# Thermal Emission Spectroscopy of Single, Isolated Carbon Nanoparticles: Effects of Particle Size, Material, Charge, Excitation Wavelength, and Thermal History

*Bryan A. Long, Daniel J. Rodriguez‡, Chris Y. Lau‡, Madeline Schultz, and Scott L. Anderson**

Department of Chemistry, University of Utah, 315 S. 1400 E., Salt Lake City, Utah 84112, USA

‡Equal contributions.

*Corresponding author: anderson@chem.utah.edu

**ABSTRACT:**

Results are presented for thermal emission from individual trapped carbon nanoparticles (NPs) in the temperature range from ~1000 to ~2100 K. We explore the effects on the magnitude and wavelength dependence of the emissivity, ϵ(λ), of the NP size and charge, and of the type of carbon material, including graphite, graphene, diamond, carbon black, and carbon dots. In addition, it is found that heating the NPs, particularly to temperatures above ~1900 K, results in significant changes in the emission properties, attributed to changes in the distribution of surface and defect sites caused by annealing and sublimation.

**KEYWORDS:** Nanoparticle, spectroscopy, single particle, trap, thermal emission, emissivity

## I. Introduction

Thermal emission spectroscopy of nanoparticles (NPs) is important for applications such as measuring temperatures and particle concentrations in hot gas flows. For example, dispersed emission has been used to monitor flame temperatures during soot particle formation, also yielding information about particle size distributions and soot volume fractions.[1-2] Similarly, laser-induced incandescence (LII) from laser-heated NPs (*e.g.* soot, metal NPs), has been used to extract particle concentrations and sizes[3-4] and to understand the energy balance in incandescing systems[5-11]

Particle temperatures can be extracted by fitting emission spectra to a model that describes the emission intensity vs. wavelength, typically using Planck's law for ideal blackbody emission, together with a function, $\epsilon(\lambda)$, that describes how the NP emissivity varies with wavelength. If the wavelength dependence of the emission is known, NP temperatures can also be extracted by two- or three-color thermometry, where calibrated intensities measured in two or three spectral regions are fit to an emission model to yield the temperature.[12-13]

In these experiments, emission is collected from ensembles of NPs allowing rapid measurement, but resulting in spectral information that is averaged over the NP ensemble, which can be quite heterogeneous. Taking soot as an example, there typically are distributions of size[14], shape[15-16], and nanostructure,[15] all of which may affect spectral properties. There have been a few experiments in which two or three color thermometry was used to measure the temperature of individual trapped particles for kinetics studies, however, the particles were in the 10-240 μm size range, and thus were quite bright, and would have had bulk-like emission properties.[17-19]

We are developing an experiment to measure surface reaction kinetics for single, isolated NPs at high temperatures, aiming to understand the effects of NP size and structural heterogeneity on their chemical properties. To provide NP temperatures for the kinetics work, we have developed experimental and calibration methods allowing measurements of dispersed emission spectra of single hot NPs in the 10 – 50 nm size range.[20] Here, we report a study of the effects of material, mass, charge, and thermal history, on the emission spectroscopy of individual carbon NPs.

Thermal emission from individual NPs is also interesting from a fundamental perspective. For NPs of black carbon materials at high temperatures, it is reasonable to expect that the emission should be "blackbody-like", i.e., the spectra should have the form:

$$\mathrm{I}(\lambda, T_{NP}) = \frac{2c}{\lambda^4 \cdot (\exp(\frac{\mathrm{hc}}{\lambda \mathrm{k} T_{NP}}) - 1)} \cdot \epsilon(\lambda) \ . \quad (1)$$

The first term is Planck's law for ideal blackbody emission, and the second is an emissivity function, $\epsilon(\lambda)$, that accounts for factors such as: 1. Coupling of the NP to light of wavelength, $\lambda$. 2. The optical/dielectric properties of the NP material. 3. The effects of surface, defect, or other localized states on the optical emission. One goal of our experiments is to examine how $\epsilon(\lambda)$ differs for different NPs of either the same, or different materials, and to measure changes $\epsilon(\lambda)$ for individual NPs as they are heated, driving sublimation and annealing.

This initial spectroscopy study is focused on carbon NPs, chosen because carbon high temperature chemistry and spectroscopy are of interest in various energy and materials applications. To obtain insight into nanostructure effects on NP emission spectra, we studied NPs made from a variety of carbon materials, including graphite and graphene, which have well ordered $sp^2$-hybridized basal planes, as well as lower coordination sites at basal plane edges or defects. The ratio of basal plane to edge/defect sites is expected to be larger for graphene platelets, compared to the more compact graphite NPs. The carbon black used in these experiments is >99% pure carbon soot produced by methane pyrolysis, and should have a variety of carbon morphologies present. In addition to possible spectral features from edge/defect sites, graphite, graphene, and carbon black NPs could potentially show band-gap emission, which has been reported for carbon materials with $sp^2$-hybridized carbon domains below ~20 nm in size.[21-23] For this reason, we also studied fluorescent carbon dot NPs synthesized by Jelinek and co-workers using a procedure discussed elsewhere.[24] Finally, carbon atoms in bulk diamond are $sp^3$ hybridized, although NPs also have a significant proportion of surface/defect atoms, and here, boron-doped diamond was used to allow laser heating.

## II. Experimental Methods

### a. NP trapping and determination of mass and charge:

The single nanoparticle mass spectrometer (NPMS) instrument has been described in detail elsewhere,[25-26] along with the methods for NP injection, and for determining the NP mass (M) and charge (Q). The trap and optical system used for NP mass determination and spectral measurements are shown in Figure S1. In brief, charged NPs are introduced into the gas phase by electrospray ionization (ESI), then pass through hexapole and quadrupole ion guides for differential pumping. The NPs are then injected into a split ring-electrode quadrupole trap, based on a design by Gerlich and Decker,[27-28] which allows lasers to be focused through the trap along several axes, and provides high solid angle optical detection along several axes.[25] NPs are trapped by applying an AC voltage between the split-ring electrode and the end cap electrodes, with amplitude $V_0$ and frequency $F_{RF.}$ Trapped NPs are detected by heating them with either 532 nm or 10.6 μm lasers loosely focused through the trap, collecting light emitted radially from the central ~50 μm of the trap, and detecting it with a fiber-coupled Si avalanche photodiode (APD - Laser Components USA). A 550 nm long pass filter (Thorlabs) is used to block 532 nm light scattering, thus the APD detects emission between 550 nm and its ~1000 nm quantum efficiency cut-off. The light can also be deflected into a CCD camera, allowing the trap center to be imaged from the radial direction for alignment purposes.

Because the APD signal is sensitive only to emission from near the trap center, it is modulated by motion of the NP. Under appropriate conditions, a trapped NP undergoes harmonic secular motion with frequencies $F_r$ and $F_z$ for radial and axial motion, respectively. The axial frequency is:

$$F_z = \left(\frac{Q}{M}\right)\left(\frac{\sqrt{2}V_0}{4\pi^2 F_{RF} z_0^2}\right), \tag{2}$$

where $z_0$ is a trap geometric parameter. For the small NPs of interest here, $F_z$ is measured by sweeping the frequency of a weak drive potential applied across the trap, with typical sweeps taking 30 to 60 seconds, depending on the frequency range covered and the desired $F_z$ resolution. When the drive potential is resonant with $F_z$, it increases the amplitude of the axial motion, such that the NP spends less time in the ~50 μm detection volume, resulting in a ~20 – 30% reduction in the APD signal. To damp the NP motion, 1.5 – 2 mTorr of Ar buffer gas is added to the trap using a mass flow controller, with pressure measured by a capacitance manometer with resolution of $10^{-5}$ Torr. By measuring $F_z$, Q/M for the trapped particle is obtained. To determine Q, a series of single electron charge steps is driven using a vacuum ultraviolet lamp, resulting in a series of steps in $F_z$ (see figure 7, below) which can be fit to extract Q. Once Q is known, M can be determined using equation 2.

In a typical experiment, NPs are injected into the trap until a step in APD signal indicates that trapping has occurred. The NP is heated at low laser intensity such that its mass is stable, and $F_z$ is measured while the vuv lamp is used to drive a series of charge steps, thus determining Q and M. Occasionally more than one NP is trapped, as shown by the presence of more than one $F_z$ peak, and in that case the particles are ejected and the experiment starts over. After Q determination, the vuv lamp isolation valve is closed, and then the laser is adjusted to vary the NP temperature ($T_{NP}$), measuring emission spectra and mass simultaneously every 30 to 60 seconds.

One important issue is the purity of the NPs studied, which could affect their emission properties. As initially trapped, the NPs have oxidized surface groups (carbonyls, etc.) from air exposure, as well as adventitious adsorbates like water. In addition, ESI was done from 2 mM ammonium acetate solutions in methanol, presumably leaving some ammonium acetate and methanol on the NP surfaces. Molecules like water or methanol should desorb at low temperatures, and ammonium acetate was chosen because its thermal decomposition generates species that also desorb well below the $T_{NP}$ range of interest here (>1500 K).[29] Similarly, oxidized surface groups have been shown to desorb as CO or $CO_2$ below ~1400 K when heated in vacuum.[30-33] In our experiments, NPs are initially trapped, held at $T_{NP} \leq 1300$ K during charge stepping (20 – 60 minutes), then heated to a series of higher $T_{NP}$ values for spectral measurements. Thus, except perhaps during initial spectral measurements at $T_{NP}$ below 1400 K, we expect that the only contaminants that could be problematic are metals or other nonvolatile elements.

The graphite NPs examined here have stated purity of 99.9999% (metal basis, Alfa Aesar), and the graphene platelet material (US Research Nanomaterials, Inc.) has stated composition of 99.7% C, <0.3% O (O should desorb as CO or $CO_2$ by 1400 K). The carbon black NPs (Plasmachem GmbH) have stated purity of >99% carbon, with < 0.02% ash content, which would include any metal or other non-volatile contaminants. Thus, the graphite, graphene, and carbon black NPs all should have metal/non-volatile contaminant levels below 0.1%. To allow diamond NPs to be laser heated at 532 nm or 10.6 μm, we used boron-doped nano diamond material (US Research Nanomaterials, Inc.), with stated composition of 98.5% carbon (98.3% diamond), 0.8% boron, ~0.15% H, 0.025% N, and a total of ~0.05% of various metals. As reviewed recently, annealing is a common approach to changing the properties of optically active impurity centers in diamond,[34] with diffusion at temperatures up to 2300 K used to interconvert between different type centers, however, it is unclear if the diffusion length scale would be large enough to bring impurities from the NP bulk to the surface. Therefore, while some contaminants may desorb at high $T_{NP}$, particularly if they are near the NP surface, we cannot assume that contaminants in the bulk are eliminated. We also did a few experiments on carbon dots, *i.e.*, small carbon aggregates prepared in the group of Raz Jelinek.[24] Carbon dots are used as non-toxic fluorescent particles for biological applications, and the question of interest was whether they would be fluorescent in the gas phase.

A related issue is whether the NP charge might have significant effects on the measured emission spectra. Experiments testing this issue are presented below.

**b. NP spectral measurements and $T_{NP}$ determination:**

For the work here, we developed the optical system illustrated in Figure S1 and associated discussion. The system is an improved version of one reported earlier,[20] redesigned to reduce chromatic effects and sensitivity to misalignment. Briefly, light emitted by the NP along the trap axis is collected with ~1 steradian acceptance and formed into a slightly convergent beam by a five element lens system. For alignment purposes, the beam can be deflected into a CCD camera allowing the central volume of the trap to be imaged from the axial direction. Normally, however, the beam passes to a beamsplitter that reflects the "visible" ($\lambda < 980$ nm) portion of the beam, and transmits the "nIR" (980 nm $< \lambda <$ 1600 nm) portion. The visible and nIR beams are then coupled into optical fibers, dispersed by spectrographs, and recorded by deep-cooled Si CCD and InGaAs photodiode cameras, respectively. Both spectrographs are wavelength-calibrated using mercury and neon emission lines.

To extract accurate $T_{NP}$ values, it is essential to know the sensitivity of the optical system vs. wavelength, $S(\lambda)$, and the calibration process is illustrated in Figure S2 and associated discussion.

As a calibration emitter, we use a thermocouple (TC) fabricated from 78 μm type C wire, with a roughly disk-shaped "bead" roughly 280 μm across and 150 μm thick. The TC bead is positioned at the trap center using the alignment cameras (Figure S1) and heated by the $CO_2$ laser with temperature read out electrically. Because emissivities of the W-Re TC materials have been well studied,[35-38] the hot TC provides an emitter with well-defined intensity vs. λ, allowing S(λ) to be measured, so that the NP spectra can be corrected. Our primary interest is in extracting $T_{NP}$ from fitting spectra, which depends only on the wavelength dependence, *i.e.*, the spectral shape. We have not attempted to quantify the absolute emissivity of the NPs, but comparing the relative intensities of different NP spectra should be accurate to within a few percent. This estimate is based on the observation of spectrum-to-spectrum repeatability better than 2 %.

Temperatures are extracted by fitting corrected NP spectra to the emission model in equation 1, and the major uncertainty is what to use for the emissivity function, ϵ(λ). For NPs much smaller than λ, as is the case here, scattering theory[39] gives the expression:

$$\epsilon(\lambda) = \frac{8\pi r}{\lambda} \cdot Im\left(\frac{(n+ik)^2-1}{(n+ik)^2+2}\right), \quad (3)$$

where r is the NP radius, and n and k are the real and imaginary parts of the index of refraction of the material. The first term in this expression accounts for coupling of the optical field to NPs much smaller than the wavelength, and the second accounts for the bulk optical properties of the NP material. If n and k are wavelength independent, $\epsilon(\lambda) \propto \lambda^{-1}$, however, that is typically not the case. For example, if n(λ) and k(λ) data for room temperature graphite are used, ϵ(λ) decreases more rapidly with increasing λ (Figure S3). The figure also shows that a simple power law expression: $\epsilon(\lambda) \propto \lambda^{-1.41}$, is a reasonable fit to the scattering theory expression, deviating by less than 3% over the wavelength range of interest.

It is impractical to use equation 3 in fitting NP spectra for several reasons. For most nanomaterials, optical (or equivalent dielectric) parameters are not available over our λ range, or for the high temperatures of interest. More fundamentally, even if *bulk* optical parameters were available over our λ and $T_{NP}$ range, the optical properties for small NPs would be modified by surface and defect states, and possibly by quantum confinement effects. Indeed, this last point is clearly illustrated by the data presented below.

Therefore, we are forced to use a model for $\epsilon(\lambda)$, and have chosen to use a power law function of the form $\epsilon(\lambda) \propto \lambda^{-n}$, where n is a fitting parameter. The power law form was chosen because it is a reasonable approximation to $\epsilon(\lambda)$ for graphite, at least at room temperature, and adds only a single fitting parameter that affects the spectral shape. With this assumption, and combining all constants into a single normalization parameter, K, the emission model (equation 1) becomes:

$$\mathrm{I}(\lambda, T_{NP}) = \frac{\mathrm{K}}{\lambda^{(4+n)} \cdot (\exp(\frac{\mathrm{hc}}{\lambda \mathrm{k} T_{NP}}) - 1)} . \qquad (4)$$

This function depends on only three parameters: $T_{NP}$, n, and K, of which only the first two affect the shape of the function. As shown below, fits based on this model are not perfect, but this approach at least allows us to treat all the NPs on a common basis, so that properties can be compared for different NPs.

Several factors contribute to the uncertainty in $T_{NP}$ values extracted by fitting spectra with equation 4. Except for a few spectra of very low brightness NPs, the signal-to-noise in the spectra below is high enough that the major sources of uncertainty result from non-idealities in the optical system and their effects on the spectral intensity calibration process, and from the nature of the fitting function used to extract $T_{NP}$. These are analyzed in some detail in the Supporting Information.

The effects of mis-positioning the TC relative to the trap center or changing the TC temperature during calibration are explored in Figure S4 and associated discussion. The effects of misalignment of various optical elements are explored in Figures S5 through S9. The effects of deliberately mismatching the intensities of the visible and nIR spectra are explored in Figure S10. The emission intensities measured range over three orders of magnitude, thus camera non-linearity is a concern. We previously tested for non-linearity using the TC emitter and neutral density filters,[20] and Figure S11 shows an additional NP-based test that also showed no significant saturation effects. These "experimental" issues contribute to uncertainties in the shape of the spectra, and given our ability to position the TC and optics, we estimate those contributions to the uncertainty in $T_{NP}$ to be ±4.8%.

There is additional $T_{NP}$ uncertainty resulting from fitting the spectra with equation 4, because this approximate emission model does not give perfect fits to the spectra. For example, a particular pair of n and $T_{NP}$ parameters might give the best overall fit to a spectrum, but fits with different n and $T_{NP}$ parameters might actually fit some spectral regions better, albeit with lower overall fit quality. Figure S12 examines how this behavior, together with the experimental signal-to-noise, introduces uncertainty in the $T_{NP}$ values extracted. For spectra with reasonable signal/noise, we estimate this contribution to the $T_{NP}$ uncertainty to be ~±4%. Taking the experimental and fitting uncertainties together, the total estimated uncertainty in $T_{NP}$ is ~±6.2 %.

This is the estimated absolute uncertainty in $T_{NP}$ values extracted using equation 4. In many cases, we are interested in comparing $T_{NP}$ values for different spectra, and because most of the experimental factors contributing to the uncertainty are identical in all spectra, the uncertainty in $T_{NP}$ *differences* is mostly due to the fitting uncertainty of ± 4%. Furthermore, for spectra with similar shapes, *i.e.*, similar n parameters, some of the $T_{NP}$ uncertainty resulting from the fitting process should also cancel when comparing $T_{NP}$ values, and in such cases, the uncertainty in the $T_{NP}$ *differences* is estimated to be < 2 %.

Note, however, that the $T_{NP}$ values here are all extracted assuming the power law approximation for $\epsilon(\lambda)$. It is not possible to estimate how much $T_{NP}$ might change if the correct form of $\epsilon(\lambda)$ were known, and we can only point out that the fits based on this assumption are often quite good, and for those where the model does not work as well, we can at least see how it fails, *i.e.*, how $\epsilon(\lambda)$ might need to be changed for different NPs at different temperatures.

This study focuses on the emission properties of carbon NPs, and the NP mass is only used to calculate mass ratios needed to approximately correct emission intensities for NP size (vide infra). We estimate that the absolute uncertainty in the mass measurements is ~±1.3 %, mostly due to the estimated assembly precision of the trap, however, this systematic error cancels exactly in mass ratios. The relative uncertainty in comparing masses mainly results from uncertainty in fitting the $F_z$ scans used to measure Q/M, which depends primarily on the emission brightness of the NPs. For the NPs studied here, the typical uncertainty in mass ratios is estimated to be ~0.2%, with the exception of one experiment using a carbon dot NP, where the emission was very weak. Mass ratios involving that NP are uncertain by up to 2%.

## III. Results

### a. Graphite NP emission spectra at different temperatures.

We begin with graphite NPs, which have the highest purity of the materials studied (99.9999%), and the best known nanostructure, at least in principle. Figure 1 shows a set of experimental emission spectra (60 second acquisition time) for a single graphite NP with initial mass of 33.7 MDa and initial charge of +50e, taken at a series of increasing 532 nm heating laser powers ranging from 6 to 48 mW, loosely focused through the trap and detected at the exit window of the vacuum system. If this NP were spherical with the bulk density of graphite, it would have a diameter of ~36 nm. Fits to the spectra based on equation 4 are shown as colored curves, and the best fit $T_{NP}$ and n values are given in the legend. Equation 4 gives reasonably good fits at low $T_{NP}$, but with increasing temperature the power law $\epsilon(\lambda)$ function is unable to capture the curvature in the experimental spectra, failing most obviously in the region above 1250 nm. Note that there is a systematic decrease in the best-fit n parameter with increasing temperature.

It is also interesting to consider how $T_{NP}$ is controlled by the balance between laser heating and cooling by thermal emission, Ar buffer gas collisions, and sublimation. Sublimation at these temperatures turns out to be too slow to result in significant cooling, and for the high temperatures and low buffer gas pressures used here, thermal emission is the dominant cooling mechanism.[40] For example, a spherical NP of this size at 2000 K would emit on the order of $10^{-10}$ W of thermal radiation, estimated as Stefan-Boltzmann emission with emissivity of ~0.02.[10] Collisions with 1 mTorr of Ar buffer gas would carry away only ~$10^{-12}$ W, assuming unit efficiency energy accommodation.

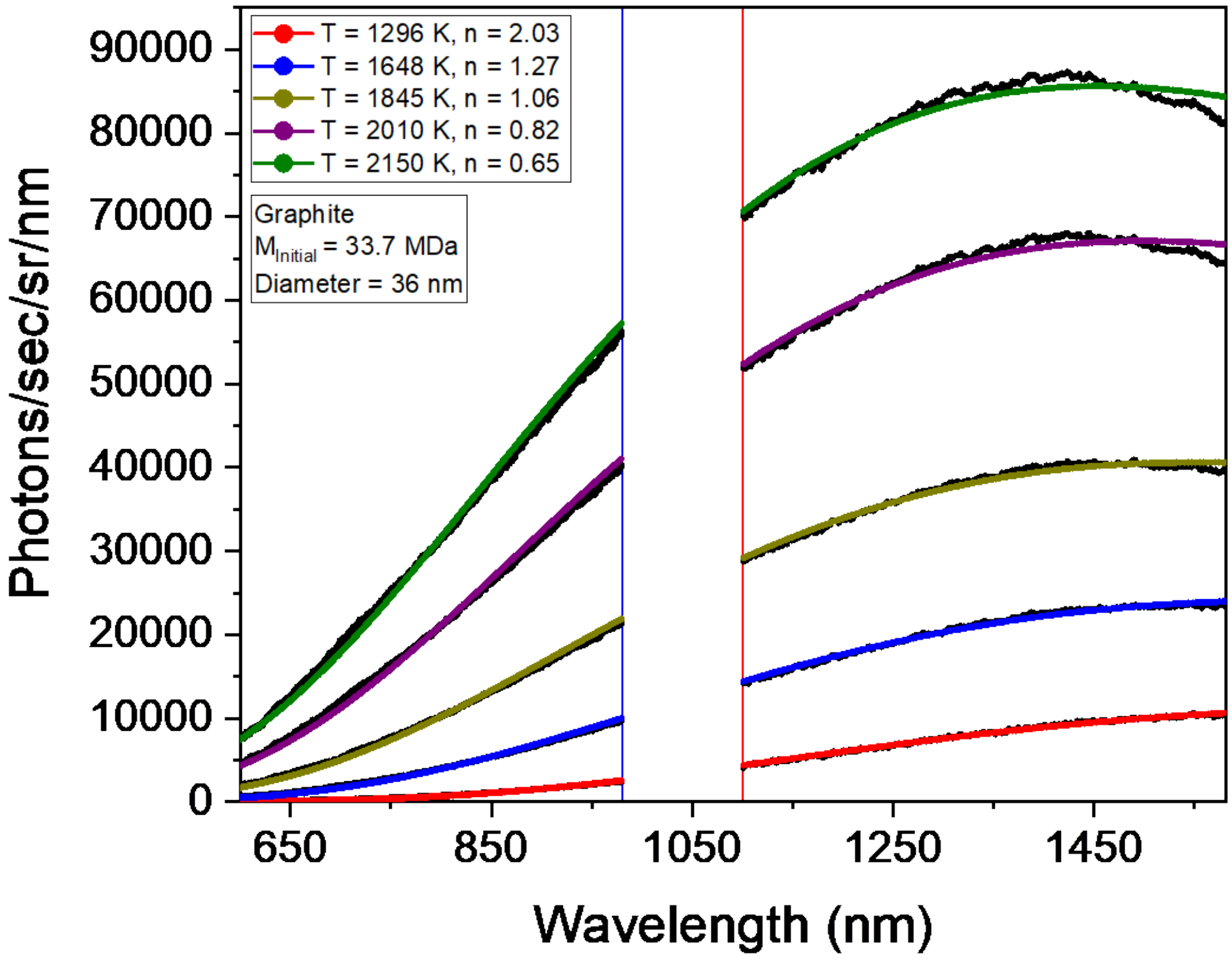

Figure 1. Emission spectra of a graphite NP being heated to five different temperatures (1296-2150 K ± 2%), along with fits based on equation 4. The emissivity parameter n, decreases as temperature increase.

The total emitted power from an ideal blackbody varies like $T^4$. Because we are measuring only the short wavelength part of the spectrum, and the spectra both increase in intensity and shift to shorter wavelength with increasing temperature, the observed emission intensity is expected to vary more strongly with temperature than the total blackbody emission. Consistent with this expectation, the integrated number of photons emitted *per* second in the 600 to 1580 nm range increased by a factor of ~19 when $T_{NP}$ was increased by a factor of 1.66, from 1296 K to 2150 K, varying like ~$T^{5.9}$. Because the cooling power varies nonlinearly with $T_{NP}$, the laser power required to heat the NP also increased nonlinearly, by a factor of about four to obtain the 66% increase in $T_{NP}$. Because our laser beam divergence is power dependent, the exact relationship between laser power and focused intensity is uncertain.

### b. Effects of carbon material structure on NP emission spectra.

Figure 2 shows analogous examples of emission spectra for NPs of four other carbon materials, with fits and extracted $T_{NP}$ and n parameters. The carbon dot emission was too weak to obtain spectra except at very high $T_{NP}$. The mass of the graphene platelet and diamond NPs were similar to that for the graphite NP in Figure 1. The carbon black NP had nearly double the mass, and the carbon dot was ~5 times less massive.

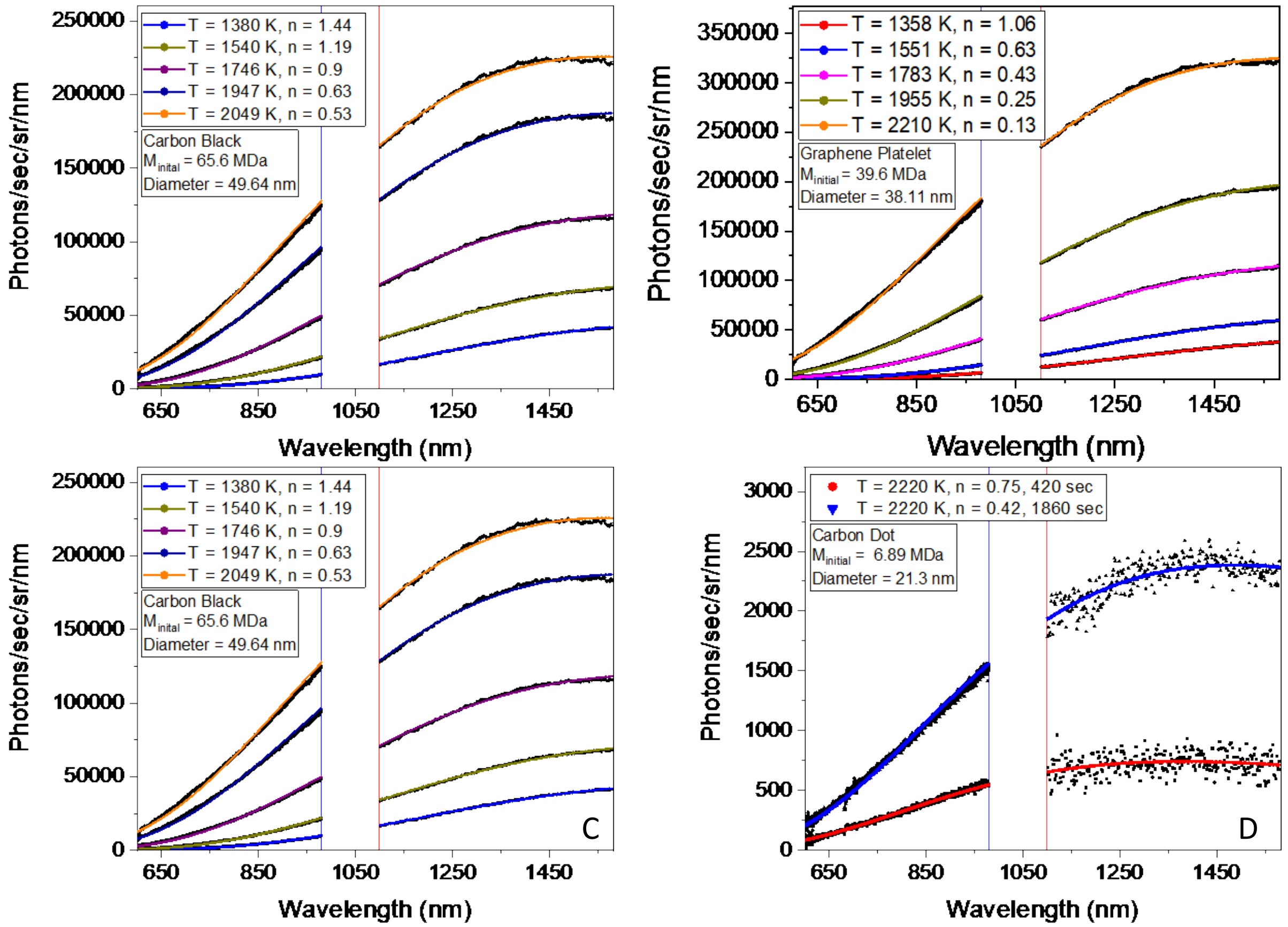

Figure 2. Emission spectra with accompanying fits and extracted $T_{NP}$ and n values, for carbon NPs of four different materials, including carbon black (2A), graphene platelet (2B), nano diamond (2C), and carbon dot (2D). The relative uncertainty in $T_{NP}$ values is ± 2%.

The spectral shapes for all the different carbon material NPs were grossly similar, and in particular, there were no obvious signs of structured emission in this wavelength range for any of the NPs, at least at the low spectral resolution used here to maximize signal – roughly 20 nm below 1000 nm, and ~70 nm at longer wavelengths. We note that carbon dot NPs are fluorescent in solution, but we observed no signs of fluorescence, even using low 532 nm laser intensities to minimize heating.

While the spectral shapes are qualitatively similar for all the carbon NPs in Figures 1 and 2, there are differences in the wavelength dependence, i.e., in the spectral shapes, which can be seen in two ways. The best-fit n parameters varied systematically between the NPs, with the graphite NP having the highest values, and the graphene platelet NP the lowest. High n corresponds to more sharply peaked spectra, with intensity falling faster at both long and short wavelengths, and low n corresponds to spectra with less sharply wavelength-dependent intensity. Furthermore, even though the graphite fit functions had the highest n values, they still were unable to capture the curvature of the experimental graphite NP spectra in the nIR.

The brightness of the different NPs in Figures 1 and 2 also varied dramatically; the carbon black and graphene NPs being >100 brighter than the carbon dot at similar $T_{NP}$. To compare quantitatively, it is necessary to correct the intensities for the differences in NP size. For an ideal blackbody of radius r, intensity scales with surface area ($r^2$), and for small particles there is an additional factor of r from the emissivity (equation 3). Thus, for spherical NPs, the intensities should scale as $r^3$, *i.e.*, with the NP volume. Because the shape of the NPs used is not known, this scaling is only approximate, but it provides at least a first order correction for the effects of varying NP mass. Using the measured NP masses and assuming bulk densities for the materials involved, we can estimate the volume ratios. For carbon black, reported densities vary for different carbon blacks and for different measurement methods, and also can change with annealing.[41] In the following we have assumed a density of 2.0 g/cm$^3$,[41] compared to 2.26 for graphite.[42] The carbon dot density is unknown, and was assumed to be equal to that of graphite, as was that for the graphene platelet NP. To compare the intensities for the NPs in Figure 2 to those for the graphite NP in Figure 1, the Figure 2 intensities should be multiplied by the following volume ratio factors: Carbon Black = 0.4, Graphene Platelet = 0.8, Nano diamond = 1.6, Carbon dot = 4.2. As noted in the Experimental section, the worst case uncertainty in the mass ratios used to calculate the volume-scaling factors is ~0.5%. This is negligible compared to the uncertainty in how brightness might scale with NP shape and how NP density might differ from the bulk values, which are the main reasons that we note that volume scaling is only an approximate correction for the effects of NP size.

The $T_{NP}$ values extracted from fitting the NP spectra varied in the five experiments, but by comparing spectra at similar $T_{NP}$, it is clear that the graphene platelet NP had the highest volume-corrected (and raw) brightness of any of the NPs. The raw graphene NP intensities were roughly 3 times higher than those for the graphite NP, or ~2.4 times brighter after correction for the larger graphene NP volume. After correction for the NP volumes and taking the differences in $T_{NP}$ values into account, the carbon black and nano diamond NPs both had brightness similar to that of the graphite NP, i.e., ~2.4 times lower than that of the graphene NP. We attribute the greater brightness of the graphene NP to its platelet shape, which would give it substantially higher emitting surface area than the more compact graphite, carbon black, and nano diamond NPs.

In terms of brightness, the real outlier was the carbon dot NP, which emitted much more weakly than any of the other materials. Even after volume scaling to correct for the small size of carbon dot NP, the intensity in the 1$^{st}$ carbon dot spectrum (red) at 2220 K was ~ 25 times lower than that of the graphite NP at 2150 K, i.e., 70 K colder. The other interesting observation was that in a 2$^{nd}$ spectrum (blue) taken for the same NP about 20 minutes later, the intensity increased by a factor of ~3.8 for the same extracted $T_{NP}$ value, while the n parameter decreased by ~40%, corresponding to a less sharply peaked spectrum. The increase in brightness in the 2$^{nd}$ spectrum is interesting, however, the extreme weakness of the emission makes detailed study difficult. The next section examines similar behavior for the other, brighter carbon NPs.

It is also interesting to compare the $T_{NP}$ dependence of the emission intensity for the different carbon NPs in Figures 1 and 2. The log of the integrated numbers of photons emitted/second from each NP, in the 600 to 1600 nm range, are plotted as a function of $T_{NP}$ in Figure 3. If intensity scales like $T_{NP}^{n}$, then plots of log I vs. log $T_{NP}$ should be linear, with slope = n. With the exception of the nano diamond NP, the intensity data are reasonably well fit by straight lines, with slopes ranging from 5.9 for the graphite NP to 6.3 for the graphene platelet NP, i.e., the black carbon NPs all have intensities that vary roughly as $T^6$. For the nano diamond NP, the three highest $T_{NP}$ points ($T_{NP}$ > 1770 K) are also fit well by a line with slope near 6, however, the intensities at the two lower $T_{NP}$ values fall well below this trend, suggesting that the emission properties of the diamond NP changed significantly as it was initially heated. As shown in the next section, all the carbon NPs show significant effects of heating on emissivity.

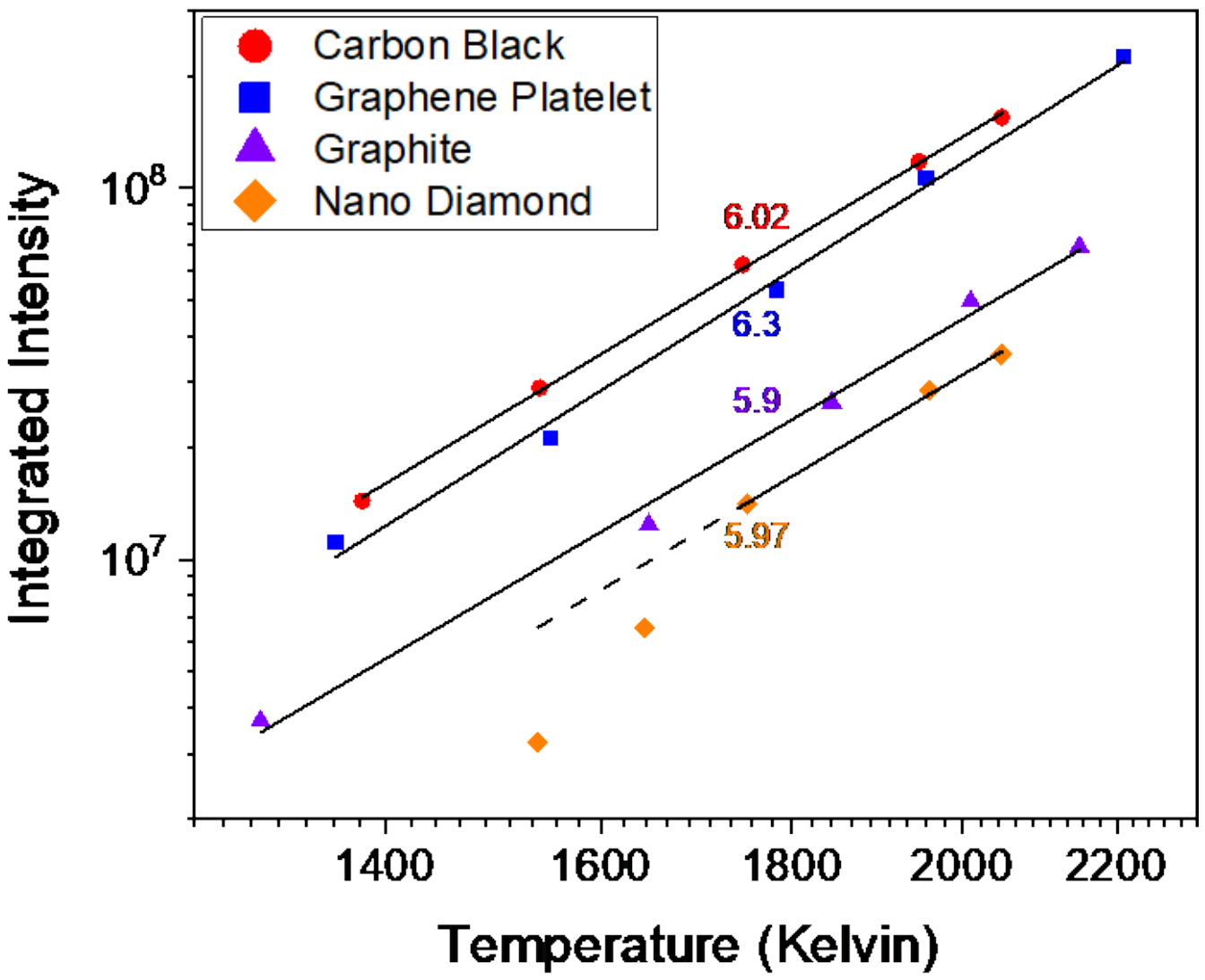

Figure 3. Emitted photons/second in the 600 to 1600 nm ranges are plotted as log I vs. log $T_{NP}$. Data are shown as points and linear fits are shown as lines, with slopes indicated.

### c. Effects of thermal history on NP emission spectra for carbon NPs

To examine the effects of heating on NP emission spectra, a series of experiments were done in which individual NPs were first heated at low $T_{NP}$ for charge determination, then a series of spectra were measured as $T_{NP}$ was stepped up, and then stepped back down. The goal was to compare spectra measured at similar $T_{NP}$ values, before and after the NP was heated into the range where sublimation caused significant mass loss. We attempted to hit the same set of $T_{NP}$ values during the heat-up and cool-down phases of the experiment, but because the emissivity properties changed, the actual $T_{NP}$ values extracted from the spectra sometimes differed. Nonetheless, comparing spectra at similar temperatures provides insights into heating effects on the brightness and shape of the spectra. It should be noted that essentially all the mass loss occurred during the times when $T_{NP}$ was above 1900 K, thus we note these times.

Figure 4 shows experiments of this type on two different carbon black NPs, both smaller than the carbon black NP in Figure 2. Figure 4A compares two spectra taken for a carbon black NP of initial mass 31.13 MDa, with $T_{NP}$ near 1670 K. The 1$^{st}$ spectrum, taken during the heat-up phase, had extracted $T_{NP}$ = 1668 K. The 2$^{nd}$ spectrum, with extracted $T_{NP}$ = 1679 K, was taken after the NP had spent ~50 minutes at temperatures ≥1900 K, during which time sublimation decreased the mass by ~15% to 26.33 MDa. The *raw* integrated emission intensity (in the 600 to 1580 nm range) after heating was 5.5 % higher than before, even though the NP volume was smaller. In Figures 4 and 5, the intensities for the 2$^{nd}$ (or 3$^{rd}$) spectra for each NP have been scaled by the volume ratio (to the 1$^{st}$ spectrum), to correct for NP mass loss. With that volume scaling, the integrated intensity in the 2$^{nd}$ spectrum was ~25% higher than in the 1$^{st}$ spectrum, however, $T_{NP}$ was also ~10 K higher. If we assume that the emission in this wavelength range scales like $T^6$, as suggested by Figure 3, then the 10 K temperature increase would have made the 2$^{nd}$ spectrum only ~4% brighter, i.e., the increase in brightness was substantially greater than would be expected from the small $T_{NP}$ difference.

It is interesting to compare the change in emission intensity to the change in laser power required to heat the NP in the 2$^{nd}$ spectrum. If the changes were identical, this would imply that the NP absorption cross section at 532 nm (heating) and its broadband emissivity (cooling) tracked together as the NP evolved. In this case, the raw emission intensity was 5.5 % higher in the 2$^{nd}$ spectrum, but the 532 nm laser power in the 2$^{nd}$ spectrum (30.28 mW) was slightly lower than in the 1$^{st}$ (30.93 mW). The implication is that the 532 nm absorption cross section increased by more than enough to compensate for the increased radiative cooling from brighter broadband emission.

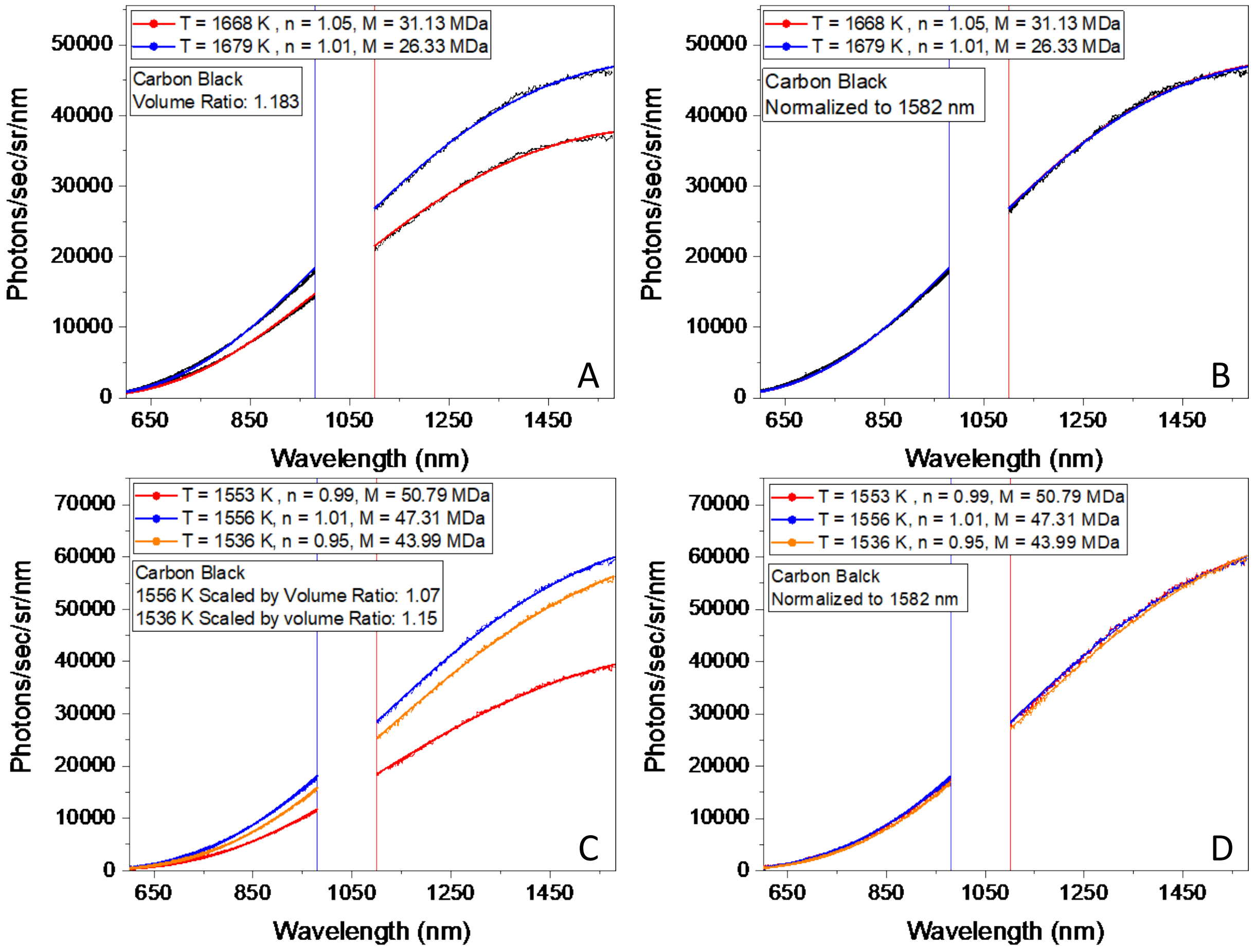

Figure 4: A and B are emission spectra for a single 31 MDa carbon black NP. C and D are spectra for a 50.8 MDa carbon black NP. Spectra are shown before and after periods in which the NP was heated above 1900 K. The order in which the spectra were taken is shown by the masses, with mass always decreasing after heating. The left-hand spectra are volume scaled, and right-hand spectra are normalized at 1580 nm to allow shape comparison.

In Figures 4C and D, a similar experiment on a larger carbon black NP is shown, with three emission spectra taken at $T_{NP}$ ≈1550 K. The first (red) was measured prior to any heating above the ~1553 K measurement temperature, at which point the mass was 50.79 MDa. The second (blue) was taken after $T_{NP}$ had been stepped up and back down, with a ~10 minute heating period at >1900 K, during which the mass had decreased by 6.85%, to 47.31 MDa. $T_{NP}$ in the 2$^{nd}$ spectrum was only 3 K higher than in the 1st, thus the emission intensity, spectral shape, and heating power should be directly comparable in the two spectra.

The raw intensity in the 2$^{nd}$ spectrum was substantially (~43 %) higher than in the 1$^{st}$, or 54% higher if the 6.8% volume (i.e. mass) change is taken into account. In this case, ~9.5% more laser power was needed to compensate for the greater radiative cooling in the 2$^{nd}$ spectrum (29.22 mW), compared to the first (26.70 mW). Note, however, that the increase in laser power required in the 2$^{nd}$ spectrum was substantially smaller than the 43% increase in raw integrated emission intensity, indicating that once again, the increase in 532 nm absorption cross section was more than enough to compensate for the increase in broadband thermal emissivity.

The 3$^{rd}$ spectrum (yellow) was taken after a second stepped heating program, including another ~10 minute period above 1900 K, leading to further decrease in the NP mass to 43.99 MDa. The volume-scaled intensity was lower than in 2$^{nd}$ spectrum, however, a ~ 7% decrease (assuming $I \propto T^6$) would be expected from the ~20 K lower 3$^{rd}$ spectrum $T_{NP}$. The intensity remained well above that in the 1$^{st}$ spectrum.

As shown by the normalized spectra in Figures 4B and D, the increases in emission brightness for the carbon black NPs after >1900 K heating were not accompanied by significant changes in spectral shape. Note also that the power-law-based emissivity model (equation 3) fit the spectra reasonably well, and with significantly lower n values, compared to the graphite NP in Figure 1.

Figure 5 shows similar pre-heating/post-heating emission spectra comparisons for graphene, graphite, and nano diamond NPs. On the left side, spectra are plotted to allow intensity comparisons, with the post-heating spectra scaled by the volume ratio to approximately account for the ~5 – 20% mass losses that occurred in the >1900 K heating periods. On the right side, spectra are normalized at 1580 nm to show the effects of >1900 K heating on spectral shapes.

Figure 5A shows spectra for two different graphene platelet NPs, with arrows indicating which intensity scale applies to each spectrum. The left-hand scale is for an NP with initial mass of 11.68 MDa, which was heated by 19.7 mW of 532 nm laser power giving an initial spectrum fit with $T_{NP}$ = 1645 K and n = 1.03. The NP was then heated above 1900 K for 94 minutes, resulting in a final mass of 10.34 MDa. In the 2$^{nd}$, post-heating spectrum, 22.63 mW of 532 nm power was needed to reach 1640 K, with best fit n = 0.90. In this example, the raw integrated intensity increased by 47% between the spectra, and after applying the 1.13 volume scaling factor, the increase was 66%. In addition to the increase in emission brightness, Figure 5B shows that there was also a significant change in the wavelength dependence, which is reflected by a drop in the best-fit n value to 0.9, implying a slower decrease in emissivity with increasing wavelength. For this NP, the 15% increase in 532 nm heating power in spectrum 2 was much smaller than the 47% increase in raw integrated intensity, again implying that the 532 nm absorption cross section increased more than the emissivity for broadband radiative cooling.

The second graphene NP in Figure 5A (right axis) was ~six times more massive than NP 1, and the emission intensity was significantly larger, even though the spectra were measured at lower $T_{NP}$, near 1290 K. In the 1$^{st}$ spectrum, $T_{NP}$ was 1288 K and n was 1.3, then the NP was heated above 1900 K for 21 minutes, during which time the mass dropped by 14.42 MDa. The 2$^{nd}$ spectrum had $T_{NP}$ = 1289 K with n = 1.02, and in this case, the raw intensity was ~16% lower than in the 1$^{st}$ spectrum. After volume scaling to account for the mass lost during the >1900 K heating, the 2nd spectrum intensity was ~4.8 % higher than in the 1st, and as shown in Figure 5B and by the drop in n, the wavelength dependence was significantly flatter after heating.

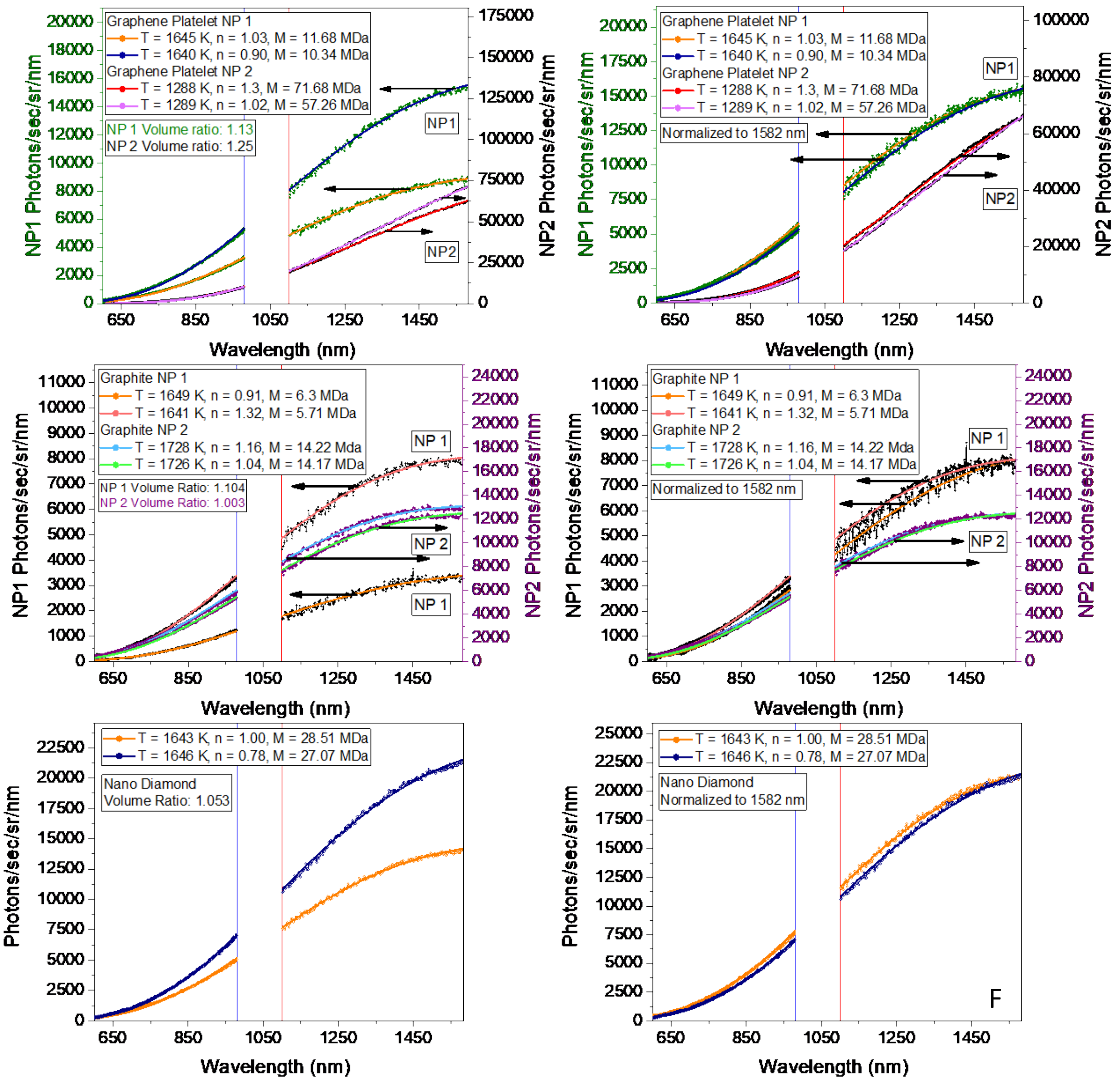

Figure 5: Emission spectra taken before and after >1900 K heating for graphene, graphite, and nano diamond NPs. The order in which spectra were taken is shown by the masses, which decreased during >1900 K heating periods. A, C, and E show volume-scaled intensities. B, D, and F are normalized to show changes in spectral shape. Note that the emission intensities changed after >1900 K heating, and that there were also significant changes in spectral shape.

Figure 5C shows data for two graphite NPs. NP 1 was a small graphite NP with initial mass of 6.3 MDa, initially heated to 1649 K with 532 nm excitation, then heated above 1900 K for ~17 minutes, during which time its mass decreased ~10% to 5.71 MDa. The post-heating spectrum was at slightly lower $T_{NP}$, (1641 K); nonetheless the integrated emission intensity increased dramatically, by factor of 2.46 (raw) and 2.7 (volume-scaled). In addition, Figure 5D shows that the spectral shape changed significantly, in this case with higher post-heating n value, implying more sharply wavelength-dependent emissivity. The laser power required to heat the NP increased by only 3.7% for the 2nd spectrum, compared to the 2.4 x increase in raw emission intensity. Thus, even though the broadband emissivity increased dramatically when the NP was heated, the increase in 532 nm absorption cross section was even larger.

The 2nd graphite NP in Figure 5 C had mass of 14.22 MDa at the point the 1st spectrum was measured, resulting in $T_{NP}$ = 1727 K, and n = 1.16. After heating to 1900 K for 15 minutes, which caused only 0.05 MDa mass loss, a 2nd spectrum was measured with $T_{NP}$ = 1726 K and n = 1.05. For this NP, the brightness of the emission *decreased* by ~9% (raw). Even after volume scaling, the post-heating emission intensity was ~8.5% lower than in the initial spectrum.

Figures 5E and F show analogous results for a diamond NP with initial mass of 28.51 MDa, with a 1st spectrum taken at 1643 K (n = 1.0) with 33.08 mW of 532 nm heating power. The NP was then heated in steps up to nearly 2200 K over a total of 3780 seconds, resulting in ~5% mass loss to 27.07 MDa. In the 2nd spectrum, 21.86 mW was required to heat the NP to 1646 K (n = 0.78). The post-heating intensity increased substantially (34% raw, 41% volume scaled), and the shape also changed, in this case with lower post-heating n value, implying a reduced dependence of emissivity on wavelength. For this NP, 33% *less* heating laser power resulted in 34% higher raw emission intensity in the 2$^{nd}$ spectrum, implying that again, the 532 nm absorption cross section increased substantially more than the broadband emissivity.

The data in Figures 4 and 5 show that heating carbon NPs into the range where sublimation becomes significant can result in substantial changes in emissivity, with emission often, but not always, substantially brighter after heating, even without correction for the mass loss. In addition, the wavelength dependence of the spectra, hence the form of $\epsilon(\lambda)$, can change significantly, with the n parameter (i.e., spectral curvature) sometimes increasing, sometimes hardly changing, and sometimes decreasing significantly.

### d. Comparisons of emission properties for different NP materials

To provide a different perspective on how the intensities and spectral curvatures vary from material-to-material, and how different materials react to heating, Figures S13 and S14 replot the data in Figures 4 and 5, comparing spectra for the different materials. Specifically, the pre- and post-heating spectra for the graphite NP1 in Figure 5C, are plotted together with the analogous spectra for the graphene, carbon black, and diamond NPs, all at similar temperatures (1660 ± 20 K). Spectra are compared both before and after >1900 K heating, and both volume-weighted intensities and spectra normalized at 1580 nm are compared.

In essence, the results are: 1. Prior to >1900 K heating, the graphite NP (Figure 5C) had volume-corrected emission intensity 30 to 50% lower than the intensities for the graphene (4A), carbon black (3C), and nano diamond NPs (4E). The carbon dot NP had much lower intensity – far too weak for spectral measurements at 1600 K. 2. The n parameter for the small graphite NP (0.91) was ~10% smaller than those for the other three NPs (1.0 to 1.05), but the spectral shapes prior to >1900 K heating were still nearly superimposable. 3. Heating above 1900 K increased the intensities for all the NPs, but with the largest effect on the graphite NP, such that after heating, the volume-scaled intensities for the graphite, graphene, carbon black, and diamond NPs were all similar. 4. The n parameter and associated curvature of the graphite NP spectrum increased after heating, while those for graphene and diamond decreased slightly, and that for carbon black remained essentially unchanged.

It is tempting to associate these differences between the NPs to the differences in the materials from which the NPs were made, however, it is important to note that the NP feedstocks are heterogeneous, with distributions of mass and shape that are expected to influence the electronic and phonon density of states, hence the emission properties. This point is made clear Figures 4 and 5, where pairs of carbon black, graphite, and graphene NPs are shown to have substantially different emission behavior. Understanding the effects of structural heterogeneity on both emission spectra and reaction kinetics is the main motivation for our single NP experiments.

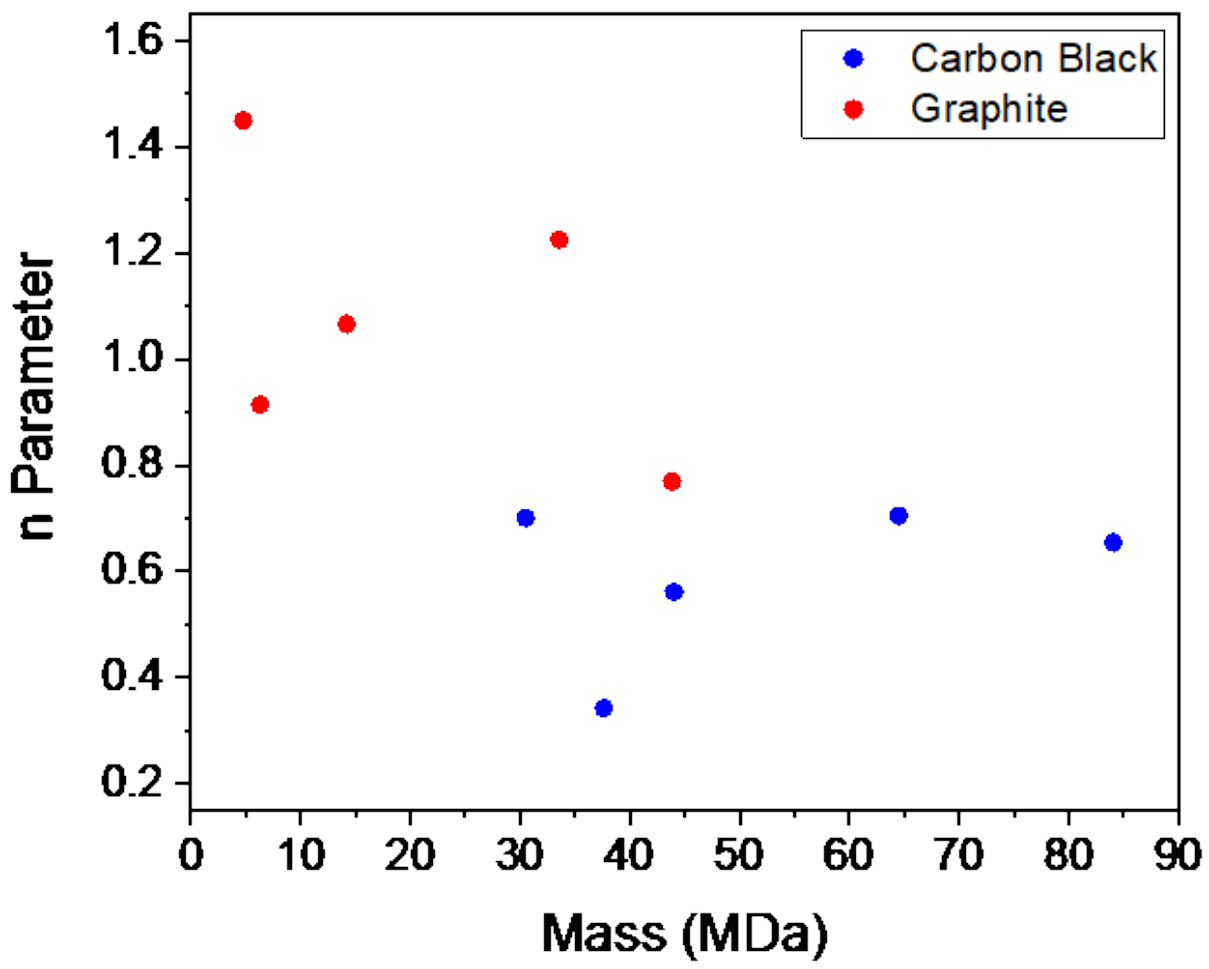

Figure 6. Emissivity properties of ten carbon NPs vs. mass. The n parameters are extracted by fitting spectra for five graphite NPs with $T_{NP}$ near 1652 K ± 2%, and five carbon black NPs near 1902 K ± 2%.

### e. The effects of NP mass on emission spectra

One obvious point to be explored is the effect of NP size on the spectral properties. As discussed above, emission intensity is expected to scale roughly with NP volume, but we are also interested in possible effects of NP size on the wavelength dependence of the spectra. Such effects can be summarized by comparing the n parameters extracted by fitting spectra for different size NPs at similar temperatures. Figure 6 shows typical data for ten different carbon NPs, with masses ranging from just under 5 MDa, to ~85 MDa, which would correspond to diameters ranging from ~20 to ~54 nm, if the NPs were spherical, with the bulk densities. The figure shows the n parameters for five graphite NPs in the 5 to 45 MDa range, obtained by fitting spectra with $T_{NP}$ close to 1652 K. Data are also included for five carbon black NPs in the mass range from ~30 to 85 MDa, taken from spectra with $T_{NP}$ close to 1902 K. (These temperatures were used because spectra were available for different NPs at very similar $T_{NP}$ values). As noted in the discussion of Figures 1 and 2, the n parameter tends to decrease with increasing $T_{NP}$, and also tends to be larger for graphite than for carbon black, and the data in Figure 6 is consistent with both tendencies.

For the graphite NPs, there appears to be a trend toward smaller n parameter (weaker dependence of $\epsilon$ on $\lambda$) with increasing mass, however, the NP-to-NP scatter in best-fit n values was quite large, presumably reflecting the structural heterogeneity of the graphite NP feedstock. For the somewhat larger carbon black NPs, the scatter was smaller, but with no obvious trend of n vs. mass. These results do not rule out systematic effects of NP mass on spectral shape, but clearly any mass effects are obscured by the effects of NP structural heterogeneity.

### f. The effects of NP charge on the NP emission spectrum and extracted $T_{NP}$ value

The NPMS method requires that the NPs be charged so they can be trapped, however, the charge-to-mass ratios are quite low – on the order of $\sim 10^{-6}$ e/Da, corresponding to $\sim 10^{-5}$ e/atom. Therefore, the effect of the charge on NP electronic and vibrational density of states should be negligible, however, there could be spectral artifacts due to the fact that the amplitude of NP motion in the trap is charge dependent.[27-28] In essence, changing the charge changes the effective "size" of the NP emitter, which could affect the intensity calibration and the laser-NP spatial overlap. Figure 7 shows an experiment in which the secular motion frequency, $F_z$, and $T_{NP}$ were monitored for a graphite NP with initial mass of 43.84 MDa, as the NP charge was varied in steps of ± e using a vacuum ultraviolet lamp to drive photoemission from both the NP and from trap surfaces (leading to electron capture by the NP). From the size of the steps in $F_z$, the absolute charge can be determined, as indicated in the figure. For this experiment, the program normally used to stabilize $T_{NP}$ was not used. Instead, the heating laser was set to constant power, giving initial $T_{NP}$ around 1650 K, which drifted slowly down over the course of the 6.5 hour-long experiment.

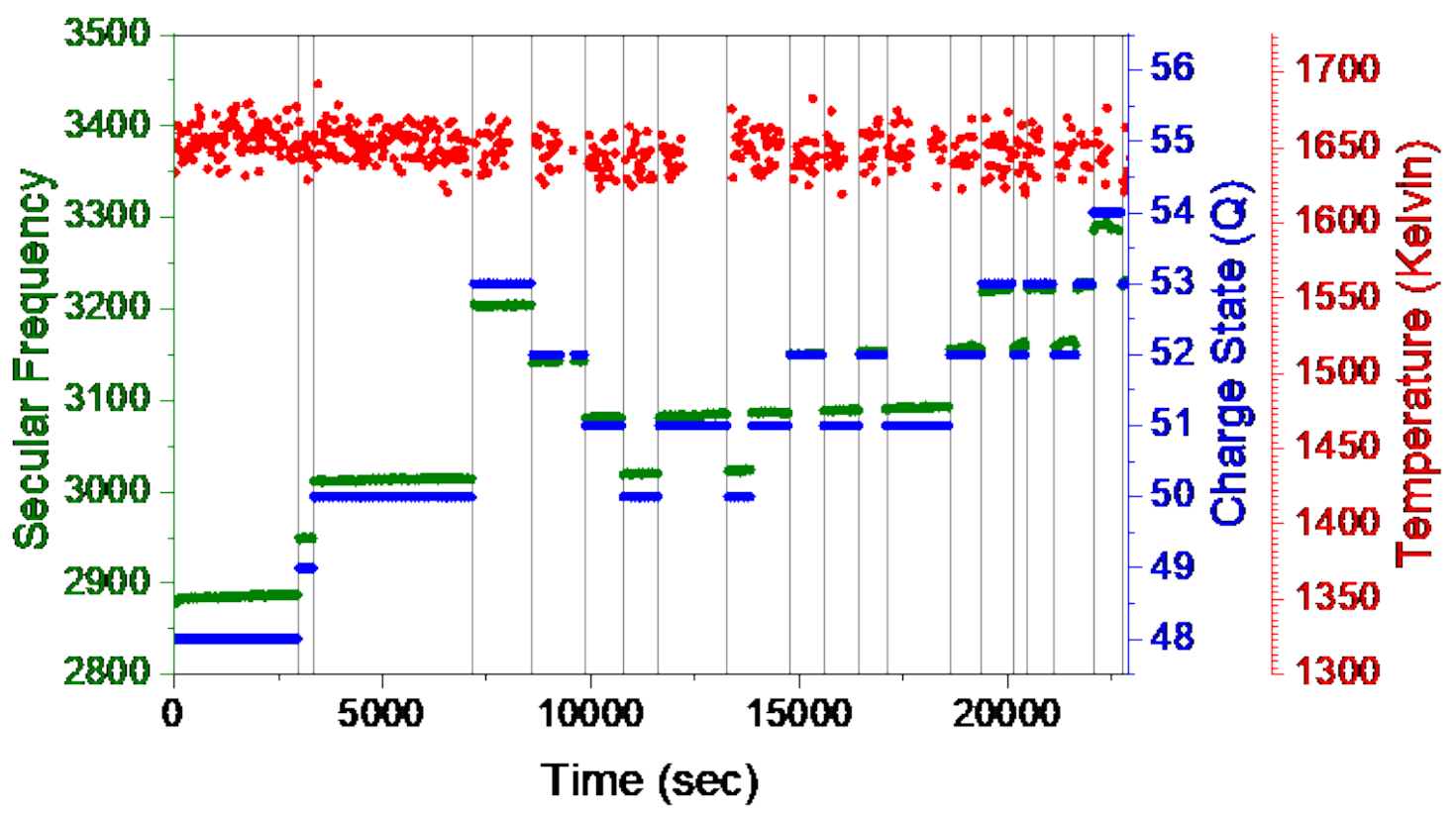

Figure 7. The effect of NP charge stepping on $T_{NP}$ measurement. Secular frequency (green), $T_{NP}$ (red), and charge state (blue, Q) measurements for a single 43.84 MDa graphite NP, as a series of charge steps was carries out.

A series of 23 charge steps was observed, including one triple step, with Q varying from +48e to +54e, *i.e.*, by ~11%. The root mean square radial amplitude[27-28] of the NP radial motion would have decreased from ~53 μm for Q = +48e, to ~47 μm for Q = +54e, *i.e.*, also by ~11%. It is clear that there were no significant effects of the NP charge on $T_{NP}$, even for the triple charge step near 7100 seconds.

A related question is whether the axial motion driven to measure $F_z$ has any effect on the spectra and extracted $T_{NP}$ values. As shown in Figure S15, the answer is “no”.

### g. Effects of heating laser wavelength on the NP emission spectrum

The fact that broadband blackbody-like emission spectra are observed when irradiating the NPs at 532 nm, shows that energy redistribution must be fast in hot carbon NPs. The obvious question is whether it is fast enough to eliminate all memory of the excitation process, or if there might still be residual effects of heating laser wavelength. In Figure 8 we compare heating using the cw 532 nm laser, presumably exciting electrons in the NP, to heating via vibrational/phonon excitation using a quasi-cw, duty-factor-modulated (at 20 kHz) 10.6 μm $CO_2$ laser.

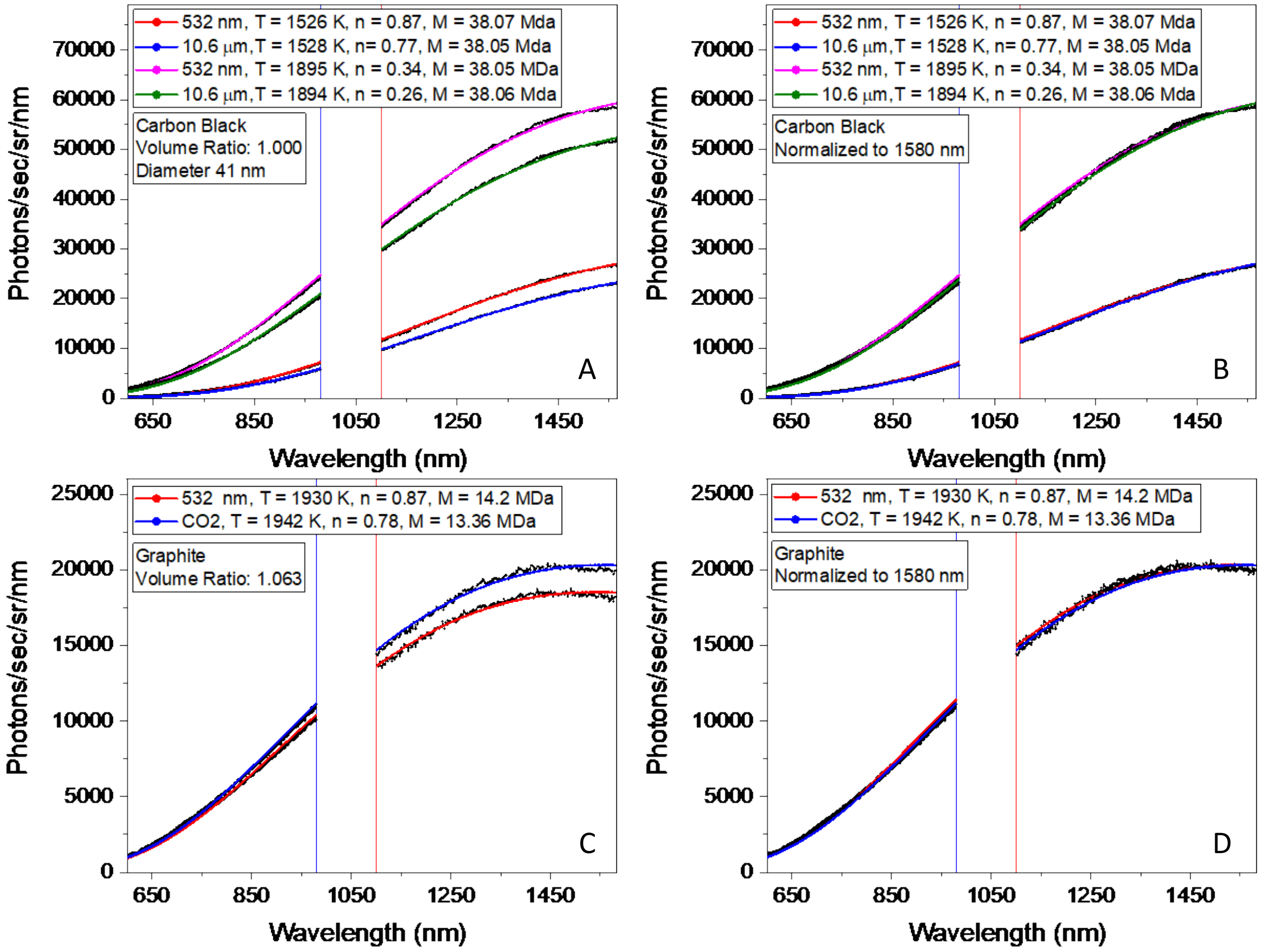

Figure 8. A: Emission spectra for a single carbon black NP with mass of 37.62 MDa, heated to similar temperatures by both 532 nm and 10.6 μm radiation. B: The same carbon black spectra normalized to allow shape comparisons. C and D: Analogous spectra for a single graphite NP with initial M = 14.2 MDa, heated at 532 nm or 10.6 μm. No significant effects are observed.

The experiment was done for both a carbon black NP of initial mass 37.62 MDa, and a graphite NP of initial mass 14.2 MDa. For the carbon black NP, a set of spectra was taken at $T_{NP}$ ~ 1550 K using first the 532 nm laser, and then the 10.6 μm laser, then the process was repeated for sets of spectra with $T_{NP}$ near 1900 K. In Figure 8A, pairs of spectra with very similar $T_{NP}$ values are compared, and it can be seen that the raw intensities and n parameters were slightly lower for the $CO_2$ laser heating, however, comparison of the normalized spectra in frame B shows that they are nearly superimposable.

For the graphite NP in frames C and D, the intensity observed with the $CO_2$ laser was slightly higher than that measured with the 532 nm laser, however, in this case, $T_{NP}$ was somewhat higher during the 10.6 μm heating. Again, although the n parameters were slightly different, the normalized spectra are nearly superimposable.

Clearly, the effects of heating laser wavelength on the shape and intensity of NP emission spectra are small. Indeed, given the ~$T^6$ dependence of intensity on $T_{NP}$ and the estimated ±2% uncertainty in comparing $T_{NP}$ values, the small intensity differences observed are not significant. In addition, the time scales of these experiments were long enough that some evolution of the NP emission properties probably occurred between the 532 nm and $CO_2$ laser experiments. We conclude, therefore, that there is no evidence for any residual effect of heating laser wavelength on the emission properties of black carbon NPs.

## IV. Discussion

The results show that NPs of black carbon materials like graphite, graphene, and carbon black have blackbody-like thermal emission spectra. The same is true for boron-doped nano diamond and carbon dot NPs, at least for the high temperatures probed here. No sign of structured emission was observed for any of the NPs.

Nonetheless, there were significant NP-to-NP variations in both the intensity (volume-corrected) and wavelength dependence of the thermal emission spectra from carbon NPs. This variability is attributed to structural variations in the NPs, including both variations in NP material (i.e., graphite, graphene, diamond, carbon black), and variations in the structure of individual NPs of the same material. Furthermore, for individual NPs, the emission brightness and wavelength dependence often change significantly if the NPs are heated hot enough to drive significant sublimation, and these changes are attributed to evolution of the NP structure at high $T_{NP}$.

One possibility, clearly, is that structural differences might include differences in the NP composition, which also might account for the evolution of the NP spectra after heating. Consider the graphite, graphene, and carbon black NPs, where the feedstocks all had stated non-volatile impurity levels well below 1 %. As initially trapped, these NPs probably had surface groups with O, H, or N heteroatoms, as well as adsorbates such as water, methanol, and ammonium acetate, however, all these should have decomposed and desorbed well below 1400 K.[29-33] Therefore, except perhaps in a few spectra at low initial $T_{NP}$, these NPs should have been near pure carbon. The nano diamond NPs would have had significant concentrations of boron and impurity atoms, and it is unclear to what extent our temperatures were high enough to allow them to diffuse in the diamond lattice, or what spectral changes that might cause.[34] Similarly, the initial composition of the carbon dot NPs is unknown, thus it is unclear why the emission was so weak. To avoid the issue of impurity effects on the spectra we focus the discussion on graphite, graphene, and carbon black NPs.

Thermal emission depends on the density and occupancy of electronic and vibrational/phonon states of the NPs. Graphite and graphene NPs have fully coordinated basal plane sites, both on the NP surface and in the interior, as well as low coordination sites such as basal plane edges on the NP surface, and defects in basal planes either on the surface or in the NP interior. These low coordination sites should make different contributions to the electronic and vibrational state distributions, compared to fully coordinated sites. Therefore, changes in the site distribution will change the state distribution, hence the spectra. For carbon black, where the initial structure includes both crystalline (graphitic) and amorphous regions, and some porosity, there may be changes in the distribution of low coordination surface and defect sites, but also larger scale structural changes such as crystallization and densification.

One structural evolution process expected at high temperatures is annealing. The melting point of graphite (at high pressures) is near 4800 K[43] thus our experimental $T_{NP}$ range corresponds to ~25 to 45% of the melting point. Given our long experimental times, it is likely that some annealing occurs, particularly during the >1900 K heating periods. Annealing should tend to reduce both the number and relative concentration of defects and other low-coordination sites, because these are less thermodynamically stable than fully coordinated sites.

The competing process that clearly occurs, again primarily at higher $T_{NP}$, is sublimation. In the experiments summarized in Figures 4 and 5, the "high" temperatures were in the 1900 to 2200 K range, and the mass losses during the heating periods ranged from 5 to 20% of the NP mass. The NPs were in the 6 to 50 MDa size range, which would have 3 to 6% of their atoms in the surface layer if the particles were spherical. The actual particles would have had larger fractions of their atoms in the surface – much larger for the graphene platelet NPs. Thus, on the order of a monolayer's worth of material would have desorbed from each NP during the >1900 K periods, corresponding to sublimation rates on the order of only $10^{-2}$ to $10^{-3}$ monolayers/second. Note, however, that sublimation is not expected to proceed in a layer-by-layer fashion. For graphite and graphene NPs, the low coordination edge and defect sites are less stable than basal plane sites, and thus should desorb at much higher rates, however, with increasing $T_{NP}$ sublimation from basal plane sites, which are more stable but also more abundant, should start to occur more frequently, which would tend to create new low coordination defect sites. For carbon black, the situation is more complex, but sublimation from low coordination sites should still dominate.

Note that if there were significant concentrations in the buffer gas of reaction species, such as $O_2$ or O atoms (possibly created in a discharge), these would tend to etch the NPs. The instrument is constructed mostly with UHV-compatible hardware, apart from a few o-rings sealing parts like windows. Because the ESI source introduces significant gas loads each time an NP is injected, the typical base pressure is in the low $10^{-7}$ Torr range, setting an upper limit on the $O_2$ concentration. In experiments in which $O_2$ is leaked into the trap chamber at >100 times higher pressure, mass loss rates are typically only a few C atoms/second at 1900 K, thus we believe the effects of reactive background gas are negligible.

The important point is that sublimation can lead to net creation or destruction of low coordination sites, depending on the NP structure. Consider an ideal graphite NP in which all the basal planes are perfect, with low coordination sites only around their edges. In that case, sublimation would occur mostly at the edges, shrinking the size of the basal planes, which would reduce the number of edge atoms, but increase the fraction of edge atoms as the NP loses mass. If the exposed basal planes also have defects like vacancies, sublimation from under-coordinated sites surrounding the defects would tend expand the defects, similar to the etch pit formation observed in high temperature reactions of graphite defects.[44-49] Thus sublimation at defects would further increase the fraction of low coordination sites.

Sublimation can also decrease the fraction of low coordination sites, however. Consider a rough NP with high surface area and many low coordination sites due to the presence of asperities on the NP surface. Because the asperities have high surface area and many low coordination sites, they should sublime more rapidly than the bulk of the NP, eventually disappearing, generating a smoother NP with fewer low coordination sites. Similar considerations apply to the case where the NP surface is decorated by smaller particles, which also would tend to sublime rapidly. In either case, both the number and fraction of low coordination surface sites should initially decrease as these small features sublime. Therefore, the structural changes that occur as an NP is heated depend on the initial NP structure, and because carbon NP feedstocks are quite heterogeneous, significant NP-to-NP variability in the effects of heating is expected, and observed.

**Conclusions**

The results presented above show that hot graphite, graphene, carbon black, nano diamond, and carbon dot NPs all have blackbody-like thermal emission spectra, regardless of heating laser wavelength and NP mass and charge, at least in the range studied. There are substantial variations from NP to NP in emission brightness and spectral shape, and while there do appear to be some effects of the carbon material on the emission properties, the magnitude of these material effects is comparable to the variation in properties between different NPs of the same material. In addition, for all the NPs studied, there were substantial changes in both the intensity and shape of the emission spectra after the NPs were heated above 1900 K. The different NPs still have different spectral properties, but in general the NP-to-NP differences are smaller after >1900 K heating. We attribute the large NP-to-NP variability to the variations in the number and type of low coordination carbon sites on the different NPs, and the evolution of the NP properties during heating to the effects of annealing and sublimation, changing the site distributions. This kind of NP-to-NP variation in emission properties, as well as the evolution in emission properties with its NP-to-NP variability, can only be studied using single NP methods of the type presented here.

Ideally, we would like to correlate the NP emission behavior with the actual distributions of carbon sites on individual NPs, and this should be possible to some extent based on correlating emission behavior to sublimation rates. This is because the sublimation rate also depends on the distribution of sites present (on the NP surface), with the less stable low coordination sites having the highest rates. At this point, the number of individual NPs and the range of heating effects studied is too small to allow correlations between emission behavior and sublimation rates to be discerned with any confidence. We are, however, beginning a systematic study of the sublimation kinetics for graphite, which will entails measuring sublimation rates and emission spectra for many particles subjected to a variety of heating regimes. One goal of this study will be to understand the factors that control graphite emissivity in more detail, by correlating emission behavior, and the temperature dependence of sublimation, and eventually oxidation kinetics, which provide means to examine the number and energetics of sites on the NP surfaces. Again, this kind of information can come only from single NP experiments.

**Supporting Information**.

The Supporting Information is available free of charge *via* the Internet at http://pubs.acs.org.

The following files are available free of charge.

Figures S1 through S15 and associated discussion illustrate the power law emissivity function, effects of varying n on fitting, comparisons of NP emission spectra for different materials, effects of drive voltage sweeps on $T_{NP}$, the optical setup and calibration method, the effects TC position and temperature on $T_{NP}$, and the effects of various optical element misalignments on $T_{NP}$ (PDF).

Data underlying Figures 1 through 8 are provided in tabular form (XLSX).

AUTHOR INFORMATION

**Corresponding Author**

*Scott L. Anderson, anderson@chem.utah.edu, (801)585-7289

**Author Contributions**

The manuscript was written through contributions of all authors. All authors have given approval to the final version of the manuscript. ‡These authors contributed equally.

ACKNOWLEDGMENT

The nIR spectrograph was purchased with funds from the Albaugh Scientific Equipment Endowment of the College of Science, University of Utah. This material is based upon work supported by the U.S. Department of Energy, Office of Science, Office of Basic Energy Sciences, under Award Number DE-SC- 0018049. We thank Prof. Joel Harris (University of Utah, Chemistry Department) and Prof. Jordan Gerton (University of Utah, Physics Department) for many helpful discussions about optics and light detection.

**Funding Sources**

U.S. Department of Energy, Office of Science, Office of Basic Energy Sciences, under Award Number DE-SC- 0018049.

ABBREVIATIONS

NP, nanoparticle; $T_{NP}$, nanoparticle temperature; VUV, vacuum ultraviolet.

TOC Graphic

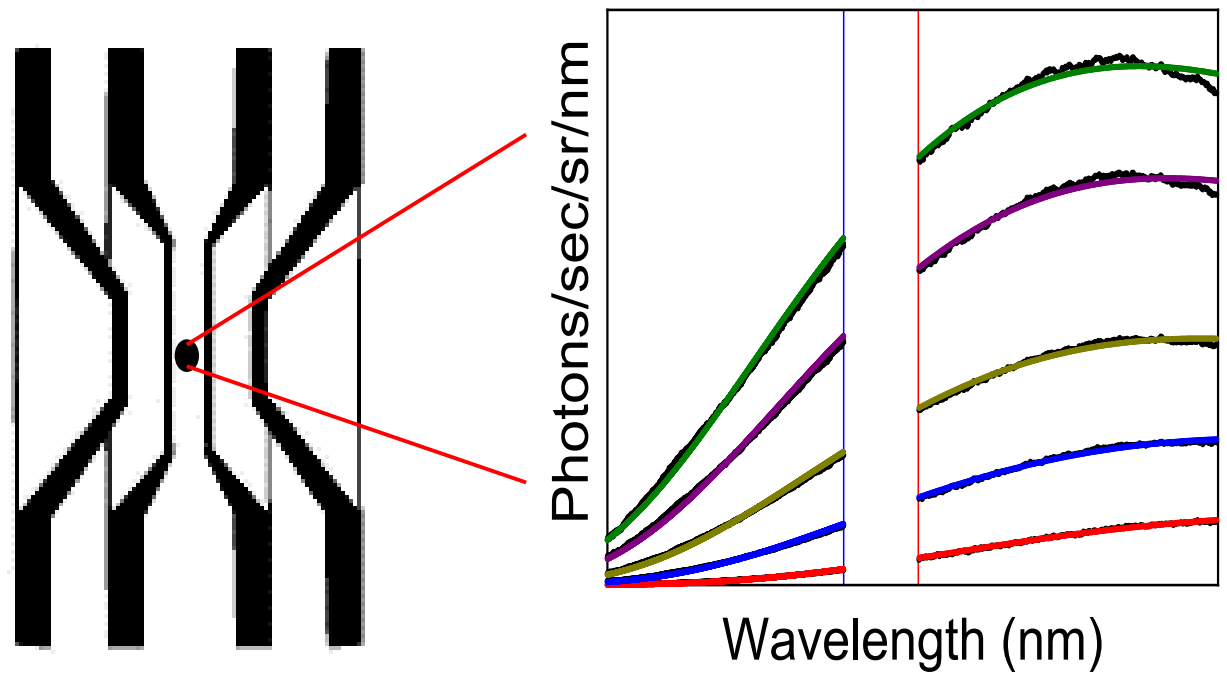

Thermal Emission Spectroscopy of Single, Isolated Carbon Nanoparticles: Effects of Particle Size, Material, Charge, Excitation Wavelength, and Thermal History

*Bryan A. Long, Daniel J. Rodriguez[‡], Chris Y. Lau[‡], Madeline Schultz, and Scott L. Anderson**

Chemistry Department, University of Utah, 315 S. 1400 E., Salt Lake City, UT 84112

[‡]*Equal contributions*

**Corresponding author: anderson@chem.utah.edu*

## Supporting Information

### Optical System Design:

Figure S1 shows the NPMS optics design. The trap is shown near the center of the dashed rectangle representing the vacuum chamber, and mirrors for focusing the $CO_2$ laser through the trap are shown, along with the laser beam path. 532 nm or other lasers are focused through the trap into the plane of the figure. The insertable thermocouple (TC) calibration emitter is shown schematically above the trap, and the coordinate system used in the following description is shown above that.

Light emitted axially (in the +z direction) from the trap center is collected by a five element lens system consisting of a 35 mm focal length achromatic doublet, a 100 mm focal length achromatic doublet, and a -75 mm focal length lens. This collects about one steradian solid angle from the trap center, and forms it into a slightly convergent beam. When the TC is used in calibration experiments, the ND filter assembly is inserted, cutting the light intensity by a factor of ~$10^8$, with transmission vs. wavelength measured as described previously.[1] For alignment purposes, a mirror can be inserted into the beam path, directing the light onto a room temperature CCD camera (The Imaging Source) that captures magnified images of the trap center from the axial direction for alignment purposes. During spectral measurements, the mirror is retracted, allowing the light to reach a 980 nm dichroic beamsplitter that reflects wavelengths ($\lambda$) < 980 nm (“visible”) and passes light with $\lambda$ > 980 (“nIR”). The visible beam is passed through a 100 mm focal length plano-convex lens, a 550 nm long pass filter used to block scattered 532 nm laser light, and then is injected into a 1 mm diameter low OH optical fiber using a fiber collimator (Thorlabs, f = 8.00 mm, NA = 0.50, F240SMA-780 nm). The visible spectrum is dispersed by an Andor Shamrock 163 spectrograph equipped with silver-coated mirrors and grating, and measured by a back-illuminated Si CCD camera with 2000 pixel x 256 pixel CCD array with pixel sizes 15 μm x 15 μm, thermoelectrically cooled to -65 °C (Andor, DU416A-LDC-DD).

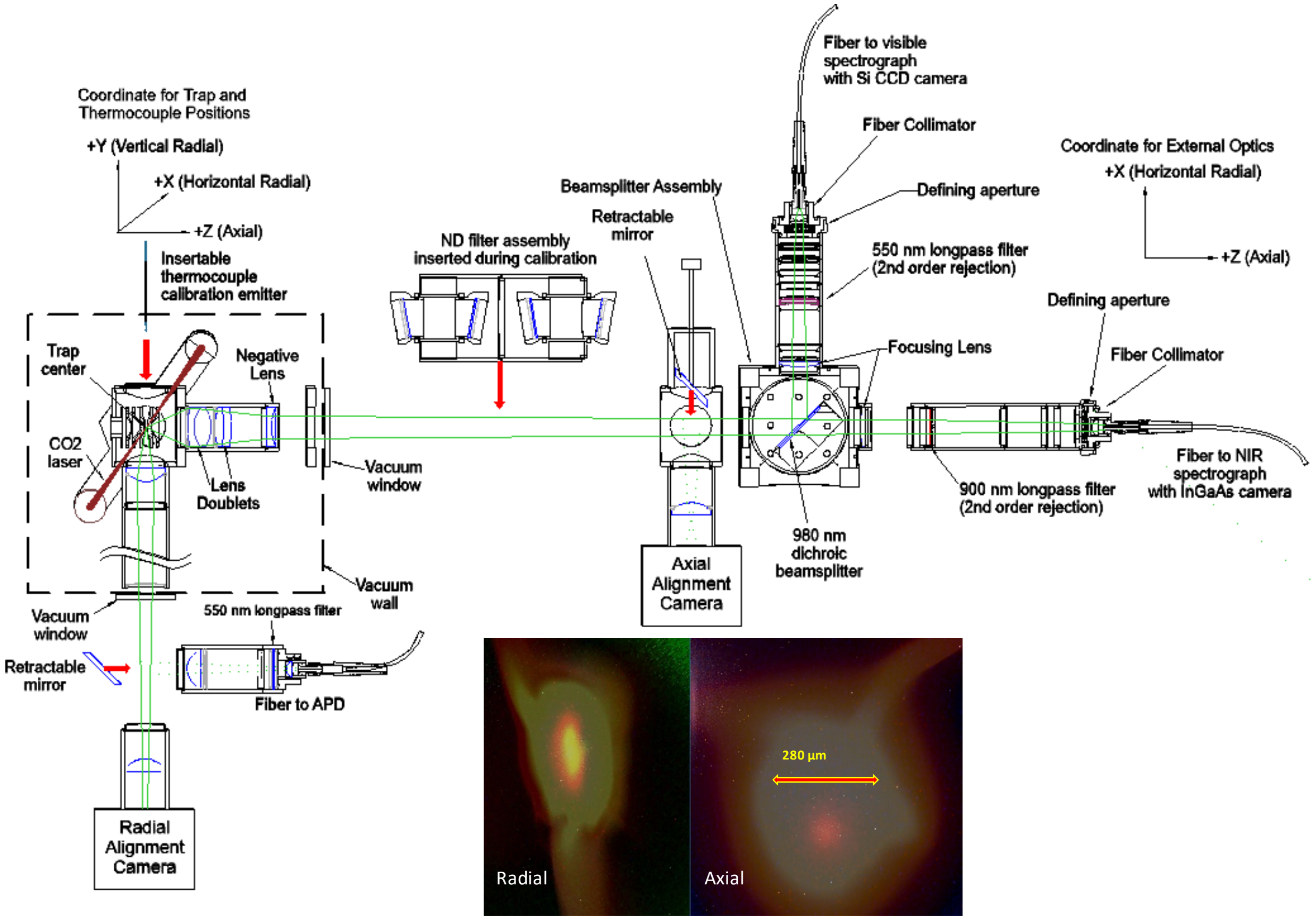

Figure S1. Alignment camera images at the bottom show views of a single NP (bright red/yellow spots) superimposed on images of the TC (larger irregular grey/green features), positioned for calibration experiments. Thermal emission is collected out of the bottom of the trap for mass and charge determination, and along the trap axis for spectra/$T_{NP}$ measurements.

The nIR beam passes through a 175 mm focal length plano-convex lens, and a 900 nm long pass filter used to block 2nd order light in the nIR spectrum. A fiber collimator (Thorlabs, f = 8.12 mm, NA = 0.49, F240SMA-C-1310 nm) is used to inject the light into a 1 mm diameter low OH fiber, which conveys it to another Andor Shamrock 163 spectrograph, equipped with silver coated mirrors and grating. The nIR spectra are acquired by an InGaAs photodiode array camera, with a 512 pixel x 1 pixel array with pixels that are 25 μm wide by 500 μm high, cooled to -85° C (Andor, DU490A-1.7).

Light emitted radially out of the bottom of the trap is collected by a 31 mm aspheric achromat, and formed into a slightly convergent beam. Normally this is reflected by a retractable mirror, passed through a 50 mm focal length achromatic lens doublet, and injected into a 400 μm diameter low OH fiber that conveys the light to an avalanche photodiode (APD) pulse counting module (Laser Components Count). The APD signal is monitored as a function of time to detect the secular motion of the NP in the trap, for mass determination. When the mirror is retracted, the light is focused onto a second room temperature CCD camera (The Imaging Source), generating a magnified view of the central volume of the trap from the radial direction, for alignment purposes.

**Optical system calibration:**

The optical system has non-idealities such as chromatic aberrations, detector quantum efficiencies, grating efficiencies, etc., that affect its sensitivity vs. wavelength, S(λ), thus distorting the measured spectra. Before fitting spectra to extract temperatures, it is essential to correct the NP spectral intensities, i.e., we need to know S(λ) quantitatively. To measure S(λ) we need a calibration light source with well-known intensity vs. λ, small enough to fit into the trap center so that calibration can be done on the system as a whole, under conditions identical to those used for the NP spectral measurements.

As a calibration emitter, we use a micro-thermocouple (TC) with a roughly disk-shaped bead about 280 μm in diameter, and ~150 μm thick. The TC is positioned so that its bead is in the trap center, where it can be heated by the $CO_2$ laser, with the bead temperature read out electrically. The TC wire manufacturer (Concept Alloys) states that the accuracy in the temperature range of interest here is ±1% of the measurement in °C. To enable the TC bead to be positioned precisely and repeatedly, the TC is mounted on a precision XYZ vacuum manipulator, and the axial and radial alignment cameras allow the position in two orthogonal planes to be imaged. The approach to determining the correct TC position is to first trap a single NP, and image it with the two cameras, thus defining the trap center precisely. Then the TC is moved into the trap, heated, and its position is adjusted until it is at the position defined by the NP.

The bottom of Figure S1 shows superimposed images of a single NP and the TC, from the radial (left) and axial (right) alignment cameras. In the radial camera image, looking from the bottom of the trap, the NP appears as a bright yellow/red oblong in the center, and the laser-heated TC appears as the larger irregular pale green shape with lead wires visible extending from the bead at the top and bottom of the image. The axial camera shows the NP as a circular red spot, and the TC bead as a roughly circular pale grey area, with lead wires visible at the top extending to either side.

The NP image makes an important point: The effective "size" of the trapped NP as a light emitter is determined by the amplitude of its secular motion in the trap, which depends on NP mass and charge and the trap operating conditions. For the hot carbon NPs of interest here, the NP effective size (FWHM) is typically in the 30 to 40 μm range along the axial direction, with double this "size" in the radial direction, hence the oblong appearance of the NP in the radial camera image. Thus, for light collection along the axis, the NP emitter "size" is ~60 - 80 μm in diameter.

Figure S2 illustrates the process of using the TC emission spectra to determine S(λ) and to correct NP spectra. Figure S2A shows a typical TC spectrum measured at a junction temperature of 2316 K. This measurement was done with the ND filter assembly inserted into the optical path, so that the intensity of the calibration spectrum was similar to those for single NPs. The scale is in counts/second/$m^2$ of TC emitter surface area (area ≈ 2.93x10-8 $m^2$). The approach to measuring the ND filter assembly transmission has been described previously, along with the transmission vs. wavelength.[1] Figure S2B shows the predicted TC emission spectrum, which is the product of three functions: 1. The Planck's law emission function for an ideal blackbody at 2316 K. 2. The emissivity of the W-Re TC at 2316 K[2-5] 3. The ND filter assembly transmission. The ratio of Figure S2A to Figure S2B is the desired S(λ) function, which is plotted in Figure S2C. Figure S2D shows a typical raw single NP spectrum, measured for a single graphite NP of mass 14.93 MDa. Figure S2E shows the result of dividing Figure S2D by Figure S2C, to obtain a sensitivity-corrected NP spectrum.

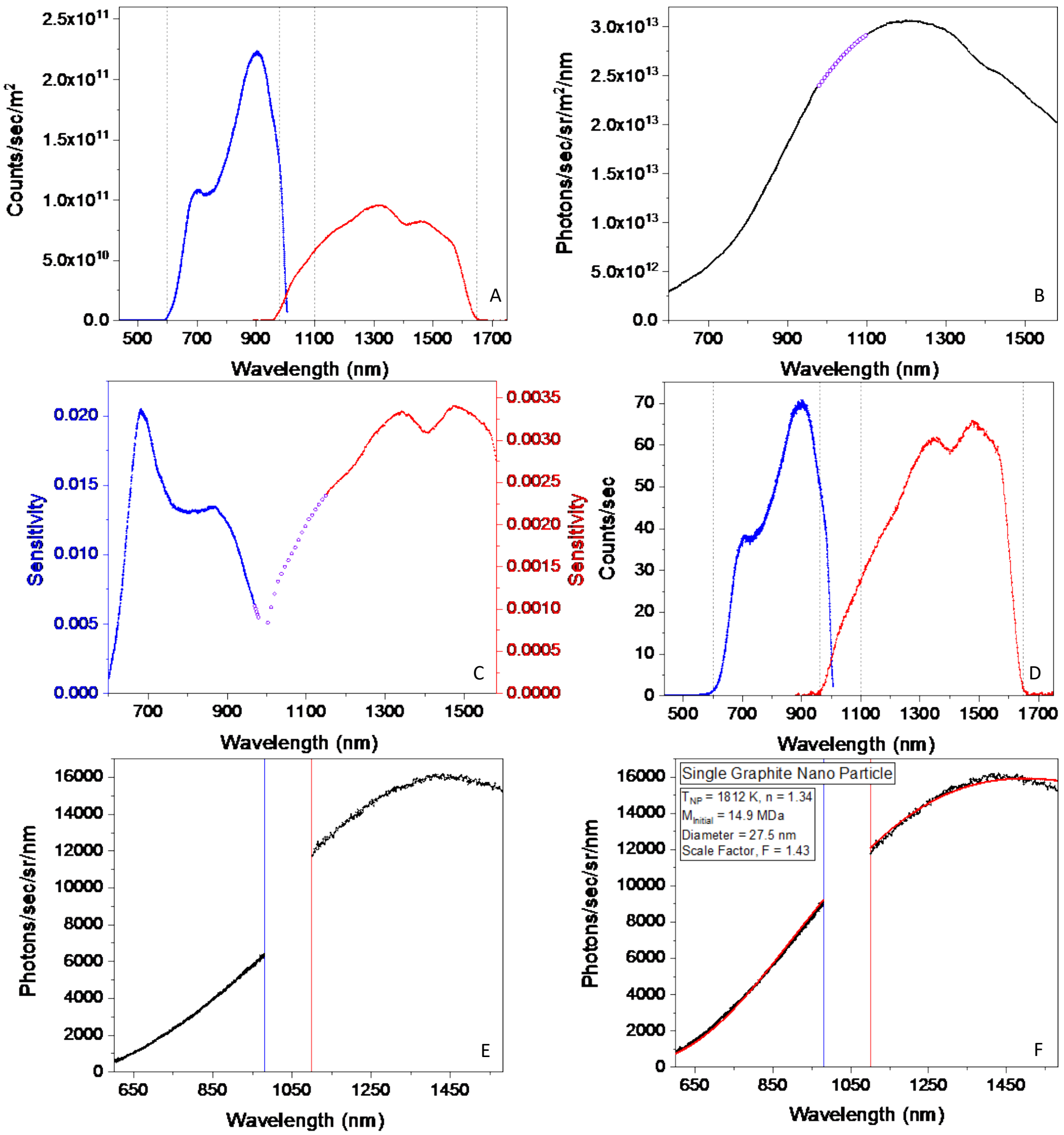

Figure S2. Data illustrating how the sensitivity function, S(λ), is extracted, and NP spectra are intensity corrected. Frame A is the raw spectrum for the TC heated to 2316 K. B is the theoretical spectrum, calculated as the product of Planck's law, the emissivity of the TC material, and the transmission of the ND filter assembly. C is the extracted S(λ) function, *i.e.* the ratio of A to B. D is a typical single raw NP spectrum. E shows the corrected NP spectrum, showing intensity mismatch between the visible and nIR. F is the final corrected spectrum with fit, after applying the visible scale factor.

Note that there is a spectral region from 980 to 1100 nm where both the visible and nIR spectrographs have low sensitivity because of the combined effects of the dichroic beamsplitter, the CCD quantum efficiency, and the grating efficiency of the nIR spectrograph. As a result, the signal in this "gap" region is low, and uncertainty is high. In the fitting process described below for $T_{NP}$ determination, this gap region is given zero weight.

Figure S2E shows one final step in the NP spectral correction process. Note that the visible and nIR spectra do not match up across the spectral gap region, i.e., the corrected visible intensity is slightly too low to match smoothly to the nIR data. This effect is believed to be a result of the fact that our visible camera uses a 2D array that is 3.84 mm tall, whereas the 1D nIR camera array is only 500 μm tall, (but equipped with a cylindrical lens to help reduce the height of the dispersed light on the array). Because the TC (~280 μm) is significantly larger than the ~80 μm effective radial emitter size for typical NPs, the solid angle of light injected into the fibers (numerical aperture 0.5) is higher for the TC than for NPs, resulting in a somewhat larger solid angle entering the spectrographs. The taller visible array captures more of this "extra" light than the nIR array, and as a result, the visible portion of S(λ) is larger than it should be for correcting spectra from the smaller NP emitters. Thus, the intensity of the visible portion of the corrected NP spectra tends to be slightly too small to match up smoothly to the nIR section. This hypothesis is supported by the observation that changing the effective NP emitter size by varying the trapping conditions, results in changes in the relative intensities of the visible and nIR spectra.

Therefore, it is necessary to scale the visible spectrum to ensure smooth matching to the nIR spectrum across the gap. The approach used is to fit the data just on either side of the gap (900 to 950 nm in the visible and 1030 to 1200 nm in the nIR), to a smooth function, with the visible spectrum scale factor as one of the fit parameters. As a smooth function, we use Planck's law times a power law model for the NP emissivity ($\epsilon \propto \lambda^{-n}$), *i.e.*, the same function used to fit spectra for $T_{NP}$ determination, but with n fixed at 1.4. The n parameter affects the curvature of the fitting function, but because only a narrow λ range is fit to extract the visible spectrum scale factor, the scale factor is not very dependent on the exponent. 1.4 was chosen as being typical for carbon NPs, and the correct value for room temperature graphite (see Figure S3). Figure S2F shows the result of applying the scale factor to the visible spectrum. The visible spectrum scale factor varies slightly from NP to NP, and in this case, it was 1.34. The effects of systematically varying this scale factor on temperatures extracted from spectral fitting are explored below.

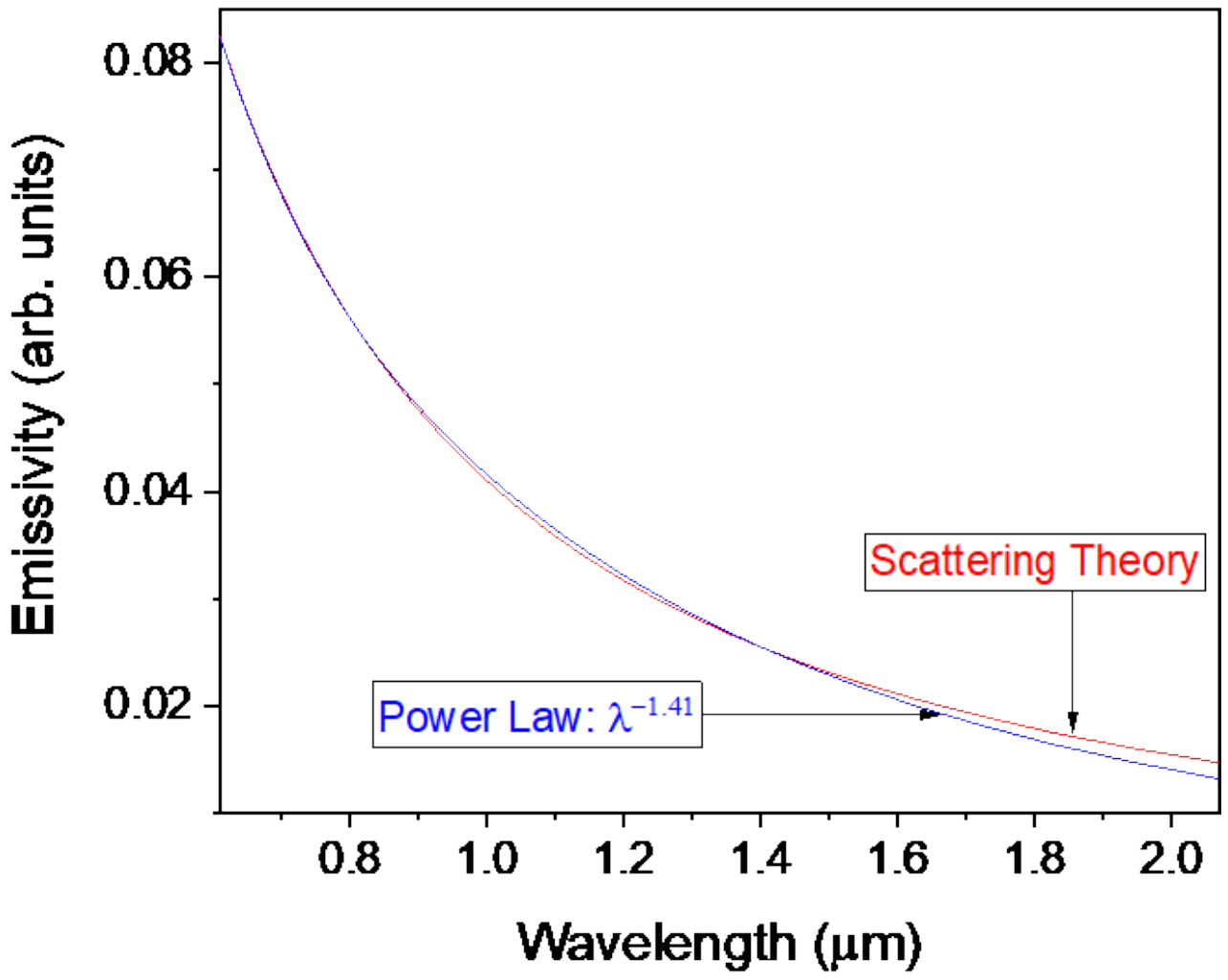

Figure S3. Comparison of the calculated emissivity function based on small particle scattering theory (Red) using room temperature graphite n and k data, to a power law model, $\lambda^{-n}$ (Blue).

**Power law approximation to the scattering theory emissivity function:**

The NP spectra are fit to a function that is the product of Planck's law times and emissivity function, $\epsilon(\lambda)$. As noted in the manuscript, scattering theory predicts that in the limit of NPs much smaller than the wavelengths emitted, as is the case here, $\epsilon(\lambda) = \frac{8\pi r}{\lambda} Im\left(\frac{(n+ik)^2-1}{(n+ik)^2+2}\right)$, where n and k are the real and imaginary components of the index of refraction of the material at the wavelengths and temperature of interest.[6] Figure S3 shows $\epsilon(\lambda)$ for graphite, calculated from the scattering theory expression, using room temperature n and k values from the literature.[7] The figure also shows the best power law fit to the scattering theory result: $\epsilon(\lambda) \propto \lambda^{-n}$, n = 1.41. The maximum difference between the two curves in the wavelength range we measure is ~3%. Ideally we would use the correct $\epsilon(\lambda)$ function in fitting NP spectra, however, the necessary index of refraction (or equivalent dielectric) parameters are not available for the materials of interest, at high temperatures, over the wavelength range of interest. More fundamentally, bulk optical properties would be inaccurate for NPs, due to the large number of surface and defect states in NPs. Therefore, we are forced to adopt a model $\epsilon(\lambda)$ function, and the power law model at least has the virtue of having roughly the correct shape and introducing a minimum number of adjustable parameters.

**Sensitivity of extracted $T_{NP}$ values to the position of the TC, and to TC temperature used in calibration:**

Figures S4A-C, show the effects of moving the TC calibration emitter to different positions in the trap, *i.e.*, deliberately positioning it so that it is not at the trap center. Each figure shows traces of NP temperature ($T_{NP}$) vs. time from an experiment in which a single NP was heated using a 532 nm laser, with its temperature stepped from ~1200 K to ~2000 K, then dropped to ~1400 K. The 532 nm laser was not being stabilized during this test, and its intensity fluctuated significantly, driving fluctuations in $T_{NP}$. All the traces shown in Figures S4A-C are based on the same raw NP emission spectra, but corrected using TC calibration spectra obtained with the TC emitter at the trap center (black trace), and deliberately misaligned by ± 0.05 and ± 0.1 mm in the X, Y, and Z directions, relative to trap center. The important point is how much the $T_{NP}$ traces obtained at different TC emitter positions vary, *i.e.*, how sensitive the $T_{NP}$ calibration is to mis-aligning the TC emitter relative to trap center. The TC temperature, $T_{TC}$, during these experiments was kept close to 2300 K, and measured electrically in conjunction with each spectral measurement.

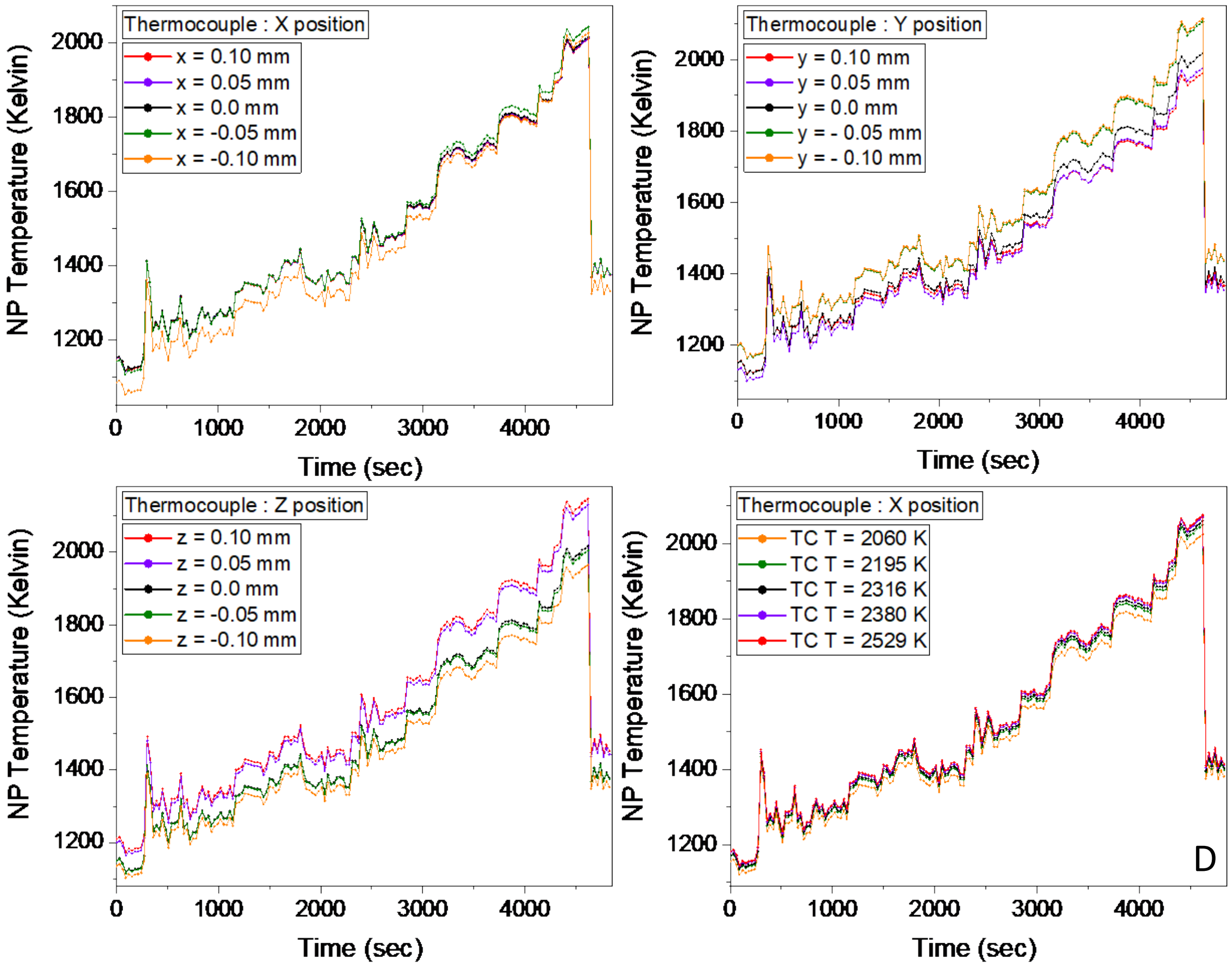

Figure S4. Temperature profile as a function of time for a single NP being heated at different laser powers. Figures A-C are correcting the same single NP raw emission spectra with different thermocouple emission spectra as a function of thermocouple position in the ion trap. Figure D is correcting the same single NP raw emission spectra but with the thermocouple in the optimal position of the trap but at different temperatures.

In Figure S4A, the TC was moved radially in the x direction, as defined in Figure S1, while keeping the y and z coordinates centered on the trap. This corresponds to horizontal motion of the right hand TC image shown in Figure S1. It can be seen that if the TC is positioned at x= -0.10 mm with respect to the trap center the $T_{NP}$ values extracted for the NP are shifted significantly to lower temperatures, particularly for low $T_{NP}$ values. For example, the shift is roughly -60 K for $T_{NP}$ near 1200 K, but only –30K for $T_{NP}$ above 1500 K, and only -10 K at temperatures greater than 1800 K. For misalignments of -0.05 mm and for both +0.05 and +0.10 mm, the variations in $T_{NP}$ relative to the values extracted for the TC at the trap center are much smaller – averaging roughly ± 6 K over the entire $T_{NP}$ range.

Figure S4B gives analogous data for deliberate misalignment of the TC emitter in the y direction, where +y corresponds to moving the right-hand TC image in Figure S1 up, and –y is downward motion. Here the effects of misalignment, particularly of having the TC positioned too high in the trap, are larger. For y = -0.1 mm and -0.05 mm, $T_{NP}$ values are similar, and shifted by ~-50 K at low $T_{NP}$ and by ~-90 K at high $T_{NP}$, relative to the $T_{NP}$ values obtained at y = 0.0. For positive y misalignments (y = 0.05 and 0.1 mm) the shifts are also similar to each other, but much smaller, remaining within 20 K of the values measured for y = 0.0 at all $T_{NP}$. The larger shifts observed for negative y misalignments are attributed to the effect of collecting light emitted from the wires leading to the TC bead. These wires connect to the bead at its top, and are significantly cooler than the laser-heated bead. Thus, for TC displacements to –y, the optical system picks up more light from the lead wires, which would tend to give TC spectra colder than the bead temperature, and therefore cause the calibrated $T_{NP}$ higher, as is observed. Knowing this, we position the TC so that the lower part of the bead is at the trap center, as shown in the left image in Figure S1.

Figure S4C shows the effects of moving the TC axially in the trap, with +z corresponding to moving closer to the collection lens. The TC positions z = 0.05 mm (purple trace, $T_{TC}$ = 2303 K) and z = 0.1 mm (red trace, $T_{TC}$ = 2302 K) cause ~+100 K shifts in the $T_{NP}$ values obtained from the TC spectra. In contrast, moving the TC to z = -0.05 mm, *i.e.*, slightly further from the collection lens, had no significant effect, and moving to z = -0.1 mm caused shifts to lower $T_{NP}$ ranging from ~-20 K at low $T_{NP}$ to ~-50 K. The larger effects from shifting the TC closer to the collection lens makes sense – the NP is a diffuse emitter extending axially ~± 0.03 mm relative to the trap center, whereas the TC emitter is the surface of the TC bead facing the collection lens. Thus, as shown in the left image in Figure S1, when z = 0, the TC surface is actually ~ 0.06 mm closer to the lens than the trap center. Thus for small negative displacements of the TC, its front surface actually moves closer to the trap center, *i.e.* to the average NP position. For positive displacements, however, the TC surface is shifted increasingly farther from the average position of the NP, hence introducing significant chromatic effects on the TC spectra that affect the $T_{NP}$ calibration. Knowing this, we are careful to position the TC with its center at, or slightly behind the trap center.

Given the ~25 μm precision with which the TC can be positioned in the X, Y, and Z directions, and the results in Figures S4A-C, we estimate that the aggregate effects of TC misalignment on the extracted $T_{NP}$ values range from ~ ±3.2% at low end of our $T_{NP}$ range, to ±3.5% at high $T_{NP}$ values.

In principle, the TC temperature used to generate the sensitivity calibration needed to correct the NP spectra should not matter, however this assumes that there is no error in reading the TC temperature ($T_{TC}$) and in calculating the emissivity used to correct the TC spectra. To test for possible $T_{TC}$ effects, Figure S4D shows the effects of correcting the same NP data using TC spectra measured at different TC temperatures, with the TC positioned at x = y = z = 0. There are effects of $T_{TC}$ on the extracted $T_{NP}$ values, but they are small, as expected. For all but the lowest $T_{TC}$ value (2060 K), the variations in $T_{NP}$ are less than 10 K at low $T_{NP}$, and less than 16 K for high $T_{NP}$, corresponding to ≤ 0.8% uncertainty in $T_{NP}$. Using the 2060 K TC data introduces larger errors (~19 K at low $T_{NP}$, 35 K at high $T_{NP}$), probably because the emitter is too cold to give good intensities at short wavelengths. Obviously, we avoid use of low TC temperatures in correcting

NP spectra, thus the uncertainty in $T_{NP}$ values due to this TC temperature effect is estimated to be ≤0.8%.

**Effects of optical system misalignment on measured and corrected NP spectra, and on the $T_{NP}$ values extracted by fitting:**

Figures S5 – S9 show the effects on raw TC and NP spectra and corrected NP spectra (and extracted $T_{NP}$ values) of deliberately misaligning various optical components. Note that misalignment generally results in substantial decreases in signal intensities,

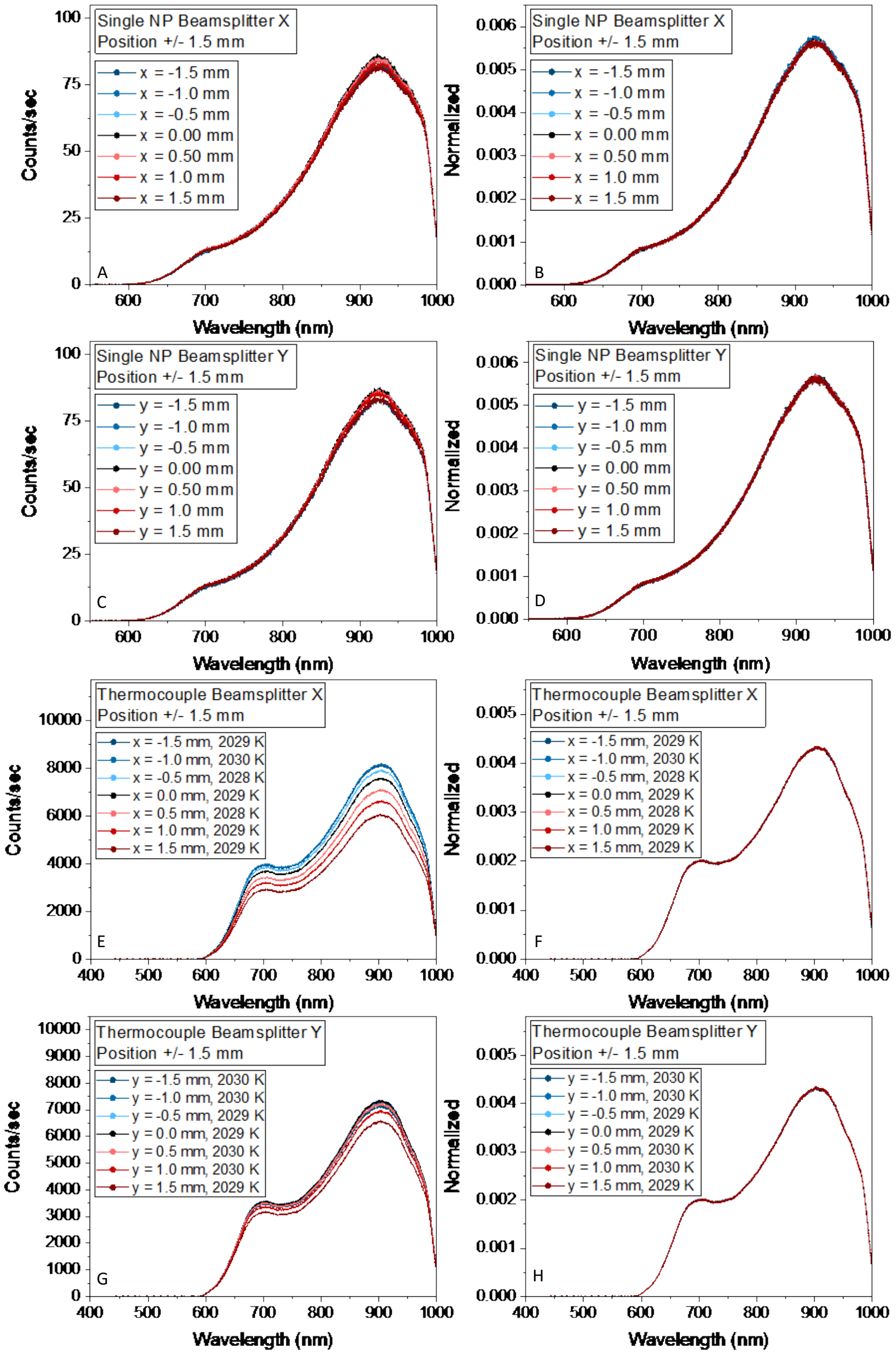

Figure S5. Raw NP visible emission spectra (A-D) and TC visible emission spectra (E-H). A, C, E, and G show the effects on the emission spectra of varying the beamsplitter's x and y positions by +/- 1.5 mm. Figures B, D, F, and H show the normalized intensities to allow shape changes observed.

however, from the $T_{NP}$ determination perspective, what is important are two factors: 1. How does the misalignment affect the *shape* of the spectra, as opposed to the intensity? To show changes in intensity and shape, we present both raw and normalized TC and NP spectra. 2. If there are changes in spectral shape from misalignments, are they the same for the TC and NP spectra? If so, they cancel in the calibration process. If not, they are apparent as changes in the corrected NP spectra.

In Figure S5 the raw visible emission spectra for a single NP and for the TC are shown as a function of displacements of the beamsplitter assembly in the x and y directions. Note that in making such comparisons, we kept the NP at relatively low $T_{NP}$, so that it did not sublime significantly on the experimental time scale, and we attempted to keep the temperature constant by holding the heating laser at constant intensity, using a PID program to stabilize the laser (unlike the measurements in Figure S4). As shown in Figure S1, beamsplitter assembly motion also moves the visible focusing lens and fiber collimator. The y motion does not move the visible beam path, therefore the lens and collimator become radially misaligned with respect to the beam. The x motion of the beamsplitter shifts the visible beam path, and again, the beam becomes radially misaligned with the lens and collimator.

The left side of Figure S5 shows raw spectra, and the right side shows the same spectra, normalized to allow changes in shape to be seen. In Figures S5A and S5E, the x position of the beamsplitter is changed by ±1.5 mm relative to the position that gives maximum intensity for the NP spectra, while measuring the NP and TC raw spectra. Figure S5C and S5G show analogous effects of ±1.5 mm misalignments in the y direction. It can be seen that radial misalignment of the focusing optics has significant effects on visible spectral intensity, but as shown in Figures S5B,

S5D, S5F, and S5H, the spectral shape, which is what is important in $T_{NP}$ determination, is relatively insensitive to this kind of misalignment.

Note that motion of the beamsplitter assembly does not shift the transmitted nIR beam path, but because the 175 mm nIR focusing lens is attached to the assembly, its motion does misalign the lens with respect to the beam path, thereby deflecting the beam slightly such that it becomes radially misaligned with respect to the nIR collimator. The effect of this radial misalignment on the nIR spectra is similar to what is observed if the beamsplitter assembly is fixed, but the nIR collimator is translated to cause radial misalignment (Figures S6 and S7A-C). Because these effects are similar and quite small, a separate figure is not presented showing the effects on nIR spectra of x and y beamsplitter assembly motion. Similarly, small displacements of the beamsplitter assembly in the z direction have negligible effects on either visible or nIR spectra, and these data are not shown.

In Figures S6-S9, we explore the misalignment of the nIR fiber collimator, which is mounted separately, allowing x, y, and z motion with respect to the nIR beam path. This motion has no effect on the visible spectra. Figure S6 shows spectra for a single NP for nIR collimator positions misaligned by up to ± 0.4 mm in the x and y (*i.e.* radial) directions. Figure S6A and S6C show raw spectra and Figures S6B and S6D show the same spectra normalized so that shape differences can be seen. It can be seen that for x and y misalignment of the nIR collimator, there are significant intensity effects, but that the spectral shape is unaffected. Figures S7A-D show analogous TC emission spectra, *i.e.*, raw and normalized spectra measured as a function of x and y misalignments up to ± 0.4 mm. Again, there is essentially no effect on the spectral shape. Figure S6E and S6F show how the integrated nIR signal from a single NP varies with x and y misalignment, showing that it is easy to find the optimum radial position for the collimator.

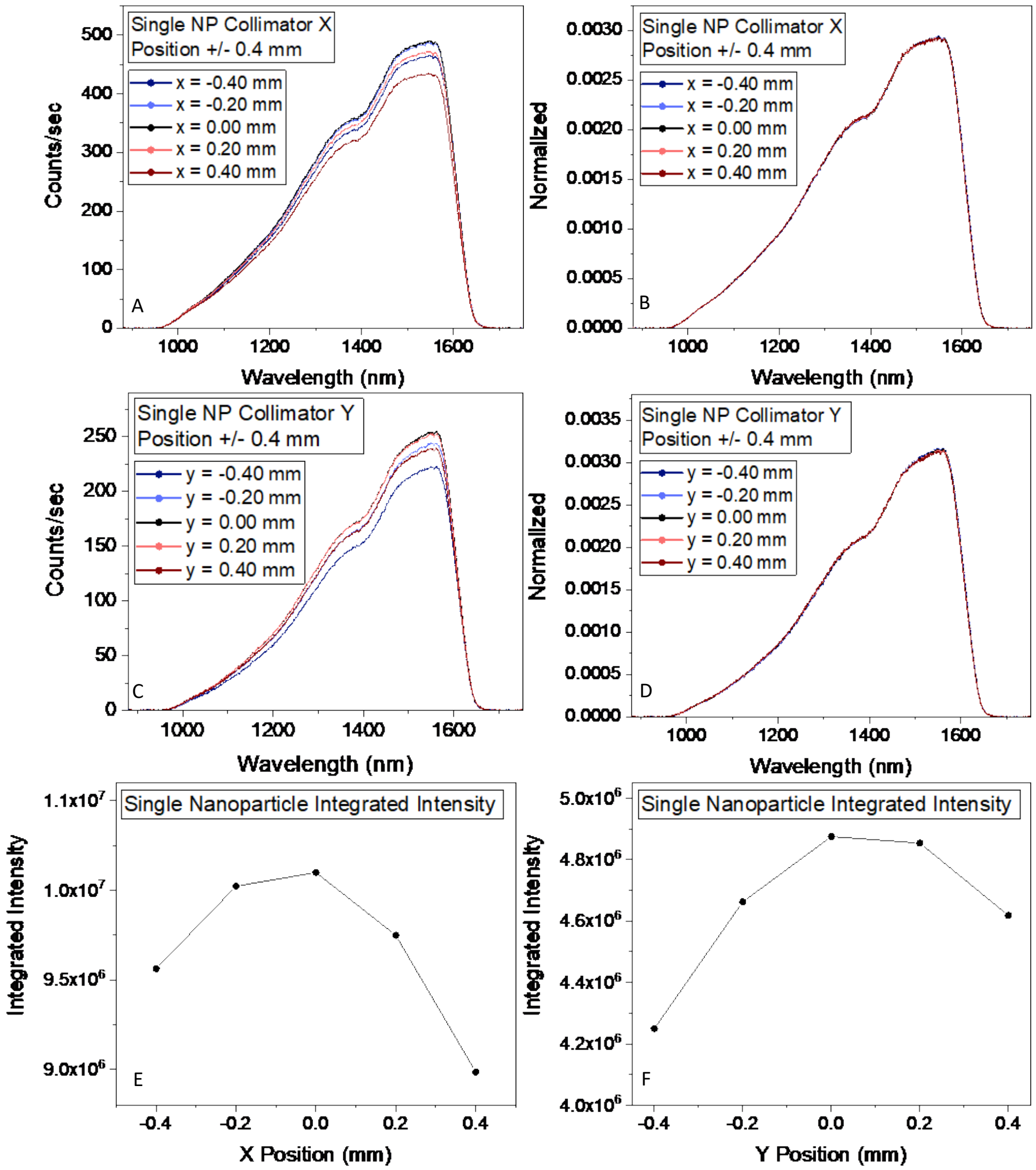

Figure S6. nIR emission spectra for a single carbon nanoparticle. Frames A and C show the effects of misalignment of the nIR fiber collimator on the raw intensities, and frames B and D show that the effects of nIR collimator on shape are negligible. Frames E and F simply show how the integrated spectral intensities vary with misalignment in the x and y directions.

The bottom half of Figure S7 shows the effects of nIR collimator x and y misalignment on the nIR part of the corrected NP spectra (there are not effects on the visible spectra). Figures S7E and S7G show that there are intensity changes in the corrected NP spectra as a function of collimator position, but Figure S7F and S7H show that the effects on spectral shape are small, as might be expected from small effects on the raw NP and TC spectral shapes. The $T_{NP}$ values extracted by fitting these spectra (combined with visible spectra taken at the same time) show ± 22 and ± 14 K variations (<±2%), however, it should be noted that the laser used to heat the NPs was not perfectly stable, thus it is almost certain that some of this $T_{NP}$ variation was real, *i.e.*, not error due to collimator position.

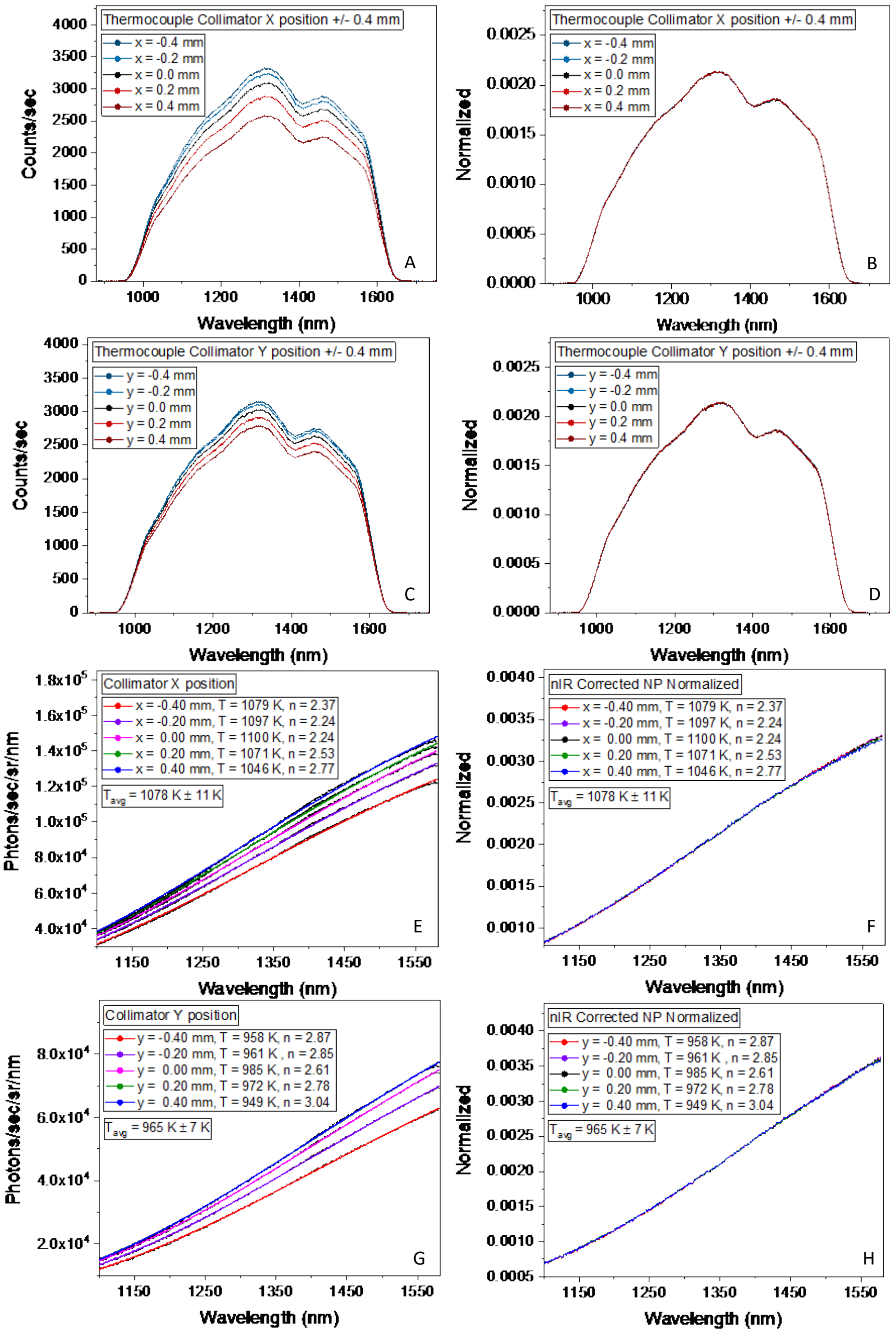

Figure S7. Frames A-D show raw TC nIR spectra at different X and Y collimator positions, taken at similar TC temperatures. Frames E and G are corrected NP spectra at different X and Y collimator positions, corrected using TC spectra taken at the same collimator positions. Images F and H are the spectra in E and G, normalized to show shape changes.

Given the relatively sharp dependence of intensity on x and y position, which makes it straightforward to align it precisely,

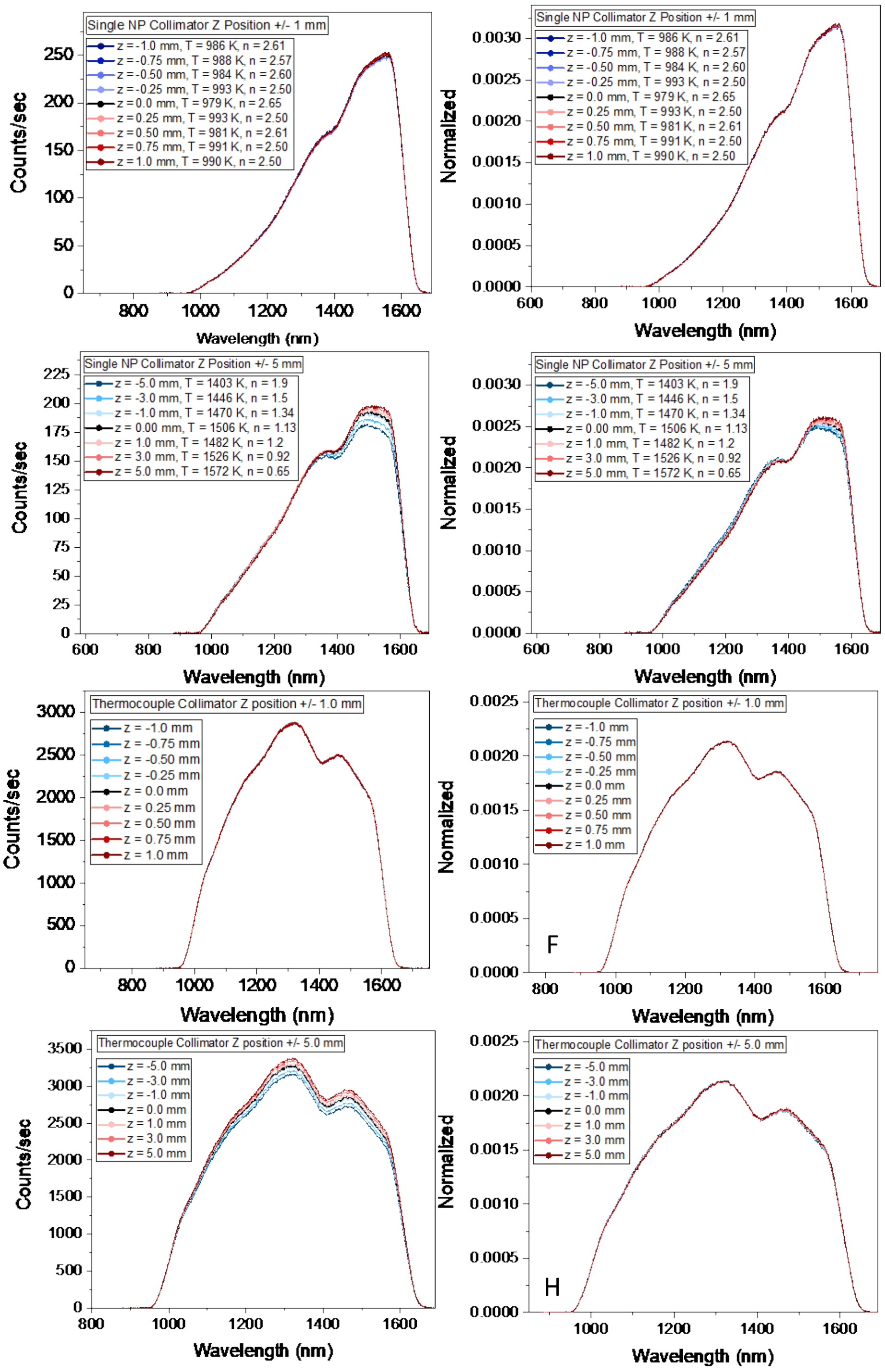

Figure S8**.** Frames A-D are raw nIR single NP emission spectra as a function of collimator Z position. Frames E-F are raw nIR TC emission spectra at the same Z positions. Frames A,B, E, and F shows small variations of collimator Z position (± 1.0 mm). Frames C, D, H, and G show effects of larger collimator misalignments up to ± 5.0 mm.

we conclude, that x and y collimator misalignment has negligible effect on the uncertainty of the $T_{NP}$ measurements.

The positioning of the collimator in z (axial) direction is more problematic. Figures S8 and S9 show the effects of changing the z position of the nIR collimator, with all other optics left in their optimal positions. Figure S8 shows raw and normalized NP and TC spectra for z position changes of up to ±1 mm, relative to the optimal z position. Because long focal length lenses are used to couple the nIR light into the collimator, the intensity varies much more slowly with z than the x and y variations shown in Figure S6E and S6F, *i.e.*, it is not as straightforward to identify the optimal position. To find the optimal z position, the collimator was translated over a ± 11.5 mm range, measuring the fall off of intensity in both directions, and choosing the mid-point as the optimal z position. For collimator displacements up to ±1 mm (Figure S8A) the changes in raw NP spectra are small, and the normalized spectra (S8B) are essentially superimposable. Certainly any differences between spectra are well within the variation expected due to real variations in the NP temperature over the experiment time. For larger ±5 mm displacements of the collimator (S8C), there are larger changes in the spectral intensity, and after normalization (S8D), it can be seen that there are net shifts to the red or blue for large positive and negative z displacements, respectively. If the TC spectra had similar red and blue shifts, there would be no problem, however, as shown in the analogous TC spectra in Figures S8E-H, the TC spectra have smaller red and blue shifts. This difference reflects the larger size of the TC emitter (~280 μm) compared to the NP effective emitter size (~80 μm), which means that chromatic effects tend to average out more in the TC spectra than in those for the NPs.

The effects of z displacements on corrected NP spectra, and the $T_{NP}$ values extracted by fitting them, are summarized in Figure S9. Note that for these experiments, relatively large NPs were used to give good signal, while keeping $T_{NP}$ low enough to minimize sublimation and other changes to the NP.

Figure S9A shows the corrected spectra for a single NP with $T_{NP} \approx 980$ K, also including the visible portion of the spectra because this is used in the fitting process.

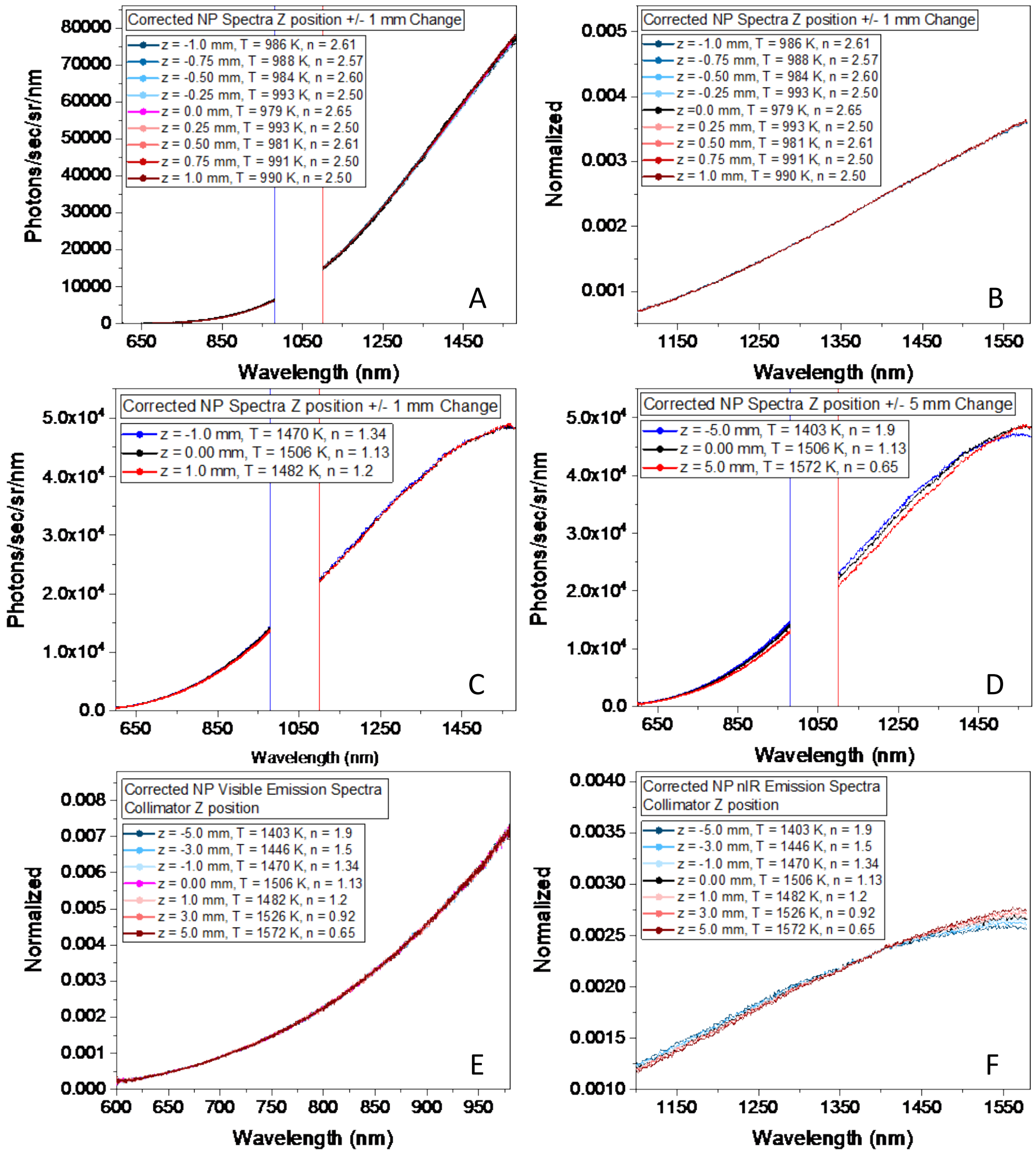

**Figure S9.** Corrected spectra for different NPs measured with different nIR collimator positions. A.) Effects of ± 1.0 mm changes in the Z position. B.) The nIR spectra, normalized to show shape changes. C.) Analogous unnormalized spectra for a different NP at higher $T_{NP}$. D.) Effects of ± 5 mm changes in the nIR collimator position. E.) Visible spectra showing that, as expected, moving the nIR collimator has no effect. F.) nIR spectra, normalized, as a function of± 5 mm misalignment of the nIR collimator.

For z displacements up to ±1 mm, $T_{NP}$ varies by ≤ 14 K (≤ 1.4 %), however, $T_{NP}$ fluctuates randomly rather than varying systematically with z. This is a sign that the $T_{NP}$ fluctuations were not due to the z displacement, but rather reflect real $T_{NP}$ variations due to laser intensity fluctuations. Figure S9B shows that the normalized nIR spectra are essentially superimposable. Figure S9C shows corrected spectra for a different NP at higher $T_{NP}$, also for ±1 mm z displacements. Again, $T_{NP}$ varies randomly with z, by about 2% in this case. The fact that $T_{NP}$ shows no systematic variation with z over this ±1 mm range is not surprising, given that the raw spectra (Figure S8B) are nearly superimposable.

Figure S9D shows the effects of ±5 mm z displacements – a range where significant, systematic red shifts were observed with increasing z in the spectra in Figure S8D. As expected, these spectral shifts result in systematic increases in the extracted $T_{NP}$ values with increasing z. Figure S9E and S9F show the visible and nIR spectra from Figure S9D, normalized to allow shape changes to be observed. As expected, there are no changes in the visible spectral shapes, but the nIR region shows significant red shifting with increasing z.

Similar axial displacement measurements were done moving the visible collimator relative to the beamsplitter assembly by up to ±8 mm in the x direction, but no effects were observed on the shape of the visible part of the spectra. Indeed, the visible spectra are generally much less affected by misalignments of the optics or of the TC position, compared to the nIR spectra. In part, this probably reflects the fact that the achromatic doublets used in the collection system are optimized for the 750-1550 nm range, which is part of our "visible" spectrum. In addition, however, the 2D CCD array used to record the visible spectra is 3.84 mm tall, whereas the 1D InGaAs array used for the nIR spectra is only 0.5 mm tall. The nIR camera has a cylindrical lens to vertically compress the dispersed light onto the array, but some light still misses the array. As a result, the visible detection efficiency is less sensitive to the divergence of light exiting the fiber, compared to that for the nIR spectrograph, and therefore, less sensitive to optical misalignments that affect how light couples into the fibers.

For x and y displacements of the nIR collimator, the intensity falls off rapidly – much faster than chromatic effects grow in. Therefore, the correct x and y positions are well defined, and we are able to position the collimator accurately enough that misalignment in these directions has a negligible effect on the uncertainty in $T_{NP}$. The same is not true for the z positioning, because the intensity maximum along this direction is much broader, and significant chromatic effects appear for ± z collimator displacements for which the intensity is only modestly affected. The method used to find the optimal z position for collimator is estimated to give the position to within ±0.5 mm, but for estimating the resulting effect on $T_{NP}$ we assume that the position is uncertain by ±2 mm. From data analogous to that in Figure S9C-D, but for ±2 mm z motion of the collimator, we find that $T_{NP}$ varied by a total of 86 K, corresponding to 5.6%. Therefore, we estimate this contribution to the uncertainty as ~±2.8 %.

**Effects of varying the visible scale factor on fit quality and $T_{NP}$:**

As discussed above, the intensity of the visible part of the spectrum must be scaled to match smoothly to the nIR part, and Figure S10 explores the effect of changing the scale factor, f, away from the best-fit value. In Figures S10A-C the spectrum is plotted with the best fit value of f (1.53) in blue, and the red and orange spectra show the effects of increasing or decreasing f by ±2%, ±5%, and ±10%. The two insets show more clearly how changing f affects both the visible spectra (the nIR is not scaled) and the resulting fits to the entire spectrum.

Consider the effect of changing f by ±10% (C). The visible inset shows how the change in scale factor shifts the spectra to increased or decreased intensity. Such a large change in f leads to gross mismatching of the visible and nIR spectra, as can be seen by the resulting fits. Note that when f is increased by 10%, shifting the visible spectrum to higher intensity, the best fit is well below the experimental spectrum in the visible, and well above it in the nIR, *i.e.*, it essentially splits the differences between the now-mismatched visible and nIR spectra. Similarly, if the visible spectrum is scaled down by 10%, the fit is obviously too high in the visible. The same effect is clear when f is changed by ±5% (Figure S10B). Clearly such large variations in f can be excluded on basis of poor fitting.

On the other hand, when f is changed by only ±2% (Figure S10A), the fits now are within the scatter in the experimental spectra, *i.e.* the precision of the process used to determine the best-fit f value is ~±2%. (Note that to keep the spectral fits in the visible inset from overlapping, they have been offset, only for the ±2% case). If we take the variation in $T_{NP}$ for the ±2% change in f as an estimate of this contribution to the uncertainty in $T_{NP}$, the result is only ±24 K or ±1.9%.

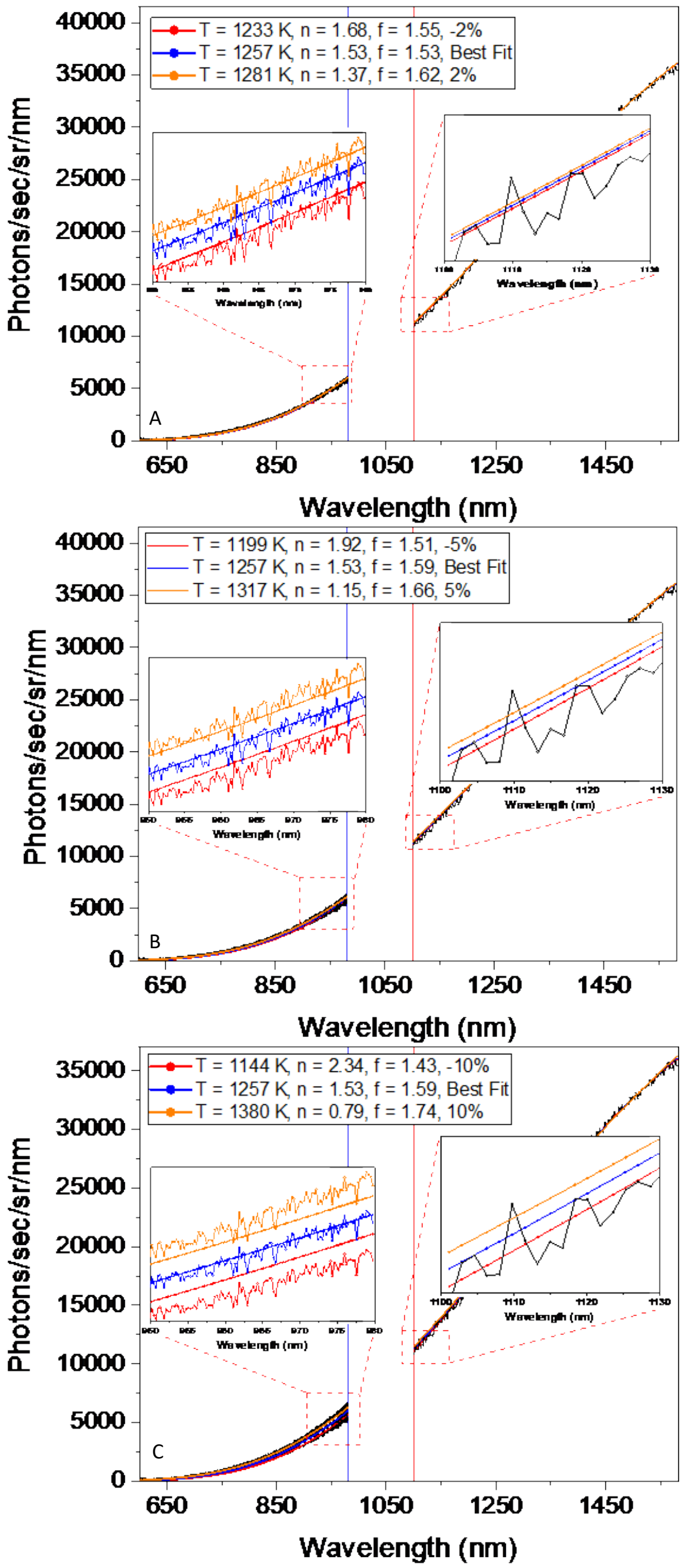

**Figures S10.** Images A-C show the effect of varying the best fit visible scale factor, f, by ±2%, ±5%, and ±10% from its best-fit value, determined as described above. Note that in the visible inset in A, the spectra and fits have been offset for clarity.

**Effects of spectral integration time:**

To test for possible saturation effects in the visible and nIR cameras, Figure S11 shows spectra of a single graphite NP with M = 14.2 MDa, with the laser stabilized to have power, measured at the vacuum exit window, of 19.6 mW ± 0.28 mW. In Figure S11A, the spectra are plotted in terms of photons/sec/sr/nm of spectral bandwidth, and have been offset vertically so that the individual spectra can be seen clearly. In Figure S11B, the spectra are plotted without the offset so that their shapes can be compared.

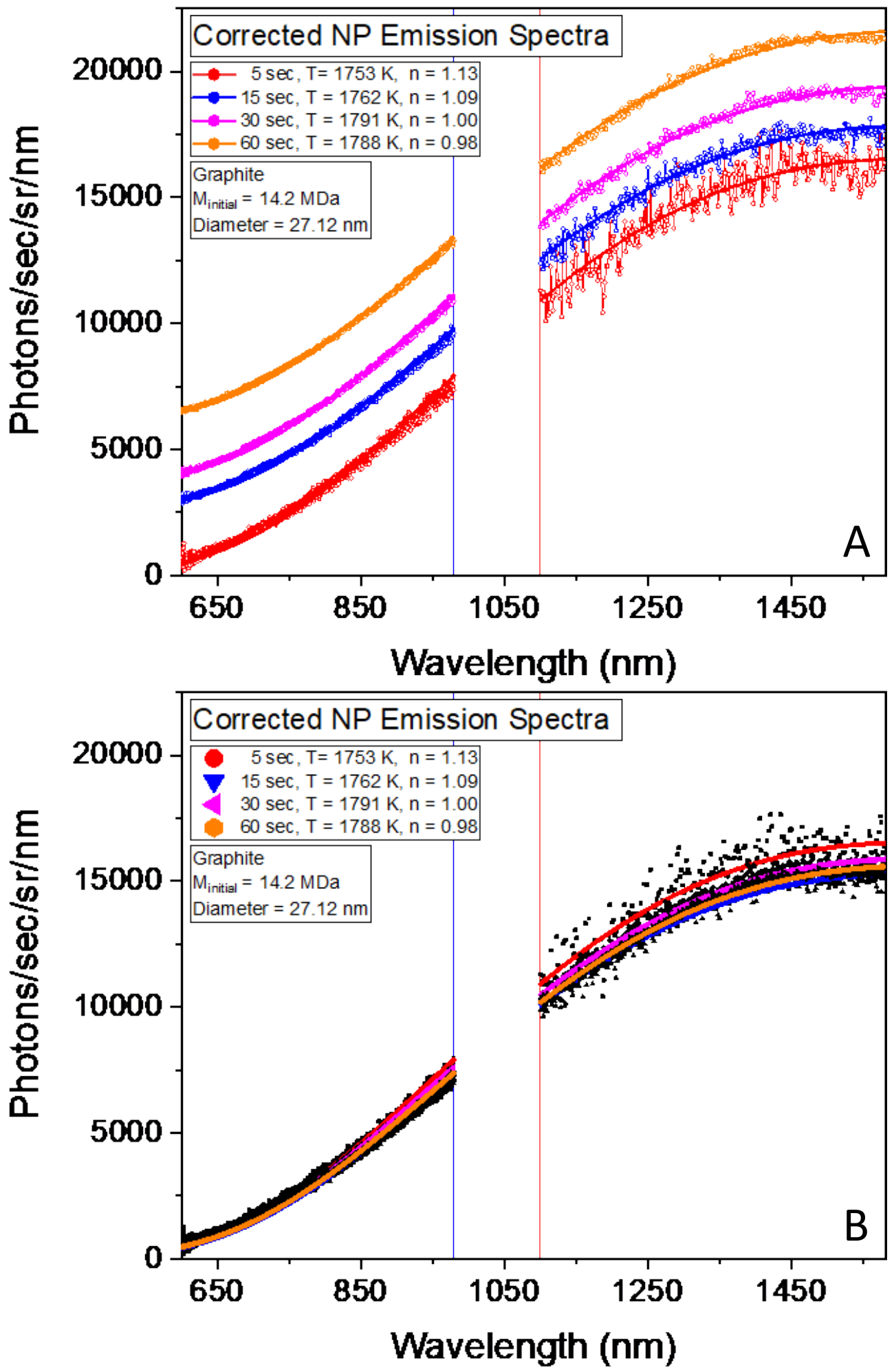

Figure S11. Emission spectra taken at different integration times of, 5, 15, 30, and 60 seconds for a single graphite NP with $M_{initial}$ = 14.2 MDa. In image A, the 15, 30, and 60 second integrations times are offset to show how the noise varies between spectra. Image B has no offset, and it showing that the overall intensity is similar between spectra.

As expected the short acquisition time spectra are relatively noisy, and the 5 second spectrum has slightly higher intensity. The 30 and 60 second spectra are essentially identical, and the $T_{NP}$ and n values extracted from fitting them also have negligible differences. There may be a slight decrease in the apparent $T_{NP}$ and increase in n for the shorter time spectra, however, the variation in $T_{NP}$ is certainly well within the fitting error for such noisy spectra.

**Summary of the effects of optical system alignment and fitting on the uncertainties in $T_{NP}$.**

As described above, the significant experimental contributions to uncertainties in the spectral shapes, hence $T_{NP}$, are: 1. Positioning of the TC during calibration (±3.2 to 3.5%), 2. The z position of the nIR collimator (±2.8 %), 3. The visible spectrum scale factor (± 1.9%). The positions of other optics are well enough defined and controlled that they do not contribute uncertainties of these magnitudes, and thus their contributions are considered negligible. Therefore, the total contribution to uncertainty in $T_{NP}$ from these experimental issues is estimated to be ±4.8 %.

In addition, given typical signal/noise level in NP spectra, we estimate ~±4 % uncertainty in $T_{NP}$ from fitting, assuming a fitting function consisting of Planck's law times the power law model for $\epsilon(\lambda)$. Therefore, the total estimated uncertainty in $T_{NP}$ including both experimental and fitting uncertainties is ±6.2 %. As emphasized in the main paper, there is additional error associated with our choice of the power law function to model $\epsilon(\lambda)$, however, because the correct form of $\epsilon(\lambda)$ is unknown, it is not possible to estimate this error. We simply note that for many of the carbon materials, the fits to the NP spectra based on the power law $\epsilon(\lambda)$ model are quite good, suggesting that the error is relatively small.

**Effects of varying n on extracted $T_{NP}$ values:**

One issue with using the power law model, however, is that the $T_{NP}$ and n parameters both affect the curvature of the emission vs. λ, and while the effects are different, they are not orthogonal. Increasing the n parameter causes the fit function to fall off more rapidly at long wavelengths, and in fitting a spectrum, this fall off is compensated by decreasing $T_{NP}$ in Planck's law, which increases the intensity at long wavelengths relative to short wavelengths. The result is a fitting function with more curvature, *i.e.*, higher at mid-wavelengths, and falling off more rapidly at both long and short wavelengths, compared to the function with best-fit n value. Conversely, decreasing n is compensated by increasing $T_{NP}$, which results in a fitting function with more gradual wavelength dependence.

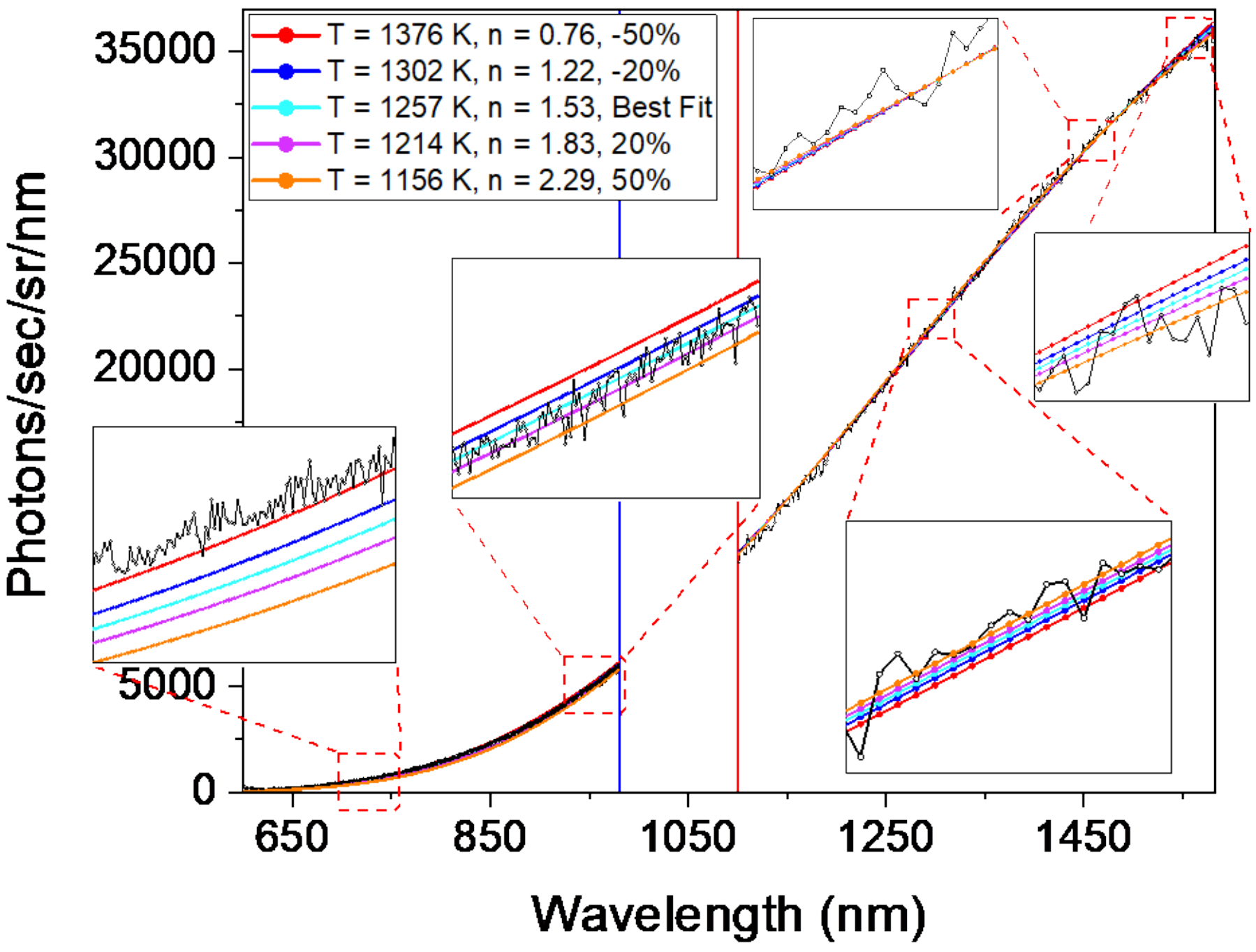

Figure S12. Effects of varying n, relative to its best fit value, on the fits and extracted $T_{NP}$ values for a typical NP spectrum.

To illustrate this behavior, and estimate the resulting uncertainty in $T_{NP}$, Figure S12 shows how forcing the n parameter away from its best fit value affects the fit quality and extracted $T_{NP}$ values.

The experimental spectrum being fit is for a carbon black NP of 85 MDa size, and the acquisition time was 30 seconds. The spectrum was first fit allowing n and $T_{NP}$ (and the normalization constant) to vary, obtaining best fit values of $T_{NP}$ = 1257 K, and n= 1.53. We then varied the n parameter by up to ±50%, refitting the spectrum allowing only $T_{NP}$ (and the normalization) to vary. For clarity, the figure shows only the results for ±20% and ±50% variations, and to help show the effects more clearly, insets with magnified intensity scales are shown for five regions across the spectrum.

Because the power law model is only an approximation to the true emissivity function, even the best-fit function over- or under-shoots the experimental data in some spectral regions. Therefore, forcing n away from its best-fit value may improve the fit in some spectral regions, even though the overall fit is worsened. For example, using n = 2.29 improves the fit in the wavelength region near 1550 nm, however, such a high n value substantially worsens the fit over the entire range below 1000 nm. Similarly, using n = 0.76 gives the best fit in the region around 700 nm, but substantially worsens the fit for most of the rest of the spectrum. Similar effects are observed for the ±20% variations in n, but in that case, the deviations are on the same order as the estimated uncertainty in the relative spectral intensities (*i.e.*, spectral shape). The temperature variations associated with ±20% n variations are roughly ±45 K (~±4%), and we take this as a rough estimate of the error associated with fitting any particular spectrum, *within the power law model*.

**NP spectra for different materials compared, before and after >1900 K heating.**

Figures S13 and S14 replot the spectra in figures 4A and for NP1 in Figure 5A and 5C and Figure 5D in the main text, to allow comparison between materials. In Figure S13, the volume-scaled intensities for NPs prior to >1900 K heating are shown on the left, and the post-heating spectra are on the right. In Figure S14, the data are replotted, normalized at 1580 nm, to allow spectral shapes to be compared.

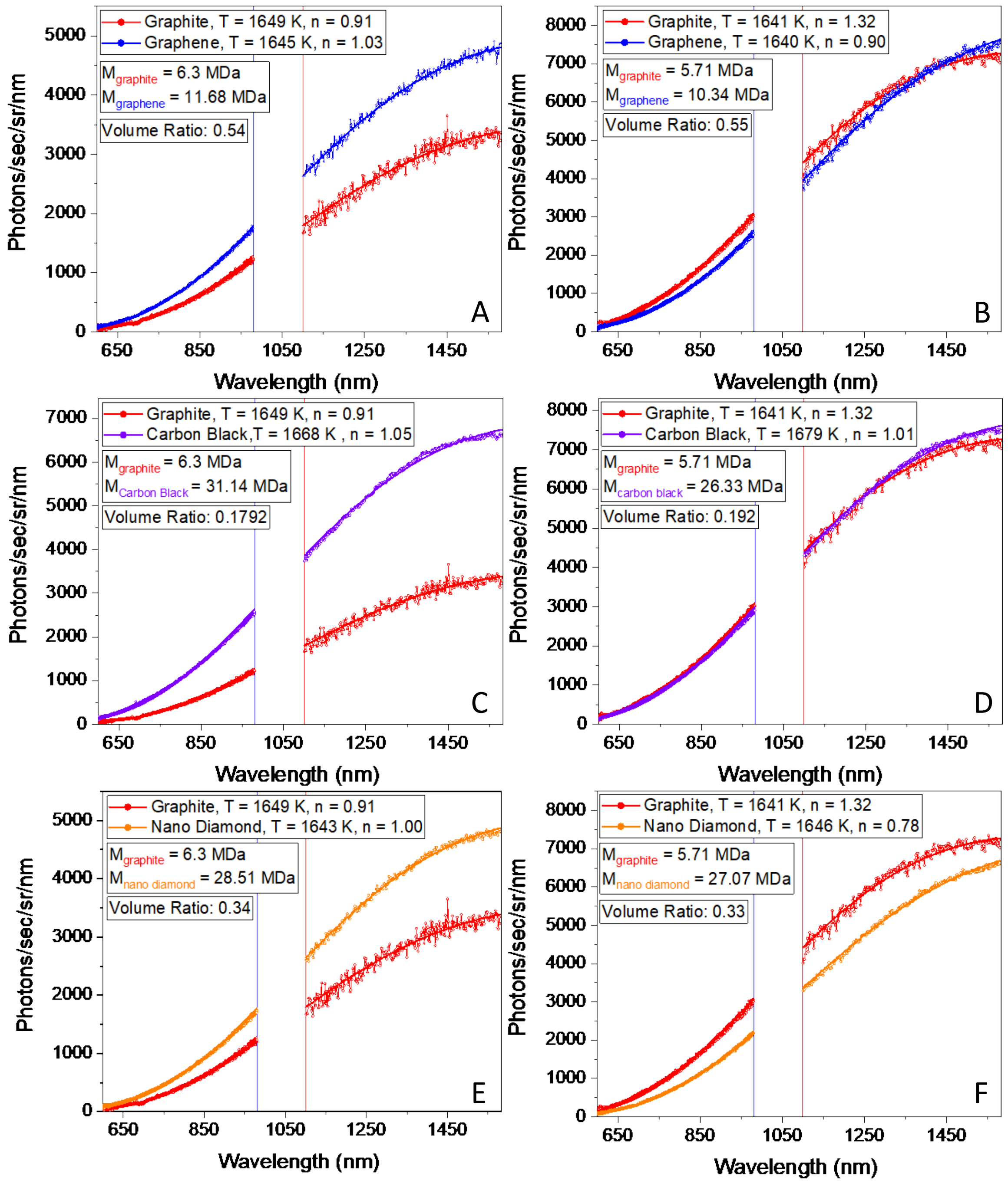

Figure S13. Emission spectra comparing the carbon NPs in figures 4 and 5, before and after heating. All spectra are volume scaled to allow direct comparison of intensities. The left column has spectra prior to heating above 1900 K. The right column shows spectra collected after >1900 K heating.

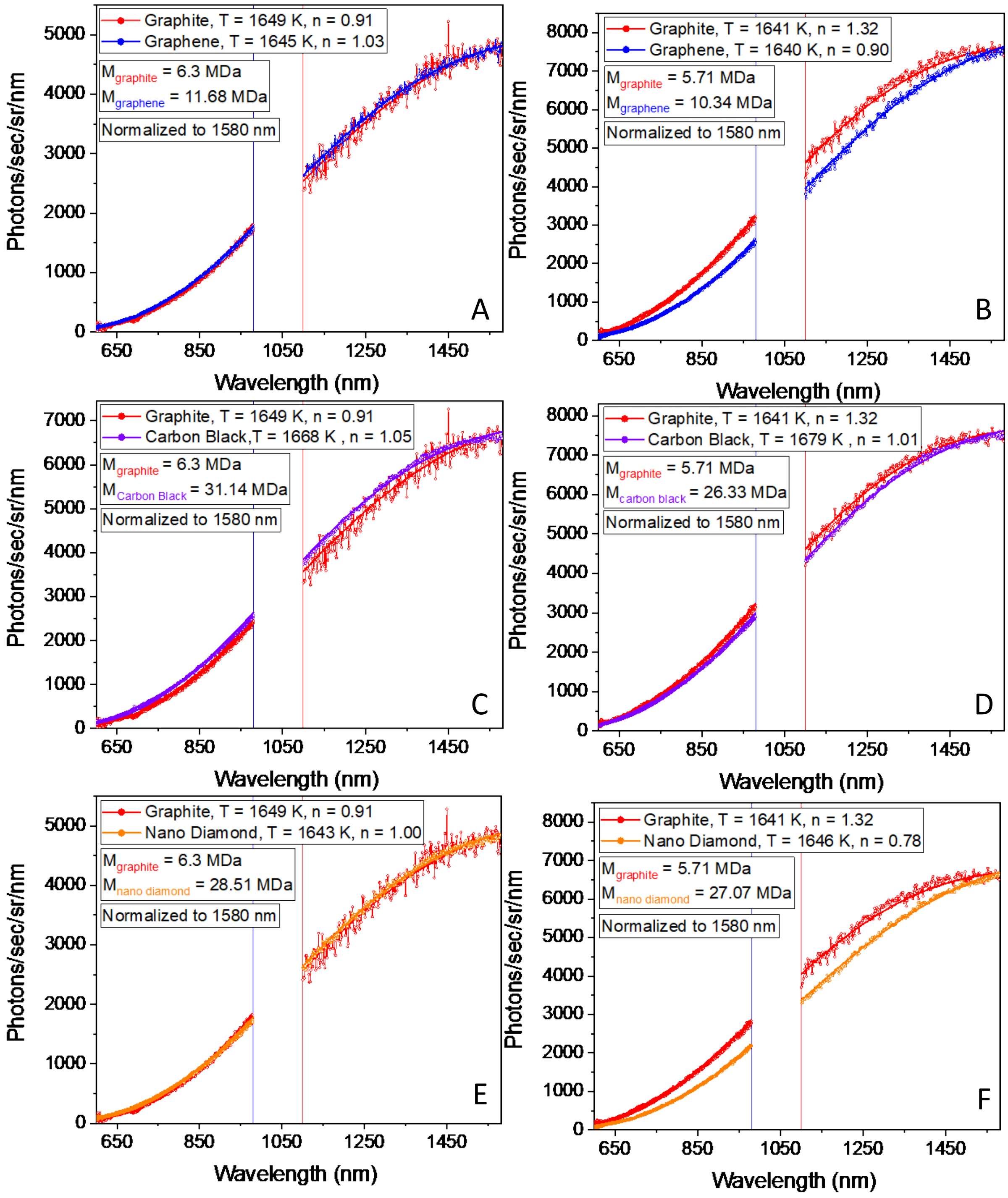

Figure S14. Emission spectra comparing the carbon NPs in figures 4 and 5, before and after heating. All spectra are normalized at 1580 nm to allow spectral shape comparison. The left column has spectra prior to heating above 1900 K. The right column shows spectra collected after >1900 K heating.

**Effects of measuring $T_{NP}$ while NP motion is being driven to measure the NP mass.**

Figure S15 addresses the question of whether $T_{NP}$ is significantly affected by the enhanced axial NP motion that occurs when the NP is being driven to measure mass, due to possible changes in chromatic effects as the NP position in the trap varies. In the figure, $T_{NP}$ is shown vs. time for a single graphite NP with initial mass of 23.22 MDa as the NP was heated with 12.2 mW ±0.05 mW of 532 nm laser power, giving $T_{NP}$ near 1500 K. The scatter in the $T_{NP}$ values partly reflects signal/noise in the spectra, taken at relatively low $T_{NP}$ to minimize sublimation, but in addition, the normal $T_{NP}$ stabilization program was not being used.

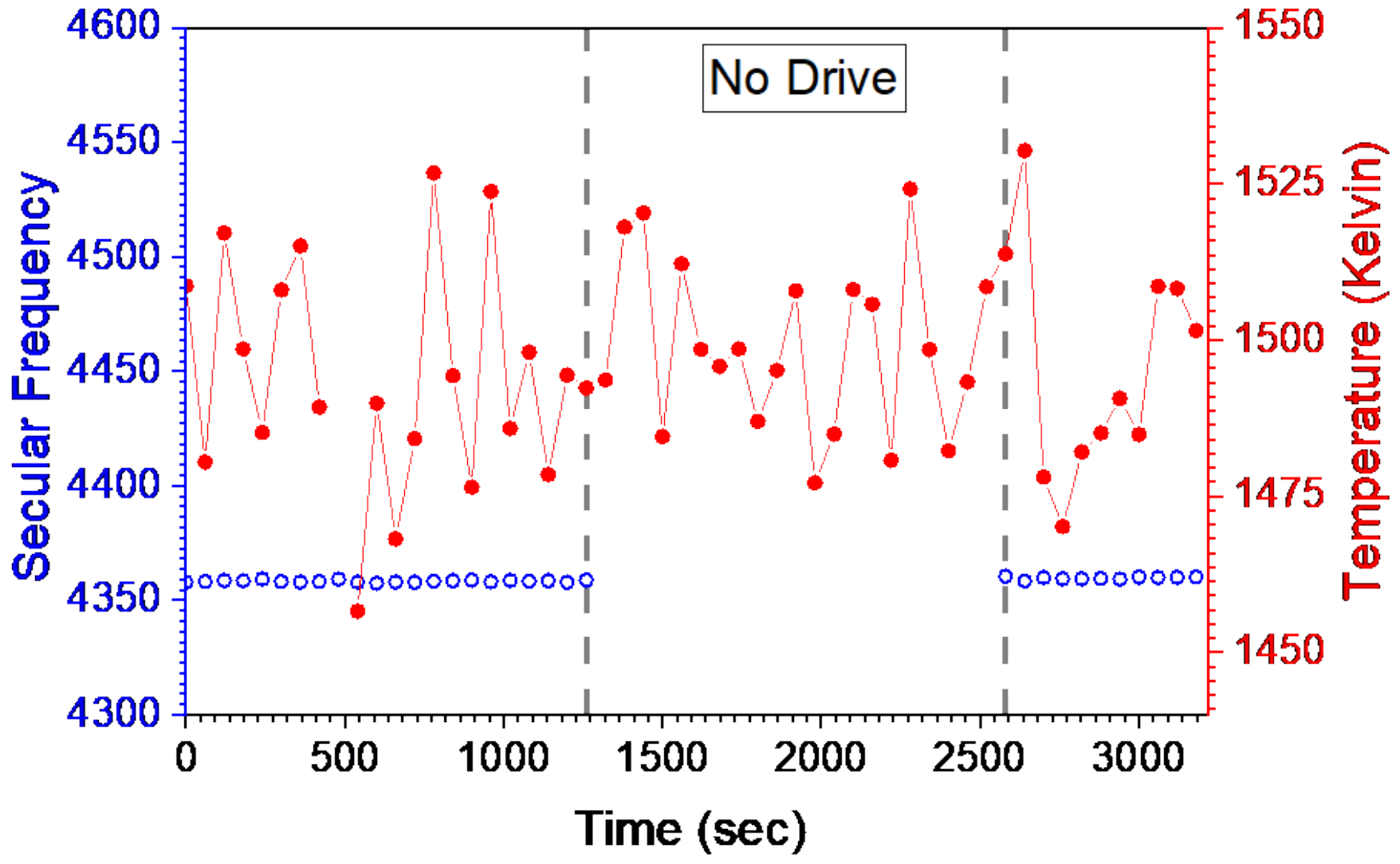

Figure S15. Single Graphite NP with initial mass of 23.22 MDa. In this experiment, the laser was stabilized by using the $T_{NP}$ for slow fluctuations and a PID for fast fluctuations. The drive frequency was turned off between ~1320-2580 seconds showing negligible changes in $T_{NP}$.

During the first ~1300 seconds, $T_{NP}$ was being measured during the AC drive frequency scan used to measure the secular frequency, hence mass. The extracted $T_{NP}$ was 1500 ± 19 K. At 1320 seconds the AC drive potential was turned off for ~1200 seconds, continuing the spectral measurements. During this "no drive" time, $T_{NP}$ was 1502 ± 14 K. At 2580 seconds the drive potential was turned back on, and eleven additional simultaneous frequency and $T_{NP}$ measurements were taken, giving $T_{NP}$ = 1498 K ± 19 K. Clearly, any effect of the AC drive potential on the spectra and extracted $T_{NP}$ values is negligible compared to the uncertainty in $T_{NP}$.

| Figure 1 | T = 1296 K, n = 2.03, 540 sec | T = 1296 K, n = 2.03, 540 sec | T = 1648 K, n = 1.27, 7740 sec | T = 1648 K, n = 1.27, 7740 sec | T = 1845 K, n = 1.06, 10320 sec | T = 1845 K, n = 1.06, 10320 sec | T = 2010 K, n = 0.82, 13980 sec | T = 2010 K, n = 0.82, 13980 sec | T = 2150 K, n = 0.65, 15060 sec | T = 2150 K, n = 0.65, 15060 sec |
|---|---|---|---|---|---|---|---|---|---|---|
| 600.015 | 22.1195 | 38.209 | 645.333 | 474.472 | 1783.07 | 1707.25 | 4524.99 | 4304.62 | 7447.01 | 7454.98 |
| 600.306 | 109.817 | 38.4397 | 640.777 | 476.604 | 1855.8 | 1713.81 | 4540.59 | 4319.43 | 7470.41 | 7478.44 |
| 600.597 | 53.2226 | 38.6716 | 719.694 | 478.743 | 1881.53 | 1720.39 | 4412.72 | 4334.27 | 7498.8 | 7501.94 |
| 600.888 | 121.514 | 38.9046 | 583.186 | 480.889 | 1898.98 | 1726.98 | 4830.55 | 4349.14 | 7756.67 | 7525.48 |
| 601.179 | 32.9083 | 39.1386 | 609.586 | 483.041 | 2045.51 | 1733.6 | 4903.62 | 4364.05 | 7904.78 | 7549.06 |
| 601.47 | 97.3672 | 39.3738 | 641.282 | 485.2 | 1912.3 | 1740.22 | 4801.04 | 4378.98 | 7821.01 | 7572.69 |
| 601.76 | 74.5391 | 39.6102 | 581.884 | 487.367 | 1944.28 | 1746.87 | 4621.81 | 4393.95 | 7845.33 | 7596.35 |
| 602.052 | 82.5042 | 39.8476 | 592.211 | 489.54 | 1923.1 | 1753.53 | 4532.04 | 4408.94 | 7574.43 | 7620.06 |
| 602.342 | 121.921 | 40.0862 | 574.021 | 491.72 | 1845.25 | 1760.21 | 4651.8 | 4423.97 | 7954.15 | 7643.81 |
| 602.633 | 30.7162 | 40.3259 | 550.013 | 493.907 | 1949.02 | 1766.9 | 4820.77 | 4439.03 | 7862.21 | 7667.61 |
| 602.924 | 60.0883 | 40.5668 | 640.01 | 496.101 | 1993.92 | 1773.61 | 4726.14 | 4454.12 | 8025.06 | 7691.44 |
| 603.215 | 69.3919 | 40.8087 | 679.424 | 498.302 | 1983.32 | 1780.34 | 4678.2 | 4469.24 | 7829.35 | 7715.32 |
| 603.506 | 88.236 | 41.0519 | 635.624 | 500.51 | 1971.01 | 1787.08 | 4754.04 | 4484.4 | 7856.2 | 7739.24 |
| 603.797 | 9.50572 | 41.2961 | 625.089 | 502.725 | 1975.66 | 1793.85 | 4743.16 | 4499.58 | 7759.66 | 7763.2 |
| 604.088 | 94.3437 | 41.5416 | 725.254 | 504.946 | 2043.3 | 1800.62 | 4758.61 | 4514.8 | 7625.28 | 7787.2 |
| 604.379 | 100.568 | 41.7881 | 652.011 | 507.175 | 1953.45 | 1807.42 | 4809.32 | 4530.04 | 8032.4 | 7811.24 |
| 604.669 | 27.3871 | 42.0359 | 600.319 | 509.411 | 2234.29 | 1814.23 | 4851.64 | 4545.32 | 7862.68 | 7835.33 |
| 604.96 | 53.5358 | 42.2848 | 661.122 | 511.654 | 1802.48 | 1821.06 | 4799.88 | 4560.63 | 7828.93 | 7859.46 |
| 605.251 | 133.289 | 42.5348 | 641.942 | 513.903 | 2019.45 | 1827.9 | 4866.7 | 4575.97 | 7786.47 | 7883.63 |
| 605.542 | 87.6995 | 42.7861 | 747.281 | 516.16 | 2010.13 | 1834.76 | 4850.65 | 4591.34 | 8149.99 | 7907.84 |
| 605.832 | 76.8268 | 43.0385 | 703.654 | 518.424 | 1898.54 | 1841.64 | 4980.13 | 4606.74 | 8119.86 | 7932.09 |
| 606.123 | 52.6426 | 43.2921 | 633.84 | 520.694 | 2019.52 | 1848.53 | 4957.22 | 4622.17 | 8365.97 | 7956.38 |
| 606.414 | 69.3261 | 43.5468 | 690.247 | 522.972 | 2053.58 | 1855.44 | 5046.27 | 4637.64 | 8073.15 | 7980.72 |
| 606.705 | 32.7087 | 43.8028 | 749.785 | 525.257 | 2009.27 | 1862.37 | 4779.45 | 4653.13 | 7919.67 | 8005.1 |
| 606.995 | 40.5757 | 44.0599 | 653.904 | 527.549 | 2041.32 | 1869.32 | 4795.9 | 4668.66 | 8187.08 | 8029.52 |
| 607.286 | 32.6726 | 44.3182 | 711.133 | 529.848 | 2205.38 | 1876.28 | 5048.26 | 4684.22 | 8184.06 | 8053.98 |
| 607.577 | 64.5955 | 44.5777 | 650.624 | 532.154 | 2187.92 | 1883.26 | 4964.12 | 4699.81 | 8360.13 | 8078.48 |
| 607.868 | 109.849 | 44.8384 | 639.515 | 534.467 | 2110.88 | 1890.26 | 5010.75 | 4715.43 | 8096.86 | 8103.02 |
| 608.158 | 47.3482 | 45.1003 | 701.652 | 536.788 | 2023.35 | 1897.27 | 4996.77 | 4731.08 | 8434.44 | 8127.61 |
| 608.449 | 98.6004 | 45.3635 | 730.58 | 539.115 | 1959.44 | 1904.3 | 4983.17 | 4746.76 | 8001.68 | 8152.23 |
| 608.74 | 92.1263 | 45.6278 | 618.615 | 541.449 | 2065.36 | 1911.34 | 5168.62 | 4762.47 | 8338.31 | 8176.9 |
| 609.03 | 0 | 45.8934 | 630.688 | 543.791 | 2122.13 | 1918.4 | 4989.26 | 4778.22 | 8187.76 | 8201.61 |
| 609.321 | 75.6907 | 46.1601 | 687.017 | 546.14 | 2117.14 | 1925.48 | 5390.76 | 4793.99 | 8561.23 | 8226.36 |
| 609.612 | 79.2879 | 46.4281 | 687.623 | 548.496 | 2114.15 | 1932.58 | 5203.81 | 4809.8 | 8443.7 | 8251.16 |
| 609.902 | 42.6906 | 46.6973 | 738.07 | 550.859 | 2093.81 | 1939.69 | 5098.24 | 4825.64 | 8096.97 | 8275.99 |
| 610.193 | 84.9278 | 46.9678 | 662.047 | 553.229 | 2117.06 | 1946.83 | 5071.17 | 4841.51 | 8380.38 | 8300.87 |
| 610.484 | 79.0783 | 47.2394 | 759.046 | 555.607 | 2290.09 | 1953.97 | 5470.18 | 4857.41 | 8670.03 | 8325.78 |
| 610.774 | 84.0812 | 47.5124 | 648.958 | 557.992 | 2238.87 | 1961.14 | 5081.29 | 4873.34 | 8506.89 | 8350.74 |
| 611.065 | 55.9232 | 47.7865 | 705.712 | 560.383 | 2335.47 | 1968.32 | 5334.24 | 4889.31 | 8688.52 | 8375.74 |
| 611.355 | 33.9571 | 48.0619 | 735.944 | 562.783 | 2090.9 | 1975.52 | 5219.15 | 4905.3 | 8511.66 | 8400.78 |
| 611.646 | 66.0772 | 48.3386 | 667.079 | 565.189 | 2220.46 | 1982.73 | 5235.33 | 4921.32 | 8487.15 | 8425.86 |
| 611.937 | 81.2142 | 48.6165 | 739.659 | 567.602 | 2228.25 | 1989.96 | 5442.78 | 4937.38 | 8802.06 | 8450.99 |
| 612.227 | 70.2212 | 48.8956 | 685.855 | 570.023 | 2256.79 | 1997.21 | 5126.97 | 4953.47 | 8538.04 | 8476.15 |
| 612.518 | 56.5436 | 49.176 | 738.998 | 572.451 | 2239.35 | 2004.48 | 5106.96 | 4969.59 | 8598.37 | 8501.35 |
| 612.808 | 43.9985 | 49.4577 | 716.051 | 574.887 | 2173.22 | 2011.76 | 5395.04 | 4985.74 | 8689.28 | 8526.6 |
| 613.099 | 88.0617 | 49.7407 | 699.099 | 577.329 | 2272.56 | 2019.06 | 5392.19 | 5001.92 | 8746.43 | 8551.89 |
| 613.389 | 61.2262 | 50.0249 | 765.543 | 579.779 | 2253.67 | 2026.38 | 5175.71 | 5018.13 | 8873.54 | 8577.22 |
| 613.68 | 101.146 | 50.3104 | 793.526 | 582.236 | 2161.16 | 2033.71 | 5164.83 | 5034.37 | 8753.37 | 8602.59 |
| 613.97 | 101.996 | 50.5971 | 711.286 | 584.701 | 2278.38 | 2041.06 | 5448.52 | 5050.65 | 8944.55 | 8628 |
| 614.26 | 58.1616 | 50.8852 | 721.839 | 587.172 | 2400.71 | 2048.43 | 5306.96 | 5066.95 | 8811.64 | 8653.45 |
| 614.551 | 129.565 | 51.1745 | 709.102 | 589.652 | 2373.72 | 2055.82 | 5399.22 | 5083.29 | 8983.61 | 8678.94 |
| 614.841 | 85.2667 | 51.4652 | 742.345 | 592.138 | 2296.66 | 2063.22 | 5377.51 | 5099.65 | 8769.38 | 8704.48 |
| 615.132 | 85.1058 | 51.7571 | 756.71 | 594.632 | 2308.85 | 2070.64 | 5422.63 | 5116.05 | 8981.79 | 8730.05 |
| 615.422 | 84.34 | 52.0503 | 760.315 | 597.133 | 2315.38 | 2078.07 | 5568.58 | 5132.48 | 9004.69 | 8755.67 |
| 615.713 | 90.7827 | 52.3449 | 751.79 | 599.641 | 2396.13 | 2085.52 | 5213.65 | 5148.94 | 8619.99 | 8781.32 |
| 616.003 | 79.3019 | 52.6407 | 798.14 | 602.157 | 2161.67 | 2092.99 | 5350.39 | 5165.44 | 8692.34 | 8807.02 |
| 616.293 | 53.2549 | 52.9378 | 799.05 | 604.68 | 2389.55 | 2100.48 | 5464.49 | 5181.96 | 9069.12 | 8832.76 |
| 616.584 | 62.6527 | 53.2363 | 759.201 | 607.211 | 2357.9 | 2107.99 | 5492.96 | 5198.51 | 9095.94 | 8858.54 |
| 616.874 | 71.147 | 53.5361 | 706.022 | 609.749 | 2334.55 | 2115.51 | 5396.09 | 5215.1 | 9011.75 | 8884.36 |
| 617.165 | 91.5171 | 53.8372 | 791.112 | 612.294 | 2320.82 | 2123.05 | 5745.29 | 5231.72 | 8974.9 | 8910.22 |
| 617.455 | 79.0576 | 54.1396 | 704.764 | 614.847 | 2322.65 | 2130.6 | 5481.88 | 5248.36 | 8858.88 | 8936.12 |
| 617.745 | 121.256 | 54.4433 | 745.709 | 617.407 | 2383.57 | 2138.17 | 5333.81 | 5265.04 | 8942.01 | 8962.06 |
| 618.036 | 102.669 | 54.7484 | 751.669 | 619.975 | 2274.42 | 2145.76 | 5672.78 | 5281.75 | 9193.87 | 8988.05 |
| 618.326 | 95.3117 | 55.0548 | 750.351 | 622.55 | 2332.71 | 2153.37 | 5565.69 | 5298.49 | 9104.66 | 9014.07 |
| 618.616 | 113.305 | 55.3626 | 743.338 | 625.132 | 2379.69 | 2160.99 | 5738.52 | 5315.26 | 9117.55 | 9040.14 |
| 618.907 | 91.8439 | 55.6717 | 765.582 | 627.722 | 2350.42 | 2168.63 | 5432.78 | 5332.06 | 9096.37 | 9066.24 |
| 619.197 | 81.8291 | 55.9821 | 804.207 | 630.32 | 2340.05 | 2176.29 | 5599.63 | 5348.9 | 8905.4 | 9092.39 |
| 619.487 | 93.4493 | 56.294 | 816.058 | 632.925 | 2359.74 | 2183.97 | 5596.86 | 5365.76 | 9127.62 | 9118.57 |
| 619.777 | 85.0158 | 56.6071 | 745.963 | 635.537 | 2368.41 | 2191.66 | 5681.01 | 5382.66 | 9312.11 | 9144.8 |
| 620.067 | 68.9526 | 56.9216 | 726.442 | 638.157 | 2398.28 | 2199.37 | 5452.89 | 5399.58 | 9107.89 | 9171.07 |
| 620.358 | 89.402 | 57.2375 | 741.284 | 640.784 | 2253.4 | 2207.09 | 5589.46 | 5416.54 | 9120.07 | 9197.37 |
| 620.648 | 92.5336 | 57.5548 | 775.226 | 643.419 | 2393.8 | 2214.84 | 5526.84 | 5433.53 | 9180.19 | 9223.72 |
| 620.938 | 86.7861 | 57.8734 | 765.525 | 646.062 | 2416.25 | 2222.6 | 5622.73 | 5450.55 | 9048.58 | 9250.11 |
| 621.228 | 85.3712 | 58.1934 | 790.697 | 648.712 | 2392.65 | 2230.38 | 5645.47 | 5467.6 | 9135.42 | 9276.54 |
| 621.519 | 80.9369 | 58.5148 | 757.124 | 651.369 | 2432.89 | 2238.17 | 5720.6 | 5484.68 | 9285.26 | 9303.01 |
| 621.809 | 55.2952 | 58.8376 | 815.761 | 654.034 | 2432 | 2245.99 | 5638.79 | 5501.79 | 9135.13 | 9329.51 |
| 622.099 | 99.2202 | 59.1617 | 827.068 | 656.707 | 2524.73 | 2253.82 | 5868.03 | 5518.94 | 9330.43 | 9356.07 |
| 622.389 | 69.1845 | 59.4873 | 758.254 | 659.387 | 2445.07 | 2261.66 | 5607.99 | 5536.11 | 9111.37 | 9382.66 |
| 622.679 | 101.322 | 59.8143 | 761.572 | 662.074 | 2482.84 | 2269.53 | 5808.25 | 5553.32 | 9543.53 | 9409.28 |
| 622.969 | 101.513 | 60.1426 | 783.503 | 664.77 | 2519.55 | 2277.41 | 5611.84 | 5570.55 | 9393.84 | 9435.95 |
| 623.259 | 115.834 | 60.4723 | 809.823 | 667.473 | 2500.4 | 2285.3 | 5767.35 | 5587.82 | 9503.98 | 9462.67 |
| 623.549 | 81.9693 | 60.8035 | 800.6 | 670.183 | 2572.84 | 2293.22 | 5920.2 | 5605.12 | 9592.73 | 9489.42 |
| 623.84 | 76.7352 | 61.1361 | 758.093 | 672.901 | 2544.54 | 2301.15 | 5964.28 | 5622.45 | 9563.7 | 9516.21 |
| 624.13 | 75.5429 | 61.4701 | 776.299 | 675.627 | 2557.41 | 2309.1 | 5933.91 | 5639.81 | 9701.28 | 9543.04 |
| 624.42 | 89.5766 | 61.8055 | 799.78 | 678.361 | 2543.82 | 2317.07 | 5991.9 | 5657.2 | 9526.62 | 9569.91 |
| 624.71 | 81.1597 | 62.1423 | 869.347 | 681.101 | 2529.2 | 2325.06 | 5942.97 | 5674.62 | 9748.19 | 9596.82 |
| 625 | 86.3307 | 62.4806 | 825.17 | 683.85 | 2498.59 | 2333.06 | 5991.56 | 5692.07 | 9639.28 | 9623.77 |
| 625.29 | 122.192 | 62.8203 | 843.947 | 686.606 | 2563.37 | 2341.08 | 6023.45 | 5709.56 | 9938.52 | 9650.76 |
| 625.58 | 94.3276 | 63.1614 | 836.208 | 689.37 | 2568.41 | 2349.11 | 6125.92 | 5727.07 | 9625.13 | 9677.79 |
| 625.87 | 91.6166 | 63.5039 | 857.969 | 692.142 | 2659.49 | 2357.17 | 6085.75 | 5744.61 | 10050.8 | 9704.86 |
| 626.16 | 86.8995 | 63.8479 | 801.757 | 694.921 | 2674.76 | 2365.24 | 6251.28 | 5762.19 | 9984.73 | 9731.97 |
| 626.45 | 98.9633 | 64.1934 | 791.319 | 697.708 | 2665.57 | 2373.32 | 6130.62 | 5779.8 | 9968.84 | 9759.12 |
| 626.74 | 121.438 | 64.5403 | 832.918 | 700.503 | 2658.48 | 2381.43 | 6298.58 | 5797.44 | 10206.1 | 9786.31 |
| 627.03 | 95.8813 | 64.8887 | 825.877 | 703.305 | 2591.87 | 2389.55 | 6136.88 | 5815.1 | 9905.27 | 9813.54 |
| 627.32 | 106.792 | 65.2385 | 896.002 | 706.115 | 2592.94 | 2397.69 | 6068.47 | 5832.8 | 9904.7 | 9840.81 |
| 627.61 | 88.1525 | 65.5898 | 868.943 | 708.933 | 2699.92 | 2405.85 | 6311.01 | 5850.53 | 10145.7 | 9868.12 |
| 627.9 | 94.0905 | 65.9425 | 859.835 | 711.759 | 2714.04 | 2414.02 | 6266.68 | 5868.29 | 10172.3 | 9895.47 |
| 628.19 | 91.0475 | 66.2968 | 891.056 | 714.592 | 2593.84 | 2422.22 | 6244.68 | 5886.08 | 10091.8 | 9922.86 |
| 628.48 | 97.02 | 66.6525 | 907.011 | 717.433 | 2726.7 | 2430.43 | 6310.62 | 5903.91 | 10127.1 | 9950.28 |
| 628.77 | 83.7278 | 67.0096 | 850.254 | 720.282 | 2741.56 | 2438.65 | 6395.58 | 5921.76 | 10219.1 | 9977.75 |
| 629.059 | 105.815 | 67.3683 | 857.542 | 723.138 | 2711.96 | 2446.89 | 6296.12 | 5939.64 | 10299.1 | 10005.3 |
| 629.349 | 90.8971 | 67.7284 | 890.761 | 726.002 | 2839.21 | 2455.16 | 6433.15 | 5957.56 | 10527.6 | 10032.8 |
| 629.639 | 98.7656 | 68.0901 | 887.85 | 728.874 | 2787.39 | 2463.43 | 6455.02 | 5975.5 | 10322.5 | 10060.4 |
| 629.929 | 95.0315 | 68.4532 | 930.846 | 731.754 | 2796.84 | 2471.73 | 6561.26 | 5993.48 | 10601.7 | 10088 |
| 630.219 | 102.043 | 68.8179 | 900.48 | 734.642 | 2807.99 | 2480.04 | 6439.84 | 6011.48 | 10474.5 | 10115.7 |
| 630.509 | 97.1007 | 69.184 | 927.6 | 737.537 | 2777.97 | 2488.37 | 6444.15 | 6029.52 | 10434.3 | 10143.4 |
| 630.799 | 107.549 | 69.5516 | 873.506 | 740.44 | 2812.19 | 2496.72 | 6554.73 | 6047.59 | 10594.3 | 10171.1 |
| 631.088 | 98.1486 | 69.9208 | 886.314 | 743.351 | 2814.68 | 2505.08 | 6291.44 | 6065.69 | 10562 | 10198.9 |
| 631.378 | 100.297 | 70.2915 | 938.978 | 746.27 | 2869.63 | 2513.47 | 6586.93 | 6083.82 | 10735.9 | 10226.7 |
| 631.668 | 116.981 | 70.6637 | 933.254 | 749.197 | 2817.09 | 2521.86 | 6552.49 | 6101.98 | 10391.7 | 10254.6 |
| 631.958 | 112.566 | 71.0374 | 915.973 | 752.131 | 2881.5 | 2530.28 | 6785.3 | 6120.17 | 10886.7 | 10282.5 |
| 632.247 | 110.525 | 71.4126 | 931.049 | 755.074 | 2847.15 | 2538.72 | 6625.24 | 6138.39 | 10771.2 | 10310.4 |
| 632.537 | 124.342 | 71.7894 | 930.908 | 758.024 | 2860.56 | 2547.17 | 6670.4 | 6156.64 | 10812.1 | 10338.4 |
| 632.827 | 114.671 | 72.1677 | 948.618 | 760.982 | 2956.36 | 2555.64 | 6877.25 | 6174.92 | 11051.9 | 10366.4 |
| 633.117 | 106.737 | 72.5476 | 946.217 | 763.948 | 3028.48 | 2564.12 | 6868.57 | 6193.23 | 10973.4 | 10394.5 |
| 633.407 | 103.904 | 72.929 | 927.142 | 766.922 | 2943.72 | 2572.62 | 6736.9 | 6211.58 | 11045 | 10422.6 |
| 633.696 | 110.709 | 73.3119 | 946.139 | 769.904 | 2956.86 | 2581.15 | 6819.4 | 6229.95 | 11238.7 | 10450.7 |
| 633.986 | 107.047 | 73.6964 | 984.54 | 772.893 | 2990.82 | 2589.68 | 6957.36 | 6248.35 | 11208.6 | 10478.9 |
| 634.276 | 101.856 | 74.0824 | 954.605 | 775.891 | 2970.06 | 2598.24 | 6830.06 | 6266.79 | 10851.9 | 10507.1 |
| 634.565 | 115.242 | 74.47 | 967.294 | 778.896 | 2984.66 | 2606.81 | 6811.13 | 6285.25 | 11003.5 | 10535.4 |
| 634.855 | 110.779 | 74.8592 | 939.034 | 781.909 | 2986.8 | 2615.4 | 6902.4 | 6303.75 | 11220.5 | 10563.6 |
| 635.144 | 100.897 | 75.2499 | 973.618 | 784.931 | 2964.85 | 2624.01 | 6946.25 | 6322.27 | 11192.4 | 10592 |
| 635.434 | 96.502 | 75.6423 | 963.549 | 787.96 | 3074.6 | 2632.64 | 7082.37 | 6340.83 | 11395.2 | 10620.3 |
| 635.724 | 98.4366 | 76.0361 | 975.315 | 790.997 | 3041.51 | 2641.28 | 6866.04 | 6359.42 | 11278.4 | 10648.8 |
| 636.013 | 113.776 | 76.4316 | 994.768 | 794.042 | 3014.46 | 2649.94 | 6918.54 | 6378.03 | 11239.7 | 10677.2 |
| 636.303 | 111.36 | 76.8287 | 959.202 | 797.095 | 2980.72 | 2658.61 | 6904.19 | 6396.68 | 11121.7 | 10705.7 |
| 636.593 | 119.772 | 77.2273 | 1004.55 | 800.156 | 3052.44 | 2667.31 | 7010.15 | 6415.36 | 11326.4 | 10734.2 |
| 636.882 | 111.356 | 77.6275 | 957.109 | 803.225 | 3125.55 | 2676.02 | 7145.03 | 6434.07 | 11486 | 10762.8 |
| 637.172 | 118.944 | 78.0293 | 963.435 | 806.302 | 3035.01 | 2684.75 | 7022.33 | 6452.81 | 11447.5 | 10791.4 |
| 637.461 | 109.796 | 78.4328 | 990.993 | 809.387 | 3123.78 | 2693.49 | 7144.45 | 6471.58 | 11539.8 | 10820 |
| 637.751 | 105.643 | 78.8378 | 960.842 | 812.479 | 3018.55 | 2702.26 | 7041.46 | 6490.38 | 11330.3 | 10848.7 |
| 638.04 | 111.624 | 79.2444 | 998.737 | 815.58 | 3064.14 | 2711.04 | 7023.53 | 6509.21 | 11493.3 | 10877.4 |
| 638.33 | 120.216 | 79.6527 | 1017.71 | 818.689 | 3111.43 | 2719.84 | 7097.24 | 6528.07 | 11723.3 | 10906.1 |
| 638.619 | 113.085 | 80.0626 | 988.34 | 821.806 | 3113.16 | 2728.65 | 7008.01 | 6546.96 | 11414.6 | 10934.9 |
| 638.909 | 117.106 | 80.474 | 1010.41 | 824.931 | 3118.72 | 2737.49 | 7189.37 | 6565.88 | 11626.1 | 10963.8 |
| 639.198 | 109.754 | 80.8871 | 988.294 | 828.064 | 3124.81 | 2746.34 | 7220.18 | 6584.83 | 11773.4 | 10992.6 |
| 639.488 | 109.719 | 81.3019 | 990.333 | 831.205 | 3101.87 | 2755.21 | 7144.9 | 6603.81 | 11511.9 | 11021.5 |
| 639.777 | 117.02 | 81.7182 | 997.486 | 834.354 | 3141.79 | 2764.09 | 7168.14 | 6622.82 | 11607.9 | 11050.5 |
| 640.067 | 113.279 | 82.1363 | 991.291 | 837.511 | 3086.2 | 2772.99 | 7162.21 | 6641.86 | 11671.5 | 11079.5 |
| 640.356 | 118.174 | 82.5559 | 1019.15 | 840.676 | 3142.11 | 2781.91 | 7223.74 | 6660.94 | 11565.9 | 11108.5 |
| 640.646 | 125.003 | 82.9772 | 1008.14 | 843.849 | 3124.77 | 2790.85 | 7204.27 | 6680.04 | 11621.1 | 11137.5 |
| 640.935 | 124.481 | 83.4001 | 992.218 | 847.03 | 3134.18 | 2799.81 | 7206.36 | 6699.17 | 11664.7 | 11166.6 |
| 641.225 | 122.427 | 83.8247 | 1046.68 | 850.22 | 3193.91 | 2808.78 | 7276 | 6718.33 | 11724.7 | 11195.8 |
| 641.514 | 114.567 | 84.251 | 1040.23 | 853.417 | 3210.38 | 2817.77 | 7367.14 | 6737.52 | 11830 | 11224.9 |
| 641.803 | 132.044 | 84.6789 | 1020.85 | 856.622 | 3154.43 | 2826.78 | 7205.7 | 6756.75 | 11706 | 11254.1 |
| 642.093 | 108.621 | 85.1085 | 1038.43 | 859.836 | 3259.59 | 2835.8 | 7365.03 | 6776 | 11925.3 | 11283.4 |
| 642.382 | 117.407 | 85.5397 | 1055.62 | 863.058 | 3159.43 | 2844.84 | 7384.3 | 6795.28 | 11685.5 | 11312.7 |
| 642.671 | 104.838 | 85.9726 | 1008.85 | 866.288 | 3144.34 | 2853.9 | 7237.39 | 6814.59 | 11813.1 | 11342 |
| 642.961 | 119.809 | 86.4072 | 1021.54 | 869.525 | 3189.64 | 2862.98 | 7296.4 | 6833.94 | 11760.1 | 11371.4 |
| 643.25 | 124.49 | 86.8435 | 1053.95 | 872.771 | 3175.93 | 2872.07 | 7323.76 | 6853.31 | 11897.3 | 11400.8 |
| 643.539 | 114.003 | 87.2814 | 1033.14 | 876.025 | 3290.98 | 2881.18 | 7466.35 | 6872.71 | 11983.7 | 11430.2 |
| 643.829 | 125.133 | 87.7211 | 1051.12 | 879.288 | 3286.39 | 2890.31 | 7430.49 | 6892.14 | 11977 | 11459.7 |
| 644.118 | 114.396 | 88.1624 | 1053.19 | 882.558 | 3175.55 | 2899.46 | 7455.1 | 6911.61 | 11955.4 | 11489.2 |
| 644.407 | 127.196 | 88.6055 | 1066.55 | 885.837 | 3266.24 | 2908.62 | 7359.16 | 6931.1 | 11859.5 | 11518.7 |
| 644.697 | 117.65 | 89.0502 | 1039.85 | 889.123 | 3240.19 | 2917.8 | 7435.01 | 6950.62 | 12245.3 | 11548.3 |
| 644.986 | 142.592 | 89.4967 | 1051.87 | 892.418 | 3318.8 | 2927 | 7488.17 | 6970.17 | 12154.5 | 11577.9 |
| 645.275 | 123.91 | 89.9448 | 1101.73 | 895.721 | 3377.63 | 2936.22 | 7503.48 | 6989.76 | 12115.1 | 11607.6 |
| 645.564 | 137.038 | 90.3947 | 1045.6 | 899.032 | 3338.42 | 2945.45 | 7598.27 | 7009.37 | 12066.5 | 11637.3 |
| 645.853 | 123.582 | 90.8463 | 1048.21 | 902.352 | 3294.36 | 2954.7 | 7490.75 | 7029.01 | 12394.6 | 11667 |
| 646.143 | 126.589 | 91.2996 | 1097.3 | 905.679 | 3379.41 | 2963.97 | 7669.51 | 7048.68 | 12226.6 | 11696.8 |
| 646.432 | 129.216 | 91.7546 | 1091.87 | 909.015 | 3315.57 | 2973.25 | 7500.69 | 7068.38 | 12360.1 | 11726.6 |
| 646.721 | 130.063 | 92.2114 | 1053.31 | 912.359 | 3358.23 | 2982.55 | 7684.42 | 7088.11 | 12251.1 | 11756.5 |
| 647.01 | 121.087 | 92.6699 | 1051.58 | 915.711 | 3395.16 | 2991.87 | 7596.52 | 7107.88 | 12345.7 | 11786.3 |
| 647.299 | 122.316 | 93.1301 | 1071.21 | 919.072 | 3290.16 | 3001.21 | 7631.84 | 7127.67 | 12390.9 | 11816.2 |
| 647.588 | 123.885 | 93.5921 | 1094.34 | 922.44 | 3370.43 | 3010.57 | 7680.09 | 7147.49 | 12334.1 | 11846.2 |
| 647.878 | 139.287 | 94.0558 | 1064.66 | 925.817 | 3355.19 | 3019.94 | 7687.23 | 7167.34 | 12348.3 | 11876.2 |
| 648.167 | 147.28 | 94.5213 | 1111.17 | 929.202 | 3437.44 | 3029.33 | 7824.56 | 7187.22 | 12452.9 | 11906.2 |
| 648.456 | 127.305 | 94.9885 | 1107.49 | 932.595 | 3344.17 | 3038.73 | 7732.79 | 7207.13 | 12567.2 | 11936.3 |
| 648.745 | 124.251 | 95.4575 | 1142.38 | 935.997 | 3386.25 | 3048.16 | 7731.41 | 7227.07 | 12608.7 | 11966.4 |
| 649.034 | 144.889 | 95.9282 | 1121.98 | 939.407 | 3351.12 | 3057.6 | 7912.42 | 7247.03 | 12549.8 | 11996.5 |
| 649.323 | 135.288 | 96.4007 | 1081.41 | 942.825 | 3386.24 | 3067.06 | 7799.19 | 7267.03 | 12565.3 | 12026.7 |
| 649.612 | 148.06 | 96.875 | 1118.77 | 946.251 | 3442.6 | 3076.53 | 7711.4 | 7287.06 | 12495.7 | 12057 |
| 649.901 | 125.352 | 97.3511 | 1148.85 | 949.685 | 3503.67 | 3086.03 | 7785.78 | 7307.12 | 12676.6 | 12087.2 |
| 650.19 | 133.504 | 97.8289 | 1131.67 | 953.128 | 3424.75 | 3095.54 | 7897.41 | 7327.21 | 12593.7 | 12117.5 |
| 650.479 | 119.747 | 98.3085 | 1117.1 | 956.579 | 3479.17 | 3105.07 | 7854.79 | 7347.32 | 12788.6 | 12147.8 |
| 650.768 | 132.062 | 98.7899 | 1148.69 | 960.039 | 3476.96 | 3114.61 | 7933.26 | 7367.47 | 12770.5 | 12178.2 |
| 651.057 | 133.738 | 99.2731 | 1157.83 | 963.506 | 3450.93 | 3124.17 | 7824.46 | 7387.65 | 12634.5 | 12208.6 |
| 651.346 | 140.188 | 99.7581 | 1140 | 966.982 | 3511.38 | 3133.76 | 7910.09 | 7407.85 | 12811.2 | 12239 |
| 651.635 | 138.476 | 100.245 | 1160.77 | 970.467 | 3477.72 | 3143.35 | 7959.01 | 7428.09 | 12845.5 | 12269.5 |
| 651.924 | 150.644 | 100.734 | 1161 | 973.959 | 3531.52 | 3152.97 | 7916.41 | 7448.35 | 12787.8 | 12300 |
| 652.213 | 137.012 | 101.224 | 1139.53 | 977.46 | 3528.32 | 3162.6 | 8027.67 | 7468.64 | 13008.1 | 12330.5 |
| 652.502 | 139.202 | 101.716 | 1171.84 | 980.969 | 3536.49 | 3172.25 | 8119.8 | 7488.97 | 12832 | 12361.1 |
| 652.791 | 142.125 | 102.21 | 1184.71 | 984.487 | 3517.2 | 3181.92 | 8153.19 | 7509.32 | 13070.1 | 12391.7 |
| 653.08 | 139.408 | 102.706 | 1159.53 | 988.013 | 3609.2 | 3191.6 | 8163.02 | 7529.7 | 13120.4 | 12422.4 |
| 653.369 | 146.996 | 103.204 | 1197.58 | 991.547 | 3648.43 | 3201.3 | 8236.71 | 7550.11 | 13230.8 | 12453.1 |
| 653.658 | 131.133 | 103.703 | 1185.74 | 995.089 | 3650.06 | 3211.02 | 8277.67 | 7570.55 | 13236.5 | 12483.8 |
| 653.947 | 137.144 | 104.204 | 1196.73 | 998.64 | 3626.32 | 3220.76 | 8267.65 | 7591.02 | 13285.6 | 12514.5 |
| 654.236 | 136.512 | 104.708 | 1198.71 | 1002.2 | 3639.63 | 3230.51 | 8169.33 | 7611.52 | 13182 | 12545.3 |
| 654.524 | 142.072 | 105.213 | 1180.55 | 1005.77 | 3627.9 | 3240.28 | 8451.93 | 7632.05 | 13361.1 | 12576.2 |
| 654.813 | 160.135 | 105.719 | 1167.07 | 1009.34 | 3616.6 | 3250.07 | 8304.11 | 7652.6 | 13398.1 | 12607 |
| 655.102 | 142.117 | 106.228 | 1222.08 | 1012.93 | 3665.91 | 3259.87 | 8357.34 | 7673.19 | 13401.8 | 12638 |
| 655.391 | 150.75 | 106.739 | 1225.31 | 1016.52 | 3638.79 | 3269.7 | 8355.37 | 7693.81 | 13479.4 | 12668.9 |
| 655.68 | 145.384 | 107.251 | 1216.64 | 1020.12 | 3693.65 | 3279.54 | 8408.37 | 7714.45 | 13470.9 | 12699.9 |
| 655.968 | 149.475 | 107.765 | 1211.37 | 1023.73 | 3718.15 | 3289.39 | 8433.96 | 7735.12 | 13572 | 12730.9 |
| 656.257 | 147.924 | 108.281 | 1217.7 | 1027.35 | 3667.2 | 3299.27 | 8382.37 | 7755.83 | 13582.7 | 12761.9 |
| 656.546 | 145.533 | 108.799 | 1223.76 | 1030.97 | 3751.81 | 3309.16 | 8543.55 | 7776.56 | 13758 | 12793 |
| 656.835 | 152.844 | 109.319 | 1196.95 | 1034.61 | 3689.38 | 3319.07 | 8369.99 | 7797.32 | 13469.2 | 12824.1 |
| 657.124 | 152.599 | 109.841 | 1214.09 | 1038.25 | 3774.39 | 3329 | 8567.2 | 7818.11 | 13612.6 | 12855.3 |
| 657.412 | 156.068 | 110.365 | 1252.75 | 1041.9 | 3762.91 | 3338.94 | 8549.5 | 7838.93 | 13561.5 | 12886.5 |
| 657.701 | 155.775 | 110.89 | 1242.38 | 1045.56 | 3748.65 | 3348.9 | 8469.23 | 7859.78 | 13658.7 | 12917.7 |
| 657.99 | 156.372 | 111.418 | 1223.13 | 1049.23 | 3777.78 | 3358.88 | 8543.4 | 7880.65 | 13791.2 | 12948.9 |
| 658.278 | 141.338 | 111.947 | 1230.73 | 1052.91 | 3766.1 | 3368.88 | 8496.38 | 7901.56 | 13536.9 | 12980.2 |
| 658.567 | 152.97 | 112.478 | 1231.6 | 1056.59 | 3774.99 | 3378.89 | 8528.93 | 7922.49 | 13641.8 | 13011.6 |
| 658.856 | 147.352 | 113.011 | 1232.76 | 1060.29 | 3787.73 | 3388.92 | 8541.96 | 7943.46 | 13821.4 | 13042.9 |
| 659.144 | 153.405 | 113.546 | 1213.24 | 1063.99 | 3784.08 | 3398.97 | 8612.18 | 7964.45 | 13924.7 | 13074.3 |
| 659.433 | 139.076 | 114.083 | 1239.66 | 1067.7 | 3761.24 | 3409.03 | 8697.49 | 7985.47 | 13846.5 | 13105.8 |
| 659.722 | 172.991 | 114.622 | 1283.97 | 1071.42 | 3820.52 | 3419.11 | 8695.23 | 8006.52 | 13832.1 | 13137.2 |
| 660.01 | 163.636 | 115.163 | 1252.8 | 1075.14 | 3839.35 | 3429.21 | 8713.11 | 8027.6 | 14067.4 | 13168.7 |
| 660.299 | 151.93 | 115.705 | 1246.03 | 1078.88 | 3795.91 | 3439.33 | 8646.6 | 8048.71 | 13840 | 13200.3 |
| 660.588 | 166.047 | 116.25 | 1287.72 | 1082.62 | 3882.89 | 3449.46 | 8814.19 | 8069.84 | 14089.1 | 13231.9 |
| 660.876 | 153.628 | 116.796 | 1289.22 | 1086.38 | 3879.34 | 3459.61 | 8761.44 | 8091.01 | 14146.4 | 13263.5 |
| 661.165 | 155.762 | 117.345 | 1290.85 | 1090.14 | 3882.43 | 3469.78 | 8902.8 | 8112.2 | 14236 | 13295.1 |
| 661.453 | 156.196 | 117.895 | 1325.58 | 1093.91 | 3947.56 | 3479.97 | 9001.72 | 8133.42 | 14227.3 | 13326.8 |
| 661.742 | 173.709 | 118.447 | 1294.51 | 1097.69 | 3936.9 | 3490.17 | 8869.11 | 8154.67 | 14286.1 | 13358.5 |
| 662.03 | 160.507 | 119.002 | 1281.4 | 1101.47 | 3879.59 | 3500.39 | 8817.41 | 8175.95 | 14063.4 | 13390.2 |
| 662.319 | 165.347 | 119.558 | 1326.88 | 1105.27 | 3904.69 | 3510.63 | 8874.9 | 8197.26 | 14219.9 | 13422 |
| 662.607 | 161.198 | 120.116 | 1286.28 | 1109.07 | 3935.84 | 3520.88 | 8928.91 | 8218.6 | 14424.3 | 13453.9 |
| 662.896 | 155.916 | 120.676 | 1311.09 | 1112.88 | 3890.27 | 3531.15 | 8850.84 | 8239.96 | 14335 | 13485.7 |
| 663.184 | 166.214 | 121.238 | 1316.86 | 1116.7 | 3941.92 | 3541.44 | 8996.58 | 8261.35 | 14432.8 | 13517.6 |
| 663.473 | 148.252 | 121.802 | 1332.11 | 1120.53 | 3939.16 | 3551.75 | 8897.49 | 8282.78 | 14331.6 | 13549.5 |
| 663.761 | 175.107 | 122.368 | 1322.95 | 1124.37 | 3931.83 | 3562.07 | 8998.62 | 8304.23 | 14418.4 | 13581.5 |
| 664.05 | 163.958 | 122.936 | 1359.57 | 1128.22 | 4027.97 | 3572.41 | 9076.93 | 8325.71 | 14509.2 | 13613.5 |
| 664.338 | 163.013 | 123.506 | 1336.53 | 1132.07 | 4020.7 | 3582.77 | 9004.5 | 8347.21 | 14447.1 | 13645.5 |
| 664.627 | 164.842 | 124.078 | 1328.57 | 1135.94 | 4008.77 | 3593.14 | 9139.71 | 8368.75 | 14620.8 | 13677.5 |
| 664.915 | 164.894 | 124.652 | 1334.32 | 1139.81 | 3997.37 | 3603.53 | 9155.27 | 8390.31 | 14559.3 | 13709.6 |
| 665.203 | 172.161 | 125.228 | 1343.39 | 1143.69 | 3981.57 | 3613.94 | 9066.81 | 8411.91 | 14536.4 | 13741.8 |
| 665.492 | 177.709 | 125.806 | 1337.44 | 1147.58 | 3994.46 | 3624.37 | 9096.89 | 8433.53 | 14396.8 | 13773.9 |
| 665.78 | 160.715 | 126.386 | 1343.59 | 1151.47 | 4032.62 | 3634.81 | 9059.41 | 8455.17 | 14402.4 | 13806.1 |
| 666.069 | 167.919 | 126.968 | 1311.13 | 1155.38 | 4033.78 | 3645.27 | 9238.35 | 8476.85 | 14601.5 | 13838.3 |
| 666.357 | 169.972 | 127.552 | 1381.41 | 1159.3 | 4040.32 | 3655.75 | 9249.36 | 8498.56 | 14821.9 | 13870.6 |
| 666.645 | 167.513 | 128.137 | 1323.61 | 1163.22 | 3971.68 | 3666.24 | 9163.11 | 8520.29 | 14629.1 | 13902.9 |
| 666.934 | 178.421 | 128.725 | 1356.8 | 1167.15 | 4088.91 | 3676.75 | 9166.99 | 8542.05 | 14691.3 | 13935.2 |
| 667.222 | 170.118 | 129.315 | 1331.58 | 1171.09 | 4076.11 | 3687.28 | 9186.42 | 8563.84 | 14749 | 13967.6 |
| 667.51 | 176.053 | 129.907 | 1378.76 | 1175.04 | 4049.01 | 3697.83 | 9364.37 | 8585.66 | 14886.5 | 14000 |
| 667.799 | 163.962 | 130.501 | 1369.59 | 1178.99 | 4012.91 | 3708.39 | 9161.9 | 8607.5 | 14709 | 14032.4 |
| 668.087 | 185.126 | 131.097 | 1373.6 | 1182.96 | 4060.42 | 3718.97 | 9221.84 | 8629.38 | 14831.1 | 14064.9 |
| 668.375 | 178.156 | 131.695 | 1365.83 | 1186.94 | 4080.64 | 3729.57 | 9270.48 | 8651.28 | 14898.9 | 14097.4 |
| 668.663 | 165.172 | 132.295 | 1370.21 | 1190.92 | 4116.37 | 3740.18 | 9387 | 8673.21 | 14772.1 | 14129.9 |
| 668.952 | 170.656 | 132.898 | 1421.59 | 1194.91 | 4160.45 | 3750.81 | 9432.65 | 8695.17 | 14960 | 14162.5 |
| 669.24 | 172.396 | 133.502 | 1391.96 | 1198.91 | 4145.46 | 3761.46 | 9359.59 | 8717.15 | 15016.6 | 14195.1 |
| 669.528 | 175.806 | 134.108 | 1382.63 | 1202.92 | 4212.36 | 3772.13 | 9492.46 | 8739.17 | 15108.9 | 14227.7 |
| 669.816 | 173.342 | 134.716 | 1408.52 | 1206.94 | 4142.24 | 3782.81 | 9429.12 | 8761.21 | 15045.4 | 14260.4 |
| 670.105 | 190.714 | 135.327 | 1411.24 | 1210.96 | 4120.74 | 3793.51 | 9385.86 | 8783.28 | 15119.2 | 14293.1 |
| 670.393 | 183.77 | 135.939 | 1385.69 | 1215 | 4210.77 | 3804.22 | 9487.14 | 8805.37 | 15170.5 | 14325.8 |
| 670.681 | 174.868 | 136.553 | 1447.52 | 1219.04 | 4178.3 | 3814.96 | 9440.04 | 8827.5 | 15181.1 | 14358.5 |
| 670.969 | 175.845 | 137.17 | 1405.98 | 1223.09 | 4242.56 | 3825.71 | 9537.04 | 8849.65 | 15120.1 | 14391.4 |
| 671.257 | 184.179 | 137.788 | 1405.94 | 1227.15 | 4244.95 | 3836.48 | 9573.19 | 8871.83 | 15257.6 | 14424.2 |
| 671.545 | 187.033 | 138.409 | 1428.67 | 1231.22 | 4192.26 | 3847.26 | 9566.96 | 8894.04 | 15254.2 | 14457 |
| 671.833 | 198.485 | 139.032 | 1443.24 | 1235.3 | 4249.6 | 3858.06 | 9563.99 | 8916.27 | 15320.5 | 14489.9 |
| 672.122 | 206.392 | 139.657 | 1438.71 | 1239.38 | 4255.49 | 3868.88 | 9648.74 | 8938.54 | 15446.2 | 14522.9 |
| 672.41 | 191.802 | 140.284 | 1458.7 | 1243.48 | 4320.95 | 3879.72 | 9834.43 | 8960.83 | 15629.5 | 14555.8 |
| 672.698 | 185.805 | 140.913 | 1434.69 | 1247.58 | 4296.75 | 3890.57 | 9647.04 | 8983.15 | 15349.1 | 14588.8 |
| 672.986 | 196.488 | 141.544 | 1481.22 | 1251.7 | 4287.53 | 3901.44 | 9738.61 | 9005.49 | 15545 | 14621.8 |
| 673.274 | 187.035 | 142.177 | 1456.21 | 1255.82 | 4289.6 | 3912.32 | 9712.01 | 9027.87 | 15541.3 | 14654.9 |
| 673.562 | 183.083 | 142.812 | 1443.59 | 1259.95 | 4352.19 | 3923.23 | 9669.28 | 9050.27 | 15369.6 | 14688 |
| 673.85 | 186.961 | 143.45 | 1474.14 | 1264.08 | 4400.68 | 3934.15 | 9800.36 | 9072.69 | 15743 | 14721.1 |
| 674.138 | 187.901 | 144.089 | 1461.86 | 1268.23 | 4291.83 | 3945.08 | 9766.45 | 9095.15 | 15574.2 | 14754.3 |
| 674.426 | 185.11 | 144.731 | 1469.41 | 1272.39 | 4372.16 | 3956.04 | 9842.82 | 9117.63 | 15697.7 | 14787.5 |
| 674.714 | 187.676 | 145.375 | 1477.5 | 1276.55 | 4328.06 | 3967.01 | 9738.16 | 9140.14 | 15561.2 | 14820.7 |
| 675.002 | 198.694 | 146.021 | 1456.3 | 1280.72 | 4340.99 | 3978 | 9795.73 | 9162.68 | 15646.5 | 14853.9 |
| 675.29 | 193.03 | 146.669 | 1481.59 | 1284.9 | 4324.88 | 3989 | 9789.84 | 9185.25 | 15603.5 | 14887.2 |
| 675.578 | 178.386 | 147.319 | 1484.32 | 1289.09 | 4305.34 | 4000.02 | 9830.69 | 9207.84 | 15661.2 | 14920.5 |
| 675.866 | 186.006 | 147.971 | 1471.86 | 1293.29 | 4391.22 | 4011.06 | 9864.13 | 9230.46 | 15722.9 | 14953.9 |
| 676.154 | 198.425 | 148.626 | 1500.13 | 1297.5 | 4413.03 | 4022.12 | 9931.92 | 9253.11 | 15715.9 | 14987.3 |
| 676.442 | 196.781 | 149.282 | 1498.55 | 1301.72 | 4358.75 | 4033.19 | 9911.89 | 9275.78 | 15856.8 | 15020.7 |
| 676.73 | 194.757 | 149.941 | 1480.89 | 1305.94 | 4377.22 | 4044.28 | 9966.31 | 9298.48 | 15918 | 15054.1 |
| 677.018 | 189.593 | 150.602 | 1488.24 | 1310.17 | 4385.52 | 4055.39 | 9949.77 | 9321.21 | 15785.6 | 15087.6 |
| 677.305 | 200.472 | 151.265 | 1499.41 | 1314.42 | 4417.5 | 4066.51 | 9955.84 | 9343.97 | 16021.8 | 15121.1 |
| 677.593 | 200.747 | 151.93 | 1503.15 | 1318.67 | 4404.64 | 4077.65 | 9928.84 | 9366.75 | 16004.4 | 15154.6 |
| 677.881 | 206.871 | 152.598 | 1507.44 | 1322.93 | 4476.11 | 4088.8 | 10118 | 9389.56 | 16023.3 | 15188.2 |
| 678.169 | 192.293 | 153.267 | 1525.2 | 1327.19 | 4482.18 | 4099.98 | 10096.6 | 9412.39 | 15990.4 | 15221.8 |
| 678.457 | 188.626 | 153.939 | 1491.09 | 1331.47 | 4461.73 | 4111.17 | 10129.7 | 9435.26 | 16105.7 | 15255.5 |
| 678.745 | 197.68 | 154.613 | 1533.04 | 1335.76 | 4541.6 | 4122.37 | 10125 | 9458.15 | 16230.5 | 15289.1 |
| 679.032 | 206.265 | 155.289 | 1537 | 1340.05 | 4504.63 | 4133.6 | 10085.2 | 9481.06 | 16054 | 15322.8 |
| 679.32 | 198.908 | 155.967 | 1548.54 | 1344.35 | 4479.64 | 4144.84 | 10183.9 | 9504.01 | 16184.6 | 15356.5 |
| 679.608 | 198.028 | 156.648 | 1526.12 | 1348.67 | 4455.28 | 4156.1 | 10141.8 | 9526.98 | 16119.5 | 15390.3 |
| 679.896 | 220.596 | 157.33 | 1544.85 | 1352.99 | 4532.6 | 4167.37 | 10150.2 | 9549.98 | 16167.5 | 15424.1 |
| 680.183 | 208.446 | 158.015 | 1534.09 | 1357.32 | 4541.85 | 4178.66 | 10259 | 9573 | 16286.7 | 15457.9 |
| 680.471 | 210.418 | 158.702 | 1554.74 | 1361.65 | 4551.58 | 4189.97 | 10338.4 | 9596.05 | 16419.4 | 15491.8 |
| 680.759 | 214.489 | 159.391 | 1564.21 | 1366 | 4591.99 | 4201.29 | 10261.5 | 9619.13 | 16354.3 | 15525.7 |
| 681.047 | 206.97 | 160.083 | 1570.68 | 1370.36 | 4608.8 | 4212.63 | 10301 | 9642.24 | 16373.3 | 15559.6 |
| 681.334 | 204.87 | 160.776 | 1531.78 | 1374.72 | 4533.85 | 4223.99 | 10306.2 | 9665.37 | 16283.6 | 15593.5 |
| 681.622 | 203.722 | 161.472 | 1549.79 | 1379.09 | 4611.87 | 4235.37 | 10314.6 | 9688.52 | 16303.4 | 15627.5 |
| 681.91 | 207.024 | 162.17 | 1566.47 | 1383.47 | 4523.05 | 4246.76 | 10319.7 | 9711.71 | 16202 | 15661.5 |
| 682.197 | 201.618 | 162.87 | 1587.68 | 1387.86 | 4490.63 | 4258.16 | 10384.4 | 9734.92 | 16277.2 | 15695.6 |
| 682.485 | 217.75 | 163.573 | 1601.35 | 1392.26 | 4494.88 | 4269.59 | 10232.3 | 9758.16 | 16355.5 | 15729.6 |
| 682.773 | 211.541 | 164.278 | 1620.3 | 1396.67 | 4525.84 | 4281.03 | 10361.9 | 9781.42 | 16600.6 | 15763.7 |
| 683.06 | 210.969 | 164.985 | 1588.36 | 1401.09 | 4578.75 | 4292.49 | 10464.5 | 9804.71 | 16495.3 | 15797.9 |
| 683.348 | 228.732 | 165.694 | 1592.77 | 1405.51 | 4569.05 | 4303.96 | 10453 | 9828.03 | 16734.7 | 15832 |
| 683.636 | 206.299 | 166.405 | 1628.01 | 1409.94 | 4613.74 | 4315.45 | 10528.3 | 9851.37 | 16708.6 | 15866.2 |
| 683.923 | 225.355 | 167.119 | 1626.65 | 1414.39 | 4618.84 | 4326.96 | 10438.7 | 9874.74 | 16809.6 | 15900.5 |
| 684.211 | 234.302 | 167.835 | 1632.77 | 1418.84 | 4643.53 | 4338.48 | 10567.2 | 9898.14 | 16775.1 | 15934.7 |
| 684.498 | 210.043 | 168.553 | 1644.7 | 1423.3 | 4679.19 | 4350.03 | 10668.6 | 9921.56 | 17037.6 | 15969 |
| 684.786 | 223.757 | 169.274 | 1671.43 | 1427.77 | 4748.99 | 4361.58 | 10804.5 | 9945.01 | 16862.8 | 16003.3 |
| 685.073 | 238.894 | 169.996 | 1682.71 | 1432.24 | 4606.38 | 4373.16 | 10788 | 9968.48 | 17010.6 | 16037.6 |
| 685.361 | 235.512 | 170.721 | 1664.8 | 1436.73 | 4677.64 | 4384.75 | 10661.3 | 9991.98 | 17028.9 | 16072 |
| 685.649 | 215.339 | 171.449 | 1710.02 | 1441.22 | 4746.34 | 4396.36 | 10715.8 | 10015.5 | 16998.8 | 16106.4 |
| 685.936 | 230.474 | 172.178 | 1677.2 | 1445.73 | 4728.95 | 4407.98 | 10826.1 | 10039.1 | 17097 | 16140.9 |
| 686.223 | 230.292 | 172.91 | 1683.5 | 1450.24 | 4719.52 | 4419.62 | 10832.9 | 10062.6 | 17163.7 | 16175.3 |
| 686.511 | 216.954 | 173.644 | 1697.72 | 1454.76 | 4771.17 | 4431.28 | 10895 | 10086.2 | 17286.7 | 16209.8 |
| 686.798 | 228.879 | 174.381 | 1685.85 | 1459.29 | 4792.96 | 4442.95 | 10863.6 | 10109.9 | 17199.8 | 16244.4 |
| 687.086 | 230.34 | 175.119 | 1696.73 | 1463.83 | 4760.86 | 4454.64 | 10923.3 | 10133.5 | 17265.8 | 16278.9 |
| 687.373 | 215.128 | 175.86 | 1750.17 | 1468.38 | 4780.64 | 4466.35 | 10967.4 | 10157.2 | 17259.9 | 16313.5 |
| 687.661 | 236.542 | 176.603 | 1749.39 | 1472.93 | 4781.37 | 4478.07 | 11028.6 | 10180.9 | 17315.8 | 16348.1 |
| 687.948 | 232.849 | 177.349 | 1787.51 | 1477.5 | 4786.6 | 4489.81 | 10908.8 | 10204.7 | 17548.1 | 16382.8 |
| 688.236 | 220.956 | 178.097 | 1709.23 | 1482.07 | 4742.09 | 4501.56 | 10926 | 10228.4 | 17446.7 | 16417.4 |
| 688.523 | 226.031 | 178.847 | 1753.61 | 1486.65 | 4862.02 | 4513.34 | 11011.4 | 10252.2 | 17410.8 | 16452.1 |
| 688.81 | 247.327 | 179.599 | 1766.31 | 1491.24 | 4839.58 | 4525.12 | 11147.1 | 10276 | 17612.5 | 16486.9 |
| 689.098 | 220.754 | 180.354 | 1775.96 | 1495.85 | 4825.7 | 4536.93 | 11116.4 | 10299.9 | 17837.6 | 16521.6 |
| 689.385 | 237.351 | 181.111 | 1747.22 | 1500.45 | 4756.05 | 4548.75 | 11203.1 | 10323.7 | 17563.9 | 16556.4 |
| 689.672 | 243.78 | 181.87 | 1762.19 | 1505.07 | 4799.06 | 4560.59 | 11145.1 | 10347.6 | 17756.1 | 16591.2 |
| 689.96 | 230.867 | 182.632 | 1786.85 | 1509.7 | 4802.37 | 4572.44 | 11209.1 | 10371.5 | 17734.2 | 16626.1 |
| 690.247 | 249.089 | 183.396 | 1763.42 | 1514.33 | 4831.87 | 4584.31 | 11152.1 | 10395.5 | 17649.9 | 16661 |
| 690.534 | 237.481 | 184.163 | 1803.43 | 1518.98 | 4753.48 | 4596.2 | 11205.5 | 10419.4 | 17684.3 | 16695.9 |
| 690.822 | 241.001 | 184.931 | 1800.63 | 1523.63 | 4844.37 | 4608.1 | 11255.8 | 10443.4 | 17710.4 | 16730.8 |
| 691.109 | 250.873 | 185.702 | 1772.59 | 1528.29 | 4825.03 | 4620.02 | 11277.2 | 10467.4 | 17903 | 16765.8 |
| 691.396 | 233.535 | 186.476 | 1777.36 | 1532.96 | 4886.7 | 4631.96 | 11252.6 | 10491.5 | 17811.1 | 16800.8 |
| 691.683 | 249.175 | 187.252 | 1780.58 | 1537.64 | 4890.3 | 4643.91 | 11426.1 | 10515.5 | 18035.5 | 16835.8 |
| 691.971 | 254.509 | 188.03 | 1795.39 | 1542.33 | 4917.63 | 4655.88 | 11330.8 | 10539.6 | 17887 | 16870.9 |
| 692.258 | 246.382 | 188.81 | 1816.66 | 1547.02 | 4856.83 | 4667.87 | 11354.8 | 10563.8 | 18046.6 | 16906 |
| 692.545 | 258.088 | 189.593 | 1845.41 | 1551.73 | 4950.51 | 4679.87 | 11411.7 | 10587.9 | 18120.8 | 16941.1 |
| 692.832 | 253.179 | 190.378 | 1819.97 | 1556.44 | 4897.32 | 4691.89 | 11452.5 | 10612.1 | 18156 | 16976.2 |
| 693.119 | 249.198 | 191.166 | 1861.58 | 1561.17 | 4957.3 | 4703.92 | 11540 | 10636.3 | 18114.6 | 17011.4 |
| 693.407 | 257.087 | 191.956 | 1808.04 | 1565.9 | 4939.69 | 4715.97 | 11384 | 10660.5 | 18068.2 | 17046.6 |
| 693.694 | 247.013 | 192.748 | 1810.66 | 1570.64 | 4968.87 | 4728.04 | 11511.2 | 10684.7 | 18143.2 | 17081.8 |
| 693.981 | 253.242 | 193.543 | 1860.28 | 1575.39 | 4950.01 | 4740.12 | 11529.1 | 10709 | 18229.5 | 17117 |
| 694.268 | 262.176 | 194.34 | 1881.66 | 1580.14 | 5102.91 | 4752.22 | 11587.6 | 10733.3 | 18254.2 | 17152.3 |
| 694.555 | 230.621 | 195.139 | 1857.41 | 1584.91 | 5008.8 | 4764.33 | 11560.5 | 10757.6 | 18097.5 | 17187.6 |
| 694.842 | 251.619 | 195.941 | 1877.76 | 1589.69 | 4982.07 | 4776.46 | 11653.6 | 10782 | 18306 | 17223 |
| 695.129 | 261.963 | 196.745 | 1862.52 | 1594.47 | 5020.62 | 4788.61 | 11552.4 | 10806.3 | 18322.3 | 17258.3 |
| 695.416 | 264.802 | 197.552 | 1863.06 | 1599.27 | 4937.84 | 4800.77 | 11646.1 | 10830.7 | 18193.4 | 17293.7 |
| 695.703 | 260.726 | 198.361 | 1867.69 | 1604.07 | 4955.09 | 4812.95 | 11580.9 | 10855.2 | 18353 | 17329.2 |
| 695.99 | 252.785 | 199.172 | 1928.39 | 1608.88 | 5060.3 | 4825.15 | 11814.7 | 10879.6 | 18470.6 | 17364.6 |
| 696.277 | 246.275 | 199.986 | 1936.13 | 1613.7 | 5007.83 | 4837.36 | 11659.6 | 10904.1 | 18519.5 | 17400.1 |
| 696.564 | 250.345 | 200.802 | 1884.71 | 1618.53 | 5030.23 | 4849.59 | 11825.7 | 10928.6 | 18651.9 | 17435.6 |
| 696.851 | 246.73 | 201.621 | 1887.53 | 1623.36 | 5033.12 | 4861.84 | 11811.5 | 10953.1 | 18644.5 | 17471.1 |
| 697.138 | 274.433 | 202.442 | 1874.44 | 1628.21 | 5057.52 | 4874.1 | 11723.2 | 10977.6 | 18555.7 | 17506.7 |
| 697.425 | 272.428 | 203.266 | 1943.49 | 1633.06 | 5077.4 | 4886.37 | 11889 | 11002.2 | 18656.1 | 17542.3 |
| 697.712 | 264.709 | 204.091 | 1901.66 | 1637.93 | 5086.76 | 4898.66 | 11865.6 | 11026.8 | 18823.7 | 17577.9 |
| 697.999 | 259.773 | 204.92 | 1899.56 | 1642.8 | 5136.61 | 4910.97 | 11817.7 | 11051.4 | 18679.9 | 17613.5 |
| 698.286 | 282.837 | 205.751 | 1927.99 | 1647.68 | 5143.16 | 4923.3 | 11952.1 | 11076 | 18690.8 | 17649.2 |
| 698.573 | 287.346 | 206.584 | 1946.8 | 1652.57 | 5172.42 | 4935.64 | 12043.5 | 11100.7 | 18868.6 | 17684.9 |
| 698.86 | 262.829 | 207.419 | 1905.8 | 1657.47 | 5172.62 | 4947.99 | 11974.7 | 11125.4 | 18732.6 | 17720.7 |
| 699.147 | 254.692 | 208.257 | 1929.04 | 1662.37 | 5157.32 | 4960.37 | 11936 | 11150.1 | 18749.8 | 17756.4 |
| 699.434 | 261.604 | 209.098 | 1904.39 | 1667.29 | 5250.84 | 4972.75 | 12112.4 | 11174.9 | 19326.9 | 17792.2 |
| 699.721 | 272.814 | 209.941 | 1913.91 | 1672.22 | 5322.33 | 4985.16 | 11998.4 | 11199.6 | 19036.6 | 17828 |
| 700.008 | 282.319 | 210.786 | 1978.6 | 1677.15 | 5301.37 | 4997.58 | 12131.7 | 11224.4 | 19110 | 17863.8 |
| 700.294 | 271.92 | 211.634 | 1926.71 | 1682.09 | 5275.12 | 5010.02 | 12127.8 | 11249.2 | 19119.6 | 17899.7 |
| 700.581 | 273.328 | 212.485 | 1950.01 | 1687.04 | 5237.05 | 5022.47 | 11994.6 | 11274.1 | 19109.6 | 17935.6 |
| 700.868 | 290.233 | 213.338 | 1983.92 | 1692 | 5353.75 | 5034.94 | 12202.4 | 11298.9 | 19284.5 | 17971.5 |
| 701.155 | 293.803 | 214.193 | 1999.51 | 1696.97 | 5369.99 | 5047.42 | 12089.7 | 11323.8 | 19289.9 | 18007.5 |
| 701.442 | 263.52 | 215.051 | 1959.95 | 1701.95 | 5381.88 | 5059.92 | 12215.2 | 11348.7 | 19187.7 | 18043.4 |
| 701.729 | 301.058 | 215.911 | 2003.56 | 1706.94 | 5294.03 | 5072.44 | 12315.8 | 11373.7 | 19272.7 | 18079.4 |
| 702.015 | 259.472 | 216.774 | 2001.12 | 1711.93 | 5432.11 | 5084.97 | 12269.8 | 11398.6 | 19388.4 | 18115.5 |
| 702.302 | 275.271 | 217.639 | 1995.89 | 1716.94 | 5465 | 5097.51 | 12271.4 | 11423.6 | 19512.2 | 18151.5 |
| 702.589 | 259.106 | 218.506 | 1952.4 | 1721.95 | 5507.45 | 5110.08 | 12396.4 | 11448.6 | 19545.8 | 18187.6 |
| 702.876 | 304.574 | 219.376 | 2015.75 | 1726.97 | 5431.77 | 5122.66 | 12405.1 | 11473.6 | 19441.7 | 18223.7 |
| 703.162 | 307.72 | 220.249 | 2013.08 | 1732 | 5505.25 | 5135.25 | 12536.3 | 11498.7 | 19622.6 | 18259.8 |
| 703.449 | 299.362 | 221.124 | 2009 | 1737.04 | 5529.56 | 5147.86 | 12490.3 | 11523.8 | 19537.8 | 18296 |
| 703.736 | 281.264 | 222.002 | 1991 | 1742.09 | 5518.27 | 5160.49 | 12505.2 | 11548.9 | 19685.1 | 18332.2 |
| 704.022 | 288.296 | 222.882 | 2002.15 | 1747.15 | 5513.72 | 5173.13 | 12378.4 | 11574 | 19615.5 | 18368.4 |
| 704.309 | 277.542 | 223.765 | 2063.98 | 1752.21 | 5542.85 | 5185.79 | 12448.7 | 11599.2 | 19545.9 | 18404.6 |
| 704.596 | 294.555 | 224.65 | 1999.69 | 1757.29 | 5413.31 | 5198.46 | 12348 | 11624.3 | 19599.7 | 18440.9 |
| 704.882 | 279.172 | 225.538 | 2054.84 | 1762.37 | 5488.77 | 5211.15 | 12539.6 | 11649.5 | 19734.7 | 18477.2 |
| 705.169 | 272.91 | 226.428 | 1968.36 | 1767.46 | 5588.37 | 5223.86 | 12490.8 | 11674.8 | 19721.5 | 18513.5 |
| 705.456 | 298.803 | 227.321 | 2047.64 | 1772.56 | 5535.43 | 5236.58 | 12600.6 | 11700 | 19679.9 | 18549.8 |
| 705.742 | 278.248 | 228.216 | 2035.25 | 1777.67 | 5484.7 | 5249.31 | 12669.1 | 11725.3 | 19628.9 | 18586.2 |
| 706.029 | 284.19 | 229.114 | 2058.36 | 1782.79 | 5505.52 | 5262.07 | 12581.8 | 11750.5 | 19627 | 18622.6 |
| 706.315 | 288.742 | 230.014 | 2045.55 | 1787.92 | 5638.44 | 5274.83 | 12557.7 | 11775.9 | 19825.9 | 18659 |
| 706.602 | 305.849 | 230.917 | 2025.98 | 1793.05 | 5655.89 | 5287.62 | 12688.2 | 11801.2 | 19899.1 | 18695.4 |
| 706.888 | 297.87 | 231.822 | 2060.06 | 1798.2 | 5710.83 | 5300.42 | 12778.1 | 11826.6 | 19828.1 | 18731.9 |
| 707.175 | 280.734 | 232.73 | 2061.8 | 1803.35 | 5713.59 | 5313.23 | 12856.7 | 11852 | 20146.3 | 18768.4 |
| 707.462 | 287.212 | 233.641 | 2034.95 | 1808.52 | 5763.17 | 5326.06 | 12810.9 | 11877.4 | 19894.2 | 18804.9 |
| 707.748 | 302.714 | 234.554 | 2062.45 | 1813.69 | 5683.67 | 5338.91 | 12775.1 | 11902.8 | 19980.5 | 18841.5 |
| 708.034 | 269.544 | 235.469 | 2105.98 | 1818.87 | 5778.61 | 5351.77 | 12851.8 | 11928.2 | 20327.7 | 18878 |
| 708.321 | 294.509 | 236.387 | 2066.5 | 1824.06 | 5806.53 | 5364.64 | 12884.9 | 11953.7 | 20232.3 | 18914.6 |
| 708.607 | 296.357 | 237.308 | 2099.5 | 1829.25 | 5847.35 | 5377.54 | 12911.6 | 11979.2 | 20356.4 | 18951.2 |
| 708.894 | 321.365 | 238.231 | 2118.3 | 1834.46 | 5788.29 | 5390.44 | 12883.6 | 12004.8 | 20254 | 18987.9 |
| 709.18 | 293.621 | 239.157 | 2112.45 | 1839.67 | 5782.8 | 5403.37 | 12954.2 | 12030.3 | 20167.4 | 19024.5 |
| 709.467 | 318.504 | 240.086 | 2096.01 | 1844.9 | 5735.93 | 5416.31 | 12976.6 | 12055.9 | 20224.3 | 19061.2 |
| 709.753 | 308.882 | 241.017 | 2149.54 | 1850.13 | 5800.49 | 5429.26 | 13105.2 | 12081.5 | 20309.5 | 19098 |
| 710.04 | 306.436 | 241.95 | 2131.08 | 1855.37 | 5880.2 | 5442.23 | 13131.8 | 12107.1 | 20300.4 | 19134.7 |
| 710.326 | 299.167 | 242.886 | 2083.56 | 1860.62 | 5932.07 | 5455.22 | 12959.4 | 12132.7 | 20331.5 | 19171.5 |
| 710.612 | 306.873 | 243.825 | 2141.07 | 1865.88 | 5833.05 | 5468.22 | 13111.4 | 12158.4 | 20424.2 | 19208.3 |
| 710.899 | 302.54 | 244.766 | 2100.58 | 1871.15 | 5898.18 | 5481.23 | 12986.4 | 12184.1 | 20348.9 | 19245.1 |
| 711.185 | 304.644 | 245.71 | 2109.75 | 1876.42 | 6022.34 | 5494.26 | 13157.3 | 12209.8 | 20617.8 | 19282 |
| 711.471 | 313.885 | 246.656 | 2159.89 | 1881.71 | 5974.4 | 5507.31 | 13276.9 | 12235.5 | 20670.9 | 19318.8 |
| 711.758 | 314.952 | 247.606 | 2125.85 | 1887 | 5982.58 | 5520.37 | 13214.5 | 12261.3 | 20706.2 | 19355.7 |
| 712.044 | 319.356 | 248.557 | 2174.11 | 1892.31 | 6034.47 | 5533.45 | 13365.8 | 12287 | 20915.6 | 19392.6 |
| 712.33 | 310.366 | 249.512 | 2175.46 | 1897.62 | 6034.93 | 5546.54 | 13326.2 | 12312.9 | 20717.8 | 19429.6 |
| 712.616 | 326.442 | 250.468 | 2166.65 | 1902.94 | 6051.23 | 5559.65 | 13341.1 | 12338.7 | 20804.9 | 19466.5 |
| 712.903 | 303.952 | 251.428 | 2160.31 | 1908.27 | 6037.36 | 5572.77 | 13401.5 | 12364.5 | 20916.7 | 19503.5 |
| 713.189 | 325.742 | 252.39 | 2171.62 | 1913.61 | 6042.79 | 5585.91 | 13409.8 | 12390.4 | 20896.2 | 19540.5 |
| 713.475 | 323.433 | 253.355 | 2172.57 | 1918.95 | 6021.21 | 5599.06 | 13257 | 12416.3 | 20869.4 | 19577.6 |
| 713.761 | 316.95 | 254.322 | 2193.57 | 1924.31 | 6027.02 | 5612.23 | 13416.1 | 12442.2 | 20898.1 | 19614.7 |
| 714.048 | 336.062 | 255.292 | 2202.27 | 1929.67 | 6143.58 | 5625.42 | 13394.3 | 12468.1 | 20933.8 | 19651.8 |
| 714.334 | 327.549 | 256.264 | 2201.23 | 1935.05 | 5970.36 | 5638.61 | 13528.4 | 12494.1 | 21056.1 | 19688.9 |
| 714.62 | 318.115 | 257.24 | 2174.2 | 1940.43 | 5962.88 | 5651.83 | 13402.5 | 12520.1 | 20909.2 | 19726 |
| 714.906 | 315.796 | 258.218 | 2204.43 | 1945.82 | 6169.62 | 5665.06 | 13516.6 | 12546.1 | 21099.2 | 19763.2 |
| 715.192 | 328.538 | 259.198 | 2197.86 | 1951.22 | 6123.97 | 5678.3 | 13600.7 | 12572.1 | 21061.8 | 19800.3 |
| 715.479 | 313.502 | 260.181 | 2197.72 | 1956.62 | 6109.99 | 5691.56 | 13496.5 | 12598.1 | 21219.1 | 19837.5 |
| 715.765 | 321.784 | 261.167 | 2204.82 | 1962.04 | 6081.73 | 5704.84 | 13523.2 | 12624.2 | 21106.7 | 19874.8 |
| 716.051 | 320.439 | 262.155 | 2244.45 | 1967.47 | 6170.26 | 5718.13 | 13629.1 | 12650.3 | 20993.9 | 19912 |
| 716.337 | 330.397 | 263.146 | 2216.76 | 1972.9 | 6180.59 | 5731.43 | 13639.1 | 12676.4 | 21208.9 | 19949.3 |
| 716.623 | 325.035 | 264.14 | 2177.41 | 1978.34 | 6125.46 | 5744.76 | 13517.1 | 12702.5 | 21131.7 | 19986.6 |
| 716.909 | 323.233 | 265.137 | 2259.86 | 1983.8 | 6178.28 | 5758.09 | 13509.8 | 12728.7 | 21351.7 | 20023.9 |
| 717.195 | 331.871 | 266.135 | 2253.39 | 1989.26 | 6154.52 | 5771.44 | 13599.1 | 12754.9 | 21356.9 | 20061.3 |
| 717.481 | 344.644 | 267.137 | 2271.99 | 1994.72 | 6204.4 | 5784.81 | 13700 | 12781.1 | 21198.3 | 20098.7 |
| 717.767 | 346.497 | 268.142 | 2270.57 | 2000.2 | 6176.17 | 5798.19 | 13736.9 | 12807.3 | 21441.5 | 20136 |
| 718.053 | 315.408 | 269.149 | 2259.75 | 2005.69 | 6230.9 | 5811.58 | 13846.2 | 12833.5 | 21303.2 | 20173.5 |
| 718.339 | 338.526 | 270.158 | 2267.61 | 2011.18 | 6263.43 | 5824.99 | 13797.9 | 12859.8 | 21405.6 | 20210.9 |
| 718.625 | 326.786 | 271.171 | 2240.79 | 2016.69 | 6314.17 | 5838.42 | 13815.7 | 12886.1 | 21538.6 | 20248.4 |
| 718.911 | 340.057 | 272.186 | 2304.72 | 2022.2 | 6334.22 | 5851.86 | 13790.3 | 12912.4 | 21789.7 | 20285.9 |
| 719.197 | 347.009 | 273.203 | 2263.6 | 2027.72 | 6346.46 | 5865.31 | 13949.3 | 12938.7 | 21683.4 | 20323.4 |
| 719.483 | 324.195 | 274.224 | 2241.34 | 2033.25 | 6366.47 | 5878.78 | 13818.4 | 12965.1 | 21491.8 | 20360.9 |
| 719.769 | 342.008 | 275.247 | 2325.07 | 2038.79 | 6383.78 | 5892.27 | 13911.2 | 12991.5 | 21539.5 | 20398.5 |
| 720.055 | 356.602 | 276.273 | 2276.8 | 2044.34 | 6288.8 | 5905.77 | 13932.6 | 13017.9 | 21584.5 | 20436 |
| 720.341 | 339.564 | 277.301 | 2262.06 | 2049.9 | 6343.99 | 5919.28 | 13770.3 | 13044.3 | 21394.4 | 20473.6 |
| 720.627 | 355.791 | 278.332 | 2337.83 | 2055.46 | 6438.24 | 5932.81 | 13962.8 | 13070.7 | 21792.6 | 20511.3 |
| 720.913 | 356.72 | 279.366 | 2312.73 | 2061.03 | 6380.57 | 5946.35 | 14010.6 | 13097.2 | 21866.9 | 20548.9 |
| 721.199 | 346.882 | 280.403 | 2322.5 | 2066.62 | 6428.25 | 5959.91 | 14043.7 | 13123.6 | 21660.6 | 20586.6 |
| 721.485 | 349.798 | 281.442 | 2351.8 | 2072.21 | 6323.91 | 5973.49 | 13984.9 | 13150.2 | 21691.8 | 20624.3 |
| 721.77 | 359.675 | 282.484 | 2289.04 | 2077.81 | 6330.47 | 5987.07 | 14006.3 | 13176.7 | 21489.8 | 20662 |
| 722.056 | 344.586 | 283.529 | 2332.51 | 2083.42 | 6367.74 | 6000.68 | 13940.4 | 13203.2 | 21764.2 | 20699.7 |
| 722.342 | 349.874 | 284.576 | 2333.37 | 2089.03 | 6352.82 | 6014.3 | 14070.1 | 13229.8 | 21864.6 | 20737.5 |
| 722.628 | 350.25 | 285.627 | 2341.54 | 2094.66 | 6449.49 | 6027.93 | 14093 | 13256.4 | 21915.5 | 20775.2 |
| 722.913 | 355.137 | 286.679 | 2320.92 | 2100.3 | 6367.82 | 6041.58 | 14193.5 | 13283 | 21841.1 | 20813 |
| 723.199 | 345.853 | 287.735 | 2302.2 | 2105.94 | 6404.89 | 6055.24 | 14293.9 | 13309.6 | 21914.4 | 20850.9 |
| 723.485 | 339.705 | 288.793 | 2384.86 | 2111.59 | 6508.02 | 6068.91 | 14096.1 | 13336.3 | 21869.9 | 20888.7 |
| 723.771 | 351.282 | 289.854 | 2360.73 | 2117.25 | 6489.12 | 6082.61 | 14125.9 | 13363 | 21999.5 | 20926.6 |
| 724.057 | 375.473 | 290.918 | 2353.13 | 2122.92 | 6465.02 | 6096.31 | 14317.8 | 13389.6 | 22030.9 | 20964.5 |
| 724.342 | 347.91 | 291.985 | 2395.28 | 2128.6 | 6475.35 | 6110.03 | 14273.8 | 13416.4 | 22132.3 | 21002.4 |
| 724.628 | 353.197 | 293.054 | 2339.12 | 2134.29 | 6433.82 | 6123.77 | 14096.3 | 13443.1 | 21896.6 | 21040.3 |
| 724.914 | 362.973 | 294.126 | 2334.59 | 2139.98 | 6523.99 | 6137.52 | 14169.2 | 13469.8 | 22094.5 | 21078.3 |
| 725.199 | 362.735 | 295.201 | 2397.99 | 2145.69 | 6537.64 | 6151.28 | 14251.4 | 13496.6 | 21896 | 21116.2 |
| 725.485 | 366.09 | 296.278 | 2374.59 | 2151.4 | 6603.42 | 6165.06 | 14384.9 | 13523.4 | 22261.4 | 21154.2 |
| 725.771 | 347.682 | 297.358 | 2377.22 | 2157.12 | 6421.54 | 6178.85 | 14204 | 13550.2 | 21976.8 | 21192.3 |
| 726.056 | 372.373 | 298.442 | 2394.31 | 2162.85 | 6539.89 | 6192.66 | 14353.5 | 13577.1 | 22166.1 | 21230.3 |
| 726.342 | 345.032 | 299.527 | 2408.42 | 2168.59 | 6528.02 | 6206.48 | 14296.8 | 13603.9 | 22442.7 | 21268.4 |
| 726.628 | 360.198 | 300.616 | 2357.52 | 2174.34 | 6667.75 | 6220.32 | 14353.8 | 13630.8 | 21976.5 | 21306.4 |
| 726.913 | 358.712 | 301.707 | 2407.3 | 2180.09 | 6676.81 | 6234.17 | 14544.9 | 13657.7 | 22505.2 | 21344.5 |
| 727.199 | 368.548 | 302.801 | 2396.08 | 2185.86 | 6583.97 | 6248.04 | 14360.8 | 13684.6 | 22277.8 | 21382.7 |
| 727.484 | 369.498 | 303.898 | 2362.34 | 2191.63 | 6685.97 | 6261.92 | 14360.3 | 13711.6 | 22200.3 | 21420.8 |
| 727.77 | 378.531 | 304.998 | 2395.15 | 2197.41 | 6599.92 | 6275.81 | 14449.3 | 13738.5 | 22305.7 | 21459 |
| 728.055 | 388.405 | 306.1 | 2393.03 | 2203.21 | 6621.33 | 6289.72 | 14449.1 | 13765.5 | 22474.8 | 21497.2 |
| 728.341 | 380.006 | 307.205 | 2422.36 | 2209.01 | 6674.83 | 6303.64 | 14446.9 | 13792.5 | 22381.4 | 21535.4 |
| 728.626 | 367.941 | 308.313 | 2434.96 | 2214.81 | 6769.58 | 6317.58 | 14571.1 | 13819.5 | 22582.3 | 21573.6 |
| 728.912 | 368.168 | 309.424 | 2401.53 | 2220.63 | 6702.07 | 6331.53 | 14614.7 | 13846.6 | 22442.3 | 21611.8 |
| 729.197 | 385.677 | 310.538 | 2430.6 | 2226.46 | 6728.16 | 6345.5 | 14562.1 | 13873.6 | 22670.3 | 21650.1 |
| 729.483 | 386.919 | 311.654 | 2480.43 | 2232.29 | 6708.69 | 6359.48 | 14547.6 | 13900.7 | 22617.5 | 21688.4 |
| 729.768 | 378.379 | 312.773 | 2400.36 | 2238.13 | 6661.87 | 6373.47 | 14543 | 13927.8 | 22586.6 | 21726.7 |
| 730.054 | 383.029 | 313.895 | 2454.96 | 2243.98 | 6789.94 | 6387.48 | 14660.1 | 13955 | 22760 | 21765 |
| 730.339 | 394.618 | 315.02 | 2412.24 | 2249.84 | 6702.89 | 6401.5 | 14611.7 | 13982.1 | 22609.5 | 21803.4 |
| 730.625 | 384.474 | 316.147 | 2442.11 | 2255.71 | 6731.91 | 6415.54 | 14525.8 | 14009.3 | 22505.9 | 21841.8 |
| 730.91 | 389.228 | 317.278 | 2455.68 | 2261.59 | 6775.98 | 6429.59 | 14685.9 | 14036.5 | 22406.4 | 21880.2 |
| 731.195 | 366.675 | 318.411 | 2464.43 | 2267.47 | 6772.19 | 6443.65 | 14693.3 | 14063.7 | 22647.8 | 21918.6 |
| 731.481 | 388.491 | 319.547 | 2485.58 | 2273.37 | 6851.42 | 6457.73 | 14726.7 | 14090.9 | 22876.9 | 21957 |
| 731.766 | 368.214 | 320.685 | 2488.08 | 2279.27 | 6768.52 | 6471.83 | 14720.5 | 14118.1 | 22744.9 | 21995.5 |
| 732.052 | 382.877 | 321.827 | 2431.33 | 2285.18 | 6822.88 | 6485.94 | 14577.5 | 14145.4 | 22679.1 | 22033.9 |
| 732.337 | 376.065 | 322.971 | 2474.33 | 2291.1 | 6797.24 | 6500.06 | 14670.9 | 14172.7 | 22794.9 | 22072.4 |
| 732.622 | 388.12 | 324.119 | 2494.69 | 2297.03 | 6767.15 | 6514.19 | 14753.2 | 14200 | 22927.1 | 22110.9 |
| 732.908 | 404.709 | 325.269 | 2450.68 | 2302.97 | 6797.88 | 6528.34 | 14686 | 14227.3 | 22907.1 | 22149.5 |
| 733.193 | 372.881 | 326.422 | 2506.41 | 2308.92 | 6908.27 | 6542.51 | 14682.8 | 14254.6 | 22851.4 | 22188 |
| 733.478 | 405.173 | 327.577 | 2505.22 | 2314.87 | 6872.57 | 6556.68 | 14955 | 14282 | 23019.4 | 22226.6 |
| 733.763 | 388.049 | 328.736 | 2552.82 | 2320.83 | 6846.82 | 6570.87 | 14980.2 | 14309.4 | 23110.2 | 22265.2 |
| 734.048 | 392.586 | 329.897 | 2522.17 | 2326.81 | 6969.64 | 6585.08 | 14942.7 | 14336.8 | 23188.5 | 22303.8 |
| 734.334 | 381.561 | 331.062 | 2531.15 | 2332.79 | 6870.12 | 6599.3 | 14875.9 | 14364.2 | 22959.5 | 22342.4 |
| 734.619 | 388.042 | 332.229 | 2506.64 | 2338.78 | 6879.7 | 6613.53 | 14780.6 | 14391.6 | 22897.6 | 22381 |
| 734.904 | 389.71 | 333.399 | 2524.32 | 2344.77 | 6914.1 | 6627.78 | 14863 | 14419.1 | 23050.4 | 22419.7 |
| 735.189 | 407.979 | 334.571 | 2511.76 | 2350.78 | 6993.66 | 6642.04 | 15161.7 | 14446.6 | 23303.3 | 22458.4 |
| 735.475 | 397.722 | 335.747 | 2493.1 | 2356.79 | 6911.73 | 6656.32 | 15212.8 | 14474.1 | 23141.2 | 22497.1 |
| 735.76 | 400.67 | 336.925 | 2546.64 | 2362.82 | 6999.58 | 6670.61 | 15045.9 | 14501.6 | 23392.9 | 22535.8 |
| 736.045 | 404.078 | 338.107 | 2586.16 | 2368.85 | 6968.04 | 6684.91 | 15114.2 | 14529.1 | 23178 | 22574.6 |
| 736.33 | 414.901 | 339.291 | 2551.78 | 2374.89 | 7004.88 | 6699.23 | 15106 | 14556.7 | 23285.8 | 22613.3 |
| 736.615 | 423.672 | 340.478 | 2544.19 | 2380.94 | 7020.72 | 6713.56 | 15194.2 | 14584.2 | 23438.7 | 22652.1 |
| 736.9 | 400.801 | 341.668 | 2607.8 | 2386.99 | 7048.09 | 6727.9 | 15158 | 14611.8 | 23201.2 | 22690.9 |
| 737.185 | 404.372 | 342.861 | 2569.33 | 2393.06 | 7055.74 | 6742.26 | 15120.5 | 14639.4 | 23474 | 22729.7 |
| 737.47 | 401.057 | 344.057 | 2549.26 | 2399.13 | 6989.37 | 6756.63 | 15120.4 | 14667.1 | 23246 | 22768.6 |
| 737.755 | 400.45 | 345.255 | 2574.15 | 2405.22 | 7094.48 | 6771.01 | 15057.6 | 14694.7 | 23373.7 | 22807.4 |
| 738.041 | 409.933 | 346.456 | 2592.51 | 2411.31 | 7091.89 | 6785.41 | 15125.5 | 14722.4 | 23409.1 | 22846.3 |
| 738.326 | 425.076 | 347.661 | 2614.88 | 2417.41 | 7116.86 | 6799.83 | 15357.6 | 14750.1 | 23763.7 | 22885.2 |
| 738.611 | 410.268 | 348.868 | 2629.4 | 2423.52 | 7135.34 | 6814.25 | 15194.2 | 14777.8 | 23593.3 | 22924.1 |
| 738.896 | 398.346 | 350.078 | 2631.8 | 2429.63 | 7079.27 | 6828.69 | 15379.5 | 14805.5 | 23505.4 | 22963 |
| 739.181 | 403.939 | 351.291 | 2629.45 | 2435.76 | 7110.95 | 6843.15 | 15216 | 14833.2 | 23720.4 | 23002 |
| 739.466 | 411.713 | 352.507 | 2586.62 | 2441.89 | 7197.5 | 6857.61 | 15423.3 | 14861 | 23654.2 | 23041 |
| 739.751 | 398.23 | 353.726 | 2616.02 | 2448.03 | 7179.26 | 6872.09 | 15367.2 | 14888.8 | 23689.7 | 23079.9 |
| 740.035 | 421.777 | 354.947 | 2638.63 | 2454.19 | 7234.41 | 6886.59 | 15325.4 | 14916.6 | 23795 | 23118.9 |
| 740.321 | 408.057 | 356.172 | 2687.03 | 2460.35 | 7155.94 | 6901.1 | 15458.6 | 14944.4 | 23766.1 | 23158 |
| 740.605 | 436.851 | 357.399 | 2693.38 | 2466.51 | 7277.95 | 6915.62 | 15680.7 | 14972.2 | 24132.1 | 23197 |
| 740.89 | 446.951 | 358.629 | 2673.54 | 2472.69 | 7294.85 | 6930.15 | 15435 | 15000.1 | 23737.6 | 23236.1 |
| 741.175 | 423.082 | 359.863 | 2719.15 | 2478.87 | 7306.32 | 6944.7 | 15512.8 | 15027.9 | 24053.8 | 23275.1 |
| 741.46 | 413.133 | 361.099 | 2627.37 | 2485.07 | 7216.1 | 6959.26 | 15480.1 | 15055.8 | 24089.1 | 23314.2 |
| 741.745 | 432.091 | 362.338 | 2739.07 | 2491.27 | 7361.5 | 6973.84 | 15670.4 | 15083.7 | 24147.6 | 23353.3 |
| 742.03 | 435.083 | 363.58 | 2680.04 | 2497.48 | 7343.68 | 6988.42 | 15714.9 | 15111.7 | 24231.7 | 23392.5 |
| 742.315 | 415.66 | 364.825 | 2671.02 | 2503.7 | 7209.01 | 7003.02 | 15548.9 | 15139.6 | 23895.8 | 23431.6 |
| 742.6 | 437.551 | 366.072 | 2741.5 | 2509.93 | 7410.16 | 7017.64 | 15869.8 | 15167.6 | 24188.6 | 23470.8 |
| 742.884 | 411.311 | 367.323 | 2720.74 | 2516.16 | 7351.77 | 7032.27 | 15889.5 | 15195.5 | 24271.3 | 23509.9 |
| 743.169 | 446.013 | 368.577 | 2754.35 | 2522.41 | 7463.33 | 7046.91 | 15908.3 | 15223.5 | 24311.8 | 23549.1 |
| 743.454 | 426.322 | 369.833 | 2739.83 | 2528.66 | 7512.41 | 7061.56 | 15934.7 | 15251.6 | 24449.5 | 23588.4 |
| 743.739 | 415.369 | 371.093 | 2726.84 | 2534.92 | 7432.67 | 7076.23 | 15816.2 | 15279.6 | 24453.3 | 23627.6 |
| 744.023 | 430.991 | 372.355 | 2692.95 | 2541.19 | 7425.76 | 7090.91 | 15742.6 | 15307.6 | 24507.1 | 23666.8 |
| 744.308 | 432.278 | 373.621 | 2769.84 | 2547.47 | 7503.59 | 7105.61 | 15914 | 15335.7 | 24567.2 | 23706.1 |
| 744.593 | 425.295 | 374.889 | 2798.65 | 2553.75 | 7470.62 | 7120.32 | 15962.4 | 15363.8 | 24507.4 | 23745.4 |
| 744.878 | 449.354 | 376.16 | 2785.53 | 2560.05 | 7530.13 | 7135.04 | 16043.2 | 15391.9 | 24511.3 | 23784.7 |
| 745.162 | 446.472 | 377.435 | 2763.27 | 2566.35 | 7541.15 | 7149.77 | 15994.3 | 15420 | 24600.3 | 23824 |
| 745.447 | 454.667 | 378.712 | 2787.38 | 2572.66 | 7522.23 | 7164.52 | 16112.8 | 15448.2 | 24824.5 | 23863.3 |
| 745.732 | 423.39 | 379.992 | 2779.17 | 2578.98 | 7577.12 | 7179.28 | 16128 | 15476.3 | 24745.8 | 23902.7 |
| 746.016 | 450.277 | 381.275 | 2756.63 | 2585.31 | 7551.06 | 7194.05 | 16286.3 | 15504.5 | 24531.7 | 23942 |
| 746.301 | 456.788 | 382.561 | 2759.74 | 2591.65 | 7498.26 | 7208.84 | 15970.1 | 15532.7 | 24777.9 | 23981.4 |
| 746.586 | 443.122 | 383.85 | 2793.97 | 2597.99 | 7526.31 | 7223.64 | 16042 | 15560.9 | 24366.7 | 24020.8 |
| 746.87 | 456.943 | 385.142 | 2803.03 | 2604.34 | 7532.84 | 7238.45 | 16129.3 | 15589.1 | 24936 | 24060.2 |
| 747.155 | 467.988 | 386.436 | 2767.19 | 2610.71 | 7629.2 | 7253.28 | 16203.7 | 15617.4 | 24711.2 | 24099.7 |
| 747.439 | 465.224 | 387.734 | 2829.76 | 2617.08 | 7625.46 | 7268.12 | 16338.7 | 15645.6 | 25095.4 | 24139.1 |
| 747.724 | 444.397 | 389.035 | 2752.54 | 2623.45 | 7520.95 | 7282.97 | 16199.7 | 15673.9 | 24774.4 | 24178.6 |
| 748.009 | 437.196 | 390.339 | 2774.35 | 2629.84 | 7618.54 | 7297.83 | 16221.2 | 15702.2 | 24762.6 | 24218.1 |
| 748.293 | 444.201 | 391.645 | 2830.47 | 2636.24 | 7670.87 | 7312.71 | 16407.7 | 15730.5 | 25043.5 | 24257.6 |
| 748.578 | 457.044 | 392.955 | 2781.74 | 2642.64 | 7630.2 | 7327.6 | 16265.9 | 15758.9 | 25200.7 | 24297.1 |
| 748.862 | 476.534 | 394.268 | 2866.36 | 2649.05 | 7740.46 | 7342.51 | 16390.9 | 15787.2 | 25093.5 | 24336.6 |
| 749.147 | 460.715 | 395.583 | 2876.56 | 2655.47 | 7757.58 | 7357.42 | 16434.4 | 15815.6 | 25208.8 | 24376.2 |
| 749.431 | 472.432 | 396.902 | 2865.96 | 2661.9 | 7730.85 | 7372.35 | 16533.7 | 15844 | 25371.6 | 24415.7 |
| 749.716 | 464.003 | 398.224 | 2899.29 | 2668.34 | 7720.71 | 7387.29 | 16331.9 | 15872.4 | 25246.2 | 24455.3 |
| 750 | 477.598 | 399.548 | 2860.99 | 2674.78 | 7736.01 | 7402.25 | 16327.9 | 15900.8 | 25080.7 | 24494.9 |
| 750.284 | 468.658 | 400.876 | 2863.59 | 2681.24 | 7681.92 | 7417.22 | 16464.5 | 15929.2 | 25242.4 | 24534.5 |
| 750.569 | 478.789 | 402.206 | 2818.89 | 2687.7 | 7839.79 | 7432.2 | 16633 | 15957.6 | 25162.7 | 24574.1 |
| 750.853 | 474.49 | 403.54 | 2874.69 | 2694.17 | 7814.43 | 7447.19 | 16563.3 | 15986.1 | 25377.7 | 24613.8 |
| 751.138 | 478.969 | 404.876 | 2923.6 | 2700.65 | 7779.29 | 7462.19 | 16613.1 | 16014.6 | 25440.7 | 24653.4 |
| 751.422 | 461.241 | 406.216 | 2946.26 | 2707.13 | 7859.84 | 7477.21 | 16713.5 | 16043.1 | 25592.5 | 24693.1 |
| 751.706 | 480.596 | 407.558 | 2878.88 | 2713.63 | 7778.48 | 7492.24 | 16628.5 | 16071.6 | 25500.5 | 24732.8 |
| 751.991 | 484.134 | 408.904 | 2893.3 | 2720.13 | 7777.95 | 7507.29 | 16664.7 | 16100.1 | 25413.7 | 24772.5 |
| 752.275 | 460.267 | 410.252 | 2922.62 | 2726.64 | 7804.4 | 7522.35 | 16437.2 | 16128.7 | 25450.9 | 24812.2 |
| 752.559 | 477.767 | 411.604 | 2904.31 | 2733.16 | 7816.34 | 7537.41 | 16565.6 | 16157.3 | 25308.5 | 24851.9 |
| 752.844 | 454.417 | 412.958 | 2893.2 | 2739.69 | 7893.83 | 7552.5 | 16703.9 | 16185.9 | 25647.8 | 24891.7 |
| 753.128 | 479.842 | 414.316 | 2946.83 | 2746.23 | 8011.19 | 7567.59 | 16845.7 | 16214.5 | 25741.9 | 24931.5 |
| 753.412 | 492.195 | 415.676 | 2973.19 | 2752.77 | 7895.35 | 7582.7 | 16693.1 | 16243.1 | 25488.7 | 24971.2 |
| 753.697 | 481.228 | 417.04 | 2946.2 | 2759.33 | 7897.85 | 7597.82 | 16766.6 | 16271.7 | 25590.5 | 25011 |
| 753.981 | 478.387 | 418.406 | 2936.49 | 2765.89 | 7925.09 | 7612.95 | 16901.9 | 16300.3 | 25928.7 | 25050.8 |
| 754.265 | 480.755 | 419.776 | 2931.46 | 2772.46 | 7919.04 | 7628.09 | 16829.9 | 16329 | 25761.8 | 25090.6 |
| 754.549 | 471.033 | 421.149 | 2960.28 | 2779.03 | 7861.91 | 7643.25 | 16758.2 | 16357.7 | 25692.3 | 25130.5 |
| 754.834 | 520.303 | 422.524 | 2990.05 | 2785.62 | 7898.19 | 7658.42 | 16804.9 | 16386.4 | 25610 | 25170.3 |
| 755.118 | 497.016 | 423.903 | 2946.11 | 2792.21 | 7889.96 | 7673.6 | 16759.4 | 16415.1 | 25724.6 | 25210.2 |
| 755.402 | 468.67 | 425.285 | 2924.64 | 2798.82 | 7987.52 | 7688.79 | 16769.8 | 16443.8 | 25825.9 | 25250.1 |
| 755.686 | 505.207 | 426.669 | 2995.68 | 2805.43 | 8010.62 | 7704 | 16996.3 | 16472.6 | 25937 | 25290 |
| 755.97 | 492.236 | 428.057 | 2976.63 | 2812.05 | 7925.83 | 7719.22 | 16894.3 | 16501.3 | 25790.9 | 25329.9 |
| 756.254 | 497.774 | 429.448 | 2960.51 | 2818.67 | 7976.17 | 7734.45 | 16726 | 16530.1 | 25931.1 | 25369.8 |
| 756.538 | 514.676 | 430.842 | 2998.88 | 2825.31 | 8073.2 | 7749.69 | 17114.7 | 16558.9 | 26038.3 | 25409.7 |
| 756.823 | 514.935 | 432.238 | 3050.79 | 2831.95 | 8019.21 | 7764.95 | 17071.9 | 16587.7 | 25983.3 | 25449.7 |
| 757.107 | 481.85 | 433.638 | 2999.74 | 2838.6 | 7941.02 | 7780.22 | 16819.9 | 16616.5 | 25762.9 | 25489.7 |
| 757.391 | 502.953 | 435.041 | 2948.97 | 2845.26 | 8069.92 | 7795.5 | 17002 | 16645.4 | 25940.5 | 25529.6 |
| 757.675 | 501.206 | 436.447 | 2993.01 | 2851.93 | 8099.07 | 7810.79 | 17144.1 | 16674.2 | 25950.5 | 25569.6 |
| 757.959 | 509.596 | 437.856 | 2999.45 | 2858.61 | 8081.1 | 7826.09 | 16982.3 | 16703.1 | 25887.5 | 25609.6 |
| 758.243 | 517.317 | 439.268 | 3053.02 | 2865.29 | 8159.33 | 7841.41 | 17234.3 | 16732 | 26331.1 | 25649.7 |
| 758.527 | 509.355 | 440.683 | 3002.68 | 2871.98 | 8028.72 | 7856.74 | 17064 | 16760.9 | 26011.7 | 25689.7 |
| 758.811 | 489.986 | 442.101 | 3064.24 | 2878.68 | 8060.48 | 7872.08 | 17037 | 16789.8 | 26251.8 | 25729.8 |
| 759.095 | 489.819 | 443.523 | 3057.41 | 2885.39 | 8161.34 | 7887.43 | 17204.3 | 16818.7 | 26237.9 | 25769.8 |
| 759.379 | 480.622 | 444.947 | 3040.11 | 2892.11 | 8085.11 | 7902.8 | 17212.3 | 16847.7 | 26338.2 | 25809.9 |
| 759.663 | 521.636 | 446.374 | 3041.66 | 2898.83 | 8223.03 | 7918.18 | 17427 | 16876.6 | 26604 | 25850 |
| 759.947 | 509.989 | 447.805 | 3083.15 | 2905.57 | 8154.99 | 7933.56 | 17282.6 | 16905.6 | 26200.2 | 25890.1 |
| 760.231 | 502.524 | 449.238 | 3089.95 | 2912.31 | 8166.83 | 7948.97 | 17352.7 | 16934.6 | 26469.1 | 25930.2 |
| 760.515 | 513.359 | 450.675 | 3109.42 | 2919.06 | 8361.14 | 7964.38 | 17471.8 | 16963.6 | 26598.1 | 25970.3 |
| 760.799 | 505.816 | 452.114 | 3085.78 | 2925.81 | 8203.82 | 7979.8 | 17334.1 | 16992.7 | 26269.7 | 26010.5 |
| 761.082 | 521.436 | 453.557 | 3144.35 | 2932.58 | 8229.4 | 7995.24 | 17336.9 | 17021.7 | 26610 | 26050.6 |
| 761.366 | 522.882 | 455.002 | 3109.36 | 2939.35 | 8301.85 | 8010.69 | 17544.7 | 17050.7 | 26708.6 | 26090.8 |
| 761.65 | 542.443 | 456.451 | 3141.64 | 2946.14 | 8261.79 | 8026.15 | 17345.4 | 17079.8 | 26510.5 | 26131 |
| 761.934 | 525.918 | 457.903 | 3061.58 | 2952.93 | 8216.04 | 8041.62 | 17393.1 | 17108.9 | 26454.2 | 26171.2 |
| 762.218 | 537.955 | 459.358 | 3142.59 | 2959.72 | 8328.2 | 8057.11 | 17506.1 | 17138 | 26517.5 | 26211.4 |
| 762.502 | 535.521 | 460.816 | 3115.82 | 2966.53 | 8340.7 | 8072.6 | 17418 | 17167.1 | 26461.7 | 26251.6 |
| 762.786 | 509.98 | 462.277 | 3098.74 | 2973.34 | 8292.18 | 8088.11 | 17460.6 | 17196.2 | 26665.2 | 26291.8 |
| 763.069 | 513.153 | 463.741 | 3172.86 | 2980.16 | 8298.59 | 8103.63 | 17534.2 | 17225.4 | 26758 | 26332.1 |
| 763.353 | 510.753 | 465.208 | 3122.41 | 2986.99 | 8304.4 | 8119.16 | 17550.5 | 17254.5 | 26738.6 | 26372.4 |
| 763.637 | 524.447 | 466.679 | 3156.64 | 2993.83 | 8317.56 | 8134.71 | 17569.9 | 17283.7 | 26807 | 26412.6 |
| 763.921 | 515.765 | 468.152 | 3093.34 | 3000.68 | 8301.9 | 8150.26 | 17515.3 | 17312.9 | 26779.6 | 26452.9 |
| 764.204 | 530.919 | 469.629 | 3132.5 | 3007.53 | 8341.56 | 8165.83 | 17510.3 | 17342.1 | 26827.4 | 26493.2 |
| 764.488 | 539.408 | 471.108 | 3132.09 | 3014.39 | 8338.25 | 8181.41 | 17738.3 | 17371.3 | 26891.8 | 26533.5 |
| 764.772 | 536.521 | 472.591 | 3140.42 | 3021.26 | 8312.63 | 8197 | 17590.6 | 17400.5 | 26794.5 | 26573.8 |
| 765.056 | 539.227 | 474.077 | 3158.24 | 3028.14 | 8345.08 | 8212.6 | 17663.4 | 17429.8 | 26831.4 | 26614.2 |
| 765.339 | 551.701 | 475.566 | 3139.05 | 3035.03 | 8405.92 | 8228.22 | 17653.4 | 17459.1 | 26951.5 | 26654.5 |
| 765.623 | 521.783 | 477.058 | 3149.08 | 3041.92 | 8502.64 | 8243.84 | 17907.8 | 17488.3 | 26979.7 | 26694.9 |
| 765.907 | 573.229 | 478.553 | 3158.11 | 3048.82 | 8420.67 | 8259.48 | 17881.9 | 17517.6 | 26776.4 | 26735.3 |
| 766.19 | 537.549 | 480.051 | 3212.28 | 3055.74 | 8516.47 | 8275.12 | 17838.9 | 17546.9 | 27179.5 | 26775.6 |
| 766.474 | 557.337 | 481.552 | 3164.18 | 3062.65 | 8480.25 | 8290.78 | 17818.2 | 17576.2 | 27049.8 | 26816 |
| 766.757 | 563.69 | 483.057 | 3235.86 | 3069.58 | 8438.33 | 8306.46 | 17875.7 | 17605.6 | 27122.7 | 26856.5 |
| 767.041 | 528.869 | 484.564 | 3218.7 | 3076.51 | 8442.76 | 8322.14 | 17877.6 | 17634.9 | 27299.5 | 26896.9 |
| 767.324 | 560.127 | 486.075 | 3247 | 3083.46 | 8648.82 | 8337.83 | 18014.3 | 17664.3 | 27317.9 | 26937.3 |
| 767.608 | 538.487 | 487.588 | 3226.06 | 3090.41 | 8592.1 | 8353.54 | 17921.1 | 17693.7 | 27203.8 | 26977.7 |
| 767.891 | 555.573 | 489.105 | 3246.12 | 3097.36 | 8661.02 | 8369.25 | 18196.5 | 17723 | 27411.1 | 27018.2 |
| 768.175 | 547.025 | 490.625 | 3235.96 | 3104.33 | 8646.71 | 8384.98 | 18104.1 | 17752.4 | 27397.5 | 27058.7 |
| 768.459 | 549.554 | 492.148 | 3258.06 | 3111.3 | 8612.46 | 8400.72 | 17932 | 17781.9 | 27135.3 | 27099.1 |
| 768.742 | 574.755 | 493.674 | 3263.65 | 3118.28 | 8563.34 | 8416.47 | 18051.8 | 17811.3 | 27183.2 | 27139.6 |
| 769.026 | 578.32 | 495.204 | 3228.17 | 3125.27 | 8486.72 | 8432.24 | 18021.7 | 17840.7 | 27187.4 | 27180.1 |
| 769.309 | 553.879 | 496.736 | 3281.33 | 3132.27 | 8628.96 | 8448.01 | 18103.3 | 17870.2 | 27380.4 | 27220.6 |
| 769.592 | 563.324 | 498.272 | 3260.9 | 3139.28 | 8709.51 | 8463.79 | 18167.3 | 17899.7 | 27723.7 | 27261.2 |
| 769.876 | 580.483 | 499.81 | 3246.8 | 3146.29 | 8544.59 | 8479.59 | 18148.8 | 17929.1 | 27333.2 | 27301.7 |
| 770.159 | 547.066 | 501.352 | 3302.01 | 3153.31 | 8691.53 | 8495.4 | 18253.2 | 17958.6 | 27632.2 | 27342.2 |
| 770.443 | 573.684 | 502.897 | 3320.87 | 3160.34 | 8687.15 | 8511.22 | 18061.2 | 17988.2 | 27391.3 | 27382.8 |
| 770.726 | 571.862 | 504.445 | 3246.25 | 3167.38 | 8792.69 | 8527.04 | 18212.9 | 18017.7 | 27699.3 | 27423.4 |
| 771.01 | 578.371 | 505.996 | 3369.39 | 3174.42 | 8750.16 | 8542.89 | 18202.3 | 18047.2 | 27722.8 | 27463.9 |
| 771.293 | 568.633 | 507.551 | 3290.64 | 3181.47 | 8731.03 | 8558.74 | 18425.8 | 18076.8 | 27812.4 | 27504.5 |
| 771.576 | 559.432 | 509.108 | 3313.71 | 3188.53 | 8751.45 | 8574.6 | 18392.7 | 18106.3 | 27668.3 | 27545.1 |
| 771.86 | 553.622 | 510.669 | 3315.75 | 3195.6 | 8759.56 | 8590.47 | 18437.2 | 18135.9 | 27862.3 | 27585.7 |
| 772.143 | 586.902 | 512.232 | 3265.44 | 3202.68 | 8815.73 | 8606.36 | 18480.9 | 18165.5 | 27862.5 | 27626.4 |
| 772.426 | 571.66 | 513.799 | 3309.81 | 3209.76 | 8739.62 | 8622.25 | 18348.7 | 18195.1 | 27743.6 | 27667 |
| 772.709 | 564.009 | 515.369 | 3346.4 | 3216.86 | 8904.78 | 8638.16 | 18566.8 | 18224.7 | 27944.5 | 27707.6 |
| 772.993 | 565.997 | 516.942 | 3375.65 | 3223.95 | 8877.15 | 8654.08 | 18536.6 | 18254.4 | 27980.5 | 27748.3 |
| 773.276 | 575.995 | 518.519 | 3340.85 | 3231.06 | 8866.15 | 8670.01 | 18376.3 | 18284 | 27930.6 | 27789 |
| 773.559 | 569.469 | 520.098 | 3378.76 | 3238.18 | 8918.52 | 8685.95 | 18405 | 18313.7 | 28077.1 | 27829.6 |
| 773.842 | 575.92 | 521.681 | 3361.54 | 3245.3 | 8902.37 | 8701.9 | 18471.1 | 18343.3 | 28060.1 | 27870.3 |
| 774.126 | 566.012 | 523.267 | 3410.85 | 3252.43 | 8991.32 | 8717.86 | 18634.9 | 18373 | 28394.1 | 27911 |
| 774.409 | 580.774 | 524.856 | 3435.95 | 3259.57 | 8927.39 | 8733.83 | 18770.1 | 18402.7 | 28332.6 | 27951.7 |
| 774.692 | 573.381 | 526.447 | 3376.11 | 3266.72 | 8941.78 | 8749.82 | 18677.1 | 18432.4 | 27946.7 | 27992.4 |
| 774.975 | 597.733 | 528.043 | 3411.74 | 3273.87 | 8916.21 | 8765.81 | 18684.9 | 18462.2 | 28253 | 28033.2 |
| 775.258 | 585.639 | 529.641 | 3455.07 | 3281.03 | 8935.8 | 8781.81 | 18660.3 | 18491.9 | 28170.9 | 28073.9 |
| 775.542 | 598.626 | 531.243 | 3382.22 | 3288.2 | 8956.79 | 8797.83 | 18685.2 | 18521.6 | 28300.2 | 28114.6 |
| 775.825 | 566.178 | 532.847 | 3428.99 | 3295.38 | 8962.89 | 8813.86 | 18591.5 | 18551.4 | 28298.8 | 28155.4 |
| 776.108 | 582.244 | 534.455 | 3445.46 | 3302.56 | 9061.65 | 8829.89 | 18839.6 | 18581.2 | 28537.7 | 28196.1 |
| 776.391 | 602.082 | 536.066 | 3381.61 | 3309.76 | 9015.7 | 8845.94 | 18849.3 | 18611 | 28389.3 | 28236.9 |
| 776.674 | 592.529 | 537.68 | 3422.04 | 3316.96 | 8973.77 | 8862 | 18762.4 | 18640.8 | 28284.6 | 28277.7 |
| 776.957 | 619.068 | 539.298 | 3434.92 | 3324.16 | 8999.17 | 8878.07 | 18835 | 18670.6 | 28375.6 | 28318.5 |
| 777.24 | 599.862 | 540.918 | 3411.37 | 3331.38 | 9023.97 | 8894.15 | 18744.3 | 18700.4 | 28508.1 | 28359.3 |
| 777.523 | 576.425 | 542.542 | 3437.75 | 3338.6 | 9165.68 | 8910.24 | 18836.6 | 18730.2 | 28428.2 | 28400.1 |
| 777.806 | 622.529 | 544.169 | 3395.7 | 3345.84 | 9087.24 | 8926.34 | 18922.7 | 18760.1 | 28604.9 | 28440.9 |
| 778.089 | 603.095 | 545.799 | 3444.57 | 3353.07 | 9094.21 | 8942.45 | 19053.4 | 18789.9 | 28597.6 | 28481.8 |
| 778.372 | 594.195 | 547.432 | 3509.89 | 3360.32 | 9127.25 | 8958.57 | 18922.2 | 18819.8 | 28824 | 28522.6 |
| 778.655 | 642.116 | 549.068 | 3483.36 | 3367.57 | 9147.06 | 8974.7 | 18953.2 | 18849.7 | 28566.4 | 28563.5 |
| 778.938 | 607.644 | 550.708 | 3492.78 | 3374.84 | 9291.39 | 8990.85 | 19103 | 18879.6 | 28827.8 | 28604.3 |
| 779.221 | 625.858 | 552.35 | 3552.94 | 3382.11 | 9220.1 | 9007 | 19221.2 | 18909.5 | 29080.9 | 28645.2 |
| 779.504 | 610.148 | 553.996 | 3457.15 | 3389.38 | 9254.06 | 9023.16 | 18976.4 | 18939.4 | 29056.7 | 28686 |
| 779.787 | 611.89 | 555.645 | 3477.33 | 3396.67 | 9266.25 | 9039.34 | 19177.2 | 18969.3 | 28829.8 | 28726.9 |
| 780.07 | 606.551 | 557.298 | 3541.48 | 3403.96 | 9215.55 | 9055.52 | 19332.3 | 18999.3 | 29051.9 | 28767.8 |
| 780.353 | 622.426 | 558.953 | 3564.56 | 3411.26 | 9304.65 | 9071.72 | 19385.8 | 19029.2 | 29208.8 | 28808.7 |
| 780.635 | 633.722 | 560.612 | 3518.99 | 3418.57 | 9294.01 | 9087.92 | 19306.3 | 19059.2 | 29186.9 | 28849.6 |
| 780.918 | 607.67 | 562.273 | 3549.35 | 3425.88 | 9276.01 | 9104.14 | 19361.7 | 19089.2 | 29030.6 | 28890.6 |
| 781.201 | 628.727 | 563.938 | 3590.74 | 3433.21 | 9359.67 | 9120.36 | 19405.2 | 19119.2 | 29390 | 28931.5 |
| 781.484 | 638.697 | 565.606 | 3520.01 | 3440.54 | 9327.01 | 9136.6 | 19268.1 | 19149.2 | 29209.2 | 28972.4 |
| 781.767 | 635.709 | 567.278 | 3610.5 | 3447.87 | 9417.24 | 9152.85 | 19418.3 | 19179.2 | 29520.6 | 29013.4 |
| 782.05 | 615.545 | 568.952 | 3492.59 | 3455.22 | 9329.15 | 9169.1 | 19414.8 | 19209.2 | 29343.8 | 29054.3 |
| 782.332 | 651.375 | 570.63 | 3610.43 | 3462.57 | 9401.18 | 9185.37 | 19530.5 | 19239.3 | 29358.2 | 29095.3 |
| 782.615 | 607.276 | 572.311 | 3562.2 | 3469.93 | 9374.93 | 9201.65 | 19356.1 | 19269.3 | 29402.7 | 29136.3 |
| 782.898 | 648.211 | 573.995 | 3572.84 | 3477.3 | 9296.37 | 9217.93 | 19392.6 | 19299.4 | 29234.2 | 29177.2 |
| 783.181 | 637.701 | 575.683 | 3559.82 | 3484.68 | 9482.11 | 9234.23 | 19501.5 | 19329.5 | 29427.1 | 29218.2 |
| 783.463 | 649.939 | 577.373 | 3591.72 | 3492.06 | 9381.46 | 9250.54 | 19689.7 | 19359.5 | 29328.7 | 29259.2 |
| 783.746 | 664.426 | 579.067 | 3572.42 | 3499.45 | 9344.26 | 9266.85 | 19556 | 19389.6 | 29487.2 | 29300.2 |
| 784.029 | 631.145 | 580.764 | 3590.86 | 3506.85 | 9325.3 | 9283.18 | 19382 | 19419.7 | 29292.9 | 29341.2 |
| 784.311 | 643.271 | 582.464 | 3594.77 | 3514.26 | 9549.19 | 9299.52 | 19703.1 | 19449.9 | 29547.6 | 29382.3 |
| 784.594 | 673.053 | 584.167 | 3661.82 | 3521.67 | 9364.73 | 9315.86 | 19675.8 | 19480 | 29371.3 | 29423.3 |
| 784.877 | 647.949 | 585.874 | 3691.82 | 3529.09 | 9544.8 | 9332.22 | 19685.8 | 19510.1 | 29456.2 | 29464.3 |
| 785.159 | 657.603 | 587.583 | 3722.35 | 3536.52 | 9528.38 | 9348.59 | 19633.9 | 19540.3 | 29651.4 | 29505.4 |
| 785.442 | 635.136 | 589.296 | 3627.12 | 3543.95 | 9510.74 | 9364.97 | 19770.8 | 19570.5 | 29724.9 | 29546.4 |
| 785.725 | 663.122 | 591.012 | 3624.84 | 3551.4 | 9525.78 | 9381.35 | 19864.1 | 19600.6 | 29675 | 29587.5 |
| 786.007 | 650.299 | 592.732 | 3645.84 | 3558.85 | 9498.89 | 9397.75 | 19767.8 | 19630.8 | 29574.2 | 29628.5 |
| 786.29 | 663.005 | 594.454 | 3622.86 | 3566.3 | 9482.03 | 9414.16 | 19683.9 | 19661 | 29634.8 | 29669.6 |
| 786.572 | 659.877 | 596.18 | 3656.3 | 3573.77 | 9589.33 | 9430.57 | 19871.9 | 19691.2 | 29882.6 | 29710.7 |
| 786.855 | 641.122 | 597.909 | 3671.12 | 3581.24 | 9567.17 | 9447 | 19852.3 | 19721.4 | 29654.8 | 29751.8 |
| 787.137 | 651.865 | 599.641 | 3682.74 | 3588.72 | 9605.99 | 9463.43 | 19985.9 | 19751.7 | 29980.4 | 29792.9 |
| 787.42 | 676.004 | 601.376 | 3719.1 | 3596.21 | 9495.27 | 9479.88 | 19812.2 | 19781.9 | 29828.4 | 29834 |
| 787.702 | 638.021 | 603.115 | 3638.73 | 3603.71 | 9587.9 | 9496.33 | 20034.7 | 19812.2 | 29824.5 | 29875.1 |
| 787.985 | 651.188 | 604.857 | 3666.77 | 3611.21 | 9619.56 | 9512.8 | 19836.8 | 19842.4 | 29677 | 29916.2 |
| 788.267 | 668.115 | 606.602 | 3723.56 | 3618.72 | 9705.01 | 9529.27 | 20018.9 | 19872.7 | 30186.8 | 29957.3 |
| 788.55 | 696.796 | 608.35 | 3745.8 | 3626.24 | 9816.92 | 9545.76 | 20038.2 | 19903 | 30120.7 | 29998.5 |
| 788.832 | 664.395 | 610.102 | 3745.75 | 3633.76 | 9687.48 | 9562.25 | 20054.2 | 19933.3 | 30035.8 | 30039.6 |
| 789.115 | 672.562 | 611.856 | 3763.01 | 3641.29 | 9676.94 | 9578.76 | 20118 | 19963.6 | 30180.9 | 30080.8 |
| 789.397 | 657.896 | 613.614 | 3751.3 | 3648.83 | 9806.73 | 9595.27 | 20092.4 | 19993.9 | 30204.8 | 30121.9 |
| 789.679 | 684.85 | 615.375 | 3748.13 | 3656.38 | 9712.66 | 9611.79 | 20160 | 20024.2 | 30201 | 30163.1 |
| 789.962 | 645.697 | 617.14 | 3724.99 | 3663.93 | 9694.89 | 9628.33 | 20150.4 | 20054.5 | 30346.8 | 30204.2 |
| 790.244 | 682.985 | 618.907 | 3683.8 | 3671.49 | 9666.97 | 9644.87 | 19779.7 | 20084.9 | 29882.6 | 30245.4 |
| 790.526 | 682.217 | 620.678 | 3788.47 | 3679.06 | 9709.8 | 9661.42 | 20301.1 | 20115.2 | 30386.8 | 30286.6 |
| 790.809 | 672.576 | 622.452 | 3740.27 | 3686.64 | 9813.67 | 9677.98 | 20259.9 | 20145.6 | 30285.1 | 30327.8 |
| 791.091 | 676.737 | 624.23 | 3742 | 3694.22 | 9812.59 | 9694.55 | 20205.2 | 20176 | 30364 | 30369 |
| 791.373 | 673.086 | 626.01 | 3718.97 | 3701.81 | 9812.68 | 9711.13 | 20261.2 | 20206.4 | 30306.4 | 30410.2 |
| 791.655 | 681.855 | 627.794 | 3763.84 | 3709.41 | 9719.64 | 9727.72 | 20161.9 | 20236.8 | 30035.3 | 30451.3 |
| 791.938 | 691.542 | 629.581 | 3816.13 | 3717.02 | 9852.05 | 9744.32 | 20337 | 20267.2 | 30549.6 | 30492.6 |
| 792.22 | 676.673 | 631.371 | 3756.35 | 3724.63 | 9810.26 | 9760.92 | 20109.5 | 20297.6 | 30198.1 | 30533.8 |
| 792.502 | 684.023 | 633.164 | 3821.29 | 3732.25 | 9950.93 | 9777.54 | 20265.6 | 20328 | 30369.6 | 30575 |
| 792.784 | 665.54 | 634.961 | 3796.31 | 3739.87 | 9823.58 | 9794.17 | 20370.8 | 20358.4 | 30564.2 | 30616.2 |
| 793.067 | 707.869 | 636.761 | 3794.57 | 3747.51 | 9801.67 | 9810.8 | 20291.4 | 20388.9 | 30461.2 | 30657.5 |
| 793.349 | 685.182 | 638.564 | 3772.83 | 3755.15 | 9914.48 | 9827.44 | 20398 | 20419.3 | 30555.2 | 30698.7 |
| 793.631 | 696.364 | 640.37 | 3869.78 | 3762.8 | 9927.28 | 9844.1 | 20405.5 | 20449.8 | 30548.5 | 30740 |
| 793.913 | 713.94 | 642.18 | 3899.05 | 3770.45 | 9971.85 | 9860.76 | 20513.3 | 20480.3 | 30533.9 | 30781.2 |
| 794.195 | 678.433 | 643.992 | 3868.15 | 3778.12 | 9920.15 | 9877.43 | 20332 | 20510.8 | 30605.4 | 30822.5 |
| 794.477 | 709.38 | 645.809 | 3921.92 | 3785.79 | 10049.2 | 9894.12 | 20635.2 | 20541.3 | 30919.5 | 30863.7 |
| 794.76 | 707.798 | 647.628 | 3878.69 | 3793.47 | 9935.48 | 9910.81 | 20428 | 20571.8 | 30548.3 | 30905 |
| 795.042 | 696.094 | 649.45 | 3872.3 | 3801.15 | 9980.6 | 9927.5 | 20574.7 | 20602.3 | 30860.4 | 30946.3 |
| 795.324 | 694.651 | 651.276 | 3892.57 | 3808.84 | 10077.5 | 9944.21 | 20599.6 | 20632.8 | 30983 | 30987.5 |
| 795.606 | 721.669 | 653.105 | 3892.93 | 3816.54 | 10007.9 | 9960.93 | 20733.7 | 20663.3 | 31028.8 | 31028.8 |
| 795.888 | 711.077 | 654.937 | 3923.26 | 3824.25 | 10014.5 | 9977.66 | 20722.5 | 20693.9 | 30943.5 | 31070.1 |
| 796.17 | 714.919 | 656.773 | 3883.09 | 3831.96 | 10003.3 | 9994.39 | 20611.1 | 20724.4 | 30929.3 | 31111.4 |
| 796.452 | 690.672 | 658.612 | 3858.27 | 3839.68 | 10113.9 | 10011.1 | 20886.6 | 20755 | 30876.8 | 31152.7 |
| 796.734 | 711.093 | 660.454 | 3903.24 | 3847.41 | 10102.8 | 10027.9 | 20674.4 | 20785.5 | 31011.6 | 31194 |
| 797.016 | 684.572 | 662.299 | 3909.92 | 3855.14 | 9940.71 | 10044.6 | 20870.7 | 20816.1 | 30766.2 | 31235.3 |
| 797.298 | 713.691 | 664.147 | 3871.06 | 3862.88 | 10064.6 | 10061.4 | 20791.5 | 20846.7 | 31032.8 | 31276.7 |
| 797.58 | 710.712 | 665.999 | 3913.32 | 3870.63 | 9965.79 | 10078.2 | 20727.8 | 20877.3 | 31003.6 | 31318 |
| 797.862 | 732.613 | 667.854 | 3922.37 | 3878.39 | 10177.9 | 10095 | 20706.5 | 20907.9 | 31247 | 31359.3 |
| 798.144 | 713.317 | 669.712 | 3912.55 | 3886.15 | 10200.1 | 10111.8 | 21042.4 | 20938.5 | 31468.7 | 31400.6 |
| 798.426 | 741.108 | 671.574 | 3960.96 | 3893.92 | 10150.2 | 10128.6 | 20952.2 | 20969.2 | 31247.5 | 31442 |
| 798.707 | 720.32 | 673.438 | 4004.26 | 3901.7 | 10177.2 | 10145.4 | 21036.7 | 20999.8 | 31279.6 | 31483.3 |
| 798.989 | 726.19 | 675.306 | 3889.96 | 3909.48 | 10254.2 | 10162.2 | 20853.9 | 21030.4 | 31337.1 | 31524.7 |
| 799.271 | 713.108 | 677.178 | 4005.96 | 3917.27 | 10172.4 | 10179.1 | 21096.9 | 21061.1 | 31468.6 | 31566 |
| 799.553 | 726.732 | 679.052 | 3978.2 | 3925.07 | 10197.7 | 10195.9 | 20960.9 | 21091.7 | 31229.8 | 31607.4 |
| 799.835 | 736.274 | 680.93 | 4043.14 | 3932.88 | 10202.9 | 10212.8 | 21081.7 | 21122.4 | 31424.9 | 31648.7 |
| 800.117 | 731.806 | 682.811 | 4073.49 | 3940.69 | 10240.6 | 10229.6 | 21093.1 | 21153.1 | 31437.7 | 31690.1 |
| 800.399 | 752.047 | 684.695 | 3911.94 | 3948.51 | 10207.2 | 10246.5 | 21280.8 | 21183.8 | 31424 | 31731.5 |
| 800.68 | 778.224 | 686.582 | 4035.84 | 3956.33 | 10437.1 | 10263.4 | 21008 | 21214.5 | 31626.7 | 31772.8 |
| 800.962 | 776.682 | 688.473 | 4035.4 | 3964.17 | 10273.2 | 10280.2 | 21037.2 | 21245.2 | 31554.4 | 31814.2 |
| 801.244 | 747.304 | 690.367 | 3988.58 | 3972.01 | 10329.1 | 10297.1 | 21258.8 | 21275.9 | 31580.3 | 31855.6 |
| 801.525 | 727.321 | 692.264 | 3985.34 | 3979.85 | 10205.2 | 10314 | 21165.9 | 21306.6 | 31638.9 | 31897 |
| 801.807 | 770.697 | 694.165 | 4018.39 | 3987.71 | 10385.8 | 10331 | 21269.5 | 21337.3 | 31863.8 | 31938.4 |
| 802.089 | 772.545 | 696.069 | 4082.52 | 3995.57 | 10282 | 10347.9 | 21316.5 | 21368 | 31669.8 | 31979.8 |
| 802.371 | 767.294 | 697.976 | 4083.92 | 4003.44 | 10508.1 | 10364.8 | 21490.4 | 21398.8 | 32045.7 | 32021.2 |
| 802.652 | 746.769 | 699.886 | 4092.03 | 4011.31 | 10403.6 | 10381.7 | 21301.8 | 21429.5 | 31703.5 | 32062.6 |
| 802.934 | 716.542 | 701.8 | 4039.8 | 4019.19 | 10476.8 | 10398.7 | 21352.7 | 21460.3 | 31865.1 | 32104 |
| 803.216 | 740.717 | 703.716 | 4050.34 | 4027.08 | 10551.1 | 10415.6 | 21276 | 21491.1 | 31653 | 32145.4 |
| 803.497 | 767.403 | 705.636 | 4124.29 | 4034.98 | 10401.8 | 10432.6 | 21288.9 | 21521.8 | 31977.3 | 32186.8 |
| 803.779 | 776.606 | 707.56 | 4156.45 | 4042.88 | 10429.3 | 10449.6 | 21331.7 | 21552.6 | 31932.6 | 32228.2 |
| 804.06 | 754.053 | 709.486 | 4080.95 | 4050.79 | 10397.3 | 10466.5 | 21447.9 | 21583.4 | 32151.5 | 32269.6 |
| 804.342 | 756.064 | 711.416 | 4151.23 | 4058.71 | 10616.7 | 10483.5 | 21588.6 | 21614.2 | 32196.6 | 32311.1 |
| 804.624 | 780.857 | 713.349 | 4085.66 | 4066.63 | 10627.2 | 10500.5 | 21457.7 | 21645 | 32309 | 32352.5 |
| 804.905 | 782.66 | 715.286 | 4133.5 | 4074.56 | 10611.9 | 10517.5 | 21600.5 | 21675.8 | 32503.2 | 32393.9 |
| 805.187 | 770.045 | 717.225 | 4124.96 | 4082.49 | 10535.6 | 10534.5 | 21904.7 | 21706.7 | 32360.3 | 32435.4 |
| 805.468 | 797.9 | 719.168 | 4122.52 | 4090.44 | 10646.8 | 10551.5 | 21467.9 | 21737.5 | 32324.8 | 32476.8 |
| 805.75 | 789.475 | 721.114 | 4102.03 | 4098.39 | 10621.1 | 10568.6 | 21797.2 | 21768.3 | 32385.6 | 32518.2 |
| 806.031 | 770.66 | 723.064 | 4173.57 | 4106.35 | 10545 | 10585.6 | 21584 | 21799.2 | 32043.5 | 32559.7 |
| 806.313 | 777.178 | 725.016 | 4193.51 | 4114.31 | 10587.1 | 10602.6 | 21798.9 | 21830 | 32247 | 32601.1 |
| 806.594 | 761.036 | 726.972 | 4156.39 | 4122.28 | 10605 | 10619.7 | 21702.2 | 21860.9 | 32540 | 32642.6 |
| 806.875 | 786.352 | 728.932 | 4130.67 | 4130.26 | 10579.3 | 10636.8 | 21700.8 | 21891.8 | 32202.9 | 32684 |
| 807.157 | 767.548 | 730.894 | 4127.51 | 4138.24 | 10640.4 | 10653.8 | 21873.6 | 21922.7 | 32559 | 32725.5 |
| 807.438 | 789.127 | 732.86 | 4191.81 | 4146.23 | 10608.6 | 10670.9 | 21758.6 | 21953.5 | 32360 | 32767 |
| 807.72 | 758.723 | 734.829 | 4143.76 | 4154.23 | 10575.6 | 10688 | 21673.9 | 21984.4 | 32335.8 | 32808.4 |
| 808.001 | 792.737 | 736.801 | 4217.42 | 4162.24 | 10753.5 | 10705.1 | 21862.1 | 22015.3 | 32425 | 32849.9 |
| 808.283 | 810.493 | 738.777 | 4164.26 | 4170.25 | 10611.9 | 10722.2 | 21985 | 22046.2 | 32551.6 | 32891.4 |
| 808.564 | 811.485 | 740.756 | 4211.04 | 4178.27 | 10902.6 | 10739.3 | 22010.3 | 22077.2 | 32551.3 | 32932.8 |
| 808.845 | 774.337 | 742.738 | 4243.88 | 4186.29 | 10795.3 | 10756.4 | 21991.1 | 22108.1 | 32598.1 | 32974.3 |
| 809.126 | 793.722 | 744.723 | 4281.16 | 4194.32 | 10680.1 | 10773.5 | 21910.5 | 22139 | 32344.8 | 33015.8 |
| 809.408 | 800.57 | 746.712 | 4225.89 | 4202.36 | 10855.8 | 10790.6 | 21936.3 | 22169.9 | 32733.1 | 33057.3 |
| 809.689 | 800.254 | 748.704 | 4254.8 | 4210.41 | 10745.2 | 10807.8 | 21957.3 | 22200.9 | 32568.2 | 33098.7 |
| 809.97 | 769.996 | 750.699 | 4248.19 | 4218.46 | 10705.3 | 10824.9 | 21961.6 | 22231.8 | 32723.5 | 33140.2 |
| 810.252 | 831.562 | 752.697 | 4313.82 | 4226.52 | 10857.9 | 10842.1 | 21910.6 | 22262.8 | 32914.2 | 33181.7 |
| 810.533 | 822.116 | 754.699 | 4291.59 | 4234.58 | 10757.3 | 10859.2 | 21983.4 | 22293.8 | 32678 | 33223.2 |
| 810.814 | 800.164 | 756.704 | 4189.82 | 4242.65 | 10839.4 | 10876.4 | 21946 | 22324.7 | 32543.7 | 33264.7 |
| 811.095 | 796.419 | 758.712 | 4264.11 | 4250.73 | 10917.2 | 10893.6 | 22000.5 | 22355.7 | 32771.3 | 33306.2 |
| 811.376 | 823.414 | 760.724 | 4188.49 | 4258.82 | 10918.6 | 10910.8 | 21995.6 | 22386.7 | 32796.2 | 33347.7 |
| 811.658 | 762.15 | 762.739 | 4311.71 | 4266.91 | 10878.4 | 10928 | 21968.1 | 22417.7 | 33115.7 | 33389.2 |
| 811.939 | 785.868 | 764.757 | 4278.86 | 4275.01 | 10845.4 | 10945.1 | 22086.2 | 22448.7 | 32999.2 | 33430.7 |
| 812.22 | 800.517 | 766.778 | 4325.67 | 4283.11 | 10765.5 | 10962.4 | 22094.1 | 22479.7 | 32833.4 | 33472.2 |
| 812.501 | 838.352 | 768.803 | 4323.56 | 4291.22 | 10992.6 | 10979.6 | 22194.9 | 22510.7 | 33006.6 | 33513.7 |
| 812.782 | 809.729 | 770.831 | 4244.89 | 4299.34 | 10928.7 | 10996.8 | 22165.2 | 22541.7 | 33021.8 | 33555.2 |
| 813.063 | 815.591 | 772.862 | 4237.1 | 4307.47 | 11011.5 | 11014 | 22089.8 | 22572.7 | 32931 | 33596.7 |
| 813.344 | 847.575 | 774.896 | 4255.6 | 4315.6 | 10874.6 | 11031.2 | 22352.5 | 22603.8 | 33017.2 | 33638.2 |
| 813.625 | 809.555 | 776.934 | 4369.15 | 4323.73 | 10819.7 | 11048.5 | 22360.8 | 22634.8 | 33098.5 | 33679.7 |
| 813.907 | 801.184 | 778.975 | 4372.2 | 4331.88 | 11042 | 11065.7 | 22197.5 | 22665.8 | 33236.9 | 33721.2 |
| 814.187 | 824.853 | 781.019 | 4386.79 | 4340.03 | 11128.4 | 11083 | 22445.5 | 22696.9 | 33248.2 | 33762.7 |
| 814.468 | 812.232 | 783.067 | 4391.87 | 4348.19 | 11014.4 | 11100.3 | 22408.9 | 22728 | 33425.5 | 33804.2 |
| 814.75 | 800.822 | 785.118 | 4356.88 | 4356.35 | 11077.6 | 11117.5 | 22296.6 | 22759 | 33211.9 | 33845.8 |
| 815.03 | 825.349 | 787.172 | 4337.53 | 4364.52 | 10903.9 | 11134.8 | 22374.1 | 22790.1 | 33152.4 | 33887.3 |
| 815.312 | 848.041 | 789.229 | 4427.17 | 4372.7 | 11109.1 | 11152.1 | 22640.6 | 22821.2 | 33638.3 | 33928.8 |
| 815.592 | 846.809 | 791.29 | 4369.12 | 4380.88 | 11006.9 | 11169.4 | 22553.6 | 22852.2 | 33084.8 | 33970.3 |
| 815.873 | 832.683 | 793.354 | 4450.17 | 4389.07 | 11139 | 11186.7 | 22472.9 | 22883.3 | 33294.2 | 34011.8 |
| 816.154 | 814.394 | 795.421 | 4399.75 | 4397.27 | 11133.7 | 11204 | 22521 | 22914.4 | 33603.4 | 34053.4 |
| 816.435 | 856.071 | 797.491 | 4473.07 | 4405.47 | 11170.6 | 11221.3 | 22444.8 | 22945.5 | 33456.8 | 34094.9 |
| 816.716 | 836.131 | 799.565 | 4412.82 | 4413.68 | 11157.4 | 11238.6 | 22644.1 | 22976.6 | 33617 | 34136.4 |
| 816.997 | 828.301 | 801.642 | 4462.15 | 4421.89 | 11169.6 | 11256 | 22770.9 | 23007.8 | 33788.1 | 34177.9 |
| 817.278 | 869.157 | 803.723 | 4420.85 | 4430.11 | 11179.2 | 11273.3 | 22674.8 | 23038.9 | 33692.4 | 34219.4 |
| 817.559 | 849.724 | 805.806 | 4447.17 | 4438.34 | 11221.9 | 11290.7 | 22897.6 | 23070 | 33939.8 | 34261 |
| 817.84 | 822.813 | 807.893 | 4493.46 | 4446.58 | 11147 | 11308 | 23008.5 | 23101.1 | 33772.8 | 34302.5 |
| 818.12 | 865.143 | 809.983 | 4454.56 | 4454.82 | 11268.2 | 11325.4 | 22749.2 | 23132.3 | 33608.6 | 34344 |
| 818.401 | 872.465 | 812.077 | 4455.75 | 4463.06 | 11340.1 | 11342.7 | 22860 | 23163.4 | 34177 | 34385.6 |
| 818.682 | 887.231 | 814.173 | 4466.22 | 4471.32 | 11311.1 | 11360.1 | 22840.5 | 23194.5 | 33791.1 | 34427.1 |
| 818.963 | 848.396 | 816.273 | 4562.3 | 4479.58 | 11381.5 | 11377.5 | 22968.2 | 23225.7 | 34017.1 | 34468.6 |
| 819.243 | 846.873 | 818.376 | 4519.46 | 4487.85 | 11426.5 | 11394.9 | 23139.8 | 23256.9 | 34235.2 | 34510.2 |
| 819.524 | 866.005 | 820.483 | 4490.59 | 4496.12 | 11344.7 | 11412.3 | 23172 | 23288 | 34224.3 | 34551.7 |
| 819.805 | 856.71 | 822.593 | 4510.38 | 4504.4 | 11336.3 | 11429.7 | 23093.2 | 23319.2 | 34031.7 | 34593.2 |
| 820.086 | 879.118 | 824.706 | 4516.48 | 4512.68 | 11400.5 | 11447.1 | 23130.2 | 23350.4 | 34295.8 | 34634.8 |
| 820.366 | 878.757 | 826.822 | 4518.52 | 4520.97 | 11373.9 | 11464.5 | 23377.4 | 23381.5 | 34320.6 | 34676.3 |
| 820.647 | 908.638 | 828.942 | 4532.66 | 4529.27 | 11499 | 11481.9 | 23169.4 | 23412.7 | 34408.7 | 34717.8 |
| 820.928 | 861.646 | 831.065 | 4591.79 | 4537.58 | 11519.6 | 11499.4 | 23433.8 | 23443.9 | 34702.7 | 34759.3 |
| 821.208 | 902.528 | 833.191 | 4515.53 | 4545.89 | 11598 | 11516.8 | 23237.9 | 23475.1 | 34503.9 | 34800.9 |
| 821.489 | 879.659 | 835.32 | 4695.1 | 4554.2 | 11382.2 | 11534.2 | 23473.8 | 23506.3 | 34510.8 | 34842.4 |
| 821.769 | 888.202 | 837.453 | 4584.25 | 4562.53 | 11561.8 | 11551.7 | 23534.1 | 23537.5 | 34411.3 | 34883.9 |
| 822.05 | 900.219 | 839.589 | 4528.29 | 4570.86 | 11496.3 | 11569.1 | 23413.3 | 23568.7 | 34388.4 | 34925.5 |
| 822.331 | 883.865 | 841.728 | 4581.07 | 4579.19 | 11461.1 | 11586.6 | 23328 | 23600 | 34593.7 | 34967 |
| 822.611 | 874.117 | 843.871 | 4653.49 | 4587.53 | 11668.5 | 11604.1 | 23705.4 | 23631.2 | 34835.5 | 35008.5 |
| 822.892 | 862.109 | 846.017 | 4619.84 | 4595.88 | 11603.5 | 11621.5 | 23526.6 | 23662.4 | 34726.3 | 35050.1 |
| 823.172 | 902.272 | 848.166 | 4617.7 | 4604.24 | 11697.6 | 11639 | 23300.4 | 23693.6 | 34559.3 | 35091.6 |
| 823.453 | 910.949 | 850.318 | 4579.75 | 4612.6 | 11645.6 | 11656.5 | 23552.1 | 23724.9 | 34824.8 | 35133.1 |
| 823.733 | 915.239 | 852.474 | 4650.8 | 4620.96 | 11715.3 | 11674 | 23520.2 | 23756.1 | 34979.6 | 35174.7 |
| 824.014 | 892.51 | 854.633 | 4639.06 | 4629.34 | 11565.5 | 11691.5 | 23476.6 | 23787.4 | 34855.8 | 35216.2 |
| 824.294 | 893.022 | 856.795 | 4660.52 | 4637.72 | 11655.2 | 11709 | 23741.2 | 23818.6 | 34829 | 35257.7 |
| 824.574 | 914.549 | 858.961 | 4717.32 | 4646.1 | 11702.2 | 11726.5 | 23771.6 | 23849.9 | 35000.1 | 35299.2 |
| 824.855 | 899.745 | 861.129 | 4680.41 | 4654.49 | 11690.8 | 11744 | 23735.3 | 23881.1 | 34853.4 | 35340.8 |
| 825.135 | 891.638 | 863.301 | 4638.04 | 4662.89 | 11758.9 | 11761.6 | 23709.9 | 23912.4 | 34931.9 | 35382.3 |
| 825.416 | 911.866 | 865.477 | 4758.04 | 4671.29 | 11769.7 | 11779.1 | 23800.9 | 23943.7 | 35169 | 35423.8 |
| 825.696 | 926.54 | 867.655 | 4693.55 | 4679.7 | 11825.4 | 11796.6 | 23752.5 | 23974.9 | 35381 | 35465.4 |
| 825.976 | 910.007 | 869.837 | 4731.72 | 4688.12 | 11860.3 | 11814.2 | 23817 | 24006.2 | 35126.5 | 35506.9 |
| 826.257 | 926.245 | 872.022 | 4799.25 | 4696.54 | 11872.8 | 11831.7 | 24055 | 24037.5 | 35284.3 | 35548.4 |
| 826.537 | 933.562 | 874.211 | 4715.93 | 4704.97 | 11801 | 11849.3 | 23860.1 | 24068.8 | 35345 | 35589.9 |
| 826.817 | 900.923 | 876.402 | 4673.04 | 4713.4 | 11814.4 | 11866.9 | 23833.9 | 24100.1 | 35220.9 | 35631.5 |
| 827.098 | 907.193 | 878.597 | 4768.95 | 4721.84 | 11910.8 | 11884.4 | 24143.2 | 24131.4 | 35308.2 | 35673 |
| 827.378 | 926.076 | 880.795 | 4742.29 | 4730.29 | 11852.9 | 11902 | 23918.8 | 24162.7 | 35351.2 | 35714.5 |
| 827.658 | 922.145 | 882.997 | 4712.16 | 4738.74 | 11755.1 | 11919.6 | 23818.7 | 24194 | 35179.7 | 35756 |
| 827.939 | 925.184 | 885.202 | 4795.93 | 4747.2 | 11785 | 11937.2 | 23983.6 | 24225.3 | 35598.6 | 35797.5 |
| 828.219 | 936.011 | 887.41 | 4743.12 | 4755.66 | 11892.8 | 11954.8 | 24121.2 | 24256.6 | 35551.8 | 35839.1 |
| 828.499 | 945.182 | 889.621 | 4749.57 | 4764.13 | 11866 | 11972.4 | 24044 | 24287.9 | 35494.1 | 35880.6 |
| 828.779 | 926.09 | 891.835 | 4762.36 | 4772.61 | 11949.4 | 11990 | 24077.6 | 24319.2 | 35547.9 | 35922.1 |
| 829.059 | 959.195 | 894.053 | 4775.29 | 4781.09 | 11911.9 | 12007.6 | 24204.1 | 24350.6 | 35532.3 | 35963.6 |
| 829.34 | 968.057 | 896.274 | 4799.38 | 4789.58 | 12047.4 | 12025.2 | 24142.1 | 24381.9 | 35608.1 | 36005.1 |
| 829.62 | 939.652 | 898.499 | 4815.45 | 4798.07 | 11994.6 | 12042.8 | 24183.9 | 24413.2 | 35623.9 | 36046.6 |
| 829.9 | 963.349 | 900.726 | 4810.35 | 4806.57 | 11934.4 | 12060.5 | 24037.1 | 24444.6 | 35560 | 36088.1 |
| 830.18 | 911.664 | 902.957 | 4876.95 | 4815.08 | 12112.9 | 12078.1 | 24082.5 | 24475.9 | 35842.3 | 36129.6 |
| 830.46 | 932.805 | 905.192 | 4843.77 | 4823.59 | 12036.6 | 12095.8 | 24313.3 | 24507.3 | 35833.1 | 36171.1 |
| 830.74 | 931.28 | 907.429 | 4862.2 | 4832.11 | 11951.1 | 12113.4 | 24416.6 | 24538.6 | 35669.7 | 36212.6 |
| 831.02 | 957.54 | 909.67 | 4832.79 | 4840.63 | 12129.1 | 12131 | 24280 | 24570 | 35811.7 | 36254.1 |
| 831.3 | 969.2 | 911.914 | 4840.09 | 4849.16 | 12095.5 | 12148.7 | 24279.3 | 24601.3 | 35910.2 | 36295.6 |
| 831.58 | 953.513 | 914.161 | 4825.21 | 4857.7 | 11994.1 | 12166.4 | 24191.2 | 24632.7 | 35537.5 | 36337.1 |
| 831.86 | 942.237 | 916.411 | 4912.91 | 4866.24 | 12163.4 | 12184 | 24590 | 24664 | 36050.3 | 36378.6 |
| 832.14 | 980.601 | 918.665 | 4922.87 | 4874.78 | 12101.3 | 12201.7 | 24399.4 | 24695.4 | 36023.1 | 36420.1 |
| 832.42 | 955.351 | 920.922 | 4891.43 | 4883.34 | 12094.4 | 12219.4 | 24522.4 | 24726.8 | 36147.5 | 36461.6 |
| 832.7 | 940.687 | 923.183 | 4850.1 | 4891.9 | 12021.4 | 12237.1 | 24369.2 | 24758.2 | 36006.7 | 36503.1 |
| 832.98 | 973.396 | 925.446 | 4934.43 | 4900.46 | 12121.6 | 12254.8 | 24496.7 | 24789.5 | 36132.2 | 36544.6 |
| 833.26 | 968.312 | 927.713 | 4924.55 | 4909.03 | 12218.6 | 12272.5 | 24568.8 | 24820.9 | 36167.7 | 36586.1 |
| 833.54 | 999.172 | 929.983 | 4932.9 | 4917.61 | 12269.3 | 12290.2 | 24513.7 | 24852.3 | 36294.1 | 36627.6 |
| 833.82 | 971.961 | 932.256 | 4914.42 | 4926.19 | 12194.6 | 12307.9 | 24706.4 | 24883.7 | 36112.5 | 36669 |
| 834.1 | 974.783 | 934.533 | 4884.35 | 4934.78 | 12241.8 | 12325.6 | 24419.6 | 24915.1 | 35938.7 | 36710.5 |
| 834.38 | 985.085 | 936.813 | 4983.27 | 4943.37 | 12328 | 12343.4 | 24714.3 | 24946.5 | 36405.2 | 36752 |
| 834.66 | 972.082 | 939.096 | 4951.81 | 4951.97 | 12299.9 | 12361.1 | 24696.2 | 24977.9 | 36420.9 | 36793.5 |
| 834.939 | 965.452 | 941.383 | 4906.72 | 4960.58 | 12179.8 | 12378.8 | 24606.3 | 25009.3 | 35959.4 | 36834.9 |
| 835.219 | 981.561 | 943.672 | 4990.74 | 4969.19 | 12364.9 | 12396.6 | 24787.3 | 25040.7 | 36454.9 | 36876.4 |
| 835.499 | 965.811 | 945.965 | 4935.8 | 4977.81 | 12301.3 | 12414.3 | 24557.1 | 25072.1 | 36144.7 | 36917.9 |
| 835.779 | 978.903 | 948.261 | 5002.61 | 4986.43 | 12369.2 | 12432.1 | 24794.7 | 25103.5 | 36671.6 | 36959.3 |
| 836.059 | 991.87 | 950.561 | 4935.11 | 4995.06 | 12367.7 | 12449.8 | 24698.7 | 25134.9 | 36402.8 | 37000.8 |
| 836.338 | 954.313 | 952.864 | 4960.92 | 5003.69 | 12499 | 12467.6 | 24770.1 | 25166.3 | 36761.4 | 37042.2 |
| 836.618 | 1018.38 | 955.17 | 4966.89 | 5012.33 | 12345.9 | 12485.3 | 24805.3 | 25197.7 | 36587.4 | 37083.7 |
| 836.898 | 1012.44 | 957.479 | 4968.25 | 5020.97 | 12284.2 | 12503.1 | 24891.6 | 25229.2 | 36422.6 | 37125.1 |
| 837.178 | 997.205 | 959.791 | 4950.7 | 5029.62 | 12402.1 | 12520.9 | 24843.9 | 25260.6 | 36452.8 | 37166.6 |
| 837.457 | 1015.36 | 962.107 | 4988.04 | 5038.28 | 12425 | 12538.7 | 24871.5 | 25292 | 36503.5 | 37208 |
| 837.737 | 1011.19 | 964.426 | 5071.77 | 5046.94 | 12392.7 | 12556.5 | 25013.1 | 25323.5 | 36484.2 | 37249.5 |
| 838.017 | 980.159 | 966.748 | 5009.48 | 5055.61 | 12393.3 | 12574.3 | 24697.4 | 25354.9 | 36507.3 | 37290.9 |
| 838.296 | 998.966 | 969.074 | 5013.55 | 5064.28 | 12488.5 | 12592.1 | 25118 | 25386.3 | 36869.1 | 37332.3 |
| 838.576 | 994.45 | 971.403 | 5049.1 | 5072.96 | 12481.6 | 12609.9 | 25049.5 | 25417.8 | 36760.5 | 37373.8 |
| 838.856 | 1020.76 | 973.735 | 5082.8 | 5081.65 | 12463.1 | 12627.7 | 25164.9 | 25449.2 | 36899.3 | 37415.2 |
| 839.135 | 1009.81 | 976.07 | 5075.68 | 5090.34 | 12537.5 | 12645.5 | 25372.5 | 25480.6 | 37119.2 | 37456.6 |
| 839.415 | 1004.81 | 978.408 | 5065.64 | 5099.03 | 12603.7 | 12663.3 | 25280.6 | 25512.1 | 36914.7 | 37498 |
| 839.694 | 998.335 | 980.75 | 5062.26 | 5107.74 | 12540.6 | 12681.1 | 25169 | 25543.5 | 36824.2 | 37539.5 |
| 839.974 | 1030.15 | 983.095 | 5069.48 | 5116.44 | 12624.5 | 12699 | 25251 | 25575 | 36828.7 | 37580.9 |
| 840.254 | 1049.19 | 985.443 | 5099.03 | 5125.16 | 12523.2 | 12716.8 | 25200.8 | 25606.4 | 37248.4 | 37622.3 |
| 840.533 | 1020.05 | 987.795 | 5132.89 | 5133.87 | 12560.4 | 12734.6 | 25472 | 25637.9 | 37083.1 | 37663.7 |
| 840.812 | 1052.98 | 990.15 | 5068.67 | 5142.6 | 12665.7 | 12752.5 | 25476.9 | 25669.3 | 37117.4 | 37705.1 |
| 841.092 | 1076.89 | 992.508 | 5161.05 | 5151.33 | 12713.5 | 12770.3 | 25419.7 | 25700.8 | 37430.9 | 37746.5 |
| 841.371 | 1027.12 | 994.869 | 5109.27 | 5160.06 | 12710.4 | 12788.2 | 25532.5 | 25732.3 | 37272.5 | 37787.9 |
| 841.651 | 1038.22 | 997.234 | 5101.34 | 5168.8 | 12758.7 | 12806 | 25400.5 | 25763.7 | 37239.8 | 37829.3 |
| 841.93 | 1038.5 | 999.601 | 5257.99 | 5177.55 | 12873 | 12823.9 | 25469.7 | 25795.2 | 37517.2 | 37870.7 |
| 842.21 | 1039.95 | 1001.97 | 5251.45 | 5186.3 | 12907.4 | 12841.8 | 25500.7 | 25826.7 | 37190.5 | 37912.1 |
| 842.489 | 1052.83 | 1004.35 | 5207.76 | 5195.05 | 12671.8 | 12859.6 | 25575.6 | 25858.1 | 37770.5 | 37953.4 |
| 842.769 | 1026.97 | 1006.72 | 5233.4 | 5203.82 | 12743.1 | 12877.5 | 25611.3 | 25889.6 | 37362.3 | 37994.8 |
| 843.048 | 1048.89 | 1009.11 | 5186.03 | 5212.58 | 12830 | 12895.4 | 25388.7 | 25921.1 | 37716.9 | 38036.2 |
| 843.327 | 1059.17 | 1011.49 | 5203.12 | 5221.36 | 12761.5 | 12913.3 | 25700.6 | 25952.5 | 37589.2 | 38077.6 |
| 843.607 | 1061.58 | 1013.88 | 5240.74 | 5230.13 | 12919.1 | 12931.2 | 25742.2 | 25984 | 37372.2 | 38118.9 |
| 843.886 | 1052.65 | 1016.27 | 5201.95 | 5238.92 | 12972.6 | 12949.1 | 25693.5 | 26015.5 | 37833.8 | 38160.3 |
| 844.165 | 1027.92 | 1018.66 | 5294.59 | 5247.71 | 12819.7 | 12967 | 25393.1 | 26047 | 37537.8 | 38201.6 |
| 844.444 | 1047.55 | 1021.06 | 5203.81 | 5256.5 | 12940 | 12984.9 | 25643 | 26078.5 | 37826.5 | 38243 |
| 844.724 | 1074.27 | 1023.46 | 5294.98 | 5265.3 | 12864.2 | 13002.8 | 25997.2 | 26110 | 37751.3 | 38284.3 |
| 845.003 | 1099.7 | 1025.86 | 5247.29 | 5274.11 | 12938.1 | 13020.7 | 25857.8 | 26141.4 | 37933.7 | 38325.7 |
| 845.282 | 1062.85 | 1028.27 | 5319.98 | 5282.92 | 12999.1 | 13038.6 | 25834.6 | 26172.9 | 37813.9 | 38367 |
| 845.562 | 1034.55 | 1030.68 | 5199.67 | 5291.73 | 12931.1 | 13056.6 | 25982.8 | 26204.4 | 37920.7 | 38408.3 |
| 845.841 | 1067.92 | 1033.09 | 5208.36 | 5300.55 | 12841.5 | 13074.5 | 25864.3 | 26235.9 | 37983.6 | 38449.7 |
| 846.12 | 1068.31 | 1035.51 | 5250.2 | 5309.38 | 13012.5 | 13092.5 | 25893.6 | 26267.4 | 37998.1 | 38491 |
| 846.399 | 1079.15 | 1037.93 | 5314.65 | 5318.21 | 13000.4 | 13110.4 | 26145.7 | 26298.9 | 38163.5 | 38532.3 |
| 846.678 | 1070.27 | 1040.35 | 5288.55 | 5327.05 | 12904.1 | 13128.3 | 25959.4 | 26330.4 | 38107.1 | 38573.6 |
| 846.957 | 1098.58 | 1042.78 | 5259.47 | 5335.89 | 13045.3 | 13146.3 | 26147.3 | 26361.9 | 38145.5 | 38614.9 |
| 847.236 | 1067.72 | 1045.2 | 5390.66 | 5344.74 | 12964.3 | 13164.2 | 26407.3 | 26393.4 | 38378.9 | 38656.2 |
| 847.516 | 1077.69 | 1047.64 | 5329.65 | 5353.59 | 13171.5 | 13182.2 | 26258.2 | 26424.9 | 38414.8 | 38697.5 |
| 847.795 | 1088.8 | 1050.07 | 5359.29 | 5362.45 | 13122.2 | 13200.1 | 25942.5 | 26456.4 | 38212.9 | 38738.8 |
| 848.074 | 1094.23 | 1052.51 | 5377.59 | 5371.31 | 13080.9 | 13218.1 | 26275.7 | 26487.9 | 38465.6 | 38780.1 |
| 848.353 | 1099.94 | 1054.95 | 5332.53 | 5380.18 | 13085 | 13236.1 | 26314.8 | 26519.4 | 38335.6 | 38821.4 |
| 848.632 | 1110.12 | 1057.4 | 5387.35 | 5389.05 | 13064.4 | 13254.1 | 26290.8 | 26550.9 | 38347.6 | 38862.7 |
| 848.911 | 1099.12 | 1059.85 | 5323.23 | 5397.93 | 13307.3 | 13272 | 26191.2 | 26582.4 | 38415.7 | 38903.9 |
| 849.19 | 1075.83 | 1062.3 | 5363.37 | 5406.82 | 13250.1 | 13290 | 26163 | 26613.9 | 38535.5 | 38945.2 |
| 849.469 | 1101.67 | 1064.75 | 5396.88 | 5415.71 | 13111.5 | 13308 | 26281.4 | 26645.4 | 38313.9 | 38986.5 |
| 849.748 | 1105.64 | 1067.21 | 5341.84 | 5424.6 | 13111.3 | 13326 | 26119 | 26676.9 | 38347.1 | 39027.7 |
| 850.027 | 1126.87 | 1069.67 | 5478.03 | 5433.5 | 13306.2 | 13344 | 26534 | 26708.4 | 38717.7 | 39069 |
| 850.306 | 1119.93 | 1072.14 | 5419.2 | 5442.4 | 13257.7 | 13362 | 26389.3 | 26739.9 | 38573.4 | 39110.2 |
| 850.585 | 1119.99 | 1074.61 | 5434.05 | 5451.31 | 13159.9 | 13380 | 26195.9 | 26771.4 | 38730.3 | 39151.5 |
| 850.864 | 1090.47 | 1077.08 | 5404.98 | 5460.23 | 13317.1 | 13398 | 26361.8 | 26802.9 | 38403.3 | 39192.7 |
| 851.143 | 1101.17 | 1079.55 | 5416.09 | 5469.15 | 13290.2 | 13416 | 26166 | 26834.4 | 38578.4 | 39233.9 |
| 851.421 | 1105.04 | 1082.03 | 5432.5 | 5478.07 | 13346.7 | 13434 | 26519.6 | 26866 | 38882.3 | 39275.2 |
| 851.7 | 1111.12 | 1084.51 | 5424.54 | 5487.01 | 13269.8 | 13452.1 | 26467.8 | 26897.5 | 38677.5 | 39316.4 |
| 851.979 | 1087.55 | 1086.99 | 5471.44 | 5495.94 | 13291.5 | 13470.1 | 26427.2 | 26929 | 38597.7 | 39357.6 |
| 852.258 | 1103.54 | 1089.48 | 5464.31 | 5504.88 | 13275.3 | 13488.1 | 26490.9 | 26960.5 | 38767.7 | 39398.8 |
| 852.537 | 1091.88 | 1091.97 | 5475.95 | 5513.83 | 13332.2 | 13506.1 | 26381.5 | 26992 | 38835.2 | 39440 |
| 852.816 | 1115.02 | 1094.46 | 5499.01 | 5522.78 | 13363.4 | 13524.2 | 26657.8 | 27023.5 | 38819.6 | 39481.2 |
| 853.094 | 1106.78 | 1096.96 | 5500.59 | 5531.73 | 13375.5 | 13542.2 | 26747.3 | 27055 | 38888.3 | 39522.4 |
| 853.373 | 1100.6 | 1099.46 | 5446.64 | 5540.69 | 13305.1 | 13560.3 | 26460.6 | 27086.5 | 38561.9 | 39563.6 |
| 853.652 | 1145.92 | 1101.96 | 5404.15 | 5549.66 | 13416 | 13578.3 | 26728.7 | 27118.1 | 38770 | 39604.8 |
| 853.93 | 1121.4 | 1104.47 | 5521.86 | 5558.63 | 13356.4 | 13596.4 | 26591.5 | 27149.6 | 38790.3 | 39645.9 |
| 854.209 | 1085.72 | 1106.98 | 5535.96 | 5567.6 | 13354.3 | 13614.4 | 26626.9 | 27181.1 | 38608.8 | 39687.1 |
| 854.488 | 1148.54 | 1109.49 | 5565.01 | 5576.59 | 13559.7 | 13632.5 | 26758.8 | 27212.6 | 39016.3 | 39728.3 |
| 854.766 | 1155.94 | 1112.01 | 5529.55 | 5585.57 | 13414 | 13650.5 | 26768.6 | 27244.1 | 38865.5 | 39769.4 |
| 855.045 | 1158.21 | 1114.53 | 5523.59 | 5594.56 | 13582.3 | 13668.6 | 26873.3 | 27275.6 | 38938.9 | 39810.6 |
| 855.324 | 1161.05 | 1117.05 | 5455.62 | 5603.56 | 13428.1 | 13686.7 | 26839.3 | 27307.2 | 39127.6 | 39851.7 |
| 855.602 | 1143.51 | 1119.58 | 5498.96 | 5612.56 | 13475.8 | 13704.8 | 26950.7 | 27338.7 | 39280.9 | 39892.9 |
| 855.881 | 1118.43 | 1122.11 | 5536.44 | 5621.56 | 13524.4 | 13722.8 | 26802.2 | 27370.2 | 38992.2 | 39934 |
| 856.16 | 1123 | 1124.64 | 5636.88 | 5630.57 | 13487.5 | 13740.9 | 26957 | 27401.7 | 39452.9 | 39975.1 |
| 856.438 | 1145.13 | 1127.17 | 5549.18 | 5639.59 | 13428.3 | 13759 | 26741.9 | 27433.2 | 39117.3 | 40016.2 |
| 856.717 | 1169.15 | 1129.71 | 5598 | 5648.61 | 13524 | 13777.1 | 27041.1 | 27464.7 | 39209.9 | 40057.3 |
| 856.995 | 1156.56 | 1132.25 | 5543.44 | 5657.63 | 13579.5 | 13795.2 | 26835.7 | 27496.3 | 39066.4 | 40098.4 |
| 857.274 | 1144.47 | 1134.8 | 5600.76 | 5666.66 | 13633 | 13813.3 | 26918 | 27527.8 | 39364.4 | 40139.5 |
| 857.552 | 1138.93 | 1137.35 | 5604.04 | 5675.69 | 13787.5 | 13831.4 | 27133.8 | 27559.3 | 39486.8 | 40180.6 |
| 857.831 | 1176.98 | 1139.9 | 5697.74 | 5684.73 | 13729.6 | 13849.5 | 27163.9 | 27590.8 | 39576.9 | 40221.7 |
| 858.109 | 1162.85 | 1142.45 | 5590 | 5693.78 | 13695.8 | 13867.6 | 27130.9 | 27622.3 | 39737.1 | 40262.8 |
| 858.388 | 1162.76 | 1145.01 | 5598.99 | 5702.83 | 13572 | 13885.7 | 27113 | 27653.8 | 39280.6 | 40303.8 |
| 858.666 | 1180.9 | 1147.57 | 5614.25 | 5711.88 | 13738.1 | 13903.8 | 27219 | 27685.3 | 39351.5 | 40344.9 |
| 858.945 | 1163.74 | 1150.13 | 5678.88 | 5720.94 | 13713.4 | 13921.9 | 27031.3 | 27716.9 | 39296.5 | 40386 |
| 859.223 | 1183.43 | 1152.7 | 5646.73 | 5730 | 13634.5 | 13940 | 27317.4 | 27748.4 | 39501.6 | 40427 |
| 859.501 | 1194.79 | 1155.27 | 5611.74 | 5739.07 | 13726.5 | 13958.2 | 27343.2 | 27779.9 | 39975.8 | 40468.1 |
| 859.78 | 1195.29 | 1157.85 | 5682.4 | 5748.14 | 13837 | 13976.3 | 27362.5 | 27811.4 | 39652.7 | 40509.1 |
| 860.058 | 1197.3 | 1160.42 | 5703.8 | 5757.22 | 13756.8 | 13994.4 | 27214.8 | 27842.9 | 40002.6 | 40550.1 |
| 860.337 | 1172.94 | 1163 | 5684.65 | 5766.3 | 13997.4 | 14012.6 | 27633.7 | 27874.4 | 40048.8 | 40591.2 |
| 860.615 | 1192.75 | 1165.58 | 5673.64 | 5775.39 | 13869.2 | 14030.7 | 27414.3 | 27905.9 | 40253.5 | 40632.2 |
| 860.893 | 1217.01 | 1168.17 | 5795.24 | 5784.48 | 13950.4 | 14048.8 | 27515.1 | 27937.5 | 40073.2 | 40673.2 |
| 861.171 | 1217.15 | 1170.76 | 5757.42 | 5793.58 | 13824.4 | 14067 | 27546.1 | 27969 | 40077.5 | 40714.2 |
| 861.45 | 1228.97 | 1173.35 | 5712.33 | 5802.68 | 14027.8 | 14085.1 | 27660.8 | 28000.5 | 40181.8 | 40755.2 |
| 861.728 | 1218.15 | 1175.95 | 5706.7 | 5811.78 | 13885.1 | 14103.3 | 27524.8 | 28032 | 40049 | 40796.1 |
| 862.006 | 1177.42 | 1178.55 | 5676.71 | 5820.9 | 13850.2 | 14121.4 | 27633 | 28063.5 | 40043 | 40837.1 |
| 862.284 | 1241.11 | 1181.15 | 5739.69 | 5830.01 | 13801 | 14139.6 | 27425.3 | 28095 | 39994.6 | 40878.1 |
| 862.563 | 1206.37 | 1183.75 | 5705.54 | 5839.13 | 13892.5 | 14157.7 | 27613.3 | 28126.5 | 40073.4 | 40919.1 |
| 862.841 | 1205.72 | 1186.36 | 5797.78 | 5848.26 | 14013.6 | 14175.9 | 27951.2 | 28158 | 40161.6 | 40960 |
| 863.119 | 1231.89 | 1188.97 | 5805.32 | 5857.38 | 13860.2 | 14194.1 | 27526.5 | 28189.5 | 40081.9 | 41000.9 |
| 863.397 | 1227.23 | 1191.59 | 5856.3 | 5866.52 | 14065 | 14212.2 | 27758.8 | 28221 | 40350.8 | 41041.9 |
| 863.675 | 1240.79 | 1194.21 | 5786.84 | 5875.65 | 13981.9 | 14230.4 | 27854.1 | 28252.5 | 40222.8 | 41082.8 |
| 863.953 | 1205.9 | 1196.83 | 5830.61 | 5884.8 | 14142.1 | 14248.6 | 27717.8 | 28284 | 40174.6 | 41123.8 |
| 864.231 | 1186.34 | 1199.45 | 5763.66 | 5893.94 | 13941.1 | 14266.8 | 27561.1 | 28315.5 | 40152.3 | 41164.7 |
| 864.51 | 1223.56 | 1202.08 | 5770.78 | 5903.1 | 14011.9 | 14284.9 | 27721.4 | 28347 | 39966.2 | 41205.6 |
| 864.788 | 1237.15 | 1204.71 | 5814.38 | 5912.25 | 14056.9 | 14303.1 | 27870.8 | 28378.5 | 40401.6 | 41246.5 |
| 865.066 | 1191.49 | 1207.34 | 5794.92 | 5921.41 | 14154.9 | 14321.3 | 27879.5 | 28410 | 40441.3 | 41287.4 |
| 865.344 | 1199.37 | 1209.98 | 5835.51 | 5930.58 | 13919.4 | 14339.5 | 27627.1 | 28441.5 | 39984.7 | 41328.2 |
| 865.622 | 1232.76 | 1212.62 | 5859.39 | 5939.75 | 13997.9 | 14357.7 | 27763.5 | 28473 | 40467.8 | 41369.1 |
| 865.9 | 1235.55 | 1215.26 | 5883 | 5948.92 | 14122.1 | 14375.9 | 27857.5 | 28504.5 | 40802 | 41410 |
| 866.178 | 1239.35 | 1217.91 | 5870.79 | 5958.1 | 14045.6 | 14394.1 | 27968.2 | 28536 | 40495.1 | 41450.8 |
| 866.456 | 1247.01 | 1220.56 | 5831.15 | 5967.29 | 14067.9 | 14412.3 | 28069.2 | 28567.5 | 40461.9 | 41491.7 |
| 866.734 | 1211.97 | 1223.21 | 5923.73 | 5976.47 | 14237.9 | 14430.5 | 28011.6 | 28598.9 | 40597.9 | 41532.6 |
| 867.011 | 1218.99 | 1225.87 | 5807.27 | 5985.67 | 14208.1 | 14448.7 | 27917.1 | 28630.4 | 40433.1 | 41573.4 |
| 867.289 | 1257.77 | 1228.52 | 5867.74 | 5994.86 | 14130.1 | 14466.9 | 27929.8 | 28661.9 | 40538.7 | 41614.2 |
| 867.567 | 1240.71 | 1231.18 | 5867.25 | 6004.06 | 14148.8 | 14485.1 | 27865.4 | 28693.4 | 40526.8 | 41655 |
| 867.845 | 1271.84 | 1233.85 | 5847.87 | 6013.27 | 14247 | 14503.3 | 27968.8 | 28724.9 | 40800.2 | 41695.8 |
| 868.123 | 1219.43 | 1236.52 | 5803.98 | 6022.48 | 14246.2 | 14521.5 | 27965.3 | 28756.3 | 40768.9 | 41736.6 |
| 868.401 | 1257.97 | 1239.19 | 5812.52 | 6031.69 | 14139.2 | 14539.8 | 27998.5 | 28787.8 | 40644.4 | 41777.4 |
| 868.679 | 1272.18 | 1241.86 | 5896.62 | 6040.91 | 14312.9 | 14558 | 28156.8 | 28819.3 | 40872.6 | 41818.2 |
| 868.957 | 1255.82 | 1244.54 | 5884.51 | 6050.14 | 14321.2 | 14576.2 | 27901 | 28850.8 | 40540.9 | 41859 |
| 869.234 | 1264.97 | 1247.22 | 5976.34 | 6059.36 | 14373.1 | 14594.4 | 28092.1 | 28882.2 | 41042.5 | 41899.8 |
| 869.512 | 1258.95 | 1249.9 | 5902.32 | 6068.6 | 14267.6 | 14612.6 | 28198.6 | 28913.7 | 41231.9 | 41940.5 |
| 869.79 | 1260.75 | 1252.59 | 5902.46 | 6077.83 | 14316.9 | 14630.9 | 28078.5 | 28945.2 | 40820.4 | 41981.3 |
| 870.068 | 1249.86 | 1255.28 | 5966.31 | 6087.07 | 14279.2 | 14649.1 | 28152 | 28976.7 | 40993 | 42022 |
| 870.345 | 1229.18 | 1257.97 | 5943.79 | 6096.32 | 14299.5 | 14667.3 | 28208.8 | 29008.1 | 40775.1 | 42062.8 |
| 870.623 | 1291.25 | 1260.67 | 5947.1 | 6105.57 | 14358.3 | 14685.6 | 28053.2 | 29039.6 | 40728.4 | 42103.5 |
| 870.901 | 1245.24 | 1263.37 | 5988.87 | 6114.82 | 14475.5 | 14703.8 | 28319.7 | 29071 | 41317.4 | 42144.2 |
| 871.178 | 1264.96 | 1266.07 | 6068.69 | 6124.08 | 14370.5 | 14722.1 | 28313.5 | 29102.5 | 40876.6 | 42184.9 |
| 871.456 | 1237.32 | 1268.78 | 5977.2 | 6133.34 | 14343.6 | 14740.3 | 28326.2 | 29134 | 41060.7 | 42225.6 |
| 871.734 | 1243.58 | 1271.49 | 5952.43 | 6142.6 | 14367.2 | 14758.6 | 28222.3 | 29165.4 | 41110.8 | 42266.3 |
| 872.011 | 1290.4 | 1274.2 | 5981.02 | 6151.87 | 14397.3 | 14776.8 | 28420.8 | 29196.8 | 41431.6 | 42307 |
| 872.289 | 1292.86 | 1276.91 | 6039.86 | 6161.15 | 14464.1 | 14795.1 | 28732.6 | 29228.3 | 41466.1 | 42347.6 |
| 872.567 | 1294.79 | 1279.63 | 6105.71 | 6170.43 | 14552.8 | 14813.3 | 28719.1 | 29259.8 | 41567.1 | 42388.3 |
| 872.844 | 1274.51 | 1282.35 | 6035.08 | 6179.71 | 14555.8 | 14831.6 | 28522.1 | 29291.2 | 41250.6 | 42428.9 |
| 873.122 | 1283.53 | 1285.08 | 6057.54 | 6189 | 14607.2 | 14849.8 | 28679.2 | 29322.6 | 41691.9 | 42469.6 |
| 873.399 | 1278.92 | 1287.81 | 6098.76 | 6198.29 | 14442 | 14868.1 | 28652 | 29354.1 | 41114.7 | 42510.2 |
| 873.677 | 1320.84 | 1290.54 | 6139.15 | 6207.59 | 14659 | 14886.4 | 28605.4 | 29385.5 | 41741.2 | 42550.8 |
| 873.954 | 1321.06 | 1293.27 | 6084.67 | 6216.89 | 14691.7 | 14904.6 | 28941 | 29417 | 41805.5 | 42591.5 |
| 874.232 | 1294.83 | 1296.01 | 6078 | 6226.19 | 14652.6 | 14922.9 | 28664.7 | 29448.4 | 41713.3 | 42632.1 |
| 874.509 | 1312.09 | 1298.74 | 6151.66 | 6235.5 | 14751.8 | 14941.1 | 28702.3 | 29479.8 | 41767.5 | 42672.7 |
| 874.787 | 1292.45 | 1301.49 | 6080.39 | 6244.81 | 14686.1 | 14959.4 | 28653.4 | 29511.2 | 41679 | 42713.2 |
| 875.064 | 1311.23 | 1304.23 | 6130.73 | 6254.13 | 14714.1 | 14977.7 | 29027 | 29542.7 | 41968.5 | 42753.8 |
| 875.342 | 1334.74 | 1306.98 | 6183.76 | 6263.45 | 14735.2 | 14996 | 28884.8 | 29574.1 | 41868.4 | 42794.4 |
| 875.619 | 1332.21 | 1309.73 | 6162.78 | 6272.77 | 14670.2 | 15014.2 | 28885.6 | 29605.5 | 41952.9 | 42835 |
| 875.896 | 1283.8 | 1312.49 | 6193.87 | 6282.1 | 14734.7 | 15032.5 | 28861 | 29636.9 | 41948.1 | 42875.5 |
| 876.174 | 1340.19 | 1315.25 | 6148.11 | 6291.43 | 14641 | 15050.8 | 28914.6 | 29668.3 | 41940.1 | 42916.1 |
| 876.451 | 1327.66 | 1318.01 | 6177.58 | 6300.77 | 14814.6 | 15069.1 | 29243.1 | 29699.7 | 42172.5 | 42956.6 |
| 876.728 | 1331.02 | 1320.77 | 6205.12 | 6310.11 | 14861.9 | 15087.4 | 29062.3 | 29731.2 | 42232.6 | 42997.1 |
| 877.006 | 1377.26 | 1323.54 | 6133.27 | 6319.45 | 14766.8 | 15105.7 | 29153.9 | 29762.5 | 42133.2 | 43037.6 |
| 877.283 | 1357.19 | 1326.31 | 6202.87 | 6328.8 | 14936 | 15124 | 29064 | 29794 | 42049 | 43078.1 |
| 877.56 | 1374.73 | 1329.09 | 6238.29 | 6338.16 | 14728.6 | 15142.2 | 29105.3 | 29825.3 | 42293.3 | 43118.6 |
| 877.838 | 1339.77 | 1331.86 | 6325.96 | 6347.51 | 14943.6 | 15160.5 | 29201.3 | 29856.8 | 42363.2 | 43159.1 |
| 878.115 | 1362.4 | 1334.64 | 6293.18 | 6356.87 | 14825.2 | 15178.8 | 29268.2 | 29888.1 | 42321.1 | 43199.5 |
| 878.392 | 1362.59 | 1337.42 | 6230.85 | 6366.24 | 14972.3 | 15197.1 | 29361.9 | 29919.5 | 42458.4 | 43240 |
| 878.669 | 1353.94 | 1340.21 | 6231.16 | 6375.61 | 15060.3 | 15215.4 | 29561.7 | 29950.9 | 42380.9 | 43280.4 |
| 878.947 | 1376.03 | 1343 | 6328.01 | 6384.98 | 14941.5 | 15233.7 | 29488.2 | 29982.3 | 42753.3 | 43320.9 |
| 879.224 | 1378.24 | 1345.79 | 6272.69 | 6394.36 | 14960.1 | 15252 | 29330 | 30013.7 | 42625.3 | 43361.3 |
| 879.501 | 1378.11 | 1348.59 | 6332.11 | 6403.74 | 15129.4 | 15270.3 | 29477.2 | 30045.1 | 42881.3 | 43401.7 |
| 879.778 | 1337.69 | 1351.38 | 6267.01 | 6413.12 | 15005.1 | 15288.6 | 29394.1 | 30076.4 | 42384.3 | 43442.2 |
| 880.055 | 1343 | 1354.19 | 6367.44 | 6422.51 | 14985.8 | 15306.9 | 29559 | 30107.8 | 42636.4 | 43482.6 |
| 880.332 | 1411.49 | 1356.99 | 6360.19 | 6431.9 | 15037.6 | 15325.2 | 29571.7 | 30139.2 | 42732.1 | 43522.9 |
| 880.609 | 1442.58 | 1359.8 | 6452.18 | 6441.3 | 15057.4 | 15343.5 | 29643.4 | 30170.5 | 42678.4 | 43563.3 |
| 880.886 | 1401.22 | 1362.61 | 6335.32 | 6450.7 | 15092.6 | 15361.9 | 29443.3 | 30201.9 | 42632.7 | 43603.7 |
| 881.163 | 1405.96 | 1365.42 | 6405.97 | 6460.1 | 15041.5 | 15380.2 | 29629.2 | 30233.2 | 42385.1 | 43644 |
| 881.441 | 1377.57 | 1368.24 | 6311.77 | 6469.51 | 15215 | 15398.5 | 29685.1 | 30264.6 | 42904.8 | 43684.4 |
| 881.717 | 1421.28 | 1371.06 | 6355.02 | 6478.92 | 15164.2 | 15416.8 | 29588.7 | 30296 | 42715.8 | 43724.7 |
| 881.995 | 1399.74 | 1373.88 | 6399.47 | 6488.34 | 15196.8 | 15435.1 | 29609.3 | 30327.3 | 42934 | 43765.1 |
| 882.271 | 1420.39 | 1376.7 | 6390.38 | 6497.76 | 15212.9 | 15453.4 | 29802.5 | 30358.7 | 43054.9 | 43805.4 |
| 882.548 | 1417.91 | 1379.53 | 6401.43 | 6507.18 | 15233.8 | 15471.8 | 29990.9 | 30390 | 43236.7 | 43845.7 |
| 882.825 | 1374.43 | 1382.37 | 6370.33 | 6516.6 | 15192.4 | 15490.1 | 29599.6 | 30421.3 | 42977.7 | 43886 |
| 883.102 | 1428.53 | 1385.2 | 6374.97 | 6526.03 | 15281.4 | 15508.4 | 29703.5 | 30452.7 | 43329.5 | 43926.2 |
| 883.379 | 1376.99 | 1388.04 | 6457.16 | 6535.47 | 15319.5 | 15526.7 | 30036.4 | 30484 | 43518.1 | 43966.5 |
| 883.656 | 1391.09 | 1390.88 | 6420.34 | 6544.91 | 15297.2 | 15545 | 30021.9 | 30515.3 | 43065.7 | 44006.8 |
| 883.933 | 1375.75 | 1393.72 | 6480.92 | 6554.35 | 15182.8 | 15563.4 | 29723.1 | 30546.6 | 43052.6 | 44047 |
| 884.21 | 1442.27 | 1396.57 | 6412.17 | 6563.8 | 15272.6 | 15581.7 | 29866 | 30578 | 43013.9 | 44087.3 |
| 884.487 | 1446.41 | 1399.42 | 6449.95 | 6573.24 | 15285.1 | 15600 | 30013.7 | 30609.3 | 43468.7 | 44127.5 |
| 884.764 | 1439.04 | 1402.27 | 6482.48 | 6582.7 | 15358 | 15618.4 | 30197.5 | 30640.6 | 43518.2 | 44167.7 |
| 885.04 | 1421.33 | 1405.13 | 6377.66 | 6592.15 | 15281.7 | 15636.7 | 29876.6 | 30671.9 | 43373.2 | 44207.9 |
| 885.317 | 1437.7 | 1407.99 | 6518.31 | 6601.61 | 15409.7 | 15655 | 29870.6 | 30703.2 | 43437.9 | 44248.1 |
| 885.594 | 1430.93 | 1410.85 | 6507.31 | 6611.08 | 15432.3 | 15673.4 | 30239.6 | 30734.5 | 43575 | 44288.3 |
| 885.871 | 1401.94 | 1413.71 | 6526.59 | 6620.55 | 15329.7 | 15691.7 | 30106.9 | 30765.8 | 43776.6 | 44328.5 |
| 886.148 | 1407.32 | 1416.58 | 6514.48 | 6630.02 | 15485.2 | 15710 | 30288.9 | 30797.1 | 43590.6 | 44368.6 |
| 886.424 | 1450.25 | 1419.45 | 6615.45 | 6639.49 | 15482 | 15728.4 | 30288.5 | 30828.3 | 43787.3 | 44408.8 |
| 886.701 | 1411.54 | 1422.33 | 6470.97 | 6648.97 | 15492.4 | 15746.7 | 30104.9 | 30859.6 | 43631.7 | 44448.9 |
| 886.978 | 1398.76 | 1425.2 | 6500.94 | 6658.45 | 15443.4 | 15765 | 30350.7 | 30890.9 | 43679.6 | 44489.1 |
| 887.254 | 1405.94 | 1428.08 | 6620.61 | 6667.94 | 15495 | 15783.4 | 30164.5 | 30922.2 | 43771.4 | 44529.2 |
| 887.531 | 1455.18 | 1430.97 | 6605.6 | 6677.43 | 15539 | 15801.7 | 30186.4 | 30953.4 | 43643.6 | 44569.3 |
| 887.808 | 1459.61 | 1433.85 | 6596.08 | 6686.92 | 15689.3 | 15820.1 | 30458.8 | 30984.7 | 43965.9 | 44609.4 |
| 888.084 | 1461.71 | 1436.74 | 6610.54 | 6696.41 | 15572.6 | 15838.4 | 30337.6 | 31016 | 43780.9 | 44649.5 |
| 888.361 | 1402.77 | 1439.63 | 6471.43 | 6705.91 | 15545.9 | 15856.7 | 30249.6 | 31047.2 | 43745.9 | 44689.5 |
| 888.638 | 1454.74 | 1442.53 | 6587.89 | 6715.42 | 15538.6 | 15875.1 | 30350.8 | 31078.5 | 43754.1 | 44729.6 |
| 888.914 | 1463.99 | 1445.43 | 6626.09 | 6724.93 | 15710 | 15893.4 | 30521.5 | 31109.7 | 43808 | 44769.6 |
| 889.191 | 1465.15 | 1448.33 | 6546.43 | 6734.44 | 15687.4 | 15911.8 | 30695.3 | 31141 | 43974.9 | 44809.7 |
| 889.467 | 1483.63 | 1451.23 | 6683.61 | 6743.95 | 15747 | 15930.1 | 30686.2 | 31172.2 | 44008.5 | 44849.7 |
| 889.744 | 1476.3 | 1454.14 | 6627.69 | 6753.47 | 15709.4 | 15948.5 | 30433.4 | 31203.4 | 43799.6 | 44889.7 |
| 890.02 | 1439.98 | 1457.05 | 6629.75 | 6762.99 | 15666.4 | 15966.8 | 30414.7 | 31234.7 | 43975.8 | 44929.7 |
| 890.297 | 1432.04 | 1459.96 | 6659.39 | 6772.51 | 15611.1 | 15985.2 | 30616.9 | 31265.9 | 43994.4 | 44969.7 |
| 890.573 | 1450.97 | 1462.88 | 6605.5 | 6782.04 | 15773.4 | 16003.5 | 30717.3 | 31297.1 | 43952.9 | 45009.7 |
| 890.85 | 1463.12 | 1465.8 | 6584.59 | 6791.57 | 15680.3 | 16021.9 | 30288.4 | 31328.3 | 44055.6 | 45049.7 |
| 891.126 | 1463.34 | 1468.72 | 6641.92 | 6801.11 | 15730.9 | 16040.2 | 30665.2 | 31359.5 | 44176.9 | 45089.6 |
| 891.402 | 1433.63 | 1471.64 | 6589.09 | 6810.64 | 15823.7 | 16058.6 | 30749.4 | 31390.7 | 44180.7 | 45129.5 |
| 891.679 | 1466.39 | 1474.57 | 6608.64 | 6820.19 | 15794.2 | 16076.9 | 30886.1 | 31421.9 | 44145.7 | 45169.5 |
| 891.955 | 1482.62 | 1477.5 | 6731.12 | 6829.73 | 15652.8 | 16095.3 | 30586.4 | 31453.1 | 44177.4 | 45209.4 |
| 892.232 | 1473.68 | 1480.43 | 6717.5 | 6839.28 | 15704.5 | 16113.6 | 30718.5 | 31484.3 | 44047.8 | 45249.3 |
| 892.508 | 1489.65 | 1483.37 | 6729.17 | 6848.83 | 15901.2 | 16132 | 30757 | 31515.5 | 44367.1 | 45289.2 |
| 892.784 | 1489.52 | 1486.31 | 6811.21 | 6858.39 | 16025.9 | 16150.4 | 30826 | 31546.7 | 44497.5 | 45329.1 |
| 893.061 | 1470.09 | 1489.25 | 6697.57 | 6867.94 | 15882.7 | 16168.7 | 30791.8 | 31577.9 | 44287.2 | 45368.9 |
| 893.337 | 1472.79 | 1492.2 | 6678.44 | 6877.51 | 15776.7 | 16187.1 | 30780.5 | 31609 | 44072.9 | 45408.8 |
| 893.613 | 1441.1 | 1495.15 | 6733.67 | 6887.07 | 15950.6 | 16205.4 | 30876.7 | 31640.2 | 44314.9 | 45448.7 |
| 893.889 | 1514.87 | 1498.1 | 6694.78 | 6896.64 | 15814.2 | 16223.8 | 30800.3 | 31671.4 | 44041.9 | 45488.5 |
| 894.166 | 1533.46 | 1501.05 | 6779 | 6906.21 | 16017.9 | 16242.1 | 31165.9 | 31702.5 | 44791.9 | 45528.3 |
| 894.442 | 1507.01 | 1504.01 | 6751.6 | 6915.78 | 15805.7 | 16260.5 | 30885.5 | 31733.7 | 44476.3 | 45568.1 |
| 894.718 | 1530.81 | 1506.97 | 6677.77 | 6925.36 | 15961.1 | 16278.9 | 30854.8 | 31764.8 | 44273.6 | 45607.9 |
| 894.994 | 1476.45 | 1509.93 | 6805.03 | 6934.94 | 15979.8 | 16297.2 | 30986.2 | 31796 | 44537 | 45647.7 |
| 895.27 | 1563.32 | 1512.9 | 6812.63 | 6944.53 | 16018 | 16315.6 | 31064.6 | 31827.1 | 44560.7 | 45687.5 |
| 895.547 | 1506.97 | 1515.87 | 6859.09 | 6954.12 | 15999.1 | 16333.9 | 31088.3 | 31858.2 | 44748.1 | 45727.2 |
| 895.823 | 1495 | 1518.84 | 6740.79 | 6963.71 | 15919.6 | 16352.3 | 31079 | 31889.4 | 44783.1 | 45767 |
| 896.099 | 1508.77 | 1521.82 | 6821.35 | 6973.3 | 15958.3 | 16370.6 | 31294.6 | 31920.5 | 44846.8 | 45806.7 |
| 896.375 | 1529.58 | 1524.8 | 6815.61 | 6982.9 | 15884.9 | 16389 | 31320.4 | 31951.6 | 44768.5 | 45846.4 |
| 896.651 | 1540.4 | 1527.78 | 6779.66 | 6992.5 | 15980.2 | 16407.4 | 31142.9 | 31982.7 | 44799.8 | 45886.1 |
| 896.927 | 1534.32 | 1530.76 | 6915.08 | 7002.11 | 16050.7 | 16425.7 | 31477.3 | 32013.8 | 44786.4 | 45925.8 |
| 897.203 | 1543.33 | 1533.75 | 6774.49 | 7011.71 | 16066 | 16444.1 | 31182.8 | 32044.9 | 44796.5 | 45965.5 |
| 897.479 | 1556.76 | 1536.74 | 6853.79 | 7021.32 | 15972.4 | 16462.5 | 31169.5 | 32076 | 45027.2 | 46005.2 |
| 897.755 | 1523.7 | 1539.73 | 6880.31 | 7030.94 | 15971.9 | 16480.8 | 31342.1 | 32107.1 | 44801.4 | 46044.8 |
| 898.031 | 1533.17 | 1542.72 | 6857.21 | 7040.55 | 16157.5 | 16499.2 | 31331 | 32138.2 | 44942.8 | 46084.5 |
| 898.307 | 1515.02 | 1545.72 | 6873.8 | 7050.17 | 16084.1 | 16517.5 | 31452.6 | 32169.2 | 45031 | 46124.1 |
| 898.583 | 1541.34 | 1548.72 | 6971.16 | 7059.8 | 16184.3 | 16535.9 | 31390.9 | 32200.3 | 45274.7 | 46163.7 |
| 898.859 | 1492.44 | 1551.73 | 6992.94 | 7069.42 | 16134.8 | 16554.3 | 31603.8 | 32231.4 | 45211.8 | 46203.3 |
| 899.135 | 1585.33 | 1554.73 | 6930.22 | 7079.05 | 16205.7 | 16572.6 | 31557.1 | 32262.4 | 45222 | 46242.9 |
| 899.411 | 1568.63 | 1557.74 | 6911.51 | 7088.68 | 16206 | 16591 | 31523.2 | 32293.5 | 45210.7 | 46282.5 |
| 899.687 | 1538.99 | 1560.76 | 6923.94 | 7098.32 | 16266.1 | 16609.3 | 31660.6 | 32324.5 | 45305.9 | 46322.1 |
| 899.963 | 1592.09 | 1563.77 | 6931.64 | 7107.96 | 16238.7 | 16627.7 | 31754.2 | 32355.6 | 45505 | 46361.6 |
| 900.239 | 1597.23 | 1566.79 | 6979.12 | 7117.6 | 16334 | 16646.1 | 31792.7 | 32386.6 | 45555.5 | 46401.2 |
| 900.514 | 1578.25 | 1569.81 | 6911.94 | 7127.24 | 16070.6 | 16664.4 | 31478 | 32417.7 | 45247.4 | 46440.7 |
| 900.79 | 1560.15 | 1572.84 | 6992.27 | 7136.89 | 16488.4 | 16682.8 | 31819.8 | 32448.7 | 45532.3 | 46480.2 |
| 901.066 | 1570.44 | 1575.87 | 7081.83 | 7146.54 | 16276.7 | 16701.2 | 31805.4 | 32479.7 | 45471 | 46519.7 |
| 901.342 | 1538.2 | 1578.9 | 7008.65 | 7156.19 | 16488.7 | 16719.5 | 31663.7 | 32510.7 | 45563.9 | 46559.2 |
| 901.617 | 1569.73 | 1581.93 | 7015.05 | 7165.85 | 16444.6 | 16737.9 | 31697.8 | 32541.7 | 45567.2 | 46598.7 |
| 901.893 | 1593.53 | 1584.96 | 7063.89 | 7175.51 | 16442.3 | 16756.2 | 31813 | 32572.7 | 45719.5 | 46638.1 |
| 902.169 | 1599.96 | 1588 | 7079.86 | 7185.17 | 16475.9 | 16774.6 | 32002.5 | 32603.7 | 45950 | 46677.6 |
| 902.445 | 1603.94 | 1591.05 | 7064.7 | 7194.84 | 16421.3 | 16793 | 32111.5 | 32634.7 | 46048.7 | 46717 |
| 902.72 | 1561.98 | 1594.09 | 7013.11 | 7204.5 | 16605.1 | 16811.3 | 32156.6 | 32665.7 | 45913.4 | 46756.4 |
| 902.996 | 1601.59 | 1597.14 | 7008.37 | 7214.17 | 16513.4 | 16829.7 | 31870.7 | 32696.7 | 45932.8 | 46795.8 |
| 903.272 | 1599.53 | 1600.19 | 7092.78 | 7223.85 | 16532 | 16848 | 31758.2 | 32727.6 | 46026.7 | 46835.2 |
| 903.547 | 1561.92 | 1603.24 | 7017.37 | 7233.52 | 16493.6 | 16866.4 | 32138.2 | 32758.6 | 45789.7 | 46874.6 |
| 903.823 | 1597.7 | 1606.3 | 7000.57 | 7243.2 | 16709.1 | 16884.8 | 32047 | 32789.6 | 45810.8 | 46914 |
| 904.098 | 1616.79 | 1609.36 | 7031.08 | 7252.89 | 16518 | 16903.1 | 31802.8 | 32820.5 | 45765.2 | 46953.3 |
| 904.374 | 1611.27 | 1612.42 | 7104.99 | 7262.57 | 16667.2 | 16921.5 | 32140.9 | 32851.4 | 46145.4 | 46992.7 |
| 904.65 | 1628.13 | 1615.48 | 7185.38 | 7272.26 | 16639.9 | 16939.8 | 32073.5 | 32882.4 | 46084.3 | 47032 |
| 904.925 | 1630.49 | 1618.55 | 6977.28 | 7281.95 | 16617.4 | 16958.2 | 32366.9 | 32913.3 | 46088.4 | 47071.3 |
| 905.201 | 1597.44 | 1621.62 | 7158.37 | 7291.65 | 16623.9 | 16976.6 | 32091.4 | 32944.2 | 46203.9 | 47110.6 |
| 905.476 | 1635.04 | 1624.69 | 7127.73 | 7301.34 | 16772.3 | 16994.9 | 32530.5 | 32975.2 | 46236.2 | 47149.9 |
| 905.752 | 1604.57 | 1627.77 | 7135.76 | 7311.04 | 16701.3 | 17013.3 | 32107.8 | 33006.1 | 46340.2 | 47189.2 |
| 906.027 | 1667.56 | 1630.85 | 7027.78 | 7320.74 | 16702.6 | 17031.6 | 32377.6 | 33037 | 46193.4 | 47228.4 |
| 906.303 | 1641.67 | 1633.93 | 7192.07 | 7330.45 | 16755.2 | 17050 | 32377.8 | 33067.9 | 46565.3 | 47267.7 |
| 906.578 | 1611.86 | 1637.02 | 7118.55 | 7340.16 | 16457.9 | 17068.4 | 32060.2 | 33098.8 | 45652.8 | 47306.9 |
| 906.854 | 1635.33 | 1640.1 | 7140.99 | 7349.87 | 16681.3 | 17086.7 | 32258.2 | 33129.7 | 46198.8 | 47346.1 |
| 907.129 | 1628.18 | 1643.19 | 7185.1 | 7359.58 | 16549.7 | 17105.1 | 32277.6 | 33160.6 | 46350.7 | 47385.3 |
| 907.404 | 1626.14 | 1646.29 | 7036.38 | 7369.3 | 16730.7 | 17123.4 | 32337 | 33191.4 | 46312.9 | 47424.5 |
| 907.68 | 1635.47 | 1649.38 | 7156.59 | 7379.02 | 16859.5 | 17141.8 | 32505 | 33222.3 | 46373.9 | 47463.7 |
| 907.955 | 1643.22 | 1652.48 | 7255.82 | 7388.74 | 16935.7 | 17160.1 | 32426.5 | 33253.2 | 46331.4 | 47502.8 |
| 908.231 | 1619.8 | 1655.58 | 7240.34 | 7398.46 | 16731.5 | 17178.5 | 32431.6 | 33284 | 46365.5 | 47542 |
| 908.506 | 1678.42 | 1658.69 | 7248.56 | 7408.19 | 16847.5 | 17196.8 | 32496.4 | 33314.9 | 46553.5 | 47581.1 |
| 908.781 | 1657.56 | 1661.79 | 7256.02 | 7417.92 | 16776.2 | 17215.2 | 32171.4 | 33345.7 | 46184.2 | 47620.2 |
| 909.056 | 1682.21 | 1664.9 | 7271.37 | 7427.65 | 16778.8 | 17233.5 | 32434.6 | 33376.5 | 46792.8 | 47659.3 |
| 909.332 | 1631.73 | 1668.02 | 7234.91 | 7437.39 | 16834.1 | 17251.9 | 32488.6 | 33407.4 | 46416.1 | 47698.4 |
| 909.607 | 1650.59 | 1671.13 | 7219.52 | 7447.12 | 16901.7 | 17270.2 | 32509.6 | 33438.2 | 46317.6 | 47737.5 |
| 909.882 | 1645.94 | 1674.25 | 7180.73 | 7456.86 | 16850.6 | 17288.6 | 32477.4 | 33469 | 46413.2 | 47776.6 |
| 910.157 | 1651.35 | 1677.37 | 7176.83 | 7466.61 | 16959.1 | 17306.9 | 32689.9 | 33499.8 | 46467.4 | 47815.6 |
| 910.433 | 1671.21 | 1680.5 | 7365.06 | 7476.35 | 17011 | 17325.3 | 33025.4 | 33530.6 | 46999 | 47854.6 |
| 910.708 | 1624.76 | 1683.62 | 7339.62 | 7486.1 | 16918.9 | 17343.6 | 32824.6 | 33561.4 | 46745.4 | 47893.6 |
| 910.983 | 1681.29 | 1686.75 | 7262.85 | 7495.85 | 17086 | 17362 | 32830.3 | 33592.2 | 47126.4 | 47932.6 |
| 911.258 | 1666.39 | 1689.89 | 7407.06 | 7505.61 | 17061.2 | 17380.3 | 32814.2 | 33623 | 47014.4 | 47971.6 |
| 911.533 | 1702.03 | 1693.02 | 7323.29 | 7515.36 | 17086.1 | 17398.7 | 32984 | 33653.7 | 46885.6 | 48010.6 |
| 911.808 | 1669.44 | 1696.16 | 7343.78 | 7525.12 | 16917.5 | 17417 | 32592.3 | 33684.5 | 46661.7 | 48049.6 |
| 912.083 | 1678.72 | 1699.3 | 7447.06 | 7534.88 | 17073.9 | 17435.3 | 32879.8 | 33715.3 | 47224 | 48088.5 |
| 912.358 | 1679.98 | 1702.44 | 7391.29 | 7544.64 | 17215.9 | 17453.7 | 33010.6 | 33746 | 47423.3 | 48127.4 |
| 912.633 | 1687 | 1705.59 | 7354.47 | 7554.41 | 17229.9 | 17472 | 33038.4 | 33776.8 | 47045.9 | 48166.3 |
| 912.909 | 1710.92 | 1708.74 | 7546.03 | 7564.18 | 17120.6 | 17490.4 | 33033.4 | 33807.5 | 47276 | 48205.2 |
| 913.183 | 1674.73 | 1711.89 | 7391.31 | 7573.95 | 17111.9 | 17508.7 | 33095.1 | 33838.2 | 47353.4 | 48244.1 |
| 913.458 | 1713.77 | 1715.05 | 7444.8 | 7583.72 | 17186.9 | 17527 | 33303.8 | 33868.9 | 47303.5 | 48283 |
| 913.734 | 1735.73 | 1718.2 | 7410.86 | 7593.5 | 17205.7 | 17545.4 | 32822.7 | 33899.7 | 47128.6 | 48321.8 |
| 914.008 | 1674.04 | 1721.36 | 7435.88 | 7603.28 | 17013.2 | 17563.7 | 32851.7 | 33930.4 | 46901.2 | 48360.7 |
| 914.283 | 1726.57 | 1724.53 | 7388.7 | 7613.06 | 17140.6 | 17582 | 33159.8 | 33961.1 | 47522.1 | 48399.5 |
| 914.558 | 1737.95 | 1727.69 | 7421.93 | 7622.84 | 17126.5 | 17600.4 | 33251.8 | 33991.8 | 47583 | 48438.3 |
| 914.833 | 1706.63 | 1730.86 | 7410.66 | 7632.63 | 17361.9 | 17618.7 | 33135.3 | 34022.4 | 47366.6 | 48477.1 |
| 915.108 | 1672.35 | 1734.03 | 7551.13 | 7642.41 | 17224 | 17637 | 33351.9 | 34053.1 | 47421.5 | 48515.9 |
| 915.383 | 1729.76 | 1737.21 | 7443.18 | 7652.2 | 17285.2 | 17655.4 | 33217.1 | 34083.8 | 47724.3 | 48554.7 |
| 915.658 | 1749.43 | 1740.38 | 7576.8 | 7662 | 17298.9 | 17673.7 | 33479.8 | 34114.4 | 47754.7 | 48593.4 |
| 915.933 | 1753.25 | 1743.56 | 7479.16 | 7671.79 | 17333.6 | 17692 | 33182.6 | 34145.1 | 47256.2 | 48632.1 |
| 916.207 | 1750.84 | 1746.74 | 7566.97 | 7681.59 | 17379.9 | 17710.3 | 33515.4 | 34175.7 | 47919.9 | 48670.9 |
| 916.482 | 1711.13 | 1749.93 | 7597.89 | 7691.39 | 17414.8 | 17728.7 | 33404.6 | 34206.4 | 47752.1 | 48709.6 |
| 916.757 | 1748.67 | 1753.12 | 7433.71 | 7701.19 | 17428.2 | 17747 | 33446.7 | 34237 | 47793 | 48748.3 |
| 917.032 | 1731.55 | 1756.31 | 7472.18 | 7711 | 17429.3 | 17765.3 | 33553.7 | 34267.6 | 47876.2 | 48786.9 |
| 917.307 | 1681.69 | 1759.5 | 7506.44 | 7720.81 | 17488.5 | 17783.6 | 33643.4 | 34298.2 | 47564.7 | 48825.6 |
| 917.581 | 1725.53 | 1762.7 | 7514.34 | 7730.61 | 17448.2 | 17802 | 33742.5 | 34328.9 | 48228.7 | 48864.2 |
| 917.856 | 1768.23 | 1765.9 | 7479.57 | 7740.43 | 17407 | 17820.3 | 33589 | 34359.5 | 48035.7 | 48902.9 |
| 918.131 | 1765.59 | 1769.1 | 7474.38 | 7750.24 | 17491.9 | 17838.6 | 33925.6 | 34390.1 | 47949.9 | 48941.5 |
| 918.405 | 1752.91 | 1772.3 | 7533.2 | 7760.06 | 17491.4 | 17856.9 | 33737.2 | 34420.6 | 47998.2 | 48980.1 |
| 918.68 | 1797.96 | 1775.51 | 7533.55 | 7769.87 | 17618 | 17875.2 | 33709.5 | 34451.2 | 47933.1 | 49018.7 |
| 918.955 | 1753.47 | 1778.72 | 7482.82 | 7779.69 | 17325.6 | 17893.5 | 33692.4 | 34481.8 | 47696.9 | 49057.2 |
| 919.229 | 1759.92 | 1781.93 | 7665.83 | 7789.52 | 17646.8 | 17911.8 | 33911.1 | 34512.3 | 48243 | 49095.8 |
| 919.504 | 1765.39 | 1785.14 | 7570.93 | 7799.34 | 17663.7 | 17930.1 | 34007.6 | 34542.9 | 48318.9 | 49134.3 |
| 919.779 | 1802.54 | 1788.36 | 7730.11 | 7809.17 | 17644.6 | 17948.5 | 33998.5 | 34573.4 | 48456.9 | 49172.8 |
| 920.053 | 1783.09 | 1791.58 | 7614.52 | 7819 | 17744.3 | 17966.8 | 33664.7 | 34604 | 48148.6 | 49211.3 |
| 920.328 | 1796.22 | 1794.81 | 7638.48 | 7828.83 | 17704.3 | 17985.1 | 33764.3 | 34634.5 | 48414.6 | 49249.8 |
| 920.602 | 1828.65 | 1798.03 | 7689.83 | 7838.66 | 17683.4 | 18003.4 | 33924.7 | 34665 | 48129.4 | 49288.3 |
| 920.877 | 1804.93 | 1801.26 | 7599.81 | 7848.5 | 17611.5 | 18021.7 | 33821.4 | 34695.5 | 48250.6 | 49326.8 |
| 921.151 | 1774.92 | 1804.49 | 7636.98 | 7858.34 | 17633.2 | 18040 | 33808.4 | 34726 | 48471.2 | 49365.2 |
| 921.426 | 1744.19 | 1807.72 | 7592.44 | 7868.18 | 17670.6 | 18058.3 | 33905.5 | 34756.5 | 48066.2 | 49403.6 |
| 921.7 | 1823.55 | 1810.96 | 7795.17 | 7878.02 | 17730.8 | 18076.6 | 33910.5 | 34787 | 48498.1 | 49442 |
| 921.975 | 1764.6 | 1814.2 | 7707.31 | 7887.87 | 17867.7 | 18094.8 | 34157.9 | 34817.5 | 48523.3 | 49480.4 |
| 922.249 | 1750.56 | 1817.44 | 7777.97 | 7897.71 | 17762.2 | 18113.1 | 33985.6 | 34848 | 48695.9 | 49518.8 |
| 922.523 | 1813.02 | 1820.69 | 7698.87 | 7907.56 | 17733.2 | 18131.4 | 34163.9 | 34878.4 | 48441.7 | 49557.2 |
| 922.798 | 1797.09 | 1823.93 | 7609.8 | 7917.41 | 17750.4 | 18149.7 | 34037.2 | 34908.9 | 48309.2 | 49595.5 |
| 923.072 | 1814.43 | 1827.18 | 7774.36 | 7927.27 | 17832.5 | 18168 | 34309.9 | 34939.3 | 48638.5 | 49633.8 |
| 923.346 | 1786.45 | 1830.44 | 7735.98 | 7937.12 | 17828.5 | 18186.3 | 33898.1 | 34969.8 | 48403.3 | 49672.2 |
| 923.621 | 1828.25 | 1833.69 | 7688.59 | 7946.98 | 17779.9 | 18204.6 | 34340.1 | 35000.2 | 48420.4 | 49710.5 |
| 923.895 | 1818.63 | 1836.95 | 7856.7 | 7956.84 | 17864.3 | 18222.9 | 34251.9 | 35030.6 | 48337.3 | 49748.7 |
| 924.169 | 1842.12 | 1840.21 | 7805.57 | 7966.7 | 17766.1 | 18241.2 | 34314.9 | 35061.1 | 48534.1 | 49787 |
| 924.444 | 1843.5 | 1843.47 | 7809.93 | 7976.56 | 17784.5 | 18259.4 | 34093.1 | 35091.5 | 48753.1 | 49825.3 |
| 924.718 | 1823.6 | 1846.74 | 7666.94 | 7986.43 | 17774.7 | 18277.7 | 34378.1 | 35121.8 | 48770.2 | 49863.5 |
| 924.992 | 1852.41 | 1850.01 | 7734.58 | 7996.29 | 17909.7 | 18296 | 34472.6 | 35152.2 | 48859 | 49901.7 |
| 925.266 | 1812.47 | 1853.28 | 7762.09 | 8006.16 | 17934 | 18314.2 | 34495.7 | 35182.6 | 48717.4 | 49939.9 |
| 925.54 | 1798.09 | 1856.55 | 7803.62 | 8016.03 | 18056.9 | 18332.5 | 34302.9 | 35213 | 48745.3 | 49978.1 |
| 925.815 | 1877.91 | 1859.83 | 7859.03 | 8025.9 | 17945.8 | 18350.8 | 34569.9 | 35243.3 | 48687.3 | 50016.3 |
| 926.089 | 1851.02 | 1863.11 | 7902.2 | 8035.78 | 18166 | 18369.1 | 34462.8 | 35273.7 | 49107.4 | 50054.4 |
| 926.363 | 1781.18 | 1866.39 | 7922.91 | 8045.66 | 18121.3 | 18387.3 | 34494.8 | 35304.1 | 49178.7 | 50092.6 |
| 926.637 | 1830.4 | 1869.68 | 7851.36 | 8055.54 | 17885.5 | 18405.6 | 34475.2 | 35334.4 | 49181.1 | 50130.7 |
| 926.911 | 1815.3 | 1872.96 | 7949.35 | 8065.42 | 18148.4 | 18423.8 | 34498.3 | 35364.7 | 49083.7 | 50168.8 |
| 927.185 | 1846.6 | 1876.25 | 7874.39 | 8075.3 | 17955.5 | 18442.1 | 34547 | 35395 | 48698.9 | 50206.9 |
| 927.459 | 1875.98 | 1879.55 | 7837.29 | 8085.18 | 18084.1 | 18460.4 | 34854.9 | 35425.3 | 49324 | 50245 |
| 927.734 | 1868.32 | 1882.84 | 7823.91 | 8095.07 | 18062.4 | 18478.6 | 34576.8 | 35455.7 | 49367.4 | 50283 |
| 928.008 | 1899.68 | 1886.14 | 7990.44 | 8104.96 | 18213.5 | 18496.9 | 34743.1 | 35485.9 | 49566.7 | 50321.1 |
| 928.282 | 1915.95 | 1889.44 | 7971.82 | 8114.85 | 18148.1 | 18515.1 | 34815.2 | 35516.2 | 49195.9 | 50359.1 |
| 928.556 | 1925.57 | 1892.74 | 7958.3 | 8124.74 | 18252.1 | 18533.4 | 34861.2 | 35546.5 | 49598.9 | 50397.1 |
| 928.83 | 1892.58 | 1896.05 | 7912.59 | 8134.64 | 18228 | 18551.6 | 35024.8 | 35576.8 | 49295.3 | 50435.1 |
| 929.104 | 1863.86 | 1899.35 | 7942.26 | 8144.53 | 18157.2 | 18569.9 | 34696.7 | 35607 | 49237 | 50473.1 |
| 929.377 | 1914.63 | 1902.66 | 8016.71 | 8154.43 | 18192.7 | 18588.1 | 34825.7 | 35637.3 | 49534.3 | 50511 |
| 929.651 | 1922.05 | 1905.98 | 8109.91 | 8164.33 | 18275.1 | 18606.4 | 35209.7 | 35667.5 | 50159.5 | 50548.9 |
| 929.925 | 1831.83 | 1909.29 | 7979.93 | 8174.23 | 18439.2 | 18624.6 | 34896.6 | 35697.7 | 49473.4 | 50586.9 |
| 930.199 | 1877.12 | 1912.61 | 7978.04 | 8184.13 | 18254.3 | 18642.8 | 34939.4 | 35728 | 49602.5 | 50624.8 |
| 930.473 | 1914.74 | 1915.93 | 8049.49 | 8194.04 | 18438.6 | 18661.1 | 34961.7 | 35758.2 | 49791.1 | 50662.7 |
| 930.747 | 1896.46 | 1919.26 | 7958.88 | 8203.94 | 18327.6 | 18679.3 | 35196.4 | 35788.4 | 50114.3 | 50700.6 |
| 931.021 | 1897.34 | 1922.58 | 8031.83 | 8213.85 | 18432.6 | 18697.5 | 35019.4 | 35818.6 | 49533.2 | 50738.4 |
| 931.294 | 1922.09 | 1925.91 | 8032.84 | 8223.76 | 18540.9 | 18715.8 | 35432.9 | 35848.8 | 50215.1 | 50776.2 |
| 931.568 | 1945.62 | 1929.24 | 8099.39 | 8233.67 | 18486.5 | 18734 | 35305.1 | 35878.9 | 50099.7 | 50814.1 |
| 931.842 | 1950.48 | 1932.58 | 8022.29 | 8243.58 | 18435.7 | 18752.2 | 35176.5 | 35909.1 | 50045.1 | 50851.9 |
| 932.116 | 1904.32 | 1935.91 | 8219.16 | 8253.5 | 18554 | 18770.4 | 35455.7 | 35939.2 | 50077.7 | 50889.7 |
| 932.39 | 1895.29 | 1939.25 | 8097.36 | 8263.42 | 18518.1 | 18788.6 | 35442.3 | 35969.4 | 50144.6 | 50927.4 |
| 932.663 | 1891.16 | 1942.59 | 8030.39 | 8273.33 | 18774.6 | 18806.9 | 35512.4 | 35999.5 | 50052.3 | 50965.2 |
| 932.937 | 1948.64 | 1945.94 | 8049.35 | 8283.25 | 18550.2 | 18825.1 | 35240.1 | 36029.7 | 50185.5 | 51002.9 |
| 933.211 | 1965.63 | 1949.28 | 8091.29 | 8293.18 | 18614.1 | 18843.3 | 35357.8 | 36059.8 | 50122.4 | 51040.7 |
| 933.484 | 1911.41 | 1952.63 | 8170.87 | 8303.1 | 18515.4 | 18861.5 | 35605.1 | 36089.9 | 50265.6 | 51078.4 |
| 933.758 | 1883.4 | 1955.98 | 8090.17 | 8313.02 | 18444.4 | 18879.7 | 35339.9 | 36120 | 49991.8 | 51116.1 |
| 934.032 | 1959.77 | 1959.34 | 8213.71 | 8322.95 | 18689.7 | 18897.9 | 35583.4 | 36150.1 | 50485 | 51153.7 |
| 934.305 | 1878.21 | 1962.69 | 8081.07 | 8332.88 | 18473.9 | 18916.1 | 35257.9 | 36180.2 | 50306.3 | 51191.4 |
| 934.579 | 1974.79 | 1966.05 | 8146.33 | 8342.81 | 18604 | 18934.3 | 35346 | 36210.2 | 50217 | 51229 |
| 934.852 | 1930.27 | 1969.42 | 8124.06 | 8352.74 | 18724.5 | 18952.5 | 35335.6 | 36240.3 | 50140.4 | 51266.6 |
| 935.126 | 1970.34 | 1972.78 | 8085.85 | 8362.67 | 18667.3 | 18970.7 | 35382.9 | 36270.3 | 50616.7 | 51304.2 |
| 935.4 | 1993.95 | 1976.15 | 8095.04 | 8372.61 | 18526.4 | 18988.9 | 35380.1 | 36300.4 | 50289.6 | 51341.8 |
| 935.673 | 1972.95 | 1979.52 | 8227.49 | 8382.54 | 18637.4 | 19007.1 | 35494.9 | 36330.4 | 50340.7 | 51379.4 |
| 935.947 | 1894.85 | 1982.89 | 8156.7 | 8392.48 | 18655.9 | 19025.2 | 35459.5 | 36360.5 | 50596.6 | 51416.9 |
| 936.22 | 1996.44 | 1986.26 | 8240.28 | 8402.42 | 18877.1 | 19043.4 | 35866.8 | 36390.5 | 50848.4 | 51454.5 |
| 936.494 | 1980.73 | 1989.64 | 8211.36 | 8412.36 | 18762 | 19061.6 | 35751.9 | 36420.5 | 50535.1 | 51492 |
| 936.767 | 2001.24 | 1993.02 | 8207.67 | 8422.3 | 18729.3 | 19079.8 | 35675.3 | 36450.5 | 50536.5 | 51529.5 |
| 937.04 | 2010.01 | 1996.4 | 8220.74 | 8432.25 | 18638.2 | 19098 | 35699.3 | 36480.5 | 50710.1 | 51567 |
| 937.314 | 1951.35 | 1999.79 | 8189.22 | 8442.19 | 18587.8 | 19116.1 | 35711.7 | 36510.4 | 50938.3 | 51604.4 |
| 937.587 | 1961.87 | 2003.17 | 8173.99 | 8452.14 | 18681 | 19134.3 | 35714.2 | 36540.4 | 50958.7 | 51641.9 |
| 937.861 | 2029.38 | 2006.56 | 8148.58 | 8462.09 | 18805.8 | 19152.5 | 35693.2 | 36570.4 | 50572.1 | 51679.3 |
| 938.134 | 1989.35 | 2009.96 | 8300.6 | 8472.04 | 18884.6 | 19170.6 | 35663.1 | 36600.3 | 50325 | 51716.7 |
| 938.407 | 2008.26 | 2013.35 | 8273.18 | 8481.99 | 18773.5 | 19188.8 | 35659.2 | 36630.2 | 50585.9 | 51754.1 |
| 938.681 | 1998.54 | 2016.75 | 8177.37 | 8491.94 | 18781.3 | 19207 | 35514.7 | 36660.2 | 50687.4 | 51791.5 |
| 938.954 | 1975.9 | 2020.15 | 8346.34 | 8501.89 | 18896.1 | 19225.1 | 35713 | 36690.1 | 50769.5 | 51828.9 |
| 939.227 | 1975.41 | 2023.55 | 8275.59 | 8511.85 | 18630.2 | 19243.2 | 35666 | 36720 | 50471.3 | 51866.2 |
| 939.5 | 1980.23 | 2026.95 | 8245.75 | 8521.81 | 18808.8 | 19261.4 | 35846.4 | 36749.9 | 50507.2 | 51903.5 |
| 939.774 | 2002.74 | 2030.36 | 8292.13 | 8531.76 | 18717 | 19279.5 | 35796.5 | 36779.8 | 50701.9 | 51940.8 |
| 940.047 | 2027.58 | 2033.77 | 8267.41 | 8541.72 | 18843.6 | 19297.7 | 35894.5 | 36809.7 | 51012.8 | 51978.1 |
| 940.32 | 1999.78 | 2037.18 | 8314.71 | 8551.68 | 18881 | 19315.8 | 36026.4 | 36839.5 | 51101.9 | 52015.4 |
| 940.593 | 2022.32 | 2040.6 | 8292.2 | 8561.65 | 18762.1 | 19334 | 35636.5 | 36869.4 | 50603.6 | 52052.7 |
| 940.867 | 1982.44 | 2044.01 | 8294.15 | 8571.61 | 18738.7 | 19352.1 | 35975.7 | 36899.2 | 50700.4 | 52089.9 |
| 941.14 | 2006.67 | 2047.43 | 8342.28 | 8581.57 | 18979.6 | 19370.2 | 36042.8 | 36929.1 | 50744.6 | 52127.1 |
| 941.413 | 1930.96 | 2050.86 | 8355.06 | 8591.54 | 19150 | 19388.4 | 36044.8 | 36958.9 | 50420.3 | 52164.3 |
| 941.686 | 1953.47 | 2054.28 | 8264.59 | 8601.51 | 19112.4 | 19406.5 | 36330 | 36988.7 | 51009.9 | 52201.5 |
| 941.959 | 1991.23 | 2057.71 | 8250.4 | 8611.48 | 18866.9 | 19424.6 | 36038.3 | 37018.5 | 51118.2 | 52238.7 |
| 942.232 | 2039.31 | 2061.14 | 8370.46 | 8621.44 | 18733.7 | 19442.7 | 36052.8 | 37048.3 | 50871.7 | 52275.8 |
| 942.505 | 2037.97 | 2064.57 | 8341.23 | 8631.42 | 18941 | 19460.8 | 36094.8 | 37078.1 | 51046.3 | 52312.9 |
| 942.778 | 1994.86 | 2068 | 8410.97 | 8641.39 | 19208.7 | 19479 | 36332.7 | 37107.9 | 51289 | 52350.1 |
| 943.051 | 2033.01 | 2071.44 | 8368.68 | 8651.36 | 18991.5 | 19497.1 | 35934.6 | 37137.6 | 51330.6 | 52387.1 |
| 943.324 | 2016.49 | 2074.88 | 8404.4 | 8661.34 | 19226.8 | 19515.2 | 36228.6 | 37167.4 | 51191.6 | 52424.2 |
| 943.597 | 2062.26 | 2078.32 | 8395.41 | 8671.31 | 18882.6 | 19533.3 | 36058.2 | 37197.1 | 51199.7 | 52461.3 |
| 943.87 | 2044.09 | 2081.76 | 8344.43 | 8681.29 | 19114.2 | 19551.4 | 36213.2 | 37226.9 | 51211 | 52498.3 |
| 944.143 | 2058.81 | 2085.21 | 8405.35 | 8691.27 | 19066.1 | 19569.5 | 36404.6 | 37256.6 | 51352.1 | 52535.3 |
| 944.416 | 1975.15 | 2088.66 | 8479.58 | 8701.25 | 19174.5 | 19587.6 | 36290 | 37286.3 | 51251.1 | 52572.3 |
| 944.689 | 2045.05 | 2092.11 | 8450.74 | 8711.23 | 19120.7 | 19605.7 | 36501.9 | 37316 | 51479.4 | 52609.3 |
| 944.962 | 2052.24 | 2095.56 | 8449.17 | 8721.21 | 19224.8 | 19623.8 | 36605.5 | 37345.7 | 51381.2 | 52646.3 |
| 945.235 | 2024.61 | 2099.02 | 8490.89 | 8731.2 | 19157.8 | 19641.8 | 36356.9 | 37375.4 | 51577.4 | 52683.2 |
| 945.508 | 2083.2 | 2102.48 | 8543.18 | 8741.18 | 19157.8 | 19659.9 | 36526.2 | 37405.1 | 51475.1 | 52720.2 |
| 945.78 | 2007.92 | 2105.94 | 8484.3 | 8751.17 | 19256.6 | 19678 | 36824.9 | 37434.7 | 51711.7 | 52757.1 |
| 946.053 | 2059.44 | 2109.4 | 8560.78 | 8761.15 | 19545.8 | 19696.1 | 36659.9 | 37464.4 | 51731.5 | 52794 |
| 946.326 | 2114.42 | 2112.87 | 8533.4 | 8771.14 | 19046 | 19714.2 | 36836.2 | 37494 | 51469.4 | 52830.9 |
| 946.599 | 2077.36 | 2116.34 | 8427.89 | 8781.13 | 19199.1 | 19732.2 | 36427.7 | 37523.7 | 51337.5 | 52867.7 |
| 946.872 | 2089.86 | 2119.81 | 8514.99 | 8791.12 | 19356.8 | 19750.3 | 36264.9 | 37553.3 | 51845.8 | 52904.6 |
| 947.144 | 2086.4 | 2123.28 | 8689.54 | 8801.11 | 19291.6 | 19768.3 | 36812.4 | 37582.9 | 51572.7 | 52941.4 |
| 947.417 | 2099.91 | 2126.75 | 8639.66 | 8811.1 | 19390.2 | 19786.4 | 36875.2 | 37612.5 | 51679.3 | 52978.2 |
| 947.69 | 2061.31 | 2130.23 | 8559.08 | 8821.1 | 19311.1 | 19804.5 | 36650.1 | 37642.1 | 51357.6 | 53015 |
| 947.963 | 2140.72 | 2133.71 | 8575.67 | 8831.09 | 19418.8 | 19822.5 | 36647.5 | 37671.7 | 51944.7 | 53051.7 |
| 948.235 | 2112.29 | 2137.19 | 8616.42 | 8841.09 | 19169.8 | 19840.6 | 36532 | 37701.2 | 51618.4 | 53088.5 |
| 948.508 | 2095.98 | 2140.68 | 8579.02 | 8851.08 | 19348.5 | 19858.6 | 36989.2 | 37730.8 | 51841.2 | 53125.2 |
| 948.78 | 2098.72 | 2144.17 | 8653.98 | 8861.08 | 19301.7 | 19876.7 | 36831 | 37760.3 | 52033.6 | 53161.9 |
| 949.053 | 2087.39 | 2147.66 | 8606.83 | 8871.08 | 19246 | 19894.7 | 36676.2 | 37789.8 | 51494.6 | 53198.6 |
| 949.326 | 2075.73 | 2151.15 | 8749.12 | 8881.08 | 19278.8 | 19912.7 | 36894.6 | 37819.4 | 51918.9 | 53235.3 |
| 949.598 | 2139.97 | 2154.64 | 8588.02 | 8891.08 | 19435.3 | 19930.8 | 36562.8 | 37848.9 | 51679.4 | 53271.9 |
| 949.871 | 2051.32 | 2158.14 | 8605.07 | 8901.08 | 19483.9 | 19948.8 | 36690.1 | 37878.4 | 51881.7 | 53308.6 |
| 950.143 | 2093.76 | 2161.64 | 8586.1 | 8911.08 | 19341.7 | 19966.8 | 36881 | 37907.9 | 51898.6 | 53345.2 |
| 950.416 | 2087.26 | 2165.14 | 8695.06 | 8921.09 | 19499.7 | 19984.8 | 37040.6 | 37937.3 | 51720.1 | 53381.8 |
| 950.688 | 2167.46 | 2168.64 | 8688.93 | 8931.09 | 19504.8 | 20002.8 | 36837.1 | 37966.8 | 52331.7 | 53418.4 |
| 950.961 | 2143.83 | 2172.15 | 8740.91 | 8941.09 | 19513 | 20020.8 | 37049 | 37996.3 | 51921.5 | 53454.9 |
| 951.233 | 2122.09 | 2175.66 | 8491.69 | 8951.1 | 19561.3 | 20038.9 | 36710.1 | 38025.7 | 51796.5 | 53491.5 |
| 951.506 | 2112.53 | 2179.17 | 8734.96 | 8961.11 | 19450.2 | 20056.9 | 37048.5 | 38055.2 | 52037 | 53528 |
| 951.778 | 2160.6 | 2182.68 | 8622.59 | 8971.12 | 19453.3 | 20074.9 | 36977.1 | 38084.6 | 51907.6 | 53564.5 |
| 952.051 | 2117.61 | 2186.2 | 8596.53 | 8981.12 | 19584 | 20092.9 | 36923.2 | 38114 | 51786.9 | 53601 |
| 952.323 | 2109.98 | 2189.71 | 8652.29 | 8991.14 | 19614.2 | 20110.9 | 37185.8 | 38143.4 | 52027.1 | 53637.5 |
| 952.595 | 2136.4 | 2193.23 | 8688.62 | 9001.15 | 19786.9 | 20128.8 | 36941.6 | 38172.8 | 52509.7 | 53673.9 |
| 952.868 | 2125.56 | 2196.75 | 8719.35 | 9011.16 | 19602.3 | 20146.8 | 36992.2 | 38202.2 | 52404.1 | 53710.4 |
| 953.14 | 2120.38 | 2200.28 | 8758.8 | 9021.17 | 19541.7 | 20164.8 | 36860.7 | 38231.5 | 52015.4 | 53746.8 |
| 953.412 | 2088.61 | 2203.81 | 8745.01 | 9031.18 | 19606.3 | 20182.8 | 36763.2 | 38260.9 | 51921.7 | 53783.2 |
| 953.685 | 2079.16 | 2207.34 | 8773.84 | 9041.2 | 19596.4 | 20200.8 | 36828.7 | 38290.2 | 52364.4 | 53819.6 |
| 953.957 | 2184.85 | 2210.87 | 8868.12 | 9051.21 | 19826.2 | 20218.8 | 37523.9 | 38319.6 | 52642.9 | 53855.9 |
| 954.229 | 2168.27 | 2214.4 | 8782.32 | 9061.23 | 19597.3 | 20236.7 | 36910.5 | 38348.9 | 51996.9 | 53892.2 |
| 954.501 | 2174.32 | 2217.94 | 8764.22 | 9071.24 | 19659 | 20254.7 | 37330.7 | 38378.2 | 52084.2 | 53928.6 |
| 954.774 | 2190.64 | 2221.47 | 8810.94 | 9081.26 | 19858.2 | 20272.6 | 37241.8 | 38407.5 | 52460.6 | 53964.9 |
| 955.046 | 2139.6 | 2225.02 | 8841.46 | 9091.28 | 19492.3 | 20290.6 | 37165.3 | 38436.8 | 52365 | 54001.1 |
| 955.318 | 2182.16 | 2228.56 | 8777.04 | 9101.3 | 19826.7 | 20308.5 | 37375.6 | 38466.1 | 52368.1 | 54037.4 |
| 955.59 | 2254.12 | 2232.1 | 8746.56 | 9111.32 | 19820.6 | 20326.5 | 37268.2 | 38495.4 | 52423.6 | 54073.6 |
| 955.862 | 2228.94 | 2235.65 | 8759.38 | 9121.34 | 19709.8 | 20344.5 | 37213.3 | 38524.6 | 52682.3 | 54109.9 |
| 956.135 | 2246.92 | 2239.2 | 8826.6 | 9131.36 | 19766.4 | 20362.4 | 37738.4 | 38553.9 | 52770.2 | 54146.1 |
| 956.407 | 2223.55 | 2242.75 | 8826.75 | 9141.38 | 19932 | 20380.3 | 37480.5 | 38583.1 | 52761.3 | 54182.3 |
| 956.679 | 2177.12 | 2246.31 | 8778.86 | 9151.4 | 20065.3 | 20398.3 | 37546.2 | 38612.3 | 52708.2 | 54218.4 |
| 956.951 | 2200.5 | 2249.86 | 8885.16 | 9161.42 | 19810.1 | 20416.2 | 37366.3 | 38641.5 | 52683.4 | 54254.6 |
| 957.223 | 2197.68 | 2253.42 | 8830.44 | 9171.45 | 19789.8 | 20434.1 | 37762.4 | 38670.7 | 52528.9 | 54290.7 |
| 957.495 | 2205.82 | 2256.99 | 8859.25 | 9181.47 | 20095 | 20452 | 37414 | 38699.9 | 52819 | 54326.8 |
| 957.767 | 2208.18 | 2260.55 | 8865.6 | 9191.5 | 19648.3 | 20470 | 37581 | 38729.1 | 52838.3 | 54362.9 |
| 958.039 | 2228.68 | 2264.11 | 8920.35 | 9201.52 | 19995.8 | 20487.9 | 37401.5 | 38758.3 | 53230.4 | 54399 |
| 958.311 | 2197.53 | 2267.68 | 9046.14 | 9211.55 | 20087.5 | 20505.8 | 37801.8 | 38787.4 | 53021.9 | 54435 |
| 958.583 | 2223.97 | 2271.25 | 9143.29 | 9221.58 | 20048.1 | 20523.7 | 37982.7 | 38816.6 | 53741.1 | 54471.1 |
| 958.855 | 2234.57 | 2274.82 | 8892.98 | 9231.6 | 20094.1 | 20541.6 | 37905.4 | 38845.7 | 53005.1 | 54507.1 |
| 959.127 | 2238.97 | 2278.4 | 8900.93 | 9241.63 | 20016 | 20559.5 | 37784.7 | 38874.8 | 53347 | 54543.1 |
| 959.399 | 2197.39 | 2281.98 | 9095.17 | 9251.66 | 19890.3 | 20577.4 | 38018.6 | 38903.9 | 52919.8 | 54579 |
| 959.67 | 2201.86 | 2285.55 | 8986.56 | 9261.69 | 20280.1 | 20595.3 | 38346.2 | 38933 | 53724.1 | 54615 |
| 959.942 | 2220.34 | 2289.14 | 9024.19 | 9271.72 | 20255.2 | 20613.1 | 38062 | 38962.1 | 53706.3 | 54650.9 |
| 960.214 | 2285.36 | 2292.72 | 9001.32 | 9281.75 | 20186.7 | 20631 | 38071.5 | 38991.2 | 53245 | 54686.8 |
| 960.486 | 2273.93 | 2296.31 | 9005.99 | 9291.78 | 20193.9 | 20648.9 | 38187.8 | 39020.2 | 53353.4 | 54722.7 |
| 960.758 | 2205.55 | 2299.89 | 9009.57 | 9301.82 | 20178.3 | 20666.8 | 38173.3 | 39049.3 | 53457.1 | 54758.6 |
| 961.03 | 2276.92 | 2303.48 | 9109.76 | 9311.85 | 20437.2 | 20684.6 | 38128.2 | 39078.3 | 53830.1 | 54794.5 |
| 961.301 | 2217.12 | 2307.08 | 9175.29 | 9321.88 | 20519.7 | 20702.5 | 38246.6 | 39107.3 | 53691.8 | 54830.3 |
| 961.573 | 2236.94 | 2310.67 | 9075.63 | 9331.91 | 20475.1 | 20720.4 | 38310.7 | 39136.3 | 53763.8 | 54866.1 |
| 961.845 | 2270.24 | 2314.27 | 9195.77 | 9341.95 | 20397.2 | 20738.2 | 38129.7 | 39165.3 | 53889.7 | 54901.9 |
| 962.117 | 2343.86 | 2317.87 | 9195.65 | 9351.98 | 20437.9 | 20756.1 | 38405.9 | 39194.3 | 53813.5 | 54937.7 |
| 962.388 | 2282.13 | 2321.47 | 9133.59 | 9362.02 | 20409.5 | 20773.9 | 38438.3 | 39223.3 | 54111.3 | 54973.5 |
| 962.66 | 2204.81 | 2325.07 | 9165.54 | 9372.05 | 20574.7 | 20791.8 | 38519.1 | 39252.3 | 54072.1 | 55009.2 |
| 962.932 | 2322.44 | 2328.68 | 9247.35 | 9382.09 | 20642.8 | 20809.6 | 38643.4 | 39281.2 | 53845.8 | 55044.9 |
| 963.203 | 2304.23 | 2332.28 | 9191.14 | 9392.12 | 20385.7 | 20827.4 | 38477.8 | 39310.2 | 53915.5 | 55080.6 |
| 963.475 | 2285.58 | 2335.89 | 9206.01 | 9402.16 | 20511.1 | 20845.3 | 38210.6 | 39339.1 | 53948.6 | 55116.3 |
| 963.746 | 2344.29 | 2339.51 | 9130.4 | 9412.2 | 20568.2 | 20863.1 | 38848 | 39368 | 54156.4 | 55152 |
| 964.018 | 2276.61 | 2343.12 | 9194.59 | 9422.23 | 20481.7 | 20880.9 | 38502.7 | 39396.9 | 53693.9 | 55187.6 |
| 964.29 | 2275.67 | 2346.74 | 9249.88 | 9432.27 | 20437.3 | 20898.7 | 38727.5 | 39425.8 | 54207.6 | 55223.2 |
| 964.561 | 2283.99 | 2350.36 | 9216.58 | 9442.31 | 20845 | 20916.5 | 38795.7 | 39454.7 | 54378 | 55258.8 |
| 964.833 | 2331.84 | 2353.98 | 9175.62 | 9452.35 | 20522.5 | 20934.3 | 38516.5 | 39483.6 | 54254.5 | 55294.4 |
| 965.104 | 2366.26 | 2357.6 | 9409.61 | 9462.39 | 20797.5 | 20952.1 | 38942.4 | 39512.4 | 54445.9 | 55330 |
| 965.376 | 2392.53 | 2361.23 | 9299.59 | 9472.43 | 20729.5 | 20969.9 | 38981.8 | 39541.3 | 54712.9 | 55365.5 |
| 965.647 | 2266.92 | 2364.85 | 9222.57 | 9482.47 | 20665.5 | 20987.7 | 38990.7 | 39570.1 | 54859.2 | 55401.1 |
| 965.919 | 2375.62 | 2368.48 | 9232.77 | 9492.51 | 20663.6 | 21005.5 | 38591.5 | 39598.9 | 54279.7 | 55436.6 |
| 966.19 | 2372.72 | 2372.11 | 9286.91 | 9502.55 | 20725.9 | 21023.3 | 39158.8 | 39627.7 | 54977.4 | 55472 |
| 966.461 | 2299.82 | 2375.75 | 9300.56 | 9512.59 | 20608.4 | 21041.1 | 38701 | 39656.5 | 54792.2 | 55507.5 |
| 966.733 | 2383.07 | 2379.39 | 9373.15 | 9522.63 | 20782.8 | 21058.9 | 39280.7 | 39685.3 | 54764.5 | 55542.9 |
| 967.004 | 2288.35 | 2383.02 | 9413.82 | 9532.67 | 20802.6 | 21076.6 | 39011.5 | 39714.1 | 54761.6 | 55578.4 |
| 967.275 | 2299 | 2386.66 | 9317.36 | 9542.71 | 20708.6 | 21094.4 | 39012.9 | 39742.8 | 54602.7 | 55613.8 |
| 967.547 | 2337.1 | 2390.3 | 9252.1 | 9552.75 | 20718.7 | 21112.2 | 38594.9 | 39771.6 | 54154.8 | 55649.2 |
| 967.818 | 2342.94 | 2393.95 | 9263.32 | 9562.8 | 20611.3 | 21129.9 | 38974.4 | 39800.3 | 54680.5 | 55684.5 |
| 968.09 | 2274.12 | 2397.6 | 9409.6 | 9572.84 | 20791.9 | 21147.7 | 38937.2 | 39829 | 54346.2 | 55719.9 |
| 968.361 | 2354.3 | 2401.24 | 9357.53 | 9582.88 | 21054.3 | 21165.4 | 39512.4 | 39857.7 | 55114.5 | 55755.2 |
| 968.632 | 2332.07 | 2404.9 | 9212.26 | 9592.92 | 20720.2 | 21183.2 | 38741.3 | 39886.4 | 54830.8 | 55790.5 |
| 968.903 | 2403.44 | 2408.55 | 9440.24 | 9602.97 | 20834.2 | 21200.9 | 38948.5 | 39915.1 | 54782.5 | 55825.8 |
| 969.174 | 2373.26 | 2412.2 | 9453.69 | 9613.01 | 21098.4 | 21218.6 | 39205.2 | 39943.8 | 54925 | 55861.1 |
| 969.446 | 2397.05 | 2415.86 | 9545.95 | 9623.06 | 21024.8 | 21236.4 | 39185.8 | 39972.5 | 55134.4 | 55896.3 |
| 969.717 | 2365.26 | 2419.52 | 9503.94 | 9633.1 | 20749.7 | 21254.1 | 38978.2 | 40001.1 | 54737.8 | 55931.5 |
| 969.988 | 2380.88 | 2423.18 | 9507.6 | 9643.14 | 20869.2 | 21271.8 | 39328.8 | 40029.7 | 55154.3 | 55966.7 |
| 970.259 | 2334.22 | 2426.84 | 9548.73 | 9653.19 | 20929.7 | 21289.5 | 39628.9 | 40058.4 | 55076.7 | 56001.9 |
| 970.53 | 2419.32 | 2430.51 | 9400.54 | 9663.23 | 21103.2 | 21307.2 | 38908.2 | 40087 | 54245 | 56037.1 |
| 970.801 | 2356.63 | 2434.18 | 9460.14 | 9673.28 | 20903.9 | 21324.9 | 38957.3 | 40115.6 | 54308.5 | 56072.2 |
| 971.072 | 2342.36 | 2437.85 | 9442.22 | 9683.32 | 20963.5 | 21342.6 | 39373.2 | 40144.2 | 55312.1 | 56107.3 |
| 971.343 | 2417.73 | 2441.52 | 9503.39 | 9693.37 | 20969.1 | 21360.3 | 39819 | 40172.7 | 55493.7 | 56142.4 |
| 971.615 | 2366.72 | 2445.19 | 9303.38 | 9703.41 | 20755.6 | 21378 | 39239.7 | 40201.3 | 55171.9 | 56177.5 |
| 971.886 | 2383.78 | 2448.87 | 9453.96 | 9713.46 | 21393.5 | 21395.7 | 39480.2 | 40229.8 | 55058.3 | 56212.6 |
| 972.157 | 2406.81 | 2452.55 | 9444.52 | 9723.5 | 21287.4 | 21413.4 | 39720.2 | 40258.4 | 55357.6 | 56247.6 |
| 972.428 | 2387.12 | 2456.23 | 9580.64 | 9733.55 | 20959.2 | 21431.1 | 39658.3 | 40286.9 | 55583.9 | 56282.7 |
| 972.698 | 2447.98 | 2459.91 | 9612.85 | 9743.59 | 21127.7 | 21448.7 | 39592.8 | 40315.4 | 55459.8 | 56317.7 |
| 972.97 | 2441.27 | 2463.59 | 9513.92 | 9753.64 | 20883 | 21466.4 | 39303.9 | 40343.9 | 55219.8 | 56352.6 |
| 973.24 | 2401.54 | 2467.28 | 9401.04 | 9763.68 | 20876.2 | 21484.1 | 39302.2 | 40372.4 | 54953.5 | 56387.6 |
| 973.511 | 2468.57 | 2470.97 | 9477.05 | 9773.73 | 21318.3 | 21501.7 | 39307.5 | 40400.9 | 54737.5 | 56422.6 |
| 973.782 | 2434.92 | 2474.66 | 9678.51 | 9783.78 | 21212.6 | 21519.4 | 39501.8 | 40429.3 | 55486.9 | 56457.5 |
| 974.053 | 2354.91 | 2478.35 | 9553.69 | 9793.82 | 21172.1 | 21537 | 39592.7 | 40457.8 | 55534.4 | 56492.4 |
| 974.324 | 2448.94 | 2482.05 | 9482.69 | 9803.87 | 21170.9 | 21554.7 | 39567.4 | 40486.2 | 55719.2 | 56527.2 |
| 974.595 | 2401.04 | 2485.74 | 9477.82 | 9813.91 | 21027.6 | 21572.3 | 39753.4 | 40514.6 | 55503.6 | 56562.1 |
| 974.866 | 2451.63 | 2489.44 | 9636.21 | 9823.96 | 21114.3 | 21589.9 | 39637.2 | 40543 | 55593 | 56596.9 |
| 975.136 | 2447.97 | 2493.14 | 9658.39 | 9834 | 21027.4 | 21607.5 | 39807.7 | 40571.4 | 55600.7 | 56631.8 |
| 975.407 | 2403.84 | 2496.84 | 9581.92 | 9844.05 | 21090.2 | 21625.2 | 39280 | 40599.8 | 55338.8 | 56666.6 |
| 975.678 | 2452.08 | 2500.55 | 9684.72 | 9854.1 | 21366.7 | 21642.8 | 39949.6 | 40628.2 | 55634.7 | 56701.3 |
| 975.949 | 2533.74 | 2504.25 | 9411.2 | 9864.14 | 21379 | 21660.4 | 39979.2 | 40656.5 | 55737 | 56736.1 |
| 976.219 | 2419.02 | 2507.96 | 9665.99 | 9874.19 | 21307.4 | 21678 | 40058.3 | 40684.9 | 55892 | 56770.8 |
| 976.49 | 2444.99 | 2511.67 | 9677.89 | 9884.23 | 21355.3 | 21695.6 | 40348.9 | 40713.2 | 55590.1 | 56805.6 |
| 976.761 | 2497.57 | 2515.39 | 9781.96 | 9894.28 | 21240.8 | 21713.2 | 40136.6 | 40741.5 | 55962.1 | 56840.2 |
| 977.031 | 2508.95 | 2519.1 | 9600.95 | 9904.33 | 21204.7 | 21730.8 | 39987.2 | 40769.8 | 56136.2 | 56874.9 |
| 977.302 | 2365.56 | 2522.82 | 9562.2 | 9914.37 | 21086 | 21748.4 | 39795.4 | 40798.1 | 56082.5 | 56909.6 |
| 977.573 | 2475.82 | 2526.53 | 9749.48 | 9924.42 | 21533 | 21766 | 40134.9 | 40826.4 | 55619.9 | 56944.2 |
| 977.843 | 2556.93 | 2530.26 | 9654.73 | 9934.46 | 21596.7 | 21783.5 | 40098.8 | 40854.7 | 55643.2 | 56978.8 |
| 978.114 | 2446.63 | 2533.98 | 9730.46 | 9944.51 | 21412.2 | 21801.1 | 40093.9 | 40882.9 | 55945.3 | 57013.4 |
| 978.385 | 2452.24 | 2537.7 | 9819.74 | 9954.55 | 21388.1 | 21818.7 | 40650.4 | 40911.2 | 56041.1 | 57048 |
| 978.655 | 2531.13 | 2541.43 | 9703.49 | 9964.6 | 21602.8 | 21836.2 | 40069.3 | 40939.4 | 56225.7 | 57082.5 |
| 978.926 | 2455.12 | 2545.16 | 9762.55 | 9974.64 | 21486.7 | 21853.8 | 39879.4 | 40967.6 | 55986 | 57117.1 |
| 979.196 | 2557.2 | 2548.89 | 9779.9 | 9984.69 | 21673.1 | 21871.3 | 40316.8 | 40995.8 | 56364.9 | 57151.6 |
| 979.466 | 2551.19 | 2552.62 | 9773.88 | 9994.73 | 21549.3 | 21888.9 | 40169.8 | 41024 | 56221.4 | 57186.1 |
| 979.737 | 2431.01 | 2556.35 | 9725.75 | 10004.8 | 21333.4 | 21906.4 | 40047.3 | 41052.1 | 55838.5 | 57220.5 |

| | | | | | | | | | | | |
|---|---|---|---|---|---|---|---|---|---|---|---|
| 1101.15 | 4090.54 | 4401.24 | 14180.7 | 14438 | 28829.5 | 29199.5 | 51899.3 | 52369.5 | 70101.2 | 70679.3 |
| 1102.87 | 4288.56 | 4429.25 | 14288.9 | 14498.3 | 28823.7 | 29292.8 | 51804.6 | 52508.6 | 69845.9 | 70839.3 |
| 1104.6 | 4447.73 | 4457.28 | 14274.8 | 14558.6 | 28990.2 | 29385.6 | 51993.9 | 52647.1 | 70225.2 | 70998.4 |
| 1106.32 | 4466.2 | 4485.33 | 14146.1 | 14618.7 | 29227.5 | 29478.1 | 52175.4 | 52784.8 | 70406.8 | 71156.5 |
| 1108.05 | 4417.68 | 4513.4 | 14568.2 | 14678.7 | 29219.1 | 29570.2 | 52102.6 | 52921.9 | 70190.1 | 71313.7 |
| 1109.77 | 4534.71 | 4541.49 | 14605.5 | 14738.6 | 29419.2 | 29661.9 | 52466.1 | 53058.3 | 70886.1 | 71469.9 |
| 1111.5 | 4653.43 | 4569.61 | 14649.6 | 14798.3 | 29474.7 | 29753.3 | 52697.3 | 53193.9 | 71118 | 71625.2 |
| 1113.22 | 4462.43 | 4597.74 | 14743.4 | 14857.9 | 29421 | 29844.3 | 52804.4 | 53328.9 | 71043.4 | 71779.6 |
| 1114.95 | 4554.38 | 4625.89 | 14759.7 | 14917.4 | 29641.3 | 29935 | 52941.9 | 53463.2 | 71565.6 | 71933 |
| 1116.67 | 4614.99 | 4654.05 | 14678.4 | 14976.7 | 29616.2 | 30025.2 | 53175.6 | 53596.8 | 71543.8 | 72085.4 |
| 1118.39 | 4529.95 | 4682.23 | 14693.3 | 15035.9 | 29648.9 | 30115.1 | 53478.2 | 53729.8 | 71590.6 | 72237 |
| 1120.12 | 4492.01 | 4710.42 | 14914.2 | 15094.9 | 29756.7 | 30204.7 | 52916.1 | 53862 | 71457.3 | 72387.6 |
| 1121.84 | 4546.61 | 4738.63 | 14791.7 | 15153.8 | 30097.8 | 30293.8 | 53257.4 | 53993.5 | 71594 | 72537.2 |
| 1123.57 | 4584.85 | 4766.85 | 14943.6 | 15212.5 | 30127.2 | 30382.6 | 53721.2 | 54124.3 | 72034.4 | 72685.9 |
| 1125.29 | 4614.12 | 4795.08 | 14891.7 | 15271.1 | 30137.1 | 30471 | 53415.9 | 54254.4 | 72373.9 | 72833.6 |
| 1127.01 | 4842.35 | 4823.32 | 14904.8 | 15329.6 | 30438.2 | 30559 | 53416.4 | 54383.8 | 72418.5 | 72980.4 |
| 1128.74 | 4803.88 | 4851.57 | 15381.6 | 15387.9 | 30196.7 | 30646.6 | 53791.2 | 54512.6 | 72953.1 | 73126.3 |
| 1130.46 | 5018.3 | 4879.83 | 15199.4 | 15446.1 | 30170.4 | 30733.9 | 54077.1 | 54640.6 | 72796 | 73271.2 |
| 1132.18 | 4894.99 | 4908.1 | 15312.1 | 15504.1 | 30506.4 | 30820.8 | 54404.1 | 54767.9 | 72565.4 | 73415.2 |
| 1133.91 | 4712.41 | 4936.38 | 15301.6 | 15562 | 30557.6 | 30907.3 | 54776.5 | 54894.5 | 73307 | 73558.2 |
| 1135.63 | 5025.35 | 4964.66 | 15353.7 | 15619.7 | 30487.4 | 30993.4 | 54902.8 | 55020.4 | 73034.3 | 73700.4 |
| 1137.35 | 4932.15 | 4992.94 | 15559 | 15677.2 | 30690.2 | 31079.1 | 54440.1 | 55145.6 | 73005.6 | 73841.5 |
| 1139.07 | 4859.53 | 5021.23 | 15237 | 15734.6 | 30554 | 31164.4 | 54511 | 55270.1 | 73287.1 | 73981.7 |
| 1140.8 | 4998.34 | 5049.53 | 15467.4 | 15791.9 | 30728.8 | 31249.4 | 54649.4 | 55393.9 | 73147.6 | 74121 |
| 1142.52 | 4985.75 | 5077.82 | 15659.4 | 15849 | 30920.9 | 31333.9 | 54474.5 | 55517 | 73469.1 | 74259.3 |
| 1144.24 | 4901.7 | 5106.12 | 15562.3 | 15905.9 | 30835.3 | 31418.1 | 54978.2 | 55639.3 | 73698 | 74396.7 |
| 1145.96 | 4930.18 | 5134.42 | 15675.3 | 15962.7 | 31238.5 | 31501.9 | 55081.3 | 55761 | 73629.6 | 74533.2 |
| 1147.69 | 5003.11 | 5162.72 | 15859.6 | 16019.3 | 31400.2 | 31585.3 | 55264.8 | 55882 | 74026.2 | 74668.7 |
| 1149.41 | 5042.95 | 5191.01 | 15814.4 | 16075.7 | 31269 | 31668.3 | 55276.2 | 56002.2 | 74275.2 | 74803.3 |
| 1151.13 | 5156.84 | 5219.31 | 16030.9 | 16132 | 31381.3 | 31750.9 | 55700.8 | 56121.8 | 74921 | 74937 |
| 1152.85 | 5199.94 | 5247.6 | 16287 | 16188.1 | 31453.3 | 31833.1 | 56197.6 | 56240.6 | 75296.2 | 75069.7 |
| 1154.57 | 4991.2 | 5275.89 | 16292.1 | 16244.1 | 31625 | 31914.9 | 56184.6 | 56358.8 | 75440.7 | 75201.5 |
| 1156.3 | 5158.55 | 5304.17 | 16365.5 | 16299.9 | 31660.8 | 31996.3 | 56330.4 | 56476.2 | 75417.8 | 75332.3 |
| 1158.02 | 5204.56 | 5332.44 | 16196.8 | 16355.5 | 31850.8 | 32077.4 | 56561.4 | 56592.9 | 75345.9 | 75462.2 |
| 1159.74 | 5327.81 | 5360.71 | 16099.1 | 16410.9 | 31740.4 | 32158 | 56582.4 | 56708.9 | 75253.6 | 75591.2 |
| 1161.46 | 5365.94 | 5388.98 | 16090.4 | 16466.2 | 31713 | 32238.2 | 56429.6 | 56824.2 | 75384.5 | 75719.3 |
| 1163.18 | 5258.82 | 5417.23 | 16209.8 | 16521.3 | 32047.3 | 32318.1 | 56599.9 | 56938.8 | 75741.7 | 75846.4 |
| 1164.9 | 5498.37 | 5445.48 | 16500.5 | 16576.2 | 31950.3 | 32397.5 | 56850.4 | 57052.7 | 75759.3 | 75972.6 |
| 1166.62 | 5523.2 | 5473.71 | 16512.6 | 16631 | 32243.8 | 32476.5 | 57140.5 | 57165.9 | 75874.7 | 76097.9 |
| 1168.34 | 5549.07 | 5501.93 | 16463.3 | 16685.6 | 32340.9 | 32555.2 | 56970.3 | 57278.4 | 75873.5 | 76222.2 |
| 1170.06 | 5487.87 | 5530.15 | 16302.5 | 16740 | 32359.1 | 32633.4 | 56602.3 | 57390.1 | 75418 | 76345.7 |
| 1171.78 | 5566.41 | 5558.35 | 16428.4 | 16794.2 | 32202.7 | 32711.2 | 56718.1 | 57501.2 | 75664.9 | 76468.2 |
| 1173.5 | 5607.54 | 5586.53 | 16624.5 | 16848.3 | 32158.7 | 32788.7 | 56774.6 | 57611.5 | 76009.5 | 76589.7 |
| 1175.22 | 5428.46 | 5614.7 | 16559 | 16902.1 | 32552.9 | 32865.7 | 56866.9 | 57721.2 | 76146.7 | 76710.4 |
| 1176.94 | 5519.75 | 5642.86 | 16774.2 | 16955.8 | 32490.9 | 32942.3 | 57198.6 | 57830.1 | 76109.6 | 76830.1 |
| 1178.66 | 5480.3 | 5671 | 16830.6 | 17009.3 | 32373.3 | 33018.5 | 57238.7 | 57938.3 | 75653.5 | 76949 |
| 1180.38 | 5288.28 | 5699.12 | 16737.6 | 17062.7 | 32606.8 | 33094.4 | 57186 | 58045.9 | 75939.8 | 77066.9 |
| 1182.11 | 5401.87 | 5727.23 | 16799.8 | 17115.8 | 32822.4 | 33169.8 | 57534.1 | 58152.7 | 76611.7 | 77183.8 |
| 1183.82 | 5590.99 | 5755.32 | 16955 | 17168.8 | 32851.6 | 33244.8 | 57754 | 58258.8 | 76798.7 | 77299.9 |
| 1185.54 | 5685.8 | 5783.39 | 16991.5 | 17221.6 | 33004.9 | 33319.4 | 57540.3 | 58364.2 | 76733.8 | 77415.1 |
| 1187.26 | 5636.25 | 5811.44 | 17327.3 | 17274.2 | 33110.3 | 33393.6 | 57832.9 | 58468.9 | 77133 | 77529.3 |
| 1188.98 | 5671.5 | 5839.46 | 17534 | 17326.6 | 33610.1 | 33467.4 | 58459.7 | 58572.9 | 77424.5 | 77642.7 |
| 1190.7 | 6010.02 | 5867.47 | 17426.9 | 17378.8 | 33578.3 | 33540.8 | 58557.5 | 58676.2 | 77361.9 | 77755.1 |
| 1192.42 | 5927.6 | 5895.45 | 17531.7 | 17430.9 | 33579.1 | 33613.8 | 58647.5 | 58778.8 | 77636.4 | 77866.6 |
| 1194.14 | 5779.31 | 5923.41 | 17367.5 | 17482.7 | 33621.8 | 33686.3 | 58483.1 | 58880.7 | 77270.5 | 77977.2 |
| 1195.86 | 5872.86 | 5951.35 | 17210.9 | 17534.4 | 33348.9 | 33758.5 | 58470.8 | 58981.9 | 77194.2 | 78086.9 |
| 1197.58 | 5813.45 | 5979.26 | 17306 | 17585.9 | 33716.7 | 33830.3 | 58734.9 | 59082.4 | 77896.8 | 78195.8 |
| 1199.3 | 6083.44 | 6007.14 | 17485.5 | 17637.2 | 33779.8 | 33901.6 | 58891.3 | 59182.2 | 78022.1 | 78303.7 |
| 1201.02 | 6162.26 | 6035 | 17467.6 | 17688.2 | 33725.6 | 33972.6 | 58960.2 | 59281.2 | 77789.4 | 78410.7 |
| 1202.73 | 5906.94 | 6062.83 | 17435.4 | 17739.1 | 33747.2 | 34043.1 | 59119.9 | 59379.6 | 78125.8 | 78516.8 |
| 1204.45 | 6010.9 | 6090.64 | 17753.8 | 17789.8 | 33808 | 34113.2 | 59267.1 | 59477.3 | 78462 | 78622 |
| 1206.17 | 6175.66 | 6118.41 | 17846 | 17840.4 | 33913.3 | 34183 | 59099.6 | 59574.3 | 78392 | 78726.4 |
| 1207.89 | 6139.52 | 6146.16 | 17783 | 17890.7 | 33752.6 | 34252.3 | 59055.1 | 59670.6 | 78460.7 | 78829.8 |
| 1209.61 | 6107.3 | 6173.87 | 17890 | 17940.8 | 33853.8 | 34321.2 | 59138.6 | 59766.2 | 78330.7 | 78932.4 |
| 1211.32 | 6044.66 | 6201.56 | 17781.7 | 17990.7 | 34342.9 | 34389.7 | 59297.5 | 59861.1 | 78566.2 | 79034 |
| 1213.04 | 6147.54 | 6229.21 | 17803.4 | 18040.5 | 34316.8 | 34457.8 | 59583.6 | 59955.3 | 78850.8 | 79134.8 |
| 1214.76 | 6065.55 | 6256.83 | 17989.2 | 18090 | 34442.3 | 34525.5 | 60128.1 | 60048.8 | 79582.9 | 79234.7 |
| 1216.48 | 6125.47 | 6284.42 | 18012.2 | 18139.4 | 34573.9 | 34592.8 | 60237.1 | 60141.6 | 79410.8 | 79333.7 |
| 1218.19 | 6452.31 | 6311.97 | 18182.5 | 18188.5 | 34703.7 | 34659.6 | 60416.1 | 60233.7 | 79371.4 | 79431.8 |
| 1219.91 | 6325.79 | 6339.49 | 18110.9 | 18237.5 | 34529.1 | 34726.1 | 60408.6 | 60325.2 | 79652.7 | 79529.1 |
| 1221.63 | 6387.35 | 6366.97 | 18142.1 | 18286.2 | 34595.4 | 34792.1 | 60285.2 | 60415.9 | 79442.5 | 79625.4 |
| 1223.35 | 6410.96 | 6394.42 | 18327.9 | 18334.7 | 34902.9 | 34857.8 | 60474.4 | 60506 | 79730.5 | 79720.9 |
| 1225.06 | 6382.13 | 6421.83 | 18552 | 18383.1 | 34907.8 | 34923 | 60849.4 | 60595.3 | 80081.6 | 79815.6 |
| 1226.78 | 6475.63 | 6449.2 | 18562.9 | 18431.2 | 35063.1 | 34987.8 | 60564.1 | 60684 | 79920.5 | 79909.3 |
| 1228.5 | 6487.96 | 6476.53 | 18262.2 | 18479.2 | 34873.5 | 35052.2 | 60577.5 | 60772 | 79939.1 | 80002.2 |
| 1230.21 | 6443.79 | 6503.83 | 18305.5 | 18526.9 | 34806 | 35116.3 | 60877.3 | 60859.3 | 80289.4 | 80094.2 |
| 1231.93 | 6320.2 | 6531.08 | 18557.3 | 18574.5 | 34922.9 | 35179.9 | 61052.1 | 60945.9 | 80373 | 80185.4 |
| 1233.64 | 6397.47 | 6558.3 | 18662.8 | 18621.8 | 35180.2 | 35243.1 | 61176.6 | 61031.8 | 80536.2 | 80275.7 |
| 1235.36 | 6524.23 | 6585.47 | 18718.4 | 18668.9 | 35410.4 | 35305.8 | 61333.7 | 61117.1 | 80860.4 | 80365.1 |
| 1237.08 | 6553.98 | 6612.6 | 18680.3 | 18715.9 | 35261.6 | 35368.2 | 61427.4 | 61201.7 | 80817 | 80453.7 |
| 1238.79 | 6499.1 | 6639.7 | 18558 | 18762.6 | 35274.7 | 35430.2 | 61169.1 | 61285.6 | 80669.5 | 80541.4 |
| 1240.51 | 6524.62 | 6666.74 | 18687.3 | 18809.1 | 35349.1 | 35491.8 | 61275.9 | 61368.8 | 80789.1 | 80628.3 |
| 1242.22 | 6611.82 | 6693.75 | 18860.9 | 18855.4 | 35684.8 | 35552.9 | 61489 | 61451.3 | 81021.7 | 80714.3 |
| 1243.94 | 6694.61 | 6720.71 | 18789.9 | 18901.5 | 35878.4 | 35613.7 | 61810.3 | 61533.2 | 81169 | 80799.5 |
| 1245.66 | 6491.87 | 6747.62 | 18856.3 | 18947.4 | 35701.9 | 35674 | 61732.1 | 61614.4 | 81097.2 | 80883.8 |
| 1247.37 | 6731.69 | 6774.49 | 18888 | 18993.1 | 35710.7 | 35733.9 | 61838.7 | 61694.9 | 81572.2 | 80967.3 |
| 1249.09 | 6788.44 | 6801.31 | 18936.8 | 19038.6 | 35926.1 | 35793.5 | 62098.5 | 61774.7 | 81534.1 | 81049.9 |
| 1250.8 | 6733.23 | 6828.09 | 18991.1 | 19083.9 | 35847.8 | 35852.6 | 62085.2 | 61853.9 | 81281.2 | 81131.7 |
| 1252.52 | 6885.11 | 6854.82 | 19038.5 | 19129 | 35731.3 | 35911.3 | 62154 | 61932.4 | 81610 | 81212.7 |
| 1254.23 | 6919.59 | 6881.5 | 19119.9 | 19173.8 | 35860.9 | 35969.7 | 62223.9 | 62010.2 | 81559.5 | 81292.8 |
| 1255.95 | 6982.04 | 6908.13 | 19074.7 | 19218.5 | 35939.2 | 36027.6 | 62259.5 | 62087.3 | 81744.8 | 81372.1 |
| 1257.66 | 6937.71 | 6934.71 | 19142.3 | 19263 | 36119.7 | 36085.1 | 62456.1 | 62163.8 | 82044.5 | 81450.6 |
| 1259.38 | 6921.85 | 6961.24 | 19211.9 | 19307.2 | 36168.3 | 36142.2 | 62468.5 | 62239.7 | 82084.1 | 81528.2 |
| 1261.09 | 7004.28 | 6987.72 | 19199.2 | 19351.2 | 36328.1 | 36198.9 | 62440.9 | 62314.8 | 82116.1 | 81605 |
| 1262.81 | 6910.86 | 7014.15 | 19380.9 | 19395 | 36380.8 | 36255.2 | 62686.1 | 62389.3 | 82022.6 | 81681 |
| 1264.52 | 6835.98 | 7040.53 | 19321.4 | 19438.7 | 36382.4 | 36311.1 | 62747.1 | 62463.2 | 82349.6 | 81756.2 |
| 1266.23 | 6966.37 | 7066.86 | 19539.5 | 19482.1 | 36529.3 | 36366.6 | 62845.4 | 62536.3 | 82602.1 | 81830.5 |
| 1267.95 | 6962.5 | 7093.13 | 19672.1 | 19525.2 | 36529.2 | 36421.7 | 62776.5 | 62608.9 | 82327.2 | 81904 |
| 1269.66 | 7102.24 | 7119.35 | 19548.1 | 19568.2 | 36766.7 | 36476.4 | 63116.9 | 62680.7 | 82865.1 | 81976.7 |
| 1271.37 | 7186.79 | 7145.52 | 19873.6 | 19611 | 36659.7 | 36530.7 | 63303.4 | 62751.9 | 82894.3 | 82048.6 |
| 1273.09 | 7132.57 | 7171.63 | 19946 | 19653.6 | 36796.8 | 36584.5 | 63367.9 | 62822.5 | 83164.8 | 82119.7 |
| 1274.8 | 7086.15 | 7197.69 | 19734.9 | 19695.9 | 37129.3 | 36638 | 63893.7 | 62892.4 | 83538.4 | 82190 |
| 1276.51 | 7288.09 | 7223.69 | 19791.3 | 19738 | 36955.5 | 36691.1 | 63643.4 | 62961.7 | 83393.9 | 82259.5 |
| 1278.23 | 7330.4 | 7249.63 | 20008.2 | 19780 | 37005.2 | 36743.8 | 63776.9 | 63030.3 | 83449.6 | 82328.2 |
| 1279.94 | 7392.58 | 7275.52 | 20267.4 | 19821.7 | 37165.3 | 36796.1 | 63924.9 | 63098.2 | 83232 | 82396.1 |
| 1281.65 | 7440.9 | 7301.35 | 20026 | 19863.2 | 37145.5 | 36848 | 63628.9 | 63165.5 | 83113.8 | 82463.1 |
| 1283.37 | 7233.87 | 7327.12 | 19993.9 | 19904.5 | 37174.4 | 36899.5 | 63731.6 | 63232.2 | 83193.1 | 82529.4 |
| 1285.08 | 7301.72 | 7352.83 | 20141.2 | 19945.5 | 37178.8 | 36950.6 | 63732.1 | 63298.2 | 83318.5 | 82594.9 |
| 1286.79 | 7537.75 | 7378.49 | 19948.8 | 19986.4 | 37309.3 | 37001.3 | 63957.6 | 63363.6 | 83356.4 | 82659.6 |
| 1288.5 | 7369.77 | 7404.08 | 20118.9 | 20027 | 37295.3 | 37051.6 | 64164.8 | 63428.4 | 83624.6 | 82723.6 |
| 1290.22 | 7409.6 | 7429.61 | 20113.9 | 20067.5 | 37108.3 | 37101.5 | 63900 | 63492.5 | 83464.9 | 82786.7 |
| 1291.93 | 7430.84 | 7455.09 | 20150.1 | 20107.7 | 37382.4 | 37151.1 | 63973.2 | 63555.9 | 83595.3 | 82849.1 |
| 1293.64 | 7401.87 | 7480.5 | 20183.1 | 20147.7 | 37412.6 | 37200.2 | 64290.7 | 63618.8 | 83893.4 | 82910.6 |
| 1295.35 | 7573.82 | 7505.85 | 20244.3 | 20187.5 | 37452.6 | 37248.9 | 64181.8 | 63681 | 83778.7 | 82971.4 |
| 1297.07 | 7562.32 | 7531.14 | 20172.7 | 20227.1 | 37418.7 | 37297.3 | 64360.4 | 63742.5 | 83785.4 | 83031.5 |
| 1298.78 | 7517.35 | 7556.37 | 20059.4 | 20266.4 | 37442.6 | 37345.2 | 64341 | 63803.5 | 83897.8 | 83090.7 |
| 1300.49 | 7677.4 | 7581.53 | 20328.1 | 20305.6 | 37808.5 | 37392.8 | 64721.9 | 63863.8 | 84349.1 | 83149.2 |
| 1302.2 | 7628.8 | 7606.63 | 20342.1 | 20344.5 | 37746.4 | 37439.9 | 64717.1 | 63923.5 | 84480.2 | 83206.9 |
| 1303.91 | 7676.97 | 7631.67 | 20621.7 | 20383.2 | 37961.2 | 37486.7 | 64798.1 | 63982.5 | 84673.6 | 83263.9 |
| 1305.62 | 7904.98 | 7656.64 | 20796.2 | 20421.7 | 38085.9 | 37533.1 | 65209.6 | 64041 | 84882.6 | 83320.1 |
| 1307.33 | 7726.54 | 7681.55 | 20648.3 | 20460 | 38064.5 | 37579.1 | 65020.9 | 64098.8 | 84573.7 | 83375.5 |
| 1309.04 | 7631.24 | 7706.39 | 20733 | 20498.1 | 38226 | 37624.7 | 65045.9 | 64156 | 84980.4 | 83430.2 |
| 1310.76 | 7802.27 | 7731.16 | 20770.6 | 20536 | 38175.1 | 37670 | 65226.2 | 64212.6 | 85104.5 | 83484.1 |
| 1312.47 | 7791.49 | 7755.87 | 20865.9 | 20573.6 | 38187.6 | 37714.8 | 65344.6 | 64268.5 | 84872.2 | 83537.3 |
| 1314.18 | 7740.18 | 7780.52 | 20856.5 | 20611.1 | 38300.5 | 37759.3 | 65379.1 | 64323.9 | 85028.1 | 83589.8 |
| 1315.89 | 7997.68 | 7805.09 | 20918.3 | 20648.3 | 38250.9 | 37803.3 | 65400.5 | 64378.6 | 84984.2 | 83641.5 |
| 1317.6 | 8079.28 | 7829.6 | 21045.2 | 20685.3 | 38237.6 | 37847 | 65579.6 | 64432.7 | 85073.9 | 83692.4 |
| 1319.31 | 7898.51 | 7854.04 | 20900 | 20722.1 | 38244.8 | 37890.3 | 65404.7 | 64486.2 | 84712.4 | 83742.6 |
| 1321.02 | 7888.72 | 7878.41 | 21007.9 | 20758.7 | 38249.3 | 37933.2 | 65290 | 64539.2 | 84678.9 | 83792.1 |
| 1322.73 | 7817.5 | 7902.71 | 20873.1 | 20795 | 38312.5 | 37975.8 | 65214.3 | 64591.5 | 84418.8 | 83840.8 |
| 1324.44 | 7834.65 | 7926.94 | 20918.6 | 20831.2 | 38439.9 | 38017.9 | 65006 | 64643.2 | 84210.6 | 83888.8 |
| 1326.15 | 7884.25 | 7951.11 | 21062.1 | 20867.1 | 38377.2 | 38059.7 | 65029.7 | 64694.3 | 84599.5 | 83936.1 |
| 1327.86 | 8062.54 | 7975.2 | 20979.4 | 20902.8 | 38307 | 38101.1 | 65052.2 | 64744.8 | 84379.3 | 83982.7 |
| 1329.57 | 8096.03 | 7999.22 | 21103.1 | 20938.3 | 38368.2 | 38142.1 | 65299.6 | 64794.7 | 84549.5 | 84028.5 |
| 1331.28 | 8106.1 | 8023.17 | 21275 | 20973.6 | 38523.7 | 38182.8 | 65542.9 | 64844 | 85305.7 | 84073.6 |
| 1332.98 | 8141.47 | 8047.05 | 21374.5 | 21008.7 | 38605.7 | 38223.1 | 65693.5 | 64892.7 | 85173.3 | 84118 |
| 1334.69 | 8085.23 | 8070.86 | 21273.6 | 21043.6 | 38744.9 | 38262.9 | 65785.7 | 64940.8 | 85232.4 | 84161.7 |
| 1336.4 | 8173.36 | 8094.6 | 21335.9 | 21078.2 | 38834 | 38302.5 | 65764 | 64988.3 | 85501.3 | 84204.7 |
| 1338.11 | 8256.86 | 8118.26 | 21425 | 21112.7 | 38878.5 | 38341.6 | 65958.9 | 65035.2 | 85375.2 | 84246.9 |
| 1339.82 | 8289.69 | 8141.85 | 21554.2 | 21146.9 | 38932.8 | 38380.4 | 66106.5 | 65081.6 | 85482.5 | 84288.5 |
| 1341.53 | 8266.06 | 8165.37 | 21635.2 | 21180.9 | 39011.9 | 38418.8 | 66136.8 | 65127.4 | 85768.1 | 84329.3 |
| 1343.24 | 8264.53 | 8188.81 | 21540.3 | 21214.7 | 38851.3 | 38456.8 | 66336.7 | 65172.5 | 85846.8 | 84369.5 |
| 1344.95 | 8299.1 | 8212.18 | 21455.8 | 21248.2 | 38827.5 | 38494.4 | 66296.9 | 65217.1 | 85651.2 | 84408.9 |
| 1346.65 | 8263.63 | 8235.48 | 21404.3 | 21281.6 | 38871.7 | 38531.7 | 66049 | 65261.2 | 85401 | 84447.7 |
| 1348.36 | 8315.71 | 8258.7 | 21458.5 | 21314.8 | 38869.8 | 38568.6 | 66128.9 | 65304.6 | 85237.9 | 84485.7 |
| 1350.07 | 8191.92 | 8281.85 | 21372.6 | 21347.7 | 38876.2 | 38605.2 | 65892.8 | 65347.5 | 85080.5 | 84523.1 |
| 1351.78 | 8100.48 | 8304.92 | 21548.7 | 21380.4 | 38904.6 | 38641.4 | 65819.2 | 65389.8 | 85284.3 | 84559.8 |
| 1353.49 | 8283.03 | 8327.91 | 21588.4 | 21412.9 | 38879.8 | 38677.2 | 65950.9 | 65431.5 | 85424.3 | 84595.8 |
| 1355.19 | 8379.01 | 8350.83 | 21564.4 | 21445.2 | 38794.2 | 38712.6 | 65893.7 | 65472.7 | 85274.8 | 84631.1 |
| 1356.9 | 8299.99 | 8373.68 | 21644.5 | 21477.3 | 38977.6 | 38747.7 | 66216.2 | 65513.2 | 85389.4 | 84665.7 |
| 1358.61 | 8219.03 | 8396.44 | 21671 | 21509.2 | 39345.7 | 38782.4 | 66398.7 | 65553.2 | 85824.7 | 84699.7 |
| 1360.32 | 8367.42 | 8419.13 | 21660 | 21540.9 | 39207.4 | 38816.8 | 66341.6 | 65592.7 | 85752.2 | 84732.9 |
| 1362.02 | 8253.94 | 8441.74 | 21638.9 | 21572.3 | 39023.8 | 38850.8 | 66345.1 | 65631.6 | 85471.9 | 84765.6 |
| 1363.73 | 8364.71 | 8464.28 | 21734.7 | 21603.6 | 39454.1 | 38884.4 | 66428.4 | 65669.9 | 85627.6 | 84797.5 |
| 1365.44 | 8725.58 | 8486.74 | 21658.5 | 21634.6 | 39168.9 | 38917.7 | 66129 | 65707.7 | 85639.3 | 84828.8 |
| 1367.14 | 8573.47 | 8509.11 | 21845.5 | 21665.4 | 39274.9 | 38950.6 | 66341.1 | 65744.9 | 85673 | 84859.4 |
| 1368.85 | 8611.9 | 8531.41 | 22024.7 | 21696 | 39645.1 | 38983.2 | 66916.8 | 65781.6 | 86150.7 | 84889.4 |
| 1370.56 | 8554.28 | 8553.64 | 21848 | 21726.4 | 39458.1 | 39015.4 | 66589 | 65817.7 | 85926.6 | 84918.7 |
| 1372.26 | 8520.68 | 8575.78 | 22106.2 | 21756.6 | 39706.6 | 39047.2 | 66876.7 | 65853.3 | 85963.4 | 84947.3 |
| 1373.97 | 8831.23 | 8597.84 | 22101.2 | 21786.5 | 39775.8 | 39078.7 | 67030.4 | 65888.3 | 86146 | 84975.3 |
| 1375.67 | 8808.71 | 8619.82 | 21964 | 21816.3 | 39790.7 | 39109.8 | 66928.2 | 65922.8 | 85975 | 85002.6 |
| 1377.38 | 8650.67 | 8641.73 | 22141.2 | 21845.9 | 39449.5 | 39140.6 | 66695.6 | 65956.7 | 85819.9 | 85029.4 |
| 1379.09 | 8649.67 | 8663.55 | 22295 | 21875.2 | 39554.6 | 39171.1 | 66845.6 | 65990.1 | 85995.3 | 85055.4 |
| 1380.79 | 8779.27 | 8685.29 | 22223.7 | 21904.3 | 39877.4 | 39201.2 | 67058.6 | 66022.9 | 86394.2 | 85080.8 |
| 1382.5 | 8918.9 | 8706.95 | 22350.1 | 21933.3 | 39942.1 | 39230.9 | 67212.6 | 66055.2 | 86612.1 | 85105.6 |
| 1384.2 | 8924.1 | 8728.54 | 22412.6 | 21962 | 40040.6 | 39260.3 | 67494.1 | 66087 | 86655.4 | 85129.7 |
| 1385.91 | 8788.17 | 8750.04 | 22303.4 | 21990.5 | 39944.5 | 39289.3 | 67483.3 | 66118.2 | 87013.5 | 85153.2 |
| 1387.61 | 8755.66 | 8771.46 | 22669.1 | 22018.8 | 39999.2 | 39318 | 67513.4 | 66148.9 | 86899.2 | 85176.1 |
| 1389.32 | 8852.9 | 8792.79 | 22623.9 | 22046.9 | 40220 | 39346.3 | 67556.5 | 66179.1 | 86887.1 | 85198.4 |
| 1391.02 | 8827.55 | 8814.05 | 22459.6 | 22074.7 | 40169.3 | 39374.3 | 67732.1 | 66208.7 | 86960.5 | 85220 |
| 1392.73 | 8932.17 | 8835.22 | 22487 | 22102.4 | 40201.9 | 39402 | 67466.5 | 66237.8 | 86841.4 | 85241 |
| 1394.43 | 9008.77 | 8856.31 | 22236.5 | 22129.9 | 40081.9 | 39429.3 | 67022.2 | 66266.4 | 86387 | 85261.4 |
| 1396.14 | 9058.5 | 8877.32 | 22280.5 | 22157.1 | 39821.2 | 39456.3 | 67179.4 | 66294.5 | 86155.8 | 85281.2 |
| 1397.84 | 8940.6 | 8898.25 | 22635.5 | 22184.2 | 39968.3 | 39482.9 | 67525.1 | 66322.1 | 86420.3 | 85300.4 |
| 1399.55 | 8952.4 | 8919.09 | 22518.8 | 22211 | 39767.9 | 39509.2 | 67266.8 | 66349.1 | 86322.5 | 85318.9 |
| 1401.25 | 9213.66 | 8939.85 | 22606.6 | 22237.7 | 39946.8 | 39535.2 | 67408.5 | 66375.6 | 86648.9 | 85336.9 |
| 1402.95 | 9038.82 | 8960.53 | 22536.1 | 22264.1 | 40253.9 | 39560.8 | 67449.1 | 66401.6 | 86633.6 | 85354.2 |
| 1404.66 | 9045 | 8981.12 | 22363.5 | 22290.3 | 40039.7 | 39586.1 | 67159.8 | 66427.1 | 86476.6 | 85370.9 |
| 1406.36 | 9134.92 | 9001.63 | 22663.9 | 22316.3 | 40230.1 | 39611 | 67539 | 66452.1 | 86785.4 | 85387.1 |
| 1408.07 | 9081.85 | 9022.05 | 22554.1 | 22342.1 | 40181.1 | 39635.6 | 67469.6 | 66476.5 | 86666.2 | 85402.6 |
| 1409.77 | 9267.83 | 9042.39 | 22666.7 | 22367.8 | 40080.4 | 39659.9 | 67356.4 | 66500.5 | 86489.7 | 85417.6 |
| 1411.47 | 9261.89 | 9062.65 | 22735.1 | 22393.2 | 40297.9 | 39683.8 | 67760.5 | 66524 | 86820.1 | 85431.9 |
| 1413.17 | 9151.35 | 9082.82 | 22751.6 | 22418.4 | 40365 | 39707.5 | 67724.2 | 66546.9 | 86902.5 | 85445.7 |
| 1414.88 | 9271.47 | 9102.91 | 22700 | 22443.4 | 40423.1 | 39730.8 | 67539 | 66569.4 | 86969.6 | 85458.9 |
| 1416.58 | 9148.64 | 9122.91 | 22760.4 | 22468.2 | 40297.2 | 39753.7 | 67643 | 66591.3 | 86978.4 | 85471.5 |
| 1418.28 | 9210.21 | 9142.83 | 22892.4 | 22492.7 | 40413.6 | 39776.3 | 67855.2 | 66612.8 | 87163 | 85483.5 |
| 1419.99 | 9386.19 | 9162.66 | 22629.3 | 22517.1 | 40182.1 | 39798.6 | 67733.6 | 66633.8 | 87006.3 | 85494.9 |
| 1421.69 | 9477.5 | 9182.41 | 23017 | 22541.3 | 40196.4 | 39820.6 | 68003.9 | 66654.3 | 87146.2 | 85505.8 |
| 1423.39 | 9363.96 | 9202.07 | 22919.7 | 22565.3 | 40693.1 | 39842.3 | 68017.7 | 66674.2 | 87351.8 | 85516.1 |
| 1425.09 | 9210.78 | 9221.65 | 22798.9 | 22589.1 | 40624.2 | 39863.6 | 67814.4 | 66693.8 | 87201.8 | 85525.8 |
| 1426.8 | 9410.09 | 9241.14 | 23080.9 | 22612.7 | 40407.8 | 39884.6 | 67654.5 | 66712.8 | 87030.3 | 85535 |
| 1428.5 | 9339.85 | 9260.54 | 22857.3 | 22636 | 40398.4 | 39905.3 | 67637.4 | 66731.3 | 86619.8 | 85543.6 |
| 1430.2 | 9287.09 | 9279.86 | 22923.6 | 22659.2 | 40478.7 | 39925.7 | 67758.4 | 66749.4 | 86833.1 | 85551.6 |
| 1431.9 | 9473.67 | 9299.09 | 22963.8 | 22682.2 | 40446.6 | 39945.7 | 67608.2 | 66766.9 | 86825.7 | 85559.1 |
| 1433.6 | 9689.67 | 9318.23 | 23027.7 | 22705 | 40477.3 | 39965.4 | 67907.4 | 66784 | 86816.1 | 85566 |
| 1435.3 | 9631.61 | 9337.29 | 23117.8 | 22727.6 | 40598.9 | 39984.9 | 67865.7 | 66800.6 | 87072.9 | 85572.4 |
| 1437.01 | 9470.25 | 9356.26 | 22921.7 | 22750 | 40540.7 | 40004 | 67530.9 | 66816.8 | 86635.5 | 85578.2 |
| 1438.71 | 9683.68 | 9375.15 | 22938 | 22772.1 | 40333 | 40022.8 | 67496.5 | 66832.4 | 86460.3 | 85583.5 |
| 1440.41 | 9466.19 | 9393.95 | 23032.7 | 22794.1 | 40333.8 | 40041.2 | 67425.4 | 66847.7 | 86452.3 | 85588.2 |
| 1442.11 | 9593.5 | 9412.66 | 22980 | 22815.9 | 40454.3 | 40059.4 | 67476.8 | 66862.4 | 86474.1 | 85592.4 |
| 1443.81 | 9656.04 | 9431.28 | 23034.8 | 22837.5 | 40572 | 40077.2 | 67983.6 | 66876.7 | 86977.5 | 85596.1 |
| 1445.51 | 9511.19 | 9449.82 | 22982.4 | 22858.9 | 40537.1 | 40094.8 | 68016.8 | 66890.5 | 87059.7 | 85599.2 |
| 1447.21 | 9622.11 | 9468.27 | 22973.8 | 22880.1 | 40616.9 | 40112 | 67619 | 66903.8 | 86738 | 85601.7 |
| 1448.91 | 9725.81 | 9486.63 | 23029.8 | 22901.1 | 40661.6 | 40128.9 | 67493.8 | 66916.7 | 86577.4 | 85603.8 |
| 1450.61 | 9664.88 | 9504.91 | 23089.7 | 22921.9 | 40689.6 | 40145.5 | 67565.9 | 66929.1 | 86506.3 | 85605.3 |
| 1452.31 | 9538.25 | 9523.09 | 22954.1 | 22942.5 | 40551.4 | 40161.8 | 67474.7 | 66941.1 | 86260.1 | 85606.3 |
| 1454.01 | 9576.03 | 9541.19 | 23076.5 | 22963 | 40443 | 40177.9 | 67491.9 | 66952.6 | 86223.1 | 85606.8 |
| 1455.71 | 9635.26 | 9559.2 | 23297.7 | 22983.2 | 40443.6 | 40193.6 | 67545.7 | 66963.7 | 85976 | 85606.7 |
| 1457.41 | 9798.97 | 9577.13 | 23226.5 | 23003.2 | 40488.4 | 40209 | 67333 | 66974.3 | 86078.6 | 85606.1 |
| 1459.11 | 9704.57 | 9594.97 | 23141.5 | 23023 | 40428.4 | 40224.1 | 67290.7 | 66984.5 | 86205.6 | 85605 |
| 1460.81 | 9760.54 | 9612.71 | 23090.3 | 23042.7 | 40317.4 | 40238.9 | 67481.6 | 66994.2 | 85876.4 | 85603.4 |
| 1462.51 | 9628.75 | 9630.37 | 23154.2 | 23062.2 | 40433.4 | 40253.4 | 67329.5 | 67003.5 | 85783.1 | 85601.3 |
| 1464.21 | 9588.07 | 9647.94 | 23097.6 | 23081.4 | 40315.2 | 40267.6 | 67240.3 | 67012.4 | 85723.8 | 85598.7 |
| 1465.91 | 9600.95 | 9665.43 | 22926.9 | 23100.5 | 40235.2 | 40281.5 | 67215 | 67020.8 | 85651.9 | 85595.6 |
| 1467.61 | 9595.41 | 9682.82 | 23141 | 23119.4 | 40379.3 | 40295.1 | 66945.6 | 67028.7 | 85666.2 | 85591.9 |
| 1469.31 | 9630.59 | 9700.13 | 23230.3 | 23138 | 40342.6 | 40308.4 | 66938.4 | 67036.3 | 85671.5 | 85587.8 |
| 1471.01 | 9630.67 | 9717.35 | 23141 | 23156.5 | 40245.6 | 40321.5 | 66896.5 | 67043.4 | 85376.7 | 85583.2 |
| 1472.7 | 9741.93 | 9734.48 | 23280.4 | 23174.8 | 40536.3 | 40334.2 | 67095.4 | 67050 | 85500.1 | 85578.1 |
| 1474.4 | 9878.93 | 9751.52 | 23216.6 | 23193 | 40584.9 | 40346.6 | 67449.9 | 67056.3 | 85797 | 85572.4 |
| 1476.1 | 9879.8 | 9768.47 | 23259.1 | 23210.9 | 40508.9 | 40358.8 | 67288.1 | 67062.1 | 85770.4 | 85566.3 |
| 1477.8 | 9851.53 | 9785.34 | 23305.9 | 23228.7 | 40475.5 | 40370.7 | 67381.3 | 67067.5 | 85675.9 | 85559.7 |
| 1479.5 | 9879.25 | 9802.11 | 23418.7 | 23246.2 | 40333 | 40382.2 | 67280.6 | 67072.4 | 85598.2 | 85552.7 |
| 1481.19 | 9764.95 | 9818.8 | 23391.1 | 23263.6 | 40267.5 | 40393.5 | 66930.6 | 67077 | 85281 | 85545.1 |
| 1482.89 | 9899.84 | 9835.4 | 23258 | 23280.8 | 40360.2 | 40404.6 | 67186.7 | 67081.1 | 85300.4 | 85537.1 |
| 1484.59 | 9856.37 | 9851.91 | 23317.6 | 23297.8 | 40421.4 | 40415.3 | 67320.8 | 67084.8 | 85479.6 | 85528.5 |
| 1486.29 | 9949.69 | 9868.33 | 23387.9 | 23314.6 | 40404.9 | 40425.7 | 67239.4 | 67088.1 | 85490.2 | 85519.6 |
| 1487.98 | 9995.8 | 9884.66 | 23483.5 | 23331.2 | 40704.6 | 40435.9 | 67355.4 | 67091 | 85675.7 | 85510.1 |
| 1489.68 | 9899.37 | 9900.9 | 23494.5 | 23347.6 | 40937.3 | 40445.8 | 67576.9 | 67093.4 | 85902.2 | 85500.2 |
| 1491.38 | 10026.3 | 9917.06 | 23575.8 | 23363.9 | 40897.6 | 40455.4 | 67370.7 | 67095.5 | 85865.7 | 85489.8 |
| 1493.08 | 9925.18 | 9933.12 | 23453.4 | 23380 | 40527.9 | 40464.7 | 66894.6 | 67097.1 | 85403.1 | 85478.9 |
| 1494.77 | 9968.84 | 9949.1 | 23265.5 | 23395.8 | 40456.5 | 40473.8 | 66775.3 | 67098.4 | 84978.8 | 85467.6 |
| 1496.47 | 10015.9 | 9964.99 | 23311.7 | 23411.5 | 40461.4 | 40482.5 | 66803.4 | 67099.2 | 84652.6 | 85455.8 |
| 1498.17 | 10055.4 | 9980.78 | 23383 | 23427.1 | 40329.6 | 40491 | 66704.9 | 67099.7 | 84827.1 | 85443.6 |
| 1499.86 | 9996.7 | 9996.49 | 23341.5 | 23442.4 | 40321.1 | 40499.2 | 66559.7 | 67099.7 | 84775.2 | 85430.9 |
| 1501.56 | 10017.3 | 10012.1 | 23442.7 | 23457.6 | 40288.3 | 40507.2 | 66749.8 | 67099.4 | 84778.3 | 85417.7 |
| 1503.25 | 10027.4 | 10027.6 | 23343 | 23472.6 | 40401.3 | 40514.9 | 66621.7 | 67098.6 | 84811.6 | 85404.1 |
| 1504.95 | 10001.5 | 10043.1 | 23334 | 23487.4 | 40258.1 | 40522.3 | 66625.1 | 67097.5 | 84586.2 | 85390.1 |
| 1506.65 | 10097.6 | 10058.4 | 23320.1 | 23502 | 40055.2 | 40529.4 | 66396.4 | 67095.9 | 84344.9 | 85375.6 |
| 1508.34 | 9965.81 | 10073.7 | 23283.7 | 23516.4 | 40158.2 | 40536.3 | 66057.5 | 67094 | 84179.8 | 85360.6 |
| 1510.04 | 9850.55 | 10088.9 | 23293.1 | 23530.7 | 39963.7 | 40542.9 | 66063.6 | 67091.7 | 84139.9 | 85345.3 |
| 1511.73 | 9997.33 | 10104 | 23414.3 | 23544.8 | 40080.7 | 40549.3 | 66171 | 67089 | 84121.2 | 85329.5 |
| 1513.43 | 10031.9 | 10119 | 23482 | 23558.7 | 40257.8 | 40555.4 | 66368.7 | 67085.9 | 84362.2 | 85313.2 |
| 1515.12 | 10050.2 | 10133.9 | 23323.5 | 23572.4 | 40012.2 | 40561.2 | 66237.2 | 67082.4 | 84398.6 | 85296.6 |
| 1516.82 | 10198.1 | 10148.7 | 23417.2 | 23586 | 40073 | 40566.7 | 66093.1 | 67078.6 | 84106.5 | 85279.5 |
| 1518.51 | 10113.6 | 10163.4 | 23530.9 | 23599.4 | 40113.8 | 40572 | 66100.9 | 67074.4 | 83938.3 | 85261.9 |
| 1520.21 | 10142.8 | 10178.1 | 23457.1 | 23612.6 | 40114.5 | 40577.1 | 66117.8 | 67069.7 | 83903.4 | 85244 |
| 1521.9 | 10256.1 | 10192.6 | 23292.9 | 23625.6 | 40009.2 | 40581.8 | 65849.6 | 67064.8 | 83734.9 | 85225.6 |
| 1523.6 | 10122.9 | 10207.1 | 23476.2 | 23638.4 | 40006.4 | 40586.4 | 65918.9 | 67059.4 | 83904.4 | 85206.8 |
| 1525.29 | 10071.6 | 10221.5 | 23312.4 | 23651.1 | 40052.8 | 40590.6 | 66049 | 67053.7 | 83816.7 | 85187.6 |
| 1526.99 | 10171.7 | 10235.8 | 23322.3 | 23663.6 | 40106.8 | 40594.6 | 65947.6 | 67047.6 | 83704.5 | 85167.9 |
| 1528.68 | 10318.3 | 10250 | 23373.9 | 23676 | 40055.8 | 40598.4 | 65934.3 | 67041.1 | 83555.2 | 85147.9 |
| 1530.37 | 10084.6 | 10264.1 | 23306.3 | 23688.1 | 39988.2 | 40601.9 | 65975.1 | 67034.3 | 83417.5 | 85127.4 |
| 1532.07 | 10057.9 | 10278.1 | 23444.2 | 23700.1 | 40001.5 | 40605.2 | 66006.1 | 67027.1 | 83692.7 | 85106.6 |
| 1533.76 | 10172.3 | 10292 | 23508.4 | 23711.9 | 40091.5 | 40608.2 | 65910.9 | 67019.6 | 83667 | 85085.3 |
| 1535.45 | 10108.2 | 10305.9 | 23523.8 | 23723.6 | 40132.3 | 40610.9 | 66006.4 | 67011.7 | 83671.5 | 85063.6 |
| 1537.15 | 10152.4 | 10319.6 | 23449 | 23735.1 | 40059.1 | 40613.4 | 65881.4 | 67003.4 | 83420.2 | 85041.6 |
| 1538.84 | 10193 | 10333.3 | 23503.5 | 23746.4 | 40085.9 | 40615.7 | 65913.4 | 66994.8 | 83547.5 | 85019.1 |
| 1540.53 | 10296.3 | 10346.9 | 23622.5 | 23757.5 | 40062.2 | 40617.7 | 65923.4 | 66985.9 | 83755.6 | 84996.2 |
| 1542.23 | 10293.1 | 10360.4 | 23545.8 | 23768.5 | 40298.7 | 40619.4 | 65912 | 66976.5 | 83554.4 | 84973 |
| 1543.92 | 10393.9 | 10373.8 | 23558.7 | 23779.3 | 40315.1 | 40620.9 | 66149.4 | 66966.9 | 83704.2 | 84949.3 |
| 1545.61 | 10464.6 | 10387.1 | 23673.1 | 23789.9 | 40083.2 | 40622.2 | 65846.7 | 66956.8 | 83405.1 | 84925.3 |
| 1547.31 | 10442 | 10400.4 | 23620.2 | 23800.4 | 40059.7 | 40623.3 | 65692.1 | 66946.5 | 83291.1 | 84900.8 |
| 1549 | 10610.8 | 10413.5 | 23545.4 | 23810.7 | 40250.3 | 40624.1 | 66093.2 | 66935.8 | 83894.2 | 84876 |
| 1550.69 | 10455.7 | 10426.5 | 23473.7 | 23820.8 | 40330.7 | 40624.6 | 66026.3 | 66924.7 | 83578.9 | 84850.8 |
| 1552.38 | 10228.1 | 10439.5 | 23421.3 | 23830.8 | 40336.7 | 40624.9 | 65853.8 | 66913.3 | 83427 | 84825.2 |
| 1554.07 | 10476.1 | 10452.4 | 23521.1 | 23840.6 | 39996.2 | 40625 | 65693.5 | 66901.6 | 83327.6 | 84799.3 |
| 1555.77 | 10470.8 | 10465.2 | 23489.3 | 23850.3 | 39719.3 | 40624.9 | 65386.8 | 66889.5 | 82990.8 | 84772.9 |
| 1557.46 | 10384.9 | 10477.9 | 23172 | 23859.8 | 39643.6 | 40624.5 | 65255.9 | 66877.1 | 82992 | 84746.2 |
| 1559.15 | 10467.6 | 10490.5 | 23584.7 | 23869.1 | 39870.5 | 40623.9 | 65466.8 | 66864.4 | 82982.8 | 84719.2 |
| 1560.84 | 10353.6 | 10503 | 23770.3 | 23878.2 | 40041.4 | 40623 | 65559.9 | 66851.3 | 82799.3 | 84691.7 |
| 1562.53 | 10410.5 | 10515.5 | 23579.2 | 23887.2 | 39786.9 | 40621.9 | 65327 | 66837.9 | 82483.7 | 84663.9 |
| 1564.22 | 10413.6 | 10527.8 | 23627 | 23896 | 39729.6 | 40620.6 | 65438.7 | 66824.2 | 82725.6 | 84635.7 |
| 1565.91 | 10414.4 | 10540.1 | 23457.8 | 23904.7 | 39812.8 | 40619.1 | 65156.5 | 66810.1 | 82395.3 | 84607.2 |
| 1567.61 | 10378.6 | 10552.3 | 23371.8 | 23913.2 | 39580.4 | 40617.3 | 64706.3 | 66795.8 | 81881.7 | 84578.3 |
| 1569.3 | 10146.8 | 10564.4 | 23470.4 | 23921.6 | 39545.8 | 40615.3 | 64905.5 | 66781.1 | 82074.1 | 84549 |
| 1570.99 | 10380.7 | 10576.4 | 23224.8 | 23929.8 | 39676.1 | 40613.1 | 64707.9 | 66766 | 81651 | 84519.4 |
| 1572.68 | 10390.2 | 10588.3 | 23293.5 | 23937.8 | 39434.7 | 40610.6 | 64319.4 | 66750.7 | 81577.5 | 84489.4 |
| 1574.37 | 10302.9 | 10600.1 | 23543 | 23945.7 | 39351.3 | 40607.9 | 64523.2 | 66735 | 81812.7 | 84459.1 |
| 1576.06 | 10394.1 | 10611.9 | 23568 | 23953.4 | 39399.1 | 40605 | 64429.9 | 66719.1 | 81167.2 | 84428.4 |
| 1577.75 | 10359.5 | 10623.6 | 23641.4 | 23961 | 39481.6 | 40601.9 | 64397.1 | 66702.8 | 81284.5 | 84397.4 |
| 1579.44 | 10407.3 | 10635.1 | 23460.5 | 23968.4 | 39702 | 40598.6 | 64412.9 | 66686.2 | 81134.4 | 84366.1 |

| Figure 2A, Carbon Black | T = 1380 K, n = 1.44, 2520 sec | T = 1380 K, n = 1.44, 2520 sec, | T = 1540 K, n = 1.19, 5100 sec | T = 1540 K, n = 1.19, 5100 sec | T = 1746 K, n = 0.9, 6660 sec | T = 1746 K, n = 0.9, 6660 sec | T = 1947 K, n = 0.63, 8760 sec | T = 1947 K, n = 0.63, 8760 secd | T = 2049 K, n = 0.53, 8880 sec | T = 2049 K, n = 0.53, 8880 sec | Figure 2B, Graphene Platelet | T = 1358 K, n = 1.06, 1290 sec | T = 1358 K, n = 1.06, 1290 sec | T = 1551 K, n = 0.63, 1440 sec | T = 1551 K, n = 0.63, 1440 sec |
|---|---|---|---|---|---|---|---|---|---|---|---|---|---|---|---|
| 600.015 | 0 | 171.575 | 796.038 | 675.883 | 2833.45 | 2718.6 | 6610.38 | 7946.14 | 10623 | 12662.4 | 600.015 | 169.478 | 90.0079 | 316.793 | 364.74 |
| 600.306 | 131.704 | 172.567 | 1021 | 679.28 | 2596.61 | 2730.22 | 7188.3 | 7975.69 | 10254.7 | 12706.3 | 600.306 | 289.202 | 90.5573 | 400.431 | 366.653 |
| 600.597 | 292.19 | 173.564 | 548.697 | 682.689 | 2511.43 | 2741.87 | 6966.09 | 8005.32 | 11324.2 | 12750.4 | 600.597 | 122.21 | 91.1094 | 507.642 | 368.574 |
| 600.888 | 166.093 | 174.565 | 797.747 | 686.112 | 2582.98 | 2753.56 | 6624.58 | 8035.02 | 11047.5 | 12794.5 | 600.888 | 121.571 | 91.6641 | 504.987 | 370.503 |
| 601.179 | 331.086 | 175.57 | 585.87 | 689.547 | 2869.86 | 2765.29 | 7167.37 | 8064.79 | 11531.2 | 12838.8 | 601.179 | 242.338 | 92.2216 | 597.688 | 372.439 |
| 601.47 | 287.14 | 176.58 | 995.471 | 692.995 | 2886.33 | 2777.05 | 7448.46 | 8094.64 | 12718.3 | 12883.1 | 601.47 | 210.172 | 92.7818 | 592.406 | 374.383 |
| 601.76 | 189.386 | 177.595 | 880.903 | 696.457 | 3244.59 | 2788.84 | 7593.4 | 8124.55 | 11981.9 | 12927.5 | 601.76 | 110.897 | 93.3447 | 403.066 | 376.335 |
| 602.052 | 182.036 | 178.614 | 920.339 | 699.931 | 2821.63 | 2800.67 | 7642.59 | 8154.54 | 12355.8 | 12972 | 602.052 | 213.185 | 93.9103 | 276.73 | 378.295 |
| 602.342 | 142.004 | 179.638 | 825.635 | 703.418 | 3185.83 | 2812.53 | 7601.4 | 8184.6 | 12308.6 | 13016.7 | 602.342 | 103.939 | 94.4787 | 539.683 | 380.262 |
| 602.633 | 103.766 | 180.666 | 734.474 | 706.918 | 2927.62 | 2824.43 | 7914.37 | 8214.73 | 12318.4 | 13061.4 | 602.633 | 151.903 | 95.0499 | 525.818 | 382.238 |
| 602.924 | 200.541 | 181.699 | 811.118 | 710.431 | 2760.82 | 2836.37 | 7858.21 | 8244.93 | 12253.5 | 13106.2 | 602.924 | 220.178 | 95.6238 | 508.102 | 384.221 |
| 603.215 | 194.315 | 182.737 | 982.42 | 713.957 | 3170.49 | 2848.34 | 7988.15 | 8275.2 | 12449.8 | 13151.1 | 603.215 | 165.933 | 96.2005 | 418.478 | 386.212 |
| 603.506 | 126.725 | 183.779 | 929.012 | 717.496 | 2875.36 | 2860.34 | 8013.89 | 8305.55 | 12212.1 | 13196.1 | 603.506 | 162.323 | 96.78 | 457.536 | 388.211 |
| 603.797 | 328.86 | 184.826 | 846.443 | 721.049 | 3079.22 | 2872.38 | 7844.81 | 8335.96 | 12494.9 | 13241.2 | 603.797 | 240.709 | 97.3623 | 409.035 | 390.218 |
| 604.088 | 265.104 | 185.878 | 833.976 | 724.615 | 3154.02 | 2884.46 | 8162.11 | 8366.45 | 12619.4 | 13286.4 | 604.088 | 150.922 | 97.9473 | 447.789 | 392.233 |
| 604.379 | 227.85 | 186.934 | 950.376 | 728.193 | 3398.18 | 2896.57 | 8101.27 | 8397.01 | 12410.6 | 13331.6 | 604.379 | 104.234 | 98.5352 | 541.214 | 394.256 |
| 604.669 | 252.15 | 187.995 | 849.885 | 731.785 | 3085.62 | 2908.71 | 8792.51 | 8427.65 | 12736.4 | 13377 | 604.669 | 123.041 | 99.126 | 404.614 | 396.287 |
| 604.96 | 282.91 | 189.061 | 829.598 | 735.39 | 3202.38 | 2920.9 | 9116.14 | 8458.35 | 13690.9 | 13422.5 | 604.96 | 207.076 | 99.7195 | 537.598 | 398.325 |
| 605.251 | 216.069 | 190.131 | 983.166 | 739.009 | 3084.76 | 2933.11 | 8986.42 | 8489.13 | 13438.1 | 13468.1 | 605.251 | 138.383 | 100.316 | 492.701 | 400.372 |
| 605.542 | 180.285 | 191.207 | 885.447 | 742.641 | 3099.17 | 2945.37 | 8434.15 | 8519.98 | 13380.9 | 13513.7 | 605.542 | 75.4055 | 100.915 | 411.103 | 402.427 |
| 605.832 | 231.8 | 192.286 | 781.292 | 746.286 | 3650.79 | 2957.66 | 9029.04 | 8550.9 | 13758.8 | 13559.5 | 605.832 | 113.11 | 101.517 | 528.572 | 404.49 |
| 606.123 | 248.113 | 193.371 | 978.441 | 749.944 | 3137.41 | 2969.98 | 8645.92 | 8581.89 | 13566.2 | 13605.3 | 606.123 | 199.766 | 102.122 | 546.91 | 406.561 |
| 606.414 | 321.808 | 194.461 | 901.108 | 753.616 | 3180.71 | 2982.34 | 9093.8 | 8612.96 | 14209.3 | 13651.3 | 606.414 | 163.071 | 102.73 | 489.211 | 408.64 |
| 606.705 | 119.112 | 195.555 | 915.358 | 757.301 | 3449.65 | 2994.74 | 8901.44 | 8644.1 | 13973.8 | 13697.3 | 606.705 | 244.115 | 103.341 | 525.114 | 410.728 |
| 606.995 | 304.576 | 196.655 | 995.001 | 761 | 3464.35 | 3007.17 | 9516.74 | 8675.31 | 13988.4 | 13743.5 | 606.995 | 257.232 | 103.954 | 498.632 | 412.823 |
| 607.286 | 202.374 | 197.758 | 841.269 | 764.712 | 3301.98 | 3019.63 | 9204.5 | 8706.59 | 14155.5 | 13789.7 | 607.286 | 164.586 | 104.571 | 564.02 | 414.926 |
| 607.577 | 429.096 | 198.867 | 981.953 | 768.438 | 3569.72 | 3032.14 | 9055 | 8737.94 | 14382.8 | 13836 | 607.577 | 181.834 | 105.191 | 480.649 | 417.038 |
| 607.868 | 215.253 | 199.981 | 914.156 | 772.177 | 3380.42 | 3044.68 | 8743.49 | 8769.37 | 14114.5 | 13882.5 | 607.868 | 173.31 | 105.813 | 556.286 | 419.158 |
| 608.158 | 282.909 | 201.1 | 924.218 | 775.93 | 3861.48 | 3057.25 | 9730.55 | 8800.87 | 14885.3 | 13929 | 608.158 | 270.79 | 106.438 | 578.95 | 421.286 |
| 608.449 | 248.109 | 202.223 | 940.796 | 779.696 | 3331.34 | 3069.86 | 9374.96 | 8832.44 | 14293.7 | 13975.6 | 608.449 | 196.737 | 107.067 | 424.321 | 423.423 |
| 608.74 | 351.157 | 203.352 | 751.929 | 783.476 | 3412.47 | 3082.51 | 9778.08 | 8864.09 | 14085.5 | 14022.3 | 608.74 | 196.552 | 107.698 | 423.922 | 425.567 |
| 609.03 | 140.26 | 204.485 | 972.522 | 787.269 | 3698.42 | 3095.19 | 9569.13 | 8895.81 | 14733.8 | 14069.1 | 609.03 | 102.663 | 108.332 | 563.518 | 427.72 |
| 609.321 | 234.601 | 205.623 | 968.651 | 791.077 | 3429.09 | 3107.91 | 9459.64 | 8927.6 | 14519 | 14116 | 609.321 | 186.026 | 108.969 | 505.24 | 429.881 |
| 609.612 | 278.171 | 206.767 | 964.375 | 794.897 | 3707.96 | 3120.66 | 9968.63 | 8959.46 | 15442.9 | 14163 | 609.612 | 189.063 | 109.609 | 649.413 | 432.051 |
| 609.902 | 252.637 | 207.915 | 1002.18 | 798.732 | 3963.55 | 3133.46 | 9729.63 | 8991.4 | 15246.7 | 14210.1 | 609.902 | 113.795 | 110.252 | 561.315 | 434.229 |
| 610.193 | 174.401 | 209.068 | 940.521 | 802.58 | 3813.85 | 3146.28 | 10128.8 | 9023.4 | 15568.3 | 14257.2 | 610.193 | 127.652 | 110.899 | 589.163 | 436.415 |
| 610.484 | 223.02 | 210.226 | 1033.59 | 806.442 | 3676.76 | 3159.15 | 9421.81 | 9055.49 | 14911.9 | 14304.5 | 610.484 | 108.826 | 111.548 | 607.437 | 438.61 |
| 610.774 | 328.938 | 211.389 | 979.352 | 810.318 | 3652.57 | 3172.05 | 9782.18 | 9087.64 | 14988 | 14351.9 | 610.774 | 147.135 | 112.2 | 500.05 | 440.812 |
| 611.065 | 291.696 | 212.557 | 995.464 | 814.208 | 3829.82 | 3184.98 | 9988.61 | 9119.86 | 15105.2 | 14399.4 | 611.065 | 186.818 | 112.855 | 623.58 | 443.024 |
| 611.355 | 373.984 | 213.731 | 1116.47 | 818.112 | 3704.8 | 3197.96 | 9888.11 | 9152.16 | 15221.7 | 14446.9 | 611.355 | 156.421 | 113.514 | 514.383 | 445.244 |
| 611.646 | 262.342 | 214.909 | 1043.4 | 822.029 | 3513.53 | 3210.97 | 10078 | 9184.53 | 15132 | 14494.6 | 611.646 | 166.418 | 114.175 | 491.867 | 447.472 |
| 611.937 | 327.858 | 216.092 | 1029.45 | 825.961 | 3765.71 | 3224.01 | 9744.02 | 9216.98 | 15291.2 | 14542.4 | 611.937 | 202.084 | 114.84 | 603.336 | 449.709 |
| 612.227 | 271.996 | 217.281 | 996.992 | 829.906 | 3588.51 | 3237.09 | 9777.94 | 9249.5 | 15242.5 | 14590.2 | 612.227 | 199.087 | 115.507 | 607.309 | 451.954 |
| 612.518 | 268.386 | 218.474 | 1051.6 | 833.865 | 3780.35 | 3250.21 | 9630.54 | 9282.09 | 15426.7 | 14638.1 | 612.518 | 171.889 | 116.178 | 573.748 | 454.208 |
| 612.808 | 279.055 | 219.673 | 1062.29 | 837.838 | 3833.34 | 3263.37 | 9924.03 | 9314.75 | 14752.6 | 14686.2 | 612.808 | 168.209 | 116.851 | 449.173 | 456.47 |
| 613.099 | 246.705 | 220.876 | 1097.62 | 841.826 | 3941.44 | 3276.56 | 9891.57 | 9347.49 | 15625.5 | 14734.3 | 613.099 | 276.882 | 117.528 | 512.554 | 458.741 |
| 613.389 | 260.117 | 222.085 | 969.889 | 845.827 | 3680.46 | 3289.79 | 9931.07 | 9380.3 | 15785.9 | 14782.6 | 613.389 | 226.091 | 118.208 | 568.429 | 461.02 |
| 613.68 | 304.998 | 223.299 | 973.904 | 849.843 | 3977.87 | 3303.06 | 9940.76 | 9413.18 | 15721.5 | 14830.9 | 613.68 | 211.493 | 118.891 | 524.664 | 463.308 |
| 613.97 | 222.523 | 224.518 | 1060.75 | 853.872 | 3906.29 | 3316.36 | 10326.7 | 9446.14 | 16232.7 | 14879.3 | 613.97 | 151.241 | 119.577 | 640.312 | 465.605 |
| 614.26 | 282.416 | 225.743 | 967.763 | 857.916 | 3983.99 | 3329.7 | 10473.7 | 9479.16 | 15645.7 | 14927.8 | 614.26 | 149.294 | 120.267 | 536.66 | 467.91 |
| 614.551 | 247.155 | 226.972 | 1062.13 | 861.974 | 3953.88 | 3343.07 | 10554.9 | 9512.26 | 15921.5 | 14976.4 | 614.551 | 169.598 | 120.959 | 645.775 | 470.224 |
| 614.841 | 314.47 | 228.207 | 1090.22 | 866.046 | 3878.77 | 3356.49 | 10232.1 | 9545.44 | 15528.7 | 15025.1 | 614.841 | 120.568 | 121.655 | 591.879 | 472.547 |
| 615.132 | 298.517 | 229.447 | 1116.84 | 870.132 | 3835.66 | 3369.94 | 10418 | 9578.69 | 16400.4 | 15074 | 615.132 | 174.799 | 122.354 | 499.184 | 474.878 |
| 615.422 | 334.759 | 230.692 | 1089.08 | 874.233 | 3725.46 | 3383.42 | 10113.2 | 9612.01 | 15703 | 15122.9 | 615.422 | 202.413 | 123.056 | 531.026 | 477.218 |
| 615.713 | 286.01 | 231.943 | 1026.67 | 878.347 | 3791.63 | 3396.95 | 10131.6 | 9645.4 | 15593.5 | 15171.8 | 615.713 | 209.345 | 123.761 | 619.578 | 479.567 |
| 616.003 | 285.213 | 233.199 | 1110.33 | 882.476 | 4086.46 | 3410.51 | 10387.8 | 9678.87 | 15759.1 | 15220.9 | 616.003 | 219.2 | 124.469 | 628.691 | 481.924 |
| 616.293 | 234.773 | 234.46 | 1075.26 | 886.62 | 3943.29 | 3424.11 | 10393.1 | 9712.41 | 16114.8 | 15270.1 | 616.293 | 151.625 | 125.181 | 587.838 | 484.291 |
| 616.584 | 288.755 | 235.726 | 1098.4 | 890.778 | 3968.24 | 3437.74 | 10304.6 | 9746.02 | 16059.3 | 15319.4 | 616.584 | 171.096 | 125.896 | 553.931 | 486.666 |
| 616.874 | 243.28 | 236.998 | 1106.98 | 894.95 | 4038.34 | 3451.41 | 10440.9 | 9779.71 | 15898.4 | 15368.8 | 616.874 | 217.639 | 126.614 | 503.384 | 489.05 |
| 617.165 | 281.852 | 238.274 | 1085.71 | 899.136 | 4119.73 | 3465.12 | 10311.9 | 9813.47 | 16251.7 | 15418.2 | 617.165 | 147.358 | 127.335 | 612.1 | 491.442 |
| 617.455 | 368.564 | 239.557 | 1091.42 | 903.337 | 3677.96 | 3478.87 | 10417.2 | 9847.31 | 16269.6 | 15467.8 | 617.455 | 173.424 | 128.06 | 590.305 | 493.844 |
| 617.745 | 318.477 | 240.844 | 1107.79 | 907.553 | 3858.29 | 3492.65 | 10322.4 | 9881.22 | 16052.5 | 15517.5 | 617.745 | 251.757 | 128.788 | 590.66 | 496.254 |
| 618.036 | 329.234 | 242.137 | 1062.75 | 911.783 | 3925.6 | 3506.48 | 10513.1 | 9915.2 | 16168.7 | 15567.2 | 618.036 | 203.908 | 129.519 | 577.5 | 498.673 |
| 618.326 | 346.741 | 243.436 | 1168.71 | 916.027 | 4020.51 | 3520.33 | 10675.6 | 9949.25 | 16465.1 | 15617.1 | 618.326 | 150.398 | 130.253 | 566.16 | 501.102 |
| 618.616 | 297.451 | 244.739 | 1152.96 | 920.286 | 4094.97 | 3534.23 | 10497.9 | 9983.38 | 16071.8 | 15667 | 618.616 | 199.576 | 130.991 | 536.969 | 503.539 |
| 618.907 | 293.162 | 246.049 | 1074.58 | 924.56 | 3836.61 | 3548.16 | 10615.8 | 10017.6 | 16070.4 | 15717.1 | 618.907 | 214.58 | 131.732 | 649.926 | 505.985 |
| 619.197 | 300.708 | 247.363 | 1082.47 | 928.848 | 4121.4 | 3562.14 | 10604.9 | 10051.9 | 16391.6 | 15767.2 | 619.197 | 211.298 | 132.476 | 603.418 | 508.44 |
| 619.487 | 285.801 | 248.683 | 1119.84 | 933.151 | 3971.01 | 3576.14 | 10611.7 | 10086.2 | 16502.6 | 15817.4 | 619.487 | 148.177 | 133.224 | 651.71 | 510.904 |
| 619.777 | 339.465 | 250.009 | 1230.98 | 937.468 | 3998.93 | 3590.19 | 10431.1 | 10120.6 | 16552.8 | 15867.8 | 619.777 | 205.631 | 133.975 | 622.824 | 513.377 |
| 620.067 | 379.71 | 251.34 | 1128.58 | 941.801 | 4083.37 | 3604.27 | 10736.5 | 10155.1 | 16391.5 | 15918.2 | 620.067 | 185.286 | 134.73 | 655.948 | 515.859 |
| 620.358 | 365.647 | 252.676 | 1178.51 | 946.148 | 3915.19 | 3618.39 | 10614 | 10189.7 | 16576.9 | 15968.7 | 620.358 | 209.09 | 135.487 | 668.763 | 518.351 |
| 620.648 | 312.897 | 254.018 | 1039.57 | 950.509 | 4125.28 | 3632.55 | 10849.5 | 10224.3 | 16633.3 | 16019.3 | 620.648 | 171.769 | 136.248 | 654.04 | 520.851 |
| 620.938 | 315.216 | 255.366 | 1066.11 | 954.886 | 4090.13 | 3646.75 | 10450.3 | 10259 | 16281.2 | 16070 | 620.938 | 175.03 | 137.013 | 636.166 | 523.36 |
| 621.228 | 315.939 | 256.719 | 1211.77 | 959.277 | 4188.39 | 3660.98 | 10828.8 | 10293.8 | 16706.5 | 16120.8 | 621.228 | 207.328 | 137.781 | 629.344 | 525.879 |
| 621.519 | 301.673 | 258.077 | 1143.9 | 963.683 | 4087.16 | 3675.26 | 10923.9 | 10328.7 | 16702.4 | 16171.7 | 621.519 | 197.151 | 138.552 | 606.009 | 528.406 |
| 621.809 | 348.75 | 259.441 | 1207.54 | 968.104 | 4019.86 | 3689.57 | 10914.9 | 10363.6 | 16914.7 | 16222.7 | 621.809 | 208.855 | 139.327 | 586.399 | 530.943 |
| 622.099 | 332.352 | 260.811 | 1270.73 | 972.539 | 4141.2 | 3703.91 | 10606.8 | 10398.6 | 16905.9 | 16273.8 | 622.099 | 205.255 | 140.105 | 647.339 | 533.489 |
| 622.389 | 355.666 | 262.186 | 1140.56 | 976.99 | 4134.74 | 3718.3 | 10815.2 | 10433.7 | 16690.2 | 16325 | 622.389 | 208.263 | 140.886 | 563.854 | 536.044 |
| 622.679 | 336.77 | 263.567 | 1093.82 | 981.456 | 4204.37 | 3732.72 | 11193.3 | 10468.9 | 16793.3 | 16376.3 | 622.679 | 194.211 | 141.671 | 550.74 | 538.608 |
| 622.969 | 373.959 | 264.953 | 1257.04 | 985.936 | 4163.94 | 3747.18 | 11181.2 | 10504.1 | 17108.2 | 16427.6 | 622.969 | 177.547 | 142.46 | 637.631 | 541.181 |
| 623.259 | 342.141 | 266.346 | 1186.15 | 990.432 | 4246.66 | 3761.68 | 10886.2 | 10539.4 | 17089.7 | 16479.1 | 623.259 | 236.119 | 143.252 | 586.994 | 543.764 |
| 623.549 | 343.746 | 267.743 | 1161.92 | 994.942 | 4486.92 | 3776.22 | 11040.4 | 10574.8 | 17067 | 16530.7 | 623.549 | 158.151 | 144.047 | 656.934 | 546.356 |
| 623.84 | 347.114 | 269.147 | 1150.46 | 999.468 | 4365.7 | 3790.8 | 11334.8 | 10610.3 | 17452.3 | 16582.3 | 623.84 | 183.495 | 144.846 | 652.273 | 548.957 |
| 624.13 | 318.305 | 270.556 | 1117.04 | 1004.01 | 4000.18 | 3805.41 | 10818.7 | 10645.8 | 16621.7 | 16634.1 | 624.13 | 191.868 | 145.648 | 640.437 | 551.568 |
| 624.42 | 336.499 | 271.97 | 1134.19 | 1008.56 | 4260.79 | 3820.06 | 11253 | 10681.4 | 17202.6 | 16685.9 | 624.42 | 198.409 | 146.454 | 632.326 | 554.188 |
| 624.71 | 341.024 | 273.391 | 1248.85 | 1013.13 | 4210.77 | 3834.75 | 11473.4 | 10717.1 | 17242.7 | 16737.9 | 624.71 | 175.403 | 147.264 | 679.554 | 556.817 |
| 625 | 355.83 | 274.817 | 1199.34 | 1017.72 | 4438.11 | 3849.48 | 11146.9 | 10752.8 | 17632.5 | 16789.9 | 625 | 186.989 | 148.077 | 658.827 | 559.456 |
| 625.29 | 373.497 | 276.249 | 1326.44 | 1022.32 | 4356.9 | 3864.24 | 11588.6 | 10788.6 | 17452.5 | 16842 | 625.29 | 220.038 | 148.893 | 567.789 | 562.104 |
| 625.58 | 354.677 | 277.687 | 1201.43 | 1026.94 | 4222.71 | 3879.05 | 11400.8 | 10824.5 | 17173.2 | 16894.3 | 625.58 | 207.684 | 149.713 | 707.672 | 564.761 |
| 625.87 | 332.534 | 279.13 | 1229.95 | 1031.57 | 4319.15 | 3893.89 | 11499.6 | 10860.5 | 17608.4 | 16946.6 | 625.87 | 230.587 | 150.537 | 625.244 | 567.428 |
| 626.16 | 324.513 | 280.579 | 1206.12 | 1036.22 | 4463.05 | 3908.77 | 11792.4 | 10896.5 | 17684.8 | 16999 | 626.16 | 224.687 | 151.364 | 653.319 | 570.105 |
| 626.45 | 301.836 | 282.034 | 1255.7 | 1040.88 | 4564.27 | 3923.69 | 11749 | 10932.7 | 17723.2 | 17051.5 | 626.45 | 201.992 | 152.194 | 721.05 | 572.79 |
| 626.74 | 423.442 | 283.495 | 1207.44 | 1045.56 | 4421.76 | 3938.65 | 11760 | 10968.9 | 17794.9 | 17104.1 | 626.74 | 198.36 | 153.029 | 688.774 | 575.486 |
| 627.03 | 340.135 | 284.962 | 1263.95 | 1050.25 | 4578.53 | 3953.64 | 11834.7 | 11005.1 | 18063.4 | 17156.8 | 627.03 | 211.617 | 153.867 | 691.583 | 578.19 |
| 627.32 | 373.769 | 286.434 | 1184.22 | 1054.96 | 4472.25 | 3968.68 | 12004.6 | 11041.5 | 17913.8 | 17209.7 | 627.32 | 231.023 | 154.708 | 656.59 | 580.904 |
| 627.61 | 446.612 | 287.913 | 1296.25 | 1059.68 | 4453.19 | 3983.75 | 11988.2 | 11077.9 | 18417.9 | 17262.5 | 627.61 | 211.878 | 155.553 | 685.224 | 583.628 |
| 627.9 | 379.827 | 289.397 | 1291.12 | 1064.42 | 4680.99 | 3998.86 | 11688.6 | 11114.4 | 18530 | 17315.5 | 627.9 | 207.032 | 156.402 | 663.41 | 586.362 |
| 628.19 | 409.662 | 290.887 | 1308.98 | 1069.17 | 4812.59 | 4014.01 | 12469.5 | 11150.9 | 18640.4 | 17368.6 | 628.19 | 233.884 | 157.254 | 722.409 | 589.105 |
| 628.48 | 389.656 | 292.383 | 1278.51 | 1073.94 | 4727.76 | 4029.2 | 12019.1 | 11187.6 | 18616.7 | 17421.8 | 628.48 | 232.823 | 158.11 | 664.887 | 591.857 |
| 628.77 | 430.711 | 293.885 | 1282.36 | 1078.73 | 4571.19 | 4044.43 | 12005.1 | 11224.3 | 18906.8 | 17475.1 | 628.77 | 239.363 | 158.97 | 721.454 | 594.619 |
| 629.059 | 424.88 | 295.392 | 1273.25 | 1083.53 | 4679.41 | 4059.7 | 11975.9 | 11261 | 18562.5 | 17528.4 | 629.059 | 214.866 | 159.833 | 687.004 | 597.391 |
| 629.349 | 339.679 | 296.906 | 1311.43 | 1088.35 | 4550.35 | 4075 | 12138.1 | 11297.9 | 18954.5 | 17581.9 | 629.349 | 214.724 | 160.701 | 680.68 | 600.173 |
| 629.639 | 434.861 | 298.426 | 1319.15 | 1093.18 | 4753.55 | 4090.34 | 11987.3 | 11334.8 | 18627.7 | 17635.5 | 629.639 | 217.782 | 161.571 | 713.265 | 602.964 |
| 629.929 | 389.455 | 299.951 | 1328.12 | 1098.03 | 4804.78 | 4105.73 | 12463.6 | 11371.9 | 19141.2 | 17689.1 | 629.929 | 223.578 | 162.446 | 760.376 | 605.765 |
| 630.219 | 372.704 | 301.483 | 1341.64 | 1102.89 | 4804.1 | 4121.15 | 12455.5 | 11408.9 | 18910.7 | 17742.8 | 630.219 | 223.696 | 163.324 | 713.895 | 608.575 |
| 630.509 | 409.18 | 303.021 | 1352.07 | 1107.77 | 4820.94 | 4136.61 | 12293.8 | 11446.1 | 18858.5 | 17796.7 | 630.509 | 240.669 | 164.206 | 760.881 | 611.396 |
| 630.799 | 383.232 | 304.564 | 1348.83 | 1112.66 | 4855.03 | 4152.1 | 12585.5 | 11483.3 | 19353.2 | 17850.6 | 630.799 | 237.351 | 165.091 | 750.642 | 614.226 |
| 631.088 | 420.307 | 306.114 | 1472.84 | 1117.57 | 5076.87 | 4167.64 | 12436 | 11520.6 | 19151.2 | 17904.7 | 631.088 | 222.776 | 165.981 | 727.08 | 617.066 |
| 631.378 | 400.298 | 307.67 | 1460.05 | 1122.5 | 4840.21 | 4183.22 | 12627.1 | 11558 | 19325.2 | 17958.8 | 631.378 | 266.837 | 166.874 | 700.897 | 619.915 |
| 631.668 | 374.566 | 309.231 | 1364.92 | 1127.44 | 4871.92 | 4198.84 | 12595.5 | 11595.4 | 19498.8 | 18013 | 631.668 | 263.817 | 167.771 | 757.428 | 622.775 |
| 631.958 | 402.884 | 310.799 | 1390.71 | 1132.4 | 4944.34 | 4214.49 | 12751.6 | 11633 | 19419.6 | 18067.3 | 631.958 | 254.216 | 168.672 | 791.977 | 625.644 |
| 632.247 | 394.751 | 312.373 | 1449.57 | 1137.37 | 4993.08 | 4230.18 | 12851.5 | 11670.6 | 19463 | 18121.7 | 632.247 | 223.039 | 169.576 | 747.491 | 628.523 |
| 632.537 | 447.811 | 313.953 | 1462.93 | 1142.36 | 5030.32 | 4245.91 | 13117.3 | 11708.2 | 19635.9 | 18176.2 | 632.537 | 237.007 | 170.484 | 811.677 | 631.412 |
| 632.827 | 415.829 | 315.539 | 1406.16 | 1147.37 | 5053.82 | 4261.69 | 13279.6 | 11746 | 19729 | 18230.8 | 632.827 | 244.49 | 171.397 | 730.589 | 634.311 |
| 633.117 | 421.666 | 317.132 | 1394.17 | 1152.39 | 4921.13 | 4277.5 | 13045.6 | 11783.8 | 19581.2 | 18285.5 | 633.117 | 235.153 | 172.312 | 839.425 | 637.22 |
| 633.407 | 460.919 | 318.73 | 1519.78 | 1157.42 | 5054.53 | 4293.35 | 12879.8 | 11821.7 | 20165.5 | 18340.3 | 633.407 | 239.581 | 173.232 | 812.391 | 640.139 |
| 633.696 | 416.869 | 320.335 | 1512.13 | 1162.47 | 5121.55 | 4309.23 | 13161.1 | 11859.7 | 20120.8 | 18395.2 | 633.696 | 256.694 | 174.156 | 824.845 | 643.067 |
| 633.986 | 391.361 | 321.946 | 1401.27 | 1167.54 | 4906.93 | 4325.16 | 12787.8 | 11897.7 | 19798 | 18450.2 | 633.986 | 239.496 | 175.083 | 750.996 | 646.006 |
| 634.276 | 469.964 | 323.563 | 1419.21 | 1172.63 | 4986.83 | 4341.13 | 13138.4 | 11935.8 | 19744.1 | 18505.2 | 634.276 | 218.48 | 176.014 | 777.191 | 648.955 |
| 634.565 | 420.673 | 325.186 | 1514.57 | 1177.73 | 5154.36 | 4357.14 | 13306.4 | 11974 | 20049.2 | 18560.4 | 634.565 | 242.597 | 176.95 | 813.916 | 651.914 |
| 634.855 | 397.427 | 326.815 | 1486.26 | 1182.85 | 5210.79 | 4373.18 | 13308.8 | 12012.3 | 20305.8 | 18615.6 | 634.855 | 217.018 | 177.889 | 791.172 | 654.882 |
| 635.144 | 338.856 | 328.451 | 1472.08 | 1187.98 | 5087.33 | 4389.27 | 13436.2 | 12050.6 | 20448.5 | 18671 | 635.144 | 252.619 | 178.832 | 767.922 | 657.861 |
| 635.434 | 402.722 | 330.093 | 1478.52 | 1193.13 | 5086.15 | 4405.39 | 13389.8 | 12089 | 20254.7 | 18726.4 | 635.434 | 258.492 | 179.779 | 842.975 | 660.85 |
| 635.724 | 450.01 | 331.741 | 1518.83 | 1198.29 | 5140.98 | 4421.56 | 13442.3 | 12127.5 | 20663.8 | 18782 | 635.724 | 267.068 | 180.73 | 818.15 | 663.849 |
| 636.013 | 433.681 | 333.395 | 1461.74 | 1203.47 | 5190.34 | 4437.76 | 13396.6 | 12166.1 | 20530.3 | 18837.6 | 636.013 | 238.074 | 181.684 | 824.1 | 666.858 |
| 636.303 | 418.824 | 335.056 | 1433.85 | 1208.67 | 5357.09 | 4454 | 13684 | 12204.7 | 20888.6 | 18893.3 | 636.303 | 232.108 | 182.643 | 832.25 | 669.877 |
| 636.593 | 462.923 | 336.723 | 1445.75 | 1213.88 | 5145 | 4470.28 | 13641.6 | 12243.5 | 20512.9 | 18949.2 | 636.593 | 244.476 | 183.606 | 832.899 | 672.907 |
| 636.882 | 400.702 | 338.397 | 1526.75 | 1219.11 | 5110.74 | 4486.6 | 13617 | 12282.2 | 20621.1 | 19005.1 | 636.882 | 272.04 | 184.572 | 825.436 | 675.946 |
| 637.172 | 497.146 | 340.076 | 1513.87 | 1224.36 | 5139.66 | 4502.97 | 13632.9 | 12321.1 | 20440.7 | 19061.1 | 637.172 | 255.138 | 185.543 | 825.254 | 678.996 |
| 637.461 | 478.603 | 341.762 | 1560.01 | 1229.62 | 5299.79 | 4519.37 | 13492.3 | 12360 | 20577.7 | 19117.2 | 637.461 | 214.309 | 186.517 | 821.726 | 682.056 |
| 637.751 | 439.175 | 343.455 | 1531.5 | 1234.9 | 5219.25 | 4535.81 | 13805.1 | 12399 | 21060.7 | 19173.4 | 637.751 | 226.666 | 187.496 | 796.023 | 685.126 |
| 638.04 | 425.863 | 345.154 | 1461.83 | 1240.19 | 5377.09 | 4552.29 | 13803.6 | 12438.1 | 21206.3 | 19229.7 | 638.04 | 241.986 | 188.478 | 847.577 | 688.206 |
| 638.33 | 484.427 | 346.859 | 1541.87 | 1245.51 | 5338.6 | 4568.81 | 13566.5 | 12477.3 | 20902.3 | 19286.1 | 638.33 | 237.727 | 189.465 | 845.214 | 691.297 |
| 638.619 | 469.979 | 348.57 | 1503.04 | 1250.83 | 5305.4 | 4585.37 | 13772 | 12516.5 | 20956.6 | 19342.6 | 638.619 | 236.001 | 190.456 | 830.768 | 694.398 |
| 638.909 | 471.594 | 350.288 | 1479.35 | 1256.18 | 5290.08 | 4601.96 | 13797.7 | 12555.8 | 20696.6 | 19399.1 | 638.909 | 243.197 | 191.45 | 802.457 | 697.509 |
| 639.198 | 451.97 | 352.013 | 1553.85 | 1261.54 | 5601.34 | 4618.6 | 13824.2 | 12595.2 | 21042.3 | 19455.8 | 639.198 | 229.627 | 192.449 | 856.833 | 700.631 |
| 639.488 | 475.747 | 353.744 | 1511.9 | 1266.91 | 5357.99 | 4635.28 | 13707.3 | 12634.6 | 21001 | 19512.6 | 639.488 | 275.517 | 193.452 | 846.416 | 703.763 |
| 639.777 | 434.707 | 355.481 | 1538.46 | 1272.31 | 5303.76 | 4652 | 13927 | 12674.2 | 20956.5 | 19569.4 | 639.777 | 238.637 | 194.459 | 826.049 | 706.905 |
| 640.067 | 465.171 | 357.225 | 1630.59 | 1277.72 | 5523.73 | 4668.76 | 14045.1 | 12713.8 | 20913.9 | 19626.4 | 640.067 | 268.602 | 195.47 | 872.152 | 710.058 |
| 640.356 | 449.199 | 358.976 | 1670.67 | 1283.14 | 5533.8 | 4685.55 | 14252.8 | 12753.5 | 21389.8 | 19683.4 | 640.356 | 264.544 | 196.485 | 883.022 | 713.221 |
| 640.646 | 481.754 | 360.733 | 1583.18 | 1288.59 | 5306.81 | 4702.39 | 13858 | 12793.2 | 21199.3 | 19740.5 | 640.646 | 264.464 | 197.504 | 820.091 | 716.394 |
| 640.935 | 501.692 | 362.496 | 1526.95 | 1294.05 | 5448.64 | 4719.27 | 14249.2 | 12833 | 21298.8 | 19797.8 | 640.935 | 235.017 | 198.527 | 861.818 | 719.578 |
| 641.225 | 451.849 | 364.266 | 1601.64 | 1299.52 | 5595.2 | 4736.19 | 14144.5 | 12872.9 | 21490.8 | 19855.1 | 641.225 | 261.677 | 199.554 | 864.286 | 722.772 |
| 641.514 | 448.902 | 366.043 | 1687.38 | 1305.01 | 5398.77 | 4753.14 | 14155.5 | 12912.9 | 21292.3 | 19912.5 | 641.514 | 282.145 | 200.586 | 827.063 | 725.977 |
| 641.803 | 486.661 | 367.826 | 1628.83 | 1310.52 | 5528.56 | 4770.14 | 14317.7 | 12953 | 21415 | 19970 | 641.803 | 267.158 | 201.621 | 854.491 | 729.192 |
| 642.093 | 491.749 | 369.616 | 1660.68 | 1316.05 | 5472.35 | 4787.18 | 14166.3 | 12993.1 | 21258.7 | 20027.6 | 642.093 | 269.078 | 202.661 | 849.164 | 732.418 |
| 642.382 | 489.136 | 371.412 | 1675.87 | 1321.59 | 5390 | 4804.25 | 14150.8 | 13033.3 | 21165.7 | 20085.3 | 642.382 | 264.172 | 203.705 | 884.358 | 735.655 |
| 642.671 | 471.878 | 373.215 | 1577.89 | 1327.15 | 5479.08 | 4821.37 | 14123 | 13073.5 | 21425.2 | 20143.1 | 642.671 | 283.836 | 204.753 | 848.743 | 738.902 |
| 642.961 | 459.725 | 375.025 | 1669.69 | 1332.73 | 5510.93 | 4838.53 | 14303.5 | 13113.9 | 21615.5 | 20201 | 642.961 | 295.291 | 205.806 | 923.509 | 742.159 |
| 643.25 | 480.994 | 376.841 | 1674.21 | 1338.32 | 5475.7 | 4855.73 | 14354.3 | 13154.3 | 21712.9 | 20259 | 643.25 | 294.514 | 206.862 | 882.367 | 745.427 |
| 643.539 | 463.361 | 378.664 | 1632.91 | 1343.93 | 5572 | 4872.96 | 14248.5 | 13194.8 | 21759.3 | 20317.1 | 643.539 | 258.565 | 207.923 | 864.808 | 748.706 |
| 643.829 | 487.944 | 380.494 | 1636.95 | 1349.56 | 5626.94 | 4890.24 | 14426.3 | 13235.4 | 21872.4 | 20375.2 | 643.829 | 307.082 | 208.988 | 901.216 | 751.995 |
| 644.118 | 483.951 | 382.33 | 1626.65 | 1355.2 | 5666.31 | 4907.56 | 14486.1 | 13276 | 21797.7 | 20433.5 | 644.118 | 255.831 | 210.057 | 888.972 | 755.295 |
| 644.407 | 473.058 | 384.173 | 1574.91 | 1360.86 | 5506.75 | 4924.91 | 14202.4 | 13316.7 | 21998.9 | 20491.8 | 644.407 | 258.058 | 211.131 | 844.652 | 758.606 |
| 644.697 | 519.839 | 386.023 | 1671.27 | 1366.54 | 5785.95 | 4942.31 | 14293.5 | 13357.5 | 21910.9 | 20550.3 | 644.697 | 276.428 | 212.208 | 884.818 | 761.927 |
| 644.986 | 463.522 | 387.88 | 1629.28 | 1372.23 | 5672.21 | 4959.75 | 14831.7 | 13398.4 | 22278.6 | 20608.8 | 644.986 | 294.038 | 213.291 | 869.061 | 765.259 |
| 645.275 | 524.496 | 389.743 | 1621.99 | 1377.94 | 5669.56 | 4977.23 | 14694.1 | 13439.3 | 22474.7 | 20667.4 | 645.275 | 262.335 | 214.377 | 903.648 | 768.602 |
| 645.564 | 487.059 | 391.613 | 1638.75 | 1383.67 | 5648.16 | 4994.75 | 14775.5 | 13480.3 | 22298.8 | 20726.2 | 645.564 | 277.63 | 215.467 | 894.407 | 771.956 |
| 645.853 | 498.229 | 393.49 | 1700.83 | 1389.41 | 5745.18 | 5012.31 | 14570.9 | 13521.4 | 22119.6 | 20785 | 645.853 | 286.755 | 216.562 | 909.534 | 775.32 |
| 646.143 | 514.288 | 395.373 | 1641.74 | 1395.17 | 5738.65 | 5029.9 | 14842.2 | 13562.6 | 22273.7 | 20843.9 | 646.143 | 301.768 | 217.661 | 862.584 | 778.695 |
| 646.432 | 503.609 | 397.264 | 1710.17 | 1400.95 | 5679.16 | 5047.55 | 14875.4 | 13603.8 | 22379.8 | 20902.9 | 646.432 | 288.749 | 218.765 | 899.562 | 782.081 |
| 646.721 | 491.492 | 399.161 | 1684.67 | 1406.75 | 5776.62 | 5065.23 | 14789.7 | 13645.1 | 22303.1 | 20962 | 646.721 | 307.919 | 219.873 | 946.618 | 785.477 |
| 647.01 | 495.064 | 401.065 | 1699.48 | 1412.56 | 5807.89 | 5082.94 | 14776.4 | 13686.5 | 22415.8 | 21021.2 | 647.01 | 267.965 | 220.985 | 913.864 | 788.885 |
| 647.299 | 525.07 | 402.976 | 1730.75 | 1418.39 | 5856.21 | 5100.71 | 14823.7 | 13727.9 | 22558.4 | 21080.5 | 647.299 | 290.513 | 222.102 | 917.626 | 792.303 |
| 647.588 | 514.214 | 404.894 | 1748.19 | 1424.24 | 6011.77 | 5118.51 | 14755.2 | 13769.5 | 22363.4 | 21139.8 | 647.588 | 290.434 | 223.223 | 895.577 | 795.732 |
| 647.878 | 540.358 | 406.819 | 1790.04 | 1430.1 | 5716.01 | 5136.35 | 14957.8 | 13811.1 | 22703.6 | 21199.3 | 647.878 | 304.015 | 224.348 | 947.12 | 799.173 |
| 648.167 | 542.115 | 408.751 | 1750.07 | 1435.98 | 5888.67 | 5154.23 | 15009.1 | 13852.8 | 23012.5 | 21258.8 | 648.167 | 332.137 | 225.478 | 937.056 | 802.624 |
| 648.456 | 518.515 | 410.689 | 1820.84 | 1441.88 | 5968.08 | 5172.15 | 15238.6 | 13894.5 | 23089.2 | 21318.5 | 648.456 | 332.454 | 226.612 | 919.619 | 806.086 |
| 648.745 | 480.759 | 412.635 | 1753.24 | 1447.8 | 5879.13 | 5190.12 | 15080.3 | 13936.3 | 23119.2 | 21378.2 | 648.745 | 314.395 | 227.75 | 946.509 | 809.559 |
| 649.034 | 521.199 | 414.587 | 1731.63 | 1453.73 | 5910.46 | 5208.12 | 15070.6 | 13978.2 | 22777.5 | 21438.1 | 649.034 | 318.387 | 228.893 | 1015.73 | 813.043 |
| 649.323 | 542.753 | 416.547 | 1736.6 | 1459.68 | 5894.58 | 5226.16 | 15153.3 | 14020.2 | 22947.7 | 21498 | 649.323 | 312.139 | 230.041 | 954.748 | 816.537 |
| 649.612 | 587.518 | 418.513 | 1805.99 | 1465.65 | 6143.09 | 5244.25 | 15301.4 | 14062.3 | 23184.4 | 21558 | 649.612 | 312.49 | 231.193 | 964.593 | 820.043 |
| 649.901 | 533.127 | 420.487 | 1718.8 | 1471.63 | 6130.02 | 5262.37 | 15485.5 | 14104.4 | 23247.8 | 21618.1 | 649.901 | 305.391 | 232.349 | 977.834 | 823.56 |
| 650.19 | 547.875 | 422.467 | 1758.15 | 1477.64 | 6060.4 | 5280.54 | 15314.7 | 14146.6 | 23003.1 | 21678.3 | 650.19 | 284.053 | 233.51 | 951.447 | 827.088 |
| 650.479 | 515.949 | 424.455 | 1800.86 | 1483.66 | 6052.65 | 5298.75 | 15540.2 | 14188.9 | 23170.8 | 21738.6 | 650.479 | 282.537 | 234.675 | 1019.65 | 830.627 |
| 650.768 | 568.53 | 426.449 | 1750.93 | 1489.69 | 5941.68 | 5316.99 | 15534.5 | 14231.2 | 22979.9 | 21799 | 650.768 | 304.622 | 235.845 | 974.43 | 834.177 |
| 651.057 | 554.887 | 428.451 | 1784.24 | 1495.75 | 6035.79 | 5335.28 | 15441.6 | 14273.6 | 23225.8 | 21859.5 | 651.057 | 308.673 | 237.019 | 961.633 | 837.739 |
| 651.346 | 557.549 | 430.459 | 1819.11 | 1501.82 | 6197.41 | 5353.61 | 15379.2 | 14316.1 | 23761.6 | 21920.1 | 651.346 | 321.036 | 238.198 | 1011.45 | 841.311 |
| 651.635 | 539.924 | 432.475 | 1765 | 1507.91 | 6128.22 | 5371.98 | 15596 | 14358.7 | 23908.2 | 21980.7 | 651.635 | 289.631 | 239.381 | 992.258 | 844.894 |
| 651.924 | 603.297 | 434.498 | 1870.77 | 1514.02 | 6126.13 | 5390.39 | 15888.1 | 14401.3 | 23324.1 | 22041.5 | 651.924 | 345.237 | 240.569 | 1028.3 | 848.489 |
| 652.213 | 535.722 | 436.528 | 1842.53 | 1520.14 | 6180.39 | 5408.84 | 15774.9 | 14444 | 23724.9 | 22102.3 | 652.213 | 317.432 | 241.761 | 963.987 | 852.095 |
| 652.502 | 587.071 | 438.565 | 1876.74 | 1526.28 | 6291.67 | 5427.34 | 15963.5 | 14486.8 | 24013.1 | 22163.3 | 652.502 | 325.616 | 242.958 | 1069.85 | 855.712 |
| 652.791 | 556.714 | 440.609 | 1819.53 | 1532.44 | 6200.95 | 5445.87 | 15871.1 | 14529.7 | 23743.2 | 22224.3 | 652.791 | 325.989 | 244.16 | 1023.77 | 859.34 |
| 653.08 | 551.146 | 442.66 | 1892.64 | 1538.62 | 6346.68 | 5464.44 | 15852 | 14572.6 | 24190.7 | 22285.4 | 653.08 | 335.302 | 245.366 | 1017.39 | 862.98 |
| 653.369 | 586.112 | 444.719 | 1914.46 | 1544.82 | 6259.53 | 5483.06 | 15894.1 | 14615.6 | 24192.7 | 22346.6 | 653.369 | 340.603 | 246.576 | 1020.61 | 866.63 |
| 653.658 | 578.788 | 446.785 | 1921.93 | 1551.03 | 6383.22 | 5501.71 | 16342.4 | 14658.7 | 24396.4 | 22407.9 | 653.658 | 332.115 | 247.792 | 1080.83 | 870.292 |
| 653.947 | 549.693 | 448.858 | 1913.34 | 1557.26 | 6317.06 | 5520.41 | 16110.5 | 14701.9 | 24367.5 | 22469.3 | 653.947 | 350.764 | 249.012 | 1039.19 | 873.966 |
| 654.236 | 555.721 | 450.938 | 1850.84 | 1563.51 | 6322.31 | 5539.14 | 16142.4 | 14745.1 | 24379 | 22530.8 | 654.236 | 345.745 | 250.236 | 1032.25 | 877.65 |
| 654.524 | 601.818 | 453.025 | 1949.44 | 1569.77 | 6361.94 | 5557.92 | 16361 | 14788.4 | 24658.3 | 22592.4 | 654.524 | 343.18 | 251.465 | 1053.18 | 881.346 |
| 654.813 | 627.857 | 455.12 | 1953.7 | 1576.05 | 6331.93 | 5576.74 | 16347.6 | 14831.8 | 24871.2 | 22654.1 | 654.813 | 335.148 | 252.699 | 1023.02 | 885.053 |
| 655.102 | 579.477 | 457.221 | 1893.06 | 1582.35 | 6349.89 | 5595.6 | 16270.8 | 14875.2 | 24599 | 22715.8 | 655.102 | 331.287 | 253.938 | 1112.88 | 888.772 |
| 655.391 | 578.128 | 459.331 | 1915.37 | 1588.67 | 6523.26 | 5614.5 | 16283.8 | 14918.8 | 24518.5 | 22777.7 | 655.391 | 328.571 | 255.181 | 1072.74 | 892.502 |
| 655.68 | 625.336 | 461.447 | 1942.12 | 1595.01 | 6429.52 | 5633.44 | 16675.7 | 14962.4 | 25250.6 | 22839.6 | 655.68 | 336.998 | 256.429 | 1075.99 | 896.243 |
| 655.968 | 617.725 | 463.571 | 1952.37 | 1601.36 | 6499.01 | 5652.43 | 16453.6 | 15006 | 24842.3 | 22901.7 | 655.968 | 340.96 | 257.681 | 1082.74 | 899.996 |
| 656.257 | 619.864 | 465.702 | 1995.55 | 1607.73 | 6570.6 | 5671.45 | 16493.8 | 15049.8 | 24823.8 | 22963.8 | 656.257 | 340.895 | 258.938 | 1087.48 | 903.76 |
| 656.546 | 604.017 | 467.84 | 1935.74 | 1614.12 | 6445.67 | 5690.52 | 16497.3 | 15093.6 | 24702.5 | 23026 | 656.546 | 343.058 | 260.2 | 1053.7 | 907.536 |
| 656.835 | 596.783 | 469.986 | 1960.37 | 1620.53 | 6596.87 | 5709.62 | 16469.4 | 15137.5 | 24859.4 | 23088.3 | 656.835 | 345.108 | 261.467 | 1042.56 | 911.323 |
| 657.124 | 639.001 | 472.139 | 1925.22 | 1626.96 | 6496.37 | 5728.77 | 16527.9 | 15181.5 | 24789.2 | 23150.7 | 657.124 | 355.56 | 262.738 | 1075.49 | 915.121 |
| 657.412 | 618.707 | 474.3 | 1969.04 | 1633.4 | 6432.24 | 5747.96 | 16894.5 | 15225.5 | 24814.5 | 23213.2 | 657.412 | 328.921 | 264.015 | 1046.99 | 918.932 |
| 657.701 | 616.655 | 476.467 | 2041.27 | 1639.86 | 6569.77 | 5767.19 | 16610.3 | 15269.6 | 24990.5 | 23275.8 | 657.701 | 337.332 | 265.296 | 1105.19 | 922.753 |
| 657.99 | 572.327 | 478.643 | 1923.67 | 1646.34 | 6566.77 | 5786.46 | 16675.7 | 15313.8 | 25011.3 | 23338.5 | 657.99 | 360.407 | 266.581 | 1098.5 | 926.587 |
| 658.278 | 623.387 | 480.825 | 1933.28 | 1652.84 | 6701.17 | 5805.77 | 16617.4 | 15358 | 25412.3 | 23401.2 | 658.278 | 364.56 | 267.872 | 1128.41 | 930.431 |
| 658.567 | 645.403 | 483.016 | 2071.46 | 1659.36 | 6653.64 | 5825.12 | 17006.2 | 15402.4 | 25181.6 | 23464.1 | 658.567 | 345.488 | 269.167 | 1108.05 | 934.288 |
| 658.856 | 598.918 | 485.213 | 1922.55 | 1665.89 | 6663.39 | 5844.52 | 16913 | 15446.8 | 25449 | 23527 | 658.856 | 375.752 | 270.467 | 1105.57 | 938.156 |
| 659.144 | 627.114 | 487.418 | 2001.63 | 1672.44 | 6698.71 | 5863.95 | 17220.1 | 15491.3 | 25575.2 | 23590.1 | 659.144 | 354.694 | 271.772 | 1126.67 | 942.035 |
| 659.433 | 675.13 | 489.631 | 2044.86 | 1679.01 | 6789.57 | 5883.43 | 16827.2 | 15535.8 | 25461.6 | 23653.2 | 659.433 | 357.526 | 273.082 | 1092.55 | 945.926 |
| 659.722 | 639.54 | 491.851 | 2027.94 | 1685.6 | 6635.79 | 5902.95 | 16960.8 | 15580.5 | 25674 | 23716.4 | 659.722 | 368.126 | 274.396 | 1097.29 | 949.829 |
| 660.01 | 595.031 | 494.079 | 2030.65 | 1692.2 | 6728.45 | 5922.51 | 17175 | 15625.2 | 25995.8 | 23779.7 | 660.01 | 365.212 | 275.716 | 1133.06 | 953.744 |
| 660.299 | 644.646 | 496.314 | 2051.75 | 1698.83 | 6755.45 | 5942.11 | 16854.3 | 15669.9 | 25707.5 | 23843.1 | 660.299 | 361.75 | 277.04 | 1119.99 | 957.67 |
| 660.588 | 659.826 | 498.556 | 2070.17 | 1705.47 | 6816.73 | 5961.75 | 17333.3 | 15714.8 | 26215.1 | 23906.6 | 660.588 | 358.834 | 278.369 | 1106.18 | 961.608 |
| 660.876 | 639.329 | 500.806 | 2063.12 | 1712.13 | 6834.73 | 5981.43 | 17366.8 | 15759.7 | 26200.7 | 23970.2 | 660.876 | 364.961 | 279.703 | 1118.39 | 965.558 |
| 661.165 | 666.641 | 503.064 | 2009.99 | 1718.81 | 6863.07 | 6001.16 | 17158.7 | 15804.7 | 25868.2 | 24033.8 | 661.165 | 348.217 | 281.042 | 1105.55 | 969.519 |
| 661.453 | 626.398 | 505.329 | 2130.78 | 1725.5 | 7014.18 | 6020.92 | 17598.5 | 15849.8 | 26280.3 | 24097.6 | 661.453 | 362.787 | 282.386 | 1139.46 | 973.492 |
| 661.742 | 636.986 | 507.602 | 2083.49 | 1732.22 | 6925.78 | 6040.73 | 17122.6 | 15894.9 | 26027.9 | 24161.5 | 661.742 | 374.307 | 283.735 | 1120.64 | 977.477 |
| 662.03 | 656.004 | 509.883 | 2132.99 | 1738.95 | 6978.17 | 6060.58 | 17576.8 | 15940.1 | 26669.3 | 24225.4 | 662.03 | 392.059 | 285.089 | 1159.65 | 981.474 |
| 662.319 | 683.274 | 512.171 | 2072.71 | 1745.7 | 7025.61 | 6080.47 | 17650.4 | 15985.4 | 26418.6 | 24289.4 | 662.319 | 395.293 | 286.447 | 1149.84 | 985.482 |
| 662.607 | 655.53 | 514.467 | 2051.53 | 1752.47 | 6848.22 | 6100.4 | 17833.1 | 16030.8 | 26474.7 | 24353.5 | 662.607 | 360.404 | 287.811 | 1165.63 | 989.503 |
| 662.896 | 680.17 | 516.77 | 2102.16 | 1759.26 | 7126.14 | 6120.37 | 17931.5 | 16076.2 | 26692.3 | 24417.7 | 662.896 | 398.28 | 289.18 | 1184.6 | 993.535 |
| 663.184 | 708.083 | 519.082 | 2118.01 | 1766.07 | 7160.37 | 6140.39 | 17933.3 | 16121.7 | 26747.5 | 24482 | 663.184 | 379.494 | 290.553 | 1211.54 | 997.579 |
| 663.473 | 658.917 | 521.4 | 2268.29 | 1772.89 | 7215.16 | 6160.44 | 17867.9 | 16167.3 | 26626.1 | 24546.4 | 663.473 | 405.127 | 291.932 | 1190.89 | 1001.63 |
| 663.761 | 674.575 | 523.727 | 2105.12 | 1779.73 | 7095.55 | 6180.54 | 17938.9 | 16213 | 26719.6 | 24610.9 | 663.761 | 383.085 | 293.315 | 1215.56 | 1005.7 |
| 664.05 | 652.354 | 526.061 | 2113.17 | 1786.6 | 7102.86 | 6200.68 | 17772.2 | 16258.7 | 26700.9 | 24675.5 | 664.05 | 382.415 | 294.704 | 1219.89 | 1009.78 |
| 664.338 | 639.959 | 528.403 | 2151.18 | 1793.48 | 7087.57 | 6220.86 | 17963.5 | 16304.5 | 26780.2 | 24740.1 | 664.338 | 403.007 | 296.097 | 1227.03 | 1013.87 |
| 664.627 | 622.185 | 530.753 | 2115.38 | 1800.37 | 7063.73 | 6241.08 | 17991.7 | 16350.4 | 26807.2 | 24804.9 | 664.627 | 376.025 | 297.496 | 1154.1 | 1017.98 |
| 664.915 | 673.467 | 533.11 | 2131.66 | 1807.29 | 7101.53 | 6261.35 | 17924.2 | 16396.3 | 26945.8 | 24869.7 | 664.915 | 381.099 | 298.899 | 1195.87 | 1022.09 |
| 665.203 | 664.248 | 535.476 | 2198.69 | 1814.23 | 7210.89 | 6281.65 | 18198.6 | 16442.3 | 26918.3 | 24934.6 | 665.203 | 365.685 | 300.308 | 1186.71 | 1026.22 |
| 665.492 | 674.565 | 537.848 | 2185.55 | 1821.18 | 7091.09 | 6302 | 18072.3 | 16488.4 | 27324.2 | 24999.6 | 665.492 | 409.399 | 301.722 | 1213.47 | 1030.36 |
| 665.78 | 650.41 | 540.229 | 2173.42 | 1828.15 | 7102.36 | 6322.38 | 18142 | 16534.6 | 27014.5 | 25064.7 | 665.78 | 408.641 | 303.141 | 1217.9 | 1034.51 |
| 666.069 | 674.828 | 542.618 | 2156.15 | 1835.14 | 7177.81 | 6342.81 | 17986.2 | 16580.8 | 27092.4 | 25129.9 | 666.069 | 368.43 | 304.565 | 1219.27 | 1038.68 |
| 666.357 | 694.111 | 545.014 | 2251.57 | 1842.15 | 7194.4 | 6363.28 | 18117.3 | 16627.1 | 27346.1 | 25195.2 | 666.357 | 375.96 | 305.994 | 1207.13 | 1042.85 |
| 666.645 | 683.131 | 547.419 | 2185.53 | 1849.18 | 7265.31 | 6383.8 | 18143.1 | 16673.5 | 27188.3 | 25260.6 | 666.645 | 412.97 | 307.428 | 1198.26 | 1047.04 |
| 666.934 | 707.441 | 549.831 | 2248.41 | 1856.23 | 7451.9 | 6404.35 | 18464.3 | 16720 | 27449.7 | 25326 | 666.934 | 410.187 | 308.867 | 1220.95 | 1051.24 |
| 667.222 | 664.272 | 552.251 | 2190.52 | 1863.3 | 7235.77 | 6424.95 | 18201.7 | 16766.5 | 26917.6 | 25391.6 | 667.222 | 410.8 | 310.312 | 1211.79 | 1055.45 |
| 667.51 | 717.318 | 554.679 | 2173.23 | 1870.38 | 7259.56 | 6445.59 | 18223.6 | 16813.1 | 27188.7 | 25457.2 | 667.51 | 427.59 | 311.761 | 1212.32 | 1059.68 |
| 667.799 | 683.567 | 557.115 | 2232.05 | 1877.48 | 7367.55 | 6466.26 | 18319.7 | 16859.8 | 27650 | 25522.9 | 667.799 | 415.299 | 313.216 | 1248.63 | 1063.91 |
| 668.087 | 709.707 | 559.559 | 2218.38 | 1884.6 | 7380.45 | 6486.98 | 18455.6 | 16906.5 | 27581.9 | 25588.7 | 668.087 | 414.785 | 314.676 | 1290.16 | 1068.16 |
| 668.375 | 705.037 | 562.01 | 2310.8 | 1891.74 | 7514.78 | 6507.75 | 18568.2 | 16953.3 | 27712.2 | 25654.6 | 668.375 | 444.871 | 316.141 | 1229.89 | 1072.42 |
| 668.663 | 721.668 | 564.47 | 2219.13 | 1898.9 | 7477.41 | 6528.55 | 18815.1 | 17000.2 | 28034 | 25720.6 | 668.663 | 411.936 | 317.611 | 1256.73 | 1076.69 |
| 668.952 | 698.94 | 566.937 | 2309.78 | 1906.08 | 7427.95 | 6549.4 | 18514.5 | 17047.2 | 27826 | 25786.7 | 668.952 | 401.822 | 319.087 | 1262 | 1080.98 |
| 669.24 | 686.095 | 569.413 | 2265.54 | 1913.28 | 7461.19 | 6570.28 | 18768.3 | 17094.2 | 28193.4 | 25852.8 | 669.24 | 411.95 | 320.567 | 1263.01 | 1085.28 |
| 669.528 | 710.003 | 571.897 | 2274.79 | 1920.49 | 7595.62 | 6591.21 | 18744.9 | 17141.3 | 27745.5 | 25919.1 | 669.528 | 406.795 | 322.053 | 1226.89 | 1089.59 |
| 669.816 | 748.235 | 574.388 | 2325.91 | 1927.73 | 7425.34 | 6612.18 | 18787.3 | 17188.5 | 27996.2 | 25985.4 | 669.816 | 425.75 | 323.545 | 1284.17 | 1093.91 |
| 670.105 | 695.627 | 576.888 | 2279.42 | 1934.98 | 7368.46 | 6633.2 | 18714.4 | 17235.8 | 27847.4 | 26051.9 | 670.105 | 403.885 | 325.041 | 1258.25 | 1098.24 |
| 670.393 | 731.01 | 579.395 | 2243.91 | 1942.25 | 7414.67 | 6654.25 | 18842 | 17283.1 | 28117.9 | 26118.4 | 670.393 | 429.961 | 326.543 | 1270.03 | 1102.59 |
| 670.681 | 740.863 | 581.911 | 2363.88 | 1949.54 | 7504.36 | 6675.35 | 18967.7 | 17330.5 | 28419.7 | 26185 | 670.681 | 463.711 | 328.049 | 1277.49 | 1106.95 |
| 670.969 | 723.606 | 584.435 | 2309.05 | 1956.85 | 7566.24 | 6696.48 | 18904.9 | 17378 | 28475.4 | 26251.7 | 670.969 | 433.692 | 329.562 | 1299.3 | 1111.32 |
| 671.257 | 720.524 | 586.967 | 2248.83 | 1964.18 | 7636.8 | 6717.66 | 19105.9 | 17425.5 | 28339 | 26318.5 | 671.257 | 399.824 | 331.079 | 1267.35 | 1115.7 |
| 671.545 | 721.511 | 589.506 | 2447.58 | 1971.53 | 7608.27 | 6738.89 | 19206.6 | 17473.1 | 28388.4 | 26385.3 | 671.545 | 435.359 | 332.602 | 1315.03 | 1120.1 |
| 671.833 | 731.694 | 592.054 | 2338.07 | 1978.89 | 7735.07 | 6760.15 | 19363.7 | 17520.8 | 28878.7 | 26452.3 | 671.833 | 430.737 | 334.13 | 1304.3 | 1124.51 |
| 672.122 | 736.802 | 594.611 | 2311.16 | 1986.28 | 7672.41 | 6781.45 | 19107.7 | 17568.6 | 28938.9 | 26519.3 | 672.122 | 440.556 | 335.663 | 1319.25 | 1128.93 |
| 672.41 | 781.107 | 597.175 | 2398.15 | 1993.68 | 7568.51 | 6802.8 | 19162.3 | 17616.4 | 28674.2 | 26586.4 | 672.41 | 443.421 | 337.202 | 1342.23 | 1133.36 |
| 672.698 | 738.196 | 599.747 | 2301.51 | 2001.11 | 7823.25 | 6824.19 | 19486 | 17664.3 | 28735.9 | 26653.7 | 672.698 | 454.021 | 338.747 | 1317.03 | 1137.81 |
| 672.986 | 730.949 | 602.328 | 2336.62 | 2008.55 | 7827.98 | 6845.62 | 19460.1 | 17712.3 | 28815.1 | 26721 | 672.986 | 450.541 | 340.296 | 1337.32 | 1142.27 |
| 673.274 | 757.695 | 604.917 | 2355.41 | 2016.01 | 7778.93 | 6867.09 | 19422.5 | 17760.3 | 29180.4 | 26788.3 | 673.274 | 430.934 | 341.851 | 1334.74 | 1146.74 |
| 673.562 | 723.767 | 607.514 | 2425.7 | 2023.49 | 7771.94 | 6888.61 | 19187.6 | 17808.5 | 28988.1 | 26855.8 | 673.562 | 438.035 | 343.411 | 1302.44 | 1151.22 |
| 673.85 | 802.673 | 610.119 | 2378.04 | 2030.99 | 7864.76 | 6910.16 | 19479.2 | 17856.7 | 29113.8 | 26923.4 | 673.85 | 432.217 | 344.977 | 1290.17 | 1155.71 |
| 674.138 | 805.624 | 612.732 | 2356.76 | 2038.51 | 7815.04 | 6931.76 | 19546.8 | 17904.9 | 29007.2 | 26991 | 674.138 | 448.747 | 346.548 | 1349.87 | 1160.22 |
| 674.426 | 799.067 | 615.354 | 2469.99 | 2046.05 | 7805.65 | 6953.4 | 19709.5 | 17953.3 | 29204.8 | 27058.8 | 674.426 | 447.911 | 348.124 | 1364.66 | 1164.74 |
| 674.714 | 769.511 | 617.983 | 2514.85 | 2053.61 | 7898.51 | 6975.08 | 19648.3 | 18001.7 | 28991 | 27126.6 | 674.714 | 460.166 | 349.706 | 1313.17 | 1169.28 |
| 675.002 | 766.64 | 620.621 | 2504.69 | 2061.18 | 7941.79 | 6996.81 | 19819.2 | 18050.2 | 29306.6 | 27194.5 | 675.002 | 448.172 | 351.294 | 1411.61 | 1173.82 |
| 675.29 | 743.399 | 623.268 | 2439.23 | 2068.78 | 7853.13 | 7018.57 | 19590.6 | 18098.7 | 29179.8 | 27262.5 | 675.29 | 454.659 | 352.887 | 1372.82 | 1178.38 |
| 675.578 | 753.439 | 625.922 | 2421.78 | 2076.39 | 7973.35 | 7040.38 | 19694.3 | 18147.3 | 29273.6 | 27330.6 | 675.578 | 454.696 | 354.485 | 1323.63 | 1182.95 |
| 675.866 | 805.046 | 628.585 | 2459.74 | 2084.03 | 7828.34 | 7062.23 | 19740.5 | 18196 | 29177.3 | 27398.8 | 675.866 | 454.253 | 356.089 | 1316.66 | 1187.54 |
| 676.154 | 807.58 | 631.256 | 2452.33 | 2091.68 | 8032.76 | 7084.12 | 19955.1 | 18244.8 | 29767.7 | 27467 | 676.154 | 443.791 | 357.698 | 1363.45 | 1192.13 |
| 676.442 | 794.748 | 633.936 | 2473.84 | 2099.35 | 8020.74 | 7106.05 | 19821.3 | 18293.6 | 29572.3 | 27535.4 | 676.442 | 466.831 | 359.312 | 1369.14 | 1196.74 |
| 676.73 | 795.764 | 636.624 | 2477.2 | 2107.05 | 7989.89 | 7128.03 | 19916.9 | 18342.5 | 29695.4 | 27603.8 | 676.73 | 457.907 | 360.933 | 1392.32 | 1201.36 |
| 677.018 | 804.39 | 639.32 | 2495.34 | 2114.76 | 8045.93 | 7150.05 | 20010.2 | 18391.5 | 29633 | 27672.3 | 677.018 | 448.848 | 362.558 | 1451.17 | 1206 |
| 677.305 | 782.769 | 642.025 | 2446.71 | 2122.49 | 8000 | 7172.1 | 19945.7 | 18440.6 | 29512.1 | 27740.9 | 677.305 | 459.439 | 364.19 | 1397.64 | 1210.64 |
| 677.593 | 830.857 | 644.738 | 2480.39 | 2130.24 | 8218.12 | 7194.2 | 19785 | 18489.7 | 29825.7 | 27809.6 | 677.593 | 468.778 | 365.827 | 1347.65 | 1215.3 |
| 677.881 | 821.811 | 647.459 | 2522.81 | 2138.01 | 8162.26 | 7216.35 | 19973.5 | 18538.9 | 29559.1 | 27878.4 | 677.881 | 457.446 | 367.469 | 1376.49 | 1219.97 |
| 678.169 | 801.137 | 650.189 | 2513.01 | 2145.79 | 8141.91 | 7238.53 | 20108.1 | 18588.2 | 30108.6 | 27947.2 | 678.169 | 482.911 | 369.117 | 1395.1 | 1224.66 |
| 678.457 | 789.401 | 652.928 | 2521.86 | 2153.6 | 8213.57 | 7260.76 | 20273.7 | 18637.5 | 29987 | 28016.2 | 678.457 | 456.824 | 370.771 | 1440.05 | 1229.36 |
| 678.745 | 790.927 | 655.674 | 2501.1 | 2161.43 | 8233.44 | 7283.03 | 20034.7 | 18686.9 | 29994.7 | 28085.2 | 678.745 | 472.843 | 372.43 | 1417.07 | 1234.07 |
| 679.032 | 792.784 | 658.429 | 2507.61 | 2169.28 | 8195.72 | 7305.34 | 20280 | 18736.4 | 30263.4 | 28154.3 | 679.032 | 471.476 | 374.094 | 1414.21 | 1238.79 |
| 679.32 | 831.547 | 661.193 | 2492.81 | 2177.14 | 8158.1 | 7327.69 | 20161.5 | 18786 | 30111.6 | 28223.5 | 679.32 | 476.179 | 375.765 | 1440.72 | 1243.53 |
| 679.608 | 769.182 | 663.965 | 2604.16 | 2185.03 | 8137.06 | 7350.09 | 20516.1 | 18835.6 | 30086 | 28292.8 | 679.608 | 458.675 | 377.441 | 1443.89 | 1248.28 |
| 679.896 | 824.378 | 666.746 | 2520.59 | 2192.93 | 8236.6 | 7372.52 | 20547.5 | 18885.3 | 30323 | 28362.2 | 679.896 | 477.694 | 379.122 | 1445.3 | 1253.04 |
| 680.183 | 805.499 | 669.535 | 2595.41 | 2200.86 | 8412.66 | 7395 | 20585.6 | 18935 | 30431.2 | 28431.7 | 680.183 | 470.585 | 380.81 | 1464.18 | 1257.82 |
| 680.471 | 806.417 | 672.333 | 2653.23 | 2208.8 | 8366.93 | 7417.52 | 20997.3 | 18984.9 | 30755.2 | 28501.2 | 680.471 | 474.732 | 382.502 | 1458.35 | 1262.6 |
| 680.759 | 838.664 | 675.139 | 2598.32 | 2216.76 | 8172.04 | 7440.08 | 20673.4 | 19034.8 | 30511.6 | 28570.8 | 680.759 | 463.964 | 384.201 | 1445.42 | 1267.4 |
| 681.047 | 800.93 | 677.954 | 2589.62 | 2224.75 | 8202.38 | 7462.69 | 20684.3 | 19084.8 | 30380.6 | 28640.5 | 681.047 | 488.236 | 385.905 | 1415.57 | 1272.22 |
| 681.334 | 826.056 | 680.777 | 2661.06 | 2232.75 | 8389.06 | 7485.34 | 20483.9 | 19134.8 | 30473.2 | 28710.3 | 681.334 | 490.482 | 387.615 | 1459.5 | 1277.04 |
| 681.622 | 846.441 | 683.609 | 2560.25 | 2240.77 | 8342.34 | 7508.02 | 20747 | 19185 | 30808.7 | 28780.2 | 681.622 | 487.046 | 389.331 | 1428.08 | 1281.88 |
| 681.91 | 817.023 | 686.449 | 2679.39 | 2248.82 | 8526.89 | 7530.76 | 20874.7 | 19235.1 | 31127.7 | 28850.2 | 681.91 | 472.31 | 391.052 | 1499.39 | 1286.74 |
| 682.197 | 802.528 | 689.298 | 2651.52 | 2256.88 | 8507.3 | 7553.53 | 20820.3 | 19285.4 | 31000.8 | 28920.2 | 682.197 | 490.996 | 392.779 | 1444.35 | 1291.6 |
| 682.485 | 834.977 | 692.156 | 2616.59 | 2264.96 | 8388.21 | 7576.34 | 20744.7 | 19335.7 | 30978.5 | 28990.4 | 682.485 | 472.179 | 394.512 | 1440.48 | 1296.48 |
| 682.773 | 856.94 | 695.023 | 2602 | 2273.06 | 8445.83 | 7599.2 | 21153.7 | 19386.2 | 31105.3 | 29060.6 | 682.773 | 519.402 | 396.251 | 1444.56 | 1301.37 |
| 683.06 | 846.319 | 697.898 | 2714.08 | 2281.18 | 8605.4 | 7622.1 | 21233.9 | 19436.6 | 31349.3 | 29130.9 | 683.06 | 478.103 | 397.995 | 1477.31 | 1306.28 |
| 683.348 | 873.261 | 700.781 | 2696.42 | 2289.32 | 8578.2 | 7645.04 | 21327.2 | 19487.2 | 31722.9 | 29201.3 | 683.348 | 513.693 | 399.745 | 1509.41 | 1311.19 |
| 683.636 | 873.382 | 703.674 | 2656.84 | 2297.48 | 8656.28 | 7668.02 | 21326.2 | 19537.8 | 31718.7 | 29271.8 | 683.636 | 492.921 | 401.501 | 1537.5 | 1316.12 |
| 683.923 | 865.041 | 706.575 | 2656.4 | 2305.66 | 8623.42 | 7691.05 | 21491.5 | 19588.5 | 31438.1 | 29342.3 | 683.923 | 557.618 | 403.263 | 1556.93 | 1321.07 |
| 684.211 | 886.511 | 709.484 | 2718.06 | 2313.86 | 8772.3 | 7714.11 | 21558.6 | 19639.2 | 32151.3 | 29413 | 684.211 | 557.221 | 405.03 | 1585.16 | 1326.03 |
| 684.498 | 856.732 | 712.403 | 2755.27 | 2322.08 | 8786.74 | 7737.22 | 21318.6 | 19690.1 | 31501.5 | 29483.7 | 684.498 | 564.196 | 406.803 | 1580.42 | 1331 |
| 684.786 | 907.475 | 715.33 | 2728.02 | 2330.32 | 8743.9 | 7760.37 | 21446.4 | 19741 | 31700.8 | 29554.5 | 684.786 | 536.766 | 408.583 | 1578.08 | 1335.98 |
| 685.073 | 913.2 | 718.266 | 2775.19 | 2338.58 | 8787.58 | 7783.57 | 21658.4 | 19791.9 | 31906.9 | 29625.4 | 685.073 | 538.899 | 410.368 | 1593.28 | 1340.97 |
| 685.361 | 890.853 | 721.211 | 2724.73 | 2346.86 | 8774.29 | 7806.8 | 21521.4 | 19843 | 32035.2 | 29696.4 | 685.361 | 564.938 | 412.158 | 1588.84 | 1345.98 |
| 685.649 | 889.005 | 724.164 | 2731.75 | 2355.15 | 8934.83 | 7830.08 | 21795.7 | 19894.1 | 32054.1 | 29767.4 | 685.649 | 562.138 | 413.955 | 1589.85 | 1351.01 |
| 685.936 | 860.534 | 727.126 | 2794.92 | 2363.47 | 8862.19 | 7853.4 | 21710.9 | 19945.3 | 32243.1 | 29838.6 | 685.936 | 546.847 | 415.758 | 1596.8 | 1356.04 |
| 686.223 | 901.111 | 730.098 | 2805.71 | 2371.81 | 8901.76 | 7876.76 | 21745.8 | 19996.5 | 32179.7 | 29909.8 | 686.223 | 592.89 | 417.566 | 1626.25 | 1361.09 |
| 686.511 | 909.124 | 733.077 | 2767.79 | 2380.17 | 8831.27 | 7900.16 | 21753.1 | 20047.8 | 32392.9 | 29981.1 | 686.511 | 542.805 | 419.38 | 1589.92 | 1366.15 |
| 686.798 | 894.114 | 736.066 | 2784.72 | 2388.55 | 9092.69 | 7923.61 | 22066.8 | 20099.2 | 32262 | 30052.5 | 686.798 | 574.9 | 421.2 | 1682.14 | 1371.23 |
| 687.086 | 912.505 | 739.064 | 2808.19 | 2396.94 | 8931.18 | 7947.1 | 21947.9 | 20150.7 | 32393.9 | 30124 | 687.086 | 568.084 | 423.027 | 1628.44 | 1376.32 |
| 687.373 | 936.29 | 742.07 | 2842.75 | 2405.36 | 9019.79 | 7970.63 | 22174.2 | 20202.2 | 32750.2 | 30195.5 | 687.373 | 605.12 | 424.859 | 1726.18 | 1381.42 |
| 687.661 | 950.566 | 745.085 | 2773.4 | 2413.8 | 9077.61 | 7994.2 | 22214.8 | 20253.8 | 32932.8 | 30267.2 | 687.661 | 556.25 | 426.697 | 1595.57 | 1386.54 |
| 687.948 | 941.19 | 748.11 | 2872.67 | 2422.26 | 9025.62 | 8017.81 | 22176.3 | 20305.5 | 32731.2 | 30338.9 | 687.948 | 568.936 | 428.541 | 1642.2 | 1391.66 |
| 688.236 | 938.649 | 751.143 | 2822.41 | 2430.74 | 8929.7 | 8041.47 | 22053.1 | 20357.2 | 32740.7 | 30410.7 | 688.236 | 582.179 | 430.391 | 1693.54 | 1396.81 |
| 688.523 | 883.601 | 754.184 | 2835.56 | 2439.23 | 9028.34 | 8065.16 | 22357 | 20409 | 32604 | 30482.6 | 688.523 | 596.025 | 432.247 | 1717.62 | 1401.96 |
| 688.81 | 918.417 | 757.235 | 2924.43 | 2447.75 | 9094.13 | 8088.9 | 22239.3 | 20460.9 | 33081.2 | 30554.6 | 688.81 | 590.254 | 434.108 | 1706.43 | 1407.13 |
| 689.098 | 917.141 | 760.295 | 2868.05 | 2456.29 | 9314.47 | 8112.69 | 22568.1 | 20512.9 | 33191.7 | 30626.7 | 689.098 | 610.938 | 435.976 | 1688.66 | 1412.31 |
| 689.385 | 953.535 | 763.364 | 2970.79 | 2464.85 | 9166.37 | 8136.51 | 22387.7 | 20564.9 | 33177.3 | 30698.8 | 689.385 | 601.105 | 437.85 | 1681.98 | 1417.51 |
| 689.672 | 904.479 | 766.442 | 2892.38 | 2473.42 | 9213.42 | 8160.38 | 22564.5 | 20617 | 33107.2 | 30771 | 689.672 | 590.905 | 439.73 | 1729.15 | 1422.72 |
| 689.96 | 946.569 | 769.528 | 2871.41 | 2482.02 | 9232.56 | 8184.28 | 22402.7 | 20669.1 | 33159 | 30843.3 | 689.96 | 617.889 | 441.616 | 1685.76 | 1427.94 |
| 690.247 | 930.373 | 772.624 | 2913.32 | 2490.64 | 9199.44 | 8208.24 | 22604.3 | 20721.4 | 33849.4 | 30915.7 | 690.247 | 624.237 | 443.508 | 1767.94 | 1433.18 |
| 690.534 | 949.5 | 775.729 | 2920.73 | 2499.28 | 9453.36 | 8232.23 | 22735.5 | 20773.7 | 33624.3 | 30988.2 | 690.534 | 625.487 | 445.406 | 1720.72 | 1438.42 |
| 690.822 | 938.441 | 778.842 | 2927.3 | 2507.94 | 9252.61 | 8256.26 | 22466.6 | 20826 | 33534.9 | 31060.7 | 690.822 | 612.189 | 447.31 | 1723.67 | 1443.69 |
| 691.109 | 1028.13 | 781.965 | 2989.2 | 2516.61 | 9295.13 | 8280.34 | 22965.2 | 20878.5 | 33830 | 31133.4 | 691.109 | 619.145 | 449.22 | 1751.17 | 1448.96 |
| 691.396 | 924.555 | 785.097 | 2965.11 | 2525.31 | 9338.39 | 8304.46 | 22970 | 20931 | 33895.4 | 31206.1 | 691.396 | 614.033 | 451.136 | 1794.22 | 1454.25 |
| 691.683 | 1001.96 | 788.237 | 2923.5 | 2534.03 | 9267.03 | 8328.62 | 22778 | 20983.5 | 33527.1 | 31278.9 | 691.683 | 662.232 | 453.059 | 1755.25 | 1459.56 |
| 691.971 | 960.818 | 791.387 | 2947.7 | 2542.77 | 9331.29 | 8352.82 | 23234.5 | 21036.2 | 33897 | 31351.7 | 691.971 | 614.901 | 454.987 | 1749.32 | 1464.87 |
| 692.258 | 953.198 | 794.546 | 2993.24 | 2551.53 | 9399.81 | 8377.07 | 23103.4 | 21088.9 | 33981.3 | 31424.7 | 692.258 | 618.535 | 456.922 | 1772.12 | 1470.2 |
| 692.545 | 1005.31 | 797.714 | 2985.74 | 2560.31 | 9576.92 | 8401.35 | 22946 | 21141.7 | 34337.1 | 31497.7 | 692.545 | 639.694 | 458.862 | 1816.25 | 1475.55 |
| 692.832 | 971.149 | 800.892 | 2994.43 | 2569.11 | 9545.91 | 8425.68 | 23229.9 | 21194.5 | 34312.9 | 31570.9 | 692.832 | 635.133 | 460.809 | 1779.27 | 1480.9 |
| 693.119 | 991.367 | 804.078 | 2994.55 | 2577.93 | 9577.14 | 8450.05 | 22950.1 | 21247.5 | 34505.8 | 31644.1 | 693.119 | 653.621 | 462.762 | 1790.85 | 1486.27 |
| 693.407 | 964.062 | 807.273 | 3088.23 | 2586.77 | 9384.79 | 8474.46 | 23467.1 | 21300.5 | 34357.8 | 31717.3 | 693.407 | 629.709 | 464.721 | 1773.29 | 1491.66 |
| 693.694 | 1022.67 | 810.478 | 3013.11 | 2595.63 | 9591.44 | 8498.92 | 23360.8 | 21353.5 | 34270.8 | 31790.7 | 693.694 | 643.45 | 466.687 | 1826.53 | 1497.06 |
| 693.981 | 973.797 | 813.692 | 3118.96 | 2604.51 | 9382.75 | 8523.42 | 23205.3 | 21406.6 | 34441.4 | 31864.2 | 693.981 | 650.31 | 468.658 | 1781.78 | 1502.47 |
| 694.268 | 1043.44 | 816.914 | 2938.44 | 2613.41 | 9535.44 | 8547.95 | 23456.7 | 21459.8 | 34343.9 | 31937.7 | 694.268 | 656.745 | 470.636 | 1793.13 | 1507.89 |
| 694.555 | 991.113 | 820.146 | 3037.05 | 2622.33 | 9586.3 | 8572.54 | 23231.9 | 21513.1 | 34409.4 | 32011.3 | 694.555 | 655.477 | 472.62 | 1808.79 | 1513.33 |
| 694.842 | 965.352 | 823.388 | 3035.15 | 2631.27 | 9467.48 | 8597.16 | 23365.1 | 21566.4 | 34451 | 32085 | 694.842 | 647.706 | 474.61 | 1831.54 | 1518.78 |
| 695.129 | 976.549 | 826.638 | 2949.59 | 2640.24 | 9729.5 | 8621.82 | 23304.5 | 21619.8 | 34583.1 | 32158.8 | 695.129 | 648.292 | 476.606 | 1792.39 | 1524.25 |
| 695.416 | 1029.89 | 829.898 | 3001.95 | 2649.22 | 9603.76 | 8646.53 | 23461.7 | 21673.3 | 34602.7 | 32232.6 | 695.416 | 641.4 | 478.609 | 1800.1 | 1529.73 |
| 695.703 | 1039.57 | 833.167 | 3008.46 | 2658.22 | 9643.77 | 8671.28 | 23518.9 | 21726.8 | 34521.1 | 32306.5 | 695.703 | 668.794 | 480.618 | 1804.21 | 1535.22 |
| 695.99 | 1027.79 | 836.445 | 3089.5 | 2667.24 | 9788.39 | 8696.07 | 23847.8 | 21780.5 | 35103.3 | 32380.5 | 695.99 | 649.875 | 482.632 | 1821.56 | 1540.73 |
| 696.277 | 988.38 | 839.732 | 3096.63 | 2676.29 | 9846.78 | 8720.91 | 23959.8 | 21834.1 | 35124.9 | 32454.6 | 696.277 | 648.475 | 484.654 | 1872.33 | 1546.25 |
| 696.564 | 1042.28 | 843.029 | 3164.3 | 2685.35 | 9913.28 | 8745.78 | 23573.8 | 21887.9 | 35130.3 | 32528.8 | 696.564 | 666.135 | 486.682 | 1818.27 | 1551.79 |
| 696.851 | 1082.86 | 846.335 | 3104.5 | 2694.43 | 9803.26 | 8770.7 | 23948.2 | 21941.7 | 35239.4 | 32603 | 696.851 | 693.756 | 488.716 | 1818.52 | 1557.34 |
| 697.138 | 991.567 | 849.65 | 3203.8 | 2703.54 | 9836.13 | 8795.66 | 23835.6 | 21995.6 | 35097.2 | 32677.4 | 697.138 | 675.401 | 490.756 | 1849.34 | 1562.9 |
| 697.425 | 1040.94 | 852.975 | 3181.03 | 2712.66 | 9825.59 | 8820.66 | 24063.6 | 22049.5 | 35694.3 | 32751.8 | 697.425 | 690.605 | 492.802 | 1844.55 | 1568.47 |
| 697.712 | 1012.53 | 856.309 | 3141.5 | 2721.81 | 9951.69 | 8845.7 | 24026.4 | 22103.5 | 35552.8 | 32826.3 | 697.712 | 701.722 | 494.855 | 1926.5 | 1574.06 |
| 697.999 | 1037.51 | 859.652 | 3113.86 | 2730.98 | 9802.24 | 8870.79 | 24120.2 | 22157.6 | 35213.1 | 32900.8 | 697.999 | 696.121 | 496.914 | 1910.16 | 1579.67 |
| 698.286 | 1023.46 | 863.005 | 3192.63 | 2740.16 | 9909.88 | 8895.91 | 24062.7 | 22211.8 | 35658.9 | 32975.5 | 698.286 | 670.465 | 498.98 | 1940.94 | 1585.29 |
| 698.573 | 1107.49 | 866.367 | 3206.27 | 2749.37 | 9876.21 | 8921.08 | 24154.7 | 22266 | 35830 | 33050.2 | 698.573 | 677.09 | 501.052 | 1882.82 | 1590.92 |
| 698.86 | 983.37 | 869.739 | 3242.66 | 2758.6 | 10077.8 | 8946.29 | 24275.6 | 22320.3 | 35683.2 | 33125 | 698.86 | 685.949 | 503.13 | 1838.39 | 1596.56 |
| 699.147 | 1075.09 | 873.119 | 3167.64 | 2767.84 | 10040.1 | 8971.55 | 24228.4 | 22374.6 | 35949.5 | 33199.9 | 699.147 | 689.015 | 505.215 | 1898.26 | 1602.22 |
| 699.434 | 1060.28 | 876.51 | 3242.45 | 2777.11 | 10120.5 | 8996.84 | 24339.8 | 22429 | 35799 | 33274.9 | 699.434 | 683.549 | 507.306 | 1858.91 | 1607.89 |
| 699.721 | 1078.21 | 879.909 | 3223.44 | 2786.4 | 10083.8 | 9022.18 | 24460.8 | 22483.5 | 36153.9 | 33349.9 | 699.721 | 699.814 | 509.404 | 1955.05 | 1613.58 |
| 700.008 | 1030.21 | 883.318 | 3287.83 | 2795.71 | 10159.1 | 9047.56 | 24442.6 | 22538.1 | 35984.5 | 33425 | 700.008 | 712.276 | 511.508 | 1909.33 | 1619.28 |
| 700.294 | 1055.49 | 886.737 | 3249.26 | 2805.04 | 10224 | 9072.98 | 24462.8 | 22592.7 | 36413.5 | 33500.2 | 700.294 | 707.708 | 513.618 | 1881.88 | 1625 |
| 700.581 | 1095.66 | 890.164 | 3284.19 | 2814.39 | 10032 | 9098.44 | 24767.4 | 22647.4 | 36101 | 33575.5 | 700.581 | 699.1 | 515.735 | 1922.78 | 1630.73 |
| 700.868 | 1070.11 | 893.602 | 3301.86 | 2823.77 | 10233.2 | 9123.95 | 24974.4 | 22702.2 | 36682.1 | 33650.9 | 700.868 | 689.196 | 517.858 | 1987.61 | 1636.47 |
| 701.155 | 1061.24 | 897.049 | 3229.8 | 2833.16 | 10251.9 | 9149.49 | 24748.4 | 22757 | 36340.1 | 33726.3 | 701.155 | 696.815 | 519.988 | 1860.44 | 1642.23 |
| 701.442 | 1151.21 | 900.505 | 3302.14 | 2842.57 | 10270.2 | 9175.08 | 24823.8 | 22811.9 | 36473.7 | 33801.8 | 701.442 | 739.448 | 522.124 | 1950.47 | 1648 |
| 701.729 | 1120.91 | 903.971 | 3245.96 | 2852.01 | 10357.4 | 9200.71 | 25038.1 | 22866.9 | 36811.1 | 33877.4 | 701.729 | 701.042 | 524.267 | 1998.01 | 1653.78 |
| 702.015 | 1116.32 | 907.447 | 3354.55 | 2861.46 | 10397.8 | 9226.38 | 25004.3 | 22921.9 | 36690.4 | 33953.1 | 702.015 | 703.923 | 526.416 | 1965.92 | 1659.58 |
| 702.302 | 1082.29 | 910.931 | 3283.11 | 2870.93 | 10419.2 | 9252.1 | 24930.5 | 22977 | 36735 | 34028.8 | 702.302 | 691.952 | 528.572 | 1947.53 | 1665.39 |
| 702.589 | 1096.69 | 914.426 | 3345.47 | 2880.43 | 10310 | 9277.85 | 25044.5 | 23032.2 | 36903.6 | 34104.7 | 702.589 | 713.961 | 530.734 | 1945.72 | 1671.22 |
| 702.876 | 1130.42 | 917.93 | 3341.62 | 2889.95 | 10230.5 | 9303.65 | 24955 | 23087.4 | 37157.1 | 34180.6 | 702.876 | 736.225 | 532.902 | 1985.61 | 1677.06 |
| 703.162 | 1127.5 | 921.443 | 3364.05 | 2899.48 | 10558.6 | 9329.49 | 25224.1 | 23142.7 | 37016.2 | 34256.6 | 703.162 | 708.067 | 535.078 | 1937.32 | 1682.92 |
| 703.449 | 1150.14 | 924.967 | 3418.95 | 2909.04 | 10526 | 9355.37 | 25468.4 | 23198.1 | 37014.9 | 34332.6 | 703.449 | 721.025 | 537.26 | 2054.01 | 1688.79 |
| 703.736 | 1123.3 | 928.499 | 3345.91 | 2918.62 | 10523 | 9381.3 | 25340.5 | 23253.5 | 36991.3 | 34408.8 | 703.736 | 667.066 | 539.448 | 2004.66 | 1694.67 |
| 704.022 | 1132.64 | 932.042 | 3337.08 | 2928.22 | 10414.1 | 9407.26 | 25411.4 | 23309 | 37315 | 34485 | 704.022 | 721.782 | 541.643 | 1959.64 | 1700.57 |
| 704.309 | 1144.86 | 935.594 | 3364.98 | 2937.84 | 10659.8 | 9433.27 | 25559.5 | 23364.6 | 37382.1 | 34561.2 | 704.309 | 710.034 | 543.844 | 2013.23 | 1706.48 |
| 704.596 | 1161.54 | 939.155 | 3437.12 | 2947.48 | 10772.2 | 9459.32 | 25596.7 | 23420.2 | 37635.8 | 34637.6 | 704.596 | 699.696 | 546.052 | 1941.33 | 1712.4 |
| 704.882 | 1159.64 | 942.727 | 3439.24 | 2957.14 | 10675.9 | 9485.41 | 25709.4 | 23475.9 | 37509.4 | 34714.1 | 704.882 | 715.501 | 548.267 | 2031.59 | 1718.34 |
| 705.169 | 1138 | 946.308 | 3476.6 | 2966.82 | 10732.5 | 9511.54 | 25967.2 | 23531.7 | 37713.5 | 34790.6 | 705.169 | 695.117 | 550.488 | 2055.12 | 1724.3 |
| 705.456 | 1188.17 | 949.898 | 3438.11 | 2976.53 | 10723.5 | 9537.72 | 25671.6 | 23587.5 | 38214.6 | 34867.2 | 705.456 | 705.996 | 552.716 | 1960.86 | 1730.27 |
| 705.742 | 1211.74 | 953.498 | 3501.55 | 2986.25 | 10646.2 | 9563.93 | 25793 | 23643.4 | 38153.9 | 34943.9 | 705.742 | 699.692 | 554.951 | 2050.86 | 1736.25 |
| 706.029 | 1187.04 | 957.108 | 3540.86 | 2996 | 10685.5 | 9590.19 | 25759.4 | 23699.4 | 37999.6 | 35020.6 | 706.029 | 697.057 | 557.192 | 2001.39 | 1742.24 |
| 706.315 | 1214.99 | 960.728 | 3513.29 | 3005.76 | 10728.9 | 9616.49 | 25894.2 | 23755.4 | 38094.9 | 35097.4 | 706.315 | 716.995 | 559.44 | 1951.92 | 1748.25 |
| 706.602 | 1203.82 | 964.357 | 3504.39 | 3015.55 | 10784.3 | 9642.83 | 25916.5 | 23811.5 | 38315 | 35174.4 | 706.602 | 703.325 | 561.695 | 2049.56 | 1754.28 |
| 706.888 | 1160.82 | 967.996 | 3601.24 | 3025.36 | 10851.2 | 9669.22 | 26109.7 | 23867.7 | 38335.7 | 35251.3 | 706.888 | 716.027 | 563.956 | 2036 | 1760.32 |
| 707.175 | 1178.8 | 971.645 | 3452 | 3035.19 | 10852.2 | 9695.64 | 26082.7 | 23923.9 | 38060.6 | 35328.4 | 707.175 | 701.415 | 566.224 | 2069.29 | 1766.37 |
| 707.462 | 1186.15 | 975.304 | 3576.01 | 3045.04 | 10978.5 | 9722.11 | 26639.2 | 23980.2 | 38561.9 | 35405.5 | 707.462 | 745.89 | 568.498 | 2013.48 | 1772.44 |
| 707.748 | 1190.72 | 978.972 | 3515.62 | 3054.91 | 10814.4 | 9748.62 | 26256.9 | 24036.6 | 38530.4 | 35482.8 | 707.748 | 706.014 | 570.78 | 2052.45 | 1778.52 |
| 708.034 | 1178.95 | 982.65 | 3554.21 | 3064.8 | 10915.1 | 9775.17 | 26014.7 | 24093 | 38520.1 | 35560 | 708.034 | 744.804 | 573.068 | 2004.31 | 1784.62 |
| 708.321 | 1173.94 | 986.338 | 3563.91 | 3074.71 | 10921.9 | 9801.76 | 26145 | 24149.5 | 38185.4 | 35637.4 | 708.321 | 699.403 | 575.363 | 2048.17 | 1790.73 |
| 708.607 | 1180.44 | 990.036 | 3558.43 | 3084.64 | 11232.4 | 9828.39 | 26556.8 | 24206 | 38742.6 | 35714.8 | 708.607 | 766.664 | 577.664 | 2095.68 | 1796.85 |
| 708.894 | 1236.33 | 993.743 | 3542.97 | 3094.6 | 10945.3 | 9855.07 | 26360.8 | 24262.7 | 38177.2 | 35792.3 | 708.894 | 744.767 | 579.972 | 2037.47 | 1802.99 |
| 709.18 | 1190.62 | 997.461 | 3572.98 | 3104.57 | 10944.1 | 9881.78 | 26309.2 | 24319.4 | 38468.9 | 35869.9 | 709.18 | 723.22 | 582.288 | 2036.73 | 1809.15 |
| 709.467 | 1217.7 | 1001.19 | 3621.28 | 3114.57 | 11125.2 | 9908.54 | 26515.4 | 24376.1 | 38938.8 | 35947.6 | 709.467 | 757.901 | 584.609 | 2084.1 | 1815.32 |
| 709.753 | 1201.54 | 1004.92 | 3597.27 | 3124.59 | 11168.8 | 9935.34 | 26402.5 | 24433 | 39172.4 | 36025.3 | 709.753 | 751.802 | 586.938 | 2106.46 | 1821.5 |
| 710.04 | 1214.12 | 1008.67 | 3609.47 | 3134.62 | 11145.9 | 9962.19 | 26843.4 | 24489.8 | 39304.1 | 36103.2 | 710.04 | 753.529 | 589.273 | 2102.99 | 1827.69 |
| 710.326 | 1227.51 | 1012.43 | 3627.93 | 3144.68 | 11047.4 | 9989.07 | 26471.9 | 24546.8 | 39380.7 | 36181.1 | 710.326 | 749.745 | 591.616 | 2113.59 | 1833.9 |
| 710.612 | 1204.11 | 1016.2 | 3689.56 | 3154.76 | 11294.5 | 10016 | 26817.5 | 24603.8 | 39352.1 | 36259 | 710.612 | 747.81 | 593.965 | 2154.77 | 1840.13 |
| 710.899 | 1248.98 | 1019.97 | 3525.56 | 3164.86 | 10980.1 | 10043 | 26614 | 24660.9 | 38995.7 | 36337.1 | 710.899 | 725.678 | 596.321 | 2172.35 | 1846.37 |
| 711.185 | 1197.1 | 1023.76 | 3665.41 | 3174.99 | 11255.6 | 10070 | 27280.6 | 24718 | 39461.8 | 36415.2 | 711.185 | 759.25 | 598.683 | 2135.21 | 1852.62 |
| 711.471 | 1187.74 | 1027.56 | 3697.15 | 3185.13 | 11323.9 | 10097 | 27207.3 | 24775.3 | 39847.4 | 36493.3 | 711.471 | 749.598 | 601.053 | 2142.28 | 1858.89 |
| 711.758 | 1248.73 | 1031.36 | 3634.8 | 3195.29 | 11271.9 | 10124.1 | 27058.8 | 24832.5 | 39346 | 36571.6 | 711.758 | 759.963 | 603.429 | 2160.67 | 1865.18 |
| 712.044 | 1270.7 | 1035.18 | 3748.31 | 3205.48 | 11334.8 | 10151.2 | 27047.6 | 24889.9 | 39841.2 | 36649.9 | 712.044 | 819.809 | 605.812 | 2208.29 | 1871.47 |
| 712.33 | 1251.78 | 1039 | 3687.93 | 3215.68 | 11410.3 | 10178.4 | 27113.4 | 24947.3 | 39851.3 | 36728.4 | 712.33 | 755.894 | 608.203 | 2153.78 | 1877.79 |
| 712.616 | 1273.59 | 1042.84 | 3730.31 | 3225.91 | 11508 | 10205.6 | 27534.3 | 25004.8 | 40036.3 | 36806.8 | 712.616 | 796.039 | 610.6 | 2184.11 | 1884.12 |
| 712.903 | 1270.15 | 1046.68 | 3806.72 | 3236.16 | 11470.6 | 10232.9 | 27590.9 | 25062.3 | 40162 | 36885.4 | 712.903 | 755.894 | 613.004 | 2152.67 | 1890.46 |
| 713.189 | 1276.87 | 1050.54 | 3766.51 | 3246.43 | 11480.1 | 10260.2 | 27454.8 | 25119.9 | 39956.9 | 36964.1 | 713.189 | 777.793 | 615.415 | 2221.19 | 1896.81 |
| 713.475 | 1226.21 | 1054.41 | 3742.97 | 3256.72 | 11615 | 10287.5 | 27613.7 | 25177.6 | 40094.1 | 37042.8 | 713.475 | 756.361 | 617.832 | 2187.88 | 1903.18 |
| 713.761 | 1276.47 | 1058.28 | 3723.58 | 3267.03 | 11697.7 | 10314.9 | 27467.2 | 25235.3 | 40217.5 | 37121.5 | 713.761 | 786.679 | 620.257 | 2181.41 | 1909.57 |
| 714.048 | 1239.6 | 1062.17 | 3699.67 | 3277.37 | 11458.5 | 10342.3 | 27637.6 | 25293.1 | 40087.1 | 37200.4 | 714.048 | 783.979 | 622.689 | 2171.01 | 1915.97 |
| 714.334 | 1244.87 | 1066.07 | 3866.2 | 3287.72 | 11542.3 | 10369.8 | 27664.7 | 25351 | 40080.1 | 37279.3 | 714.334 | 774.712 | 625.128 | 2112.64 | 1922.38 |
| 714.62 | 1313.47 | 1069.97 | 3795.07 | 3298.1 | 11502.6 | 10397.3 | 27463.5 | 25408.9 | 40230.5 | 37358.3 | 714.62 | 792.726 | 627.573 | 2174.07 | 1928.81 |
| 714.906 | 1311.07 | 1073.89 | 3789.82 | 3308.49 | 11666.9 | 10424.9 | 27406.5 | 25466.9 | 40513.9 | 37437.4 | 714.906 | 783.808 | 630.026 | 2169.07 | 1935.25 |
| 715.192 | 1294.58 | 1077.82 | 3757.26 | 3318.91 | 11618.7 | 10452.5 | 27495.1 | 25525 | 40681 | 37516.5 | 715.192 | 803.42 | 632.485 | 2252 | 1941.71 |
| 715.479 | 1326.73 | 1081.75 | 3785.81 | 3329.35 | 11674.1 | 10480.1 | 27850.8 | 25583.1 | 40448.2 | 37595.8 | 715.479 | 823.836 | 634.952 | 2227.45 | 1948.18 |
| 715.765 | 1310.17 | 1085.7 | 3885.89 | 3339.81 | 11771.7 | 10507.8 | 27760.5 | 25641.3 | 40629.9 | 37675.1 | 715.765 | 790.252 | 637.426 | 2195.78 | 1954.67 |
| 716.051 | 1315.55 | 1089.65 | 3893.15 | 3350.29 | 11725.2 | 10535.5 | 27984.2 | 25699.5 | 40883.5 | 37754.4 | 716.051 | 832.386 | 639.906 | 2315.43 | 1961.17 |
| 716.337 | 1303.88 | 1093.62 | 3863.68 | 3360.8 | 11609.1 | 10563.2 | 27880.5 | 25757.8 | 40483.2 | 37833.9 | 716.337 | 793.886 | 642.394 | 2279.92 | 1967.69 |
| 716.623 | 1325.38 | 1097.6 | 3925.87 | 3371.32 | 11692.8 | 10591 | 27616.7 | 25816.2 | 40524.1 | 37913.4 | 716.623 | 779.516 | 644.888 | 2223.89 | 1974.22 |
| 716.909 | 1255.28 | 1101.59 | 3813.79 | 3381.87 | 11699.8 | 10618.9 | 27984.7 | 25874.6 | 40769.9 | 37993 | 716.909 | 805.287 | 647.39 | 2304.15 | 1980.77 |
| 717.195 | 1275.18 | 1105.58 | 3826.76 | 3392.43 | 11655.7 | 10646.7 | 27919.4 | 25933.1 | 40541.2 | 38072.6 | 717.195 | 815.357 | 649.899 | 2266.07 | 1987.33 |
| 717.481 | 1309.28 | 1109.59 | 3870.09 | 3403.02 | 11865.8 | 10674.6 | 27987.1 | 25991.7 | 41130.6 | 38152.3 | 717.481 | 804.737 | 652.415 | 2345.31 | 1993.9 |
| 717.767 | 1316.29 | 1113.61 | 3855.37 | 3413.63 | 11803.3 | 10702.6 | 28119 | 26050.3 | 41127.7 | 38232.2 | 717.767 | 831.388 | 654.938 | 2210.12 | 2000.49 |
| 718.053 | 1331.12 | 1117.63 | 3929.79 | 3424.26 | 11810.9 | 10730.6 | 28292.3 | 26109 | 40964 | 38312 | 718.053 | 828.602 | 657.468 | 2312.65 | 2007.1 |
| 718.339 | 1366.06 | 1121.67 | 3941.43 | 3434.91 | 11970.6 | 10758.7 | 28535.2 | 26167.8 | 41114.2 | 38392 | 718.339 | 831.417 | 660.005 | 2219.82 | 2013.72 |
| 718.625 | 1396.44 | 1125.72 | 3975.49 | 3445.59 | 11936.3 | 10786.8 | 28389.3 | 26226.6 | 41499.2 | 38472 | 718.625 | 833.489 | 662.549 | 2318.74 | 2020.35 |
| 718.911 | 1357.59 | 1129.78 | 4005.86 | 3456.28 | 11979 | 10814.9 | 28596.3 | 26285.5 | 41726.5 | 38552.1 | 718.911 | 815.614 | 665.101 | 2308.08 | 2027 |
| 719.197 | 1312.37 | 1133.85 | 3965.97 | 3466.99 | 12008.3 | 10843 | 28547.8 | 26344.4 | 41155.4 | 38632.2 | 719.197 | 807.764 | 667.659 | 2269.37 | 2033.66 |
| 719.483 | 1381.6 | 1137.93 | 3926.29 | 3477.73 | 11974.2 | 10871.2 | 28481.4 | 26403.5 | 41608.9 | 38712.5 | 719.483 | 830.602 | 670.225 | 2283.34 | 2040.34 |
| 719.769 | 1356.86 | 1142.02 | 3942.63 | 3488.49 | 11996 | 10899.5 | 28865.4 | 26462.5 | 41305.1 | 38792.8 | 719.769 | 869.837 | 672.798 | 2318.82 | 2047.04 |
| 720.055 | 1371.77 | 1146.12 | 3952.57 | 3499.27 | 12057.6 | 10927.8 | 28672 | 26521.7 | 41768.1 | 38873.1 | 720.055 | 853.01 | 675.378 | 2304.5 | 2053.74 |
| 720.341 | 1339.73 | 1150.22 | 3984.02 | 3510.07 | 12081.7 | 10956.1 | 28945 | 26580.9 | 42154 | 38953.6 | 720.341 | 842.438 | 677.965 | 2358.36 | 2060.47 |
| 720.627 | 1330.78 | 1154.34 | 4052.26 | 3520.89 | 12077.4 | 10984.5 | 29185.8 | 26640.1 | 41993.4 | 39034.1 | 720.627 | 833.956 | 680.559 | 2353.29 | 2067.2 |
| 720.913 | 1424.99 | 1158.47 | 4089.69 | 3531.74 | 12039.5 | 11012.9 | 28933.2 | 26699.5 | 41803.1 | 39114.7 | 720.913 | 827.757 | 683.161 | 2380.58 | 2073.96 |
| 721.199 | 1305.16 | 1162.61 | 4107.2 | 3542.6 | 12298.9 | 11041.4 | 28966.3 | 26758.8 | 41901.3 | 39195.3 | 721.199 | 830.317 | 685.77 | 2401.32 | 2080.73 |
| 721.485 | 1411.84 | 1166.77 | 4050.03 | 3553.49 | 12416.2 | 11069.9 | 28902.8 | 26818.3 | 42018.7 | 39276.1 | 721.485 | 819.974 | 688.385 | 2309.57 | 2087.51 |
| 721.77 | 1339.71 | 1170.93 | 3956 | 3564.4 | 12445.1 | 11098.4 | 28911.2 | 26877.8 | 42147 | 39356.9 | 721.77 | 824.24 | 691.009 | 2340.49 | 2094.31 |
| 722.056 | 1405.25 | 1175.1 | 4147.26 | 3575.32 | 12261 | 11127 | 28915.5 | 26937.4 | 42495.7 | 39437.7 | 722.056 | 847.723 | 693.639 | 2408.57 | 2101.12 |
| 722.342 | 1395.77 | 1179.28 | 4047.95 | 3586.28 | 12247.2 | 11155.6 | 29029.6 | 26997 | 42270.2 | 39518.7 | 722.342 | 833.024 | 696.277 | 2357.35 | 2107.95 |
| 722.628 | 1403.84 | 1183.47 | 4174.67 | 3597.25 | 12251.5 | 11184.3 | 29072.9 | 27056.7 | 42212.6 | 39599.7 | 722.628 | 807.192 | 698.922 | 2362.93 | 2114.79 |
| 722.913 | 1355.56 | 1187.67 | 4099.6 | 3608.24 | 12431.8 | 11213 | 29036.1 | 27116.5 | 42264 | 39680.8 | 722.913 | 838.512 | 701.574 | 2349.41 | 2121.64 |
| 723.199 | 1409.28 | 1191.89 | 4140.35 | 3619.26 | 12450.9 | 11241.7 | 29580.3 | 27176.3 | 42396.9 | 39761.9 | 723.199 | 841.083 | 704.234 | 2460.2 | 2128.52 |
| 723.485 | 1359.71 | 1196.11 | 4143.45 | 3630.29 | 12365.6 | 11270.5 | 28916.9 | 27236.2 | 42435.4 | 39843.1 | 723.485 | 861.48 | 706.9 | 2408.39 | 2135.4 |
| 723.771 | 1371.12 | 1200.34 | 4102.81 | 3641.35 | 12454.3 | 11299.3 | 29134.1 | 27296.1 | 42400.3 | 39924.4 | 723.771 | 840.106 | 709.574 | 2360.24 | 2142.3 |
| 724.057 | 1434.1 | 1204.59 | 4181.2 | 3652.43 | 12429.2 | 11328.2 | 29278.9 | 27356.2 | 42469.6 | 40005.8 | 724.057 | 900.056 | 712.256 | 2429.67 | 2149.22 |
| 724.342 | 1455.37 | 1208.84 | 4194.28 | 3663.53 | 12299 | 11357.1 | 29403.1 | 27416.2 | 42737 | 40087.2 | 724.342 | 855.401 | 714.945 | 2409.06 | 2156.15 |
| 724.628 | 1423.01 | 1213.11 | 4112.03 | 3674.65 | 12520.4 | 11386 | 29424.1 | 27476.4 | 42932.6 | 40168.7 | 724.628 | 888.868 | 717.641 | 2447.27 | 2163.1 |
| 724.914 | 1406.76 | 1217.38 | 4213.66 | 3685.8 | 12522.2 | 11415 | 29543.9 | 27536.5 | 42568.8 | 40250.3 | 724.914 | 867.218 | 720.344 | 2468.84 | 2170.06 |
| 725.199 | 1380.25 | 1221.67 | 4168.15 | 3696.96 | 12462.2 | 11444 | 29150 | 27596.8 | 42346.2 | 40331.9 | 725.199 | 871.784 | 723.055 | 2409.45 | 2177.04 |
| 725.485 | 1426.67 | 1225.96 | 4213.65 | 3708.15 | 12598.2 | 11473.1 | 29471.5 | 27657.1 | 42893.9 | 40413.6 | 725.485 | 872.121 | 725.773 | 2380.93 | 2184.03 |
| 725.771 | 1409.46 | 1230.27 | 4231.46 | 3719.36 | 12513 | 11502.2 | 29260.2 | 27717.5 | 42644.8 | 40495.4 | 725.771 | 883.298 | 728.498 | 2476.86 | 2191.04 |
| 726.056 | 1447.43 | 1234.59 | 4270.12 | 3730.59 | 12506.3 | 11531.4 | 29474.9 | 27778 | 42974.7 | 40577.2 | 726.056 | 870.997 | 731.231 | 2443.8 | 2198.06 |
| 726.342 | 1412.02 | 1238.92 | 4175 | 3741.84 | 12708.7 | 11560.6 | 29753.7 | 27838.5 | 43159.8 | 40659.1 | 726.342 | 874.346 | 733.971 | 2481.95 | 2205.09 |
| 726.628 | 1462.23 | 1243.25 | 4187.65 | 3753.11 | 12666.6 | 11589.8 | 29738 | 27899 | 43109.8 | 40741.1 | 726.628 | 888.446 | 736.719 | 2471.46 | 2212.15 |
| 726.913 | 1468.32 | 1247.6 | 4217.34 | 3764.41 | 12666.1 | 11619.1 | 30069 | 27959.7 | 42865.4 | 40823.2 | 726.913 | 877.852 | 739.474 | 2496.71 | 2219.21 |
| 727.199 | 1423.38 | 1251.96 | 4194.17 | 3775.72 | 12561.7 | 11648.4 | 29649.2 | 28020.3 | 43266.8 | 40905.3 | 727.199 | 869.73 | 742.237 | 2487.91 | 2226.3 |
| 727.484 | 1443.42 | 1256.33 | 4219.27 | 3787.06 | 12694.8 | 11677.7 | 29790.9 | 28081.1 | 43304.7 | 40987.4 | 727.484 | 920.409 | 745.006 | 2508.09 | 2233.39 |
| 727.77 | 1441.94 | 1260.71 | 4300.4 | 3798.42 | 12652.2 | 11707.1 | 30129.5 | 28141.9 | 43236.7 | 41069.7 | 727.77 | 869.855 | 747.784 | 2461.81 | 2240.51 |
| 728.055 | 1570.35 | 1265.1 | 4236.48 | 3809.8 | 12869.5 | 11736.6 | 30255.5 | 28202.8 | 43686 | 41152 | 728.055 | 895.798 | 750.569 | 2486.3 | 2247.63 |
| 728.341 | 1457.35 | 1269.51 | 4253.56 | 3821.2 | 12846.5 | 11766 | 29971.6 | 28263.7 | 43366.1 | 41234.4 | 728.341 | 922.307 | 753.361 | 2454.91 | 2254.78 |
| 728.626 | 1501.76 | 1273.92 | 4321.71 | 3832.62 | 12655.4 | 11795.6 | 29855.6 | 28324.7 | 43524.2 | 41316.9 | 728.626 | 911.371 | 756.161 | 2465.99 | 2261.93 |
| 728.912 | 1543.26 | 1278.34 | 4359.72 | 3844.07 | 12810.8 | 11825.1 | 30085 | 28385.8 | 43702.6 | 41399.4 | 728.912 | 873.393 | 758.968 | 2466.99 | 2269.11 |
| 729.197 | 1513.09 | 1282.77 | 4308.52 | 3855.53 | 12852.8 | 11854.7 | 30057.1 | 28446.9 | 43871.4 | 41482 | 729.197 | 887.405 | 761.783 | 2584.67 | 2276.29 |
| 729.483 | 1513.92 | 1287.22 | 4303.03 | 3867.02 | 12814.6 | 11884.4 | 30398.3 | 28508 | 43562.4 | 41564.6 | 729.483 | 889.286 | 764.605 | 2516.62 | 2283.5 |
| 729.768 | 1509.33 | 1291.67 | 4275.24 | 3878.53 | 12944.6 | 11914 | 30387.7 | 28569.3 | 43582 | 41647.3 | 729.768 | 882.868 | 767.434 | 2544.3 | 2290.72 |
| 730.054 | 1479.27 | 1296.14 | 4336.16 | 3890.06 | 12927.8 | 11943.8 | 30545.3 | 28630.6 | 43505.5 | 41730.1 | 730.054 | 938.381 | 770.272 | 2493 | 2297.95 |
| 730.339 | 1517.59 | 1300.61 | 4235.3 | 3901.62 | 12807.5 | 11973.5 | 30291.2 | 28691.9 | 43860.2 | 41813 | 730.339 | 944.762 | 773.116 | 2540.16 | 2305.2 |
| 730.625 | 1582.98 | 1305.1 | 4363.06 | 3913.19 | 13031.3 | 12003.3 | 30295.5 | 28753.3 | 44127.3 | 41895.9 | 730.625 | 932.066 | 775.969 | 2583.52 | 2312.46 |
| 730.91 | 1566.36 | 1309.6 | 4401.73 | 3924.79 | 13062.8 | 12033.2 | 30631.8 | 28814.8 | 43705.3 | 41978.9 | 730.91 | 888.592 | 778.829 | 2510.21 | 2319.74 |
| 731.195 | 1559.61 | 1314.1 | 4349.74 | 3936.41 | 13026.6 | 12063 | 30414.8 | 28876.3 | 43940.9 | 42062 | 731.195 | 924.005 | 781.696 | 2521.53 | 2327.03 |
| 731.481 | 1468.75 | 1318.62 | 4400.78 | 3948.05 | 12912.7 | 12093 | 30424.9 | 28938 | 43639.8 | 42145.1 | 731.481 | 935.431 | 784.571 | 2513.09 | 2334.34 |
| 731.766 | 1549.84 | 1323.15 | 4384.1 | 3959.71 | 13025.8 | 12122.9 | 30447 | 28999.6 | 44009.6 | 42228.3 | 731.766 | 928.147 | 787.453 | 2506.96 | 2341.67 |
| 732.052 | 1544.94 | 1327.69 | 4309.81 | 3971.39 | 13219.6 | 12152.9 | 30597.4 | 29061.3 | 44544.4 | 42311.5 | 732.052 | 949.79 | 790.344 | 2514.62 | 2349.01 |
| 732.337 | 1554.78 | 1332.24 | 4363.82 | 3983.09 | 12889.7 | 12183 | 30233.8 | 29123.1 | 43958.4 | 42394.9 | 732.337 | 964.743 | 793.241 | 2582.43 | 2356.36 |
| 732.622 | 1494.21 | 1336.8 | 4350.34 | 3994.82 | 13007.7 | 12213.1 | 30639.9 | 29184.9 | 44265.8 | 42478.2 | 732.622 | 890.979 | 796.147 | 2660.69 | 2363.74 |
| 732.908 | 1518.82 | 1341.37 | 4332.49 | 4006.57 | 12876.5 | 12243.2 | 30356 | 29246.8 | 44002.8 | 42561.7 | 732.908 | 915.101 | 799.06 | 2600.55 | 2371.12 |
| 733.193 | 1535.78 | 1345.96 | 4344.31 | 4018.34 | 13199.4 | 12273.4 | 30354.7 | 29308.8 | 44223.7 | 42645.2 | 733.193 | 948.106 | 801.98 | 2522.34 | 2378.52 |
| 733.478 | 1597.73 | 1350.55 | 4438.71 | 4030.13 | 13296.7 | 12303.5 | 30974.7 | 29370.8 | 45026.3 | 42728.8 | 733.478 | 916.99 | 804.909 | 2550.07 | 2385.94 |
| 733.763 | 1603.86 | 1355.15 | 4531.79 | 4041.94 | 13249.4 | 12333.8 | 30827.5 | 29432.9 | 44242 | 42812.5 | 733.763 | 975.132 | 807.845 | 2561.08 | 2393.37 |
| 734.048 | 1563.27 | 1359.77 | 4498.46 | 4053.78 | 13145.2 | 12364.1 | 31001 | 29495 | 44897.3 | 42896.2 | 734.048 | 951.141 | 810.788 | 2668.58 | 2400.82 |
| 734.334 | 1555.63 | 1364.39 | 4544.21 | 4065.63 | 13293 | 12394.4 | 30971.7 | 29557.2 | 44337.3 | 42979.9 | 734.334 | 938.125 | 813.739 | 2652.41 | 2408.28 |
| 734.619 | 1563.47 | 1369.03 | 4506.12 | 4077.51 | 13113 | 12424.8 | 31128.1 | 29619.4 | 44632.7 | 43063.8 | 734.619 | 946.907 | 816.698 | 2599.13 | 2415.76 |
| 734.904 | 1579.92 | 1373.68 | 4590.07 | 4089.41 | 13529 | 12455.2 | 31190.2 | 29681.7 | 45449.9 | 43147.7 | 734.904 | 956.871 | 819.665 | 2636.47 | 2423.25 |
| 735.189 | 1527.36 | 1378.34 | 4550.38 | 4101.33 | 13563.4 | 12485.6 | 31195.5 | 29744.1 | 44857.8 | 43231.7 | 735.189 | 942.82 | 822.639 | 2690.6 | 2430.76 |
| 735.475 | 1515.96 | 1383.01 | 4575.58 | 4113.27 | 13447.9 | 12516.1 | 30977.9 | 29806.5 | 44659.8 | 43315.7 | 735.475 | 954.499 | 825.621 | 2651.48 | 2438.28 |
| 735.76 | 1472.47 | 1387.69 | 4526.01 | 4125.24 | 13424.5 | 12546.6 | 31094.7 | 29869 | 44875.1 | 43399.8 | 735.76 | 953.779 | 828.611 | 2671.76 | 2445.82 |
| 736.045 | 1587.2 | 1392.38 | 4545.61 | 4137.22 | 13407.7 | 12577.2 | 31314.6 | 29931.6 | 45239.6 | 43484 | 736.045 | 962.111 | 831.608 | 2660.13 | 2453.38 |
| 736.33 | 1600.19 | 1397.08 | 4493.48 | 4149.23 | 13415 | 12607.8 | 30974 | 29994.2 | 45033.1 | 43568.2 | 736.33 | 967.663 | 834.613 | 2676.36 | 2460.95 |
| 736.615 | 1605.09 | 1401.79 | 4549.08 | 4161.26 | 13642.4 | 12638.5 | 31462.7 | 30056.8 | 45359.9 | 43652.5 | 736.615 | 993.911 | 837.626 | 2683.05 | 2468.53 |
| 736.9 | 1604.51 | 1406.51 | 4600.76 | 4173.31 | 13536.7 | 12669.1 | 31565.6 | 30119.5 | 45512.8 | 43736.9 | 736.9 | 962.205 | 840.647 | 2697.11 | 2476.13 |
| 737.185 | 1635.03 | 1411.25 | 4504.45 | 4185.39 | 13115.8 | 12699.9 | 31140.3 | 30182.3 | 44901.2 | 43821.3 | 737.185 | 969.827 | 843.675 | 2711.3 | 2483.75 |
| 737.47 | 1627.55 | 1415.99 | 4589.99 | 4197.48 | 13537 | 12730.6 | 31403.9 | 30245.2 | 45332.9 | 43905.8 | 737.47 | 952.06 | 846.711 | 2684.98 | 2491.38 |
| 737.755 | 1622.79 | 1420.75 | 4628.62 | 4209.6 | 13754.6 | 12761.4 | 31498.6 | 30308.1 | 45822.4 | 43990.4 | 737.755 | 970.076 | 849.755 | 2647.83 | 2499.03 |
| 738.041 | 1598.59 | 1425.52 | 4595.48 | 4221.74 | 13634.3 | 12792.3 | 31598.9 | 30371 | 45560 | 44075 | 738.041 | 969.43 | 852.807 | 2733.94 | 2506.69 |
| 738.326 | 1616.26 | 1430.29 | 4571.36 | 4233.9 | 13844.3 | 12823.1 | 31950.6 | 30434 | 45856.2 | 44159.7 | 738.326 | 974.537 | 855.867 | 2701.23 | 2514.37 |
| 738.611 | 1596.9 | 1435.08 | 4621.23 | 4246.08 | 13688.7 | 12854.1 | 31867.1 | 30497.1 | 46072.8 | 44244.4 | 738.611 | 990.344 | 858.934 | 2765.34 | 2522.06 |
| 738.896 | 1643.62 | 1439.88 | 4612.14 | 4258.28 | 13691.6 | 12885 | 31544.6 | 30560.2 | 45547.4 | 44329.2 | 738.896 | 954.225 | 862.009 | 2777.09 | 2529.77 |
| 739.181 | 1673.72 | 1444.69 | 4608.61 | 4270.51 | 13858.5 | 12916 | 31859.3 | 30623.4 | 46102.8 | 44414.1 | 739.181 | 998.123 | 865.092 | 2815.55 | 2537.49 |
| 739.466 | 1638.83 | 1449.51 | 4731.75 | 4282.75 | 13747.6 | 12947.1 | 32009.6 | 30686.6 | 46180.9 | 44499 | 739.466 | 1001.65 | 868.183 | 2696.84 | 2545.24 |
| 739.751 | 1672.18 | 1454.35 | 4616.02 | 4295.02 | 13798.8 | 12978.2 | 32090.7 | 30749.9 | 46290 | 44584 | 739.751 | 964.473 | 871.282 | 2821.67 | 2552.99 |
| 740.035 | 1661 | 1459.19 | 4729.25 | 4307.31 | 14054.7 | 13009.3 | 32139.8 | 30813.3 | 46824.3 | 44669.1 | 740.035 | 1038.57 | 874.389 | 2780.62 | 2560.76 |
| 740.321 | 1702.08 | 1464.04 | 4750.73 | 4319.63 | 13885.8 | 13040.5 | 32297.5 | 30876.7 | 45937.8 | 44754.2 | 740.321 | 1025.55 | 877.503 | 2798.46 | 2568.55 |
| 740.605 | 1660.48 | 1468.91 | 4596.1 | 4331.96 | 13980.9 | 13071.7 | 32501.3 | 30940.2 | 46585 | 44839.4 | 740.605 | 1009.14 | 880.625 | 2820.03 | 2576.35 |
| 740.89 | 1664.2 | 1473.78 | 4802.09 | 4344.32 | 14101.1 | 13102.9 | 32769.3 | 31003.7 | 47023.9 | 44924.6 | 740.89 | 1015.09 | 883.755 | 2774.17 | 2584.17 |
| 741.175 | 1766.7 | 1478.67 | 4834.25 | 4356.69 | 14162 | 13134.2 | 32581.7 | 31067.3 | 46574.8 | 45009.9 | 741.175 | 1004.4 | 886.894 | 2813.86 | 2592 |
| 741.46 | 1676.69 | 1483.57 | 4953.17 | 4369.09 | 14008.9 | 13165.5 | 32796.3 | 31130.9 | 46525.3 | 45095.3 | 741.46 | 990.195 | 890.04 | 2841.22 | 2599.85 |
| 741.745 | 1636.25 | 1488.48 | 4871.24 | 4381.51 | 14041.7 | 13196.9 | 32559.8 | 31194.6 | 47238.9 | 45180.7 | 741.745 | 1039.28 | 893.193 | 2937.1 | 2607.71 |
| 742.03 | 1708.44 | 1493.4 | 4756.68 | 4393.96 | 14024.3 | 13228.3 | 32794.2 | 31258.4 | 47054.5 | 45266.2 | 742.03 | 1031.72 | 896.355 | 2912.13 | 2615.59 |
| 742.315 | 1682.51 | 1498.33 | 4849.56 | 4406.42 | 14138.6 | 13259.7 | 32939.8 | 31322.2 | 47555.7 | 45351.8 | 742.315 | 1079.75 | 899.525 | 2844.53 | 2623.49 |
| 742.6 | 1606.8 | 1503.27 | 4864.01 | 4418.91 | 14054.6 | 13291.2 | 32315.1 | 31386.1 | 46904.1 | 45437.4 | 742.6 | 1075.22 | 902.703 | 2800.36 | 2631.4 |
| 742.884 | 1695 | 1508.23 | 4926.67 | 4431.42 | 14223.5 | 13322.7 | 33273.1 | 31450 | 47675 | 45523.1 | 742.884 | 1034.72 | 905.889 | 2947.08 | 2639.32 |
| 743.169 | 1743.26 | 1513.19 | 4832.82 | 4443.95 | 14294.8 | 13354.2 | 33089.1 | 31514 | 47440.4 | 45608.8 | 743.169 | 1038.76 | 909.082 | 2880.88 | 2647.27 |
| 743.454 | 1670.14 | 1518.17 | 5024.57 | 4456.5 | 14471.5 | 13385.9 | 33187.3 | 31578 | 48121.6 | 45694.6 | 743.454 | 1106.51 | 912.284 | 2853.03 | 2655.22 |
| 743.739 | 1728.83 | 1523.15 | 4791.82 | 4469.07 | 14166.8 | 13417.5 | 33333.1 | 31642.1 | 47616.5 | 45780.5 | 743.739 | 1077.86 | 915.493 | 2791.77 | 2663.2 |
| 744.023 | 1707.49 | 1528.15 | 5018.88 | 4481.67 | 14490.4 | 13449.2 | 33315 | 31706.3 | 48210.8 | 45866.4 | 744.023 | 1070.17 | 918.711 | 2924.12 | 2671.19 |
| 744.308 | 1788.49 | 1533.16 | 4941.03 | 4494.28 | 14450.2 | 13480.9 | 33207.4 | 31770.5 | 48152.2 | 45952.4 | 744.308 | 1036.67 | 921.936 | 2907.45 | 2679.19 |
| 744.593 | 1757.98 | 1538.18 | 4825.92 | 4506.93 | 14404.9 | 13512.6 | 33292.5 | 31834.8 | 47762 | 46038.4 | 744.593 | 1065.59 | 925.17 | 2941.29 | 2687.21 |
| 744.878 | 1749.5 | 1543.21 | 5001.72 | 4519.59 | 14428.4 | 13544.4 | 33814.4 | 31899.1 | 48413.2 | 46124.5 | 744.878 | 1143.34 | 928.412 | 2978.83 | 2695.25 |
| 745.162 | 1823.93 | 1548.25 | 4971.56 | 4532.27 | 14545.8 | 13576.2 | 33536.3 | 31963.5 | 48437.9 | 46210.7 | 745.162 | 1105.37 | 931.661 | 2979.76 | 2703.3 |
| 745.447 | 1769.72 | 1553.31 | 4922.81 | 4544.97 | 14683.4 | 13608.1 | 33478.4 | 32027.9 | 47868.7 | 46296.9 | 745.447 | 1118.94 | 934.919 | 2965.12 | 2711.36 |
| 745.732 | 1704.37 | 1558.37 | 4919.53 | 4557.7 | 14375 | 13640 | 33212.4 | 32092.4 | 48090.7 | 46383.2 | 745.732 | 1073.62 | 938.184 | 2941.68 | 2719.45 |
| 746.016 | 1809.38 | 1563.44 | 4975.47 | 4570.45 | 14416.1 | 13672 | 33981.3 | 32156.9 | 48227.9 | 46469.5 | 746.016 | 1086.79 | 941.458 | 3007.82 | 2727.55 |
| 746.301 | 1780.22 | 1568.53 | 4897.37 | 4583.22 | 14446.3 | 13704 | 33565 | 32221.5 | 48073.2 | 46555.9 | 746.301 | 1122.26 | 944.74 | 2873.15 | 2735.66 |
| 746.586 | 1753.98 | 1573.63 | 4945.11 | 4596.01 | 14582.8 | 13736 | 33686.3 | 32286.2 | 48063.6 | 46642.4 | 746.586 | 1092.51 | 948.03 | 2885.98 | 2743.79 |
| 746.87 | 1785.61 | 1578.74 | 5030.99 | 4608.82 | 14648.6 | 13768.1 | 33778.9 | 32350.9 | 48520.7 | 46728.9 | 746.87 | 1048.12 | 951.328 | 2985.93 | 2751.93 |
| 747.155 | 1759.31 | 1583.86 | 5025.18 | 4621.66 | 14546.9 | 13800.2 | 33859.9 | 32415.7 | 48439.2 | 46815.5 | 747.155 | 1166.05 | 954.634 | 3011.43 | 2760.1 |
| 747.439 | 1785.02 | 1588.99 | 5127.44 | 4634.52 | 14774.5 | 13832.3 | 34153.3 | 32480.5 | 48827.6 | 46902.1 | 747.439 | 1113.73 | 957.948 | 2995.47 | 2768.27 |
| 747.724 | 1820.09 | 1594.13 | 5008.44 | 4647.4 | 14852.1 | 13864.5 | 33839.8 | 32545.3 | 48404.3 | 46988.8 | 747.724 | 1065.26 | 961.27 | 3007.15 | 2776.47 |
| 748.009 | 1833.42 | 1599.29 | 5032.49 | 4660.3 | 14448.4 | 13896.7 | 33856.4 | 32610.3 | 48686.5 | 47075.6 | 748.009 | 1090.83 | 964.601 | 3034.79 | 2784.67 |
| 748.293 | 1849.17 | 1604.45 | 5133.17 | 4673.22 | 14811.8 | 13929 | 34174.3 | 32675.3 | 48971.8 | 47162.4 | 748.293 | 1098.12 | 967.939 | 2983.48 | 2792.9 |
| 748.578 | 1840.35 | 1609.63 | 5030.4 | 4686.16 | 14758.4 | 13961.3 | 34104.9 | 32740.3 | 49180.2 | 47249.3 | 748.578 | 1120.41 | 971.286 | 3048.91 | 2801.14 |
| 748.862 | 1834.55 | 1614.82 | 5118.19 | 4699.13 | 15045.2 | 13993.6 | 34380.1 | 32805.4 | 49021.8 | 47336.2 | 748.862 | 1101.45 | 974.64 | 2978.18 | 2809.39 |
| 749.147 | 1844.97 | 1620.02 | 5083.05 | 4712.12 | 15137.9 | 14026 | 34534.6 | 32870.6 | 49481.3 | 47423.2 | 749.147 | 1136.11 | 978.003 | 3123.8 | 2817.66 |
| 749.431 | 1813.02 | 1625.22 | 5136.66 | 4725.13 | 14933.1 | 14058.4 | 34282.4 | 32935.8 | 49003.6 | 47510.3 | 749.431 | 1122.88 | 981.374 | 3084.77 | 2825.95 |
| 749.716 | 1877.87 | 1630.45 | 5175.38 | 4738.16 | 15100.3 | 14090.9 | 34663.1 | 33001 | 50032.8 | 47597.4 | 749.716 | 1121.3 | 984.754 | 3037.17 | 2834.25 |
| 750 | 1887.48 | 1635.68 | 5188 | 4751.22 | 15035.7 | 14123.4 | 34344.1 | 33066.3 | 49552.4 | 47684.6 | 750 | 1149.57 | 988.141 | 3040.65 | 2842.57 |
| 750.284 | 1801.56 | 1640.92 | 5326.51 | 4764.29 | 15048.3 | 14155.9 | 34449.3 | 33131.7 | 49542 | 47771.8 | 750.284 | 1161.85 | 991.537 | 3072.38 | 2850.91 |
| 750.569 | 1859.35 | 1646.18 | 5289.78 | 4777.39 | 14940.3 | 14188.5 | 34253.6 | 33197.1 | 49273.3 | 47859.1 | 750.569 | 1145.65 | 994.94 | 3071.43 | 2859.26 |
| 750.853 | 1829.63 | 1651.44 | 5248.34 | 4790.51 | 14858.3 | 14221.1 | 34382.6 | 33262.6 | 49493.7 | 47946.4 | 750.853 | 1143.84 | 998.352 | 3099.04 | 2867.62 |
| 751.138 | 1943.54 | 1656.72 | 5116.77 | 4803.65 | 15167.4 | 14253.7 | 34860.8 | 33328.1 | 49836.3 | 48033.8 | 751.138 | 1119.95 | 1001.77 | 3094.24 | 2876 |
| 751.422 | 1920.32 | 1662.01 | 5195.9 | 4816.81 | 15215.4 | 14286.4 | 34969.9 | 33393.7 | 49831.1 | 48121.3 | 751.422 | 1167.43 | 1005.2 | 3029.47 | 2884.4 |
| 751.706 | 1882.95 | 1667.31 | 5300.35 | 4830 | 15209.8 | 14319.1 | 34927.9 | 33459.4 | 49964.4 | 48208.8 | 751.706 | 1142.48 | 1008.64 | 3131.48 | 2892.82 |
| 751.991 | 1937.66 | 1672.62 | 5206.85 | 4843.2 | 15456.7 | 14351.9 | 34566.7 | 33525.1 | 50086.8 | 48296.4 | 751.991 | 1132.54 | 1012.08 | 3134.45 | 2901.24 |
| 752.275 | 1854.87 | 1677.94 | 5308.73 | 4856.43 | 15449.4 | 14384.7 | 35617.7 | 33590.8 | 50710.5 | 48384 | 752.275 | 1158.86 | 1015.54 | 3156.62 | 2909.69 |
| 752.559 | 1910 | 1683.28 | 5372.72 | 4869.68 | 15150 | 14417.5 | 34792 | 33656.6 | 49735.5 | 48471.7 | 752.559 | 1139.61 | 1019 | 3091.42 | 2918.15 |
| 752.844 | 1945.91 | 1688.62 | 5362.98 | 4882.95 | 15487.6 | 14450.4 | 35473.6 | 33722.5 | 50540.2 | 48559.4 | 752.844 | 1130.53 | 1022.47 | 3189.7 | 2926.63 |
| 753.128 | 1826.35 | 1693.98 | 5264.63 | 4896.25 | 15285.8 | 14483.3 | 34920.9 | 33788.4 | 50046.7 | 48647.2 | 753.128 | 1126.95 | 1025.94 | 3077.72 | 2935.12 |
| 753.412 | 1894.65 | 1699.35 | 5297.05 | 4909.56 | 15370.3 | 14516.3 | 34979 | 33854.3 | 49899.6 | 48735.1 | 753.412 | 1172.04 | 1029.43 | 3095.18 | 2943.63 |
| 753.697 | 1932.21 | 1704.73 | 5257.11 | 4922.9 | 15478.6 | 14549.3 | 35329.3 | 33920.3 | 50871.7 | 48823 | 753.697 | 1215.23 | 1032.92 | 3119.56 | 2952.15 |
| 753.981 | 1989.14 | 1710.12 | 5414.33 | 4936.26 | 15544.2 | 14582.3 | 35203 | 33986.4 | 50358.3 | 48911 | 753.981 | 1156.89 | 1036.43 | 3179.16 | 2960.69 |
| 754.265 | 1908.22 | 1715.52 | 5391.12 | 4949.64 | 15611.3 | 14615.4 | 35222.8 | 34052.5 | 50827.2 | 48999 | 754.265 | 1171.78 | 1039.94 | 3196.94 | 2969.25 |
| 754.549 | 1988.96 | 1720.94 | 5353.4 | 4963.05 | 15503.6 | 14648.5 | 35463 | 34118.7 | 50912.5 | 49087.1 | 754.549 | 1184.66 | 1043.46 | 3236.84 | 2977.82 |
| 754.834 | 1916.31 | 1726.36 | 5412.59 | 4976.47 | 15371.5 | 14681.6 | 35391.8 | 34184.9 | 50488.4 | 49175.2 | 754.834 | 1193.68 | 1046.99 | 3184.42 | 2986.41 |
| 755.118 | 1930.6 | 1731.8 | 5329.62 | 4989.92 | 15404.4 | 14714.8 | 35425.7 | 34251.2 | 50675.5 | 49263.4 | 755.118 | 1131 | 1050.52 | 3132.9 | 2995.01 |
| 755.402 | 1937.67 | 1737.25 | 5487.5 | 5003.39 | 15545.3 | 14748 | 35501.1 | 34317.6 | 50640.1 | 49351.7 | 755.402 | 1240.34 | 1054.07 | 3233.68 | 3003.63 |
| 755.686 | 1903.72 | 1742.7 | 5346.35 | 5016.88 | 15450.3 | 14781.3 | 35393.7 | 34383.9 | 50928.2 | 49440 | 755.686 | 1219.91 | 1057.62 | 3199.82 | 3012.26 |
| 755.97 | 1912.95 | 1748.18 | 5414.6 | 5030.39 | 15635.4 | 14814.6 | 35489.3 | 34450.4 | 50841.8 | 49528.4 | 755.97 | 1192.65 | 1061.18 | 3172.67 | 3020.92 |
| 756.254 | 2021.95 | 1753.66 | 5470.39 | 5043.93 | 15723.1 | 14848 | 35929.8 | 34516.9 | 51289.6 | 49616.8 | 756.254 | 1164.35 | 1064.75 | 3247.28 | 3029.58 |
| 756.538 | 1931.06 | 1759.15 | 5462.99 | 5057.48 | 15691.8 | 14881.3 | 35543.1 | 34583.4 | 51131 | 49705.3 | 756.538 | 1211.52 | 1068.33 | 3177.95 | 3038.27 |
| 756.823 | 1999.71 | 1764.66 | 5548.67 | 5071.06 | 15758.8 | 14914.7 | 36237.4 | 34650 | 51495.2 | 49793.8 | 756.823 | 1229.92 | 1071.91 | 3266.16 | 3046.97 |
| 757.107 | 1910.61 | 1770.17 | 5478.7 | 5084.66 | 15726.7 | 14948.2 | 36157.4 | 34716.6 | 51339.1 | 49882.4 | 757.107 | 1166.27 | 1075.51 | 3331.96 | 3055.68 |
| 757.391 | 2008.48 | 1775.7 | 5498.84 | 5098.28 | 15871.6 | 14981.7 | 35765.5 | 34783.3 | 51342.3 | 49971 | 757.391 | 1228.63 | 1079.11 | 3257.21 | 3064.41 |
| 757.675 | 2094.75 | 1781.24 | 5607.71 | 5111.93 | 15855.3 | 15015.2 | 35995.8 | 34850.1 | 51645.7 | 50059.7 | 757.675 | 1225.53 | 1082.72 | 3259.46 | 3073.16 |
| 757.959 | 2025.17 | 1786.79 | 5522.01 | 5125.59 | 15843.1 | 15048.8 | 36089.7 | 34916.9 | 51920.3 | 50148.4 | 757.959 | 1205.55 | 1086.34 | 3360.54 | 3081.92 |
| 758.243 | 2066.81 | 1792.35 | 5627.19 | 5139.28 | 16019.8 | 15082.4 | 36031.6 | 34983.8 | 52015.4 | 50237.2 | 758.243 | 1253.54 | 1089.97 | 3361.23 | 3090.7 |
| 758.527 | 1940.64 | 1797.93 | 5592.39 | 5152.99 | 15843.1 | 15116 | 36274.4 | 35050.7 | 51650.3 | 50326.1 | 758.527 | 1220.57 | 1093.61 | 3279.49 | 3099.49 |
| 758.811 | 1943.89 | 1803.51 | 5498.5 | 5166.72 | 15988.7 | 15149.7 | 36002.1 | 35117.6 | 52061.6 | 50415 | 758.811 | 1235.97 | 1097.25 | 3332.11 | 3108.3 |
| 759.095 | 2016.18 | 1809.11 | 5606.4 | 5180.47 | 15993.9 | 15183.4 | 36235.9 | 35184.6 | 52211.8 | 50504 | 759.095 | 1240.48 | 1100.91 | 3314.86 | 3117.13 |
| 759.379 | 2047.52 | 1814.72 | 5532.09 | 5194.24 | 16004.2 | 15217.1 | 36358.2 | 35251.7 | 51767.4 | 50593 | 759.379 | 1268.94 | 1104.57 | 3362.31 | 3125.97 |
| 759.663 | 1999.17 | 1820.34 | 5657.95 | 5208.04 | 16003.9 | 15250.9 | 36408.1 | 35318.8 | 51769.7 | 50682.1 | 759.663 | 1284.06 | 1108.24 | 3355.82 | 3134.83 |
| 759.947 | 2063.1 | 1825.97 | 5758.16 | 5221.86 | 16210.7 | 15284.8 | 36635.4 | 35386 | 52256.1 | 50771.2 | 759.947 | 1247.23 | 1111.92 | 3275.58 | 3143.71 |
| 760.231 | 2044.71 | 1831.61 | 5624.58 | 5235.7 | 16023.3 | 15318.6 | 36753.4 | 35453.2 | 52361.9 | 50860.4 | 760.231 | 1260.6 | 1115.61 | 3439.8 | 3152.6 |
| 760.515 | 2057.09 | 1837.27 | 5656.43 | 5249.56 | 15932.5 | 15352.5 | 36785.7 | 35520.4 | 51866.8 | 50949.6 | 760.515 | 1278.36 | 1119.3 | 3327.15 | 3161.5 |
| 760.799 | 2031.37 | 1842.93 | 5732.83 | 5263.45 | 16087 | 15386.5 | 36595.8 | 35587.8 | 52270.9 | 51038.9 | 760.799 | 1256.46 | 1123.01 | 3466.44 | 3170.43 |
| 761.082 | 2044.29 | 1848.61 | 5751.47 | 5277.35 | 16043.9 | 15420.5 | 36966.7 | 35655.1 | 52269.8 | 51128.2 | 761.082 | 1324.08 | 1126.72 | 3325.7 | 3179.36 |
| 761.366 | 2045.16 | 1854.3 | 5706.46 | 5291.28 | 16145.5 | 15454.5 | 37016.3 | 35722.6 | 52675.1 | 51217.6 | 761.366 | 1265.82 | 1130.44 | 3435.57 | 3188.32 |
| 761.65 | 2009.15 | 1860 | 5650.32 | 5305.23 | 16368.3 | 15488.5 | 37121.1 | 35790 | 52535.5 | 51307 | 761.65 | 1343.54 | 1134.17 | 3413.62 | 3197.29 |
| 761.934 | 2045.32 | 1865.71 | 5814.6 | 5319.2 | 16176.8 | 15522.6 | 36988.5 | 35857.5 | 52264 | 51396.5 | 761.934 | 1241.74 | 1137.91 | 3541.91 | 3206.28 |
| 762.218 | 2081.51 | 1871.44 | 5794.65 | 5333.19 | 16404.1 | 15556.7 | 37602.1 | 35925.1 | 53427.5 | 51486.1 | 762.218 | 1297.34 | 1141.66 | 3396.39 | 3215.28 |
| 762.502 | 2085.49 | 1877.17 | 5691.81 | 5347.21 | 16577 | 15590.9 | 37397.8 | 35992.7 | 53115.7 | 51575.7 | 762.502 | 1269.81 | 1145.41 | 3420.05 | 3224.3 |
| 762.786 | 2093.54 | 1882.92 | 5718.27 | 5361.24 | 16250.2 | 15625.1 | 36880.5 | 36060.4 | 52934.4 | 51665.3 | 762.786 | 1228.05 | 1149.18 | 3339.2 | 3233.33 |
| 763.069 | 2131.2 | 1888.68 | 5752.86 | 5375.3 | 16303.7 | 15659.4 | 37280.5 | 36128.1 | 53086.1 | 51755.1 | 763.069 | 1281.95 | 1152.95 | 3438.83 | 3242.38 |
| 763.353 | 2107.44 | 1894.45 | 5811.06 | 5389.38 | 16179.9 | 15693.6 | 37207.3 | 36195.9 | 52599.3 | 51844.8 | 763.353 | 1255.49 | 1156.73 | 3443.3 | 3251.45 |
| 763.637 | 2123.84 | 1900.23 | 5710.61 | 5403.48 | 16531.8 | 15727.9 | 37341.7 | 36263.7 | 52931.2 | 51934.6 | 763.637 | 1262.56 | 1160.52 | 3470.1 | 3260.53 |
| 763.921 | 2068.15 | 1906.02 | 5750.9 | 5417.61 | 16322.4 | 15762.3 | 37379.2 | 36331.6 | 53105.7 | 52024.5 | 763.921 | 1284.01 | 1164.32 | 3475.24 | 3269.63 |
| 764.204 | 2143.62 | 1911.83 | 5782.61 | 5431.75 | 16341 | 15796.7 | 37091.6 | 36399.5 | 52477.5 | 52114.4 | 764.204 | 1293.71 | 1168.13 | 3441.93 | 3278.75 |
| 764.488 | 2116.46 | 1917.64 | 5826.43 | 5445.92 | 16385 | 15831.1 | 37353.6 | 36467.5 | 52834.8 | 52204.3 | 764.488 | 1308.53 | 1171.94 | 3472.93 | 3287.88 |
| 764.772 | 2076.93 | 1923.47 | 5842.83 | 5460.11 | 16527.9 | 15865.6 | 37748.2 | 36535.5 | 53335.2 | 52294.4 | 764.772 | 1278.61 | 1175.77 | 3512.55 | 3297.02 |
| 765.056 | 2130.7 | 1929.31 | 5990.07 | 5474.32 | 16615.7 | 15900 | 37372.4 | 36603.6 | 53046.7 | 52384.4 | 765.056 | 1295.14 | 1179.6 | 3508.07 | 3306.19 |
| 765.339 | 2227.79 | 1935.16 | 5868.1 | 5488.55 | 16484.8 | 15934.6 | 37265.8 | 36671.7 | 53031.1 | 52474.5 | 765.339 | 1312.59 | 1183.44 | 3532.23 | 3315.37 |
| 765.623 | 2186.09 | 1941.03 | 5729.23 | 5502.81 | 16685.6 | 15969.2 | 37582.5 | 36739.9 | 53905 | 52564.7 | 765.623 | 1314.37 | 1187.29 | 3571.96 | 3324.56 |
| 765.907 | 2144.87 | 1946.9 | 5923.12 | 5517.08 | 16819.4 | 16003.8 | 37681.4 | 36808.1 | 53242.5 | 52654.9 | 765.907 | 1315.5 | 1191.15 | 3471.44 | 3333.77 |
| 766.19 | 2085.41 | 1952.79 | 5908.07 | 5531.38 | 16508.7 | 16038.4 | 37640.5 | 36876.3 | 53558.4 | 52745.2 | 766.19 | 1318.39 | 1195.02 | 3529.34 | 3343 |
| 766.474 | 2190.03 | 1958.69 | 6057.92 | 5545.7 | 16583.4 | 16073.1 | 37857.8 | 36944.7 | 54071.9 | 52835.5 | 766.474 | 1304.13 | 1198.89 | 3456.24 | 3352.24 |
| 766.757 | 2212.7 | 1964.6 | 5911.54 | 5560.04 | 16527.1 | 16107.8 | 37516.6 | 37013 | 53823.3 | 52925.9 | 766.757 | 1305.42 | 1202.78 | 3597.18 | 3361.5 |
| 767.041 | 2208.34 | 1970.52 | 5958.4 | 5574.4 | 16673.8 | 16142.6 | 37729.8 | 37081.4 | 53615.5 | 53016.3 | 767.041 | 1314.34 | 1206.67 | 3492.75 | 3370.78 |
| 767.324 | 2134.89 | 1976.45 | 5951.59 | 5588.79 | 16941 | 16177.4 | 38136.4 | 37149.9 | 54345.8 | 53106.8 | 767.324 | 1344.08 | 1210.57 | 3580.21 | 3380.07 |
| 767.608 | 2276.35 | 1982.4 | 6007.23 | 5603.2 | 16885 | 16212.2 | 38106.8 | 37218.4 | 53981.3 | 53197.3 | 767.608 | 1406.69 | 1214.48 | 3596.65 | 3389.38 |
| 767.891 | 2238.1 | 1988.35 | 5964.61 | 5617.63 | 16941 | 16247.1 | 38406.6 | 37287 | 54235.7 | 53287.9 | 767.891 | 1338.34 | 1218.4 | 3575.02 | 3398.7 |
| 768.175 | 2175.43 | 1994.32 | 5999.23 | 5632.08 | 16937 | 16282 | 38464.2 | 37355.6 | 54841 | 53378.5 | 768.175 | 1321.91 | 1222.33 | 3661.61 | 3408.04 |
| 768.459 | 2216.63 | 2000.3 | 5905.68 | 5646.55 | 16853 | 16316.9 | 38047.6 | 37424.2 | 54476 | 53469.2 | 768.459 | 1307.21 | 1226.27 | 3598.32 | 3417.4 |
| 768.742 | 2169.45 | 2006.3 | 6012.01 | 5661.04 | 16785.1 | 16351.9 | 37594.8 | 37492.9 | 53723.9 | 53559.9 | 768.742 | 1319.66 | 1230.21 | 3655.37 | 3426.77 |
| 769.026 | 2185.98 | 2012.3 | 5945.82 | 5675.56 | 16736 | 16386.9 | 38305.3 | 37561.7 | 54454.5 | 53650.7 | 769.026 | 1371.06 | 1234.17 | 3662.79 | 3436.16 |
| 769.309 | 2177.07 | 2018.31 | 6095.53 | 5690.1 | 17447 | 16421.9 | 38810.5 | 37630.5 | 54985.4 | 53741.5 | 769.309 | 1305.03 | 1238.13 | 3566.35 | 3445.57 |
| 769.592 | 2200.33 | 2024.34 | 6082.39 | 5704.65 | 17069.9 | 16457 | 38642.2 | 37699.4 | 55188.6 | 53832.3 | 769.592 | 1365.93 | 1242.1 | 3650.31 | 3454.99 |
| 769.876 | 2220.06 | 2030.38 | 6176.15 | 5719.23 | 17240.2 | 16492.1 | 38504.7 | 37768.3 | 54915.5 | 53923.2 | 769.876 | 1390.1 | 1246.08 | 3590.47 | 3464.42 |
| 770.159 | 2154.8 | 2036.43 | 6099.3 | 5733.84 | 17233.2 | 16527.3 | 38515.2 | 37837.2 | 54740.2 | 54014.2 | 770.159 | 1465.33 | 1250.07 | 3595.93 | 3473.88 |
| 770.443 | 2254.59 | 2042.49 | 6204.75 | 5748.46 | 17301.2 | 16562.5 | 39140.6 | 37906.2 | 55277.4 | 54105.2 | 770.443 | 1344.95 | 1254.07 | 3701.61 | 3483.35 |
| 770.726 | 2300.03 | 2048.57 | 6155.4 | 5763.11 | 17235.8 | 16597.7 | 39158 | 37975.2 | 55216.5 | 54196.3 | 770.726 | 1375.18 | 1258.08 | 3680.67 | 3492.83 |
| 771.01 | 2210.01 | 2054.65 | 6135.9 | 5777.77 | 17212.1 | 16633 | 38777.1 | 38044.3 | 55122.3 | 54287.4 | 771.01 | 1401.02 | 1262.09 | 3647.21 | 3502.33 |
| 771.293 | 2251.62 | 2060.75 | 6167.31 | 5792.46 | 17314.5 | 16668.3 | 39024 | 38113.5 | 55434.4 | 54378.6 | 771.293 | 1360.14 | 1266.12 | 3719.06 | 3511.85 |
| 771.576 | 2304.35 | 2066.86 | 6240.06 | 5807.18 | 17301.4 | 16703.6 | 38682.7 | 38182.7 | 55315.6 | 54469.8 | 771.576 | 1374.02 | 1270.15 | 3693.88 | 3521.39 |
| 771.86 | 2319.82 | 2072.98 | 6187.98 | 5821.91 | 17443 | 16739 | 39562.9 | 38251.9 | 55569 | 54561 | 771.86 | 1430.61 | 1274.19 | 3732.68 | 3530.93 |
| 772.143 | 2307.12 | 2079.11 | 6307.13 | 5836.66 | 17346.6 | 16774.4 | 39742.7 | 38321.2 | 55507.8 | 54652.3 | 772.143 | 1399.91 | 1278.24 | 3758.61 | 3540.5 |
| 772.426 | 2256 | 2085.26 | 6186.92 | 5851.44 | 17451.6 | 16809.8 | 39155.1 | 38390.5 | 56124.3 | 54743.7 | 772.426 | 1425.98 | 1282.31 | 3771.95 | 3550.08 |
| 772.709 | 2301.73 | 2091.41 | 6168.44 | 5866.24 | 17278.2 | 16845.3 | 39562.3 | 38459.9 | 56023.9 | 54835.1 | 772.709 | 1368.17 | 1286.37 | 3733.49 | 3559.68 |
| 772.993 | 2259.76 | 2097.58 | 6344.24 | 5881.06 | 17527.4 | 16880.8 | 39500.6 | 38529.3 | 55946.3 | 54926.5 | 772.993 | 1465.07 | 1290.45 | 3807.81 | 3569.3 |
| 773.276 | 2341.86 | 2103.76 | 6318.5 | 5895.9 | 17751.6 | 16916.4 | 39565.9 | 38598.8 | 56233.9 | 55018 | 773.276 | 1449.14 | 1294.54 | 3820.94 | 3578.93 |
| 773.559 | 2321.12 | 2109.95 | 6254.97 | 5910.76 | 17530.4 | 16952 | 39867.1 | 38668.3 | 55908.1 | 55109.5 | 773.559 | 1457.02 | 1298.63 | 3819.76 | 3588.58 |
| 773.842 | 2395.61 | 2116.16 | 6297.36 | 5925.65 | 17476.6 | 16987.6 | 39749.4 | 38737.8 | 56272.6 | 55201.1 | 773.842 | 1425.89 | 1302.73 | 3759 | 3598.24 |
| 774.126 | 2322.2 | 2122.37 | 6252.79 | 5940.55 | 17766.5 | 17023.3 | 39871.2 | 38807.4 | 56119.3 | 55292.8 | 774.126 | 1435.27 | 1306.85 | 3864.91 | 3607.92 |
| 774.409 | 2393.6 | 2128.6 | 6282.93 | 5955.48 | 17804.1 | 17059 | 39833.7 | 38877.1 | 56191.5 | 55384.4 | 774.409 | 1451.73 | 1310.97 | 3901.91 | 3617.61 |
| 774.692 | 2320.56 | 2134.84 | 6306.36 | 5970.43 | 17620.6 | 17094.7 | 39515.8 | 38946.8 | 55968.9 | 55476.2 | 774.692 | 1422.28 | 1315.1 | 3831.62 | 3627.33 |
| 774.975 | 2324.38 | 2141.09 | 6284.01 | 5985.4 | 17706.6 | 17130.5 | 39693 | 39016.5 | 56470.4 | 55567.9 | 774.975 | 1442.97 | 1319.24 | 3841.77 | 3637.05 |
| 775.258 | 2316.05 | 2147.35 | 6390.12 | 6000.39 | 17566.8 | 17166.3 | 39700 | 39086.3 | 56640.3 | 55659.8 | 775.258 | 1500.79 | 1323.39 | 3773.97 | 3646.8 |
| 775.542 | 2446.84 | 2153.63 | 6544.3 | 6015.41 | 17712.9 | 17202.1 | 39914.2 | 39156.2 | 56820.7 | 55751.6 | 775.542 | 1506.29 | 1327.55 | 3894.76 | 3656.56 |
| 775.825 | 2378.22 | 2159.91 | 6476.14 | 6030.44 | 17920.6 | 17238 | 40199.6 | 39226 | 56936.8 | 55843.5 | 775.825 | 1418.48 | 1331.71 | 3829.88 | 3666.34 |
| 776.108 | 2382.62 | 2166.21 | 6385.19 | 6045.5 | 17873 | 17273.9 | 40155.2 | 39296 | 56698.5 | 55935.5 | 776.108 | 1456.98 | 1335.89 | 3796.85 | 3676.13 |
| 776.391 | 2374.7 | 2172.52 | 6409.53 | 6060.58 | 17607.4 | 17309.8 | 39907.3 | 39365.9 | 56521.2 | 56027.5 | 776.391 | 1450.75 | 1340.07 | 3794.78 | 3685.94 |
| 776.674 | 2354.32 | 2178.84 | 6333.09 | 6075.68 | 17820.7 | 17345.8 | 39860.7 | 39436 | 56692.1 | 56119.5 | 776.674 | 1462.14 | 1344.26 | 3887.27 | 3695.76 |
| 776.957 | 2361.67 | 2185.18 | 6531.87 | 6090.81 | 17834.1 | 17381.8 | 40242.1 | 39506 | 56778.3 | 56211.6 | 776.957 | 1488.77 | 1348.46 | 3802.06 | 3705.61 |
| 777.24 | 2407.28 | 2191.52 | 6369.75 | 6105.95 | 17845.6 | 17417.9 | 40230.2 | 39576.1 | 57261.1 | 56303.8 | 777.24 | 1465.59 | 1352.67 | 3856.2 | 3715.46 |
| 777.523 | 2356.07 | 2197.88 | 6379.45 | 6121.11 | 17977.8 | 17454 | 40183.1 | 39646.3 | 57126 | 56395.9 | 777.523 | 1450.79 | 1356.89 | 3973.96 | 3725.34 |
| 777.806 | 2405.1 | 2204.25 | 6464.86 | 6136.3 | 17995 | 17490.1 | 40433.8 | 39716.5 | 57386.2 | 56488.2 | 777.806 | 1487.24 | 1361.12 | 3957.08 | 3735.23 |
| 778.089 | 2412.29 | 2210.63 | 6659.04 | 6151.51 | 18154.4 | 17526.2 | 41055.7 | 39786.8 | 57561 | 56580.4 | 778.089 | 1471.86 | 1365.36 | 4070.14 | 3745.14 |
| 778.372 | 2378.23 | 2217.03 | 6564.4 | 6166.74 | 18021.6 | 17562.4 | 40466.9 | 39857 | 57917.4 | 56672.8 | 778.372 | 1552.55 | 1369.61 | 3997.6 | 3755.06 |
| 778.655 | 2522.01 | 2223.43 | 6635.79 | 6181.99 | 18254 | 17598.7 | 40843.3 | 39927.4 | 58105.9 | 56765.1 | 778.655 | 1530 | 1373.86 | 3920.3 | 3765 |
| 778.938 | 2453.43 | 2229.85 | 6592.76 | 6197.27 | 18183 | 17634.9 | 40615.3 | 39997.8 | 56867.6 | 56857.5 | 778.938 | 1477.27 | 1378.13 | 3949.26 | 3774.95 |
| 779.221 | 2450.54 | 2236.28 | 6553.09 | 6212.56 | 18050.3 | 17671.2 | 40888.3 | 40068.2 | 57979.4 | 56950 | 779.221 | 1487.37 | 1382.4 | 3991.8 | 3784.93 |
| 779.504 | 2478.2 | 2242.72 | 6659.43 | 6227.88 | 18309.6 | 17707.5 | 40972.6 | 40138.7 | 57918.3 | 57042.5 | 779.504 | 1507.45 | 1386.68 | 3973.78 | 3794.91 |
| 779.787 | 2418.52 | 2249.17 | 6521.38 | 6243.22 | 18248.5 | 17743.9 | 41001 | 40209.2 | 58187 | 57135 | 779.787 | 1518.13 | 1390.97 | 4030.47 | 3804.92 |
| 780.07 | 2379.37 | 2255.64 | 6693.78 | 6258.58 | 18292.5 | 17780.3 | 41403.5 | 40279.7 | 58115.7 | 57227.6 | 780.07 | 1539.46 | 1395.27 | 4088.66 | 3814.94 |
| 780.353 | 2471.86 | 2262.11 | 6786.42 | 6273.96 | 18440.2 | 17816.8 | 41245.2 | 40350.3 | 58100.1 | 57320.2 | 780.353 | 1476.79 | 1399.58 | 3987.89 | 3824.97 |
| 780.635 | 2482 | 2268.6 | 6727.09 | 6289.36 | 18568.8 | 17853.2 | 41287.4 | 40421 | 58475.8 | 57412.9 | 780.635 | 1497.25 | 1403.9 | 3963.51 | 3835.03 |
| 780.918 | 2528.68 | 2275.1 | 6595.15 | 6304.79 | 18317 | 17889.7 | 41563.8 | 40491.7 | 58378.7 | 57505.6 | 780.918 | 1490.67 | 1408.23 | 4131.83 | 3845.1 |
| 781.201 | 2508.49 | 2281.61 | 6826.67 | 6320.23 | 18603.2 | 17926.3 | 41573 | 40562.4 | 58832 | 57598.4 | 781.201 | 1519.43 | 1412.57 | 4119.17 | 3855.18 |
| 781.484 | 2443.94 | 2288.14 | 6805.42 | 6335.7 | 18433.4 | 17962.8 | 41253 | 40633.2 | 58486.3 | 57691.2 | 781.484 | 1559 | 1416.91 | 3988.44 | 3865.28 |
| 781.767 | 2386.22 | 2294.68 | 6637.29 | 6351.19 | 18605.8 | 17999.5 | 41297.8 | 40704 | 58809.4 | 57784.1 | 781.767 | 1594.11 | 1421.27 | 4062.26 | 3875.4 |
| 782.05 | 2468.66 | 2301.22 | 6721.18 | 6366.7 | 18388.6 | 18036.1 | 41276.3 | 40774.9 | 58336.9 | 57876.9 | 782.05 | 1493.4 | 1425.63 | 4124.14 | 3885.53 |
| 782.332 | 2491.89 | 2307.78 | 6701.34 | 6382.23 | 18599 | 18072.8 | 41577.5 | 40845.8 | 58701 | 57969.9 | 782.332 | 1566.14 | 1430 | 4101.91 | 3895.68 |
| 782.615 | 2502.59 | 2314.36 | 6751.9 | 6397.79 | 18649.3 | 18109.5 | 41705.1 | 40916.8 | 58508.3 | 58062.8 | 782.615 | 1540.35 | 1434.38 | 4086.87 | 3905.85 |
| 782.898 | 2510.92 | 2320.94 | 6766.73 | 6413.36 | 18423.7 | 18146.2 | 41530.4 | 40987.8 | 58719.3 | 58155.9 | 782.898 | 1572.95 | 1438.77 | 4043.48 | 3916.03 |
| 783.181 | 2492.6 | 2327.54 | 6764.56 | 6428.96 | 18511.7 | 18183 | 41645.1 | 41058.8 | 59301.2 | 58248.9 | 783.181 | 1569.37 | 1443.17 | 4057.03 | 3926.23 |
| 783.463 | 2591.36 | 2334.15 | 6802.74 | 6444.58 | 18587.3 | 18219.8 | 42104.4 | 41129.9 | 59544.5 | 58342 | 783.463 | 1601.51 | 1447.58 | 4214.14 | 3936.45 |
| 783.746 | 2468.45 | 2340.77 | 6846.85 | 6460.22 | 18964 | 18256.7 | 41767.2 | 41201 | 59034.4 | 58435.2 | 783.746 | 1618.92 | 1452 | 4119.75 | 3946.68 |
| 784.029 | 2578.03 | 2347.4 | 6786.9 | 6475.88 | 18804.4 | 18293.6 | 41852.1 | 41272.2 | 59337.4 | 58528.4 | 784.029 | 1510.14 | 1456.43 | 4157.94 | 3956.93 |
| 784.311 | 2503.29 | 2354.04 | 6786.75 | 6491.56 | 18921.2 | 18330.5 | 41837.7 | 41343.4 | 58877.2 | 58621.6 | 784.311 | 1584.3 | 1460.86 | 4154.56 | 3967.19 |
| 784.594 | 2525.03 | 2360.7 | 6685.01 | 6507.27 | 18811 | 18367.5 | 41841 | 41414.7 | 59034.8 | 58714.9 | 784.594 | 1610.25 | 1465.31 | 4079.89 | 3977.47 |
| 784.877 | 2536.31 | 2367.37 | 6898.62 | 6522.99 | 18834.2 | 18404.4 | 42119.9 | 41486 | 58459.3 | 58808.2 | 784.877 | 1592.03 | 1469.76 | 4070.22 | 3987.77 |
| 785.159 | 2502.87 | 2374.05 | 6844.69 | 6538.74 | 18949.2 | 18441.5 | 42095.9 | 41557.3 | 59027.4 | 58901.5 | 785.159 | 1618.24 | 1474.23 | 4211.28 | 3998.08 |
| 785.442 | 2534.66 | 2380.74 | 6930.34 | 6554.51 | 18931.8 | 18478.5 | 42204.3 | 41628.7 | 59584.2 | 58994.9 | 785.442 | 1616.75 | 1478.7 | 4144.05 | 4008.41 |
| 785.725 | 2597.29 | 2387.44 | 6938.45 | 6570.3 | 19030.1 | 18515.6 | 41895.8 | 41700.1 | 59757.7 | 59088.4 | 785.725 | 1616.06 | 1483.18 | 4163.92 | 4018.76 |
| 786.007 | 2546.79 | 2394.16 | 6913.24 | 6586.11 | 19034 | 18552.7 | 42069.7 | 41771.6 | 59899.7 | 59181.8 | 786.007 | 1581.17 | 1487.67 | 4202.9 | 4029.12 |
| 786.29 | 2508.58 | 2400.89 | 6940.22 | 6601.94 | 18713.6 | 18589.9 | 42167.8 | 41843.1 | 59794 | 59275.4 | 786.29 | 1622.9 | 1492.17 | 4233.42 | 4039.5 |
| 786.572 | 2630.82 | 2407.63 | 6934.81 | 6617.79 | 18822.1 | 18627.1 | 42204.1 | 41914.7 | 59628.9 | 59368.9 | 786.572 | 1607.91 | 1496.68 | 4157.15 | 4049.89 |
| 786.855 | 2687.78 | 2414.38 | 6846.5 | 6633.67 | 19201.9 | 18664.3 | 42896.7 | 41986.2 | 60304.6 | 59462.5 | 786.855 | 1651.99 | 1501.2 | 4297.49 | 4060.31 |
| 787.137 | 2636.76 | 2421.14 | 6926.73 | 6649.56 | 18984.9 | 18701.6 | 42865.1 | 42057.9 | 60113.5 | 59556.2 | 787.137 | 1657.44 | 1505.73 | 4300.07 | 4070.73 |
| 787.42 | 2668.76 | 2427.92 | 6988.57 | 6665.48 | 19279.3 | 18738.9 | 42627.5 | 42129.6 | 60173.6 | 59649.8 | 787.42 | 1652.87 | 1510.27 | 4175.52 | 4081.18 |
| 787.702 | 2666.03 | 2434.71 | 6951.29 | 6681.42 | 19345.7 | 18776.2 | 42825.5 | 42201.3 | 60287.7 | 59743.6 | 787.702 | 1625.23 | 1514.81 | 4272.9 | 4091.64 |
| 787.985 | 2609.36 | 2441.51 | 7033.43 | 6697.38 | 19276.4 | 18813.5 | 42622.4 | 42273.1 | 60007.7 | 59837.3 | 787.985 | 1685.06 | 1519.37 | 4286.74 | 4102.11 |
| 788.267 | 2680.96 | 2448.32 | 7003.13 | 6713.36 | 19234.2 | 18851 | 42860.9 | 42344.9 | 60130.7 | 59931.1 | 788.267 | 1588.15 | 1523.93 | 4271.31 | 4112.6 |
| 788.55 | 2692.84 | 2455.15 | 7095.69 | 6729.37 | 19528.9 | 18888.4 | 42739.4 | 42416.7 | 60297.5 | 60025 | 788.55 | 1645.3 | 1528.5 | 4313.36 | 4123.11 |
| 788.832 | 2627.67 | 2461.98 | 7044.51 | 6745.39 | 19224.5 | 18925.8 | 43027.3 | 42488.6 | 60191.8 | 60118.9 | 788.832 | 1618.47 | 1533.09 | 4204.66 | 4133.64 |
| 789.115 | 2630.61 | 2468.83 | 7117.64 | 6761.44 | 19477 | 18963.3 | 43282.3 | 42560.5 | 60819.7 | 60212.8 | 789.115 | 1652.39 | 1537.68 | 4331.81 | 4144.18 |
| 789.397 | 2634.32 | 2475.69 | 7047.87 | 6777.51 | 19537.8 | 19000.9 | 43352.9 | 42632.5 | 60950.3 | 60306.7 | 789.397 | 1676.32 | 1542.28 | 4251.71 | 4154.74 |
| 789.679 | 2689.27 | 2482.56 | 7169.39 | 6793.6 | 19530.6 | 19038.4 | 42819 | 42704.5 | 60884.5 | 60400.7 | 789.679 | 1653.8 | 1546.89 | 4322.41 | 4165.31 |
| 789.962 | 2684.53 | 2489.45 | 7167.98 | 6809.7 | 19408.2 | 19076 | 43582.1 | 42776.6 | 61000.3 | 60494.8 | 789.962 | 1699.79 | 1551.51 | 4402.74 | 4175.9 |
| 790.244 | 2615.2 | 2496.34 | 7128.36 | 6825.84 | 19394.9 | 19113.7 | 42816.7 | 42848.7 | 60467.8 | 60588.9 | 790.244 | 1714.45 | 1556.14 | 4229.13 | 4186.51 |
| 790.526 | 2616.92 | 2503.25 | 7060.23 | 6841.99 | 19905.6 | 19151.3 | 43730.7 | 42920.8 | 61226.8 | 60683 | 790.526 | 1683.02 | 1560.78 | 4330.11 | 4197.13 |
| 790.809 | 2623.22 | 2510.17 | 7159.08 | 6858.16 | 19623.7 | 19189 | 42830.1 | 42993 | 60879.7 | 60777.1 | 790.809 | 1639.42 | 1565.42 | 4374.47 | 4207.77 |
| 791.091 | 2688.58 | 2517.11 | 7009.9 | 6874.36 | 19704 | 19226.7 | 43418.7 | 43065.2 | 60701.7 | 60871.3 | 791.091 | 1750.8 | 1570.08 | 4376.22 | 4218.42 |
| 791.373 | 2678.51 | 2524.05 | 7207.07 | 6890.57 | 19697.5 | 19264.5 | 43506.4 | 43137.4 | 61370.1 | 60965.6 | 791.373 | 1660.2 | 1574.75 | 4427.07 | 4229.1 |
| 791.655 | 2785.66 | 2531.01 | 7153.2 | 6906.81 | 19718.7 | 19302.3 | 43164.4 | 43209.7 | 61009.4 | 61059.8 | 791.655 | 1666.73 | 1579.42 | 4373.22 | 4239.78 |
| 791.938 | 2747.62 | 2537.98 | 7276.67 | 6923.07 | 19687.3 | 19340.1 | 43845.8 | 43282.1 | 61295.9 | 61154.2 | 791.938 | 1707.5 | 1584.11 | 4376.51 | 4250.49 |
| 792.22 | 2780.8 | 2544.96 | 7112.09 | 6939.35 | 19718.8 | 19378 | 43406.7 | 43354.4 | 61037.6 | 61248.5 | 792.22 | 1724.4 | 1588.8 | 4400.38 | 4261.21 |
| 792.502 | 2717.89 | 2551.95 | 7310.67 | 6955.65 | 19897.4 | 19415.9 | 43704.1 | 43426.9 | 61564.6 | 61342.9 | 792.502 | 1636 | 1593.5 | 4452.43 | 4271.94 |
| 792.784 | 2711.7 | 2558.96 | 7438.08 | 6971.97 | 19921.8 | 19453.8 | 43867.6 | 43499.3 | 61324.9 | 61437.3 | 792.784 | 1744.41 | 1598.21 | 4453.26 | 4282.7 |
| 793.067 | 2818.96 | 2565.98 | 7242.26 | 6988.32 | 20146.2 | 19491.8 | 43875.2 | 43571.8 | 62449.4 | 61531.8 | 793.067 | 1684.87 | 1602.94 | 4509.54 | 4293.47 |
| 793.349 | 2721.8 | 2573.01 | 7211.57 | 7004.68 | 19757.3 | 19529.8 | 43579.2 | 43644.4 | 61353.9 | 61626.3 | 793.349 | 1715.14 | 1607.67 | 4411.02 | 4304.25 |
| 793.631 | 2785.07 | 2580.05 | 7225.72 | 7021.07 | 19880.8 | 19567.8 | 43760.9 | 43716.9 | 61397.1 | 61720.8 | 793.631 | 1764.69 | 1612.41 | 4411.93 | 4315.06 |
| 793.913 | 2679.67 | 2587.1 | 7272.7 | 7037.47 | 19849.2 | 19605.8 | 43815.3 | 43789.6 | 61398.7 | 61815.4 | 793.913 | 1784.33 | 1617.16 | 4355.16 | 4325.87 |
| 794.195 | 2811.9 | 2594.17 | 7287.15 | 7053.9 | 19723.2 | 19644 | 43811.4 | 43862.2 | 62100 | 61910 | 794.195 | 1710.04 | 1621.91 | 4479.99 | 4336.71 |
| 794.477 | 2800.27 | 2601.24 | 7266.84 | 7070.35 | 20055.2 | 19682.1 | 44286.7 | 43934.9 | 61744.1 | 62004.7 | 794.477 | 1738.6 | 1626.68 | 4444.43 | 4347.56 |
| 794.76 | 2705.04 | 2608.33 | 7182.88 | 7086.82 | 19835 | 19720.2 | 44246.4 | 44007.6 | 62046 | 62099.3 | 794.76 | 1755.34 | 1631.46 | 4561.43 | 4358.43 |
| 795.042 | 2778.88 | 2615.44 | 7428.92 | 7103.31 | 19910.4 | 19758.4 | 44310.1 | 44080.4 | 61868.1 | 62194.1 | 795.042 | 1660.63 | 1636.25 | 4490.64 | 4369.31 |
| 795.324 | 2770.35 | 2622.55 | 7356.24 | 7119.83 | 20031.2 | 19796.6 | 43803.7 | 44153.2 | 62206.9 | 62288.8 | 795.324 | 1755.54 | 1641.04 | 4609.55 | 4380.21 |
| 795.606 | 2732.18 | 2629.68 | 7422.52 | 7136.36 | 20500.9 | 19834.9 | 44367.1 | 44226.1 | 62388.6 | 62383.6 | 795.606 | 1742.49 | 1645.85 | 4475.84 | 4391.13 |
| 795.888 | 2792.11 | 2636.82 | 7480.91 | 7152.91 | 20373 | 19873.2 | 44844.6 | 44298.9 | 62550.3 | 62478.5 | 795.888 | 1726.46 | 1650.66 | 4533.16 | 4402.06 |
| 796.17 | 2841.15 | 2643.97 | 7470.23 | 7169.49 | 20056.5 | 19911.5 | 44236.8 | 44371.9 | 62066.6 | 62573.3 | 796.17 | 1821.02 | 1655.49 | 4486.56 | 4413.01 |
| 796.452 | 2814.06 | 2651.13 | 7353.7 | 7186.09 | 20103.3 | 19949.9 | 44555.8 | 44444.8 | 62497.1 | 62668.2 | 796.452 | 1762.32 | 1660.32 | 4595.45 | 4423.97 |
| 796.734 | 2810.74 | 2658.3 | 7438.87 | 7202.71 | 20340.2 | 19988.2 | 44732.7 | 44517.8 | 62702.2 | 62763.2 | 796.734 | 1761.42 | 1665.16 | 4626.53 | 4434.95 |
| 797.016 | 2796.65 | 2665.49 | 7480.55 | 7219.35 | 20232 | 20026.7 | 44787.4 | 44590.9 | 62872.2 | 62858.2 | 797.016 | 1778.91 | 1670.02 | 4642.96 | 4445.95 |
| 797.298 | 2884.72 | 2672.69 | 7455.21 | 7236.01 | 20340.9 | 20065.1 | 44712.8 | 44664 | 62362.7 | 62953.2 | 797.298 | 1797.94 | 1674.88 | 4543.81 | 4456.97 |
| 797.58 | 2857.63 | 2679.9 | 7546.82 | 7252.69 | 20632.9 | 20103.6 | 45010.6 | 44737.1 | 63083.1 | 63048.2 | 797.58 | 1789.23 | 1679.75 | 4649.47 | 4468 |
| 797.862 | 2880.37 | 2687.12 | 7341.1 | 7269.39 | 20432.7 | 20142.1 | 44884 | 44810.3 | 63290.9 | 63143.3 | 797.862 | 1725.72 | 1684.63 | 4633.9 | 4479.04 |
| 798.144 | 2855.09 | 2694.36 | 7471.91 | 7286.11 | 20486.5 | 20180.7 | 45599.9 | 44883.4 | 63921.8 | 63238.4 | 798.144 | 1802.89 | 1689.52 | 4752.13 | 4490.11 |
| 798.426 | 2945.1 | 2701.61 | 7481.55 | 7302.86 | 20040.2 | 20219.2 | 45060.1 | 44956.7 | 63178.5 | 63333.6 | 798.426 | 1821.02 | 1694.42 | 4717.49 | 4501.19 |
| 798.707 | 2877.99 | 2708.87 | 7633.75 | 7319.62 | 20700.5 | 20257.8 | 45476.1 | 45030 | 63511.4 | 63428.8 | 798.707 | 1826.05 | 1699.33 | 4576.1 | 4512.28 |
| 798.989 | 2802.58 | 2716.14 | 7670.38 | 7336.41 | 20399.6 | 20296.5 | 45054.9 | 45103.3 | 63413.9 | 63524 | 798.989 | 1830.26 | 1704.24 | 4580.15 | 4523.39 |
| 799.271 | 2839.27 | 2723.42 | 7548.3 | 7353.22 | 20686.4 | 20335.2 | 44960.8 | 45176.6 | 63247.1 | 63619.3 | 799.271 | 1774.96 | 1709.17 | 4634.04 | 4534.52 |
| 799.553 | 2932.02 | 2730.72 | 7609.28 | 7370.04 | 20590.3 | 20373.8 | 45077.1 | 45250 | 63528.9 | 63714.6 | 799.553 | 1863.12 | 1714.11 | 4698.73 | 4545.66 |
| 799.835 | 2884.86 | 2738.03 | 7587.47 | 7386.89 | 20566 | 20412.6 | 45171.4 | 45323.4 | 63322.1 | 63809.9 | 799.835 | 1854.95 | 1719.05 | 4675.23 | 4556.82 |
| 800.117 | 2905.95 | 2745.35 | 7647.6 | 7403.76 | 20967.5 | 20451.4 | 45626.1 | 45396.9 | 63945.8 | 63905.2 | 800.117 | 1839.65 | 1724.01 | 4774.53 | 4568 |
| 800.399 | 2856.05 | 2752.68 | 7578.41 | 7420.65 | 20895.8 | 20490.2 | 45637.7 | 45470.4 | 63823.9 | 64000.6 | 800.399 | 1863.2 | 1728.97 | 4720.57 | 4579.19 |
| 800.68 | 2852.87 | 2760.03 | 7546.73 | 7437.57 | 20825.6 | 20529 | 45281.3 | 45543.9 | 64518.8 | 64096.1 | 800.68 | 1919.98 | 1733.95 | 4770.61 | 4590.4 |
| 800.962 | 3016.98 | 2767.38 | 7729.37 | 7454.5 | 20727.1 | 20567.9 | 46171.5 | 45617.5 | 63681.5 | 64191.5 | 800.962 | 1815.44 | 1738.93 | 4662.16 | 4601.63 |
| 801.244 | 2847.61 | 2774.75 | 7709.35 | 7471.45 | 20721.5 | 20606.8 | 45600 | 45691.1 | 64334 | 64287 | 801.244 | 1921.82 | 1743.92 | 4680.93 | 4612.87 |
| 801.525 | 2991.26 | 2782.13 | 7706.88 | 7488.43 | 20684.9 | 20645.7 | 45828.2 | 45764.7 | 64305.8 | 64382.6 | 801.525 | 1850.24 | 1748.93 | 4838.43 | 4624.13 |
| 801.807 | 2963.46 | 2789.53 | 7797.2 | 7505.42 | 20834.1 | 20684.6 | 46250.8 | 45838.4 | 64208 | 64478.1 | 801.807 | 1830 | 1753.94 | 4694.2 | 4635.4 |
| 802.089 | 2934.87 | 2796.93 | 7682.75 | 7522.44 | 20915.2 | 20723.6 | 45295.7 | 45912.1 | 63764.3 | 64573.7 | 802.089 | 1780.39 | 1758.96 | 4802.99 | 4646.7 |
| 802.371 | 2927.26 | 2804.35 | 7846.74 | 7539.48 | 21032.4 | 20762.7 | 46219.3 | 45985.9 | 64706.9 | 64669.4 | 802.371 | 1842.92 | 1763.99 | 4863.01 | 4658 |
| 802.652 | 2970.26 | 2811.78 | 7851.53 | 7556.54 | 21124.7 | 20801.7 | 46015.7 | 46059.7 | 64738.2 | 64765 | 802.652 | 1885.14 | 1769.03 | 4803.79 | 4669.33 |
| 802.934 | 3061.68 | 2819.22 | 7817.05 | 7573.61 | 21016.3 | 20840.8 | 45977.8 | 46133.5 | 64729.5 | 64860.7 | 802.934 | 1932.86 | 1774.08 | 4808.53 | 4680.67 |
| 803.216 | 2915.9 | 2826.68 | 8014.05 | 7590.72 | 21147.8 | 20879.9 | 46831.4 | 46207.4 | 64864.8 | 64956.5 | 803.216 | 1869.96 | 1779.14 | 4929.16 | 4692.02 |
| 803.497 | 2941.09 | 2834.14 | 7859.38 | 7607.84 | 20943.8 | 20919 | 46151.7 | 46281.3 | 65011.9 | 65052.2 | 803.497 | 1914.16 | 1784.21 | 4847.02 | 4703.4 |
| 803.779 | 2952.59 | 2841.62 | 7790.06 | 7624.98 | 20782.5 | 20958.2 | 46119.1 | 46355.2 | 64546 | 65148 | 803.779 | 1861.14 | 1789.29 | 4875.06 | 4714.78 |
| 804.06 | 3054.08 | 2849.11 | 8011.4 | 7642.14 | 21409.9 | 20997.4 | 46537 | 46429.2 | 65078 | 65243.9 | 804.06 | 1932.13 | 1794.38 | 4884.85 | 4726.19 |
| 804.342 | 2914.37 | 2856.61 | 7934.28 | 7659.32 | 20932.2 | 21036.7 | 46612.6 | 46503.2 | 64770.1 | 65339.7 | 804.342 | 1934.93 | 1799.48 | 4865.34 | 4737.61 |
| 804.624 | 3058.67 | 2864.13 | 7943 | 7676.53 | 21580.2 | 21076 | 46950.4 | 46577.2 | 64883.1 | 65435.6 | 804.624 | 1863.32 | 1804.58 | 4859.28 | 4749.05 |
| 804.905 | 3084.82 | 2871.66 | 7881.45 | 7693.75 | 20981.1 | 21115.2 | 46640.8 | 46651.3 | 65865.2 | 65531.6 | 804.905 | 1873.66 | 1809.7 | 4884.68 | 4760.5 |
| 805.187 | 3087.13 | 2879.2 | 7933.89 | 7711 | 21262.7 | 21154.6 | 46736.1 | 46725.4 | 65878.4 | 65627.5 | 805.187 | 1872.26 | 1814.83 | 4940.81 | 4771.97 |
| 805.468 | 2952.47 | 2886.75 | 7862.93 | 7728.26 | 21275.2 | 21194 | 46805.7 | 46799.6 | 66105.7 | 65723.5 | 805.468 | 1947.76 | 1819.96 | 4931.34 | 4783.46 |
| 805.75 | 3033.88 | 2894.31 | 8023.35 | 7745.55 | 21415.1 | 21233.3 | 46701.4 | 46873.8 | 65353.2 | 65819.5 | 805.75 | 1977.02 | 1825.11 | 4926.16 | 4794.96 |
| 806.031 | 3075.12 | 2901.89 | 8090.8 | 7762.86 | 21657 | 21272.8 | 47154.4 | 46948 | 66195.6 | 65915.6 | 806.031 | 1904.33 | 1830.26 | 5046.31 | 4806.48 |
| 806.313 | 3118.13 | 2909.47 | 7920.89 | 7780.19 | 21430.4 | 21312.2 | 47283.4 | 47022.2 | 65107.1 | 66011.7 | 806.313 | 1883.32 | 1835.42 | 4922.7 | 4818.01 |
| 806.594 | 3031.37 | 2917.07 | 7994.35 | 7797.53 | 21607.8 | 21351.7 | 47146.7 | 47096.5 | 65973.9 | 66107.8 | 806.594 | 1899.03 | 1840.6 | 4981.13 | 4829.56 |
| 806.875 | 3098.27 | 2924.69 | 7942.24 | 7814.9 | 21528.3 | 21391.2 | 47259.5 | 47170.8 | 65747 | 66203.9 | 806.875 | 1932.22 | 1845.78 | 4950.99 | 4841.13 |
| 807.157 | 3071.35 | 2932.31 | 7968.31 | 7832.29 | 21508.6 | 21430.8 | 46715.5 | 47245.2 | 65923.4 | 66300.1 | 807.157 | 2019.36 | 1850.97 | 4863.11 | 4852.72 |
| 807.438 | 3082.73 | 2939.95 | 8044.88 | 7849.7 | 21896.2 | 21470.4 | 47564.7 | 47319.6 | 65822.2 | 66396.3 | 807.438 | 1975.76 | 1856.18 | 5044.83 | 4864.32 |
| 807.72 | 3039.85 | 2947.6 | 8198.04 | 7867.14 | 21761.8 | 21510 | 47174 | 47394 | 65874.8 | 66492.6 | 807.72 | 1953.87 | 1861.39 | 5036.25 | 4875.93 |
| 808.001 | 3032.04 | 2955.26 | 8054.71 | 7884.59 | 21686.6 | 21549.6 | 47271.5 | 47468.5 | 66123.1 | 66588.8 | 808.001 | 2037.62 | 1866.61 | 4995.33 | 4887.56 |
| 808.283 | 3122.45 | 2962.93 | 8173.67 | 7902.06 | 21568.7 | 21589.3 | 46939.2 | 47543 | 65926.6 | 66685.1 | 808.283 | 1932.37 | 1871.84 | 5058.57 | 4899.21 |
| 808.564 | 3150.14 | 2970.62 | 8150.98 | 7919.55 | 21835 | 21629 | 47595.2 | 47617.5 | 66394.3 | 66781.5 | 808.564 | 1969.55 | 1877.08 | 5062.31 | 4910.88 |
| 808.845 | 3253.36 | 2978.32 | 8134 | 7937.07 | 21744.9 | 21668.7 | 47336.1 | 47692.1 | 65770 | 66877.8 | 808.845 | 1964.5 | 1882.33 | 4896.7 | 4922.56 |
| 809.126 | 3138.6 | 2986.03 | 8198.24 | 7954.6 | 21805.6 | 21708.5 | 47496.8 | 47766.7 | 66334.7 | 66974.2 | 809.126 | 1990.24 | 1887.59 | 5052.46 | 4934.26 |
| 809.408 | 3137.74 | 2993.75 | 8144.85 | 7972.16 | 21810.2 | 21748.3 | 47720.1 | 47841.3 | 66198.6 | 67070.6 | 809.408 | 1949.53 | 1892.86 | 5078.56 | 4945.97 |
| 809.689 | 3183.24 | 3001.48 | 8129.83 | 7989.73 | 21941.8 | 21788.1 | 47532.2 | 47916 | 66483.1 | 67167.1 | 809.689 | 1992.33 | 1898.14 | 5108.64 | 4957.7 |
| 809.97 | 3144.34 | 3009.23 | 8197.77 | 8007.33 | 21713 | 21827.9 | 47820.2 | 47990.7 | 66728.1 | 67263.6 | 809.97 | 1968.71 | 1903.43 | 5099.45 | 4969.45 |
| 810.252 | 3157.01 | 3016.99 | 8022.85 | 8024.94 | 22029 | 21867.8 | 47872 | 48065.4 | 67214.4 | 67360.1 | 810.252 | 2002.48 | 1908.73 | 5087.69 | 4981.21 |
| 810.533 | 3167.55 | 3024.76 | 8265.2 | 8042.58 | 21838.5 | 21907.7 | 47668.1 | 48140.2 | 66752.8 | 67456.6 | 810.533 | 1933.93 | 1914.04 | 5095.99 | 4992.99 |
| 810.814 | 3186.47 | 3032.54 | 8223.87 | 8060.24 | 22069.2 | 21947.7 | 47551.9 | 48215 | 66301.4 | 67553.2 | 810.814 | 2035.24 | 1919.35 | 5126.62 | 5004.78 |
| 811.095 | 3049.01 | 3040.34 | 8166.43 | 8077.92 | 22024.4 | 21987.6 | 47672.8 | 48289.8 | 66992.8 | 67649.8 | 811.095 | 1945.46 | 1924.68 | 5079.55 | 5016.59 |
| 811.376 | 3094.42 | 3048.15 | 8078.08 | 8095.62 | 22035.3 | 22027.6 | 47498.7 | 48364.7 | 66292.6 | 67746.4 | 811.376 | 2005.32 | 1930.02 | 5140.12 | 5028.42 |
| 811.658 | 3161.68 | 3055.97 | 8266.2 | 8113.34 | 21905.1 | 22067.7 | 48199.8 | 48439.6 | 66921.3 | 67843.1 | 811.658 | 2006.56 | 1935.36 | 5051.01 | 5040.27 |
| 811.939 | 3203.99 | 3063.8 | 8108.99 | 8131.07 | 21796.3 | 22107.7 | 47741.4 | 48514.5 | 66974.6 | 67939.8 | 811.939 | 2062.07 | 1940.72 | 5150.26 | 5052.12 |
| 812.22 | 3217.01 | 3071.64 | 8390.33 | 8148.84 | 21900.2 | 22147.8 | 47903 | 48589.5 | 66748.8 | 68036.5 | 812.22 | 2053.09 | 1946.08 | 5195.07 | 5064 |
| 812.501 | 3167.3 | 3079.5 | 8263.32 | 8166.61 | 22180.5 | 22187.9 | 47861.8 | 48664.5 | 67071.1 | 68133.2 | 812.501 | 2079.82 | 1951.46 | 5180.64 | 5075.89 |
| 812.782 | 3206.44 | 3087.37 | 8233.82 | 8184.42 | 22155.2 | 22228.1 | 48082.2 | 48739.5 | 66986.3 | 68230 | 812.782 | 2003.36 | 1956.84 | 5147.77 | 5087.8 |
| 813.063 | 3240.86 | 3095.25 | 8423.37 | 8202.24 | 22478.9 | 22268.3 | 48173.4 | 48814.6 | 67328 | 68326.8 | 813.063 | 2015.76 | 1962.24 | 5195.62 | 5099.73 |
| 813.344 | 3149.44 | 3103.14 | 8420.82 | 8220.08 | 22385.4 | 22308.5 | 48094.9 | 48889.7 | 67318.9 | 68423.6 | 813.344 | 2005.99 | 1967.64 | 5104.27 | 5111.67 |
| 813.625 | 3209.27 | 3111.05 | 8203 | 8237.94 | 22510 | 22348.7 | 48541.6 | 48964.8 | 67729.4 | 68520.5 | 813.625 | 2084.93 | 1973.06 | 5311.74 | 5123.62 |
| 813.907 | 3313.97 | 3118.96 | 8415.77 | 8255.82 | 22473.2 | 22389 | 48645.2 | 49039.9 | 67503.2 | 68617.4 | 813.907 | 2053.76 | 1978.48 | 5138.76 | 5135.6 |
| 814.187 | 3248.79 | 3126.89 | 8591.11 | 8273.73 | 22524.6 | 22429.3 | 48788 | 49115.2 | 67753.6 | 68714.3 | 814.187 | 2049.76 | 1983.91 | 5221.84 | 5147.59 |
| 814.468 | 3214.41 | 3134.84 | 8538.27 | 8291.65 | 22295.5 | 22469.6 | 48616.1 | 49190.4 | 67705.5 | 68811.2 | 814.468 | 2069.45 | 1989.35 | 5255.77 | 5159.59 |
| 814.75 | 3279.06 | 3142.79 | 8469.12 | 8309.59 | 22646.5 | 22510 | 48625.1 | 49265.6 | 67373.4 | 68908.2 | 814.75 | 2089.7 | 1994.81 | 5246.72 | 5171.61 |
| 815.03 | 3290.16 | 3150.76 | 8415.29 | 8327.56 | 22539.9 | 22550.3 | 48877.9 | 49340.9 | 67675.5 | 69005.2 | 815.03 | 2096.41 | 2000.27 | 5308.14 | 5183.65 |
| 815.312 | 3295.11 | 3158.73 | 8377.78 | 8345.54 | 22576 | 22590.8 | 48714.9 | 49416.2 | 68429.2 | 69102.2 | 815.312 | 2150.36 | 2005.74 | 5248.98 | 5195.7 |
| 815.592 | 3338.57 | 3166.72 | 8418.52 | 8363.54 | 22668.7 | 22631.2 | 48819.7 | 49491.6 | 67964 | 69199.2 | 815.592 | 2069.21 | 2011.22 | 5268.26 | 5207.77 |
| 815.873 | 3299.39 | 3174.73 | 8654.16 | 8381.57 | 22554.9 | 22671.7 | 48937.9 | 49567 | 68478.2 | 69296.3 | 815.873 | 2114.15 | 2016.71 | 5319.43 | 5219.86 |
| 816.154 | 3362.23 | 3182.74 | 8636.1 | 8399.61 | 22707.8 | 22712.2 | 48862.5 | 49642.4 | 69112.3 | 69393.4 | 816.154 | 2126.53 | 2022.21 | 5383.31 | 5231.96 |
| 816.435 | 3297.39 | 3190.77 | 8652.78 | 8417.68 | 22816.6 | 22752.7 | 49650.8 | 49717.8 | 68839.1 | 69490.5 | 816.435 | 2157.42 | 2027.72 | 5287.31 | 5244.08 |
| 816.716 | 3362.95 | 3198.81 | 8485.87 | 8435.76 | 22809.9 | 22793.3 | 49468.3 | 49793.3 | 68515.3 | 69587.7 | 816.716 | 2144.79 | 2033.24 | 5322.95 | 5256.22 |
| 816.997 | 3367.5 | 3206.86 | 8445.69 | 8453.87 | 22528.1 | 22833.9 | 49431.7 | 49868.8 | 68674 | 69684.9 | 816.997 | 2184.8 | 2038.77 | 5329.44 | 5268.37 |
| 817.278 | 3353.48 | 3214.93 | 8739.77 | 8471.99 | 22956.9 | 22874.5 | 49468.4 | 49944.4 | 69094.3 | 69782.1 | 817.278 | 2105.11 | 2044.31 | 5374.45 | 5280.54 |
| 817.559 | 3355.13 | 3223 | 8709.56 | 8490.14 | 22956.6 | 22915.1 | 50093.6 | 50019.9 | 68909.8 | 69879.3 | 817.559 | 2064.52 | 2049.86 | 5330.99 | 5292.72 |
| 817.84 | 3420.87 | 3231.09 | 8695.75 | 8508.3 | 23045.8 | 22955.8 | 50503.4 | 50095.6 | 69249.6 | 69976.6 | 817.84 | 2097.26 | 2055.42 | 5373.94 | 5304.92 |
| 818.12 | 3399.2 | 3239.19 | 8734.8 | 8526.49 | 23038.3 | 22996.5 | 49734.2 | 50171.2 | 69323.4 | 70073.8 | 818.12 | 2138.16 | 2060.99 | 5352 | 5317.14 |
| 818.401 | 3363.22 | 3247.3 | 8755.26 | 8544.69 | 23164.1 | 23037.2 | 49866.8 | 50246.8 | 69375.8 | 70171.1 | 818.401 | 2131.07 | 2066.57 | 5513.82 | 5329.37 |
| 818.682 | 3401.52 | 3255.43 | 8779.95 | 8562.92 | 23503.5 | 23078 | 49939.3 | 50322.5 | 69613.8 | 70268.5 | 818.682 | 2193.87 | 2072.15 | 5382.56 | 5341.61 |
| 818.963 | 3360.39 | 3263.57 | 8711.19 | 8581.17 | 22987.1 | 23118.8 | 49923.7 | 50398.3 | 69540.7 | 70365.8 | 818.963 | 2154.6 | 2077.75 | 5525.99 | 5353.88 |
| 819.243 | 3388.72 | 3271.71 | 8599 | 8599.43 | 23187.7 | 23159.6 | 49926.8 | 50474 | 69441.3 | 70463.2 | 819.243 | 2174.84 | 2083.36 | 5451.5 | 5366.16 |
| 819.524 | 3344.01 | 3279.87 | 8758.48 | 8617.72 | 23243.7 | 23200.5 | 50162.2 | 50549.8 | 69693 | 70560.6 | 819.524 | 2183.71 | 2088.98 | 5415.34 | 5378.45 |
| 819.805 | 3382.14 | 3288.05 | 8782.47 | 8636.03 | 23161.4 | 23241.3 | 50353.4 | 50625.6 | 70180 | 70658.1 | 819.805 | 2166.45 | 2094.6 | 5450.93 | 5390.77 |
| 820.086 | 3482.15 | 3296.24 | 8806.37 | 8654.35 | 23177.5 | 23282.2 | 50096.2 | 50701.5 | 69439.4 | 70755.5 | 820.086 | 2154.31 | 2100.24 | 5373.76 | 5403.1 |
| 820.366 | 3501.17 | 3304.43 | 8927.44 | 8672.7 | 23420.4 | 23323.2 | 50507 | 50777.3 | 70304.3 | 70853 | 820.366 | 2181.89 | 2105.89 | 5544.72 | 5415.44 |
| 820.647 | 3484.54 | 3312.64 | 8885.54 | 8691.07 | 23178.7 | 23364.1 | 50245.1 | 50853.2 | 70384.9 | 70950.5 | 820.647 | 2180.18 | 2111.54 | 5473.57 | 5427.8 |
| 820.928 | 3401.51 | 3320.87 | 8683.16 | 8709.45 | 23460.5 | 23405.1 | 50369.7 | 50929.2 | 70523.7 | 71048.1 | 820.928 | 2186.85 | 2117.21 | 5410.46 | 5440.18 |
| 821.208 | 3549.18 | 3329.1 | 9029.9 | 8727.86 | 23576.3 | 23446.1 | 51019.6 | 51005.1 | 71039.1 | 71145.6 | 821.208 | 2209.54 | 2122.88 | 5662.32 | 5452.57 |
| 821.489 | 3502.91 | 3337.35 | 8868.43 | 8746.28 | 23587.8 | 23487.2 | 50606.7 | 51081.1 | 70059.5 | 71243.2 | 821.489 | 2159.26 | 2128.57 | 5515.08 | 5464.98 |
| 821.769 | 3390.36 | 3345.61 | 8890.24 | 8764.73 | 23840.4 | 23528.2 | 51234.8 | 51157.2 | 70846.3 | 71340.8 | 821.769 | 2280.82 | 2134.26 | 5625.98 | 5477.41 |
| 822.05 | 3464.16 | 3353.88 | 8934.53 | 8783.2 | 23643 | 23569.3 | 50563.4 | 51233.2 | 70985.2 | 71438.4 | 822.05 | 2192.11 | 2139.97 | 5611.28 | 5489.85 |
| 822.331 | 3463.37 | 3362.16 | 8918.81 | 8801.68 | 23720.4 | 23610.5 | 51116.1 | 51309.3 | 70471.9 | 71536.1 | 822.331 | 2261.71 | 2145.68 | 5679.13 | 5502.3 |
| 822.611 | 3502.01 | 3370.46 | 9012.68 | 8820.19 | 23743.5 | 23651.6 | 50808.1 | 51385.4 | 71628.5 | 71633.8 | 822.611 | 2289.95 | 2151.41 | 5735.05 | 5514.78 |
| 822.892 | 3490.2 | 3378.76 | 8996.28 | 8838.72 | 23888.5 | 23692.8 | 51186.4 | 51461.6 | 71409.1 | 71731.4 | 822.892 | 2175.35 | 2157.14 | 5630.15 | 5527.27 |
| 823.172 | 3549.66 | 3387.09 | 9007.89 | 8857.26 | 23664.8 | 23734 | 51152.9 | 51537.7 | 71035.9 | 71829.2 | 823.172 | 2277.58 | 2162.89 | 5601.73 | 5539.77 |
| 823.453 | 3534.74 | 3395.42 | 8932.61 | 8875.83 | 23811.8 | 23775.3 | 51448.9 | 51613.9 | 71830.9 | 71926.9 | 823.453 | 2267.85 | 2168.64 | 5683.45 | 5552.29 |
| 823.733 | 3546.04 | 3403.76 | 9194.71 | 8894.41 | 23831.3 | 23816.5 | 51443.1 | 51690.1 | 72078.6 | 72024.7 | 823.733 | 2286.86 | 2174.4 | 5702.46 | 5564.83 |
| 824.014 | 3571.24 | 3412.12 | 9156.89 | 8913.02 | 23712.8 | 23857.8 | 51925.3 | 51766.4 | 71345.9 | 72122.5 | 824.014 | 2251.3 | 2180.18 | 5741.87 | 5577.39 |
| 824.294 | 3595.38 | 3420.49 | 8909.79 | 8931.64 | 23697.7 | 23899.1 | 51134.9 | 51842.7 | 71461.5 | 72220.3 | 824.294 | 2353.09 | 2185.96 | 5658.52 | 5589.96 |
| 824.574 | 3505.7 | 3428.87 | 9242.04 | 8950.29 | 23928 | 23940.5 | 51486.2 | 51919 | 71911.7 | 72318.1 | 824.574 | 2194.64 | 2191.75 | 5746.89 | 5602.54 |
| 824.855 | 3611.16 | 3437.26 | 9124.84 | 8968.95 | 24045.3 | 23981.9 | 51596.6 | 51995.3 | 71645.4 | 72416 | 824.855 | 2283.51 | 2197.55 | 5602.59 | 5615.14 |
| 825.135 | 3616.75 | 3445.67 | 9235.61 | 8987.64 | 24230.9 | 24023.3 | 52030.5 | 52071.7 | 71819.9 | 72513.9 | 825.135 | 2297.48 | 2203.37 | 5826.55 | 5627.76 |
| 825.416 | 3698.88 | 3454.09 | 9274.7 | 9006.34 | 24230.1 | 24064.7 | 52643.5 | 52148.1 | 72673.2 | 72611.8 | 825.416 | 2254.75 | 2209.19 | 5728.13 | 5640.4 |
| 825.696 | 3607.55 | 3462.51 | 9173.42 | 9025.07 | 24382.1 | 24106.2 | 51892.4 | 52224.5 | 72410.8 | 72709.7 | 825.696 | 2281.23 | 2215.02 | 5718.47 | 5653.05 |
| 825.976 | 3626.56 | 3470.96 | 9183.01 | 9043.81 | 24279.4 | 24147.6 | 51870.7 | 52301 | 72402.7 | 72807.7 | 825.976 | 2333.8 | 2220.86 | 5811.31 | 5665.71 |
| 826.257 | 3604.58 | 3479.41 | 9210.62 | 9062.57 | 24050.8 | 24189.2 | 52429.6 | 52377.4 | 72710.8 | 72905.6 | 826.257 | 2344.59 | 2226.71 | 5764.4 | 5678.39 |
| 826.537 | 3631.58 | 3487.88 | 9178.39 | 9081.36 | 24221.2 | 24230.7 | 52014.9 | 52453.9 | 71733.5 | 73003.6 | 826.537 | 2265.67 | 2232.57 | 5734.6 | 5691.09 |
| 826.817 | 3577.19 | 3496.36 | 9040.9 | 9100.16 | 24302.6 | 24272.3 | 52177.8 | 52530.5 | 71889.7 | 73101.7 | 826.817 | 2317.81 | 2238.45 | 5764.55 | 5703.81 |
| 827.098 | 3728.5 | 3504.85 | 9374.99 | 9118.98 | 24858.7 | 24313.9 | 52276.9 | 52607.1 | 72808.7 | 73199.7 | 827.098 | 2381.63 | 2244.33 | 5848.45 | 5716.54 |
| 827.378 | 3624.12 | 3513.35 | 9343.13 | 9137.83 | 24455.2 | 24355.5 | 52419.7 | 52683.6 | 73031.8 | 73297.8 | 827.378 | 2406.43 | 2250.22 | 5805.76 | 5729.28 |
| 827.658 | 3650.8 | 3521.86 | 9242.3 | 9156.69 | 24174.1 | 24397.1 | 51692.6 | 52760.3 | 72264.6 | 73395.8 | 827.658 | 2389.16 | 2256.12 | 5858.26 | 5742.04 |
| 827.939 | 3593.3 | 3530.39 | 9310.15 | 9175.57 | 24317.8 | 24438.8 | 52378.9 | 52836.9 | 72758.8 | 73493.9 | 827.939 | 2365.99 | 2262.03 | 5765.68 | 5754.82 |
| 828.219 | 3702.73 | 3538.93 | 9377.48 | 9194.47 | 24568.7 | 24480.5 | 52588.1 | 52913.6 | 72882 | 73592.1 | 828.219 | 2309.11 | 2267.95 | 5964.41 | 5767.62 |
| 828.499 | 3663.03 | 3547.48 | 9275.12 | 9213.39 | 24477.2 | 24522.2 | 52201.6 | 52990.3 | 72080.7 | 73690.2 | 828.499 | 2351.55 | 2273.88 | 5919.44 | 5780.43 |
| 828.779 | 3594.7 | 3556.05 | 9318.09 | 9232.33 | 24639 | 24564 | 52458.2 | 53067 | 73088.8 | 73788.4 | 828.779 | 2397.01 | 2279.82 | 5907.49 | 5793.25 |
| 829.059 | 3770.64 | 3564.62 | 9287.53 | 9251.29 | 24623.1 | 24605.8 | 52943.3 | 53143.8 | 72615.6 | 73886.6 | 829.059 | 2361.66 | 2285.77 | 5862.25 | 5806.09 |
| 829.34 | 3732.47 | 3573.21 | 9473.94 | 9270.27 | 24760.1 | 24647.6 | 52827.6 | 53220.5 | 73575.6 | 73984.8 | 829.34 | 2366.04 | 2291.73 | 5885.85 | 5818.95 |
| 829.62 | 3638.57 | 3581.81 | 9449.19 | 9289.27 | 24858.8 | 24689.4 | 53153.8 | 53297.3 | 74071.8 | 74083 | 829.62 | 2383.5 | 2297.7 | 5923.54 | 5831.83 |
| 829.9 | 3724.63 | 3590.42 | 9425.18 | 9308.29 | 24737.3 | 24731.3 | 53018.7 | 53374.2 | 73580.1 | 74181.2 | 829.9 | 2388.6 | 2303.68 | 5995.41 | 5844.72 |
| 830.18 | 3680.85 | 3599.05 | 9412.75 | 9327.32 | 24911.4 | 24773.2 | 53100.2 | 53451 | 73168.1 | 74279.5 | 830.18 | 2367.88 | 2309.66 | 5948.97 | 5857.62 |
| 830.46 | 3851.19 | 3607.68 | 9640.91 | 9346.38 | 24782.2 | 24815.1 | 53470.2 | 53527.9 | 73967.2 | 74377.8 | 830.46 | 2369.08 | 2315.66 | 6051.84 | 5870.54 |
| 830.74 | 3746.71 | 3616.33 | 9415.57 | 9365.46 | 25143.4 | 24857 | 53526.6 | 53604.8 | 73877.2 | 74476.1 | 830.74 | 2405.51 | 2321.67 | 5921.25 | 5883.48 |
| 831.02 | 3757.21 | 3624.99 | 9619.88 | 9384.55 | 25012 | 24899 | 53738.7 | 53681.8 | 74402.4 | 74574.4 | 831.02 | 2486.57 | 2327.69 | 6026.1 | 5896.43 |
| 831.3 | 3807.32 | 3633.67 | 9564.72 | 9403.67 | 25398.4 | 24941 | 53665.8 | 53758.7 | 74184.9 | 74672.8 | 831.3 | 2428.99 | 2333.72 | 6095.56 | 5909.4 |
| 831.58 | 3789.01 | 3642.35 | 9805.23 | 9422.81 | 24933.1 | 24983 | 53593 | 53835.7 | 74535.7 | 74771.1 | 831.58 | 2439.22 | 2339.76 | 6072.33 | 5922.39 |
| 831.86 | 3762.58 | 3651.05 | 9659.26 | 9441.96 | 24911.2 | 25025 | 53539.4 | 53912.7 | 74067 | 74869.5 | 831.86 | 2431.65 | 2345.81 | 6019.99 | 5935.39 |
| 832.14 | 3844.71 | 3659.76 | 9600.23 | 9461.13 | 24773.2 | 25067.1 | 53375.7 | 53989.8 | 73645.4 | 74967.9 | 832.14 | 2420.04 | 2351.86 | 6046.45 | 5948.41 |
| 832.42 | 3919.03 | 3668.48 | 9709.78 | 9480.33 | 24843.9 | 25109.2 | 53613.6 | 54066.8 | 74533.5 | 75066.3 | 832.42 | 2438.94 | 2357.93 | 6044.04 | 5961.44 |
| 832.7 | 3750.22 | 3677.22 | 9589.71 | 9499.54 | 24960.9 | 25151.3 | 53627.7 | 54143.9 | 74389.6 | 75164.8 | 832.7 | 2441.82 | 2364.01 | 6001.43 | 5974.49 |
| 832.98 | 3785.69 | 3685.96 | 9635.22 | 9518.77 | 25108.9 | 25193.5 | 53825.3 | 54221 | 74749.1 | 75263.2 | 832.98 | 2503.86 | 2370.1 | 6055.45 | 5987.56 |
| 833.26 | 3886.61 | 3694.72 | 9738.38 | 9538.02 | 25298.4 | 25235.7 | 54166.7 | 54298.2 | 74739 | 75361.7 | 833.26 | 2451.93 | 2376.2 | 6177.49 | 6000.64 |
| 833.54 | 3797.76 | 3703.49 | 9500.75 | 9557.29 | 25347.3 | 25277.9 | 53840.6 | 54375.3 | 74411.3 | 75460.2 | 833.54 | 2488.47 | 2382.3 | 6078.71 | 6013.73 |
| 833.82 | 3869.28 | 3712.27 | 9765.86 | 9576.58 | 25233.6 | 25320.1 | 53970.2 | 54452.5 | 74800.7 | 75558.7 | 833.82 | 2457.28 | 2388.42 | 6130.04 | 6026.85 |
| 834.1 | 3878.68 | 3721.07 | 9785.87 | 9595.89 | 25167.3 | 25362.4 | 53407 | 54529.7 | 74687.3 | 75657.2 | 834.1 | 2466.17 | 2394.55 | 5968.05 | 6039.97 |
| 834.38 | 3879.64 | 3729.87 | 9705.55 | 9615.22 | 25315.1 | 25404.6 | 54711.7 | 54607 | 75301.5 | 75755.8 | 834.38 | 2492.03 | 2400.68 | 6201.08 | 6053.12 |
| 834.66 | 3779.8 | 3738.69 | 9705.51 | 9634.57 | 25398.3 | 25446.9 | 53497 | 54684.2 | 74652.7 | 75854.4 | 834.66 | 2452.17 | 2406.83 | 6115.55 | 6066.28 |
| 834.939 | 3895.34 | 3747.52 | 9662.14 | 9653.93 | 25022.3 | 25489.3 | 54006.1 | 54761.5 | 74514.7 | 75953 | 834.939 | 2481.34 | 2412.99 | 6096.52 | 6079.45 |
| 835.219 | 3919.12 | 3756.37 | 9729.49 | 9673.32 | 25104.9 | 25531.6 | 53912 | 54838.8 | 75278.2 | 76051.6 | 835.219 | 2542.18 | 2419.15 | 6179.02 | 6092.65 |
| 835.499 | 3842.66 | 3765.22 | 9723.48 | 9692.72 | 25650.6 | 25574 | 54498.8 | 54916.2 | 75643.4 | 76150.2 | 835.499 | 2524.44 | 2425.33 | 6204.19 | 6105.85 |
| 835.779 | 3898.68 | 3774.09 | 9760.37 | 9712.15 | 25527.7 | 25616.4 | 54452.8 | 54993.5 | 74938.1 | 76248.8 | 835.779 | 2456.99 | 2431.52 | 6129.84 | 6119.08 |
| 836.059 | 3810.23 | 3782.97 | 9913.12 | 9731.59 | 25632.3 | 25658.8 | 54297.3 | 55070.9 | 75941.6 | 76347.5 | 836.059 | 2462.09 | 2437.71 | 6249.58 | 6132.32 |
| 836.338 | 3892.23 | 3791.86 | 9908.11 | 9751.05 | 25452.1 | 25701.3 | 54372.5 | 55148.3 | 75721.6 | 76446.2 | 836.338 | 2513.58 | 2443.92 | 6239.27 | 6145.57 |
| 836.618 | 3989.28 | 3800.76 | 9879.97 | 9770.53 | 25642.7 | 25743.8 | 54639.3 | 55225.8 | 75540.2 | 76544.9 | 836.618 | 2493.72 | 2450.14 | 6116.21 | 6158.84 |
| 836.898 | 3968.31 | 3809.68 | 9918.69 | 9790.03 | 25989.5 | 25786.3 | 54885.4 | 55303.2 | 75874.2 | 76643.6 | 836.898 | 2528.03 | 2456.36 | 6256.89 | 6172.13 |
| 837.178 | 3943.18 | 3818.61 | 9870.34 | 9809.55 | 25868.9 | 25828.8 | 54766.1 | 55380.7 | 75818.3 | 76742.3 | 837.178 | 2523.01 | 2462.6 | 6339.91 | 6185.43 |
| 837.457 | 3980.48 | 3827.55 | 9732.92 | 9829.09 | 25702.9 | 25871.4 | 54747.8 | 55458.2 | 75633.4 | 76841 | 837.457 | 2461.46 | 2468.84 | 6294.42 | 6198.75 |
| 837.737 | 3931.11 | 3836.5 | 10047.9 | 9848.65 | 25930.6 | 25914 | 55030.5 | 55535.7 | 75128.5 | 76939.8 | 837.737 | 2532.86 | 2475.1 | 6253.69 | 6212.09 |
| 838.017 | 3955.83 | 3845.47 | 9996.17 | 9868.22 | 25848.5 | 25956.6 | 54775.4 | 55613.3 | 75634.5 | 77038.6 | 838.017 | 2559.04 | 2481.36 | 6308.59 | 6225.44 |
| 838.296 | 3948.75 | 3854.44 | 10037.5 | 9887.82 | 25860.2 | 25999.2 | 54748.6 | 55690.9 | 76042.5 | 77137.4 | 838.296 | 2552.86 | 2487.64 | 6244.14 | 6238.8 |
| 838.576 | 3990.86 | 3863.43 | 10094.2 | 9907.43 | 26020.2 | 26041.9 | 55009.3 | 55768.5 | 76407.9 | 77236.2 | 838.576 | 2561.84 | 2493.93 | 6394.05 | 6252.18 |
| 838.856 | 4011 | 3872.43 | 10093.5 | 9927.07 | 26010.8 | 26084.6 | 55437.7 | 55846.1 | 76439.1 | 77335 | 838.856 | 2566.79 | 2500.22 | 6385.67 | 6265.58 |
| 839.135 | 3964.82 | 3881.44 | 10091.7 | 9946.72 | 26188 | 26127.3 | 55420.1 | 55923.7 | 76910.2 | 77433.8 | 839.135 | 2519.18 | 2506.53 | 6398.97 | 6278.99 |
| 839.415 | 4034.67 | 3890.47 | 10191.6 | 9966.39 | 26262.2 | 26170 | 55143.5 | 56001.4 | 76423.5 | 77532.7 | 839.415 | 2590.94 | 2512.84 | 6435 | 6292.42 |
| 839.694 | 4034.33 | 3899.51 | 10038 | 9986.08 | 26086.7 | 26212.8 | 55329.2 | 56079.1 | 76432.9 | 77631.6 | 839.694 | 2654.31 | 2519.17 | 6336.83 | 6305.87 |
| 839.974 | 3948.26 | 3908.55 | 10051.6 | 10005.8 | 26041.6 | 26255.5 | 55617.5 | 56156.8 | 76797.3 | 77730.4 | 839.974 | 2560.2 | 2525.5 | 6442.35 | 6319.33 |
| 840.254 | 4043.51 | 3917.61 | 10147 | 10025.5 | 26179.4 | 26298.3 | 55816.1 | 56234.6 | 76705.4 | 77829.4 | 840.254 | 2544.35 | 2531.84 | 6402.8 | 6332.81 |
| 840.533 | 4030.18 | 3926.69 | 10142 | 10045.3 | 26293.7 | 26341.2 | 55816.4 | 56312.3 | 77000.4 | 77928.3 | 840.533 | 2606.74 | 2538.2 | 6500.24 | 6346.3 |
| 840.812 | 4132.89 | 3935.77 | 10068 | 10065 | 26277.4 | 26384 | 56225.4 | 56390.1 | 77589.5 | 78027.2 | 840.812 | 2567.14 | 2544.57 | 6432.67 | 6359.8 |
| 841.092 | 3997.03 | 3944.87 | 9924.24 | 10084.8 | 26116 | 26426.9 | 55965 | 56467.9 | 77181.4 | 78126.2 | 841.092 | 2546.07 | 2550.94 | 6467.54 | 6373.33 |
| 841.371 | 4069.19 | 3953.98 | 10338.4 | 10104.6 | 26439.8 | 26469.8 | 55922.1 | 56545.7 | 77657.4 | 78225.1 | 841.371 | 2619.69 | 2557.32 | 6385.23 | 6386.87 |
| 841.651 | 4138.57 | 3963.1 | 10280.4 | 10124.4 | 26625 | 26512.8 | 55696.5 | 56623.6 | 77361.2 | 78324.1 | 841.651 | 2625.33 | 2563.72 | 6429.33 | 6400.42 |
| 841.93 | 4181.66 | 3972.24 | 10254 | 10144.3 | 26414.8 | 26555.7 | 56475.9 | 56701.4 | 77947.5 | 78423.1 | 841.93 | 2677.12 | 2570.12 | 6408.9 | 6413.99 |
| 842.21 | 4031.31 | 3981.38 | 10265.4 | 10164.1 | 26677.1 | 26598.7 | 56571.8 | 56779.3 | 77677.2 | 78522.1 | 842.21 | 2654.26 | 2576.54 | 6631.89 | 6427.58 |
| 842.489 | 4165.31 | 3990.54 | 10315.7 | 10184 | 26834.4 | 26641.7 | 56834.9 | 56857.2 | 78231.1 | 78621.1 | 842.489 | 2666.65 | 2582.96 | 6633.35 | 6441.18 |
| 842.769 | 4090.2 | 3999.71 | 10216.9 | 10203.9 | 26875.8 | 26684.7 | 56128.7 | 56935.2 | 78215.2 | 78720.2 | 842.769 | 2718.88 | 2589.4 | 6486.12 | 6454.8 |
| 843.048 | 4174.49 | 4008.89 | 10529.7 | 10223.8 | 26426.6 | 26727.8 | 56868.6 | 57013.1 | 77977.3 | 78819.2 | 843.048 | 2681.2 | 2595.84 | 6478.49 | 6468.43 |
| 843.327 | 4133.57 | 4018.08 | 10130.7 | 10243.8 | 26561.6 | 26770.8 | 56094.7 | 57091.1 | 77299.4 | 78918.3 | 843.327 | 2602.75 | 2602.29 | 6557.54 | 6482.08 |
| 843.607 | 4130.01 | 4027.29 | 10280.3 | 10263.7 | 26557.2 | 26813.9 | 56392.8 | 57169.1 | 77908.3 | 79017.4 | 843.607 | 2758.17 | 2608.76 | 6524.75 | 6495.74 |
| 843.886 | 4139.14 | 4036.51 | 10450 | 10283.7 | 26813.7 | 26857 | 56501.6 | 57247.1 | 78088.5 | 79116.4 | 843.886 | 2708.69 | 2615.23 | 6602.61 | 6509.42 |
| 844.165 | 4241.35 | 4045.74 | 10520.2 | 10303.7 | 26934.7 | 26900.2 | 56868.1 | 57325.2 | 78774.9 | 79215.6 | 844.165 | 2659.77 | 2621.72 | 6713.21 | 6523.12 |
| 844.444 | 4299.04 | 4054.98 | 10272 | 10323.7 | 27011 | 26943.4 | 56890.4 | 57403.2 | 78584 | 79314.7 | 844.444 | 2739.13 | 2628.21 | 6564.17 | 6536.83 |
| 844.724 | 4189.41 | 4064.23 | 10405.2 | 10343.7 | 26904.6 | 26986.5 | 57001.2 | 57481.3 | 78612.9 | 79413.8 | 844.724 | 2726.33 | 2634.71 | 6647.25 | 6550.56 |
| 845.003 | 4258.59 | 4073.5 | 10580.7 | 10363.8 | 26594.2 | 27029.8 | 56858.8 | 57559.4 | 78431.6 | 79513 | 845.003 | 2645.79 | 2641.23 | 6677.1 | 6564.3 |
| 845.282 | 4169.14 | 4082.78 | 10486 | 10383.8 | 26817.8 | 27073 | 56974.1 | 57637.6 | 78748.6 | 79612.1 | 845.282 | 2718.35 | 2647.75 | 6658.24 | 6578.06 |
| 845.562 | 4333.21 | 4092.07 | 10385.8 | 10403.9 | 27022.9 | 27116.3 | 56940.5 | 57715.7 | 78456.2 | 79711.3 | 845.562 | 2715.05 | 2654.28 | 6690.16 | 6591.83 |
| 845.841 | 4246.6 | 4101.37 | 10480 | 10424 | 26728 | 27159.6 | 56675.6 | 57793.9 | 78753.9 | 79810.5 | 845.841 | 2719.76 | 2660.83 | 6713.2 | 6605.62 |
| 846.12 | 4209.09 | 4110.68 | 10588.7 | 10444.1 | 27339.8 | 27202.9 | 57135.1 | 57872.1 | 78695.6 | 79909.7 | 846.12 | 2774.14 | 2667.38 | 6702.8 | 6619.43 |
| 846.399 | 4350.73 | 4120.01 | 10520.9 | 10464.3 | 27279 | 27246.2 | 57184.7 | 57950.3 | 78948.7 | 80008.9 | 846.399 | 2703.52 | 2673.95 | 6576.64 | 6633.25 |
| 846.678 | 4252.74 | 4129.35 | 10521.8 | 10484.4 | 27339.5 | 27289.6 | 57588.6 | 58028.5 | 79057.3 | 80108.1 | 846.678 | 2784.4 | 2680.52 | 6737.24 | 6647.08 |
| 846.957 | 4251.04 | 4138.7 | 10438.1 | 10504.6 | 26935 | 27332.9 | 57155.7 | 58106.8 | 79208.9 | 80207.4 | 846.957 | 2734.3 | 2687.1 | 6845.58 | 6660.93 |
| 847.236 | 4262.04 | 4148.06 | 10602.1 | 10524.8 | 27295.3 | 27376.3 | 57916.8 | 58185 | 79202.6 | 80306.6 | 847.236 | 2765.6 | 2693.7 | 6800.8 | 6674.8 |
| 847.516 | 4301.46 | 4157.43 | 10756.8 | 10545 | 27536.3 | 27419.8 | 57980.7 | 58263.3 | 79466 | 80405.9 | 847.516 | 2787.84 | 2700.3 | 6763.76 | 6688.68 |
| 847.795 | 4250.58 | 4166.82 | 10609 | 10565.2 | 27403.5 | 27463.2 | 58018.2 | 58341.6 | 79736.9 | 80505.1 | 847.795 | 2841.99 | 2706.91 | 6873.85 | 6702.58 |
| 848.074 | 4326.76 | 4176.22 | 10669.5 | 10585.5 | 27428 | 27506.7 | 58326 | 58419.9 | 79764.8 | 80604.4 | 848.074 | 2839.73 | 2713.53 | 6828.07 | 6716.49 |
| 848.353 | 4383.42 | 4185.63 | 10631.6 | 10605.8 | 27450.2 | 27550.2 | 57957.2 | 58498.3 | 79873.8 | 80703.7 | 848.353 | 2783.96 | 2720.17 | 6832.54 | 6730.43 |
| 848.632 | 4283.1 | 4195.05 | 10680.6 | 10626 | 27278.4 | 27593.7 | 57991.4 | 58576.7 | 79528.1 | 80803 | 848.632 | 2787.59 | 2726.81 | 6826.11 | 6744.37 |
| 848.911 | 4319.87 | 4204.48 | 10764.2 | 10646.3 | 27298.8 | 27637.2 | 58102.6 | 58655.1 | 79765.6 | 80902.3 | 848.911 | 2861.71 | 2733.47 | 6873.1 | 6758.33 |
| 849.19 | 4406.27 | 4213.93 | 10782.3 | 10666.7 | 27459.9 | 27680.8 | 58065.4 | 58733.5 | 79851.8 | 81001.7 | 849.19 | 2858.23 | 2740.13 | 6876.02 | 6772.31 |
| 849.469 | 4397.29 | 4223.39 | 10755.9 | 10687 | 27324.1 | 27724.4 | 58398.4 | 58811.9 | 80382.2 | 81101 | 849.469 | 2860.05 | 2746.8 | 6982.6 | 6786.3 |
| 849.748 | 4255.12 | 4232.85 | 10790.7 | 10707.4 | 27527.7 | 27768 | 58277 | 58890.3 | 79775.8 | 81200.4 | 849.748 | 2837.14 | 2753.48 | 6850.81 | 6800.3 |
| 850.027 | 4330.6 | 4242.34 | 10654.6 | 10727.7 | 27632.7 | 27811.7 | 57839.2 | 58968.8 | 80007.1 | 81299.7 | 850.027 | 2770.48 | 2760.18 | 6892.93 | 6814.33 |
| 850.306 | 4414.69 | 4251.83 | 10800.3 | 10748.1 | 27533.7 | 27855.3 | 58345.4 | 59047.3 | 80634.4 | 81399.1 | 850.306 | 2823.62 | 2766.88 | 6865.67 | 6828.36 |
| 850.585 | 4403.46 | 4261.34 | 10915.4 | 10768.5 | 27654.5 | 27899 | 58344.3 | 59125.8 | 80097.6 | 81498.5 | 850.585 | 2819.53 | 2773.59 | 6962.12 | 6842.42 |
| 850.864 | 4359.74 | 4270.85 | 10672.2 | 10789 | 27645.3 | 27942.7 | 57910.4 | 59204.3 | 79614.3 | 81597.9 | 850.864 | 2792.56 | 2780.32 | 6820.43 | 6856.48 |
| 851.143 | 4352 | 4280.38 | 10905 | 10809.4 | 27788.2 | 27986.4 | 58227.5 | 59282.8 | 80023.9 | 81697.3 | 851.143 | 2804.72 | 2787.05 | 6857.66 | 6870.57 |
| 851.421 | 4350.9 | 4289.93 | 10712.6 | 10829.9 | 27680 | 28030.2 | 58741.7 | 59361.4 | 80475.1 | 81796.7 | 851.421 | 2874.67 | 2793.79 | 6987.36 | 6884.67 |
| 851.7 | 4407.34 | 4299.48 | 10833.8 | 10850.4 | 27651.8 | 28073.9 | 58210.2 | 59440 | 80443.5 | 81896.1 | 851.7 | 2808.02 | 2800.54 | 6847.52 | 6898.78 |
| 851.979 | 4448.02 | 4309.04 | 10833 | 10870.9 | 27759.1 | 28117.7 | 58280.6 | 59518.6 | 80629.4 | 81995.6 | 851.979 | 2841.06 | 2807.31 | 6915.35 | 6912.91 |
| 852.258 | 4626.14 | 4318.62 | 10827.1 | 10891.4 | 28490.1 | 28161.5 | 59569.2 | 59597.2 | 81133.4 | 82095 | 852.258 | 2867.28 | 2814.08 | 6969.59 | 6927.06 |
| 852.537 | 4406.79 | 4328.21 | 11015.1 | 10911.9 | 28161.2 | 28205.4 | 58963.6 | 59675.8 | 80865.5 | 82194.5 | 852.537 | 2831.25 | 2820.86 | 7029.02 | 6941.22 |
| 852.816 | 4390.57 | 4337.81 | 10932.6 | 10932.5 | 27890.1 | 28249.2 | 58891.9 | 59754.5 | 81610.3 | 82294 | 852.816 | 2871.5 | 2827.65 | 6867.83 | 6955.4 |
| 853.094 | 4463.47 | 4347.43 | 10854.6 | 10953.1 | 27706.3 | 28293.1 | 58798 | 59833.2 | 80900.9 | 82393.4 | 853.094 | 2889.75 | 2834.46 | 6967.05 | 6969.59 |
| 853.373 | 4440.73 | 4357.05 | 10780.5 | 10973.7 | 27835.2 | 28337 | 58312.9 | 59911.8 | 80126.2 | 82492.9 | 853.373 | 2852.21 | 2841.27 | 6899.85 | 6983.8 |
| 853.652 | 4517.18 | 4366.69 | 11041.9 | 10994.3 | 28032.5 | 28380.9 | 59161.1 | 59990.6 | 81218.4 | 82592.4 | 853.652 | 2889.28 | 2848.09 | 7117.73 | 6998.02 |
| 853.93 | 4316.05 | 4376.34 | 11129.5 | 11014.9 | 28140.6 | 28424.9 | 59013.2 | 60069.3 | 81028.5 | 82691.9 | 853.93 | 2959.63 | 2854.92 | 7086.26 | 7012.26 |
| 854.209 | 4412.77 | 4386 | 11068.4 | 11035.5 | 27782.9 | 28468.8 | 59186.5 | 60148 | 80889 | 82791.4 | 854.209 | 2897.11 | 2861.77 | 6932.07 | 7026.51 |
| 854.488 | 4540.65 | 4395.67 | 11101.1 | 11056.2 | 28206.9 | 28512.8 | 59233 | 60226.8 | 81546.4 | 82891 | 854.488 | 2878.93 | 2868.62 | 7070.83 | 7040.78 |
| 854.766 | 4530.48 | 4405.35 | 11134.1 | 11076.9 | 28332.1 | 28556.8 | 59404.5 | 60305.5 | 81428.2 | 82990.5 | 854.766 | 2918.04 | 2875.48 | 7166.58 | 7055.06 |
| 855.045 | 4482.73 | 4415.05 | 11075.9 | 11097.6 | 28320.3 | 28600.9 | 59032.3 | 60384.3 | 81204.3 | 83090 | 855.045 | 2910.51 | 2882.35 | 7074.99 | 7069.36 |
| 855.324 | 4563.76 | 4424.76 | 11230.9 | 11118.3 | 28293.3 | 28644.9 | 59445.4 | 60463.1 | 81423.1 | 83189.6 | 855.324 | 2938.6 | 2889.24 | 7105.29 | 7083.68 |
| 855.602 | 4487.09 | 4434.48 | 10989.5 | 11139 | 28382.7 | 28689 | 59152.2 | 60542 | 81160.6 | 83289.2 | 855.602 | 2851.1 | 2896.13 | 6998.39 | 7098.01 |
| 855.881 | 4436.69 | 4444.21 | 11089.4 | 11159.8 | 28764.6 | 28733.1 | 59599.4 | 60620.8 | 81407.4 | 83388.7 | 855.881 | 2977.49 | 2903.03 | 7053.27 | 7112.35 |
| 856.16 | 4474.15 | 4453.95 | 11355.5 | 11180.6 | 28472.4 | 28777.2 | 59578.9 | 60699.7 | 82045.4 | 83488.3 | 856.16 | 2949.55 | 2909.94 | 7250.1 | 7126.71 |
| 856.438 | 4528.27 | 4463.71 | 11136.1 | 11201.4 | 28322.4 | 28821.3 | 59585 | 60778.6 | 81805.7 | 83587.9 | 856.438 | 2872.53 | 2916.87 | 7107.65 | 7141.09 |
| 856.717 | 4618.11 | 4473.48 | 11244.7 | 11222.2 | 28525 | 28865.5 | 59295.2 | 60857.4 | 81616.7 | 83687.5 | 856.717 | 2994.69 | 2923.8 | 7130.4 | 7155.48 |
| 856.995 | 4594.98 | 4483.26 | 11310.2 | 11243 | 28658.8 | 28909.7 | 60114 | 60936.3 | 82603.9 | 83787.1 | 856.995 | 2981.1 | 2930.74 | 7175.22 | 7169.89 |
| 857.274 | 4571.97 | 4493.05 | 11387.6 | 11263.8 | 28686.4 | 28953.9 | 60149 | 61015.3 | 82520 | 83886.7 | 857.274 | 3001.96 | 2937.69 | 7279.64 | 7184.31 |
| 857.552 | 4595.1 | 4502.85 | 11495.1 | 11284.7 | 28831.8 | 28998.1 | 59852.5 | 61094.2 | 82677.6 | 83986.3 | 857.552 | 2986.02 | 2944.66 | 7246.69 | 7198.75 |
| 857.831 | 4554.44 | 4512.67 | 11347.7 | 11305.6 | 28394.3 | 29042.4 | 59910.6 | 61173.2 | 82273.8 | 84085.9 | 857.831 | 3036.75 | 2951.63 | 7107.48 | 7213.2 |
| 858.109 | 4620.66 | 4522.49 | 11403.5 | 11326.5 | 28728.4 | 29086.6 | 60139.7 | 61252.2 | 82740.8 | 84185.6 | 858.109 | 2931.69 | 2958.61 | 7106.52 | 7227.67 |
| 858.388 | 4532.16 | 4532.33 | 11345.8 | 11347.4 | 28896.3 | 29130.9 | 59702.4 | 61331.2 | 82470.9 | 84285.2 | 858.388 | 3042 | 2965.6 | 7303.04 | 7242.15 |
| 858.666 | 4689.7 | 4542.18 | 11375.6 | 11368.3 | 29048.5 | 29175.2 | 60639.3 | 61410.2 | 82887.3 | 84384.8 | 858.666 | 2948.88 | 2972.6 | 7319.41 | 7256.65 |
| 858.945 | 4597.91 | 4552.05 | 11339.6 | 11389.3 | 29085.8 | 29219.6 | 60823 | 61489.2 | 83159.1 | 84484.5 | 858.945 | 3017.95 | 2979.62 | 7245.45 | 7271.16 |
| 859.223 | 4576.35 | 4561.92 | 11117.1 | 11410.2 | 28371.2 | 29263.9 | 60016.5 | 61568.2 | 82558.1 | 84584.2 | 859.223 | 3045.88 | 2986.64 | 7120.31 | 7285.69 |
| 859.501 | 4691.46 | 4571.81 | 11458.3 | 11431.2 | 29024.4 | 29308.3 | 60619.2 | 61647.3 | 83021.4 | 84683.8 | 859.501 | 3060.24 | 2993.67 | 7216.91 | 7300.23 |
| 859.78 | 4671.68 | 4581.7 | 11367.3 | 11452.2 | 28971.3 | 29352.7 | 60348.9 | 61726.3 | 82802.2 | 84783.5 | 859.78 | 3076.12 | 3000.71 | 7311.24 | 7314.79 |
| 860.058 | 4642.34 | 4591.61 | 11403.3 | 11473.2 | 29161.9 | 29397.1 | 61189.5 | 61805.4 | 83545.3 | 84883.2 | 860.058 | 3048.32 | 3007.76 | 7406.15 | 7329.36 |
| 860.337 | 4696.05 | 4601.53 | 11457.3 | 11494.3 | 29203.8 | 29441.5 | 61014.7 | 61884.5 | 83139 | 84982.9 | 860.337 | 3069.58 | 3014.83 | 7328.43 | 7343.95 |
| 860.615 | 4659.41 | 4611.47 | 11548.3 | 11515.3 | 29339.1 | 29486 | 61119.8 | 61963.7 | 83990.6 | 85082.6 | 860.615 | 3096.19 | 3021.9 | 7389.71 | 7358.56 |
| 860.893 | 4625.48 | 4621.41 | 11553.5 | 11536.4 | 29334.5 | 29530.5 | 60729.4 | 62042.8 | 83227.8 | 85182.3 | 860.893 | 3080.31 | 3028.98 | 7357.18 | 7373.18 |
| 861.171 | 4649.75 | 4631.37 | 11408.1 | 11557.5 | 29003.8 | 29575 | 60891.5 | 62121.9 | 83584.8 | 85282 | 861.171 | 3003.14 | 3036.07 | 7424.78 | 7387.81 |
| 861.45 | 4778.35 | 4641.34 | 11663.4 | 11578.6 | 29351.2 | 29619.5 | 61255.7 | 62201.1 | 83865.9 | 85381.7 | 861.45 | 3040.48 | 3043.17 | 7542.55 | 7402.46 |
| 861.728 | 4704.97 | 4651.32 | 11592.2 | 11599.7 | 29206.1 | 29664 | 61036.7 | 62280.3 | 84148.2 | 85481.4 | 861.728 | 3050.84 | 3050.29 | 7387.68 | 7417.12 |
| 862.006 | 4698.97 | 4661.31 | 11504.2 | 11620.8 | 29267.1 | 29708.6 | 60988.6 | 62359.5 | 83670.3 | 85581.1 | 862.006 | 3044.54 | 3057.41 | 7474.22 | 7431.81 |
| 862.284 | 4520.91 | 4671.31 | 11535.3 | 11642 | 29427.9 | 29753.1 | 61009.7 | 62438.7 | 83436.6 | 85680.9 | 862.284 | 3151.63 | 3064.54 | 7394.18 | 7446.5 |
| 862.563 | 4823.54 | 4681.33 | 11602.5 | 11663.2 | 29494.2 | 29797.7 | 61875.9 | 62517.9 | 84974.6 | 85780.6 | 862.563 | 3202.38 | 3071.68 | 7523.81 | 7461.21 |
| 862.841 | 4756 | 4691.36 | 11610.5 | 11684.4 | 29476.8 | 29842.4 | 61919.9 | 62597.1 | 84035 | 85880.3 | 862.841 | 3149.28 | 3078.83 | 7590.27 | 7475.93 |
| 863.119 | 4846.83 | 4701.4 | 11529.6 | 11705.6 | 29431.8 | 29887 | 61320.3 | 62676.4 | 84045.6 | 85980.1 | 863.119 | 3138.92 | 3085.99 | 7486.97 | 7490.67 |
| 863.397 | 4865.75 | 4711.45 | 11740.6 | 11726.8 | 29290.5 | 29931.7 | 61331.9 | 62755.7 | 84060.7 | 86079.8 | 863.397 | 3110.18 | 3093.17 | 7504.17 | 7505.43 |
| 863.675 | 4866.63 | 4721.51 | 11576.3 | 11748 | 29302.7 | 29976.3 | 61624.3 | 62834.9 | 84368.1 | 86179.6 | 863.675 | 3123.67 | 3100.35 | 7613.16 | 7520.2 |
| 863.953 | 4734.72 | 4731.58 | 11525 | 11769.3 | 29800.6 | 30021 | 62082.3 | 62914.2 | 84627.9 | 86279.3 | 863.953 | 3189.31 | 3107.54 | 7566.96 | 7534.98 |
| 864.231 | 4825.85 | 4741.67 | 11681.2 | 11790.6 | 29584.5 | 30065.8 | 61424.1 | 62993.5 | 84364.5 | 86379.1 | 864.231 | 3090.4 | 3114.74 | 7463.71 | 7549.78 |
| 864.51 | 4909.26 | 4751.77 | 11698.1 | 11811.9 | 29658.8 | 30110.5 | 61817.6 | 63072.9 | 84655.1 | 86478.9 | 864.51 | 3190.75 | 3121.95 | 7494.11 | 7564.6 |
| 864.788 | 4690.32 | 4761.87 | 11745.8 | 11833.2 | 29430.3 | 30155.2 | 61812.3 | 63152.2 | 84685.1 | 86578.6 | 864.788 | 3139.05 | 3129.18 | 7531.67 | 7579.43 |
| 865.066 | 4855.61 | 4772 | 12097.9 | 11854.5 | 29629.9 | 30200 | 61754.5 | 63231.5 | 84399.5 | 86678.4 | 865.066 | 3175.79 | 3136.41 | 7647.12 | 7594.27 |
| 865.344 | 4889.31 | 4782.13 | 11685.9 | 11875.9 | 29753 | 30244.8 | 61745.6 | 63310.9 | 84777.1 | 86778.2 | 865.344 | 3203.74 | 3143.65 | 7559.69 | 7609.14 |
| 865.622 | 4859.65 | 4792.27 | 11951.5 | 11897.2 | 29920.3 | 30289.6 | 62251.9 | 63390.3 | 85283.3 | 86878 | 865.622 | 3110 | 3150.9 | 7670.67 | 7624.01 |
| 865.9 | 4794.96 | 4802.43 | 11954.4 | 11918.6 | 29603.2 | 30334.5 | 61713 | 63469.7 | 84956.2 | 86977.8 | 865.9 | 3232.8 | 3158.16 | 7602.16 | 7638.9 |
| 866.178 | 4770.21 | 4812.6 | 11889.6 | 11940 | 30031.8 | 30379.3 | 61976.9 | 63549.1 | 85363 | 87077.6 | 866.178 | 3170.67 | 3165.43 | 7646.99 | 7653.81 |
| 866.456 | 4948.15 | 4822.78 | 11923.3 | 11961.4 | 30317.8 | 30424.2 | 62531.4 | 63628.5 | 85635.6 | 87177.4 | 866.456 | 3161.19 | 3172.72 | 7692.99 | 7668.73 |
| 866.734 | 5010.74 | 4832.97 | 12029.6 | 11982.9 | 29727.3 | 30469.1 | 62454 | 63707.9 | 85590.2 | 87277.2 | 866.734 | 3100 | 3180.01 | 7617.31 | 7683.66 |
| 867.011 | 4787.7 | 4843.17 | 11911.2 | 12004.3 | 30237.9 | 30514 | 62490.8 | 63787.4 | 85117.9 | 87377 | 867.011 | 3254.04 | 3187.31 | 7727.48 | 7698.61 |
| 867.289 | 4929.96 | 4853.39 | 11887.7 | 12025.8 | 29791.4 | 30559 | 62619 | 63866.8 | 85388.3 | 87476.8 | 867.289 | 3228.64 | 3194.62 | 7732.88 | 7713.58 |
| 867.567 | 4938.13 | 4863.61 | 12087.8 | 12047.2 | 29937.2 | 30603.9 | 62568.1 | 63946.3 | 85338.8 | 87576.6 | 867.567 | 3185.14 | 3201.94 | 7792.82 | 7728.56 |
| 867.845 | 5027.86 | 4873.85 | 12041 | 12068.7 | 30037.6 | 30648.9 | 62575.9 | 64025.8 | 85568.3 | 87676.4 | 867.845 | 3265.48 | 3209.27 | 7688.67 | 7743.55 |
| 868.123 | 4947.3 | 4884.1 | 12123.7 | 12090.3 | 30084.2 | 30693.9 | 62304.4 | 64105.3 | 85741.4 | 87776.3 | 868.123 | 3277.33 | 3216.61 | 7798.57 | 7758.56 |
| 868.401 | 4970.98 | 4894.36 | 12040.2 | 12111.8 | 30078.2 | 30738.9 | 62513 | 64184.8 | 85548.6 | 87876.1 | 868.401 | 3233.32 | 3223.97 | 7616.16 | 7773.58 |
| 868.679 | 4835.78 | 4904.63 | 11979 | 12133.3 | 30117.3 | 30783.9 | 62486.8 | 64264.3 | 85454.5 | 87975.9 | 868.679 | 3260.18 | 3231.33 | 7647.08 | 7788.62 |
| 868.957 | 5013.94 | 4914.92 | 11990.9 | 12154.9 | 29918.7 | 30829 | 63011.6 | 64343.8 | 85467.1 | 88075.8 | 868.957 | 3160.01 | 3238.7 | 7667.24 | 7803.68 |
| 869.234 | 4813.13 | 4925.21 | 12093.8 | 12176.5 | 30159.5 | 30874 | 62681.4 | 64423.4 | 85644.7 | 88175.6 | 869.234 | 3239.6 | 3246.08 | 7795.07 | 7818.75 |
| 869.512 | 4922.48 | 4935.52 | 12217.7 | 12198.1 | 30357.9 | 30919.1 | 63176.1 | 64502.9 | 86355.5 | 88275.4 | 869.512 | 3286.86 | 3253.47 | 7785.97 | 7833.83 |
| 869.79 | 4906.99 | 4945.84 | 11976.8 | 12219.7 | 30263.3 | 30964.2 | 63341.5 | 64582.5 | 85762.8 | 88375.2 | 869.79 | 3278.05 | 3260.87 | 7688.99 | 7848.93 |
| 870.068 | 4921.95 | 4956.17 | 12202.6 | 12241.3 | 30473.2 | 31009.3 | 62964.3 | 64662.1 | 86362.8 | 88475.1 | 870.068 | 3244.81 | 3268.28 | 7800 | 7864.04 |
| 870.345 | 4994.92 | 4966.51 | 12200.8 | 12263 | 30432.7 | 31054.5 | 63252.4 | 64741.7 | 86234.4 | 88574.9 | 870.345 | 3235.85 | 3275.7 | 7899.25 | 7879.17 |
| 870.623 | 4947.55 | 4976.87 | 12173.7 | 12284.6 | 30336.3 | 31099.6 | 63073.4 | 64821.3 | 85866.4 | 88674.8 | 870.623 | 3247.29 | 3283.13 | 7709.84 | 7894.31 |
| 870.901 | 5148.42 | 4987.23 | 12093.2 | 12306.3 | 30393.5 | 31144.8 | 63381.3 | 64900.9 | 86721.2 | 88774.6 | 870.901 | 3237.43 | 3290.57 | 7888.82 | 7909.47 |
| 871.178 | 4976.26 | 4997.61 | 12389.5 | 12328 | 30614.8 | 31190 | 63243.5 | 64980.5 | 86970.6 | 88874.5 | 871.178 | 3271.86 | 3298.03 | 7799.7 | 7924.64 |
| 871.456 | 5072.73 | 5008 | 12263.1 | 12349.8 | 30431.2 | 31235.2 | 63587.6 | 65060.2 | 86629 | 88974.3 | 871.456 | 3282.88 | 3305.49 | 7779.39 | 7939.83 |
| 871.734 | 5083.66 | 5018.4 | 12343.8 | 12371.5 | 30591.4 | 31280.5 | 64041.5 | 65139.8 | 86801.2 | 89074.2 | 871.734 | 3344.24 | 3312.96 | 7954.95 | 7955.03 |
| 872.011 | 5129.77 | 5028.81 | 12167.1 | 12393.2 | 30729.8 | 31325.7 | 63292.8 | 65219.5 | 86807.2 | 89174 | 872.011 | 3315.72 | 3320.44 | 7891.69 | 7970.25 |
| 872.289 | 4953.44 | 5039.23 | 12331.2 | 12415 | 30669.8 | 31371 | 64356.5 | 65299.2 | 87713 | 89273.9 | 872.289 | 3360.64 | 3327.93 | 7904.73 | 7985.48 |
| 872.567 | 5146.49 | 5049.67 | 12160.1 | 12436.8 | 30626.6 | 31416.3 | 63718.1 | 65378.8 | 87072.1 | 89373.7 | 872.567 | 3314.17 | 3335.43 | 7928.48 | 8000.73 |
| 872.844 | 5092.84 | 5060.11 | 12310.9 | 12458.6 | 30554.8 | 31461.6 | 63917.8 | 65458.5 | 87589.8 | 89473.6 | 872.844 | 3330.85 | 3342.94 | 7912.65 | 8015.99 |
| 873.122 | 5062.02 | 5070.57 | 12361.9 | 12480.4 | 30838.6 | 31506.9 | 64044.8 | 65538.2 | 87938 | 89573.5 | 873.122 | 3284.91 | 3350.46 | 7906.33 | 8031.27 |
| 873.399 | 5049.42 | 5081.04 | 12405.7 | 12502.2 | 30659.3 | 31552.2 | 63810.2 | 65617.9 | 87907.8 | 89673.3 | 873.399 | 3291.34 | 3357.99 | 7971.35 | 8046.56 |
| 873.677 | 5162.34 | 5091.52 | 12587.1 | 12524.1 | 31188.4 | 31597.6 | 64248.9 | 65697.7 | 87877.7 | 89773.2 | 873.677 | 3369.4 | 3365.53 | 8033.12 | 8061.86 |
| 873.954 | 5106.29 | 5102.01 | 12589.3 | 12545.9 | 31124.7 | 31643 | 64799.6 | 65777.4 | 88496.9 | 89873 | 873.954 | 3391.27 | 3373.08 | 8027.97 | 8077.18 |
| 874.232 | 5080.86 | 5112.52 | 12422.8 | 12567.8 | 31183.4 | 31688.4 | 64616.8 | 65857.2 | 88033.7 | 89972.9 | 874.232 | 3337.15 | 3380.64 | 8071.88 | 8092.52 |
| 874.509 | 5218.03 | 5123.03 | 12506.5 | 12589.7 | 31306 | 31733.8 | 64593.7 | 65936.9 | 88144.4 | 90072.8 | 874.509 | 3359.8 | 3388.22 | 8035.29 | 8107.86 |
| 874.787 | 5107.35 | 5133.56 | 12697.6 | 12611.6 | 31276.8 | 31779.2 | 64494.2 | 66016.7 | 88208.3 | 90172.6 | 874.787 | 3426.8 | 3395.8 | 7998.94 | 8123.23 |
| 875.064 | 5085.71 | 5144.1 | 12461.2 | 12633.6 | 31335.8 | 31824.7 | 63988.6 | 66096.5 | 87799 | 90272.5 | 875.064 | 3442.06 | 3403.39 | 8116.7 | 8138.61 |
| 875.342 | 5143.18 | 5154.65 | 12726.1 | 12655.5 | 30944.2 | 31870.2 | 64755.8 | 66176.3 | 88243.2 | 90372.4 | 875.342 | 3508.81 | 3410.99 | 8035.66 | 8154 |
| 875.619 | 5213.05 | 5165.21 | 12610 | 12677.5 | 31133.8 | 31915.7 | 64906 | 66256.1 | 88757.6 | 90472.2 | 875.619 | 3422.29 | 3418.6 | 8204.83 | 8169.41 |
| 875.896 | 5216.32 | 5175.78 | 12652.9 | 12699.5 | 31338.3 | 31961.2 | 64510.3 | 66335.9 | 88378.7 | 90572.1 | 875.896 | 3323.3 | 3426.22 | 8212.43 | 8184.83 |
| 876.174 | 5230.44 | 5186.37 | 12591 | 12721.4 | 31164.6 | 32006.7 | 65189.8 | 66415.7 | 88374.3 | 90671.9 | 876.174 | 3411.62 | 3433.85 | 8180.6 | 8200.26 |
| 876.451 | 5203.31 | 5196.96 | 12721.7 | 12743.5 | 31334.4 | 32052.2 | 65359.2 | 66495.5 | 88706.2 | 90771.8 | 876.451 | 3389.98 | 3441.49 | 8093.08 | 8215.72 |
| 876.728 | 5207.07 | 5207.57 | 12669.3 | 12765.5 | 31217.9 | 32097.8 | 64804.6 | 66575.4 | 88211.7 | 90871.7 | 876.728 | 3408.38 | 3449.14 | 8165.11 | 8231.18 |
| 877.006 | 5315.64 | 5218.19 | 12693.1 | 12787.5 | 31173 | 32143.3 | 65691.6 | 66655.2 | 88533.7 | 90971.5 | 877.006 | 3462.46 | 3456.8 | 8235.78 | 8246.66 |
| 877.283 | 5195.69 | 5228.82 | 12595.1 | 12809.6 | 31575.2 | 32188.9 | 65025.6 | 66735.1 | 88916 | 91071.4 | 877.283 | 3458.26 | 3464.47 | 8156.21 | 8262.16 |
| 877.56 | 5347.76 | 5239.46 | 12695.6 | 12831.7 | 31813.2 | 32234.5 | 66070.7 | 66814.9 | 89765.1 | 91171.2 | 877.56 | 3524.53 | 3472.15 | 8242.4 | 8277.67 |
| 877.838 | 5350.29 | 5250.11 | 12915.6 | 12853.8 | 32108.1 | 32280.2 | 65795.5 | 66894.8 | 89329.4 | 91271.1 | 877.838 | 3459.67 | 3479.84 | 8243.32 | 8293.19 |
| 878.115 | 5274.83 | 5260.78 | 12860 | 12875.9 | 31879.6 | 32325.8 | 65528.8 | 66974.7 | 89540.9 | 91371 | 878.115 | 3488.43 | 3487.53 | 8313.2 | 8308.73 |
| 878.392 | 5420.69 | 5271.45 | 12867.2 | 12898 | 31959.6 | 32371.5 | 65491.5 | 67054.6 | 89516.5 | 91470.8 | 878.392 | 3532.97 | 3495.24 | 8220.38 | 8324.28 |
| 878.669 | 5338.43 | 5282.14 | 12842.9 | 12920.1 | 31961.4 | 32417.2 | 65848.8 | 67134.5 | 89664.4 | 91570.7 | 878.669 | 3587.17 | 3502.96 | 8309.23 | 8339.85 |
| 878.947 | 5279.97 | 5292.84 | 12984.6 | 12942.3 | 31857 | 32462.8 | 65789.2 | 67214.4 | 89594.8 | 91670.5 | 878.947 | 3475.7 | 3510.69 | 8378.58 | 8355.43 |
| 879.224 | 5452.07 | 5303.55 | 12880.2 | 12964.5 | 32095.5 | 32508.6 | 65912.8 | 67294.4 | 89544.8 | 91770.4 | 879.224 | 3452.04 | 3518.43 | 8288.22 | 8371.03 |
| 879.501 | 5274.25 | 5314.27 | 12788.8 | 12986.6 | 31953.1 | 32554.3 | 65887.5 | 67374.3 | 89708.1 | 91870.2 | 879.501 | 3465.25 | 3526.18 | 8358.96 | 8386.64 |
| 879.778 | 5307.1 | 5325 | 12816.6 | 13008.8 | 32101.6 | 32600 | 66462.8 | 67454.2 | 90402 | 91970.1 | 879.778 | 3557.09 | 3533.94 | 8445.65 | 8402.26 |
| 880.055 | 5370.52 | 5335.75 | 12946.6 | 13031.1 | 32229 | 32645.8 | 66416 | 67534.2 | 90199.1 | 92069.9 | 880.055 | 3637.38 | 3541.71 | 8387.8 | 8417.9 |
| 880.332 | 5344.7 | 5346.5 | 12996.8 | 13053.3 | 32164 | 32691.6 | 66069.6 | 67614.1 | 90343.1 | 92169.8 | 880.332 | 3584.64 | 3549.49 | 8450.45 | 8433.56 |
| 880.609 | 5400 | 5357.27 | 12912.5 | 13075.5 | 32079 | 32737.4 | 66252.3 | 67694.1 | 91071.2 | 92269.6 | 880.609 | 3528.94 | 3557.28 | 8477.06 | 8449.23 |
| 880.886 | 5364.62 | 5368.05 | 13072.2 | 13097.8 | 32410.7 | 32783.2 | 66468.9 | 67774.1 | 90813.4 | 92369.5 | 880.886 | 3551.6 | 3565.08 | 8451.73 | 8464.91 |
| 881.163 | 5428.95 | 5378.84 | 12859.3 | 13120.1 | 32191.5 | 32829 | 66226.4 | 67854.1 | 90216.9 | 92469.3 | 881.163 | 3592.87 | 3572.89 | 8465.36 | 8480.61 |
| 881.441 | 5351.3 | 5389.64 | 13001.3 | 13142.4 | 32471.2 | 32874.8 | 67275.3 | 67934.1 | 90985.3 | 92569.1 | 881.441 | 3535.91 | 3580.7 | 8501.33 | 8496.32 |
| 881.717 | 5435.48 | 5400.45 | 13150.3 | 13164.7 | 32342.2 | 32920.7 | 66876.1 | 68014.1 | 91140.2 | 92669 | 881.717 | 3645.78 | 3588.53 | 8541.17 | 8512.05 |
| 881.995 | 5411.46 | 5411.27 | 13101.5 | 13187 | 32257.4 | 32966.6 | 66587.5 | 68094.1 | 91207.8 | 92768.8 | 881.995 | 3680 | 3596.37 | 8579.47 | 8527.78 |
| 882.271 | 5611.54 | 5422.11 | 13094 | 13209.4 | 32590.1 | 33012.5 | 67382.9 | 68174.1 | 91979.2 | 92868.6 | 882.271 | 3649.41 | 3604.22 | 8507.15 | 8543.54 |
| 882.548 | 5426.18 | 5432.95 | 13022.3 | 13231.7 | 31948.4 | 33058.4 | 66491.1 | 68254.1 | 90549.9 | 92968.5 | 882.548 | 3682.37 | 3612.08 | 8380.11 | 8559.31 |
| 882.825 | 5472.16 | 5443.81 | 13108.4 | 13254.1 | 32607.6 | 33104.3 | 67252.6 | 68334.1 | 90959.8 | 93068.3 | 882.825 | 3559.68 | 3619.95 | 8535.52 | 8575.09 |
| 883.102 | 5507.86 | 5454.68 | 13198.4 | 13276.5 | 32671.7 | 33150.3 | 67027.8 | 68414.2 | 91422.1 | 93168.1 | 883.102 | 3612.76 | 3627.82 | 8575.74 | 8590.89 |
| 883.379 | 5457.86 | 5465.56 | 13280.2 | 13298.9 | 32828.1 | 33196.2 | 67066.8 | 68494.2 | 90998.1 | 93267.9 | 883.379 | 3644.79 | 3635.71 | 8593.12 | 8606.7 |
| 883.656 | 5693.45 | 5476.45 | 13231 | 13321.3 | 32678.4 | 33242.2 | 67463.8 | 68574.3 | 91193.6 | 93367.8 | 883.656 | 3720.01 | 3643.61 | 8746.3 | 8622.53 |
| 883.933 | 5467.02 | 5487.35 | 13246.3 | 13343.8 | 32473.6 | 33288.2 | 66637.3 | 68654.3 | 91326.1 | 93467.6 | 883.933 | 3638.31 | 3651.52 | 8547.81 | 8638.37 |
| 884.21 | 5494.14 | 5498.27 | 13219.3 | 13366.2 | 32848.3 | 33334.2 | 67569.7 | 68734.4 | 92286.3 | 93567.4 | 884.21 | 3651.97 | 3659.44 | 8647.3 | 8654.22 |
| 884.487 | 5427.25 | 5509.19 | 13248 | 13388.7 | 32959.6 | 33380.2 | 66933.3 | 68814.5 | 91628.1 | 93667.2 | 884.487 | 3571.84 | 3667.36 | 8555.19 | 8670.09 |
| 884.764 | 5553.05 | 5520.13 | 13164.5 | 13411.2 | 32813.5 | 33426.2 | 67889.1 | 68894.6 | 91562.2 | 93767 | 884.764 | 3692.46 | 3675.3 | 8506.62 | 8685.98 |
| 885.04 | 5568.83 | 5531.07 | 13170.2 | 13433.7 | 32735.4 | 33472.3 | 67341.6 | 68974.6 | 91281.1 | 93866.8 | 885.04 | 3655.21 | 3683.25 | 8588.92 | 8701.87 |
| 885.317 | 5517.26 | 5542.03 | 13190.6 | 13456.2 | 33016.7 | 33518.4 | 67311.8 | 69054.7 | 92138.2 | 93966.6 | 885.317 | 3701.59 | 3691.21 | 8551.88 | 8717.78 |
| 885.594 | 5546.75 | 5553 | 13077.1 | 13478.7 | 32775.9 | 33564.5 | 67287.2 | 69134.8 | 91754.4 | 94066.4 | 885.594 | 3690.85 | 3699.17 | 8652.42 | 8733.71 |
| 885.871 | 5727.59 | 5563.98 | 13255.7 | 13501.3 | 33013.4 | 33610.6 | 67673.2 | 69214.9 | 92489.2 | 94166.1 | 885.871 | 3652.27 | 3707.15 | 8736.57 | 8749.65 |
| 886.148 | 5571.45 | 5574.98 | 13394.9 | 13523.9 | 33086 | 33656.7 | 67973.2 | 69295.1 | 92201.2 | 94265.9 | 886.148 | 3729.28 | 3715.14 | 8848.09 | 8765.6 |
| 886.424 | 5545.65 | 5585.98 | 13256 | 13546.4 | 33162.3 | 33702.8 | 67497.6 | 69375.2 | 92070.5 | 94365.7 | 886.424 | 3706.17 | 3723.14 | 8767.47 | 8781.57 |
| 886.701 | 5624.06 | 5596.99 | 13472.8 | 13569 | 32925.9 | 33749 | 68117.7 | 69455.3 | 92257.2 | 94465.5 | 886.701 | 3680.35 | 3731.14 | 8872.35 | 8797.56 |
| 886.978 | 5754.98 | 5608.02 | 13536.1 | 13591.6 | 33211.1 | 33795.1 | 68484.1 | 69535.4 | 93103.2 | 94565.2 | 886.978 | 3786.83 | 3739.16 | 8698.27 | 8813.55 |
| 887.254 | 5596.58 | 5619.06 | 13625.9 | 13614.2 | 33079.6 | 33841.3 | 68192.8 | 69615.6 | 92992.1 | 94665 | 887.254 | 3768.59 | 3747.18 | 8872.02 | 8829.56 |
| 887.531 | 5618.98 | 5630.1 | 13598.9 | 13636.9 | 33518.4 | 33887.5 | 68497 | 69695.7 | 92736.6 | 94764.8 | 887.531 | 3740.48 | 3755.22 | 8851.27 | 8845.59 |
| 887.808 | 5658.92 | 5641.16 | 13464.5 | 13659.5 | 33242.6 | 33933.7 | 67629 | 69775.9 | 92502.9 | 94864.5 | 887.808 | 3817.06 | 3763.27 | 8875.02 | 8861.63 |
| 888.084 | 5624.35 | 5652.23 | 13491.6 | 13682.2 | 33239.7 | 33979.9 | 68490 | 69856 | 93055.7 | 94964.3 | 888.084 | 3798.74 | 3771.32 | 8826.78 | 8877.68 |
| 888.361 | 5669.52 | 5663.32 | 13435.8 | 13704.9 | 33419.3 | 34026.2 | 68556.4 | 69936.2 | 93087.6 | 95064 | 888.361 | 3795.41 | 3779.39 | 8779.04 | 8893.75 |
| 888.638 | 5652.29 | 5674.41 | 13608.2 | 13727.6 | 33560.1 | 34072.4 | 68779.2 | 70016.4 | 93340.6 | 95163.8 | 888.638 | 3841.47 | 3787.46 | 8816.19 | 8909.83 |
| 888.914 | 5684.2 | 5685.51 | 13662.1 | 13750.3 | 33413.1 | 34118.7 | 68801.8 | 70096.5 | 93036 | 95263.5 | 888.914 | 3848 | 3795.55 | 8834.64 | 8925.92 |
| 889.191 | 5528.62 | 5696.63 | 13685.1 | 13773 | 33635.9 | 34165 | 68373.4 | 70176.7 | 93148.5 | 95363.2 | 889.191 | 3783.19 | 3803.64 | 8990.47 | 8942.03 |
| 889.467 | 5636.61 | 5707.75 | 13525.3 | 13795.8 | 33338.2 | 34211.3 | 68639.4 | 70256.9 | 92645.5 | 95462.9 | 889.467 | 3791.82 | 3811.75 | 8997.34 | 8958.16 |
| 889.744 | 5765.27 | 5718.89 | 13933.9 | 13818.5 | 33490.8 | 34257.6 | 69081.4 | 70337.1 | 93648.6 | 95562.7 | 889.744 | 3894.61 | 3819.86 | 8952.65 | 8974.29 |
| 890.02 | 5859.66 | 5730.04 | 13723.3 | 13841.3 | 33067.3 | 34303.9 | 68477.3 | 70417.3 | 93181.1 | 95662.4 | 890.02 | 3797.74 | 3827.99 | 8889.94 | 8990.44 |
| 890.297 | 5585.82 | 5741.2 | 13795.6 | 13864.1 | 33514.3 | 34350.2 | 68245.4 | 70497.5 | 93020.4 | 95762.1 | 890.297 | 3741.61 | 3836.12 | 8964.94 | 9006.61 |
| 890.573 | 5716.44 | 5752.37 | 13705.3 | 13886.9 | 33449.3 | 34396.6 | 68975.2 | 70577.7 | 93126.9 | 95861.8 | 890.573 | 3793.86 | 3844.27 | 8830.18 | 9022.79 |
| 890.85 | 5691.92 | 5763.55 | 13483.5 | 13909.7 | 33604.2 | 34443 | 68512.6 | 70657.9 | 93296.7 | 95961.5 | 890.85 | 3698.82 | 3852.42 | 8885.08 | 9038.98 |
| 891.126 | 5769.02 | 5774.74 | 13549.4 | 13932.6 | 33431.7 | 34489.4 | 68441.9 | 70738.1 | 92937.1 | 96061.2 | 891.126 | 3778.9 | 3860.59 | 8838.42 | 9055.19 |
| 891.402 | 5848.84 | 5785.95 | 13564.2 | 13955.4 | 34060.2 | 34535.8 | 69454.7 | 70818.4 | 93629.1 | 96160.8 | 891.402 | 3743.76 | 3868.76 | 8945.38 | 9071.41 |
| 891.679 | 5731.77 | 5797.16 | 13731.2 | 13978.3 | 33545.1 | 34582.2 | 68653.2 | 70898.6 | 93080 | 96260.5 | 891.679 | 3827.55 | 3876.95 | 8928.26 | 9087.65 |
| 891.955 | 5712.97 | 5808.39 | 13792.5 | 14001.2 | 33360.3 | 34628.6 | 69226.7 | 70978.8 | 93237.1 | 96360.2 | 891.955 | 3858.61 | 3885.14 | 9098.11 | 9103.89 |
| 892.232 | 5659.21 | 5819.62 | 13866.5 | 14024.1 | 33519.6 | 34675.1 | 68517.4 | 71059.1 | 92942.1 | 96459.8 | 892.232 | 3920.6 | 3893.34 | 8890.08 | 9120.16 |
| 892.508 | 5782.43 | 5830.87 | 13656.3 | 14047 | 33987 | 34721.5 | 69318.2 | 71139.3 | 93728.5 | 96559.5 | 892.508 | 3832.39 | 3901.56 | 9211.88 | 9136.43 |
| 892.784 | 5851.61 | 5842.13 | 13834.2 | 14069.9 | 33996.1 | 34768 | 68797.9 | 71219.5 | 94099.6 | 96659.1 | 892.784 | 3819.73 | 3909.78 | 9030.92 | 9152.72 |
| 893.061 | 5917.27 | 5853.4 | 13886.1 | 14092.8 | 34091.4 | 34814.5 | 69514.8 | 71299.8 | 93820.8 | 96758.8 | 893.061 | 3868.96 | 3918.01 | 9092.01 | 9169.03 |
| 893.337 | 5802.04 | 5864.68 | 13834.3 | 14115.8 | 33744.3 | 34861 | 69645.4 | 71380 | 94210.7 | 96858.4 | 893.337 | 3853.63 | 3926.26 | 9072.41 | 9185.35 |
| 893.613 | 5903.36 | 5875.97 | 14012.4 | 14138.8 | 34032.9 | 34907.5 | 68906 | 71460.3 | 94163.7 | 96958.1 | 893.613 | 3877.56 | 3934.51 | 9100.67 | 9201.68 |
| 893.889 | 5934.66 | 5887.27 | 13849.5 | 14161.7 | 34063.6 | 34954 | 69649.2 | 71540.6 | 94164.9 | 97057.7 | 893.889 | 3855.4 | 3942.77 | 9224.14 | 9218.02 |
| 894.166 | 5890.67 | 5898.59 | 13835.3 | 14184.7 | 33933.5 | 35000.6 | 69798.2 | 71620.8 | 94208.2 | 97157.3 | 894.166 | 3883.7 | 3951.05 | 9196.84 | 9234.39 |
| 894.442 | 5794.79 | 5909.91 | 14145.7 | 14207.8 | 34254.7 | 35047.1 | 68863.6 | 71701.1 | 94341.1 | 97256.9 | 894.442 | 3859.87 | 3959.33 | 9166.18 | 9250.76 |
| 894.718 | 5871.35 | 5921.25 | 14057.7 | 14230.8 | 34346.1 | 35093.7 | 69878.9 | 71781.4 | 95019.9 | 97356.5 | 894.718 | 3852.55 | 3967.62 | 9208.33 | 9267.15 |
| 894.994 | 5806.78 | 5932.59 | 14001.2 | 14253.8 | 34219.4 | 35140.3 | 69806.8 | 71861.6 | 95486.7 | 97456.1 | 894.994 | 3974.92 | 3975.92 | 9353.69 | 9283.55 |
| 895.27 | 5817.06 | 5943.95 | 14142.6 | 14276.9 | 34112.3 | 35186.9 | 70049 | 71941.9 | 94663.4 | 97555.6 | 895.27 | 3939.69 | 3984.23 | 9173.39 | 9299.96 |
| 895.547 | 5884.12 | 5955.32 | 14087.8 | 14300 | 34387.2 | 35233.5 | 69888.1 | 72022.2 | 94688.7 | 97655.2 | 895.547 | 3949.68 | 3992.55 | 9142.65 | 9316.39 |
| 895.823 | 5987.82 | 5966.7 | 14087.1 | 14323 | 34438.6 | 35280.1 | 69897.2 | 72102.5 | 94612.8 | 97754.8 | 895.823 | 3958.26 | 4000.89 | 9099.63 | 9332.83 |
| 896.099 | 5876.43 | 5978.09 | 14263.3 | 14346.1 | 34606.6 | 35326.8 | 70304 | 72182.8 | 94921 | 97854.4 | 896.099 | 3978.5 | 4009.22 | 9362.92 | 9349.29 |
| 896.375 | 5976.72 | 5989.49 | 14228.5 | 14369.2 | 34532.1 | 35373.4 | 70613.2 | 72263.1 | 94872.6 | 97953.9 | 896.375 | 3937.19 | 4017.58 | 9290.06 | 9365.76 |
| 896.651 | 6010.4 | 6000.9 | 14372.4 | 14392.4 | 34292.3 | 35420.1 | 70680 | 72343.4 | 96031.5 | 98053.4 | 896.651 | 4048.19 | 4025.94 | 9274.87 | 9382.24 |
| 896.927 | 5968.46 | 6012.33 | 14044 | 14415.5 | 34348.1 | 35466.8 | 70554.5 | 72423.7 | 95313.1 | 98153 | 896.927 | 3940.26 | 4034.31 | 9374.86 | 9398.74 |
| 897.203 | 6090.1 | 6023.76 | 14214.8 | 14438.7 | 34675.6 | 35513.4 | 71112.5 | 72504 | 95768.2 | 98252.5 | 897.203 | 4001.51 | 4042.69 | 9310.44 | 9415.25 |
| 897.479 | 6002.51 | 6035.2 | 13977 | 14461.9 | 34337.4 | 35560.2 | 70406.1 | 72584.3 | 95358 | 98352 | 897.479 | 4026.43 | 4051.08 | 9307.99 | 9431.77 |
| 897.755 | 6019.11 | 6046.66 | 14186.4 | 14485 | 34909.1 | 35606.9 | 71210.8 | 72664.6 | 96382.7 | 98451.5 | 897.755 | 4050.15 | 4059.48 | 9384.06 | 9448.31 |
| 898.031 | 6218.28 | 6058.13 | 14162.6 | 14508.2 | 34660.6 | 35653.6 | 70303.1 | 72744.9 | 95592.5 | 98551 | 898.031 | 4002.1 | 4067.89 | 9342.59 | 9464.86 |
| 898.307 | 6020.65 | 6069.6 | 14390.2 | 14531.5 | 34794.2 | 35700.3 | 70903.2 | 72825.2 | 95747.7 | 98650.5 | 898.307 | 4013.45 | 4076.3 | 9257 | 9481.43 |
| 898.583 | 6036.25 | 6081.09 | 14104.2 | 14554.7 | 34818.3 | 35747.1 | 71024.6 | 72905.5 | 96349.1 | 98750 | 898.583 | 4077.91 | 4084.73 | 9422.12 | 9498.01 |
| 898.859 | 6015.38 | 6092.59 | 14271.8 | 14577.9 | 34970.4 | 35793.9 | 70834 | 72985.8 | 96292.7 | 98849.5 | 898.859 | 4035.05 | 4093.17 | 9384.88 | 9514.6 |
| 899.135 | 6005.35 | 6104.1 | 14353.3 | 14601.2 | 35174.1 | 35840.7 | 71402.6 | 73066.1 | 96261.9 | 98948.9 | 899.135 | 4087.15 | 4101.62 | 9289.48 | 9531.21 |
| 899.411 | 6064.6 | 6115.62 | 14222.2 | 14624.5 | 34866.7 | 35887.5 | 71508.3 | 73146.5 | 96849.6 | 99048.4 | 899.411 | 4098.9 | 4110.08 | 9615.92 | 9547.83 |
| 899.687 | 6073.2 | 6127.15 | 14463.2 | 14647.7 | 35009.5 | 35934.3 | 71065.4 | 73226.8 | 96103 | 99147.8 | 899.687 | 4055.76 | 4118.55 | 9470.22 | 9564.46 |
| 899.963 | 6100.5 | 6138.69 | 14229.2 | 14671 | 35025.2 | 35981.1 | 71541.7 | 73307.1 | 97011.8 | 99247.3 | 899.963 | 4034.48 | 4127.02 | 9545.45 | 9581.11 |
| 900.239 | 6197.93 | 6150.25 | 14119.5 | 14694.4 | 35150.8 | 36027.9 | 71490.8 | 73387.4 | 96473.8 | 99346.7 | 900.239 | 4029.54 | 4135.51 | 9588.34 | 9597.77 |
| 900.514 | 6139.54 | 6161.81 | 14591.3 | 14717.7 | 35068.5 | 36074.8 | 72065 | 73467.8 | 96588.2 | 99446.1 | 900.514 | 4103.19 | 4144.01 | 9447.88 | 9614.44 |
| 900.79 | 6175.12 | 6173.38 | 14358.9 | 14741 | 35102.1 | 36121.7 | 71593.7 | 73548.1 | 96598.5 | 99545.5 | 900.79 | 4121.94 | 4152.52 | 9541.59 | 9631.13 |
| 901.066 | 6213.75 | 6184.97 | 14618 | 14764.4 | 35338.3 | 36168.5 | 71977.3 | 73628.4 | 97403.8 | 99644.9 | 901.066 | 4007.27 | 4161.03 | 9508.19 | 9647.83 |
| 901.342 | 6095.14 | 6196.56 | 14305.2 | 14787.8 | 35364.6 | 36215.4 | 71741.4 | 73708.8 | 96800.2 | 99744.3 | 901.342 | 4118.15 | 4169.56 | 9560.2 | 9664.55 |
| 901.617 | 6293.59 | 6208.17 | 14644.7 | 14811.1 | 35514 | 36262.3 | 72031.5 | 73789.1 | 97713.9 | 99843.7 | 901.617 | 4104.56 | 4178.09 | 9492.79 | 9681.27 |
| 901.893 | 6170.62 | 6219.79 | 14635.2 | 14834.5 | 35409 | 36309.2 | 72665.1 | 73869.4 | 97780.6 | 99943 | 901.893 | 4165.56 | 4186.64 | 9593.32 | 9698.01 |
| 902.169 | 6276.59 | 6231.42 | 14526.2 | 14857.9 | 35282.9 | 36356.2 | 72093.4 | 73949.8 | 97323.4 | 100042 | 902.169 | 4120.95 | 4195.19 | 9718.92 | 9714.77 |
| 902.445 | 6250.43 | 6243.06 | 14572 | 14881.4 | 35607.5 | 36403.1 | 72122.1 | 74030.1 | 97418.3 | 100142 | 902.445 | 4152.83 | 4203.76 | 9598.26 | 9731.54 |
| 902.72 | 6240.13 | 6254.7 | 14866.9 | 14904.8 | 35441.1 | 36450 | 72153.9 | 74110.5 | 97950.2 | 100241 | 902.72 | 4169.04 | 4212.34 | 9801.59 | 9748.32 |
| 902.996 | 6116.52 | 6266.36 | 14598.6 | 14928.3 | 35632.9 | 36497 | 72661.3 | 74190.8 | 97435.7 | 100340 | 902.996 | 4098.39 | 4220.92 | 9650.89 | 9765.11 |
| 903.272 | 6191.62 | 6278.04 | 14452.6 | 14951.7 | 35235.8 | 36544 | 72374.6 | 74271.1 | 96901.5 | 100440 | 903.272 | 4139.82 | 4229.51 | 9555.47 | 9781.92 |
| 903.547 | 6264.05 | 6289.72 | 14936.7 | 14975.2 | 35685.3 | 36590.9 | 72396 | 74351.5 | 98321.2 | 100539 | 903.547 | 4231.11 | 4238.12 | 9610.52 | 9798.74 |
| 903.823 | 6382.14 | 6301.41 | 14794.8 | 14998.7 | 35831.4 | 36637.9 | 72705.9 | 74431.8 | 98125.7 | 100638 | 903.823 | 4239.03 | 4246.73 | 9538.83 | 9815.58 |
| 904.098 | 6327.81 | 6313.11 | 14808.4 | 15022.2 | 36161.5 | 36685 | 72576.5 | 74512.2 | 98261.4 | 100738 | 904.098 | 4207.82 | 4255.35 | 9616.36 | 9832.43 |
| 904.374 | 6302.64 | 6324.83 | 14894 | 15045.7 | 35837.7 | 36732 | 72386.8 | 74592.5 | 97959.9 | 100837 | 904.374 | 4163.36 | 4263.99 | 9825.79 | 9849.29 |
| 904.65 | 6425.88 | 6336.55 | 14908.5 | 15069.3 | 35867.8 | 36779 | 72313.4 | 74672.9 | 98339.6 | 100936 | 904.65 | 4206.4 | 4272.63 | 9812.66 | 9866.16 |
| 904.925 | 6392.73 | 6348.28 | 14899.4 | 15092.8 | 35927.1 | 36826.1 | 73472 | 74753.2 | 98818.6 | 101035 | 904.925 | 4183.25 | 4281.28 | 9831.94 | 9883.05 |
| 905.201 | 6408.35 | 6360.03 | 14877.3 | 15116.4 | 36190.7 | 36873.1 | 73543.1 | 74833.6 | 98401.1 | 101135 | 905.201 | 4297.93 | 4289.94 | 9836.77 | 9899.95 |
| 905.476 | 6230.83 | 6371.79 | 14779.1 | 15140 | 36087.7 | 36920.2 | 73104.1 | 74913.9 | 98345.8 | 101234 | 905.476 | 4302.9 | 4298.62 | 9739.75 | 9916.87 |
| 905.752 | 6335.91 | 6383.55 | 14965.6 | 15163.6 | 36252.8 | 36967.3 | 72575.9 | 74994.3 | 98713.6 | 101333 | 905.752 | 4175.02 | 4307.3 | 9802.46 | 9933.8 |
| 906.027 | 6362.44 | 6395.33 | 15015.7 | 15187.2 | 36583.1 | 37014.3 | 73484.7 | 75074.6 | 99660 | 101432 | 906.027 | 4347.74 | 4315.99 | 9942.42 | 9950.74 |
| 906.303 | 6405.2 | 6407.12 | 14858.6 | 15210.8 | 36255.8 | 37061.5 | 73188.1 | 75154.9 | 98717.3 | 101531 | 906.303 | 4195.1 | 4324.69 | 9930.01 | 9967.69 |
| 906.578 | 6158.01 | 6418.92 | 14748 | 15234.4 | 36468.3 | 37108.6 | 73331.1 | 75235.3 | 98093.5 | 101631 | 906.578 | 4220.88 | 4333.4 | 9826.6 | 9984.66 |
| 906.854 | 6393.15 | 6430.72 | 14820.2 | 15258.1 | 35934.2 | 37155.7 | 73575.6 | 75315.6 | 99335.9 | 101730 | 906.854 | 4296.65 | 4342.12 | 9925.55 | 10001.6 |
| 907.129 | 6413.85 | 6442.54 | 14969 | 15281.7 | 36630.9 | 37202.8 | 73876 | 75396 | 99248.3 | 101829 | 907.129 | 4391.24 | 4350.85 | 10036.7 | 10018.6 |
| 907.404 | 6405.73 | 6454.37 | 14823.3 | 15305.4 | 36324.2 | 37250 | 73861.1 | 75476.3 | 99907.7 | 101928 | 907.404 | 4273.6 | 4359.59 | 9859.29 | 10035.6 |
| 907.68 | 6364.38 | 6466.21 | 15035.9 | 15329.1 | 36478.9 | 37297.1 | 74112.1 | 75556.7 | 98917.7 | 102027 | 907.68 | 4344.35 | 4368.34 | 9975.66 | 10052.7 |
| 907.955 | 6428.21 | 6478.06 | 14849.8 | 15352.8 | 36233.7 | 37344.3 | 73279.6 | 75637 | 99132.1 | 102126 | 907.955 | 4275.15 | 4377.1 | 10059.2 | 10069.7 |
| 908.231 | 6392.7 | 6489.93 | 14825.1 | 15376.5 | 36055.1 | 37391.5 | 73040.2 | 75717.4 | 99059.1 | 102225 | 908.231 | 4211.22 | 4385.87 | 9711.81 | 10086.7 |
| 908.506 | 6321.31 | 6501.8 | 15014.8 | 15400.2 | 36431.9 | 37438.7 | 73362.1 | 75797.7 | 98793.2 | 102324 | 908.506 | 4396.61 | 4394.64 | 9947.6 | 10103.8 |
| 908.781 | 6307.77 | 6513.68 | 15034 | 15424 | 36250.6 | 37485.9 | 73585.7 | 75878.1 | 99116.4 | 102424 | 908.781 | 4266.31 | 4403.43 | 10058.6 | 10120.9 |
| 909.056 | 6541.11 | 6525.57 | 15254.1 | 15447.7 | 37070.3 | 37533.1 | 74233.6 | 75958.4 | 100297 | 102523 | 909.056 | 4368.32 | 4412.23 | 10095.4 | 10138 |
| 909.332 | 6465.84 | 6537.48 | 15245.4 | 15471.5 | 36721.2 | 37580.3 | 74612.1 | 76038.8 | 100081 | 102622 | 909.332 | 4403.71 | 4421.04 | 9951.81 | 10155.1 |
| 909.607 | 6529.22 | 6549.39 | 15148 | 15495.3 | 36590.3 | 37627.6 | 74055 | 76119.1 | 100134 | 102721 | 909.607 | 4462.93 | 4429.85 | 10112.4 | 10172.2 |
| 909.882 | 6667.81 | 6561.32 | 15306.5 | 15519 | 36580.9 | 37674.8 | 74352.2 | 76199.4 | 100478 | 102820 | 909.882 | 4413.91 | 4438.68 | 10075 | 10189.3 |
| 910.157 | 6413.28 | 6573.25 | 15448.4 | 15542.9 | 37209.3 | 37722.1 | 74509.8 | 76279.8 | 99579.8 | 102919 | 910.157 | 4345.32 | 4447.51 | 10037.3 | 10206.4 |
| 910.433 | 6620.1 | 6585.2 | 15135 | 15566.7 | 36947.1 | 37769.3 | 74929.4 | 76360.1 | 100275 | 103018 | 910.433 | 4369.78 | 4456.36 | 10132.1 | 10223.6 |
| 910.708 | 6463.47 | 6597.15 | 15318.6 | 15590.5 | 36579 | 37816.6 | 74526.6 | 76440.4 | 99844.8 | 103117 | 910.708 | 4317.25 | 4465.21 | 9993.71 | 10240.8 |
| 910.983 | 6692.67 | 6609.12 | 15264.5 | 15614.4 | 36802.5 | 37863.9 | 74434.2 | 76520.8 | 100326 | 103216 | 910.983 | 4504.29 | 4474.08 | 10174.2 | 10257.9 |
| 911.258 | 6688.88 | 6621.1 | 15460.8 | 15638.2 | 37061.3 | 37911.2 | 74869 | 76601.1 | 100568 | 103315 | 911.258 | 4434.57 | 4482.95 | 10262.9 | 10275.1 |
| 911.533 | 6538.6 | 6633.08 | 15327.3 | 15662.1 | 37027.6 | 37958.5 | 74699.6 | 76681.4 | 100685 | 103414 | 911.533 | 4453.05 | 4491.84 | 10122.6 | 10292.3 |
| 911.808 | 6771.72 | 6645.08 | 15379.4 | 15686 | 37118 | 38005.9 | 75306.4 | 76761.8 | 100637 | 103512 | 911.808 | 4373.43 | 4500.73 | 10245.5 | 10309.5 |
| 912.083 | 6631.19 | 6657.09 | 15302.5 | 15709.9 | 37271.6 | 38053.2 | 75873.2 | 76842.1 | 101267 | 103611 | 912.083 | 4490.77 | 4509.63 | 10359.3 | 10326.8 |
| 912.358 | 6695.77 | 6669.1 | 15523.6 | 15733.8 | 37340.6 | 38100.5 | 74976.2 | 76922.4 | 100956 | 103710 | 912.358 | 4444.76 | 4518.54 | 10313.8 | 10344 |
| 912.633 | 6660.12 | 6681.13 | 15395 | 15757.7 | 37407.6 | 38147.9 | 75362.9 | 77002.8 | 101377 | 103809 | 912.633 | 4500.24 | 4527.47 | 10236.3 | 10361.3 |
| 912.909 | 6721.13 | 6693.17 | 15544 | 15781.6 | 36945.6 | 38195.2 | 75478.7 | 77083.1 | 101085 | 103908 | 912.909 | 4441.76 | 4536.4 | 10125.7 | 10378.5 |
| 913.183 | 6669.3 | 6705.22 | 15505.8 | 15805.6 | 37223.8 | 38242.6 | 75236.4 | 77163.4 | 100864 | 104007 | 913.183 | 4430.35 | 4545.34 | 10305.8 | 10395.8 |
| 913.458 | 6657.36 | 6717.28 | 15554.6 | 15829.6 | 37372.6 | 38290 | 75775.6 | 77243.7 | 102343 | 104106 | 913.458 | 4442.61 | 4554.29 | 10229.7 | 10413.1 |
| 913.734 | 6603.34 | 6729.35 | 15672 | 15853.5 | 37587.5 | 38337.4 | 75515.1 | 77324 | 101769 | 104205 | 913.734 | 4591.8 | 4563.25 | 10328.8 | 10430.4 |
| 914.008 | 6640.3 | 6741.43 | 15437.9 | 15877.5 | 37632 | 38384.8 | 75444.1 | 77404.3 | 101721 | 104303 | 914.008 | 4483.32 | 4572.22 | 10222.9 | 10447.7 |
| 914.283 | 6816.12 | 6753.52 | 15450 | 15901.5 | 37681.6 | 38432.2 | 76132.7 | 77484.6 | 102610 | 104402 | 914.283 | 4506.14 | 4581.2 | 10232.3 | 10465.1 |
| 914.558 | 6782.66 | 6765.62 | 15680.1 | 15925.5 | 37482.6 | 38479.7 | 75979.6 | 77564.9 | 101731 | 104501 | 914.558 | 4578.99 | 4590.19 | 10284.5 | 10482.4 |
| 914.833 | 6781.18 | 6777.73 | 15776.6 | 15949.6 | 37777 | 38527.1 | 75816.2 | 77645.2 | 103262 | 104600 | 914.833 | 4607.8 | 4599.19 | 10538.3 | 10499.8 |
| 915.108 | 6771.91 | 6789.85 | 15885.9 | 15973.6 | 37552.4 | 38574.5 | 76303 | 77725.6 | 102123 | 104698 | 915.108 | 4555.41 | 4608.19 | 10621.3 | 10517.1 |
| 915.383 | 6750.06 | 6801.99 | 15806.2 | 15997.6 | 37928.5 | 38622 | 76142.8 | 77805.9 | 102533 | 104797 | 915.383 | 4587.56 | 4617.21 | 10438.1 | 10534.5 |
| 915.658 | 6773.56 | 6814.13 | 15832.1 | 16021.7 | 38111.8 | 38669.4 | 76459.7 | 77886.1 | 103297 | 104896 | 915.658 | 4575.78 | 4626.24 | 10403.8 | 10551.9 |
| 915.933 | 6830.48 | 6826.28 | 15854.8 | 16045.8 | 38133.1 | 38716.9 | 76535.4 | 77966.4 | 103045 | 104995 | 915.933 | 4539.2 | 4635.27 | 10407.1 | 10569.3 |
| 916.207 | 6943.16 | 6838.44 | 15855.6 | 16069.9 | 38121.8 | 38764.4 | 75938.6 | 78046.7 | 102446 | 105093 | 916.207 | 4504.46 | 4644.32 | 10484.6 | 10586.8 |
| 916.482 | 6770.84 | 6850.61 | 15811.5 | 16094 | 38067.9 | 38811.9 | 76159.5 | 78127 | 102627 | 105192 | 916.482 | 4623.36 | 4653.38 | 10504.2 | 10604.2 |
| 916.757 | 7033.77 | 6862.8 | 15975.9 | 16118.1 | 37787.6 | 38859.4 | 76093.2 | 78207.3 | 102713 | 105291 | 916.757 | 4570.95 | 4662.44 | 10538.6 | 10621.6 |
| 917.032 | 6780.41 | 6874.99 | 15888.4 | 16142.2 | 38160.7 | 38906.9 | 77115 | 78287.6 | 103052 | 105389 | 917.032 | 4647.86 | 4671.52 | 10540.3 | 10639.1 |
| 917.307 | 6957.51 | 6887.19 | 15738.6 | 16166.4 | 37942.4 | 38954.4 | 76461.5 | 78367.9 | 103775 | 105488 | 917.307 | 4661.96 | 4680.6 | 10482.6 | 10656.6 |
| 917.581 | 6789.59 | 6899.41 | 15677.6 | 16190.5 | 38360.7 | 39002 | 76750 | 78448.1 | 103748 | 105587 | 917.581 | 4509.6 | 4689.69 | 10631 | 10674 |
| 917.856 | 6980.35 | 6911.63 | 16272.8 | 16214.7 | 38324.6 | 39049.5 | 77102.1 | 78528.4 | 103567 | 105685 | 917.856 | 4636.78 | 4698.8 | 10621.7 | 10691.5 |
| 918.131 | 6780.61 | 6923.86 | 15813.3 | 16238.9 | 38075.6 | 39097.1 | 76425.9 | 78608.7 | 102467 | 105784 | 918.131 | 4603.94 | 4707.91 | 10657.3 | 10709 |
| 918.405 | 6912.06 | 6936.11 | 16046.8 | 16263.1 | 38461 | 39144.6 | 77109.7 | 78688.9 | 103505 | 105882 | 918.405 | 4645.01 | 4717.03 | 10714.3 | 10726.6 |
| 918.68 | 6812.11 | 6948.36 | 16066.4 | 16287.3 | 38472.7 | 39192.2 | 76950.1 | 78769.2 | 103788 | 105981 | 918.68 | 4650.67 | 4726.16 | 10562 | 10744.1 |
| 918.955 | 6842.19 | 6960.63 | 16004.9 | 16311.5 | 38330.4 | 39239.8 | 77426.9 | 78849.5 | 103481 | 106079 | 918.955 | 4628.93 | 4735.3 | 10642.5 | 10761.7 |
| 919.229 | 6861.45 | 6972.9 | 16146.4 | 16335.7 | 38845.7 | 39287.3 | 77381.5 | 78929.7 | 103884 | 106178 | 919.229 | 4631.1 | 4744.45 | 10625.3 | 10779.2 |
| 919.504 | 6847.63 | 6985.19 | 16222.8 | 16360 | 38569 | 39334.9 | 77509.5 | 79009.9 | 103935 | 106276 | 919.504 | 4584.85 | 4753.61 | 10877.3 | 10796.8 |
| 919.779 | 7005.35 | 6997.48 | 16110.2 | 16384.2 | 38718.3 | 39382.5 | 77192.8 | 79090.2 | 103718 | 106375 | 919.779 | 4770.97 | 4762.78 | 10622.1 | 10814.4 |
| 920.053 | 7077.4 | 7009.78 | 16025.6 | 16408.5 | 38820 | 39430.2 | 77542.4 | 79170.4 | 104329 | 106473 | 920.053 | 4718.65 | 4771.96 | 10735.1 | 10832 |
| 920.328 | 7062.02 | 7022.1 | 16128.8 | 16432.8 | 38975.6 | 39477.8 | 77867.3 | 79250.7 | 104131 | 106572 | 920.328 | 4751 | 4781.15 | 10635.7 | 10849.6 |
| 920.602 | 7076.53 | 7034.42 | 15999 | 16457 | 38515.3 | 39525.4 | 77779.6 | 79330.9 | 104430 | 106670 | 920.602 | 4691.57 | 4790.35 | 10747.4 | 10867.2 |
| 920.877 | 7162.51 | 7046.76 | 16261.2 | 16481.3 | 38556.5 | 39573 | 77382.7 | 79411.1 | 103451 | 106769 | 920.877 | 4742.19 | 4799.56 | 10699.1 | 10884.9 |
| 921.151 | 6867.46 | 7059.1 | 16072 | 16505.7 | 38361 | 39620.7 | 77074.5 | 79491.4 | 104054 | 106867 | 921.151 | 4702.55 | 4808.77 | 10608.2 | 10902.5 |
| 921.426 | 7130.45 | 7071.46 | 16160.6 | 16530 | 38547.7 | 39668.3 | 77339.2 | 79571.6 | 103585 | 106966 | 921.426 | 4793.62 | 4818 | 10680.9 | 10920.2 |
| 921.7 | 7120.47 | 7083.82 | 16417.4 | 16554.3 | 39176.1 | 39716 | 78156.4 | 79651.8 | 104998 | 107064 | 921.7 | 4762.99 | 4827.23 | 10928.9 | 10937.8 |
| 921.975 | 6954.83 | 7096.2 | 16410.6 | 16578.7 | 38914.7 | 39763.7 | 77834.2 | 79732 | 104730 | 107162 | 921.975 | 4773.67 | 4836.48 | 10763.2 | 10955.5 |
| 922.249 | 7070.79 | 7108.58 | 16246 | 16603 | 38817.1 | 39811.4 | 78275.4 | 79812.2 | 104816 | 107261 | 922.249 | 4896.34 | 4845.73 | 10948 | 10973.2 |
| 922.523 | 7097.13 | 7120.98 | 16342.8 | 16627.4 | 38975.4 | 39859.1 | 78261.8 | 79892.4 | 105426 | 107359 | 922.523 | 4667.54 | 4855 | 10921.5 | 10990.9 |
| 922.798 | 7113.66 | 7133.38 | 16385.5 | 16651.8 | 39009.9 | 39906.8 | 77691.2 | 79972.6 | 104641 | 107457 | 922.798 | 4742.36 | 4864.27 | 10891 | 11008.6 |
| 923.072 | 7097.83 | 7145.8 | 16348 | 16676.2 | 39124.5 | 39954.4 | 78405.6 | 80052.8 | 105097 | 107555 | 923.072 | 4877.52 | 4873.55 | 10821.7 | 11026.4 |
| 923.346 | 6989.92 | 7158.22 | 16288 | 16700.6 | 38987.9 | 40002.2 | 78435 | 80132.9 | 105202 | 107654 | 923.346 | 4822.26 | 4882.84 | 10882.2 | 11044.1 |
| 923.621 | 7111.17 | 7170.65 | 16379.5 | 16725 | 38956.9 | 40049.9 | 77992.8 | 80213.1 | 105361 | 107752 | 923.621 | 4829.85 | 4892.15 | 10719.1 | 11061.9 |
| 923.895 | 7112.25 | 7183.1 | 16752.4 | 16749.5 | 38965.1 | 40097.6 | 78514 | 80293.3 | 104826 | 107850 | 923.895 | 4816.05 | 4901.45 | 10812 | 11079.7 |
| 924.169 | 7167.52 | 7195.55 | 16303.4 | 16773.9 | 38955.5 | 40145.4 | 78286.5 | 80373.5 | 105245 | 107948 | 924.169 | 4704.95 | 4910.77 | 10970.3 | 11097.4 |
| 924.444 | 7249.6 | 7208.02 | 16584.6 | 16798.4 | 39154.3 | 40193.1 | 78527 | 80453.6 | 105437 | 108047 | 924.444 | 4834.81 | 4920.1 | 11031.5 | 11115.2 |
| 924.718 | 7280.25 | 7220.49 | 16211.4 | 16822.8 | 39348.5 | 40240.9 | 78442.5 | 80533.8 | 105478 | 108145 | 924.718 | 4844.65 | 4929.44 | 10936.4 | 11133 |
| 924.992 | 7148.67 | 7232.98 | 16477.2 | 16847.3 | 39308.9 | 40288.6 | 78332 | 80613.9 | 104905 | 108243 | 924.992 | 4727.97 | 4938.79 | 10881.5 | 11150.9 |
| 925.266 | 7238 | 7245.47 | 16722.6 | 16871.8 | 39065.6 | 40336.4 | 79209.1 | 80694.1 | 106004 | 108341 | 925.266 | 4854.35 | 4948.15 | 11117.4 | 11168.7 |
| 925.54 | 7238.61 | 7257.98 | 16515.3 | 16896.3 | 39725.4 | 40384.2 | 78603.3 | 80774.2 | 106374 | 108439 | 925.54 | 4831.71 | 4957.52 | 11115.4 | 11186.5 |
| 925.815 | 7323.02 | 7270.49 | 16465.7 | 16920.8 | 39849 | 40432 | 79207.6 | 80854.3 | 106704 | 108537 | 925.815 | 4968.55 | 4966.89 | 11121.6 | 11204.4 |
| 926.089 | 7290.69 | 7283.01 | 16605.1 | 16945.4 | 39489.9 | 40479.8 | 78961.4 | 80934.5 | 105918 | 108635 | 926.089 | 4865.45 | 4976.28 | 11029.3 | 11222.3 |
| 926.363 | 7237.35 | 7295.55 | 16686.2 | 16969.9 | 39582.8 | 40527.6 | 79140.1 | 81014.6 | 106316 | 108733 | 926.363 | 4905.62 | 4985.67 | 11092.2 | 11240.2 |
| 926.637 | 7303.48 | 7308.09 | 16840.2 | 16994.5 | 39822.2 | 40575.4 | 79885.6 | 81094.7 | 106778 | 108831 | 926.637 | 5000.77 | 4995.08 | 11120.8 | 11258 |
| 926.911 | 7252.88 | 7320.64 | 16431.5 | 17019 | 39870 | 40623.2 | 79204.3 | 81174.8 | 106450 | 108929 | 926.911 | 4946.94 | 5004.49 | 11163.1 | 11276 |
| 927.185 | 7296.71 | 7333.21 | 16838.1 | 17043.6 | 40104.9 | 40671 | 80109.1 | 81254.9 | 106546 | 109027 | 927.185 | 4920.28 | 5013.91 | 11223.4 | 11293.9 |
| 927.459 | 7374.52 | 7345.78 | 16823 | 17068.2 | 39910.1 | 40718.8 | 79951.1 | 81335 | 107140 | 109125 | 927.459 | 4921.81 | 5023.35 | 11305.9 | 11311.8 |
| 927.734 | 7312.48 | 7358.36 | 16875.2 | 17092.8 | 40112.7 | 40766.7 | 79506.4 | 81415.1 | 106256 | 109223 | 927.734 | 4837.11 | 5032.79 | 11109.1 | 11329.8 |
| 928.008 | 7106.99 | 7370.95 | 16563 | 17117.4 | 39687.7 | 40814.5 | 79796.1 | 81495.2 | 106289 | 109321 | 928.008 | 4987.99 | 5042.24 | 11131.9 | 11347.7 |
| 928.282 | 7378.7 | 7383.56 | 16999.5 | 17142 | 39968.7 | 40862.4 | 79198.3 | 81575.2 | 106917 | 109419 | 928.282 | 4945.47 | 5051.7 | 11191.4 | 11365.7 |
| 928.556 | 7412.23 | 7396.17 | 16984 | 17166.6 | 39690.9 | 40910.2 | 79705.6 | 81655.3 | 107298 | 109517 | 928.556 | 5002.3 | 5061.17 | 11330.8 | 11383.6 |
| 928.83 | 7415.9 | 7408.79 | 17068.3 | 17191.3 | 40074.2 | 40958.1 | 80588.1 | 81735.4 | 107571 | 109615 | 928.83 | 5119 | 5070.65 | 11362.1 | 11401.6 |
| 929.104 | 7348.97 | 7421.42 | 16925.6 | 17215.9 | 39987.4 | 41006 | 80147 | 81815.4 | 107134 | 109713 | 929.104 | 5045.54 | 5080.13 | 11175.5 | 11419.6 |
| 929.377 | 7421.64 | 7434.06 | 16996.3 | 17240.6 | 40429.3 | 41053.9 | 79786.1 | 81895.5 | 107345 | 109811 | 929.377 | 5060.09 | 5089.63 | 11401.3 | 11437.7 |
| 929.651 | 7326.53 | 7446.72 | 17172.5 | 17265.3 | 40240.7 | 41101.8 | 80698.2 | 81975.5 | 107660 | 109909 | 929.651 | 4972.96 | 5099.14 | 11330.8 | 11455.7 |
| 929.925 | 7408.93 | 7459.38 | 16905.5 | 17290 | 40471 | 41149.7 | 79948.8 | 82055.6 | 107280 | 110006 | 929.925 | 5044.78 | 5108.65 | 11218.6 | 11473.7 |
| 930.199 | 7355.07 | 7472.05 | 16687.8 | 17314.7 | 40490.5 | 41197.6 | 80279.7 | 82135.6 | 107386 | 110104 | 930.199 | 5075.09 | 5118.18 | 11434.4 | 11491.8 |
| 930.473 | 7592.76 | 7484.73 | 17096.1 | 17339.4 | 40359.4 | 41245.5 | 80537.7 | 82215.6 | 108233 | 110202 | 930.473 | 4990.56 | 5127.71 | 11456.2 | 11509.8 |
| 930.747 | 7514.41 | 7497.42 | 17298 | 17364.1 | 40904.5 | 41293.4 | 80375.9 | 82295.6 | 108200 | 110300 | 930.747 | 5082.66 | 5137.26 | 11491.2 | 11527.9 |
| 931.021 | 7548.38 | 7510.11 | 17108 | 17388.8 | 40671.3 | 41341.3 | 81229.4 | 82375.6 | 109181 | 110397 | 931.021 | 5178.8 | 5146.81 | 11630.2 | 11546 |
| 931.294 | 7471.96 | 7522.82 | 17027.3 | 17413.6 | 40629.7 | 41389.2 | 81302.8 | 82455.6 | 108374 | 110495 | 931.294 | 5046.71 | 5156.37 | 11419.3 | 11564.1 |
| 931.568 | 7608.09 | 7535.54 | 17286.5 | 17438.3 | 40581.6 | 41437.2 | 81402.3 | 82535.6 | 108605 | 110593 | 931.568 | 5173.58 | 5165.94 | 11638 | 11582.2 |
| 931.842 | 7503.86 | 7548.27 | 17361.2 | 17463.1 | 40618.2 | 41485.1 | 81422.7 | 82615.6 | 108451 | 110690 | 931.842 | 5097.78 | 5175.52 | 11433.9 | 11600.3 |
| 932.116 | 7548.53 | 7561.01 | 17258.5 | 17487.9 | 40850.6 | 41533.1 | 80904.3 | 82695.6 | 108322 | 110788 | 932.116 | 5184.62 | 5185.11 | 11494.2 | 11618.5 |
| 932.39 | 7591.87 | 7573.76 | 17042.8 | 17512.7 | 40608.1 | 41581 | 81269.7 | 82775.5 | 108472 | 110886 | 932.39 | 5112.46 | 5194.71 | 11529.7 | 11636.6 |
| 932.663 | 7567.45 | 7586.51 | 17077.1 | 17537.5 | 40691.2 | 41629 | 81289.8 | 82855.5 | 109220 | 110983 | 932.663 | 5060.85 | 5204.32 | 11538.4 | 11654.8 |
| 932.937 | 7666.63 | 7599.28 | 17100.6 | 17562.3 | 40680.2 | 41677 | 80785.7 | 82935.4 | 108336 | 111081 | 932.937 | 5118.5 | 5213.94 | 11765.6 | 11672.9 |
| 933.211 | 7576.82 | 7612.06 | 17149.2 | 17587.1 | 40567.2 | 41724.9 | 81020.9 | 83015.4 | 108381 | 111178 | 933.211 | 5081.85 | 5223.56 | 11747.7 | 11691.1 |
| 933.484 | 7524.94 | 7624.84 | 17261 | 17612 | 40728.4 | 41772.9 | 81522.5 | 83095.3 | 108960 | 111276 | 933.484 | 5066.8 | 5233.2 | 11531.8 | 11709.3 |
| 933.758 | 7539.3 | 7637.64 | 17318.6 | 17636.8 | 40857.5 | 41820.9 | 81491 | 83175.2 | 108899 | 111374 | 933.758 | 5149.99 | 5242.85 | 11577.1 | 11727.5 |
| 934.032 | 7702.69 | 7650.44 | 17449.9 | 17661.7 | 40898.7 | 41868.9 | 81397.5 | 83255.1 | 109125 | 111471 | 934.032 | 5101.73 | 5252.5 | 11725.1 | 11745.7 |
| 934.305 | 7623.27 | 7663.26 | 17603 | 17686.5 | 41183.8 | 41916.9 | 81063 | 83335.1 | 108640 | 111568 | 934.305 | 5258.86 | 5262.16 | 11542.4 | 11764 |
| 934.579 | 7645.82 | 7676.08 | 17488.4 | 17711.4 | 41101.3 | 41964.9 | 80912.2 | 83414.9 | 109460 | 111666 | 934.579 | 5157.5 | 5271.84 | 11666 | 11782.2 |
| 934.852 | 7673.13 | 7688.91 | 17552 | 17736.3 | 41112.9 | 42012.9 | 81544.7 | 83494.9 | 109174 | 111763 | 934.852 | 5139.23 | 5281.52 | 11770 | 11800.5 |
| 935.126 | 7729.42 | 7701.76 | 17559.4 | 17761.2 | 40978.1 | 42060.9 | 81538 | 83574.7 | 108940 | 111861 | 935.126 | 5217.43 | 5291.21 | 11832.3 | 11818.7 |
| 935.4 | 7554.67 | 7714.61 | 17414.4 | 17786.1 | 41667.8 | 42109 | 81771.1 | 83654.6 | 109555 | 111958 | 935.4 | 5224.1 | 5300.91 | 11664.8 | 11837 |
| 935.673 | 7792.55 | 7727.47 | 17619.5 | 17811 | 41115.2 | 42157 | 81763.6 | 83734.5 | 108622 | 112056 | 935.673 | 5281.25 | 5310.62 | 11850.1 | 11855.3 |
| 935.947 | 7674.46 | 7740.34 | 17732.2 | 17836 | 41340.4 | 42205 | 82040.1 | 83814.3 | 109669 | 112153 | 935.947 | 5141.91 | 5320.34 | 11789.1 | 11873.6 |
| 936.22 | 7675.26 | 7753.23 | 17312.5 | 17860.9 | 41454.4 | 42253.1 | 81753.3 | 83894.2 | 109715 | 112250 | 936.22 | 5123.77 | 5330.06 | 11837.6 | 11891.9 |
| 936.494 | 7693.7 | 7766.11 | 17378.5 | 17885.9 | 41426.4 | 42301.1 | 81900.5 | 83974 | 109302 | 112348 | 936.494 | 5225.6 | 5339.8 | 11692 | 11910.2 |
| 936.767 | 7639.96 | 7779.02 | 17455.7 | 17910.8 | 41126.1 | 42349.2 | 82144.5 | 84053.9 | 109445 | 112445 | 936.767 | 5259.74 | 5349.55 | 11784.8 | 11928.5 |
| 937.04 | 7855.99 | 7791.92 | 17780.5 | 17935.8 | 41666.2 | 42397.2 | 81917.1 | 84133.7 | 109228 | 112542 | 937.04 | 5132.17 | 5359.3 | 11707.5 | 11946.9 |
| 937.314 | 7679.8 | 7804.84 | 17796.2 | 17960.8 | 41716.6 | 42445.3 | 81658.5 | 84213.5 | 110120 | 112639 | 937.314 | 5250.85 | 5369.07 | 11923.2 | 11965.2 |
| 937.587 | 7575.88 | 7817.77 | 17655.4 | 17985.8 | 41602.1 | 42493.4 | 82599.6 | 84293.3 | 110095 | 112736 | 937.587 | 5170.07 | 5378.84 | 11874.2 | 11983.6 |
| 937.861 | 7756.96 | 7830.71 | 17733.5 | 18010.8 | 41367.4 | 42541.4 | 81561.3 | 84373.1 | 110315 | 112834 | 937.861 | 5267.73 | 5388.62 | 11820.2 | 12002 |
| 938.134 | 7833.68 | 7843.65 | 17635.9 | 18035.8 | 41995.3 | 42589.5 | 82333.9 | 84452.9 | 110088 | 112931 | 938.134 | 5240.16 | 5398.41 | 11949.1 | 12020.4 |
| 938.407 | 7792.42 | 7856.61 | 17704.5 | 18060.9 | 41282.5 | 42637.6 | 82654.1 | 84532.6 | 110041 | 113028 | 938.407 | 5248.45 | 5408.21 | 11987.4 | 12038.8 |
| 938.681 | 7547.95 | 7869.57 | 17835.5 | 18085.9 | 41721.5 | 42685.7 | 82834.1 | 84612.4 | 110956 | 113125 | 938.681 | 5290.12 | 5418.02 | 11847.9 | 12057.2 |
| 938.954 | 7697.79 | 7882.55 | 17775.1 | 18111 | 41722.4 | 42733.8 | 82992 | 84692.2 | 110367 | 113222 | 938.954 | 5277.09 | 5427.84 | 11893.8 | 12075.6 |
| 939.227 | 7602.19 | 7895.53 | 17514.2 | 18136 | 41399 | 42781.9 | 81560.9 | 84771.9 | 109578 | 113319 | 939.227 | 5318.58 | 5437.67 | 11864.8 | 12094.1 |
| 939.5 | 7848.98 | 7908.53 | 17903.8 | 18161.1 | 41749.2 | 42830 | 82694.6 | 84851.6 | 110702 | 113416 | 939.5 | 5369.69 | 5447.51 | 12001.7 | 12112.5 |
| 939.774 | 7992.88 | 7921.53 | 17848.5 | 18186.2 | 41978.7 | 42878.1 | 82570.8 | 84931.4 | 109831 | 113513 | 939.774 | 5383.3 | 5457.35 | 12146.6 | 12131 |
| 940.047 | 7610.6 | 7934.54 | 17978.2 | 18211.2 | 42011.3 | 42926.2 | 82424.2 | 85011.1 | 110345 | 113610 | 940.047 | 5249.19 | 5467.21 | 12088.8 | 12149.4 |
| 940.32 | 7766.3 | 7947.56 | 17786.5 | 18236.3 | 41719.6 | 42974.4 | 83199.4 | 85090.8 | 110705 | 113707 | 940.32 | 5376.94 | 5477.07 | 11901.7 | 12167.9 |
| 940.593 | 7777.9 | 7960.59 | 17536.7 | 18261.5 | 41804.2 | 43022.5 | 82273.7 | 85170.5 | 110796 | 113804 | 940.593 | 5284.95 | 5486.95 | 11956.9 | 12186.4 |
| 940.867 | 7787.94 | 7973.63 | 17922.1 | 18286.6 | 41805.3 | 43070.6 | 82409.6 | 85250.2 | 110814 | 113901 | 940.867 | 5371.43 | 5496.83 | 12055.8 | 12204.9 |
| 941.14 | 7962.56 | 7986.68 | 17961.8 | 18311.7 | 42090.5 | 43118.8 | 83941.3 | 85329.8 | 111257 | 113998 | 941.14 | 5443.16 | 5506.72 | 12028.9 | 12223.4 |
| 941.413 | 7852.32 | 7999.74 | 17750.5 | 18336.9 | 41907.2 | 43166.9 | 83426.5 | 85409.5 | 111108 | 114095 | 941.413 | 5416.07 | 5516.62 | 11916.1 | 12242 |
| 941.686 | 8002.69 | 8012.81 | 18297.2 | 18362 | 42134.2 | 43215.1 | 83458.2 | 85489.2 | 112038 | 114192 | 941.686 | 5392.54 | 5526.53 | 12263.1 | 12260.5 |
| 941.959 | 7959.61 | 8025.88 | 18217.3 | 18387.2 | 41912 | 43263.2 | 83421.5 | 85568.8 | 111463 | 114289 | 941.959 | 5512.54 | 5536.45 | 12087.5 | 12279 |
| 942.232 | 8125.81 | 8038.97 | 17953.2 | 18412.4 | 42355.4 | 43311.4 | 82948.1 | 85648.4 | 111575 | 114386 | 942.232 | 5354.55 | 5546.37 | 12140.4 | 12297.6 |
| 942.505 | 7963.24 | 8052.06 | 18223.9 | 18437.5 | 42265.2 | 43359.5 | 83570.5 | 85728.1 | 112139 | 114482 | 942.505 | 5470.82 | 5556.31 | 12315 | 12316.2 |
| 942.778 | 8188.14 | 8065.17 | 18210.5 | 18462.7 | 42536.7 | 43407.7 | 83206.1 | 85807.7 | 111764 | 114579 | 942.778 | 5494.54 | 5566.26 | 12340.6 | 12334.8 |
| 943.051 | 7984.05 | 8078.28 | 18124.8 | 18487.9 | 42371.1 | 43455.9 | 83546.2 | 85887.3 | 111761 | 114676 | 943.051 | 5443.59 | 5576.21 | 12064.5 | 12353.4 |
| 943.324 | 7873.72 | 8091.4 | 17992.2 | 18513.2 | 42331.4 | 43504.1 | 83627.4 | 85966.9 | 111616 | 114773 | 943.324 | 5490.1 | 5586.17 | 12167.2 | 12372 |
| 943.597 | 8035.24 | 8104.54 | 18166.4 | 18538.4 | 42064.1 | 43552.2 | 83200.4 | 86046.4 | 111532 | 114869 | 943.597 | 5366.35 | 5596.15 | 12301.4 | 12390.6 |
| 943.87 | 7949.27 | 8117.68 | 18065.4 | 18563.6 | 42886.3 | 43600.4 | 83537.1 | 86126 | 111313 | 114966 | 943.87 | 5418.91 | 5606.13 | 12287.6 | 12409.2 |
| 944.143 | 7945.72 | 8130.83 | 18018.3 | 18588.9 | 42311.3 | 43648.6 | 83961.1 | 86205.6 | 111316 | 115063 | 944.143 | 5335.73 | 5616.12 | 12278.5 | 12427.9 |
| 944.416 | 7981.85 | 8143.99 | 18344.1 | 18614.1 | 42491.3 | 43696.8 | 84615.9 | 86285.1 | 112602 | 115159 | 944.416 | 5501.61 | 5626.12 | 12186.4 | 12446.5 |
| 944.689 | 8086.53 | 8157.15 | 18199 | 18639.4 | 42583.8 | 43745 | 84032.9 | 86364.7 | 112237 | 115256 | 944.689 | 5389.62 | 5636.12 | 12245.1 | 12465.2 |
| 944.962 | 8090.87 | 8170.33 | 18359.5 | 18664.7 | 42561.4 | 43793.2 | 83964.1 | 86444.2 | 112291 | 115352 | 944.962 | 5516.31 | 5646.14 | 12273.4 | 12483.8 |
| 945.235 | 7979.73 | 8183.52 | 18194.3 | 18689.9 | 42796.8 | 43841.4 | 83985.8 | 86523.7 | 112788 | 115449 | 945.235 | 5519.32 | 5656.17 | 12363.5 | 12502.5 |
| 945.508 | 8037.22 | 8196.71 | 18260.7 | 18715.2 | 42667 | 43889.6 | 84090.8 | 86603.2 | 112817 | 115546 | 945.508 | 5475.07 | 5666.2 | 12421 | 12521.2 |
| 945.78 | 8107.95 | 8209.92 | 18580 | 18740.5 | 42926.5 | 43937.8 | 84534.6 | 86682.7 | 113270 | 115642 | 945.78 | 5526.65 | 5676.25 | 12637.8 | 12539.9 |
| 946.053 | 7926.03 | 8223.13 | 18489.1 | 18765.8 | 43151.6 | 43986 | 84526.9 | 86762.2 | 112836 | 115738 | 946.053 | 5621.22 | 5686.3 | 12338.8 | 12558.6 |
| 946.326 | 8179.94 | 8236.35 | 18355.6 | 18791.2 | 42753.9 | 44034.3 | 84313.1 | 86841.6 | 112393 | 115835 | 946.326 | 5541.25 | 5696.36 | 12546.5 | 12577.4 |
| 946.599 | 8105.96 | 8249.58 | 18354.1 | 18816.5 | 43056.3 | 44082.5 | 84597.6 | 86921.1 | 113195 | 115931 | 946.599 | 5656.98 | 5706.43 | 12389.9 | 12596.1 |
| 946.872 | 8105.57 | 8262.82 | 18412.7 | 18841.8 | 42860.3 | 44130.7 | 85274.7 | 87000.5 | 113390 | 116028 | 946.872 | 5599.65 | 5716.51 | 12335.5 | 12614.9 |
| 947.144 | 8286.46 | 8276.07 | 18506.8 | 18867.2 | 43103 | 44178.9 | 84805.1 | 87080 | 114090 | 116124 | 947.144 | 5549.35 | 5726.6 | 12714.1 | 12633.6 |
| 947.417 | 8160.97 | 8289.33 | 18265.6 | 18892.6 | 43150.6 | 44227.2 | 85275.7 | 87159.4 | 113516 | 116220 | 947.417 | 5562.96 | 5736.7 | 12438 | 12652.4 |
| 947.69 | 8155.96 | 8302.6 | 18679.5 | 18917.9 | 43022.8 | 44275.4 | 84483.3 | 87238.8 | 113149 | 116317 | 947.69 | 5611.39 | 5746.8 | 12580.2 | 12671.2 |
| 947.963 | 8293.45 | 8315.87 | 18713.9 | 18943.3 | 43596.6 | 44323.7 | 85026.3 | 87318.2 | 113750 | 116413 | 947.963 | 5683.85 | 5756.92 | 12543.8 | 12690 |
| 948.235 | 8319.12 | 8329.16 | 18317.7 | 18968.7 | 42966.7 | 44371.9 | 84555.6 | 87397.6 | 113088 | 116509 | 948.235 | 5673.31 | 5767.04 | 12555.8 | 12708.8 |
| 948.508 | 8109.19 | 8342.45 | 18400.8 | 18994.1 | 43402.6 | 44420.2 | 84874.9 | 87476.9 | 112308 | 116606 | 948.508 | 5737.95 | 5777.18 | 12560.1 | 12727.6 |
| 948.78 | 8095.93 | 8355.76 | 18776 | 19019.5 | 43162.3 | 44468.4 | 85753.8 | 87556.3 | 113282 | 116702 | 948.78 | 5634.37 | 5787.32 | 12582.5 | 12746.4 |
| 949.053 | 8133.54 | 8369.07 | 18572.1 | 19044.9 | 43526.7 | 44516.7 | 85007.7 | 87635.6 | 113449 | 116798 | 949.053 | 5731.86 | 5797.47 | 12820 | 12765.3 |
| 949.326 | 8243.59 | 8382.39 | 18499.1 | 19070.4 | 43424.5 | 44564.9 | 86085.5 | 87715 | 113137 | 116894 | 949.326 | 5665.37 | 5807.63 | 12531.9 | 12784.1 |
| 949.598 | 8302.83 | 8395.72 | 18401.1 | 19095.8 | 43174.6 | 44613.2 | 85501 | 87794.3 | 113648 | 116990 | 949.598 | 5746.82 | 5817.8 | 12557 | 12803 |
| 949.871 | 8261.29 | 8409.06 | 18494 | 19121.2 | 43002.2 | 44661.5 | 85658 | 87873.6 | 113596 | 117086 | 949.871 | 5541.08 | 5827.97 | 12498.4 | 12821.8 |
| 950.143 | 8123.3 | 8422.41 | 18237.5 | 19146.7 | 43300.3 | 44709.7 | 84704.5 | 87952.9 | 113629 | 117183 | 950.143 | 5735.72 | 5838.16 | 12650.8 | 12840.7 |
| 950.416 | 8202.29 | 8435.76 | 18641 | 19172.2 | 43367.4 | 44758 | 85344.4 | 88032.2 | 113355 | 117279 | 950.416 | 5720.76 | 5848.35 | 12581 | 12859.6 |
| 950.688 | 8485.83 | 8449.13 | 18837.4 | 19197.6 | 43211.3 | 44806.3 | 85173.5 | 88111.5 | 113732 | 117375 | 950.688 | 5701.49 | 5858.56 | 12656.5 | 12878.5 |
| 950.961 | 8481.69 | 8462.5 | 18885.6 | 19223.1 | 43208.9 | 44854.6 | 85387.7 | 88190.7 | 114161 | 117471 | 950.961 | 5719.12 | 5868.77 | 12621.1 | 12897.4 |
| 951.233 | 8388.85 | 8475.88 | 18398.6 | 19248.6 | 43215.3 | 44902.9 | 85482.8 | 88270 | 113603 | 117567 | 951.233 | 5800.35 | 5878.99 | 12795.1 | 12916.4 |
| 951.506 | 8363.62 | 8489.27 | 18476.8 | 19274.1 | 43604.4 | 44951.1 | 85292.4 | 88349.2 | 113666 | 117663 | 951.506 | 5702.07 | 5889.22 | 12619.6 | 12935.3 |
| 951.778 | 8381.22 | 8502.67 | 18637 | 19299.6 | 43274.2 | 44999.4 | 85629.6 | 88428.4 | 113202 | 117759 | 951.778 | 5770.01 | 5899.46 | 12769.5 | 12954.2 |
| 952.051 | 8297.03 | 8516.08 | 18967.1 | 19325.2 | 43584.2 | 45047.7 | 85611.2 | 88507.6 | 113973 | 117854 | 952.051 | 5641.51 | 5909.71 | 12751.3 | 12973.2 |
| 952.323 | 8320.86 | 8529.5 | 19195.1 | 19350.7 | 43755.4 | 45096 | 85248.3 | 88586.8 | 114196 | 117950 | 952.323 | 5758.74 | 5919.96 | 12682.8 | 12992.2 |
| 952.595 | 8414.04 | 8542.93 | 18807.8 | 19376.2 | 43696.4 | 45144.3 | 85266.7 | 88666 | 112995 | 118046 | 952.595 | 5776.89 | 5930.23 | 12709.8 | 13011.1 |
| 952.868 | 8352.54 | 8556.36 | 19125 | 19401.8 | 43876.2 | 45192.6 | 85703.1 | 88745.2 | 114282 | 118142 | 952.868 | 5797.73 | 5940.5 | 12967.7 | 13030.1 |
| 953.14 | 8358.78 | 8569.81 | 18847.8 | 19427.3 | 43206.8 | 45240.9 | 85311.6 | 88824.3 | 113782 | 118238 | 953.14 | 5831.27 | 5950.79 | 12764.7 | 13049.1 |
| 953.412 | 8423.37 | 8583.26 | 18974.1 | 19452.9 | 43950 | 45289.2 | 85663.1 | 88903.4 | 114061 | 118334 | 953.412 | 5806.37 | 5961.08 | 12950.5 | 13068.1 |
| 953.685 | 8305.81 | 8596.72 | 18804.7 | 19478.5 | 43704.1 | 45337.5 | 85703.7 | 88982.6 | 113201 | 118429 | 953.685 | 5677.56 | 5971.38 | 12820.6 | 13087.2 |
| 953.957 | 8292.7 | 8610.19 | 18891.1 | 19504.1 | 43502.4 | 45385.8 | 85671.2 | 89061.7 | 114014 | 118525 | 953.957 | 5737.56 | 5981.69 | 12932.1 | 13106.2 |
| 954.229 | 8480.59 | 8623.67 | 19372.5 | 19529.7 | 44042.9 | 45434.2 | 86557.1 | 89140.8 | 114131 | 118621 | 954.229 | 5947.94 | 5992 | 12843.2 | 13125.2 |
| 954.501 | 8381.89 | 8637.16 | 18895.8 | 19555.3 | 44200.7 | 45482.5 | 85782 | 89219.9 | 114258 | 118716 | 954.501 | 5763.14 | 6002.33 | 12923 | 13144.3 |
| 954.774 | 8387.07 | 8650.65 | 18909.4 | 19580.9 | 44173.2 | 45530.8 | 86709.9 | 89298.9 | 114391 | 118812 | 954.774 | 5882.52 | 6012.67 | 12984.2 | 13163.4 |
| 955.046 | 8724.64 | 8664.16 | 19185 | 19606.5 | 44565.6 | 45579.1 | 86892.3 | 89378 | 114132 | 118908 | 955.046 | 5907.28 | 6023.01 | 12932.6 | 13182.4 |
| 955.318 | 8418.12 | 8677.67 | 19109.5 | 19632.1 | 43877.9 | 45627.4 | 86642.8 | 89457 | 114358 | 119003 | 955.318 | 5804.5 | 6033.36 | 12938.1 | 13201.5 |
| 955.59 | 8519.49 | 8691.19 | 18794.3 | 19657.8 | 44369.1 | 45675.8 | 86529.5 | 89536.1 | 115550 | 119099 | 955.59 | 5866.19 | 6043.72 | 12951.3 | 13220.6 |
| 955.862 | 8533.38 | 8704.72 | 19149.6 | 19683.4 | 44713.1 | 45724.1 | 86965.9 | 89615.1 | 115537 | 119194 | 955.862 | 5932.49 | 6054.09 | 13027.5 | 13239.7 |
| 956.135 | 8549.16 | 8718.26 | 19223.6 | 19709.1 | 44525.8 | 45772.4 | 86796.4 | 89694.1 | 115648 | 119290 | 956.135 | 5919.56 | 6064.47 | 13046.5 | 13258.8 |
| 956.407 | 8370.7 | 8731.81 | 19009.3 | 19734.7 | 43952.2 | 45820.8 | 86744.3 | 89773.1 | 114613 | 119385 | 956.407 | 5779.3 | 6074.86 | 13056 | 13277.9 |
| 956.679 | 8612.69 | 8745.36 | 19070.7 | 19760.4 | 44592.4 | 45869.1 | 87315.8 | 89852 | 115094 | 119481 | 956.679 | 5983.01 | 6085.26 | 13098 | 13297.1 |
| 956.951 | 8512.18 | 8758.93 | 19238 | 19786.1 | 44497.4 | 45917.4 | 87089.3 | 89931 | 115946 | 119576 | 956.951 | 5986.34 | 6095.66 | 13087.2 | 13316.2 |
| 957.223 | 8687.1 | 8772.5 | 19650.9 | 19811.8 | 44751 | 45965.8 | 87143.9 | 90009.9 | 115470 | 119672 | 957.223 | 5898.39 | 6106.07 | 13165.4 | 13335.4 |
| 957.495 | 8531.92 | 8786.08 | 19667.8 | 19837.5 | 44764.4 | 46014.1 | 88069.2 | 90088.9 | 116673 | 119767 | 957.495 | 5964.42 | 6116.49 | 13266.6 | 13354.5 |
| 957.767 | 8664.31 | 8799.67 | 19419.8 | 19863.2 | 45035.4 | 46062.4 | 87248 | 90167.8 | 116122 | 119862 | 957.767 | 6041.48 | 6126.93 | 13064.1 | 13373.7 |
| 958.039 | 8504.6 | 8813.27 | 19781.7 | 19888.9 | 44834.2 | 46110.8 | 87078.7 | 90246.7 | 115257 | 119958 | 958.039 | 5978.3 | 6137.36 | 13138.7 | 13392.9 |
| 958.311 | 8613.34 | 8826.88 | 19302.5 | 19914.7 | 45075.7 | 46159.1 | 88015.3 | 90325.6 | 116299 | 120053 | 958.311 | 5991.04 | 6147.81 | 13362.7 | 13412.1 |
| 958.583 | 8613.53 | 8840.5 | 19194.1 | 19940.4 | 44786.3 | 46207.5 | 86900.5 | 90404.4 | 116506 | 120148 | 958.583 | 5958.79 | 6158.27 | 13207.1 | 13431.3 |
| 958.855 | 8387.64 | 8854.12 | 19285.6 | 19966.1 | 44943.7 | 46255.8 | 87765.7 | 90483.3 | 115817 | 120243 | 958.855 | 5915.44 | 6168.73 | 13314 | 13450.5 |
| 959.127 | 8703.81 | 8867.75 | 19369.1 | 19991.9 | 45228.6 | 46304.2 | 87686.8 | 90562.1 | 116340 | 120338 | 959.127 | 6070.94 | 6179.21 | 13402.7 | 13469.8 |
| 959.399 | 8728.74 | 8881.39 | 19843.3 | 20017.6 | 44842.1 | 46352.5 | 87953.8 | 90640.9 | 117122 | 120434 | 959.399 | 5952.35 | 6189.69 | 13253.8 | 13489 |
| 959.67 | 8684.92 | 8895.04 | 19358.9 | 20043.4 | 44989 | 46400.9 | 87818.6 | 90719.8 | 116992 | 120529 | 959.67 | 5920.41 | 6200.18 | 13484.1 | 13508.2 |
| 959.942 | 8731.27 | 8908.7 | 19624.8 | 20069.2 | 45017.1 | 46449.2 | 88728.3 | 90798.6 | 117065 | 120624 | 959.942 | 6043.63 | 6210.68 | 13210.5 | 13527.5 |
| 960.214 | 8902.54 | 8922.37 | 19607.6 | 20095 | 44998.8 | 46497.6 | 87593 | 90877.3 | 116864 | 120719 | 960.214 | 6097.27 | 6221.19 | 13455 | 13546.8 |
| 960.486 | 8677.5 | 8936.04 | 19473 | 20120.8 | 45077.8 | 46545.9 | 88333.7 | 90956.1 | 117249 | 120814 | 960.486 | 6072.92 | 6231.7 | 13522.9 | 13566.1 |
| 960.758 | 8756.45 | 8949.72 | 19876.1 | 20146.6 | 45388.3 | 46594.3 | 88946.9 | 91034.9 | 118008 | 120909 | 960.758 | 5908.39 | 6242.23 | 13433.6 | 13585.4 |
| 961.03 | 8468.61 | 8963.42 | 19460.4 | 20172.4 | 45434.8 | 46642.7 | 88740.6 | 91113.6 | 116774 | 121004 | 961.03 | 6015.38 | 6252.76 | 13534.5 | 13604.6 |
| 961.301 | 8766.55 | 8977.11 | 19493.3 | 20198.2 | 45493.2 | 46691 | 88029.7 | 91192.3 | 117299 | 121099 | 961.301 | 6116.44 | 6263.3 | 13173.6 | 13624 |
| 961.573 | 8830.28 | 8990.82 | 19791.4 | 20224 | 45698.1 | 46739.4 | 89266.2 | 91271 | 117754 | 121194 | 961.573 | 6221.46 | 6273.85 | 13397 | 13643.3 |
| 961.845 | 9101.13 | 9004.54 | 20215.8 | 20249.9 | 45854.6 | 46787.8 | 90037.2 | 91349.7 | 118706 | 121289 | 961.845 | 6131.14 | 6284.41 | 13808.5 | 13662.6 |
| 962.117 | 8604.75 | 9018.26 | 19747.4 | 20275.7 | 45293.7 | 46836.1 | 88884.1 | 91428.4 | 117665 | 121384 | 962.117 | 6169.6 | 6294.98 | 13300 | 13682 |
| 962.388 | 9043.15 | 9032 | 19794.2 | 20301.6 | 45847.9 | 46884.5 | 89471.8 | 91507 | 119185 | 121478 | 962.388 | 6175.76 | 6305.56 | 13704 | 13701.3 |
| 962.66 | 8932.43 | 9045.74 | 20029.1 | 20327.4 | 45378.1 | 46932.8 | 89175.5 | 91585.6 | 118159 | 121573 | 962.66 | 6209.88 | 6316.14 | 13811.7 | 13720.7 |
| 962.932 | 8945.68 | 9059.49 | 19826.5 | 20353.3 | 45875.1 | 46981.2 | 89856 | 91664.3 | 118001 | 121668 | 962.932 | 6228.5 | 6326.74 | 13610.1 | 13740 |
| 963.203 | 8809.91 | 9073.24 | 20079.9 | 20379.2 | 45812.8 | 47029.6 | 89170.8 | 91742.9 | 118124 | 121763 | 963.203 | 6201.6 | 6337.34 | 13643.6 | 13759.4 |
| 963.475 | 8678.62 | 9087.01 | 19846.3 | 20405 | 46251.6 | 47077.9 | 88415.3 | 91821.5 | 118350 | 121858 | 963.475 | 6430.92 | 6347.95 | 13519.8 | 13778.8 |
| 963.746 | 9043.74 | 9100.78 | 19839.2 | 20430.9 | 45964.4 | 47126.3 | 89601.4 | 91900.1 | 119680 | 121952 | 963.746 | 6143.69 | 6358.57 | 13637.8 | 13798.2 |
| 964.018 | 9050.93 | 9114.57 | 19928 | 20456.8 | 45736 | 47174.7 | 89500.8 | 91978.6 | 118806 | 122047 | 964.018 | 6243.53 | 6369.19 | 13511.7 | 13817.6 |
| 964.29 | 8998.42 | 9128.36 | 19961.7 | 20482.8 | 46229.8 | 47223.1 | 89168.2 | 92057.1 | 119106 | 122142 | 964.29 | 6017.12 | 6379.83 | 13729.1 | 13837 |
| 964.561 | 8942.79 | 9142.16 | 19718 | 20508.7 | 45851.2 | 47271.4 | 90075.4 | 92135.7 | 118882 | 122236 | 964.561 | 6139.24 | 6390.48 | 13428.2 | 13856.5 |
| 964.833 | 9233.12 | 9155.96 | 20029.5 | 20534.6 | 46210.7 | 47319.8 | 89975.6 | 92214.2 | 119298 | 122331 | 964.833 | 6237.49 | 6401.13 | 13629.2 | 13875.9 |
| 965.104 | 8920.35 | 9169.78 | 20320.9 | 20560.5 | 46622.3 | 47368.2 | 89502 | 92292.7 | 118883 | 122425 | 965.104 | 6163.52 | 6411.79 | 13806.4 | 13895.4 |
| 965.376 | 9167.71 | 9183.6 | 19979 | 20586.5 | 46026.7 | 47416.6 | 90413.6 | 92371.2 | 119489 | 122520 | 965.376 | 6238.35 | 6422.46 | 13610 | 13914.8 |
| 965.647 | 9097.15 | 9197.43 | 20175.6 | 20612.4 | 46104.9 | 47464.9 | 90071 | 92449.6 | 120045 | 122614 | 965.647 | 6273.03 | 6433.14 | 13739.2 | 13934.3 |
| 965.919 | 8992.05 | 9211.27 | 20288.9 | 20638.4 | 46341.6 | 47513.3 | 90371.4 | 92528.1 | 118172 | 122709 | 965.919 | 6309.75 | 6443.82 | 13839.1 | 13953.8 |
| 966.19 | 9178.18 | 9225.12 | 20245.2 | 20664.3 | 46237.1 | 47561.7 | 91240.7 | 92606.5 | 120662 | 122803 | 966.19 | 6320.84 | 6454.52 | 13941.2 | 13973.2 |
| 966.461 | 9059.33 | 9238.98 | 20203.3 | 20690.3 | 46199.7 | 47610.1 | 90024 | 92684.9 | 120204 | 122898 | 966.461 | 6261.35 | 6465.22 | 13880.8 | 13992.8 |
| 966.733 | 9034.13 | 9252.84 | 20146 | 20716.3 | 46874.4 | 47658.4 | 91150.2 | 92763.3 | 120194 | 122992 | 966.733 | 6352.01 | 6475.93 | 14096.3 | 14012.3 |
| 967.004 | 9177.01 | 9266.71 | 20402.6 | 20742.3 | 46992.7 | 47706.8 | 90604.6 | 92841.7 | 120150 | 123086 | 967.004 | 6224.26 | 6486.65 | 13899.1 | 14031.8 |
| 967.275 | 9021.16 | 9280.59 | 20329.8 | 20768.3 | 46207 | 47755.2 | 91050.5 | 92920 | 120222 | 123180 | 967.275 | 6352.28 | 6497.38 | 13985.9 | 14051.3 |
| 967.547 | 9249.22 | 9294.48 | 20174 | 20794.3 | 46361.3 | 47803.6 | 90796.9 | 92998.4 | 119398 | 123275 | 967.547 | 6090.73 | 6508.12 | 13771.5 | 14070.8 |
| 967.818 | 9264.63 | 9308.37 | 20077.7 | 20820.3 | 46433.3 | 47851.9 | 91051.4 | 93076.7 | 120301 | 123369 | 967.818 | 6371.44 | 6518.86 | 13856.7 | 14090.4 |
| 968.09 | 9107.16 | 9322.28 | 20189.5 | 20846.3 | 46906.2 | 47900.3 | 90176.2 | 93155 | 119799 | 123463 | 968.09 | 6339.71 | 6529.62 | 14107.5 | 14110 |
| 968.361 | 9315.88 | 9336.19 | 20667.5 | 20872.3 | 46822.7 | 47948.7 | 91187.9 | 93233.3 | 120479 | 123557 | 968.361 | 6360.77 | 6540.38 | 14091.5 | 14129.5 |
| 968.632 | 9205.89 | 9350.11 | 20466.5 | 20898.4 | 46765.5 | 47997.1 | 90598.3 | 93311.6 | 120081 | 123652 | 968.632 | 6284.49 | 6551.15 | 13977.1 | 14149.1 |
| 968.903 | 9243.94 | 9364.04 | 20705.7 | 20924.4 | 47316.3 | 48045.4 | 91276.7 | 93389.8 | 120907 | 123746 | 968.903 | 6417.79 | 6561.93 | 13738.3 | 14168.7 |
| 969.174 | 9190.91 | 9377.98 | 20088.5 | 20950.5 | 46703.6 | 48093.8 | 90143.4 | 93468.1 | 120209 | 123840 | 969.174 | 6487.02 | 6572.71 | 14162.1 | 14188.3 |
| 969.446 | 9323.95 | 9391.92 | 20477.1 | 20976.5 | 46551.6 | 48142.2 | 91010.1 | 93546.3 | 120754 | 123934 | 969.446 | 6338.62 | 6583.51 | 14060.7 | 14207.9 |
| 969.717 | 9097.43 | 9405.87 | 20549.8 | 21002.6 | 46765.5 | 48190.6 | 91335.2 | 93624.5 | 121660 | 124028 | 969.717 | 6178.26 | 6594.31 | 14064.4 | 14227.5 |
| 969.988 | 8891.73 | 9419.83 | 20506.2 | 21028.6 | 46856.8 | 48238.9 | 91722.3 | 93702.7 | 121345 | 124122 | 969.988 | 6386.48 | 6605.12 | 14222 | 14247.1 |
| 970.259 | 9317.35 | 9433.8 | 20605.2 | 21054.7 | 47152.1 | 48287.3 | 92389.7 | 93780.9 | 119821 | 124216 | 970.259 | 6454.48 | 6615.94 | 14013.6 | 14266.8 |
| 970.53 | 9359.85 | 9447.77 | 20746 | 21080.8 | 47351.3 | 48335.7 | 91029.5 | 93859 | 120865 | 124310 | 970.53 | 6565.72 | 6626.77 | 13860.1 | 14286.4 |
| 970.801 | 9382.3 | 9461.76 | 20363.2 | 21106.9 | 47467.9 | 48384.1 | 92751.3 | 93937.1 | 121690 | 124404 | 970.801 | 6518.37 | 6637.61 | 14275.2 | 14306 |
| 971.072 | 9359.39 | 9475.75 | 20661 | 21133 | 47516.8 | 48432.4 | 91588.7 | 94015.3 | 120864 | 124497 | 971.072 | 6532.51 | 6648.45 | 14515.6 | 14325.7 |
| 971.343 | 9231.27 | 9489.75 | 20950.9 | 21159.1 | 47101.8 | 48480.8 | 92289.8 | 94093.4 | 122412 | 124591 | 971.343 | 6447.8 | 6659.31 | 14323.4 | 14345.4 |
| 971.615 | 9478.82 | 9503.76 | 20177.8 | 21185.2 | 47189.8 | 48529.2 | 91634.4 | 94171.4 | 121693 | 124685 | 971.615 | 6391.99 | 6670.17 | 14261.7 | 14365.1 |
| 971.886 | 9356.4 | 9517.77 | 21045.2 | 21211.3 | 47923.4 | 48577.6 | 91943.8 | 94249.5 | 122451 | 124779 | 971.886 | 6752.71 | 6681.04 | 14354 | 14384.8 |
| 972.157 | 9591.04 | 9531.79 | 20737.9 | 21237.5 | 47533.4 | 48625.9 | 92389.1 | 94327.5 | 120899 | 124872 | 972.157 | 6636.47 | 6691.92 | 14219.2 | 14404.5 |
| 972.428 | 9321.42 | 9545.83 | 20694.7 | 21263.6 | 46919.7 | 48674.3 | 91564.1 | 94405.6 | 120472 | 124966 | 972.428 | 6409.84 | 6702.8 | 14046.1 | 14424.2 |
| 972.698 | 9435.78 | 9559.86 | 20552.7 | 21289.8 | 47740.8 | 48722.7 | 91666.6 | 94483.6 | 122005 | 125060 | 972.698 | 6555.74 | 6713.7 | 14111.9 | 14443.9 |
| 972.97 | 9522.61 | 9573.91 | 20410.3 | 21315.9 | 46802.4 | 48771 | 92667.8 | 94561.6 | 121824 | 125153 | 972.97 | 6353.46 | 6724.6 | 14319.3 | 14463.6 |
| 973.24 | 9478.76 | 9587.96 | 20603.5 | 21342.1 | 48177.5 | 48819.4 | 92684.3 | 94639.5 | 122553 | 125247 | 973.24 | 6492.13 | 6735.51 | 14390.5 | 14483.4 |
| 973.511 | 9448.03 | 9602.02 | 20963.7 | 21368.2 | 47501.6 | 48867.8 | 92438.7 | 94717.5 | 121800 | 125341 | 973.511 | 6512.27 | 6746.43 | 14166.2 | 14503.1 |
| 973.782 | 9218.75 | 9616.09 | 21073.2 | 21394.4 | 47337.9 | 48916.1 | 91935.3 | 94795.4 | 122226 | 125434 | 973.782 | 6618.95 | 6757.36 | 14408.9 | 14522.9 |
| 974.053 | 9254.48 | 9630.17 | 20777.2 | 21420.6 | 48034.1 | 48964.5 | 91780.8 | 94873.3 | 122038 | 125528 | 974.053 | 6736.95 | 6768.29 | 14170.6 | 14542.6 |
| 974.324 | 9463.79 | 9644.26 | 20542.6 | 21446.8 | 47693.2 | 49012.9 | 92775.1 | 94951.2 | 121554 | 125621 | 974.324 | 6666.78 | 6779.24 | 14438.3 | 14562.4 |
| 974.595 | 9559.5 | 9658.35 | 20950.6 | 21473 | 47576.6 | 49061.2 | 91520.5 | 95029.1 | 121614 | 125715 | 974.595 | 6606.62 | 6790.19 | 14229.8 | 14582.2 |
| 974.866 | 9588.87 | 9672.45 | 20668.5 | 21499.2 | 47848.1 | 49109.6 | 92550.3 | 95107 | 121886 | 125808 | 974.866 | 6631.42 | 6801.15 | 14466.7 | 14602 |
| 975.136 | 9783.44 | 9686.56 | 21326.7 | 21525.4 | 47969.6 | 49158 | 93683.9 | 95184.8 | 124029 | 125901 | 975.136 | 6770.15 | 6812.12 | 14663.3 | 14621.8 |
| 975.407 | 9610.63 | 9700.67 | 21035.2 | 21551.6 | 47924.8 | 49206.3 | 92401 | 95262.6 | 123143 | 125995 | 975.407 | 6600.34 | 6823.09 | 14433.6 | 14641.6 |
| 975.678 | 9579.91 | 9714.8 | 20963.5 | 21577.8 | 48044.8 | 49254.7 | 93224.9 | 95340.4 | 123429 | 126088 | 975.678 | 6468.24 | 6834.08 | 14675 | 14661.4 |
| 975.949 | 9546.76 | 9728.93 | 21370.8 | 21604 | 48464.7 | 49303 | 94258.2 | 95418.2 | 122738 | 126181 | 975.949 | 6707.98 | 6845.07 | 14744 | 14681.3 |
| 976.219 | 9284.14 | 9743.07 | 21422.5 | 21630.3 | 47560.7 | 49351.4 | 93462.5 | 95496 | 123388 | 126274 | 976.219 | 6833.84 | 6856.07 | 14617.8 | 14701.1 |
| 976.49 | 9618.03 | 9757.22 | 21517.8 | 21656.5 | 47018.5 | 49399.8 | 93762.4 | 95573.7 | 122606 | 126368 | 976.49 | 6707.41 | 6867.08 | 14508.4 | 14721 |
| 976.761 | 9632.77 | 9771.37 | 21364.7 | 21682.8 | 48085.2 | 49448.1 | 93172.2 | 95651.5 | 124044 | 126461 | 976.761 | 6684.01 | 6878.1 | 14683.8 | 14740.8 |
| 977.031 | 9604.98 | 9785.53 | 20903.5 | 21709 | 48381.1 | 49496.5 | 92887.6 | 95729.2 | 122946 | 126554 | 977.031 | 6643.36 | 6889.12 | 14715.3 | 14760.7 |
| 977.302 | 9527.15 | 9799.7 | 21314 | 21735.3 | 47807.8 | 49544.8 | 93948.8 | 95806.9 | 124468 | 126647 | 977.302 | 6784.92 | 6900.16 | 14638.4 | 14780.5 |
| 977.573 | 9759.38 | 9813.88 | 21264.5 | 21761.5 | 48021.2 | 49593.2 | 93021.4 | 95884.5 | 122941 | 126740 | 977.573 | 6570.85 | 6911.2 | 14734.8 | 14800.4 |
| 977.843 | 9670.02 | 9828.06 | 21537 | 21787.8 | 48728.5 | 49641.5 | 94393.8 | 95962.2 | 123535 | 126833 | 977.843 | 6742.48 | 6922.25 | 14693.5 | 14820.3 |
| 978.114 | 9768.06 | 9842.25 | 21484.1 | 21814.1 | 48314.3 | 49689.8 | 94177.7 | 96039.8 | 124454 | 126926 | 978.114 | 6845.76 | 6933.31 | 14657.2 | 14840.2 |
| 978.385 | 9478.31 | 9856.45 | 21251.9 | 21840.4 | 48181.2 | 49738.2 | 93379 | 96117.4 | 123276 | 127019 | 978.385 | 6666.99 | 6944.37 | 14895.6 | 14860.1 |
| 978.655 | 9565.26 | 9870.66 | 21421.9 | 21866.7 | 48957.1 | 49786.5 | 94041.1 | 96195 | 124327 | 127112 | 978.655 | 6735.07 | 6955.44 | 14741.5 | 14880.1 |
| 978.926 | 9842.83 | 9884.87 | 21266.8 | 21893 | 48340.9 | 49834.9 | 93418.6 | 96272.6 | 123758 | 127205 | 978.926 | 6871.52 | 6966.53 | 14741.8 | 14900 |
| 979.196 | 9473.81 | 9899.1 | 21681.5 | 21919.3 | 48492.4 | 49883.2 | 93642.3 | 96350.1 | 123645 | 127298 | 979.196 | 6759.81 | 6977.62 | 14834.2 | 14919.9 |
| 979.466 | 9748.57 | 9913.33 | 21413.7 | 21945.6 | 48812 | 49931.6 | 94511.3 | 96427.7 | 123493 | 127391 | 979.466 | 6831.39 | 6988.71 | 14903.9 | 14939.9 |
| 979.737 | 9731.35 | 9927.56 | 21657.4 | 21971.9 | 49229.2 | 49979.9 | 95239.8 | 96505.2 | 125314 | 127484 | 979.737 | 6809.31 | 6999.82 | 14779.4 | 14959.9 |
| 1101.15 | 16716 | 16976.8 | 33667.2 | 34266.8 | 70298.5 | 71247.7 | 127997 | 129123 | 164189 | 165662 | 1101.15 | 12969.2 | 12748.1 | 24677.1 | 24716.7 |
| 1102.87 | 16623.7 | 17084.3 | 33793.5 | 34444.2 | 70584.4 | 71537 | 129019 | 129546 | 165402 | 166145 | 1102.87 | 12471.5 | 12839.5 | 24349.8 | 24864.1 |
| 1104.6 | 16954.9 | 17191.8 | 34388.8 | 34621.6 | 71238.2 | 71825.6 | 128554 | 129968 | 166022 | 166627 | 1104.6 | 12927.7 | 12931.1 | 24678.1 | 25011.7 |
| 1106.32 | 17122.8 | 17299.5 | 34760.7 | 34798.8 | 71056.8 | 72113.7 | 128796 | 130389 | 165857 | 167106 | 1106.32 | 13302.5 | 13022.9 | 24568.7 | 25159.3 |
| 1108.05 | 16961.2 | 17407.3 | 34541.7 | 34976 | 70829.8 | 72401.1 | 129799 | 130808 | 166529 | 167584 | 1108.05 | 13050.3 | 13114.9 | 25195.2 | 25307.1 |
| 1109.77 | 17242.4 | 17515.2 | 34537.8 | 35153 | 71242.8 | 72687.9 | 130485 | 131226 | 167049 | 168060 | 1109.77 | 13566.9 | 13207.1 | 25418.1 | 25455 |
| 1111.5 | 17377 | 17623.2 | 34812.2 | 35330 | 71989.6 | 72974.1 | 130420 | 131642 | 167150 | 168533 | 1111.5 | 12904.1 | 13299.5 | 25333.6 | 25603.1 |
| 1113.22 | 17133.9 | 17731.2 | 35282.1 | 35506.8 | 72516.9 | 73259.6 | 130840 | 132057 | 168116 | 169005 | 1113.22 | 12725.4 | 13392.1 | 25806.3 | 25751.2 |
| 1114.95 | 17748.5 | 17839.4 | 35823.5 | 35683.5 | 72422 | 73544.5 | 131821 | 132470 | 169180 | 169474 | 1114.95 | 13279.9 | 13484.8 | 25789.9 | 25899.4 |
| 1116.67 | 17742.9 | 17947.6 | 35750.4 | 35860.2 | 72962.6 | 73828.7 | 131561 | 132882 | 169302 | 169942 | 1116.67 | 13317.8 | 13577.8 | 25674 | 26047.7 |
| 1118.39 | 17992.7 | 18056 | 35715.8 | 36036.6 | 72930.5 | 74112.3 | 132126 | 133293 | 169111 | 170408 | 1118.39 | 13412.2 | 13670.9 | 25617.5 | 26196.1 |
| 1120.12 | 18283.6 | 18164.4 | 35751.3 | 36213 | 72826.1 | 74395.2 | 132603 | 133702 | 169378 | 170872 | 1120.12 | 13365.1 | 13764.2 | 25683.4 | 26344.6 |
| 1121.84 | 18044 | 18272.8 | 35944.2 | 36389.2 | 73503.1 | 74677.4 | 132676 | 134109 | 169716 | 171333 | 1121.84 | 13668.5 | 13857.7 | 26003.3 | 26493.2 |
| 1123.57 | 18691.7 | 18381.4 | 36001.5 | 36565.2 | 74233.5 | 74959 | 133746 | 134515 | 170560 | 171793 | 1123.57 | 13442.3 | 13951.3 | 26528.3 | 26641.9 |
| 1125.29 | 18396 | 18490 | 36420.7 | 36741.1 | 74383.3 | 75239.9 | 133877 | 134920 | 171213 | 172251 | 1125.29 | 13741.4 | 14045.2 | 26368.8 | 26790.6 |
| 1127.01 | 18283.7 | 18598.7 | 36468.7 | 36916.9 | 74304.5 | 75520.1 | 133973 | 135323 | 171687 | 172706 | 1127.01 | 14292.9 | 14139.2 | 26162.8 | 26939.5 |
| 1128.74 | 18433.1 | 18707.4 | 36450.4 | 37092.5 | 74730.3 | 75799.6 | 134461 | 135724 | 172422 | 173160 | 1128.74 | 14099.5 | 14233.3 | 26714.7 | 27088.3 |
| 1130.46 | 18208.5 | 18816.2 | 36487.8 | 37267.9 | 75171.9 | 76078.4 | 134663 | 136124 | 172505 | 173612 | 1130.46 | 14461 | 14327.6 | 26769.9 | 27237.3 |
| 1132.18 | 18538.8 | 18925.1 | 36655.7 | 37443.2 | 75352.7 | 76356.5 | 134971 | 136522 | 172639 | 174061 | 1132.18 | 14312.2 | 14422.1 | 26876.3 | 27386.3 |
| 1133.91 | 18536.7 | 19034 | 37067.9 | 37618.3 | 75748.6 | 76633.8 | 135973 | 136919 | 173312 | 174509 | 1133.91 | 14254.4 | 14516.8 | 27409.6 | 27535.4 |
| 1135.63 | 18688.9 | 19143 | 36968.8 | 37793.2 | 76301.7 | 76910.5 | 136406 | 137315 | 173830 | 174954 | 1135.63 | 14727.4 | 14611.6 | 27013.9 | 27684.5 |
| 1137.35 | 18990.8 | 19252 | 37375.8 | 37968 | 75873.6 | 77186.4 | 135848 | 137709 | 173792 | 175398 | 1137.35 | 14514.5 | 14706.6 | 27298.6 | 27833.7 |
| 1139.07 | 18763.5 | 19361 | 37511.5 | 38142.6 | 75801 | 77461.6 | 136726 | 138101 | 174134 | 175839 | 1139.07 | 14415.2 | 14801.7 | 27909.2 | 27982.9 |
| 1140.8 | 18989 | 19470.1 | 37356.2 | 38316.9 | 76497.8 | 77736.1 | 137256 | 138492 | 175314 | 176278 | 1140.8 | 15005.1 | 14896.9 | 27383.5 | 28132.2 |
| 1142.52 | 19368.7 | 19579.2 | 37999.8 | 38491.1 | 77134.8 | 78009.8 | 137010 | 138881 | 175440 | 176716 | 1142.52 | 15143.7 | 14992.3 | 27696.2 | 28281.5 |
| 1144.24 | 19470.8 | 19688.3 | 38015.2 | 38665.1 | 77653.3 | 78282.8 | 137727 | 139269 | 175852 | 177151 | 1144.24 | 15042.2 | 15087.9 | 28217 | 28430.8 |
| 1145.96 | 19625.3 | 19797.5 | 38613.2 | 38838.8 | 77955.5 | 78555.1 | 139014 | 139655 | 177180 | 177584 | 1145.96 | 15357.8 | 15183.6 | 28366 | 28580.2 |
| 1147.69 | 19298.7 | 19906.7 | 38889.3 | 39012.4 | 78193.4 | 78826.6 | 139011 | 140039 | 177048 | 178015 | 1147.69 | 15357.8 | 15279.4 | 28715.8 | 28729.5 |
| 1149.41 | 19360.1 | 20015.9 | 38354.9 | 39185.8 | 78471.4 | 79097.3 | 139141 | 140422 | 177461 | 178444 | 1149.41 | 15141.5 | 15375.4 | 28647.7 | 28878.9 |
| 1151.13 | 19776.9 | 20125.2 | 38989.6 | 39358.9 | 78902.1 | 79367.2 | 140143 | 140804 | 179159 | 178871 | 1151.13 | 15441 | 15471.5 | 29096.9 | 29028.4 |
| 1152.85 | 20128 | 20234.4 | 38775.7 | 39531.8 | 79517.5 | 79636.4 | 140778 | 141184 | 179898 | 179296 | 1152.85 | 15916 | 15567.7 | 29267.1 | 29177.8 |
| 1154.57 | 20405.8 | 20343.7 | 38781.1 | 39704.5 | 79525 | 79904.8 | 141228 | 141562 | 179955 | 179719 | 1154.57 | 15560.2 | 15664 | 28883.9 | 29327.3 |
| 1156.3 | 20653.1 | 20453 | 39373.3 | 39876.9 | 79829.9 | 80172.4 | 141618 | 141939 | 180376 | 180140 | 1156.3 | 15867.6 | 15760.5 | 29385.8 | 29476.7 |
| 1158.02 | 20351.3 | 20562.3 | 39630 | 40049.2 | 80405.2 | 80439.3 | 142123 | 142314 | 180886 | 180559 | 1158.02 | 16201.9 | 15857.1 | 29720.4 | 29626.2 |
| 1159.74 | 20347.8 | 20671.6 | 40190.6 | 40221.1 | 80133.6 | 80705.3 | 142568 | 142687 | 180305 | 180975 | 1159.74 | 15694.6 | 15953.8 | 29628.9 | 29775.7 |
| 1161.46 | 20725.1 | 20780.9 | 40140.3 | 40392.9 | 80343.3 | 80970.6 | 142382 | 143059 | 180708 | 181390 | 1161.46 | 16165.1 | 16050.7 | 29838.3 | 29925.1 |
| 1163.18 | 20803.6 | 20890.2 | 39872 | 40564.4 | 80795.5 | 81235 | 142270 | 143430 | 181055 | 181802 | 1163.18 | 16207.3 | 16147.6 | 29792 | 30074.6 |
| 1164.9 | 20588.9 | 20999.5 | 40518.2 | 40735.6 | 80692.2 | 81498.7 | 143159 | 143798 | 181460 | 182213 | 1164.9 | 16255.9 | 16244.7 | 29942.1 | 30224 |
| 1166.62 | 20321 | 21108.7 | 40438.4 | 40906.6 | 81012.1 | 81761.5 | 143306 | 144166 | 182301 | 182621 | 1166.62 | 16172.2 | 16341.9 | 30438.7 | 30373.4 |
| 1168.34 | 21000 | 21218 | 40507.4 | 41077.3 | 81372.3 | 82023.6 | 143741 | 144531 | 182242 | 183027 | 1168.34 | 16230.4 | 16439.1 | 30646.9 | 30522.8 |
| 1170.06 | 21069.2 | 21327.2 | 40772.3 | 41247.8 | 80964.2 | 82284.8 | 142709 | 144895 | 180770 | 183431 | 1170.06 | 16007.9 | 16536.5 | 30159.3 | 30672.2 |
| 1171.78 | 20692.4 | 21436.5 | 40457.7 | 41417.9 | 80884.8 | 82545.2 | 143192 | 145257 | 181658 | 183833 | 1171.78 | 16304.8 | 16634 | 29855.3 | 30821.6 |
| 1173.5 | 21155.7 | 21545.7 | 40967.1 | 41587.9 | 81261 | 82804.7 | 143862 | 145618 | 182510 | 184233 | 1173.5 | 16679.2 | 16731.6 | 30633.3 | 30970.9 |
| 1175.22 | 21112.6 | 21654.9 | 40874.3 | 41757.5 | 82226.9 | 83063.5 | 144262 | 145977 | 182502 | 184631 | 1175.22 | 16611.7 | 16829.3 | 30609.5 | 31120.2 |
| 1176.94 | 20776.7 | 21764 | 41060.3 | 41926.8 | 82064.8 | 83321.4 | 145555 | 146334 | 183514 | 185027 | 1176.94 | 16801 | 16927.1 | 30511.9 | 31269.5 |
| 1178.66 | 21334.7 | 21873.2 | 41200.7 | 42095.9 | 81832.5 | 83578.4 | 144873 | 146690 | 183144 | 185421 | 1178.66 | 16730.7 | 17024.9 | 30876.5 | 31418.7 |
| 1180.38 | 21807.4 | 21982.2 | 41371.7 | 42264.7 | 82665.4 | 83834.7 | 145447 | 147044 | 184482 | 185812 | 1180.38 | 17045.8 | 17122.9 | 31038.3 | 31567.9 |
| 1182.11 | 21908 | 22091.3 | 41931.1 | 42433.2 | 83276.7 | 84090 | 146248 | 147397 | 186054 | 186202 | 1182.11 | 17458.5 | 17220.9 | 31179.9 | 31717 |
| 1183.82 | 21898 | 22200.3 | 42243.4 | 42601.4 | 83581.1 | 84344.6 | 146473 | 147748 | 185511 | 186589 | 1183.82 | 17365.5 | 17319.1 | 31142.5 | 31866.1 |
| 1185.54 | 21558.3 | 22309.3 | 42061.1 | 42769.3 | 83409.9 | 84598.2 | 146541 | 148097 | 185325 | 186974 | 1185.54 | 17452.1 | 17417.3 | 31330.6 | 32015.1 |
| 1187.26 | 21834.3 | 22418.2 | 41845.2 | 42936.9 | 84415.8 | 84851.1 | 147384 | 148445 | 186854 | 187358 | 1187.26 | 17625.2 | 17515.5 | 31951 | 32164.1 |
| 1188.98 | 22186.4 | 22527.1 | 42885.4 | 43104.2 | 84734.2 | 85103 | 148470 | 148791 | 188263 | 187739 | 1188.98 | 17538.4 | 17613.9 | 31779.1 | 32313 |
| 1190.7 | 22272.4 | 22636 | 43277.5 | 43271.2 | 84857.5 | 85354.1 | 149457 | 149136 | 189396 | 188118 | 1190.7 | 17593.2 | 17712.4 | 32253.9 | 32461.8 |
| 1192.42 | 22544.7 | 22744.7 | 42825.8 | 43437.8 | 84997.4 | 85604.3 | 149618 | 149478 | 189022 | 188494 | 1192.42 | 17638.8 | 17810.9 | 32672.6 | 32610.6 |
| 1194.14 | 22428.9 | 22853.5 | 43242.5 | 43604.2 | 85224.6 | 85853.7 | 149440 | 149819 | 188706 | 188869 | 1194.14 | 17446.6 | 17909.4 | 32680.7 | 32759.3 |
| 1195.86 | 22306.7 | 22962.1 | 43176 | 43770.2 | 85701.3 | 86102.1 | 149321 | 150159 | 188606 | 189242 | 1195.86 | 17509.7 | 18008.1 | 32628.8 | 32907.9 |
| 1197.58 | 22936.5 | 23070.7 | 43008.9 | 43935.9 | 85448 | 86349.7 | 149501 | 150497 | 188433 | 189613 | 1197.58 | 18153.5 | 18106.8 | 32647.4 | 33056.5 |
| 1199.3 | 23079 | 23179.3 | 43494.9 | 44101.2 | 85693.6 | 86596.4 | 149888 | 150833 | 188980 | 189981 | 1199.3 | 18201 | 18205.6 | 32696.9 | 33204.9 |
| 1201.02 | 22681.1 | 23287.8 | 43445.4 | 44266.3 | 86562.4 | 86842.3 | 150165 | 151167 | 189835 | 190348 | 1201.02 | 17970.6 | 18304.4 | 33022.1 | 33353.3 |
| 1202.73 | 22939.7 | 23396.2 | 43966.3 | 44431 | 86609 | 87087.2 | 150787 | 151500 | 190549 | 190712 | 1202.73 | 17772.4 | 18403.3 | 33060.1 | 33501.6 |
| 1204.45 | 22984.4 | 23504.5 | 44531.9 | 44595.3 | 86356.7 | 87331.3 | 151495 | 151832 | 191068 | 191074 | 1204.45 | 18333.8 | 18502.2 | 33213.6 | 33649.8 |
| 1206.17 | 23204.8 | 23612.7 | 44538.2 | 44759.3 | 86536.6 | 87574.4 | 151869 | 152161 | 190872 | 191434 | 1206.17 | 18282.5 | 18601.2 | 33543 | 33797.9 |
| 1207.89 | 23433.7 | 23720.9 | 44569.2 | 44923 | 86757.3 | 87816.7 | 151496 | 152489 | 190752 | 191792 | 1207.89 | 18524.8 | 18700.2 | 33675.7 | 33945.9 |
| 1209.61 | 23457.6 | 23829 | 44638.9 | 45086.3 | 86980.7 | 88058 | 151463 | 152815 | 191075 | 192148 | 1209.61 | 18886.2 | 18799.3 | 33398.5 | 34093.8 |
| 1211.32 | 23681.6 | 23937 | 44751.4 | 45249.3 | 87358.5 | 88298.5 | 152280 | 153140 | 191837 | 192502 | 1211.32 | 18870.1 | 18898.5 | 33539 | 34241.7 |
| 1213.04 | 23898.6 | 24044.9 | 44509.8 | 45411.9 | 88042.7 | 88538 | 153388 | 153463 | 193158 | 192854 | 1213.04 | 19109.3 | 18997.6 | 34408.2 | 34389.3 |
| 1214.76 | 23834.7 | 24152.8 | 45002.9 | 45574.2 | 88986.9 | 88776.7 | 154183 | 153784 | 193803 | 193203 | 1214.76 | 18970.1 | 19096.8 | 34459.1 | 34536.9 |
| 1216.48 | 24058.3 | 24260.5 | 45287.5 | 45736 | 88843.4 | 89014.4 | 154382 | 154104 | 193563 | 193551 | 1216.48 | 19113.3 | 19196.1 | 34695.9 | 34684.4 |
| 1218.19 | 24109 | 24368.2 | 45383 | 45897.6 | 88657.8 | 89251.2 | 154558 | 154422 | 194139 | 193896 | 1218.19 | 19266.7 | 19295.3 | 34813.7 | 34831.8 |
| 1219.91 | 24437.6 | 24475.7 | 45887.8 | 46058.7 | 89193.4 | 89487.1 | 154776 | 154738 | 194457 | 194239 | 1219.91 | 19356 | 19394.7 | 34770 | 34979 |
| 1221.63 | 24518.7 | 24583.2 | 46126.8 | 46219.5 | 89338.9 | 89722.1 | 154950 | 155053 | 195488 | 194581 | 1221.63 | 19586.9 | 19494 | 35100.7 | 35126.1 |
| 1223.35 | 24185.2 | 24690.5 | 45897 | 46379.9 | 89264.2 | 89956.1 | 155143 | 155366 | 195239 | 194920 | 1223.35 | 19587.5 | 19593.4 | 35204.2 | 35273.1 |
| 1225.06 | 24450 | 24797.7 | 46078.7 | 46539.9 | 90055.1 | 90189.3 | 155926 | 155677 | 195169 | 195257 | 1225.06 | 19777.4 | 19692.8 | 35184.6 | 35420 |
| 1226.78 | 24557 | 24904.9 | 46588.5 | 46699.6 | 90488.8 | 90421.5 | 156406 | 155986 | 196319 | 195592 | 1226.78 | 20284.4 | 19792.2 | 35344.1 | 35566.7 |
| 1228.5 | 25120.7 | 25011.9 | 46704.5 | 46858.9 | 90258.7 | 90652.8 | 156646 | 156294 | 196602 | 195925 | 1228.5 | 20402.8 | 19891.6 | 35653.9 | 35713.3 |
| 1230.21 | 25134.1 | 25118.8 | 46637.8 | 47017.7 | 90553.2 | 90883.1 | 156872 | 156601 | 197188 | 196256 | 1230.21 | 19881.9 | 19991.1 | 35807.7 | 35859.8 |
| 1231.93 | 25335.1 | 25225.6 | 46775.2 | 47176.2 | 90760.1 | 91112.5 | 156767 | 156906 | 197149 | 196584 | 1231.93 | 20229.5 | 20090.5 | 35516.3 | 36006.1 |
| 1233.64 | 25220.4 | 25332.3 | 46901.4 | 47334.3 | 91079.7 | 91341 | 157043 | 157208 | 197002 | 196911 | 1233.64 | 20105.6 | 20190 | 35506.6 | 36152.2 |
| 1235.36 | 25464.7 | 25438.8 | 47412.3 | 47492 | 91697.8 | 91568.5 | 158215 | 157510 | 198167 | 197236 | 1235.36 | 20180.2 | 20289.5 | 35936.3 | 36298.2 |
| 1237.08 | 25545.9 | 25545.3 | 47418.8 | 47649.3 | 91851.9 | 91795.1 | 157880 | 157809 | 198024 | 197558 | 1237.08 | 20299.2 | 20389 | 36346.7 | 36444.1 |
| 1238.79 | 25729.2 | 25651.6 | 46980.7 | 47806.2 | 91656 | 92020.8 | 157626 | 158107 | 197672 | 197879 | 1238.79 | 19980.1 | 20488.5 | 36457.1 | 36589.8 |
| 1240.51 | 25522.7 | 25757.8 | 47728 | 47962.7 | 91855.4 | 92245.5 | 158142 | 158404 | 198509 | 198197 | 1240.51 | 20263.6 | 20588 | 36653.5 | 36735.4 |
| 1242.22 | 25462.7 | 25863.8 | 48407.1 | 48118.8 | 92302.4 | 92469.2 | 159226 | 158698 | 199269 | 198513 | 1242.22 | 20341.3 | 20687.5 | 36944.8 | 36880.8 |
| 1243.94 | 26102.7 | 25969.7 | 48168.4 | 48274.5 | 92421.4 | 92692.1 | 159516 | 158991 | 199174 | 198827 | 1243.94 | 20456.9 | 20787 | 36952.2 | 37026.1 |
| 1245.66 | 25751.9 | 26075.5 | 48280.1 | 48429.8 | 93027.9 | 92913.9 | 159525 | 159283 | 200019 | 199140 | 1245.66 | 20744.7 | 20886.5 | 37243.7 | 37171.1 |
| 1247.37 | 25868.3 | 26181.2 | 47892.4 | 48584.6 | 93102.5 | 93134.8 | 159488 | 159572 | 200439 | 199450 | 1247.37 | 20941.6 | 20986 | 37243.2 | 37316 |
| 1249.09 | 26084.2 | 26286.7 | 48244.4 | 48739.1 | 93319.1 | 93354.8 | 159790 | 159860 | 200032 | 199758 | 1249.09 | 21041.8 | 21085.5 | 37279.8 | 37460.8 |
| 1250.8 | 26209 | 26392 | 48917 | 48893.1 | 93732.9 | 93573.8 | 160936 | 160147 | 201276 | 200064 | 1250.8 | 21055.4 | 21185 | 37746.4 | 37605.3 |
| 1252.52 | 26395.5 | 26497.2 | 49070.3 | 49046.7 | 93773.2 | 93791.9 | 160941 | 160432 | 201459 | 200368 | 1252.52 | 21100.3 | 21284.4 | 37589.4 | 37749.7 |
| 1254.23 | 26316.4 | 26602.3 | 49192.7 | 49199.8 | 94341 | 94008.9 | 161309 | 160714 | 201170 | 200670 | 1254.23 | 21593.4 | 21383.9 | 37716.1 | 37894 |
| 1255.95 | 26705.2 | 26707.2 | 48948.3 | 49352.6 | 94244.6 | 94225.1 | 161550 | 160996 | 201766 | 200969 | 1255.95 | 21497.5 | 21483.3 | 38311.4 | 38038 |
| 1257.66 | 26916.3 | 26812 | 49561.9 | 49504.9 | 94180.2 | 94440.2 | 161842 | 161276 | 202485 | 201267 | 1257.66 | 21410.8 | 21582.7 | 38184.7 | 38181.8 |
| 1259.38 | 26784.8 | 26916.6 | 49794.1 | 49656.8 | 94267.7 | 94654.4 | 162308 | 161554 | 202543 | 201563 | 1259.38 | 21684 | 21682.1 | 38590.2 | 38325.5 |
| 1261.09 | 26959.5 | 27021.1 | 49641 | 49808.2 | 94786.9 | 94867.7 | 161950 | 161830 | 202610 | 201856 | 1261.09 | 21922 | 21781.5 | 38585.6 | 38469 |
| 1262.81 | 27123.5 | 27125.4 | 49781.8 | 49959.2 | 94811.4 | 95080 | 162282 | 162105 | 202859 | 202148 | 1262.81 | 21882.4 | 21880.8 | 38511.7 | 38612.3 |
| 1264.52 | 27185.6 | 27229.5 | 50255.3 | 50109.8 | 95024.2 | 95291.3 | 163028 | 162378 | 203319 | 202438 | 1264.52 | 21780.6 | 21980.1 | 38739 | 38755.4 |
| 1266.23 | 26895.5 | 27333.5 | 50016.9 | 50259.9 | 95807.8 | 95501.6 | 163082 | 162649 | 203214 | 202725 | 1266.23 | 21683.7 | 22079.4 | 38612.4 | 38898.2 |
| 1267.95 | 27163.4 | 27437.3 | 49795.4 | 50409.6 | 95478.4 | 95711 | 162811 | 162919 | 203574 | 203011 | 1267.95 | 22003.5 | 22178.7 | 38855 | 39040.9 |
| 1269.66 | 27353.5 | 27541 | 50734.7 | 50558.8 | 96151.7 | 95919.4 | 163920 | 163187 | 204914 | 203294 | 1269.66 | 22409.6 | 22277.9 | 39430.3 | 39183.4 |
| 1271.37 | 27537.7 | 27644.5 | 50600.2 | 50707.7 | 96624 | 96126.8 | 164593 | 163454 | 205051 | 203576 | 1271.37 | 22503.3 | 22377 | 39536.7 | 39325.7 |
| 1273.09 | 27691.3 | 27747.8 | 51157.9 | 50856 | 97274.8 | 96333.3 | 164982 | 163718 | 206371 | 203855 | 1273.09 | 22629.8 | 22476.2 | 39358.8 | 39467.8 |
| 1274.8 | 27942.4 | 27850.9 | 51622.1 | 51003.9 | 97484 | 96538.8 | 165848 | 163981 | 206979 | 204133 | 1274.8 | 22531.2 | 22575.3 | 39725.4 | 39609.7 |
| 1276.51 | 28192.2 | 27953.9 | 51530.9 | 51151.3 | 97506.9 | 96743.3 | 166139 | 164243 | 206754 | 204408 | 1276.51 | 22553.4 | 22674.3 | 39663.7 | 39751.3 |
| 1278.23 | 28235.2 | 28056.7 | 51935.7 | 51298.3 | 98063.1 | 96946.8 | 166671 | 164503 | 207519 | 204681 | 1278.23 | 23006.7 | 22773.3 | 39764.1 | 39892.8 |
| 1279.94 | 28294.6 | 28159.3 | 51956.8 | 51444.8 | 97913.6 | 97149.4 | 166926 | 164761 | 207449 | 204953 | 1279.94 | 23020.5 | 22872.3 | 40181.8 | 40034 |
| 1281.65 | 28185.7 | 28261.7 | 51722.5 | 51590.9 | 97701.4 | 97351 | 166720 | 165018 | 207295 | 205222 | 1281.65 | 23171.6 | 22971.2 | 40163.6 | 40175 |
| 1283.37 | 28466.6 | 28363.9 | 52046.1 | 51736.5 | 98400 | 97551.6 | 167188 | 165272 | 208207 | 205490 | 1283.37 | 23428.5 | 23070 | 40671.6 | 40315.8 |
| 1285.08 | 28345.2 | 28466 | 51997.3 | 51881.6 | 98414.8 | 97751.2 | 166632 | 165526 | 207787 | 205755 | 1285.08 | 23102.2 | 23168.9 | 40969.7 | 40456.4 |
| 1286.79 | 28384.1 | 28567.9 | 52051.4 | 52026.3 | 98310.1 | 97949.9 | 166316 | 165777 | 207494 | 206018 | 1286.79 | 23340.2 | 23267.6 | 40678.2 | 40596.7 |
| 1288.5 | 28641.5 | 28669.5 | 52537.8 | 52170.5 | 98830.8 | 98147.5 | 167400 | 166027 | 208138 | 206280 | 1288.5 | 23233 | 23366.3 | 40827.2 | 40736.8 |
| 1290.22 | 28603 | 28771 | 52310.7 | 52314.2 | 98348.3 | 98344.2 | 167190 | 166276 | 207530 | 206539 | 1290.22 | 23094.9 | 23464.9 | 41035.7 | 40876.7 |
| 1291.93 | 28780.8 | 28872.3 | 52522.8 | 52457.5 | 98469.8 | 98539.9 | 167012 | 166522 | 207181 | 206796 | 1291.93 | 23284.2 | 23563.5 | 41129.5 | 41016.4 |
| 1293.64 | 29189.9 | 28973.4 | 52729.5 | 52600.2 | 99073.1 | 98734.6 | 167583 | 166768 | 207786 | 207052 | 1293.64 | 23548.5 | 23662 | 41174.6 | 41155.8 |
| 1295.35 | 29338.5 | 29074.3 | 52600.9 | 52742.6 | 99161.8 | 98928.4 | 167966 | 167011 | 208359 | 207305 | 1295.35 | 23727.6 | 23760.4 | 41203.3 | 41294.9 |
| 1297.07 | 29174.9 | 29175 | 52799.6 | 52884.4 | 99868.8 | 99121.1 | 168505 | 167253 | 209093 | 207557 | 1297.07 | 23981.9 | 23858.8 | 41264.8 | 41433.9 |
| 1298.78 | 29194 | 29275.5 | 52852.6 | 53025.7 | 99891.1 | 99312.9 | 168394 | 167493 | 209228 | 207806 | 1298.78 | 23969 | 23957.1 | 41679 | 41572.6 |
| 1300.49 | 29773.5 | 29375.9 | 53185.6 | 53166.6 | 100361 | 99503.7 | 169494 | 167732 | 210195 | 208054 | 1300.49 | 23983.9 | 24055.3 | 42130.9 | 41711 |
| 1302.2 | 29786.9 | 29476 | 53680.4 | 53307 | 101086 | 99693.5 | 170223 | 167969 | 210855 | 208300 | 1302.2 | 24137.5 | 24153.5 | 42190.3 | 41849.2 |
| 1303.91 | 29700.4 | 29575.8 | 54347.8 | 53446.9 | 101075 | 99882.4 | 170695 | 168204 | 211463 | 208543 | 1303.91 | 24395.2 | 24251.5 | 42340.9 | 41987.2 |
| 1305.62 | 29888.4 | 29675.5 | 54580.1 | 53586.3 | 101621 | 100070 | 170884 | 168438 | 212074 | 208785 | 1305.62 | 24662.7 | 24349.5 | 42631.9 | 42124.9 |
| 1307.33 | 29740.7 | 29775 | 54472 | 53725.2 | 101435 | 100257 | 170640 | 168670 | 211557 | 209024 | 1307.33 | 24421.6 | 24447.5 | 42509.8 | 42262.3 |
| 1309.04 | 29883.7 | 29874.3 | 54570.4 | 53863.7 | 101471 | 100443 | 171012 | 168900 | 211931 | 209262 | 1309.04 | 24819 | 24545.3 | 42655.7 | 42399.5 |
| 1310.76 | 30147.1 | 29973.3 | 54476.5 | 54001.6 | 102464 | 100628 | 171656 | 169129 | 212882 | 209498 | 1310.76 | 25018.1 | 24643 | 43032.2 | 42536.5 |
| 1312.47 | 30667.5 | 30072.2 | 54615.8 | 54139.1 | 101892 | 100812 | 171765 | 169357 | 212686 | 209732 | 1312.47 | 24818 | 24740.7 | 42828.1 | 42673.2 |
| 1314.18 | 30574.7 | 30170.8 | 54748.3 | 54276.1 | 102146 | 100995 | 172298 | 169582 | 213416 | 209964 | 1314.18 | 24966.5 | 24838.3 | 43191.7 | 42809.6 |
| 1315.89 | 30580.8 | 30269.2 | 54800 | 54412.5 | 102767 | 101176 | 172427 | 169806 | 213682 | 210194 | 1315.89 | 24965.7 | 24935.8 | 43529.6 | 42945.7 |
| 1317.6 | 31133 | 30367.3 | 54976.4 | 54548.5 | 102363 | 101358 | 172077 | 170029 | 213416 | 210422 | 1317.6 | 25208 | 25033.2 | 43543.8 | 43081.6 |
| 1319.31 | 30519.6 | 30465.3 | 54857.1 | 54684 | 102267 | 101537 | 171856 | 170250 | 213058 | 210648 | 1319.31 | 25347.6 | 25130.5 | 43580.3 | 43217.2 |
| 1321.02 | 30479.8 | 30563 | 55240.8 | 54819 | 102368 | 101716 | 171699 | 170469 | 212711 | 210872 | 1321.02 | 25417.1 | 25227.7 | 43403.6 | 43352.6 |
| 1322.73 | 30794.3 | 30660.5 | 55371 | 54953.4 | 102436 | 101894 | 171875 | 170687 | 212111 | 211095 | 1322.73 | 25273.5 | 25324.8 | 43537.6 | 43487.7 |
| 1324.44 | 30889.1 | 30757.8 | 55112.1 | 55087.4 | 102989 | 102071 | 171605 | 170903 | 212313 | 211315 | 1324.44 | 25404.7 | 25421.8 | 43826.3 | 43622.5 |
| 1326.15 | 30860.9 | 30854.9 | 55251.9 | 55220.9 | 102757 | 102247 | 171660 | 171117 | 212303 | 211534 | 1326.15 | 25575 | 25518.7 | 43907.7 | 43757 |
| 1327.86 | 30752.2 | 30951.7 | 55294.6 | 55353.9 | 102675 | 102422 | 172375 | 171330 | 212734 | 211750 | 1327.86 | 25743.8 | 25615.5 | 43756.6 | 43891.3 |
| 1329.57 | 31096.3 | 31048.3 | 55729.3 | 55486.3 | 103728 | 102596 | 173429 | 171541 | 214505 | 211965 | 1329.57 | 25812.1 | 25712.2 | 44232.5 | 44025.3 |
| 1331.28 | 31197.3 | 31144.7 | 55845.3 | 55618.3 | 103828 | 102770 | 173564 | 171751 | 214193 | 212178 | 1331.28 | 25875.6 | 25808.8 | 44558.3 | 44159 |
| 1332.98 | 31305.2 | 31240.8 | 56142 | 55749.8 | 103822 | 102942 | 173543 | 171959 | 214422 | 212389 | 1332.98 | 26096.5 | 25905.3 | 44165.7 | 44292.4 |
| 1334.69 | 31246.2 | 31336.7 | 56487.9 | 55880.7 | 104064 | 103113 | 173473 | 172166 | 214594 | 212598 | 1334.69 | 26101.6 | 26001.7 | 44344.4 | 44425.5 |
| 1336.4 | 31524 | 31432.3 | 56405.7 | 56011.1 | 104280 | 103283 | 173604 | 172371 | 214510 | 212805 | 1336.4 | 26270.2 | 26098 | 44522.9 | 44558.4 |
| 1338.11 | 31776.8 | 31527.7 | 56673.8 | 56141 | 104317 | 103452 | 174849 | 172574 | 215492 | 213011 | 1338.11 | 26514.7 | 26194.2 | 44745.5 | 44690.9 |
| 1339.82 | 31563 | 31622.9 | 56492.4 | 56270.4 | 104381 | 103620 | 174176 | 172776 | 215374 | 213214 | 1339.82 | 26256.5 | 26290.2 | 44787.5 | 44823.2 |
| 1341.53 | 32253.3 | 31717.8 | 56711.1 | 56399.3 | 104555 | 103787 | 174449 | 172977 | 215348 | 213416 | 1341.53 | 26379.5 | 26386.1 | 45178.5 | 44955.2 |
| 1343.24 | 32093.2 | 31812.5 | 57402.7 | 56527.7 | 104862 | 103953 | 175296 | 173175 | 216066 | 213615 | 1343.24 | 26688.3 | 26481.9 | 45658.8 | 45086.9 |
| 1344.95 | 32185.4 | 31907 | 57234.2 | 56655.6 | 105041 | 104119 | 174948 | 173373 | 215790 | 213813 | 1344.95 | 26564.4 | 26577.6 | 45597.7 | 45218.3 |
| 1346.65 | 32280.3 | 32001.2 | 57274.6 | 56782.9 | 105144 | 104283 | 174880 | 173568 | 215715 | 214009 | 1346.65 | 26700.9 | 26673.2 | 45719.7 | 45349.4 |
| 1348.36 | 32041 | 32095.1 | 57175.3 | 56909.8 | 105477 | 104446 | 174309 | 173762 | 215489 | 214204 | 1348.36 | 26766.5 | 26768.6 | 45674.6 | 45480.2 |
| 1350.07 | 32366 | 32188.8 | 57353.8 | 57036.1 | 105474 | 104608 | 174766 | 173955 | 215306 | 214396 | 1350.07 | 26989.4 | 26864 | 45703.8 | 45610.7 |
| 1351.78 | 32133.2 | 32282.3 | 57685.1 | 57161.9 | 105876 | 104770 | 175348 | 174146 | 216107 | 214586 | 1351.78 | 27114.3 | 26959.2 | 46132.8 | 45740.9 |
| 1353.49 | 32421.8 | 32375.5 | 57189.7 | 57287.2 | 105629 | 104930 | 175220 | 174335 | 216017 | 214775 | 1353.49 | 27015 | 27054.2 | 46016.7 | 45870.8 |
| 1355.19 | 32801.2 | 32468.4 | 57454.2 | 57411.9 | 105948 | 105089 | 175990 | 174523 | 216331 | 214962 | 1355.19 | 27221.1 | 27149.2 | 46180.9 | 46000.3 |
| 1356.9 | 32611.7 | 32561.1 | 57750.3 | 57536.2 | 106356 | 105248 | 175983 | 174710 | 216693 | 215147 | 1356.9 | 27177.8 | 27244 | 46112.6 | 46129.6 |
| 1358.61 | 32720.6 | 32653.5 | 57878.9 | 57659.9 | 106371 | 105405 | 176037 | 174894 | 216630 | 215330 | 1358.61 | 27356 | 27338.6 | 46191.4 | 46258.6 |
| 1360.32 | 32956.7 | 32745.7 | 58075.5 | 57783.1 | 106516 | 105561 | 176536 | 175078 | 216969 | 215512 | 1360.32 | 27026.9 | 27433.2 | 46441.6 | 46387.3 |
| 1362.02 | 33102.5 | 32837.6 | 58108.9 | 57905.7 | 106619 | 105717 | 176515 | 175260 | 217384 | 215691 | 1362.02 | 27070.3 | 27527.6 | 46428.3 | 46515.7 |
| 1363.73 | 32991.4 | 32929.3 | 58486.3 | 58027.8 | 106602 | 105871 | 176899 | 175440 | 217551 | 215869 | 1363.73 | 27836.3 | 27621.8 | 46766 | 46643.7 |
| 1365.44 | 32991.6 | 33020.7 | 58559.6 | 58149.5 | 106779 | 106025 | 177172 | 175619 | 217495 | 216045 | 1365.44 | 27631.3 | 27716 | 46919.4 | 46771.4 |
| 1367.14 | 33548 | 33111.8 | 58687.1 | 58270.6 | 107005 | 106177 | 177482 | 175796 | 217701 | 216219 | 1367.14 | 27742.7 | 27810 | 47015.6 | 46898.8 |
| 1368.85 | 33495.2 | 33202.7 | 58847.3 | 58391.1 | 107141 | 106329 | 177597 | 175972 | 218765 | 216392 | 1368.85 | 27928.5 | 27903.8 | 47056.9 | 47026 |
| 1370.56 | 33627.1 | 33293.3 | 58903.6 | 58511.1 | 107025 | 106479 | 177448 | 176146 | 218495 | 216563 | 1370.56 | 27929.2 | 27997.5 | 47256.8 | 47152.7 |
| 1372.26 | 33714.4 | 33383.7 | 59152.2 | 58630.6 | 107190 | 106629 | 177452 | 176319 | 218678 | 216732 | 1372.26 | 28175.2 | 28091 | 47461.7 | 47279.2 |
| 1373.97 | 33613.3 | 33473.8 | 59157.7 | 58749.6 | 108151 | 106777 | 177850 | 176490 | 219540 | 216899 | 1373.97 | 28640.3 | 28184.5 | 47582.7 | 47405.4 |
| 1375.67 | 34029.8 | 33563.6 | 59469.2 | 58868.1 | 107979 | 106925 | 178664 | 176660 | 219190 | 217064 | 1375.67 | 28600.8 | 28277.7 | 47753.7 | 47531.2 |
| 1377.38 | 33961.8 | 33653.1 | 59369.2 | 58986 | 107745 | 107072 | 178575 | 176828 | 219056 | 217228 | 1377.38 | 28373.8 | 28370.8 | 47859.1 | 47656.7 |
| 1379.09 | 34189.9 | 33742.4 | 59908.1 | 59103.4 | 108417 | 107218 | 178698 | 176994 | 219584 | 217390 | 1379.09 | 28390.2 | 28463.8 | 48251.9 | 47781.9 |
| 1380.79 | 34349.6 | 33831.4 | 59948.6 | 59220.2 | 108891 | 107362 | 179540 | 177160 | 220488 | 217550 | 1380.79 | 28567.5 | 28556.6 | 48385.1 | 47906.7 |
| 1382.5 | 34144.2 | 33920.1 | 60046.9 | 59336.5 | 108923 | 107506 | 179550 | 177324 | 220309 | 217708 | 1382.5 | 28906.3 | 28649.2 | 48206.6 | 48031.2 |
| 1384.2 | 34644.1 | 34008.6 | 60414.7 | 59452.3 | 109414 | 107649 | 180019 | 177486 | 220929 | 217865 | 1384.2 | 28919.5 | 28741.7 | 48607.8 | 48155.4 |
| 1385.91 | 34684 | 34096.8 | 60494.9 | 59567.6 | 109710 | 107791 | 180963 | 177646 | 222087 | 218020 | 1385.91 | 29334.4 | 28834.1 | 49110 | 48279.3 |
| 1387.61 | 34185.5 | 34184.7 | 60720.8 | 59682.3 | 109316 | 107932 | 180728 | 177806 | 221627 | 218173 | 1387.61 | 29379.4 | 28926.3 | 48898 | 48402.8 |
| 1389.32 | 34535.1 | 34272.3 | 60555.4 | 59796.5 | 109914 | 108072 | 180406 | 177964 | 221347 | 218324 | 1389.32 | 29247.2 | 29018.3 | 48877.2 | 48526 |
| 1391.02 | 34997.2 | 34359.7 | 60639 | 59910.2 | 109930 | 108211 | 180144 | 178120 | 221581 | 218474 | 1391.02 | 29383.7 | 29110.1 | 49217.2 | 48648.9 |
| 1392.73 | 35124.4 | 34446.8 | 60492.1 | 60023.3 | 109752 | 108349 | 180240 | 178275 | 221199 | 218622 | 1392.73 | 29244.8 | 29201.8 | 49242.2 | 48771.4 |
| 1394.43 | 34992.6 | 34533.6 | 60715.5 | 60135.9 | 109927 | 108486 | 180362 | 178428 | 221165 | 218768 | 1394.43 | 29215.3 | 29293.4 | 49092.6 | 48893.6 |
| 1396.14 | 35224.9 | 34620.1 | 61113.1 | 60247.9 | 110373 | 108622 | 181275 | 178580 | 221860 | 218913 | 1396.14 | 29387.1 | 29384.8 | 49608.5 | 49015.5 |
| 1397.84 | 34784.4 | 34706.3 | 61010.9 | 60359.4 | 109975 | 108757 | 181110 | 178731 | 221480 | 219056 | 1397.84 | 29388.8 | 29476 | 49737.6 | 49137 |
| 1399.55 | 34554.7 | 34792.3 | 60929.3 | 60470.4 | 109694 | 108891 | 180532 | 178880 | 221438 | 219197 | 1399.55 | 29751.3 | 29567 | 49573.4 | 49258.2 |
| 1401.25 | 35080.6 | 34877.9 | 60944.5 | 60580.9 | 110546 | 109025 | 181219 | 179028 | 221945 | 219336 | 1401.25 | 29823 | 29657.9 | 49932.4 | 49379 |
| 1402.95 | 35040.3 | 34963.3 | 61212.5 | 60690.8 | 110167 | 109157 | 181390 | 179174 | 221775 | 219474 | 1402.95 | 29826.7 | 29748.5 | 50202.3 | 49499.5 |
| 1404.66 | 35078.3 | 35048.4 | 61466.2 | 60800.2 | 110136 | 109288 | 181140 | 179319 | 221644 | 219611 | 1404.66 | 29742.2 | 29839.1 | 49986.4 | 49619.7 |
| 1406.36 | 35330.1 | 35133.3 | 61424.8 | 60909 | 110638 | 109419 | 181232 | 179462 | 221865 | 219745 | 1406.36 | 29881.2 | 29929.4 | 49787.2 | 49739.5 |
| 1408.07 | 35524.1 | 35217.8 | 61888.7 | 61017.3 | 110710 | 109548 | 181517 | 179604 | 221918 | 219878 | 1408.07 | 30058.9 | 30019.6 | 50025.4 | 49859 |
| 1409.77 | 35690.9 | 35302.1 | 62244.3 | 61125.1 | 111013 | 109677 | 181374 | 179745 | 222269 | 220009 | 1409.77 | 29984.9 | 30109.6 | 50435.6 | 49978.1 |
| 1411.47 | 35624.1 | 35386 | 61960.7 | 61232.3 | 110787 | 109804 | 181616 | 179884 | 222628 | 220139 | 1411.47 | 30216.9 | 30199.4 | 50446.1 | 50096.9 |
| 1413.17 | 35791.9 | 35469.7 | 61776.7 | 61339 | 111038 | 109931 | 181791 | 180022 | 222777 | 220267 | 1413.17 | 30245.6 | 30289.1 | 50551.9 | 50215.3 |
| 1414.88 | 36025.1 | 35553.1 | 62076.8 | 61445.1 | 111387 | 110057 | 182134 | 180158 | 222236 | 220393 | 1414.88 | 30565 | 30378.6 | 50573 | 50333.4 |
| 1416.58 | 35822.6 | 35636.2 | 62077.6 | 61550.7 | 110918 | 110182 | 182040 | 180293 | 222097 | 220518 | 1416.58 | 30848 | 30467.8 | 50282.7 | 50451.1 |
| 1418.28 | 35716.7 | 35719 | 62324.9 | 61655.8 | 111546 | 110306 | 182533 | 180426 | 223347 | 220641 | 1418.28 | 30821.8 | 30557 | 50647.2 | 50568.5 |
| 1419.99 | 35833 | 35801.5 | 62695.1 | 61760.3 | 111905 | 110428 | 182745 | 180558 | 223179 | 220762 | 1419.99 | 30766.8 | 30645.9 | 50952.5 | 50685.5 |
| 1421.69 | 36454.7 | 35883.7 | 62760.1 | 61864.3 | 112490 | 110550 | 183253 | 180689 | 223810 | 220882 | 1421.69 | 30961.9 | 30734.6 | 51318.3 | 50802.2 |
| 1423.39 | 36642 | 35965.6 | 63079.1 | 61967.8 | 112580 | 110672 | 183807 | 180818 | 224710 | 221000 | 1423.39 | 31150.3 | 30823.2 | 51571.3 | 50918.5 |
| 1425.09 | 36421.2 | 36047.2 | 62994.2 | 62070.7 | 111966 | 110792 | 182510 | 180946 | 223192 | 221116 | 1425.09 | 30707.2 | 30911.6 | 51048.7 | 51034.5 |
| 1426.8 | 36473.2 | 36128.6 | 62937.1 | 62173.1 | 112279 | 110911 | 182461 | 181072 | 222582 | 221231 | 1426.8 | 30929.2 | 30999.8 | 51332.4 | 51150.1 |
| 1428.5 | 36395.5 | 36209.6 | 62842.3 | 62275 | 112008 | 111029 | 182303 | 181198 | 222654 | 221345 | 1428.5 | 31358.2 | 31087.7 | 51421.4 | 51265.4 |
| 1430.2 | 36595.1 | 36290.4 | 62918.9 | 62376.3 | 112062 | 111147 | 182557 | 181321 | 222962 | 221456 | 1430.2 | 31145.9 | 31175.5 | 51291 | 51380.3 |
| 1431.9 | 36854.7 | 36370.8 | 63228.6 | 62477.1 | 112484 | 111263 | 183225 | 181444 | 223170 | 221567 | 1431.9 | 31414 | 31263.2 | 51727.2 | 51494.8 |
| 1433.6 | 37160.8 | 36451 | 63064.6 | 62577.3 | 112471 | 111379 | 183062 | 181565 | 223623 | 221675 | 1433.6 | 31432.6 | 31350.6 | 51732 | 51609 |
| 1435.3 | 36950.7 | 36530.8 | 63137.9 | 62677 | 112621 | 111493 | 183588 | 181684 | 223822 | 221782 | 1435.3 | 31508.2 | 31437.8 | 51887.3 | 51722.8 |
| 1437.01 | 36817.9 | 36610.4 | 63664.8 | 62776.1 | 112769 | 111607 | 183495 | 181803 | 223743 | 221888 | 1437.01 | 31646 | 31524.9 | 52054.1 | 51836.3 |
| 1438.71 | 37201.3 | 36689.6 | 63482.8 | 62874.8 | 112970 | 111720 | 183025 | 181920 | 223692 | 221992 | 1438.71 | 31494.1 | 31611.7 | 52132.8 | 51949.4 |
| 1440.41 | 37130.3 | 36768.6 | 63235.2 | 62972.8 | 112880 | 111832 | 183081 | 182035 | 223461 | 222094 | 1440.41 | 31564.3 | 31698.3 | 52173.4 | 52062.1 |
| 1442.11 | 37165.8 | 36847.2 | 63426.7 | 63070.4 | 113184 | 111943 | 183669 | 182149 | 223988 | 222195 | 1442.11 | 31828.1 | 31784.8 | 52328 | 52174.5 |
| 1443.81 | 37533.3 | 36925.6 | 63874.7 | 63167.4 | 113323 | 112053 | 184591 | 182262 | 224619 | 222294 | 1443.81 | 32016.4 | 31871.1 | 52848.1 | 52286.5 |
| 1445.51 | 37320.4 | 37003.6 | 63726.2 | 63263.9 | 113071 | 112162 | 183854 | 182374 | 223989 | 222392 | 1445.51 | 32093 | 31957.1 | 52728.9 | 52398.2 |
| 1447.21 | 37416.2 | 37081.4 | 64076.2 | 63359.8 | 113365 | 112270 | 184028 | 182484 | 224480 | 222488 | 1447.21 | 32280.7 | 32043 | 52797.3 | 52509.4 |
| 1448.91 | 37777.4 | 37158.8 | 64492.9 | 63455.2 | 113237 | 112378 | 184576 | 182593 | 224319 | 222583 | 1448.91 | 32296.1 | 32128.6 | 52825.3 | 52620.4 |
| 1450.61 | 37578.7 | 37236 | 64200.2 | 63550.1 | 113485 | 112484 | 183902 | 182700 | 223701 | 222676 | 1450.61 | 32308.1 | 32214.1 | 52976.8 | 52730.9 |
| 1452.31 | 37625.4 | 37312.8 | 64267 | 63644.4 | 113361 | 112590 | 183561 | 182807 | 223790 | 222767 | 1452.31 | 32469.5 | 32299.3 | 53119.6 | 52841.1 |
| 1454.01 | 37498.5 | 37389.3 | 64355 | 63738.2 | 112900 | 112694 | 183227 | 182912 | 223045 | 222857 | 1454.01 | 32526.5 | 32384.4 | 53176.2 | 52950.9 |
| 1455.71 | 37584.7 | 37465.6 | 64327 | 63831.5 | 113082 | 112798 | 183235 | 183015 | 223441 | 222946 | 1455.71 | 32534.8 | 32469.2 | 53531.6 | 53060.4 |
| 1457.41 | 37625.8 | 37541.5 | 64251.2 | 63924.2 | 113158 | 112901 | 183523 | 183118 | 223056 | 223033 | 1457.41 | 32621.2 | 32553.8 | 53118.1 | 53169.5 |
| 1459.11 | 37816.3 | 37617.1 | 64369 | 64016.4 | 113364 | 113003 | 183866 | 183218 | 223106 | 223119 | 1459.11 | 32657.1 | 32638.3 | 53089.6 | 53278.2 |
| 1460.81 | 37947.2 | 37692.4 | 64041.7 | 64108 | 113456 | 113104 | 183750 | 183318 | 223416 | 223203 | 1460.81 | 32522.6 | 32722.5 | 53221.8 | 53386.5 |
| 1462.51 | 37852.5 | 37767.4 | 63927.2 | 64199.1 | 113452 | 113204 | 183574 | 183416 | 222889 | 223285 | 1462.51 | 32896 | 32806.5 | 53367.6 | 53494.5 |
| 1464.21 | 38136.8 | 37842.2 | 64657.1 | 64289.7 | 113788 | 113304 | 183844 | 183514 | 223632 | 223366 | 1464.21 | 32693.8 | 32890.3 | 53605.9 | 53602.1 |
| 1465.91 | 37955.7 | 37916.6 | 64584.8 | 64379.8 | 113152 | 113402 | 182911 | 183610 | 223052 | 223446 | 1465.91 | 32755.7 | 32973.9 | 53586 | 53709.3 |
| 1467.61 | 38193.1 | 37990.6 | 64575.9 | 64469.3 | 113322 | 113500 | 183227 | 183704 | 223120 | 223524 | 1467.61 | 33384.2 | 33057.3 | 53948.1 | 53816.2 |
| 1469.31 | 38283.3 | 38064.4 | 64573.3 | 64558.3 | 113771 | 113596 | 184143 | 183797 | 223478 | 223601 | 1469.31 | 33043.9 | 33140.4 | 53812.7 | 53922.7 |
| 1471.01 | 37852 | 38137.9 | 64462 | 64646.7 | 113081 | 113692 | 182893 | 183889 | 222454 | 223676 | 1471.01 | 33128.8 | 33223.4 | 53487.7 | 54028.8 |
| 1472.7 | 38261.7 | 38211 | 64762.9 | 64734.7 | 113804 | 113787 | 182945 | 183980 | 222880 | 223750 | 1472.7 | 33260.9 | 33306.1 | 54009.9 | 54134.5 |
| 1474.4 | 38724.5 | 38283.9 | 65192.2 | 64822 | 114708 | 113881 | 184794 | 184069 | 224179 | 223823 | 1474.4 | 33622.7 | 33388.7 | 54516.9 | 54239.9 |
| 1476.1 | 38389.3 | 38356.4 | 65188.3 | 64908.9 | 114273 | 113974 | 184632 | 184158 | 224215 | 223894 | 1476.1 | 33690.7 | 33471 | 54618.9 | 54344.9 |
| 1477.8 | 38570.5 | 38428.7 | 64981.7 | 64995.2 | 114449 | 114066 | 184363 | 184244 | 224135 | 223963 | 1477.8 | 33641.4 | 33553.1 | 54787.7 | 54449.5 |
| 1479.5 | 38893.8 | 38500.6 | 65272.9 | 65081 | 114802 | 114158 | 184924 | 184330 | 224010 | 224031 | 1479.5 | 33803.8 | 33635 | 54956.7 | 54553.7 |
| 1481.19 | 38732.3 | 38572.2 | 65177.7 | 65166.3 | 113840 | 114248 | 184036 | 184415 | 222984 | 224098 | 1481.19 | 33626.9 | 33716.6 | 54869.6 | 54657.6 |
| 1482.89 | 38612.5 | 38643.5 | 65374.2 | 65251 | 114044 | 114338 | 184028 | 184498 | 223712 | 224163 | 1482.89 | 33798.8 | 33798.1 | 54800.1 | 54761.1 |
| 1484.59 | 38917.3 | 38714.5 | 65675.6 | 65335.2 | 114859 | 114427 | 185126 | 184580 | 224509 | 224227 | 1484.59 | 34202.1 | 33879.3 | 55432.4 | 54864.2 |
| 1486.29 | 38947 | 38785.2 | 65274.5 | 65418.9 | 114264 | 114515 | 184458 | 184660 | 223825 | 224289 | 1486.29 | 33970.4 | 33960.3 | 55247.9 | 54966.9 |
| 1487.98 | 38936.8 | 38855.6 | 65757.6 | 65502.1 | 115080 | 114602 | 184979 | 184740 | 224500 | 224350 | 1487.98 | 34151.1 | 34041.1 | 55233 | 55069.2 |
| 1489.68 | 39349.3 | 38925.6 | 65915.7 | 65584.7 | 115577 | 114688 | 185596 | 184818 | 225494 | 224410 | 1489.68 | 34655.5 | 34121.6 | 55577.3 | 55171.2 |
| 1491.38 | 39346.8 | 38995.3 | 65806.3 | 65666.8 | 115302 | 114774 | 185142 | 184895 | 224704 | 224468 | 1491.38 | 34588.7 | 34202 | 55722.8 | 55272.8 |
| 1493.08 | 39099.1 | 39064.8 | 65972.4 | 65748.4 | 115052 | 114858 | 185433 | 184971 | 224100 | 224525 | 1493.08 | 34285 | 34282.1 | 55733.5 | 55374 |
| 1494.77 | 39386.8 | 39133.9 | 65787.8 | 65829.4 | 114659 | 114942 | 184660 | 185046 | 223620 | 224581 | 1494.77 | 34537 | 34362 | 55721.4 | 55474.9 |
| 1496.47 | 39165.2 | 39202.7 | 65660 | 65909.9 | 114833 | 115025 | 183989 | 185119 | 223153 | 224635 | 1496.47 | 34650.6 | 34441.7 | 55899.5 | 55575.3 |
| 1498.17 | 39128.6 | 39271.2 | 65764 | 65989.9 | 114588 | 115107 | 184589 | 185191 | 224036 | 224688 | 1498.17 | 34855.9 | 34521.1 | 55774.1 | 55675.4 |
| 1499.86 | 39666 | 39339.4 | 66163.2 | 66069.4 | 114722 | 115188 | 184705 | 185262 | 223720 | 224739 | 1499.86 | 34844.8 | 34600.3 | 55848.9 | 55775.1 |
| 1501.56 | 39517.7 | 39407.2 | 65901.3 | 66148.3 | 115015 | 115269 | 184534 | 185332 | 223242 | 224789 | 1501.56 | 34489 | 34679.3 | 56324.1 | 55874.4 |
| 1503.25 | 39519.1 | 39474.8 | 65988.2 | 66226.8 | 114718 | 115348 | 184757 | 185400 | 223718 | 224838 | 1503.25 | 34571.5 | 34758.1 | 56142.4 | 55973.4 |
| 1504.95 | 39811.9 | 39542.1 | 66362.2 | 66304.7 | 115247 | 115427 | 185236 | 185468 | 223991 | 224885 | 1504.95 | 34819.4 | 34836.6 | 55959 | 56071.9 |
| 1506.65 | 39513.8 | 39609 | 66301.4 | 66382.1 | 115064 | 115505 | 184562 | 185534 | 223653 | 224931 | 1506.65 | 35168.2 | 34914.9 | 56182.8 | 56170.1 |
| 1508.34 | 39460.2 | 39675.6 | 66323.7 | 66458.9 | 114407 | 115582 | 184151 | 185599 | 223065 | 224976 | 1508.34 | 34943 | 34993 | 56088.5 | 56267.9 |
| 1510.04 | 39611.1 | 39741.9 | 66399.1 | 66535.3 | 114564 | 115658 | 184556 | 185663 | 223425 | 225019 | 1510.04 | 35191.5 | 35070.8 | 55979.2 | 56365.3 |
| 1511.73 | 39595.5 | 39807.9 | 66395 | 66611.1 | 114821 | 115734 | 184366 | 185726 | 222864 | 225061 | 1511.73 | 35191.7 | 35148.4 | 56368.7 | 56462.3 |
| 1513.43 | 39637.5 | 39873.6 | 66280.1 | 66686.4 | 114763 | 115808 | 184096 | 185787 | 222153 | 225102 | 1513.43 | 34745 | 35225.8 | 56238.1 | 56559 |
| 1515.12 | 39775.4 | 39938.9 | 66086 | 66761.2 | 114089 | 115882 | 183598 | 185847 | 222014 | 225142 | 1515.12 | 34907.4 | 35303 | 56398.2 | 56655.2 |
| 1516.82 | 39735.3 | 40003.9 | 66263.7 | 66835.5 | 114946 | 115955 | 183643 | 185907 | 222075 | 225180 | 1516.82 | 34927.4 | 35379.9 | 56823.3 | 56751.1 |
| 1518.51 | 39689.6 | 40068.7 | 66373.1 | 66909.2 | 115345 | 116027 | 183328 | 185965 | 222915 | 225216 | 1518.51 | 35360.3 | 35456.6 | 56357.5 | 56846.6 |
| 1520.21 | 39524.6 | 40133.1 | 66286.2 | 66982.4 | 114662 | 116098 | 183385 | 186022 | 222119 | 225252 | 1520.21 | 35370 | 35533.1 | 56285.5 | 56941.7 |
| 1521.9 | 39810.8 | 40197.2 | 66356.2 | 67055.2 | 115153 | 116169 | 184220 | 186077 | 222568 | 225286 | 1521.9 | 35312.4 | 35609.3 | 56850.2 | 57036.5 |
| 1523.6 | 40085.8 | 40261 | 66446.2 | 67127.4 | 115303 | 116239 | 184404 | 186132 | 223436 | 225319 | 1523.6 | 35872.3 | 35685.3 | 56990.3 | 57130.8 |
| 1525.29 | 40036.7 | 40324.4 | 66592.5 | 67199.1 | 115194 | 116308 | 184320 | 186185 | 222424 | 225351 | 1525.29 | 35762 | 35761 | 56870.2 | 57224.8 |
| 1526.99 | 40320.1 | 40387.6 | 66610.9 | 67270.3 | 115075 | 116376 | 184433 | 186238 | 223051 | 225381 | 1526.99 | 35670 | 35836.6 | 57336.6 | 57318.4 |
| 1528.68 | 40152 | 40450.4 | 66809.6 | 67340.9 | 115477 | 116443 | 184513 | 186289 | 223239 | 225410 | 1528.68 | 35978.4 | 35911.8 | 57322.2 | 57411.6 |
| 1530.37 | 39944.7 | 40513 | 66862.6 | 67411.1 | 115662 | 116509 | 183950 | 186339 | 222615 | 225438 | 1530.37 | 35860.1 | 35986.9 | 56841.4 | 57504.4 |
| 1532.07 | 40462.7 | 40575.2 | 67158.4 | 67480.8 | 114868 | 116575 | 183855 | 186388 | 222555 | 225465 | 1532.07 | 35900.1 | 36061.7 | 57116.9 | 57596.8 |
| 1533.76 | 40598.9 | 40637.1 | 67194.7 | 67549.9 | 115508 | 116640 | 184623 | 186436 | 222807 | 225490 | 1533.76 | 36050.3 | 36136.3 | 57071 | 57688.9 |
| 1535.45 | 40357.2 | 40698.7 | 67078.8 | 67618.5 | 115735 | 116704 | 184414 | 186483 | 222766 | 225514 | 1535.45 | 36042.8 | 36210.6 | 57165.8 | 57780.6 |
| 1537.15 | 40385.8 | 40759.9 | 66979.1 | 67686.7 | 115326 | 116767 | 183993 | 186528 | 222206 | 225538 | 1537.15 | 36370.4 | 36284.7 | 57151.6 | 57871.8 |
| 1538.84 | 40400.5 | 40820.9 | 67103.4 | 67754.3 | 115762 | 116830 | 184739 | 186573 | 222497 | 225559 | 1538.84 | 36315.1 | 36358.6 | 57476.9 | 57962.7 |
| 1540.53 | 40582.8 | 40881.5 | 67453.6 | 67821.4 | 115655 | 116892 | 184686 | 186616 | 222656 | 225580 | 1540.53 | 36100.2 | 36432.2 | 57878.4 | 58053.2 |
| 1542.23 | 40647.7 | 40941.8 | 67667.4 | 67888 | 115962 | 116953 | 185166 | 186659 | 223544 | 225599 | 1542.23 | 36626.1 | 36505.5 | 57664.3 | 58143.4 |
| 1543.92 | 41074.7 | 41001.8 | 67643.3 | 67954.1 | 115945 | 117013 | 184920 | 186700 | 223699 | 225617 | 1543.92 | 36498.3 | 36578.7 | 58127.9 | 58233.1 |
| 1545.61 | 41141.1 | 41061.5 | 67602.9 | 68019.7 | 115718 | 117072 | 184669 | 186740 | 223089 | 225634 | 1545.61 | 36340.7 | 36651.6 | 58080.8 | 58322.5 |
| 1547.31 | 41144.5 | 41120.9 | 67871.3 | 68084.8 | 116329 | 117131 | 185290 | 186779 | 223524 | 225649 | 1547.31 | 36552 | 36724.2 | 58139 | 58411.5 |
| 1549 | 41087.8 | 41180 | 67775.5 | 68149.4 | 116266 | 117189 | 184591 | 186818 | 223381 | 225664 | 1549 | 36614.6 | 36796.6 | 58444.3 | 58500.1 |
| 1550.69 | 41336.7 | 41238.7 | 68141.9 | 68213.5 | 116469 | 117246 | 185142 | 186854 | 223908 | 225677 | 1550.69 | 37060.9 | 36868.8 | 58503.1 | 58588.3 |
| 1552.38 | 41535.4 | 41297.2 | 67613.9 | 68277.1 | 116382 | 117302 | 185308 | 186890 | 223498 | 225689 | 1552.38 | 36723.7 | 36940.7 | 58529.9 | 58676.1 |
| 1554.07 | 41138.3 | 41355.3 | 67495.8 | 68340.2 | 116121 | 117358 | 184614 | 186925 | 222825 | 225700 | 1554.07 | 36639.5 | 37012.4 | 58243 | 58763.5 |
| 1555.77 | 40980 | 41413.1 | 67872.7 | 68402.8 | 116074 | 117413 | 184631 | 186959 | 222800 | 225710 | 1555.77 | 37014.9 | 37083.9 | 58262.9 | 58850.6 |
| 1557.46 | 40855.5 | 41470.6 | 67104.4 | 68464.9 | 115439 | 117467 | 183906 | 186992 | 222207 | 225718 | 1557.46 | 36824.2 | 37155.1 | 58470.9 | 58937.3 |
| 1559.15 | 41384.8 | 41527.8 | 67909.2 | 68526.5 | 116774 | 117520 | 185173 | 187024 | 223445 | 225725 | 1559.15 | 37132.6 | 37226 | 58623.8 | 59023.6 |
| 1560.84 | 41495.5 | 41584.6 | 68166.2 | 68587.6 | 116757 | 117572 | 185340 | 187054 | 223446 | 225732 | 1560.84 | 37664 | 37296.7 | 58636.5 | 59109.4 |
| 1562.53 | 41087.5 | 41641.2 | 68264.1 | 68648.2 | 116012 | 117624 | 184873 | 187084 | 223467 | 225736 | 1562.53 | 37261.1 | 37367.2 | 58943 | 59195 |
| 1564.22 | 41548.8 | 41697.4 | 68309.5 | 68708.3 | 116726 | 117675 | 185383 | 187112 | 223535 | 225740 | 1564.22 | 37206.8 | 37437.4 | 59091.5 | 59280.1 |
| 1565.91 | 41399.1 | 41753.4 | 67802.4 | 68768 | 116236 | 117726 | 183738 | 187140 | 221716 | 225743 | 1565.91 | 37601.7 | 37507.4 | 58710.6 | 59364.9 |
| 1567.61 | 40973 | 41809 | 67783.3 | 68827.1 | 115709 | 117775 | 183368 | 187166 | 221344 | 225745 | 1567.61 | 37110.2 | 37577.1 | 58732 | 59449.2 |
| 1569.3 | 41063.9 | 41864.3 | 67660.5 | 68885.8 | 115232 | 117824 | 182593 | 187192 | 220650 | 225745 | 1569.3 | 37183.2 | 37646.6 | 59161.9 | 59533.2 |
| 1570.99 | 41116.3 | 41919.3 | 67796.6 | 68943.9 | 115272 | 117872 | 182719 | 187216 | 220479 | 225744 | 1570.99 | 37619.2 | 37715.8 | 59154.4 | 59616.8 |
| 1572.68 | 41208.2 | 41973.9 | 67810.4 | 69001.6 | 115290 | 117919 | 183222 | 187240 | 220594 | 225743 | 1572.68 | 37642.8 | 37784.8 | 59198.6 | 59700.1 |
| 1574.37 | 41022.5 | 42028.3 | 67601.9 | 69058.8 | 115196 | 117966 | 182312 | 187262 | 219740 | 225740 | 1574.37 | 37675.2 | 37853.6 | 59458.1 | 59782.9 |
| 1576.06 | 41346.2 | 42082.4 | 67958.8 | 69115.5 | 115794 | 118012 | 183779 | 187284 | 221029 | 225736 | 1576.06 | 37856.1 | 37922.1 | 59494.7 | 59865.3 |
| 1577.75 | 41637.7 | 42136.1 | 68391.2 | 69171.7 | 116049 | 118057 | 184753 | 187304 | 222056 | 225731 | 1577.75 | 37991.9 | 37990.3 | 60076.1 | 59947.4 |
| 1579.44 | 41253.7 | 42189.5 | 68174 | 69227.4 | 115899 | 118101 | 182948 | 187324 | 220637 | 225724 | 1579.44 | 37938.1 | 38058.3 | 60335.5 | 60029.1 |

| T = 1783 K, n = 0.43, 2730 sec | T = 1783 K, n = 0.43, 2730 sec | T = 1955 K, n = 0.25, 4260 sec | T = 1955 K, n = 0.25, 4260 sec | T = 2210 K, n = 0.13, 6690 sec | T = 2210 K, n = 0.13, 6690 sec | Figure 2C, Nano Diamond | T = 1538 K, n = 0.92, 1380 sec | T = 1538 K, n = 0.92, 1380 sec | T = 1643 K, n = 1.00, 2040 sec | T = 1643 K, n = 1.00, 2040 sec | T = 1751 K, n = 0.82, 3900 sec | T = 1751 K, n = 0.82, 3900 sec | T = 1959 K, n = 0.63, 10380 sec | T = 1959 K, n = 0.63, 10380 sec | T = 2049 K, n = 0.51, 11700 sec |
|---|---|---|---|---|---|---|---|---|---|---|---|---|---|---|---|
| 2266.93 | 1992.26 | 5495.23 | 5917.95 | 16274.4 | 20804.8 | 600.015 | 18.3979 | 63.1154 | 214.365 | 211.425 | 529.211 | 599.135 | 1961.28 | 1986.52 | 2685.66 |
| 1569.17 | 2000.95 | 5399.68 | 5940.93 | 16442.8 | 20872.5 | 600.306 | 52.3244 | 63.4414 | 304.831 | 212.406 | 553.598 | 601.709 | 1824.24 | 1993.84 | 2654.38 |
| 1654.14 | 2009.68 | 5604.05 | 5963.96 | 15875 | 20940.3 | 600.597 | 49.7502 | 63.7687 | 177.121 | 213.391 | 674.403 | 604.291 | 1934.62 | 2001.17 | 2540.96 |
| 2419.84 | 2018.43 | 5646.67 | 5987.06 | 16631.5 | 21008.2 | 600.888 | 82.4833 | 64.0973 | 224.248 | 214.379 | 752.689 | 606.88 | 1957.69 | 2008.52 | 2681.38 |
| 2251.04 | 2027.2 | 5843.07 | 6010.21 | 18289.9 | 21076.3 | 601.179 | 65.7683 | 64.4272 | 239.471 | 215.371 | 733.891 | 609.478 | 2232.32 | 2015.89 | 2910.82 |
| 2103.66 | 2036.01 | 5542.72 | 6033.43 | 18957.7 | 21144.5 | 601.47 | -32.5935 | 64.7584 | 221.531 | 216.365 | 743.569 | 612.083 | 2048.69 | 2023.28 | 3053.82 |
| 2266.23 | 2044.84 | 5642.99 | 6056.7 | 17724.1 | 21212.9 | 601.76 | 45.1446 | 65.0909 | 321.449 | 217.364 | 806.009 | 614.696 | 2254.94 | 2030.68 | 2866.59 |
| 2121.69 | 2053.7 | 5329.36 | 6080.04 | 17456.8 | 21281.4 | 602.052 | 43.3923 | 65.4247 | 238.751 | 218.365 | 674.297 | 617.317 | 1963.77 | 2038.1 | 2860.14 |
| 1958.54 | 2062.59 | 5504.2 | 6103.43 | 17227.3 | 21350 | 602.342 | -28.2081 | 65.7597 | 342.365 | 219.37 | 657.513 | 619.945 | 2184.39 | 2045.54 | 2978.77 |
| 2257.61 | 2071.5 | 6111.78 | 6126.89 | 18916.1 | 21418.7 | 602.633 | 96.192 | 66.0961 | 320.226 | 220.378 | 654.251 | 622.582 | 2266.47 | 2053 | 3016.06 |
| 2129.61 | 2080.44 | 5992.72 | 6150.41 | 18471.9 | 21487.6 | 602.924 | 26.5575 | 66.4338 | 206.291 | 221.39 | 737.576 | 625.226 | 2337.01 | 2060.47 | 2831.96 |
| 2390.63 | 2089.41 | 6031.08 | 6173.98 | 19457.5 | 21556.6 | 603.215 | 0 | 66.7728 | 287.337 | 222.405 | 663.629 | 627.879 | 1992.72 | 2067.96 | 2903.88 |
| 2240.15 | 2098.41 | 5982.2 | 6197.62 | 18912.2 | 21625.8 | 603.506 | 50.3463 | 67.1131 | 195.538 | 223.423 | 736.582 | 630.539 | 2075.95 | 2075.47 | 3010.11 |
| 2415.94 | 2107.43 | 6318.45 | 6221.32 | 18451.2 | 21695 | 603.797 | 59.3873 | 67.4547 | 299.848 | 224.445 | 624.398 | 633.207 | 2174 | 2083 | 2815.93 |
| 2494.8 | 2116.48 | 5919.22 | 6245.08 | 20760.3 | 21764.4 | 604.088 | 70.2151 | 67.7977 | 261.344 | 225.47 | 557.164 | 635.883 | 2200.83 | 2090.54 | 2931.96 |
| 2279.46 | 2125.56 | 6315.42 | 6268.9 | 18757 | 21834 | 604.379 | 11.3153 | 68.1419 | 241.709 | 226.498 | 718.302 | 638.567 | 2128 | 2098.1 | 3022.37 |
| 2002.81 | 2134.67 | 6091.07 | 6292.78 | 19799.8 | 21903.6 | 604.669 | 33.3921 | 68.4875 | 248.574 | 227.53 | 662.424 | 641.258 | 2238.81 | 2105.68 | 3065.26 |
| 2330.18 | 2143.8 | 6518.27 | 6316.73 | 20701.9 | 21973.4 | 604.96 | 44.9587 | 68.8344 | 371.054 | 228.565 | 746.948 | 643.958 | 2294.64 | 2113.28 | 3037.09 |
| 2371.46 | 2152.96 | 6386.57 | 6340.73 | 20491.7 | 22043.4 | 605.251 | 96.5719 | 69.1827 | 322.979 | 229.604 | 755.664 | 646.666 | 2201.42 | 2120.89 | 3188.26 |
| 2261.39 | 2162.15 | 6536.3 | 6364.79 | 20036.5 | 22113.4 | 605.542 | 0 | 69.5323 | 228.507 | 230.646 | 771.335 | 649.381 | 2284.47 | 2128.52 | 3209.76 |
| 2401.52 | 2171.37 | 6625.66 | 6388.92 | 20532.8 | 22183.6 | 605.832 | 71.6266 | 69.8832 | 337.8 | 231.691 | 832.246 | 652.105 | 2222.77 | 2136.17 | 3146.27 |
| 2390.59 | 2180.62 | 6167.84 | 6413.1 | 20926.7 | 22254 | 606.123 | 88.7149 | 70.2355 | 296.702 | 232.74 | 713.738 | 654.836 | 2280.06 | 2143.84 | 3041.12 |
| 2423.58 | 2189.89 | 6496.78 | 6437.35 | 21617.6 | 22324.4 | 606.414 | 98.3464 | 70.5891 | 286.473 | 233.793 | 799.897 | 657.576 | 2413.31 | 2151.52 | 3115.61 |
| 2572.97 | 2199.19 | 6499.78 | 6461.66 | 21377.2 | 22395 | 606.705 | 37.8575 | 70.944 | 266.498 | 234.849 | 732.231 | 660.323 | 2331.97 | 2159.22 | 3164.88 |
| 2566.89 | 2208.52 | 6798.3 | 6486.03 | 21475.2 | 22465.7 | 606.995 | 83.7724 | 71.3003 | 253.059 | 235.908 | 720.137 | 663.079 | 2358.98 | 2166.94 | 3237.88 |
| 2358.74 | 2217.88 | 6797.34 | 6510.46 | 21585.1 | 22536.6 | 607.286 | 44.667 | 71.658 | 329.613 | 236.97 | 797.483 | 665.843 | 2237.08 | 2174.68 | 3200.04 |
| 2562.05 | 2227.26 | 6944.35 | 6534.95 | 22244.6 | 22607.5 | 607.577 | 71.7788 | 72.017 | 357.186 | 238.037 | 783.161 | 668.614 | 2400.23 | 2182.43 | 3269.73 |
| 2341.6 | 2236.68 | 6413.73 | 6559.51 | 21782.1 | 22678.7 | 607.868 | 76.9657 | 72.3774 | 257.408 | 239.106 | 737.966 | 671.394 | 2330.72 | 2190.2 | 3196.14 |
| 2637.93 | 2246.12 | 6974.41 | 6584.12 | 22126.6 | 22749.9 | 608.158 | 60.5211 | 72.7391 | 293.82 | 240.179 | 746.088 | 674.182 | 2312.89 | 2197.99 | 3329.78 |
| 2586.55 | 2255.59 | 6715.76 | 6608.8 | 22296 | 22821.3 | 608.449 | 82.1423 | 73.1022 | 358.908 | 241.256 | 790.315 | 676.977 | 2494.82 | 2205.8 | 3350.62 |
| 2680.43 | 2265.09 | 7264.09 | 6633.54 | 22294.9 | 22892.8 | 608.74 | 139.511 | 73.4667 | 247.016 | 242.336 | 789.572 | 679.781 | 2475.97 | 2213.62 | 3220.03 |
| 2709.07 | 2274.61 | 6994.3 | 6658.34 | 22340.5 | 22964.4 | 609.03 | 63.6843 | 73.8325 | 324.635 | 243.419 | 781.699 | 682.593 | 2409.76 | 2221.47 | 3288.35 |
| 2506.5 | 2284.17 | 7112.15 | 6683.2 | 22644.5 | 23036.2 | 609.321 | 62.1362 | 74.1997 | 286.578 | 244.506 | 785.809 | 685.413 | 2382.43 | 2229.32 | 3240.58 |
| 2686.38 | 2293.75 | 7090.6 | 6708.12 | 23607.3 | 23108.1 | 609.612 | 71.0446 | 74.5683 | 367.904 | 245.597 | 790.81 | 688.241 | 2365.76 | 2237.2 | 3456.94 |
| 2642.57 | 2303.36 | 7221.27 | 6733.1 | 23426.4 | 23180.1 | 609.902 | 38.6037 | 74.9383 | 292.367 | 246.691 | 895.998 | 691.077 | 2562.35 | 2245.1 | 3493.04 |
| 2499.47 | 2313 | 7603.37 | 6758.15 | 24310.7 | 23252.3 | 610.193 | 100.082 | 75.3096 | 276.578 | 247.788 | 817.066 | 693.921 | 2671.13 | 2253.01 | 3514.9 |
| 2570.5 | 2322.67 | 7356.68 | 6783.26 | 23298.2 | 23324.6 | 610.484 | 103.371 | 75.6824 | 243.754 | 248.889 | 739.694 | 696.774 | 2457.89 | 2260.94 | 3485.75 |
| 2612.72 | 2332.37 | 7360.32 | 6808.43 | 23489.3 | 23397 | 610.774 | 101.642 | 76.0565 | 274.926 | 249.994 | 806.542 | 699.634 | 2497.11 | 2268.89 | 3427.48 |
| 2549.86 | 2342.09 | 7674.48 | 6833.66 | 23661.8 | 23469.6 | 611.065 | 130.373 | 76.432 | 365.699 | 251.102 | 840.55 | 702.503 | 2418.35 | 2276.86 | 3554.32 |
| 2753.73 | 2351.85 | 7111.1 | 6858.96 | 22993.8 | 23542.3 | 611.355 | 91.9776 | 76.809 | 322.88 | 252.213 | 807.047 | 705.38 | 2447.72 | 2284.84 | 3493.98 |
| 2690.77 | 2361.63 | 7710.75 | 6884.31 | 23760.3 | 23615.1 | 611.646 | 20.8451 | 77.1873 | 350.824 | 253.328 | 868.389 | 708.265 | 2382.88 | 2292.84 | 3488.89 |
| 2855.92 | 2371.44 | 7473.16 | 6909.73 | 23774.9 | 23688.1 | 611.937 | 95.9766 | 77.567 | 332.822 | 254.447 | 802.382 | 711.158 | 2550.96 | 2300.86 | 3463.54 |
| 2668.26 | 2381.28 | 7377.04 | 6935.21 | 23520.5 | 23761.1 | 612.227 | 67.5379 | 77.9481 | 255.751 | 255.569 | 823.977 | 714.059 | 2553.88 | 2308.9 | 3405.18 |
| 2567.68 | 2391.15 | 7511.59 | 6960.75 | 24032 | 23834.4 | 612.518 | 66.6415 | 78.3307 | 304.121 | 256.695 | 865.92 | 716.969 | 2459.66 | 2316.95 | 3401.37 |
| 2716.79 | 2401.05 | 7208.61 | 6986.36 | 23422.7 | 23907.7 | 612.808 | 97.8224 | 78.7146 | 291.279 | 257.824 | 730.949 | 719.886 | 2511.95 | 2325.02 | 3436.59 |
| 2798.75 | 2410.97 | 7479.07 | 7012.02 | 24307.6 | 23981.2 | 613.099 | 124.15 | 79.1 | 222.057 | 258.957 | 816.628 | 722.812 | 2575.98 | 2333.11 | 3335.04 |
| 2640.16 | 2420.93 | 7618.13 | 7037.75 | 23761.2 | 24054.8 | 613.389 | 77.506 | 79.4868 | 294.752 | 260.094 | 820.023 | 725.747 | 2643.71 | 2341.22 | 3497.19 |
| 2818.94 | 2430.91 | 7605.59 | 7063.54 | 24451.1 | 24128.5 | 613.68 | 82.9072 | 79.875 | 328.193 | 261.234 | 847.648 | 728.689 | 2494.96 | 2349.34 | 3651.21 |
| 2778.84 | 2440.92 | 7503.17 | 7089.4 | 24286.8 | 24202.4 | 613.97 | 69.4615 | 80.2646 | 312.699 | 262.377 | 845.566 | 731.64 | 2654.56 | 2357.49 | 3621.8 |
| 2913.74 | 2450.97 | 7895.8 | 7115.31 | 24638.8 | 24276.4 | 614.26 | 81.0338 | 80.6557 | 344.987 | 263.524 | 828.495 | 734.598 | 2582.77 | 2365.65 | 3730.04 |
| 2796.66 | 2461.03 | 7626.52 | 7141.29 | 24510.6 | 24350.5 | 614.551 | 98.1916 | 81.0482 | 321.775 | 264.675 | 785.247 | 737.566 | 2616.89 | 2373.83 | 3507.17 |
| 2722.76 | 2471.14 | 7548.92 | 7167.33 | 24481.9 | 24424.8 | 614.841 | 65.4422 | 81.4421 | 248.392 | 265.829 | 861.549 | 740.541 | 2680.45 | 2382.03 | 3486.14 |
| 2795.05 | 2481.26 | 7679.4 | 7193.43 | 24876.2 | 24499.2 | 615.132 | 142.316 | 81.8375 | 310.915 | 266.987 | 817.562 | 743.525 | 2641.88 | 2390.24 | 3474.75 |
| 2849.97 | 2491.42 | 7639.76 | 7219.6 | 24538 | 24573.7 | 615.422 | 127.213 | 82.2343 | 325.644 | 268.149 | 797.236 | 746.517 | 2721.58 | 2398.47 | 3484.14 |
| 2677.96 | 2501.61 | 7828.37 | 7245.83 | 24315.9 | 24648.4 | 615.713 | 85.2214 | 82.6325 | 259.274 | 269.314 | 901.653 | 749.517 | 2736.89 | 2406.72 | 3646.8 |
| 2770.22 | 2511.82 | 7695.38 | 7272.12 | 24742.4 | 24723.2 | 616.003 | 124.643 | 83.0322 | 313.563 | 270.483 | 730.551 | 752.525 | 2672.29 | 2414.99 | 3736.35 |
| 2714.91 | 2522.07 | 7799.19 | 7298.47 | 24639 | 24798.1 | 616.293 | 104.246 | 83.4333 | 346.277 | 271.655 | 832.643 | 755.542 | 2593.39 | 2423.27 | 3538.8 |
| 2724.49 | 2532.34 | 7765.34 | 7324.89 | 24214.3 | 24873.1 | 616.584 | 103.793 | 83.8359 | 334.165 | 272.831 | 910.306 | 758.567 | 2681.03 | 2431.58 | 3591.31 |
| 2867.02 | 2542.65 | 7820.11 | 7351.37 | 24464.7 | 24948.3 | 616.874 | 85.9131 | 84.2399 | 276.325 | 274.011 | 868.139 | 761.601 | 2619.07 | 2439.9 | 3602.28 |
| 3118.22 | 2552.98 | 7765.69 | 7377.91 | 24901.8 | 25023.6 | 617.165 | 69.3187 | 84.6455 | 336.531 | 275.194 | 936.143 | 764.643 | 2718.82 | 2448.24 | 3654.5 |
| 2771.78 | 2563.34 | 7889.75 | 7404.52 | 25068.5 | 25099.1 | 617.455 | 73.213 | 85.0524 | 294.505 | 276.381 | 897.364 | 767.693 | 2519.19 | 2456.59 | 3632.84 |
| 2722.11 | 2573.73 | 7812.2 | 7431.18 | 24432.9 | 25174.6 | 617.745 | 75.9162 | 85.4608 | 314.505 | 277.572 | 958.824 | 770.751 | 2524.59 | 2464.97 | 3583.95 |
| 2991.16 | 2584.16 | 8105.47 | 7457.91 | 25091.4 | 25250.3 | 618.036 | 90.5544 | 85.8708 | 346.817 | 278.766 | 948.097 | 773.818 | 2641.03 | 2473.36 | 3671.89 |
| 3033.54 | 2594.61 | 8164.7 | 7484.71 | 25731.3 | 25326.2 | 618.326 | 86.7354 | 86.2821 | 307.145 | 279.964 | 895.741 | 776.894 | 2737.45 | 2481.77 | 3713.35 |
| 3033.54 | 2605.09 | 7901.05 | 7511.56 | 24796.9 | 25402.1 | 618.616 | 118.174 | 86.695 | 325.105 | 281.165 | 893.765 | 779.977 | 2706.24 | 2490.2 | 3833.46 |
| 2913.86 | 2615.59 | 7988.14 | 7538.48 | 25168.5 | 25478.2 | 618.907 | 92.2052 | 87.1093 | 325.129 | 282.371 | 880.876 | 783.069 | 2708.69 | 2498.64 | 3637.5 |
| 2757.15 | 2626.13 | 8074.37 | 7565.47 | 25223.8 | 25554.4 | 619.197 | 76.4592 | 87.5251 | 306.237 | 283.579 | 872.147 | 786.17 | 2652.86 | 2507.11 | 3873.78 |
| 3044.27 | 2636.7 | 8045.43 | 7592.51 | 25408 | 25630.8 | 619.487 | 66.2347 | 87.9424 | 344.527 | 284.792 | 877.528 | 789.279 | 2702.53 | 2515.59 | 3908.62 |
| 3010.65 | 2647.3 | 8111.29 | 7619.62 | 25437.7 | 25707.3 | 619.777 | 93.0107 | 88.3611 | 316.086 | 286.008 | 899.497 | 792.396 | 2719.68 | 2524.09 | 3745.81 |
| 2914.68 | 2657.93 | 8212.36 | 7646.79 | 25325.8 | 25783.9 | 620.067 | 109.712 | 88.7814 | 359.529 | 287.228 | 879.645 | 795.522 | 2735.43 | 2532.6 | 3753.01 |
| 2894.44 | 2668.59 | 8006.88 | 7674.03 | 25579 | 25860.6 | 620.358 | 63.5544 | 89.2031 | 334.993 | 288.452 | 905.057 | 798.656 | 2748.4 | 2541.14 | 3759.85 |
| 2900.18 | 2679.27 | 8091.94 | 7701.33 | 25403.2 | 25937.5 | 620.648 | 93.2328 | 89.6264 | 357.793 | 289.679 | 841.095 | 801.798 | 2777.19 | 2549.69 | 3736.83 |
| 2939.16 | 2689.99 | 8087.34 | 7728.69 | 25253.1 | 26014.5 | 620.938 | 77.7299 | 90.0511 | 331.243 | 290.91 | 822.392 | 804.949 | 2683.92 | 2558.26 | 3791.18 |
| 2844.32 | 2700.74 | 7992.64 | 7756.12 | 25657.2 | 26091.7 | 621.228 | 112.534 | 90.4773 | 361.421 | 292.145 | 897.26 | 808.109 | 2772.78 | 2566.85 | 3701.29 |
| 2971.95 | 2711.51 | 8174.94 | 7783.6 | 25902.7 | 26168.9 | 621.519 | 128.412 | 90.9051 | 357.426 | 293.384 | 861.869 | 811.277 | 2767.96 | 2575.46 | 3908.54 |
| 3062.98 | 2722.32 | 8119.46 | 7811.16 | 25519.8 | 26246.3 | 621.809 | 104.966 | 91.3343 | 346.521 | 294.626 | 945.354 | 814.453 | 2769.97 | 2584.08 | 3946.89 |
| 2881.05 | 2733.16 | 8177.38 | 7838.77 | 25699.8 | 26323.8 | 622.099 | 57.7672 | 91.7651 | 384.618 | 295.872 | 945.426 | 817.638 | 2747.11 | 2592.72 | 3848.94 |
| 3190 | 2744.02 | 8159.36 | 7866.45 | 25751.8 | 26401.5 | 622.389 | 109.004 | 92.1974 | 341.038 | 297.122 | 844.936 | 820.832 | 2789.34 | 2601.38 | 3882.92 |
| 3076.7 | 2754.92 | 8273.65 | 7894.19 | 26431.3 | 26479.3 | 622.679 | 97.3051 | 92.6312 | 377.92 | 298.375 | 904.834 | 824.034 | 2841.99 | 2610.06 | 3870.07 |
| 3015.7 | 2765.85 | 8500.48 | 7922 | 26605.3 | 26557.2 | 622.969 | 100.385 | 93.0665 | 311.905 | 299.632 | 987.736 | 827.244 | 2855.04 | 2618.76 | 3828.69 |
| 3030.71 | 2776.8 | 8348.65 | 7949.87 | 26297.3 | 26635.2 | 623.259 | 69.9061 | 93.5033 | 343.154 | 300.893 | 943.78 | 830.463 | 2683.27 | 2627.47 | 3715.16 |
| 2999.14 | 2787.79 | 8421.84 | 7977.8 | 26458.5 | 26713.4 | 623.549 | 124.86 | 93.9417 | 329.608 | 302.158 | 924.985 | 833.691 | 2844.97 | 2636.2 | 3987.08 |
| 3289.04 | 2798.81 | 8485.27 | 8005.8 | 26458.3 | 26791.7 | 623.84 | 91.9361 | 94.3816 | 349.629 | 303.426 | 949.898 | 836.927 | 2727.55 | 2644.95 | 3878.62 |
| 3135.28 | 2809.85 | 8433.35 | 8033.86 | 26401.7 | 26870.2 | 624.13 | 107.862 | 94.823 | 325.025 | 304.698 | 988.705 | 840.171 | 2734.34 | 2653.72 | 3912.25 |
| 3006.88 | 2820.93 | 8565.83 | 8061.99 | 26396.2 | 26948.7 | 624.42 | 115.119 | 95.2659 | 374.994 | 305.974 | 961.367 | 843.424 | 2823.41 | 2662.5 | 3956.08 |
| 3122.51 | 2832.04 | 8622.02 | 8090.18 | 26986.5 | 27027.4 | 624.71 | 87.8819 | 95.7104 | 366.21 | 307.254 | 915.273 | 846.686 | 2813.5 | 2671.3 | 3908.5 |
| 3168.99 | 2843.18 | 8519.21 | 8118.43 | 27092 | 27106.2 | 625 | 79.7455 | 96.1565 | 373.073 | 308.537 | 931.206 | 849.956 | 2850.73 | 2680.13 | 3951.62 |
| 3142.83 | 2854.34 | 8458.65 | 8146.74 | 27304.3 | 27185.2 | 625.29 | 97.7176 | 96.6041 | 365.466 | 309.825 | 983.607 | 853.235 | 2944.58 | 2688.97 | 3982.95 |
| 3203.76 | 2865.54 | 8586.52 | 8175.12 | 26534 | 27264.3 | 625.58 | 116.25 | 97.0532 | 400.192 | 311.116 | 1009.81 | 856.523 | 2859.02 | 2697.82 | 4015.4 |
| 3155.04 | 2876.77 | 8519.64 | 8203.57 | 26919.8 | 27343.5 | 625.87 | 114.729 | 97.5038 | 337.569 | 312.411 | 944.866 | 859.819 | 2895.03 | 2706.7 | 3984.45 |
| 3107.63 | 2888.03 | 8903.48 | 8232.08 | 27360.4 | 27422.8 | 626.16 | 87.1115 | 97.9561 | 399.229 | 313.709 | 991.926 | 863.123 | 3024.21 | 2715.59 | 4080.01 |
| 3236.57 | 2899.32 | 8791.88 | 8260.65 | 27176.7 | 27502.3 | 626.45 | 113.064 | 98.4099 | 372.59 | 315.012 | 1005.92 | 866.437 | 3011.51 | 2724.5 | 4079.15 |
| 3231.02 | 2910.64 | 8780.52 | 8289.28 | 27772.6 | 27581.9 | 626.74 | 121.124 | 98.8652 | 369.158 | 316.318 | 1001.18 | 869.758 | 3018.28 | 2733.43 | 4131.21 |
| 3330.09 | 2921.99 | 8912.07 | 8317.99 | 27836.7 | 27661.6 | 627.03 | 128.375 | 99.3221 | 387.064 | 317.628 | 961.702 | 873.089 | 2989.82 | 2742.37 | 4165.54 |
| 3246.34 | 2933.37 | 8928.23 | 8346.75 | 27558.4 | 27741.5 | 627.32 | 92.3963 | 99.7806 | 387.692 | 318.942 | 981.928 | 876.428 | 2903.82 | 2751.34 | 4113.26 |
| 3303.19 | 2944.78 | 8890.18 | 8375.58 | 28460.3 | 27821.4 | 627.61 | 82.1451 | 100.241 | 386.039 | 320.26 | 1023.37 | 879.776 | 3053.37 | 2760.32 | 4116.14 |
| 3108.34 | 2956.22 | 9106.85 | 8404.47 | 28400.7 | 27901.5 | 627.9 | 128.426 | 100.702 | 427.09 | 321.581 | 964.936 | 883.132 | 3048.11 | 2769.32 | 4125.05 |
| 3361.43 | 2967.7 | 9197.36 | 8433.43 | 28493.9 | 27981.8 | 628.19 | 100.907 | 101.165 | 366.626 | 322.907 | 1033.17 | 886.497 | 3054.27 | 2778.34 | 4296.69 |
| 3447.9 | 2979.2 | 9002.51 | 8462.45 | 28398.2 | 28062.2 | 628.48 | 116.894 | 101.63 | 383.447 | 324.236 | 1015.31 | 889.871 | 2996.18 | 2787.37 | 4167 |
| 3303.34 | 2990.73 | 9278.35 | 8491.53 | 29174.9 | 28142.7 | 628.77 | 104.571 | 102.097 | 412.294 | 325.569 | 1018.37 | 893.253 | 3081.69 | 2796.43 | 4264.85 |
| 3205.38 | 3002.3 | 8993.02 | 8520.68 | 28643.5 | 28223.3 | 629.059 | 128.902 | 102.564 | 375.478 | 326.906 | 1031.98 | 896.644 | 3021.74 | 2805.5 | 4197.35 |
| 3371.22 | 3013.89 | 9475.19 | 8549.9 | 28446.1 | 28304 | 629.349 | 92.0116 | 103.034 | 413.943 | 328.247 | 1049.55 | 900.044 | 3093.76 | 2814.59 | 4251.72 |
| 3378.98 | 3025.52 | 9357.11 | 8579.17 | 28956.1 | 28384.9 | 629.639 | 115.177 | 103.505 | 391.415 | 329.591 | 1076.28 | 903.453 | 3161.01 | 2823.7 | 4277.01 |
| 3667 | 3037.18 | 9418.89 | 8608.51 | 29373.3 | 28466 | 629.929 | 127.421 | 103.978 | 415.352 | 330.94 | 1041.19 | 906.87 | 3185.36 | 2832.82 | 4369.01 |
| 3475.2 | 3048.86 | 9064.91 | 8637.92 | 29102.7 | 28547.1 | 630.219 | 121.418 | 104.452 | 393.935 | 332.292 | 1042.77 | 910.296 | 3133.14 | 2841.97 | 4292.31 |
| 3429.24 | 3060.58 | 9271.85 | 8667.39 | 28879.4 | 28628.4 | 630.509 | 136.436 | 104.928 | 408.7 | 333.649 | 1062.48 | 913.731 | 3205.53 | 2851.13 | 4273.61 |
| 3590.55 | 3072.33 | 9645.51 | 8696.93 | 29838.2 | 28709.8 | 630.799 | 146.398 | 105.405 | 432.127 | 335.009 | 1132.64 | 917.174 | 3180.19 | 2860.31 | 4443.86 |
| 3457.33 | 3084.11 | 9415.25 | 8726.53 | 29709.3 | 28791.3 | 631.088 | 132.435 | 105.885 | 410.928 | 336.373 | 1068.02 | 920.626 | 3199.42 | 2869.51 | 4384.49 |
| 3610.3 | 3095.93 | 9640.2 | 8756.19 | 29904.2 | 28873 | 631.378 | 127.795 | 106.365 | 405.343 | 337.741 | 1109.84 | 924.087 | 3247.33 | 2878.72 | 4448.39 |
| 3470.6 | 3107.77 | 9555.6 | 8785.92 | 30348.4 | 28954.7 | 631.668 | 117.925 | 106.847 | 392.577 | 339.113 | 1058.29 | 927.556 | 3156.93 | 2887.96 | 4467.81 |
| 3481.34 | 3119.64 | 9536.37 | 8815.72 | 29828.7 | 29036.7 | 631.958 | 113.146 | 107.331 | 383.175 | 340.488 | 1018.27 | 931.035 | 3202.79 | 2897.21 | 4448.45 |
| 3530.1 | 3131.55 | 9765.74 | 8845.57 | 30132.7 | 29118.7 | 632.247 | 129.316 | 107.817 | 414.085 | 341.868 | 1072.53 | 934.522 | 3184.89 | 2906.48 | 4463.6 |
| 3581.32 | 3143.48 | 9816.35 | 8875.5 | 30572.9 | 29200.9 | 632.537 | 106.746 | 108.304 | 414.587 | 343.251 | 1121.24 | 938.018 | 3256.39 | 2915.76 | 4451.7 |
| 3548.9 | 3155.45 | 9996.4 | 8905.49 | 30657.9 | 29283.2 | 632.827 | 143.538 | 108.793 | 426 | 344.639 | 1106.75 | 941.522 | 3311.99 | 2925.07 | 4631.94 |
| 3682.11 | 3167.45 | 9936.68 | 8935.54 | 30307.6 | 29365.6 | 633.117 | 117 | 109.283 | 433.758 | 346.03 | 1126.22 | 945.036 | 3243.84 | 2934.39 | 4503.81 |
| 3752.74 | 3179.48 | 9824.62 | 8965.65 | 31071.4 | 29448.1 | 633.407 | 138.002 | 109.775 | 414.87 | 347.425 | 1137.17 | 948.558 | 3304.2 | 2943.73 | 4478.49 |
| 3665.92 | 3191.54 | 10030 | 8995.84 | 31192.3 | 29530.8 | 633.696 | 128.813 | 110.269 | 431.375 | 348.825 | 1089.94 | 952.089 | 3291.54 | 2953.09 | 4569.6 |
| 3609.28 | 3203.63 | 9830.55 | 9026.08 | 30972.7 | 29613.6 | 633.986 | 132.543 | 110.764 | 430.632 | 350.228 | 1107.36 | 955.629 | 3327.31 | 2962.47 | 4438.56 |
| 3676.41 | 3215.75 | 9852.14 | 9056.39 | 30469.9 | 29696.6 | 634.276 | 118.587 | 111.261 | 431.176 | 351.635 | 1138.7 | 959.177 | 3288.58 | 2971.86 | 4584.35 |
| 3763.99 | 3227.91 | 9998.19 | 9086.77 | 31285.9 | 29779.6 | 634.565 | 134.209 | 111.76 | 408.149 | 353.046 | 1102.64 | 962.735 | 3295.38 | 2981.28 | 4700.56 |
| 3715.53 | 3240.09 | 10086.8 | 9117.21 | 31486.9 | 29862.8 | 634.855 | 117.793 | 112.26 | 433.159 | 354.461 | 1098.77 | 966.301 | 3364.88 | 2990.71 | 4597.79 |
| 3744.7 | 3252.31 | 10261.9 | 9147.72 | 31780.1 | 29946.1 | 635.144 | 119.665 | 112.762 | 421.193 | 355.88 | 1144.91 | 969.876 | 3409.82 | 3000.16 | 4679.38 |
| 3750.28 | 3264.56 | 10212.5 | 9178.29 | 31443.7 | 30029.6 | 635.434 | 118.151 | 113.265 | 434.984 | 357.302 | 1101.12 | 973.46 | 3398.86 | 3009.62 | 4737.4 |
| 3732.95 | 3276.84 | 10250.3 | 9208.92 | 31653 | 30113.1 | 635.724 | 140.128 | 113.771 | 438.674 | 358.729 | 1179.02 | 977.053 | 3445.38 | 3019.11 | 4527.28 |
| 3767.63 | 3289.15 | 10340.6 | 9239.62 | 32076.7 | 30196.8 | 636.013 | 129.222 | 114.277 | 432.176 | 360.16 | 1129.83 | 980.655 | 3352.43 | 3028.61 | 4613.03 |
| 3928.48 | 3301.49 | 10199.1 | 9270.39 | 31448.1 | 30280.7 | 636.303 | 128.361 | 114.786 | 424.679 | 361.595 | 1150.59 | 984.265 | 3315.74 | 3038.13 | 4661.02 |
| 3724.51 | 3313.87 | 10196.8 | 9301.22 | 31324.1 | 30364.6 | 636.593 | 144.337 | 115.296 | 436.263 | 363.033 | 1147.64 | 987.885 | 3394.84 | 3047.67 | 4808.31 |
| 3749.8 | 3326.27 | 10276.5 | 9332.11 | 31882.1 | 30448.7 | 636.882 | 119.972 | 115.808 | 425.633 | 364.476 | 1141.93 | 991.513 | 3434.03 | 3057.23 | 4650.57 |
| 3769.7 | 3338.71 | 10265.4 | 9363.08 | 31757 | 30532.9 | 637.172 | 124.863 | 116.321 | 421.025 | 365.923 | 1179.95 | 995.15 | 3397.33 | 3066.8 | 4698.36 |
| 3793.24 | 3351.18 | 10222.3 | 9394.1 | 31633.3 | 30617.2 | 637.461 | 134.219 | 116.837 | 456.127 | 367.373 | 1182.63 | 998.796 | 3417.31 | 3076.39 | 4675.85 |
| 3981.24 | 3363.68 | 10329.3 | 9425.19 | 31919.7 | 30701.7 | 637.751 | 116.319 | 117.353 | 460.457 | 368.828 | 1164.85 | 1002.45 | 3439.7 | 3086 | 4712.77 |
| 3997 | 3376.21 | 10566.2 | 9456.35 | 32441.1 | 30786.3 | 638.04 | 126.893 | 117.872 | 466.896 | 370.287 | 1172.52 | 1006.12 | 3403.07 | 3095.63 | 4796.2 |
| 3811.17 | 3388.78 | 10537.6 | 9487.57 | 32230.6 | 30871 | 638.33 | 126.847 | 118.392 | 439.568 | 371.749 | 1175.74 | 1009.79 | 3433.36 | 3105.28 | 4748.02 |
| 3898.14 | 3401.37 | 10409 | 9518.85 | 32338.6 | 30955.8 | 638.619 | 136.781 | 118.914 | 455.349 | 373.216 | 1160.74 | 1013.47 | 3458.65 | 3114.94 | 4747.26 |
| 3951.73 | 3414 | 10383.8 | 9550.21 | 32321.1 | 31040.8 | 638.909 | 132.003 | 119.437 | 429.991 | 374.687 | 1138.26 | 1017.16 | 3511.56 | 3124.62 | 4747.88 |
| 3809.4 | 3426.66 | 10418 | 9581.62 | 32182 | 31125.9 | 639.198 | 143.65 | 119.963 | 438.95 | 376.161 | 1148.26 | 1020.86 | 3490.59 | 3134.32 | 4713.1 |
| 3920.1 | 3439.35 | 10356.3 | 9613.1 | 32084.5 | 31211.1 | 639.488 | 153.699 | 120.489 | 445.694 | 377.64 | 1159.87 | 1024.57 | 3408.99 | 3144.04 | 4750.06 |
| 3791.95 | 3452.07 | 10377 | 9644.65 | 32118.5 | 31296.5 | 639.777 | 133.64 | 121.018 | 473.124 | 379.123 | 1211.35 | 1028.29 | 3469.6 | 3153.78 | 4810.54 |
| 3971.92 | 3464.83 | 10556.4 | 9676.26 | 32650.3 | 31382 | 640.067 | 133.471 | 121.548 | 450.597 | 380.61 | 1207.79 | 1032.01 | 3531.32 | 3163.53 | 4895.9 |
| 3951.76 | 3477.62 | 10894.3 | 9707.94 | 33387.1 | 31467.5 | 640.356 | 137.436 | 122.08 | 460.088 | 382.1 | 1208.57 | 1035.75 | 3471.95 | 3173.3 | 4967.27 |
| 3895.42 | 3490.43 | 10466.8 | 9739.68 | 32556.1 | 31553.3 | 640.646 | 133.578 | 122.614 | 441.366 | 383.595 | 1166.74 | 1039.49 | 3496.8 | 3183.09 | 4844.35 |
| 3952.85 | 3503.29 | 10763.8 | 9771.49 | 32779.3 | 31639.1 | 640.935 | 143.508 | 123.149 | 448.989 | 385.094 | 1231.67 | 1043.24 | 3539.96 | 3192.9 | 4807.97 |
| 4020.26 | 3516.17 | 10657.5 | 9803.36 | 33054.4 | 31725.1 | 641.225 | 140.061 | 123.686 | 475.022 | 386.597 | 1177.92 | 1047.01 | 3559.14 | 3202.73 | 4832.1 |
| 4094.7 | 3529.08 | 10756.1 | 9835.3 | 32838.9 | 31811.2 | 641.514 | 127.942 | 124.225 | 447.972 | 388.104 | 1226.74 | 1050.78 | 3550.13 | 3212.57 | 4862.8 |
| 3906.26 | 3542.03 | 10909.2 | 9867.31 | 33085.7 | 31897.4 | 641.803 | 152.742 | 124.766 | 467.45 | 389.615 | 1212.02 | 1054.56 | 3556.4 | 3222.43 | 4914.12 |
| 4065.85 | 3555.01 | 10669.1 | 9899.38 | 32917.6 | 31983.8 | 642.093 | 155.534 | 125.308 | 462.264 | 391.13 | 1204.08 | 1058.35 | 3515.62 | 3232.31 | 4783.57 |
| 3999.97 | 3568.02 | 10814 | 9931.51 | 33666.3 | 32070.3 | 642.382 | 152.821 | 125.852 | 479.959 | 392.649 | 1223.88 | 1062.15 | 3569.04 | 3242.21 | 4945.66 |
| 4018.75 | 3581.06 | 10724 | 9963.71 | 33036.1 | 32156.9 | 642.671 | 141.068 | 126.398 | 466.787 | 394.172 | 1237.22 | 1065.95 | 3541.17 | 3252.12 | 4856.04 |
| 4049.68 | 3594.14 | 10934.2 | 9995.98 | 33229.2 | 32243.6 | 642.961 | 147.233 | 126.945 | 468.685 | 395.7 | 1227.46 | 1069.77 | 3683.03 | 3262.06 | 4995.42 |
| 4060.87 | 3607.25 | 10920.3 | 10028.3 | 33313 | 32330.5 | 643.25 | 152.507 | 127.494 | 485.272 | 397.231 | 1228.38 | 1073.6 | 3682.85 | 3272.01 | 4959.25 |
| 4185.05 | 3620.39 | 10856.3 | 10060.7 | 33314.5 | 32417.4 | 643.539 | 134.876 | 128.045 | 468.98 | 398.766 | 1207.65 | 1077.43 | 3645.89 | 3281.98 | 4951.43 |
| 4110.35 | 3633.56 | 11016.3 | 10093.2 | 33461.6 | 32504.5 | 643.829 | 146.75 | 128.598 | 487.277 | 400.306 | 1207.6 | 1081.27 | 3631.32 | 3291.97 | 5023.02 |
| 4056.38 | 3646.76 | 10980.3 | 10125.7 | 33303.3 | 32591.8 | 644.118 | 129.959 | 129.152 | 456.343 | 401.849 | 1216.65 | 1085.12 | 3614.8 | 3301.97 | 4959.75 |
| 4147.39 | 3660 | 11159.8 | 10158.3 | 34057.4 | 32679.1 | 644.407 | 156.026 | 129.708 | 435.551 | 403.397 | 1271.49 | 1088.99 | 3578.71 | 3311.99 | 4917.57 |
| 4229.16 | 3673.27 | 11210.5 | 10191 | 33979.4 | 32766.6 | 644.697 | 157.101 | 130.266 | 479.9 | 404.949 | 1279.88 | 1092.86 | 3662.29 | 3322.03 | 5058.45 |
| 4195.1 | 3686.57 | 11165.2 | 10223.7 | 34075.4 | 32854.2 | 644.986 | 143.814 | 130.826 | 466.596 | 406.505 | 1276.86 | 1096.74 | 3700.47 | 3332.09 | 5071.31 |
| 4187.54 | 3699.9 | 11175.8 | 10256.5 | 34213.7 | 32941.9 | 645.275 | 147.6 | 131.387 | 494.013 | 408.065 | 1250.45 | 1100.62 | 3728.46 | 3342.17 | 4988.76 |
| 4169.72 | 3713.27 | 11050.9 | 10289.3 | 33951.4 | 33029.8 | 645.564 | 160.967 | 131.95 | 488.832 | 409.629 | 1229.73 | 1104.52 | 3688.87 | 3352.26 | 4987.02 |
| 4126.14 | 3726.67 | 11102.3 | 10322.2 | 33694.3 | 33117.8 | 645.853 | 150.57 | 132.515 | 501.017 | 411.197 | 1271.98 | 1108.43 | 3624.04 | 3362.38 | 4991.79 |
| 4184.39 | 3740.1 | 11147.5 | 10355.2 | 34739.6 | 33205.9 | 646.143 | 163.794 | 133.082 | 486.953 | 412.769 | 1271.25 | 1112.35 | 3708.89 | 3372.51 | 5045.28 |
| 4147.96 | 3753.56 | 11257.9 | 10388.3 | 34370.5 | 33294.1 | 646.432 | 178.403 | 133.65 | 487.292 | 414.346 | 1227.11 | 1116.27 | 3725.86 | 3382.66 | 5081.81 |
| 4155.6 | 3767.06 | 11292.3 | 10421.4 | 34545.1 | 33382.4 | 646.721 | 157.204 | 134.22 | 535.042 | 415.926 | 1263.84 | 1120.2 | 3751.1 | 3392.82 | 5083 |
| 4189.42 | 3780.59 | 11404.9 | 10454.6 | 34900.3 | 33470.9 | 647.01 | 166.933 | 134.792 | 476.629 | 417.511 | 1254.13 | 1124.15 | 3726.66 | 3403.01 | 5013.6 |
| 4311.22 | 3794.15 | 11545.5 | 10487.8 | 34926.8 | 33559.5 | 647.299 | 156.042 | 135.365 | 510.356 | 419.1 | 1269.16 | 1128.1 | 3748.16 | 3413.21 | 5162.78 |
| 4209.5 | 3807.74 | 11482.1 | 10521.1 | 34656.9 | 33648.2 | 647.588 | 156.033 | 135.941 | 488.871 | 420.693 | 1268.45 | 1132.06 | 3714.35 | 3423.43 | 5125.98 |
| 4292.71 | 3821.37 | 11442.6 | 10554.5 | 34943.6 | 33737.1 | 647.878 | 158.605 | 136.518 | 472.89 | 422.29 | 1298.27 | 1136.03 | 3739.58 | 3433.66 | 5178.19 |
| 4287.25 | 3835.03 | 11554.7 | 10588 | 35276.9 | 33826.1 | 648.167 | 165.919 | 137.097 | 492.6 | 423.892 | 1286.52 | 1140.01 | 3821.82 | 3443.92 | 5169.76 |
| 4472.49 | 3848.72 | 11631.9 | 10621.5 | 35438.3 | 33915.2 | 648.456 | 156.496 | 137.678 | 514.778 | 425.497 | 1308.34 | 1144 | 3804.59 | 3454.19 | 5225.95 |
| 4341.89 | 3862.44 | 11519.8 | 10655.1 | 35388.1 | 34004.4 | 648.745 | 148.729 | 138.26 | 500.119 | 427.107 | 1284.22 | 1148 | 3807.1 | 3464.48 | 5207.71 |
| 4293.44 | 3876.2 | 11700.3 | 10688.7 | 35422 | 34093.7 | 649.034 | 165.03 | 138.845 | 486.763 | 428.72 | 1304.9 | 1152.01 | 3737.45 | 3474.79 | 5222.35 |
| 4455.3 | 3889.99 | 11467.5 | 10722.4 | 35475.5 | 34183.2 | 649.323 | 169.423 | 139.431 | 526.413 | 430.338 | 1309.25 | 1156.02 | 3810.48 | 3485.12 | 5233.37 |
| 4495.15 | 3903.81 | 12020 | 10756.2 | 36407 | 34272.8 | 649.612 | 158.721 | 140.018 | 510.689 | 431.96 | 1316.58 | 1160.05 | 3852.9 | 3495.47 | 5316.33 |
| 4475.73 | 3917.67 | 11902.5 | 10790 | 35872.1 | 34362.5 | 649.901 | 162.691 | 140.608 | 524.57 | 433.587 | 1370.13 | 1164.08 | 3837.28 | 3505.83 | 5367.59 |
| 4425.63 | 3931.56 | 11844 | 10824 | 35874.2 | 34452.3 | 650.19 | 163.248 | 141.2 | 525.427 | 435.217 | 1326.87 | 1168.12 | 3811.04 | 3516.21 | 5303.46 |
| 4552.5 | 3945.48 | 12072.9 | 10857.9 | 36175.8 | 34542.3 | 650.479 | 171.576 | 141.793 | 511.579 | 436.852 | 1338.88 | 1172.18 | 3898.47 | 3526.61 | 5357.24 |
| 4417.62 | 3959.43 | 11715.6 | 10892 | 35849.1 | 34632.4 | 650.768 | 177.153 | 142.388 | 507.429 | 438.49 | 1345.69 | 1176.24 | 3818.6 | 3537.02 | 5249.98 |
| 4397.84 | 3973.42 | 11996.4 | 10926.1 | 35994.9 | 34722.6 | 651.057 | 166.072 | 142.985 | 526.562 | 440.134 | 1326.55 | 1180.31 | 3863.58 | 3547.46 | 5247.77 |
| 4410.27 | 3987.44 | 11970.2 | 10960.2 | 36257 | 34812.9 | 651.346 | 172.776 | 143.583 | 510.448 | 441.781 | 1356.35 | 1184.39 | 3880.62 | 3557.91 | 5383.02 |
| 4482.7 | 4001.49 | 11896.6 | 10994.5 | 36642.6 | 34903.3 | 651.635 | 160.145 | 144.184 | 533.534 | 443.432 | 1352.37 | 1188.48 | 3875.69 | 3568.38 | 5504.74 |
| 4469 | 4015.57 | 11939.6 | 11028.8 | 36253.4 | 34993.9 | 651.924 | 155.431 | 144.786 | 504.941 | 445.088 | 1363.03 | 1192.58 | 3995.54 | 3578.86 | 5404.98 |
| 4615.77 | 4029.69 | 12131 | 11063.2 | 36734.6 | 35084.6 | 652.213 | 152.026 | 145.39 | 542.65 | 446.747 | 1364.31 | 1196.68 | 3953.33 | 3589.37 | 5400.76 |
| 4573.02 | 4043.84 | 12260.9 | 11097.6 | 37253.7 | 35175.4 | 652.502 | 159.354 | 145.996 | 526.073 | 448.411 | 1373.69 | 1200.8 | 4012.35 | 3599.89 | 5508.76 |
| 4563.02 | 4058.03 | 12052.1 | 11132.1 | 37172 | 35266.4 | 652.791 | 171.233 | 146.604 | 532.036 | 450.08 | 1326.19 | 1204.93 | 3896.19 | 3610.43 | 5473.4 |
| 4624.58 | 4072.25 | 12201.2 | 11166.7 | 36977.2 | 35357.4 | 653.08 | 164.934 | 147.214 | 520.472 | 451.752 | 1377.87 | 1209.06 | 4102.48 | 3620.99 | 5490.64 |
| 4534.91 | 4086.5 | 12396.4 | 11201.3 | 37221.5 | 35448.6 | 653.369 | 166.527 | 147.825 | 516.592 | 453.429 | 1415.19 | 1213.21 | 4086.1 | 3631.57 | 5483.32 |
| 4722.19 | 4100.78 | 12403.2 | 11236 | 37523 | 35539.9 | 653.658 | 177.427 | 148.438 | 533.366 | 455.11 | 1396.64 | 1217.36 | 4026.97 | 3642.16 | 5532.71 |
| 4720.27 | 4115.1 | 12357.9 | 11270.8 | 37560.4 | 35631.4 | 653.947 | 163.79 | 149.053 | 539.63 | 456.795 | 1402.44 | 1221.53 | 4015.28 | 3652.78 | 5508.86 |
| 4542.09 | 4129.45 | 12421.8 | 11305.6 | 37237 | 35722.9 | 654.236 | 165.586 | 149.67 | 527.889 | 458.484 | 1367.32 | 1225.7 | 3998.07 | 3663.41 | 5484.33 |
| 4757.84 | 4143.83 | 12447 | 11340.5 | 38135.7 | 35814.6 | 654.524 | 172.371 | 150.289 | 570.936 | 460.177 | 1400.87 | 1229.88 | 4089.15 | 3674.05 | 5562.31 |
| 4587.5 | 4158.25 | 12523.1 | 11375.5 | 37804.1 | 35906.4 | 654.813 | 188.803 | 150.91 | 559.33 | 461.875 | 1401.11 | 1234.07 | 4078.9 | 3684.72 | 5618.58 |
| 4686.52 | 4172.7 | 12527.3 | 11410.5 | 37810.9 | 35998.3 | 655.102 | 159.383 | 151.532 | 544.95 | 463.577 | 1378.22 | 1238.27 | 4078.49 | 3695.4 | 5665.26 |
| 4674.52 | 4187.18 | 12604.3 | 11445.6 | 37969.2 | 36090.3 | 655.391 | 185.098 | 152.156 | 549.666 | 465.284 | 1358.88 | 1242.48 | 4087.18 | 3706.1 | 5621.62 |
| 4824.3 | 4201.7 | 12844.8 | 11480.7 | 38669.4 | 36182.5 | 655.68 | 176.091 | 152.782 | 556.674 | 466.994 | 1405.42 | 1246.7 | 4148.65 | 3716.82 | 5682.56 |
| 4702.82 | 4216.25 | 12695.1 | 11516 | 38217.5 | 36274.8 | 655.968 | 187.749 | 153.411 | 526.06 | 468.709 | 1429.95 | 1250.92 | 4117.57 | 3727.56 | 5668.81 |
| 4764.43 | 4230.83 | 12752.3 | 11551.3 | 38441.8 | 36367.2 | 656.257 | 171.72 | 154.04 | 557.073 | 470.428 | 1416.75 | 1255.16 | 4120.64 | 3738.31 | 5629.7 |
| 4693.36 | 4245.45 | 12776.5 | 11586.6 | 38262.4 | 36459.7 | 656.546 | 171.781 | 154.672 | 552.583 | 472.151 | 1442.44 | 1259.41 | 4134.34 | 3749.09 | 5642.12 |
| 4875.41 | 4260.1 | 12785.8 | 11622 | 38545.7 | 36552.3 | 656.835 | 170.289 | 155.305 | 553.27 | 473.879 | 1420.12 | 1263.66 | 4193.2 | 3759.88 | 5712.09 |
| 4813.2 | 4274.78 | 12614.3 | 11657.5 | 38365 | 36645.1 | 657.124 | 187.81 | 155.941 | 555.876 | 475.611 | 1454.32 | 1267.93 | 4150.13 | 3770.68 | 5653.46 |
| 4784.73 | 4289.49 | 12819.4 | 11693.1 | 38862.1 | 36738 | 657.412 | 187.588 | 156.578 | 567.78 | 477.347 | 1439.77 | 1272.2 | 4145.22 | 3781.51 | 5729.82 |
| 4746.16 | 4304.24 | 12931.5 | 11728.7 | 38995.7 | 36831 | 657.701 | 192.124 | 157.217 | 584.676 | 479.087 | 1479.75 | 1276.49 | 4150.93 | 3792.35 | 5724.16 |
| 4879.41 | 4319.03 | 12714.5 | 11764.4 | 38573.6 | 36924.1 | 657.99 | 190.541 | 157.858 | 558.727 | 480.832 | 1441.4 | 1280.78 | 4137.83 | 3803.21 | 5669.41 |
| 4946.24 | 4333.84 | 13056.4 | 11800.2 | 39007.7 | 37017.3 | 658.278 | 188.94 | 158.501 | 578.873 | 482.58 | 1457.47 | 1285.08 | 4203.45 | 3814.09 | 5755.9 |
| 4900.17 | 4348.69 | 12985.6 | 11836 | 39550.8 | 37110.7 | 658.567 | 187.524 | 159.146 | 566.058 | 484.334 | 1463.98 | 1289.39 | 4205.47 | 3824.99 | 5785.97 |
| 4826.1 | 4363.58 | 12944.4 | 11871.9 | 39340.6 | 37204.2 | 658.856 | 177.513 | 159.793 | 575.753 | 486.091 | 1459.78 | 1293.71 | 4232.65 | 3835.9 | 5837.88 |
| 4894.98 | 4378.49 | 13354.7 | 11907.8 | 39762.5 | 37297.8 | 659.144 | 173.647 | 160.441 | 565.682 | 487.853 | 1492.73 | 1298.04 | 4264.64 | 3846.83 | 5877.86 |
| 5001.76 | 4393.44 | 12940.5 | 11943.9 | 39196.7 | 37391.5 | 659.433 | 176.754 | 161.091 | 580.875 | 489.619 | 1462.63 | 1302.38 | 4216.52 | 3857.78 | 5759.77 |
| 4928.37 | 4408.43 | 13031.3 | 11979.9 | 39193.8 | 37485.4 | 659.722 | 191.178 | 161.744 | 565.265 | 491.389 | 1470.53 | 1306.73 | 4267.07 | 3868.75 | 5746.21 |
| 4912.5 | 4423.45 | 13247.3 | 12016.1 | 39686 | 37579.3 | 660.01 | 200.692 | 162.398 | 584.597 | 493.164 | 1492.37 | 1311.09 | 4174.15 | 3879.73 | 5790.94 |
| 4951.81 | 4438.5 | 13039.3 | 12052.3 | 39611.3 | 37673.4 | 660.299 | 182.936 | 163.054 | 594.45 | 494.943 | 1460.08 | 1315.45 | 4292.8 | 3890.74 | 5834.54 |
| 5002.43 | 4453.58 | 13323.9 | 12088.6 | 39919.9 | 37767.6 | 660.588 | 180.069 | 163.712 | 574.477 | 496.726 | 1534.57 | 1319.83 | 4316.69 | 3901.76 | 5874.26 |
| 5050.93 | 4468.7 | 13266.5 | 12124.9 | 39985.1 | 37861.9 | 660.876 | 188.371 | 164.372 | 562.867 | 498.514 | 1494.74 | 1324.22 | 4387.76 | 3912.79 | 5893.36 |
| 5031.63 | 4483.85 | 13099.6 | 12161.4 | 39649.4 | 37956.4 | 661.165 | 175.763 | 165.034 | 583.284 | 500.306 | 1524.85 | 1328.61 | 4275.01 | 3923.85 | 5850.34 |
| 5075.19 | 4499.04 | 13556.3 | 12197.8 | 40613.5 | 38051 | 661.453 | 189.666 | 165.697 | 598.223 | 502.102 | 1538.57 | 1333.02 | 4395.64 | 3934.92 | 6019.6 |
| 5035.52 | 4514.26 | 13314.2 | 12234.4 | 40020.5 | 38145.6 | 661.742 | 190.097 | 166.363 | 590.65 | 503.903 | 1555.58 | 1337.43 | 4320.65 | 3946.01 | 5917.72 |
| 5155.77 | 4529.51 | 13535.4 | 12271 | 40374.3 | 38240.4 | 662.03 | 204.433 | 167.03 | 602.459 | 505.708 | 1508.36 | 1341.85 | 4390.88 | 3957.12 | 5996.7 |
| 5112.14 | 4544.8 | 13720.6 | 12307.7 | 40552.6 | 38335.4 | 662.319 | 194.406 | 167.7 | 596.21 | 507.517 | 1499.12 | 1346.29 | 4393.06 | 3968.24 | 6019.59 |
| 5148.94 | 4560.12 | 13691.4 | 12344.4 | 40902.7 | 38430.4 | 662.607 | 197.977 | 168.371 | 591.563 | 509.331 | 1528.89 | 1350.73 | 4343.55 | 3979.39 | 5976.9 |
| 5200.07 | 4575.48 | 13729.6 | 12381.2 | 41138.2 | 38525.6 | 662.896 | 205.605 | 169.044 | 598.906 | 511.149 | 1553.41 | 1355.18 | 4397.81 | 3990.55 | 6104.85 |
| 5225.73 | 4590.86 | 13741.9 | 12418.1 | 41410.6 | 38620.8 | 663.184 | 189.504 | 169.719 | 612.577 | 512.971 | 1576.11 | 1359.64 | 4409.44 | 4001.73 | 6129.47 |
| 5176.83 | 4606.29 | 13958.9 | 12455.1 | 41147.2 | 38716.2 | 663.473 | 203.607 | 170.396 | 619.068 | 514.797 | 1540.62 | 1364.11 | 4461.55 | 4012.92 | 6047.93 |
| 5191.91 | 4621.74 | 13819.1 | 12492.1 | 41153.6 | 38811.7 | 663.761 | 197.535 | 171.075 | 593.346 | 516.628 | 1574.33 | 1368.59 | 4435.56 | 4024.14 | 6047.89 |
| 5261.84 | 4637.23 | 13701.2 | 12529.2 | 41165.8 | 38907.4 | 664.05 | 197.246 | 171.756 | 612.418 | 518.464 | 1570.86 | 1373.08 | 4474.84 | 4035.37 | 6032.44 |
| 5189.91 | 4652.76 | 13967.7 | 12566.3 | 41233.4 | 39003.1 | 664.338 | 176.37 | 172.439 | 590.478 | 520.303 | 1534.69 | 1377.58 | 4413.62 | 4046.62 | 6110.99 |
| 5215.97 | 4668.32 | 13712.7 | 12603.5 | 41554.7 | 39099 | 664.627 | 183.689 | 173.124 | 615.439 | 522.147 | 1595.93 | 1382.08 | 4438.2 | 4057.89 | 6019.77 |
| 5173.64 | 4683.91 | 13843.2 | 12640.8 | 41281.7 | 39195 | 664.915 | 197.86 | 173.81 | 602.544 | 523.996 | 1582.3 | 1386.6 | 4439.88 | 4069.17 | 6080.1 |
| 5220.92 | 4699.53 | 14000.2 | 12678.1 | 41576.6 | 39291.1 | 665.203 | 207.509 | 174.499 | 619.784 | 525.848 | 1560.47 | 1391.13 | 4456.22 | 4080.47 | 6064.35 |
| 5248.09 | 4715.19 | 13908.5 | 12715.6 | 41693.1 | 39387.3 | 665.492 | 202.114 | 175.19 | 619.097 | 527.705 | 1568.35 | 1395.66 | 4490.91 | 4091.79 | 6078.57 |
| 5194.83 | 4730.89 | 13801.2 | 12753 | 41812.4 | 39483.6 | 665.78 | 204.058 | 175.882 | 619.168 | 529.567 | 1537.82 | 1400.21 | 4411.11 | 4103.13 | 6173.61 |
| 5224.24 | 4746.61 | 13817.5 | 12790.6 | 41913.7 | 39580.1 | 666.069 | 201.076 | 176.577 | 597.449 | 531.433 | 1577.03 | 1404.76 | 4474.27 | 4114.48 | 6098.42 |
| 5343.79 | 4762.38 | 14049.2 | 12828.2 | 42041.9 | 39676.7 | 666.357 | 206.27 | 177.273 | 606.199 | 533.303 | 1553.62 | 1409.33 | 4492.77 | 4125.85 | 6149.54 |
| 5389.76 | 4778.17 | 14129.1 | 12865.9 | 41927.4 | 39773.3 | 666.645 | 205.473 | 177.971 | 605.99 | 535.178 | 1606.47 | 1413.9 | 4490.88 | 4137.24 | 6163.32 |
| 5341.78 | 4794 | 14305 | 12903.6 | 42449.9 | 39870.1 | 666.934 | 210.518 | 178.672 | 606.797 | 537.056 | 1603.79 | 1418.48 | 4508.11 | 4148.65 | 6250.87 |
| 5340.63 | 4809.87 | 14050.8 | 12941.4 | 41509.9 | 39967.1 | 667.222 | 189.582 | 179.374 | 617.08 | 538.94 | 1603.72 | 1423.07 | 4482.71 | 4160.07 | 6121.27 |
| 5438.62 | 4825.77 | 14127.9 | 12979.3 | 41912.7 | 40064.1 | 667.51 | 191.067 | 180.078 | 648.797 | 540.827 | 1590.02 | 1427.67 | 4537.9 | 4171.51 | 6189.13 |
| 5277.89 | 4841.7 | 14186.6 | 13017.2 | 42314.5 | 40161.3 | 667.799 | 199.655 | 180.784 | 629.517 | 542.72 | 1621.54 | 1432.29 | 4612.8 | 4182.97 | 6238.81 |
| 5353.14 | 4857.66 | 14243.9 | 13055.2 | 42355.7 | 40258.5 | 668.087 | 212.273 | 181.492 | 617.288 | 544.616 | 1604.65 | 1436.9 | 4567.2 | 4194.45 | 6259.95 |
| 5457.32 | 4873.66 | 14289 | 13093.3 | 42901.4 | 40355.9 | 668.375 | 209.272 | 182.202 | 629.387 | 546.517 | 1640.37 | 1441.53 | 4610.77 | 4205.94 | 6326.4 |
| 5486.18 | 4889.7 | 14374.6 | 13131.4 | 42810.5 | 40453.4 | 668.663 | 202.195 | 182.914 | 630.524 | 548.423 | 1636.27 | 1446.17 | 4652.21 | 4217.45 | 6307.61 |
| 5474.63 | 4905.77 | 14306.9 | 13169.6 | 42785.7 | 40551.1 | 668.952 | 219.165 | 183.628 | 645.638 | 550.332 | 1623.02 | 1450.82 | 4616.9 | 4228.98 | 6348.8 |
| 5497.72 | 4921.87 | 14645.6 | 13207.9 | 43072.6 | 40648.8 | 669.24 | 222.534 | 184.344 | 633.745 | 552.246 | 1617.97 | 1455.48 | 4654.84 | 4240.53 | 6350.54 |
| 5535.49 | 4938.01 | 14499.3 | 13246.2 | 43238.9 | 40746.6 | 669.528 | 220.801 | 185.062 | 627.784 | 554.165 | 1617.94 | 1460.15 | 4644.09 | 4252.09 | 6326.24 |
| 5565.38 | 4954.18 | 14391 | 13284.6 | 43334.7 | 40844.6 | 669.816 | 222.686 | 185.782 | 654.782 | 556.088 | 1618.04 | 1464.82 | 4663.94 | 4263.68 | 6370.44 |
| 5498.67 | 4970.39 | 14388.2 | 13323.1 | 43101.8 | 40942.7 | 670.105 | 203.637 | 186.504 | 605.28 | 558.015 | 1633.4 | 1469.51 | 4620.45 | 4275.27 | 6274.16 |
| 5499.58 | 4986.63 | 14606.1 | 13361.6 | 43032 | 41040.9 | 670.393 | 202.258 | 187.228 | 651.598 | 559.947 | 1665.63 | 1474.21 | 4647.12 | 4286.89 | 6351.68 |
| 5557.34 | 5002.9 | 14655.1 | 13400.2 | 43502.6 | 41139.2 | 670.681 | 220.492 | 187.954 | 652.37 | 561.883 | 1645.43 | 1478.91 | 4667.15 | 4298.52 | 6356.13 |
| 5677.59 | 5019.21 | 14783.5 | 13438.9 | 44089.4 | 41237.6 | 670.969 | 219.776 | 188.682 | 641.198 | 563.824 | 1755.31 | 1483.62 | 4698.14 | 4310.18 | 6334.25 |
| 5558.23 | 5035.55 | 14703.4 | 13477.6 | 43597.9 | 41336.2 | 671.257 | 213.917 | 189.412 | 654.224 | 565.769 | 1645.17 | 1488.35 | 4671.64 | 4321.84 | 6412.9 |
| 5654.54 | 5051.93 | 14521.7 | 13516.5 | 43596.2 | 41434.8 | 671.545 | 216.784 | 190.144 | 639.45 | 567.718 | 1677.4 | 1493.08 | 4665.17 | 4333.53 | 6399.05 |
| 5768.39 | 5068.34 | 14876.7 | 13555.3 | 44340.5 | 41533.6 | 671.833 | 214.141 | 190.877 | 654.908 | 569.672 | 1684.86 | 1497.83 | 4742.07 | 4345.23 | 6487.07 |
| 5694.89 | 5084.79 | 14959.4 | 13594.2 | 44233.4 | 41632.5 | 672.122 | 212.327 | 191.613 | 648.511 | 571.631 | 1685.86 | 1502.58 | 4797.3 | 4356.95 | 6482.57 |
| 5678.78 | 5101.27 | 14891.2 | 13633.2 | 44392 | 41731.5 | 672.41 | 208.931 | 192.351 | 652.351 | 573.594 | 1672.11 | 1507.34 | 4777.15 | 4368.69 | 6503.99 |
| 5648.33 | 5117.78 | 14979.3 | 13672.3 | 44034.7 | 41830.7 | 672.698 | 220.976 | 193.091 | 664.444 | 575.561 | 1716.09 | 1512.11 | 4736.55 | 4380.44 | 6442.58 |
| 5761.36 | 5134.33 | 14950.6 | 13711.4 | 44340.4 | 41929.9 | 672.986 | 203.787 | 193.833 | 666.825 | 577.533 | 1737.34 | 1516.89 | 4800.9 | 4392.22 | 6560.41 |
| 5692.55 | 5150.91 | 15065.9 | 13750.6 | 44996.6 | 42029.2 | 673.274 | 210.513 | 194.577 | 667.512 | 579.509 | 1694.65 | 1521.68 | 4826.4 | 4404.01 | 6565.64 |
| 5737.12 | 5167.53 | 15028 | 13789.9 | 44953.8 | 42128.7 | 673.562 | 222.516 | 195.323 | 650.141 | 581.489 | 1706.23 | 1526.48 | 4829.22 | 4415.81 | 6610.24 |
| 5790.05 | 5184.19 | 15262.3 | 13829.2 | 45314.2 | 42228.3 | 673.85 | 226.475 | 196.071 | 667.587 | 583.474 | 1719.53 | 1531.29 | 4839.21 | 4427.64 | 6575.6 |
| 5659.49 | 5200.87 | 15029.2 | 13868.6 | 44677 | 42328 | 674.138 | 224.448 | 196.82 | 669.431 | 585.464 | 1709.13 | 1536.11 | 4854.62 | 4439.48 | 6541.76 |
| 5741.31 | 5217.59 | 15137 | 13908.1 | 44786.1 | 42427.8 | 674.426 | 228.048 | 197.572 | 690.62 | 587.458 | 1746.81 | 1540.94 | 4856.69 | 4451.34 | 6539.99 |
| 5762.4 | 5234.35 | 15068.7 | 13947.6 | 44977.3 | 42527.7 | 674.714 | 208.807 | 198.326 | 662.559 | 589.456 | 1701.5 | 1545.78 | 4792.73 | 4463.21 | 6569.07 |
| 5821.33 | 5251.14 | 15362.8 | 13987.2 | 45409.1 | 42627.7 | 675.002 | 219.135 | 199.082 | 681.262 | 591.459 | 1746.83 | 1550.62 | 4860.53 | 4475.11 | 6670.76 |
| 5714.36 | 5267.97 | 15190.2 | 14026.9 | 44615.1 | 42727.9 | 675.29 | 222.003 | 199.84 | 672.655 | 593.466 | 1685.92 | 1555.48 | 4838.11 | 4487.02 | 6540.12 |
| 5827.26 | 5284.82 | 15208.7 | 14066.6 | 44839.8 | 42828.1 | 675.578 | 218.057 | 200.6 | 684.259 | 595.478 | 1728.33 | 1560.34 | 4856.45 | 4498.94 | 6566.31 |
| 5937.79 | 5301.72 | 15288.9 | 14106.4 | 45283.3 | 42928.5 | 675.866 | 229.727 | 201.362 | 681.67 | 597.495 | 1712.9 | 1565.22 | 4872.64 | 4510.89 | 6662.44 |
| 5807.88 | 5318.65 | 15550.3 | 14146.2 | 45936.6 | 43029 | 676.154 | 231.886 | 202.126 | 693.9 | 599.515 | 1771.63 | 1570.11 | 4939.54 | 4522.85 | 6655.72 |
| 5894.69 | 5335.61 | 15567.5 | 14186.1 | 45665.2 | 43129.6 | 676.442 | 216.765 | 202.892 | 677.545 | 601.54 | 1776.01 | 1575 | 4892.52 | 4534.83 | 6674.03 |
| 5965.5 | 5352.61 | 15380.6 | 14226.1 | 45609.5 | 43230.3 | 676.73 | 227.666 | 203.661 | 684.405 | 603.57 | 1762.17 | 1579.9 | 4915.18 | 4546.83 | 6708.72 |
| 5876.06 | 5369.64 | 15678.7 | 14266.2 | 45996 | 43331.1 | 677.018 | 223.899 | 204.431 | 701.997 | 605.605 | 1728.72 | 1584.82 | 4981.6 | 4558.84 | 6689.56 |
| 5894.87 | 5386.71 | 15321.3 | 14306.3 | 45535.5 | 43432.1 | 677.305 | 237.639 | 205.203 | 681.775 | 607.643 | 1769.29 | 1589.74 | 4882.13 | 4570.87 | 6683.21 |
| 5956.39 | 5403.81 | 15558.4 | 14346.5 | 46053.1 | 43533.1 | 677.593 | 224.972 | 205.977 | 699.199 | 609.686 | 1758.86 | 1594.67 | 4906.41 | 4582.92 | 6743.24 |
| 5854.17 | 5420.95 | 15543 | 14386.7 | 46011.6 | 43634.3 | 677.881 | 235.585 | 206.753 | 691.931 | 611.734 | 1739.46 | 1599.61 | 4910.57 | 4594.99 | 6721.91 |
| 6032.31 | 5438.12 | 15586.3 | 14427 | 46264.6 | 43735.6 | 678.169 | 237.48 | 207.532 | 695.582 | 613.786 | 1809.16 | 1604.57 | 5008.53 | 4607.07 | 6717.82 |
| 6022.66 | 5455.32 | 15570.4 | 14467.4 | 46296.9 | 43836.9 | 678.457 | 228.354 | 208.312 | 698.48 | 615.843 | 1777.99 | 1609.53 | 4959.74 | 4619.17 | 6711.91 |
| 6109.43 | 5472.57 | 15701.4 | 14507.9 | 46190.5 | 43938.4 | 678.745 | 233.23 | 209.095 | 711.591 | 617.904 | 1766.49 | 1614.5 | 4995.39 | 4631.28 | 6745.75 |
| 6133.18 | 5489.84 | 15873.1 | 14548.4 | 46836.4 | 44040.1 | 679.032 | 230.317 | 209.879 | 711.03 | 619.97 | 1832.47 | 1619.48 | 5127.5 | 4643.42 | 6849.2 |
| 6010.93 | 5507.15 | 15593.6 | 14589 | 46206.1 | 44141.8 | 679.32 | 235.141 | 210.666 | 713.246 | 622.04 | 1818.64 | 1624.47 | 4999.3 | 4655.57 | 6753.44 |
| 6161.95 | 5524.5 | 15768.4 | 14629.6 | 46541.1 | 44243.6 | 679.608 | 234.316 | 211.454 | 695.81 | 624.114 | 1818.64 | 1629.47 | 5110.65 | 4667.74 | 6817.65 |
| 6133 | 5541.88 | 15921.6 | 14670.3 | 46759.4 | 44345.6 | 679.896 | 225.167 | 212.245 | 694.693 | 626.194 | 1769.32 | 1634.47 | 5059.31 | 4679.92 | 6859.19 |
| 6270.41 | 5559.29 | 16000.1 | 14711.1 | 47261.4 | 44447.6 | 680.183 | 233.895 | 213.038 | 702.217 | 628.277 | 1819.09 | 1639.49 | 5071.63 | 4692.12 | 6855.67 |
| 6164.5 | 5576.74 | 16202 | 14752 | 47308.2 | 44549.8 | 680.471 | 239.06 | 213.832 | 716.336 | 630.365 | 1854.21 | 1644.52 | 5130.63 | 4704.34 | 6874.64 |
| 6145.33 | 5594.22 | 15958.1 | 14792.9 | 46752.2 | 44652.1 | 680.759 | 234.396 | 214.629 | 710.987 | 632.458 | 1822.49 | 1649.56 | 5056.52 | 4716.58 | 6943.13 |
| 6238.56 | 5611.74 | 15884.3 | 14833.9 | 46609.3 | 44754.4 | 681.047 | 244.695 | 215.428 | 727.797 | 634.555 | 1817.92 | 1654.6 | 5050.15 | 4728.83 | 6885.99 |
| 6170.62 | 5629.3 | 15943.7 | 14874.9 | 47040.5 | 44856.9 | 681.334 | 237.182 | 216.229 | 727.546 | 636.657 | 1807.16 | 1659.66 | 5109.45 | 4741.1 | 6873.46 |
| 6200.68 | 5646.88 | 16195.2 | 14916 | 47256.8 | 44959.6 | 681.622 | 240.062 | 217.032 | 730.418 | 638.763 | 1821.04 | 1664.72 | 5141.35 | 4753.39 | 7010.27 |
| 6323.69 | 5664.51 | 15983.4 | 14957.2 | 47433 | 45062.3 | 681.91 | 256.361 | 217.837 | 732.556 | 640.874 | 1865.05 | 1669.8 | 5123.07 | 4765.69 | 6794.4 |
| 6146.76 | 5682.16 | 15922.4 | 14998.4 | 47355.2 | 45165.1 | 682.197 | 190.429 | 218.645 | 733.954 | 642.989 | 1836.93 | 1674.88 | 5043.17 | 4778.01 | 6812.69 |
| 6189.11 | 5699.86 | 15472.5 | 15039.7 | 47546 | 45268.1 | 682.485 | 186.656 | 219.454 | 736.218 | 645.109 | 1877.32 | 1679.98 | 5094.67 | 4790.35 | 6828.12 |
| 6189.28 | 5717.59 | 15895.7 | 15081.1 | 47921.1 | 45371.1 | 682.773 | 211.197 | 220.265 | 758.695 | 647.234 | 1907.12 | 1685.08 | 5101.02 | 4802.71 | 6927.14 |
| 6316.12 | 5735.35 | 16333.9 | 15122.5 | 48331.3 | 45474.3 | 683.06 | 207.702 | 221.079 | 776.317 | 649.362 | 1883.72 | 1690.2 | 5211.78 | 4815.08 | 6944.02 |
| 6354.17 | 5753.15 | 16449.4 | 15164 | 48278.2 | 45577.5 | 683.348 | 225.411 | 221.894 | 758.419 | 651.496 | 1858.64 | 1695.32 | 5216.91 | 4827.47 | 6980.93 |
| 6304.1 | 5770.98 | 16390 | 15205.6 | 48612.4 | 45680.9 | 683.636 | 213.941 | 222.712 | 750.476 | 653.634 | 1902.6 | 1700.45 | 5168.22 | 4839.87 | 7128.59 |
| 6307.21 | 5788.85 | 16556.3 | 15247.2 | 48383 | 45784.4 | 683.923 | 236.762 | 223.532 | 736.588 | 655.777 | 1912.62 | 1705.59 | 5292.41 | 4852.3 | 7104.87 |
| 6449.09 | 5806.75 | 16754.5 | 15288.9 | 48880.7 | 45888 | 684.211 | 250.727 | 224.354 | 740.387 | 657.923 | 1929.93 | 1710.74 | 5371.03 | 4864.74 | 7244.15 |
| 6484.39 | 5824.69 | 16586.1 | 15330.7 | 48968.3 | 45991.7 | 684.498 | 242.842 | 225.178 | 737.678 | 660.075 | 1905.69 | 1715.91 | 5226.31 | 4877.19 | 7286.43 |
| 6450.03 | 5842.66 | 16575.6 | 15372.5 | 48850.4 | 46095.5 | 684.786 | 257.729 | 226.004 | 745.537 | 662.231 | 1932.99 | 1721.08 | 5304.46 | 4889.67 | 7052.33 |
| 6553.51 | 5860.66 | 16664.3 | 15414.4 | 49349 | 46199.5 | 685.073 | 239.165 | 226.832 | 748.455 | 664.392 | 1965.09 | 1726.26 | 5345.72 | 4902.16 | 7270.02 |
| 6464.48 | 5878.71 | 16446.5 | 15456.4 | 49094.8 | 46303.5 | 685.361 | 246.188 | 227.662 | 757.355 | 666.557 | 1945.79 | 1731.45 | 5320.47 | 4914.66 | 7201.39 |
| 6507.71 | 5896.78 | 16734.2 | 15498.4 | 49346 | 46407.7 | 685.649 | 255.818 | 228.495 | 744.935 | 668.727 | 1929.71 | 1736.64 | 5370.2 | 4927.19 | 7281.64 |
| 6653.03 | 5914.9 | 16900 | 15540.5 | 49334.6 | 46511.9 | 685.936 | 226.287 | 229.329 | 751.414 | 670.901 | 1941.36 | 1741.85 | 5438.16 | 4939.73 | 7257.02 |
| 6682.37 | 5933.04 | 17108.9 | 15582.6 | 49461.7 | 46616.3 | 686.223 | 232.798 | 230.166 | 751.249 | 673.08 | 1948.18 | 1747.07 | 5397.13 | 4952.29 | 7284.22 |
| 6586.48 | 5951.23 | 16803.3 | 15624.9 | 49527.2 | 46720.7 | 686.511 | 232.959 | 231.004 | 755.569 | 675.264 | 1983.5 | 1752.3 | 5448.85 | 4964.86 | 7366.39 |
| 6579.96 | 5969.44 | 16874.3 | 15667.2 | 49827.4 | 46825.3 | 686.798 | 228.637 | 231.845 | 779.377 | 677.452 | 1968.04 | 1757.54 | 5441.05 | 4977.45 | 7355.66 |
| 6645.31 | 5987.69 | 17311.6 | 15709.5 | 50608.1 | 46930 | 687.086 | 263.768 | 232.688 | 754.7 | 679.644 | 1976.74 | 1762.78 | 5463.91 | 4990.06 | 7421.53 |
| 6764.87 | 6005.98 | 17203.8 | 15751.9 | 50482 | 47034.8 | 687.373 | 253.261 | 233.533 | 775.186 | 681.841 | 1968.43 | 1768.04 | 5436.31 | 5002.69 | 7327.49 |
| 6723.97 | 6024.31 | 17190.1 | 15794.4 | 50428.8 | 47139.7 | 687.661 | 254.47 | 234.38 | 755.331 | 684.043 | 2007.54 | 1773.31 | 5438.83 | 5015.33 | 7430.02 |
| 6753.69 | 6042.66 | 17102.6 | 15837 | 50848.4 | 47244.7 | 687.948 | 240.732 | 235.23 | 777.865 | 686.249 | 2060.93 | 1778.58 | 5460.2 | 5027.98 | 7421.23 |
| 6769.55 | 6061.06 | 17289.7 | 15879.6 | 50383.5 | 47349.8 | 688.236 | 215.898 | 236.081 | 764.195 | 688.46 | 2006.16 | 1783.87 | 5377.83 | 5040.66 | 7431.56 |
| 6723.48 | 6079.48 | 17144.2 | 15922.3 | 50499.5 | 47455 | 688.523 | 251.729 | 236.935 | 754.267 | 690.676 | 2049.19 | 1789.16 | 5441.99 | 5053.35 | 7324.65 |
| 6875.24 | 6097.94 | 17423.5 | 15965 | 50818 | 47560.4 | 688.81 | 227.429 | 237.79 | 764.252 | 692.896 | 2028.3 | 1794.46 | 5520.2 | 5066.06 | 7434.27 |
| 6829.31 | 6116.44 | 17509.2 | 16007.8 | 51553.9 | 47665.8 | 689.098 | 255.158 | 238.648 | 801.089 | 695.12 | 2100.53 | 1799.78 | 5619.46 | 5078.79 | 7469.53 |
| 6922.36 | 6134.98 | 17484.9 | 16050.7 | 51062.1 | 47771.3 | 689.385 | 246.932 | 239.508 | 790.545 | 697.349 | 2037.14 | 1805.1 | 5572.15 | 5091.53 | 7509.25 |
| 6826.67 | 6153.54 | 17589.4 | 16093.6 | 50756.8 | 47877 | 689.672 | 248.963 | 240.37 | 785.279 | 699.583 | 2005.99 | 1810.43 | 5482.49 | 5104.29 | 7549.05 |
| 6939.77 | 6172.15 | 17639.5 | 16136.6 | 51472.6 | 47982.8 | 689.96 | 247.937 | 241.234 | 790.981 | 701.822 | 2059.92 | 1815.78 | 5565.75 | 5117.06 | 7552.44 |
| 6986.31 | 6190.79 | 17650.9 | 16179.7 | 51662.5 | 48088.6 | 690.247 | 255.857 | 242.101 | 806.548 | 704.064 | 2044.04 | 1821.13 | 5571.95 | 5129.85 | 7495 |
| 6888.58 | 6209.46 | 17563.4 | 16222.9 | 51605.9 | 48194.6 | 690.534 | 226.335 | 242.969 | 810.62 | 706.312 | 2044.12 | 1826.49 | 5556.81 | 5142.66 | 7579.52 |
| 6936.02 | 6228.17 | 17554.9 | 16266.1 | 51375.9 | 48300.7 | 690.822 | 230.424 | 243.84 | 796.031 | 708.564 | 2058.95 | 1831.86 | 5678.01 | 5155.49 | 7545.73 |
| 6977.06 | 6246.91 | 17915.3 | 16309.3 | 52071 | 48406.8 | 691.109 | 248.675 | 244.713 | 798.061 | 710.821 | 2108.97 | 1837.24 | 5650.63 | 5168.33 | 7602.56 |
| 7027.42 | 6265.69 | 17797 | 16352.7 | 51951 | 48513.1 | 691.396 | 252.216 | 245.588 | 795.904 | 713.082 | 2111.55 | 1842.63 | 5663.88 | 5181.19 | 7627.32 |
| 6958.49 | 6284.51 | 17642.2 | 16396.1 | 51367.2 | 48619.5 | 691.683 | 245.573 | 246.465 | 775.901 | 715.348 | 2055.7 | 1848.03 | 5637.5 | 5194.06 | 7662.04 |
| 7076.87 | 6303.35 | 18040 | 16439.5 | 52319.1 | 48726 | 691.971 | 228.833 | 247.344 | 802.404 | 717.618 | 2094.33 | 1853.44 | 5649.88 | 5206.95 | 7714.55 |
| 7161.33 | 6322.24 | 17872 | 16483 | 52367.1 | 48832.6 | 692.258 | 253.795 | 248.225 | 783.909 | 719.893 | 2109.03 | 1858.86 | 5658.49 | 5219.86 | 7686.53 |
| 7171.65 | 6341.16 | 18014.8 | 16526.6 | 52730.6 | 48939.3 | 692.545 | 256.898 | 249.109 | 811.653 | 722.173 | 2141.03 | 1864.29 | 5733.38 | 5232.78 | 7684.41 |
| 7122.7 | 6360.11 | 17887.6 | 16570.3 | 52384.6 | 49046.2 | 692.832 | 236.506 | 249.995 | 823.2 | 724.457 | 2101.35 | 1869.72 | 5690.32 | 5245.73 | 7789.65 |
| 7150.42 | 6379.1 | 18093.7 | 16614 | 52238.7 | 49153.1 | 693.119 | 257.561 | 250.883 | 801.824 | 726.746 | 2092.48 | 1875.17 | 5626.03 | 5258.68 | 7729.25 |
| 7117.43 | 6398.13 | 18059.7 | 16657.8 | 52723.7 | 49260.1 | 693.407 | 257.351 | 251.773 | 809.179 | 729.039 | 2119.88 | 1880.63 | 5718.21 | 5271.66 | 7785.37 |
| 7252.54 | 6417.19 | 18137.1 | 16701.7 | 52792.8 | 49367.2 | 693.694 | 236.171 | 252.665 | 848.267 | 731.337 | 2133.11 | 1886.09 | 5755.72 | 5284.65 | 7767.2 |
| 7141.5 | 6436.28 | 17899.9 | 16745.6 | 52287.9 | 49474.5 | 693.981 | 254.262 | 253.559 | 832.616 | 733.64 | 2153.09 | 1891.57 | 5777.03 | 5297.65 | 7700.94 |
| 7166.75 | 6455.42 | 18156.6 | 16789.6 | 53091.1 | 49581.8 | 694.268 | 245.322 | 254.456 | 840.02 | 735.947 | 2115.5 | 1897.06 | 5697.7 | 5310.68 | 7785.28 |
| 7195.92 | 6474.58 | 18156.4 | 16833.6 | 52797.2 | 49689.2 | 694.555 | 260.839 | 255.355 | 805.406 | 738.259 | 2127.28 | 1902.55 | 5781.68 | 5323.72 | 7773.36 |
| 7176.75 | 6493.78 | 18081.9 | 16877.7 | 52885.2 | 49796.8 | 694.842 | 237.704 | 256.255 | 816.538 | 740.576 | 2158.63 | 1908.06 | 5752.44 | 5336.77 | 7739.53 |
| 7222.84 | 6513.02 | 18158.6 | 16921.9 | 52938.1 | 49904.4 | 695.129 | 239.6 | 257.159 | 820.579 | 742.897 | 2129.95 | 1913.57 | 5746.83 | 5349.85 | 7773.8 |
| 7249.23 | 6532.29 | 18250.2 | 16966.2 | 52981.9 | 50012.2 | 695.416 | 244.098 | 258.064 | 838.767 | 745.222 | 2153.26 | 1919.09 | 5801.15 | 5362.94 | 7808.2 |
| 7229.86 | 6551.6 | 18266.2 | 17010.5 | 53078.3 | 50120.1 | 695.703 | 279.007 | 258.971 | 847.189 | 747.553 | 2165.34 | 1924.63 | 5788.9 | 5376.04 | 7752.45 |
| 7304.2 | 6570.94 | 18595.9 | 17054.8 | 53782.9 | 50228 | 695.99 | 248.636 | 259.881 | 829.259 | 749.888 | 2179.47 | 1930.17 | 5878.24 | 5389.16 | 7865.68 |
| 7365.59 | 6590.32 | 18501.6 | 17099.3 | 54174.3 | 50336.1 | 696.277 | 237.027 | 260.793 | 831.685 | 752.227 | 2170.42 | 1935.72 | 5872.45 | 5402.3 | 7973.67 |
| 7327.87 | 6609.73 | 18498.3 | 17143.8 | 53527.3 | 50444.2 | 696.564 | 260.57 | 261.707 | 878.651 | 754.571 | 2178.84 | 1941.28 | 5920.57 | 5415.46 | 7908.94 |
| 7430.12 | 6629.18 | 18555.7 | 17188.3 | 53857.3 | 50552.5 | 696.851 | 267.234 | 262.623 | 886.542 | 756.92 | 2188.8 | 1946.86 | 5845.43 | 5428.63 | 8023.3 |
| 7425.44 | 6648.67 | 18526.3 | 17233 | 53877.1 | 50660.9 | 697.138 | 270.389 | 263.541 | 861.363 | 759.274 | 2178.71 | 1952.44 | 5905.02 | 5441.81 | 7949.22 |
| 7546.5 | 6668.19 | 18536.8 | 17277.7 | 54147.3 | 50769.4 | 697.425 | 275.024 | 264.462 | 843.646 | 761.631 | 2239.93 | 1958.03 | 5872.91 | 5455.02 | 7969.28 |
| 7411.52 | 6687.74 | 18963.7 | 17322.4 | 54265.9 | 50877.9 | 697.712 | 246.453 | 265.385 | 852.374 | 763.994 | 2237.45 | 1963.63 | 5967.99 | 5468.24 | 8062.55 |
| 7437.33 | 6707.33 | 18609.7 | 17367.3 | 54197.6 | 50986.6 | 697.999 | 240.445 | 266.31 | 845.571 | 766.361 | 2201.56 | 1969.24 | 5909.19 | 5481.47 | 7939.63 |
| 7399.86 | 6726.95 | 18810.8 | 17412.2 | 54271.9 | 51095.4 | 698.286 | 271.412 | 267.237 | 841.921 | 768.733 | 2209.14 | 1974.86 | 5911 | 5494.73 | 8003.5 |
| 7487.36 | 6746.61 | 19010.8 | 17457.1 | 54523.4 | 51204.3 | 698.573 | 277.675 | 268.166 | 844.525 | 771.11 | 2171.99 | 1980.48 | 5959.86 | 5507.99 | 8030.28 |
| 7617.08 | 6766.31 | 18842.2 | 17502.1 | 54559.2 | 51313.3 | 698.86 | 270.315 | 269.098 | 841.873 | 773.491 | 2185.44 | 1986.12 | 5996.16 | 5521.28 | 8057.7 |
| 7512.31 | 6786.04 | 18768.9 | 17547.2 | 54279.4 | 51422.3 | 699.147 | 285.597 | 270.031 | 867.138 | 775.877 | 2189.21 | 1991.77 | 5930.46 | 5534.58 | 8076.89 |
| 7511.77 | 6805.81 | 19047 | 17592.3 | 54635.1 | 51531.5 | 699.434 | 257.252 | 270.967 | 867.774 | 778.267 | 2280.22 | 1997.43 | 6051.5 | 5547.9 | 8140.62 |
| 7661.26 | 6825.61 | 19292.5 | 17637.6 | 55335.8 | 51640.8 | 699.721 | 275.594 | 271.905 | 896.988 | 780.662 | 2226.8 | 2003.1 | 6023.89 | 5561.23 | 8059.51 |
| 7670.28 | 6845.45 | 19163.9 | 17682.8 | 55499.8 | 51750.2 | 700.008 | 271.142 | 272.846 | 871.895 | 783.062 | 2260.96 | 2008.77 | 6005.31 | 5574.58 | 8135.3 |
| 7683.01 | 6865.32 | 19391.9 | 17728.2 | 55527 | 51859.7 | 700.294 | 273.344 | 273.788 | 873.633 | 785.466 | 2263.49 | 2014.46 | 6090.48 | 5587.94 | 8224.95 |
| 7621.73 | 6885.23 | 19388.5 | 17773.6 | 55701.9 | 51969.3 | 700.581 | 286.403 | 274.733 | 895.504 | 787.875 | 2292.12 | 2020.15 | 6083.02 | 5601.32 | 8210.79 |
| 7835.05 | 6905.17 | 19526.1 | 17819.1 | 55789.2 | 52079 | 700.868 | 279.259 | 275.68 | 900.466 | 790.289 | 2335.84 | 2025.86 | 6233.17 | 5614.72 | 8306.66 |
| 7670.11 | 6925.15 | 19360.7 | 17864.6 | 55844.3 | 52188.8 | 701.155 | 264.546 | 276.629 | 886.988 | 792.707 | 2247.82 | 2031.57 | 6128.54 | 5628.14 | 8293.5 |
| 7711.93 | 6945.16 | 19418.3 | 17910.2 | 55926 | 52298.7 | 701.442 | 269.647 | 277.581 | 879.104 | 795.13 | 2293.97 | 2037.3 | 6185.03 | 5641.57 | 8322.28 |
| 7812.2 | 6965.21 | 19709.1 | 17955.9 | 55994 | 52408.7 | 701.729 | 281.203 | 278.534 | 890.167 | 797.558 | 2342.32 | 2043.03 | 6149.14 | 5655.01 | 8392.9 |
| 7765.93 | 6985.3 | 19555.2 | 18001.6 | 56611 | 52518.8 | 702.015 | 284.995 | 279.49 | 901.681 | 799.99 | 2304.84 | 2048.77 | 6076.82 | 5668.48 | 8369.28 |
| 7849.38 | 7005.42 | 19692 | 18047.4 | 56579 | 52629 | 702.302 | 318.039 | 280.449 | 920.32 | 802.427 | 2325.47 | 2054.53 | 6201.28 | 5681.95 | 8366.96 |
| 7914.21 | 7025.57 | 19659.5 | 18093.2 | 56516.1 | 52739.3 | 702.589 | 314.733 | 281.409 | 891.87 | 804.868 | 2310.44 | 2060.29 | 6226.2 | 5695.45 | 8437.15 |
| 7799.34 | 7045.77 | 19611.2 | 18139.2 | 56292.7 | 52849.7 | 702.876 | 308.955 | 282.371 | 880.225 | 807.314 | 2267.61 | 2066.06 | 6138.81 | 5708.96 | 8327.64 |
| 7934.19 | 7065.99 | 20022.1 | 18185.1 | 56669.2 | 52960.2 | 703.162 | 301.782 | 283.336 | 903.564 | 809.765 | 2271.6 | 2071.84 | 6292.6 | 5722.48 | 8449.29 |
| 7980.1 | 7086.25 | 19943.9 | 18231.2 | 57091.3 | 53070.8 | 703.449 | 290.875 | 284.303 | 901.721 | 812.221 | 2312.32 | 2077.63 | 6304.83 | 5736.02 | 8507.22 |
| 7940.4 | 7106.55 | 19930.9 | 18277.3 | 57019.5 | 53181.5 | 703.736 | 301.55 | 285.272 | 934.084 | 814.681 | 2271.73 | 2083.43 | 6259.58 | 5749.58 | 8375.27 |
| 7874.64 | 7126.88 | 19878.8 | 18323.5 | 56848.3 | 53292.3 | 704.022 | 293.696 | 286.244 | 899.301 | 817.146 | 2321.14 | 2089.24 | 6201.67 | 5763.16 | 8332.36 |
| 7945.07 | 7147.25 | 19961.1 | 18369.7 | 57564.9 | 53403.2 | 704.309 | 295.804 | 287.218 | 957.386 | 819.615 | 2299.91 | 2095.06 | 6315.48 | 5776.75 | 8492.5 |
| 8060.67 | 7167.65 | 20081 | 18416 | 57737.1 | 53514.2 | 704.596 | 303.931 | 288.194 | 914.678 | 822.089 | 2353.85 | 2100.89 | 6278.06 | 5790.35 | 8458.97 |
| 8079.47 | 7188.09 | 20212.5 | 18462.4 | 57673.7 | 53625.2 | 704.882 | 299.383 | 289.172 | 956.66 | 824.568 | 2341.86 | 2106.73 | 6275.35 | 5803.98 | 8424.67 |
| 7966.85 | 7208.57 | 20297.2 | 18508.8 | 57757.9 | 53736.4 | 705.169 | 294.462 | 290.152 | 946.471 | 827.051 | 2408.68 | 2112.58 | 6314.04 | 5817.61 | 8554.19 |
| 7936.98 | 7229.08 | 20297.8 | 18555.3 | 57666.2 | 53847.7 | 705.456 | 279.641 | 291.135 | 914.6 | 829.539 | 2338.95 | 2118.43 | 6304.45 | 5831.27 | 8587.22 |
| 8105.73 | 7249.62 | 20349.5 | 18601.9 | 58094.4 | 53959.1 | 705.742 | 316.661 | 292.12 | 983.688 | 832.032 | 2345.08 | 2124.3 | 6246.09 | 5844.94 | 8521.82 |
| 7955.35 | 7270.2 | 20263 | 18648.5 | 58196.2 | 54070.6 | 706.029 | 324.759 | 293.107 | 924.136 | 834.529 | 2340.98 | 2130.18 | 6399.3 | 5858.62 | 8541.72 |
| 8071.97 | 7290.82 | 20386.8 | 18695.2 | 58204.4 | 54182.2 | 706.315 | 312.842 | 294.097 | 934.243 | 837.031 | 2389.66 | 2136.06 | 6455.72 | 5872.32 | 8609.06 |
| 8152.2 | 7311.47 | 20454.4 | 18741.9 | 58076.1 | 54293.8 | 706.602 | 316.339 | 295.088 | 917.3 | 839.538 | 2358.08 | 2141.96 | 6293.78 | 5886.04 | 8688.28 |
| 8213.13 | 7332.16 | 20654.5 | 18788.7 | 58527.9 | 54405.6 | 706.888 | 321.526 | 296.082 | 954.443 | 842.049 | 2449.33 | 2147.86 | 6454.06 | 5899.77 | 8626.69 |
| 8025.72 | 7352.88 | 20434.9 | 18835.6 | 58835.4 | 54517.5 | 707.175 | 319.065 | 297.078 | 946.208 | 844.565 | 2417.02 | 2153.78 | 6413.28 | 5913.52 | 8672.45 |
| 8173.66 | 7373.64 | 20640.6 | 18882.6 | 58996.5 | 54629.5 | 707.462 | 327.584 | 298.077 | 970.071 | 847.086 | 2448.28 | 2159.7 | 6428.08 | 5927.28 | 8702.85 |
| 8197.84 | 7394.43 | 20646.2 | 18929.6 | 58831.5 | 54741.5 | 707.748 | 319.343 | 299.078 | 976.461 | 849.611 | 2441.67 | 2165.64 | 6437.36 | 5941.06 | 8693.44 |
| 8174.52 | 7415.26 | 20757.5 | 18976.6 | 58844.8 | 54853.7 | 708.034 | 312.98 | 300.081 | 954.942 | 852.141 | 2390.82 | 2171.58 | 6510.53 | 5954.86 | 8752.07 |
| 8048.42 | 7436.12 | 20305.9 | 19023.8 | 58607.6 | 54966 | 708.321 | 302.614 | 301.086 | 969.948 | 854.676 | 2437.85 | 2177.53 | 6404.15 | 5968.67 | 8644.21 |
| 8220.59 | 7457.02 | 20958 | 19071 | 59586.4 | 55078.3 | 708.607 | 324.758 | 302.093 | 997.297 | 857.215 | 2448.14 | 2183.49 | 6522.17 | 5982.5 | 8705.24 |
| 8284.63 | 7477.96 | 20768.3 | 19118.2 | 58897.1 | 55190.8 | 708.894 | 320.571 | 303.103 | 961.225 | 859.759 | 2446.75 | 2189.47 | 6491.63 | 5996.34 | 8665 |
| 8273.07 | 7498.93 | 20672.9 | 19165.5 | 58906.9 | 55303.3 | 709.18 | 342.53 | 304.115 | 992.476 | 862.308 | 2379.33 | 2195.45 | 6442.32 | 6010.2 | 8657.19 |
| 8309.65 | 7519.93 | 21027.6 | 19212.9 | 59363.3 | 55416 | 709.467 | 326.906 | 305.129 | 965.025 | 864.861 | 2441.69 | 2201.44 | 6566.5 | 6024.07 | 8846.08 |
| 8355.31 | 7540.97 | 21006.6 | 19260.4 | 59256.5 | 55528.7 | 709.753 | 339.87 | 306.146 | 986.803 | 867.419 | 2440.52 | 2207.44 | 6498.71 | 6037.96 | 8737.41 |
| 8431.91 | 7562.05 | 21054.5 | 19307.9 | 59955 | 55641.6 | 710.04 | 320.088 | 307.165 | 965.839 | 869.982 | 2509.06 | 2213.45 | 6614.81 | 6051.87 | 8877.2 |
| 8370.09 | 7583.16 | 20874 | 19355.5 | 59629.7 | 55754.5 | 710.326 | 337.28 | 308.186 | 988.904 | 872.549 | 2470.14 | 2219.47 | 6634.97 | 6065.79 | 8933.36 |
| 8366.09 | 7604.31 | 21241 | 19403.1 | 60313.2 | 55867.5 | 710.612 | 345.682 | 309.209 | 983.116 | 875.121 | 2516.28 | 2225.49 | 6646.83 | 6079.72 | 8896.24 |
| 8396.58 | 7625.49 | 21094.5 | 19450.8 | 59258.5 | 55980.7 | 710.899 | 327.87 | 310.235 | 973.215 | 877.698 | 2468.54 | 2231.53 | 6540.52 | 6093.68 | 8744.61 |
| 8521.89 | 7646.71 | 21250.2 | 19498.6 | 60485.8 | 56093.9 | 711.185 | 331.913 | 311.263 | 977.644 | 880.279 | 2494.56 | 2237.58 | 6583.07 | 6107.64 | 8913.54 |
| 8507.84 | 7667.96 | 21269.7 | 19546.4 | 60433.8 | 56207.2 | 711.471 | 326.166 | 312.293 | 979.475 | 882.865 | 2491.44 | 2243.64 | 6626.97 | 6121.63 | 8959.48 |
| 8514.62 | 7689.25 | 21265.2 | 19594.3 | 60485.8 | 56320.6 | 711.758 | 327.765 | 313.326 | 1016.45 | 885.456 | 2483.63 | 2249.7 | 6593.74 | 6135.62 | 8796.1 |
| 8674.24 | 7710.58 | 21574.5 | 19642.2 | 60781.8 | 56434.1 | 712.044 | 334.298 | 314.36 | 981.452 | 888.051 | 2488.01 | 2255.78 | 6701.29 | 6149.64 | 9082.67 |
| 8574.87 | 7731.94 | 21736.7 | 19690.3 | 61040 | 56547.7 | 712.33 | 327.776 | 315.397 | 996.481 | 890.651 | 2515.73 | 2261.86 | 6671.26 | 6163.67 | 8970.38 |
| 8730.82 | 7753.33 | 21839.8 | 19738.3 | 61538.3 | 56661.4 | 712.616 | 336.564 | 316.437 | 980.377 | 893.256 | 2516.16 | 2267.96 | 6775.31 | 6177.71 | 9103.05 |
| 8593.49 | 7774.77 | 21577.1 | 19786.5 | 61405.7 | 56775.2 | 712.903 | 351.186 | 317.478 | 1004.21 | 895.865 | 2546.58 | 2274.07 | 6795.09 | 6191.77 | 9015.37 |
| 8538.8 | 7796.23 | 21590.3 | 19834.7 | 61115.6 | 56889.1 | 713.189 | 346.135 | 318.522 | 1028.09 | 898.479 | 2510.22 | 2280.18 | 6705.31 | 6205.85 | 9065.69 |
| 8742.88 | 7817.73 | 21768.1 | 19883 | 61316.2 | 57003 | 713.475 | 354.504 | 319.569 | 1008.21 | 901.098 | 2536.26 | 2286.3 | 6702.61 | 6219.94 | 9105.53 |
| 8662.46 | 7839.27 | 21923.1 | 19931.3 | 61519.1 | 57117.1 | 713.761 | 352.585 | 320.617 | 1000.37 | 903.722 | 2554.8 | 2292.43 | 6741.86 | 6234.05 | 9160.09 |
| 8697.65 | 7860.84 | 21535.9 | 19979.7 | 60716.9 | 57231.3 | 714.048 | 323.402 | 321.668 | 1013.65 | 906.35 | 2486.31 | 2298.58 | 6669.19 | 6248.17 | 8981.28 |
| 8681.11 | 7882.45 | 21833.6 | 20028.1 | 61575.5 | 57345.5 | 714.334 | 338.45 | 322.721 | 1014.64 | 908.983 | 2575.99 | 2304.73 | 6707.83 | 6262.3 | 9084.32 |
| 8654.92 | 7904.09 | 21907.9 | 20076.7 | 61348.9 | 57459.9 | 714.62 | 361.615 | 323.776 | 1005.57 | 911.62 | 2539.15 | 2310.89 | 6730.27 | 6276.46 | 9083.4 |
| 8725.56 | 7925.77 | 21782.1 | 20125.2 | 61689.8 | 57574.3 | 714.906 | 356.447 | 324.834 | 1031.6 | 914.262 | 2566.12 | 2317.06 | 6838.79 | 6290.62 | 9088.76 |
| 8765.25 | 7947.49 | 21829.5 | 20173.9 | 61623 | 57688.8 | 715.192 | 355.538 | 325.894 | 1047.94 | 916.909 | 2562.16 | 2323.24 | 6792.52 | 6304.81 | 9109.24 |
| 8904.3 | 7969.24 | 22127.3 | 20222.6 | 62096.4 | 57803.5 | 715.479 | 357.962 | 326.957 | 1013.46 | 919.561 | 2570.43 | 2329.43 | 6823.68 | 6319.01 | 9192.91 |
| 8867.59 | 7991.02 | 21996.5 | 20271.4 | 62031.2 | 57918.2 | 715.765 | 362.855 | 328.021 | 1009.68 | 922.217 | 2587.22 | 2335.63 | 6787.75 | 6333.22 | 9204.29 |
| 8847.82 | 8012.84 | 22326.3 | 20320.2 | 62276.7 | 58033 | 716.051 | 350.757 | 329.088 | 1034.13 | 924.878 | 2582.84 | 2341.84 | 6836.65 | 6347.45 | 9195.04 |
| 8952.5 | 8034.7 | 22086.2 | 20369.1 | 62047.2 | 58147.9 | 716.337 | 353.088 | 330.157 | 1009.34 | 927.543 | 2599.02 | 2348.05 | 6868.33 | 6361.69 | 9263.78 |
| 8881.07 | 8056.59 | 21974.3 | 20418.1 | 61808.4 | 58262.9 | 716.623 | 353.364 | 331.229 | 1049.63 | 930.213 | 2602.21 | 2354.28 | 6839.81 | 6375.95 | 9257.79 |
| 8836.4 | 8078.52 | 21928.1 | 20467.1 | 62143.6 | 58378 | 716.909 | 353.396 | 332.303 | 1028.28 | 932.888 | 2577.08 | 2360.52 | 6805.35 | 6390.23 | 9135.86 |
| 8778.69 | 8100.48 | 22235.2 | 20516.2 | 61925.3 | 58493.2 | 717.195 | 349.39 | 333.379 | 1038.09 | 935.568 | 2595.7 | 2366.76 | 6809.72 | 6404.52 | 9191.32 |
| 8995.36 | 8122.48 | 22166.7 | 20565.3 | 62714.8 | 58608.4 | 717.481 | 356.952 | 334.457 | 1052.31 | 938.252 | 2628.63 | 2373.02 | 6995.45 | 6418.82 | 9230.22 |
| 8927 | 8144.51 | 22221 | 20614.5 | 62821.2 | 58723.8 | 717.767 | 356.074 | 335.538 | 1041.77 | 940.941 | 2636.65 | 2379.28 | 6957.56 | 6433.14 | 9306.04 |
| 8950.98 | 8166.58 | 22305.5 | 20663.8 | 62663.1 | 58839.3 | 718.053 | 373.021 | 336.621 | 1041.87 | 943.635 | 2641.48 | 2385.56 | 6967.47 | 6447.48 | 9250.12 |
| 9035.22 | 8188.68 | 22673.2 | 20713.1 | 63222.7 | 58954.8 | 718.339 | 359.834 | 337.707 | 1056.23 | 946.333 | 2635.04 | 2391.84 | 6984.45 | 6461.83 | 9412.93 |
| 9057.73 | 8210.82 | 22483.1 | 20762.5 | 63296.6 | 59070.4 | 718.625 | 341.683 | 338.794 | 1047.31 | 949.036 | 2614.45 | 2398.14 | 6995.87 | 6476.19 | 9438.85 |
| 9031.82 | 8233 | 22511.2 | 20812 | 63557.4 | 59186.2 | 718.911 | 348.432 | 339.884 | 1003.36 | 951.744 | 2673.71 | 2404.44 | 7037.11 | 6490.57 | 9374.01 |
| 8892.59 | 8255.21 | 22633.8 | 20861.5 | 62868.8 | 59302 | 719.197 | 354.306 | 340.977 | 1053.92 | 954.456 | 2684.51 | 2410.75 | 7010.86 | 6504.97 | 9393.53 |
| 8997.61 | 8277.45 | 22826 | 20911.1 | 63347.4 | 59417.9 | 719.483 | 357.559 | 342.071 | 1056.63 | 957.173 | 2660.53 | 2417.07 | 6987.43 | 6519.38 | 9403.01 |
| 9128.84 | 8299.73 | 22598.6 | 20960.7 | 63664.3 | 59533.9 | 719.769 | 362.166 | 343.168 | 1058.44 | 959.895 | 2605.88 | 2423.4 | 7071.79 | 6533.8 | 9418.25 |
| 9192.19 | 8322.05 | 22628 | 21010.4 | 63686.5 | 59650 | 720.055 | 359.306 | 344.268 | 1089.94 | 962.622 | 2699.24 | 2429.74 | 7074.23 | 6548.24 | 9463.18 |
| 9189.24 | 8344.4 | 22863.2 | 21060.2 | 63772.4 | 59766.2 | 720.341 | 377.421 | 345.369 | 1101.74 | 965.353 | 2696.1 | 2436.09 | 7015.94 | 6562.7 | 9437.89 |
| 8985.34 | 8366.79 | 22774.6 | 21110 | 63862 | 59882.4 | 720.627 | 377.817 | 346.473 | 1045.46 | 968.089 | 2656.78 | 2442.45 | 7016.65 | 6577.17 | 9440.1 |
| 9128.93 | 8389.21 | 22731.8 | 21159.9 | 63778.3 | 59998.8 | 720.913 | 357.747 | 347.58 | 1053.78 | 970.829 | 2675.07 | 2448.82 | 7109.65 | 6591.65 | 9536.15 |
| 9210.2 | 8411.67 | 22963.7 | 21209.8 | 63933.2 | 60115.2 | 721.199 | 358.606 | 348.688 | 1077.52 | 973.574 | 2710.98 | 2455.2 | 7058.2 | 6606.15 | 9463.41 |
| 9262.76 | 8434.16 | 23141.8 | 21259.9 | 64282.6 | 60231.8 | 721.485 | 386.537 | 349.799 | 1093.98 | 976.324 | 2741.86 | 2461.59 | 7230.27 | 6620.67 | 9524.76 |
| 9307.24 | 8456.69 | 22856.8 | 21309.9 | 64119.4 | 60348.4 | 721.77 | 390.399 | 350.913 | 1108.94 | 979.079 | 2680.55 | 2467.98 | 7096.44 | 6635.2 | 9549.61 |
| 9392.85 | 8479.25 | 23391.3 | 21360.1 | 64852.3 | 60465.1 | 722.056 | 377.918 | 352.028 | 1100.84 | 981.838 | 2731.07 | 2474.39 | 7123.84 | 6649.74 | 9575.61 |
| 9136.42 | 8501.85 | 22983.8 | 21410.3 | 64447.4 | 60581.9 | 722.342 | 369.276 | 353.146 | 1089.86 | 984.602 | 2675.17 | 2480.8 | 7086.82 | 6664.3 | 9545.09 |
| 9237.34 | 8524.49 | 23012.8 | 21460.5 | 64314.5 | 60698.8 | 722.628 | 375.887 | 354.267 | 1096.11 | 987.371 | 2700.65 | 2487.23 | 7135.87 | 6678.88 | 9558.26 |
| 9225.41 | 8547.16 | 23285 | 21510.8 | 64809.6 | 60815.8 | 722.913 | 381.523 | 355.389 | 1111.34 | 990.144 | 2704.97 | 2493.66 | 7065.66 | 6693.47 | 9529.24 |
| 9448.64 | 8569.86 | 23423.4 | 21561.2 | 64722 | 60932.8 | 723.199 | 381.462 | 356.514 | 1108.77 | 992.922 | 2702.89 | 2500.1 | 7149.2 | 6708.07 | 9752.1 |
| 9347.48 | 8592.6 | 23208.6 | 21611.7 | 64960.1 | 61050 | 723.485 | 385.151 | 357.642 | 1071.72 | 995.705 | 2752.91 | 2506.56 | 7100.88 | 6722.69 | 9709.98 |
| 9443.89 | 8615.38 | 23354.7 | 21662.2 | 64782.7 | 61167.2 | 723.771 | 395.605 | 358.772 | 1102.1 | 998.492 | 2694.2 | 2513.02 | 7219.68 | 6737.32 | 9578.26 |
| 9297.31 | 8638.19 | 23185.4 | 21712.8 | 64440.8 | 61284.5 | 724.057 | 375.322 | 359.904 | 1117.17 | 1001.28 | 2744.47 | 2519.49 | 7139.52 | 6751.97 | 9589.34 |
| 9417.39 | 8661.03 | 23595.1 | 21763.4 | 64956.6 | 61401.9 | 724.342 | 411.056 | 361.038 | 1109.61 | 1004.08 | 2691.33 | 2525.97 | 7259.34 | 6766.63 | 9739.25 |
| 9426.39 | 8683.92 | 23135.6 | 21814.1 | 64868.9 | 61519.4 | 724.628 | 398.339 | 362.175 | 1119.48 | 1006.88 | 2779.25 | 2532.46 | 7174.84 | 6781.31 | 9699.03 |
| 9383.56 | 8706.83 | 23519.6 | 21864.8 | 65029.5 | 61637 | 724.914 | 398.675 | 363.314 | 1115.47 | 1009.69 | 2749.6 | 2538.96 | 7310.67 | 6796 | 9648.71 |
| 9495.71 | 8729.79 | 23381.3 | 21915.6 | 64492.3 | 61754.7 | 725.199 | 405.413 | 364.455 | 1116.32 | 1012.5 | 2784.31 | 2545.47 | 7246.03 | 6810.71 | 9667.62 |
| 9404.47 | 8752.77 | 23498 | 21966.5 | 65228.1 | 61872.5 | 725.485 | 384.925 | 365.599 | 1112.78 | 1015.32 | 2780.11 | 2551.99 | 7280.04 | 6825.43 | 9648.07 |
| 9544.13 | 8775.8 | 23452.4 | 22017.4 | 64764 | 61990.3 | 725.771 | 408.823 | 366.745 | 1101.85 | 1018.13 | 2783.24 | 2558.51 | 7228.76 | 6840.17 | 9606.52 |
| 9446.41 | 8798.85 | 23529.7 | 22068.4 | 65037.8 | 62108.3 | 726.056 | 399.165 | 367.894 | 1096.11 | 1020.96 | 2798.7 | 2565.05 | 7360.1 | 6854.92 | 9734.67 |
| 9536.61 | 8821.95 | 23707.5 | 22119.5 | 65776.5 | 62226.3 | 726.342 | 420.735 | 369.045 | 1130.72 | 1023.79 | 2794.56 | 2571.6 | 7287.65 | 6869.68 | 9763.03 |
| 9666.15 | 8845.08 | 23486.8 | 22170.6 | 65340.5 | 62344.4 | 726.628 | 411.021 | 370.198 | 1134.19 | 1026.62 | 2772.02 | 2578.15 | 7314.85 | 6884.46 | 9706.93 |
| 9579.77 | 8868.24 | 23805.3 | 22221.8 | 66073.7 | 62462.6 | 726.913 | 402.306 | 371.353 | 1148.68 | 1029.46 | 2802.03 | 2584.72 | 7301.7 | 6899.26 | 9819.1 |
| 9581.27 | 8891.44 | 23479.1 | 22273 | 65229.9 | 62580.9 | 727.199 | 425.987 | 372.511 | 1121.12 | 1032.31 | 2763.74 | 2591.29 | 7351.89 | 6914.07 | 9818.51 |
| 9477.04 | 8914.67 | 23795.6 | 22324.3 | 65819.5 | 62699.2 | 727.484 | 401.918 | 373.672 | 1129.41 | 1035.15 | 2769.49 | 2597.88 | 7255.56 | 6928.89 | 9785.55 |
| 9527.26 | 8937.94 | 23829 | 22375.7 | 66112.2 | 62817.7 | 727.77 | 394.08 | 374.834 | 1114.91 | 1038.01 | 2823.6 | 2604.47 | 7342.74 | 6943.73 | 9791.71 |
| 9746.76 | 8961.25 | 23795 | 22427.1 | 66254.3 | 62936.2 | 728.055 | 405.396 | 375.999 | 1141.64 | 1040.87 | 2836.04 | 2611.07 | 7332.31 | 6958.58 | 9949.97 |
| 9692.15 | 8984.59 | 24050.1 | 22478.6 | 66124 | 63054.8 | 728.341 | 399.477 | 377.167 | 1131.72 | 1043.73 | 2787.45 | 2617.68 | 7351.03 | 6973.45 | 9822.43 |
| 9642.95 | 9007.96 | 23935.4 | 22530.1 | 65966.9 | 63173.6 | 728.626 | 405.306 | 378.336 | 1142.73 | 1046.59 | 2820.37 | 2624.3 | 7263.62 | 6988.33 | 9794.38 |
| 9778.65 | 9031.37 | 23948 | 22581.7 | 66191.2 | 63292.3 | 728.912 | 414.646 | 379.509 | 1174.68 | 1049.47 | 2870.19 | 2630.93 | 7408.21 | 7003.23 | 9900.91 |
| 9641.76 | 9054.82 | 24150.2 | 22633.4 | 66826.6 | 63411.2 | 729.197 | 425.748 | 380.683 | 1118 | 1052.34 | 2850.16 | 2637.57 | 7427.17 | 7018.14 | 9878.58 |
| 9759.3 | 9078.3 | 23945.2 | 22685.1 | 66543.4 | 63530.2 | 729.483 | 406.873 | 381.86 | 1161.87 | 1055.22 | 2851.32 | 2644.22 | 7339.98 | 7033.06 | 9896.34 |
| 9704.62 | 9101.82 | 24192.6 | 22736.9 | 66848.2 | 63649.2 | 729.768 | 413.282 | 383.039 | 1173.08 | 1058.11 | 2860.7 | 2650.88 | 7407.47 | 7048 | 9970.88 |
| 9670.01 | 9125.37 | 23969.8 | 22788.8 | 66552 | 63768.4 | 730.054 | 389.269 | 384.221 | 1168.26 | 1061 | 2818.11 | 2657.54 | 7393.72 | 7062.95 | 9891.99 |
| 9684.36 | 9148.95 | 24255.8 | 22840.7 | 67057.1 | 63887.6 | 730.339 | 430.294 | 385.405 | 1157.27 | 1063.89 | 2839.06 | 2664.22 | 7449.83 | 7077.92 | 9980.6 |
| 9716.1 | 9172.58 | 24218.6 | 22892.6 | 66398.1 | 64006.9 | 730.625 | 408.276 | 386.591 | 1189.27 | 1066.79 | 2823.43 | 2670.91 | 7358.93 | 7092.9 | 9854.34 |
| 9945.86 | 9196.23 | 24202 | 22944.6 | 66749.5 | 64126.2 | 730.91 | 411.046 | 387.78 | 1171.39 | 1069.7 | 2821.16 | 2677.6 | 7512.61 | 7107.9 | 10056.5 |
| 9786.73 | 9219.92 | 24365.8 | 22996.7 | 66668.9 | 64245.7 | 731.195 | 405.035 | 388.971 | 1155.17 | 1072.61 | 2909.22 | 2684.3 | 7415.14 | 7122.9 | 9946.92 |
| 9806.86 | 9243.65 | 24190.7 | 23048.9 | 66351.7 | 64365.2 | 731.481 | 401.641 | 390.164 | 1156.45 | 1075.52 | 2876.37 | 2691.02 | 7434.54 | 7137.93 | 9814.81 |
| 9905.3 | 9267.41 | 24310.6 | 23101.1 | 66564.3 | 64484.9 | 731.766 | 426.178 | 391.36 | 1153.71 | 1078.44 | 2866.93 | 2697.74 | 7452.59 | 7152.97 | 10025.5 |
| 9785.81 | 9291.21 | 24551.8 | 23153.3 | 67098.5 | 64604.6 | 732.052 | 415.995 | 392.558 | 1158.48 | 1081.36 | 2868.09 | 2704.47 | 7495.89 | 7168.02 | 10066.7 |
| 9809.01 | 9315.04 | 24124.3 | 23205.7 | 66566.6 | 64724.4 | 732.337 | 419.424 | 393.759 | 1164.97 | 1084.29 | 2914.67 | 2711.21 | 7436.66 | 7183.09 | 10071.4 |
| 9906.37 | 9338.91 | 24523.6 | 23258.1 | 67474.8 | 64844.3 | 732.622 | 427.313 | 394.962 | 1160.25 | 1087.22 | 2893.32 | 2717.96 | 7569.45 | 7198.17 | 10090.3 |
| 9918.28 | 9362.81 | 24455.4 | 23310.5 | 66717.3 | 64964.2 | 732.908 | 416.668 | 396.167 | 1184.11 | 1090.16 | 2808.59 | 2724.72 | 7476.29 | 7213.26 | 10003.1 |
| 9858.15 | 9386.75 | 24449.9 | 23363 | 66803.6 | 65084.3 | 733.193 | 405.067 | 397.375 | 1163.84 | 1093.1 | 2889.54 | 2731.49 | 7506.94 | 7228.37 | 9963.17 |
| 9886.32 | 9410.72 | 24983.6 | 23415.6 | 67912.6 | 65204.4 | 733.478 | 435.671 | 398.585 | 1164.88 | 1096.05 | 2981.35 | 2738.27 | 7629.29 | 7243.5 | 10252.3 |
| 10090.5 | 9434.73 | 24582.9 | 23468.2 | 67719.3 | 65324.6 | 733.763 | 395.678 | 399.798 | 1175.02 | 1099 | 2914.2 | 2745.06 | 7488.32 | 7258.63 | 10158.6 |
| 10044 | 9458.77 | 24937.1 | 23520.9 | 68001.5 | 65444.9 | 734.048 | 430.865 | 401.013 | 1180.94 | 1101.96 | 2914.58 | 2751.85 | 7544.67 | 7273.78 | 10172.6 |
| 10123.7 | 9482.85 | 24835.8 | 23573.6 | 67959.9 | 65565.2 | 734.334 | 430.161 | 402.23 | 1180.05 | 1104.92 | 2905.73 | 2758.66 | 7659.28 | 7288.95 | 10206.7 |
| 10097.7 | 9506.96 | 24826.5 | 23626.4 | 67703 | 65685.7 | 734.619 | 424.859 | 403.45 | 1176.13 | 1107.88 | 2949.88 | 2765.47 | 7489.69 | 7304.13 | 10150 |
| 10114.6 | 9531.11 | 25102 | 23679.3 | 68326.9 | 65806.2 | 734.904 | 472.393 | 404.672 | 1211.32 | 1110.85 | 2955.04 | 2772.3 | 7689.25 | 7319.32 | 10171.7 |
| 10278.8 | 9555.29 | 24928.8 | 23732.2 | 68325.3 | 65926.8 | 735.189 | 444.033 | 405.896 | 1189.75 | 1113.83 | 2956.44 | 2779.13 | 7714.63 | 7334.53 | 10210.5 |
| 10102.4 | 9579.51 | 25027.5 | 23785.2 | 68017.7 | 66047.5 | 735.475 | 430.01 | 407.123 | 1164.54 | 1116.81 | 2958.67 | 2785.97 | 7678.73 | 7349.75 | 10243.3 |
| 10125.1 | 9603.76 | 25084.8 | 23838.2 | 68314.4 | 66168.3 | 735.76 | 452.981 | 408.352 | 1196.34 | 1119.79 | 3011.63 | 2792.82 | 7626.07 | 7364.99 | 10113.6 |
| 10351.5 | 9628.05 | 24894.1 | 23891.3 | 68626.5 | 66289.1 | 736.045 | 436.05 | 409.584 | 1175.1 | 1122.78 | 2965.43 | 2799.68 | 7732.08 | 7380.24 | 10239.8 |
| 10142.9 | 9652.38 | 24965.6 | 23944.5 | 68254.1 | 66410.1 | 736.33 | 422.524 | 410.818 | 1204.26 | 1125.77 | 2951.72 | 2806.55 | 7686.63 | 7395.5 | 10353.1 |
| 10320.7 | 9676.73 | 25316.1 | 23997.7 | 68782.3 | 66531.1 | 736.615 | 434.724 | 412.054 | 1252.32 | 1128.77 | 3018.37 | 2813.43 | 7707.59 | 7410.77 | 10374.6 |
| 10201.8 | 9701.13 | 25409.8 | 24051 | 69179.6 | 66652.2 | 736.9 | 448.966 | 413.293 | 1202.31 | 1131.77 | 3003.01 | 2820.32 | 7764.07 | 7426.07 | 10323.4 |
| 10120.5 | 9725.56 | 24963.2 | 24104.3 | 68383.9 | 66773.3 | 737.185 | 434.348 | 414.534 | 1203.53 | 1134.78 | 2964.32 | 2827.22 | 7712.81 | 7441.37 | 10184.2 |
| 10237.7 | 9750.02 | 25541 | 24157.7 | 68969.2 | 66894.6 | 737.47 | 440.689 | 415.778 | 1209.82 | 1137.79 | 2988.24 | 2834.12 | 7795.53 | 7456.69 | 10231.5 |
| 10382.9 | 9774.51 | 25438.4 | 24211.2 | 69368.9 | 67015.9 | 737.755 | 434.624 | 417.024 | 1230.32 | 1140.81 | 3013.78 | 2841.04 | 7727.79 | 7472.02 | 10240.4 |
| 10288.7 | 9799.05 | 25452.9 | 24264.7 | 69205.1 | 67137.3 | 738.041 | 440.896 | 418.273 | 1187.45 | 1143.83 | 2976.63 | 2847.96 | 7830.86 | 7487.37 | 10364 |
| 10417.6 | 9823.61 | 25649.6 | 24318.3 | 68733.3 | 67258.8 | 738.326 | 432.905 | 419.523 | 1225.24 | 1146.85 | 3054.04 | 2854.9 | 7857.83 | 7502.73 | 10339.5 |
| 10609.4 | 9848.22 | 25696 | 24371.9 | 69951.6 | 67380.4 | 738.611 | 468.523 | 420.777 | 1234.6 | 1149.88 | 3102.97 | 2861.84 | 7883.42 | 7518.1 | 10483 |
| 10209.4 | 9872.85 | 25368.5 | 24425.6 | 68814.4 | 67502 | 738.896 | 445.818 | 422.032 | 1240.06 | 1152.92 | 2988.76 | 2868.79 | 7711.72 | 7533.49 | 10348.3 |
| 10450.8 | 9897.53 | 25638.8 | 24479.4 | 70055.5 | 67623.7 | 739.181 | 458.312 | 423.29 | 1235.98 | 1155.96 | 3069.17 | 2875.75 | 7918.25 | 7548.89 | 10496.3 |
| 10532.2 | 9922.23 | 25767.9 | 24533.2 | 69746.1 | 67745.5 | 739.466 | 453.507 | 424.551 | 1247.63 | 1159 | 3021.21 | 2882.72 | 7920.2 | 7564.3 | 10441.8 |
| 10524.5 | 9946.98 | 25668.4 | 24587 | 69820.1 | 67867.4 | 739.751 | 458.393 | 425.814 | 1220.43 | 1162.05 | 3017.98 | 2889.7 | 7879.79 | 7579.73 | 10367.6 |
| 10607.3 | 9971.75 | 25992.8 | 24641 | 70555.1 | 67989.4 | 740.035 | 463.53 | 427.079 | 1259.42 | 1165.11 | 3055.41 | 2896.69 | 7926.22 | 7595.17 | 10611.5 |
| 10494.7 | 9996.56 | 25905.8 | 24694.9 | 69979.3 | 68111.4 | 740.321 | 451.696 | 428.347 | 1252.54 | 1168.16 | 3043.98 | 2903.69 | 7910.94 | 7610.62 | 10530.1 |
| 10530.4 | 10021.4 | 25868.2 | 24749 | 69860.7 | 68233.5 | 740.605 | 482.439 | 429.617 | 1229.31 | 1171.23 | 3134.2 | 2910.7 | 7977.41 | 7626.09 | 10581.2 |
| 10626.2 | 10046.3 | 26251.4 | 24803.1 | 71144.6 | 68355.7 | 740.89 | 447.498 | 430.889 | 1255.24 | 1174.3 | 3076.41 | 2917.71 | 8015.68 | 7641.57 | 10651.5 |
| 10618 | 10071.2 | 25915.1 | 24857.2 | 70839.2 | 68478 | 741.175 | 460.782 | 432.164 | 1278.48 | 1177.37 | 3120.91 | 2924.74 | 7943.89 | 7657.07 | 10601.7 |
| 10684.8 | 10096.2 | 26167.6 | 24911.5 | 70775.1 | 68600.3 | 741.46 | 491.889 | 433.442 | 1283.16 | 1180.44 | 3092.24 | 2931.77 | 8034.99 | 7672.58 | 10680.4 |
| 10743.9 | 10121.1 | 26321.9 | 24965.7 | 71349.7 | 68722.7 | 741.745 | 453.969 | 434.721 | 1275.42 | 1183.53 | 3093.38 | 2938.81 | 8012.63 | 7688.1 | 10726.3 |
| 10722.9 | 10146.2 | 26245.1 | 25020.1 | 71363.3 | 68845.2 | 742.03 | 470.938 | 436.003 | 1324.63 | 1186.61 | 3119.91 | 2945.86 | 8002.66 | 7703.64 | 10692.8 |
| 10849.7 | 10171.2 | 26452.9 | 25074.5 | 71861.1 | 68967.8 | 742.315 | 476.687 | 437.288 | 1278.4 | 1189.7 | 3157.04 | 2952.93 | 8072.4 | 7719.19 | 10788.7 |
| 10683 | 10196.3 | 26015.4 | 25128.9 | 71077.7 | 69090.4 | 742.6 | 459.407 | 438.575 | 1277.26 | 1192.8 | 3108.95 | 2959.99 | 7889.29 | 7734.75 | 10554.9 |
| 10848.4 | 10221.4 | 26634.9 | 25183.4 | 72171.1 | 69213.2 | 742.884 | 488.014 | 439.864 | 1253.44 | 1195.9 | 3205.84 | 2967.07 | 8137.81 | 7750.33 | 10782.1 |
| 10701.2 | 10246.6 | 26718.2 | 25238 | 71733.5 | 69336 | 743.169 | 494.697 | 441.156 | 1314.67 | 1199.01 | 3114.83 | 2974.16 | 8111.51 | 7765.92 | 10709 |
| 10994.7 | 10271.8 | 27026.5 | 25292.6 | 72442 | 69458.8 | 743.454 | 492.514 | 442.45 | 1332.36 | 1202.11 | 3236.43 | 2981.26 | 8145.23 | 7781.52 | 10826.8 |
| 10829.1 | 10297 | 26480.8 | 25347.3 | 71836.2 | 69581.8 | 743.739 | 480.519 | 443.747 | 1260 | 1205.23 | 3112.85 | 2988.37 | 8043.33 | 7797.14 | 10836.3 |
| 11043.8 | 10322.3 | 26950.8 | 25402 | 72659.4 | 69704.8 | 744.023 | 481.996 | 445.046 | 1312.01 | 1208.35 | 3244.44 | 2995.48 | 8250.24 | 7812.77 | 10958.2 |
| 10951.6 | 10347.6 | 26830.8 | 25456.8 | 72815.4 | 69827.9 | 744.308 | 473.698 | 446.348 | 1334.64 | 1211.47 | 3214.3 | 3002.61 | 8266.68 | 7828.41 | 11030.1 |
| 10973.3 | 10372.9 | 26962.6 | 25511.7 | 72373.1 | 69951.1 | 744.593 | 466.526 | 447.652 | 1294.04 | 1214.6 | 3152.59 | 3009.74 | 8168.14 | 7844.06 | 10810.8 |
| 11153 | 10398.3 | 27060 | 25566.6 | 72936.5 | 70074.3 | 744.878 | 492.33 | 448.958 | 1324.31 | 1217.73 | 3191.28 | 3016.88 | 8245.27 | 7859.73 | 11063.6 |
| 11027.1 | 10423.7 | 27063.1 | 25621.6 | 72904.1 | 70197.7 | 745.162 | 497.238 | 450.267 | 1326.04 | 1220.87 | 3256.77 | 3024.03 | 8185.69 | 7875.42 | 10909.8 |
| 11022.4 | 10449.1 | 26542.1 | 25676.6 | 73103.5 | 70321.1 | 745.447 | 459.608 | 451.578 | 1294.96 | 1224.01 | 3202 | 3031.19 | 8227.86 | 7891.11 | 10920 |
| 11199.4 | 10474.6 | 26915.9 | 25731.7 | 73461.2 | 70444.5 | 745.732 | 489.721 | 452.892 | 1285.72 | 1227.16 | 3154.64 | 3038.36 | 8287.35 | 7906.82 | 10958.6 |
| 11016.7 | 10500.1 | 27166.2 | 25786.8 | 73112.4 | 70568.1 | 746.016 | 471.913 | 454.208 | 1292.05 | 1230.31 | 3216.72 | 3045.54 | 8251.68 | 7922.54 | 10943.1 |
| 10983.6 | 10525.6 | 27039.5 | 25842 | 72475.3 | 70691.7 | 746.301 | 500.125 | 455.526 | 1319.2 | 1233.46 | 3231.87 | 3052.73 | 8195.56 | 7938.28 | 10942.4 |
| 11266.5 | 10551.2 | 27045.5 | 25897.3 | 72932.1 | 70815.4 | 746.586 | 487.785 | 456.847 | 1318.71 | 1236.63 | 3163.17 | 3059.92 | 8236.49 | 7954.03 | 10899.2 |
| 11215.6 | 10576.8 | 27259.9 | 25952.6 | 73549.9 | 70939.2 | 746.87 | 500.026 | 458.17 | 1342.84 | 1239.79 | 3187.59 | 3067.13 | 8343.8 | 7969.79 | 11026.6 |
| 11312.9 | 10602.5 | 27207.4 | 26008 | 73241.5 | 71063 | 747.155 | 473.306 | 459.496 | 1335.94 | 1242.96 | 3197.55 | 3074.34 | 8291.26 | 7985.56 | 10874.2 |
| 11214.5 | 10628.1 | 27001.6 | 26063.4 | 73682.1 | 71186.9 | 747.439 | 491.596 | 460.824 | 1333.03 | 1246.13 | 3260.28 | 3081.57 | 8266.22 | 8001.35 | 11105.6 |
| 11243.9 | 10653.9 | 27447.4 | 26118.9 | 73709.1 | 71310.9 | 747.724 | 485.746 | 462.155 | 1307.74 | 1249.31 | 3289.14 | 3088.8 | 8343.55 | 8017.15 | 11065 |
| 11073.8 | 10679.6 | 27546.3 | 26174.5 | 73273.7 | 71435 | 748.009 | 480.55 | 463.488 | 1324.92 | 1252.5 | 3245.05 | 3096.04 | 8303.08 | 8032.97 | 10979.4 |
| 11340.3 | 10705.4 | 27464.5 | 26230 | 74119.6 | 71559.1 | 748.293 | 499.01 | 464.824 | 1367.43 | 1255.69 | 3273.62 | 3103.29 | 8368.35 | 8048.79 | 11025.5 |
| 11321 | 10731.2 | 27743.8 | 26285.7 | 73848.6 | 71683.3 | 748.578 | 511.39 | 466.161 | 1319.19 | 1258.88 | 3231.26 | 3110.55 | 8241.33 | 8064.63 | 11046.6 |
| 11420.3 | 10757 | 27703.2 | 26341.4 | 74666.5 | 71807.6 | 748.862 | 511.446 | 467.502 | 1386.95 | 1262.08 | 3302.27 | 3117.82 | 8494.61 | 8080.48 | 11162.7 |
| 11624.9 | 10782.9 | 27891.1 | 26397.2 | 74985.3 | 71931.9 | 749.147 | 504.54 | 468.845 | 1363.53 | 1265.28 | 3257.07 | 3125.09 | 8492.11 | 8096.35 | 11226.1 |
| 11462.5 | 10808.9 | 27518 | 26453 | 74306.6 | 72056.4 | 749.431 | 488.968 | 470.19 | 1350.34 | 1268.48 | 3287.84 | 3132.38 | 8312.4 | 8112.23 | 11188.6 |
| 11503.1 | 10834.8 | 28027.4 | 26508.9 | 75224.7 | 72180.9 | 749.716 | 513.262 | 471.538 | 1363 | 1271.7 | 3374.51 | 3139.67 | 8521.27 | 8128.12 | 11227.1 |
| 11434.7 | 10860.8 | 27963.8 | 26564.8 | 74754.3 | 72305.4 | 750 | 514.84 | 472.888 | 1381.5 | 1274.91 | 3319.31 | 3146.98 | 8489.75 | 8144.02 | 11196.1 |
| 11545.4 | 10886.8 | 27914.7 | 26620.8 | 75118 | 72430.1 | 750.284 | 474.367 | 474.24 | 1361.46 | 1278.13 | 3354.52 | 3154.29 | 8404.8 | 8159.94 | 11195.8 |
| 11566.2 | 10912.9 | 27760.8 | 26676.9 | 74498.4 | 72554.8 | 750.569 | 518.896 | 475.595 | 1371.01 | 1281.36 | 3348.91 | 3161.61 | 8433.53 | 8175.87 | 11248.3 |
| 11457.9 | 10939 | 28054.4 | 26733 | 75019.2 | 72679.6 | 750.853 | 488.314 | 476.952 | 1380.41 | 1284.59 | 3390.47 | 3168.94 | 8460.84 | 8191.81 | 11202.9 |
| 11642.1 | 10965.1 | 27984.8 | 26789.2 | 75236.3 | 72804.4 | 751.138 | 515.232 | 478.312 | 1393.13 | 1287.82 | 3296.08 | 3176.28 | 8513.73 | 8207.77 | 11363.1 |
| 11618.3 | 10991.2 | 28402 | 26845.4 | 75339.8 | 72929.3 | 751.422 | 504.398 | 479.675 | 1399.69 | 1291.06 | 3379.5 | 3183.63 | 8550.1 | 8223.73 | 11350.2 |
| 11542.5 | 11017.4 | 28130.9 | 26901.7 | 75639.8 | 73054.3 | 751.706 | 511.84 | 481.039 | 1347.58 | 1294.31 | 3340.42 | 3190.98 | 8644.75 | 8239.71 | 11295.4 |
| 11711.2 | 11043.7 | 28301.3 | 26958 | 75811.1 | 73179.4 | 751.991 | 514.617 | 482.406 | 1399.09 | 1297.55 | 3360.53 | 3198.35 | 8589.86 | 8255.71 | 11426.3 |
| 11841.3 | 11069.9 | 28383.8 | 27014.4 | 76378.2 | 73304.5 | 752.275 | 525.356 | 483.776 | 1391.07 | 1300.81 | 3354.72 | 3205.72 | 8570.64 | 8271.71 | 11427.7 |
| 11694.6 | 11096.2 | 27953.2 | 27070.9 | 75365 | 73429.7 | 752.559 | 518.971 | 485.148 | 1355.09 | 1304.06 | 3273.6 | 3213.11 | 8546.88 | 8287.73 | 11355.6 |
| 11948.6 | 11122.6 | 28299.4 | 27127.4 | 76440.4 | 73555 | 752.844 | 531.052 | 486.523 | 1414.19 | 1307.33 | 3441.48 | 3220.5 | 8642.56 | 8303.76 | 11485.2 |
| 11581.2 | 11148.9 | 28082.2 | 27183.9 | 75542 | 73680.3 | 753.128 | 532.94 | 487.9 | 1374.91 | 1310.59 | 3315.36 | 3227.9 | 8592.55 | 8319.81 | 11182 |
| 11744.5 | 11175.3 | 28103.6 | 27240.5 | 75007.3 | 73805.7 | 753.412 | 506.963 | 489.279 | 1344.47 | 1313.86 | 3337.48 | 3235.31 | 8483.83 | 8335.86 | 11338.9 |
| 11796.9 | 11201.8 | 28634.4 | 27297.2 | 76689 | 73931.2 | 753.697 | 534.507 | 490.661 | 1413.42 | 1317.14 | 3424.96 | 3242.73 | 8676.57 | 8351.93 | 11464.8 |
| 11835.8 | 11228.2 | 28526.2 | 27353.9 | 76818.4 | 74056.8 | 753.981 | 545.352 | 492.045 | 1404.68 | 1320.42 | 3439.05 | 3250.16 | 8730.91 | 8368.01 | 11437.9 |
| 11864.7 | 11254.7 | 28596.7 | 27410.7 | 76533.1 | 74182.4 | 754.265 | 526.705 | 493.432 | 1372.96 | 1323.7 | 3378.21 | 3257.59 | 8626.6 | 8384.11 | 11358.6 |
| 11908.2 | 11281.3 | 28667.2 | 27467.6 | 76916 | 74308.1 | 754.549 | 521.553 | 494.821 | 1412.4 | 1326.99 | 3402.99 | 3265.04 | 8833.53 | 8400.22 | 11466.4 |
| 11762.4 | 11307.8 | 28588.4 | 27524.5 | 76242.6 | 74433.8 | 754.834 | 527.401 | 496.212 | 1408.24 | 1330.29 | 3414.35 | 3272.49 | 8668.95 | 8416.33 | 11607.2 |
| 11834.6 | 11334.4 | 28464.9 | 27581.4 | 75947.5 | 74559.6 | 755.118 | 514.644 | 497.606 | 1415.13 | 1333.59 | 3446.72 | 3279.95 | 8554.93 | 8432.47 | 11520.7 |
| 11828.3 | 11361.1 | 28758 | 27638.4 | 76215.9 | 74685.5 | 755.402 | 541.142 | 499.003 | 1402.53 | 1336.89 | 3457.19 | 3287.42 | 8741.84 | 8448.61 | 11504.8 |
| 11953.9 | 11387.8 | 29080.6 | 27695.5 | 77015.3 | 74811.5 | 755.686 | 523.721 | 500.402 | 1453.49 | 1340.2 | 3432.48 | 3294.9 | 8781.72 | 8464.76 | 11587.2 |
| 11935.9 | 11414.5 | 28587.6 | 27752.6 | 76828.4 | 74937.5 | 755.97 | 506.186 | 501.803 | 1453.7 | 1343.51 | 3407.06 | 3302.39 | 8712.93 | 8480.93 | 11539.4 |
| 12005.7 | 11441.2 | 28969.5 | 27809.8 | 77515.8 | 75063.6 | 756.254 | 538.411 | 503.207 | 1422.97 | 1346.83 | 3471.28 | 3309.89 | 8788.13 | 8497.11 | 11689.3 |
| 12042.8 | 11468 | 29058.4 | 27867 | 76696 | 75189.8 | 756.538 | 525.217 | 504.613 | 1399.99 | 1350.15 | 3406.69 | 3317.4 | 8714.7 | 8513.31 | 11443.9 |
| 11906.1 | 11494.8 | 29338.6 | 27924.3 | 77783.7 | 75316 | 756.823 | 527.718 | 506.022 | 1458.79 | 1353.47 | 3496.73 | 3324.91 | 8848.27 | 8529.51 | 11649.3 |
| 11977.6 | 11521.6 | 29048.4 | 27981.6 | 77405.1 | 75442.3 | 757.107 | 525.616 | 507.433 | 1411.47 | 1356.81 | 3439.21 | 3332.44 | 8850.94 | 8545.73 | 11672.1 |
| 12107.6 | 11548.5 | 29087.9 | 28039 | 77844.9 | 75568.6 | 757.391 | 512.759 | 508.847 | 1397.12 | 1360.14 | 3491.62 | 3339.97 | 8891.1 | 8561.96 | 11723.6 |
| 12118.1 | 11575.4 | 29559.4 | 28096.4 | 78134 | 75695.1 | 757.675 | 551.909 | 510.263 | 1442.66 | 1363.48 | 3473.31 | 3347.51 | 8899.8 | 8578.2 | 11714.9 |
| 12012.5 | 11602.4 | 29222 | 28153.9 | 78079.4 | 75821.6 | 757.959 | 522.902 | 511.682 | 1471.27 | 1366.83 | 3503.1 | 3355.06 | 8950.09 | 8594.46 | 11856 |
| 12206.5 | 11629.3 | 29427.8 | 28211.5 | 78435.6 | 75948.1 | 758.243 | 546.93 | 513.103 | 1445.25 | 1370.17 | 3561.46 | 3362.62 | 8891.15 | 8610.73 | 11824.8 |
| 12199.2 | 11656.4 | 29373.5 | 28269.1 | 77923.1 | 76074.7 | 758.527 | 523.639 | 514.526 | 1500.03 | 1373.53 | 3519.52 | 3370.19 | 8812.88 | 8627 | 11723 |
| 12188.5 | 11683.4 | 29732.4 | 28326.8 | 78390.8 | 76201.4 | 758.811 | 543.353 | 515.952 | 1474.4 | 1376.89 | 3580.03 | 3377.76 | 8944.23 | 8643.3 | 11832.6 |
| 12249.2 | 11710.5 | 29434.9 | 28384.5 | 78570.8 | 76328.2 | 759.095 | 563.026 | 517.38 | 1444.19 | 1380.25 | 3509.07 | 3385.35 | 8959 | 8659.6 | 11787.1 |
| 12305.2 | 11737.6 | 29390.8 | 28442.3 | 78708.7 | 76455 | 759.379 | 584.354 | 518.811 | 1452.96 | 1383.62 | 3568.29 | 3392.94 | 8915.83 | 8675.92 | 11802.3 |
| 12160.3 | 11764.7 | 29526.1 | 28500.1 | 78461.5 | 76581.9 | 759.663 | 545.953 | 520.244 | 1446.5 | 1386.99 | 3587.59 | 3400.54 | 8932.46 | 8692.24 | 11796 |
| 12374.7 | 11791.9 | 29699.6 | 28558 | 78637.6 | 76708.9 | 759.947 | 550.312 | 521.68 | 1491.33 | 1390.36 | 3604.32 | 3408.15 | 8932.7 | 8708.58 | 11884.5 |
| 12217.7 | 11819.1 | 29787.5 | 28615.9 | 78645.6 | 76835.9 | 760.231 | 543.014 | 523.118 | 1484.21 | 1393.74 | 3594.09 | 3415.77 | 8961.69 | 8724.93 | 11941.4 |
| 12276.2 | 11846.4 | 29701.8 | 28673.9 | 78612.8 | 76962.9 | 760.515 | 538.549 | 524.559 | 1509.54 | 1397.13 | 3570.79 | 3423.4 | 8984.08 | 8741.3 | 11869.5 |
| 12317.4 | 11873.7 | 29588.3 | 28732 | 78752.4 | 77090.1 | 760.799 | 562.747 | 526.002 | 1485.33 | 1400.52 | 3553.89 | 3431.04 | 9043.5 | 8757.67 | 11822.3 |
| 12396.3 | 11901 | 29821.6 | 28790.1 | 79082.1 | 77217.3 | 761.082 | 556.541 | 527.448 | 1482.15 | 1403.91 | 3648.29 | 3438.68 | 9093.43 | 8774.06 | 11965.2 |
| 12459.2 | 11928.3 | 29853.9 | 28848.2 | 78864.2 | 77344.6 | 761.366 | 565.99 | 528.896 | 1494.29 | 1407.31 | 3591.21 | 3446.34 | 8984.15 | 8790.46 | 11950.4 |
| 12601.2 | 11955.7 | 30112.2 | 28906.4 | 79455.1 | 77471.9 | 761.65 | 551.704 | 530.346 | 1501.63 | 1410.72 | 3586.09 | 3454 | 9078.2 | 8806.87 | 12012.9 |
| 12311.8 | 11983.1 | 29837.2 | 28964.7 | 79181.7 | 77599.3 | 761.934 | 549.988 | 531.799 | 1474.58 | 1414.12 | 3676.87 | 3461.67 | 9058.24 | 8823.29 | 11901.5 |
| 12617.9 | 12010.6 | 30301.7 | 29023 | 80312.4 | 77726.8 | 762.218 | 569.552 | 533.254 | 1473.75 | 1417.54 | 3691.86 | 3469.35 | 9066.51 | 8839.73 | 12177.5 |
| 12513.5 | 12038.1 | 30197 | 29081.4 | 79710.7 | 77854.3 | 762.502 | 560.18 | 534.712 | 1480.82 | 1420.95 | 3626.17 | 3477.04 | 9210.05 | 8856.18 | 11978.5 |
| 12456.4 | 12065.6 | 30161 | 29139.8 | 79470.2 | 77981.9 | 762.786 | 559.852 | 536.172 | 1477.51 | 1424.38 | 3610.24 | 3484.74 | 9063.01 | 8872.64 | 11952.4 |
| 12523.1 | 12093.1 | 30001.9 | 29198.3 | 79920.5 | 78109.6 | 763.069 | 572.767 | 537.635 | 1531.87 | 1427.8 | 3616.5 | 3492.44 | 9049.97 | 8889.11 | 12070.4 |
| 12536.4 | 12120.7 | 30044.5 | 29256.8 | 79050.2 | 78237.3 | 763.353 | 528.566 | 539.1 | 1467.56 | 1431.23 | 3602.06 | 3500.16 | 9148.07 | 8905.59 | 12004.2 |
| 12725.8 | 12148.3 | 30115.1 | 29315.4 | 80266.3 | 78365.1 | 763.637 | 559.093 | 540.568 | 1541.67 | 1434.67 | 3665.02 | 3507.88 | 9149.89 | 8922.08 | 12029.2 |
| 12594.9 | 12176 | 30210.1 | 29374 | 79643.6 | 78492.9 | 763.921 | 572.221 | 542.038 | 1550 | 1438.11 | 3645.6 | 3515.61 | 9135.35 | 8938.59 | 12080.7 |
| 12442.7 | 12203.7 | 30106.4 | 29432.7 | 78943 | 78620.8 | 764.204 | 562.921 | 543.511 | 1474.91 | 1441.55 | 3672.47 | 3523.35 | 9122.21 | 8955.11 | 11924.6 |
| 12642.6 | 12231.4 | 30393.6 | 29491.5 | 80092.5 | 78748.8 | 764.488 | 534.15 | 544.986 | 1524.58 | 1445 | 3641.76 | 3531.1 | 9208.98 | 8971.64 | 12194.9 |
| 12757.9 | 12259.1 | 30508.7 | 29550.3 | 80642.5 | 78876.8 | 764.772 | 605.595 | 546.463 | 1507.94 | 1448.45 | 3702.01 | 3538.85 | 9228.98 | 8988.18 | 12139.7 |
| 12694.9 | 12286.9 | 30157.7 | 29609.1 | 79997.6 | 79004.9 | 765.056 | 568.24 | 547.943 | 1531.51 | 1451.91 | 3663.59 | 3546.62 | 9287.93 | 9004.73 | 12128.3 |
| 12726.8 | 12314.8 | 30593.6 | 29668 | 80236.4 | 79133.1 | 765.339 | 583.711 | 549.426 | 1518.47 | 1455.37 | 3668.29 | 3554.39 | 9152.72 | 9021.29 | 12177.1 |
| 12852.5 | 12342.6 | 30492.7 | 29727 | 80874.6 | 79261.3 | 765.623 | 574.573 | 550.911 | 1501.57 | 1458.84 | 3672.92 | 3562.18 | 9246.98 | 9037.87 | 12257.9 |
| 12775.5 | 12370.5 | 30561.2 | 29786 | 80544 | 79389.6 | 765.907 | 564.17 | 552.398 | 1574.9 | 1462.31 | 3745.19 | 3569.97 | 9206.66 | 9054.46 | 12259.6 |
| 12716.7 | 12398.4 | 30707.4 | 29845.1 | 80457.7 | 79517.9 | 766.19 | 566.026 | 553.888 | 1527.75 | 1465.79 | 3669.68 | 3577.77 | 9212.67 | 9071.06 | 12249.9 |
| 12892.6 | 12426.4 | 30746.4 | 29904.2 | 80349.1 | 79646.3 | 766.474 | 575.133 | 555.38 | 1542.14 | 1469.27 | 3700.72 | 3585.57 | 9253.03 | 9087.67 | 12272 |
| 12768.4 | 12454.4 | 30613.9 | 29963.4 | 80611.5 | 79774.8 | 766.757 | 568.903 | 556.875 | 1597.21 | 1472.75 | 3716.72 | 3593.39 | 9281.41 | 9104.29 | 12274.2 |
| 12886.8 | 12482.4 | 30752.8 | 30022.6 | 81252 | 79903.3 | 767.041 | 610.007 | 558.372 | 1574.68 | 1476.24 | 3710.92 | 3601.22 | 9322.51 | 9120.92 | 12326.4 |
| 12902.6 | 12510.4 | 31035.1 | 30081.9 | 81423.4 | 80031.9 | 767.324 | 607.952 | 559.872 | 1595.25 | 1479.74 | 3693.13 | 3609.05 | 9391.9 | 9137.57 | 12245.8 |
| 13054.2 | 12538.5 | 30910.5 | 30141.2 | 81780.3 | 80160.5 | 767.608 | 619.714 | 561.374 | 1554.69 | 1483.24 | 3746.94 | 3616.89 | 9339.68 | 9154.22 | 12444.9 |
| 13116.7 | 12566.6 | 31213.9 | 30200.6 | 81674.7 | 80289.2 | 767.891 | 587.356 | 562.878 | 1561.51 | 1486.74 | 3756.75 | 3624.74 | 9332.15 | 9170.89 | 12350.1 |
| 13131.5 | 12594.8 | 31321.9 | 30260 | 82352.6 | 80418 | 768.175 | 570.746 | 564.385 | 1589.12 | 1490.25 | 3776.06 | 3632.6 | 9412.03 | 9187.57 | 12474.7 |
| 12956.8 | 12623 | 30809.7 | 30319.5 | 80843.5 | 80546.8 | 768.459 | 564.968 | 565.895 | 1523.95 | 1493.76 | 3744.19 | 3640.47 | 9306.05 | 9204.26 | 12357.9 |
| 12919 | 12651.2 | 30934.8 | 30379 | 81249.6 | 80675.7 | 768.742 | 580.972 | 567.407 | 1588.06 | 1497.27 | 3756.62 | 3648.34 | 9342.55 | 9220.96 | 12155.9 |
| 13140 | 12679.4 | 31381.9 | 30438.6 | 81754.2 | 80804.6 | 769.026 | 605.114 | 568.921 | 1548.59 | 1500.8 | 3769.99 | 3656.23 | 9420.01 | 9237.67 | 12506.4 |
| 13174.4 | 12707.7 | 31397.1 | 30498.3 | 82494 | 80933.6 | 769.309 | 633.781 | 570.438 | 1569.58 | 1504.32 | 3826.56 | 3664.12 | 9490.34 | 9254.4 | 12519.3 |
| 13385.9 | 12736 | 31612.5 | 30558 | 82413.2 | 81062.7 | 769.592 | 614.6 | 571.958 | 1567.21 | 1507.85 | 3761.26 | 3672.02 | 9472.51 | 9271.13 | 12480.3 |
| 13217.6 | 12764.4 | 31474.9 | 30617.7 | 82684.8 | 81191.8 | 769.876 | 598.873 | 573.48 | 1578.94 | 1511.38 | 3844.92 | 3679.93 | 9465.07 | 9287.88 | 12441.6 |
| 13313.6 | 12792.8 | 31467.6 | 30677.5 | 82349.2 | 81321 | 770.159 | 586.517 | 575.004 | 1573.28 | 1514.92 | 3819.63 | 3687.85 | 9556.85 | 9304.64 | 12510.8 |
| 13311.3 | 12821.2 | 31886.9 | 30737.4 | 83448.4 | 81450.2 | 770.443 | 615.062 | 576.531 | 1527.8 | 1518.46 | 3823.32 | 3695.77 | 9602.75 | 9321.41 | 12611.7 |
| 13185.2 | 12849.6 | 31494.4 | 30797.3 | 82666.2 | 81579.5 | 770.726 | 613.133 | 578.06 | 1606.24 | 1522.01 | 3806.13 | 3703.71 | 9504.03 | 9338.19 | 12617.2 |
| 13260.5 | 12878.1 | 31604.4 | 30857.2 | 83404.7 | 81708.9 | 771.01 | 569.963 | 579.591 | 1567.77 | 1525.57 | 3829.69 | 3711.65 | 9672.68 | 9354.98 | 12591.9 |
| 13347.3 | 12906.6 | 31808.1 | 30917.2 | 82994.1 | 81838.2 | 771.293 | 616.206 | 581.126 | 1599.84 | 1529.12 | 3811.18 | 3719.6 | 9580.12 | 9371.78 | 12612.7 |
| 13215 | 12935.2 | 31912.6 | 30977.3 | 83361.4 | 81967.7 | 771.576 | 592.464 | 582.662 | 1626.06 | 1532.68 | 3838.97 | 3727.56 | 9557.34 | 9388.6 | 12699 |
| 13624.5 | 12963.8 | 32159.5 | 31037.4 | 83773.9 | 82097.2 | 771.86 | 604.45 | 584.201 | 1598 | 1536.25 | 3856.24 | 3735.53 | 9720.24 | 9405.42 | 12664.4 |
| 13465.6 | 12992.4 | 32123.4 | 31097.6 | 83872.5 | 82226.8 | 772.143 | 628.477 | 585.743 | 1582.83 | 1539.82 | 3831.33 | 3743.51 | 9657.98 | 9422.26 | 12555.7 |
| 13603.8 | 13021 | 31935.1 | 31157.8 | 83600.3 | 82356.4 | 772.426 | 615.915 | 587.287 | 1618.89 | 1543.39 | 3863.72 | 3751.49 | 9594.28 | 9439.1 | 12725.2 |
| 13410.8 | 13049.7 | 32114 | 31218 | 83779.7 | 82486.1 | 772.709 | 633.54 | 588.833 | 1598.8 | 1546.97 | 3831.16 | 3759.48 | 9605.28 | 9455.96 | 12766.8 |
| 13410.4 | 13078.4 | 32062.6 | 31278.4 | 84213.2 | 82615.9 | 772.993 | 648.063 | 590.382 | 1570.23 | 1550.55 | 3883.37 | 3767.48 | 9671.64 | 9472.83 | 12710.8 |
| 13520.2 | 13107.2 | 32304.9 | 31338.7 | 84775 | 82745.7 | 773.276 | 636.743 | 591.934 | 1658.38 | 1554.14 | 3930.64 | 3775.49 | 9652.23 | 9489.71 | 12858.6 |
| 13593.9 | 13136 | 32418.3 | 31399.2 | 84623.8 | 82875.6 | 773.559 | 608.801 | 593.488 | 1603.86 | 1557.73 | 3914.74 | 3783.51 | 9685.22 | 9506.6 | 12820.5 |
| 13550.3 | 13164.8 | 32200 | 31459.6 | 84835.6 | 83005.5 | 773.842 | 606.608 | 595.044 | 1627.72 | 1561.33 | 3893.18 | 3791.54 | 9809.06 | 9523.5 | 12817.1 |
| 13689.5 | 13193.6 | 32208.8 | 31520.2 | 84539.4 | 83135.5 | 774.126 | 632.5 | 596.603 | 1655.12 | 1564.93 | 3942.19 | 3799.57 | 9810.69 | 9540.42 | 12901.4 |
| 13756.1 | 13222.5 | 32272.4 | 31580.7 | 84948.3 | 83265.5 | 774.409 | 622.003 | 598.164 | 1627.46 | 1568.54 | 3876.74 | 3807.61 | 9766.6 | 9557.34 | 12817.2 |
| 13632.4 | 13251.4 | 32467.4 | 31641.3 | 84176.4 | 83395.6 | 774.692 | 619.074 | 599.728 | 1646.18 | 1572.15 | 3939.05 | 3815.66 | 9743.58 | 9574.27 | 12861.5 |
| 13885.7 | 13280.4 | 32830.9 | 31702 | 84902.2 | 83525.7 | 774.975 | 607.179 | 601.294 | 1673.05 | 1575.76 | 3977.53 | 3823.72 | 9740.17 | 9591.22 | 12867.2 |
| 13696.4 | 13309.3 | 32463.7 | 31762.8 | 84460.9 | 83655.9 | 775.258 | 649.598 | 602.863 | 1609.32 | 1579.38 | 3930.84 | 3831.79 | 9812.89 | 9608.18 | 12842.6 |
| 13860.4 | 13338.4 | 32912.5 | 31823.5 | 85735.2 | 83786.2 | 775.542 | 633.066 | 604.434 | 1656.15 | 1583 | 3963.99 | 3839.87 | 9916.65 | 9625.14 | 13083.3 |
| 13877 | 13367.4 | 32740.7 | 31884.3 | 85302.8 | 83916.5 | 775.825 | 615.939 | 606.008 | 1673.69 | 1586.63 | 3997.81 | 3847.95 | 9835.95 | 9642.12 | 13063.1 |
| 13896.9 | 13396.5 | 32458.3 | 31945.2 | 85841.7 | 84046.8 | 776.108 | 635.052 | 607.584 | 1637.52 | 1590.26 | 3942.81 | 3856.04 | 9808.95 | 9659.11 | 13006.6 |
| 13776.5 | 13425.6 | 32557.2 | 32006.2 | 84666.1 | 84177.2 | 776.391 | 638.865 | 609.162 | 1634.46 | 1593.9 | 3961.25 | 3864.14 | 9819.82 | 9676.11 | 12928.3 |
| 13820.9 | 13454.7 | 32842.5 | 32067.1 | 85104.5 | 84307.7 | 776.674 | 628.038 | 610.743 | 1623.73 | 1597.54 | 4002.51 | 3872.25 | 9811.75 | 9693.12 | 12980.9 |
| 13916.7 | 13483.9 | 32980.9 | 32128.2 | 85362.9 | 84438.2 | 776.957 | 653.941 | 612.327 | 1643.33 | 1601.18 | 4001.65 | 3880.37 | 10018.5 | 9710.14 | 12988.9 |
| 13874.5 | 13513.1 | 32874.6 | 32189.2 | 85137.4 | 84568.8 | 777.24 | 652.486 | 613.913 | 1660.52 | 1604.83 | 3998.41 | 3888.49 | 9859.48 | 9727.17 | 12882.6 |
| 14026.8 | 13542.3 | 32992.1 | 32250.4 | 85464.2 | 84699.4 | 777.523 | 628.482 | 615.501 | 1682.11 | 1608.49 | 4001.14 | 3896.62 | 9912.72 | 9744.21 | 12990.2 |
| 13890.2 | 13571.6 | 33233.6 | 32311.6 | 86036.3 | 84830.1 | 777.806 | 635.014 | 617.092 | 1689.77 | 1612.14 | 4056.96 | 3904.77 | 9840 | 9761.26 | 13050.8 |
| 14253.8 | 13600.9 | 33113.1 | 32372.8 | 86432.8 | 84960.9 | 778.089 | 633.4 | 618.685 | 1663.89 | 1615.81 | 4082.95 | 3912.91 | 9921.25 | 9778.32 | 13082.4 |
| 14096.1 | 13630.2 | 33411.1 | 32434.1 | 86689.2 | 85091.7 | 778.372 | 653.428 | 620.281 | 1716.68 | 1619.47 | 4070.53 | 3921.07 | 9923.68 | 9795.4 | 13133.2 |
| 14291.4 | 13659.6 | 33331.4 | 32495.4 | 86341.6 | 85222.5 | 778.655 | 677.003 | 621.88 | 1666.74 | 1623.14 | 4027.57 | 3929.24 | 10057 | 9812.48 | 13259.4 |
| 13971.3 | 13689 | 33396.9 | 32556.8 | 86340.4 | 85353.4 | 778.938 | 654.283 | 623.48 | 1681.56 | 1626.82 | 3955.94 | 3937.41 | 9855.13 | 9829.57 | 13055.3 |
| 14131.6 | 13718.4 | 33527 | 32618.2 | 86678.8 | 85484.4 | 779.221 | 654.539 | 625.083 | 1662.28 | 1630.5 | 4017.19 | 3945.59 | 10074.3 | 9846.68 | 13176 |
| 14004 | 13747.9 | 33400.4 | 32679.7 | 86701.4 | 85615.4 | 779.504 | 633.894 | 626.689 | 1734.43 | 1634.19 | 4071.28 | 3953.78 | 10020.7 | 9863.79 | 13207.5 |
| 14260.5 | 13777.4 | 33569.9 | 32741.2 | 86985.8 | 85746.4 | 779.787 | 678.157 | 628.297 | 1681.94 | 1637.87 | 4110.5 | 3961.98 | 10029.1 | 9880.92 | 13188.7 |
| 14144 | 13806.9 | 33476.3 | 32802.8 | 86916.6 | 85877.5 | 780.07 | 661.256 | 629.908 | 1718.96 | 1641.57 | 4079.94 | 3970.19 | 9994.95 | 9898.05 | 13259.3 |
| 14189 | 13836.5 | 33535.4 | 32864.4 | 87330 | 86008.7 | 780.353 | 661.711 | 631.521 | 1723.07 | 1645.27 | 4139.43 | 3978.4 | 10133.4 | 9915.2 | 13372.8 |
| 14119.3 | 13866.1 | 33729.9 | 32926.1 | 87853.7 | 86139.9 | 780.635 | 661.902 | 633.136 | 1704.07 | 1648.97 | 4124.63 | 3986.62 | 10183.7 | 9932.36 | 13237 |
| 14295.2 | 13895.7 | 33529.4 | 32987.8 | 87557.5 | 86271.1 | 780.918 | 646.684 | 634.755 | 1709.96 | 1652.67 | 4099.23 | 3994.85 | 10215.3 | 9949.52 | 13304.5 |
| 14363.6 | 13925.3 | 33643.5 | 33049.6 | 88194.8 | 86402.5 | 781.201 | 658.868 | 636.375 | 1714.27 | 1656.38 | 4127.51 | 4003.09 | 10133.4 | 9966.7 | 13338 |
| 14313 | 13955 | 33807.5 | 33111.4 | 87683.4 | 86533.8 | 781.484 | 674.853 | 637.998 | 1720.6 | 1660.1 | 4095.28 | 4011.34 | 10070.1 | 9983.89 | 13332.3 |
| 14406.5 | 13984.8 | 34054.2 | 33173.3 | 88089.8 | 86665.2 | 781.767 | 677.156 | 639.623 | 1718.26 | 1663.82 | 4143.41 | 4019.59 | 10262.1 | 10001.1 | 13276.4 |
| 14227.5 | 14014.5 | 33610.3 | 33235.2 | 87116.5 | 86796.7 | 782.05 | 673.253 | 641.251 | 1687.17 | 1667.54 | 4067.45 | 4027.86 | 10252.4 | 10018.3 | 13357.6 |
| 14304.2 | 14044.3 | 33957 | 33297.2 | 88302 | 86928.2 | 782.332 | 698.113 | 642.881 | 1681.46 | 1671.27 | 4135.29 | 4036.13 | 10295.3 | 10035.5 | 13380.2 |
| 14610.8 | 14074.1 | 34364 | 33359.2 | 87919.5 | 87059.8 | 782.615 | 661.325 | 644.514 | 1724.52 | 1675 | 4127.02 | 4044.41 | 10141.5 | 10052.7 | 13526.4 |
| 14351.6 | 14104 | 33994.8 | 33421.3 | 87838.8 | 87191.4 | 782.898 | 684.512 | 646.15 | 1753.95 | 1678.74 | 4104.95 | 4052.69 | 10315.6 | 10070 | 13394.9 |
| 14543.8 | 14133.8 | 34229.2 | 33483.4 | 88637.4 | 87323.1 | 783.181 | 672.729 | 647.787 | 1756.47 | 1682.48 | 4166.34 | 4060.99 | 10360.3 | 10087.2 | 13345.1 |
| 14709.8 | 14163.8 | 34449.4 | 33545.6 | 88542.3 | 87454.8 | 783.463 | 678.784 | 649.428 | 1758.52 | 1686.23 | 4144.63 | 4069.29 | 10332.5 | 10104.5 | 13480.4 |
| 14482.1 | 14193.7 | 34104.9 | 33607.8 | 87966.6 | 87586.6 | 783.746 | 697.277 | 651.07 | 1732.62 | 1689.98 | 4167.56 | 4077.6 | 10260.8 | 10121.8 | 13302.9 |
| 14478.3 | 14223.7 | 34560.8 | 33670.1 | 88972.7 | 87718.4 | 784.029 | 703.87 | 652.715 | 1768.46 | 1693.73 | 4132.27 | 4085.92 | 10276.3 | 10139 | 13473.4 |
| 14560.6 | 14253.7 | 33968.4 | 33732.4 | 88214.4 | 87850.2 | 784.311 | 680.465 | 654.363 | 1773.02 | 1697.49 | 4192.07 | 4094.24 | 10171.8 | 10156.3 | 13454.8 |
| 14671 | 14283.7 | 34587.7 | 33794.8 | 88450.1 | 87982.2 | 784.594 | 678.788 | 656.013 | 1790.6 | 1701.25 | 4154.38 | 4102.58 | 10336.4 | 10173.6 | 13569.8 |
| 14655.1 | 14313.8 | 34250.4 | 33857.2 | 88619.6 | 88114.1 | 784.877 | 689.206 | 657.665 | 1744.85 | 1705.02 | 4201.04 | 4110.92 | 10188 | 10190.9 | 13558.6 |
| 14627.2 | 14343.9 | 34634 | 33919.6 | 88853.3 | 88246.1 | 785.159 | 685.506 | 659.32 | 1713.02 | 1708.79 | 4254.51 | 4119.27 | 10348.7 | 10208.2 | 13626.2 |
| 14746.2 | 14374 | 34398.6 | 33982.1 | 89176.4 | 88378.2 | 785.442 | 684.875 | 660.978 | 1784.51 | 1712.57 | 4195.35 | 4127.63 | 10274.1 | 10225.6 | 13624.6 |
| 14694.1 | 14404.2 | 34642.7 | 34044.7 | 89400.6 | 88510.3 | 785.725 | 699.932 | 662.638 | 1773.4 | 1716.35 | 4207.23 | 4135.99 | 10383.5 | 10242.9 | 13525.5 |
| 14688.8 | 14434.4 | 34628.8 | 34107.3 | 89688.3 | 88642.5 | 786.007 | 688.09 | 664.3 | 1757.27 | 1720.13 | 4247.38 | 4144.37 | 10380.1 | 10260.2 | 13662.9 |
| 14638.3 | 14464.6 | 34680.4 | 34169.9 | 89665.8 | 88774.7 | 786.29 | 703.503 | 665.965 | 1757.32 | 1723.92 | 4221.06 | 4152.75 | 10351.1 | 10277.6 | 13636 |
| 14775.8 | 14494.9 | 34749.8 | 34232.6 | 89117.3 | 88906.9 | 786.572 | 710.791 | 667.632 | 1757.42 | 1727.71 | 4216.76 | 4161.14 | 10395.3 | 10295 | 13529.9 |
| 14942.2 | 14525.2 | 35300.8 | 34295.4 | 90255.9 | 89039.2 | 786.855 | 692.191 | 669.302 | 1791.27 | 1731.51 | 4262.17 | 4169.54 | 10426.7 | 10312.4 | 13782 |
| 14958.8 | 14555.5 | 34975.6 | 34358.2 | 90012 | 89171.6 | 787.137 | 720.004 | 670.974 | 1783.66 | 1735.31 | 4213.11 | 4177.94 | 10506.3 | 10329.7 | 13597.2 |
| 14852.6 | 14585.9 | 34897.9 | 34421 | 90798.2 | 89304 | 787.42 | 714.396 | 672.649 | 1824.33 | 1739.12 | 4331 | 4186.36 | 10574.9 | 10347.1 | 13833.7 |
| 14820.6 | 14616.3 | 35103.6 | 34483.9 | 90484 | 89436.4 | 787.702 | 717.217 | 674.326 | 1805.63 | 1742.93 | 4326.92 | 4194.78 | 10522.8 | 10364.5 | 13895 |
| 14867.6 | 14646.7 | 35295.1 | 34546.9 | 90200.4 | 89568.9 | 787.985 | 712.71 | 676.005 | 1820.02 | 1746.74 | 4240.07 | 4203.21 | 10522.7 | 10382 | 13618.5 |
| 14952.9 | 14677.1 | 35383.5 | 34609.8 | 90083 | 89701.4 | 788.267 | 702.554 | 677.687 | 1785 | 1750.56 | 4221.41 | 4211.65 | 10456.1 | 10399.4 | 13793.1 |
| 15172.9 | 14707.6 | 35059.2 | 34672.9 | 90595.6 | 89834 | 788.55 | 720.777 | 679.372 | 1787.04 | 1754.39 | 4360.02 | 4220.09 | 10558.6 | 10416.8 | 13830.8 |
| 15045.5 | 14738.1 | 35117.4 | 34736 | 90179.2 | 89966.7 | 788.832 | 716.018 | 681.059 | 1809.64 | 1758.21 | 4310.49 | 4228.54 | 10489.8 | 10434.3 | 13611 |
| 15251.9 | 14768.7 | 35408.2 | 34799.1 | 91049.4 | 90099.3 | 789.115 | 704.805 | 682.748 | 1857.71 | 1762.05 | 4308.53 | 4237 | 10487.8 | 10451.7 | 13917.1 |
| 15199 | 14799.2 | 35518.1 | 34862.3 | 91159.3 | 90232.1 | 789.397 | 738.23 | 684.44 | 1868.69 | 1765.88 | 4384.4 | 4245.47 | 10552.9 | 10469.2 | 13897.1 |
| 14964.3 | 14829.9 | 35513.7 | 34925.5 | 91475.2 | 90364.8 | 789.679 | 696.661 | 686.134 | 1837.09 | 1769.72 | 4328.92 | 4253.95 | 10491.2 | 10486.7 | 13857 |
| 15265.3 | 14860.5 | 35849.9 | 34988.8 | 91280.9 | 90497.6 | 789.962 | 728.697 | 687.831 | 1832.84 | 1773.57 | 4419.35 | 4262.43 | 10688 | 10504.1 | 13780.2 |
| 15225.8 | 14891.2 | 35645.7 | 35052.1 | 90585.3 | 90630.5 | 790.244 | 749.877 | 689.53 | 1815.88 | 1777.41 | 4365.68 | 4270.92 | 10514.1 | 10521.6 | 13827.6 |
| 15369.5 | 14921.9 | 35627.5 | 35115.4 | 91782.7 | 90763.4 | 790.526 | 714.394 | 691.232 | 1816.78 | 1781.27 | 4354.11 | 4279.42 | 10653.1 | 10539.1 | 13978.6 |
| 15156 | 14952.6 | 35838.2 | 35178.8 | 91700.3 | 90896.4 | 790.809 | 713.384 | 692.936 | 1830.53 | 1785.12 | 4374.23 | 4287.93 | 10684.3 | 10556.6 | 13892.1 |
| 15164.4 | 14983.4 | 35781.3 | 35242.3 | 91114.9 | 91029.4 | 791.091 | 720.656 | 694.643 | 1858.62 | 1788.98 | 4359.28 | 4296.44 | 10579.6 | 10574.2 | 14069.3 |
| 15171.1 | 15014.2 | 35784.6 | 35305.8 | 91407.9 | 91162.4 | 791.373 | 688.295 | 696.352 | 1831.99 | 1792.85 | 4412.17 | 4304.97 | 10617 | 10591.7 | 13965.8 |
| 15148.8 | 15045 | 35837.8 | 35369.4 | 91714.3 | 91295.5 | 791.655 | 723.735 | 698.063 | 1835.86 | 1796.72 | 4396.91 | 4313.5 | 10638.8 | 10609.2 | 13846 |
| 15219.8 | 15075.9 | 36094.1 | 35433 | 91584.3 | 91428.6 | 791.938 | 719.668 | 699.777 | 1842.36 | 1800.6 | 4377.45 | 4322.03 | 10744 | 10626.8 | 14092.1 |
| 15305.1 | 15106.8 | 35574.3 | 35496.6 | 90893.7 | 91561.8 | 792.22 | 734.007 | 701.494 | 1855.64 | 1804.47 | 4363.82 | 4330.58 | 10718.1 | 10644.4 | 14091.9 |
| 15415.1 | 15137.7 | 35947.7 | 35560.3 | 91959.4 | 91695 | 792.502 | 720.359 | 703.213 | 1849.06 | 1808.36 | 4384.45 | 4339.13 | 10794 | 10661.9 | 13990.1 |
| 15684.2 | 15168.6 | 36184.5 | 35624 | 92105.7 | 91828.3 | 792.784 | 743.506 | 704.934 | 1885.83 | 1812.24 | 4466.98 | 4347.69 | 10819.3 | 10679.5 | 14120.9 |
| 15547.1 | 15199.6 | 36489.4 | 35687.8 | 92709.5 | 91961.6 | 793.067 | 757.785 | 706.658 | 1843.89 | 1816.13 | 4441.9 | 4356.26 | 10802.5 | 10697.1 | 14126.2 |
| 15413.2 | 15230.6 | 35631.9 | 35751.6 | 92166.4 | 92094.9 | 793.349 | 694.824 | 708.384 | 1820.97 | 1820.03 | 4382.75 | 4364.84 | 10691.9 | 10714.7 | 13985.2 |
| 15389.2 | 15261.7 | 35978.8 | 35815.5 | 91870.8 | 92228.3 | 793.631 | 722.135 | 710.113 | 1867.36 | 1823.93 | 4366.16 | 4373.42 | 10760.8 | 10732.3 | 14016.7 |
| 15429.8 | 15292.8 | 36069.5 | 35879.4 | 91957.8 | 92361.7 | 793.913 | 735.761 | 711.844 | 1860.36 | 1827.83 | 4404.07 | 4382.02 | 10801.8 | 10749.9 | 13994 |
| 15581.2 | 15323.9 | 36233.8 | 35943.4 | 92706.5 | 92495.2 | 794.195 | 725.609 | 713.577 | 1843.55 | 1831.74 | 4472.77 | 4390.61 | 10813.8 | 10767.5 | 14287.2 |
| 15549.6 | 15355 | 36300.5 | 36007.4 | 92411.6 | 92628.7 | 794.477 | 731.124 | 715.313 | 1889.64 | 1835.65 | 4487.22 | 4399.22 | 10759.4 | 10785.1 | 14039.8 |
| 15596.3 | 15386.2 | 36687.7 | 36071.5 | 91999.6 | 92762.3 | 794.76 | 752.578 | 717.052 | 1867.74 | 1839.57 | 4478.83 | 4407.84 | 10753.5 | 10802.8 | 14172.5 |
| 15591 | 15417.4 | 36543.2 | 36135.6 | 92710.6 | 92895.9 | 795.042 | 748.613 | 718.793 | 1863.45 | 1843.49 | 4521.24 | 4416.46 | 10826.2 | 10820.5 | 14179.2 |
| 15790.9 | 15448.6 | 36417.9 | 36199.7 | 92784.2 | 93029.6 | 795.324 | 731.239 | 720.536 | 1932.39 | 1847.41 | 4489.42 | 4425.09 | 10821.8 | 10838.1 | 14026.4 |
| 15689.2 | 15479.9 | 36606.7 | 36263.9 | 93447.4 | 93163.2 | 795.606 | 731.737 | 722.282 | 1936.9 | 1851.34 | 4526.47 | 4433.72 | 10939.3 | 10855.8 | 14250.6 |
| 15756 | 15511.1 | 36718.4 | 36328.2 | 94094.7 | 93297 | 795.888 | 769.481 | 724.03 | 1896.7 | 1855.28 | 4531.06 | 4442.37 | 11059.3 | 10873.5 | 14243.2 |
| 15589.3 | 15542.5 | 36555.1 | 36392.4 | 93608 | 93430.7 | 796.17 | 745.462 | 725.781 | 1894.53 | 1859.21 | 4523.29 | 4451.02 | 10726.5 | 10891.1 | 14301.2 |
| 15728.1 | 15573.8 | 36486.1 | 36456.8 | 92808.8 | 93564.6 | 796.452 | 752.873 | 727.534 | 1883.94 | 1863.15 | 4550.92 | 4459.68 | 10806.6 | 10908.8 | 14234.6 |
| 15750.4 | 15605.2 | 36971.2 | 36521.2 | 93786.3 | 93698.4 | 796.734 | 759.096 | 729.29 | 1943.42 | 1867.1 | 4561.2 | 4468.35 | 10861.5 | 10926.5 | 14326.3 |
| 15940.8 | 15636.6 | 36970.1 | 36585.6 | 94251.5 | 93832.3 | 797.016 | 706.36 | 731.048 | 1920.68 | 1871.05 | 4525.52 | 4477.02 | 10943.8 | 10944.2 | 14219.3 |
| 15722.9 | 15668.1 | 36862.5 | 36650.1 | 92857.8 | 93966.2 | 797.298 | 762.035 | 732.809 | 1899.49 | 1875 | 4493.29 | 4485.71 | 10968.5 | 10962 | 14440.8 |
| 15890.9 | 15699.5 | 37094.7 | 36714.6 | 94357.9 | 94100.2 | 797.58 | 744.709 | 734.572 | 1881.5 | 1878.96 | 4581.24 | 4494.4 | 11092.1 | 10979.7 | 14488.4 |
| 15995.9 | 15731 | 37102.7 | 36779.1 | 94097.3 | 94234.2 | 797.862 | 744.197 | 736.337 | 1926.42 | 1882.92 | 4570.27 | 4503.09 | 11016.1 | 10997.5 | 14224.5 |
| 16048.9 | 15762.6 | 37514.2 | 36843.7 | 94759.4 | 94368.3 | 798.144 | 741.947 | 738.105 | 1928.49 | 1886.89 | 4668.44 | 4511.8 | 11016 | 11015.2 | 14566 |
| 15941.9 | 15794.2 | 37112.6 | 36908.4 | 94364.1 | 94502.4 | 798.426 | 770.449 | 739.875 | 1981.16 | 1890.86 | 4564.17 | 4520.51 | 11194.8 | 11033 | 14462.5 |
| 16178 | 15825.8 | 37123.6 | 36973.1 | 94631.2 | 94636.5 | 798.707 | 770.382 | 741.648 | 1896.65 | 1894.84 | 4630.09 | 4529.23 | 11111.6 | 11050.7 | 14461.9 |
| 16066.6 | 15857.4 | 37462.4 | 37037.8 | 94524.8 | 94770.7 | 798.989 | 791.402 | 743.423 | 1963.1 | 1898.82 | 4652.63 | 4537.96 | 11055.6 | 11068.5 | 14497.9 |
| 16039.9 | 15889 | 37086.8 | 37102.6 | 94589.8 | 94904.9 | 799.271 | 779.186 | 745.201 | 1931 | 1902.8 | 4622.21 | 4546.69 | 10984.4 | 11086.3 | 14512.5 |
| 16322.5 | 15920.7 | 37719.8 | 37167.4 | 95386 | 95039.2 | 799.553 | 795.327 | 746.981 | 1954.95 | 1906.79 | 4576.38 | 4555.44 | 11167.4 | 11104.1 | 14437 |
| 16129.5 | 15952.5 | 37600.5 | 37232.3 | 94736.4 | 95173.5 | 799.835 | 755.502 | 748.763 | 1969.9 | 1910.78 | 4614.87 | 4564.19 | 11247.8 | 11121.9 | 14453.6 |
| 16404.1 | 15984.2 | 37875.6 | 37297.2 | 95859.4 | 95307.8 | 800.117 | 796.68 | 750.548 | 1996.98 | 1914.77 | 4651.81 | 4572.94 | 11238.5 | 11139.7 | 14700.7 |
| 16231.6 | 16016 | 37695.1 | 37362.1 | 95574.1 | 95442.2 | 800.399 | 767.62 | 752.336 | 1966.85 | 1918.77 | 4584.97 | 4581.71 | 11183.1 | 11157.5 | 14650.1 |
| 16332.5 | 16047.8 | 37722.7 | 37427.1 | 95876.7 | 95576.6 | 800.68 | 754.592 | 754.126 | 1977.95 | 1922.78 | 4605.5 | 4590.48 | 11210.4 | 11175.3 | 14718.8 |
| 16255 | 16079.6 | 37826.8 | 37492.2 | 95552.9 | 95711.1 | 800.962 | 801.103 | 755.918 | 2000.81 | 1926.78 | 4628.64 | 4599.26 | 11245.6 | 11193.2 | 14703.3 |
| 16196.4 | 16111.5 | 37998.8 | 37557.3 | 95935.4 | 95845.6 | 801.244 | 790.707 | 757.712 | 1981.39 | 1930.8 | 4576.04 | 4608.05 | 11322.5 | 11211 | 14579.2 |
| 16356.3 | 16143.4 | 37885.4 | 37622.4 | 96329.2 | 95980.1 | 801.525 | 798.856 | 759.51 | 1939.9 | 1934.81 | 4680.29 | 4616.84 | 11290.5 | 11228.9 | 14628.5 |
| 16306.2 | 16175.3 | 38148 | 37687.6 | 96060.9 | 96114.7 | 801.807 | 783.376 | 761.309 | 1976.17 | 1938.83 | 4668.16 | 4625.64 | 11245.4 | 11246.7 | 14644.1 |
| 16282.7 | 16207.3 | 37597.2 | 37752.8 | 95643.7 | 96249.3 | 802.089 | 768.252 | 763.111 | 1938.29 | 1942.86 | 4630.61 | 4634.45 | 11209.1 | 11264.6 | 14579.6 |
| 16400.5 | 16239.3 | 37926.4 | 37818.1 | 95908.1 | 96383.9 | 802.371 | 775.31 | 764.916 | 2023.7 | 1946.88 | 4650.31 | 4643.27 | 11324.3 | 11282.5 | 14703.1 |
| 16471.4 | 16271.3 | 38501.6 | 37883.4 | 96559.6 | 96518.6 | 802.652 | 797.866 | 766.723 | 1975.19 | 1950.92 | 4749.87 | 4652.09 | 11249 | 11300.4 | 14799.1 |
| 16504.3 | 16303.4 | 38260.7 | 37948.7 | 96774.8 | 96653.3 | 802.934 | 796.722 | 768.532 | 1984.17 | 1954.95 | 4750.59 | 4660.92 | 11291 | 11318.2 | 14817.2 |
| 16567.6 | 16335.4 | 38350.8 | 38014.1 | 96738.8 | 96788.1 | 803.216 | 818.091 | 770.344 | 1973.99 | 1958.99 | 4677.97 | 4669.76 | 11280.5 | 11336.1 | 14782.4 |
| 16512.6 | 16367.6 | 38282 | 38079.6 | 96818.9 | 96922.9 | 803.497 | 789.13 | 772.158 | 1985.07 | 1963.04 | 4820.27 | 4678.61 | 11322.1 | 11354.1 | 14853.9 |
| 16578.6 | 16399.7 | 38019.7 | 38145.1 | 95881.6 | 97057.7 | 803.779 | 806.645 | 773.975 | 1993.93 | 1967.09 | 4695.92 | 4687.46 | 11234.5 | 11372 | 14685.5 |
| 16644.4 | 16431.9 | 38374.1 | 38210.6 | 97418.5 | 97192.6 | 804.06 | 810.931 | 775.794 | 2049.87 | 1971.14 | 4774.7 | 4696.32 | 11544.9 | 11389.9 | 14804.2 |
| 16406 | 16464.1 | 38007.1 | 38276.1 | 96316 | 97327.5 | 804.342 | 795.64 | 777.615 | 1977.7 | 1975.2 | 4762.16 | 4705.19 | 11347.1 | 11407.8 | 14761 |
| 16553.3 | 16496.3 | 38589.3 | 38341.8 | 97102.9 | 97462.4 | 804.624 | 796.934 | 779.439 | 2037.86 | 1979.26 | 4776 | 4714.06 | 11488.3 | 11425.8 | 14970.8 |
| 16638.9 | 16528.6 | 38685.5 | 38407.4 | 97420.7 | 97597.4 | 804.905 | 847.997 | 781.265 | 2044.4 | 1983.32 | 4779.41 | 4722.94 | 11458.8 | 11443.7 | 14897.9 |
| 16749.7 | 16560.8 | 38694.6 | 38473.1 | 97652.5 | 97732.4 | 805.187 | 789.213 | 783.094 | 2000.07 | 1987.39 | 4764.86 | 4731.83 | 11389.2 | 11461.7 | 15029 |
| 16747.2 | 16593.2 | 38921.3 | 38538.8 | 97660.3 | 97867.4 | 805.468 | 830.227 | 784.925 | 2030.84 | 1991.47 | 4804.96 | 4740.73 | 11456.5 | 11479.7 | 15017.3 |
| 16724.8 | 16625.5 | 38705.3 | 38604.6 | 97391 | 98002.5 | 805.75 | 788.857 | 786.759 | 2067.63 | 1995.54 | 4806.31 | 4749.63 | 11484.4 | 11497.6 | 15010.6 |
| 16753 | 16657.9 | 38820.2 | 38670.4 | 98092.1 | 98137.6 | 806.031 | 814.982 | 788.595 | 2031.01 | 1999.62 | 4816.85 | 4758.54 | 11570 | 11515.6 | 15052.9 |
| 16770.5 | 16690.3 | 39185.1 | 38736.3 | 97605 | 98272.8 | 806.313 | 801.128 | 790.433 | 2064.51 | 2003.71 | 4694.16 | 4767.46 | 11358.8 | 11533.6 | 14966.9 |
| 17036.1 | 16722.8 | 38917.1 | 38802.2 | 97774.1 | 98408 | 806.594 | 834.351 | 792.274 | 2029.75 | 2007.8 | 4820.51 | 4776.39 | 11451.9 | 11551.6 | 15053.4 |
| 16816.4 | 16755.2 | 38960.5 | 38868.2 | 97592.5 | 98543.2 | 806.875 | 801.379 | 794.118 | 2038.12 | 2011.89 | 4799.37 | 4785.32 | 11545.7 | 11569.6 | 15098.9 |
| 17040.2 | 16787.7 | 39099.2 | 38934.2 | 98430.7 | 98678.5 | 807.157 | 809.941 | 795.963 | 2057.94 | 2015.99 | 4829.23 | 4794.26 | 11525.5 | 11587.6 | 14954 |
| 16862.8 | 16820.2 | 39391.6 | 39000.2 | 98539.2 | 98813.8 | 807.438 | 834.768 | 797.812 | 2079.58 | 2020.09 | 4828.97 | 4803.2 | 11566.4 | 11605.7 | 15021.5 |
| 16825 | 16852.8 | 39535 | 39066.3 | 98277.2 | 98949.1 | 807.72 | 851.152 | 799.662 | 2071.26 | 2024.2 | 4851.78 | 4812.16 | 11573.5 | 11623.7 | 15170.3 |
| 17049.6 | 16885.4 | 39063.1 | 39132.4 | 98790.1 | 99084.4 | 808.001 | 807.107 | 801.515 | 2091.76 | 2028.31 | 4883.11 | 4821.12 | 11745.3 | 11641.8 | 15329.6 |
| 17024.9 | 16918 | 39265.6 | 39198.6 | 98427.8 | 99219.8 | 808.283 | 837.572 | 803.371 | 2093.95 | 2032.42 | 4811.02 | 4830.09 | 11553 | 11659.8 | 15024 |
| 17044.9 | 16950.7 | 39496.1 | 39264.8 | 99275.8 | 99355.3 | 808.564 | 821.163 | 805.228 | 2095.18 | 2036.54 | 4874.94 | 4839.06 | 11693.2 | 11677.9 | 15323 |
| 16993.6 | 16983.3 | 39402.1 | 39331 | 98720.7 | 99490.7 | 808.845 | 806.884 | 807.089 | 2054.56 | 2040.66 | 4831.94 | 4848.05 | 11703.5 | 11695.9 | 15225.4 |
| 17275.8 | 17016 | 39569.3 | 39397.3 | 99488.5 | 99626.2 | 809.126 | 803.003 | 808.951 | 2081.62 | 2044.79 | 4912.68 | 4857.03 | 11613.8 | 11714 | 15230.8 |
| 16913.4 | 17048.8 | 39262 | 39463.7 | 99173.9 | 99761.7 | 809.408 | 850.151 | 810.817 | 2081.24 | 2048.91 | 4886.09 | 4866.03 | 11615.1 | 11732.1 | 15196.7 |
| 17098 | 17081.5 | 39644.6 | 39530 | 99493.9 | 99897.3 | 809.689 | 823.925 | 812.684 | 2110.3 | 2053.05 | 4867.39 | 4875.03 | 11609.8 | 11750.2 | 15241.2 |
| 17087.7 | 17114.3 | 40005.7 | 39596.4 | 99784.4 | 100033 | 809.97 | 830.275 | 814.554 | 2141.43 | 2057.19 | 4951.77 | 4884.05 | 11822.8 | 11768.2 | 15356.1 |
| 17254.1 | 17147.1 | 39633.9 | 39662.9 | 99336.8 | 100168 | 810.252 | 821.59 | 816.426 | 2081.43 | 2061.33 | 4911.97 | 4893.06 | 11781.5 | 11786.4 | 15316.8 |
| 17228.8 | 17180 | 39475.9 | 39729.4 | 99532.6 | 100304 | 810.533 | 839.457 | 818.301 | 2120.69 | 2065.47 | 4881.85 | 4902.09 | 11702.2 | 11804.5 | 15250.5 |
| 17134.2 | 17212.9 | 39708.1 | 39795.9 | 99410.9 | 100440 | 810.814 | 833.414 | 820.178 | 2082.31 | 2069.62 | 4895.63 | 4911.12 | 11781.2 | 11822.6 | 15262.5 |
| 17454.4 | 17245.8 | 39818.2 | 39862.5 | 99531.1 | 100576 | 811.095 | 873.428 | 822.058 | 2093.08 | 2073.78 | 4903.28 | 4920.16 | 11756 | 11840.7 | 15252 |
| 17227 | 17278.7 | 39674.8 | 39929.1 | 99141.1 | 100711 | 811.376 | 872.558 | 823.94 | 2040.91 | 2077.93 | 4890.99 | 4929.2 | 11704.9 | 11858.9 | 15220.4 |
| 17461.7 | 17311.7 | 39635.2 | 39995.8 | 99450.8 | 100847 | 811.658 | 822.857 | 825.824 | 2094.55 | 2082.09 | 4871.72 | 4938.26 | 11776.2 | 11877 | 15231.4 |
| 17365.2 | 17344.7 | 39946.4 | 40062.5 | 99793.4 | 100983 | 811.939 | 866.18 | 827.711 | 2011.16 | 2086.26 | 4929.29 | 4947.32 | 11768.3 | 11895.1 | 15232.3 |
| 17224.7 | 17377.7 | 39839.1 | 40129.2 | 99343.2 | 101119 | 812.22 | 869.575 | 829.6 | 2105.72 | 2090.43 | 4886.12 | 4956.38 | 11714.7 | 11913.3 | 15306 |
| 17368.8 | 17410.7 | 40261.9 | 40196 | 99646.9 | 101255 | 812.501 | 824.838 | 831.492 | 2130.84 | 2094.6 | 4973.07 | 4965.46 | 11839 | 11931.5 | 15344.6 |
| 17390.3 | 17443.8 | 40023.5 | 40262.9 | 99775.8 | 101391 | 812.782 | 821.178 | 833.386 | 2170.77 | 2098.78 | 4952.73 | 4974.54 | 11793.4 | 11949.6 | 15430.5 |
| 17485.7 | 17476.9 | 40189.4 | 40329.7 | 100735 | 101526 | 813.063 | 843.257 | 835.282 | 2131.31 | 2102.96 | 5025.98 | 4983.62 | 11889.5 | 11967.8 | 15526.6 |
| 17359.8 | 17510.1 | 40309.3 | 40396.6 | 100582 | 101662 | 813.344 | 843.681 | 837.181 | 2089.58 | 2107.15 | 5029.97 | 4992.72 | 11860.6 | 11986 | 15287.8 |
| 17624 | 17543.2 | 40390.5 | 40463.6 | 100532 | 101798 | 813.625 | 835.92 | 839.082 | 2090.66 | 2111.34 | 4973.33 | 5001.82 | 11921.2 | 12004.2 | 15463.6 |
| 17686.5 | 17576.4 | 40475.5 | 40530.6 | 100382 | 101935 | 813.907 | 842.083 | 840.986 | 2145.01 | 2115.53 | 4977.14 | 5010.93 | 11898.7 | 12022.4 | 15418 |
| 17700.3 | 17609.7 | 40623.7 | 40597.6 | 100587 | 102071 | 814.187 | 845.375 | 842.892 | 2175.93 | 2119.73 | 4977.19 | 5020.04 | 11920.8 | 12040.6 | 15449 |
| 17702 | 17642.9 | 40737.7 | 40664.7 | 100689 | 102207 | 814.468 | 871.468 | 844.801 | 2141.77 | 2123.93 | 5086.19 | 5029.17 | 11956.1 | 12058.8 | 15547 |
| 17626.5 | 17676.2 | 40314.1 | 40731.8 | 100566 | 102343 | 814.75 | 880.021 | 846.712 | 2098.2 | 2128.13 | 5080.62 | 5038.3 | 11992.5 | 12077 | 15591.1 |
| 17700.7 | 17709.5 | 40974.9 | 40798.9 | 100895 | 102479 | 815.03 | 890.064 | 848.625 | 2156.15 | 2132.34 | 5097.7 | 5047.43 | 11984.7 | 12095.3 | 15574.4 |
| 17958.1 | 17742.8 | 40872.8 | 40866.1 | 101721 | 102615 | 815.312 | 863.679 | 850.54 | 2110.44 | 2136.55 | 5008.23 | 5056.57 | 12013 | 12113.5 | 15691.6 |
| 17776.2 | 17776.2 | 40248.3 | 40933.3 | 101965 | 102751 | 815.592 | 880.736 | 852.458 | 2175.19 | 2140.77 | 5011.57 | 5065.72 | 11911.4 | 12131.8 | 15658.4 |
| 17916.3 | 17809.6 | 41176.4 | 41000.6 | 102066 | 102888 | 815.873 | 836.074 | 854.379 | 2190.24 | 2144.99 | 5047.72 | 5074.88 | 12065.1 | 12150 | 15683.8 |
| 17713.8 | 17843 | 41170.7 | 41067.9 | 102276 | 103024 | 816.154 | 865.303 | 856.302 | 2200.33 | 2149.22 | 5094.18 | 5084.04 | 12175.6 | 12168.3 | 15617.7 |
| 17979.4 | 17876.5 | 40874.8 | 41135.3 | 102326 | 103160 | 816.435 | 912.628 | 858.227 | 2178.65 | 2153.44 | 5103.12 | 5093.21 | 12088.2 | 12186.5 | 15718.1 |
| 17917.1 | 17910 | 41158.4 | 41202.7 | 101808 | 103297 | 816.716 | 879.144 | 860.155 | 2161.14 | 2157.68 | 5106.43 | 5102.39 | 12147.6 | 12204.8 | 15755.5 |
| 18036.8 | 17943.5 | 40979.8 | 41270.1 | 101803 | 103433 | 816.997 | 878.868 | 862.085 | 2146.53 | 2161.91 | 5078.17 | 5111.57 | 12032.1 | 12223.1 | 15644.1 |
| 18116.5 | 17977 | 41098.8 | 41337.6 | 102585 | 103569 | 817.278 | 865.618 | 864.017 | 2167.72 | 2166.15 | 5167.99 | 5120.77 | 12154.3 | 12241.4 | 15755.4 |
| 17998.7 | 18010.6 | 41325.9 | 41405.1 | 102205 | 103706 | 817.559 | 870.562 | 865.952 | 2161.48 | 2170.4 | 5118.25 | 5129.96 | 12116.4 | 12259.7 | 15746 |
| 17965.1 | 18044.2 | 41565.1 | 41472.6 | 102537 | 103842 | 817.84 | 881.356 | 867.889 | 2182.72 | 2174.64 | 5134.27 | 5139.17 | 12219.2 | 12278 | 15901.5 |
| 18155.7 | 18077.8 | 41614.2 | 41540.2 | 102342 | 103979 | 818.12 | 898.298 | 869.829 | 2184.93 | 2178.9 | 5183.1 | 5148.38 | 12215.8 | 12296.3 | 15920.8 |
| 18130.9 | 18111.5 | 41536.6 | 41607.8 | 102712 | 104115 | 818.401 | 864.136 | 871.771 | 2167.17 | 2183.15 | 5093.39 | 5157.59 | 12153.1 | 12314.6 | 15958.2 |
| 18175.5 | 18145.2 | 41736.8 | 41675.5 | 103374 | 104252 | 818.682 | 871.126 | 873.715 | 2212.41 | 2187.41 | 5188.86 | 5166.82 | 12335.3 | 12332.9 | 16056.4 |
| 18207.1 | 18178.9 | 41549.9 | 41743.2 | 103159 | 104388 | 818.963 | 878.99 | 875.662 | 2177.52 | 2191.67 | 5253.54 | 5176.05 | 12260.9 | 12351.3 | 15890.9 |
| 18272.6 | 18212.6 | 42013.2 | 41810.9 | 103933 | 104525 | 819.243 | 894.015 | 877.611 | 2222.11 | 2195.94 | 5166.55 | 5185.29 | 12316.1 | 12369.6 | 15929.7 |
| 18294.8 | 18246.4 | 41783.6 | 41878.7 | 104036 | 104661 | 819.524 | 931.935 | 879.562 | 2243.16 | 2200.21 | 5085.55 | 5194.53 | 12354 | 12387.9 | 16027.3 |
| 18345.8 | 18280.2 | 41882.9 | 41946.5 | 105070 | 104798 | 819.805 | 885.712 | 881.516 | 2207.23 | 2204.49 | 5269.65 | 5203.78 | 12333.8 | 12406.3 | 16208.2 |
| 18301.7 | 18314 | 41589.3 | 42014.4 | 103364 | 104935 | 820.086 | 896.825 | 883.473 | 2241.66 | 2208.76 | 5190.27 | 5213.04 | 12152.7 | 12424.6 | 16002.6 |
| 18444.5 | 18347.8 | 42183.4 | 42082.3 | 104319 | 105071 | 820.366 | 902.171 | 885.431 | 2217.78 | 2213.05 | 5235.93 | 5222.3 | 12497.5 | 12443 | 15875.1 |
| 18506.7 | 18381.7 | 41912.1 | 42150.2 | 104481 | 105208 | 820.647 | 887.144 | 887.392 | 2231.98 | 2217.33 | 5180.74 | 5231.57 | 12357 | 12461.4 | 16147.2 |
| 18554.5 | 18415.6 | 42300.2 | 42218.2 | 104491 | 105345 | 820.928 | 895.892 | 889.356 | 2262.57 | 2221.62 | 5198.6 | 5240.85 | 12434.9 | 12479.8 | 16289.2 |
| 18420.9 | 18449.5 | 42427.1 | 42286.2 | 104983 | 105482 | 821.208 | 927.902 | 891.322 | 2258.3 | 2225.92 | 5275.61 | 5250.14 | 12480.6 | 12498.1 | 16326.3 |
| 18405.5 | 18483.5 | 42228.2 | 42354.2 | 104039 | 105618 | 821.489 | 908.718 | 893.29 | 2220.45 | 2230.22 | 5355.84 | 5259.43 | 12408.6 | 12516.5 | 16119.6 |
| 18719 | 18517.5 | 42294 | 42422.3 | 104869 | 105755 | 821.769 | 914.11 | 895.26 | 2217.46 | 2234.52 | 5257.02 | 5268.73 | 12462.5 | 12535 | 16215.2 |
| 18616.8 | 18551.5 | 42734.6 | 42490.5 | 105579 | 105892 | 822.05 | 929.132 | 897.233 | 2266.43 | 2238.82 | 5246.92 | 5278.03 | 12444.7 | 12553.4 | 16387.2 |
| 18849.7 | 18585.6 | 42625.1 | 42558.6 | 105306 | 106029 | 822.331 | 935.467 | 899.208 | 2291.35 | 2243.13 | 5328.54 | 5287.34 | 12496.7 | 12571.8 | 16081.8 |
| 18621.7 | 18619.6 | 42856.2 | 42626.8 | 105690 | 106166 | 822.611 | 941.669 | 901.186 | 2263.77 | 2247.44 | 5326.37 | 5296.66 | 12643.3 | 12590.2 | 16385.3 |
| 18886.1 | 18653.7 | 42755.7 | 42695.1 | 106408 | 106303 | 822.892 | 905.235 | 903.166 | 2303.21 | 2251.76 | 5367.82 | 5305.98 | 12533.8 | 12608.6 | 16422.6 |
| 18916.6 | 18687.8 | 42787.2 | 42763.3 | 106191 | 106440 | 823.172 | 910.905 | 905.149 | 2299.3 | 2256.08 | 5394.07 | 5315.31 | 12651.8 | 12627 | 16368.7 |
| 18647.1 | 18722 | 42899.8 | 42831.7 | 106114 | 106577 | 823.453 | 926.617 | 907.133 | 2293.39 | 2260.41 | 5323.86 | 5324.65 | 12630.8 | 12645.5 | 16349.2 |
| 18879.4 | 18756.2 | 43003.8 | 42900 | 106992 | 106714 | 823.733 | 936.661 | 909.12 | 2278.84 | 2264.73 | 5388.56 | 5333.99 | 12690.1 | 12663.9 | 16407.9 |
| 18927.1 | 18790.4 | 43203.8 | 42968.4 | 106138 | 106851 | 824.014 | 944.866 | 911.11 | 2286.23 | 2269.07 | 5283.76 | 5343.34 | 12590.7 | 12682.4 | 16319.2 |
| 18665.8 | 18824.6 | 42847.7 | 43036.8 | 105937 | 106988 | 824.294 | 935.719 | 913.102 | 2252.39 | 2273.4 | 5295.69 | 5352.7 | 12702.6 | 12700.8 | 16174.3 |
| 18994 | 18858.9 | 42864.9 | 43105.3 | 106301 | 107125 | 824.574 | 945.692 | 915.096 | 2313.25 | 2277.74 | 5392.1 | 5362.06 | 12499.6 | 12719.3 | 16471.5 |
| 18999.4 | 18893.2 | 43259 | 43173.8 | 106453 | 107262 | 824.855 | 929.207 | 917.093 | 2270.27 | 2282.08 | 5423.43 | 5371.43 | 12720.7 | 12737.8 | 16343 |
| 18977 | 18927.5 | 43487.2 | 43242.3 | 106306 | 107399 | 825.135 | 914.386 | 919.092 | 2328.73 | 2286.43 | 5468.92 | 5380.81 | 12794.8 | 12756.2 | 16491 |
| 19325 | 18961.9 | 43509.9 | 43310.9 | 107748 | 107536 | 825.416 | 938.479 | 921.093 | 2355.87 | 2290.78 | 5413.4 | 5390.19 | 12831.4 | 12774.7 | 16604.1 |
| 19276.1 | 18996.2 | 43387.1 | 43379.5 | 106636 | 107673 | 825.696 | 972.12 | 923.097 | 2335.52 | 2295.14 | 5400.01 | 5399.58 | 12676 | 12793.2 | 16483.2 |
| 19157.1 | 19030.6 | 43359.9 | 43448.2 | 107434 | 107810 | 825.976 | 939.841 | 925.103 | 2320.33 | 2299.49 | 5407.9 | 5408.97 | 12739.5 | 12811.7 | 16585 |
| 19136.7 | 19065.1 | 44074.6 | 43516.9 | 107364 | 107948 | 826.257 | 962.804 | 927.111 | 2315.01 | 2303.86 | 5452.04 | 5418.37 | 12847.3 | 12830.2 | 16618.3 |
| 19031.8 | 19099.5 | 43609.7 | 43585.6 | 106893 | 108085 | 826.537 | 942.719 | 929.122 | 2292.77 | 2308.22 | 5480.05 | 5427.78 | 12717.1 | 12848.7 | 16513.6 |
| 19338.8 | 19134 | 43378.3 | 43654.4 | 107170 | 108222 | 826.817 | 952.989 | 931.135 | 2365.36 | 2312.59 | 5386.91 | 5437.2 | 12723.7 | 12867.2 | 16491.1 |
| 19238.9 | 19168.5 | 43926.6 | 43723.2 | 108534 | 108359 | 827.098 | 977.893 | 933.15 | 2343.48 | 2316.96 | 5480.34 | 5446.62 | 12830.9 | 12885.8 | 16716.5 |
| 19444.9 | 19203 | 43738.8 | 43792 | 108208 | 108497 | 827.378 | 941.654 | 935.168 | 2374.27 | 2321.34 | 5486.58 | 5456.05 | 12787.8 | 12904.3 | 16692.6 |
| 19221 | 19237.6 | 43660.1 | 43860.9 | 106967 | 108634 | 827.658 | 935.316 | 937.188 | 2339.87 | 2325.72 | 5446.1 | 5465.48 | 12649.4 | 12922.8 | 16489.6 |
| 19402.9 | 19272.2 | 43557.2 | 43929.8 | 107508 | 108771 | 827.939 | 988.438 | 939.211 | 2364.92 | 2330.11 | 5535.25 | 5474.92 | 12772.6 | 12941.4 | 16718.4 |
| 19468.5 | 19306.8 | 43852.4 | 43998.7 | 108568 | 108909 | 828.219 | 979.692 | 941.236 | 2341.42 | 2334.49 | 5404.57 | 5484.36 | 12933.2 | 12959.9 | 16633.7 |
| 19221.1 | 19341.5 | 43599.8 | 44067.7 | 107548 | 109046 | 828.499 | 947.134 | 943.263 | 2328.28 | 2338.89 | 5416 | 5493.82 | 12821.3 | 12978.5 | 16682 |
| 19194.1 | 19376.2 | 44124.9 | 44136.7 | 108536 | 109183 | 828.779 | 963.685 | 945.292 | 2403.16 | 2343.28 | 5409.57 | 5503.28 | 12911.1 | 12997 | 16713.4 |
| 19607.4 | 19410.8 | 44227.9 | 44205.8 | 108749 | 109321 | 829.059 | 955.353 | 947.324 | 2417.38 | 2347.68 | 5547.67 | 5512.74 | 12931.1 | 13015.6 | 16820.4 |
| 19578 | 19445.6 | 44524.1 | 44274.8 | 108719 | 109458 | 829.34 | 955.221 | 949.358 | 2402.64 | 2352.08 | 5620.15 | 5522.21 | 12924.7 | 13034.1 | 16854.2 |
| 19767.1 | 19480.3 | 44645.7 | 44344 | 109214 | 109596 | 829.62 | 968.769 | 951.395 | 2373.72 | 2356.49 | 5538.03 | 5531.69 | 12975.7 | 13052.7 | 16718.6 |
| 19678.3 | 19515.1 | 44324.2 | 44413.1 | 108712 | 109733 | 829.9 | 984.482 | 953.434 | 2389.75 | 2360.9 | 5584.05 | 5541.18 | 12964 | 13071.3 | 16865.8 |
| 19514.5 | 19549.9 | 44285.5 | 44482.3 | 109098 | 109871 | 830.18 | 952.199 | 955.475 | 2413.54 | 2365.31 | 5552.72 | 5550.67 | 12948.4 | 13089.9 | 17051.7 |
| 19790.6 | 19584.7 | 44942.1 | 44551.6 | 109418 | 110008 | 830.46 | 978.068 | 957.519 | 2404.36 | 2369.73 | 5525.99 | 5560.16 | 13074 | 13108.5 | 16926.7 |
| 19643 | 19619.6 | 44682.7 | 44620.8 | 109398 | 110146 | 830.74 | 989.83 | 959.565 | 2378.52 | 2374.15 | 5597 | 5569.66 | 13062.6 | 13127.1 | 17014.2 |
| 19873.6 | 19654.5 | 44980.7 | 44690.1 | 110204 | 110283 | 831.02 | 972.306 | 961.613 | 2424.46 | 2378.58 | 5637.08 | 5579.17 | 13126.6 | 13145.7 | 17099.9 |
| 19785.9 | 19689.4 | 44789.7 | 44759.5 | 109988 | 110421 | 831.3 | 996.771 | 963.663 | 2495.45 | 2383 | 5649.19 | 5588.69 | 13080.6 | 13164.3 | 16963.6 |
| 19941 | 19724.3 | 44734.3 | 44828.8 | 109298 | 110558 | 831.58 | 1011.11 | 965.716 | 2399.35 | 2387.44 | 5648.65 | 5598.21 | 13096.1 | 13182.9 | 17130.5 |
| 19816.3 | 19759.3 | 45142.9 | 44898.2 | 109679 | 110696 | 831.86 | 945.769 | 967.772 | 2412.21 | 2391.87 | 5502.89 | 5607.74 | 13079.5 | 13201.5 | 17093.7 |
| 19816 | 19794.3 | 44632.1 | 44967.7 | 109539 | 110834 | 832.14 | 1001.74 | 969.829 | 2436.75 | 2396.31 | 5601.56 | 5617.27 | 13117.9 | 13220.2 | 16914.7 |
| 19633.3 | 19829.3 | 44601.3 | 45037.2 | 109296 | 110971 | 832.42 | 974.808 | 971.889 | 2442.22 | 2400.75 | 5580.56 | 5626.81 | 13184.9 | 13238.8 | 16989.9 |
| 19890.3 | 19864.4 | 45236.2 | 45106.7 | 109908 | 111109 | 832.7 | 982.711 | 973.951 | 2462.4 | 2405.2 | 5628.32 | 5636.36 | 13248.8 | 13257.5 | 16984.7 |
| 20015.1 | 19899.4 | 45462.2 | 45176.2 | 110139 | 111247 | 832.98 | 987.573 | 976.016 | 2422.18 | 2409.65 | 5759.03 | 5645.91 | 13291.2 | 13276.1 | 17110.8 |
| 20131.3 | 19934.5 | 45270.8 | 45245.8 | 110845 | 111384 | 833.26 | 1001.17 | 978.083 | 2506.62 | 2414.11 | 5728.78 | 5655.47 | 13274.9 | 13294.8 | 17169.7 |
| 20091.2 | 19969.7 | 45114.1 | 45315.4 | 110402 | 111522 | 833.54 | 1013.4 | 980.152 | 2405.11 | 2418.56 | 5697.99 | 5665.03 | 13271.9 | 13313.4 | 17132.7 |
| 20196.8 | 20004.8 | 45376 | 45385.1 | 110715 | 111660 | 833.82 | 1026.32 | 982.224 | 2478.87 | 2423.02 | 5638.58 | 5674.61 | 13180.2 | 13332.1 | 17017.8 |
| 19950.5 | 20040 | 44782.5 | 45454.8 | 110233 | 111798 | 834.1 | 987.829 | 984.298 | 2411.7 | 2427.49 | 5651.4 | 5684.18 | 13040 | 13350.8 | 16892.2 |
| 20236.9 | 20075.2 | 45531.9 | 45524.5 | 111375 | 111936 | 834.38 | 1032.59 | 986.374 | 2409.78 | 2431.96 | 5682.57 | 5693.77 | 13416.5 | 13369.4 | 17212.3 |
| 19941.1 | 20110.4 | 45061.6 | 45594.2 | 110565 | 112073 | 834.66 | 982.149 | 988.452 | 2464.04 | 2436.43 | 5695.09 | 5703.35 | 13188.4 | 13388.1 | 17031.6 |
| 20150.2 | 20145.7 | 45488.9 | 45664 | 110323 | 112211 | 834.939 | 999.255 | 990.533 | 2433.61 | 2440.9 | 5695.76 | 5712.95 | 13171.2 | 13406.8 | 17071.9 |
| 19998.2 | 20181 | 45550.8 | 45733.8 | 111438 | 112349 | 835.219 | 974.451 | 992.616 | 2440.48 | 2445.38 | 5684.6 | 5722.55 | 13208.1 | 13425.5 | 17113 |
| 20233.1 | 20216.3 | 46017.1 | 45803.7 | 111097 | 112487 | 835.499 | 991.191 | 994.702 | 2525.06 | 2449.86 | 5672.13 | 5732.16 | 13190.3 | 13444.2 | 17306.3 |
| 20112 | 20251.6 | 45555.7 | 45873.6 | 110968 | 112625 | 835.779 | 997.215 | 996.789 | 2379.28 | 2454.35 | 5736.03 | 5741.77 | 13242.1 | 13462.9 | 17159.3 |
| 20359.9 | 20287 | 45915.8 | 45943.5 | 111686 | 112763 | 836.059 | 1031.22 | 998.879 | 2452.11 | 2458.84 | 5698.75 | 5751.39 | 13311.3 | 13481.6 | 17240.6 |
| 20314.4 | 20322.3 | 45745.7 | 46013.5 | 111963 | 112900 | 836.338 | 1036.41 | 1000.97 | 2478.55 | 2463.33 | 5686.06 | 5761.02 | 13356.3 | 13500.3 | 17099.2 |
| 20476.1 | 20357.8 | 46167.9 | 46083.5 | 111945 | 113038 | 836.618 | 1006.52 | 1003.07 | 2481.16 | 2467.83 | 5769.58 | 5770.65 | 13502.2 | 13519 | 17310.3 |
| 20358.6 | 20393.2 | 46004.4 | 46153.5 | 112275 | 113176 | 836.898 | 985.915 | 1005.16 | 2491 | 2472.33 | 5727.37 | 5780.28 | 13387.1 | 13537.7 | 17465.7 |
| 20223.2 | 20428.7 | 45595.2 | 46223.6 | 112680 | 113314 | 837.178 | 999.24 | 1007.26 | 2497.17 | 2476.84 | 5734.83 | 5789.93 | 13361.5 | 13556.5 | 17348.2 |
| 20333.8 | 20464.2 | 46213.8 | 46293.7 | 111930 | 113452 | 837.457 | 1026.34 | 1009.36 | 2444.86 | 2481.34 | 5732.62 | 5799.58 | 13418.3 | 13575.2 | 17365 |
| 20355 | 20499.7 | 45941.5 | 46363.8 | 112433 | 113590 | 837.737 | 1036.7 | 1011.47 | 2482.37 | 2485.85 | 5732.01 | 5809.23 | 13372.5 | 13593.9 | 17290.3 |
| 20553.3 | 20535.2 | 46266.4 | 46433.9 | 112178 | 113728 | 838.017 | 1026.78 | 1013.57 | 2476.44 | 2490.37 | 5782.6 | 5818.89 | 13511.5 | 13612.7 | 17422 |
| 20544.2 | 20570.8 | 46221.3 | 46504.1 | 112647 | 113866 | 838.296 | 1031.35 | 1015.68 | 2508.62 | 2494.89 | 5790.94 | 5828.56 | 13547.3 | 13631.4 | 17432.1 |
| 20512.9 | 20606.4 | 46193.6 | 46574.4 | 113036 | 114004 | 838.576 | 1039.51 | 1017.79 | 2465.8 | 2499.41 | 5760.95 | 5838.23 | 13376.4 | 13650.2 | 17447.5 |
| 20882.7 | 20642 | 46750.7 | 46644.6 | 113492 | 114142 | 838.856 | 1040.76 | 1019.91 | 2540.25 | 2503.93 | 5916.02 | 5847.91 | 13544.8 | 13669 | 17575.1 |
| 20490.3 | 20677.6 | 46614.5 | 46714.9 | 112399 | 114280 | 839.135 | 1046.6 | 1022.02 | 2506.18 | 2508.46 | 5879.15 | 5857.6 | 13622.7 | 13687.7 | 17551.8 |
| 20624.3 | 20713.3 | 46592.8 | 46785.3 | 112801 | 114418 | 839.415 | 1061.11 | 1024.14 | 2538.6 | 2512.99 | 5869.9 | 5867.29 | 13423.2 | 13706.5 | 17538.2 |
| 20669.5 | 20749 | 46840.8 | 46855.6 | 112926 | 114557 | 839.694 | 1043.02 | 1026.26 | 2513.62 | 2517.53 | 5876.85 | 5876.98 | 13604.7 | 13725.3 | 17435.7 |
| 21013.4 | 20784.7 | 46910.7 | 46926 | 113345 | 114695 | 839.974 | 1051.24 | 1028.38 | 2527.71 | 2522.07 | 5863.97 | 5886.69 | 13630.5 | 13744 | 17617.5 |
| 20717.7 | 20820.5 | 46752.7 | 46996.4 | 113220 | 114833 | 840.254 | 1048.92 | 1030.51 | 2560.75 | 2526.61 | 5856.39 | 5896.4 | 13542.8 | 13762.8 | 17550.8 |
| 20766.3 | 20856.2 | 46739.6 | 47066.9 | 113917 | 114971 | 840.533 | 1045.4 | 1032.63 | 2501.16 | 2531.16 | 5920.55 | 5906.11 | 13647.9 | 13781.6 | 17542.2 |
| 20888.5 | 20892 | 46957.5 | 47137.4 | 114179 | 115109 | 840.812 | 1040.9 | 1034.76 | 2508.43 | 2535.71 | 5852.59 | 5915.83 | 13625.5 | 13800.4 | 17697.6 |
| 20872.9 | 20927.8 | 47249.8 | 47207.9 | 114130 | 115247 | 841.092 | 1063.16 | 1036.89 | 2556.64 | 2540.26 | 5900.61 | 5925.56 | 13734.9 | 13819.2 | 17825.7 |
| 21025.9 | 20963.7 | 47425.8 | 47278.5 | 114801 | 115386 | 841.371 | 1074.74 | 1039.03 | 2556.08 | 2544.82 | 5878.09 | 5935.29 | 13884.2 | 13838 | 17884 |
| 20821.7 | 20999.6 | 47134.7 | 47349.1 | 113868 | 115524 | 841.651 | 1058.6 | 1041.16 | 2541.23 | 2549.38 | 5891.96 | 5945.02 | 13734.4 | 13856.8 | 17631.7 |
| 20962.2 | 21035.5 | 47076.4 | 47419.7 | 114891 | 115662 | 841.93 | 1062.28 | 1043.3 | 2552.23 | 2553.94 | 5947.8 | 5954.77 | 13804.8 | 13875.6 | 17938.5 |
| 21256.4 | 21071.4 | 47101.5 | 47490.3 | 114599 | 115800 | 842.21 | 1048.18 | 1045.44 | 2590.36 | 2558.51 | 5915.85 | 5964.52 | 13716 | 13894.5 | 17794 |
| 21453.6 | 21107.3 | 47532.7 | 47561 | 115702 | 115938 | 842.489 | 1088.58 | 1047.59 | 2585.85 | 2563.08 | 6014.47 | 5974.27 | 13959.6 | 13913.3 | 17856.5 |
| 21156.5 | 21143.3 | 47578.1 | 47631.7 | 114999 | 116076 | 842.769 | 1068.75 | 1049.73 | 2577.48 | 2567.65 | 6046.51 | 5984.03 | 13875.9 | 13932.1 | 17867.4 |
| 21431.2 | 21179.3 | 47241.7 | 47702.5 | 115225 | 116215 | 843.048 | 1106.15 | 1051.88 | 2614.22 | 2572.23 | 5972.52 | 5993.8 | 13924.5 | 13951 | 17922.6 |
| 21101.7 | 21215.3 | 47519.4 | 47773.3 | 114146 | 116353 | 843.327 | 1086.37 | 1054.03 | 2498.88 | 2576.81 | 6061.78 | 6003.57 | 13764 | 13969.8 | 17632.8 |
| 21046.7 | 21251.4 | 47594.8 | 47844.1 | 115659 | 116491 | 843.607 | 1058.44 | 1056.18 | 2610.7 | 2581.4 | 5973.9 | 6013.35 | 13881.7 | 13988.6 | 17889.2 |
| 21191.8 | 21287.4 | 47615.9 | 47914.9 | 115409 | 116630 | 843.886 | 1051.23 | 1058.33 | 2583.86 | 2585.98 | 5991.08 | 6023.13 | 13825.4 | 14007.5 | 17951.3 |
| 21358.2 | 21323.5 | 48294.5 | 47985.8 | 116211 | 116768 | 844.165 | 1055.98 | 1060.49 | 2616.03 | 2590.57 | 5995.67 | 6032.92 | 13945.4 | 14026.4 | 17957.3 |
| 21393.6 | 21359.7 | 48066.8 | 48056.7 | 116117 | 116906 | 844.444 | 1074.43 | 1062.65 | 2619.78 | 2595.17 | 6050.99 | 6042.71 | 13968.4 | 14045.2 | 17833.5 |
| 21477.8 | 21395.8 | 48079.9 | 48127.6 | 116249 | 117045 | 844.724 | 1101.6 | 1064.81 | 2633.23 | 2599.77 | 6005.16 | 6052.51 | 13902.7 | 14064.1 | 17944.6 |
| 21437 | 21432 | 48086.3 | 48198.6 | 115979 | 117183 | 845.003 | 1059.11 | 1066.97 | 2598.51 | 2604.37 | 5981.14 | 6062.32 | 13908.5 | 14083 | 17955.8 |
| 21378.5 | 21468.2 | 47694.9 | 48269.6 | 115885 | 117321 | 845.282 | 1097.26 | 1069.14 | 2640.28 | 2608.97 | 6025.98 | 6072.13 | 13972.7 | 14101.8 | 17955.9 |
| 21300 | 21504.4 | 48154.7 | 48340.6 | 115810 | 117460 | 845.562 | 1067.56 | 1071.31 | 2576.19 | 2613.58 | 5966.85 | 6081.95 | 13958.8 | 14120.7 | 17895.5 |
| 21547.6 | 21540.6 | 48189.2 | 48411.7 | 116061 | 117598 | 845.841 | 1070.9 | 1073.48 | 2571.94 | 2618.19 | 6042.46 | 6091.77 | 13955.9 | 14139.6 | 17904.8 |
| 21612.2 | 21576.9 | 48346.7 | 48482.8 | 116988 | 117736 | 846.12 | 1133.81 | 1075.65 | 2631.37 | 2622.81 | 6096.11 | 6101.6 | 14223.2 | 14158.5 | 18000.2 |
| 21580.8 | 21613.2 | 48427.4 | 48553.9 | 116972 | 117875 | 846.399 | 1122.32 | 1077.82 | 2585.93 | 2627.43 | 6130.11 | 6111.43 | 13997.2 | 14177.4 | 18093.8 |
| 21449.8 | 21649.5 | 48412.4 | 48625.1 | 116523 | 118013 | 846.678 | 1078.44 | 1080 | 2639.65 | 2632.05 | 6095.77 | 6121.27 | 14066.4 | 14196.2 | 18078.1 |
| 21707.3 | 21685.9 | 48437.7 | 48696.3 | 116836 | 118152 | 846.957 | 1108.98 | 1082.18 | 2623.76 | 2636.67 | 6186.32 | 6131.12 | 14145.6 | 14215.1 | 18198.2 |
| 21617.1 | 21722.2 | 48616.5 | 48767.5 | 117477 | 118290 | 847.236 | 1110.83 | 1084.36 | 2622.1 | 2641.3 | 6171.64 | 6140.97 | 14159.1 | 14234.1 | 18179.8 |
| 21847.4 | 21758.6 | 48770.1 | 48838.7 | 117974 | 118428 | 847.516 | 1127.74 | 1086.54 | 2692.9 | 2645.93 | 6091.74 | 6150.82 | 14065.7 | 14253 | 18312.7 |
| 21932.2 | 21795 | 48788.5 | 48910 | 118403 | 118567 | 847.795 | 1126.81 | 1088.73 | 2632.84 | 2650.57 | 6145.66 | 6160.69 | 14251.6 | 14271.9 | 18364.2 |
| 21688.1 | 21831.5 | 49166.4 | 48981.3 | 118349 | 118705 | 848.074 | 1106.34 | 1090.92 | 2714.05 | 2655.21 | 6167.65 | 6170.55 | 14229.7 | 14290.8 | 18441.5 |
| 21694.4 | 21867.9 | 48780.8 | 49052.7 | 117187 | 118844 | 848.353 | 1134.06 | 1093.11 | 2648.24 | 2659.85 | 6152.58 | 6180.43 | 14121 | 14309.7 | 18187.4 |
| 21809.6 | 21904.4 | 48926.8 | 49124 | 117337 | 118982 | 848.632 | 1114.96 | 1095.3 | 2627.28 | 2664.5 | 6118.9 | 6190.3 | 14230.8 | 14328.6 | 18248.9 |
| 21941 | 21940.9 | 48864.2 | 49195.4 | 117547 | 119121 | 848.911 | 1122.73 | 1097.5 | 2675.25 | 2669.14 | 6155.3 | 6200.19 | 14114.7 | 14347.6 | 18195.2 |
| 21913.3 | 21977.5 | 49066.9 | 49266.8 | 117922 | 119259 | 849.19 | 1115.62 | 1099.69 | 2655.01 | 2673.8 | 6195.32 | 6210.08 | 14292 | 14366.5 | 18326.5 |
| 22122.8 | 22014 | 48926.8 | 49338.3 | 118348 | 119398 | 849.469 | 1110.77 | 1101.89 | 2702.11 | 2678.45 | 6123.51 | 6219.97 | 14099.1 | 14385.5 | 18198 |
| 21800.2 | 22050.6 | 48882 | 49409.8 | 118237 | 119536 | 849.748 | 1124.03 | 1104.09 | 2696.63 | 2683.11 | 6160.07 | 6229.87 | 14149.8 | 14404.4 | 18265.7 |
| 22109 | 22087.2 | 48989.9 | 49481.3 | 117559 | 119675 | 850.027 | 1095.54 | 1106.3 | 2732.83 | 2687.77 | 6153.25 | 6239.78 | 14202.1 | 14423.4 | 18182.5 |
| 22281 | 22123.9 | 49250.6 | 49552.9 | 118168 | 119813 | 850.306 | 1164.78 | 1108.5 | 2719.25 | 2692.44 | 6274.98 | 6249.69 | 14348.7 | 14442.3 | 18303.5 |
| 21817.8 | 22160.5 | 49252.4 | 49624.4 | 118826 | 119952 | 850.585 | 1116.13 | 1110.71 | 2660.83 | 2697.11 | 6206.4 | 6259.6 | 14331.5 | 14461.2 | 18326.3 |
| 21771.8 | 22197.2 | 48859 | 49696.1 | 117973 | 120090 | 850.864 | 1109.77 | 1112.92 | 2734.88 | 2701.78 | 6063.78 | 6269.53 | 14160.5 | 14480.2 | 18312.5 |
| 21958.7 | 22233.9 | 49062.7 | 49767.7 | 117957 | 120229 | 851.143 | 1122.44 | 1115.13 | 2653.17 | 2706.45 | 6230.22 | 6279.45 | 14404.1 | 14499.2 | 18423.5 |
| 22116.1 | 22270.7 | 49436.6 | 49839.4 | 118516 | 120368 | 851.421 | 1147.91 | 1117.35 | 2739.46 | 2711.13 | 6246.75 | 6289.39 | 14441.7 | 14518.1 | 18299.5 |
| 21913.4 | 22307.4 | 49191.1 | 49911.1 | 118218 | 120506 | 851.7 | 1098.63 | 1119.57 | 2676.18 | 2715.81 | 6208.87 | 6299.32 | 14320.2 | 14537.1 | 18292 |
| 22094.2 | 22344.2 | 48969.7 | 49982.8 | 118644 | 120645 | 851.979 | 1137.98 | 1121.79 | 2675.82 | 2720.5 | 6203.72 | 6309.27 | 14308.8 | 14556.1 | 18450.5 |
| 22185.9 | 22381 | 49762.8 | 50054.5 | 119163 | 120783 | 852.258 | 1144.39 | 1124.01 | 2693.55 | 2725.19 | 6301.01 | 6319.22 | 14541.4 | 14575.1 | 18590.4 |
| 22254.3 | 22417.8 | 49122.2 | 50126.3 | 119044 | 120922 | 852.537 | 1142.9 | 1126.23 | 2701.42 | 2729.88 | 6237.16 | 6329.17 | 14332.5 | 14594.1 | 18451.7 |
| 22178.2 | 22454.7 | 49785.3 | 50198.1 | 118560 | 121060 | 852.816 | 1115.83 | 1128.46 | 2754.03 | 2734.57 | 6325.29 | 6339.13 | 14340.3 | 14613 | 18620.3 |
| 22449.1 | 22491.5 | 49437.8 | 50270 | 118587 | 121199 | 853.094 | 1141.13 | 1130.69 | 2737.59 | 2739.27 | 6321.72 | 6349.09 | 14407.3 | 14632 | 18487.7 |
| 22362.5 | 22528.4 | 49618.8 | 50341.8 | 118264 | 121338 | 853.373 | 1092.36 | 1132.92 | 2675.18 | 2743.97 | 6244.34 | 6359.06 | 14179.5 | 14651 | 18219.5 |
| 22539.4 | 22565.3 | 49851.2 | 50413.7 | 119715 | 121476 | 853.652 | 1145.28 | 1135.15 | 2759.1 | 2748.68 | 6287.08 | 6369.04 | 14504.6 | 14670 | 18531.8 |
| 22344.3 | 22602.3 | 49715.3 | 50485.7 | 118934 | 121615 | 853.93 | 1133.26 | 1137.38 | 2749.44 | 2753.39 | 6277.96 | 6379.02 | 14373.1 | 14689 | 18538.6 |
| 22449.6 | 22639.3 | 49498.2 | 50557.6 | 119689 | 121754 | 854.209 | 1126.81 | 1139.62 | 2697.13 | 2758.1 | 6259.51 | 6389 | 14356.6 | 14708 | 18582.4 |
| 22567.7 | 22676.2 | 50017.9 | 50629.6 | 120116 | 121892 | 854.488 | 1148.51 | 1141.86 | 2764.04 | 2762.81 | 6328.34 | 6398.99 | 14465.8 | 14727.1 | 18645.3 |
| 22701.2 | 22713.3 | 50227.9 | 50701.6 | 120867 | 122031 | 854.766 | 1162.19 | 1144.1 | 2758.91 | 2767.53 | 6371.22 | 6408.99 | 14565.1 | 14746.1 | 18844.1 |
| 22588.7 | 22750.3 | 50076.5 | 50773.7 | 119671 | 122170 | 855.045 | 1159.68 | 1146.35 | 2736.03 | 2772.25 | 6316.22 | 6418.99 | 14487.8 | 14765.1 | 18663 |
| 22701.8 | 22787.4 | 50201.6 | 50845.7 | 119918 | 122308 | 855.324 | 1139.51 | 1148.59 | 2783.36 | 2776.97 | 6465.18 | 6429 | 14448.8 | 14784.1 | 18675.9 |
| 22336.5 | 22824.4 | 50070.5 | 50917.8 | 118788 | 122447 | 855.602 | 1166.89 | 1150.84 | 2716.95 | 2781.7 | 6308.97 | 6439.01 | 14530.2 | 14803.1 | 18608 |
| 22589.1 | 22861.5 | 50167.7 | 50990 | 119946 | 122586 | 855.881 | 1142.54 | 1153.09 | 2769.1 | 2786.43 | 6400.31 | 6449.03 | 14520.8 | 14822.2 | 18654.5 |
| 22510.3 | 22898.7 | 50609.4 | 51062.1 | 120278 | 122724 | 856.16 | 1133.58 | 1155.34 | 2816.5 | 2791.16 | 6371.01 | 6459.05 | 14578.7 | 14841.2 | 18734.4 |
| 22556.5 | 22935.8 | 50914.9 | 51134.3 | 120560 | 122863 | 856.438 | 1151.97 | 1157.6 | 2813.56 | 2795.9 | 6416.61 | 6469.07 | 14560.5 | 14860.2 | 18841.3 |
| 22859 | 22973 | 50423.3 | 51206.5 | 119365 | 123002 | 856.717 | 1158.31 | 1159.85 | 2767.34 | 2800.64 | 6345.31 | 6479.11 | 14543.5 | 14879.3 | 18688.1 |
| 22703.4 | 23010.2 | 50706.6 | 51278.8 | 120957 | 123140 | 856.995 | 1200.23 | 1162.11 | 2758.07 | 2805.38 | 6412.48 | 6489.14 | 14538.4 | 14898.3 | 18686.4 |
| 22795.4 | 23047.4 | 50885.9 | 51351 | 120976 | 123279 | 857.274 | 1175.31 | 1164.37 | 2838.56 | 2810.13 | 6422.06 | 6499.19 | 14728.8 | 14917.4 | 18829.5 |
| 22955.7 | 23084.7 | 50855 | 51423.3 | 121747 | 123418 | 857.552 | 1172.53 | 1166.64 | 2815.94 | 2814.88 | 6409.81 | 6509.23 | 14677.3 | 14936.4 | 18701.8 |
| 22722.5 | 23122 | 50755.3 | 51495.6 | 120799 | 123556 | 857.831 | 1166.37 | 1168.9 | 2751.6 | 2819.63 | 6441.43 | 6519.29 | 14660.4 | 14955.5 | 18908.8 |
| 23034.6 | 23159.2 | 50831.1 | 51568 | 121574 | 123695 | 858.109 | 1158.63 | 1171.17 | 2796.66 | 2824.39 | 6448.62 | 6529.35 | 14635.4 | 14974.6 | 19052.8 |
| 22848.4 | 23196.5 | 51091 | 51640.3 | 121136 | 123834 | 858.388 | 1185.09 | 1173.44 | 2788.37 | 2829.15 | 6426.24 | 6539.41 | 14776.7 | 14993.6 | 19027.2 |
| 23027.5 | 23233.9 | 50934.3 | 51712.8 | 122472 | 123972 | 858.666 | 1168.93 | 1175.71 | 2861.96 | 2833.91 | 6462.27 | 6549.48 | 14790.2 | 15012.7 | 19041.9 |
| 23257.3 | 23271.2 | 51507.9 | 51785.2 | 121696 | 124111 | 858.945 | 1189.58 | 1177.99 | 2834.97 | 2838.67 | 6477.09 | 6559.55 | 14837.1 | 15031.8 | 19066.9 |
| 22919.2 | 23308.6 | 51061.9 | 51857.6 | 121302 | 124250 | 859.223 | 1199.82 | 1180.26 | 2781.66 | 2843.44 | 6464.41 | 6569.63 | 14852.9 | 15050.8 | 18888.5 |
| 23157.9 | 23346 | 51164.5 | 51930.1 | 122416 | 124388 | 859.501 | 1193.23 | 1182.54 | 2821.71 | 2848.21 | 6522.72 | 6579.71 | 14748.6 | 15069.9 | 19077.2 |
| 23052.7 | 23383.5 | 51240.1 | 52002.6 | 122238 | 124527 | 859.78 | 1155.01 | 1184.82 | 2807.28 | 2852.99 | 6477.77 | 6589.8 | 14724.3 | 15089 | 19068 |
| 23040.9 | 23420.9 | 51429.9 | 52075.2 | 122595 | 124666 | 860.058 | 1166.8 | 1187.11 | 2834.47 | 2857.77 | 6598.42 | 6599.89 | 14770.2 | 15108.1 | 19207.9 |
| 23275.8 | 23458.4 | 51373.3 | 52147.7 | 122682 | 124805 | 860.337 | 1197.47 | 1189.39 | 2859.39 | 2862.55 | 6585.43 | 6609.99 | 14811.2 | 15127.2 | 19163.8 |
| 23466.4 | 23495.9 | 51676.8 | 52220.3 | 122734 | 124943 | 860.615 | 1210.35 | 1191.68 | 2850.18 | 2867.33 | 6574.98 | 6620.09 | 14808.3 | 15146.2 | 19063.2 |
| 23280.7 | 23533.4 | 51454.8 | 52292.9 | 123024 | 125082 | 860.893 | 1232.5 | 1193.97 | 2827.29 | 2872.12 | 6472.05 | 6630.2 | 14962 | 15165.3 | 19216.5 |
| 23279.1 | 23570.9 | 51841.5 | 52365.6 | 122846 | 125221 | 861.171 | 1211.82 | 1196.26 | 2875.59 | 2876.91 | 6477.55 | 6640.31 | 14948.6 | 15184.4 | 19150.8 |
| 23427.8 | 23608.5 | 52088.7 | 52438.2 | 123039 | 125360 | 861.45 | 1192.8 | 1198.55 | 2875.96 | 2881.7 | 6608.11 | 6650.43 | 14967.3 | 15203.5 | 19161.7 |
| 23673.4 | 23646.1 | 51632.3 | 52510.9 | 123464 | 125498 | 861.728 | 1227.88 | 1200.85 | 2830.3 | 2886.5 | 6608.22 | 6660.55 | 14979.4 | 15222.6 | 19141.9 |
| 23271.8 | 23683.7 | 51622.2 | 52583.7 | 123012 | 125637 | 862.006 | 1191.35 | 1203.15 | 2871.81 | 2891.3 | 6587.23 | 6670.68 | 14930.6 | 15241.8 | 19120.7 |
| 23469 | 23721.3 | 52101.3 | 52656.4 | 123090 | 125776 | 862.284 | 1208.21 | 1205.45 | 2894.2 | 2896.1 | 6475.77 | 6680.81 | 14966.3 | 15260.9 | 19305.8 |
| 23759.2 | 23759 | 52115.2 | 52729.2 | 124341 | 125915 | 862.563 | 1225.38 | 1207.75 | 2922.4 | 2900.91 | 6662.73 | 6690.95 | 15157.1 | 15280 | 19436.8 |
| 23553.3 | 23796.7 | 52313.9 | 52802 | 124190 | 126053 | 862.841 | 1261.64 | 1210.06 | 2927.94 | 2905.72 | 6589.37 | 6701.09 | 14993.3 | 15299.1 | 19374.6 |
| 23588.8 | 23834.4 | 52028.9 | 52874.8 | 123577 | 126192 | 863.119 | 1227.53 | 1212.36 | 2838.99 | 2910.53 | 6555.78 | 6711.24 | 14988.2 | 15318.2 | 19292.4 |
| 23547.8 | 23872.1 | 51964.5 | 52947.7 | 122987 | 126331 | 863.397 | 1217.41 | 1214.67 | 2821.2 | 2915.34 | 6568.99 | 6721.39 | 15023 | 15337.3 | 19253.2 |
| 23622.7 | 23909.8 | 51824.6 | 53020.6 | 123505 | 126470 | 863.675 | 1222.45 | 1216.98 | 2890.61 | 2920.16 | 6613.95 | 6731.55 | 14943.2 | 15356.5 | 19364.2 |
| 23578.1 | 23947.6 | 52449.6 | 53093.5 | 124449 | 126608 | 863.953 | 1248.59 | 1219.3 | 2893.18 | 2924.98 | 6695.92 | 6741.71 | 15131 | 15375.6 | 19330.6 |
| 23576 | 23985.4 | 52448.6 | 53166.4 | 124174 | 126747 | 864.231 | 1240.66 | 1221.61 | 2921.04 | 2929.81 | 6636.12 | 6751.88 | 15138 | 15394.7 | 19410.8 |
| 23765.5 | 24023.2 | 52176.1 | 53239.3 | 123558 | 126886 | 864.51 | 1215.12 | 1223.93 | 2831.33 | 2934.63 | 6584.26 | 6762.05 | 14949.5 | 15413.8 | 19268.2 |
| 23887.2 | 24061 | 52188.9 | 53312.3 | 124116 | 127025 | 864.788 | 1233.9 | 1226.25 | 2902.64 | 2939.46 | 6661.76 | 6772.22 | 15201.7 | 15433 | 19371.3 |
| 24012.7 | 24098.9 | 52208.3 | 53385.3 | 123869 | 127163 | 865.066 | 1248.14 | 1228.57 | 2902.82 | 2944.3 | 6639.84 | 6782.4 | 15092.6 | 15452.1 | 19372.1 |
| 23827.4 | 24136.7 | 52411.3 | 53458.4 | 123873 | 127302 | 865.344 | 1269.9 | 1230.89 | 2901.43 | 2949.13 | 6652.03 | 6792.59 | 15014.1 | 15471.3 | 19172.6 |
| 23648.9 | 24174.6 | 52653 | 53531.4 | 125028 | 127441 | 865.622 | 1259.01 | 1233.22 | 2970.67 | 2953.97 | 6723.83 | 6802.78 | 15241.7 | 15490.4 | 19539.4 |
| 23870.5 | 24212.5 | 52244.8 | 53604.5 | 123705 | 127580 | 865.9 | 1233.15 | 1235.55 | 2924 | 2958.81 | 6709.05 | 6812.98 | 15104.3 | 15509.6 | 19386.5 |
| 23752 | 24250.5 | 52657.9 | 53677.6 | 124959 | 127718 | 866.178 | 1239.82 | 1237.88 | 2939.33 | 2963.66 | 6773.9 | 6823.18 | 15188.1 | 15528.7 | 19440 |
| 23895.6 | 24288.4 | 53241.6 | 53750.7 | 125907 | 127857 | 866.456 | 1202.42 | 1240.21 | 2996.81 | 2968.51 | 6773.69 | 6833.38 | 15302.9 | 15547.9 | 19712.8 |
| 23954.7 | 24326.4 | 52850.3 | 53823.9 | 125375 | 127996 | 866.734 | 1244.19 | 1242.55 | 2893.62 | 2973.36 | 6637.69 | 6843.59 | 15201.5 | 15567 | 19631.3 |
| 24062.6 | 24364.4 | 52612.8 | 53897.1 | 124663 | 128135 | 867.011 | 1240.5 | 1244.88 | 2966.75 | 2978.21 | 6811.32 | 6853.8 | 15196.5 | 15586.2 | 19476.7 |
| 24150.4 | 24402.5 | 53285.8 | 53970.3 | 125222 | 128273 | 867.289 | 1234.05 | 1247.22 | 2965.32 | 2983.07 | 6819.02 | 6864.02 | 15358.3 | 15605.3 | 19640.1 |
| 23997.5 | 24440.5 | 53036.6 | 54043.5 | 124838 | 128412 | 867.567 | 1242.81 | 1249.56 | 2914.81 | 2987.93 | 6790.25 | 6874.24 | 15416.7 | 15624.5 | 19558.9 |
| 24263.1 | 24478.6 | 53163.7 | 54116.8 | 125833 | 128551 | 867.845 | 1251.22 | 1251.91 | 2977.57 | 2992.79 | 6690.02 | 6884.47 | 15302.6 | 15643.7 | 19676 |
| 24183.2 | 24516.7 | 53111.2 | 54190 | 125478 | 128690 | 868.123 | 1261.95 | 1254.25 | 2942.47 | 2997.66 | 6809.48 | 6894.7 | 15415.8 | 15662.8 | 19557.4 |
| 24379.2 | 24554.8 | 53155.2 | 54263.3 | 124920 | 128828 | 868.401 | 1231 | 1256.6 | 2952.32 | 3002.53 | 6794.61 | 6904.94 | 15237.6 | 15682 | 19536.9 |
| 24076.9 | 24592.9 | 53058.8 | 54336.7 | 125523 | 128967 | 868.679 | 1252.83 | 1258.95 | 2926.07 | 3007.4 | 6778.43 | 6915.18 | 15289.8 | 15701.2 | 19525.8 |
| 24153.2 | 24631.1 | 53119.7 | 54410 | 124806 | 129106 | 868.957 | 1251.43 | 1261.3 | 2954.82 | 3012.28 | 6679.51 | 6925.43 | 15318.8 | 15720.4 | 19635.1 |
| 24310.9 | 24669.2 | 53306.8 | 54483.4 | 126059 | 129244 | 869.234 | 1274.8 | 1263.65 | 2946.57 | 3017.15 | 6775.55 | 6935.68 | 15374.2 | 15739.5 | 19663.6 |
| 24531.9 | 24707.4 | 53689.7 | 54556.8 | 125894 | 129383 | 869.512 | 1241.14 | 1266.01 | 2984.75 | 3022.03 | 6787.74 | 6945.94 | 15256.6 | 15758.7 | 19630.9 |
| 24530.2 | 24745.7 | 53701.4 | 54630.2 | 126068 | 129522 | 869.79 | 1240.45 | 1268.37 | 2972.28 | 3026.92 | 6842 | 6956.2 | 15387 | 15777.9 | 19796.7 |
| 24489.4 | 24783.9 | 53741.3 | 54703.7 | 126796 | 129661 | 870.068 | 1255.69 | 1270.73 | 2991.22 | 3031.8 | 6855.39 | 6966.46 | 15494 | 15797.1 | 19850 |
| 24208.1 | 24822.1 | 53712.4 | 54777.1 | 126189 | 129800 | 870.345 | 1286.91 | 1273.09 | 2933.35 | 3036.7 | 6730.48 | 6976.73 | 15443 | 15816.3 | 19851.6 |
| 24162.5 | 24860.4 | 53285.4 | 54850.6 | 126297 | 129938 | 870.623 | 1252.04 | 1275.45 | 2957.09 | 3041.59 | 6780.01 | 6987.01 | 15358.6 | 15835.5 | 19790.7 |
| 24280.2 | 24898.7 | 53905.9 | 54924.2 | 127035 | 130077 | 870.901 | 1266.56 | 1277.82 | 2964.68 | 3046.48 | 6807.66 | 6997.28 | 15459.1 | 15854.7 | 19644.3 |
| 24339.5 | 24937 | 53825.3 | 54997.7 | 127142 | 130216 | 871.178 | 1255.41 | 1280.19 | 3005.04 | 3051.38 | 6915.45 | 7007.57 | 15595.3 | 15873.9 | 19896.8 |
| 24406.9 | 24975.4 | 53898.6 | 55071.2 | 126580 | 130354 | 871.456 | 1307.6 | 1282.56 | 2995.49 | 3056.28 | 6864.54 | 7017.85 | 15293.2 | 15893.1 | 19801.5 |
| 24451.4 | 25013.8 | 53963.9 | 55144.8 | 127409 | 130493 | 871.734 | 1290.64 | 1284.93 | 2996.95 | 3061.19 | 6832.94 | 7028.15 | 15580.3 | 15912.3 | 19910.9 |
| 24558.3 | 25052.2 | 54157.9 | 55218.4 | 127304 | 130632 | 872.011 | 1327.58 | 1287.31 | 3006.32 | 3066.09 | 6909.96 | 7038.44 | 15537.9 | 15931.5 | 19822.8 |
| 24645.5 | 25090.5 | 54444.6 | 55292.1 | 127657 | 130770 | 872.289 | 1299.11 | 1289.68 | 2960.53 | 3071 | 6994.46 | 7048.74 | 15688.5 | 15950.7 | 20096.1 |
| 24567 | 25129 | 54106 | 55365.8 | 127912 | 130909 | 872.567 | 1311.77 | 1292.06 | 2984.38 | 3075.92 | 6903.59 | 7059.05 | 15503.2 | 15969.9 | 19855.9 |
| 24888.3 | 25167.4 | 54253.3 | 55439.4 | 127913 | 131048 | 872.844 | 1293.92 | 1294.44 | 3044.96 | 3080.83 | 6985.87 | 7069.36 | 15602 | 15989.1 | 19981.5 |
| 24702 | 25205.9 | 54493.4 | 55513.1 | 128205 | 131187 | 873.122 | 1308.92 | 1296.83 | 3012.27 | 3085.75 | 6954.32 | 7079.67 | 15624 | 16008.3 | 20075.1 |
| 24757.1 | 25244.4 | 54363.3 | 55586.9 | 127638 | 131325 | 873.399 | 1305.73 | 1299.21 | 3117.68 | 3090.67 | 6964.3 | 7089.99 | 15632.6 | 16027.5 | 20088.2 |
| 25035.4 | 25282.9 | 54651.8 | 55660.6 | 128249 | 131464 | 873.677 | 1284.82 | 1301.6 | 3043.53 | 3095.6 | 7018.02 | 7100.31 | 15835.3 | 16046.8 | 20131.4 |
| 24976.2 | 25321.4 | 55130.2 | 55734.4 | 129527 | 131603 | 873.954 | 1305.21 | 1303.99 | 3095.15 | 3100.52 | 6980.59 | 7110.64 | 15621.2 | 16066 | 20159.5 |
| 25183.5 | 25360 | 54777 | 55808.2 | 128308 | 131741 | 874.232 | 1294.03 | 1306.38 | 3047.06 | 3105.45 | 7053.5 | 7120.97 | 15787.7 | 16085.2 | 20198.9 |
| 25162.7 | 25398.6 | 55102.5 | 55882 | 128906 | 131880 | 874.509 | 1305.37 | 1308.77 | 3097.62 | 3110.39 | 7016.1 | 7131.31 | 15918.8 | 16104.4 | 20427.8 |
| 24984.8 | 25437.2 | 55271.8 | 55955.8 | 129523 | 132019 | 874.787 | 1312.33 | 1311.17 | 3116.55 | 3115.32 | 7030.84 | 7141.65 | 15830 | 16123.7 | 20288.2 |
| 25239.7 | 25475.8 | 55219.7 | 56029.7 | 129165 | 132157 | 875.064 | 1299.74 | 1313.56 | 3041.27 | 3120.26 | 6956.62 | 7151.99 | 15817.7 | 16142.9 | 20247.5 |
| 25324 | 25514.4 | 55095.1 | 56103.6 | 129226 | 132296 | 875.342 | 1338.83 | 1315.96 | 3150.63 | 3125.2 | 7039.57 | 7162.34 | 15833.2 | 16162.1 | 20247.3 |
| 25057.5 | 25553 | 55284.9 | 56177.5 | 128956 | 132435 | 875.619 | 1342.4 | 1318.37 | 3107.04 | 3130.15 | 6933 | 7172.69 | 15892.6 | 16181.4 | 20260.7 |
| 25442.7 | 25591.7 | 55132.6 | 56251.4 | 129368 | 132573 | 875.896 | 1329.37 | 1320.77 | 3032.09 | 3135.09 | 6998.41 | 7183.05 | 15881.2 | 16200.6 | 20204 |
| 25231.9 | 25630.4 | 55260.2 | 56325.4 | 130181 | 132712 | 876.174 | 1367.85 | 1323.17 | 3101.41 | 3140.04 | 7094.33 | 7193.41 | 15914 | 16219.8 | 20229 |
| 25311 | 25669.1 | 55602 | 56399.3 | 129558 | 132850 | 876.451 | 1298.92 | 1325.58 | 3059.41 | 3144.99 | 7040.89 | 7203.78 | 15863.7 | 16239.1 | 20414.4 |
| 25284 | 25707.9 | 55517 | 56473.3 | 129687 | 132989 | 876.728 | 1306.24 | 1327.99 | 3084.97 | 3149.95 | 7137.9 | 7214.15 | 15856 | 16258.3 | 20402.9 |
| 25553.8 | 25746.6 | 55980.1 | 56547.3 | 130165 | 133128 | 877.006 | 1322.9 | 1330.4 | 3134.05 | 3154.91 | 7039.88 | 7224.52 | 16007.4 | 16277.6 | 20441.3 |
| 25338.5 | 25785.4 | 55580.3 | 56621.4 | 129769 | 133266 | 877.283 | 1368.14 | 1332.82 | 3028.91 | 3159.87 | 7110.35 | 7234.9 | 16006.8 | 16296.8 | 20338.2 |
| 25626.6 | 25824.2 | 56046 | 56695.4 | 131416 | 133405 | 877.56 | 1329.77 | 1335.23 | 3133.99 | 3164.83 | 7122.47 | 7245.29 | 16022.2 | 16316 | 20634.8 |
| 25552.5 | 25863 | 56026 | 56769.5 | 130594 | 133544 | 877.838 | 1349.09 | 1337.65 | 3117.11 | 3169.8 | 7239.15 | 7255.67 | 16117.1 | 16335.3 | 20425 |
| 25627.7 | 25901.8 | 56142.4 | 56843.6 | 130753 | 133682 | 878.115 | 1363.89 | 1340.07 | 3136.41 | 3174.77 | 7099.34 | 7266.07 | 16077.5 | 16354.5 | 20499.4 |
| 25651.2 | 25940.7 | 55733.3 | 56917.7 | 131314 | 133821 | 878.392 | 1360.07 | 1342.49 | 3156.28 | 3179.74 | 7173.01 | 7276.46 | 16040.5 | 16373.8 | 20512.8 |
| 25881.9 | 25979.6 | 56268.8 | 56991.9 | 131397 | 133959 | 878.669 | 1328.77 | 1344.92 | 3192.45 | 3184.72 | 7203.21 | 7286.86 | 16078.6 | 16393.1 | 20804 |
| 25586.3 | 26018.5 | 56040.7 | 57066.1 | 131737 | 134098 | 878.947 | 1357.23 | 1347.35 | 3159.84 | 3189.69 | 7218.7 | 7297.27 | 16118.6 | 16412.3 | 20555.4 |
| 25845 | 26057.4 | 56365.1 | 57140.2 | 131239 | 134236 | 879.224 | 1332.95 | 1349.77 | 3109.14 | 3194.67 | 7219.16 | 7307.67 | 16182.4 | 16431.6 | 20529 |
| 25735.8 | 26096.3 | 55943.8 | 57214.4 | 131557 | 134375 | 879.501 | 1359.67 | 1352.2 | 3089.25 | 3199.66 | 7149.26 | 7318.09 | 16029 | 16450.8 | 20692.5 |
| 25696.2 | 26135.3 | 56722 | 57288.7 | 132141 | 134514 | 879.778 | 1351.96 | 1354.64 | 3168.2 | 3204.64 | 7242.85 | 7328.5 | 16387.7 | 16470.1 | 20740.3 |
| 25863.2 | 26174.2 | 56929.6 | 57362.9 | 132854 | 134652 | 880.055 | 1370.33 | 1357.07 | 3169.43 | 3209.63 | 7112.14 | 7338.92 | 16101 | 16489.4 | 20749.3 |
| 26052.5 | 26213.2 | 56549.6 | 57437.2 | 132000 | 134791 | 880.332 | 1391.29 | 1359.51 | 3142.74 | 3214.62 | 7260.06 | 7349.35 | 16161.1 | 16508.6 | 20791.3 |
| 25994.5 | 26252.3 | 56967.6 | 57511.5 | 132941 | 134929 | 880.609 | 1339.9 | 1361.95 | 3172.65 | 3219.61 | 7194.85 | 7359.78 | 16225.4 | 16527.9 | 20626.5 |
| 25688.1 | 26291.3 | 56333 | 57585.8 | 132075 | 135068 | 880.886 | 1371.75 | 1364.39 | 3138.8 | 3224.61 | 7178.29 | 7370.21 | 16252.2 | 16547.2 | 20731.5 |
| 25893.2 | 26330.3 | 56700.4 | 57660.1 | 132122 | 135206 | 881.163 | 1349.75 | 1366.83 | 3187.85 | 3229.61 | 7304.58 | 7380.65 | 16218.8 | 16566.4 | 20704.2 |
| 26278.6 | 26369.4 | 57358 | 57734.5 | 133152 | 135345 | 881.441 | 1372.41 | 1369.27 | 3209.45 | 3234.61 | 7309.24 | 7391.09 | 16351.6 | 16585.7 | 20730.7 |
| 26089.3 | 26408.5 | 57420.8 | 57808.9 | 133284 | 135483 | 881.717 | 1366.68 | 1371.72 | 3202.35 | 3239.62 | 7280.44 | 7401.53 | 16188.8 | 16605 | 20878.3 |
| 26034.2 | 26447.6 | 57048 | 57883.2 | 132778 | 135622 | 881.995 | 1360.09 | 1374.17 | 3242.27 | 3244.62 | 7276.1 | 7411.98 | 16272.8 | 16624.2 | 20663.5 |
| 26227.8 | 26486.8 | 57202.5 | 57957.7 | 133266 | 135760 | 882.271 | 1356.36 | 1376.62 | 3174.75 | 3249.63 | 7347.69 | 7422.43 | 16443 | 16643.5 | 20947.4 |
| 26257.8 | 26525.9 | 56728.9 | 58032.1 | 133072 | 135899 | 882.548 | 1359.92 | 1379.07 | 3188.6 | 3254.65 | 7284.31 | 7432.89 | 16316.9 | 16662.8 | 20889.9 |
| 26266.6 | 26565.1 | 57642.2 | 58106.6 | 133722 | 136037 | 882.825 | 1390.15 | 1381.53 | 3179.74 | 3259.66 | 7301.34 | 7443.35 | 16408.8 | 16682.1 | 20927.2 |
| 26198 | 26604.3 | 57369.3 | 58181 | 133843 | 136176 | 883.102 | 1403.52 | 1383.98 | 3195.22 | 3264.68 | 7362.78 | 7453.82 | 16549 | 16701.4 | 20902.4 |
| 26002.8 | 26643.5 | 57297 | 58255.5 | 133554 | 136314 | 883.379 | 1399.08 | 1386.44 | 3207.93 | 3269.7 | 7371.82 | 7464.29 | 16443.7 | 16720.6 | 20992.3 |
| 26435.5 | 26682.7 | 57398.9 | 58330 | 133455 | 136452 | 883.656 | 1404.49 | 1388.9 | 3242.18 | 3274.72 | 7380.06 | 7474.76 | 16302.3 | 16739.9 | 20969.3 |
| 26044.1 | 26721.9 | 57124.2 | 58404.6 | 133050 | 136591 | 883.933 | 1387.87 | 1391.36 | 3235.36 | 3279.75 | 7310.57 | 7485.24 | 16308 | 16759.2 | 20729.8 |
| 26383.8 | 26761.2 | 57415.6 | 58479.1 | 134827 | 136729 | 884.21 | 1426.89 | 1393.83 | 3281.68 | 3284.78 | 7319.85 | 7495.72 | 16509.6 | 16778.5 | 21047.2 |
| 26264.1 | 26800.5 | 57232.3 | 58553.7 | 133451 | 136868 | 884.487 | 1410.4 | 1396.3 | 3223.64 | 3289.81 | 7328.96 | 7506.2 | 16318.9 | 16797.8 | 21068.7 |
| 26398.7 | 26839.8 | 58226.1 | 58628.3 | 133894 | 137006 | 884.764 | 1364.4 | 1398.76 | 3249.84 | 3294.84 | 7355.37 | 7516.69 | 16560.5 | 16817.1 | 21098.9 |
| 26304.5 | 26879.1 | 57478.1 | 58702.9 | 133860 | 137144 | 885.04 | 1381.43 | 1401.23 | 3284.07 | 3299.88 | 7430.51 | 7527.18 | 16407.8 | 16836.3 | 21052.7 |
| 26410.2 | 26918.5 | 57574.2 | 58777.5 | 134055 | 137283 | 885.317 | 1408.56 | 1403.71 | 3261.8 | 3304.92 | 7386.34 | 7537.68 | 16507.6 | 16855.7 | 21050.2 |
| 26595.3 | 26957.8 | 57987.7 | 58852.2 | 133966 | 137421 | 885.594 | 1393.88 | 1406.18 | 3271.75 | 3309.96 | 7436.67 | 7548.18 | 16425 | 16874.9 | 21207.5 |
| 26476.5 | 26997.2 | 57689.5 | 58926.8 | 134296 | 137560 | 885.871 | 1394.61 | 1408.66 | 3251.99 | 3315.01 | 7378.6 | 7558.68 | 16385.3 | 16894.2 | 21092.1 |
| 26902.7 | 27036.6 | 57953.5 | 59001.5 | 134986 | 137698 | 886.148 | 1403.02 | 1411.13 | 3213.68 | 3320.05 | 7447.11 | 7569.19 | 16483 | 16913.5 | 21237.8 |
| 26453 | 27076 | 58180.6 | 59076.2 | 134794 | 137836 | 886.424 | 1427.61 | 1413.62 | 3199.88 | 3325.1 | 7536.84 | 7579.7 | 16531.2 | 16932.8 | 21115.8 |
| 26795.2 | 27115.4 | 58045 | 59151 | 134932 | 137974 | 886.701 | 1443.68 | 1416.1 | 3315.85 | 3330.15 | 7486.73 | 7590.22 | 16607.3 | 16952.1 | 21176.3 |
| 26830.5 | 27154.9 | 58222.2 | 59225.7 | 135528 | 138113 | 886.978 | 1413.68 | 1418.58 | 3291.62 | 3335.21 | 7444.64 | 7600.74 | 16522.9 | 16971.4 | 21181.7 |
| 26794.2 | 27194.4 | 58666.9 | 59300.5 | 135276 | 138251 | 887.254 | 1409.58 | 1421.07 | 3275.16 | 3340.27 | 7480.3 | 7611.26 | 16514.9 | 16990.7 | 21068.2 |
| 26795.5 | 27233.8 | 58434.3 | 59375.2 | 135791 | 138389 | 887.531 | 1464.4 | 1423.56 | 3292.07 | 3345.32 | 7567.89 | 7621.79 | 16730.1 | 17010 | 21247.7 |
| 26600.6 | 27273.4 | 58184.8 | 59450.1 | 135160 | 138528 | 887.808 | 1438.99 | 1426.05 | 3309.73 | 3350.39 | 7500.29 | 7632.32 | 16617.6 | 17029.3 | 21199.6 |
| 26781.1 | 27312.9 | 58269.7 | 59524.9 | 135159 | 138666 | 888.084 | 1421.24 | 1428.54 | 3309.53 | 3355.45 | 7451.09 | 7642.85 | 16809.5 | 17048.6 | 21158.2 |
| 26881.1 | 27352.4 | 58486 | 59599.7 | 135616 | 138804 | 888.361 | 1486.11 | 1431.03 | 3345.32 | 3360.52 | 7565.66 | 7653.39 | 16778.8 | 17067.9 | 21298.9 |
| 27189.9 | 27392 | 58824.7 | 59674.6 | 135552 | 138942 | 888.638 | 1429.01 | 1433.53 | 3321.27 | 3365.59 | 7507.51 | 7663.93 | 16651.7 | 17087.2 | 21401.1 |
| 27124.6 | 27431.6 | 58355.5 | 59749.4 | 135823 | 139081 | 888.914 | 1434.95 | 1436.03 | 3301.89 | 3370.66 | 7596.66 | 7674.48 | 16622.5 | 17106.5 | 21410.6 |
| 26973.1 | 27471.2 | 58328.8 | 59824.3 | 135522 | 139219 | 889.191 | 1428.55 | 1438.53 | 3341.66 | 3375.74 | 7445.59 | 7685.03 | 16611.7 | 17125.8 | 21416.9 |
| 26963.3 | 27510.8 | 58800.7 | 59899.2 | 135875 | 139357 | 889.467 | 1424.98 | 1441.03 | 3285.3 | 3380.82 | 7546.95 | 7695.58 | 16732.4 | 17145.1 | 21283.5 |
| 27132.1 | 27550.4 | 59096.6 | 59974.2 | 136782 | 139495 | 889.744 | 1448.34 | 1443.53 | 3343.54 | 3385.9 | 7537.83 | 7706.14 | 16724.8 | 17164.5 | 21443.7 |
| 27141.7 | 27590.1 | 58808 | 60049.1 | 136145 | 139633 | 890.02 | 1475.06 | 1446.04 | 3321.69 | 3390.98 | 7661.9 | 7716.7 | 16553.2 | 17183.8 | 21313.5 |
| 27028 | 27629.8 | 58978.6 | 60124.1 | 136004 | 139772 | 890.297 | 1442.62 | 1448.55 | 3268.39 | 3396.06 | 7521.96 | 7727.26 | 16774.6 | 17203.1 | 21291.3 |
| 27315 | 27669.5 | 59084 | 60199.1 | 135696 | 139910 | 890.573 | 1422.15 | 1451.06 | 3357.62 | 3401.15 | 7561.36 | 7737.83 | 16655.4 | 17222.4 | 21296.5 |
| 26841 | 27709.2 | 58764.9 | 60274.1 | 136209 | 140048 | 890.85 | 1444.26 | 1453.57 | 3356.23 | 3406.24 | 7540.93 | 7748.4 | 16730.2 | 17241.7 | 21320 |
| 27029.8 | 27748.9 | 58808.1 | 60349.1 | 136221 | 140186 | 891.126 | 1465.3 | 1456.08 | 3310.22 | 3411.33 | 7515.22 | 7758.98 | 16657 | 17261 | 21278.5 |
| 27277.4 | 27788.6 | 59569.2 | 60424.1 | 137084 | 140324 | 891.402 | 1455.04 | 1458.6 | 3299.06 | 3416.43 | 7632.97 | 7769.56 | 16796.7 | 17280.3 | 21430.4 |
| 27109.8 | 27828.4 | 58977.6 | 60499.2 | 136202 | 140462 | 891.679 | 1461.44 | 1461.12 | 3360.49 | 3421.53 | 7589.81 | 7780.14 | 16862.2 | 17299.6 | 21464.8 |
| 27243.9 | 27868.2 | 59042.9 | 60574.2 | 136180 | 140600 | 891.955 | 1455.73 | 1463.64 | 3389.29 | 3426.63 | 7590.69 | 7790.72 | 16817.4 | 17318.9 | 21409.1 |
| 27337.5 | 27908 | 58913.9 | 60649.3 | 136403 | 140738 | 892.232 | 1417.12 | 1466.16 | 3343.62 | 3431.73 | 7680.57 | 7801.31 | 16794.3 | 17338.2 | 21447.4 |
| 27414 | 27947.8 | 59406.3 | 60724.4 | 137551 | 140876 | 892.508 | 1449.18 | 1468.68 | 3399.36 | 3436.84 | 7588.22 | 7811.91 | 16811.9 | 17357.6 | 21457.4 |
| 27462.4 | 27987.6 | 59512.7 | 60799.5 | 136726 | 141014 | 892.784 | 1441.39 | 1471.21 | 3366.84 | 3441.94 | 7691.83 | 7822.5 | 16936.7 | 17376.9 | 21640.2 |
| 27460.7 | 28027.5 | 59729.3 | 60874.7 | 137370 | 141152 | 893.061 | 1481.52 | 1473.73 | 3374.89 | 3447.05 | 7608.04 | 7833.1 | 16933.7 | 17396.2 | 21674.1 |
| 27513.4 | 28067.4 | 59458.1 | 60949.8 | 137297 | 141290 | 893.337 | 1509.52 | 1476.26 | 3301.62 | 3452.16 | 7607.77 | 7843.7 | 16809.7 | 17415.5 | 21500.6 |
| 27454.7 | 28107.2 | 59281.4 | 61025 | 136982 | 141428 | 893.613 | 1445.58 | 1478.79 | 3375.39 | 3457.28 | 7603.01 | 7854.31 | 16926.9 | 17434.8 | 21499.6 |
| 27362.7 | 28147.2 | 59668.7 | 61100.2 | 137653 | 141566 | 893.889 | 1470.84 | 1481.33 | 3331.31 | 3462.4 | 7682.76 | 7864.92 | 16866.6 | 17454.2 | 21411.7 |
| 27688.8 | 28187.1 | 59915.1 | 61175.4 | 137790 | 141704 | 894.166 | 1475.92 | 1483.86 | 3418.49 | 3467.52 | 7722.53 | 7875.54 | 16983.8 | 17473.5 | 21688.2 |
| 27572.8 | 28227 | 59734.8 | 61250.6 | 138147 | 141842 | 894.442 | 1476.82 | 1486.4 | 3362.25 | 3472.64 | 7633.17 | 7886.15 | 17081.7 | 17492.8 | 21678.6 |
| 27792.7 | 28267 | 59804.1 | 61325.8 | 138389 | 141980 | 894.718 | 1533.36 | 1488.94 | 3364.94 | 3477.76 | 7686.57 | 7896.77 | 17100.4 | 17512.1 | 21887.5 |
| 27787.3 | 28307 | 60283 | 61401.1 | 139480 | 142118 | 894.994 | 1464.31 | 1491.48 | 3394.72 | 3482.89 | 7739.99 | 7907.4 | 17001.8 | 17531.4 | 21702.8 |
| 27748.9 | 28346.9 | 59533.6 | 61476.4 | 137548 | 142256 | 895.27 | 1466.78 | 1494.02 | 3355.93 | 3488.02 | 7626.32 | 7918.03 | 16901.8 | 17550.8 | 21604 |
| 27824.2 | 28387 | 60175.7 | 61551.7 | 138199 | 142394 | 895.547 | 1487.48 | 1496.57 | 3374.64 | 3493.15 | 7742.74 | 7928.66 | 17041.9 | 17570.1 | 21757.1 |
| 27543.2 | 28427 | 60075.7 | 61627 | 137954 | 142532 | 895.823 | 1508.06 | 1499.11 | 3400.7 | 3498.29 | 7665.12 | 7939.29 | 17114.6 | 17589.4 | 21815.2 |
| 27976.1 | 28467 | 60222.1 | 61702.3 | 138362 | 142670 | 896.099 | 1501.75 | 1501.66 | 3469.86 | 3503.42 | 7708.44 | 7949.93 | 17202.6 | 17608.7 | 21834 |
| 27883 | 28507.1 | 60079.2 | 61777.6 | 138619 | 142807 | 896.375 | 1439.03 | 1504.21 | 3391.77 | 3508.56 | 7775.36 | 7960.57 | 17094.2 | 17628 | 21715.3 |
| 27921.1 | 28547.2 | 60251.6 | 61853 | 139706 | 142945 | 896.651 | 1539.06 | 1506.76 | 3449.03 | 3513.7 | 7739.28 | 7971.22 | 17237 | 17647.4 | 21937.5 |
| 27938 | 28587.3 | 60394.1 | 61928.3 | 139570 | 143083 | 896.927 | 1492.95 | 1509.32 | 3401.73 | 3518.85 | 7750.88 | 7981.87 | 16992.3 | 17666.7 | 21670.2 |
| 28078.5 | 28627.4 | 61103 | 62003.7 | 140156 | 143221 | 897.203 | 1548.17 | 1511.87 | 3466.94 | 3523.99 | 7848.69 | 7992.52 | 17296.6 | 17686 | 21858.5 |
| 27838.1 | 28667.5 | 60524 | 62079.1 | 138991 | 143359 | 897.479 | 1458.57 | 1514.43 | 3429.9 | 3529.14 | 7847.18 | 8003.17 | 16983.7 | 17705.3 | 21799.2 |
| 28297.9 | 28707.7 | 60469.6 | 62154.5 | 140410 | 143496 | 897.755 | 1511.66 | 1516.99 | 3521.1 | 3534.29 | 7855.89 | 8013.83 | 17376.3 | 17724.7 | 22289.8 |
| 28134.6 | 28747.8 | 60714.9 | 62230 | 139829 | 143634 | 898.031 | 1531.01 | 1519.55 | 3427.64 | 3539.45 | 7810.77 | 8024.49 | 17365.8 | 17744 | 22037.8 |
| 28398.8 | 28788 | 60352.8 | 62305.4 | 139930 | 143772 | 898.307 | 1528.27 | 1522.11 | 3518.02 | 3544.6 | 7848.44 | 8035.16 | 17410.4 | 17763.3 | 21951.4 |
| 28364.9 | 28828.2 | 60685.2 | 62380.9 | 139648 | 143910 | 898.583 | 1508.27 | 1524.68 | 3467.02 | 3549.76 | 7795.6 | 8045.82 | 17354.2 | 17782.6 | 22240.1 |
| 28113.9 | 28868.4 | 61003.7 | 62456.3 | 140610 | 144047 | 898.859 | 1560.48 | 1527.25 | 3485.37 | 3554.92 | 7870.08 | 8056.5 | 17319.7 | 17802 | 22055.9 |
| 28199.3 | 28908.7 | 61198.3 | 62531.8 | 141049 | 144185 | 899.135 | 1497.2 | 1529.82 | 3471.75 | 3560.08 | 7864.37 | 8067.17 | 17328.7 | 17821.3 | 21963.8 |
| 28419.8 | 28948.9 | 61496.4 | 62607.3 | 141215 | 144323 | 899.411 | 1535.02 | 1532.39 | 3529.6 | 3565.25 | 7913.15 | 8077.85 | 17438.9 | 17840.6 | 22295.6 |
| 28213.2 | 28989.2 | 61576.8 | 62682.9 | 140440 | 144460 | 899.687 | 1512.24 | 1534.96 | 3533.73 | 3570.41 | 7903.28 | 8088.53 | 17425.7 | 17859.9 | 22189.3 |
| 28526.6 | 29029.5 | 61218.5 | 62758.4 | 140731 | 144598 | 899.963 | 1528.69 | 1537.53 | 3468.09 | 3575.58 | 7904.55 | 8099.21 | 17310.3 | 17879.3 | 22120.3 |
| 28288.1 | 29069.8 | 61228.4 | 62834 | 140046 | 144735 | 900.239 | 1508 | 1540.11 | 3498.8 | 3580.75 | 7841.17 | 8109.9 | 17425.3 | 17898.6 | 22095.3 |
| 28626.6 | 29110.1 | 61340.8 | 62909.5 | 141824 | 144873 | 900.514 | 1555.43 | 1542.69 | 3470.02 | 3585.93 | 7943.65 | 8120.59 | 17448.1 | 17917.9 | 22209.5 |
| 28420.5 | 29150.4 | 61369.4 | 62985.1 | 140487 | 145011 | 900.79 | 1555.48 | 1545.27 | 3408.78 | 3591.11 | 7842.19 | 8131.29 | 17376.2 | 17937.2 | 21932.8 |
| 28652.2 | 29190.7 | 61520 | 63060.7 | 141232 | 145148 | 901.066 | 1561.96 | 1547.85 | 3486.8 | 3596.28 | 7884.81 | 8141.98 | 17511 | 17956.6 | 22246.3 |
| 28552.6 | 29231.1 | 61597.9 | 63136.3 | 141273 | 145286 | 901.342 | 1579.26 | 1550.43 | 3505.6 | 3601.46 | 7946.28 | 8152.68 | 17495.1 | 17975.9 | 22519.3 |
| 28803.9 | 29271.5 | 61993.1 | 63211.9 | 141871 | 145423 | 901.617 | 1556.83 | 1553.02 | 3558.73 | 3606.65 | 7959.81 | 8163.39 | 17643.1 | 17995.2 | 22342.5 |
| 29187.3 | 29311.9 | 62754.1 | 63287.6 | 142681 | 145561 | 901.893 | 1578.19 | 1555.61 | 3564.95 | 3611.83 | 8096.22 | 8174.09 | 17526.4 | 18014.5 | 22391.9 |
| 28756.9 | 29352.3 | 62109.6 | 63363.2 | 141824 | 145698 | 902.169 | 1508.52 | 1558.2 | 3538.7 | 3617.02 | 8058.65 | 8184.8 | 17453.3 | 18033.9 | 22307.3 |
| 28730.4 | 29392.7 | 61859.8 | 63438.9 | 142181 | 145836 | 902.445 | 1542.26 | 1560.79 | 3522.01 | 3622.21 | 7990.48 | 8195.52 | 17586.1 | 18053.2 | 22197.2 |
| 29092.4 | 29433.2 | 62177.1 | 63514.6 | 141937 | 145973 | 902.72 | 1583.12 | 1563.38 | 3532.28 | 3627.4 | 7909.64 | 8206.23 | 17662.8 | 18072.5 | 22361.2 |
| 29029.7 | 29473.6 | 62064.1 | 63590.3 | 142086 | 146110 | 902.996 | 1575.44 | 1565.98 | 3564.54 | 3632.6 | 8031.65 | 8216.95 | 17743.3 | 18091.8 | 22466.4 |
| 28669.2 | 29514.1 | 61869.2 | 63666 | 141716 | 146248 | 903.272 | 1567.87 | 1568.57 | 3563.73 | 3637.79 | 8017.5 | 8227.67 | 17504.9 | 18111.2 | 22359.9 |
| 28995.2 | 29554.6 | 62342.1 | 63741.7 | 143185 | 146385 | 903.547 | 1594.79 | 1571.17 | 3594.28 | 3642.99 | 8046.86 | 8238.4 | 17679.7 | 18130.5 | 22674.9 |
| 29025.8 | 29595.1 | 62199.9 | 63817.4 | 143719 | 146523 | 903.823 | 1573.85 | 1573.77 | 3563.03 | 3648.19 | 8116.29 | 8249.13 | 17747.5 | 18149.8 | 22628 |
| 29112.3 | 29635.7 | 62472 | 63893.2 | 143311 | 146660 | 904.098 | 1490.3 | 1576.37 | 3552.97 | 3653.4 | 8186.39 | 8259.86 | 17721 | 18169.2 | 22629.8 |
| 29111.4 | 29676.2 | 62465.8 | 63969 | 142868 | 146797 | 904.374 | 1530.56 | 1578.98 | 3538.7 | 3658.6 | 8092.98 | 8270.59 | 17749.5 | 18188.5 | 22520 |
| 29121.9 | 29716.8 | 62702.4 | 64044.8 | 142811 | 146935 | 904.65 | 1580.77 | 1581.58 | 3585.64 | 3663.81 | 8117.11 | 8281.33 | 17669.5 | 18207.8 | 22591.7 |
| 29225.7 | 29757.3 | 62961.1 | 64120.6 | 143880 | 147072 | 904.925 | 1573.72 | 1584.19 | 3628.48 | 3669.02 | 8106.15 | 8292.07 | 17668.6 | 18227.1 | 22758.4 |
| 29243.3 | 29797.9 | 62999.3 | 64196.3 | 143851 | 147209 | 905.201 | 1594.32 | 1586.8 | 3633.47 | 3674.23 | 8200.27 | 8302.81 | 17813.5 | 18246.4 | 22764.5 |
| 29242.2 | 29838.5 | 62366.3 | 64272.2 | 142980 | 147346 | 905.476 | 1608.05 | 1589.41 | 3597.36 | 3679.45 | 8156.2 | 8313.56 | 17726.3 | 18265.8 | 22702.2 |
| 29393.5 | 29879.2 | 62866.7 | 64348 | 143779 | 147484 | 905.752 | 1567.46 | 1592.03 | 3589.17 | 3684.66 | 8132.48 | 8324.31 | 17860.1 | 18285.1 | 22696.6 |
| 29483.9 | 29919.8 | 63026.3 | 64423.8 | 144231 | 147621 | 906.027 | 1575.46 | 1594.64 | 3648.42 | 3689.88 | 8120.01 | 8335.06 | 17778.1 | 18304.4 | 22900.3 |
| 29399.8 | 29960.4 | 62883.2 | 64499.7 | 143693 | 147758 | 906.303 | 1599.53 | 1597.26 | 3619.61 | 3695.1 | 8037.67 | 8345.82 | 17834.6 | 18323.7 | 22781.8 |
| 29634.8 | 30001.1 | 63177 | 64575.6 | 143621 | 147895 | 906.578 | 1617.77 | 1599.88 | 3668.41 | 3700.33 | 8083.92 | 8356.58 | 17825.2 | 18343.1 | 22835 |
| 29506.7 | 30041.8 | 63138.1 | 64651.5 | 144175 | 148032 | 906.854 | 1600.75 | 1602.5 | 3639.55 | 3705.55 | 8187.77 | 8367.34 | 17919.1 | 18362.4 | 22834.8 |
| 29713.6 | 30082.5 | 63336.8 | 64727.4 | 144111 | 148169 | 907.129 | 1551.84 | 1605.12 | 3612.73 | 3710.78 | 8171.39 | 8378.1 | 17878.3 | 18381.7 | 22947.1 |
| 29760 | 30123.2 | 63252.2 | 64803.3 | 144221 | 148306 | 907.404 | 1592.04 | 1607.74 | 3673.33 | 3716.01 | 8114.09 | 8388.87 | 18006.5 | 18401 | 22821.1 |
| 29475.6 | 30163.9 | 63575.6 | 64879.2 | 145111 | 148444 | 907.68 | 1629.4 | 1610.37 | 3685.06 | 3721.24 | 8223.45 | 8399.64 | 17859 | 18420.3 | 23057.8 |
| 29457.4 | 30204.7 | 63133.3 | 64955.2 | 144365 | 148581 | 907.955 | 1571.98 | 1613 | 3617.58 | 3726.47 | 8176.45 | 8410.41 | 17884.8 | 18439.7 | 22725.8 |
| 29459.6 | 30245.5 | 62740 | 65031.1 | 143514 | 148718 | 908.231 | 1593.57 | 1615.63 | 3550.73 | 3731.71 | 8118.87 | 8421.18 | 17826 | 18459 | 22722.6 |
| 29611 | 30286.2 | 63516.6 | 65107.1 | 144432 | 148855 | 908.506 | 1584.82 | 1618.26 | 3653.66 | 3736.95 | 8220.15 | 8431.96 | 17884.2 | 18478.3 | 22810.1 |
| 29407.7 | 30327 | 63473.7 | 65183 | 144581 | 148992 | 908.781 | 1612.79 | 1620.89 | 3702.55 | 3742.19 | 8238.64 | 8442.74 | 17954.2 | 18497.6 | 22750.2 |
| 29954.7 | 30367.8 | 63723.5 | 65259 | 145335 | 149129 | 909.056 | 1610.49 | 1623.53 | 3702.2 | 3747.43 | 8316.63 | 8453.53 | 18016.3 | 18516.9 | 23082.4 |
| 29736.2 | 30408.7 | 63820.1 | 65335 | 145660 | 149266 | 909.332 | 1588.45 | 1626.16 | 3730.7 | 3752.68 | 8270.43 | 8464.32 | 18068.7 | 18536.2 | 23031.8 |
| 29798.6 | 30449.5 | 63878.2 | 65411 | 144810 | 149402 | 909.607 | 1614.93 | 1628.8 | 3663.67 | 3757.92 | 8289.45 | 8475.1 | 18071.8 | 18555.6 | 22989 |
| 30330.1 | 30490.3 | 64019.5 | 65487.1 | 145625 | 149540 | 909.882 | 1630.96 | 1631.44 | 3726.23 | 3763.17 | 8261.11 | 8485.9 | 18115.7 | 18574.9 | 22978.5 |
| 30078.2 | 30531.2 | 64072.1 | 65563.1 | 145378 | 149676 | 910.157 | 1612.95 | 1634.08 | 3646.07 | 3768.42 | 8269.32 | 8496.69 | 18081.5 | 18594.2 | 22947.7 |
| 30083 | 30572.1 | 64612.6 | 65639.1 | 145540 | 149813 | 910.433 | 1622.57 | 1636.73 | 3665.62 | 3773.68 | 8355.41 | 8507.49 | 18207.8 | 18613.5 | 23209.8 |
| 29744 | 30613 | 64512.9 | 65715.2 | 145486 | 149950 | 910.708 | 1652.82 | 1639.37 | 3676.85 | 3778.93 | 8272.39 | 8518.29 | 18090.9 | 18632.8 | 22910.9 |
| 29939.5 | 30653.9 | 64289 | 65791.3 | 146747 | 150087 | 910.983 | 1675.2 | 1642.02 | 3706.79 | 3784.19 | 8301.31 | 8529.09 | 18120.3 | 18652.1 | 23035.2 |
| 30438.2 | 30694.8 | 64146.6 | 65867.4 | 146825 | 150224 | 911.258 | 1677.92 | 1644.67 | 3734.76 | 3789.45 | 8306.31 | 8539.9 | 18113.3 | 18671.5 | 22974.3 |
| 30087.2 | 30735.8 | 64550.9 | 65943.5 | 147156 | 150360 | 911.533 | 1677.26 | 1647.32 | 3798.33 | 3794.71 | 8358.84 | 8550.71 | 18340.6 | 18690.8 | 23052.2 |
| 30199.5 | 30776.8 | 64783.3 | 66019.6 | 146087 | 150497 | 911.808 | 1623.38 | 1649.97 | 3745.66 | 3799.97 | 8278.92 | 8561.52 | 18148.8 | 18710.1 | 23015 |
| 30540.3 | 30817.7 | 64846.6 | 66095.7 | 146810 | 150634 | 912.083 | 1632.42 | 1652.62 | 3688.98 | 3805.24 | 8413.62 | 8572.34 | 18332.8 | 18729.4 | 23217.2 |
| 30480.1 | 30858.7 | 65100.9 | 66171.8 | 147382 | 150771 | 912.358 | 1659.37 | 1655.28 | 3743.31 | 3810.5 | 8497.31 | 8583.15 | 18325.8 | 18748.7 | 23330 |
| 30580.6 | 30899.7 | 64724.9 | 66248 | 147775 | 150907 | 912.633 | 1642.16 | 1657.94 | 3762.11 | 3815.77 | 8407.99 | 8593.97 | 18465.1 | 18768 | 23382.3 |
| 30287 | 30940.7 | 64728.3 | 66324.1 | 146865 | 151044 | 912.909 | 1574.75 | 1660.6 | 3741.13 | 3821.04 | 8422.76 | 8604.79 | 18302.2 | 18787.3 | 23104.5 |
| 30443.4 | 30981.8 | 65205.8 | 66400.3 | 146729 | 151181 | 913.183 | 1623.69 | 1663.26 | 3711.88 | 3826.32 | 8377.08 | 8615.62 | 18313.7 | 18806.6 | 23138.5 |
| 30553.8 | 31022.8 | 65285.8 | 66476.4 | 148199 | 151317 | 913.458 | 1657.16 | 1665.92 | 3807.64 | 3831.59 | 8505.13 | 8626.45 | 18322.1 | 18825.9 | 23274.1 |
| 30377.6 | 31063.9 | 65202 | 66552.6 | 147970 | 151454 | 913.734 | 1650.64 | 1668.59 | 3756.42 | 3836.87 | 8369.29 | 8637.28 | 18451.2 | 18845.2 | 23412 |
| 30261.8 | 31105 | 64842.6 | 66628.8 | 147846 | 151590 | 914.008 | 1623.45 | 1671.25 | 3750.44 | 3842.15 | 8522.21 | 8648.11 | 18324.6 | 18864.5 | 23322.2 |
| 30711.7 | 31146 | 65313.9 | 66705 | 149302 | 151727 | 914.283 | 1673.38 | 1673.92 | 3810.86 | 3847.43 | 8552.13 | 8658.94 | 18538.2 | 18883.8 | 23547.9 |
| 30664.4 | 31187.2 | 65374.6 | 66781.3 | 147922 | 151863 | 914.558 | 1667.31 | 1676.59 | 3763.43 | 3852.71 | 8361.52 | 8669.78 | 18403 | 18903.1 | 23578.7 |
| 30825.6 | 31228.3 | 65770.4 | 66857.5 | 149027 | 152000 | 914.833 | 1709.67 | 1679.26 | 3813.18 | 3858 | 8475.55 | 8680.62 | 18514.6 | 18922.4 | 23615.1 |
| 30815.1 | 31269.4 | 65650.1 | 66933.7 | 148630 | 152136 | 915.108 | 1691.95 | 1681.94 | 3767.85 | 3863.29 | 8617.55 | 8691.47 | 18675.4 | 18941.7 | 23771.1 |
| 30675.7 | 31310.5 | 65521.7 | 67010 | 148382 | 152273 | 915.383 | 1691.97 | 1684.61 | 3808.51 | 3868.58 | 8484.57 | 8702.31 | 18535.4 | 18961 | 23567.5 |
| 30941.9 | 31351.7 | 65787.6 | 67086.2 | 149160 | 152409 | 915.658 | 1666.22 | 1687.29 | 3785.69 | 3873.87 | 8528.02 | 8713.16 | 18590.5 | 18980.3 | 23444.8 |
| 30925.4 | 31392.9 | 65990.3 | 67162.5 | 149017 | 152546 | 915.933 | 1677.48 | 1689.97 | 3781.81 | 3879.16 | 8504.78 | 8724.01 | 18481.7 | 18999.6 | 23535.6 |
| 31167.7 | 31434.1 | 65718.1 | 67238.8 | 149015 | 152682 | 916.207 | 1714.61 | 1692.65 | 3834.72 | 3884.46 | 8630.3 | 8734.86 | 18596.4 | 19018.9 | 23575.2 |
| 30915.4 | 31475.3 | 66131.5 | 67315.1 | 150586 | 152818 | 916.482 | 1736.65 | 1695.33 | 3785.94 | 3889.75 | 8638.89 | 8745.72 | 18573.9 | 19038.2 | 23703.5 |
| 30891.9 | 31516.5 | 65696 | 67391.4 | 149725 | 152955 | 916.757 | 1684.37 | 1698.01 | 3823.45 | 3895.05 | 8541.06 | 8756.58 | 18469.7 | 19057.5 | 23681 |
| 31111.3 | 31557.7 | 66593.8 | 67467.7 | 150383 | 153091 | 917.032 | 1701.24 | 1700.7 | 3837.76 | 3900.35 | 8668 | 8767.44 | 18747.5 | 19076.8 | 23734.4 |
| 31041.6 | 31599 | 66543 | 67544 | 149424 | 153227 | 917.307 | 1685.19 | 1703.39 | 3784.39 | 3905.66 | 8497.03 | 8778.3 | 18603.7 | 19096.1 | 23539.3 |
| 30945.6 | 31640.2 | 66258.2 | 67620.4 | 149980 | 153363 | 917.581 | 1696.88 | 1706.08 | 3821.09 | 3910.96 | 8635.32 | 8789.17 | 18697.6 | 19115.4 | 23651.7 |
| 31599.1 | 31681.5 | 66307.3 | 67696.7 | 150349 | 153500 | 917.856 | 1689.04 | 1708.77 | 3831.25 | 3916.27 | 8681.93 | 8800.04 | 18729.7 | 19134.7 | 23726.7 |
| 31227.5 | 31722.8 | 66270.1 | 67773.1 | 149067 | 153636 | 918.131 | 1701.66 | 1711.46 | 3827.98 | 3921.58 | 8552.49 | 8810.91 | 18816 | 19153.9 | 23535 |
| 31168.9 | 31764.1 | 66575 | 67849.4 | 150626 | 153772 | 918.405 | 1650.12 | 1714.15 | 3945.37 | 3926.89 | 8687.7 | 8821.78 | 18835.2 | 19173.2 | 23640.3 |
| 31143.7 | 31805.4 | 66519.6 | 67925.8 | 149770 | 153908 | 918.68 | 1714.68 | 1716.85 | 3824.12 | 3932.2 | 8680.94 | 8832.65 | 18611.5 | 19192.5 | 23741.9 |
| 31285 | 31846.8 | 66259.1 | 68002.2 | 150045 | 154044 | 918.955 | 1721.13 | 1719.55 | 3905.69 | 3937.51 | 8667.82 | 8843.53 | 18674.4 | 19211.8 | 23825.7 |
| 31371.5 | 31888.1 | 67132.9 | 68078.6 | 150823 | 154180 | 919.229 | 1718 | 1722.24 | 3862.84 | 3942.83 | 8663.66 | 8854.41 | 18871.1 | 19231.1 | 23826 |
| 31540.9 | 31929.5 | 66767.8 | 68155 | 151157 | 154316 | 919.504 | 1725.05 | 1724.95 | 3914.61 | 3948.15 | 8650.83 | 8865.3 | 18660.2 | 19250.4 | 23902.3 |
| 31198.3 | 31970.8 | 67191.3 | 68231.4 | 151411 | 154452 | 919.779 | 1698.92 | 1727.65 | 3875.15 | 3953.47 | 8752.73 | 8876.18 | 18828.7 | 19269.6 | 23965.6 |
| 31609.7 | 32012.2 | 66975.9 | 68307.8 | 151138 | 154588 | 920.053 | 1718.2 | 1730.35 | 3885.79 | 3958.79 | 8682.86 | 8887.07 | 18908.8 | 19288.9 | 23855 |
| 31469.6 | 32053.6 | 67368.4 | 68384.2 | 152208 | 154724 | 920.328 | 1719.77 | 1733.06 | 3888.72 | 3964.12 | 8711.28 | 8897.96 | 18874.8 | 19308.2 | 24177 |
| 31423.3 | 32095 | 66740.4 | 68460.7 | 151731 | 154860 | 920.602 | 1712.08 | 1735.77 | 3919.7 | 3969.44 | 8657.2 | 8908.85 | 18894.8 | 19327.5 | 23934.4 |
| 31595 | 32136.4 | 67321.7 | 68537.1 | 150798 | 154996 | 920.877 | 1733.97 | 1738.47 | 3894.69 | 3974.77 | 8649.45 | 8919.74 | 18889.2 | 19346.7 | 23966.5 |
| 31620.4 | 32177.9 | 66786.2 | 68613.6 | 150362 | 155132 | 921.151 | 1689.07 | 1741.19 | 3858.69 | 3980.1 | 8636.77 | 8930.64 | 18737.9 | 19366 | 23734.5 |
| 31545.8 | 32219.3 | 66997.7 | 68690.1 | 151244 | 155268 | 921.426 | 1737.6 | 1743.9 | 3847.33 | 3985.43 | 8855.53 | 8941.54 | 18961.1 | 19385.3 | 24006.2 |
| 31706.8 | 32260.8 | 67704.1 | 68766.5 | 152682 | 155404 | 921.7 | 1792.59 | 1746.61 | 3871.22 | 3990.76 | 8861.39 | 8952.44 | 19020.8 | 19404.5 | 23979.3 |
| 31776.4 | 32302.3 | 67413.3 | 68843 | 151421 | 155540 | 921.975 | 1736.42 | 1749.33 | 3841.43 | 3996.1 | 8778.71 | 8963.34 | 18885.3 | 19423.8 | 24093.8 |
| 31994 | 32343.8 | 67815.1 | 68919.5 | 152394 | 155675 | 922.249 | 1766.89 | 1752.05 | 3893.76 | 4001.44 | 8856.93 | 8974.25 | 19078.3 | 19443.1 | 24088.3 |
| 31569.3 | 32385.3 | 67942.2 | 68996 | 151481 | 155811 | 922.523 | 1793.46 | 1754.77 | 3925.22 | 4006.78 | 8787.9 | 8985.15 | 18973.4 | 19462.3 | 24007.7 |
| 31935.8 | 32426.8 | 68176.2 | 69072.5 | 151635 | 155947 | 922.798 | 1746.49 | 1757.49 | 3925.49 | 4012.12 | 8801.12 | 8996.06 | 19048.8 | 19481.6 | 24131.7 |
| 31903.4 | 32468.3 | 67787.1 | 69149 | 153531 | 156083 | 923.072 | 1761.28 | 1760.21 | 3974.3 | 4017.46 | 8802.28 | 9006.98 | 18956.7 | 19500.9 | 24145.4 |
| 32032 | 32509.9 | 68098.5 | 69225.6 | 152529 | 156218 | 923.346 | 1748.01 | 1762.93 | 3948.41 | 4022.8 | 8854.72 | 9017.89 | 19066.4 | 19520.1 | 24074.9 |
| 32069.7 | 32551.4 | 67630.2 | 69302.1 | 151666 | 156354 | 923.621 | 1717.12 | 1765.66 | 3941.95 | 4028.15 | 8839.86 | 9028.81 | 19101 | 19539.4 | 24100 |
| 31852.3 | 32593 | 67856.8 | 69378.6 | 152748 | 156490 | 923.895 | 1767.85 | 1768.39 | 3977.83 | 4033.5 | 8723.73 | 9039.73 | 19029.8 | 19558.6 | 24275 |
| 32295.7 | 32634.6 | 68279 | 69455.2 | 152979 | 156625 | 924.169 | 1740.7 | 1771.12 | 4028.03 | 4038.85 | 8885.8 | 9050.65 | 19161.9 | 19577.9 | 24202.5 |
| 32101.1 | 32676.2 | 68207.8 | 69531.7 | 153167 | 156761 | 924.444 | 1765.57 | 1773.85 | 3974.25 | 4044.2 | 8815.09 | 9061.57 | 19004 | 19597.2 | 24230.9 |
| 32142.8 | 32717.8 | 67807.9 | 69608.3 | 152646 | 156896 | 924.718 | 1767.24 | 1776.58 | 3996.06 | 4049.55 | 8795.73 | 9072.49 | 18964.1 | 19616.4 | 24110.7 |
| 31833.6 | 32759.4 | 67728.3 | 69684.9 | 151980 | 157032 | 924.992 | 1750.31 | 1779.31 | 3979.24 | 4054.91 | 8915.42 | 9083.42 | 19142.5 | 19635.7 | 24217.8 |
| 32306.6 | 32801 | 68531.3 | 69761.5 | 152995 | 157167 | 925.266 | 1746.61 | 1782.05 | 3958.92 | 4060.26 | 8862.11 | 9094.35 | 19160.7 | 19654.9 | 24268.5 |
| 32133.1 | 32842.7 | 68755.8 | 69838.1 | 153757 | 157303 | 925.54 | 1795.27 | 1784.79 | 4092.85 | 4065.62 | 8821.67 | 9105.28 | 19253.9 | 19674.1 | 24422.5 |
| 32335 | 32884.3 | 68542.6 | 69914.6 | 154128 | 157438 | 925.815 | 1747.11 | 1787.52 | 3988.07 | 4070.98 | 8912.78 | 9116.22 | 19308.7 | 19693.4 | 24477.8 |
| 32372.7 | 32926 | 68968.8 | 69991.3 | 153850 | 157573 | 926.089 | 1751.81 | 1790.27 | 3985.92 | 4076.34 | 8932.23 | 9127.15 | 19162.2 | 19712.6 | 24363.5 |
| 32276.8 | 32967.7 | 68053.9 | 70067.9 | 154343 | 157709 | 926.363 | 1823.41 | 1793.01 | 3953.88 | 4081.71 | 9020.71 | 9138.09 | 19219.8 | 19731.9 | 24324.7 |
| 32518.3 | 33009.4 | 68691.6 | 70144.5 | 154080 | 157844 | 926.637 | 1781.81 | 1795.75 | 4058.88 | 4087.07 | 8936.19 | 9149.03 | 19363.3 | 19751.1 | 24325.7 |
| 32635.9 | 33051.1 | 68392.2 | 70221.1 | 153728 | 157979 | 926.911 | 1822.17 | 1798.5 | 3973.06 | 4092.44 | 8958.26 | 9159.97 | 19271.1 | 19770.3 | 24408.9 |
| 32682.7 | 33092.8 | 68933.1 | 70297.8 | 154592 | 158115 | 927.185 | 1765.21 | 1801.24 | 4042.94 | 4097.81 | 9039.32 | 9170.91 | 19340.3 | 19789.6 | 24488.2 |
| 32835.3 | 33134.6 | 69558.2 | 70374.4 | 155782 | 158250 | 927.459 | 1777.79 | 1803.99 | 4026.5 | 4103.18 | 9141.39 | 9181.86 | 19470.7 | 19808.8 | 24659.9 |
| 32562.5 | 33176.3 | 68701.7 | 70451.1 | 153751 | 158385 | 927.734 | 1788.38 | 1806.74 | 3966.14 | 4108.55 | 9046.73 | 9192.81 | 19238.8 | 19828 | 24384.6 |
| 32452.4 | 33218.1 | 68897.4 | 70527.7 | 154916 | 158520 | 928.008 | 1740.19 | 1809.49 | 4018.97 | 4113.93 | 8936.29 | 9203.76 | 19339.1 | 19847.3 | 24459.6 |
| 32594.3 | 33259.8 | 69020.7 | 70604.4 | 155331 | 158656 | 928.282 | 1766.55 | 1812.25 | 4070.48 | 4119.3 | 9108.84 | 9214.71 | 19236.2 | 19866.5 | 24436.1 |
| 32579.2 | 33301.6 | 69625.5 | 70681.1 | 156264 | 158791 | 928.556 | 1815.89 | 1815 | 4101.57 | 4124.68 | 9095.75 | 9225.66 | 19506.3 | 19885.7 | 24479.7 |
| 32963.7 | 33343.4 | 69514.1 | 70757.7 | 155468 | 158926 | 928.83 | 1770.06 | 1817.76 | 4057.04 | 4130.06 | 8942.98 | 9236.62 | 19484.5 | 19904.9 | 24561.5 |
| 32732.7 | 33385.2 | 69663 | 70834.4 | 154631 | 159061 | 929.104 | 1810.34 | 1820.52 | 4031.68 | 4135.44 | 9152.1 | 9247.58 | 19351.1 | 19924.2 | 24550.5 |
| 32822.9 | 33427.1 | 69471.5 | 70911.1 | 156209 | 159196 | 929.377 | 1868.57 | 1823.28 | 4088.82 | 4140.82 | 9074.37 | 9258.54 | 19632 | 19943.4 | 24688.2 |
| 32662 | 33468.9 | 69823.1 | 70987.8 | 155451 | 159331 | 929.651 | 1814.36 | 1826.04 | 4022.68 | 4146.2 | 9215.71 | 9269.5 | 19601.3 | 19962.6 | 24555.8 |
| 32844.7 | 33510.7 | 69109.3 | 71064.5 | 155271 | 159466 | 929.925 | 1812.56 | 1828.8 | 4040.72 | 4151.59 | 9000.81 | 9280.46 | 19315.5 | 19981.8 | 24569.3 |
| 33064.4 | 33552.6 | 69585.7 | 71141.2 | 155862 | 159601 | 930.199 | 1818.27 | 1831.57 | 4018.64 | 4156.98 | 9079.38 | 9291.43 | 19455 | 20001 | 24544.3 |
| 33148.3 | 33594.5 | 69919.7 | 71217.9 | 157235 | 159736 | 930.473 | 1877.72 | 1834.33 | 4124.1 | 4162.37 | 9158.86 | 9302.39 | 19716 | 20020.2 | 24831.6 |
| 33125.5 | 33636.3 | 69821.9 | 71294.7 | 157194 | 159870 | 930.747 | 1844.44 | 1837.1 | 4115.6 | 4167.76 | 9147.36 | 9313.36 | 19661.1 | 20039.5 | 24713.7 |
| 33299.7 | 33678.2 | 70699.3 | 71371.4 | 157729 | 160005 | 931.021 | 1855.51 | 1839.87 | 4104.58 | 4173.15 | 9210.11 | 9324.33 | 19828.5 | 20058.7 | 24911.5 |
| 33497.9 | 33720.2 | 70252.4 | 71448.1 | 157449 | 160140 | 931.294 | 1788.62 | 1842.64 | 4114.33 | 4178.55 | 9115.6 | 9335.31 | 19661.6 | 20077.9 | 24817.5 |
| 33240.5 | 33762.1 | 70458.1 | 71524.9 | 157835 | 160275 | 931.568 | 1898.56 | 1845.41 | 4124.6 | 4183.94 | 9118.33 | 9346.28 | 19686.8 | 20097.1 | 24888.2 |
| 33149.2 | 33804 | 69921.6 | 71601.6 | 157631 | 160410 | 931.842 | 1830.77 | 1848.19 | 4140.71 | 4189.34 | 9168.1 | 9357.26 | 19812.2 | 20116.3 | 24845.3 |
| 33132.3 | 33845.9 | 70071.1 | 71678.4 | 157313 | 160544 | 932.116 | 1872.1 | 1850.96 | 4137.21 | 4194.74 | 9207.68 | 9368.24 | 19773 | 20135.5 | 25113.9 |
| 33733.5 | 33887.9 | 70578.4 | 71755.1 | 157658 | 160679 | 932.39 | 1855.95 | 1853.74 | 4072.54 | 4200.14 | 9277.6 | 9379.22 | 19803.8 | 20154.7 | 24873.5 |
| 33160.4 | 33929.8 | 70547.7 | 71831.9 | 157116 | 160814 | 932.663 | 1880.53 | 1856.52 | 4088 | 4205.54 | 9179.84 | 9390.2 | 19805.8 | 20173.8 | 24958.4 |
| 33516.3 | 33971.8 | 70320.6 | 71908.7 | 157367 | 160948 | 932.937 | 1869.46 | 1859.3 | 4121.01 | 4210.94 | 9126.84 | 9401.18 | 19688.8 | 20193 | 24883.6 |
| 33541.6 | 34013.8 | 70378 | 71985.4 | 157064 | 161083 | 933.211 | 1858.88 | 1862.08 | 4073.66 | 4216.35 | 9159.5 | 9412.17 | 19737.9 | 20212.2 | 24757.5 |
| 33355 | 34055.8 | 70687.7 | 72062.2 | 157834 | 161218 | 933.484 | 1834.78 | 1864.87 | 4137.33 | 4221.76 | 9269.1 | 9423.16 | 19844.3 | 20231.4 | 24729.4 |
| 33509.9 | 34097.8 | 71001 | 72139 | 158181 | 161352 | 933.758 | 1864.22 | 1867.65 | 4114.39 | 4227.16 | 9219.42 | 9434.15 | 19753.8 | 20250.6 | 25056.7 |
| 33519.4 | 34139.8 | 71085.4 | 72215.8 | 157326 | 161487 | 934.032 | 1879.77 | 1870.44 | 4133.18 | 4232.58 | 9188.27 | 9445.14 | 19494.2 | 20269.8 | 25027.4 |
| 33852.9 | 34181.9 | 71480.7 | 72292.6 | 157535 | 161621 | 934.305 | 1867.83 | 1873.22 | 4162.23 | 4237.99 | 9335.7 | 9456.13 | 19905.2 | 20289 | 24983.5 |
| 33459.9 | 34223.9 | 71179.3 | 72369.4 | 158477 | 161756 | 934.579 | 1903.59 | 1876.02 | 4175.03 | 4243.4 | 9265.02 | 9467.12 | 19987.8 | 20308.1 | 25272.2 |
| 33356.5 | 34266 | 71405.8 | 72446.2 | 158793 | 161890 | 934.852 | 1871.89 | 1878.81 | 4185.49 | 4248.82 | 9257.2 | 9478.12 | 20000.5 | 20327.3 | 25106.8 |
| 33755.5 | 34308 | 71245.4 | 72523.1 | 158619 | 162024 | 935.126 | 1845.75 | 1881.6 | 4238.54 | 4254.23 | 9182.23 | 9489.12 | 19880.8 | 20346.5 | 25065.7 |
| 33735.4 | 34350.1 | 70917.4 | 72599.9 | 158561 | 162159 | 935.4 | 1842.59 | 1884.39 | 4152.69 | 4259.65 | 9184.92 | 9500.12 | 19963.9 | 20365.6 | 25329 |
| 33654.9 | 34392.2 | 70773.2 | 72676.7 | 158255 | 162293 | 935.673 | 1853.03 | 1887.19 | 4149.16 | 4265.07 | 9224.49 | 9511.12 | 19749.8 | 20384.8 | 24810.2 |
| 33651.8 | 34434.3 | 71793.5 | 72753.5 | 158890 | 162427 | 935.947 | 1855.81 | 1889.99 | 4247.25 | 4270.49 | 9279.5 | 9522.12 | 20036 | 20404 | 25269.7 |
| 33979.5 | 34476.4 | 71428.5 | 72830.4 | 158394 | 162562 | 936.22 | 1893.94 | 1892.79 | 4172.71 | 4275.91 | 9306.93 | 9533.12 | 19756.8 | 20423.1 | 25212.5 |
| 33711.2 | 34518.5 | 71524.1 | 72907.2 | 160028 | 162696 | 936.494 | 1877.4 | 1895.59 | 4168.34 | 4281.34 | 9443.76 | 9544.13 | 19846.3 | 20442.3 | 25214.8 |
| 33972.3 | 34560.7 | 71101.9 | 72984.1 | 158831 | 162830 | 936.767 | 1879.42 | 1898.39 | 4211.35 | 4286.76 | 9284.85 | 9555.14 | 19960.4 | 20461.5 | 25205.6 |
| 33838.5 | 34602.8 | 71262.2 | 73060.9 | 159126 | 162964 | 937.04 | 1886.61 | 1901.19 | 4254.4 | 4292.19 | 9436.1 | 9566.15 | 20050.5 | 20480.6 | 25358.1 |
| 34080.1 | 34645 | 71377.8 | 73137.8 | 159391 | 163098 | 937.314 | 1923.42 | 1904 | 4176.39 | 4297.62 | 9340.73 | 9577.16 | 20274.6 | 20499.8 | 25198.6 |
| 34482.4 | 34687.1 | 71834.1 | 73214.6 | 159670 | 163232 | 937.587 | 1861.71 | 1906.8 | 4296 | 4303.05 | 9380.86 | 9588.17 | 20097.3 | 20518.9 | 25147.7 |
| 33951 | 34729.3 | 72085.5 | 73291.5 | 158708 | 163366 | 937.861 | 1878.86 | 1909.61 | 4215.27 | 4308.48 | 9361.79 | 9599.18 | 19983.9 | 20538.1 | 25252.4 |
| 34268.6 | 34771.5 | 71427.2 | 73368.4 | 159987 | 163500 | 938.134 | 1861.01 | 1912.42 | 4198.86 | 4313.91 | 9346.74 | 9610.2 | 20122.9 | 20557.2 | 25295.9 |
| 34202.3 | 34813.7 | 71666.8 | 73445.2 | 159477 | 163634 | 938.407 | 1925.92 | 1915.23 | 4199.31 | 4319.35 | 9394.39 | 9621.22 | 19840.3 | 20576.3 | 25483.3 |
| 33902.9 | 34855.9 | 71971.3 | 73522.1 | 159748 | 163768 | 938.681 | 1867.56 | 1918.04 | 4123.29 | 4324.78 | 9478.86 | 9632.24 | 20201.2 | 20595.5 | 25365.8 |
| 34546.1 | 34898.1 | 71875.2 | 73599 | 160128 | 163902 | 938.954 | 1909.54 | 1920.86 | 4283.46 | 4330.22 | 9499.81 | 9643.26 | 20181.4 | 20614.6 | 25386.8 |
| 34282.6 | 34940.4 | 71735.3 | 73675.9 | 158541 | 164036 | 939.227 | 1885.01 | 1923.67 | 4133.57 | 4335.66 | 9344.4 | 9654.28 | 20056.5 | 20633.8 | 25385.4 |
| 34256.7 | 34982.6 | 72226.3 | 73752.8 | 160214 | 164170 | 939.5 | 1889.43 | 1926.49 | 4260.54 | 4341.1 | 9383.25 | 9665.3 | 20132 | 20652.9 | 25518.2 |
| 34217.5 | 35024.8 | 71854.9 | 73829.7 | 159539 | 164304 | 939.774 | 1914.31 | 1929.31 | 4178.49 | 4346.55 | 9496.02 | 9676.33 | 20034.2 | 20672 | 25386.2 |
| 34226.6 | 35067.1 | 71390.5 | 73906.6 | 160186 | 164437 | 940.047 | 1951.13 | 1932.13 | 4188.35 | 4351.99 | 9441.51 | 9687.35 | 20272.1 | 20691.1 | 25347.9 |
| 34757.3 | 35109.4 | 72403.1 | 73983.5 | 160750 | 164571 | 940.32 | 1921.09 | 1934.95 | 4204.34 | 4357.43 | 9447.96 | 9698.38 | 20253.6 | 20710.2 | 25350.5 |
| 34363.2 | 35151.7 | 72279.2 | 74060.4 | 159716 | 164705 | 940.593 | 1900.08 | 1937.77 | 4252.78 | 4362.88 | 9408.22 | 9709.41 | 20174.6 | 20729.4 | 25287.2 |
| 34690.2 | 35194 | 72536.1 | 74137.3 | 159963 | 164838 | 940.867 | 1916.09 | 1940.6 | 4245.24 | 4368.33 | 9353.71 | 9720.44 | 20316.7 | 20748.5 | 25335.2 |
| 34592.7 | 35236.3 | 72523.2 | 74214.2 | 161990 | 164972 | 941.14 | 1955.7 | 1943.42 | 4293.39 | 4373.78 | 9542.03 | 9731.47 | 20201.5 | 20767.6 | 25701.8 |
| 34484.7 | 35278.6 | 72347.7 | 74291.1 | 159258 | 165106 | 941.413 | 1961.29 | 1946.25 | 4211.29 | 4379.23 | 9404.82 | 9742.5 | 20128.3 | 20786.7 | 25626.6 |
| 34897.7 | 35320.9 | 72695.7 | 74368 | 162094 | 165239 | 941.686 | 1912.94 | 1949.08 | 4284.48 | 4384.68 | 9594.69 | 9753.54 | 20204.6 | 20805.8 | 25844.4 |
| 34679.9 | 35363.2 | 72378.7 | 74444.9 | 160815 | 165373 | 941.959 | 1878.35 | 1951.91 | 4289.18 | 4390.13 | 9497.48 | 9764.58 | 20279.4 | 20824.9 | 25623.5 |
| 34813.6 | 35405.6 | 72222.6 | 74521.9 | 161285 | 165506 | 942.232 | 1938.3 | 1954.74 | 4339.66 | 4395.59 | 9438.91 | 9775.61 | 20446.8 | 20844 | 25386.2 |
| 34807.1 | 35447.9 | 72927.6 | 74598.8 | 161574 | 165640 | 942.505 | 1960.29 | 1957.58 | 4238.15 | 4401.04 | 9531.47 | 9786.65 | 20401.6 | 20863.1 | 25553.3 |
| 34739.3 | 35490.3 | 72323 | 74675.7 | 161731 | 165773 | 942.778 | 1973.31 | 1960.41 | 4201.33 | 4406.5 | 9537.94 | 9797.69 | 20396.7 | 20882.2 | 25680.5 |
| 34817.5 | 35532.7 | 72705.9 | 74752.7 | 161139 | 165906 | 943.051 | 1901.84 | 1963.25 | 4347.93 | 4411.96 | 9518.68 | 9808.73 | 20424.3 | 20901.3 | 25726.3 |
| 34587.1 | 35575.1 | 73051.4 | 74829.6 | 161574 | 166040 | 943.324 | 1951.43 | 1966.08 | 4315.03 | 4417.42 | 9566.64 | 9819.78 | 20321 | 20920.4 | 25613 |
| 34621.9 | 35617.5 | 73167.1 | 74906.6 | 161233 | 166173 | 943.597 | 1925.3 | 1968.92 | 4244.47 | 4422.88 | 9597.6 | 9830.82 | 20416.9 | 20939.5 | 25720.9 |
| 35023.8 | 35659.9 | 73336 | 74983.5 | 162273 | 166306 | 943.87 | 1972.15 | 1971.76 | 4275.06 | 4428.35 | 9552.19 | 9841.87 | 20317.7 | 20958.6 | 25728.9 |
| 34841.3 | 35702.3 | 73101.9 | 75060.4 | 161926 | 166440 | 944.143 | 1937.55 | 1974.61 | 4290.68 | 4433.81 | 9575.44 | 9852.91 | 20463.5 | 20977.7 | 25575.1 |
| 35100.5 | 35744.7 | 73377 | 75137.4 | 162478 | 166573 | 944.416 | 2006.76 | 1977.45 | 4333.69 | 4439.28 | 9791.61 | 9863.96 | 20386.2 | 20996.8 | 25888.2 |
| 34850.9 | 35787.1 | 73209.5 | 75214.3 | 162521 | 166706 | 944.689 | 1999.66 | 1980.29 | 4277.28 | 4444.74 | 9727.2 | 9875.01 | 20523.2 | 21015.8 | 25624.1 |
| 35189.9 | 35829.6 | 73412 | 75291.3 | 161640 | 166839 | 944.962 | 1959.39 | 1983.14 | 4319.61 | 4450.21 | 9568.39 | 9886.06 | 20526.9 | 21034.9 | 25838.9 |
| 34872.4 | 35872 | 73241.9 | 75368.3 | 161618 | 166972 | 945.235 | 1969.64 | 1985.99 | 4368.78 | 4455.68 | 9602.08 | 9897.11 | 20476.9 | 21054 | 25798.4 |
| 35576.9 | 35914.5 | 73818 | 75445.2 | 162457 | 167105 | 945.508 | 1957.35 | 1988.84 | 4313.68 | 4461.15 | 9540.43 | 9908.17 | 20243.7 | 21073 | 25925.9 |
| 35863 | 35957 | 74192.2 | 75522.2 | 163587 | 167238 | 945.78 | 1999.84 | 1991.69 | 4370.14 | 4466.63 | 9876.47 | 9919.22 | 20599.8 | 21092.1 | 26090.2 |
| 35663.2 | 35999.4 | 73771.2 | 75599.1 | 163070 | 167371 | 946.053 | 2028.63 | 1994.54 | 4364.62 | 4472.1 | 9645.01 | 9930.28 | 20570.4 | 21111.2 | 25894.7 |
| 35386.1 | 36041.9 | 73927.7 | 75676.1 | 163638 | 167504 | 946.326 | 1974.09 | 1997.39 | | 4477.58 | 9522.5 | 9941.34 | 20522.8 | 21130.2 | 25918.1 |
| 35583.2 | 36084.4 | 74071.6 | 75753.1 | 164363 | 167637 | 946.599 | 1932.24 | 2000.25 | | 4483.05 | 9666.44 | 9952.4 | 20530 | 21149.3 | 26005.6 |
| 35680.8 | 36127 | 73802.8 | 75830.1 | 162734 | 167770 | 946.872 | 1935.44 | 2003.11 | | 4488.53 | 9801.35 | 9963.45 | 20707 | 21168.3 | 26026.8 |
| 35549.1 | 36169.5 | 73426.7 | 75907 | 163321 | 167903 | 947.144 | 1936.67 | 2005.96 | 4385.81 | 4494.01 | 9707.51 | 9974.51 | 20731.3 | 21187.4 | 26055.6 |
| 35553.8 | 36212 | 74532.2 | 75984 | 164223 | 168036 | 947.417 | 1979.01 | 2008.82 | 4396.56 | 4499.49 | 9723.07 | 9985.58 | 20732.2 | 21206.4 | 26146.8 |
| 35616.4 | 36254.6 | 74068 | 76061 | 163925 | 168168 | 947.69 | 2009.15 | 2011.68 | 4403.75 | 4504.97 | 9679.57 | 9996.64 | 20609.1 | 21225.5 | 25913.6 |
| 35746.3 | 36297.1 | 74371.9 | 76138 | 164435 | 168301 | 947.963 | 2031.29 | 2014.55 | 4269.88 | 4510.45 | 9823.39 | 10007.7 | 20749.8 | 21244.5 | 26171.3 |
| 35745.7 | 36339.7 | 73704.9 | 76214.9 | 163186 | 168434 | 948.235 | 1981.63 | 2017.41 | 4298.07 | 4515.94 | 9650.99 | 10018.8 | 20602.9 | 21263.5 | 26003.7 |
| 35415.4 | 36382.2 | 74187.6 | 76291.9 | 163547 | 168566 | 948.508 | 1989.72 | 2020.27 | 4403.01 | 4521.43 | 9747.35 | 10029.8 | 20501.8 | 21282.5 | 25886.2 |
| 35714.7 | 36424.8 | 74205.4 | 76368.9 | 164043 | 168699 | 948.78 | 1970.11 | 2023.14 | 4397.15 | 4526.91 | 9758.76 | 10040.9 | 20705.1 | 21301.6 | 26088.6 |
| 35436.9 | 36467.4 | 74135.2 | 76445.9 | 163907 | 168832 | 949.053 | 2017.19 | 2026.01 | | 4532.4 | 9763.28 | 10052 | 20833.2 | 21320.6 | 26111 |
| 36000.7 | 36510 | 74261.6 | 76522.9 | 163520 | 168964 | 949.326 | 2043.61 | 2028.88 | | 4537.89 | 9788.55 | 10063 | 20729.4 | 21339.6 | 26102.4 |
| 35670.5 | 36552.6 | 74569.1 | 76599.9 | 164765 | 169097 | 949.598 | 2021.34 | 2031.75 | | 4543.38 | 9832.39 | 10074.1 | 20847.7 | 21358.6 | 26038 |
| 35997.1 | 36595.2 | 74930.6 | 76676.9 | 164342 | 169229 | 949.871 | 1994.53 | 2034.62 | 4388.07 | 4548.87 | 9807.42 | 10085.2 | 20860.2 | 21377.7 | 25953.7 |
| 35575 | 36637.8 | 73916.5 | 76753.9 | 163178 | 169361 | 950.143 | 2021.3 | 2037.49 | 4449.46 | 4554.36 | 9827.15 | 10096.3 | 20777.1 | 21396.7 | 25970.8 |
| 35675.7 | 36680.4 | 74579.3 | 76830.8 | 163961 | 169494 | 950.416 | 2020.52 | 2040.37 | 4503.52 | 4559.86 | 9706.36 | 10107.3 | 20790.9 | 21415.7 | 26058.3 |
| 35836.8 | 36723.1 | 74617.2 | 76907.8 | 163853 | 169626 | 950.688 | 2029.65 | 2043.24 | 4466.81 | 4565.35 | 9762.34 | 10118.4 | 20722.2 | 21434.7 | 26249.5 |
| 35726.1 | 36765.7 | 74307.1 | 76984.8 | 164072 | 169758 | 950.961 | 2000.67 | 2046.12 | 4414.96 | 4570.85 | 9805.16 | 10129.5 | 20879.4 | 21453.7 | 26231.9 |
| 35891.8 | 36808.4 | 74452.7 | 77061.8 | 163632 | 169891 | 951.233 | 1980 | 2049 | 4456.41 | 4576.35 | 9719.64 | 10140.6 | 20785.9 | 21472.7 | 26075.5 |
| 35801.2 | 36851.1 | 74510.6 | 77138.8 | 164841 | 170023 | 951.506 | 2022.88 | 2051.88 | 4404.3 | 4581.85 | 9848.13 | 10151.6 | 20814.5 | 21491.7 | 26164.3 |
| 35982.8 | 36893.7 | 74520.9 | 77215.8 | 163827 | 170155 | 951.778 | 2018.63 | 2054.76 | 4386.04 | 4587.35 | 9839.52 | 10162.7 | 20868.5 | 21510.7 | 26079.9 |
| 36026.2 | 36936.4 | 74794.1 | 77292.8 | 164723 | 170287 | 952.051 | 2053.03 | 2057.64 | 4402.95 | 4592.85 | 9828.47 | 10173.8 | 20749.9 | 21529.7 | 26192 |
| 36186.3 | 36979.1 | 74869.5 | 77369.8 | 164576 | 170419 | 952.323 | 1992.97 | 2060.53 | 4394.66 | 4598.35 | 9809.59 | 10184.9 | 20993.8 | 21548.6 | 26328.5 |
| 35936.5 | 37021.8 | 74534.4 | 77446.8 | 164439 | 170552 | 952.595 | 2054.61 | 2063.41 | 4414.11 | 4603.86 | 9865.31 | 10196 | 20876.5 | 21567.6 | 26003.6 |
| 35952.1 | 37064.5 | 75282.7 | 77523.8 | 164929 | 170684 | 952.868 | 2067.67 | 2066.3 | 4495.16 | 4609.36 | 9986.87 | 10207 | 21002.4 | 21586.6 | 26350.8 |
| 36110.5 | 37107.2 | 74712.3 | 77600.8 | 165180 | 170816 | 953.14 | 2077.41 | 2069.19 | 4390.3 | 4614.87 | 9846.82 | 10218.1 | 20902.5 | 21605.6 | 26240.5 |
| 36384.6 | 37150 | 75416.5 | 77677.8 | 165616 | 170948 | 953.412 | 2022.58 | 2072.08 | 4534.83 | 4620.38 | 9915.37 | 10229.2 | 21066.5 | 21624.5 | 26346.2 |
| 36535.3 | 37192.7 | 75061.1 | 77754.8 | 164785 | 171079 | 953.685 | 2069.67 | 2074.97 | 4496.92 | 4625.89 | 10010.3 | 10240.3 | 20967.5 | 21643.5 | 26169.4 |
| 36257.2 | 37235.4 | 75290 | 77831.8 | 165168 | 171211 | 953.957 | 2110.87 | 2077.86 | 4533.27 | 4631.4 | 9871.61 | 10251.4 | 20866.2 | 21662.5 | 26369.1 |
| 36618.5 | 37278.2 | 75555 | 77908.8 | 166093 | 171343 | 954.229 | 2030.25 | 2080.76 | 4484.53 | 4636.91 | 9969.84 | 10262.5 | 20938.2 | 21681.4 | 26381.1 |
| 36148.5 | 37321 | 75526.5 | 77985.8 | 164782 | 171475 | 954.501 | 2044.53 | 2083.65 | 4444.29 | 4642.42 | 9921.69 | 10273.6 | 20813.5 | 21700.4 | 26260.1 |
| 36526.2 | 37363.7 | 76112.7 | 78062.8 | 166065 | 171607 | 954.774 | 2048.73 | 2086.55 | 4541.49 | 4647.93 | 9991.82 | 10284.6 | 20933 | 21719.3 | 26567.6 |
| 36615.1 | 37406.5 | 75972.3 | 78139.8 | 167226 | 171738 | 955.046 | 2083.87 | 2089.45 | 4490.78 | 4653.45 | 9958.56 | 10295.7 | 21203 | 21738.3 | 26653 |
| 36016.4 | 37449.3 | 75546.5 | 78216.8 | 165699 | 171870 | 955.318 | 2093.49 | 2092.34 | 4487.02 | 4658.96 | 9944.16 | 10306.8 | 20994.3 | 21757.2 | 26546.3 |
| 36686.9 | 37492.1 | 75741.6 | 78293.8 | 167081 | 172002 | 955.59 | 1949 | 2095.24 | 4488.16 | 4664.48 | 10017.4 | 10317.9 | 21223 | 21776.2 | 26608.8 |
| 36681 | 37534.9 | 76059.5 | 78370.8 | 166176 | 172133 | 955.862 | 2094.34 | 2098.15 | 4527.15 | 4670 | 9903.74 | 10329 | 21191.6 | 21795.1 | 26626.1 |
| 36689.8 | 37577.7 | 75661.4 | 78447.8 | 166336 | 172265 | 956.135 | 2070.39 | 2101.05 | 4529.48 | 4675.52 | 10083.4 | 10340.1 | 21169.6 | 21814 | 26456.7 |
| 36713.8 | 37620.5 | 76001 | 78524.9 | 165554 | 172396 | 956.407 | 2072.92 | 2103.95 | 4590.36 | 4681.03 | 9991.97 | 10351.2 | 21105.5 | 21833 | 26493.6 |
| 37003.7 | 37663.3 | 76367.5 | 78601.9 | 167303 | 172528 | 956.679 | 2141.21 | 2106.86 | 4583.65 | 4686.56 | 10092.8 | 10362.3 | 21202.1 | 21851.9 | 26768.1 |
| 36854.8 | 37706.2 | 76156.6 | 78678.9 | 166458 | 172659 | 956.951 | 2088.44 | 2109.77 | 4562.56 | 4692.08 | 10031.5 | 10373.4 | 21349.8 | 21870.8 | 26917 |
| 36728 | 37749 | 76623.6 | 78755.9 | 167914 | 172791 | 957.223 | 2114.87 | 2112.68 | 4563.56 | 4697.6 | 10238.2 | 10384.5 | 21408.1 | 21889.7 | 26612.5 |
| 36731.3 | 37791.9 | 76637.9 | 78832.9 | 166651 | 172922 | 957.495 | 2110.17 | 2115.59 | 4541.31 | 4703.13 | 10136.7 | 10395.6 | 21412.6 | 21908.6 | 26835.2 |
| 36934 | 37834.7 | 76405.2 | 78909.9 | 167628 | 173053 | 957.767 | 2087.25 | 2118.5 | 4613.81 | 4708.65 | 10131.6 | 10406.7 | 21249.3 | 21927.5 | 26755.5 |
| 37031.5 | 37877.6 | 76541.8 | 78986.8 | 167130 | 173184 | 958.039 | 2125.79 | 2121.41 | 4564.87 | 4714.18 | 10011.6 | 10417.8 | 21403.1 | 21946.4 | 26742.4 |
| 36828.7 | 37920.5 | 76890.7 | 79063.8 | 168468 | 173316 | 958.311 | 2106.66 | 2124.32 | 4578.77 | 4719.71 | 10190.7 | 10428.9 | 21561.4 | 21965.3 | 26918.7 |
| 36826.8 | 37963.3 | 76022.6 | 79140.8 | 167914 | 173447 | 958.583 | 2164.26 | 2127.24 | 4533.56 | 4725.23 | 10089.8 | 10440 | 21264.1 | 21984.2 | 26632.2 |
| 36997.1 | 38006.2 | 77025.5 | 79217.8 | 167466 | 173578 | 958.855 | 2103.62 | 2130.15 | 4591.14 | 4730.76 | 10173 | 10451.1 | 21365.1 | 22003.1 | 27160.4 |
| 36838.9 | 38049.1 | 77009.8 | 79294.8 | 168446 | 173709 | 959.127 | 2050.34 | 2133.07 | 4695.27 | 4736.3 | 10187.4 | 10462.2 | 21498.4 | 22022 | 26817.4 |
| 36911.8 | 38092 | 77428.9 | 79371.8 | 169498 | 173840 | 959.399 | 2081.24 | 2135.99 | 4583.86 | 4741.83 | 10181.3 | 10473.3 | 21563.7 | 22040.9 | 26846.2 |
| 37188.1 | 38134.9 | 77339.7 | 79448.8 | 168647 | 173971 | 959.67 | 2162.3 | 2138.91 | 4590.38 | 4747.36 | 10267.2 | 10484.4 | 21559.8 | 22059.8 | 26694.1 |
| 37068.5 | 38177.9 | 77521.2 | 79525.8 | 168915 | 174102 | 959.942 | 2122.27 | 2141.83 | 4638.46 | 4752.89 | 10178.5 | 10495.5 | 21618.2 | 22078.7 | 26807.2 |
| 37336.2 | 38220.8 | 77338 | 79602.8 | 169360 | 174233 | 960.214 | 2082.12 | 2144.76 | 4564.73 | 4758.43 | 10097.9 | 10506.6 | 21421.9 | 22097.5 | 27072.7 |
| 37213.5 | 38263.7 | 77889 | 79679.8 | 169217 | 174364 | 960.486 | 2119.62 | 2147.68 | 4674.05 | 4763.96 | 10389.4 | 10517.7 | 21477.2 | 22116.4 | 26947.8 |
| 37812.5 | 38306.7 | 78029.1 | 79756.8 | 171564 | 174495 | 960.758 | 2150 | 2150.61 | 4724.66 | 4769.5 | 10228 | 10528.8 | 21756.3 | 22135.3 | 27181.5 |
| 37434.3 | 38349.6 | 77266.9 | 79833.8 | 169509 | 174626 | 961.03 | 2201.55 | 2153.53 | 4672.21 | 4775.04 | 10280 | 10539.9 | 21379.6 | 22154.2 | 27145.1 |
| 37408.3 | 38392.6 | 77546.9 | 79910.7 | 168940 | 174757 | 961.301 | 2217.86 | 2156.46 | 4636.8 | 4780.58 | 10103.5 | 10551 | 21518.8 | 22173 | 26910.3 |
| 37344.1 | 38435.5 | 77934.5 | 79987.7 | 169574 | 174887 | 961.573 | 2133.95 | 2159.39 | 4683.68 | 4786.12 | 10311.7 | 10562.1 | 21475 | 22191.9 | 27315 |
| 38262.8 | 38478.5 | 78948.1 | 80064.7 | 172951 | 175018 | 961.845 | 2173.41 | 2162.32 | 4727.65 | 4791.66 | 10403.7 | 10573.2 | 21630.8 | 22210.7 | 27583 |
| 37560.2 | 38521.5 | 78319.1 | 80141.7 | 169664 | 175149 | 962.117 | 2147.78 | 2165.25 | 4702.36 | 4797.2 | 10297.1 | 10584.4 | 21420.2 | 22229.5 | 27070.4 |
| 38240.1 | 38564.4 | 79088.1 | 80218.7 | 171992 | 175279 | 962.388 | 2120.53 | 2168.18 | 4785.93 | 4802.74 | 10497 | 10595.5 | 21781.6 | 22248.4 | 27311.7 |
| 37789.6 | 38607.4 | 78239.2 | 80295.6 | 171021 | 175410 | 962.66 | 2123.59 | 2171.12 | 4694.96 | 4808.29 | 10450.5 | 10606.6 | 21618.5 | 22267.2 | 27219.7 |
| 37927.7 | 38650.4 | 78048.3 | 80372.6 | 171678 | 175540 | 962.932 | 2213.09 | 2174.05 | 4716.96 | 4813.83 | 10395.2 | 10617.7 | 21802.6 | 22286.1 | 27370.8 |
| 38029.4 | 38693.4 | 78626.5 | 80449.6 | 170816 | 175671 | 963.203 | 2226.02 | 2176.99 | 4796.03 | 4819.38 | 10395.2 | 10628.8 | 21767.8 | 22304.9 | 27289.1 |
| 37467.6 | 38736.5 | 77745.4 | 80526.6 | 170409 | 175801 | 963.475 | 2159.21 | 2179.93 | 4753.78 | 4824.93 | 10485.5 | 10639.9 | 21586.3 | 22323.7 | 27269.7 |
| 38291.2 | 38779.5 | 79038.7 | 80603.5 | 171521 | 175932 | 963.746 | 2152.89 | 2182.87 | 4861.7 | 4830.47 | 10468.2 | 10651 | 21811.1 | 22342.5 | 27400.8 |
| 38233.4 | 38822.5 | 78766 | 80680.5 | 171224 | 176062 | 964.018 | 2141.4 | 2185.81 | 4738.97 | 4836.02 | 10477.6 | 10662.1 | 21668.9 | 22361.4 | 27127.4 |
| 38433.4 | 38865.5 | 78866.9 | 80757.4 | 171186 | 176192 | 964.29 | 2168.66 | 2188.75 | 4686.64 | 4841.57 | 10425.9 | 10673.2 | 21789.7 | 22380.2 | 27241.3 |
| 38377.2 | 38908.6 | 78573.4 | 80834.4 | 172003 | 176322 | 964.561 | 2217.64 | 2191.69 | 4689.81 | 4847.12 | 10497.3 | 10684.4 | 21712.9 | 22399 | 27466.7 |
| 38142.6 | 38951.6 | 78748.4 | 80911.4 | 171415 | 176453 | 964.833 | 2196.85 | 2194.64 | 4582.98 | 4852.67 | 10421.2 | 10695.5 | 21976.3 | 22417.8 | 27521.8 |
| 37991 | 38994.7 | 78856.4 | 80988.3 | 172270 | 176583 | 965.104 | 2180.67 | 2197.58 | 4671.22 | 4858.23 | 10377.6 | 10706.6 | 21737 | 22436.6 | 27372.6 |
| 37834.8 | 39037.7 | 78688.8 | 81065.3 | 172414 | 176713 | 965.376 | 2190.63 | 2200.53 | 4811.55 | 4863.78 | 10455.4 | 10717.7 | 22128.7 | 22455.4 | 27565.9 |
| 38093.8 | 39080.8 | 78893.1 | 81142.2 | 172208 | 176843 | 965.647 | 2187.38 | 2203.48 | 4785.87 | 4869.33 | 10383.2 | 10728.8 | 22042.8 | 22474.2 | 27455.7 |
| 38324.8 | 39123.9 | 79090 | 81219.2 | 170565 | 176973 | 965.919 | 2196.6 | 2206.43 | 4744.4 | 4874.89 | 10457.6 | 10739.9 | 21904 | 22493 | 27613.5 |
| 39013.4 | 39167 | 79700.8 | 81296.1 | 173381 | 177103 | 966.19 | 2188.43 | 2209.38 | 4877.65 | 4880.44 | 10672.2 | 10751 | 22129.3 | 22511.7 | 27897.7 |
| 38142.3 | 39210.1 | 79715.9 | 81373.1 | 172688 | 177233 | 966.461 | 2200.82 | 2212.33 | 4744.84 | 4886 | 10540.6 | 10762.2 | 21886.5 | 22530.5 | 27856.5 |
| 38364.3 | 39253.1 | 80041.4 | 81450 | 174608 | 177363 | 966.733 | 2211.25 | 2215.28 | 4790.7 | 4891.56 | 10521.9 | 10773.3 | 21984.2 | 22549.3 | 27561.8 |
| 38293.3 | 39296.2 | 79593.2 | 81526.9 | 173163 | 177493 | 967.004 | 2191.16 | 2218.23 | 4722.03 | 4897.12 | 10562.8 | 10784.4 | 21773.3 | 22568.1 | 27411.7 |
| 38627.1 | 39339.3 | 79780.3 | 81603.9 | 173848 | 177623 | 967.275 | 2244.16 | 2221.19 | 4819.06 | 4902.68 | 10685.8 | 10795.5 | 22034 | 22586.8 | 27789.5 |
| 38487.6 | 39382.5 | 79415.4 | 81680.8 | 172652 | 177753 | 967.547 | 2154.6 | 2224.15 | 4759.18 | 4908.24 | 10574.7 | 10806.6 | 22057.6 | 22605.6 | 27615.5 |
| 38461.4 | 39425.6 | 79177.8 | 81757.7 | 173487 | 177882 | 967.818 | 2184.67 | 2227.1 | 4884.48 | 4913.8 | 10544.6 | 10817.8 | 22271.4 | 22624.3 | 27876.3 |
| 38416.1 | 39468.7 | 79728.8 | 81834.7 | 173619 | 178012 | 968.09 | 2238.07 | 2230.06 | 4832.39 | 4919.36 | 10509 | 10828.9 | 22145.8 | 22643.1 | 27615 |
| 39146.4 | 39511.8 | 80109.1 | 81911.6 | 173456 | 178142 | 968.361 | 2267.91 | 2233.02 | 4845.15 | 4924.93 | 10510.9 | 10840 | 22021.2 | 22661.8 | 27829.1 |
| 38264.8 | 39555 | 80080.8 | 81988.5 | 174311 | 178271 | 968.632 | 2231.99 | 2235.98 | 4842.55 | 4930.49 | 10703 | 10851.1 | 21957.8 | 22680.6 | 28134 |
| 39126.4 | 39598.1 | 79608.6 | 82065.4 | 174544 | 178401 | 968.903 | 2219 | 2238.95 | 4935.06 | 4936.05 | 10525.4 | 10862.2 | 22123.1 | 22699.3 | 27853.3 |
| 39072.1 | 39641.3 | 80184.2 | 82142.3 | 174103 | 178530 | 969.174 | 2192.11 | 2241.91 | 4757.39 | 4941.62 | 10563.6 | 10873.4 | 22127 | 22718 | 27564.7 |
| 39201.4 | 39684.4 | 80713.5 | 82219.2 | 174576 | 178660 | 969.446 | 2179.18 | 2244.87 | 4895.7 | 4947.19 | 10553.6 | 10884.5 | 22292.6 | 22736.8 | 28165.5 |
| 39155.7 | 39727.6 | 80930.1 | 82296.1 | 175855 | 178789 | 969.717 | 2250.3 | 2247.84 | 4803.26 | 4952.75 | 10664.5 | 10895.6 | 22217 | 22755.5 | 28129 |
| 38991.4 | 39770.8 | 80010.2 | 82373 | 173917 | 178919 | 969.988 | 2260.6 | 2250.81 | 4762.89 | 4958.32 | 10658.5 | 10906.7 | 22015.9 | 22774.2 | 27673.8 |
| 38864.7 | 39813.9 | 80511.8 | 82449.9 | 173920 | 179048 | 970.259 | 2219.11 | 2253.78 | 4862.56 | 4963.89 | 10721.4 | 10917.8 | 22235.2 | 22792.9 | 27647.4 |
| 38891.4 | 39857.1 | 80617.4 | 82526.8 | 173822 | 179177 | 970.53 | 2227.34 | 2256.75 | 4837.65 | 4969.46 | 10548.1 | 10929 | 21974.5 | 22811.7 | 27720.5 |
| 39352.4 | 39900.3 | 81224.5 | 82603.7 | 175714 | 179306 | 970.801 | 2279.51 | 2259.72 | 4866.2 | 4975.03 | 10801.9 | 10940.1 | 22589 | 22830.4 | 28189.8 |
| 39201.3 | 39943.5 | 80712.6 | 82680.6 | 175782 | 179436 | 971.072 | 2238.06 | 2262.69 | 4799.4 | 4980.6 | 10722.1 | 10951.2 | 22382.6 | 22849.1 | 27811.2 |
| 39492.1 | 39986.7 | 80535 | 82757.5 | 175386 | 179565 | 971.343 | 2306.28 | 2265.66 | 4838.82 | 4986.17 | 10631.6 | 10962.3 | 22347.4 | 22867.8 | 27927.5 |
| 39296.2 | 40029.9 | 80992.8 | 82834.4 | 174535 | 179694 | 971.615 | 2245.32 | 2268.64 | 4882.61 | 4991.75 | 10659.4 | 10973.5 | 22114.5 | 22886.5 | 27904.7 |
| 39004.8 | 40073.1 | 80596.3 | 82911.2 | 175404 | 179823 | 971.886 | 2343.94 | 2271.61 | 4898.51 | 4997.32 | 10835.7 | 10984.6 | 22681.5 | 22905.1 | 28185.6 |
| 39763.5 | 40116.3 | 81251.4 | 82988.1 | 176221 | 179952 | 972.157 | 2323.34 | 2274.59 | 4853 | 5002.9 | 10676.8 | 10995.7 | 22501.7 | 22923.8 | 28109.4 |
| 39347.1 | 40159.5 | 80873.8 | 83065 | 174342 | 180081 | 972.428 | 2290.19 | 2277.57 | 4838.52 | 5008.47 | 10604.1 | 11006.8 | 22463.8 | 22942.5 | 27971.3 |
| 39785.8 | 40202.8 | 81243.1 | 83141.8 | 175503 | 180210 | 972.698 | 2287.97 | 2280.55 | 4844.68 | 5014.05 | 10936.8 | 11017.9 | 22531.2 | 22961.2 | 27989.6 |
| 39557.1 | 40246 | 81523.3 | 83218.7 | 175929 | 180339 | 972.97 | 2257.46 | 2283.53 | 4798.02 | 5019.62 | 10515.5 | 11029.1 | 22478.8 | 22979.8 | 28015 |
| 39665.9 | 40289.2 | 81718.9 | 83295.5 | 177081 | 180468 | 973.24 | 2230.97 | 2286.51 | 4908.55 | 5025.2 | 10978.9 | 11040.2 | 22630 | 22998.5 | 28306.8 |
| 39450 | 40332.5 | 81467.1 | 83372.4 | 176287 | 180596 | 973.511 | 2271.39 | 2289.49 | 4977.53 | 5030.78 | 10801 | 11051.3 | 22489.7 | 23017.2 | 28153.3 |
| 39739.1 | 40375.7 | 81416.4 | 83449.2 | 176667 | 180725 | 973.782 | 2288.65 | 2292.47 | 4941.81 | 5036.36 | 10792.8 | 11062.4 | 22357.7 | 23035.8 | 28057.6 |
| 39416.6 | 40419 | 81743.6 | 83526.1 | 174412 | 180854 | 974.053 | 2270.02 | 2295.46 | 4874.29 | 5041.94 | 10838.1 | 11073.5 | 22363.4 | 23054.5 | 28067 |
| 39412.1 | 40462.2 | 81576.4 | 83602.9 | 177154 | 180983 | 974.324 | 2263.31 | 2298.44 | 4878.47 | 5047.52 | 10874.5 | 11084.7 | 22450.8 | 23073.1 | 28250.5 |
| 39528.1 | 40505.5 | 81950.6 | 83679.7 | 176419 | 181111 | 974.595 | 2280.74 | 2301.43 | 4856.93 | 5053.1 | 10907.5 | 11095.8 | 22565.1 | 23091.8 | 28028.4 |
| 40084.4 | 40548.8 | 81097.6 | 83756.6 | 176853 | 181240 | 974.866 | 2257.26 | 2304.42 | 4946.18 | 5058.68 | 10791.3 | 11106.9 | 22588.6 | 23110.4 | 28246.5 |
| 40070.9 | 40592.1 | 82368.9 | 83833.4 | 177483 | 181368 | 975.136 | 2309.52 | 2307.41 | 4883.6 | 5064.26 | 11114.5 | 11118 | 22593.6 | 23129 | 28380.7 |
| 39435.8 | 40635.3 | 81658.3 | 83910.2 | 176228 | 181497 | 975.407 | 2236.95 | 2310.4 | 4887.85 | 5069.85 | 10714.6 | 11129.2 | 22421.3 | 23147.6 | 28328.2 |
| 39930.9 | 40678.6 | 82529.3 | 83987 | 177473 | 181625 | 975.678 | 2254.76 | 2313.39 | 5018.4 | 5075.43 | 10845.3 | 11140.3 | 22627.7 | 23166.3 | 28328.3 |
| 40053.5 | 40721.9 | 82420.1 | 84063.8 | 177824 | 181754 | 975.949 | 2322.62 | 2316.38 | 4902.44 | 5081.02 | 10960.9 | 11151.4 | 22626.1 | 23184.9 | 28419.2 |
| 40158.4 | 40765.2 | 82623.2 | 84140.6 | 178469 | 181882 | 976.219 | 2312.23 | 2319.37 | 4924.41 | 5086.6 | 10841.5 | 11162.5 | 22699.6 | 23203.5 | 28650.7 |
| 39885.5 | 40808.5 | 81817.7 | 84217.4 | 177606 | 182010 | 976.49 | 2304.48 | 2322.37 | 5017.68 | 5092.19 | 10874.7 | 11173.7 | 22631.4 | 23222.1 | 28529.1 |
| 40805 | 40851.8 | 83101.1 | 84294.2 | 177725 | 182138 | 976.761 | 2276.64 | 2325.37 | 4973.71 | 5097.77 | 10864.8 | 11184.8 | 22562.4 | 23240.7 | 28533.2 |
| 40206 | 40895.1 | 81943.6 | 84371 | 177823 | 182267 | 977.031 | 2275.56 | 2328.36 | 4945.87 | 5103.36 | 10993.8 | 11195.9 | 22790.1 | 23259.3 | 28183.5 |
| 40394.8 | 40938.5 | 82052.1 | 84447.7 | 178236 | 182395 | 977.302 | 2331.69 | 2331.36 | 4928.72 | 5108.95 | 10853.6 | 11207 | 22874.9 | 23277.9 | 28441.8 |
| 40306 | 40981.8 | 81684.1 | 84524.5 | 178047 | 182523 | 977.573 | 2369.45 | 2334.36 | 4981.9 | 5114.54 | 10896.5 | 11218.2 | 22799.5 | 23296.5 | 28270.7 |
| 40928.1 | 41025.1 | 82497.6 | 84601.3 | 178596 | 182651 | 977.843 | 2317.21 | 2337.36 | 5019.88 | 5120.13 | 11208.7 | 11229.3 | 22930.6 | 23315 | 28994.9 |
| 40300.1 | 41068.4 | 83786.8 | 84678 | 178277 | 182779 | 978.114 | 2377.94 | 2340.36 | 4979.97 | 5125.72 | 11067.5 | 11240.4 | 22774.9 | 23333.6 | 28636.3 |
| 39935.7 | 41111.8 | 82223.1 | 84754.8 | 177518 | 182907 | 978.385 | 2239.31 | 2343.37 | 5065.23 | 5131.31 | 11091.5 | 11251.5 | 22805.2 | 23352.2 | 28569.3 |
| 40101.3 | 41155.1 | 82882.8 | 84831.5 | 178976 | 183035 | 978.655 | 2322.72 | 2346.37 | 5082.29 | 5136.9 | 11021.4 | 11262.7 | 22774.5 | 23370.8 | 28460 |
| 40573.2 | 41198.5 | 82434.6 | 84908.3 | 178380 | 183163 | 978.926 | 2320.24 | 2349.37 | 4940.89 | 5142.49 | 10937.1 | 11273.8 | 22622.1 | 23389.3 | 28391.5 |
| 40469.8 | 41241.8 | 83012.4 | 84985 | 178680 | 183290 | 979.196 | 2284.53 | 2352.38 | 5062.6 | 5148.09 | 10983 | 11284.9 | 22935 | 23407.9 | 28570.5 |
| 40399.3 | 41285.2 | 83455.5 | 85061.8 | 178968 | 183418 | 979.466 | 2321.59 | 2355.39 | 4957.08 | 5153.68 | 11051.8 | 11296 | 23130.5 | 23426.4 | 28379.1 |
| 40738 | 41328.5 | 83394.6 | 85138.5 | 180766 | 183546 | 979.737 | 2419.58 | 2358.39 | 5061.48 | 5159.27 | 11082.3 | 11307.2 | 23176.4 | 23445 | 28845.5 |
| 60394.9 | 61055.3 | 117731 | 118740 | 235038 | 236116 | 1101.15 | 3648.72 | 3799.7 | 7602.33 | 7700.15 | 16002 | 16234.6 | 30950.1 | 31219.1 | 37681.6 |
| 61045.6 | 61333.5 | 118307 | 119196 | 236478 | 236784 | 1102.87 | 3844.03 | 3821.04 | 7563.48 | 7735.84 | 16238.3 | 16302.2 | 30994.7 | 31319.6 | 37881 |
| 60754.9 | 61611.5 | 118919 | 119651 | 236925 | 237449 | 1104.6 | 3558.65 | 3842.39 | 7738.38 | 7771.49 | 16172.6 | 16369.6 | 31439.1 | 31419.8 | 38013.6 |
| 60928.9 | 61889.2 | 119035 | 120106 | 237023 | 238111 | 1106.32 | 3795.46 | 3863.75 | 7697.3 | 7807.09 | 15926.4 | 16436.8 | 31142.8 | 31519.6 | 38216.8 |
| 61782.3 | 62166.6 | 119699 | 120559 | 238141 | 238771 | 1108.05 | 4070.02 | 3885.11 | 7510.64 | 7842.65 | 16344 | 16504 | 31322.8 | 31619 | 38288.9 |
| 62114.8 | 62443.7 | 119896 | 121012 | 238564 | 239429 | 1109.77 | 3617.57 | 3906.47 | 7793.42 | 7878.15 | 16225.2 | 16571 | 31671.7 | 31718.1 | 38339.2 |
| 62128.6 | 62720.5 | 119836 | 121463 | 238547 | 240084 | 1111.5 | 4059.53 | 3927.84 | 7958.79 | 7913.62 | 16442.6 | 16637.8 | 31670 | 31816.9 | 38653.2 |
| 62493.1 | 62996.9 | 121044 | 121914 | 239700 | 240736 | 1113.22 | 3906.72 | 3949.2 | 7971.19 | 7949.03 | 16499.2 | 16704.6 | 31861 | 31915.2 | 38581.2 |
| 63562.3 | 63273.1 | 121728 | 122363 | 241485 | 241385 | 1114.95 | 3676.95 | 3970.57 | 8093.13 | 7984.4 | 16388.5 | 16771.2 | 31919 | 32013.3 | 38577 |
| 63403.1 | 63548.9 | 121604 | 122812 | 241741 | 242032 | 1116.67 | 4016.95 | 3991.95 | 7854.21 | 8019.71 | 16627.9 | 16837.6 | 31787 | 32110.9 | 38794.9 |
| 63198.6 | 63824.3 | 121884 | 123259 | 241817 | 242677 | 1118.39 | 3959.56 | 4013.32 | 7693.66 | 8054.98 | 16953.2 | 16904 | 31737.9 | 32208.2 | 38766.3 |
| 63007.8 | 64099.5 | 122433 | 123706 | 241873 | 243318 | 1120.12 | 3814.03 | 4034.69 | 7931.03 | 8090.19 | 16859.4 | 16970.1 | 31864 | 32305.2 | 38848.4 |
| 63500.8 | 64374.2 | 122952 | 124151 | 242548 | 243957 | 1121.84 | 3869.47 | 4056.06 | 7995.3 | 8125.35 | 16917.7 | 17036.2 | 32339.8 | 32401.8 | 38874.8 |
| 64159.8 | 64648.7 | 123428 | 124596 | 243956 | 244594 | 1123.57 | 4013.75 | 4077.43 | 8253.6 | 8160.46 | 17037.3 | 17102.1 | 32507.2 | 32498 | 39012.7 |
| 64319.4 | 64922.7 | 124276 | 125039 | 244705 | 245227 | 1125.29 | 4201.01 | 4098.8 | 8410.7 | 8195.51 | 17167 | 17167.8 | 32470 | 32593.8 | 39478.3 |
| 64745.4 | 65196.4 | 124709 | 125482 | 245272 | 245858 | 1127.01 | 4139.06 | 4120.17 | 8421.99 | 8230.51 | 17312.2 | 17233.4 | 32462.6 | 32689.3 | 39247.5 |
| 65372.7 | 65469.7 | 124937 | 125923 | 246136 | 246487 | 1128.74 | 4071.92 | 4141.53 | 8205.78 | 8265.45 | 17149.5 | 17298.9 | 32405.2 | 32784.4 | 39725.2 |
| 65554.6 | 65742.7 | 125372 | 126364 | 246683 | 247112 | 1130.46 | 3949.65 | 4162.89 | 8166.55 | 8300.34 | 17339.7 | 17364.2 | 32622.6 | 32879.1 | 39771.8 |
| 65072.5 | 66015.2 | 125564 | 126803 | 246141 | 247735 | 1132.18 | 3992.66 | 4184.25 | 8283.37 | 8335.16 | 17393.1 | 17429.3 | 32770.8 | 32973.5 | 39397.6 |
| 65720 | 66287.4 | 125922 | 127241 | 247029 | 248356 | 1133.91 | 4213.72 | 4205.6 | 8320.11 | 8369.93 | 17110.2 | 17494.4 | 32985.2 | 33067.5 | 39886.8 |
| 66021.1 | 66559.1 | 126949 | 127678 | 248559 | 248973 | 1135.63 | 4268.11 | 4226.95 | 8345.1 | 8404.64 | 17196.6 | 17559.2 | 33043.3 | 33161.1 | 40045.4 |
| 65787.9 | 66830.5 | 126938 | 128114 | 247955 | 249588 | 1137.35 | 4417.42 | 4248.29 | 8558.7 | 8439.29 | 17473.8 | 17623.9 | 33171.9 | 33254.4 | 40241.9 |
| 66284.2 | 67101.4 | 126659 | 128549 | 248665 | 250200 | 1139.07 | 4035.92 | 4269.62 | 8569.62 | 8473.87 | 17498.8 | 17688.5 | 33364.3 | 33347.2 | 40086.8 |
| 66393.4 | 67372 | 127599 | 128983 | 249784 | 250810 | 1140.8 | 4094.98 | 4290.95 | 8401.81 | 8508.4 | 17722.5 | 17752.8 | 32973.7 | 33439.7 | 40196.4 |
| 66939 | 67642.1 | 128152 | 129416 | 250276 | 251417 | 1142.52 | 4382.89 | 4312.27 | 8425.23 | 8542.86 | 17871.8 | 17817.1 | 32891.2 | 33531.8 | 40438.1 |
| 67244 | 67911.8 | 128428 | 129847 | 251099 | 252021 | 1144.24 | 4295.93 | 4333.58 | 8510.15 | 8577.26 | 17922.6 | 17881.1 | 33249 | 33623.6 | 40383.2 |
| 67332.6 | 68181 | 129379 | 130278 | 252059 | 252622 | 1145.96 | 4287.44 | 4354.88 | 8477.69 | 8611.59 | 17884.5 | 17945 | 33459.8 | 33714.9 | 40814.8 |
| 67922.1 | 68449.8 | 129881 | 130707 | 252822 | 253221 | 1147.69 | 4335.91 | 4376.18 | 8414.95 | 8645.86 | 17708.1 | 18008.8 | 33468.2 | 33805.9 | 40773.3 |
| 68324.5 | 68718.2 | 130248 | 131135 | 253315 | 253817 | 1149.41 | 4302.84 | 4397.46 | 8913.24 | 8680.06 | 17873.6 | 18072.4 | 33665.8 | 33896.5 | 40914.2 |
| 68616.4 | 68986.1 | 131036 | 131562 | 254780 | 254410 | 1151.13 | 4327.53 | 4418.73 | 8716.75 | 8714.2 | 18148.3 | 18135.8 | 33753 | 33986.7 | 41039.7 |
| 69471.4 | 69253.6 | 131538 | 131988 | 256171 | 255001 | 1152.85 | 4427.01 | 4440 | 8683.94 | 8748.26 | 18354.1 | 18199 | 33570.4 | 34076.6 | 41097.6 |
| 69490.5 | 69520.6 | 132148 | 132413 | 256944 | 255589 | 1154.57 | 4488.13 | 4461.25 | 8860.82 | 8782.26 | 18490.3 | 18262.1 | 33801.2 | 34166 | 41470.6 |
| 68983.1 | 69787.1 | 132594 | 132837 | 257110 | 256174 | 1156.3 | 4354.78 | 4482.49 | 8590.1 | 8816.19 | 18395.7 | 18325 | 34317.5 | 34255.1 | 41323.3 |
| 69824.9 | 70053.2 | 133066 | 133259 | 257729 | 256756 | 1158.02 | 4568.68 | 4503.71 | 8944.02 | 8850.05 | 18287.7 | 18387.8 | 34234.4 | 34343.8 | 41412.9 |
| 69978.8 | 70318.8 | 133485 | 133680 | 257832 | 257336 | 1159.74 | 4341.61 | 4524.93 | 9033.03 | 8883.84 | 18292.5 | 18450.3 | 34132.3 | 34432 | 41746.1 |
| 69944.4 | 70583.9 | 133478 | 134100 | 257941 | 257913 | 1161.46 | 4309.43 | 4546.13 | 8869.36 | 8917.56 | 18193 | 18512.7 | 34512.8 | 34519.9 | 41831.7 |
| 70345.7 | 70848.5 | 133663 | 134519 | 258334 | 258487 | 1163.18 | 4403.11 | 4567.31 | 8972.75 | 8951.21 | 18397.3 | 18575 | 34305.2 | 34607.4 | 41884.4 |
| 70725.9 | 71112.6 | 134117 | 134937 | 259675 | 259058 | 1164.9 | 4456.8 | 4588.48 | 9044.4 | 8984.78 | 18627.9 | 18637 | 34428.6 | 34694.6 | 41876.1 |
| 71341.8 | 71376.2 | 134775 | 135354 | 260102 | 259627 | 1166.62 | 4536.7 | 4609.64 | 8942.19 | 9018.28 | 18491.1 | 18698.9 | 34767.8 | 34781.3 | 41836.4 |
| 71301 | 71639.4 | 134781 | 135769 | 259307 | 260193 | 1168.34 | 4523.17 | 4630.78 | 8827.7 | 9051.71 | 18439.4 | 18760.6 | 34848.1 | 34867.7 | 42088.3 |
| 70143.7 | 71902 | 133791 | 136183 | 258253 | 260756 | 1170.06 | 4890.83 | 4651.9 | 8833.52 | 9085.06 | 18423.3 | 18822.1 | 34423.5 | 34953.6 | 41900.4 |
| 70772.6 | 72164.1 | 134648 | 136596 | 258661 | 261316 | 1171.78 | 4513.77 | 4673.01 | 8976.96 | 9118.34 | 18543.7 | 18883.5 | 34345.4 | 35039.2 | 41901.8 |
| 71510.4 | 72425.7 | 135580 | 137007 | 260424 | 261874 | 1173.5 | 4536.16 | 4694.1 | 9088.2 | 9151.54 | 18760.5 | 18944.6 | 34749.8 | 35124.3 | 42169.4 |
| 71433.3 | 72686.7 | 135886 | 137418 | 261067 | 262429 | 1175.22 | 4834.21 | 4715.18 | 9005.47 | 9184.66 | 18613 | 19005.6 | 34951.8 | 35209.1 | 42013 |
| 72073.4 | 72947.3 | 136689 | 137827 | 261777 | 262981 | 1176.94 | 4501.77 | 4736.23 | 9016.27 | 9217.7 | 18836.3 | 19066.4 | 35227 | 35293.5 | 42364.8 |
| 72297.4 | 73207.3 | 136076 | 138235 | 261672 | 263531 | 1178.66 | 4711.87 | 4757.27 | 9137.04 | 9250.67 | 19097.8 | 19127 | 35063.6 | 35377.5 | 42290.7 |
| 72736.3 | 73466.8 | 137206 | 138642 | 263000 | 264078 | 1180.38 | 4675.73 | 4778.28 | 9057.73 | 9283.56 | 18970.8 | 19187.5 | 35006.8 | 35461.1 | 42468.2 |
| 73089 | 73725.7 | 138245 | 139047 | 264584 | 264622 | 1182.11 | 4491.18 | 4799.28 | 9193.42 | 9316.37 | 19254.3 | 19247.7 | 35151.8 | 35544.2 | 42811 |
| 73163.7 | 73984.1 | 138088 | 139451 | 264038 | 265163 | 1183.82 | 4801.9 | 4820.26 | 9347.84 | 9349.09 | 19390 | 19307.8 | 35512.8 | 35627 | 42698.4 |
| 73250.8 | 74241.9 | 138330 | 139854 | 264380 | 265701 | 1185.54 | 4943.02 | 4841.21 | 9107.46 | 9381.74 | 19220.9 | 19367.7 | 35555.2 | 35709.4 | 42646.8 |
| 74270.8 | 74499.2 | 139575 | 140256 | 266591 | 266237 | 1187.26 | 4771.14 | 4862.15 | 9250.14 | 9414.31 | 19346.4 | 19427.4 | 35652.2 | 35791.4 | 42968.1 |
| 74818.2 | 74755.9 | 140383 | 140656 | 267779 | 266770 | 1188.98 | 4792 | 4883.06 | 9516.37 | 9446.79 | 19558.1 | 19486.9 | 35771.2 | 35873.1 | 43335.2 |
| 74667 | 75012.1 | 140716 | 141055 | 268559 | 267300 | 1190.7 | 4889.12 | 4903.95 | 9553.21 | 9479.19 | 19562.5 | 19546.2 | 35852.4 | 35954.3 | 43469.7 |
| 75274.5 | 75267.7 | 140775 | 141453 | 268755 | 267828 | 1192.42 | 4650.3 | 4924.81 | 9439.65 | 9511.51 | 19516.2 | 19605.3 | 36112.5 | 36035.1 | 43722.1 |
| 75652.8 | 75522.8 | 140700 | 141850 | 268262 | 268352 | 1194.14 | 4730.52 | 4945.66 | 9486.55 | 9543.75 | 19428.1 | 19664.3 | 35944.8 | 36115.5 | 43464.7 |
| 75013.5 | 75777.2 | 141156 | 142245 | 268050 | 268874 | 1195.86 | 4721.23 | 4966.48 | 9489.74 | 9575.9 | 19493.6 | 19723 | 35831.4 | 36195.5 | 43438.2 |
| 75434.9 | 76031.1 | 141275 | 142639 | 268446 | 269394 | 1197.58 | 5116.82 | 4987.28 | 9616.08 | 9607.96 | 19614.1 | 19781.6 | 36165.1 | 36275.1 | 43369.6 |
| 75819.8 | 76284.4 | 142056 | 143032 | 269544 | 269910 | 1199.3 | 5129.97 | 5008.05 | 9688.01 | 9639.94 | 19490.6 | 19840 | 36162.1 | 36354.3 | 43550.2 |
| 76233.4 | 76537.1 | 142131 | 143423 | 269883 | 270424 | 1201.02 | 4940.43 | 5028.8 | 9725.56 | 9671.84 | 19774.7 | 19898.1 | 36201.4 | 36433.2 | 43598.4 |
| 76113.9 | 76789.2 | 142550 | 143813 | 270507 | 270935 | 1202.73 | 4954.8 | 5049.52 | 9909.17 | 9703.64 | 19701.4 | 19956.1 | 36624.7 | 36511.6 | 43604 |
| 76397.8 | 77040.7 | 143440 | 144202 | 271856 | 271443 | 1204.45 | 4996.06 | 5070.21 | 9818.19 | 9735.36 | 19661.5 | 20013.9 | 36609.8 | 36589.6 | 44065.9 |
| 76993 | 77291.6 | 143671 | 144590 | 272143 | 271949 | 1206.17 | 5176.06 | 5090.88 | 9712.56 | 9766.99 | 19953.1 | 20071.5 | 36477.2 | 36667.2 | 44187.3 |
| 76960.1 | 77541.9 | 144053 | 144976 | 272166 | 272451 | 1207.89 | 5228.09 | 5111.52 | 9631.94 | 9798.54 | 19897.9 | 20128.8 | 36289 | 36744.4 | 44155.4 |
| 77202.3 | 77791.6 | 144441 | 145361 | 271810 | 272951 | 1209.61 | 5158.54 | 5132.14 | 9638.5 | 9829.99 | 20118.2 | 20186 | 36609.1 | 36821.2 | 43965.8 |
| 77301.6 | 78040.7 | 144558 | 145744 | 273182 | 273449 | 1211.32 | 5161.85 | 5152.73 | 9801.73 | 9861.36 | 20026.5 | 20243 | 36808.5 | 36897.7 | 44079.1 |
| 78002.1 | 78289.2 | 145287 | 146126 | 275060 | 273943 | 1213.04 | 5259.29 | 5173.29 | 9848.77 | 9892.63 | 20079.1 | 20299.8 | 37173.2 | 36973.7 | 44367 |
| 78288.6 | 78537 | 146312 | 146507 | 275463 | 274435 | 1214.76 | 5320.47 | 5193.82 | 9910.97 | 9923.82 | 20477.3 | 20356.4 | 37235.5 | 37049.3 | 44725 |
| 78500.9 | 78784.3 | 146362 | 146887 | 275932 | 274924 | 1216.48 | 5241.49 | 5214.32 | 9967.82 | 9954.91 | 20378.8 | 20412.8 | 37192.8 | 37124.5 | 44958.3 |
| 78640.7 | 79030.9 | 146885 | 147265 | 276738 | 275410 | 1218.19 | 5177.76 | 5234.79 | 9976.02 | 9985.91 | 20355.7 | 20469 | 37320.1 | 37199.3 | 45086.4 |
| 78782.4 | 79276.9 | 147058 | 147642 | 276967 | 275894 | 1219.91 | 5236.42 | 5255.23 | 10120.1 | 10016.8 | 20659.6 | 20524.9 | 37220.7 | 37273.7 | 45096.1 |
| 79490 | 79522.2 | 147288 | 148017 | 277064 | 276375 | 1221.63 | 5397.24 | 5275.65 | 10270.1 | 10047.6 | 20559.2 | 20580.7 | 37365 | 37347.7 | 44913.5 |
| 79343.6 | 79766.9 | 147842 | 148391 | 277786 | 276853 | 1223.35 | 5468.3 | 5296.03 | 10155.8 | 10078.4 | 20377.2 | 20636.3 | 37204.4 | 37421.3 | 44952.2 |
| 79645 | 80011 | 148525 | 148764 | 278521 | 277328 | 1225.06 | 5310.73 | 5316.38 | 10024 | 10109 | 20685.7 | 20691.7 | 37593.5 | 37494.5 | 45121 |
| 80057.9 | 80254.4 | 148792 | 149136 | 278795 | 277801 | 1226.78 | 5368.06 | 5336.69 | 10221.9 | 10139.5 | 20697.4 | 20746.8 | 37737.3 | 37567.3 | 45071.7 |
| 80114.1 | 80497.2 | 149242 | 149506 | 279859 | 278271 | 1228.5 | 5578.07 | 5356.98 | 10019.7 | 10170 | 20717 | 20801.8 | 37753.2 | 37639.7 | 45212.5 |
| 80500.4 | 80739.3 | 149877 | 149875 | 280113 | 278738 | 1230.21 | 5523.73 | 5377.24 | 10065.4 | 10200.3 | 20932.5 | 20856.6 | 37827.2 | 37711.7 | 45478.7 |
| 80657.6 | 80980.8 | 149661 | 150242 | 279381 | 279202 | 1231.93 | 5303.09 | 5397.46 | 10239.2 | 10230.6 | 20858.3 | 20911.1 | 37831.9 | 37783.3 | 45421.1 |
| 80919.2 | 81221.6 | 149828 | 150608 | 279741 | 279664 | 1233.64 | 5209.41 | 5417.65 | 10071.8 | 10260.8 | 20834.6 | 20965.5 | 37772.2 | 37854.4 | 45474.4 |
| 81403.6 | 81461.7 | 151001 | 150973 | 281273 | 280123 | 1235.36 | 5367.05 | 5437.8 | 10311.3 | 10290.8 | 21070.2 | 21019.6 | 38048 | 37925.2 | 45894.1 |
| 81439.8 | 81701.2 | 150905 | 151336 | 281176 | 280580 | 1237.08 | 5338.61 | 5457.92 | 10410 | 10320.8 | 20959.6 | 21073.5 | 37911.8 | 37995.6 | 45700.9 |
| 81654.3 | 81940 | 150712 | 151698 | 280727 | 281033 | 1238.79 | 5251.25 | 5478.01 | 10385.1 | 10350.6 | 20854.7 | 21127.2 | 37759.2 | 38065.6 | 45464.6 |
| 81753.3 | 82178.2 | 151478 | 152058 | 281910 | 281484 | 1240.51 | 5570.28 | 5498.06 | 10282.4 | 10380.4 | 20983.4 | 21180.8 | 38185.7 | 38135.2 | 45934.4 |
| 82204.8 | 82415.7 | 152169 | 152418 | 283392 | 281932 | 1242.22 | 5546.73 | 5518.07 | 10385.1 | 10410.1 | 21384.1 | 21234 | 38252.1 | 38204.4 | 45952.1 |
| 82547.8 | 82652.5 | 152399 | 152775 | 283144 | 282378 | 1243.94 | 5427.04 | 5538.06 | 10436.4 | 10439.7 | 21619.3 | 21287.1 | 38427.5 | 38273.1 | 45721.3 |
| 82514 | 82888.6 | 153126 | 153132 | 283780 | 282820 | 1245.66 | 5675.5 | 5558 | 10305.1 | 10469.1 | 21459.7 | 21340 | 38508.8 | 38341.5 | 46173.1 |
| 82622.2 | 83124.1 | 153238 | 153487 | 283483 | 283261 | 1247.37 | 5591.73 | 5577.91 | 10281.3 | 10498.5 | 21542.8 | 21392.7 | 38421.7 | 38409.5 | 46073.1 |
| 82665 | 83358.8 | 153361 | 153840 | 283899 | 283698 | 1249.09 | 5530.33 | 5597.78 | 10348.5 | 10527.8 | 21478.2 | 21445.2 | 38587.4 | 38477.1 | 45966.8 |
| 82973.8 | 83592.9 | 154062 | 154193 | 285271 | 284133 | 1250.8 | 5696.51 | 5617.61 | 10570 | 10557 | 21444.8 | 21497.4 | 38625.7 | 38544.2 | 46251.3 |
| 83569.9 | 83826.3 | 154636 | 154544 | 285528 | 284565 | 1252.52 | 5697.02 | 5637.41 | 10614.1 | 10586.1 | 21610 | 21549.4 | 38637.1 | 38611 | 46385.3 |
| 83535.3 | 84059 | 155039 | 154893 | 286130 | 284994 | 1254.23 | 5720.84 | 5657.17 | 10694 | 10615 | 21631.3 | 21601.2 | 38726.9 | 38677.4 | 46531.9 |
| 83959 | 84291 | 155085 | 155241 | 287031 | 285421 | 1255.95 | 5768.42 | 5676.9 | 10828.8 | 10643.9 | 21712.1 | 21652.8 | 38902.2 | 38743.3 | 46601.8 |
| 84604.9 | 84522.3 | 155566 | 155588 | 287240 | 285845 | 1257.66 | 5622.73 | 5696.58 | 10588.6 | 10672.7 | 21721.8 | 21704.2 | 39116.5 | 38808.9 | 46477.4 |
| 84485.8 | 84752.9 | 155942 | 155933 | 287076 | 286266 | 1259.38 | 5562.07 | 5716.22 | 10678.2 | 10701.4 | 21878.6 | 21755.4 | 39142.4 | 38874.1 | 46412.1 |
| 84495.3 | 84982.8 | 156114 | 156277 | 287542 | 286685 | 1261.09 | 5747.26 | 5735.83 | 10587.4 | 10730 | 22018.9 | 21806.4 | 38949.4 | 38938.8 | 46628.8 |
| 84683.1 | 85212 | 156285 | 156619 | 286863 | 287101 | 1262.81 | 5586.97 | 5755.39 | 10610.9 | 10758.4 | 22008.1 | 21857.1 | 38989.4 | 39003.2 | 46531.1 |
| 84753 | 85440.4 | 156992 | 156960 | 288304 | 287514 | 1264.52 | 5662.87 | 5774.92 | 10862.1 | 10786.8 | 22106.7 | 21907.7 | 39116.4 | 39067.2 | 46747.6 |
| 85111.4 | 85668.2 | 157187 | 157300 | 288924 | 287924 | 1266.23 | 5982.26 | 5794.41 | 10904.2 | 10815.1 | 21771.7 | 21958 | 39077.7 | 39130.8 | 47161.8 |
| 85881.9 | 85895.3 | 157107 | 157638 | 288050 | 288332 | 1267.95 | 5775.2 | 5813.85 | 10887.8 | 10843.2 | 21728.9 | 22008.1 | 39163.6 | 39193.9 | 46951.2 |
| 86578.6 | 86121.6 | 158523 | 157975 | 290542 | 288738 | 1269.66 | 5912.09 | 5833.26 | 10887.3 | 10871.3 | 22077.8 | 22058 | 39505.4 | 39256.7 | 47254.2 |
| 86507.4 | 86347.3 | 158687 | 158310 | 291014 | 289140 | 1271.37 | 5875.73 | 5852.62 | 10896.1 | 10899.3 | 22123.9 | 22107.6 | 39553.3 | 39319.1 | 47522.6 |
| 86842.6 | 86572.2 | 159555 | 158644 | 292115 | 289541 | 1273.09 | 5877.58 | 5871.94 | 10861.8 | 10927.1 | 22360.2 | 22157.1 | 39859 | 39381.1 | 47633.5 |
| 87542 | 86796.4 | 160470 | 158977 | 293954 | 289938 | 1274.8 | 6036.12 | 5891.22 | 11135.1 | 10954.9 | 22554.3 | 22206.3 | 39979.6 | 39442.7 | 47872.7 |
| 87421 | 87019.8 | 160153 | 159308 | 293592 | 290333 | 1276.51 | 5760.16 | 5910.46 | 11288.6 | 10982.5 | 22516 | 22255.3 | 39878.6 | 39503.8 | 47809.2 |
| 87778.9 | 87242.6 | 160792 | 159638 | 294594 | 290725 | 1278.23 | 5998.49 | 5929.66 | 11134.4 | 11010 | 22382.2 | 22304.1 | 40047.2 | 39564.6 | 47909.2 |
| 87985.8 | 87464.6 | 160906 | 159966 | 294493 | 291114 | 1279.94 | 6052.76 | 5948.81 | 11014.8 | 11037.5 | 22516.5 | 22352.7 | 40135.9 | 39625 | 47915.5 |
| 88326.5 | 87685.9 | 161122 | 160293 | 294031 | 291501 | 1281.65 | 6061.21 | 5967.92 | 10968.7 | 11064.8 | 22573.2 | 22401.1 | 39964.2 | 39685 | 47641.8 |
| 88689.7 | 87906.4 | 161952 | 160619 | 295318 | 291886 | 1283.37 | 5958.08 | 5986.99 | 11073.9 | 11092 | 22615 | 22449.2 | 40170.6 | 39744.6 | 47873.2 |
| 88269.2 | 88126.2 | 162013 | 160943 | 295238 | 292267 | 1285.08 | 6006.29 | 6006.01 | 11120.5 | 11119.2 | 22692.3 | 22497.2 | 40372 | 39803.8 | 47928.7 |
| 88584.2 | 88345.3 | 162113 | 161266 | 294854 | 292646 | 1286.79 | 5908.49 | 6024.99 | 11167.8 | 11146.2 | 22623 | 22544.9 | 40290.1 | 39862.7 | 47909.6 |
| 89025.4 | 88563.7 | 162677 | 161587 | 295551 | 293023 | 1288.5 | 6062.26 | 6043.93 | 11261.8 | 11173.1 | 22645.1 | 22592.3 | 40327.8 | 39921.1 | 48240.7 |
| 89232.1 | 88781.3 | 162784 | 161907 | 295459 | 293396 | 1290.22 | 6270.6 | 6062.82 | 11120.5 | 11199.9 | 22814.5 | 22639.6 | 40297.2 | 39979.1 | 48312.3 |
| 89409.2 | 88998.1 | 162923 | 162225 | 294935 | 293768 | 1291.93 | 6113.48 | 6081.66 | 11327.7 | 11226.6 | 22812.6 | 22686.7 | 40297.7 | 40036.8 | 48280.3 |
| 89459.7 | 89214.2 | 163157 | 162542 | 295464 | 294136 | 1293.64 | 6192.23 | 6100.47 | 11410.4 | 11253.2 | 22699.2 | 22733.5 | 40214.3 | 40094 | 48013.8 |
| 89715.8 | 89429.6 | 163070 | 162858 | 295455 | 294502 | 1295.35 | 6148.2 | 6119.22 | 11364.3 | 11279.6 | 23030 | 22780.1 | 40496.5 | 40150.8 | 48061.1 |
| 90212.5 | 89644.2 | 164076 | 163172 | 297119 | 294866 | 1297.07 | 6168.37 | 6137.93 | 11312.6 | 11306 | 23240 | 22826.5 | 40624.3 | 40207.3 | 48228.8 |
| 90233.9 | 89858.1 | 164578 | 163485 | 297698 | 295227 | 1298.78 | 6145.59 | 6156.6 | 11137.4 | 11332.3 | 23140.5 | 22872.6 | 40498.1 | 40263.4 | 48197.4 |
| 90810 | 90071.2 | 165245 | 163796 | 298788 | 295585 | 1300.49 | 6132.44 | 6175.21 | 11358.3 | 11358.5 | 23207 | 22918.6 | 40781.3 | 40319.1 | 48739.3 |
| 91006 | 90283.6 | 165768 | 164106 | 300167 | 295941 | 1302.2 | 6142.33 | 6193.78 | 11355.1 | 11384.5 | 23156.2 | 22964.3 | 40857.2 | 40374.3 | 48806.8 |
| 91791.3 | 90495.2 | 166792 | 164414 | 300985 | 296294 | 1303.91 | 6286.59 | 6212.31 | 11506.6 | 11410.4 | 23189.4 | 23009.8 | 40968.4 | 40429.2 | 48883.8 |
| 92379.6 | 90706.1 | 167269 | 164721 | 301510 | 296645 | 1305.62 | 6234.8 | 6230.79 | 11761.2 | 11436.3 | 23354 | 23055.1 | 41180.6 | 40483.8 | 48958.1 |
| 91708.6 | 90916.2 | 166668 | 165027 | 300380 | 296993 | 1307.33 | 6318.7 | 6249.22 | 11560.5 | 11462 | 23280.2 | 23100.2 | 41060.8 | 40537.9 | 48917.7 |
| 91869.6 | 91125.5 | 167253 | 165331 | 301068 | 297339 | 1309.04 | 6516.68 | 6267.6 | 11644.6 | 11487.6 | 23312.7 | 23145 | 41059.4 | 40591.6 | 49141.8 |
| 92519.1 | 91334.1 | 167969 | 165634 | 302344 | 297682 | 1310.76 | 6312.35 | 6285.94 | 11688.1 | 11513.1 | 23467.6 | 23189.6 | 41329.6 | 40645 | 49196.1 |
| 92452.5 | 91541.9 | 167770 | 165935 | 301793 | 298022 | 1312.47 | 6365.11 | 6304.22 | 11584.8 | 11538.5 | 23606 | 23234 | 41343.8 | 40697.9 | 49284 |
| 92787.7 | 91749 | 168424 | 166235 | 302838 | 298360 | 1314.18 | 6516.06 | 6322.46 | 11638.5 | 11563.8 | 23661.3 | 23278.2 | 41391.9 | 40750.5 | 49341.3 |
| 92877.4 | 91955.3 | 168937 | 166534 | 303466 | 298696 | 1315.89 | 6234.09 | 6340.65 | 11640.7 | 11589 | 23579.8 | 23322.1 | 41433.2 | 40802.7 | 49223.4 |
| 92908.2 | 92160.9 | 168530 | 166831 | 302660 | 299028 | 1317.6 | 6226.3 | 6358.8 | 11623.6 | 11614 | 23460.8 | 23365.9 | 41432.5 | 40854.5 | 49327.5 |
| 93364.3 | 92365.6 | 167976 | 167126 | 302154 | 299359 | 1319.31 | 6302.96 | 6376.89 | 11881.7 | 11639 | 23412.3 | 23409.4 | 41312.8 | 40906 | 49232.9 |
| 93261.4 | 92569.6 | 168667 | 167420 | 302456 | 299687 | 1321.02 | 6322.06 | 6394.93 | 11732.3 | 11663.9 | 23682.7 | 23452.7 | 41355.3 | 40957 | 49069.3 |
| 93247 | 92772.9 | 168852 | 167713 | 301792 | 300012 | 1322.73 | 6469.94 | 6412.93 | 11621.7 | 11688.6 | 23537.2 | 23495.7 | 41300.4 | 41007.7 | 49021.9 |
| 93223.5 | 92975.3 | 168513 | 168005 | 301560 | 300335 | 1324.44 | 6490.15 | 6430.87 | 11770.4 | 11713.2 | 23557.5 | 23538.6 | 41158.8 | 41057.9 | 48967.6 |
| 93375.9 | 93177 | 169034 | 168294 | 302225 | 300655 | 1326.15 | 6436.28 | 6448.77 | 11669.2 | 11737.8 | 23717.7 | 23581.2 | 41396.3 | 41107.8 | 49231.1 |
| 93959.5 | 93377.9 | 169408 | 168583 | 302134 | 300973 | 1327.86 | 6480.9 | 6466.61 | 11798.8 | 11762.1 | 23707.8 | 23623.6 | 41500.4 | 41157.4 | 49465.6 |
| 94649.3 | 93578.1 | 170141 | 168870 | 304054 | 301289 | 1329.57 | 6643.32 | 6484.41 | 11859.5 | 11786.5 | 23645.7 | 23665.8 | 41638.2 | 41206.5 | 49677.1 |
| 94589.7 | 93777.4 | 170552 | 169156 | 304494 | 301602 | 1331.28 | 6494.26 | 6502.15 | 11883.3 | 11810.6 | 23917.5 | 23707.7 | 41655 | 41255.3 | 49696.5 |
| 94453 | 93976 | 170789 | 169440 | 304215 | 301912 | 1332.98 | 6336.64 | 6519.84 | 11914.7 | 11834.7 | 24024.2 | 23749.5 | 41751.4 | 41303.7 | 49760.3 |
| 94778.4 | 94173.9 | 170887 | 169723 | 304912 | 302220 | 1334.69 | 6445.66 | 6537.48 | 11956.3 | 11858.7 | 23896.1 | 23791 | 41740.6 | 41351.7 | 49826.9 |
| 94471.3 | 94370.9 | 171002 | 170004 | 304224 | 302526 | 1336.4 | 6444.55 | 6555.08 | 11986.4 | 11882.5 | 24065.8 | 23832.3 | 41734 | 41399.3 | 49624.3 |
| 95220.8 | 94567.2 | 171952 | 170284 | 305693 | 302829 | 1338.11 | 6581.58 | 6572.62 | 11877.6 | 11906.3 | 24120.6 | 23873.3 | 42118.8 | 41446.5 | 49477.2 |
| 95690.1 | 94762.6 | 171864 | 170563 | 305808 | 303130 | 1339.82 | 6454.69 | 6590.1 | 11874.9 | 11929.9 | 23925.6 | 23914.2 | 42003.8 | 41493.4 | 49571.9 |
| 95452.1 | 94957.4 | 172243 | 170840 | 304991 | 303428 | 1341.53 | 6576.01 | 6607.54 | 11990.5 | 11953.4 | 23997.8 | 23954.8 | 41958.6 | 41539.9 | 49776.6 |
| 96293.8 | 95151.3 | 173354 | 171115 | 306737 | 303724 | 1343.24 | 6609.08 | 6624.92 | 11841 | 11976.8 | 24245.5 | 23995.2 | 42137.2 | 41586.1 | 50040.8 |
| 96527.6 | 95344.4 | 173247 | 171390 | 307204 | 304017 | 1344.95 | 6711.68 | 6642.26 | 11969.1 | 12000.1 | 24286.5 | 24035.4 | 41962.4 | 41631.8 | 50081.5 |
| 95874.6 | 95536.8 | 173008 | 171662 | 306193 | 304308 | 1346.65 | 6703 | 6659.54 | 12040.7 | 12023.3 | 24418.9 | 24075.3 | 42122.5 | 41677.2 | 49834.2 |
| 95728.4 | 95728.4 | 172870 | 171934 | 306051 | 304596 | 1348.36 | 6739.62 | 6676.76 | 11997.2 | 12046.4 | 24118.8 | 24115.1 | 42075.6 | 41722.2 | 49753.6 |
| 96210.2 | 95919.1 | 173040 | 172204 | 306938 | 304883 | 1350.07 | 6765.71 | 6693.94 | 12189.1 | 12069.4 | 24143.1 | 24154.6 | 41889.5 | 41766.8 | 49938.7 |
| 96437.9 | 96109.2 | 174083 | 172472 | 307653 | 305166 | 1351.78 | 6729.91 | 6711.06 | 12189.9 | 12092.2 | 24285.9 | 24193.9 | 42141 | 41811.1 | 50165.8 |
| 96533.7 | 96298.4 | 174097 | 172740 | 307241 | 305448 | 1353.49 | 6826.15 | 6728.13 | 12108.2 | 12115 | 24353.7 | 24232.9 | 42101.9 | 41855 | 50170.8 |
| 97319.2 | 96486.8 | 174022 | 173005 | 307769 | 305727 | 1355.19 | 6646.32 | 6745.14 | 12279.2 | 12137.6 | 24600.7 | 24271.8 | 42125.5 | 41898.6 | 50303.8 |
| 97417.8 | 96674.5 | 174357 | 173270 | 308375 | 306003 | 1356.9 | 6885.23 | 6762.1 | 12241.9 | 12160.1 | 24364.4 | 24310.4 | 42104.6 | 41941.7 | 50362.9 |
| 97212.1 | 96861.3 | 174696 | 173533 | 308312 | 306277 | 1358.61 | 7001.99 | 6779.01 | 12235 | 12182.5 | 24384.4 | 24348.8 | 42103.8 | 41984.5 | 50284.2 |
| 97650.1 | 97047.4 | 175159 | 173794 | 308885 | 306549 | 1360.32 | 6838.84 | 6795.87 | 12237.4 | 12204.8 | 24575.3 | 24387 | 42287.6 | 42027 | 50378.5 |
| 97950 | 97232.7 | 175142 | 174054 | 309196 | 306819 | 1362.02 | 6707.26 | 6812.67 | 12055.7 | 12227 | 24567.6 | 24425 | 42394.2 | 42069 | 50134.9 |
| 97864.9 | 97417.2 | 175547 | 174313 | 309242 | 307086 | 1363.73 | 6792.4 | 6829.41 | 12318.7 | 12249.1 | 24711.2 | 24462.7 | 42689.6 | 42110.8 | 50235 |
| 98299.7 | 97600.9 | 175780 | 174570 | 309296 | 307350 | 1365.44 | 6871.94 | 6846.11 | 12326.4 | 12271 | 24744.7 | 24500.2 | 42565.5 | 42152.1 | 50493.9 |
| 98672.4 | 97783.8 | 176397 | 174826 | 309750 | 307613 | 1367.14 | 6927.8 | 6862.74 | 12227.6 | 12292.9 | 24699.1 | 24537.5 | 42487.5 | 42193.1 | 50454.7 |
| 98806.7 | 97965.9 | 176638 | 175081 | 310427 | 307873 | 1368.85 | 6978.56 | 6879.33 | 12363.2 | 12314.6 | 24640.9 | 24574.6 | 42585.2 | 42233.7 | 50641.1 |
| 98890.7 | 98147.3 | 176576 | 175334 | 310326 | 308131 | 1370.56 | 6914.2 | 6895.86 | 12354.9 | 12336.2 | 24655.9 | 24611.5 | 42548.8 | 42273.9 | 50507.6 |
| 98764.6 | 98327.8 | 177507 | 175585 | 311027 | 308386 | 1372.26 | 6970.36 | 6912.33 | 12468.9 | 12357.7 | 24774 | 24648.1 | 42738.7 | 42313.8 | 50519.6 |
| 99287.5 | 98507.6 | 178005 | 175836 | 311610 | 308639 | 1373.97 | 7068.9 | 6928.75 | 12469.8 | 12379.1 | 24801.7 | 24684.6 | 42925.9 | 42353.4 | 50807.4 |
| 99351.4 | 98686.6 | 177871 | 176084 | 312011 | 308890 | 1375.67 | 7036.93 | 6945.11 | 12531.8 | 12400.4 | 24938.8 | 24720.8 | 42858.5 | 42392.6 | 50679.9 |
| 99141.9 | 98864.7 | 177951 | 176332 | 311392 | 309138 | 1377.38 | 6893.6 | 6961.42 | 12560 | 12421.6 | 24848.4 | 24756.7 | 42763 | 42431.4 | 50604 |
| 100173 | 99042.1 | 178810 | 176578 | 312106 | 309384 | 1379.09 | 7055.28 | 6977.67 | 12457 | 12442.6 | 24921.5 | 24792.5 | 43058.8 | 42469.9 | 50791.9 |
| 100548 | 99218.7 | 179420 | 176823 | 313462 | 309628 | 1380.79 | 7067.99 | 6993.87 | 12582.1 | 12463.6 | 25035.9 | 24828 | 42972 | 42508 | 50954.5 |
| 100292 | 99394.5 | 179402 | 177066 | 313314 | 309870 | 1382.5 | 6955.19 | 7010.01 | 12492.4 | 12484.4 | 25006.1 | 24863.4 | 43031.7 | 42545.7 | 51111.2 |
| 100882 | 99569.5 | 180043 | 177308 | 314791 | 310109 | 1384.2 | 7212.89 | 7026.1 | 12602 | 12505.1 | 25260.2 | 24898.5 | 43293.5 | 42583.2 | 51279.6 |
| 101673 | 99743.8 | 180730 | 177548 | 315780 | 310346 | 1385.91 | 7338.36 | 7042.12 | 12708.9 | 12525.8 | 25475.5 | 24933.4 | 43139.5 | 42620.2 | 51288.6 |
| 101624 | 99917.2 | 180791 | 177788 | 315313 | 310581 | 1387.61 | 7232.09 | 7058.1 | 12587.9 | 12546.3 | 25323.7 | 24968 | 43197.8 | 42656.9 | 51110.4 |
| 101727 | 100090 | 181141 | 178025 | 315382 | 310813 | 1389.32 | 7123.55 | 7074.02 | 12694.6 | 12566.6 | 25198.5 | 25002.5 | 43357.6 | 42693.2 | 51170.5 |
| 101643 | 100262 | 180981 | 178262 | 315155 | 311044 | 1391.02 | 7198.65 | 7089.88 | 12628.2 | 12586.9 | 25254.1 | 25036.7 | 43303.7 | 42729.2 | 51252 |
| 101262 | 100433 | 180590 | 178496 | 314656 | 311272 | 1392.73 | 7184.2 | 7105.68 | 12777.1 | 12607.1 | 25313.5 | 25070.8 | 43053.5 | 42764.9 | 51025 |
| 101499 | 100603 | 180569 | 178730 | 314710 | 311497 | 1394.43 | 7167.42 | 7121.43 | 12891.5 | 12627.1 | 25364.4 | 25104.5 | 43030.4 | 42800.2 | 50867.7 |
| 101960 | 100772 | 181673 | 178962 | 315599 | 311721 | 1396.14 | 7154 | 7137.12 | 12756.4 | 12647.1 | 25471.3 | 25138.1 | 43382.7 | 42835.2 | 51190.8 |
| 101755 | 100941 | 181438 | 179193 | 315037 | 311942 | 1397.84 | 7175.91 | 7152.75 | 12684 | 12666.9 | 25513.1 | 25171.5 | 43137.2 | 42869.8 | 51267.1 |
| 101756 | 101109 | 180850 | 179422 | 314164 | 312161 | 1399.55 | 7298.48 | 7168.33 | 12757 | 12686.6 | 25237.9 | 25204.7 | 43330.1 | 42904 | 51166.4 |
| 102306 | 101276 | 181682 | 179650 | 315453 | 312378 | 1401.25 | 7079.79 | 7183.85 | 12868 | 12706.2 | 25507 | 25237.6 | 43508.2 | 42938 | 51328.4 |
| 102379 | 101442 | 181550 | 179877 | 316111 | 312592 | 1402.95 | 7393.22 | 7199.31 | 12850.4 | 12725.7 | 25544 | 25270.3 | 43386.6 | 42971.5 | 51359 |
| 102392 | 101608 | 181261 | 180102 | 314848 | 312805 | 1404.66 | 7354.15 | 7214.72 | 12952.4 | 12745.1 | 25334.2 | 25302.8 | 43319.5 | 43004.8 | 51177.9 |
| 102469 | 101772 | 181496 | 180326 | 314673 | 313015 | 1406.36 | 7097.3 | 7230.07 | 12989 | 12764.4 | 25602.2 | 25335.1 | 43126.1 | 43037.7 | 51160.6 |
| 102270 | 101936 | 182023 | 180549 | 315385 | 313223 | 1408.07 | 7470.8 | 7245.36 | 12902.6 | 12783.5 | 25467.6 | 25367.2 | 43481.8 | 43070.2 | 51469.4 |
| 102492 | 102100 | 182503 | 180770 | 315520 | 313429 | 1409.77 | 7411.67 | 7260.59 | 12945.1 | 12802.6 | 25452.8 | 25399 | 43566 | 43102.4 | 51765.8 |
| 103029 | 102262 | 182733 | 180990 | 316356 | 313633 | 1411.47 | 7269.33 | 7275.77 | 12974 | 12821.5 | 25589.4 | 25430.7 | 43606.4 | 43134.3 | 51543.7 |
| 103223 | 102423 | 182873 | 181208 | 316906 | 313834 | 1413.17 | 7351.87 | 7290.89 | 12927.1 | 12840.4 | 25714.3 | 25462.1 | 43555.1 | 43165.8 | 51177.7 |
| 102951 | 102584 | 182573 | 181426 | 315862 | 314034 | 1414.88 | 7372.87 | 7305.95 | 12966.6 | 12859.1 | 25563 | 25493.3 | 43638.2 | 43197 | 51335.6 |
| 102528 | 102744 | 182672 | 181641 | 315192 | 314231 | 1416.58 | 7266.84 | 7320.95 | 13102 | 12877.7 | 25486.5 | 25524.3 | 43552.4 | 43227.9 | 51551.3 |
| 103274 | 102903 | 183371 | 181856 | 316898 | 314426 | 1418.28 | 7356.79 | 7335.89 | 13081.7 | 12896.2 | 25720.4 | 25555.1 | 43518 | 43258.4 | 51538.1 |
| 103803 | 103062 | 183409 | 182069 | 316666 | 314619 | 1419.99 | 7455.28 | 7350.78 | 12987.6 | 12914.5 | 25634.9 | 25585.6 | 43690.1 | 43288.6 | 51667.7 |
| 104362 | 103219 | 184243 | 182280 | 317810 | 314810 | 1421.69 | 7415.55 | 7365.61 | 13209.3 | 12932.8 | 25889.2 | 25616 | 43753.1 | 43318.5 | 51898.7 |
| 104205 | 103376 | 184601 | 182491 | 319114 | 314998 | 1423.39 | 7329.66 | 7380.38 | 13205.8 | 12951 | 26121.5 | 25646.1 | 43949.6 | 43348 | 51949.5 |
| 103310 | 103532 | 183537 | 182700 | 316422 | 315185 | 1425.09 | 7345.48 | 7395.09 | 12904.7 | 12969 | 25862.9 | 25676.1 | 43507.8 | 43377.2 | 51500.4 |
| 103575 | 103687 | 183720 | 182907 | 315583 | 315370 | 1426.8 | 7523.55 | 7409.74 | 12936.6 | 12987 | 25777.8 | 25705.8 | 43196.4 | 43406.1 | 51290.6 |
| 103772 | 103842 | 183440 | 183114 | 316207 | 315552 | 1428.5 | 7615.54 | 7424.34 | 13047.4 | 13004.8 | 25723 | 25735.3 | 43431.5 | 43434.7 | 51250.7 |
| 104241 | 103995 | 183716 | 183319 | 317155 | 315732 | 1430.2 | 7472.27 | 7438.88 | 13056.8 | 13022.5 | 25811.6 | 25764.6 | 43685.2 | 43462.9 | 51528.2 |
| 104688 | 104148 | 184355 | 183522 | 317090 | 315910 | 1431.9 | 7405.47 | 7453.35 | 13079.6 | 13040.1 | 26020.3 | 25793.7 | 43846.2 | 43490.8 | 51552.7 |
| 104844 | 104300 | 184400 | 183725 | 317563 | 316086 | 1433.6 | 7616.18 | 7467.77 | 13050.9 | 13057.6 | 26057.5 | 25822.6 | 43958.4 | 43518.3 | 51488.5 |
| 105121 | 104452 | 184750 | 183926 | 317705 | 316260 | 1435.3 | 7682.44 | 7482.14 | 13158.5 | 13075 | 25978.1 | 25851.2 | 43846.2 | 43545.6 | 51873.9 |
| 105083 | 104602 | 185043 | 184125 | 316986 | 316432 | 1437.01 | 7637.51 | 7496.44 | 13267.2 | 13092.3 | 25979.6 | 25879.7 | 43863.7 | 43572.5 | 51778.2 |
| 105093 | 104752 | 185161 | 184324 | 316984 | 316602 | 1438.71 | 7615.47 | 7510.68 | 13316.4 | 13109.4 | 26158 | 25907.9 | 43854.3 | 43599.1 | 51881.3 |
| 105171 | 104901 | 185033 | 184520 | 316606 | 316770 | 1440.41 | 7410.21 | 7524.86 | 13287.7 | 13126.5 | 26081.6 | 25936 | 43706.2 | 43625.3 | 51736.9 |
| 105590 | 105049 | 186078 | 184716 | 317466 | 316936 | 1442.11 | 7472.44 | 7538.99 | 13243.7 | 13143.4 | 26098.7 | 25963.8 | 43828 | 43651.3 | 51698.9 |
| 106019 | 105196 | 186695 | 184910 | 318530 | 317100 | 1443.81 | 7585.54 | 7553.06 | 13346.1 | 13160.3 | 26174.7 | 25991.4 | 44074.2 | 43676.9 | 51985.8 |
| 105684 | 105343 | 185804 | 185104 | 317728 | 317261 | 1445.51 | 7663.08 | 7567.06 | 13303.1 | 13177 | 26189.6 | 26018.8 | 43881.7 | 43702.2 | 51727.4 |
| 105997 | 105489 | 186678 | 185295 | 318109 | 317421 | 1447.21 | 7770.54 | 7581.01 | 13351.1 | 13193.6 | 26285.1 | 26046.1 | 43849.2 | 43727.2 | 51704.8 |
| 105912 | 105634 | 186799 | 185486 | 318170 | 317578 | 1448.91 | 7662.69 | 7594.9 | 13200.6 | 13210.1 | 26163.1 | 26073 | 44102.2 | 43751.9 | 51800 |
| 105901 | 105778 | 186165 | 185675 | 317272 | 317734 | 1450.61 | 7619.89 | 7608.73 | 13131.9 | 13226.5 | 26154.4 | 26099.8 | 44088.6 | 43776.3 | 51860.5 |
| 106396 | 105921 | 186603 | 185862 | 317476 | 317888 | 1452.31 | 7725.72 | 7622.51 | 13276.5 | 13242.8 | 26231.4 | 26126.4 | 43920.3 | 43800.3 | 51960.5 |
| 105916 | 106064 | 186066 | 186049 | 316460 | 318039 | 1454.01 | 7651.2 | 7636.22 | 13267.7 | 13259 | 26239.1 | 26152.8 | 43622.7 | 43824.1 | 51684.3 |
| 106381 | 106206 | 186422 | 186234 | 317051 | 318189 | 1455.71 | 7653.73 | 7649.87 | 13482 | 13275.1 | 26208.7 | 26179 | 43714 | 43847.5 | 51629.9 |
| 106907 | 106347 | 186781 | 186418 | 318123 | 318337 | 1457.41 | 7614.16 | 7663.46 | 13489.5 | 13291 | 26176 | 26205 | 44066.8 | 43870.6 | 51739.2 |
| 106609 | 106487 | 186899 | 186600 | 318318 | 318482 | 1459.11 | 7561.3 | 7677 | 13337.6 | 13306.9 | 26165.5 | 26230.7 | 43947.2 | 43893.4 | 51807.4 |
| 106483 | 106627 | 187202 | 186782 | 317547 | 318626 | 1460.81 | 7616.07 | 7690.47 | 13225.7 | 13322.6 | 26263.5 | 26256.3 | 43915.1 | 43915.9 | 51861.9 |
| 106473 | 106766 | 186597 | 186962 | 317430 | 318768 | 1462.51 | 7668.95 | 7703.89 | 13199 | 13338.3 | 26130.7 | 26281.6 | 44020.3 | 43938.1 | 51682.6 |
| 106917 | 106904 | 187366 | 187140 | 318080 | 318908 | 1464.21 | 7587.42 | 7717.24 | 13475.3 | 13353.8 | 26362.6 | 26306.8 | 44045.5 | 43959.9 | 51910.4 |
| 106455 | 107041 | 187212 | 187318 | 316998 | 319046 | 1465.91 | 7603.51 | 7730.54 | 13455 | 13369.2 | 26360 | 26331.7 | 43901.2 | 43981.5 | 51769.4 |
| 106840 | 107177 | 187140 | 187494 | 318036 | 319181 | 1467.61 | 7834.79 | 7743.78 | 13482 | 13384.5 | 26267.5 | 26356.5 | 43890.9 | 44002.8 | 51550.3 |
| 107388 | 107313 | 187855 | 187669 | 318544 | 319315 | 1469.31 | 7788.62 | 7756.96 | 13536.6 | 13399.8 | 26438.1 | 26381 | 43922.6 | 44023.8 | 51632.9 |
| 106991 | 107448 | 186961 | 187842 | 317127 | 319447 | 1471.01 | 7688.14 | 7770.07 | 13259.4 | 13414.9 | 26252 | 26405.4 | 43876.8 | 44044.4 | 51460.7 |
| 107274 | 107582 | 187430 | 188014 | 317754 | 319578 | 1472.7 | 7675.91 | 7783.13 | 13437.5 | 13429.9 | 26339.3 | 26429.5 | 44031.6 | 44064.8 | 51672.9 |
| 108381 | 107716 | 189190 | 188185 | 320408 | 319706 | 1474.4 | 7673.66 | 7796.13 | 13560 | 13444.8 | 26738.5 | 26453.5 | 44237.2 | 44084.8 | 52060.2 |
| 108093 | 107848 | 189138 | 188355 | 320172 | 319832 | 1476.1 | 7814.92 | 7809.07 | 13443.1 | 13459.5 | 26752.6 | 26477.2 | 44124.8 | 44104.6 | 51837.7 |
| 108286 | 107980 | 189321 | 188524 | 320178 | 319956 | 1477.8 | 7920.04 | 7821.95 | 13465 | 13474.2 | 26598.1 | 26500.7 | 44208.8 | 44124.1 | 51966.8 |
| 109018 | 108111 | 189640 | 188691 | 320631 | 320079 | 1479.5 | 7799.99 | 7834.77 | 13548 | 13488.8 | 26618 | 26524.1 | 44223.7 | 44143.2 | 52278.7 |
| 108516 | 108241 | 188727 | 188857 | 318859 | 320200 | 1481.19 | 7683.79 | 7847.53 | 13484.3 | 13503.2 | 26452.6 | 26547.2 | 43985.6 | 44162.1 | 51988.5 |
| 108362 | 108371 | 189042 | 189021 | 319359 | 320318 | 1482.89 | 7805.33 | 7860.24 | 13567.6 | 13517.6 | 26367.9 | 26570.2 | 44001.2 | 44180.6 | 51842.6 |
| 108802 | 108500 | 190095 | 189185 | 321027 | 320435 | 1484.59 | 8067.27 | 7872.88 | 13619.3 | 13531.9 | 26651.3 | 26592.9 | 44371.2 | 44198.9 | 52289.2 |
| 109176 | 108628 | 189867 | 189347 | 320424 | 320550 | 1486.29 | 7869.2 | 7885.46 | 13603 | 13546 | 26600 | 26615.4 | 44424.7 | 44216.9 | 52108.3 |
| 109550 | 108755 | 190711 | 189508 | 321942 | 320663 | 1487.98 | 7947.11 | 7897.98 | 13632.2 | 13560 | 26757.4 | 26637.8 | 44473.1 | 44234.6 | 52204.1 |
| 110112 | 108882 | 191688 | 189667 | 322906 | 320775 | 1489.68 | 7994.17 | 7910.45 | 13437.7 | 13574 | 27130.1 | 26659.9 | 44558.4 | 44252 | 52465.4 |
| 110008 | 109007 | 191243 | 189826 | 321487 | 320884 | 1491.38 | 7960.24 | 7922.85 | 13502.5 | 13587.8 | 27015.4 | 26681.9 | 44324.4 | 44269.1 | 52321.5 |
| 109649 | 109132 | 190567 | 189983 | 321085 | 320992 | 1493.08 | 8147.04 | 7935.2 | 13643.6 | 13601.5 | 26751.5 | 26703.7 | 44213.4 | 44285.9 | 52108.4 |
| 109409 | 109256 | 189965 | 190139 | 320463 | 321098 | 1494.77 | 8003.15 | 7947.48 | 13716.1 | 13615.1 | 26782.9 | 26725.2 | 44212.5 | 44302.4 | 51828.5 |
| 109062 | 109380 | 190127 | 190294 | 319463 | 321202 | 1496.47 | 7983.89 | 7959.71 | 13836.1 | 13628.7 | 26843.4 | 26746.6 | 44286.8 | 44318.7 | 51891 |
| 109220 | 109503 | 190690 | 190447 | 320392 | 321304 | 1498.17 | 8033.96 | 7971.87 | 13776.3 | 13642.1 | 26837.5 | 26767.8 | 44350.5 | 44334.6 | 51934.8 |
| 109873 | 109624 | 190424 | 190599 | 320785 | 321404 | 1499.86 | 8020.82 | 7983.98 | 13610.5 | 13655.4 | 26807 | 26788.7 | 44368.8 | 44350.3 | 52014.3 |
| 110064 | 109746 | 190249 | 190750 | 319970 | 321503 | 1501.56 | 7993.42 | 7996.02 | 13600.4 | 13668.6 | 26602.4 | 26809.5 | 44320.8 | 44365.7 | 51891.3 |
| 109890 | 109866 | 190565 | 190900 | 320215 | 321599 | 1503.25 | 7940.34 | 8008.01 | 13512 | 13681.7 | 26619.9 | 26830.1 | 44381.4 | 44380.8 | 51781.5 |
| 109649 | 109986 | 190921 | 191048 | 320691 | 321694 | 1504.95 | 8072.75 | 8019.94 | 13695.7 | 13694.7 | 26691.8 | 26850.5 | 44331.2 | 44395.6 | 51841 |
| 109554 | 110105 | 190708 | 191196 | 319511 | 321788 | 1506.65 | 8026.04 | 8031.81 | 13790.9 | 13707.6 | 26724.9 | 26870.7 | 44151.8 | 44410.2 | 51826.6 |
| 109981 | 110223 | 190436 | 191342 | 318437 | 321879 | 1508.34 | 8011.41 | 8043.61 | 13674.5 | 13720.4 | 26794.3 | 26890.7 | 44127.4 | 44424.4 | 51783.4 |
| 109858 | 110340 | 190470 | 191487 | 319699 | 321969 | 1510.04 | 8023.02 | 8055.36 | 13569.6 | 13733 | 26674.2 | 26910.5 | 44123.1 | 44438.4 | 51907.8 |
| 110272 | 110457 | 190822 | 191630 | 319648 | 322056 | 1511.73 | 7878.42 | 8067.05 | 13600.6 | 13745.6 | 26757.2 | 26930.2 | 44080.8 | 44452.1 | 51820.3 |
| 110330 | 110573 | 190379 | 191773 | 318334 | 322143 | 1513.43 | 7820.88 | 8078.68 | 13792.8 | 13758.1 | 26783.8 | 26949.6 | 44004.1 | 44465.6 | 51533.4 |
| 109664 | 110688 | 189812 | 191914 | 318614 | 322227 | 1515.12 | 7932.81 | 8090.26 | 13583.6 | 13770.5 | 26713.3 | 26968.9 | 43893 | 44478.7 | 51553.2 |
| 110007 | 110803 | 190133 | 192054 | 318722 | 322310 | 1516.82 | 8163.88 | 8101.77 | 13650.6 | 13782.8 | 26773.8 | 26987.9 | 43989.8 | 44491.6 | 51575.6 |
| 110043 | 110916 | 190332 | 192193 | 318257 | 322391 | 1518.51 | 8226.52 | 8113.22 | 13701.5 | 13794.9 | 26826.4 | 27006.8 | 43977.2 | 44504.2 | 51702.2 |
| 109925 | 111030 | 190147 | 192331 | 318329 | 322470 | 1520.21 | 8351.15 | 8124.61 | 13682.5 | 13807 | 26743.3 | 27025.5 | 44091.3 | 44516.6 | 51638.6 |
| 110526 | 111142 | 190802 | 192467 | 319101 | 322547 | 1521.9 | 8132.45 | 8135.94 | 13924.1 | 13819 | 26887.3 | 27044 | 44173.7 | 44528.6 | 51853.5 |
| 110375 | 111253 | 191086 | 192603 | 319703 | 322623 | 1523.6 | 7970.28 | 8147.22 | 13761 | 13830.9 | 26886.2 | 27062.3 | 44132.2 | 44540.4 | 52020.8 |
| 110252 | 111364 | 190973 | 192737 | 318927 | 322697 | 1525.29 | 8115.64 | 8158.44 | 13861.5 | 13842.6 | 26602.6 | 27080.4 | 44161.2 | 44551.9 | 51744.5 |
| 110849 | 111474 | 191398 | 192870 | 318739 | 322770 | 1526.99 | 8107.79 | 8169.59 | 13948.6 | 13854.3 | 27165.8 | 27098.4 | 44115.9 | 44563.2 | 51772.3 |
| 110877 | 111583 | 191302 | 193001 | 319623 | 322840 | 1528.68 | 8046.68 | 8180.69 | 13866.5 | 13865.9 | 26960.4 | 27116.1 | 44221.8 | 44574.2 | 51832.2 |
| 110692 | 111692 | 191445 | 193132 | 319366 | 322909 | 1530.37 | 8061.22 | 8191.73 | 13704.9 | 13877.3 | 26820.8 | 27133.7 | 43926.2 | 44584.9 | 51833.6 |
| 110511 | 111800 | 191339 | 193261 | 318668 | 322977 | 1532.07 | 7963.18 | 8202.7 | 13782.5 | 13888.7 | 27239 | 27151.1 | 43992.5 | 44595.4 | 51705.2 |
| 110954 | 111907 | 191282 | 193390 | 319696 | 323042 | 1533.76 | 8109.69 | 8213.62 | 13942.6 | 13900 | 26900 | 27168.3 | 44290.7 | 44605.6 | 51851.2 |
| 111181 | 112013 | 191426 | 193517 | 319578 | 323106 | 1535.45 | 8059.46 | 8224.49 | 13711.6 | 13911.1 | 26781.3 | 27185.3 | 44149.2 | 44615.5 | 51647.4 |
| 110849 | 112119 | 191049 | 193643 | 318515 | 323169 | 1537.15 | 8056.14 | 8235.29 | 13606 | 13922.2 | 26901.8 | 27202.1 | 44201.4 | 44625.2 | 51610 |
| 111486 | 112224 | 191664 | 193767 | 319607 | 323229 | 1538.84 | 8315.35 | 8246.03 | 13751.3 | 13933.2 | 26930.4 | 27218.8 | 44076.3 | 44634.6 | 51758.8 |
| 111236 | 112328 | 191762 | 193891 | 319677 | 323288 | 1540.53 | 8156.15 | 8256.71 | 13787.9 | 13944 | 26962.8 | 27235.2 | 44111.9 | 44643.8 | 51445.9 |
| 111676 | 112432 | 192308 | 194014 | 320881 | 323346 | 1542.23 | 8190.9 | 8267.34 | 13911 | 13954.8 | 27136.7 | 27251.5 | 44347.3 | 44652.7 | 51869.9 |
| 111865 | 112534 | 192250 | 194135 | 320656 | 323402 | 1543.92 | 8170.86 | 8277.91 | 13890.8 | 13965.5 | 26982.6 | 27267.6 | 44020.6 | 44661.3 | 51953.7 |
| 111518 | 112636 | 192229 | 194255 | 320380 | 323456 | 1545.61 | 8316.06 | 8288.41 | 14012.6 | 13976 | 26927.9 | 27283.5 | 44161.1 | 44669.7 | 51599.2 |
| 112423 | 112738 | 193045 | 194374 | 321440 | 323508 | 1547.31 | 8300.94 | 8298.86 | 14081.3 | 13986.5 | 27297.2 | 27299.3 | 44126.2 | 44677.8 | 51693.8 |
| 112257 | 112838 | 192696 | 194492 | 320662 | 323559 | 1549 | 8259.33 | 8309.25 | 13829.5 | 13996.9 | 27402.4 | 27314.8 | 43929.6 | 44685.7 | 51863.7 |
| 112223 | 112938 | 193212 | 194608 | 321519 | 323609 | 1550.69 | 8439.88 | 8319.58 | 14013.1 | 14007.2 | 27256 | 27330.2 | 44265.1 | 44693.3 | 52039.7 |
| 112588 | 113037 | 193163 | 194724 | 320994 | 323656 | 1552.38 | 8396.61 | 8329.86 | 13873.9 | 14017.4 | 27078.7 | 27345.4 | 44167.1 | 44700.7 | 51844 |
| 111812 | 113136 | 192374 | 194838 | 320271 | 323703 | 1554.07 | 8344.26 | 8340.07 | 13815.3 | 14027.4 | 26912.1 | 27360.4 | 44239.9 | 44707.8 | 51591.2 |
| 111897 | 113234 | 192582 | 194952 | 320550 | 323747 | 1555.77 | 8295.37 | 8350.23 | 13841.4 | 14037.4 | 26952.6 | 27375.3 | 44249.8 | 44714.7 | 51366.9 |
| 111878 | 113331 | 192107 | 195064 | 319168 | 323790 | 1557.46 | 8052.84 | 8360.33 | 13712.1 | 14047.3 | 26890 | 27390 | 43893.1 | 44721.3 | 51330.2 |
| 112870 | 113427 | 193295 | 195175 | 321512 | 323832 | 1559.15 | 8575.06 | 8370.37 | 13987.8 | 14057.1 | 27169.1 | 27404.5 | 44226.5 | 44727.7 | 51815.7 |
| 113021 | 113523 | 193656 | 195285 | 321436 | 323872 | 1560.84 | 8519.02 | 8380.35 | 13994.5 | 14066.8 | 27363.7 | 27418.8 | 44108.5 | 44733.8 | 51680.7 |
| 112845 | 113618 | 193577 | 195394 | 320880 | 323910 | 1562.53 | 8320.42 | 8390.27 | 13839.7 | 14076.4 | 27275.8 | 27432.9 | 43857.7 | 44739.7 | 51522.3 |
| 113248 | 113712 | 193710 | 195502 | 321737 | 323947 | 1564.22 | 8500.22 | 8400.14 | 14009.3 | 14085.9 | 27394.6 | 27446.9 | 44108.8 | 44745.3 | 51793.8 |
| 112334 | 113805 | 192718 | 195608 | 319772 | 323982 | 1565.91 | 8341.05 | 8409.95 | 14036.1 | 14095.3 | 27087.2 | 27460.7 | 43927.1 | 44750.7 | 51499.7 |
| 112249 | 113898 | 192626 | 195714 | 319357 | 324016 | 1567.61 | 8269.33 | 8419.7 | 13995.5 | 14104.6 | 27071.5 | 27474.3 | 43913 | 44755.8 | 51331.3 |
| 112280 | 113990 | 192191 | 195818 | 318434 | 324048 | 1569.3 | 8114.18 | 8429.39 | 13845.4 | 14113.9 | 27131.4 | 27487.7 | 43786.5 | 44760.8 | 51416.5 |
| 112760 | 114082 | 192616 | 195922 | 318774 | 324079 | 1570.99 | 8470.15 | 8439.02 | 13931.6 | 14123 | 27045.2 | 27501 | 43775.8 | 44765.4 | 51191.3 |
| 113117 | 114172 | 193210 | 196024 | 319622 | 324108 | 1572.68 | 8406.74 | 8448.6 | 13889.3 | 14132 | 27253.2 | 27514.1 | 44010.8 | 44769.8 | 51436.2 |
| 112895 | 114262 | 192669 | 196125 | 319186 | 324136 | 1574.37 | 8315.3 | 8458.12 | 13816.3 | 14141 | 27101.2 | 27527 | 43915.4 | 44774 | 51208 |
| 113573 | 114352 | 193915 | 196225 | 321162 | 324162 | 1576.06 | 8570.09 | 8467.58 | 13987.5 | 14149.8 | 27284.1 | 27539.8 | 44168.2 | 44778 | 51284.8 |
| 114959 | 114440 | 195626 | 196324 | 323359 | 324186 | 1577.75 | 8634.42 | 8476.98 | 13972.8 | 14158.6 | 27334.7 | 27552.4 | 44638.2 | 44781.7 | 52009.2 |
| 115030 | 114528 | 195571 | 196422 | 322629 | 324210 | 1579.44 | 8349.56 | 8486.33 | 14113.8 | 14167.2 | 27370.8 | 27564.8 | 44601.9 | 44785.2 | 52020.1 |

| T = 2049 K, n = 0.51, 11700 sec | Figure 2D, Carbon Dot | T = 2220 K, n = 0.75, 420 sec | T = 2220 K, n = 0.75, 420 sec | T = 2220 K, n = 0.42, 1860 sec | T = 2220 K, n = 0.42, 1860 sec |
|---|---|---|---|---|---|
| 2869.58 | 600.015 | 59.7845 | 85.549 | 273.436 | 207.771 |
| 2879.58 | 600.306 | 96.9647 | 85.7998 | 244.805 | 208.412 |
| 2889.6 | 600.597 | 127.997 | 86.0509 | 162.116 | 209.055 |
| 2899.64 | 600.888 | 94.7722 | 86.3025 | 270.479 | 209.699 |
| 2909.7 | 601.179 | 93.3024 | 86.5544 | 225.317 | 210.344 |
| 2919.78 | 601.47 | 57.0036 | 86.8068 | 260.717 | 210.99 |
| 2929.89 | 601.76 | 64.33 | 87.0595 | 191.667 | 211.638 |
| 2940.01 | 602.052 | 46.3114 | 87.3126 | 264.344 | 212.286 |
| 2950.16 | 602.342 | 83.1911 | 87.5661 | 263.807 | 212.936 |
| 2960.33 | 602.633 | 81.7761 | 87.8199 | 239.372 | 213.586 |
| 2970.52 | 602.924 | 17.5377 | 88.0742 | 231.011 | 214.238 |
| 2980.73 | 603.215 | 67.8613 | 88.3288 | 251.405 | 214.891 |
| 2990.97 | 603.506 | 43.2016 | 88.5838 | 189.688 | 215.545 |
| 3001.23 | 603.797 | 42.2745 | 88.8392 | 176.336 | 216.2 |
| 3011.51 | 604.088 | 81.9103 | 89.095 | 188.815 | 216.856 |
| 3021.8 | 604.379 | 72.3009 | 89.3512 | 211.637 | 217.513 |
| 3032.13 | 604.669 | 65.7702 | 89.6077 | 234.634 | 218.172 |
| 3042.47 | 604.96 | 56.814 | 89.8646 | 285.093 | 218.831 |
| 3052.84 | 605.251 | 70.8173 | 90.122 | 198.657 | 219.492 |
| 3063.22 | 605.542 | 93.7876 | 90.3797 | 205.899 | 220.154 |
| 3073.63 | 605.832 | 122.144 | 90.6377 | 209.494 | 220.816 |
| 3084.06 | 606.123 | 82.4606 | 90.8962 | 246.862 | 221.481 |
| 3094.52 | 606.414 | 80.2635 | 91.155 | 232.275 | 222.146 |
| 3104.99 | 606.705 | 81.1719 | 91.4142 | 259.204 | 222.812 |
| 3115.49 | 606.995 | 87.1441 | 91.6738 | 318.856 | 223.479 |
| 3126.01 | 607.286 | 103.89 | 91.9338 | 235.679 | 224.147 |
| 3136.55 | 607.577 | 77.9959 | 92.1941 | 295.762 | 224.817 |
| 3147.11 | 607.868 | 70.3156 | 92.4549 | 200.68 | 225.488 |
| 3157.69 | 608.158 | 54.3468 | 92.716 | 156.596 | 226.159 |
| 3168.3 | 608.449 | 92.9131 | 92.9775 | 247.689 | 226.832 |
| 3178.92 | 608.74 | 47.4138 | 93.2393 | 252.793 | 227.506 |
| 3189.57 | 609.03 | 105.738 | 93.5015 | 246.643 | 228.181 |
| 3200.24 | 609.321 | 128.911 | 93.7642 | 219.331 | 228.857 |
| 3210.93 | 609.612 | 64.4186 | 94.0271 | 254.561 | 229.534 |
| 3221.65 | 609.902 | 74.4949 | 94.2905 | 245.316 | 230.213 |
| 3232.39 | 610.193 | 68.5536 | 94.5542 | 273.638 | 230.892 |
| 3243.14 | 610.484 | 61.4222 | 94.8183 | 256.205 | 231.573 |
| 3253.92 | 610.774 | 131.936 | 95.0828 | 197.488 | 232.254 |
| 3264.72 | 611.065 | 70.2272 | 95.3477 | 244.11 | 232.937 |
| 3275.55 | 611.355 | 110.892 | 95.6129 | 243.451 | 233.621 |
| 3286.39 | 611.646 | 107.11 | 95.8785 | 228.615 | 234.306 |
| 3297.26 | 611.937 | 85.7202 | 96.1445 | 263.463 | 234.992 |
| 3308.15 | 612.227 | 97.0114 | 96.4108 | 238.032 | 235.679 |
| 3319.06 | 612.518 | 117.714 | 96.6776 | 270.733 | 236.367 |
| 3329.99 | 612.808 | 94.2344 | 96.9447 | 206.88 | 237.056 |
| 3340.95 | 613.099 | 144.495 | 97.2121 | 231.815 | 237.747 |
| 3351.92 | 613.389 | 123.282 | 97.48 | 205.929 | 238.438 |
| 3362.92 | 613.68 | 106.529 | 97.7482 | 268.951 | 239.131 |
| 3373.94 | 613.97 | 115.874 | 98.0167 | 213.917 | 239.825 |
| 3384.98 | 614.26 | 88.8094 | 98.2857 | 280.986 | 240.519 |
| 3396.05 | 614.551 | 45.2061 | 98.555 | 259.138 | 241.215 |
| 3407.13 | 614.841 | 74.794 | 98.8247 | 246.301 | 241.912 |
| 3418.24 | 615.132 | 119.191 | 99.0947 | 256.217 | 242.61 |
| 3429.37 | 615.422 | 72.7873 | 99.3651 | 260.999 | 243.309 |
| 3440.52 | 615.713 | 70.0797 | 99.6359 | 246.162 | 244.01 |
| 3451.69 | 616.003 | 93.8403 | 99.9071 | 169.962 | 244.711 |
| 3462.89 | 616.293 | 100.596 | 100.179 | 235.903 | 245.413 |
| 3474.11 | 616.584 | 82.4206 | 100.451 | 236.232 | 246.117 |
| 3485.34 | 616.874 | 108.632 | 100.723 | 248.425 | 246.821 |
| 3496.6 | 617.165 | 97.2918 | 100.995 | 237.864 | 247.527 |
| 3507.89 | 617.455 | 112.817 | 101.268 | 204.808 | 248.234 |
| 3519.19 | 617.745 | 99.2695 | 101.542 | 284.261 | 248.941 |
| 3530.52 | 618.036 | 109.243 | 101.815 | 253.665 | 249.65 |
| 3541.86 | 618.326 | 105.734 | 102.089 | 241.412 | 250.36 |
| 3553.23 | 618.616 | 94.6336 | 102.364 | 257.437 | 251.071 |
| 3564.62 | 618.907 | 81.7363 | 102.639 | 237.76 | 251.784 |
| 3576.04 | 619.197 | 72.2967 | 102.914 | 273.348 | 252.497 |
| 3587.47 | 619.487 | 103.8 | 103.189 | 227.88 | 253.211 |
| 3598.93 | 619.777 | 112.678 | 103.465 | 302.815 | 253.926 |
| 3610.41 | 620.067 | 104.199 | 103.741 | 249.937 | 254.643 |
| 3621.91 | 620.358 | 99.316 | 104.018 | 280.931 | 255.36 |
| 3633.43 | 620.648 | 111.511 | 104.295 | 252.968 | 256.079 |
| 3644.98 | 620.938 | 102.04 | 104.572 | 236.016 | 256.799 |
| 3656.54 | 621.228 | 82.3727 | 104.85 | 247.671 | 257.519 |
| 3668.13 | 621.519 | 112.147 | 105.128 | 246.205 | 258.241 |
| 3679.74 | 621.809 | 93.4527 | 105.406 | 228.837 | 258.964 |
| 3691.37 | 622.099 | 79.1634 | 105.685 | 245.577 | 259.688 |
| 3703.02 | 622.389 | 115.005 | 105.964 | 252.479 | 260.413 |
| 3714.7 | 622.679 | 108.258 | 106.244 | 271.089 | 261.139 |
| 3726.4 | 622.969 | 92.0088 | 106.523 | 284.955 | 261.866 |
| 3738.12 | 623.259 | 81.3526 | 106.804 | 275.043 | 262.594 |
| 3749.86 | 623.549 | 94.4022 | 107.084 | 253.7 | 263.324 |
| 3761.62 | 623.84 | 129.25 | 107.365 | 283.752 | 264.054 |
| 3773.4 | 624.13 | 89.2577 | 107.646 | 248.442 | 264.786 |
| 3785.21 | 624.42 | 106.845 | 107.928 | 313.902 | 265.518 |
| 3797.04 | 624.71 | 116.634 | 108.21 | 239.209 | 266.252 |
| 3808.89 | 625 | 110.507 | 108.492 | 259.452 | 266.986 |
| 3820.76 | 625.29 | 118.054 | 108.775 | 295.723 | 267.722 |
| 3832.65 | 625.58 | 106.358 | 109.058 | 291.868 | 268.459 |
| 3844.57 | 625.87 | 98.9326 | 109.341 | 284.269 | 269.197 |
| 3856.5 | 626.16 | 116.873 | 109.625 | 253.45 | 269.935 |
| 3868.46 | 626.45 | 118.358 | 109.909 | 304.202 | 270.675 |
| 3880.44 | 626.74 | 84.9783 | 110.193 | 279.839 | 271.416 |
| 3892.45 | 627.03 | 114.981 | 110.478 | 270.897 | 272.158 |
| 3904.47 | 627.32 | 110.853 | 110.763 | 295.079 | 272.901 |
| 3916.52 | 627.61 | 101.161 | 111.049 | 297.116 | 273.646 |
| 3928.58 | 627.9 | 110.665 | 111.335 | 290.357 | 274.391 |
| 3940.67 | 628.19 | 107.241 | 111.621 | 261.92 | 275.137 |
| 3952.78 | 628.48 | 116.894 | 111.907 | 265.375 | 275.884 |
| 3964.92 | 628.77 | 106.016 | 112.194 | 320.023 | 276.632 |
| 3977.07 | 629.059 | 113.145 | 112.481 | 276.623 | 277.382 |
| 3989.25 | 629.349 | 97.0252 | 112.769 | 291.483 | 278.132 |
| 4001.45 | 629.639 | 105.512 | 113.057 | 297.822 | 278.884 |
| 4013.67 | 629.929 | 121.673 | 113.345 | 306.073 | 279.636 |
| 4025.91 | 630.219 | 129.175 | 113.633 | 310.857 | 280.39 |
| 4038.17 | 630.509 | 93.8438 | 113.922 | 292.769 | 281.145 |
| 4050.46 | 630.799 | 92.9749 | 114.212 | 308.858 | 281.9 |
| 4062.76 | 631.088 | 105.972 | 114.501 | 264.373 | 282.657 |
| 4075.09 | 631.378 | 112.379 | 114.791 | 304.456 | 283.415 |
| 4087.44 | 631.668 | 127.665 | 115.081 | 298.438 | 284.174 |
| 4099.81 | 631.958 | 98.5847 | 115.372 | 283.421 | 284.933 |
| 4112.21 | 632.247 | 97.2344 | 115.663 | 299.869 | 285.694 |
| 4124.62 | 632.537 | 129.487 | 115.954 | 258.89 | 286.456 |
| 4137.06 | 632.827 | 130.052 | 116.246 | 330.595 | 287.219 |
| 4149.52 | 633.117 | 111.715 | 116.538 | 316.379 | 287.983 |
| 4162 | 633.407 | 127.122 | 116.83 | 281.528 | 288.748 |
| 4174.5 | 633.696 | 118.148 | 117.123 | 293.003 | 289.514 |
| 4187.03 | 633.986 | 124.099 | 117.416 | 308.293 | 290.281 |
| 4199.57 | 634.276 | 124.991 | 117.709 | 305.153 | 291.049 |
| 4212.14 | 634.565 | 112.434 | 118.003 | 316.695 | 291.819 |
| 4224.73 | 634.855 | 120.957 | 118.297 | 307.475 | 292.589 |
| 4237.34 | 635.144 | 133.797 | 118.591 | 323.567 | 293.36 |
| 4249.97 | 635.434 | 131.525 | 118.886 | 322.584 | 294.132 |
| 4262.62 | 635.724 | 129.418 | 119.181 | 293 | 294.906 |
| 4275.3 | 636.013 | 128.94 | 119.476 | 296.341 | 295.68 |
| 4288 | 636.303 | 115.117 | 119.772 | 322.441 | 296.455 |
| 4300.72 | 636.593 | 124.661 | 120.068 | 301.915 | 297.231 |
| 4313.46 | 636.882 | 115.728 | 120.364 | 336.637 | 298.009 |
| 4326.22 | 637.172 | 139.506 | 120.661 | 291.584 | 298.787 |
| 4339 | 637.461 | 127.381 | 120.958 | 315.126 | 299.567 |
| 4351.81 | 637.751 | 124.489 | 121.255 | 322.992 | 300.347 |
| 4364.63 | 638.04 | 100.787 | 121.553 | 321.654 | 301.128 |
| 4377.48 | 638.33 | 134.81 | 121.851 | 318.258 | 301.911 |
| 4390.35 | 638.619 | 129.582 | 122.149 | 322.544 | 302.694 |
| 4403.24 | 638.909 | 128.203 | 122.448 | 302.955 | 303.479 |
| 4416.16 | 639.198 | 111.574 | 122.747 | 343.716 | 304.264 |
| 4429.09 | 639.488 | 117.302 | 123.046 | 328.63 | 305.051 |
| 4442.05 | 639.777 | 121.944 | 123.346 | 355.676 | 305.839 |
| 4455.03 | 640.067 | 118.576 | 123.646 | 318.797 | 306.627 |
| 4468.02 | 640.356 | 116.736 | 123.946 | 331.655 | 307.417 |
| 4481.05 | 640.646 | 114.308 | 124.247 | 297.081 | 308.207 |
| 4494.09 | 640.935 | 129.976 | 124.548 | 332.902 | 308.999 |
| 4507.15 | 641.225 | 116.829 | 124.849 | 332.346 | 309.791 |
| 4520.24 | 641.514 | 131.78 | 125.15 | 312.813 | 310.585 |
| 4533.34 | 641.803 | 121.489 | 125.452 | 332.504 | 311.38 |
| 4546.47 | 642.093 | 127.115 | 125.755 | 340.885 | 312.175 |
| 4559.62 | 642.382 | 124.323 | 126.057 | 332.421 | 312.972 |
| 4572.79 | 642.671 | 121.213 | 126.36 | 308.755 | 313.769 |
| 4585.98 | 642.961 | 120.948 | 126.663 | 319.994 | 314.568 |
| 4599.2 | 643.25 | 129.742 | 126.967 | 333.999 | 315.368 |
| 4612.44 | 643.539 | 129.64 | 127.271 | 323.115 | 316.168 |
| 4625.69 | 643.829 | 135.939 | 127.575 | 305.069 | 316.97 |
| 4638.97 | 644.118 | 124.95 | 127.879 | 341.256 | 317.772 |
| 4652.27 | 644.407 | 124.689 | 128.184 | 327.51 | 318.576 |
| 4665.59 | 644.697 | 135.102 | 128.489 | 333.676 | 319.38 |
| 4678.93 | 644.986 | 133.572 | 128.795 | 349.634 | 320.186 |
| 4692.3 | 645.275 | 120.797 | 129.1 | 333.073 | 320.993 |
| 4705.68 | 645.564 | 126.575 | 129.406 | 331.561 | 321.8 |
| 4719.09 | 645.853 | 121.058 | 129.713 | 317.357 | 322.609 |
| 4732.52 | 646.143 | 136.724 | 130.019 | 338.84 | 323.418 |
| 4745.97 | 646.432 | 132.545 | 130.327 | 337.42 | 324.229 |
| 4759.44 | 646.721 | 133.391 | 130.634 | 323.348 | 325.04 |
| 4772.93 | 647.01 | 131.828 | 130.941 | 347.295 | 325.853 |
| 4786.44 | 647.299 | 128.637 | 131.249 | 337.987 | 326.666 |
| 4799.98 | 647.588 | 125.242 | 131.558 | 334.727 | 327.481 |
| 4813.53 | 647.878 | 104.861 | 131.866 | 342.361 | 328.296 |
| 4827.11 | 648.167 | 127.388 | 132.175 | 323.821 | 329.112 |
| 4840.71 | 648.456 | 126.894 | 132.484 | 344.56 | 329.93 |
| 4854.33 | 648.745 | 138.622 | 132.794 | 342.005 | 330.748 |
| 4867.97 | 649.034 | 129.853 | 133.104 | 349.072 | 331.567 |
| 4881.63 | 649.323 | 132.997 | 133.414 | 367.1 | 332.388 |
| 4895.32 | 649.612 | 126.466 | 133.724 | 349.884 | 333.209 |
| 4909.02 | 649.901 | 121.338 | 134.035 | 347.7 | 334.031 |
| 4922.75 | 650.19 | 124.275 | 134.346 | 368.877 | 334.855 |
| 4936.49 | 650.479 | 124.61 | 134.657 | 346.793 | 335.679 |
| 4950.26 | 650.768 | 126.079 | 134.969 | 339.823 | 336.504 |
| 4964.05 | 651.057 | 140.073 | 135.281 | 352.415 | 337.33 |
| 4977.86 | 651.346 | 149.693 | 135.593 | 366.656 | 338.157 |
| 4991.69 | 651.635 | 127.647 | 135.906 | 360.493 | 338.985 |
| 5005.55 | 651.924 | 132.875 | 136.219 | 341.222 | 339.814 |
| 5019.42 | 652.213 | 146.688 | 136.532 | 359.607 | 340.644 |
| 5033.32 | 652.502 | 135.146 | 136.845 | 369.546 | 341.475 |
| 5047.23 | 652.791 | 140.495 | 137.159 | 351.937 | 342.307 |
| 5061.17 | 653.08 | 138.483 | 137.473 | 359.89 | 343.14 |
| 5075.13 | 653.369 | 125.206 | 137.787 | 370.82 | 343.974 |
| 5089.11 | 653.658 | 130.335 | 138.102 | 351.164 | 344.809 |
| 5103.11 | 653.947 | 139.719 | 138.417 | 372.464 | 345.644 |
| 5117.13 | 654.236 | 143.723 | 138.732 | 351.44 | 346.481 |
| 5131.18 | 654.524 | 158.143 | 139.048 | 380.616 | 347.319 |
| 5145.24 | 654.813 | 138.122 | 139.364 | 375.247 | 348.157 |
| 5159.32 | 655.102 | 142.991 | 139.68 | 361.635 | 348.997 |
| 5173.43 | 655.391 | 136.306 | 139.996 | 359.091 | 349.837 |
| 5187.56 | 655.68 | 131.029 | 140.313 | 369.49 | 350.679 |
| 5201.71 | 655.968 | 137.184 | 140.63 | 371.526 | 351.521 |
| 5215.87 | 656.257 | 128.636 | 140.948 | 373.264 | 352.365 |
| 5230.06 | 656.546 | 136.273 | 141.265 | 374.57 | 353.209 |
| 5244.28 | 656.835 | 137.437 | 141.583 | 371.726 | 354.054 |
| 5258.51 | 657.124 | 137.707 | 141.901 | 358.703 | 354.9 |
| 5272.76 | 657.412 | 140.768 | 142.22 | 373.722 | 355.747 |
| 5287.03 | 657.701 | 149.77 | 142.539 | 395.515 | 356.595 |
| 5301.33 | 657.99 | 140.02 | 142.858 | 384.245 | 357.444 |
| 5315.65 | 658.278 | 143.776 | 143.177 | 408.198 | 358.294 |
| 5329.98 | 658.567 | 141.515 | 143.497 | 391.267 | 359.145 |
| 5344.34 | 658.856 | 144.896 | 143.817 | 398.796 | 359.997 |
| 5358.72 | 659.144 | 150.91 | 144.137 | 373.866 | 360.849 |
| 5373.11 | 659.433 | 145.204 | 144.458 | 385.767 | 361.703 |
| 5387.53 | 659.722 | 147.872 | 144.779 | 381.788 | 362.558 |
| 5401.98 | 660.01 | 142.805 | 145.1 | 398.18 | 363.413 |
| 5416.44 | 660.299 | 144.857 | 145.421 | 382.754 | 364.27 |
| 5430.92 | 660.588 | 146.303 | 145.743 | 386.785 | 365.127 |
| 5445.42 | 660.876 | 137.659 | 146.065 | 405.957 | 365.985 |
| 5459.95 | 661.165 | 151.712 | 146.387 | 400.577 | 366.844 |
| 5474.49 | 661.453 | 140.554 | 146.71 | 393.536 | 367.705 |
| 5489.06 | 661.742 | 146.067 | 147.032 | 401.952 | 368.566 |
| 5503.64 | 662.03 | 146.637 | 147.356 | 393.585 | 369.428 |
| 5518.25 | 662.319 | 154.174 | 147.679 | 379.129 | 370.291 |
| 5532.87 | 662.607 | 136.348 | 148.003 | 404.212 | 371.154 |
| 5547.52 | 662.896 | 149.123 | 148.327 | 394.591 | 372.019 |
| 5562.19 | 663.184 | 151.453 | 148.651 | 406.384 | 372.885 |
| 5576.88 | 663.473 | 138.611 | 148.975 | 399.664 | 373.751 |
| 5591.59 | 663.761 | 152.891 | 149.3 | 397.654 | 374.619 |
| 5606.32 | 664.05 | 150.749 | 149.625 | 375.5 | 375.487 |
| 5621.07 | 664.338 | 142.143 | 149.951 | 411.832 | 376.356 |
| 5635.84 | 664.627 | 157.014 | 150.276 | 411.964 | 377.226 |
| 5650.64 | 664.915 | 149.419 | 150.602 | 389.667 | 378.097 |
| 5665.45 | 665.203 | 144.614 | 150.928 | 405.208 | 378.969 |
| 5680.28 | 665.492 | 156.368 | 151.255 | 383.857 | 379.842 |
| 5695.14 | 665.78 | 157.125 | 151.582 | 410.652 | 380.716 |
| 5710.01 | 666.069 | 160.561 | 151.909 | 375.166 | 381.591 |
| 5724.9 | 666.357 | 152.161 | 152.236 | 408.284 | 382.466 |
| 5739.82 | 666.645 | 154.477 | 152.563 | 400.057 | 383.343 |
| 5754.76 | 666.934 | 152.51 | 152.891 | 419.529 | 384.22 |
| 5769.71 | 667.222 | 154.594 | 153.219 | 403.656 | 385.098 |
| 5784.69 | 667.51 | 148.023 | 153.548 | 392.751 | 385.978 |
| 5799.69 | 667.799 | 162.594 | 153.876 | 387.034 | 386.858 |
| 5814.7 | 668.087 | 155.101 | 154.205 | 417.168 | 387.738 |
| 5829.74 | 668.375 | 159.02 | 154.534 | 392.871 | 388.62 |
| 5844.8 | 668.663 | 143.158 | 154.864 | 413.743 | 389.503 |
| 5859.88 | 668.952 | 143.566 | 155.193 | 394.471 | 390.387 |
| 5874.98 | 669.24 | 150.177 | 155.523 | 405.702 | 391.271 |
| 5890.1 | 669.528 | 150.02 | 155.854 | 397.389 | 392.156 |
| 5905.23 | 669.816 | 161.104 | 156.184 | 419.508 | 393.043 |
| 5920.39 | 670.105 | 151.597 | 156.515 | 419.421 | 393.93 |
| 5935.57 | 670.393 | 162.865 | 156.846 | 393.5 | 394.818 |
| 5950.77 | 670.681 | 151.448 | 157.177 | 414.64 | 395.707 |
| 5965.99 | 670.969 | 154.29 | 157.509 | 443.845 | 396.596 |
| 5981.24 | 671.257 | 153.098 | 157.841 | 423.999 | 397.487 |
| 5996.5 | 671.545 | 168.258 | 158.173 | 437.688 | 398.378 |
| 6011.78 | 671.833 | 152.581 | 158.505 | 401.967 | 399.271 |
| 6027.08 | 672.122 | 164.967 | 158.838 | 409.818 | 400.164 |
| 6042.4 | 672.41 | 164.788 | 159.171 | 427.463 | 401.058 |
| 6057.74 | 672.698 | 162.491 | 159.504 | 418.398 | 401.953 |
| 6073.1 | 672.986 | 161.811 | 159.837 | 443.571 | 402.849 |
| 6088.49 | 673.274 | 158.531 | 160.171 | 422.344 | 403.746 |
| 6103.89 | 673.562 | 171.422 | 160.505 | 413.212 | 404.643 |
| 6119.31 | 673.85 | 166.161 | 160.839 | 436.24 | 405.542 |
| 6134.75 | 674.138 | 159.397 | 161.173 | 436.951 | 406.441 |
| 6150.22 | 674.426 | 161.398 | 161.508 | 438.694 | 407.341 |
| 6165.7 | 674.714 | 173.187 | 161.843 | 425.874 | 408.242 |
| 6181.2 | 675.002 | 162.013 | 162.178 | 424.223 | 409.144 |
| 6196.72 | 675.29 | 148.963 | 162.513 | 430.495 | 410.047 |
| 6212.26 | 675.578 | 168.53 | 162.849 | 435.06 | 410.95 |
| 6227.82 | 675.866 | 164.061 | 163.185 | 438.153 | 411.855 |
| 6243.41 | 676.154 | 154.214 | 163.521 | 445.86 | 412.76 |
| 6259.01 | 676.442 | 157.956 | 163.858 | 433.93 | 413.666 |
| 6274.63 | 676.73 | 165.155 | 164.194 | 440.404 | 414.573 |
| 6290.27 | 677.018 | 160.215 | 164.531 | 431.421 | 415.481 |
| 6305.94 | 677.305 | 157.594 | 164.868 | 448.859 | 416.389 |
| 6321.62 | 677.593 | 162.325 | 165.206 | 445.912 | 417.299 |
| 6337.32 | 677.881 | 156.684 | 165.544 | 433.159 | 418.209 |
| 6353.04 | 678.169 | 163.623 | 165.881 | 438.533 | 419.12 |
| 6368.78 | 678.457 | 168.469 | 166.22 | 419.957 | 420.033 |
| 6384.54 | 678.745 | 157.025 | 166.558 | 455.079 | 420.945 |
| 6400.32 | 679.032 | 165.732 | 166.897 | 431.492 | 421.859 |
| 6416.12 | 679.32 | 170.133 | 167.236 | 452.236 | 422.774 |
| 6431.94 | 679.608 | 176.891 | 167.575 | 421.159 | 423.689 |
| 6447.78 | 679.896 | 171.89 | 167.914 | 442.081 | 424.605 |
| 6463.64 | 680.183 | 170.582 | 168.254 | 438.118 | 425.522 |
| 6479.52 | 680.471 | 176.954 | 168.594 | 454.739 | 426.44 |
| 6495.42 | 680.759 | 168.712 | 168.934 | 451.182 | 427.358 |
| 6511.34 | 681.047 | 161.075 | 169.274 | 436.192 | 428.278 |
| 6527.27 | 681.334 | 176.48 | 169.615 | 443.303 | 429.198 |
| 6543.23 | 681.622 | 160.279 | 169.956 | 339.754 | 430.119 |
| 6559.21 | 681.91 | 167.359 | 170.297 | 356.159 | 431.041 |
| 6575.2 | 682.197 | 167.529 | 170.638 | 357.825 | 431.964 |
| 6591.22 | 682.485 | 136.157 | 170.98 | 361.416 | 432.888 |
| 6607.26 | 682.773 | 168.273 | 171.321 | 345.995 | 433.812 |
| 6623.31 | 683.06 | 178.257 | 171.663 | 452.347 | 434.737 |
| 6639.39 | 683.348 | 182.569 | 172.006 | 446.985 | 435.663 |
| 6655.48 | 683.636 | 176.204 | 172.348 | 454.347 | 436.59 |
| 6671.59 | 683.923 | 168.508 | 172.691 | 455.865 | 437.518 |
| 6687.73 | 684.211 | 181.24 | 173.034 | 444.674 | 438.446 |
| 6703.88 | 684.498 | 175.616 | 173.377 | 389.851 | 439.375 |
| 6720.05 | 684.786 | 179.524 | 173.721 | 407.456 | 440.305 |
| 6736.24 | 685.073 | 203.065 | 174.064 | 408.654 | 441.236 |
| 6752.45 | 685.361 | 185.008 | 174.408 | 418.485 | 442.168 |
| 6768.68 | 685.649 | 185.501 | 174.752 | 480.178 | 443.1 |
| 6784.93 | 685.936 | 193.009 | 175.097 | 506.61 | 444.034 |
| 6801.2 | 686.223 | 202.888 | 175.441 | 451.791 | 444.968 |
| 6817.48 | 686.511 | 200.091 | 175.786 | 498.302 | 445.903 |
| 6833.79 | 686.798 | 192.286 | 176.131 | 514.782 | 446.838 |
| 6850.12 | 687.086 | 186.453 | 176.476 | 514.881 | 447.775 |
| 6866.46 | 687.373 | 189.864 | 176.822 | 473.135 | 448.712 |
| 6882.83 | 687.661 | 213.175 | 177.167 | 461.612 | 449.65 |
| 6899.21 | 687.948 | 217.87 | 177.513 | 462.788 | 450.589 |
| 6915.61 | 688.236 | 180.537 | 177.86 | 471.43 | 451.528 |
| 6932.03 | 688.523 | 211.485 | 178.206 | 488.387 | 452.469 |
| 6948.48 | 688.81 | 215.834 | 178.553 | 478.18 | 453.41 |
| 6964.94 | 689.098 | 200.758 | 178.899 | 468.714 | 454.352 |
| 6981.41 | 689.385 | 174.547 | 179.246 | 509.623 | 455.294 |
| 6997.91 | 689.672 | 175.106 | 179.594 | 509.82 | 456.238 |
| 7014.43 | 689.96 | 184.453 | 179.941 | 462.004 | 457.182 |
| 7030.97 | 690.247 | 184.483 | 180.289 | 469.152 | 458.127 |
| 7047.52 | 690.534 | 197.164 | 180.637 | 512.05 | 459.073 |
| 7064.09 | 690.822 | 180.01 | 180.985 | 472.39 | 460.02 |
| 7080.69 | 691.109 | 191.491 | 181.333 | 474.886 | 460.967 |
| 7097.3 | 691.396 | 198.668 | 181.682 | 459.155 | 461.915 |
| 7113.93 | 691.683 | 175.186 | 182.031 | 460.411 | 462.864 |
| 7130.58 | 691.971 | 188.886 | 182.38 | 461.883 | 463.814 |
| 7147.25 | 692.258 | 187.967 | 182.729 | 434.863 | 464.764 |
| 7163.93 | 692.545 | 193.46 | 183.078 | 480.601 | 465.715 |
| 7180.64 | 692.832 | 184.213 | 183.428 | 488.122 | 466.667 |
| 7197.36 | 693.119 | 179.761 | 183.778 | 470.607 | 467.62 |
| 7214.11 | 693.407 | 198.274 | 184.128 | 496.153 | 468.573 |
| 7230.87 | 693.694 | 201.431 | 184.478 | 480.933 | 469.527 |
| 7247.65 | 693.981 | 195.867 | 184.829 | 478.219 | 470.483 |
| 7264.45 | 694.268 | 208.455 | 185.18 | 501.089 | 471.438 |
| 7281.27 | 694.555 | 194.99 | 185.53 | 486.832 | 472.395 |
| 7298.1 | 694.842 | 191.837 | 185.882 | 489.801 | 473.352 |
| 7314.96 | 695.129 | 187.661 | 186.233 | 492.147 | 474.31 |
| 7331.84 | 695.416 | 188.421 | 186.584 | 524.52 | 475.269 |
| 7348.73 | 695.703 | 177.894 | 186.936 | 521.657 | 476.228 |
| 7365.64 | 695.99 | 200.258 | 187.288 | 485.01 | 477.188 |
| 7382.57 | 696.277 | 204.389 | 187.64 | 501.499 | 478.149 |
| 7399.52 | 696.564 | 191.716 | 187.993 | 493.154 | 479.111 |
| 7416.48 | 696.851 | 183.802 | 188.345 | 476.005 | 480.073 |
| 7433.47 | 697.138 | 187.758 | 188.698 | 504.006 | 481.036 |
| 7450.47 | 697.425 | 207.217 | 189.051 | 506.144 | 482 |
| 7467.5 | 697.712 | 213.401 | 189.404 | 521.819 | 482.965 |
| 7484.54 | 697.999 | 194.571 | 189.758 | 504.046 | 483.93 |
| 7501.6 | 698.286 | 196.075 | 190.111 | 479.545 | 484.896 |
| 7518.67 | 698.573 | 192.416 | 190.465 | 504.353 | 485.863 |
| 7535.77 | 698.86 | 194.533 | 190.819 | 533.84 | 486.831 |
| 7552.88 | 699.147 | 211.655 | 191.173 | 522.268 | 487.799 |
| 7570.02 | 699.434 | 197.524 | 191.528 | 538.587 | 488.768 |
| 7587.17 | 699.721 | 195.478 | 191.882 | 515.097 | 489.738 |
| 7604.33 | 700.008 | 195.025 | 192.237 | 506.044 | 490.708 |
| 7621.52 | 700.294 | 182.43 | 192.592 | 539.673 | 491.679 |
| 7638.73 | 700.581 | 187.69 | 192.947 | 551.957 | 492.651 |
| 7655.95 | 700.868 | 196.43 | 193.302 | 547.266 | 493.624 |
| 7673.19 | 701.155 | 214.322 | 193.658 | 560.37 | 494.597 |
| 7690.45 | 701.442 | 211.644 | 194.014 | 569.735 | 495.571 |
| 7707.73 | 701.729 | 213.762 | 194.37 | 518.534 | 496.546 |
| 7725.03 | 702.015 | 215.142 | 194.726 | 606.333 | 497.521 |
| 7742.34 | 702.302 | 210.8 | 195.082 | 579.392 | 498.497 |
| 7759.67 | 702.589 | 207.814 | 195.439 | 524.723 | 499.474 |
| 7777.02 | 702.876 | 218.113 | 195.795 | 575.828 | 500.452 |
| 7794.39 | 703.162 | 226.257 | 196.152 | 569.258 | 501.43 |
| 7811.78 | 703.449 | 217.074 | 196.509 | 599.353 | 502.409 |
| 7829.18 | 703.736 | 216.681 | 196.867 | 534.297 | 503.389 |
| 7846.6 | 704.022 | 207.854 | 197.224 | 558.047 | 504.369 |
| 7864.04 | 704.309 | 205.251 | 197.582 | 565.962 | 505.35 |
| 7881.5 | 704.596 | 209.187 | 197.939 | 524.286 | 506.332 |
| 7898.98 | 704.882 | 208.561 | 198.297 | 526.254 | 507.314 |
| 7916.47 | 705.169 | 204.431 | 198.656 | 551.398 | 508.297 |
| 7933.98 | 705.456 | 211.332 | 199.014 | 544.853 | 509.281 |
| 7951.51 | 705.742 | 212.716 | 199.373 | 473.589 | 510.266 |
| 7969.06 | 706.029 | 194.129 | 199.731 | 482.26 | 511.251 |
| 7986.62 | 706.315 | 190.429 | 200.09 | 494.839 | 512.237 |
| 8004.2 | 706.602 | 206.042 | 200.449 | 519.39 | 513.223 |
| 8021.8 | 706.888 | 209.244 | 200.809 | 560.232 | 514.21 |
| 8039.42 | 707.175 | 173.214 | 201.168 | 560.344 | 515.198 |
| 8057.06 | 707.462 | 211.859 | 201.528 | 530.548 | 516.187 |
| 8074.71 | 707.748 | 202.21 | 201.888 | 527.668 | 517.176 |
| 8092.38 | 708.034 | 201.852 | 202.248 | 533.185 | 518.166 |
| 8110.07 | 708.321 | 216.158 | 202.608 | 568.849 | 519.157 |
| 8127.77 | 708.607 | 207.033 | 202.968 | 549.882 | 520.148 |
| 8145.49 | 708.894 | 199.3 | 203.329 | 549.962 | 521.14 |
| 8163.23 | 709.18 | 219.013 | 203.689 | 546.011 | 522.133 |
| 8180.99 | 709.467 | 196.334 | 204.05 | 524.551 | 523.126 |
| 8198.77 | 709.753 | 207.68 | 204.411 | 560.207 | 524.12 |
| 8216.56 | 710.04 | 202.688 | 204.773 | 533.56 | 525.114 |
| 8234.37 | 710.326 | 206.713 | 205.134 | 574.419 | 526.11 |
| 8252.2 | 710.612 | 221.044 | 205.495 | 550.881 | 527.106 |
| 8270.04 | 710.899 | 197.86 | 205.857 | 559.519 | 528.102 |
| 8287.91 | 711.185 | 215.524 | 206.219 | 582.588 | 529.1 |
| 8305.78 | 711.471 | 214.304 | 206.581 | 575.029 | 530.097 |
| 8323.68 | 711.758 | 217.465 | 206.943 | 561.85 | 531.096 |
| 8341.6 | 712.044 | 221.893 | 207.306 | 548.031 | 532.095 |
| 8359.53 | 712.33 | 211.216 | 207.668 | 588.539 | 533.095 |
| 8377.48 | 712.616 | 214.56 | 208.031 | 597.136 | 534.096 |
| 8395.44 | 712.903 | 219.075 | 208.394 | 556.89 | 535.097 |
| 8413.42 | 713.189 | 214.832 | 208.757 | 561.928 | 536.098 |
| 8431.42 | 713.475 | 205.648 | 209.12 | 542.616 | 537.101 |
| 8449.44 | 713.761 | 222.005 | 209.484 | 582.511 | 538.104 |
| 8467.48 | 714.048 | 219.679 | 209.847 | 580.441 | 539.108 |
| 8485.53 | 714.334 | 204.206 | 210.211 | 574.729 | 540.112 |
| 8503.59 | 714.62 | 215.728 | 210.575 | 592.54 | 541.117 |
| 8521.68 | 714.906 | 221.024 | 210.939 | 561.846 | 542.123 |
| 8539.78 | 715.192 | 234.016 | 211.303 | 571.551 | 543.129 |
| 8557.9 | 715.479 | 202.35 | 211.668 | 560.121 | 544.136 |
| 8576.04 | 715.765 | 230.927 | 212.032 | 592.917 | 545.143 |
| 8594.19 | 716.051 | 222.997 | 212.397 | 595.237 | 546.151 |
| 8612.36 | 716.337 | 216.818 | 212.762 | 571.842 | 547.16 |
| 8630.55 | 716.623 | 217.292 | 213.127 | 576.329 | 548.169 |
| 8648.75 | 716.909 | 236.306 | 213.492 | 574.13 | 549.179 |
| 8666.97 | 717.195 | 223.465 | 213.857 | 566.903 | 550.19 |
| 8685.21 | 717.481 | 226.975 | 214.223 | 571.833 | 551.201 |
| 8703.46 | 717.767 | 212.75 | 214.588 | 584.933 | 552.213 |
| 8721.73 | 718.053 | 213.029 | 214.954 | 568.053 | 553.226 |
| 8740.02 | 718.339 | 232.535 | 215.32 | 574.869 | 554.239 |
| 8758.32 | 718.625 | 205.768 | 215.686 | 596.075 | 555.253 |
| 8776.64 | 718.911 | 215.591 | 216.052 | 569.964 | 556.267 |
| 8794.98 | 719.197 | 222.737 | 216.419 | 601.236 | 557.282 |
| 8813.33 | 719.483 | 215.087 | 216.785 | 589.133 | 558.297 |
| 8831.7 | 719.769 | 228.761 | 217.152 | 587.035 | 559.313 |
| 8850.09 | 720.055 | 214.529 | 217.519 | 588.155 | 560.33 |
| 8868.49 | 720.341 | 218.28 | 217.886 | 596.77 | 561.348 |
| 8886.91 | 720.627 | 207.219 | 218.253 | 570.906 | 562.366 |
| 8905.35 | 720.913 | 230.808 | 218.62 | 600.169 | 563.384 |
| 8923.8 | 721.199 | 215.39 | 218.988 | 588.827 | 564.403 |
| 8942.27 | 721.485 | 231.939 | 219.355 | 593.682 | 565.423 |
| 8960.75 | 721.77 | 213.026 | 219.723 | 595.942 | 566.443 |
| 8979.26 | 722.056 | 224.82 | 220.091 | 570.897 | 567.464 |
| 8997.77 | 722.342 | 228.616 | 220.459 | 587.84 | 568.486 |
| 9016.31 | 722.628 | 217.348 | 220.827 | 608.437 | 569.508 |
| 9034.86 | 722.913 | 231.161 | 221.195 | 608.068 | 570.531 |
| 9053.43 | 723.199 | 218.459 | 221.564 | 600.074 | 571.554 |
| 9072.01 | 723.485 | 222.165 | 221.932 | 594.214 | 572.578 |
| 9090.61 | 723.771 | 227.177 | 222.301 | 579.181 | 573.602 |
| 9109.22 | 724.057 | 222.44 | 222.67 | 615.584 | 574.627 |
| 9127.86 | 724.342 | 223.578 | 223.039 | 611.819 | 575.653 |
| 9146.5 | 724.628 | 228.294 | 223.408 | 601.428 | 576.679 |
| 9165.17 | 724.914 | 217.514 | 223.777 | 596.906 | 577.706 |
| 9183.85 | 725.199 | 226.594 | 224.147 | 613.686 | 578.733 |
| 9202.54 | 725.485 | 220.33 | 224.516 | 587.854 | 579.761 |
| 9221.25 | 725.771 | 231.095 | 224.886 | 583.789 | 580.789 |
| 9239.98 | 726.056 | 219.596 | 225.256 | 621.832 | 581.818 |
| 9258.73 | 726.342 | 234.812 | 225.626 | 616.258 | 582.848 |
| 9277.49 | 726.628 | 244.123 | 225.996 | 602.642 | 583.878 |
| 9296.26 | 726.913 | 238.267 | 226.366 | 612.592 | 584.909 |
| 9315.05 | 727.199 | 212.726 | 226.736 | 623.863 | 585.94 |
| 9333.86 | 727.484 | 219.668 | 227.107 | 622.717 | 586.972 |
| 9352.68 | 727.77 | 222.191 | 227.477 | 590.979 | 588.004 |
| 9371.52 | 728.055 | 204.929 | 227.848 | 625.637 | 589.037 |
| 9390.37 | 728.341 | 221.591 | 228.219 | 591.055 | 590.071 |
| 9409.25 | 728.626 | 241.159 | 228.59 | 623.557 | 591.105 |
| 9428.13 | 728.912 | 219.673 | 228.961 | 614.652 | 592.139 |
| 9447.03 | 729.197 | 221.951 | 229.333 | 621.441 | 593.175 |
| 9465.95 | 729.483 | 219.806 | 229.704 | 645.39 | 594.21 |
| 9484.88 | 729.768 | 231.303 | 230.075 | 630.612 | 595.247 |
| 9503.83 | 730.054 | 230.406 | 230.447 | 634.05 | 596.284 |
| 9522.8 | 730.339 | 234.427 | 230.819 | 619.712 | 597.321 |
| 9541.77 | 730.625 | 243.453 | 231.191 | 643.741 | 598.359 |
| 9560.77 | 730.91 | 228.842 | 231.563 | 623.207 | 599.397 |
| 9579.78 | 731.195 | 233.128 | 231.935 | 632.647 | 600.436 |
| 9598.81 | 731.481 | 242.453 | 232.307 | 617.819 | 601.476 |
| 9617.85 | 731.766 | 236.891 | 232.679 | 628.611 | 602.516 |
| 9636.91 | 732.052 | 229.521 | 233.052 | 622.726 | 603.556 |
| 9655.98 | 732.337 | 243.447 | 233.425 | 596.493 | 604.597 |
| 9675.07 | 732.622 | 225.693 | 233.797 | 605.588 | 605.639 |
| 9694.17 | 732.908 | 225.213 | 234.17 | 635.35 | 606.681 |
| 9713.29 | 733.193 | 221.045 | 234.543 | 611.401 | 607.724 |
| 9732.42 | 733.478 | 236.617 | 234.916 | 618.064 | 608.767 |
| 9751.57 | 733.763 | 232.913 | 235.29 | 620.559 | 609.811 |
| 9770.73 | 734.048 | 226.911 | 235.663 | 616.707 | 610.855 |
| 9789.91 | 734.334 | 241.312 | 236.036 | 644.152 | 611.9 |
| 9809.1 | 734.619 | 233.31 | 236.41 | 649.272 | 612.945 |
| 9828.31 | 734.904 | 248.57 | 236.783 | 661.605 | 613.991 |
| 9847.54 | 735.189 | 228.772 | 237.157 | 642.289 | 615.037 |
| 9866.78 | 735.475 | 231.494 | 237.531 | 605.019 | 616.084 |
| 9886.03 | 735.76 | 232.143 | 237.905 | 654.038 | 617.131 |
| 9905.3 | 736.045 | 230.663 | 238.279 | 674.786 | 618.179 |
| 9924.59 | 736.33 | 245.871 | 238.654 | 621.232 | 619.227 |
| 9943.89 | 736.615 | 244.63 | 239.028 | 671.317 | 620.276 |
| 9963.2 | 736.9 | 253.518 | 239.402 | 623.697 | 621.325 |
| 9982.53 | 737.185 | 256.738 | 239.777 | 630.737 | 622.375 |
| 10001.9 | 737.47 | 240.837 | 240.151 | 631.806 | 623.426 |
| 10021.2 | 737.755 | 231.616 | 240.526 | 636.212 | 624.476 |
| 10040.6 | 738.041 | 245.264 | 240.901 | 651.031 | 625.528 |
| 10060 | 738.326 | 240.438 | 241.276 | 667.793 | 626.58 |
| 10079.4 | 738.611 | 254.267 | 241.651 | 681.71 | 627.632 |
| 10098.8 | 738.896 | 244.504 | 242.026 | 646.349 | 628.685 |
| 10118.2 | 739.181 | 230.202 | 242.402 | 641.536 | 629.738 |
| 10137.7 | 739.466 | 229.577 | 242.777 | 652.75 | 630.792 |
| 10157.2 | 739.751 | 231.276 | 243.153 | 648.913 | 631.846 |
| 10176.6 | 740.035 | 249.33 | 243.528 | 660.579 | 632.901 |
| 10196.1 | 740.321 | 239.634 | 243.904 | 673.19 | 633.956 |
| 10215.6 | 740.605 | 249.615 | 244.28 | 635.179 | 635.012 |
| 10235.1 | 740.89 | 227.849 | 244.656 | 666.951 | 636.068 |
| 10254.7 | 741.175 | 246.957 | 245.032 | 682.037 | 637.125 |
| 10274.2 | 741.46 | 241.456 | 245.408 | 647.745 | 638.182 |
| 10293.8 | 741.745 | 239.415 | 245.784 | 684.041 | 639.239 |
| 10313.4 | 742.03 | 238.679 | 246.16 | 659.977 | 640.298 |
| 10333 | 742.315 | 223.756 | 246.537 | 638.347 | 641.356 |
| 10352.6 | 742.6 | 246.562 | 246.913 | 643.262 | 642.415 |
| 10372.2 | 742.884 | 264.918 | 247.29 | 666.306 | 643.475 |
| 10391.8 | 743.169 | 247.84 | 247.666 | 679.496 | 644.534 |
| 10411.5 | 743.454 | 260.053 | 248.043 | 669.874 | 645.595 |
| 10431.1 | 743.739 | 260.134 | 248.42 | 666.693 | 646.656 |
| 10450.8 | 744.023 | 263.121 | 248.797 | 663.789 | 647.717 |
| 10470.5 | 744.308 | 230.629 | 249.174 | 665.807 | 648.779 |
| 10490.2 | 744.593 | 248.426 | 249.551 | 665.169 | 649.841 |
| 10509.9 | 744.878 | 236.564 | 249.928 | 670.084 | 650.904 |
| 10529.7 | 745.162 | 251.011 | 250.306 | 689.463 | 651.967 |
| 10549.4 | 745.447 | 222.732 | 250.683 | 672.348 | 653.031 |
| 10569.2 | 745.732 | 257.563 | 251.061 | 677.518 | 654.095 |
| 10589 | 746.016 | 253.24 | 251.438 | 659.242 | 655.159 |
| 10608.8 | 746.301 | 253.693 | 251.816 | 671.917 | 656.224 |
| 10628.6 | 746.586 | 274.988 | 252.194 | 694.129 | 657.29 |
| 10648.4 | 746.87 | 254.323 | 252.572 | 679.435 | 658.356 |
| 10668.2 | 747.155 | 247.355 | 252.95 | 675.609 | 659.422 |
| 10688.1 | 747.439 | 242.993 | 253.328 | 651.437 | 660.489 |
| 10707.9 | 747.724 | 245.118 | 253.706 | 678.452 | 661.556 |
| 10727.8 | 748.009 | 247.772 | 254.084 | 666.405 | 662.623 |
| 10747.7 | 748.293 | 255.105 | 254.462 | 682.643 | 663.691 |
| 10767.6 | 748.578 | 244.738 | 254.84 | 707.779 | 664.76 |
| 10787.5 | 748.862 | 254.858 | 255.219 | 690.404 | 665.829 |
| 10807.4 | 749.147 | 258.45 | 255.597 | 722.844 | 666.898 |
| 10827.4 | 749.431 | 260.101 | 255.976 | 695.768 | 667.968 |
| 10847.3 | 749.716 | 261.758 | 256.355 | 717.672 | 669.038 |
| 10867.3 | 750 | 244.257 | 256.733 | 685.19 | 670.109 |
| 10887.3 | 750.284 | 251.747 | 257.112 | 705.77 | 671.18 |
| 10907.3 | 750.569 | 260.787 | 257.491 | 682.627 | 672.251 |
| 10927.3 | 750.853 | 243.897 | 257.87 | 688.244 | 673.323 |
| 10947.3 | 751.138 | 260.155 | 258.249 | 705.447 | 674.395 |
| 10967.4 | 751.422 | 259.03 | 258.628 | 689.965 | 675.468 |
| 10987.4 | 751.706 | 256.573 | 259.008 | 719.156 | 676.541 |
| 11007.5 | 751.991 | 261.113 | 259.387 | 667.473 | 677.615 |
| 11027.5 | 752.275 | 246.261 | 259.766 | 707.789 | 678.688 |
| 11047.6 | 752.559 | 260.372 | 260.146 | 697.474 | 679.763 |
| 11067.7 | 752.844 | 265.476 | 260.525 | 715.398 | 680.837 |
| 11087.9 | 753.128 | 271.055 | 260.905 | 696.441 | 681.913 |
| 11108 | 753.412 | 245.603 | 261.285 | 703.578 | 682.988 |
| 11128.1 | 753.697 | 255.107 | 261.664 | 689.503 | 684.064 |
| 11148.3 | 753.981 | 257.134 | 262.044 | 688.486 | 685.14 |
| 11168.5 | 754.265 | 269.719 | 262.424 | 672.761 | 686.217 |
| 11188.6 | 754.549 | 281.956 | 262.804 | 688.072 | 687.294 |
| 11208.8 | 754.834 | 265.045 | 263.184 | 709.568 | 688.372 |
| 11229 | 755.118 | 259.92 | 263.564 | 692.898 | 689.45 |
| 11249.3 | 755.402 | 262.143 | 263.944 | 705.455 | 690.528 |
| 11269.5 | 755.686 | 273.108 | 264.324 | 704.368 | 691.607 |
| 11289.8 | 755.97 | 278.149 | 264.705 | 699.97 | 692.686 |
| 11310 | 756.254 | 252.46 | 265.085 | 703.363 | 693.765 |
| 11330.3 | 756.538 | 258.449 | 265.465 | 695.254 | 694.845 |
| 11350.6 | 756.823 | 264.567 | 265.846 | 701.496 | 695.926 |
| 11370.9 | 757.107 | 266.044 | 266.226 | 736.084 | 697.006 |
| 11391.2 | 757.391 | 258.722 | 266.607 | 694.064 | 698.087 |
| 11411.5 | 757.675 | 263.347 | 266.988 | 721.344 | 699.169 |
| 11431.9 | 757.959 | 256.492 | 267.368 | 740.564 | 700.25 |
| 11452.2 | 758.243 | 282.825 | 267.749 | 740.002 | 701.332 |
| 11472.6 | 758.527 | 273.74 | 268.13 | 707.411 | 702.415 |
| 11492.9 | 758.811 | 267.986 | 268.511 | 705.19 | 703.498 |
| 11513.3 | 759.095 | 274.939 | 268.892 | 741.019 | 704.581 |
| 11533.7 | 759.379 | 269.016 | 269.273 | 750.319 | 705.665 |
| 11554.1 | 759.663 | 261.15 | 269.654 | 693.101 | 706.749 |
| 11574.6 | 759.947 | 252.574 | 270.035 | 756.006 | 707.833 |
| 11595 | 760.231 | 254.357 | 270.416 | 716.219 | 708.918 |
| 11615.5 | 760.515 | 260.441 | 270.797 | 740.879 | 710.003 |
| 11635.9 | 760.799 | 276.155 | 271.179 | 726.705 | 711.088 |
| 11656.4 | 761.082 | 267.121 | 271.56 | 718.851 | 712.174 |
| 11676.9 | 761.366 | 262.316 | 271.942 | 731.343 | 713.26 |
| 11697.4 | 761.65 | 282.943 | 272.323 | 751.366 | 714.347 |
| 11717.9 | 761.934 | 268.303 | 272.704 | 718.555 | 715.433 |
| 11738.4 | 762.218 | 262.045 | 273.086 | 717.751 | 716.521 |
| 11759 | 762.502 | 261.375 | 273.468 | 720.67 | 717.608 |
| 11779.5 | 762.786 | 284.532 | 273.849 | 749.045 | 718.696 |
| 11800.1 | 763.069 | 265.142 | 274.231 | 741.081 | 719.784 |
| 11820.7 | 763.353 | 281.589 | 274.613 | 731.212 | 720.873 |
| 11841.2 | 763.637 | 285.413 | 274.995 | 743.733 | 721.962 |
| 11861.9 | 763.921 | 282.719 | 275.376 | 732.778 | 723.051 |
| 11882.5 | 764.204 | 268.652 | 275.758 | 739.969 | 724.141 |
| 11903.1 | 764.488 | 270.781 | 276.14 | 749.189 | 725.231 |
| 11923.7 | 764.772 | 286.428 | 276.522 | 725.452 | 726.321 |
| 11944.4 | 765.056 | 268.024 | 276.904 | 742.328 | 727.411 |
| 11965 | 765.339 | 282.423 | 277.286 | 721.558 | 728.502 |
| 11985.7 | 765.623 | 268.144 | 277.668 | 717.535 | 729.594 |
| 12006.4 | 765.907 | 259.372 | 278.051 | 719.986 | 730.685 |
| 12027.1 | 766.19 | 273.133 | 278.433 | 749.536 | 731.777 |
| 12047.8 | 766.474 | 276.845 | 278.815 | 766.568 | 732.869 |
| 12068.5 | 766.757 | 262.059 | 279.197 | 778.048 | 733.962 |
| 12089.2 | 767.041 | 276.005 | 279.58 | 743.024 | 735.055 |
| 12110 | 767.324 | 288.467 | 279.962 | 765.116 | 736.148 |
| 12130.7 | 767.608 | 260.874 | 280.344 | 752.203 | 737.241 |
| 12151.5 | 767.891 | 258.662 | 280.727 | 776.713 | 738.335 |
| 12172.3 | 768.175 | 274.435 | 281.109 | 725.165 | 739.429 |
| 12193.1 | 768.459 | 270.786 | 281.492 | 747.891 | 740.524 |
| 12213.9 | 768.742 | 286.041 | 281.875 | 726.687 | 741.619 |
| 12234.7 | 769.026 | 275.291 | 282.257 | 763.701 | 742.714 |
| 12255.5 | 769.309 | 261.024 | 282.64 | 752.635 | 743.809 |
| 12276.3 | 769.592 | 289.114 | 283.022 | 761.93 | 744.905 |
| 12297.2 | 769.876 | 278.828 | 283.405 | 726.048 | 746.001 |
| 12318 | 770.159 | 281.66 | 283.788 | 776.372 | 747.097 |
| 12338.9 | 770.443 | 289.898 | 284.171 | 772.226 | 748.194 |
| 12359.8 | 770.726 | 259.296 | 284.553 | 742.504 | 749.291 |
| 12380.7 | 771.01 | 276.959 | 284.936 | 740.091 | 750.388 |
| 12401.6 | 771.293 | 292.469 | 285.319 | 749.091 | 751.485 |
| 12422.5 | 771.576 | 279.38 | 285.702 | 776.993 | 752.583 |
| 12443.4 | 771.86 | 281.957 | 286.085 | 770.988 | 753.681 |
| 12464.4 | 772.143 | 280.672 | 286.468 | 758.533 | 754.779 |
| 12485.3 | 772.426 | 288.423 | 286.851 | 784.584 | 755.878 |
| 12506.3 | 772.709 | 306.695 | 287.234 | 742.016 | 756.977 |
| 12527.2 | 772.993 | 278.435 | 287.617 | 766.152 | 758.076 |
| 12548.2 | 773.276 | 286.305 | 288 | 783.626 | 759.176 |
| 12569.2 | 773.559 | 286.637 | 288.383 | 802.395 | 760.276 |
| 12590.2 | 773.842 | 297.317 | 288.766 | 770.268 | 761.376 |
| 12611.2 | 774.126 | 281.433 | 289.149 | 755.765 | 762.476 |
| 12632.3 | 774.409 | 301.53 | 289.533 | 788.849 | 763.577 |
| 12653.3 | 774.692 | 280.28 | 289.916 | 757.422 | 764.678 |
| 12674.4 | 774.975 | 274.309 | 290.299 | 769.949 | 765.779 |
| 12695.4 | 775.258 | 295.802 | 290.682 | 800.001 | 766.88 |
| 12716.5 | 775.542 | 291.519 | 291.066 | 764.13 | 767.982 |
| 12737.6 | 775.825 | 284.156 | 291.449 | 759.084 | 769.084 |
| 12758.7 | 776.108 | 282.275 | 291.832 | 799.522 | 770.186 |
| 12779.8 | 776.391 | 269.203 | 292.216 | 754.558 | 771.289 |
| 12800.9 | 776.674 | 282.871 | 292.599 | 797.053 | 772.392 |
| 12822 | 776.957 | 291.673 | 292.982 | 773.967 | 773.495 |
| 12843.1 | 777.24 | 294.33 | 293.366 | 760.762 | 774.598 |
| 12864.3 | 777.523 | 286.137 | 293.749 | 773.462 | 775.702 |
| 12885.4 | 777.806 | 275.746 | 294.132 | 769.836 | 776.805 |
| 12906.6 | 778.089 | 298.267 | 294.516 | 762.562 | 777.909 |
| 12927.8 | 778.372 | 285.826 | 294.899 | 807.971 | 779.014 |
| 12949 | 778.655 | 277.765 | 295.283 | 788.58 | 780.118 |
| 12970.2 | 778.938 | 295.964 | 295.666 | 803.86 | 781.223 |
| 12991.4 | 779.221 | 291.266 | 296.05 | 808.295 | 782.328 |
| 13012.6 | 779.504 | 300.46 | 296.433 | 807.132 | 783.434 |
| 13033.8 | 779.787 | 309.524 | 296.817 | 794.173 | 784.539 |
| 13055.1 | 780.07 | 296.048 | 297.201 | 797.145 | 785.645 |
| 13076.3 | 780.353 | 287.556 | 297.584 | 796.697 | 786.751 |
| 13097.6 | 780.635 | 288.354 | 297.968 | 778.813 | 787.857 |
| 13118.9 | 780.918 | 286.275 | 298.351 | 796.724 | 788.964 |
| 13140.1 | 781.201 | 296.85 | 298.735 | 786.195 | 790.071 |
| 13161.4 | 781.484 | 291.953 | 299.118 | 787.25 | 791.178 |
| 13182.7 | 781.767 | 287.576 | 299.502 | 791.939 | 792.285 |
| 13204 | 782.05 | 281.791 | 299.886 | 795.391 | 793.393 |
| 13225.4 | 782.332 | 284.657 | 300.269 | 780.466 | 794.5 |
| 13246.7 | 782.615 | 288.225 | 300.653 | 825.644 | 795.608 |
| 13268 | 782.898 | 293.617 | 301.037 | 834.351 | 796.716 |
| 13289.4 | 783.181 | 294.315 | 301.42 | 794.493 | 797.825 |
| 13310.7 | 783.463 | 289.171 | 301.804 | 803.943 | 798.933 |
| 13332.1 | 783.746 | 290.775 | 302.188 | 800.726 | 800.042 |
| 13353.5 | 784.029 | 283.05 | 302.571 | 793.396 | 801.152 |
| 13374.9 | 784.311 | 295.56 | 302.955 | 783.587 | 802.261 |
| 13396.3 | 784.594 | 298.678 | 303.339 | 803.728 | 803.37 |
| 13417.7 | 784.877 | 298.209 | 303.722 | 807.991 | 804.48 |
| 13439.1 | 785.159 | 292.522 | 304.106 | 825.681 | 805.59 |
| 13460.6 | 785.442 | 298.259 | 304.49 | 810.89 | 806.7 |
| 13482 | 785.725 | 291.573 | 304.873 | 823.797 | 807.811 |
| 13503.5 | 786.007 | 303.574 | 305.257 | 827.52 | 808.921 |
| 13524.9 | 786.29 | 305.911 | 305.641 | 785.37 | 810.032 |
| 13546.4 | 786.572 | 291.553 | 306.024 | 818.787 | 811.143 |
| 13567.9 | 786.855 | 302.748 | 306.408 | 830.112 | 812.254 |
| 13589.4 | 787.137 | 302.625 | 306.792 | 808.136 | 813.365 |
| 13610.9 | 787.42 | 298.527 | 307.175 | 809.505 | 814.477 |
| 13632.4 | 787.702 | 314.725 | 307.559 | 841.386 | 815.589 |
| 13653.9 | 787.985 | 304.437 | 307.943 | 815.949 | 816.701 |
| 13675.4 | 788.267 | 307.872 | 308.327 | 817.168 | 817.813 |
| 13696.9 | 788.55 | 295.679 | 308.71 | 818.403 | 818.925 |
| 13718.5 | 788.832 | 305.188 | 309.094 | 822.148 | 820.038 |
| 13740 | 789.115 | 302.075 | 309.477 | 823.387 | 821.151 |
| 13761.6 | 789.397 | 300.357 | 309.861 | 831.23 | 822.264 |
| 13783.2 | 789.679 | 308.249 | 310.245 | 831.921 | 823.377 |
| 13804.8 | 789.962 | 288.311 | 310.628 | 832.829 | 824.49 |
| 13826.4 | 790.244 | 321.838 | 311.012 | 827.56 | 825.604 |
| 13848 | 790.526 | 313.117 | 311.396 | 843.894 | 826.717 |
| 13869.6 | 790.809 | 318.843 | 311.779 | 834.377 | 827.831 |
| 13891.2 | 791.091 | 302.357 | 312.163 | 836.649 | 828.945 |
| 13912.8 | 791.373 | 318.908 | 312.546 | 847.958 | 830.06 |
| 13934.5 | 791.655 | 297.382 | 312.93 | 815.388 | 831.174 |
| 13956.1 | 791.938 | 313.684 | 313.313 | 855.219 | 832.289 |
| 13977.8 | 792.22 | 291.354 | 313.697 | 847.721 | 833.404 |
| 13999.4 | 792.502 | 324.675 | 314.081 | 826.939 | 834.519 |
| 14021.1 | 792.784 | 305.082 | 314.464 | 814.284 | 835.634 |
| 14042.8 | 793.067 | 302.35 | 314.848 | 814.376 | 836.749 |
| 14064.5 | 793.349 | 322.298 | 315.231 | 852.421 | 837.864 |
| 14086.2 | 793.631 | 308.422 | 315.615 | 855.434 | 838.98 |
| 14107.9 | 793.913 | 299.046 | 315.998 | 866.044 | 840.096 |
| 14129.6 | 794.195 | 320.562 | 316.382 | 847.575 | 841.212 |
| 14151.4 | 794.477 | 318.071 | 316.765 | 837.943 | 842.328 |
| 14173.1 | 794.76 | 317.011 | 317.148 | 858.069 | 843.444 |
| 14194.8 | 795.042 | 325.209 | 317.532 | 848.346 | 844.561 |
| 14216.6 | 795.324 | 324.476 | 317.915 | 831.997 | 845.677 |
| 14238.4 | 795.606 | 299.696 | 318.298 | 814.067 | 846.794 |
| 14260.1 | 795.888 | 313.675 | 318.682 | 835.005 | 847.911 |
| 14281.9 | 796.17 | 319.486 | 319.065 | 852.29 | 849.028 |
| 14303.7 | 796.452 | 317.713 | 319.449 | 828.628 | 850.145 |
| 14325.5 | 796.734 | 316.16 | 319.832 | 861.339 | 851.262 |
| 14347.3 | 797.016 | 320.035 | 320.215 | 841.157 | 852.38 |
| 14369.1 | 797.298 | 318.207 | 320.598 | 859.258 | 853.497 |
| 14390.9 | 797.58 | 311.046 | 320.981 | 853.578 | 854.615 |
| 14412.8 | 797.862 | 306.27 | 321.365 | 853.079 | 855.733 |
| 14434.6 | 798.144 | 317.333 | 321.748 | 885.521 | 856.851 |
| 14456.5 | 798.426 | 289.067 | 322.131 | 834.499 | 857.969 |
| 14478.3 | 798.707 | 318.354 | 322.514 | 851.269 | 859.088 |
| 14500.2 | 798.989 | 324.974 | 322.897 | 860.903 | 860.206 |
| 14522 | 799.271 | 302.26 | 323.28 | 848.368 | 861.325 |
| 14543.9 | 799.553 | 317.749 | 323.663 | 811.981 | 862.444 |
| 14565.8 | 799.835 | 316.915 | 324.046 | 857.059 | 863.562 |
| 14587.7 | 800.117 | 336.2 | 324.429 | 838.093 | 864.682 |
| 14609.6 | 800.399 | 319.353 | 324.812 | 874.971 | 865.801 |
| 14631.5 | 800.68 | 321.386 | 325.195 | 855.355 | 866.92 |
| 14653.5 | 800.962 | 312.365 | 325.578 | 840.472 | 868.039 |
| 14675.4 | 801.244 | 302.902 | 325.961 | 836.141 | 869.159 |
| 14697.3 | 801.525 | 330.093 | 326.344 | 847.789 | 870.278 |
| 14719.3 | 801.807 | 319.906 | 326.726 | 895.381 | 871.398 |
| 14741.2 | 802.089 | 318.757 | 327.109 | 884.534 | 872.518 |
| 14763.2 | 802.371 | 326.467 | 327.492 | 885.357 | 873.638 |
| 14785.2 | 802.652 | 345.796 | 327.875 | 864.589 | 874.758 |
| 14807.1 | 802.934 | 322.364 | 328.257 | 881.829 | 875.878 |
| 14829.1 | 803.216 | 318.837 | 328.64 | 874.958 | 876.999 |
| 14851.1 | 803.497 | 327.237 | 329.023 | 912.802 | 878.119 |
| 14873.1 | 803.779 | 312.352 | 329.405 | 906.345 | 879.24 |
| 14895.1 | 804.06 | 320.814 | 329.788 | 895.114 | 880.36 |
| 14917.1 | 804.342 | 340.464 | 330.17 | 903.988 | 881.481 |
| 14939.2 | 804.624 | 331.52 | 330.553 | 855.177 | 882.602 |
| 14961.2 | 804.905 | 333.2 | 330.935 | 872.716 | 883.723 |
| 14983.2 | 805.187 | 337.757 | 331.317 | 887.558 | 884.844 |
| 15005.3 | 805.468 | 329.997 | 331.7 | 917.564 | 885.965 |
| 15027.3 | 805.75 | 319.596 | 332.082 | 884.056 | 887.086 |
| 15049.4 | 806.031 | 335.306 | 332.464 | 880.269 | 888.208 |
| 15071.5 | 806.313 | 319.12 | 332.847 | 882.017 | 889.329 |
| 15093.5 | 806.594 | 326.397 | 333.229 | 884.51 | 890.451 |
| 15115.6 | 806.875 | 345.945 | 333.611 | 926.227 | 891.572 |
| 15137.7 | 807.157 | 314.542 | 333.993 | 924.927 | 892.694 |
| 15159.8 | 807.438 | 317.101 | 334.375 | 875.088 | 893.816 |
| 15181.9 | 807.72 | 338.956 | 334.757 | 885.408 | 894.938 |
| 15204 | 808.001 | 330.093 | 335.139 | 897.42 | 896.06 |
| 15226.2 | 808.283 | 338.308 | 335.521 | 888.453 | 897.182 |
| 15248.3 | 808.564 | 324.467 | 335.903 | 919.855 | 898.304 |
| 15270.4 | 808.845 | 355.063 | 336.285 | 898.237 | 899.426 |
| 15292.6 | 809.126 | 338.647 | 336.667 | 917.399 | 900.549 |
| 15314.7 | 809.408 | 345.799 | 337.048 | 909.874 | 901.671 |
| 15336.9 | 809.689 | 326.385 | 337.43 | 907.613 | 902.793 |
| 15359 | 809.97 | 313.999 | 337.812 | 907.918 | 903.916 |
| 15381.2 | 810.252 | 349.082 | 338.193 | 896.426 | 905.039 |
| 15403.4 | 810.533 | 317.142 | 338.575 | 917.143 | 906.161 |
| 15425.6 | 810.814 | 336.129 | 338.957 | 908.434 | 907.284 |
| 15447.8 | 811.095 | 336.795 | 339.338 | 897.833 | 908.407 |
| 15470 | 811.376 | 319.018 | 339.719 | 896.909 | 909.53 |
| 15492.2 | 811.658 | 328.523 | 340.101 | 907.129 | 910.653 |
| 15514.4 | 811.939 | 338.496 | 340.482 | 887.854 | 911.776 |
| 15536.6 | 812.22 | 328.034 | 340.863 | 900.197 | 912.899 |
| 15558.9 | 812.501 | 356.276 | 341.245 | 873.689 | 914.022 |
| 15581.1 | 812.782 | 320.671 | 341.626 | 921.033 | 915.145 |
| 15603.3 | 813.063 | 321.774 | 342.007 | 894.144 | 916.269 |
| 15625.6 | 813.344 | 338.349 | 342.388 | 922.917 | 917.392 |
| 15647.8 | 813.625 | 353.575 | 342.769 | 920.904 | 918.515 |
| 15670.1 | 813.907 | 325.385 | 343.15 | 882.503 | 919.639 |
| 15692.4 | 814.187 | 322.29 | 343.531 | 897.613 | 920.762 |
| 15714.6 | 814.468 | 345.152 | 343.912 | 891.457 | 921.886 |
| 15736.9 | 814.75 | 323.528 | 344.293 | 882.957 | 923.01 |
| 15759.2 | 815.03 | 346.898 | 344.673 | 919.185 | 924.133 |
| 15781.5 | 815.312 | 327.026 | 345.054 | 894.649 | 925.257 |
| 15803.8 | 815.592 | 332.62 | 345.435 | 918.357 | 926.381 |
| 15826.1 | 815.873 | 349.163 | 345.815 | 904.523 | 927.504 |
| 15848.4 | 816.154 | 345.385 | 346.196 | 937.396 | 928.628 |
| 15870.7 | 816.435 | 314.858 | 346.576 | 901.121 | 929.752 |
| 15893.1 | 816.716 | 345.608 | 346.957 | 900.481 | 930.876 |
| 15915.4 | 816.997 | 328.508 | 347.337 | 952.814 | 932 |
| 15937.7 | 817.278 | 341.891 | 347.717 | 907.643 | 933.124 |
| 15960.1 | 817.559 | 355.758 | 348.098 | 949.177 | 934.248 |
| 15982.4 | 817.84 | 343.781 | 348.478 | 903.726 | 935.372 |
| 16004.8 | 818.12 | 356.63 | 348.858 | 913.426 | 936.496 |
| 16027.1 | 818.401 | 354.396 | 349.238 | 904.916 | 937.62 |
| 16049.5 | 818.682 | 330.181 | 349.618 | 944.276 | 938.744 |
| 16071.9 | 818.963 | 333.989 | 349.998 | 956.81 | 939.868 |
| 16094.3 | 819.243 | 350.918 | 350.377 | 915.021 | 940.993 |
| 16116.7 | 819.524 | 340.881 | 350.757 | 941.711 | 942.117 |
| 16139 | 819.805 | 343.479 | 351.137 | 897.365 | 943.241 |
| 16161.5 | 820.086 | 333.94 | 351.517 | 938.619 | 944.365 |
| 16183.9 | 820.366 | 341.22 | 351.896 | 901.702 | 945.49 |
| 16206.3 | 820.647 | 334.648 | 352.276 | 943.394 | 946.614 |
| 16228.7 | 820.928 | 355.506 | 352.655 | 930.988 | 947.738 |
| 16251.1 | 821.208 | 373.101 | 353.035 | 928.498 | 948.863 |
| 16273.5 | 821.489 | 350.72 | 353.414 | 940.911 | 949.987 |
| 16296 | 821.769 | 333.276 | 353.793 | 982.045 | 951.111 |
| 16318.4 | 822.05 | 339.992 | 354.172 | 953.281 | 952.236 |
| 16340.9 | 822.331 | 336.409 | 354.551 | 921.09 | 953.36 |
| 16363.3 | 822.611 | 361.314 | 354.93 | 945.978 | 954.484 |
| 16385.8 | 822.892 | 358.935 | 355.31 | 953.448 | 955.609 |
| 16408.2 | 823.172 | 340.703 | 355.688 | 967.881 | 956.733 |
| 16430.7 | 823.453 | 364.26 | 356.067 | 932.968 | 957.858 |
| 16453.2 | 823.733 | 384.023 | 356.446 | 963.514 | 958.982 |
| 16475.7 | 824.014 | 362.466 | 356.825 | 954.057 | 960.106 |
| 16498.2 | 824.294 | 329.347 | 357.203 | 949.165 | 961.231 |
| 16520.7 | 824.574 | 354.636 | 357.582 | 937.619 | 962.355 |
| 16543.1 | 824.855 | 368.711 | 357.96 | 964.909 | 963.48 |
| 16565.7 | 825.135 | 339.706 | 358.339 | 971.478 | 964.604 |
| 16588.2 | 825.416 | 375.923 | 358.717 | 959.577 | 965.728 |
| 16610.7 | 825.696 | 364.548 | 359.095 | 972.416 | 966.853 |
| 16633.2 | 825.976 | 361.242 | 359.474 | 951.393 | 967.977 |
| 16655.7 | 826.257 | 348.807 | 359.852 | 966.251 | 969.101 |
| 16678.2 | 826.537 | 341.897 | 360.23 | 959.246 | 970.226 |
| 16700.8 | 826.817 | 357.087 | 360.608 | 968.068 | 971.35 |
| 16723.3 | 827.098 | 346.148 | 360.986 | 970.086 | 972.475 |
| 16745.9 | 827.378 | 354.991 | 361.363 | 942.748 | 973.599 |
| 16768.4 | 827.658 | 352.656 | 361.741 | 961.477 | 974.723 |
| 16791 | 827.939 | 376.491 | 362.119 | 961.97 | 975.847 |
| 16813.5 | 828.219 | 347.011 | 362.496 | 993.434 | 976.972 |
| 16836.1 | 828.499 | 337.548 | 362.874 | 924.912 | 978.096 |
| 16858.7 | 828.779 | 361.934 | 363.251 | 967.436 | 979.22 |
| 16881.2 | 829.059 | 343.868 | 363.628 | 964.62 | 980.344 |
| 16903.8 | 829.34 | 362.711 | 364.006 | 999.535 | 981.468 |
| 16926.4 | 829.62 | 366.644 | 364.383 | 951.144 | 982.592 |
| 16949 | 829.9 | 352.65 | 364.76 | 972.639 | 983.716 |
| 16971.6 | 830.18 | 361.014 | 365.137 | 956.635 | 984.841 |
| 16994.2 | 830.46 | 344.134 | 365.514 | 979.924 | 985.965 |
| 17016.8 | 830.74 | 338.272 | 365.891 | 991.568 | 987.089 |
| 17039.4 | 831.02 | 365.653 | 366.267 | 983.153 | 988.212 |
| 17062 | 831.3 | 395.544 | 366.644 | 973.075 | 989.337 |
| 17084.7 | 831.58 | 353.841 | 367.021 | 1000.95 | 990.46 |
| 17107.3 | 831.86 | 366.975 | 367.397 | 966.497 | 991.584 |
| 17129.9 | 832.14 | 374.357 | 367.774 | 982.591 | 992.708 |
| 17152.5 | 832.42 | 368.613 | 368.15 | 983.649 | 993.832 |
| 17175.2 | 832.7 | 370.119 | 368.526 | 967.521 | 994.956 |
| 17197.8 | 832.98 | 367.538 | 368.902 | 972.433 | 996.079 |
| 17220.5 | 833.26 | 365.538 | 369.278 | 1006.59 | 997.203 |
| 17243.1 | 833.54 | 378.62 | 369.654 | 962.309 | 998.326 |
| 17265.8 | 833.82 | 369.433 | 370.03 | 1016.6 | 999.45 |
| 17288.4 | 834.1 | 356.045 | 370.406 | 970.134 | 1000.57 |
| 17311.1 | 834.38 | 362.899 | 370.782 | 986.853 | 1001.7 |
| 17333.8 | 834.66 | 358.258 | 371.157 | 1009.45 | 1002.82 |
| 17356.5 | 834.939 | 382.119 | 371.533 | 993.062 | 1003.94 |
| 17379.1 | 835.219 | 382.007 | 371.908 | 971.066 | 1005.07 |
| 17401.8 | 835.499 | 372.892 | 372.284 | 1005.32 | 1006.19 |
| 17424.5 | 835.779 | 373.134 | 372.659 | 1003.21 | 1007.31 |
| 17447.2 | 836.059 | 364.753 | 373.034 | 997.499 | 1008.44 |
| 17469.9 | 836.338 | 364.443 | 373.409 | 1001.5 | 1009.56 |
| 17492.6 | 836.618 | 372.752 | 373.784 | 1005.58 | 1010.68 |
| 17515.3 | 836.898 | 371.149 | 374.159 | 994.956 | 1011.8 |
| 17538 | 837.178 | 371.358 | 374.534 | 1004.48 | 1012.93 |
| 17560.7 | 837.457 | 384.755 | 374.908 | 1004.45 | 1014.05 |
| 17583.4 | 837.737 | 390.06 | 375.283 | 1013.92 | 1015.17 |
| 17606.2 | 838.017 | 364.413 | 375.658 | 1015.31 | 1016.3 |
| 17628.9 | 838.296 | 371.651 | 376.032 | 1006.75 | 1017.42 |
| 17651.6 | 838.576 | 352.671 | 376.406 | 987.164 | 1018.54 |
| 17674.3 | 838.856 | 384.424 | 376.78 | 1006.36 | 1019.66 |
| 17697.1 | 839.135 | 377.284 | 377.154 | 1009.73 | 1020.79 |
| 17719.8 | 839.415 | 367.78 | 377.529 | 1018.29 | 1021.91 |
| 17742.6 | 839.694 | 362.016 | 377.902 | 1032.36 | 1023.03 |
| 17765.3 | 839.974 | 387.827 | 378.276 | 1037.24 | 1024.15 |
| 17788.1 | 840.254 | 386.384 | 378.65 | 1000.72 | 1025.27 |
| 17810.8 | 840.533 | 366.596 | 379.024 | 993.443 | 1026.4 |
| 17833.6 | 840.812 | 390.43 | 379.397 | 993.505 | 1027.52 |
| 17856.4 | 841.092 | 382.513 | 379.771 | 1014.94 | 1028.64 |
| 17879.1 | 841.371 | 362.281 | 380.144 | 1021.99 | 1029.76 |
| 17901.9 | 841.651 | 374.336 | 380.517 | 1036.98 | 1030.88 |
| 17924.7 | 841.93 | 389.629 | 380.89 | 1044.23 | 1032.01 |
| 17947.5 | 842.21 | 369.767 | 381.263 | 1034.18 | 1033.13 |
| 17970.2 | 842.489 | 376.537 | 381.636 | 1042.35 | 1034.25 |
| 17993 | 842.769 | 379.139 | 382.009 | 1023.04 | 1035.37 |
| 18015.8 | 843.048 | 395.873 | 382.382 | 1030.65 | 1036.49 |
| 18038.6 | 843.327 | 369.299 | 382.754 | 1038.43 | 1037.61 |
| 18061.4 | 843.607 | 384.465 | 383.127 | 1030.84 | 1038.73 |
| 18084.2 | 843.886 | 370.632 | 383.499 | 1054.46 | 1039.85 |
| 18107 | 844.165 | 365.599 | 383.872 | 975.606 | 1040.97 |
| 18129.8 | 844.444 | 389.93 | 384.244 | 1038.89 | 1042.09 |
| 18152.6 | 844.724 | 379.744 | 384.616 | 1021.26 | 1043.21 |
| 18175.5 | 845.003 | 371.026 | 384.988 | 1029.26 | 1044.34 |
| 18198.3 | 845.282 | 378.08 | 385.36 | 1031.29 | 1045.46 |
| 18221.1 | 845.562 | 382.911 | 385.732 | 1035.48 | 1046.58 |
| 18243.9 | 845.841 | 390.527 | 386.103 | 1056.38 | 1047.7 |
| 18266.7 | 846.12 | 397.924 | 386.475 | 1030.28 | 1048.82 |
| 18289.6 | 846.399 | 384.062 | 386.846 | 1028.99 | 1049.94 |
| 18312.4 | 846.678 | 390.887 | 387.218 | 1029.77 | 1051.06 |
| 18335.2 | 846.957 | 376.44 | 387.589 | 1020.55 | 1052.18 |
| 18358.1 | 847.236 | 381.962 | 387.96 | 1026.12 | 1053.3 |
| 18380.9 | 847.516 | 370.481 | 388.331 | 1018.76 | 1054.41 |
| 18403.8 | 847.795 | 410.721 | 388.702 | 1029.7 | 1055.53 |
| 18426.6 | 848.074 | 389.038 | 389.073 | 1050.97 | 1056.65 |
| 18449.5 | 848.353 | 385.026 | 389.443 | 1055.9 | 1057.77 |
| 18472.3 | 848.632 | 388.1 | 389.814 | 1065.72 | 1058.89 |
| 18495.2 | 848.911 | 376.921 | 390.185 | 1056.42 | 1060.01 |
| 18518 | 849.19 | 409.415 | 390.555 | 1050.46 | 1061.13 |
| 18540.9 | 849.469 | 382.012 | 390.925 | 1046.25 | 1062.25 |
| 18563.8 | 849.748 | 387.927 | 391.295 | 1064.56 | 1063.37 |
| 18586.7 | 850.027 | 393.86 | 391.665 | 1071.2 | 1064.48 |
| 18609.5 | 850.306 | 382.582 | 392.035 | 1049.89 | 1065.6 |
| 18632.4 | 850.585 | 383.721 | 392.405 | 1052.33 | 1066.72 |
| 18655.3 | 850.864 | 393.133 | 392.774 | 1067.85 | 1067.84 |
| 18678.2 | 851.143 | 392.903 | 393.144 | 1058.23 | 1068.96 |
| 18701 | 851.421 | 394.634 | 393.514 | 1047.09 | 1070.07 |
| 18723.9 | 851.7 | 395.689 | 393.883 | 1077.61 | 1071.19 |
| 18746.8 | 851.979 | 388.929 | 394.252 | 1079.62 | 1072.31 |
| 18769.7 | 852.258 | 392.01 | 394.621 | 1076.45 | 1073.43 |
| 18792.6 | 852.537 | 398.985 | 394.99 | 1054.4 | 1074.54 |
| 18815.5 | 852.816 | 382.056 | 395.359 | 1074.48 | 1075.66 |
| 18838.4 | 853.094 | 399.463 | 395.728 | 1072.16 | 1076.78 |
| 18861.3 | 853.373 | 395.261 | 396.096 | 1065.46 | 1077.89 |
| 18884.2 | 853.652 | 402.083 | 396.465 | 1045.33 | 1079.01 |
| 18907.1 | 853.93 | 407.196 | 396.833 | 1063.96 | 1080.13 |
| 18930 | 854.209 | 403.005 | 397.202 | 1066.34 | 1081.24 |
| 18952.9 | 854.488 | 384.731 | 397.57 | 1055.79 | 1082.36 |
| 18975.8 | 854.766 | 387.982 | 397.938 | 1045.47 | 1083.48 |
| 18998.8 | 855.045 | 376.414 | 398.305 | 1078.11 | 1084.59 |
| 19021.7 | 855.324 | 382.789 | 398.673 | 1073.04 | 1085.71 |
| 19044.6 | 855.602 | 407.794 | 399.041 | 1074.04 | 1086.82 |
| 19067.5 | 855.881 | 380.933 | 399.409 | 1077.58 | 1087.94 |
| 19090.5 | 856.16 | 399.37 | 399.776 | 1059.2 | 1089.05 |
| 19113.4 | 856.438 | 402.955 | 400.143 | 1057.18 | 1090.17 |
| 19136.3 | 856.717 | 392.038 | 400.51 | 1077.89 | 1091.28 |
| 19159.2 | 856.995 | 399.138 | 400.877 | 1094.64 | 1092.4 |
| 19182.2 | 857.274 | 395.492 | 401.244 | 1079.86 | 1093.51 |
| 19205.1 | 857.552 | 402.097 | 401.611 | 1063.13 | 1094.63 |
| 19228 | 857.831 | 414.871 | 401.978 | 1085.32 | 1095.74 |
| 19251 | 858.109 | 413.235 | 402.344 | 1069.79 | 1096.86 |
| 19273.9 | 858.388 | 400.246 | 402.711 | 1042.42 | 1097.97 |
| 19296.9 | 858.666 | 400.35 | 403.077 | 1102.05 | 1099.08 |
| 19319.8 | 858.945 | 399.7 | 403.443 | 1048.34 | 1100.2 |
| 19342.8 | 859.223 | 401.628 | 403.809 | 1091.92 | 1101.31 |
| 19365.7 | 859.501 | 422.809 | 404.175 | 1055.17 | 1102.42 |
| 19388.7 | 859.78 | 397.35 | 404.541 | 1077.42 | 1103.54 |
| 19411.6 | 860.058 | 409.211 | 404.907 | 1067.67 | 1104.65 |
| 19434.6 | 860.337 | 411.096 | 405.272 | 1086.56 | 1105.76 |
| 19457.5 | 860.615 | 392.659 | 405.638 | 1096.64 | 1106.88 |
| 19480.5 | 860.893 | 385.38 | 406.003 | 1088.17 | 1107.99 |
| 19503.5 | 861.171 | 386.89 | 406.368 | 1075.96 | 1109.1 |
| 19526.4 | 861.45 | 413.438 | 406.733 | 1113.44 | 1110.21 |
| 19549.4 | 861.728 | 403.903 | 407.098 | 1085.89 | 1111.33 |
| 19572.3 | 862.006 | 413.526 | 407.463 | 1100.83 | 1112.44 |
| 19595.3 | 862.284 | 413.269 | 407.827 | 1101.5 | 1113.55 |
| 19618.3 | 862.563 | 395.448 | 408.192 | 1094.04 | 1114.66 |
| 19641.2 | 862.841 | 415.106 | 408.556 | 1135.05 | 1115.77 |
| 19664.2 | 863.119 | 423.492 | 408.92 | 1105.91 | 1116.88 |
| 19687.2 | 863.397 | 395.76 | 409.284 | 1084.01 | 1117.99 |
| 19710.2 | 863.675 | 410.01 | 409.649 | 1125.16 | 1119.1 |
| 19733.2 | 863.953 | 397.588 | 410.012 | 1084.95 | 1120.21 |
| 19756.1 | 864.231 | 392.348 | 410.376 | 1096.39 | 1121.32 |
| 19779.1 | 864.51 | 403.449 | 410.74 | 1100.36 | 1122.43 |
| 19802.1 | 864.788 | 412.106 | 411.103 | 1103.66 | 1123.54 |
| 19825.1 | 865.066 | 403.81 | 411.466 | 1082.31 | 1124.65 |
| 19848 | 865.344 | 404.057 | 411.829 | 1081.61 | 1125.76 |
| 19871 | 865.622 | 414.829 | 412.192 | 1115.34 | 1126.87 |
| 19894 | 865.9 | 413.819 | 412.555 | 1110.37 | 1127.98 |
| 19917 | 866.178 | 395.906 | 412.918 | 1108.79 | 1129.09 |
| 19940 | 866.456 | 413.744 | 413.281 | 1085.92 | 1130.2 |
| 19963 | 866.734 | 399.733 | 413.643 | 1107.14 | 1131.31 |
| 19986 | 867.011 | 418.68 | 414.006 | 1102.72 | 1132.41 |
| 20009 | 867.289 | 410.055 | 414.368 | 1114.43 | 1133.52 |
| 20032 | 867.567 | 413.661 | 414.73 | 1116.14 | 1134.63 |
| 20055 | 867.845 | 390.155 | 415.092 | 1108.62 | 1135.74 |
| 20078 | 868.123 | 426.336 | 415.454 | 1118.26 | 1136.85 |
| 20100.9 | 868.401 | 426.274 | 415.815 | 1090.9 | 1137.95 |
| 20123.9 | 868.679 | 396.441 | 416.177 | 1092.66 | 1139.06 |
| 20146.9 | 868.957 | 409.345 | 416.538 | 1078.67 | 1140.17 |
| 20169.9 | 869.234 | 416.99 | 416.899 | 1096.91 | 1141.27 |
| 20192.9 | 869.512 | 422.032 | 417.26 | 1131.66 | 1142.38 |
| 20215.9 | 869.79 | 418.509 | 417.621 | 1098.19 | 1143.49 |
| 20238.9 | 870.068 | 409.219 | 417.982 | 1108.01 | 1144.59 |
| 20261.9 | 870.345 | 409.774 | 418.343 | 1072.51 | 1145.7 |
| 20284.9 | 870.623 | 412.824 | 418.704 | 1151.25 | 1146.8 |
| 20307.9 | 870.901 | 419.1 | 419.064 | 1116.86 | 1147.91 |
| 20331 | 871.178 | 412.996 | 419.424 | 1111.02 | 1149.02 |
| 20354 | 871.456 | 421.738 | 419.784 | 1124.85 | 1150.12 |
| 20377 | 871.734 | 420.33 | 420.144 | 1116.05 | 1151.22 |
| 20400 | 872.011 | 407.241 | 420.504 | 1115.53 | 1152.33 |
| 20423 | 872.289 | 434.718 | 420.864 | 1132.97 | 1153.43 |
| 20446 | 872.567 | 413.236 | 421.223 | 1110.92 | 1154.54 |
| 20469 | 872.844 | 409.616 | 421.583 | 1131.59 | 1155.64 |
| 20492 | 873.122 | 407.815 | 421.942 | 1138.98 | 1156.74 |
| 20515 | 873.399 | 423.834 | 422.301 | 1126.94 | 1157.85 |
| 20538 | 873.677 | 417.36 | 422.66 | 1120.79 | 1158.95 |
| 20561 | 873.954 | 422.688 | 423.019 | 1131.89 | 1160.05 |
| 20584 | 874.232 | 411.488 | 423.377 | 1134.02 | 1161.16 |
| 20607 | 874.509 | 417.332 | 423.736 | 1127.53 | 1162.26 |
| 20630.1 | 874.787 | 445.54 | 424.094 | 1125.12 | 1163.36 |
| 20653.1 | 875.064 | 413.638 | 424.452 | 1150.89 | 1164.46 |
| 20676.1 | 875.342 | 402.852 | 424.81 | 1116.43 | 1165.56 |
| 20699.1 | 875.619 | 426.452 | 425.168 | 1122.37 | 1166.66 |
| 20722.1 | 875.896 | 424.041 | 425.526 | 1156.11 | 1167.77 |
| 20745.1 | 876.174 | 403.979 | 425.884 | 1175.19 | 1168.87 |
| 20768.1 | 876.451 | 434.594 | 426.241 | 1127.82 | 1169.97 |
| 20791.2 | 876.728 | 419.943 | 426.599 | 1159.98 | 1171.07 |
| 20814.2 | 877.006 | 429.075 | 426.956 | 1143.64 | 1172.17 |
| 20837.2 | 877.283 | 419.297 | 427.313 | 1131.23 | 1173.27 |
| 20860.2 | 877.56 | 418.971 | 427.67 | 1167.12 | 1174.37 |
| 20883.2 | 877.838 | 414.436 | 428.026 | 1115.11 | 1175.47 |
| 20906.2 | 878.115 | 444.305 | 428.383 | 1166.62 | 1176.57 |
| 20929.2 | 878.392 | 422.376 | 428.739 | 1153.54 | 1177.67 |
| 20952.2 | 878.669 | 442.822 | 429.096 | 1190.2 | 1178.77 |
| 20975.3 | 878.947 | 420.804 | 429.452 | 1160.34 | 1179.86 |
| 20998.3 | 879.224 | 437.976 | 429.808 | 1155.22 | 1180.96 |
| 21021.3 | 879.501 | 431.325 | 430.164 | 1163.4 | 1182.06 |
| 21044.3 | 879.778 | 406.154 | 430.519 | 1175.37 | 1183.16 |
| 21067.3 | 880.055 | 443.133 | 430.875 | 1159.61 | 1184.26 |
| 21090.3 | 880.332 | 424.549 | 431.23 | 1144.13 | 1185.35 |
| 21113.3 | 880.609 | 423.375 | 431.585 | 1186.88 | 1186.45 |
| 21136.3 | 880.886 | 438.868 | 431.94 | 1144.66 | 1187.55 |
| 21159.4 | 881.163 | 430.286 | 432.295 | 1217.94 | 1188.64 |
| 21182.4 | 881.441 | 440.973 | 432.65 | 1165.1 | 1189.74 |
| 21205.4 | 881.717 | 435.293 | 433.005 | 1190.32 | 1190.83 |
| 21228.4 | 881.995 | 430.791 | 433.359 | 1184.3 | 1191.93 |
| 21251.4 | 882.271 | 430.4 | 433.713 | 1170.53 | 1193.03 |
| 21274.4 | 882.548 | 464.225 | 434.067 | 1190.42 | 1194.12 |
| 21297.4 | 882.825 | 423.744 | 434.421 | 1185.4 | 1195.22 |
| 21320.5 | 883.102 | 444.575 | 434.775 | 1194.47 | 1196.31 |
| 21343.5 | 883.379 | 441.467 | 435.129 | 1160.47 | 1197.4 |
| 21366.5 | 883.656 | 441.931 | 435.482 | 1208.5 | 1198.5 |
| 21389.5 | 883.933 | 434.995 | 435.836 | 1167.51 | 1199.59 |
| 21412.5 | 884.21 | 426.432 | 436.189 | 1183.74 | 1200.69 |
| 21435.5 | 884.487 | 450.372 | 436.542 | 1204.67 | 1201.78 |
| 21458.5 | 884.764 | 413.765 | 436.895 | 1166.69 | 1202.87 |
| 21481.5 | 885.04 | 456.032 | 437.247 | 1193.14 | 1203.96 |
| 21504.5 | 885.317 | 452.84 | 437.6 | 1190.7 | 1205.06 |
| 21527.5 | 885.594 | 447.434 | 437.952 | 1147.89 | 1206.15 |
| 21550.5 | 885.871 | 448.968 | 438.305 | 1141.36 | 1207.24 |
| 21573.5 | 886.148 | 439.739 | 438.657 | 1178.97 | 1208.33 |
| 21596.5 | 886.424 | 427.89 | 439.009 | 1212.89 | 1209.42 |
| 21619.5 | 886.701 | 429.81 | 439.36 | 1173.77 | 1210.52 |
| 21642.5 | 886.978 | 440.15 | 439.712 | 1207.87 | 1211.61 |
| 21665.6 | 887.254 | 452.461 | 440.063 | 1210.88 | 1212.7 |
| 21688.6 | 887.531 | 433.256 | 440.414 | 1163.81 | 1213.79 |
| 21711.6 | 887.808 | 473.119 | 440.765 | 1198.36 | 1214.88 |
| 21734.6 | 888.084 | 446.92 | 441.116 | 1177.67 | 1215.97 |
| 21757.6 | 888.361 | 449.753 | 441.467 | 1184.57 | 1217.06 |
| 21780.5 | 888.638 | 422.078 | 441.818 | 1218.17 | 1218.14 |
| 21803.5 | 888.914 | 458.755 | 442.168 | 1203.37 | 1219.23 |
| 21826.5 | 889.191 | 430.912 | 442.519 | 1191.17 | 1220.32 |
| 21849.5 | 889.467 | 442.954 | 442.869 | 1224.94 | 1221.41 |
| 21872.5 | 889.744 | 473.33 | 443.219 | 1208.97 | 1222.5 |
| 21895.5 | 890.02 | 460.705 | 443.568 | 1202.14 | 1223.59 |
| 21918.5 | 890.297 | 436.742 | 443.918 | 1204.29 | 1224.67 |
| 21941.5 | 890.573 | 468.976 | 444.267 | 1217.56 | 1225.76 |
| 21964.5 | 890.85 | 436.469 | 444.617 | 1210.07 | 1226.85 |
| 21987.5 | 891.126 | 459.566 | 444.966 | 1230.71 | 1227.93 |
| 22010.5 | 891.402 | 457.088 | 445.315 | 1209.39 | 1229.02 |
| 22033.5 | 891.679 | 442.784 | 445.664 | 1224.53 | 1230.1 |
| 22056.5 | 891.955 | 454.603 | 446.012 | 1200.25 | 1231.19 |
| 22079.4 | 892.232 | 440.303 | 446.361 | 1200.31 | 1232.28 |
| 22102.4 | 892.508 | 437.404 | 446.709 | 1242.53 | 1233.36 |
| 22125.4 | 892.784 | 464.774 | 447.057 | 1185.07 | 1234.44 |
| 22148.4 | 893.061 | 441.569 | 447.405 | 1243.88 | 1235.53 |
| 22171.4 | 893.337 | 453.704 | 447.753 | 1206.36 | 1236.61 |
| 22194.3 | 893.613 | 440.548 | 448.1 | 1190.81 | 1237.7 |
| 22217.3 | 893.889 | 465.108 | 448.448 | 1241.98 | 1238.78 |
| 22240.3 | 894.166 | 461.843 | 448.795 | 1239.56 | 1239.86 |
| 22263.3 | 894.442 | 440.602 | 449.142 | 1191.52 | 1240.94 |
| 22286.2 | 894.718 | 447.709 | 449.489 | 1197.9 | 1242.03 |
| 22309.2 | 894.994 | 454.623 | 449.836 | 1212.36 | 1243.11 |
| 22332.2 | 895.27 | 459.677 | 450.183 | 1230.47 | 1244.19 |
| 22355.1 | 895.547 | 454.327 | 450.529 | 1241.59 | 1245.27 |
| 22378.1 | 895.823 | 448.042 | 450.875 | 1214.8 | 1246.35 |
| 22401.1 | 896.099 | 472.048 | 451.221 | 1243.59 | 1247.43 |
| 22424 | 896.375 | 454.385 | 451.567 | 1260.99 | 1248.52 |
| 22447 | 896.651 | 441.482 | 451.913 | 1241.07 | 1249.6 |
| 22470 | 896.927 | 459.437 | 452.259 | 1234.91 | 1250.68 |
| 22492.9 | 897.203 | 445.957 | 452.604 | 1218.63 | 1251.76 |
| 22515.9 | 897.479 | 452.488 | 452.949 | 1212.82 | 1252.84 |
| 22538.8 | 897.755 | 451.291 | 453.294 | 1237.7 | 1253.91 |
| 22561.8 | 898.031 | 459.076 | 453.639 | 1265.73 | 1254.99 |
| 22584.7 | 898.307 | 453.548 | 453.984 | 1206.74 | 1256.07 |
| 22607.7 | 898.583 | 471.853 | 454.328 | 1198.12 | 1257.15 |
| 22630.6 | 898.859 | 465.474 | 454.673 | 1267.68 | 1258.23 |
| 22653.6 | 899.135 | 464 | 455.017 | 1265.87 | 1259.3 |
| 22676.5 | 899.411 | 445.565 | 455.361 | 1221.24 | 1260.38 |
| 22699.5 | 899.687 | 461.429 | 455.705 | 1194.76 | 1261.46 |
| 22722.4 | 899.963 | 463.578 | 456.048 | 1247.54 | 1262.53 |
| 22745.3 | 900.239 | 463.213 | 456.392 | 1218.22 | 1263.61 |
| 22768.3 | 900.514 | 453.81 | 456.735 | 1225.85 | 1264.68 |
| 22791.2 | 900.79 | 458.993 | 457.079 | 1264.05 | 1265.76 |
| 22814.1 | 901.066 | 456.927 | 457.421 | 1224.71 | 1266.84 |
| 22837.1 | 901.342 | 452.347 | 457.764 | 1257.79 | 1267.91 |
| 22860 | 901.617 | 473.229 | 458.107 | 1237.44 | 1268.98 |
| 22882.9 | 901.893 | 454.23 | 458.449 | 1235.99 | 1270.06 |
| 22905.8 | 902.169 | 445.169 | 458.791 | 1259.98 | 1271.13 |
| 22928.8 | 902.445 | 464.707 | 459.134 | 1249.6 | 1272.21 |
| 22951.7 | 902.72 | 475.799 | 459.476 | 1273.34 | 1273.28 |
| 22974.6 | 902.996 | 454.108 | 459.817 | 1264.3 | 1274.35 |
| 22997.5 | 903.272 | 473.95 | 460.159 | 1236.53 | 1275.42 |
| 23020.4 | 903.547 | 480.034 | 460.5 | 1250.13 | 1276.49 |
| 23043.3 | 903.823 | 469.024 | 460.841 | 1286.35 | 1277.57 |
| 23066.3 | 904.098 | 467.283 | 461.182 | 1262.6 | 1278.64 |
| 23089.2 | 904.374 | 490.365 | 461.523 | 1273.09 | 1279.71 |
| 23112.1 | 904.65 | 461.123 | 461.864 | 1242.69 | 1280.78 |
| 23135 | 904.925 | 474.227 | 462.204 | 1242.95 | 1281.85 |
| 23157.9 | 905.201 | 467.805 | 462.545 | 1244.26 | 1282.92 |
| 23180.8 | 905.476 | 444.613 | 462.885 | 1268.01 | 1283.99 |
| 23203.7 | 905.752 | 476.002 | 463.225 | 1248.54 | 1285.06 |
| 23226.6 | 906.027 | 462.98 | 463.564 | 1285.73 | 1286.13 |
| 23249.5 | 906.303 | 465.577 | 463.904 | 1270.04 | 1287.2 |
| 23272.3 | 906.578 | 473.127 | 464.243 | 1251.01 | 1288.26 |
| 23295.2 | 906.854 | 478.13 | 464.583 | 1287.54 | 1289.33 |
| 23318.1 | 907.129 | 477.15 | 464.922 | 1270.24 | 1290.4 |
| 23341 | 907.404 | 446.008 | 465.26 | 1236.91 | 1291.47 |
| 23363.9 | 907.68 | 463.864 | 465.599 | 1241.41 | 1292.53 |
| 23386.7 | 907.955 | 472.629 | 465.938 | 1267.84 | 1293.6 |
| 23409.6 | 908.231 | 494.892 | 466.276 | 1233.79 | 1294.67 |
| 23432.5 | 908.506 | 473.265 | 466.614 | 1290.31 | 1295.73 |
| 23455.3 | 908.781 | 478.452 | 466.952 | 1268.31 | 1296.8 |
| 23478.2 | 909.056 | 448.415 | 467.29 | 1236.72 | 1297.86 |
| 23501.1 | 909.332 | 446.466 | 467.627 | 1248.17 | 1298.93 |
| 23523.9 | 909.607 | 467.134 | 467.965 | 1247.06 | 1299.99 |
| 23546.8 | 909.882 | 463.437 | 468.302 | 1227.71 | 1301.05 |
| 23569.7 | 910.157 | 459.824 | 468.639 | 1261.86 | 1302.12 |
| 23592.5 | 910.433 | 479.607 | 468.976 | 1272.53 | 1303.18 |
| 23615.3 | 910.708 | 482.599 | 469.312 | 1231.42 | 1304.24 |
| 23638.2 | 910.983 | 464.629 | 469.649 | 1269.58 | 1305.31 |
| 23661 | 911.258 | 469.423 | 469.985 | 1256.29 | 1306.37 |
| 23683.9 | 911.533 | 491.766 | 470.321 | 1263.71 | 1307.43 |
| 23706.7 | 911.808 | 481.909 | 470.657 | 1258.3 | 1308.49 |
| 23729.5 | 912.083 | 470.975 | 470.993 | 1294.03 | 1309.55 |
| 23752.4 | 912.358 | 473.697 | 471.328 | 1294.43 | 1310.61 |
| 23775.2 | 912.633 | 470.477 | 471.664 | 1266.05 | 1311.67 |
| 23798 | 912.909 | 491.389 | 471.999 | 1265.38 | 1312.73 |
| 23820.8 | 913.183 | 471.875 | 472.334 | 1316.18 | 1313.79 |
| 23843.7 | 913.458 | 474.762 | 472.669 | 1259.55 | 1314.85 |
| 23866.5 | 913.734 | 474.434 | 473.003 | 1310.61 | 1315.91 |
| 23889.3 | 914.008 | 486.287 | 473.338 | 1279.83 | 1316.97 |
| 23912.1 | 914.283 | 473.788 | 473.672 | 1305.71 | 1318.03 |
| 23934.9 | 914.558 | 478.244 | 474.006 | 1282.36 | 1319.08 |
| 23957.7 | 914.833 | 456.63 | 474.34 | 1291.83 | 1320.14 |
| 23980.5 | 915.108 | 484.181 | 474.673 | 1340.6 | 1321.2 |
| 24003.3 | 915.383 | 471.508 | 475.007 | 1279.63 | 1322.25 |
| 24026.1 | 915.658 | 475.381 | 475.34 | 1345.91 | 1323.31 |
| 24048.9 | 915.933 | 493.42 | 475.673 | 1306.92 | 1324.37 |
| 24071.7 | 916.207 | 450.607 | 476.006 | 1283.48 | 1325.42 |
| 24094.5 | 916.482 | 472.074 | 476.339 | 1252.29 | 1326.48 |
| 24117.2 | 916.757 | 487.568 | 476.671 | 1291.53 | 1327.53 |
| 24140 | 917.032 | 490.222 | 477.004 | 1315.56 | 1328.58 |
| 24162.8 | 917.307 | 474.226 | 477.336 | 1292.52 | 1329.64 |
| 24185.6 | 917.581 | 489.586 | 477.668 | 1298.21 | 1330.69 |
| 24208.3 | 917.856 | 473.081 | 478 | 1286.95 | 1331.74 |
| 24231.1 | 918.131 | 465.98 | 478.331 | 1313.66 | 1332.8 |
| 24253.9 | 918.405 | 479.923 | 478.663 | 1329.17 | 1333.85 |
| 24276.6 | 918.68 | 473.498 | 478.994 | 1311.6 | 1334.9 |
| 24299.4 | 918.955 | 491.722 | 479.325 | 1289.24 | 1335.95 |
| 24322.1 | 919.229 | 475.402 | 479.656 | 1277.61 | 1337 |
| 24344.9 | 919.504 | 493.414 | 479.986 | 1297.92 | 1338.05 |
| 24367.6 | 919.779 | 519.368 | 480.317 | 1304.85 | 1339.1 |
| 24390.3 | 920.053 | 484.372 | 480.647 | 1305.43 | 1340.15 |
| 24413.1 | 920.328 | 463.368 | 480.977 | 1284.67 | 1341.2 |
| 24435.8 | 920.602 | 493.384 | 481.307 | 1276.82 | 1342.25 |
| 24458.5 | 920.877 | 481.267 | 481.636 | 1331.4 | 1343.3 |
| 24481.3 | 921.151 | 493.178 | 481.966 | 1334.35 | 1344.35 |
| 24504 | 921.426 | 468.462 | 482.295 | 1307.88 | 1345.4 |
| 24526.7 | 921.7 | 477.546 | 482.624 | 1299.67 | 1346.44 |
| 24549.4 | 921.975 | 476.596 | 482.953 | 1312.04 | 1347.49 |
| 24572.1 | 922.249 | 503.363 | 483.282 | 1325.42 | 1348.54 |
| 24594.8 | 922.523 | 490.314 | 483.61 | 1299.42 | 1349.59 |
| 24617.5 | 922.798 | 506.745 | 483.938 | 1314.62 | 1350.63 |
| 24640.2 | 923.072 | 497.263 | 484.266 | 1294.9 | 1351.68 |
| 24662.9 | 923.346 | 502.919 | 484.594 | 1327.26 | 1352.72 |
| 24685.6 | 923.621 | 505.858 | 484.922 | 1325.28 | 1353.77 |
| 24708.3 | 923.895 | 497.574 | 485.25 | 1327.64 | 1354.81 |
| 24731 | 924.169 | 483.933 | 485.577 | 1361.85 | 1355.85 |
| 24753.7 | 924.444 | 477.108 | 485.904 | 1323.69 | 1356.9 |
| 24776.4 | 924.718 | 463.362 | 486.231 | 1358.03 | 1357.94 |
| 24799 | 924.992 | 485.124 | 486.558 | 1330.44 | 1358.98 |
| 24821.7 | 925.266 | 478.732 | 486.884 | 1349.58 | 1360.03 |
| 24844.4 | 925.54 | 485.684 | 487.21 | 1359.39 | 1361.07 |
| 24867 | 925.815 | 493.287 | 487.536 | 1331.41 | 1362.11 |
| 24889.7 | 926.089 | 473.411 | 487.863 | 1295.7 | 1363.15 |
| 24912.3 | 926.363 | 492.651 | 488.188 | 1329.55 | 1364.19 |
| 24935 | 926.637 | 488.221 | 488.514 | 1325.95 | 1365.23 |
| 24957.6 | 926.911 | 482.434 | 488.839 | 1317.87 | 1366.27 |
| 24980.3 | 927.185 | 496.374 | 489.164 | 1341.34 | 1367.31 |
| 25002.9 | 927.459 | 477.58 | 489.489 | 1341.83 | 1368.35 |
| 25025.5 | 927.734 | 486.144 | 489.814 | 1321.88 | 1369.39 |
| 25048.2 | 928.008 | 467.486 | 490.139 | 1324.04 | 1370.42 |
| 25070.8 | 928.282 | 493.896 | 490.463 | 1340.44 | 1371.46 |
| 25093.4 | 928.556 | 501.838 | 490.787 | 1331.54 | 1372.5 |
| 25116 | 928.83 | 508.48 | 491.111 | 1327.59 | 1373.54 |
| 25138.6 | 929.104 | 527.731 | 491.435 | 1364.06 | 1374.57 |
| 25161.2 | 929.377 | 499.788 | 491.758 | 1335.13 | 1375.61 |
| 25183.8 | 929.651 | 500.37 | 492.082 | 1285.14 | 1376.64 |
| 25206.4 | 929.925 | 502.285 | 492.405 | 1373.92 | 1377.68 |
| 25229 | 930.199 | 502.431 | 492.728 | 1378.78 | 1378.71 |
| 25251.6 | 930.473 | 505.836 | 493.05 | 1386.33 | 1379.75 |
| 25274.2 | 930.747 | 506.392 | 493.373 | 1376.16 | 1380.78 |
| 25296.8 | 931.021 | 514.454 | 493.695 | 1380.2 | 1381.82 |
| 25319.4 | 931.294 | 504.344 | 494.017 | 1371.37 | 1382.85 |
| 25341.9 | 931.568 | 506.123 | 494.339 | 1366.14 | 1383.88 |
| 25364.5 | 931.842 | 501.32 | 494.661 | 1372.09 | 1384.91 |
| 25387.1 | 932.116 | 513.17 | 494.983 | 1353.74 | 1385.94 |
| 25409.6 | 932.39 | 510.686 | 495.304 | 1368.67 | 1386.98 |
| 25432.2 | 932.663 | 506.655 | 495.625 | 1382.17 | 1388.01 |
| 25454.7 | 932.937 | 506.345 | 495.946 | 1365.75 | 1389.04 |
| 25477.3 | 933.211 | 512.718 | 496.267 | 1394.13 | 1390.07 |
| 25499.8 | 933.484 | 506.823 | 496.587 | 1355.03 | 1391.1 |
| 25522.3 | 933.758 | 507.423 | 496.908 | 1350.06 | 1392.13 |
| 25544.9 | 934.032 | 513.66 | 497.228 | 1363.53 | 1393.15 |
| 25567.4 | 934.305 | 503.869 | 497.548 | 1331.86 | 1394.18 |
| 25589.9 | 934.579 | 530.222 | 497.867 | 1364.86 | 1395.21 |
| 25612.4 | 934.852 | 499.898 | 498.187 | 1361.6 | 1396.24 |
| 25634.9 | 935.126 | 510.959 | 498.506 | 1413.4 | 1397.27 |
| 25657.5 | 935.4 | 509.142 | 498.825 | 1354.06 | 1398.29 |
| 25680 | 935.673 | 511.348 | 499.144 | 1372.28 | 1399.32 |
| 25702.5 | 935.947 | 495.964 | 499.463 | 1369.51 | 1400.34 |
| 25724.9 | 936.22 | 486.041 | 499.781 | 1359.25 | 1401.37 |
| 25747.4 | 936.494 | 499.102 | 500.1 | 1354.53 | 1402.39 |
| 25769.9 | 936.767 | 509.605 | 500.418 | 1419.4 | 1403.42 |
| 25792.4 | 937.04 | 502.996 | 500.736 | 1411.85 | 1404.44 |
| 25814.9 | 937.314 | 489.674 | 501.053 | 1342.82 | 1405.46 |
| 25837.3 | 937.587 | 478.274 | 501.371 | 1378.35 | 1406.49 |
| 25859.8 | 937.861 | 524.181 | 501.688 | 1372.08 | 1407.51 |
| 25882.2 | 938.134 | 503.203 | 502.005 | 1398.28 | 1408.53 |
| 25904.7 | 938.407 | 523.488 | 502.322 | 1402.71 | 1409.55 |
| 25927.2 | 938.681 | 507.359 | 502.639 | 1381.63 | 1410.58 |
| 25949.6 | 938.954 | 522.272 | 502.955 | 1372.01 | 1411.6 |
| 25972 | 939.227 | 521.951 | 503.271 | 1374.66 | 1412.62 |
| 25994.5 | 939.5 | 502.268 | 503.587 | 1444.06 | 1413.64 |
| 26016.9 | 939.774 | 512.184 | 503.903 | 1411.44 | 1414.65 |
| 26039.3 | 940.047 | 498.882 | 504.219 | 1399.68 | 1415.67 |
| 26061.8 | 940.32 | 507.538 | 504.534 | 1350.17 | 1416.69 |
| 26084.2 | 940.593 | 492.152 | 504.849 | 1362.49 | 1417.71 |
| 26106.6 | 940.867 | 523.937 | 505.164 | 1396.72 | 1418.73 |
| 26129 | 941.14 | 501.525 | 505.479 | 1397.31 | 1419.74 |
| 26151.4 | 941.413 | 538.025 | 505.794 | 1372 | 1420.76 |
| 26173.8 | 941.686 | 547.132 | 506.108 | 1353.82 | 1421.78 |
| 26196.2 | 941.959 | 496.32 | 506.422 | 1427.35 | 1422.79 |
| 26218.5 | 942.232 | 519.962 | 506.736 | 1382.36 | 1423.81 |
| 26240.9 | 942.505 | 511.598 | 507.05 | 1376.57 | 1424.82 |
| 26263.3 | 942.778 | 500.507 | 507.363 | 1412.45 | 1425.84 |
| 26285.7 | 943.051 | 545.393 | 507.677 | 1410.57 | 1426.85 |
| 26308 | 943.324 | 501.632 | 507.99 | 1410.36 | 1427.87 |
| 26330.4 | 943.597 | 530.99 | 508.303 | 1394.34 | 1428.88 |
| 26352.7 | 943.87 | 551.603 | 508.615 | 1355.14 | 1429.89 |
| 26375.1 | 944.143 | 506.322 | 508.928 | 1417.35 | 1430.9 |
| 26397.4 | 944.416 | 501.637 | 509.24 | 1381.8 | 1431.92 |
| 26419.7 | 944.689 | 520.329 | 509.552 | 1422.67 | 1432.93 |
| 26442.1 | 944.962 | 535.593 | 509.864 | 1399.69 | 1433.94 |
| 26464.4 | 945.235 | 533.916 | 510.176 | 1396.91 | 1434.95 |
| 26486.7 | 945.508 | 517.523 | 510.487 | 1411.72 | 1435.96 |
| 26509 | 945.78 | 505.228 | 510.798 | 1392.81 | 1436.97 |
| 26531.3 | 946.053 | 533.369 | 511.11 | 1404.7 | 1437.98 |
| 26553.6 | 946.326 | 499.224 | 511.42 | 1439.28 | 1438.98 |
| 26575.9 | 946.599 | 518.032 | 511.731 | 1441.83 | 1439.99 |
| 26598.2 | 946.872 | 537.631 | 512.041 | 1375.59 | 1441 |
| 26620.5 | 947.144 | 492.582 | 512.351 | 1426.07 | 1442.01 |
| 26642.8 | 947.417 | 520.814 | 512.662 | 1446.42 | 1443.01 |
| 26665.1 | 947.69 | 525.232 | 512.971 | 1364.7 | 1444.02 |
| 26687.3 | 947.963 | 536.23 | 513.281 | 1414.85 | 1445.03 |
| 26709.6 | 948.235 | 516.213 | 513.59 | 1369.92 | 1446.03 |
| 26731.9 | 948.508 | 515.957 | 513.899 | 1433.33 | 1447.04 |
| 26754.1 | 948.78 | 508.36 | 514.208 | 1418.03 | 1448.04 |
| 26776.4 | 949.053 | 511.916 | 514.517 | 1385.05 | 1449.04 |
| 26798.6 | 949.326 | 529.869 | 514.825 | 1449.67 | 1450.05 |
| 26820.8 | 949.598 | 493.607 | 515.134 | 1461.15 | 1451.05 |
| 26843.1 | 949.871 | 539.948 | 515.442 | 1422.02 | 1452.05 |
| 26865.3 | 950.143 | 525.978 | 515.749 | 1418.97 | 1453.05 |
| 26887.5 | 950.416 | 502.815 | 516.057 | 1421.17 | 1454.06 |
| 26909.7 | 950.688 | 513.331 | 516.365 | 1425.53 | 1455.06 |
| 26931.9 | 950.961 | 506.502 | 516.672 | 1452.72 | 1456.06 |
| 26954.1 | 951.233 | 470.437 | 516.979 | 1412.22 | 1457.06 |
| 26976.3 | 951.506 | 525.667 | 517.285 | 1387.54 | 1458.06 |
| 26998.5 | 951.778 | 507.092 | 517.592 | 1442.69 | 1459.06 |
| 27020.7 | 952.051 | 515.103 | 517.898 | 1429.39 | 1460.05 |
| 27042.9 | 952.323 | 561.596 | 518.205 | 1426.98 | 1461.05 |
| 27065 | 952.595 | 535.831 | 518.51 | 1430.39 | 1462.05 |
| 27087.2 | 952.868 | 537.303 | 518.816 | 1418.25 | 1463.05 |
| 27109.4 | 953.14 | 529.995 | 519.122 | 1378.24 | 1464.04 |
| 27131.5 | 953.412 | 516.911 | 519.427 | 1442.64 | 1465.04 |
| 27153.7 | 953.685 | 521.848 | 519.732 | 1377.57 | 1466.03 |
| 27175.8 | 953.957 | 510.939 | 520.037 | 1466.13 | 1467.03 |
| 27198 | 954.229 | 528.624 | 520.341 | 1427.26 | 1468.02 |
| 27220.1 | 954.501 | 518.359 | 520.646 | 1404.93 | 1469.02 |
| 27242.2 | 954.774 | 518.028 | 520.95 | 1443.81 | 1470.01 |
| 27264.3 | 955.046 | 548.807 | 521.254 | 1498.93 | 1471.01 |
| 27286.4 | 955.318 | 517.505 | 521.558 | 1422.51 | 1472 |
| 27308.5 | 955.59 | 524.749 | 521.861 | 1410.15 | 1472.99 |
| 27330.6 | 955.862 | 553.481 | 522.165 | 1444.72 | 1473.98 |
| 27352.7 | 956.135 | 522.44 | 522.468 | 1444.49 | 1474.97 |
| 27374.8 | 956.407 | 540.559 | 522.771 | 1439.99 | 1475.96 |
| 27396.9 | 956.679 | 543.452 | 523.073 | 1436.03 | 1476.95 |
| 27419 | 956.951 | 536.396 | 523.376 | 1450.27 | 1477.94 |
| 27441 | 957.223 | 524.087 | 523.678 | 1428.54 | 1478.93 |
| 27463.1 | 957.495 | 543.268 | 523.98 | 1422.97 | 1479.92 |
| 27485.2 | 957.767 | 505.799 | 524.282 | 1465.07 | 1480.91 |
| 27507.2 | 958.039 | 544.303 | 524.584 | 1463.2 | 1481.9 |
| 27529.2 | 958.311 | 560.757 | 524.885 | 1471.4 | 1482.89 |
| 27551.3 | 958.583 | 545.594 | 525.186 | 1426.99 | 1483.87 |
| 27573.3 | 958.855 | 550.835 | 525.487 | 1444.7 | 1484.86 |
| 27595.3 | 959.127 | 508.128 | 525.788 | 1465.07 | 1485.85 |
| 27617.4 | 959.399 | 540.856 | 526.089 | 1444.32 | 1486.83 |
| 27639.4 | 959.67 | 533.938 | 526.389 | 1461.5 | 1487.82 |
| 27661.4 | 959.942 | 519.335 | 526.689 | 1495.18 | 1488.8 |
| 27683.4 | 960.214 | 554.67 | 526.989 | 1471.82 | 1489.78 |
| 27705.4 | 960.486 | 537.258 | 527.288 | 1440.18 | 1490.77 |
| 27727.3 | 960.758 | 554.639 | 527.588 | 1447.53 | 1491.75 |
| 27749.3 | 961.03 | 566.932 | 527.887 | 1480.85 | 1492.73 |
| 27771.3 | 961.301 | 536.156 | 528.186 | 1421.65 | 1493.71 |
| 27793.3 | 961.573 | 517.431 | 528.485 | 1520.55 | 1494.69 |
| 27815.2 | 961.845 | 535.582 | 528.784 | 1445.36 | 1495.68 |
| 27837.2 | 962.117 | 571.869 | 529.082 | 1454.38 | 1496.66 |
| 27859.1 | 962.388 | 559.077 | 529.38 | 1468.2 | 1497.64 |
| 27881.1 | 962.66 | 530.418 | 529.678 | 1446.53 | 1498.62 |
| 27903 | 962.932 | 565.619 | 529.976 | 1526.1 | 1499.6 |
| 27924.9 | 963.203 | 553.559 | 530.273 | 1486.49 | 1500.57 |
| 27946.9 | 963.475 | 542.51 | 530.57 | 1478.32 | 1501.55 |
| 27968.8 | 963.746 | 513.116 | 530.867 | 1463.37 | 1502.53 |
| 27990.7 | 964.018 | 572.763 | 531.164 | 1475.77 | 1503.51 |
| 28012.6 | 964.29 | 535.931 | 531.461 | 1481.31 | 1504.48 |
| 28034.5 | 964.561 | 552.796 | 531.757 | 1475.56 | 1505.46 |
| 28056.4 | 964.833 | 525.139 | 532.053 | 1525.55 | 1506.43 |
| 28078.2 | 965.104 | 529.127 | 532.349 | 1503.13 | 1507.41 |
| 28100.1 | 965.376 | 542.965 | 532.645 | 1517.11 | 1508.38 |
| 28122 | 965.647 | 570.31 | 532.94 | 1490.53 | 1509.36 |
| 28143.8 | 965.919 | 588.547 | 533.236 | 1531.96 | 1510.33 |
| 28165.7 | 966.19 | 546.238 | 533.531 | 1501.72 | 1511.3 |
| 28187.5 | 966.461 | 556.218 | 533.826 | 1534.63 | 1512.28 |
| 28209.4 | 966.733 | 531.892 | 534.12 | 1492.68 | 1513.25 |
| 28231.2 | 967.004 | 588.906 | 534.415 | 1483.49 | 1514.22 |
| 28253 | 967.275 | 545.106 | 534.709 | 1448.8 | 1515.19 |
| 28274.9 | 967.547 | 543.819 | 535.003 | 1514.67 | 1516.16 |
| 28296.7 | 967.818 | 546.995 | 535.297 | 1483.09 | 1517.13 |
| 28318.5 | 968.09 | 555.326 | 535.59 | 1474.78 | 1518.1 |
| 28340.3 | 968.361 | 546.637 | 535.883 | 1521.22 | 1519.07 |
| 28362.1 | 968.632 | 534.019 | 536.176 | 1486.7 | 1520.04 |
| 28383.9 | 968.903 | 567.998 | 536.469 | 1554.44 | 1521 |
| 28405.7 | 969.174 | 551.766 | 536.762 | 1472.94 | 1521.97 |
| 28427.4 | 969.446 | 555.24 | 537.054 | 1539.44 | 1522.94 |
| 28449.2 | 969.717 | 549.221 | 537.346 | 1498.57 | 1523.9 |
| 28471 | 969.988 | 532.583 | 537.638 | 1519.41 | 1524.87 |
| 28492.7 | 970.259 | 566.451 | 537.93 | 1550.13 | 1525.83 |
| 28514.4 | 970.53 | 565.068 | 538.222 | 1514.27 | 1526.8 |
| 28536.2 | 970.801 | 564.247 | 538.513 | 1484.14 | 1527.76 |
| 28557.9 | 971.072 | 562.324 | 538.804 | 1512.42 | 1528.73 |
| 28579.6 | 971.343 | 525.807 | 539.095 | 1504.77 | 1529.69 |
| 28601.3 | 971.615 | 557.668 | 539.385 | 1474.44 | 1530.65 |
| 28623 | 971.886 | 553.01 | 539.676 | 1496.95 | 1531.61 |
| 28644.8 | 972.157 | 573.091 | 539.966 | 1538.4 | 1532.58 |
| 28666.5 | 972.428 | 572.766 | 540.256 | 1481.98 | 1533.54 |
| 28688.1 | 972.698 | 566.59 | 540.546 | 1529.31 | 1534.5 |
| 28709.8 | 972.97 | 577.811 | 540.835 | 1484.94 | 1535.46 |
| 28731.5 | 973.24 | 540.064 | 541.125 | 1547.42 | 1536.42 |
| 28753.2 | 973.511 | 552.678 | 541.414 | 1516.67 | 1537.37 |
| 28774.8 | 973.782 | 543.341 | 541.702 | 1515.39 | 1538.33 |
| 28796.5 | 974.053 | 537.406 | 541.991 | 1500.74 | 1539.29 |
| 28818.1 | 974.324 | 587.664 | 542.279 | 1536.61 | 1540.25 |
| 28839.8 | 974.595 | 549.105 | 542.568 | 1500.71 | 1541.21 |
| 28861.4 | 974.866 | 582.038 | 542.856 | 1509.28 | 1542.16 |
| 28883 | 975.136 | 568.685 | 543.143 | 1560.6 | 1543.12 |
| 28904.6 | 975.407 | 542.638 | 543.431 | 1512.51 | 1544.07 |
| 28926.2 | 975.678 | 581.338 | 543.718 | 1505.73 | 1545.03 |
| 28947.8 | 975.949 | 572.38 | 544.005 | 1517.33 | 1545.98 |
| 28969.4 | 976.219 | 554.482 | 544.292 | 1482.2 | 1546.94 |
| 28991 | 976.49 | 556.655 | 544.578 | 1538.66 | 1547.89 |
| 29012.6 | 976.761 | 577.763 | 544.865 | 1414.27 | 1548.84 |
| 29034.2 | 977.031 | 544.83 | 545.151 | 1487.14 | 1549.79 |
| 29055.7 | 977.302 | 546.073 | 545.437 | 1471.3 | 1550.74 |
| 29077.3 | 977.573 | 548.777 | 545.723 | 1539.79 | 1551.7 |
| 29098.8 | 977.843 | 562.645 | 546.008 | 1552.11 | 1552.65 |
| 29120.4 | 978.114 | 546.828 | 546.293 | 1549.55 | 1553.6 |
| 29141.9 | 978.385 | 544.731 | 546.578 | 1519.37 | 1554.55 |
| 29163.4 | 978.655 | 548.041 | 546.863 | 1546.45 | 1555.49 |
| 29184.9 | 978.926 | 538.266 | 547.148 | 1544.79 | 1556.44 |
| 29206.5 | 979.196 | 540.155 | 547.432 | 1507.15 | 1557.39 |
| 29228 | 979.466 | 552.259 | 547.716 | 1532.17 | 1558.34 |
| 29249.5 | 979.737 | 551.284 | 548 | 1531.23 | 1559.29 |
| 38120.1 | 1101.15 | 695.102 | 654.201 | 1789.74 | 1932.7 |
| 38232.9 | 1102.87 | 632.991 | 655.394 | 1863.71 | 1937.21 |
| 38345.2 | 1104.6 | 659.102 | 656.577 | 1795.73 | 1941.68 |
| 38457 | 1106.32 | 588.991 | 657.752 | 1818.32 | 1946.14 |
| 38568.5 | 1108.05 | 473.575 | 658.918 | 2082.07 | 1950.57 |
| 38679.4 | 1109.77 | 498.664 | 660.074 | 2131.72 | 1954.97 |
| 38790 | 1111.5 | 758.704 | 661.222 | 1801.12 | 1959.35 |
| 38900 | 1113.22 | 718.777 | 662.36 | 1898.97 | 1963.71 |
| 39009.7 | 1114.95 | 751.917 | 663.49 | 2074.54 | 1968.04 |
| 39118.8 | 1116.67 | 594.392 | 664.61 | 2052.01 | 1972.35 |
| 39227.6 | 1118.39 | 625.452 | 665.721 | 2140.16 | 1976.63 |
| 39335.8 | 1120.12 | 964.754 | 666.824 | 2111.41 | 1980.89 |
| 39443.6 | 1121.84 | 750.79 | 667.917 | 1983.38 | 1985.12 |
| 39551 | 1123.57 | 562.415 | 669.001 | 1894.91 | 1989.33 |
| 39657.9 | 1125.29 | 752.456 | 670.077 | 1906.23 | 1993.51 |
| 39764.3 | 1127.01 | 605.541 | 671.144 | 2150.55 | 1997.67 |
| 39870.3 | 1128.74 | 572.739 | 672.201 | 2091.89 | 2001.81 |
| 39975.8 | 1130.46 | 799.621 | 673.25 | 1915.23 | 2005.92 |
| 40080.8 | 1132.18 | 656.986 | 674.29 | 2063.03 | 2010 |
| 40185.4 | 1133.91 | 616.081 | 675.321 | 2010.43 | 2014.07 |
| 40289.6 | 1135.63 | 656.424 | 676.342 | 2101.12 | 2018.1 |
| 40393.2 | 1137.35 | 826.11 | 677.356 | 2125.96 | 2022.12 |
| 40496.4 | 1139.07 | 770.789 | 678.36 | 1863.27 | 2026.1 |
| 40599.1 | 1140.8 | 688.719 | 679.356 | 2010.87 | 2030.07 |
| 40701.4 | 1142.52 | 721.148 | 680.342 | 2007.54 | 2034.01 |
| 40803.2 | 1144.24 | 513.064 | 681.32 | 2098.78 | 2037.92 |
| 40904.5 | 1145.96 | 703.367 | 682.289 | 2122.66 | 2041.81 |
| 41005.4 | 1147.69 | 834.972 | 683.25 | 1992.3 | 2045.68 |
| 41105.8 | 1149.41 | 746.453 | 684.202 | 2037.26 | 2049.52 |
| 41205.7 | 1151.13 | 581.466 | 685.145 | 2015.38 | 2053.33 |
| 41305.1 | 1152.85 | 665.858 | 686.079 | 2079.64 | 2057.13 |
| 41404.1 | 1154.57 | 585.484 | 687.005 | 2185.22 | 2060.89 |
| 41502.6 | 1156.3 | 551.918 | 687.921 | 2150.35 | 2064.64 |
| 41600.6 | 1158.02 | 621.986 | 688.83 | 2035.64 | 2068.36 |
| 41698.1 | 1159.74 | 578.815 | 689.729 | 2040.46 | 2072.05 |
| 41795.2 | 1161.46 | 696.525 | 690.62 | 1850.7 | 2075.72 |
| 41891.8 | 1163.18 | 709.463 | 691.503 | 1918.67 | 2079.37 |
| 41987.9 | 1164.9 | 754.189 | 692.377 | 2168.39 | 2082.99 |
| 42083.6 | 1166.62 | 647.509 | 693.242 | 1940.12 | 2086.59 |
| 42178.8 | 1168.34 | 741.025 | 694.099 | 2119.4 | 2090.16 |
| 42273.4 | 1170.06 | 629.226 | 694.948 | 2087.95 | 2093.71 |
| 42367.7 | 1171.78 | 741.157 | 695.788 | 2010.46 | 2097.24 |
| 42461.4 | 1173.5 | 781.932 | 696.619 | 2127.11 | 2100.74 |
| 42554.7 | 1175.22 | 642.223 | 697.442 | 1947.22 | 2104.22 |
| 42647.4 | 1176.94 | 696.997 | 698.257 | 2036.21 | 2107.67 |
| 42739.8 | 1178.66 | 477.181 | 699.063 | 2023.57 | 2111.1 |
| 42831.6 | 1180.38 | 524.708 | 699.861 | 1827.46 | 2114.5 |
| 42922.9 | 1182.11 | 671.714 | 700.65 | 1985.67 | 2117.88 |
| 43013.8 | 1183.82 | 664.782 | 701.432 | 2151.56 | 2121.24 |
| 43104.2 | 1185.54 | 656.46 | 702.205 | 2112.07 | 2124.57 |
| 43194.1 | 1187.26 | 679.092 | 702.969 | 2258.81 | 2127.88 |
| 43283.5 | 1188.98 | 625.012 | 703.726 | 2254.91 | 2131.17 |
| 43372.5 | 1190.7 | 711.398 | 704.474 | 2164.31 | 2134.43 |
| 43460.9 | 1192.42 | 809.604 | 705.214 | 1958.3 | 2137.67 |
| 43548.9 | 1194.14 | 806.784 | 705.946 | 1935.08 | 2140.88 |
| 43636.4 | 1195.86 | 810.749 | 706.669 | 2022.99 | 2144.07 |
| 43723.5 | 1197.58 | 632.366 | 707.385 | 1871.14 | 2147.24 |
| 43810 | 1199.3 | 563.977 | 708.092 | 2084.89 | 2150.38 |
| 43896.1 | 1201.02 | 598.922 | 708.791 | 2253.69 | 2153.5 |
| 43981.7 | 1202.73 | 645.897 | 709.483 | 2169.33 | 2156.59 |
| 44066.8 | 1204.45 | 821.099 | 710.166 | 2006.01 | 2159.67 |
| 44151.4 | 1206.17 | 770.801 | 710.841 | 2115.62 | 2162.72 |
| 44235.6 | 1207.89 | 782.88 | 711.508 | 2282.89 | 2165.74 |
| 44319.2 | 1209.61 | 735.392 | 712.167 | 2047.72 | 2168.74 |
| 44402.4 | 1211.32 | 566.2 | 712.819 | 2029.72 | 2171.72 |
| 44485.1 | 1213.04 | 672.573 | 713.462 | 2055.27 | 2174.68 |
| 44567.4 | 1214.76 | 659.986 | 714.097 | 2060.07 | 2177.61 |
| 44649.1 | 1216.48 | 821.397 | 714.725 | 2223.42 | 2180.52 |
| 44730.4 | 1218.19 | 855.953 | 715.345 | 2181.75 | 2183.4 |
| 44811.2 | 1219.91 | 776.049 | 715.957 | 2165.14 | 2186.26 |
| 44891.5 | 1221.63 | 726.397 | 716.561 | 2294.51 | 2189.1 |
| 44971.3 | 1223.35 | 597.414 | 717.157 | 2095.58 | 2191.92 |
| 45050.7 | 1225.06 | 755.875 | 717.746 | 2116.27 | 2194.71 |
| 45129.6 | 1226.78 | 698.521 | 718.327 | 2112.16 | 2197.48 |
| 45208 | 1228.5 | 563.501 | 718.9 | 2136.29 | 2200.23 |
| 45285.9 | 1230.21 | 812.615 | 719.466 | 2218.59 | 2202.95 |
| 45363.3 | 1231.93 | 816.181 | 720.024 | 2174.15 | 2205.66 |
| 45440.3 | 1233.64 | 712.41 | 720.575 | 2238.81 | 2208.33 |
| 45516.8 | 1235.36 | 753.863 | 721.117 | 2299.96 | 2210.99 |
| 45592.8 | 1237.08 | 708.874 | 721.653 | 2344.48 | 2213.62 |
| 45668.3 | 1238.79 | 739.021 | 722.181 | 2441.29 | 2216.23 |
| 45743.4 | 1240.51 | 648.917 | 722.701 | 2414.76 | 2218.82 |
| 45818 | 1242.22 | 653.979 | 723.214 | 2147.87 | 2221.39 |
| 45892.1 | 1243.94 | 833.165 | 723.719 | 2250.25 | 2223.93 |
| 45965.8 | 1245.66 | 713.953 | 724.217 | 2324 | 2226.45 |
| 46038.9 | 1247.37 | 686.796 | 724.708 | 2293.48 | 2228.95 |
| 46111.6 | 1249.09 | 713.019 | 725.191 | 2413.72 | 2231.43 |
| 46183.8 | 1250.8 | 629.63 | 725.667 | 2331.56 | 2233.88 |
| 46255.6 | 1252.52 | 827.04 | 726.135 | 2139.48 | 2236.31 |
| 46326.9 | 1254.23 | 872.908 | 726.597 | 2248.05 | 2238.72 |
| 46397.7 | 1255.95 | 740.062 | 727.051 | 2316.76 | 2241.11 |
| 46468 | 1257.66 | 827.163 | 727.498 | 2347.66 | 2243.48 |
| 46537.9 | 1259.38 | 808.744 | 727.937 | 2142.26 | 2245.82 |
| 46607.3 | 1261.09 | 767.668 | 728.37 | 2128.91 | 2248.14 |
| 46676.2 | 1262.81 | 652.676 | 728.795 | 2278.09 | 2250.44 |
| 46744.7 | 1264.52 | 641.861 | 729.214 | 2234.04 | 2252.72 |
| 46812.7 | 1266.23 | 703.172 | 729.625 | 2443.78 | 2254.98 |
| 46880.2 | 1267.95 | 643.783 | 730.029 | 2359.13 | 2257.21 |
| 46947.2 | 1269.66 | 614.393 | 730.426 | 2127.64 | 2259.43 |
| 47013.8 | 1271.37 | 717.187 | 730.816 | 2245.55 | 2261.62 |
| 47080 | 1273.09 | 754.108 | 731.199 | 2177.74 | 2263.79 |
| 47145.6 | 1274.8 | 719.732 | 731.576 | 2142.29 | 2265.94 |
| 47210.8 | 1276.51 | 718.132 | 731.945 | 2391.94 | 2268.07 |
| 47275.6 | 1278.23 | 680.023 | 732.307 | 2226.87 | 2270.17 |
| 47339.8 | 1279.94 | 695.511 | 732.663 | 2296.03 | 2272.26 |
| 47403.7 | 1281.65 | 785.252 | 733.012 | 2396.85 | 2274.32 |
| 47467 | 1283.37 | 867.223 | 733.354 | 2399.53 | 2276.36 |
| 47529.9 | 1285.08 | 786.217 | 733.689 | 2259.57 | 2278.39 |
| 47592.3 | 1286.79 | 710.623 | 734.017 | 2207.32 | 2280.39 |
| 47654.3 | 1288.5 | 710.48 | 734.339 | 2375.11 | 2282.37 |
| 47715.8 | 1290.22 | 698.853 | 734.654 | 2335.96 | 2284.33 |
| 47776.9 | 1291.93 | 580.634 | 734.963 | 2429.03 | 2286.26 |
| 47837.5 | 1293.64 | 689.576 | 735.264 | 2352.8 | 2288.18 |
| 47897.7 | 1295.35 | 707.982 | 735.56 | 2383.91 | 2290.08 |
| 47957.3 | 1297.07 | 749.785 | 735.848 | 2494.07 | 2291.95 |
| 48016.6 | 1298.78 | 802.115 | 736.13 | 2301.18 | 2293.81 |
| 48075.4 | 1300.49 | 760.046 | 736.406 | 2310.92 | 2295.64 |
| 48133.7 | 1302.2 | 916.373 | 736.675 | 2334.88 | 2297.46 |
| 48191.6 | 1303.91 | 714.601 | 736.938 | 2187.64 | 2299.25 |
| 48249 | 1305.62 | 667.266 | 737.194 | 2266.87 | 2301.03 |
| 48306 | 1307.33 | 840.789 | 737.444 | 2345.75 | 2302.78 |
| 48362.6 | 1309.04 | 688.955 | 737.687 | 2284.95 | 2304.51 |
| 48418.6 | 1310.76 | 706.045 | 737.924 | 2395.45 | 2306.23 |
| 48474.3 | 1312.47 | 772.426 | 738.155 | 2446.22 | 2307.92 |
| 48529.5 | 1314.18 | 715.742 | 738.379 | 2312.45 | 2309.59 |
| 48584.2 | 1315.89 | 649.457 | 738.598 | 2215.98 | 2311.25 |
| 48638.6 | 1317.6 | 610.894 | 738.81 | 2385.4 | 2312.88 |
| 48692.4 | 1319.31 | 704.959 | 739.015 | 2317.03 | 2314.5 |
| 48745.8 | 1321.02 | 754.562 | 739.215 | 2151.18 | 2316.09 |
| 48798.8 | 1322.73 | 681.466 | 739.409 | 2352.78 | 2317.67 |
| 48851.4 | 1324.44 | 799.716 | 739.596 | 2365.09 | 2319.22 |
| 48903.5 | 1326.15 | 843.94 | 739.778 | 2309.98 | 2320.76 |
| 48955.1 | 1327.86 | 743.479 | 739.953 | 2422.19 | 2322.27 |
| 49006.4 | 1329.57 | 884.31 | 740.122 | 2497.66 | 2323.77 |
| 49057.2 | 1331.28 | 803.139 | 740.285 | 2295.47 | 2325.25 |
| 49107.5 | 1332.98 | 605.668 | 740.443 | 2266.08 | 2326.7 |
| 49157.4 | 1334.69 | 729.303 | 740.594 | 2316 | 2328.14 |
| 49206.9 | 1336.4 | 803.636 | 740.739 | 2330.76 | 2329.56 |
| 49256 | 1338.11 | 689.151 | 740.879 | 2442.93 | 2330.97 |
| 49304.6 | 1339.82 | 669.085 | 741.013 | 2297.95 | 2332.35 |
| 49352.8 | 1341.53 | 796.228 | 741.14 | 2391.2 | 2333.71 |
| 49400.6 | 1343.24 | 832.723 | 741.263 | 2285.28 | 2335.05 |
| 49447.9 | 1344.95 | 680.455 | 741.379 | 2218.28 | 2336.38 |
| 49494.8 | 1346.65 | 557.585 | 741.489 | 2342.73 | 2337.69 |
| 49541.3 | 1348.36 | 659.191 | 741.594 | 2389.08 | 2338.97 |
| 49587.4 | 1350.07 | 744.35 | 741.693 | 2367.07 | 2340.24 |
| 49633.1 | 1351.78 | 710.655 | 741.787 | 2282.78 | 2341.49 |
| 49678.3 | 1353.49 | 736.632 | 741.875 | 2454.17 | 2342.73 |
| 49723.1 | 1355.19 | 800.082 | 741.957 | 2350.41 | 2343.94 |
| 49767.4 | 1356.9 | 757.546 | 742.034 | 2348.6 | 2345.14 |
| 49811.4 | 1358.61 | 687.449 | 742.105 | 2442.4 | 2346.32 |
| 49854.9 | 1360.32 | 753.276 | 742.171 | 2396.09 | 2347.48 |
| 49898.1 | 1362.02 | 684.77 | 742.231 | 2462.93 | 2348.62 |
| 49940.8 | 1363.73 | 586.196 | 742.286 | 2473.05 | 2349.74 |
| 49983.1 | 1365.44 | 645.314 | 742.335 | 2453.57 | 2350.85 |
| 50024.9 | 1367.14 | 861.681 | 742.379 | 2463.88 | 2351.93 |
| 50066.4 | 1368.85 | 766.376 | 742.417 | 2282.51 | 2353.01 |
| 50107.5 | 1370.56 | 730.195 | 742.451 | 2304.71 | 2354.06 |
| 50148.1 | 1372.26 | 820.295 | 742.478 | 2421.5 | 2355.09 |
| 50188.4 | 1373.97 | 686.766 | 742.501 | 2406.31 | 2356.11 |
| 50228.2 | 1375.67 | 776.772 | 742.518 | 2393.16 | 2357.11 |
| 50267.6 | 1377.38 | 756.799 | 742.53 | 2372.24 | 2358.09 |
| 50306.6 | 1379.09 | 715.439 | 742.537 | 2503.92 | 2359.06 |
| 50345.2 | 1380.79 | 734.112 | 742.539 | 2482.38 | 2360 |
| 50383.5 | 1382.5 | 780.125 | 742.535 | 2446.45 | 2360.93 |
| 50421.3 | 1384.2 | 862.019 | 742.527 | 2586.7 | 2361.85 |
| 50458.7 | 1385.91 | 760.878 | 742.513 | 2367 | 2362.74 |
| 50495.7 | 1387.61 | 831.566 | 742.494 | 2308.89 | 2363.62 |
| 50532.3 | 1389.32 | 933.581 | 742.47 | 2457.02 | 2364.48 |
| 50568.5 | 1391.02 | 728.258 | 742.441 | 2369.97 | 2365.33 |
| 50604.3 | 1392.73 | 753.92 | 742.407 | 2344.17 | 2366.16 |
| 50639.8 | 1394.43 | 778.75 | 742.369 | 2448.87 | 2366.97 |
| 50674.8 | 1396.14 | 836.797 | 742.325 | 2516.4 | 2367.76 |
| 50709.4 | 1397.84 | 870.988 | 742.276 | 2448.54 | 2368.54 |
| 50743.7 | 1399.55 | 866.117 | 742.223 | 2456.29 | 2369.3 |
| 50777.5 | 1401.25 | 842.497 | 742.164 | 2347.63 | 2370.05 |
| 50811 | 1402.95 | 702.676 | 742.101 | 2397.94 | 2370.78 |
| 50844.1 | 1404.66 | 711.649 | 742.033 | 2299.54 | 2371.49 |
| 50876.8 | 1406.36 | 669.193 | 741.96 | 2246.77 | 2372.19 |
| 50909.1 | 1408.07 | 742.714 | 741.883 | 2379.35 | 2372.87 |
| 50941 | 1409.77 | 796.354 | 741.8 | 2249.3 | 2373.53 |
| 50972.6 | 1411.47 | 841.925 | 741.713 | 2317.22 | 2374.18 |
| 51003.7 | 1413.17 | 802.253 | 741.622 | 2420.35 | 2374.81 |
| 51034.5 | 1414.88 | 703.998 | 741.525 | 2478.47 | 2375.43 |
| 51064.9 | 1416.58 | 802.03 | 741.425 | 2511.77 | 2376.03 |
| 51094.9 | 1418.28 | 832.372 | 741.319 | 2351.43 | 2376.61 |
| 51124.6 | 1419.99 | 798.613 | 741.209 | 2348.99 | 2377.18 |
| 51153.9 | 1421.69 | 804.989 | 741.094 | 2494.01 | 2377.73 |
| 51182.8 | 1423.39 | 691.513 | 740.975 | 2353.39 | 2378.27 |
| 51211.3 | 1425.09 | 770.039 | 740.852 | 2366.14 | 2378.79 |
| 51239.5 | 1426.8 | 713.234 | 740.724 | 2432.45 | 2379.3 |
| 51267.3 | 1428.5 | 598.511 | 740.591 | 2467.07 | 2379.79 |
| 51294.7 | 1430.2 | 679.038 | 740.454 | 2404.36 | 2380.26 |
| 51321.7 | 1431.9 | 727.266 | 740.313 | 2447.05 | 2380.72 |
| 51348.4 | 1433.6 | 726.056 | 740.167 | 2585.11 | 2381.17 |
| 51374.8 | 1435.3 | 707.396 | 740.017 | 2595.83 | 2381.6 |
| 51400.7 | 1437.01 | 669.654 | 739.863 | 2539.55 | 2382.01 |
| 51426.3 | 1438.71 | 666.177 | 739.705 | 2470.28 | 2382.41 |
| 51451.5 | 1440.41 | 852.205 | 739.542 | 2439.02 | 2382.8 |
| 51476.4 | 1442.11 | 851.372 | 739.375 | 2199.26 | 2383.17 |
| 51500.9 | 1443.81 | 696.752 | 739.204 | 2380.97 | 2383.52 |
| 51525.1 | 1445.51 | 566.956 | 739.028 | 2466.5 | 2383.86 |
| 51548.9 | 1447.21 | 708.1 | 738.849 | 2428.03 | 2384.19 |
| 51572.3 | 1448.91 | 660.045 | 738.665 | 2462.69 | 2384.5 |
| 51595.4 | 1450.61 | 685.189 | 738.477 | 2405.29 | 2384.8 |
| 51618.2 | 1452.31 | 694.818 | 738.285 | 2407.05 | 2385.08 |
| 51640.6 | 1454.01 | 844.291 | 738.089 | 2383.18 | 2385.35 |
| 51662.6 | 1455.71 | 909.542 | 737.889 | 2371 | 2385.61 |
| 51684.3 | 1457.41 | 740.825 | 737.685 | 2338.48 | 2385.84 |
| 51705.7 | 1459.11 | 853.019 | 737.477 | 2451.55 | 2386.07 |
| 51726.7 | 1460.81 | 852.888 | 737.265 | 2460.36 | 2386.28 |
| 51747.3 | 1462.51 | 856.411 | 737.05 | 2392.51 | 2386.48 |
| 51767.6 | 1464.21 | 708.934 | 736.83 | 2432.47 | 2386.66 |
| 51787.6 | 1465.91 | 750.122 | 736.606 | 2380.28 | 2386.83 |
| 51807.2 | 1467.61 | 843.954 | 736.379 | 2441.71 | 2386.99 |
| 51826.5 | 1469.31 | 668.917 | 736.147 | 2416.02 | 2387.13 |
| 51845.5 | 1471.01 | 830.101 | 735.912 | 2357.88 | 2387.26 |
| 51864.1 | 1472.7 | 784.203 | 735.673 | 2509.1 | 2387.38 |
| 51882.4 | 1474.4 | 603.343 | 735.431 | 2335.62 | 2387.48 |
| 51900.3 | 1476.1 | 660.092 | 735.184 | 2299.6 | 2387.57 |
| 51917.9 | 1477.8 | 887.634 | 734.934 | 2470.56 | 2387.64 |
| 51935.2 | 1479.5 | 900.276 | 734.68 | 2396.91 | 2387.7 |
| 51952.2 | 1481.19 | 691.401 | 734.423 | 2320.76 | 2387.75 |
| 51968.8 | 1482.89 | 858.358 | 734.162 | 2171.91 | 2387.78 |
| 51985.1 | 1484.59 | 778.158 | 733.897 | 2316.03 | 2387.8 |
| 52001.1 | 1486.29 | 682.191 | 733.629 | 2357.3 | 2387.81 |
| 52016.7 | 1487.98 | 718.272 | 733.357 | 2288.99 | 2387.81 |
| 52032 | 1489.68 | 780.577 | 733.081 | 2457.97 | 2387.79 |
| 52047 | 1491.38 | 685.139 | 732.802 | 2293.38 | 2387.76 |
| 52061.7 | 1493.08 | 779.14 | 732.52 | 2509.4 | 2387.72 |
| 52076.1 | 1494.77 | 690.334 | 732.234 | 2568.86 | 2387.66 |
| 52090.1 | 1496.47 | 561.994 | 731.944 | 2336.53 | 2387.59 |
| 52103.8 | 1498.17 | 842.192 | 731.651 | 2404.32 | 2387.51 |
| 52117.2 | 1499.86 | 756.147 | 731.355 | 2315.65 | 2387.42 |
| 52130.3 | 1501.56 | 714.624 | 731.055 | 2447.08 | 2387.31 |
| 52143.1 | 1503.25 | 764.482 | 730.752 | 2379.59 | 2387.19 |
| 52155.5 | 1504.95 | 631.905 | 730.446 | 2354.92 | 2387.06 |
| 52167.7 | 1506.65 | 687.332 | 730.136 | 2375.89 | 2386.91 |
| 52179.5 | 1508.34 | 670.069 | 729.823 | 2321.35 | 2386.76 |
| 52191.1 | 1510.04 | 637.186 | 729.507 | 2405.47 | 2386.59 |
| 52202.3 | 1511.73 | 888.472 | 729.187 | 2354.07 | 2386.41 |
| 52213.2 | 1513.43 | 763.654 | 728.864 | 2368.32 | 2386.22 |
| 52223.8 | 1515.12 | 687.55 | 728.538 | 2368.31 | 2386.01 |
| 52234.1 | 1516.82 | 648.366 | 728.209 | 2233.7 | 2385.8 |
| 52244.1 | 1518.51 | 694.211 | 727.876 | 2303.83 | 2385.57 |
| 52253.8 | 1520.21 | 790.53 | 727.541 | 2419.26 | 2385.33 |
| 52263.2 | 1521.9 | 698.992 | 727.202 | 2362.24 | 2385.07 |
| 52272.3 | 1523.6 | 766.124 | 726.861 | 2338.62 | 2384.81 |
| 52281.2 | 1525.29 | 631.08 | 726.516 | 2403.36 | 2384.54 |
| 52289.7 | 1526.99 | 800.87 | 726.168 | 2494.41 | 2384.25 |
| 52297.9 | 1528.68 | 853.072 | 725.817 | 2315.01 | 2383.95 |
| 52305.8 | 1530.37 | 631.947 | 725.463 | 2213.78 | 2383.64 |
| 52313.5 | 1532.07 | 763.759 | 725.106 | 2383.77 | 2383.32 |
| 52320.8 | 1533.76 | 669.044 | 724.746 | 2335.8 | 2382.99 |
| 52327.9 | 1535.45 | 539.886 | 724.383 | 2388.51 | 2382.64 |
| 52334.6 | 1537.15 | 699.757 | 724.017 | 2364.8 | 2382.29 |
| 52341.1 | 1538.84 | 663.769 | 723.648 | 2282.61 | 2381.92 |
| 52347.3 | 1540.53 | 540.298 | 723.277 | 2317.95 | 2381.54 |
| 52353.2 | 1542.23 | 738.137 | 722.902 | 2274.72 | 2381.16 |
| 52358.8 | 1543.92 | 704.767 | 722.524 | 2290.11 | 2380.76 |
| 52364.2 | 1545.61 | 797.859 | 722.144 | 2334.95 | 2380.35 |
| 52369.2 | 1547.31 | 799.605 | 721.761 | 2287.78 | 2379.93 |
| 52374 | 1549 | 582.775 | 721.375 | 2257.48 | 2379.49 |
| 52378.5 | 1550.69 | 542.724 | 720.987 | 2386.43 | 2379.05 |
| 52382.8 | 1552.38 | 567.838 | 720.595 | 2339.19 | 2378.6 |
| 52386.7 | 1554.07 | 805.667 | 720.201 | 2320.12 | 2378.13 |
| 52390.4 | 1555.77 | 789.338 | 719.804 | 2329.08 | 2377.66 |
| 52393.8 | 1557.46 | 574.234 | 719.405 | 2373.62 | 2377.17 |
| 52396.9 | 1559.15 | 711.301 | 719.003 | 2347.84 | 2376.68 |
| 52399.8 | 1560.84 | 836.449 | 718.598 | 2226.65 | 2376.17 |
| 52402.4 | 1562.53 | 745.804 | 718.191 | 2206.22 | 2375.66 |
| 52404.7 | 1564.22 | 790.206 | 717.78 | 2316.64 | 2375.13 |
| 52406.8 | 1565.91 | 882.647 | 717.368 | 2382.48 | 2374.59 |
| 52408.5 | 1567.61 | 629.215 | 716.953 | 2373.3 | 2374.05 |
| 52410.1 | 1569.3 | 628.008 | 716.535 | 2324.23 | 2373.49 |
| 52411.3 | 1570.99 | 719.745 | 716.114 | 2221.96 | 2372.92 |
| 52412.3 | 1572.68 | 599.035 | 715.692 | 2336.23 | 2372.34 |
| 52413.1 | 1574.37 | 786.802 | 715.266 | 2234.16 | 2371.76 |
| 52413.5 | 1576.06 | 689.935 | 714.838 | 2214.32 | 2371.16 |
| 52413.8 | 1577.75 | 677.044 | 714.408 | 2352.7 | 2370.55 |
| 52413.7 | 1579.44 | 744.617 | 713.975 | 2415.24 | 2369.94 |

Figure 3 data

| Temperature | Intensity Cblack | Fit Carbon Black | Temperature | Intensity Graphene Platelet | Fit Graphene Platelet | Temperature | Intensity Graphite | Fit Graphite | Temperature | Intensity ND | Fit Nano Diamond |
|---|---|---|---|---|---|---|---|---|---|---|---|
| 1380 | 1.44E+07 | 1.47E+07 | 1357.69 | 1.11E+07 | 1.01E+07 | 1296 | 3.67E+06 | 3.42E+06 | 1538.45 | 3.22E+06 | 6.56E+06 |
| 1540 | 2.89E+07 | 2.84E+07 | 1551.04 | 2.12E+07 | 2.35E+07 | 1647.64 | 1.24E+07 | 1.41E+07 | 1643.21 | 6.56E+06 | 9.72E+06 |
| 1746 | 6.21E+07 | 6.06E+07 | 1783.43 | 5.31E+07 | 5.67E+07 | 1844.7 | 2.64E+07 | 2.75E+07 | 1751.22 | 1.41E+07 | 1.42E+07 |
| 1947 | 1.18E+08 | 1.17E+08 | 1955.04 | 1.07E+08 | 1.01E+08 | 2010.09 | 4.95E+07 | 4.57E+07 | 1959.4 | 2.86E+07 | 2.78E+07 |
| 2049 | 1.55E+08 | 1.59E+08 | 2209.65 | 2.26E+08 | 2.19E+08 | 2149.62 | 6.95E+07 | 6.80E+07 | 2048.91 | 3.56E+07 | 3.63E+07 |

Figure 4 Data

| Figures 4 A and B | T = 1668 K , n = 1.05, 7560 sec | T = 1668 K , n = 1.05, 7560 sec | T = 1679 K , n = 1.01, 14760 sec | T = 1679 K , n = 1.01, 14760 sec | Figure 4 C and D | T = 1553 K , n = 0.99, 1560 sec | T = 1553 K , n = 0.99, 1560 sec | T = 1556 K, n = 1.01, 5940 sec | T = 1556 K, n = 1.01, 5940 sec | T = 1536 K, n = 0.95, 12990 sec | T = 1536 K, n = 0.95, 12990 sec |
|---|---|---|---|---|---|---|---|---|---|---|---|
| 600.015 | 791.878 | 671.536 | 718.1705372 | 852.2893007 | 600.015 | 614.858 | 343.085 | 730.4580148 | 543.6074728 | 464.6284234 | 425.6070806 |
| 600.306 | 667.298 | 674.568 | 916.5069928 | 856.1160621 | 600.306 | 466.314 | 344.821 | 494.6312152 | 546.345076 | 440.4734536 | 427.8042465 |
| 600.597 | 594.815 | 677.609 | 771.8279761 | 859.9546491 | 600.597 | 591.164 | 346.564 | 376.2379297 | 549.0934149 | 314.1023551 | 430.0094944 |
| 600.888 | 769.215 | 680.661 | 842.0920932 | 863.8050617 | 600.888 | 294.036 | 348.314 | 561.4040406 | 551.8524894 | 468.6890815 | 432.223979 |
| 601.179 | 982.911 | 683.722 | 814.6212441 | 867.6684825 | 601.179 | 402.962 | 350.071 | 699.4339927 | 554.6212261 | 519.0437824 | 434.4465456 |
| 601.47 | 584.534 | 686.793 | 978.6930483 | 871.5449114 | 601.47 | 472.019 | 351.834 | 647.0362679 | 557.4006985 | 411.5650777 | 436.6783489 |
| 601.76 | 809.625 | 689.874 | 994.0816905 | 875.4319833 | 601.76 | 335.273 | 353.604 | 512.1110799 | 560.1909066 | 475.0415811 | 438.9193889 |
| 602.052 | 743.613 | 692.965 | 825.2016008 | 879.3332459 | 602.052 | 515.616 | 355.381 | 451.2139024 | 562.9907769 | 456.6029373 | 441.1685108 |
| 602.342 | 809.418 | 696.066 | 783.4856442 | 883.2463342 | 602.342 | 408.51 | 357.164 | 799.9695169 | 565.8013828 | 667.8552993 | 443.4257149 |
| 602.633 | 805.053 | 699.176 | 804.6191588 | 887.171248 | 602.633 | 398.015 | 358.954 | 584.5630899 | 568.6227244 | 433.7988262 | 445.6921556 |
| 602.924 | 777.929 | 702.297 | 797.4469376 | 891.1091699 | 602.924 | 591.7 | 360.751 | 602.526061 | 571.4548017 | 461.1011813 | 447.9678329 |
| 603.215 | 769.16 | 705.428 | 869.2755804 | 895.0589175 | 603.215 | 515.997 | 362.554 | 328.3985826 | 574.2976148 | 365.5515979 | 450.2527469 |
| 603.506 | 722.33 | 708.568 | 907.0547975 | 899.0216732 | 603.506 | 392.6 | 364.365 | 678.2030749 | 577.1511635 | 516.5325707 | 452.5457429 |
| 603.797 | 766.84 | 711.719 | 766.7914567 | 902.9974371 | 603.797 | 344.019 | 366.182 | 808.4142176 | 580.0143743 | 524.9263679 | 454.846821 |
| 604.088 | 825.503 | 714.879 | 878.485351 | 906.9850265 | 604.088 | 469.318 | 368.006 | 663.7560451 | 582.8893944 | 443.3102569 | 457.1582903 |
| 604.379 | 906.415 | 718.05 | 968.3338302 | 910.9856241 | 604.379 | 630.26 | 369.837 | 545.5227215 | 585.7740767 | 392.9209186 | 459.4778417 |
| 604.669 | 851.707 | 721.231 | 1036.095652 | 914.9980473 | 604.669 | 421.586 | 371.674 | 726.0209506 | 588.6705682 | 527.0611865 | 461.8066297 |
| 604.96 | 792.854 | 724.422 | 944.98893 | 919.0234787 | 604.96 | 425.714 | 373.519 | 796.8797829 | 591.5767218 | 461.2582042 | 464.1434998 |
| 605.251 | 769.745 | 727.622 | 1047.145485 | 923.0619182 | 605.251 | 621.579 | 375.37 | 639.0370988 | 594.4946848 | 237.1114919 | 466.4907611 |
| 605.542 | 844.12 | 730.833 | 998.5411211 | 927.1133658 | 605.542 | 455.945 | 377.229 | 696.4301442 | 597.4223098 | 452.2120693 | 468.8461044 |
| 605.832 | 831.902 | 734.054 | 1060.010546 | 931.1766391 | 605.832 | 524.347 | 379.094 | 783.498808 | 600.3617441 | 613.7285853 | 471.2106844 |
| 606.123 | 907.469 | 737.285 | 1080.343468 | 935.2517379 | 606.123 | 549.046 | 380.966 | 754.7732989 | 603.3108406 | 497.8757113 | 473.5833465 |
| 606.414 | 917.149 | 740.527 | 974.5127017 | 939.3410275 | 606.414 | 438.231 | 382.845 | 641.4826909 | 606.2717463 | 496.7361408 | 475.9663998 |
| 606.705 | 769.462 | 743.778 | 923.6129908 | 943.4421426 | 606.705 | 442.818 | 384.731 | 644.1709099 | 609.2433877 | 478.0330981 | 478.3575351 |
| 606.995 | 1012.71 | 747.04 | 1020.155937 | 947.5574484 | 606.995 | 539.196 | 386.624 | 580.7379604 | 612.2257648 | 587.6720221 | 480.7590617 |
| 607.286 | 939.911 | 750.312 | 1059.568269 | 951.6833973 | 607.286 | 557.301 | 388.524 | 532.0290217 | 615.2188776 | 338.4108841 | 483.1686703 |
| 607.577 | 847.463 | 753.593 | 1037.247465 | 955.8235369 | 607.577 | 419.798 | 390.431 | 763.3543429 | 618.2227261 | 623.1257125 | 485.5875155 |
| 607.868 | 838.409 | 756.886 | 1078.498675 | 959.9766846 | 607.868 | 609.705 | 392.345 | 751.8231289 | 621.2383839 | 593.9148366 | 488.0155974 |
| 608.158 | 940.67 | 760.188 | 947.5822822 | 964.141658 | 608.158 | 597.153 | 394.266 | 612.981558 | 624.2647773 | 491.2795953 | 490.452916 |
| 608.449 | 923.17 | 763.501 | 1122.259277 | 968.3196394 | 608.449 | 439.231 | 396.194 | 792.034762 | 627.3019065 | 700.1281215 | 492.8994712 |
| 608.74 | 941.925 | 766.824 | 1170.487587 | 972.5106291 | 608.74 | 475.386 | 398.129 | 884.3832425 | 630.3497714 | 595.8441399 | 495.355263 |
| 609.03 | 932.729 | 770.157 | 1063.6954 | 976.7146269 | 609.03 | 585.29 | 400.071 | 812.7246006 | 633.4094455 | 552.8556599 | 497.8191369 |
| 609.321 | 928.629 | 773.5 | 1049.49996 | 980.9316328 | 609.321 | 536.453 | 402.02 | 572.7001428 | 636.4798553 | 416.8218808 | 500.293402 |
| 609.612 | 934.353 | 776.854 | 1066.636425 | 985.1604643 | 609.612 | 562.799 | 403.977 | 850.689252 | 639.5610009 | 523.3053375 | 502.7769037 |
| 609.902 | 913.864 | 780.218 | 1089.612366 | 989.4034865 | 609.902 | 688.072 | 405.94 | 919.6178057 | 642.6539557 | 682.4399548 | 505.2696421 |
| 610.193 | 948.06 | 783.593 | 1155.836861 | 993.6583344 | 610.193 | 531.726 | 407.911 | 807.8151656 | 645.7576462 | 534.6640963 | 507.7716171 |
| 610.484 | 900.442 | 786.977 | 1008.776171 | 997.9273729 | 610.484 | 575.773 | 409.888 | 816.6420571 | 648.8720724 | 652.6390884 | 510.2828288 |
| 610.774 | 911.43 | 790.372 | 1122.71338 | 1002.208237 | 610.774 | 501.445 | 411.873 | 782.400546 | 651.9983078 | 618.8098922 | 512.8032771 |
| 611.065 | 1021.84 | 793.778 | 1109.176625 | 1006.503292 | 611.065 | 500.256 | 413.865 | 842.1672544 | 655.135279 | 640.2077248 | 515.3341166 |
| 611.355 | 972.8 | 797.194 | 1136.600172 | 1010.810172 | 611.355 | 520.197 | 415.865 | 702.2725114 | 658.2840594 | 580.7099043 | 517.8730382 |
| 611.646 | 996.897 | 800.62 | 1064.063176 | 1015.130061 | 611.646 | 510.87 | 417.871 | 768.5021104 | 661.4446492 | 526.4296312 | 520.422351 |
| 611.937 | 999.964 | 804.057 | 1204.223634 | 1019.46414 | 611.937 | 534.589 | 419.885 | 777.6714707 | 664.614901 | 497.7510166 | 522.9809005 |
| 612.227 | 977.058 | 807.504 | 1084.967274 | 1023.810045 | 612.227 | 451.423 | 421.906 | 632.0631889 | 667.7980357 | 533.009583 | 525.547532 |
| 612.518 | 908.315 | 810.962 | 1160.613218 | 1028.170141 | 612.518 | 534.518 | 423.934 | 737.0690581 | 670.991906 | 673.1963867 | 528.1257093 |
| 612.808 | 998.031 | 814.43 | 1184.723234 | 1032.542062 | 612.808 | 566.665 | 425.969 | 702.7953399 | 674.1965121 | 391.1532565 | 530.7119687 |
| 613.099 | 984.349 | 817.908 | 1216.463121 | 1036.928174 | 613.099 | 538.652 | 428.012 | 722.6950311 | 677.4129274 | 598.1867829 | 533.3086192 |
| 613.389 | 957.553 | 821.397 | 1221.83194 | 1041.327294 | 613.389 | 590.001 | 430.062 | 824.262256 | 680.641152 | 672.8442398 | 535.9133519 |
| 613.68 | 975.991 | 824.897 | 1158.56266 | 1045.739422 | 613.68 | 483.106 | 432.119 | 777.7068985 | 683.8801123 | 724.7668604 | 538.5296303 |
| 613.97 | 951.282 | 828.407 | 1194.562126 | 1050.164558 | 613.97 | 604.971 | 434.184 | 698.417322 | 687.1308819 | 598.0251417 | 541.1539909 |
| 614.26 | 1013.56 | 831.927 | 1179.175849 | 1054.602703 | 614.26 | 652.733 | 436.256 | 883.8765175 | 690.3934608 | 590.3252471 | 543.7887426 |
| 614.551 | 1019.9 | 835.458 | 1197.790512 | 1059.053855 | 614.551 | 628.966 | 438.335 | 748.3791167 | 693.6667753 | 581.1959825 | 546.432731 |
| 614.841 | 953.141 | 839 | 1187.963446 | 1063.519199 | 614.841 | 556.71 | 440.422 | 641.1316336 | 696.9518991 | 544.6408227 | 549.085956 |
| 615.132 | 1028.01 | 842.553 | 1299.80989 | 1067.996368 | 615.132 | 620.95 | 442.516 | 891.2895174 | 700.2488322 | 580.3000283 | 551.7495722 |
| 615.422 | 1009.37 | 846.115 | 1172.003628 | 1072.487727 | 615.422 | 566.862 | 444.618 | 770.7383565 | 703.5575745 | 638.8880397 | 554.4235797 |
| 615.713 | 1032.49 | 849.689 | 1211.23621 | 1076.992095 | 615.713 | 544.303 | 446.726 | 757.2757903 | 706.8770526 | 555.9880357 | 557.1056692 |
| 616.003 | 1022.84 | 853.273 | 1267.40777 | 1081.510654 | 616.003 | 618.523 | 448.843 | 803.3673656 | 710.2094135 | 554.4385892 | 559.7981499 |
| 616.293 | 957.735 | 856.868 | 1243.851191 | 1086.041038 | 616.293 | 550.089 | 450.967 | 793.5495691 | 713.5525101 | 484.9651969 | 562.5010219 |
| 616.584 | 1136.45 | 860.473 | 1140.028411 | 1090.585613 | 616.584 | 620.727 | 453.098 | 929.5354442 | 716.9074159 | 586.3292455 | 565.2131305 |
| 616.874 | 1027.18 | 864.089 | 1193.131229 | 1095.143196 | 616.874 | 598.17 | 455.237 | 883.2141249 | 720.2730575 | 559.3721099 | 567.9344757 |
| 617.165 | 1026.41 | 867.716 | 1280.888944 | 1099.714969 | 617.165 | 558.367 | 457.383 | 831.7031688 | 723.6515818 | 589.1441117 | 570.6662122 |
| 617.455 | 1081.67 | 871.353 | 1193.403218 | 1104.299751 | 617.455 | 629.172 | 459.536 | 860.1753154 | 727.0419155 | 643.830797 | 573.4083399 |
| 617.745 | 962.116 | 875.001 | 1284.152807 | 1108.897541 | 617.745 | 575.08 | 461.698 | 846.825474 | 730.4429849 | 623.0956934 | 576.1597042 |
| 618.036 | 1028.54 | 878.66 | 1170.721733 | 1113.508339 | 618.036 | 560.431 | 463.866 | 784.6904705 | 733.8569371 | 587.6050565 | 578.9203052 |
| 618.326 | 1043.11 | 882.33 | 1210.290163 | 1118.133328 | 618.326 | 477.432 | 466.043 | 810.2779349 | 737.2816249 | 676.4615391 | 581.6912973 |
| 618.616 | 1053.78 | 886.01 | 1315.880869 | 1122.771325 | 618.616 | 526.582 | 468.227 | 809.910774 | 740.7191957 | 668.3829424 | 584.4726807 |
| 618.907 | 1067.6 | 889.701 | 1194.893242 | 1127.42233 | 618.907 | 637.923 | 470.418 | 853.2808497 | 744.1675021 | 720.0238454 | 587.2633008 |
| 619.197 | 1005.56 | 893.403 | 1355.993276 | 1132.087526 | 619.197 | 574.935 | 472.617 | 799.5776639 | 747.6286914 | 633.586208 | 590.064312 |
| 619.487 | 1091.7 | 897.115 | 1299.845367 | 1136.76573 | 619.487 | 516.498 | 474.824 | 952.6086079 | 751.1016899 | 716.8787693 | 592.8757145 |
| 619.777 | 1050.88 | 900.839 | 1200.924294 | 1141.458125 | 619.777 | 549.153 | 477.038 | 883.6424792 | 754.5854242 | 645.9552244 | 595.6963537 |
| 620.067 | 983.793 | 904.573 | 1304.019801 | 1146.163528 | 620.067 | 631.467 | 479.26 | 985.2719712 | 758.0820412 | 649.3866364 | 598.527384 |
| 620.358 | 1080.09 | 908.318 | 1342.665834 | 1150.883121 | 620.358 | 627.082 | 481.49 | 862.5640084 | 761.5904676 | 616.2167054 | 601.367651 |
| 620.648 | 1040.38 | 912.074 | 1213.128305 | 1155.615723 | 620.648 | 573.71 | 483.727 | 868.7552858 | 765.1107032 | 602.6492349 | 604.2183092 |
| 620.938 | 970.645 | 915.841 | 1218.875542 | 1160.361333 | 620.938 | 625.38 | 485.972 | 747.0403751 | 768.6438216 | 736.1325466 | 607.0805132 |
| 621.228 | 1086.72 | 919.618 | 1273.651682 | 1165.122316 | 621.228 | 588.24 | 488.224 | 969.6966195 | 772.1876758 | 669.5040396 | 609.9507993 |
| 621.519 | 1054.24 | 923.407 | 1285.288064 | 1169.895125 | 621.519 | 553.129 | 490.485 | 898.2838252 | 775.7444128 | 635.0802346 | 612.8326312 |
| 621.809 | 1159.59 | 927.206 | 1260.72631 | 1174.682124 | 621.809 | 598.689 | 492.753 | 928.7517381 | 779.312959 | 689.2184944 | 615.7236997 |
| 622.099 | 1090.27 | 931.017 | 1344.309591 | 1179.483314 | 622.099 | 652.72 | 495.029 | 854.230959 | 782.8933146 | 755.4890816 | 618.6251594 |
| 622.389 | 1119.84 | 934.838 | 1321.356118 | 1184.297512 | 622.389 | 521.702 | 497.312 | 927.3925987 | 786.4854793 | 726.4421561 | 621.5370103 |
| 622.679 | 1119.76 | 938.67 | 1229.589528 | 1189.122354 | 622.679 | 632.321 | 499.603 | 942.8380484 | 790.090527 | 716.7367559 | 624.4592525 |
| 622.969 | 1089.78 | 942.513 | 1392.593481 | 1193.970846 | 622.969 | 563.615 | 501.902 | 888.2201812 | 793.7073839 | 621.1121249 | 627.3907313 |
| 623.259 | 1128.33 | 946.367 | 1323.591155 | 1198.819339 | 623.259 | 597.044 | 504.209 | 925.1681619 | 797.3371236 | 637.5187078 | 630.3337559 |
| 623.549 | 1105.63 | 950.232 | 1341.518752 | 1203.691483 | 623.549 | 625.925 | 506.524 | 995.900313 | 800.9775991 | 665.1431907 | 633.2860172 |
| 623.84 | 1117.51 | 954.108 | 1242.266562 | 1208.575452 | 623.84 | 674.244 | 508.847 | 847.3611854 | 804.6309573 | 689.2819963 | 636.2486697 |
| 624.13 | 1027.23 | 957.995 | 1334.60078 | 1213.471247 | 624.13 | 604.933 | 511.177 | 854.3844795 | 808.2971984 | 751.4445876 | 639.2217134 |
| 624.42 | 1109.98 | 961.894 | 1326.937797 | 1218.378867 | 624.42 | 661.899 | 513.515 | 989.9505887 | 811.9752488 | 738.5410007 | 642.2063029 |
| 624.71 | 1116.39 | 965.803 | 1396.389496 | 1223.310139 | 624.71 | 620.036 | 515.861 | 965.7619859 | 815.6651085 | 693.5643338 | 645.200129 |
| 625 | 1087.79 | 969.723 | 1311.552703 | 1228.24141 | 625 | 654.159 | 518.215 | 945.7366871 | 819.367851 | 709.4513557 | 648.2043464 |
| 625.29 | 1094.76 | 973.654 | 1439.919498 | 1233.196333 | 625.29 | 693.462 | 520.577 | 995.5879041 | 823.0834763 | 662.6492977 | 651.2189549 |
| 625.58 | 1149.82 | 977.596 | 1380.401296 | 1238.163082 | 625.58 | 557.252 | 522.947 | 939.0859217 | 826.8109109 | 633.8702347 | 654.2439547 |
| 625.87 | 1076.58 | 981.55 | 1477.170112 | 1243.141656 | 625.87 | 573.198 | 525.325 | 838.0608496 | 830.5501548 | 636.5500151 | 657.2793458 |
| 626.16 | 1128.99 | 985.514 | 1464.800543 | 1248.132055 | 626.16 | 644.36 | 527.711 | 968.4126299 | 834.3022815 | 758.9816864 | 660.325128 |
| 626.45 | 1122.4 | 989.49 | 1476.306843 | 1253.146106 | 626.45 | 564.879 | 530.104 | 971.6440753 | 838.067291 | 724.6537115 | 663.3813015 |
| 626.74 | 1134.4 | 993.476 | 1394.189936 | 1258.160157 | 626.74 | 667.166 | 532.506 | 992.3414288 | 841.8441098 | 828.4585424 | 666.4478661 |
| 627.03 | 1199.61 | 997.474 | 1430.317118 | 1263.197859 | 627.03 | 669.887 | 534.915 | 1005.969325 | 845.6327379 | 725.1905913 | 669.5259766 |
| 627.32 | 1171.76 | 1001.48 | 1402.06578 | 1268.247387 | 627.32 | 624.928 | 537.333 | 973.2587243 | 849.4353224 | 781.2766276 | 672.6133238 |
| 627.61 | 1178.56 | 1005.5 | 1445.418398 | 1273.30874 | 627.61 | 666.19 | 539.759 | 1015.703383 | 853.2497161 | 694.965994 | 675.7122167 |
| 627.9 | 1143.92 | 1009.53 | 1475.02968 | 1278.381918 | 627.9 | 658.107 | 542.192 | 928.7377817 | 857.0759192 | 709.477911 | 678.8203462 |
| 628.19 | 1222.02 | 1013.58 | 1451.44945 | 1283.478748 | 628.19 | 630.95 | 544.634 | 1070.823681 | 860.9160786 | 698.7426255 | 681.9400216 |
| 628.48 | 1136.96 | 1017.63 | 1451.437625 | 1288.575578 | 628.48 | 647.581 | 547.084 | 1039.316552 | 864.7680473 | 728.0493315 | 685.0700882 |
| 628.77 | 1174.48 | 1021.7 | 1489.113959 | 1293.696059 | 628.77 | 783.674 | 549.542 | 970.5544018 | 868.6328989 | 800.2660061 | 688.2117006 |
| 629.059 | 1218.24 | 1025.77 | 1492.933625 | 1298.840192 | 629.059 | 663.278 | 552.008 | 1035.750153 | 872.5095597 | 794.4561592 | 691.3625496 |
| 629.349 | 1213.77 | 1029.86 | 1505.752566 | 1303.984324 | 629.349 | 656.005 | 554.481 | 1008.968879 | 876.4001769 | 726.1546656 | 694.5249444 |
| 629.639 | 1184.99 | 1033.96 | 1474.391098 | 1309.140282 | 629.639 | 682.053 | 556.964 | 1014.290565 | 880.3026034 | 841.998303 | 697.6977305 |
| 629.929 | 1294.94 | 1038.07 | 1557.773343 | 1314.319891 | 629.929 | 682.699 | 559.454 | 1144.307391 | 884.2179127 | 842.7961178 | 700.8809078 |
| 630.219 | 1193.21 | 1042.19 | 1569.492505 | 1319.511326 | 630.219 | 679.597 | 561.952 | 1024.608645 | 888.1461049 | 841.3736752 | 704.0756308 |
| 630.509 | 1249.46 | 1046.32 | 1568.984004 | 1324.70276 | 630.509 | 730.845 | 564.459 | 1061.990348 | 892.0861063 | 806.4210725 | 707.2807451 |
| 630.799 | 1263.74 | 1050.46 | 1551.754117 | 1329.929672 | 630.799 | 698.011 | 566.974 | 1046.244299 | 896.0400641 | 739.4358002 | 710.4962506 |
| 631.088 | 1294.25 | 1054.62 | 1577.687639 | 1335.156583 | 631.088 | 724.832 | 569.496 | 1053.252563 | 900.0058312 | 772.5202923 | 713.7221474 |
| 631.378 | 1239.31 | 1058.79 | 1543.452552 | 1340.407146 | 631.378 | 702.325 | 572.028 | 1095.309662 | 903.9844812 | 815.8101177 | 716.9595899 |
| 631.668 | 1285.71 | 1062.96 | 1530.20789 | 1345.657709 | 631.668 | 688.122 | 574.567 | 1066.993183 | 907.9770875 | 842.0317858 | 720.2085782 |
| 631.958 | 1260.4 | 1067.15 | 1533.010555 | 1350.931922 | 631.958 | 670.191 | 577.115 | 1040.895773 | 911.9815032 | 879.883539 | 723.4668032 |
| 632.247 | 1230.3 | 1071.35 | 1594.503631 | 1356.229787 | 632.247 | 754.004 | 579.67 | 990.9597444 | 915.9988016 | 799.0767887 | 726.736574 |
| 632.537 | 1351.53 | 1075.56 | 1627.295996 | 1361.527652 | 632.537 | 829.356 | 582.235 | 1047.899743 | 920.0289829 | 898.6015912 | 730.0178906 |
| 632.827 | 1350.25 | 1079.79 | 1642.692916 | 1366.849169 | 632.827 | 730.113 | 584.807 | 1121.354468 | 924.0720471 | 812.1951275 | 733.3095984 |
| 633.117 | 1328.91 | 1084.02 | 1568.960353 | 1372.170685 | 633.117 | 710.935 | 587.388 | 1055.748613 | 928.1290676 | 898.1824787 | 736.6116974 |
| 633.407 | 1332.65 | 1088.27 | 1549.944802 | 1377.515852 | 633.407 | 715.453 | 589.976 | 1046.151972 | 932.1978974 | 787.5102055 | 739.9253422 |
| 633.696 | 1332.65 | 1092.53 | 1614.252368 | 1382.872845 | 633.696 | 714.56 | 592.574 | 1177.9316 | 936.2796101 | 904.560379 | 743.2493782 |
| 633.986 | 1295.18 | 1096.79 | 1576.646987 | 1388.253489 | 633.986 | 783.694 | 595.179 | 1011.998493 | 940.3752792 | 796.5898232 | 746.5849601 |
| 634.276 | 1321.29 | 1101.07 | 1640.256844 | 1393.634133 | 634.276 | 736.418 | 597.793 | 1137.715672 | 944.4838311 | 828.358094 | 749.9320877 |
| 634.565 | 1338.18 | 1105.37 | 1676.596887 | 1399.038429 | 634.565 | 727.798 | 600.416 | 1206.467087 | 948.6041922 | 855.336012 | 753.2896066 |
| 634.855 | 1309.46 | 1109.67 | 1655.618287 | 1404.45455 | 634.855 | 764.994 | 603.046 | 1236.720286 | 952.7385098 | 775.3467044 | 756.6575167 |
| 635.144 | 1391.97 | 1113.98 | 1658.125313 | 1409.894322 | 635.144 | 783.179 | 605.685 | 1074.654178 | 956.8857102 | 857.8287504 | 760.0369725 |
| 635.434 | 1312.56 | 1118.31 | 1589.099335 | 1415.334094 | 635.434 | 767.787 | 608.333 | 1123.888092 | 961.0457935 | 916.9097675 | 763.4279742 |
| 635.724 | 1357.63 | 1122.65 | 1610.515481 | 1420.797517 | 635.724 | 753.596 | 610.989 | 1178.47912 | 965.2198331 | 915.2159985 | 766.8293671 |
| 636.013 | 1298.93 | 1127 | 1728.120989 | 1426.272766 | 636.013 | 773.083 | 613.653 | 1187.625936 | 969.4067556 | 883.8449033 | 770.2423058 |
| 636.303 | 1332.9 | 1131.36 | 1727.305023 | 1431.75984 | 636.303 | 773.226 | 616.326 | 1186.455745 | 973.606561 | 967.9999359 | 773.6667903 |
| 636.593 | 1310.98 | 1135.73 | 1670.707742 | 1437.25874 | 636.593 | 809.144 | 619.007 | 1148.773442 | 977.8192492 | 970.0804892 | 777.1016661 |
| 636.882 | 1348.88 | 1140.11 | 1690.373701 | 1442.781292 | 636.882 | 796.754 | 621.697 | 1099.217457 | 982.0448202 | 873.9894077 | 780.5480876 |
| 637.172 | 1403.29 | 1144.51 | 1666.722518 | 1448.303843 | 637.172 | 743.537 | 624.395 | 1139.572948 | 986.2843476 | 859.999361 | 784.0060549 |
| 637.461 | 1337.27 | 1148.91 | 1736.351601 | 1453.850045 | 637.461 | 777.501 | 627.101 | 1199.005776 | 990.5378314 | 918.0193189 | 787.4744135 |
| 637.751 | 1326.53 | 1153.33 | 1655.701067 | 1459.419899 | 637.751 | 817.348 | 629.817 | 1148.225921 | 994.8031246 | 925.0530208 | 790.9554724 |
| 638.04 | 1402.68 | 1157.76 | 1724.63244 | 1464.989752 | 638.04 | 788.631 | 632.54 | 1243.730697 | 999.0823741 | 878.4541689 | 794.445768 |
| 638.33 | 1406.76 | 1162.2 | 1717.265096 | 1470.583257 | 638.33 | 784.498 | 635.272 | 1203.235642 | 1003.374506 | 959.6500128 | 797.9487639 |
| 638.619 | 1417.31 | 1166.66 | 1740.656116 | 1476.188588 | 638.619 | 817.499 | 638.013 | 1182.182937 | 1007.680595 | 938.9737927 | 801.4633057 |
| 638.909 | 1422.95 | 1171.12 | 1729.315374 | 1481.805744 | 638.909 | 811.152 | 640.763 | 1183.439013 | 1011.999567 | 974.554487 | 804.9882386 |
| 639.198 | 1401.77 | 1175.6 | 1712.676766 | 1487.434725 | 639.198 | 776.596 | 643.52 | 1162.246744 | 1016.332495 | 840.26066 | 808.5247174 |
| 639.488 | 1365.81 | 1180.09 | 1683.905102 | 1493.087358 | 639.488 | 828.339 | 646.287 | 1142.718508 | 1020.678306 | 852.3745141 | 812.072742 |
| 639.777 | 1408.52 | 1184.59 | 1747.124715 | 1498.739991 | 639.777 | 792.471 | 649.062 | 1206.950193 | 1025.038073 | 850.238541 | 815.6323124 |
| 640.067 | 1428.85 | 1189.1 | 1757.259246 | 1504.416274 | 640.067 | 828.073 | 651.846 | 1094.794349 | 1029.410723 | 926.9546142 | 819.202274 |
| 640.356 | 1432.27 | 1193.62 | 1764.673892 | 1510.104384 | 640.356 | 781.512 | 654.638 | 1273.994632 | 1033.796256 | 913.0419242 | 822.7849359 |
| 640.646 | 1375.38 | 1198.15 | 1769.013884 | 1515.816145 | 640.646 | 746.249 | 657.439 | 1226.929329 | 1038.196819 | 899.9997877 | 826.3779891 |
| 640.935 | 1391.69 | 1202.7 | 1741.613989 | 1521.539731 | 640.935 | 799.338 | 660.249 | 1243.558926 | 1042.609191 | 887.1747122 | 829.9837427 |
| 641.225 | 1429.28 | 1207.26 | 1794.793674 | 1527.275143 | 641.225 | 830.681 | 663.067 | 1230.772709 | 1047.036593 | 940.3327335 | 833.5998875 |
| 641.514 | 1399.89 | 1211.83 | 1766.625115 | 1533.02238 | 641.514 | 803.336 | 665.894 | 1258.932447 | 1051.476878 | 911.8065236 | 837.2275781 |
| 641.803 | 1447.08 | 1216.41 | 1820.053137 | 1538.781443 | 641.803 | 784.003 | 668.73 | 1167.915193 | 1055.930046 | 970.4557277 | 840.8668145 |
| 642.093 | 1408.28 | 1221 | 1831.086414 | 1544.564158 | 642.093 | 857.872 | 671.575 | 1215.678317 | 1060.398244 | 952.04133 | 844.5187513 |
| 642.382 | 1391.78 | 1225.61 | 1812.851352 | 1550.358697 | 642.382 | 849.111 | 674.428 | 1193.165556 | 1064.878251 | 988.6738467 | 848.1810793 |
| 642.671 | 1444.72 | 1230.23 | 1744.511259 | 1556.165063 | 642.671 | 760.934 | 677.29 | 1252.823834 | 1069.373288 | 931.6618371 | 851.8549531 |
| 642.961 | 1459.51 | 1234.86 | 1855.139667 | 1561.99508 | 642.961 | 859.533 | 680.16 | 1279.061882 | 1073.881208 | 864.8509064 | 855.5403727 |
| 643.25 | 1405.98 | 1239.5 | 1842.781924 | 1567.825096 | 643.25 | 814.666 | 683.04 | 1297.505812 | 1078.400937 | 939.6665408 | 859.2384927 |
| 643.539 | 1438.25 | 1244.15 | 1795.112965 | 1573.678764 | 643.539 | 820.295 | 685.928 | 1230.214453 | 1082.942138 | 960.8761768 | 862.9470039 |
| 643.829 | 1481.61 | 1248.81 | 1855.068714 | 1579.544257 | 643.829 | 843.631 | 688.825 | 1238.244755 | 1087.494074 | 989.4289421 | 866.6682155 |
| 644.118 | 1426.09 | 1253.49 | 1849.581639 | 1585.433402 | 644.118 | 781.384 | 691.731 | 1257.128849 | 1092.056746 | 1062.165178 | 870.3998183 |
| 644.407 | 1428.77 | 1258.18 | 1828.756773 | 1591.322546 | 644.407 | 829.562 | 694.646 | 1216.827036 | 1096.630153 | 1080.236665 | 874.1441215 |
| 644.697 | 1513.19 | 1262.87 | 1828.614866 | 1597.235342 | 644.697 | 814.093 | 697.569 | 1331.58092 | 1101.214297 | 969.5828651 | 877.8999705 |
| 644.986 | 1469.91 | 1267.59 | 1864.245373 | 1603.159964 | 644.986 | 773.689 | 700.502 | 1283.227333 | 1105.819911 | 952.2734006 | 881.6673652 |
| 645.275 | 1488.58 | 1272.31 | 1856.227622 | 1609.108236 | 645.275 | 897.572 | 703.443 | 1285.299323 | 1110.446998 | 981.2198777 | 885.4463058 |
| 645.564 | 1486.39 | 1277.05 | 1876.792325 | 1615.056509 | 645.564 | 801.203 | 706.393 | 1272.35207 | 1115.074084 | 951.4109293 | 889.2379468 |
| 645.853 | 1506.93 | 1281.79 | 1866.894305 | 1621.028432 | 645.853 | 848.098 | 709.352 | 1309.819659 | 1119.722641 | 966.660623 | 893.039979 |
| 646.143 | 1504.08 | 1286.55 | 1919.139769 | 1627.012182 | 646.143 | 850.255 | 712.32 | 1249.871517 | 1124.381935 | 991.4840946 | 896.8547116 |
| 646.432 | 1495.09 | 1291.32 | 1884.940158 | 1633.019582 | 646.432 | 854.399 | 715.297 | 1271.944114 | 1129.051964 | 1031.623071 | 900.6809899 |
| 646.721 | 1519.45 | 1296.1 | 1875.739848 | 1639.026983 | 646.721 | 903.276 | 718.283 | 1314.006581 | 1133.743464 | 1034.312088 | 904.5199687 |
| 647.01 | 1576.92 | 1300.9 | 1873.492985 | 1645.058034 | 647.01 | 824.864 | 721.278 | 1288.992403 | 1138.4457 | 1017.421736 | 908.3693387 |
| 647.299 | 1527.87 | 1305.7 | 1903.801976 | 1651.100912 | 647.299 | 834.389 | 724.281 | 1346.213677 | 1143.158672 | 1073.336895 | 912.231409 |
| 647.588 | 1532.82 | 1310.52 | 1895.831528 | 1657.16744 | 647.588 | 860.145 | 727.294 | 1309.261402 | 1147.882379 | 1000.366279 | 916.1050252 |
| 647.878 | 1564.92 | 1315.35 | 1955.503462 | 1663.245794 | 647.878 | 906.631 | 730.316 | 1372.108183 | 1152.627558 | 1046.88662 | 919.9913417 |
| 648.167 | 1613.69 | 1320.2 | 1940.142019 | 1669.335974 | 648.167 | 888.624 | 733.346 | 1334.705008 | 1157.383472 | 1022.364493 | 923.8892041 |
| 648.456 | 1550.32 | 1325.05 | 1944.387406 | 1675.437979 | 648.456 | 914.377 | 736.386 | 1335.971821 | 1162.160858 | 1073.752543 | 927.7986122 |
| 648.745 | 1561.09 | 1329.92 | 1922.699272 | 1681.55181 | 648.745 | 868.535 | 739.435 | 1309.76598 | 1166.938244 | 1027.976907 | 931.7195661 |
| 649.034 | 1548.7 | 1334.8 | 1979.816879 | 1687.689292 | 649.034 | 915.749 | 742.492 | 1381.97429 | 1171.737102 | 1027.190638 | 935.6532204 |
| 649.323 | 1548.68 | 1339.69 | 1960.919583 | 1693.838599 | 649.323 | 868.191 | 745.559 | 1419.592178 | 1176.55743 | 1162.050209 | 939.5995751 |
| 649.612 | 1579.54 | 1344.59 | 2004.508714 | 1699.999732 | 649.612 | 904.879 | 748.635 | 1434.224936 | 1181.377759 | 1090.522819 | 943.5574756 |
| 649.901 | 1636.85 | 1349.5 | 1905.67042 | 1706.184517 | 649.901 | 871.99 | 751.72 | 1310.152465 | 1186.219559 | 1085.304117 | 947.526922 |
| 650.19 | 1574.09 | 1354.43 | 2012.952186 | 1712.381127 | 650.19 | 892.453 | 754.814 | 1384.604536 | 1191.082831 | 1140.431852 | 951.5079141 |
| 650.479 | 1612.05 | 1359.37 | 1994.669822 | 1718.589562 | 650.479 | 947.221 | 757.917 | 1352.09684 | 1195.946102 | 1064.089863 | 955.5027611 |
| 650.768 | 1581.48 | 1364.32 | 1996.999463 | 1724.809823 | 650.768 | 888.068 | 761.029 | 1352.290083 | 1200.830845 | 1085.851388 | 959.5079994 |
| 651.057 | 1586.7 | 1369.28 | 1954.959485 | 1731.053736 | 651.057 | 920.111 | 764.151 | 1362.907689 | 1205.73706 | 1002.120086 | 963.5259381 |
| 651.346 | 1581.95 | 1374.26 | 1939.95281 | 1737.309474 | 651.346 | 921.234 | 767.281 | 1402.941109 | 1210.643274 | 1058.204968 | 967.5565771 |
| 651.635 | 1547.56 | 1379.24 | 2011.710499 | 1743.577037 | 651.635 | 910.008 | 770.421 | 1466.646746 | 1215.57096 | 1178.052689 | 971.598762 |
| 651.924 | 1642.98 | 1384.24 | 2002.072642 | 1749.856426 | 651.924 | 915.913 | 773.57 | 1396.531897 | 1220.509381 | 1105.138648 | 975.6536472 |
| 652.213 | 1630.67 | 1389.25 | 2075.947112 | 1756.159466 | 652.213 | 893.557 | 776.728 | 1437.123574 | 1225.469274 | 1106.088868 | 979.7200782 |
| 652.502 | 1676.61 | 1394.28 | 2037.951486 | 1762.474332 | 652.502 | 981.202 | 779.895 | 1384.51865 | 1230.439902 | 1088.413401 | 983.7992096 |
| 652.791 | 1600.28 | 1399.31 | 2037.372032 | 1768.801024 | 652.791 | 921.974 | 783.071 | 1400.171299 | 1235.421267 | 1108.122083 | 987.8898868 |
| 653.08 | 1650.66 | 1404.36 | 2042.906409 | 1775.151366 | 653.08 | 937.687 | 786.257 | 1471.77841 | 1240.424102 | 1090.698315 | 991.994419 |
| 653.369 | 1648.48 | 1409.42 | 2050.983288 | 1781.513535 | 653.369 | 946.419 | 789.452 | 1448.804014 | 1245.437673 | 1109.2986 | 996.1093424 |
| 653.658 | 1668.21 | 1414.49 | 2018.108144 | 1787.887528 | 653.658 | 945.575 | 792.656 | 1529.654561 | 1250.46198 | 1129.16546 | 1000.238121 |
| 653.947 | 1660.35 | 1419.57 | 2040.825105 | 1794.273348 | 653.947 | 932.582 | 795.869 | 1437.166517 | 1255.507759 | 1073.876083 | 1004.378445 |
| 654.236 | 1684.44 | 1424.67 | 2079.991464 | 1800.682818 | 654.236 | 949.982 | 799.092 | 1459.657806 | 1260.564273 | 1071.579624 | 1008.530315 |
| 654.524 | 1690.25 | 1429.78 | 2091.202125 | 1807.104115 | 654.524 | 975.591 | 802.324 | 1470.447183 | 1265.631523 | 1180.431124 | 1012.694885 |
| 654.813 | 1692.17 | 1434.89 | 2176.653849 | 1813.537236 | 654.813 | 970.263 | 805.565 | 1457.789794 | 1270.720244 | 1183.317574 | 1016.87331 |
| 655.102 | 1688.97 | 1440.03 | 2041.132571 | 1819.994009 | 655.102 | 1004.61 | 808.815 | 1444.864013 | 1275.819701 | 1126.686576 | 1021.062127 |
| 655.391 | 1713.89 | 1445.17 | 2115.681099 | 1826.462608 | 655.391 | 915.101 | 812.075 | 1463.672957 | 1280.929893 | 1134.51232 | 1025.264798 |
| 655.68 | 1723.45 | 1450.33 | 2164.189676 | 1832.943032 | 655.68 | 988.429 | 815.344 | 1494.280434 | 1286.061557 | 1197.946104 | 1029.479016 |
| 655.968 | 1688.36 | 1455.5 | 2140.266504 | 1839.435282 | 655.968 | 1009.91 | 818.623 | 1441.407118 | 1291.203957 | 1244.637322 | 1033.705933 |
| 656.257 | 1680.67 | 1460.68 | 2138.457189 | 1845.951183 | 656.257 | 966.858 | 821.911 | 1502.493244 | 1296.357092 | 1159.810324 | 1037.945551 |
| 656.546 | 1727.72 | 1465.87 | 2100.650773 | 1852.478909 | 656.546 | 961.201 | 825.208 | 1573.048256 | 1301.531699 | 1233.576445 | 1042.19787 |
| 656.835 | 1689.87 | 1471.08 | 2096.452688 | 1859.018461 | 656.835 | 983.531 | 828.515 | 1541.667808 | 1306.717042 | 1199.19305 | 1046.461734 |
| 657.124 | 1664.74 | 1476.3 | 2139.083945 | 1865.581665 | 657.124 | 1012.91 | 831.831 | 1498.682071 | 1311.91312 | 1206.201351 | 1050.739453 |
| 657.412 | 1672.05 | 1481.53 | 2212.308008 | 1872.144868 | 657.412 | 988.658 | 835.156 | 1551.941872 | 1317.13067 | 1184.356696 | 1055.028718 |
| 657.701 | 1723.55 | 1486.77 | 2193.351584 | 1878.743548 | 657.701 | 965.269 | 838.491 | 1491.940052 | 1322.358955 | 1233.345529 | 1059.330684 |
| 657.99 | 1738.96 | 1492.02 | 2160.783905 | 1885.342228 | 657.99 | 987.727 | 841.835 | 1534.635926 | 1327.608711 | 1190.97244 | 1063.645349 |
| 658.278 | 1723.23 | 1497.29 | 2208.003492 | 1891.964559 | 658.278 | 1023.95 | 845.189 | 1545.919145 | 1332.869204 | 1180.80059 | 1067.971561 |
| 658.567 | 1804.31 | 1502.57 | 2194.877086 | 1898.598716 | 658.567 | 1008.98 | 848.552 | 1566.488744 | 1338.140432 | 1220.194862 | 1072.311627 |
| 658.856 | 1761.1 | 1507.86 | 2173.6738 | 1905.244699 | 658.856 | 1004.17 | 851.925 | 1581.658286 | 1343.433131 | 1216.13074 | 1076.664394 |
| 659.144 | 1767.71 | 1513.16 | 2217.889687 | 1911.914332 | 659.144 | 1012.07 | 855.307 | 1591.545865 | 1348.736567 | 1302.885887 | 1081.028707 |
| 659.433 | 1771.9 | 1518.48 | 2202.764755 | 1918.595791 | 659.433 | 1093.3 | 858.698 | 1580.917523 | 1354.061473 | 1178.387517 | 1085.406875 |
| 659.722 | 1778.44 | 1523.81 | 2220.290282 | 1925.289076 | 659.722 | 1000.28 | 862.1 | 1619.523096 | 1359.38638 | 1234.280739 | 1089.796588 |
| 660.01 | 1778.27 | 1529.15 | 2177.576245 | 1931.994187 | 660.01 | 1045.16 | 865.51 | 1578.265805 | 1364.732758 | 1236.001063 | 1094.200157 |
| 660.299 | 1783.29 | 1534.5 | 2193.552619 | 1938.722948 | 660.299 | 1048.84 | 868.93 | 1566.767872 | 1370.100607 | 1201.178928 | 1098.615271 |
| 660.588 | 1811.67 | 1539.87 | 2276.816609 | 1945.463535 | 660.588 | 1058.93 | 872.36 | 1601.476386 | 1375.479192 | 1264.496099 | 1103.04424 |
| 660.876 | 1822.08 | 1545.24 | 2300.82256 | 1952.227774 | 660.876 | 1083.07 | 875.799 | 1578.512726 | 1380.868513 | 1208.48742 | 1107.484755 |
| 661.165 | 1793.41 | 1550.63 | 2286.395338 | 1959.003838 | 661.165 | 1072.88 | 879.248 | 1587.541449 | 1386.279305 | 1333.920999 | 1111.939125 |
| 661.453 | 1844.45 | 1556.04 | 2285.307384 | 1965.791727 | 661.453 | 1038.94 | 882.707 | 1620.500044 | 1391.700833 | 1292.806403 | 1116.405041 |
| 661.742 | 1758.59 | 1561.45 | 2270.750081 | 1972.591442 | 661.742 | 1019.13 | 886.175 | 1651.0109 | 1397.143832 | 1237.698296 | 1120.884811 |
| 662.03 | 1821.01 | 1566.88 | 2281.32216 | 1979.414808 | 662.03 | 1092.08 | 889.652 | 1691.817291 | 1402.597567 | 1343.550197 | 1125.377282 |
| 662.319 | 1853.79 | 1572.32 | 2301.165502 | 1986.238175 | 662.319 | 1080.2 | 893.14 | 1653.973953 | 1408.062038 | 1268.560221 | 1129.882454 |
| 662.607 | 1838.67 | 1577.77 | 2342.495944 | 1993.097018 | 662.607 | 1047.6 | 896.636 | 1664.312431 | 1413.537244 | 1253.666138 | 1134.400326 |
| 662.896 | 1879.48 | 1583.23 | 2259.586722 | 1999.955861 | 662.896 | 1068 | 900.143 | 1609.313446 | 1419.044658 | 1331.484835 | 1138.930898 |
| 663.184 | 1909.67 | 1588.71 | 2357.39619 | 2006.838355 | 663.184 | 1059.47 | 903.659 | 1632.298577 | 1424.552071 | 1337.650293 | 1143.47417 |
| 663.473 | 1887.65 | 1594.2 | 2380.869989 | 2013.732675 | 663.473 | 1099.09 | 907.185 | 1669.562187 | 1430.080956 | 1274.471671 | 1148.031298 |
| 663.761 | 1825.86 | 1599.7 | 2379.510046 | 2020.650646 | 663.761 | 1122.14 | 910.72 | 1713.353102 | 1435.620576 | 1276.330545 | 1152.599971 |
| 664.05 | 1823.56 | 1605.22 | 2317.449342 | 2027.580443 | 664.05 | 1055.23 | 914.266 | 1710.669178 | 1441.181668 | 1317.733787 | 1157.177881 |
| 664.338 | 1892.34 | 1610.74 | 2333.295634 | 2034.522065 | 664.338 | 1076.79 | 917.821 | 1705.151029 | 1446.753496 | 1323.218042 | 1161.77311 |
| 664.627 | 1834.23 | 1616.28 | 2402.037798 | 2041.475513 | 664.627 | 1103.99 | 921.385 | 1704.292173 | 1452.336059 | 1399.558877 | 1166.39143 |
| 664.915 | 1884.43 | 1621.83 | 2339.338512 | 2048.452612 | 664.915 | 1074.53 | 924.959 | 1673.802788 | 1457.940094 | 1284.701251 | 1171.009751 |
| 665.203 | 1911.98 | 1627.4 | 2377.227707 | 2055.441536 | 665.203 | 1082.96 | 928.543 | 1679.106223 | 1463.5656 | 1317.248863 | 1175.639617 |
| 665.492 | 1853.08 | 1632.97 | 2365.544022 | 2062.442286 | 665.492 | 1134.48 | 932.137 | 1668.885838 | 1469.191106 | 1325.411744 | 1180.292574 |
| 665.78 | 1849.68 | 1638.56 | 2375.997845 | 2069.466688 | 665.78 | 1116.84 | 935.741 | 1726.665369 | 1474.838084 | 1295.346479 | 1184.945532 |
| 666.069 | 1898.3 | 1644.16 | 2350.762033 | 2076.502915 | 666.069 | 1099.18 | 939.354 | 1701.372063 | 1480.506533 | 1352.763746 | 1189.621581 |
| 666.357 | 1893.81 | 1649.78 | 2323.468568 | 2083.550967 | 666.357 | 1113.29 | 942.977 | 1717.378989 | 1486.185717 | 1309.270715 | 1194.309177 |
| 666.645 | 1915.39 | 1655.4 | 2387.078424 | 2090.622671 | 666.645 | 1133.47 | 946.61 | 1701.393534 | 1491.875638 | 1342.36098 | 1199.008318 |
| 666.934 | 1950.31 | 1661.04 | 2435.835338 | 2097.7062 | 666.934 | 1136.98 | 950.253 | 1762.919823 | 1497.587029 | 1402.191319 | 1203.719004 |
| 667.222 | 1922.79 | 1666.69 | 2325.573523 | 2104.801555 | 667.222 | 1132.77 | 953.905 | 1680.147586 | 1503.309157 | 1298.960315 | 1208.452783 |
| 667.51 | 1934.65 | 1672.36 | 2427.78211 | 2111.908736 | 667.51 | 1106.33 | 957.568 | 1720.481606 | 1509.052756 | 1305.148864 | 1213.186561 |
| 667.799 | 1917.36 | 1678.04 | 2414.147203 | 2119.039567 | 667.799 | 1119.25 | 961.24 | 1805.175534 | 1514.80709 | 1402.860976 | 1217.943431 |
| 668.087 | 1943.19 | 1683.72 | 2391.844138 | 2126.182225 | 668.087 | 1139.36 | 964.922 | 1717.808417 | 1520.57216 | 1397.73464 | 1222.700301 |
| 668.375 | 1943.91 | 1689.42 | 2452.320213 | 2133.348533 | 668.375 | 1197.92 | 968.614 | 1725.666949 | 1526.358702 | 1419.51002 | 1227.480262 |
| 668.663 | 1990.24 | 1695.14 | 2442.989821 | 2140.526667 | 668.663 | 1165.55 | 972.316 | 1765.754048 | 1532.166715 | 1425.179009 | 1232.271769 |
| 668.952 | 1995.79 | 1700.87 | 2472.636579 | 2147.716627 | 668.952 | 1166.23 | 976.027 | 1813.334665 | 1537.974728 | 1397.15735 | 1237.074823 |
| 669.24 | 1996.1 | 1706.61 | 2538.552426 | 2154.930238 | 669.24 | 1179.02 | 979.749 | 1799.678856 | 1543.814948 | 1394.905919 | 1241.900967 |
| 669.528 | 1945.19 | 1712.36 | 2498.156205 | 2162.143848 | 669.528 | 1151.01 | 983.48 | 1725.742099 | 1549.655168 | 1484.09723 | 1246.727112 |
| 669.816 | 1929.01 | 1718.12 | 2445.981696 | 2169.392936 | 669.816 | 1216.96 | 987.222 | 1811.166053 | 1555.51686 | 1445.74208 | 1251.576348 |
| 670.105 | 2004.89 | 1723.9 | 2434.936593 | 2176.642024 | 670.105 | 1173.61 | 990.973 | 1761.975082 | 1561.400023 | 1407.02901 | 1256.425585 |
| 670.393 | 1961.85 | 1729.69 | 2474.45772 | 2183.914762 | 670.393 | 1199.37 | 994.734 | 1853.153371 | 1567.293921 | 1401.394659 | 1261.297913 |
| 670.681 | 1969.7 | 1735.49 | 2514.014324 | 2191.199327 | 670.681 | 1193.38 | 998.506 | 1760.880041 | 1573.198556 | 1405.227865 | 1266.181786 |
| 670.969 | 2019.93 | 1741.3 | 2531.800013 | 2198.495717 | 670.969 | 1195.14 | 1002.29 | 1781.202718 | 1579.124661 | 1407.306109 | 1271.088752 |
| 671.257 | 2022.6 | 1747.13 | 2511.921194 | 2205.815758 | 671.257 | 1165.04 | 1006.08 | 1784.799177 | 1585.061503 | 1412.570994 | 1275.995717 |
| 671.545 | 2019.45 | 1752.97 | 2481.907843 | 2213.147625 | 671.545 | 1210.92 | 1009.88 | 1783.178087 | 1591.019815 | 1365.441035 | 1280.914228 |
| 671.833 | 1996.27 | 1758.82 | 2550.271587 | 2220.503142 | 671.833 | 1196.22 | 1013.69 | 1865.88591 | 1596.988864 | 1505.480053 | 1285.855831 |
| 672.122 | 2067.84 | 1764.69 | 2593.565078 | 2227.85866 | 672.122 | 1203.33 | 1017.51 | 1844.446719 | 1602.979384 | 1428.400287 | 1290.808979 |
| 672.41 | 2066.96 | 1770.57 | 2575.566527 | 2235.23783 | 672.41 | 1184.28 | 1021.34 | 1794.987355 | 1608.980639 | 1458.130724 | 1295.773674 |
| 672.698 | 2050.28 | 1776.45 | 2525.686182 | 2242.64065 | 672.698 | 1218.36 | 1025.18 | 1865.58531 | 1614.992631 | 1462.656678 | 1300.749914 |
| 672.986 | 2046.65 | 1782.36 | 2576.169633 | 2250.055296 | 672.986 | 1223.63 | 1029.04 | 1860.936753 | 1621.026093 | 1486.060016 | 1305.749246 |
| 673.274 | 2085.26 | 1788.27 | 2613.952397 | 2257.481767 | 673.274 | 1228.07 | 1032.9 | 1860.228197 | 1627.070292 | 1486.394844 | 1310.748577 |
| 673.562 | 2099.18 | 1794.2 | 2637.532627 | 2264.920064 | 673.562 | 1229.23 | 1036.77 | 1910.428324 | 1633.135961 | 1555.635011 | 1315.771001 |
| 673.85 | 2016.84 | 1800.14 | 2646.969449 | 2272.382012 | 673.85 | 1217.34 | 1040.65 | 1883.610548 | 1639.223103 | 1487.572516 | 1320.80497 |
| 674.138 | 2058.96 | 1806.1 | 2656.524527 | 2279.855786 | 674.138 | 1242.33 | 1044.54 | 1855.332718 | 1645.310244 | 1461.559827 | 1325.850485 |
| 674.426 | 2115.18 | 1812.06 | 2600.305665 | 2287.341386 | 674.426 | 1172.86 | 1048.44 | 1900.315296 | 1651.429592 | 1525.431197 | 1330.907545 |
| 674.714 | 2105.9 | 1818.04 | 2615.962748 | 2294.850636 | 674.714 | 1244.3 | 1052.35 | 1819.003114 | 1657.54894 | 1472.863166 | 1335.976152 |
| 675.002 | 2071.95 | 1824.03 | 2656.146108 | 2302.371712 | 675.002 | 1213.86 | 1056.28 | 1895.720417 | 1663.68976 | 1510.456293 | 1341.06785 |
| 675.29 | 2080.76 | 1830.04 | 2574.194759 | 2309.91644 | 675.29 | 1227.72 | 1060.21 | 1888.77442 | 1669.852051 | 1548.684439 | 1346.159548 |
| 675.578 | 2053.68 | 1836.05 | 2605.662658 | 2317.461167 | 675.578 | 1258.74 | 1064.15 | 1886.111966 | 1676.025078 | 1514.393411 | 1351.274338 |
| 675.866 | 2145.21 | 1842.08 | 2650.69451 | 2325.041371 | 675.866 | 1281.78 | 1068.1 | 1920.788273 | 1682.219576 | 1506.66927 | 1356.400673 |
| 676.154 | 2088.89 | 1848.12 | 2665.109906 | 2332.621576 | 676.154 | 1233.7 | 1072.07 | 1884.984718 | 1688.42481 | 1530.291979 | 1361.550101 |
| 676.442 | 2115.92 | 1854.18 | 2645.148308 | 2340.225431 | 676.442 | 1296.69 | 1076.04 | 1847.012551 | 1694.640779 | 1584.164685 | 1366.699528 |
| 676.73 | 2106.27 | 1860.25 | 2658.345668 | 2347.841112 | 676.73 | 1302.43 | 1080.02 | 1925.673016 | 1700.87822 | 1525.269555 | 1371.872047 |
| 677.018 | 2095.57 | 1866.33 | 2705.517952 | 2355.468618 | 677.018 | 1272.39 | 1084.01 | 1930.074653 | 1707.137133 | 1519.473563 | 1377.044565 |
| 677.305 | 2108.12 | 1872.42 | 2667.794316 | 2363.119776 | 677.305 | 1281.16 | 1088.02 | 1927.508821 | 1713.396045 | 1540.486921 | 1382.240176 |
| 677.593 | 2124.8 | 1878.53 | 2663.761789 | 2370.782759 | 677.593 | 1260.75 | 1092.03 | 1955.818858 | 1719.687165 | 1550.63568 | 1387.458878 |
| 677.881 | 2166.85 | 1884.65 | 2709.243014 | 2378.469394 | 677.881 | 1267.56 | 1096.05 | 1998.783124 | 1725.98902 | 1552.194363 | 1392.677579 |
| 678.169 | 2165.42 | 1890.78 | 2695.525328 | 2386.167854 | 678.169 | 1286.54 | 1100.09 | 2003.678603 | 1732.301611 | 1539.794173 | 1397.919373 |
| 678.457 | 2143.16 | 1896.92 | 2720.666535 | 2393.87814 | 678.457 | 1259.92 | 1104.13 | 1945.587737 | 1738.635673 | 1531.458105 | 1403.161167 |
| 678.745 | 2139.8 | 1903.08 | 2719.247464 | 2401.600251 | 678.745 | 1295.83 | 1108.18 | 1984.289931 | 1744.980471 | 1586.496937 | 1408.426052 |
| 679.032 | 2184.12 | 1909.25 | 2747.096732 | 2409.346013 | 679.032 | 1317.95 | 1112.25 | 1992.996583 | 1751.34674 | 1587.732338 | 1413.702483 |
| 679.32 | 2167.76 | 1915.43 | 2717.75744 | 2417.103601 | 679.32 | 1264.27 | 1116.33 | 2025.343243 | 1757.723745 | 1595.006192 | 1419.002005 |
| 679.608 | 2221.34 | 1921.62 | 2757.172136 | 2424.88484 | 679.608 | 1285.56 | 1120.41 | 2001.832063 | 1764.122222 | 1544.100756 | 1424.301528 |
| 679.896 | 2190.97 | 1927.83 | 2685.426273 | 2432.66608 | 679.896 | 1331.28 | 1124.51 | 2009.669123 | 1770.531434 | 1587.778521 | 1429.624142 |
| 680.183 | 2264.07 | 1934.05 | 2771.079032 | 2440.482796 | 680.183 | 1321.33 | 1128.61 | 2112.055482 | 1776.962117 | 1541.595318 | 1434.958302 |
| 680.471 | 2223.32 | 1940.29 | 2775.714664 | 2448.299512 | 680.471 | 1301 | 1132.73 | 2056.122491 | 1783.403537 | 1660.886531 | 1440.304008 |
| 680.759 | 2217.64 | 1946.53 | 2770.227589 | 2456.139879 | 680.759 | 1329.32 | 1136.85 | 1983.216361 | 1789.866427 | 1602.18768 | 1445.661259 |
| 681.047 | 2186.67 | 1952.79 | 2787.989628 | 2463.992072 | 681.047 | 1295.07 | 1140.99 | 2029.712672 | 1796.340054 | 1596.807337 | 1451.041602 |
| 681.334 | 2213.05 | 1959.06 | 2819.694038 | 2471.867916 | 681.334 | 1348.06 | 1145.14 | 2078.302445 | 1802.835151 | 1595.2602 | 1456.421945 |
| 681.622 | 2236.89 | 1965.35 | 2791.430875 | 2479.755585 | 681.622 | 1409.68 | 1149.3 | 2053.438567 | 1809.340985 | 1598.493024 | 1461.82538 |
| 681.91 | 2210.8 | 1971.65 | 2847.898074 | 2487.65508 | 681.91 | 1348.66 | 1153.46 | 2076.037212 | 1815.868289 | 1618.559626 | 1467.251907 |
| 682.197 | 2246.65 | 1977.96 | 2825.559532 | 2495.578226 | 682.197 | 1355.95 | 1157.64 | 2080.481792 | 1822.40633 | 1621.411438 | 1472.678433 |
| 682.485 | 2147.26 | 1984.28 | 2734.869071 | 2503.513198 | 682.485 | 1391.98 | 1161.83 | 2046.084613 | 1828.965842 | 1593.689971 | 1478.116505 |
| 682.773 | 2223.01 | 1990.62 | 2761.476651 | 2511.459996 | 682.773 | 1358.39 | 1166.03 | 2019.53523 | 1835.53609 | 1545.901901 | 1483.577669 |
| 683.06 | 2246.65 | 1996.96 | 2737.541654 | 2519.430445 | 683.06 | 1369.77 | 1170.24 | 2031.816869 | 1842.127809 | 1656.926322 | 1489.050378 |
| 683.348 | 2289.59 | 2003.33 | 2791.986678 | 2527.412719 | 683.348 | 1404.02 | 1174.46 | 2099.806049 | 1848.730263 | 1597.950371 | 1494.534634 |
| 683.636 | 2275.88 | 2009.7 | 2828.634186 | 2535.406819 | 683.636 | 1411.52 | 1178.69 | 2055.134807 | 1855.354189 | 1669.187962 | 1500.041981 |
| 683.923 | 2311.86 | 2016.09 | 2890.623936 | 2543.42457 | 683.923 | 1444.38 | 1182.93 | 2130.62824 | 1861.988851 | 1681.287961 | 1505.549328 |
| 684.211 | 2352.81 | 2022.49 | 2862.810145 | 2551.454146 | 684.211 | 1426.73 | 1187.18 | 2149.909555 | 1868.644984 | 1628.084911 | 1511.079766 |
| 684.498 | 2322.74 | 2028.9 | 2900.640212 | 2559.495549 | 684.498 | 1032.13 | 1191.44 | 2135.212384 | 1875.311853 | 1653.277848 | 1516.621751 |
| 684.786 | 2383.01 | 2035.32 | 2879.046682 | 2567.560602 | 684.786 | 1267.85 | 1195.71 | 2187.194637 | 1882.000194 | 1730.242156 | 1522.186827 |
| 685.073 | 2377.5 | 2041.76 | 2950.816197 | 2575.637481 | 685.073 | 1430.45 | 1200 | 2158.165308 | 1888.69927 | 1778.468966 | 1527.751903 |
| 685.361 | 2331.43 | 2048.21 | 2864.347472 | 2583.738011 | 685.361 | 1479.34 | 1204.29 | 2188.536599 | 1895.419817 | 1780.558756 | 1533.34007 |
| 685.649 | 2365.92 | 2054.68 | 2888.116911 | 2591.850367 | 685.649 | 1523.54 | 1208.59 | 2172.991308 | 1902.1511 | 1669.338057 | 1538.939784 |
| 685.936 | 2345.94 | 2061.16 | 2899.173839 | 2599.974548 | 685.936 | 1436.11 | 1212.91 | 2196.083796 | 1908.903855 | 1736.384522 | 1544.551043 |
| 686.223 | 2340.11 | 2067.64 | 2898.547083 | 2608.110555 | 686.223 | 1484.11 | 1217.23 | 2205.305761 | 1915.667345 | 1673.586912 | 1550.173848 |
| 686.511 | 2374.5 | 2074.15 | 2933.976555 | 2616.270213 | 686.511 | 1527.65 | 1221.57 | 2151.305196 | 1922.452306 | 1718.003608 | 1555.819744 |
| 686.798 | 2389.84 | 2080.66 | 2946.488031 | 2624.453523 | 686.798 | 1517.28 | 1225.91 | 2223.492034 | 1929.248004 | 1765.676219 | 1561.477187 |
| 687.086 | 2414.54 | 2087.19 | 2941.793271 | 2632.636832 | 687.086 | 1540.8 | 1230.27 | 2253.884797 | 1936.065172 | 1890.012947 | 1567.146175 |
| 687.373 | 2375.08 | 2093.73 | 2991.389802 | 2640.843793 | 687.373 | 1346.74 | 1234.64 | 2272.543442 | 1942.903813 | 1808.23404 | 1572.826709 |
| 687.661 | 2453.95 | 2100.29 | 2953.051234 | 2649.062579 | 687.661 | 1519.56 | 1239.01 | 2256.332537 | 1949.753188 | 1732.355038 | 1578.530334 |
| 687.948 | 2447.65 | 2106.85 | 2949.17244 | 2657.305016 | 687.948 | 1483.76 | 1243.4 | 2249.569047 | 1956.6133 | 1731.673836 | 1584.245506 |
| 688.236 | 2428.75 | 2113.43 | 3052.811924 | 2665.559279 | 688.236 | 1532.7 | 1247.8 | 2184.703955 | 1963.494883 | 1852.569915 | 1589.972223 |
| 688.523 | 2458.3 | 2120.03 | 2983.632213 | 2673.825367 | 688.523 | 1507.29 | 1252.21 | 2208.60162 | 1970.397937 | 1819.017818 | 1595.710486 |
| 688.81 | 2411.77 | 2126.63 | 3058.23987 | 2682.115107 | 688.81 | 1526.75 | 1256.63 | 2287.57342 | 1977.311727 | 1897.102069 | 1601.460295 |
| 689.098 | 2489.21 | 2133.25 | 3040.820774 | 2690.416672 | 689.098 | 1546.21 | 1261.06 | 2266.574393 | 1984.236253 | 1814.688143 | 1607.233195 |
| 689.385 | 2493.48 | 2139.88 | 2992.678791 | 2698.741889 | 689.385 | 1579.79 | 1265.5 | 2252.736078 | 1991.18225 | 1828.266005 | 1613.017641 |
| 689.672 | 2434.55 | 2146.53 | 3037.828899 | 2707.078931 | 689.672 | 1559.31 | 1269.95 | 2299.221653 | 1998.149718 | 1812.482895 | 1618.813633 |
| 689.96 | 2460.45 | 2153.18 | 3041.258321 | 2715.427798 | 689.96 | 1602.77 | 1274.41 | 2327.145206 | 2005.127922 | 1832.410947 | 1624.632717 |
| 690.247 | 2443.07 | 2159.85 | 3096.933206 | 2723.788491 | 690.247 | 1560.71 | 1278.89 | 2271.190744 | 2012.127598 | 1765.941772 | 1630.4518 |
| 690.534 | 2488.87 | 2166.54 | 2989.734219 | 2732.172836 | 690.534 | 1523.88 | 1283.37 | 2263.450305 | 2019.138009 | 1867.718006 | 1636.293976 |
| 690.822 | 2467.04 | 2173.23 | 3086.810499 | 2740.569006 | 690.822 | 1544.56 | 1287.86 | 2296.956421 | 2026.169892 | 1854.405698 | 1642.159242 |
| 691.109 | 2669.87 | 2179.94 | 3078.260597 | 2748.988827 | 691.109 | 1602.06 | 1292.37 | 2376.250292 | 2033.21251 | 1891.606268 | 1648.024509 |
| 691.396 | 2861.14 | 2186.66 | 3069.911729 | 2757.420474 | 691.396 | 1616.68 | 1296.88 | 2358.697424 | 2040.2766 | 1927.294338 | 1653.912868 |
| 691.683 | 2457.78 | 2193.4 | 3068.480833 | 2765.863946 | 691.683 | 1612.73 | 1301.41 | 2378.536995 | 2047.362161 | 1775.513241 | 1659.812772 |
| 691.971 | 2532.24 | 2200.15 | 3074.050686 | 2774.331069 | 691.971 | 1582.95 | 1305.95 | 2315.282258 | 2054.458458 | 1861.148445 | 1665.724222 |
| 692.258 | 2551.15 | 2206.91 | 3123.3161 | 2782.810019 | 692.258 | 1600.64 | 1310.49 | 2371.773505 | 2061.565491 | 1813.695204 | 1671.647217 |
| 692.545 | 2528.57 | 2213.68 | 3106.665668 | 2791.300793 | 692.545 | 1645.56 | 1315.05 | 2411.903547 | 2068.693994 | 1876.966192 | 1677.593305 |
| 692.832 | 2528.14 | 2220.47 | 3102.443931 | 2799.815219 | 692.832 | 1627.69 | 1319.62 | 2387.308061 | 2075.84397 | 1898.141191 | 1683.550938 |
| 693.119 | 2582.15 | 2227.27 | 3113.098789 | 2808.341471 | 693.119 | 1672.41 | 1324.2 | 2346.190335 | 2083.004681 | 1890.301592 | 1689.520117 |
| 693.407 | 2563.68 | 2234.08 | 3196.658419 | 2816.891373 | 693.407 | 1657.42 | 1328.79 | 2400.470028 | 2090.186863 | 1958.017714 | 1695.512388 |
| 693.694 | 2541.3 | 2240.9 | 3116.078838 | 2825.441276 | 693.694 | 1640.99 | 1333.39 | 2469.060407 | 2097.379781 | 1873.340811 | 1701.504658 |
| 693.981 | 2538.08 | 2247.74 | 3106.3109 | 2834.026655 | 693.981 | 1635.06 | 1338.01 | 2290.472059 | 2104.594171 | 1907.481744 | 1707.52002 |
| 694.268 | 2589.49 | 2254.59 | 3117.923631 | 2842.612035 | 694.268 | 1633.04 | 1342.63 | 2391.011877 | 2111.819296 | 1906.06161 | 1713.558474 |
| 694.555 | 2554.62 | 2261.46 | 3125.40923 | 2851.221065 | 694.555 | 1658.84 | 1347.26 | 2397.732425 | 2119.065893 | 1974.943858 | 1719.596928 |
| 694.842 | 2544.66 | 2268.33 | 3201.424132 | 2859.841922 | 694.842 | 1611.07 | 1351.91 | 2418.892487 | 2126.333961 | 1916.9493 | 1725.658473 |
| 695.129 | 2585.39 | 2275.22 | 3148.729296 | 2868.486429 | 695.129 | 1646.15 | 1356.56 | 2476.307004 | 2133.612765 | 1870.350449 | 1731.731565 |
| 695.416 | 2588.32 | 2282.13 | 3172.143968 | 2877.142762 | 695.416 | 1711.79 | 1361.23 | 2422.875431 | 2140.91304 | 1958.006168 | 1737.816202 |
| 695.703 | 2613.65 | 2289.04 | 3119.898504 | 2885.810921 | 695.703 | 1648.76 | 1365.9 | 2395.016293 | 2148.224051 | 1920.955693 | 1743.92393 |
| 695.99 | 2564.26 | 2295.97 | 3215.472935 | 2894.50273 | 695.99 | 1700.16 | 1370.59 | 2503.028158 | 2155.556533 | 1934.314185 | 1750.043205 |
| 696.277 | 2652.68 | 2302.91 | 3294.408758 | 2903.206366 | 696.277 | 1688.55 | 1375.29 | 2457.991902 | 2162.899751 | 2023.159122 | 1756.174025 |
| 696.564 | 2617.88 | 2309.87 | 3254.012538 | 2911.933652 | 696.564 | 1656.14 | 1380 | 2553.432263 | 2170.26444 | 1940.040902 | 1762.316391 |
| 696.851 | 2645.61 | 2316.84 | 3215.804052 | 2920.660939 | 696.851 | 1641.85 | 1384.72 | 2440.095492 | 2177.639865 | 1933.263517 | 1768.481848 |
| 697.138 | 2652.82 | 2323.82 | 3219.48181 | 2929.423702 | 697.138 | 1723.8 | 1389.45 | 2492.850715 | 2185.036762 | 2026.819141 | 1774.658852 |
| 697.425 | 2656.13 | 2330.81 | 3250.523988 | 2938.186465 | 697.425 | 1661.24 | 1394.19 | 2414.973957 | 2192.455129 | 1993.636509 | 1780.847401 |
| 697.712 | 2675.07 | 2337.82 | 3235.328103 | 2946.97288 | 697.712 | 1688.13 | 1398.95 | 2506.903745 | 2199.884233 | 1967.508362 | 1787.047496 |
| 697.999 | 2685.13 | 2344.84 | 3261.084241 | 2955.77112 | 697.999 | 1665.65 | 1403.71 | 2541.322395 | 2207.324072 | 1951.806073 | 1793.270683 |
| 698.286 | 2665.28 | 2351.87 | 3276.682197 | 2964.593011 | 698.286 | 1687.29 | 1408.48 | 2571.468237 | 2214.796118 | 1979.931644 | 1799.505415 |
| 698.573 | 2705.96 | 2358.91 | 3295.283852 | 2973.426728 | 698.573 | 1667.2 | 1413.27 | 2579.552218 | 2222.2789 | 2091.510262 | 1805.751693 |
| 698.86 | 2720.89 | 2365.97 | 3222.213522 | 2982.27227 | 698.86 | 1727.24 | 1418.07 | 2539.894547 | 2229.772418 | 2025.44519 | 1812.021063 |
| 699.147 | 2699.92 | 2373.04 | 3331.328255 | 2991.141464 | 699.147 | 1705.17 | 1422.87 | 2541.386809 | 2237.287407 | 2012.814084 | 1818.301979 |
| 699.434 | 2716.68 | 2380.13 | 3326.03039 | 3000.022483 | 699.434 | 1705.77 | 1427.69 | 2599.220018 | 2244.823868 | 2032.661316 | 1824.59444 |
| 699.721 | 2669.33 | 2387.23 | 3438.763754 | 3008.927154 | 699.721 | 1724.79 | 1432.52 | 2575.966495 | 2252.371064 | 2016.959027 | 1830.898447 |
| 700.008 | 2775.86 | 2394.34 | 3406.420761 | 3017.831824 | 700.008 | 1731.92 | 1437.36 | 2572.906821 | 2259.928996 | 2099.176674 | 1837.225546 |
| 700.294 | 2771.7 | 2401.46 | 3442.051268 | 3026.771971 | 700.294 | 1727.83 | 1442.21 | 2622.140735 | 2267.519135 | 2046.100628 | 1843.564191 |
| 700.581 | 2765.36 | 2408.6 | 3376.312805 | 3035.712118 | 700.581 | 1801.44 | 1447.08 | 2633.134091 | 2275.109274 | 2137.889744 | 1849.914381 |
| 700.868 | 2785.68 | 2415.75 | 3448.011366 | 3044.675917 | 700.868 | 1759.77 | 1451.95 | 2714.25303 | 2282.73162 | 2110.168276 | 1856.287663 |
| 701.155 | 2789.79 | 2422.91 | 3403.949212 | 3053.651541 | 701.155 | 1765.92 | 1456.83 | 2586.272766 | 2290.364702 | 2128.572283 | 1862.672491 |
| 701.442 | 2763.88 | 2430.08 | 3460.58197 | 3062.650816 | 701.442 | 1846.22 | 1461.73 | 2570.58791 | 2298.019255 | 2121.506253 | 1869.068865 |
| 701.729 | 2777.15 | 2437.27 | 3463.822182 | 3071.661917 | 701.729 | 1774.75 | 1466.63 | 2642.935784 | 2305.684544 | 2176.348806 | 1875.476784 |
| 702.015 | 2803.11 | 2444.47 | 3483.618222 | 3080.684843 | 702.015 | 1732.68 | 1471.55 | 2645.394259 | 2313.371304 | 2129.634496 | 1881.907795 |
| 702.302 | 2829.35 | 2451.69 | 3444.144398 | 3089.73142 | 702.302 | 1780.8 | 1476.48 | 2719.54573 | 2321.0688 | 2128.930202 | 1888.350352 |
| 702.589 | 2836.3 | 2458.91 | 3489.306332 | 3098.789824 | 702.589 | 1841.15 | 1481.42 | 2740.072386 | 2328.787767 | 2190.434683 | 1894.804455 |
| 702.876 | 2805.6 | 2466.16 | 3553.401038 | 3107.871878 | 702.876 | 1738.87 | 1486.37 | 2674.348438 | 2336.528206 | 2144.69022 | 1901.281649 |
| 703.162 | 2779.63 | 2473.41 | 3460.487365 | 3116.965758 | 703.162 | 1752.38 | 1491.33 | 2701.702998 | 2344.279381 | 2150.95959 | 1907.770389 |
| 703.449 | 2870.69 | 2480.68 | 3592.118024 | 3126.071463 | 703.449 | 1827.55 | 1496.3 | 2714.693194 | 2352.041291 | 2167.031345 | 1914.270675 |
| 703.736 | 2855.33 | 2487.95 | 3477.291531 | 3135.188994 | 703.736 | 1795.13 | 1501.28 | 2740.90977 | 2359.835408 | 2173.831822 | 1920.782506 |
| 704.022 | 2869.97 | 2495.25 | 3497.631548 | 3144.330176 | 704.022 | 1797.14 | 1506.28 | 2683.892474 | 2367.640261 | 2136.561977 | 1927.31743 |
| 704.309 | 2883.82 | 2502.55 | 3592.259931 | 3153.49501 | 704.309 | 1788.18 | 1511.28 | 2614.378825 | 2375.45585 | 2200.606534 | 1933.863899 |
| 704.596 | 2844.9 | 2509.87 | 3563.677477 | 3162.659843 | 704.596 | 1827.03 | 1516.3 | 2651.234479 | 2383.29291 | 2140.764648 | 1940.433459 |
| 704.882 | 2850.19 | 2517.2 | 3542.143075 | 3171.848328 | 704.882 | 1794.54 | 1521.33 | 2676.484842 | 2391.151441 | 2213.468556 | 1947.00302 |
| 705.169 | 2849.71 | 2524.55 | 3567.840085 | 3181.060464 | 705.169 | 1859.83 | 1526.36 | 2698.664796 | 2399.020709 | 2207.695655 | 1953.595672 |
| 705.456 | 2882.04 | 2531.9 | 3583.721854 | 3190.284425 | 705.456 | 1817.25 | 1531.41 | 2650.826522 | 2406.911447 | 2190.446229 | 1960.211416 |
| 705.742 | 2867.65 | 2539.27 | 3583.639075 | 3199.520212 | 705.742 | 1842.46 | 1536.47 | 2736.583284 | 2414.823657 | 2208.653957 | 1966.82716 |
| 706.029 | 2914.04 | 2546.66 | 3506.394312 | 3208.767825 | 706.029 | 1833.98 | 1541.55 | 2766.042041 | 2422.746603 | 2199.267221 | 1973.465995 |
| 706.315 | 2897.17 | 2554.05 | 3675.192805 | 3218.039088 | 706.315 | 1791.63 | 1546.63 | 2737.83936 | 2430.680284 | 2127.879535 | 1980.116376 |
| 706.602 | 2887.26 | 2561.46 | 3627.251857 | 3227.322178 | 706.602 | 1830.1 | 1551.72 | 2736.980504 | 2438.646173 | 2180.031917 | 1986.789849 |
| 706.888 | 2903.2 | 2568.88 | 3650.524621 | 3236.628918 | 706.888 | 1828.29 | 1556.83 | 2849.093406 | 2446.622797 | 2200.860541 | 1993.474868 |
| 707.175 | 2965.16 | 2576.32 | 3640.638426 | 3245.947484 | 707.175 | 1780.81 | 1561.94 | 2791.238726 | 2454.610157 | 2209.069606 | 2000.171432 |
| 707.462 | 2952.42 | 2583.77 | 3745.034748 | 3255.277876 | 707.462 | 1884.05 | 1567.07 | 2780.868041 | 2462.618988 | 2329.342212 | 2006.879543 |
| 707.748 | 2930.21 | 2591.23 | 3684.440417 | 3264.631919 | 707.748 | 1820.94 | 1572.21 | 2790.594584 | 2470.64929 | 2241.744222 | 2013.610744 |
| 708.034 | 2949.42 | 2598.7 | 3698.796685 | 3273.997787 | 708.034 | 1840.15 | 1577.36 | 2819.967456 | 2478.690329 | 2250.472847 | 2020.353492 |
| 708.321 | 2913.9 | 2606.19 | 3634.654677 | 3283.375481 | 708.321 | 1819.68 | 1582.52 | 2857.585344 | 2486.752839 | 2249.549183 | 2027.107786 |
| 708.607 | 3016.76 | 2613.69 | 3665.933367 | 3292.776827 | 708.607 | 1868.99 | 1587.69 | 2897.243015 | 2494.83682 | 2356.220836 | 2033.885171 |
| 708.894 | 2975.26 | 2621.2 | 3692.079749 | 3302.189997 | 708.894 | 1820.68 | 1592.87 | 2804.411428 | 2502.931537 | 2281.242406 | 2040.674101 |
| 709.18 | 2973.32 | 2628.73 | 3684.724231 | 3311.626819 | 709.18 | 1814.62 | 1598.07 | 2827.847458 | 2511.047725 | 2265.574754 | 2047.474578 |
| 709.467 | 2995.64 | 2636.27 | 3748.546949 | 3321.075467 | 709.467 | 1779.31 | 1603.28 | 2890.146718 | 2519.174649 | 2302.902328 | 2054.298146 |
| 709.753 | 2981.21 | 2643.82 | 3791.497497 | 3330.53594 | 709.753 | 1808.53 | 1608.49 | 2866.614067 | 2527.323044 | 2281.219314 | 2061.13326 |
| 710.04 | 3003.12 | 2651.39 | 3742.752409 | 3340.020065 | 710.04 | 1844.8 | 1613.72 | 2896.620344 | 2535.492911 | 2241.9405 | 2067.97992 |
| 710.326 | 3026.38 | 2658.97 | 3704.035422 | 3349.516015 | 710.326 | 1877.41 | 1618.96 | 2904.017241 | 2543.673513 | 2290.02876 | 2074.849672 |
| 710.612 | 3029.08 | 2666.56 | 3776.928368 | 3359.02379 | 710.612 | 1885.69 | 1624.21 | 2956.826143 | 2551.864851 | 2369.059767 | 2081.730969 |
| 710.899 | 3012.37 | 2674.16 | 3713.365814 | 3368.555217 | 710.899 | 1912.03 | 1629.47 | 2904.91904 | 2560.088397 | 2327.056144 | 2088.623812 |
| 711.185 | 3096.12 | 2681.78 | 3755.796036 | 3378.098469 | 711.185 | 1873.57 | 1634.74 | 2918.628527 | 2568.322677 | 2394.425891 | 2095.528201 |
| 711.471 | 3036.64 | 2689.41 | 3851.92627 | 3387.653547 | 711.471 | 1920.38 | 1640.03 | 2895.761489 | 2576.57843 | 2282.200708 | 2102.455681 |
| 711.758 | 2996.39 | 2697.05 | 3806.196707 | 3397.232276 | 711.758 | 1865.39 | 1645.32 | 2845.486211 | 2584.844918 | 2308.69832 | 2109.394708 |
| 712.044 | 3069.22 | 2704.71 | 3865.998724 | 3406.822831 | 712.044 | 1965.2 | 1650.63 | 3017.107091 | 2593.132877 | 2431.487911 | 2116.356825 |
| 712.33 | 3091.88 | 2712.38 | 3882.578203 | 3416.437037 | 712.33 | 1944.07 | 1655.95 | 2916.878608 | 2601.431572 | 2379.866636 | 2123.330489 |
| 712.616 | 3162.08 | 2720.06 | 3844.405194 | 3426.063068 | 712.616 | 1930.39 | 1661.28 | 3027.885733 | 2609.762475 | 2458.747547 | 2130.315699 |
| 712.903 | 3147.05 | 2727.76 | 3839.249236 | 3435.700926 | 712.903 | 1942.18 | 1666.62 | 2978.168712 | 2618.093377 | 2357.167592 | 2137.312454 |
| 713.189 | 3120.76 | 2735.46 | 3845.587753 | 3445.362434 | 713.189 | 1919.12 | 1671.97 | 3002.796405 | 2626.456486 | 2403.893448 | 2144.332301 |
| 713.475 | 3197.99 | 2743.18 | 3917.960373 | 3455.035768 | 713.475 | 1921.08 | 1677.33 | 2905.799367 | 2634.830331 | 2397.035242 | 2151.363693 |
| 713.761 | 3091.84 | 2750.92 | 3856.751111 | 3464.720927 | 713.761 | 2025.11 | 1682.7 | 3022.464205 | 2643.214912 | 2421.223695 | 2158.418178 |
| 714.048 | 3079.72 | 2758.67 | 3809.377792 | 3474.429738 | 714.048 | 1906.27 | 1688.09 | 3024.99783 | 2651.6317 | 2396.457952 | 2165.484208 |
| 714.334 | 3130.95 | 2766.43 | 3919.663258 | 3484.150374 | 714.334 | 1972.76 | 1693.49 | 3011.986163 | 2660.059223 | 2370.664633 | 2172.561784 |
| 714.62 | 3134.53 | 2774.2 | 3953.318891 | 3493.882836 | 714.62 | 1965.75 | 1698.89 | 3021.401371 | 2668.497482 | 2384.242495 | 2179.650905 |
| 714.906 | 3168.77 | 2781.99 | 3903.178383 | 3503.638949 | 714.906 | 2000.78 | 1704.31 | 3019.554831 | 2676.957213 | 2431.949743 | 2186.763118 |
| 715.192 | 3155.79 | 2789.78 | 3889.42522 | 3513.406887 | 715.192 | 1989.32 | 1709.74 | 2972.661299 | 2685.438415 | 2466.286955 | 2193.886878 |
| 715.479 | 3162.07 | 2797.6 | 3950.717261 | 3523.198477 | 715.479 | 1928.03 | 1715.19 | 3029.055924 | 2693.930353 | 2526.98323 | 2201.033728 |
| 715.765 | 3149.08 | 2805.42 | 3891.754862 | 3533.001892 | 715.765 | 1981.07 | 1720.64 | 3060.812121 | 2702.454497 | 2488.535713 | 2208.192125 |
| 716.051 | 3188.3 | 2813.26 | 3923.447447 | 3542.817133 | 716.051 | 2028.76 | 1726.11 | 3023.720282 | 2710.978642 | 2485.857087 | 2215.362067 |
| 716.337 | 3193.92 | 2821.11 | 3989.789016 | 3552.656026 | 716.337 | 1961.52 | 1731.58 | 2977.664134 | 2719.524258 | 2500.577983 | 2222.543555 |
| 716.623 | 3153.34 | 2828.98 | 3916.127406 | 3562.506743 | 716.623 | 1939.75 | 1737.07 | 3042.625847 | 2728.091346 | 2414.469401 | 2229.748135 |
| 716.909 | 3198.1 | 2836.85 | 3949.641132 | 3572.369287 | 716.909 | 2027.98 | 1742.57 | 3049.496694 | 2736.679905 | 2513.763288 | 2236.96426 |
| 717.195 | 3176.15 | 2844.74 | 4030.374446 | 3582.255481 | 717.195 | 2031.73 | 1748.08 | 3048.519746 | 2745.2792 | 2444.88104 | 2244.203477 |
| 717.481 | 3181.14 | 2852.65 | 4056.367096 | 3592.153501 | 717.481 | 1956.83 | 1753.6 | 3056.872119 | 2753.899966 | 2479.749359 | 2251.45424 |
| 717.767 | 3178.17 | 2860.56 | 4008.57988 | 3602.063347 | 717.767 | 2000.35 | 1759.13 | 3099.406958 | 2762.531467 | 2485.533805 | 2258.716549 |
| 718.053 | 3172.7 | 2868.49 | 4033.697437 | 3611.996844 | 718.053 | 1980.41 | 1764.67 | 3029.657123 | 2771.195176 | 2582.414619 | 2265.990403 |
| 718.339 | 3240.13 | 2876.43 | 4013.475675 | 3621.953992 | 718.339 | 2037.35 | 1770.23 | 3209.608905 | 2779.858885 | 2515.529795 | 2273.287349 |
| 718.625 | 3212.62 | 2884.39 | 4057.431399 | 3631.91114 | 718.625 | 2090.17 | 1775.8 | 3072.417411 | 2788.554801 | 2495.486285 | 2280.607387 |
| 718.911 | 3274.2 | 2892.36 | 4102.569682 | 3641.891939 | 718.911 | 2118.89 | 1781.38 | 3180.987532 | 2797.261453 | 2546.414812 | 2287.927425 |
| 719.197 | 3218.87 | 2900.34 | 3967.320392 | 3651.884564 | 719.197 | 2059.29 | 1786.97 | 3165.603276 | 2805.97884 | 2446.428178 | 2295.270554 |
| 719.483 | 3294.17 | 2908.33 | 4133.008754 | 3661.90084 | 719.483 | 1974.28 | 1792.57 | 3072.578447 | 2814.728435 | 2529.280844 | 2302.625229 |
| 719.769 | 3255.41 | 2916.34 | 4039.480151 | 3671.928941 | 719.769 | 2093.16 | 1798.18 | 3074.478665 | 2823.488765 | 2524.281512 | 2310.002996 |
| 720.055 | 3265.15 | 2924.36 | 4156.281518 | 3681.968869 | 720.055 | 2052.37 | 1803.8 | 3142.403431 | 2832.259831 | 2561.59754 | 2317.392308 |
| 720.341 | 3262.85 | 2932.39 | 4044.612458 | 3692.032447 | 720.341 | 2048.33 | 1809.44 | 3255.632846 | 2841.052368 | 2573.812997 | 2324.793167 |
| 720.627 | 3255.84 | 2940.44 | 4141.026505 | 3702.107851 | 720.627 | 2078.89 | 1815.09 | 3190.445683 | 2849.866377 | 2598.775019 | 2332.217116 |
| 720.913 | 3254.73 | 2948.49 | 4164.831421 | 3712.206906 | 720.913 | 2053.03 | 1820.74 | 3204.208849 | 2858.701857 | 2536.242962 | 2339.652612 |
| 721.199 | 3290.95 | 2956.57 | 4141.653262 | 3722.317787 | 721.199 | 2094.61 | 1826.41 | 3260.549796 | 2867.548072 | 2638.896676 | 2347.111199 |
| 721.485 | 3335.61 | 2964.65 | 4179.483329 | 3732.440493 | 721.485 | 2097.32 | 1832.1 | 3302.429756 | 2876.415759 | 2640.605455 | 2354.569787 |
| 721.77 | 3309.37 | 2972.75 | 4128.384948 | 3742.575025 | 721.77 | 2112.39 | 1837.79 | 3321.453414 | 2885.294182 | 2617.917956 | 2362.051466 |
| 722.056 | 3358 | 2980.86 | 4207.545457 | 3752.733208 | 722.056 | 2135.07 | 1843.49 | 3239.636654 | 2894.194077 | 2719.139992 | 2369.556236 |
| 722.342 | 3338.26 | 2988.98 | 4160.869848 | 3762.903217 | 722.342 | 2077.23 | 1849.21 | 3159.022293 | 2903.115442 | 2550.756033 | 2377.072552 |
| 722.628 | 3313.68 | 2997.12 | 4164.772293 | 3773.096877 | 722.628 | 2115.44 | 1854.93 | 3253.367613 | 2912.058279 | 2581.236947 | 2384.600415 |
| 722.913 | 3325.05 | 3005.27 | 4258.265419 | 3783.302362 | 722.913 | 2049.92 | 1860.67 | 3207.676479 | 2921.011852 | 2687.60841 | 2392.139822 |
| 723.199 | 3398.51 | 3013.43 | 4171.560183 | 3793.519673 | 723.199 | 2111.03 | 1866.42 | 3248.912298 | 2929.986896 | 2587.090668 | 2399.702322 |
| 723.485 | 3410.26 | 3021.61 | 4276.831598 | 3803.760636 | 723.485 | 2116.5 | 1872.18 | 3332.640012 | 2938.972676 | 2672.529594 | 2407.276367 |
| 723.771 | 3391.95 | 3029.8 | 4207.580934 | 3814.013423 | 723.771 | 2103.3 | 1877.96 | 3173.526222 | 2947.979927 | 2645.535512 | 2414.873504 |
| 724.057 | 3344.18 | 3038 | 4149.529106 | 3824.278037 | 724.057 | 2130.29 | 1883.74 | 3217.607001 | 2957.00865 | 2692.053543 | 2422.482187 |
| 724.342 | 3386.93 | 3046.21 | 4284.683791 | 3834.566301 | 724.342 | 2157.17 | 1889.54 | 3405.911156 | 2966.048108 | 2638.238566 | 2430.102415 |
| 724.628 | 3323.44 | 3054.44 | 4268.08066 | 3844.866392 | 724.628 | 2158.11 | 1895.35 | 3378.492182 | 2975.109038 | 2639.924253 | 2437.745735 |
| 724.914 | 3488.06 | 3062.68 | 4223.450878 | 3855.190133 | 724.914 | 2020 | 1901.16 | 3300.282617 | 2984.180703 | 2764.076248 | 2445.400601 |
| 725.199 | 3339.96 | 3070.93 | 4184.852147 | 3865.5257 | 725.199 | 2092.05 | 1906.99 | 3302.419021 | 2993.284576 | 2637.418814 | 2453.067013 |
| 725.485 | 3385.36 | 3079.2 | 4266.07031 | 3875.873093 | 725.485 | 2131.54 | 1912.84 | 3359.672502 | 3002.399184 | 2701.60192 | 2460.756516 |
| 725.771 | 3397.83 | 3087.47 | 4318.847925 | 3886.232311 | 725.771 | 2144.65 | 1918.69 | 3277.941627 | 3011.524528 | 2624.164235 | 2468.457565 |
| 726.056 | 3416.76 | 3095.76 | 4331.749645 | 3896.61518 | 726.056 | 2117.74 | 1924.56 | 3318.211233 | 3020.682079 | 2740.684456 | 2476.181706 |
| 726.342 | 3377.5 | 3104.07 | 4256.27872 | 3907.009875 | 726.342 | 2131.46 | 1930.43 | 3291.908772 | 3029.850366 | 2814.462123 | 2483.917393 |
| 726.628 | 3357.78 | 3112.39 | 4250.235843 | 3917.428221 | 726.628 | 2098.72 | 1936.32 | 3379.984444 | 3039.029388 | 2697.56089 | 2491.664625 |
| 726.913 | 3527.83 | 3120.72 | 4296.000882 | 3927.858393 | 726.913 | 2207.23 | 1942.22 | 3390.6772 | 3048.229882 | 2690.806597 | 2499.434949 |
| 727.199 | 3379.03 | 3129.06 | 4342.274422 | 3938.30039 | 727.199 | 2109.11 | 1948.13 | 3295.719945 | 3057.451847 | 2795.677105 | 2507.205273 |
| 727.484 | 3431.11 | 3137.42 | 4282.247719 | 3948.766039 | 727.484 | 2214.97 | 1954.05 | 3412.363311 | 3066.695284 | 2750.994856 | 2515.010234 |
| 727.77 | 3388.46 | 3145.79 | 4406.22722 | 3959.243513 | 727.77 | 2173.99 | 1959.99 | 3388.476382 | 3075.949456 | 2762.309741 | 2522.826741 |
| 728.055 | 3495.52 | 3154.17 | 4372.382377 | 3969.732812 | 728.055 | 2242.58 | 1965.93 | 3431.140048 | 3085.2251 | 2850.54275 | 2530.654794 |
| 728.341 | 3421.91 | 3162.56 | 4295.113963 | 3980.245763 | 728.341 | 2118.06 | 1971.89 | 3413.03966 | 3094.511479 | 2765.577202 | 2538.494393 |
| 728.626 | 3501.96 | 3170.97 | 4440.994459 | 3990.77054 | 728.626 | 2235.12 | 1977.86 | 3398.310282 | 3103.830066 | 2781.452678 | 2546.357083 |
| 728.912 | 3491.42 | 3179.39 | 4359.63439 | 4001.318967 | 728.912 | 2216.63 | 1983.84 | 3413.018189 | 3113.148652 | 2721.668522 | 2554.231319 |
| 729.197 | 3484.16 | 3187.82 | 4434.336651 | 4011.867395 | 729.197 | 2262.4 | 1989.83 | 3427.575796 | 3122.499446 | 2744.171288 | 2562.128647 |
| 729.483 | 3497.36 | 3196.27 | 4315.879701 | 4022.451299 | 729.483 | 2235.31 | 1995.84 | 3501.029447 | 3131.860975 | 2704.453733 | 2570.037521 |
| 729.768 | 3508.86 | 3204.73 | 4375.752671 | 4033.035204 | 729.768 | 2217.24 | 2001.85 | 3426.255305 | 3141.243976 | 2765.369378 | 2577.95794 |
| 730.054 | 3523.83 | 3213.2 | 4406.01436 | 4043.642759 | 730.054 | 2199.2 | 2007.88 | 3440.877327 | 3150.648448 | 2792.825292 | 2585.901451 |
| 730.339 | 3449.47 | 3221.69 | 4337.34315 | 4054.26214 | 730.339 | 2256.75 | 2013.92 | 3473.696358 | 3160.063655 | 2784.824052 | 2593.856508 |
| 730.625 | 3567.03 | 3230.18 | 4437.364003 | 4064.905173 | 730.625 | 2237.37 | 2019.97 | 3437.592203 | 3169.500335 | 2745.799246 | 2601.82311 |
| 730.91 | 3596.85 | 3238.7 | 4434.537686 | 4075.560031 | 730.91 | 2254.08 | 2026.03 | 3486.289333 | 3178.94775 | 2836.214411 | 2609.812804 |
| 731.195 | 3534.05 | 3247.22 | 4416.858427 | 4086.226714 | 731.195 | 2229.17 | 2032.1 | 3485.591512 | 3188.416636 | 2873.876813 | 2617.814044 |
| 731.481 | 3529.04 | 3255.76 | 4423.953782 | 4096.905223 | 731.481 | 2172.41 | 2038.19 | 3384.858451 | 3197.906993 | 2751.098768 | 2625.838376 |
| 731.766 | 3492.25 | 3264.3 | 4452.193295 | 4107.607384 | 731.766 | 2344.05 | 2044.29 | 3405.782328 | 3207.418823 | 2847.344563 | 2633.874253 |
| 732.052 | 3504.44 | 3272.87 | 4515.602116 | 4118.333195 | 732.052 | 2189.15 | 2050.39 | 3463.229052 | 3216.941387 | 2880.203912 | 2641.921676 |
| 732.337 | 3568 | 3281.44 | 4476.636792 | 4129.059007 | 732.337 | 2231.41 | 2056.51 | 3521.534632 | 3226.485423 | 2868.750478 | 2649.992191 |
| 732.622 | 3530.39 | 3290.03 | 4437.080189 | 4139.808469 | 732.622 | 2231.91 | 2062.64 | 3602.868285 | 3236.040195 | 2827.127866 | 2658.074251 |
| 732.908 | 3543.39 | 3298.63 | 4424.757922 | 4150.581583 | 732.908 | 2193.48 | 2068.79 | 3562.963693 | 3245.616439 | 2867.272615 | 2666.179403 |
| 733.193 | 3551.53 | 3307.24 | 4476.99156 | 4161.354697 | 733.193 | 2233.52 | 2074.94 | 3471.538483 | 3255.214153 | 2875.100668 | 2674.284556 |
| 733.478 | 3606.04 | 3315.87 | 4532.110642 | 4172.151462 | 733.478 | 2301.39 | 2081.11 | 3603.963326 | 3264.833339 | 2795.538555 | 2682.424345 |
| 733.763 | 3658.64 | 3324.51 | 4509.417332 | 4182.971878 | 733.763 | 2229.68 | 2087.29 | 3486.589932 | 3274.463261 | 2895.582919 | 2690.564135 |
| 734.048 | 3585.89 | 3333.16 | 4528.184546 | 4193.792295 | 734.048 | 2415.77 | 2093.48 | 3643.717619 | 3284.114654 | 2883.690744 | 2698.727016 |
| 734.334 | 3559.84 | 3341.82 | 4594.703498 | 4204.636362 | 734.334 | 2332.49 | 2099.68 | 3527.471473 | 3293.787519 | 2891.738167 | 2706.912989 |
| 734.619 | 3626.76 | 3350.5 | 4542.623593 | 4215.504081 | 734.619 | 2235.55 | 2105.89 | 3541.395674 | 3303.471119 | 2849.52672 | 2715.098961 |
| 734.904 | 3650.89 | 3359.19 | 4562.987261 | 4226.383625 | 734.904 | 2325.95 | 2112.11 | 3708.18549 | 3313.176191 | 2933.072134 | 2723.319571 |
| 735.189 | 3694.42 | 3367.9 | 4572.743374 | 4237.274995 | 735.189 | 2323.85 | 2118.35 | 3508.190158 | 3322.891998 | 2922.657822 | 2731.540182 |
| 735.475 | 3634.47 | 3376.61 | 4596.985837 | 4248.17819 | 735.475 | 2305.7 | 2124.6 | 3596.018909 | 3332.640012 | 2980.67547 | 2739.783883 |
| 735.76 | 3631.72 | 3385.34 | 4591.569716 | 4259.105036 | 735.76 | 2324.14 | 2130.86 | 3571.283859 | 3342.398763 | 3056.046458 | 2748.050677 |
| 736.045 | 3725.88 | 3394.08 | 4590.517238 | 4270.043709 | 736.045 | 2353.17 | 2137.13 | 3739.469316 | 3352.168248 | 2963.229765 | 2756.31747 |
| 736.33 | 3699.4 | 3402.84 | 4657.178098 | 4280.994206 | 736.33 | 2288.28 | 2143.41 | 3613.206763 | 3361.959206 | 3008.28148 | 2764.607355 |
| 736.615 | 3682.13 | 3411.6 | 4619.856531 | 4291.968355 | 736.615 | 2403.9 | 2149.71 | 3709.699223 | 3371.771634 | 3067.084243 | 2772.920332 |
| 736.9 | 3672.99 | 3420.38 | 4622.127044 | 4302.95433 | 736.9 | 2368.05 | 2156.01 | 3563.607835 | 3381.605534 | 3004.009534 | 2781.244854 |
| 737.185 | 3689.41 | 3429.18 | 4555.052289 | 4313.963955 | 737.185 | 2351.43 | 2162.33 | 3607.119622 | 3391.45017 | 2893.735591 | 2789.580922 |
| 737.47 | 3647.45 | 3437.98 | 4682.626771 | 4324.985407 | 737.47 | 2387.43 | 2168.66 | 3663.771904 | 3401.316277 | 3018.337873 | 2797.940082 |
| 737.755 | 3701.83 | 3446.8 | 4697.81083 | 4336.018684 | 737.755 | 2410.62 | 2175 | 3712.619333 | 3411.203856 | 3092.230997 | 2806.310787 |
| 738.041 | 3741.73 | 3455.63 | 4696.651922 | 4347.063786 | 738.041 | 2391.47 | 2181.36 | 3691.534421 | 3421.10217 | 3057.027851 | 2814.693039 |
| 738.326 | 3744.22 | 3464.47 | 4670.138946 | 4358.13254 | 738.326 | 2398.09 | 2187.72 | 3780.114671 | 3431.032691 | 3032.262108 | 2823.098382 |
| 738.611 | 3729.16 | 3473.33 | 4750.588445 | 4369.213119 | 738.611 | 2389.36 | 2194.1 | 3715.110015 | 3440.963213 | 3163.364676 | 2831.526816 |
| 738.896 | 3756.23 | 3482.2 | 4665.645221 | 4380.317349 | 738.896 | 2375.11 | 2200.49 | 3621.902679 | 3450.925941 | 2972.004574 | 2839.955251 |
| 739.181 | 3734.36 | 3491.08 | 4818.632899 | 4391.433405 | 739.181 | 2419.99 | 2206.89 | 3865.442005 | 3460.899405 | 3035.772032 | 2848.406777 |
| 739.466 | 3816.11 | 3499.97 | 4722.337107 | 4402.561287 | 739.466 | 2422.62 | 2213.3 | 3726.747513 | 3470.883605 | 3030.703425 | 2856.881395 |
| 739.751 | 3783.97 | 3508.88 | 4809.515367 | 4413.700994 | 739.751 | 2442.23 | 2219.72 | 3719.071488 | 3480.900012 | 3087.046933 | 2865.356012 |
| 740.035 | 3817.1 | 3517.8 | 4805.305457 | 4424.864353 | 740.035 | 2431.56 | 2226.16 | 3848.941236 | 3490.927155 | 3137.386624 | 2873.865268 |
| 740.321 | 3820.06 | 3526.74 | 4659.354007 | 4436.039537 | 740.321 | 2370.91 | 2232.6 | 3726.167785 | 3500.975769 | 3011.191022 | 2882.374523 |
| 740.605 | 3836.87 | 3535.68 | 4756.063694 | 4447.238372 | 740.605 | 2414.01 | 2239.06 | 3843.734422 | 3511.035118 | 3184.759045 | 2890.90687 |
| 740.89 | 3857.68 | 3544.64 | 4793.787331 | 4458.449032 | 740.89 | 2446.16 | 2245.53 | 3792.009825 | 3521.126675 | 3092.588917 | 2899.462308 |
| 741.175 | 3848.38 | 3553.61 | 4794.331308 | 4469.671519 | 741.175 | 2465.08 | 2252.01 | 3791.108027 | 3531.228968 | 3095.244451 | 2908.017746 |
| 741.46 | 3812.26 | 3562.59 | 4761.030442 | 4480.905831 | 741.46 | 2481.51 | 2258.51 | 3926.45298 | 3541.341996 | 3126.741396 | 2916.607822 |
| 741.745 | 3911.77 | 3571.59 | 4857.704653 | 4492.163794 | 741.745 | 2435.84 | 2265.01 | 3927.05418 | 3551.476495 | 3213.842917 | 2925.197898 |
| 742.03 | 3863.94 | 3580.6 | 4922.993744 | 4503.433583 | 742.03 | 2477.31 | 2271.53 | 3914.826219 | 3561.632466 | 3186.294636 | 2933.811065 |
| 742.315 | 3912.01 | 3589.62 | 4880.007718 | 4514.727022 | 742.315 | 2536.3 | 2278.06 | 3959.841003 | 3571.809909 | 3180.129179 | 2942.447324 |
| 742.6 | 3846.47 | 3598.66 | 4806.085946 | 4526.032288 | 742.6 | 2469.63 | 2284.6 | 3957.801221 | 3581.998087 | 3145.029944 | 2951.095129 |
| 742.884 | 3949.02 | 3607.7 | 4915.945691 | 4537.349379 | 742.884 | 2536.01 | 2291.15 | 4039.832695 | 3592.207736 | 3111.246931 | 2959.754479 |
| 743.169 | 3908.51 | 3616.76 | 5005.725582 | 4548.678296 | 743.169 | 2455 | 2297.72 | 3901.32071 | 3602.438857 | 3217.445207 | 2968.436922 |
| 743.454 | 3976.39 | 3625.83 | 4995.342712 | 4560.030863 | 743.454 | 2566.51 | 2304.29 | 3984.60826 | 3612.680714 | 3113.856282 | 2977.119364 |
| 743.739 | 3907.97 | 3634.92 | 4939.987119 | 4571.395257 | 743.739 | 2549.47 | 2310.88 | 4018.898082 | 3622.944042 | 3218.06868 | 2985.836443 |
| 744.023 | 3983.03 | 3644.02 | 4912.989293 | 4582.783302 | 744.023 | 2508.79 | 2317.48 | 4015.076173 | 3633.228841 | 3186.179178 | 2994.565069 |
| 744.308 | 3964.49 | 3653.13 | 4908.460092 | 4594.183172 | 744.308 | 2525.64 | 2324.09 | 3940.849552 | 3643.524376 | 3154.970879 | 3003.30524 |
| 744.593 | 3972.96 | 3662.25 | 4952.120175 | 4605.594868 | 744.593 | 2531.68 | 2330.72 | 3992.348699 | 3653.841383 | 3147.84712 | 3012.068503 |
| 744.878 | 4023.14 | 3671.39 | 4944.327111 | 4617.018389 | 744.878 | 2528.73 | 2337.35 | 3985.295345 | 3664.179861 | 3217.214291 | 3020.843311 |
| 745.162 | 3992.8 | 3680.53 | 5018.012371 | 4628.465562 | 745.162 | 2575.81 | 2344 | 3884.691112 | 3674.529074 | 3139.360956 | 3029.629666 |
| 745.447 | 4039.54 | 3689.7 | 4982.228131 | 4639.92456 | 745.447 | 2575.25 | 2350.66 | 4030.524844 | 3684.910495 | 3251.539956 | 3038.439112 |
| 745.732 | 3985.73 | 3698.87 | 4970.1069 | 4651.407209 | 745.732 | 2514.4 | 2357.32 | 4053.97161 | 3695.302651 | 3307.814189 | 3047.260103 |
| 746.016 | 3979.07 | 3708.05 | 4986.355263 | 4662.889858 | 746.016 | 2579.65 | 2364.01 | 3976.062644 | 3705.705543 | 3304.269628 | 3056.104187 |
| 746.301 | 3949.41 | 3717.25 | 4969.633876 | 4674.396159 | 746.301 | 2553.42 | 2370.7 | 3996.052515 | 3716.129907 | 3265.129364 | 3064.959816 |
| 746.586 | 3974.56 | 3726.46 | 4988.945067 | 4685.926111 | 746.586 | 2578.45 | 2377.41 | 3905.883382 | 3726.575742 | 3213.392631 | 3073.826991 |
| 746.87 | 4030.02 | 3735.69 | 5042.018322 | 4697.456062 | 746.87 | 2590.26 | 2384.12 | 4086.16797 | 3737.043048 | 3252.625262 | 3082.717258 |
| 747.155 | 4010.55 | 3744.93 | 5050.651004 | 4709.009665 | 747.155 | 2608.72 | 2390.85 | 4019.016175 | 3747.52109 | 3292.296633 | 3091.61907 |
| 747.439 | 3992.45 | 3754.17 | 5073.403442 | 4720.586919 | 747.439 | 2626.21 | 2397.59 | 4057.460712 | 3758.020603 | 3412.372961 | 3100.543974 |
| 747.724 | 4009.22 | 3763.43 | 5102.884642 | 4732.164173 | 747.724 | 2604.14 | 2404.34 | 4062.216626 | 3768.541588 | 3332.533749 | 3109.480424 |
| 748.009 | 3969.14 | 3772.71 | 5044.146929 | 4743.765079 | 748.009 | 2512.53 | 2411.11 | 4131.397469 | 3779.073308 | 3273.061329 | 3118.439965 |
| 748.293 | 4085.17 | 3782 | 5113.24386 | 4755.389635 | 748.293 | 2628.15 | 2417.89 | 4076.462899 | 3789.6265 | 3294.940621 | 3127.411052 |
| 748.578 | 4081.64 | 3791.3 | 5162.804914 | 4767.014192 | 748.578 | 2595.93 | 2424.67 | 4052.940983 | 3800.201163 | 3403.240233 | 3136.393685 |
| 748.862 | 4047.14 | 3800.61 | 5123.981497 | 4778.662399 | 748.862 | 2648.49 | 2431.47 | 4153.706251 | 3810.786562 | 3418.238228 | 3145.39941 |
| 749.147 | 4090.24 | 3809.93 | 5100.484046 | 4790.334258 | 749.147 | 2697.88 | 2438.28 | 4069.978537 | 3821.404168 | 3499.58994 | 3154.41668 |
| 749.431 | 4027.69 | 3819.27 | 5145.054701 | 4802.006117 | 749.431 | 2657.2 | 2445.11 | 4054.003817 | 3832.021774 | 3253.306464 | 3163.457042 |
| 749.716 | 4114.53 | 3828.62 | 5213.075503 | 4813.701627 | 749.716 | 2687.02 | 2451.94 | 4163.937372 | 3842.671588 | 3368.914567 | 3172.50895 |
| 750 | 4123.14 | 3837.98 | 5179.680033 | 4825.408962 | 750 | 2758.56 | 2458.79 | 4082.614454 | 3853.332136 | 3374.15636 | 3181.572404 |
| 750.284 | 4180.27 | 3847.36 | 5184.362967 | 4837.139949 | 750.284 | 2763.45 | 2465.65 | 4166.417318 | 3864.014157 | 3439.805784 | 3190.658949 |
| 750.569 | 4058.34 | 3856.74 | 5154.680732 | 4848.870936 | 750.569 | 2724.43 | 2472.52 | 4197.701144 | 3874.706913 | 3372.713135 | 3199.75704 |
| 750.853 | 4125.18 | 3866.14 | 5169.651931 | 4860.6374 | 750.853 | 2620.44 | 2479.4 | 4194.072478 | 3885.431876 | 3422.152254 | 3208.878223 |
| 751.138 | 4166.07 | 3875.55 | 5145.054701 | 4872.403863 | 751.138 | 2668.09 | 2486.3 | 4176.487404 | 3896.167574 | 3328.307986 | 3218.010951 |
| 751.422 | 4162.62 | 3884.98 | 5191.292764 | 4884.193978 | 751.422 | 2685.84 | 2493.2 | 4295.331589 | 3906.914009 | 3361.975541 | 3227.155225 |
| 751.706 | 4192.92 | 3894.41 | 5265.190885 | 4895.995918 | 751.706 | 2700.57 | 2500.12 | 4155.273663 | 3917.69265 | 3449.065516 | 3236.322591 |
| 751.991 | 4219.29 | 3903.86 | 5167.073952 | 4907.809684 | 751.991 | 2736.05 | 2507.05 | 4242.34018 | 3928.482027 | 3431.573628 | 3245.501503 |
| 752.275 | 4196.51 | 3913.33 | 5369.894672 | 4919.647101 | 752.275 | 2730.1 | 2513.99 | 4288.450006 | 3939.292876 | 3532.045186 | 3254.703506 |
| 752.559 | 4105.26 | 3922.8 | 5188.809389 | 4931.496344 | 752.559 | 2656.29 | 2520.95 | 4192.580216 | 3950.11446 | 3466.973053 | 3263.917055 |
| 752.844 | 4231.72 | 3932.29 | 5322.462225 | 4943.357412 | 752.844 | 2734.33 | 2527.91 | 4195.103105 | 3960.957516 | 3442.738417 | 3273.153695 |
| 753.128 | 4298.9 | 3941.79 | 5161.82339 | 4955.242132 | 753.128 | 2750.73 | 2534.89 | 4290.919216 | 3971.822043 | 3555.644803 | 3282.390336 |
| 753.412 | 4175.73 | 3951.3 | 5169.817489 | 4967.126851 | 753.412 | 2715.84 | 2541.88 | 4245.464268 | 3982.697306 | 3449.019333 | 3291.661614 |
| 753.697 | 4186.17 | 3960.82 | 5361.309293 | 4979.047047 | 753.697 | 2736.49 | 2548.88 | 4390.278108 | 3993.604776 | 3612.507871 | 3300.944438 |
| 753.981 | 4242.41 | 3970.36 | 5382.926474 | 4990.967244 | 753.981 | 2753.67 | 2555.89 | 4320.764459 | 4004.522982 | 3465.610648 | 3310.238807 |
| 754.265 | 4280.81 | 3979.91 | 5407.53553 | 5002.911091 | 754.265 | 2767.68 | 2562.91 | 4159.041893 | 4015.451923 | 3567.421519 | 3319.544723 |
| 754.549 | 4267.62 | 3989.47 | 5391.180737 | 5014.866764 | 754.549 | 2702.89 | 2569.95 | 4417.042205 | 4026.402336 | 3468.797289 | 3328.87373 |
| 754.834 | 4261.38 | 3999.04 | 5402.24949 | 5026.834263 | 754.834 | 2782.84 | 2577 | 4406.628577 | 4037.37422 | 3428.259983 | 3338.225828 |
| 755.118 | 4274.96 | 4008.63 | 5310.754889 | 5038.825412 | 755.118 | 2792.33 | 2584.06 | 4298.101399 | 4048.367575 | 3593.872949 | 3347.589473 |
| 755.402 | 4267.43 | 4018.23 | 5384.345545 | 5050.828388 | 755.402 | 2825.87 | 2591.13 | 4330.050838 | 4059.382402 | 3506.228775 | 3356.964663 |
| 755.686 | 4255.17 | 4027.84 | 5373.371396 | 5062.843189 | 755.686 | 2797.89 | 2598.21 | 4382.891947 | 4070.407965 | 3522.796999 | 3366.362945 |
| 755.97 | 4287.52 | 4037.46 | 5439.653836 | 5074.881641 | 755.97 | 2827.39 | 2605.31 | 4278.251092 | 4081.444263 | 3560.447856 | 3375.772773 |
| 756.254 | 4351.18 | 4047.1 | 5319.115582 | 5086.931919 | 756.254 | 2838.58 | 2612.41 | 4458.245817 | 4092.512768 | 3531.09843 | 3385.194146 |
| 756.538 | 4237.4 | 4056.75 | 5372.330744 | 5098.994022 | 756.538 | 2765.31 | 2619.53 | 4403.890974 | 4103.592009 | 3653.01054 | 3394.638611 |
| 756.823 | 4316.61 | 4066.41 | 5435.668612 | 5111.079776 | 756.823 | 2843.02 | 2626.66 | 4440.682214 | 4114.692722 | 3545.72696 | 3404.094622 |
| 757.107 | 4274 | 4076.08 | 5442.267292 | 5123.165531 | 757.107 | 2839.93 | 2633.8 | 4366.26235 | 4125.80417 | 3648.576953 | 3413.573724 |
| 757.391 | 4329.28 | 4085.77 | 5386.698838 | 5135.274937 | 757.391 | 2849.65 | 2640.96 | 4480.941084 | 4136.947825 | 3673.850711 | 3423.075918 |
| 757.675 | 4417.62 | 4095.47 | 5556.076785 | 5147.407994 | 757.675 | 2902.63 | 2648.13 | 4506.094826 | 4148.102216 | 3768.79183 | 3432.578112 |
| 757.959 | 4428.09 | 4105.18 | 5497.61106 | 5159.54105 | 757.959 | 2816.01 | 2655.3 | 4448.766195 | 4159.267343 | 3606.908158 | 3442.103398 |
| 758.243 | 4388.32 | 4114.9 | 5645.608338 | 5171.697758 | 758.243 | 2918.72 | 2662.49 | 4488.627844 | 4170.453941 | 3663.759681 | 3451.651775 |
| 758.527 | 4449.9 | 4124.63 | 5542.288145 | 5183.878118 | 758.527 | 2865.1 | 2669.7 | 4422.227548 | 4181.672746 | 3808.347744 | 3461.200152 |
| 758.811 | 4386.33 | 4134.38 | 5560.27487 | 5196.058477 | 758.811 | 2895.73 | 2676.91 | 4495.380599 | 4192.891551 | 3741.543741 | 3470.783167 |
| 759.095 | 4373.78 | 4144.14 | 5509.732292 | 5208.262487 | 759.095 | 2883.87 | 2684.13 | 4543.229608 | 4204.142564 | 3756.276182 | 3480.377727 |
| 759.379 | 4391.3 | 4153.91 | 5499.077434 | 5220.478323 | 759.379 | 2820.33 | 2691.37 | 4494.586157 | 4215.404312 | 3780.926467 | 3489.983834 |
| 759.663 | 4421.64 | 4163.69 | 5556.774495 | 5232.705985 | 759.663 | 2840.38 | 2698.62 | 4525.472762 | 4226.687531 | 3772.371029 | 3499.601486 |
| 759.947 | 4395.07 | 4173.49 | 5564.934153 | 5244.957298 | 759.947 | 2880.35 | 2705.88 | 4558.109286 | 4237.981486 | 3630.704053 | 3509.242229 |
| 760.231 | 4396.66 | 4183.3 | 5547.266719 | 5257.220436 | 760.231 | 2935.39 | 2713.15 | 4520.802733 | 4249.296912 | 3782.242688 | 3518.906065 |
| 760.515 | 4465.58 | 4193.12 | 5609.76497 | 5269.4954 | 760.515 | 2888.13 | 2720.44 | 4577.777086 | 4260.63381 | 3817.180281 | 3528.581446 |
| 760.799 | 4419.41 | 4202.95 | 5634.752445 | 5281.794015 | 760.799 | 2932.01 | 2727.74 | 4527.963444 | 4271.99218 | 3718.209677 | 3538.268372 |
| 761.082 | 4456.97 | 4212.8 | 5577.031733 | 5294.104456 | 761.082 | 3023.45 | 2735.04 | 4507.125453 | 4283.361285 | 3688.998801 | 3547.966845 |
| 761.366 | 4384.59 | 4222.65 | 5665.759146 | 5306.426723 | 761.366 | 2830.8 | 2742.36 | 4629.201083 | 4294.751861 | 3808.278469 | 3557.699955 |
| 761.65 | 4540.14 | 4232.52 | 5553.617062 | 5318.760814 | 761.65 | 3010.87 | 2749.7 | 4539.847863 | 4306.163909 | 3724.098036 | 3567.433065 |
| 761.934 | 4449.48 | 4242.4 | 5637.74432 | 5331.118558 | 761.934 | 2879.79 | 2757.04 | 4600.611918 | 4317.586692 | 3725.702902 | 3577.189267 |
| 762.218 | 4531.47 | 4252.3 | 5819.148893 | 5343.488126 | 762.218 | 2933.44 | 2764.39 | 4531.012383 | 4329.030947 | 3838.759383 | 3586.957014 |
| 762.502 | 4512.95 | 4262.2 | 5708.898904 | 5355.869521 | 762.502 | 2930.7 | 2771.76 | 4528.886714 | 4340.496673 | 3860.800317 | 3596.747853 |
| 762.786 | 4451.34 | 4272.12 | 5692.851576 | 5368.26274 | 762.786 | 2927.85 | 2779.14 | 4584.884119 | 4351.973135 | 3830.007666 | 3606.550238 |
| 763.069 | 4481.87 | 4282.05 | 5676.42583 | 5380.679612 | 763.069 | 2986.33 | 2786.53 | 4723.460518 | 4363.471068 | 3810.44908 | 3616.375715 |
| 763.353 | 4493.33 | 4292 | 5633.688142 | 5393.108308 | 763.353 | 2939.25 | 2793.94 | 4538.258979 | 4374.990473 | 3874.967014 | 3626.212737 |
| 763.637 | 4512.5 | 4301.95 | 5644.473081 | 5405.560656 | 763.637 | 2978.46 | 2801.35 | 4723.385368 | 4386.531349 | 3784.667306 | 3636.072851 |
| 763.921 | 4529.42 | 4311.92 | 5645.903978 | 5418.013004 | 763.921 | 3007.47 | 2808.78 | 4726.895942 | 4398.082961 | 3867.519973 | 3645.932965 |
| 764.204 | 4431.94 | 4321.9 | 5589.306697 | 5430.489003 | 764.204 | 2915.66 | 2816.22 | 4612.335301 | 4409.656044 | 3673.134871 | 3655.827716 |
| 764.488 | 4570.43 | 4331.89 | 5785.505086 | 5442.976828 | 764.488 | 2955.59 | 2823.67 | 4677.640556 | 4421.250599 | 3831.023697 | 3665.734013 |
| 764.772 | 4494.05 | 4341.89 | 5722.971358 | 5455.488303 | 764.772 | 3007.12 | 2831.13 | 4538.044265 | 4432.855889 | 3990.933037 | 3675.651856 |
| 765.056 | 4570.12 | 4351.91 | 5782.832502 | 5467.999779 | 765.056 | 2975.84 | 2838.6 | 4689.138489 | 4444.482651 | 3869.655946 | 3685.581244 |
| 765.339 | 4565.12 | 4361.94 | 5693.052611 | 5480.534906 | 765.339 | 2992.14 | 2846.09 | 4592.731915 | 4456.130884 | 3939.323308 | 3695.533725 |
| 765.623 | 4632.13 | 4371.98 | 5885.478636 | 5493.093684 | 765.623 | 3100 | 2853.59 | 4750.149465 | 4467.789853 | 3972.66758 | 3705.509296 |
| 765.907 | 4605.73 | 4382.03 | 5682.113939 | 5505.652463 | 765.907 | 3021.31 | 2861.1 | 4770.708328 | 4479.470293 | 3881.96377 | 3715.496414 |
| 766.19 | 4658.3 | 4392.09 | 5792.352103 | 5518.234892 | 766.19 | 3005.74 | 2868.62 | 4695.064595 | 4491.172205 | 3874.551365 | 3725.495078 |
| 766.474 | 4596.55 | 4402.17 | 5849.540664 | 5530.829147 | 766.474 | 3058.92 | 2876.15 | 4721.678392 | 4502.884852 | 3947.717105 | 3735.516833 |
| 766.757 | 4552.13 | 4412.26 | 5760.41118 | 5543.435227 | 766.757 | 3055.79 | 2883.7 | 4760.938842 | 4514.629706 | 3879.539151 | 3745.550134 |
| 767.041 | 4570.21 | 4422.36 | 5942.986488 | 5556.064959 | 767.041 | 2965 | 2891.25 | 4879.879648 | 4526.374561 | 3944.588193 | 3755.606526 |
| 767.324 | 4639.56 | 4432.47 | 5861.67372 | 5568.694691 | 767.324 | 3065.83 | 2898.82 | 4768.668545 | 4538.151622 | 3922.616534 | 3765.674464 |
| 767.608 | 4679.57 | 4442.6 | 5918.590292 | 5581.359899 | 767.608 | 3114.88 | 2906.4 | 4780.510021 | 4549.93942 | 3893.902127 | 3775.753948 |
| 767.891 | 4703.44 | 4452.73 | 5870.743949 | 5594.025108 | 767.891 | 3141.36 | 2913.99 | 4829.303772 | 4561.748688 | 3899.813577 | 3785.856524 |
| 768.175 | 4643.77 | 4462.88 | 5884.757275 | 5606.702142 | 768.175 | 3045.28 | 2921.6 | 4788.561795 | 4573.579428 | 4052.726163 | 3795.970645 |
| 768.459 | 4645.45 | 4473.04 | 5926.477962 | 5619.402827 | 768.459 | 3076.21 | 2929.22 | 4802.485996 | 4585.420904 | 3930.202125 | 3806.107859 |
| 768.742 | 4689.95 | 4483.22 | 5809.842153 | 5632.115338 | 768.742 | 3047.59 | 2936.84 | 4671.703714 | 4597.283851 | 3979.110137 | 3816.256617 |
| 769.026 | 4676.95 | 4493.4 | 5924.089192 | 5644.8515 | 769.026 | 3095.24 | 2944.48 | 4791.578526 | 4609.168269 | 4011.934849 | 3826.428468 |
| 769.309 | 4764.02 | 4503.6 | 5930.853431 | 5657.587662 | 769.309 | 3026.8 | 2952.14 | 4901.544289 | 4621.063424 | 3996.729029 | 3836.611864 |
| 769.592 | 4785.23 | 4513.81 | 5984.352407 | 5670.347475 | 769.592 | 3167.4 | 2959.8 | 4924.325441 | 4632.980049 | 4111.090186 | 3846.806806 |
| 769.876 | 4802.97 | 4524.03 | 5871.015938 | 5683.119114 | 769.876 | 3068.2 | 2967.47 | 4855.498877 | 4644.918146 | 4029.057271 | 3857.02484 |
| 770.159 | 4777.65 | 4534.26 | 5977.221575 | 5695.914404 | 770.159 | 3095.3 | 2975.16 | 4800.982999 | 4656.877715 | 3992.918915 | 3867.254419 |
| 770.443 | 4697.8 | 4544.5 | 6020.231251 | 5708.72152 | 770.443 | 3219.68 | 2982.86 | 4813.726273 | 4668.848019 | 3939.808231 | 3877.50709 |
| 770.726 | 4777.31 | 4554.76 | 6003.474388 | 5721.528636 | 770.726 | 3137.97 | 2990.57 | 4926.644352 | 4680.839794 | 3992.63027 | 3887.771307 |
| 771.01 | 4738.12 | 4565.03 | 6025.789279 | 5734.371228 | 771.01 | 3126.61 | 2998.29 | 4950.445396 | 4692.842305 | 3997.82588 | 3898.058616 |
| 771.293 | 4788.8 | 4575.31 | 5995.586719 | 5747.21382 | 771.293 | 3140.44 | 3006.03 | 4947.11733 | 4704.877024 | 4125.406979 | 3908.35747 |
| 771.576 | 4794.72 | 4585.6 | 6018.670273 | 5760.080064 | 771.576 | 3227.06 | 3013.77 | 5027.93567 | 4716.922478 | 4205.800389 | 3918.66787 |
| 771.86 | 4910.31 | 4595.91 | 6026.770803 | 5772.958133 | 771.86 | 3243.45 | 3021.53 | 4913.450178 | 4728.978667 | 4113.814995 | 3929.001362 |
| 772.143 | 4886.84 | 4606.22 | 6054.773804 | 5785.848028 | 772.143 | 3106.49 | 3029.3 | 5050.641672 | 4741.067064 | 4175.307929 | 3939.346399 |
| 772.426 | 4844.48 | 4616.55 | 6162.871536 | 5798.749748 | 772.426 | 3206.36 | 3037.08 | 5007.505635 | 4753.166196 | 4072.250112 | 3949.714528 |
| 772.709 | 4805.6 | 4626.89 | 6181.496843 | 5811.67512 | 772.709 | 3235.86 | 3044.88 | 5097.674769 | 4765.276065 | 4004.637903 | 3960.094203 |
| 772.993 | 4821.34 | 4637.24 | 6007.459613 | 5824.612317 | 772.993 | 3268.61 | 3052.68 | 4986.710586 | 4777.41814 | 4260.527485 | 3970.49697 |
| 773.276 | 4890.19 | 4647.61 | 6166.6439 | 5837.561339 | 773.276 | 3247.9 | 3060.5 | 5077.405769 | 4789.570951 | 4199.865848 | 3980.911282 |
| 773.559 | 4898.95 | 4657.98 | 6129.381461 | 5850.522188 | 773.559 | 3244.74 | 3068.33 | 5014.558989 | 4801.745233 | 4253.53073 | 3991.33714 |
| 773.842 | 4928.57 | 4668.37 | 6028.591945 | 5863.506687 | 773.842 | 3145.78 | 3076.17 | 5131.277506 | 4813.930251 | 4254.373573 | 4001.78609 |
| 774.126 | 4984.26 | 4678.77 | 6118.596521 | 5876.503012 | 774.126 | 3151.57 | 3084.02 | 4986.850151 | 4826.136741 | 4116.077972 | 4012.246585 |
| 774.409 | 4967.92 | 4689.18 | 6233.907864 | 5889.511163 | 774.409 | 3254.68 | 3091.89 | 5037.490441 | 4838.364702 | 4233.879777 | 4022.730172 |
| 774.692 | 4840.37 | 4699.6 | 6049.251253 | 5902.531139 | 774.692 | 3231.59 | 3099.76 | 5031.220793 | 4850.614134 | 4152.285603 | 4033.225305 |
| 774.975 | 4969.25 | 4710.04 | 6163.912188 | 5915.574767 | 774.975 | 3244.05 | 3107.65 | 5038.864611 | 4862.874302 | 4262.051531 | 4043.731984 |
| 775.258 | 4887.46 | 4720.48 | 6146.280231 | 5928.618394 | 775.258 | 3255.6 | 3115.55 | 5109.333737 | 4875.155941 | 4321.847233 | 4054.261754 |
| 775.542 | 5005.91 | 4730.94 | 6243.309209 | 5941.697498 | 775.542 | 3372.42 | 3123.46 | 5135.024264 | 4887.448316 | 4176.878158 | 4064.814616 |
| 775.825 | 4974.4 | 4741.41 | 6316.107551 | 5954.776602 | 775.825 | 3310.87 | 3131.39 | 5062.246963 | 4899.772898 | 4233.302487 | 4075.379024 |
| 776.108 | 4940.62 | 4751.9 | 6258.079373 | 5967.867532 | 776.108 | 3400.42 | 3139.32 | 5097.095041 | 4912.108216 | 4307.334161 | 4085.954977 |
| 776.391 | 4911.6 | 4762.39 | 6207.820609 | 5980.982113 | 776.391 | 3323.45 | 3147.27 | 5241.554603 | 4924.45427 | 4202.382832 | 4096.554022 |
| 776.674 | 5024.32 | 4772.89 | 6159.1583 | 5994.10852 | 776.674 | 3213.99 | 3155.23 | 5097.878747 | 4936.83253 | 4191.218043 | 4107.164613 |
| 776.957 | 4927.29 | 4783.41 | 6245.780758 | 6007.246752 | 776.957 | 3307.38 | 3163.2 | 5130.762192 | 4949.221527 | 4270.433782 | 4117.78675 |
| 777.24 | 5028 | 4793.94 | 6260.148852 | 6020.408635 | 777.24 | 3395.37 | 3171.18 | 5285.098597 | 4961.621259 | 4345.262117 | 4128.431978 |
| 777.523 | 4945.48 | 4804.48 | 6161.145 | 6033.570519 | 777.523 | 3257.33 | 3179.18 | 5047.861126 | 4974.053198 | 4296.296376 | 4139.088752 |
| 777.806 | 5081.79 | 4815.03 | 6271.20578 | 6046.756053 | 777.806 | 3486.97 | 3187.18 | 5330.403245 | 4986.495873 | 4448.920316 | 4149.768618 |
| 778.089 | 5029.45 | 4825.6 | 6349.538498 | 6059.953413 | 778.089 | 3452.61 | 3195.2 | 5312.077408 | 4998.960019 | 4369.496752 | 4160.471575 |
| 778.372 | 5091.53 | 4836.17 | 6280.855463 | 6073.162599 | 778.372 | 3309.94 | 3203.23 | 5120.488128 | 5011.434901 | 4391.029671 | 4171.174533 |
| 778.655 | 5104.78 | 4846.76 | 6322.162254 | 6086.395436 | 778.655 | 3418.95 | 3211.27 | 5350.253552 | 5023.931254 | 4388.304862 | 4181.900582 |
| 778.938 | 5060.65 | 4857.36 | 6303.371389 | 6099.640098 | 778.938 | 3337.2 | 3219.33 | 5135.421485 | 5036.449079 | 4300.256585 | 4192.649722 |
| 779.221 | 5091.18 | 4867.97 | 6429.14838 | 6112.896586 | 779.221 | 3333.76 | 3227.39 | 5279.881047 | 5048.988375 | 4222.333976 | 4203.410409 |
| 779.504 | 5084.83 | 4878.6 | 6373.946519 | 6126.1649 | 779.504 | 3428.95 | 3235.47 | 5252.837822 | 5061.538407 | 4451.564304 | 4214.182641 |
| 779.787 | 5091.36 | 4889.23 | 6400.175681 | 6139.445039 | 779.787 | 3507.64 | 3243.56 | 5287.793257 | 5074.10991 | 4580.657903 | 4224.977964 |
| 780.07 | 5081.4 | 4899.87 | 6365.100976 | 6152.74883 | 780.07 | 3348.36 | 3251.66 | 5310.477789 | 5086.692149 | 4364.670608 | 4235.784834 |
| 780.353 | 5106.48 | 4910.53 | 6396.580701 | 6166.064446 | 780.353 | 3484.69 | 3259.77 | 5450.610864 | 5099.295859 | 4495.865542 | 4246.614795 |
| 780.635 | 5130.4 | 4921.2 | 6419.782512 | 6179.391887 | 780.635 | 3372.7 | 3267.89 | 5511.643311 | 5111.921041 | 4397.922514 | 4257.456302 |
| 780.918 | 5113.76 | 4931.88 | 6387.652379 | 6192.731155 | 780.918 | 3581.92 | 3276.03 | 5267.309544 | 5124.567694 | 4415.229669 | 4268.309355 |
| 781.201 | 5168.11 | 4942.57 | 6464.873492 | 6206.094073 | 781.201 | 3436.43 | 3284.18 | 5378.939339 | 5137.225083 | 4497.759053 | 4279.185499 |
| 781.484 | 5182.57 | 4953.28 | 6609.476825 | 6219.468817 | 781.484 | 3439.12 | 3292.34 | 5374.977867 | 5149.903943 | 4464.7727 | 4290.084735 |
| 781.767 | 5167.11 | 4963.99 | 6486.265987 | 6232.855387 | 781.767 | 3510.25 | 3300.51 | 5402.675969 | 5162.604274 | 4474.806001 | 4300.995517 |
| 782.05 | 5152.71 | 4974.72 | 6423.365666 | 6246.253782 | 782.05 | 3458.99 | 3308.69 | 5385.026481 | 5175.315342 | 4486.270981 | 4311.917844 |
| 782.332 | 5153.7 | 4985.46 | 6516.858792 | 6259.664003 | 782.332 | 3452.71 | 3316.89 | 5320.751852 | 5188.04788 | 4426.602283 | 4322.851717 |
| 782.615 | 5230.39 | 4996.21 | 6558.934247 | 6273.097874 | 782.615 | 3584.57 | 3325.09 | 5536.743375 | 5200.80189 | 4441.646461 | 4333.808682 |
| 782.898 | 5179.05 | 5006.97 | 6465.287388 | 6286.543572 | 782.898 | 3423.95 | 3333.31 | 5462.656318 | 5213.566636 | 4473.039494 | 4344.788739 |
| 783.181 | 5189.38 | 5017.74 | 6575.229912 | 6300.001095 | 783.181 | 3514.06 | 3341.54 | 5471.663569 | 5226.352853 | 4560.903037 | 4355.780341 |
| 783.463 | 5227.22 | 5028.53 | 6629.828668 | 6313.470444 | 783.463 | 3587.15 | 3349.78 | 5483.462102 | 5239.160542 | 4610.088149 | 4366.783489 |
| 783.746 | 5218.93 | 5039.33 | 6497.311089 | 6326.951618 | 783.746 | 3578.13 | 3358.04 | 5426.273035 | 5251.989702 | 4563.269927 | 4377.809729 |
| 784.029 | 5189.55 | 5050.13 | 6512.187684 | 6340.456444 | 784.029 | 3488.28 | 3366.3 | 5382.750512 | 5264.829598 | 4543.653611 | 4388.847515 |
| 784.311 | 5258.68 | 5060.95 | 6567.578754 | 6353.961269 | 784.311 | 3511.96 | 3374.58 | 5564.505892 | 5277.680229 | 4603.056756 | 4399.908392 |
| 784.594 | 5258.99 | 5071.78 | 6603.895146 | 6367.489746 | 784.594 | 3539.94 | 3382.87 | 5493.940144 | 5290.563068 | 4664.849882 | 4410.980815 |
| 784.877 | 5247.97 | 5082.63 | 6587.812341 | 6381.041874 | 784.877 | 3587.39 | 3391.17 | 5460.884928 | 5303.456642 | 4488.233768 | 4422.064783 |
| 785.159 | 5221.98 | 5093.48 | 6677.308418 | 6394.594002 | 785.159 | 3578.26 | 3399.48 | 5601.458167 | 5316.371687 | 4660.935855 | 4433.171844 |
| 785.442 | 5292.76 | 5104.35 | 6625.961699 | 6408.169781 | 785.442 | 3528.01 | 3407.8 | 5442.623504 | 5329.297469 | 4514.084815 | 4444.29045 |
| 785.725 | 5288.62 | 5115.22 | 6654.626933 | 6421.74556 | 785.725 | 3530.51 | 3416.14 | 5528.959993 | 5342.244721 | 4585.045307 | 4455.432148 |
| 786.007 | 5312.34 | 5126.11 | 6655.06448 | 6435.34499 | 786.007 | 3612.93 | 3424.49 | 5615.049561 | 5355.213445 | 4667.540053 | 4466.585391 |
| 786.29 | 5391.73 | 5137.01 | 6698.369796 | 6448.968071 | 786.29 | 3620.43 | 3432.85 | 5669.833832 | 5368.203641 | 4721.609038 | 4477.761726 |
| 786.572 | 5276.44 | 5147.92 | 6569.529977 | 6462.591153 | 786.572 | 3531.45 | 3441.22 | 5532.728224 | 5381.204572 | 4662.794729 | 4488.949607 |
| 786.855 | 5329.51 | 5158.84 | 6662.893022 | 6476.22606 | 786.855 | 3668.21 | 3449.6 | 5601.264924 | 5394.226974 | 4767.076402 | 4500.149034 |
| 787.137 | 5276.93 | 5169.77 | 6660.421473 | 6489.884618 | 787.137 | 3665.71 | 3457.99 | 5582.037288 | 5407.260113 | 4707.823352 | 4511.371552 |
| 787.42 | 5332.47 | 5180.72 | 6748.33292 | 6503.555002 | 787.42 | 3637.63 | 3466.4 | 5619.665912 | 5420.314722 | 4691.589956 | 4522.605616 |
| 787.702 | 5329.57 | 5191.68 | 6660.717113 | 6517.237211 | 787.702 | 3644.29 | 3474.82 | 5625.656432 | 5433.390803 | 4719.288332 | 4533.862772 |
| 787.985 | 5409.96 | 5202.64 | 6714.298868 | 6530.943072 | 787.985 | 3575.33 | 3483.25 | 5700.409102 | 5446.488356 | 4628.191964 | 4545.131474 |
| 788.267 | 5316.85 | 5213.62 | 6660.740764 | 6544.648932 | 788.267 | 3583.07 | 3491.69 | 5588.983285 | 5459.596644 | 4720.581462 | 4556.423267 |
| 788.55 | 5376.54 | 5224.61 | 6783.265718 | 6558.378444 | 788.55 | 3683.11 | 3500.14 | 5725.154888 | 5472.726403 | 4875.064276 | 4567.726606 |
| 788.832 | 5429.59 | 5235.61 | 6680.35942 | 6572.107956 | 788.832 | 3636.32 | 3508.61 | 5622.521608 | 5485.866899 | 4758.082223 | 4579.04149 |
| 789.115 | 5385.11 | 5246.63 | 6891.706392 | 6585.872944 | 789.115 | 3625.85 | 3517.08 | 5739.02541 | 5499.028865 | 4712.453218 | 4590.379467 |
| 789.397 | 5504.63 | 5257.65 | 6846.225167 | 6599.637933 | 789.397 | 3668.59 | 3525.57 | 5832.737325 | 5512.212303 | 4857.491567 | 4601.728989 |
| 789.679 | 5405.91 | 5268.69 | 6874.618412 | 6613.414747 | 789.679 | 3643.03 | 3534.07 | 5712.808834 | 5525.417213 | 4756.223349 | 4613.101603 |
| 789.962 | 5432.07 | 5279.73 | 6898.565235 | 6627.215212 | 789.962 | 3675.74 | 3542.58 | 5701.267958 | 5538.632858 | 4763.300925 | 4624.485762 |
| 790.244 | 5435.15 | 5290.79 | 6809.282019 | 6641.027503 | 790.244 | 3700.42 | 3551.11 | 5828.97983 | 5551.869974 | 4758.186135 | 4635.881468 |
| 790.526 | 5479.17 | 5301.86 | 6919.177241 | 6654.839794 | 790.526 | 3823.39 | 3559.64 | 5957.260694 | 5565.117827 | 4759.906459 | 4647.300265 |
| 790.809 | 5461.96 | 5312.94 | 6856.513431 | 6668.675736 | 790.809 | 3639.2 | 3568.19 | 5846.95139 | 5578.397886 | 4732.554457 | 4658.742153 |
| 791.091 | 5361.18 | 5324.03 | 6909.149139 | 6682.535329 | 791.091 | 3615.17 | 3576.74 | 5848.443652 | 5591.677945 | 4721.620584 | 4670.184042 |
| 791.373 | 5412.2 | 5335.13 | 6871.070735 | 6696.394922 | 791.373 | 3658.93 | 3585.31 | 5838.792259 | 5604.990212 | 4769.697298 | 4681.649022 |
| 791.655 | 5548.74 | 5346.25 | 6861.350098 | 6710.278167 | 791.655 | 3722.14 | 3593.89 | 5755.483237 | 5618.313214 | 4802.602831 | 4693.137094 |
| 791.938 | 5399.43 | 5357.37 | 6989.078312 | 6724.173237 | 791.938 | 3685.17 | 3602.49 | 5839.844357 | 5631.657687 | 4887.337462 | 4704.636711 |
| 792.22 | 5454.7 | 5368.51 | 6836.078809 | 6738.080132 | 792.22 | 3710.64 | 3611.09 | 5714.805674 | 5645.023632 | 4943.496237 | 4716.147875 |
| 792.502 | 5505.23 | 5379.66 | 6967.520259 | 6751.998854 | 792.502 | 3731.44 | 3619.71 | 5790.363522 | 5658.400313 | 4733.986137 | 4727.68213 |
| 792.784 | 5558.58 | 5390.82 | 6964.161791 | 6765.9294 | 792.784 | 3786.35 | 3628.34 | 5796.354042 | 5671.798465 | 4770.840333 | 4739.239476 |
| 793.067 | 5530.66 | 5401.98 | 7036.427981 | 6779.883598 | 793.067 | 3814.03 | 3636.98 | 5994.008991 | 5685.207353 | 5096.58201 | 4750.796823 |
| 793.349 | 5446.52 | 5413.17 | 6913.903027 | 6793.837796 | 793.349 | 3843.36 | 3645.63 | 5838.953294 | 5698.637712 | 4994.540223 | 4762.377261 |
| 793.631 | 5505.22 | 5424.36 | 6906.322823 | 6807.815646 | 793.631 | 3773.09 | 3654.29 | 5769.149781 | 5712.089542 | 4900.499675 | 4773.980791 |
| 793.913 | 5477.35 | 5435.56 | 6956.995482 | 6821.80532 | 793.913 | 3669.97 | 3662.97 | 5740.238544 | 5725.552108 | 4953.41408 | 4785.595866 |
| 794.195 | 5505.27 | 5446.77 | 6963.085662 | 6835.806821 | 794.195 | 3775.57 | 3671.65 | 6023.328184 | 5739.046882 | 4962.974003 | 4797.222487 |
| 794.477 | 5553.01 | 5458 | 6998.006634 | 6849.820147 | 794.477 | 3785.2 | 3680.35 | 6015.093903 | 5752.552391 | 4939.616848 | 4808.8722 |
| 794.76 | 5564.27 | 5469.24 | 6885.557084 | 6863.857124 | 794.76 | 3826.32 | 3689.06 | 6039.989988 | 5766.068635 | 4841.754641 | 4820.533459 |
| 795.042 | 5563.74 | 5480.48 | 7086.485709 | 6877.894101 | 795.042 | 3743.53 | 3697.78 | 5961.640859 | 5779.606352 | 4926.951105 | 4832.21781 |
| 795.324 | 5660.23 | 5491.74 | 7001.365102 | 6891.954729 | 795.324 | 3869.77 | 3706.51 | 5896.314132 | 5793.165539 | 5002.275909 | 4843.913706 |
| 795.606 | 5653.97 | 5503.01 | 7054.296449 | 6906.027183 | 795.606 | 3872.83 | 3715.26 | 6044.692224 | 5806.735462 | 4917.021716 | 4855.621148 |
| 795.888 | 5756.79 | 5514.29 | 7089.453933 | 6920.111463 | 795.888 | 3815.82 | 3724.01 | 5971.904187 | 5820.326857 | 4959.198526 | 4867.351681 |
| 796.17 | 5681.44 | 5525.58 | 7066.405855 | 6934.207568 | 796.17 | 3839.54 | 3732.78 | 6021.309873 | 5833.939723 | 4973.319041 | 4879.093761 |
| 796.452 | 5590.23 | 5536.89 | 7056.093939 | 6948.315498 | 796.452 | 3754.82 | 3741.56 | 5947.501944 | 5847.563324 | 5062.926 | 4890.858932 |
| 796.734 | 5630.24 | 5548.2 | 7115.58849 | 6962.44708 | 796.734 | 3821.38 | 3750.35 | 6054.365089 | 5861.208398 | 4964.821331 | 4902.635648 |
| 797.016 | 5697.83 | 5559.52 | 7214.131144 | 6976.578662 | 797.016 | 3829.73 | 3759.15 | 6117.072305 | 5874.874942 | 4919.030686 | 4914.423911 |
| 797.298 | 5621.4 | 5570.86 | 7201.430459 | 6990.733895 | 797.298 | 3912.35 | 3767.97 | 6013.483548 | 5888.552222 | 4953.806637 | 4926.235265 |
| 797.58 | 5649.35 | 5582.21 | 7144.289201 | 7004.900954 | 797.58 | 3934.7 | 3776.79 | 6236.743139 | 5902.250974 | 5047.235257 | 4938.069711 |
| 797.862 | 5769.88 | 5593.56 | 7252.256851 | 7019.079838 | 797.862 | 3886.84 | 3785.63 | 6043.522033 | 5915.971197 | 5023.91274 | 4949.904157 |
| 798.144 | 5734.95 | 5604.93 | 7233.667021 | 7033.270548 | 798.144 | 3911.55 | 3794.48 | 6173.831944 | 5929.702155 | 5200.020839 | 4961.77324 |
| 798.426 | 5856.28 | 5616.31 | 7088.046687 | 7047.484909 | 798.426 | 3858.04 | 3803.34 | 6138.479289 | 5943.454585 | 5079.332583 | 4973.642323 |
| 798.707 | 5708.46 | 5627.7 | 7239.165921 | 7061.69927 | 798.707 | 3887.36 | 3812.21 | 6165.705021 | 5957.228487 | 5161.861967 | 4985.534498 |
| 798.989 | 5769.84 | 5639.1 | 7306.962037 | 7075.937282 | 798.989 | 3875.07 | 3821.09 | 6177.417668 | 5971.013124 | 5154.830574 | 4997.438218 |
| 799.271 | 5795.4 | 5650.51 | 7232.153346 | 7090.175294 | 799.271 | 3879.14 | 3829.99 | 6213.800951 | 5984.819233 | 4995.740986 | 5009.365031 |
| 799.553 | 5699.92 | 5661.94 | 7331.109895 | 7104.436957 | 799.553 | 3991.21 | 3838.9 | 6184.320722 | 5998.646813 | 5122.190596 | 5021.303389 |
| 799.835 | 5872.23 | 5673.37 | 7269.403959 | 7118.710446 | 799.835 | 3987.19 | 3847.81 | 6191.545847 | 6012.485128 | 5042.362929 | 5033.264838 |
| 800.117 | 5808.32 | 5684.81 | 7234.021789 | 7133.007586 | 800.117 | 3917.95 | 3856.74 | 6261.993502 | 6026.344915 | 5140.940976 | 5045.237834 |
| 800.399 | 5858.06 | 5696.27 | 7331.015291 | 7147.304727 | 800.399 | 3946.48 | 3865.68 | 6314.834612 | 6040.215438 | 5183.002328 | 5057.233921 |
| 800.68 | 5835.15 | 5707.73 | 7284.008564 | 7161.613692 | 800.68 | 3942.08 | 3874.64 | 6243.066466 | 6054.107432 | 5182.020935 | 5069.230008 |
| 800.962 | 5832.39 | 5719.21 | 7365.096645 | 7175.946309 | 800.962 | 3953.15 | 3883.6 | 6236.893439 | 6068.020898 | 5192.458339 | 5081.260732 |
| 801.244 | 5844.94 | 5730.7 | 7298.601344 | 7190.278926 | 801.244 | 4034.96 | 3892.58 | 6257.173174 | 6081.945099 | 5208.218357 | 5093.291456 |
| 801.525 | 5843.48 | 5742.2 | 7481.791582 | 7204.635194 | 801.525 | 4051.28 | 3901.57 | 6140.626428 | 6095.890772 | 5163.743933 | 5105.345272 |
| 801.807 | 5830.61 | 5753.71 | 7381.073019 | 7219.003288 | 801.807 | 4127.86 | 3910.56 | 6302.488558 | 6109.857916 | 5267.644594 | 5117.42218 |
| 802.089 | 5805.87 | 5765.23 | 7324.688599 | 7233.383207 | 802.089 | 3975.91 | 3919.57 | 6055.825144 | 6123.835795 | 5132.096893 | 5129.510633 |
| 802.371 | 5827.14 | 5776.76 | 7312.780228 | 7247.774952 | 802.371 | 4023.8 | 3928.6 | 6350.477131 | 6137.835147 | 5149.346319 | 5141.610633 |
| 802.652 | 5931.13 | 5788.3 | 7322.193399 | 7262.190348 | 802.652 | 4121.83 | 3937.63 | 6252.674916 | 6151.855969 | 5224.867402 | 5153.733723 |
| 802.934 | 5998.97 | 5799.85 | 7444.079771 | 7276.605744 | 802.934 | 4041.39 | 3946.68 | 6476.557177 | 6165.887527 | 5198.173511 | 5165.86836 |
| 803.216 | 5850.42 | 5811.41 | 7415.319932 | 7291.044791 | 803.216 | 4090.88 | 3955.73 | 6319.57979 | 6179.940557 | 5275.911387 | 5178.014542 |
| 803.497 | 5899.52 | 5822.98 | 7447.118948 | 7305.483838 | 803.497 | 4137.49 | 3964.8 | 6489.139416 | 6194.015058 | 5318.884858 | 5190.183816 |
| 803.779 | 5945.2 | 5834.57 | 7377.312481 | 7319.946537 | 803.779 | 4105.02 | 3973.88 | 6289.841905 | 6208.100295 | 5078.559015 | 5202.376182 |
| 804.06 | 6037.83 | 5846.16 | 7421.717577 | 7334.421061 | 804.06 | 4201.04 | 3982.97 | 6410.801022 | 6222.207003 | 5403.792676 | 5214.568548 |
| 804.342 | 5937.73 | 5857.77 | 7517.646776 | 7348.90741 | 804.342 | 4159.9 | 3992.07 | 6223.355723 | 6236.324447 | 5300.827225 | 5226.795551 |
| 804.624 | 5927.84 | 5869.39 | 7375.290305 | 7363.405586 | 804.624 | 4032.3 | 4001.18 | 6366.140516 | 6250.463362 | 5252.369499 | 5239.022554 |
| 804.905 | 5952.15 | 5881.01 | 7485.753155 | 7377.915586 | 804.905 | 4012.72 | 4010.31 | 6402.931755 | 6264.623749 | 5286.62589 | 5251.272648 |
| 805.187 | 6003.97 | 5892.65 | 7477.6408 | 7392.449238 | 805.187 | 4117.56 | 4019.45 | 6364.519425 | 6278.794871 | 5329.068254 | 5263.534289 |
| 805.468 | 5986.7 | 5904.3 | 7572.600299 | 7406.98289 | 805.468 | 4144.11 | 4028.59 | 6445.670572 | 6292.987465 | 5434.077312 | 5275.819021 |
| 805.75 | 5973.35 | 5915.95 | 7480.407988 | 7421.528368 | 805.75 | 4063.75 | 4037.76 | 6448.537003 | 6307.190794 | 5378.784472 | 5288.115298 |
| 806.031 | 6063.92 | 5927.62 | 7612.381589 | 7436.097497 | 806.031 | 4156.19 | 4046.93 | 6638.880942 | 6321.426331 | 5526.801638 | 5300.434668 |
| 806.313 | 6062.04 | 5939.3 | 7537.277258 | 7450.678451 | 806.313 | 4183.91 | 4056.11 | 6382.1045 | 6335.661867 | 5421.330748 | 5312.765583 |
| 806.594 | 6079.97 | 5950.99 | 7579.979468 | 7465.271231 | 806.594 | 4230.09 | 4065.3 | 6541.529626 | 6349.929611 | 5435.982369 | 5325.108044 |
| 806.875 | 6073.61 | 5962.69 | 7560.762882 | 7479.875836 | 806.875 | 4189.77 | 4074.51 | 6473.669274 | 6364.20809 | 5399.763192 | 5337.473597 |
| 807.157 | 6005.69 | 5974.4 | 7459.736854 | 7494.492268 | 807.157 | 4219.6 | 4083.72 | 6560.274156 | 6378.508041 | 5371.718442 | 5349.850695 |
| 807.438 | 6110.33 | 5986.13 | 7675.246434 | 7509.120524 | 807.438 | 4133.78 | 4092.95 | 6501.603562 | 6392.818727 | 5563.725108 | 5362.250885 |
| 807.72 | 6035.09 | 5997.86 | 7630.864989 | 7523.760606 | 807.72 | 4198.37 | 4102.19 | 6639.095656 | 6407.150885 | 5531.88179 | 5374.662621 |
| 808.001 | 6136.33 | 6009.6 | 7764.825289 | 7538.412514 | 808.001 | 4204.89 | 4111.44 | 6634.414891 | 6421.504514 | 5402.257085 | 5387.085903 |
| 808.283 | 6138.34 | 6021.35 | 7547.258057 | 7553.088073 | 808.283 | 4297.1 | 4120.71 | 6505.361057 | 6435.868879 | 5469.130363 | 5399.532276 |
| 808.564 | 6099.8 | 6033.12 | 7625.484345 | 7567.763632 | 808.564 | 4235.07 | 4129.98 | 6537.82581 | 6450.254715 | 5482.119389 | 5411.990195 |
| 808.845 | 6139.9 | 6044.89 | 7535.739931 | 7582.462843 | 808.845 | 4180.16 | 4139.27 | 6591.117818 | 6464.651287 | 5489.508701 | 5424.45966 |
| 809.126 | 6124.54 | 6056.68 | 7604.363838 | 7597.173879 | 809.126 | 4307.31 | 4148.56 | 6690.036546 | 6479.069331 | 5514.643909 | 5436.952216 |
| 809.408 | 6082.44 | 6068.47 | 7649.253784 | 7611.884914 | 809.408 | 4201.35 | 4157.87 | 6583.194873 | 6493.508845 | 5515.024921 | 5449.467864 |
| 809.689 | 6164.04 | 6080.28 | 7747.216984 | 7626.619601 | 809.689 | 4295.68 | 4167.19 | 6481.51707 | 6507.959096 | 5622.677967 | 5461.983512 |
| 809.97 | 6275.47 | 6092.09 | 7745.880692 | 7641.366114 | 809.97 | 4315.19 | 4176.52 | 6796.964106 | 6522.430818 | 5615.762032 | 5474.522252 |
| 810.252 | 6138.89 | 6103.92 | 7763.146055 | 7656.124452 | 810.252 | 4336.1 | 4185.86 | 6626.567095 | 6536.924011 | 5520.312898 | 5487.084083 |
| 810.533 | 6117.79 | 6115.75 | 7684.588651 | 7670.894616 | 810.533 | 4337.09 | 4195.22 | 6725.89378 | 6551.42794 | 5500.534941 | 5499.65746 |
| 810.814 | 6209.23 | 6127.6 | 7797.144631 | 7685.676605 | 810.814 | 4361.74 | 4204.58 | 6590.505883 | 6565.95334 | 5533.013278 | 5512.242383 |
| 811.095 | 6182.68 | 6139.46 | 7616.5915 | 7700.482246 | 811.095 | 4285.22 | 4213.96 | 6699.451754 | 6580.489476 | 5557.328735 | 5524.850397 |
| 811.376 | 6200.25 | 6151.32 | 7677.824413 | 7715.287886 | 811.376 | 4192.08 | 4223.35 | 6581.745553 | 6595.047084 | 5461.648684 | 5537.469958 |
| 811.658 | 6124.09 | 6163.2 | 7737.661906 | 7730.117178 | 811.658 | 4176.09 | 4232.74 | 6599.727849 | 6609.626163 | 5584.58837 | 5550.101064 |
| 811.939 | 6222.65 | 6175.09 | 7777.809789 | 7744.94647 | 811.939 | 4306.08 | 4242.15 | 6655.897024 | 6624.215977 | 5611.374628 | 5562.755261 |
| 812.22 | 6139.24 | 6186.99 | 7650.637378 | 7759.799413 | 812.22 | 4252.94 | 4251.58 | 6623.045786 | 6638.827263 | 5437.991338 | 5575.43255 |
| 812.501 | 6192.41 | 6198.9 | 7715.015898 | 7774.652356 | 812.501 | 4295.88 | 4261.01 | 6714.524675 | 6653.449285 | 5706.604393 | 5588.10984 |
| 812.782 | 6147.96 | 6210.82 | 7812.789888 | 7789.52895 | 812.782 | 4269.03 | 4270.45 | 6760.398316 | 6668.092778 | 5512.184654 | 5600.810221 |
| 813.063 | 6287.7 | 6222.74 | 7907.028027 | 7804.41737 | 813.063 | 4393.91 | 4279.91 | 6814.302259 | 6682.757743 | 5758.11021 | 5613.533693 |
| 813.344 | 6260.82 | 6234.68 | 7870.959973 | 7819.317615 | 813.344 | 4248.64 | 4289.37 | 6738.626319 | 6697.433443 | 5568.782169 | 5626.268711 |
| 813.625 | 6290.63 | 6246.63 | 7951.4686 | 7834.229686 | 813.625 | 4427.47 | 4298.85 | 6765.283059 | 6712.130614 | 5663.480827 | 5639.015275 |
| 813.907 | 6249.71 | 6258.59 | 7807.267337 | 7849.153582 | 813.907 | 4504.13 | 4308.34 | 6869.548165 | 6726.838521 | 5820.653813 | 5651.773385 |
| 814.187 | 6324.3 | 6270.56 | 7966.581706 | 7864.089304 | 814.187 | 4345.49 | 4317.84 | 6725.679066 | 6741.578636 | 5742.419467 | 5664.554587 |
| 814.468 | 6294.85 | 6282.54 | 7888.816616 | 7879.036852 | 814.468 | 4293.08 | 4327.36 | 6794.323124 | 6756.31875 | 5625.853062 | 5677.35888 |
| 814.75 | 6293.15 | 6294.53 | 8056.113259 | 7893.996225 | 814.75 | 4323.56 | 4336.88 | 6783.179468 | 6771.091072 | 5793.544273 | 5690.174718 |
| 815.03 | 6318.13 | 6306.53 | 7949.77754 | 7908.967424 | 815.03 | 4464.41 | 4346.41 | 6916.849653 | 6785.874129 | 5800.78349 | 5703.002103 |
| 815.312 | 6379.84 | 6318.54 | 8011.388872 | 7923.962274 | 815.312 | 4437.27 | 4355.96 | 6847.04614 | 6800.667922 | 5853.305337 | 5715.841033 |
| 815.592 | 6390.06 | 6330.56 | 7942.859569 | 7938.957124 | 815.592 | 4420.99 | 4365.52 | 6886.886318 | 6815.483186 | 5669.923384 | 5728.703056 |
| 815.873 | 6397.16 | 6342.6 | 7974.114608 | 7953.9638 | 815.873 | 4496.92 | 4375.09 | 6890.235856 | 6830.319921 | 5798.855341 | 5741.576623 |
| 816.154 | 6462.38 | 6354.64 | 8087.782193 | 7968.994127 | 816.154 | 4408.54 | 4384.66 | 6864.212522 | 6845.167393 | 5849.991693 | 5754.473283 |
| 816.435 | 6429.42 | 6366.69 | 8032.816844 | 7984.024453 | 816.435 | 4433.61 | 4394.26 | 6916.065947 | 6860.036336 | 5871.674706 | 5767.381488 |
| 816.716 | 6459.9 | 6378.75 | 7973.818968 | 7999.078431 | 816.716 | 4471.82 | 4403.86 | 7120.409237 | 6874.92675 | 5869.504096 | 5780.301239 |
| 816.997 | 6335.25 | 6390.82 | 7957.653384 | 8014.144235 | 816.997 | 4503 | 4413.47 | 7105.980458 | 6889.8279 | 5738.851814 | 5793.244082 |
| 817.278 | 6583.99 | 6402.9 | 8086.930751 | 8029.210039 | 817.278 | 4514.59 | 4423.1 | 6967.50068 | 6904.739785 | 5940.395313 | 5806.210016 |
| 817.559 | 6457.57 | 6414.99 | 8099.950727 | 8044.299493 | 817.559 | 4479.9 | 4432.73 | 6842.784068 | 6919.673142 | 5848.225185 | 5819.17595 |
| 817.84 | 6492.41 | 6427.09 | 8225.384776 | 8059.400774 | 817.84 | 4536.96 | 4442.38 | 7099.732281 | 6934.62797 | 5955.85514 | 5832.164976 |
| 818.12 | 6436.84 | 6439.2 | 8114.638112 | 8074.51388 | 818.12 | 4550.18 | 4452.04 | 7052.720656 | 6949.60427 | 5861.814593 | 5845.165548 |
| 818.401 | 6456.82 | 6451.32 | 8142.026182 | 8089.626986 | 818.401 | 4499.07 | 4461.7 | 7056.843165 | 6964.591305 | 5879.514305 | 5858.189211 |
| 818.682 | 6587.85 | 6463.45 | 8206.629388 | 8104.763743 | 818.682 | 4516.31 | 4471.39 | 7089.179089 | 6979.599812 | 6016.459048 | 5871.22442 |
| 818.963 | 6498.09 | 6475.6 | 8193.77497 | 8119.912325 | 818.963 | 4567.34 | 4481.08 | 7085.625573 | 6994.619055 | 5877.436061 | 5884.282721 |
| 819.243 | 6577.35 | 6487.75 | 8286.570387 | 8135.072734 | 819.243 | 4548.03 | 4490.78 | 7118.734468 | 7009.659769 | 5944.286247 | 5897.341021 |
| 819.524 | 6487.02 | 6499.91 | 8114.283344 | 8150.244968 | 819.524 | 4495.4 | 4500.49 | 7060.622131 | 7024.711218 | 5845.777475 | 5910.43396 |
| 819.805 | 6597.72 | 6512.08 | 8303.13804 | 8165.429027 | 819.805 | 4600.03 | 4510.22 | 7184.018252 | 7039.784139 | 5992.824794 | 5923.526898 |
| 820.086 | 6551.63 | 6524.26 | 8150.682515 | 8180.624912 | 820.086 | 4529.97 | 4519.95 | 7035.618688 | 7054.878532 | 5907.963158 | 5936.642927 |
| 820.366 | 6565.63 | 6536.45 | 8333.02131 | 8195.844448 | 820.366 | 4635.18 | 4529.7 | 7230.578977 | 7069.98366 | 6086.472784 | 5949.782049 |
| 820.647 | 6635.07 | 6548.65 | 8302.393028 | 8211.063985 | 820.647 | 4586.15 | 4539.46 | 7131.939377 | 7085.099524 | 5944.574892 | 5962.92117 |
| 820.928 | 6623.34 | 6560.86 | 8265.083287 | 8226.295347 | 820.928 | 4590.23 | 4549.23 | 7164.447073 | 7100.247595 | 6141.938811 | 5976.083383 |
| 821.208 | 6681.8 | 6573.08 | 8388.294125 | 8241.538534 | 821.208 | 4651.57 | 4559.01 | 7344.076784 | 7115.406401 | 6131.224307 | 5989.268687 |
| 821.489 | 6637.52 | 6585.31 | 8368.852852 | 8256.793547 | 821.489 | 4568.81 | 4568.8 | 7214.969271 | 7130.575943 | 5993.748458 | 6002.465537 |
| 821.769 | 6667.86 | 6597.55 | 8291.253321 | 8272.072211 | 821.769 | 4666.73 | 4578.61 | 7347.909428 | 7145.766957 | 6120.08261 | 6015.673933 |
| 822.05 | 6732.29 | 6609.8 | 8328.811399 | 8287.350876 | 822.05 | 4633.25 | 4588.42 | 7161.280042 | 7160.979442 | 6105.407897 | 6028.905421 |
| 822.331 | 6650.96 | 6622.06 | 8260.94433 | 8302.641365 | 822.331 | 4765.37 | 4598.25 | 7383.584155 | 7176.202663 | 6192.867338 | 6042.136909 |
| 822.611 | 6775.09 | 6634.33 | 8448.616467 | 8317.943681 | 822.611 | 4725.64 | 4608.08 | 7264.514521 | 7191.447355 | 5978.773554 | 6055.403034 |
| 822.892 | 6776.15 | 6646.61 | 8438.588365 | 8333.269647 | 822.892 | 4650.92 | 4617.93 | 7405.388359 | 7206.702783 | 6312.620383 | 6068.669159 |
| 823.172 | 6782.34 | 6658.9 | 8437.311201 | 8348.595614 | 823.172 | 4726.5 | 4627.79 | 7307.854536 | 7221.979682 | 6128.44177 | 6081.958376 |
| 823.453 | 6744.95 | 6671.19 | 8310.741896 | 8363.933406 | 823.453 | 4749.7 | 4637.66 | 7454.128432 | 7237.278053 | 6089.266868 | 6095.270684 |
| 823.733 | 6837.61 | 6683.5 | 8439.877355 | 8379.294849 | 823.733 | 4757.4 | 4647.54 | 7429.071311 | 7252.587159 | 6182.937949 | 6108.594538 |
| 824.014 | 6770.87 | 6695.82 | 8468.317902 | 8394.656293 | 824.014 | 4733.23 | 4657.43 | 7433.29044 | 7267.907002 | 6233.185274 | 6121.929938 |
| 824.294 | 6667.52 | 6708.15 | 8451.691121 | 8410.041387 | 824.294 | 4725.95 | 4667.33 | 7329.540648 | 7283.259051 | 6069.211812 | 6135.276884 |
| 824.574 | 6794.9 | 6720.48 | 8416.557288 | 8425.426482 | 824.574 | 4747.5 | 4677.25 | 7413.99839 | 7298.6111 | 6300.832121 | 6148.646921 |
| 824.855 | 6725.49 | 6732.83 | 8579.490288 | 8440.835228 | 824.855 | 4747.46 | 4687.17 | 7493.539182 | 7313.995356 | 6186.724972 | 6162.028504 |
| 825.135 | 6813.7 | 6745.19 | 8553.923359 | 8456.243973 | 825.135 | 4744.56 | 4697.11 | 7361.629651 | 7329.390348 | 6190.223349 | 6175.433179 |
| 825.416 | 6832.64 | 6757.55 | 8529.574466 | 8471.664545 | 825.416 | 4892.44 | 4707.06 | 7512.734611 | 7344.796076 | 6098.595875 | 6188.849399 |
| 825.696 | 6868.78 | 6769.93 | 8561.811029 | 8487.108767 | 825.696 | 4661.87 | 4717.01 | 7413.15027 | 7360.223275 | 6352.349484 | 6202.277165 |
| 825.976 | 6933.87 | 6782.31 | 8502.600292 | 8502.55299 | 825.976 | 4815.11 | 4726.98 | 7566.831797 | 7375.671945 | 6363.768281 | 6215.728023 |
| 826.257 | 6941.29 | 6794.71 | 8679.936862 | 8518.020863 | 826.257 | 4818.07 | 4736.96 | 7403.262691 | 7391.131352 | 6342.420095 | 6229.190427 |
| 826.537 | 6836.77 | 6807.11 | 8580.625545 | 8533.488737 | 826.537 | 4729.21 | 4746.95 | 7532.091076 | 7406.612229 | 6242.872201 | 6242.664376 |
| 826.817 | 6881.52 | 6819.53 | 8650.432012 | 8548.980262 | 826.817 | 4724.33 | 4756.96 | 7529.675544 | 7422.103842 | 6379.124196 | 6256.161418 |
| 827.098 | 6878.66 | 6831.95 | 8683.437237 | 8564.471787 | 827.098 | 4835.25 | 4766.97 | 7517.55494 | 7437.616927 | 6380.463509 | 6269.670004 |
| 827.378 | 6793.2 | 6844.38 | 8694.647898 | 8579.986963 | 827.378 | 4886.32 | 4776.99 | 7623.80615 | 7453.140747 | 6285.637847 | 6283.201683 |
| 827.658 | 6884.28 | 6856.83 | 8504.634294 | 8595.502139 | 827.658 | 4734.88 | 4787.03 | 7526.122027 | 7468.686039 | 6304.596052 | 6296.733361 |
| 827.939 | 6917.47 | 6869.28 | 8573.23455 | 8611.029141 | 827.939 | 4868.46 | 4797.07 | 7592.071425 | 7484.252802 | 6445.443275 | 6310.299677 |
| 828.219 | 6886.95 | 6881.74 | 8695.842283 | 8626.579793 | 828.219 | 4911.19 | 4807.13 | 7683.367807 | 7499.830301 | 6481.292987 | 6323.865993 |
| 828.499 | 6959.74 | 6894.21 | 8588.371307 | 8642.130446 | 828.499 | 4816.64 | 4817.2 | 7593.348973 | 7515.418536 | 6302.540899 | 6337.455401 |
| 828.779 | 6940.24 | 6906.69 | 8760.528268 | 8657.692925 | 828.779 | 4830.11 | 4827.28 | 7787.020978 | 7531.038977 | 6409.074003 | 6351.056354 |
| 829.059 | 6903.94 | 6919.18 | 8705.834908 | 8673.279054 | 829.059 | 4856.54 | 4837.37 | 7712.171686 | 7546.659419 | 6380.059406 | 6364.680399 |
| 829.34 | 6999.23 | 6931.68 | 8884.779758 | 8688.865184 | 829.34 | 5016.54 | 4847.47 | 7722.681935 | 7562.312068 | 6566.962828 | 6378.31599 |
| 829.62 | 6962.29 | 6944.19 | 8806.624424 | 8704.463139 | 829.62 | 5017.13 | 4857.58 | 7622.045495 | 7577.964716 | 6519.070847 | 6391.963126 |
| 829.9 | 6966.33 | 6956.71 | 8827.449291 | 8720.084745 | 829.9 | 5058.37 | 4867.7 | 7658.514664 | 7593.649572 | 6378.188986 | 6405.633354 |
| 830.18 | 7060.57 | 6969.23 | 8725.299831 | 8735.706352 | 830.18 | 4846.73 | 4877.83 | 7757.71252 | 7609.345164 | 6666.499177 | 6419.315128 |
| 830.46 | 7132.32 | 6981.77 | 8899.384364 | 8751.339784 | 830.46 | 4969.78 | 4887.98 | 7865.41305 | 7625.051491 | 6534.149662 | 6433.008448 |
| 830.74 | 7015.74 | 6994.32 | 8986.881915 | 8766.996867 | 830.74 | 5060.49 | 4898.13 | 7765.613994 | 7640.77929 | 6487.90873 | 6446.724859 |
| 831.02 | 7153.24 | 7006.87 | 8968.859714 | 8782.65395 | 831.02 | 5010.62 | 4908.3 | 7857.973211 | 7656.52856 | 6590.262254 | 6460.452816 |
| 831.3 | 7068.83 | 7019.44 | 8815.079722 | 8798.322859 | 831.3 | 5056.81 | 4918.48 | 7847.903125 | 7672.288565 | 6585.366835 | 6474.192319 |
| 831.58 | 7088.87 | 7032.01 | 8765.601447 | 8814.003593 | 831.58 | 4936.52 | 4928.66 | 7809.426381 | 7688.070043 | 6469.620182 | 6487.954914 |
| 831.86 | 7091.17 | 7044.6 | 8811.780382 | 8829.696153 | 831.86 | 5020.91 | 4938.86 | 7704.238005 | 7703.862255 | 6675.932096 | 6501.729054 |
| 832.14 | 7045.59 | 7057.19 | 8811.200928 | 8845.400539 | 832.14 | 5124.9 | 4949.07 | 7847.634733 | 7719.665204 | 6557.564546 | 6515.51474 |
| 832.42 | 7072.04 | 7069.79 | 8776.705678 | 8861.128575 | 832.42 | 4990.61 | 4959.29 | 7777.30517 | 7735.50036 | 6548.697371 | 6529.323518 |
| 832.7 | 7114.57 | 7082.4 | 8877.353287 | 8876.856612 | 832.7 | 5041.52 | 4969.52 | 7757.143528 | 7751.335515 | 6604.729143 | 6543.143841 |
| 832.98 | 7196.41 | 7095.02 | 8941.554423 | 8892.584649 | 832.98 | 5086.95 | 4979.76 | 7738.34532 | 7767.202878 | 6678.460626 | 6556.987256 |
| 833.26 | 7178.4 | 7107.65 | 8957.412541 | 8908.336337 | 833.26 | 5007.07 | 4990.02 | 7801.567849 | 7783.080976 | 6664.732669 | 6570.830671 |
| 833.54 | 7176.12 | 7120.29 | 8966.0807 | 8924.09985 | 833.54 | 5106.16 | 5000.28 | 8006.909559 | 7798.96981 | 6688.147553 | 6584.697178 |
| 833.82 | 7117.38 | 7132.94 | 8953.380015 | 8939.875189 | 833.82 | 4982.93 | 5010.55 | 7735.940523 | 7814.880116 | 6603.47065 | 6598.586777 |
| 834.1 | 7107.81 | 7145.6 | 8885.01627 | 8955.662354 | 834.1 | 5143.18 | 5020.84 | 7923.933344 | 7830.801157 | 6714.679803 | 6612.487921 |
| 834.38 | 7232.32 | 7158.26 | 8948.153103 | 8971.461344 | 834.38 | 5099 | 5031.14 | 7941.700925 | 7846.74367 | 6738.556519 | 6626.400611 |
| 834.66 | 7063.47 | 7170.94 | 9018.196082 | 8987.27216 | 834.66 | 5043.6 | 5041.44 | 7842.223941 | 7862.707654 | 6597.905574 | 6640.324846 |
| 834.939 | 7222.02 | 7183.62 | 8999.204182 | 9003.082976 | 834.939 | 5050.15 | 5051.76 | 7919.456557 | 7878.682373 | 6696.702991 | 6654.272174 |
| 835.219 | 7175.24 | 7196.32 | 8942.58325 | 9018.917443 | 835.219 | 4910.85 | 5062.09 | 7932.811766 | 7894.667829 | 6721.076177 | 6668.231047 |
| 835.499 | 7247.1 | 7209.02 | 9086.879117 | 9034.763735 | 835.499 | 5153.98 | 5072.43 | 8027.683136 | 7910.674756 | 6804.009664 | 6682.213011 |
| 835.779 | 7210.85 | 7221.73 | 9039.198332 | 9050.610028 | 835.779 | 5073.15 | 5082.78 | 7899.047994 | 7926.703154 | 6791.574836 | 6696.206522 |
| 836.059 | 7275.36 | 7234.45 | 9054.678032 | 9066.479972 | 836.059 | 5194.62 | 5093.14 | 8163.844003 | 7942.742288 | 6814.41243 | 6710.211578 |
| 836.338 | 7328.68 | 7247.19 | 9035.757085 | 9082.349916 | 836.338 | 5097.47 | 5103.51 | 7934.55095 | 7958.802893 | 6828.359757 | 6724.22818 |
| 836.618 | 7304.8 | 7259.93 | 9199.955423 | 9098.231685 | 836.618 | 5174.53 | 5113.89 | 8184.145209 | 7974.874234 | 6843.219203 | 6738.267874 |
| 836.898 | 7237.9 | 7272.67 | 9077.596028 | 9114.137106 | 836.898 | 5153.35 | 5124.28 | 7953.274008 | 7990.956311 | 6779.763482 | 6752.319114 |
| 837.178 | 7279.72 | 7285.43 | 9195.615431 | 9130.042526 | 837.178 | 5294.18 | 5134.69 | 8053.266306 | 8007.070594 | 6988.130548 | 6766.393445 |
| 837.457 | 7301.74 | 7298.2 | 9240.931098 | 9145.959772 | 837.457 | 5186.69 | 5145.1 | 8176.114907 | 8023.184878 | 6810.498404 | 6780.467776 |
| 837.737 | 7179.35 | 7310.97 | 9058.0365 | 9161.888844 | 837.737 | 5142.82 | 5155.52 | 8238.886537 | 8039.331369 | 6710.604136 | 6794.576744 |
| 838.017 | 7303.11 | 7323.76 | 9242.66946 | 9177.829741 | 838.017 | 5268.97 | 5165.96 | 8264.534121 | 8055.47786 | 6818.15327 | 6808.685713 |
| 838.296 | 7315.95 | 7336.55 | 9252.248189 | 9193.782464 | 838.296 | 5191.06 | 5176.4 | 8089.12354 | 8071.656558 | 6833.786284 | 6822.817773 |
| 838.576 | 7379.95 | 7349.35 | 9159.168958 | 9209.747013 | 838.576 | 5253.69 | 5186.86 | 8112.344856 | 8087.835256 | 7003.451826 | 6836.961379 |
| 838.856 | 7395.79 | 7362.16 | 9192.174184 | 9225.723387 | 838.856 | 5201.39 | 5197.33 | 8196.845541 | 8104.046161 | 6989.354403 | 6851.116531 |
| 839.135 | 7396.31 | 7374.98 | 9391.801994 | 9241.711587 | 839.135 | 5197.13 | 5207.81 | 8357.698515 | 8120.257066 | 6745.691824 | 6865.294774 |
| 839.415 | 7337.47 | 7387.81 | 9295.045004 | 9257.711612 | 839.415 | 5183.09 | 5218.29 | 8248.881473 | 8136.500178 | 6950.548967 | 6879.484563 |
| 839.694 | 7408.63 | 7400.65 | 9330.308918 | 9273.711637 | 839.694 | 5229.47 | 5228.79 | 8316.280189 | 8152.74329 | 6812.576648 | 6893.685898 |
| 839.974 | 7454.92 | 7413.49 | 9262.051604 | 9289.735314 | 839.974 | 5331.03 | 5239.3 | 8321.132725 | 8169.01861 | 6898.246489 | 6907.910325 |
| 840.254 | 7445.48 | 7426.35 | 9367.595008 | 9305.75899 | 840.254 | 5233.11 | 5249.82 | 8318.867493 | 8185.293929 | 6954.509177 | 6922.146297 |
| 840.533 | 7459.78 | 7439.22 | 9322.409423 | 9321.806318 | 840.533 | 5387.82 | 5260.35 | 8272.188674 | 8201.601455 | 7003.255547 | 6936.405361 |
| 840.812 | 7449.04 | 7452.09 | 9408.298694 | 9337.853646 | 840.812 | 5292.73 | 5270.9 | 8304.77152 | 8217.919717 | 7037.869858 | 6950.664425 |
| 841.092 | 7526.02 | 7464.97 | 9353.605334 | 9353.912799 | 841.092 | 5357.25 | 5281.45 | 8311.696046 | 8234.248715 | 7011.418429 | 6964.946581 |
| 841.371 | 7453.52 | 7477.86 | 9436.69194 | 9369.983778 | 841.371 | 5367.5 | 5292.01 | 8253.336788 | 8250.588449 | 6876.020823 | 6979.240282 |
| 841.651 | 7391.11 | 7490.76 | 9490.935928 | 9386.078408 | 841.651 | 5309.25 | 5302.58 | 8227.109476 | 8266.960389 | 6986.929785 | 6993.557075 |
| 841.93 | 7481.48 | 7503.67 | 9440.074059 | 9402.173038 | 841.93 | 5469.77 | 5313.17 | 8528.503482 | 8283.343065 | 7111.705254 | 7007.885414 |
| 842.21 | 7526.42 | 7516.59 | 9525.691341 | 9418.267668 | 842.21 | 5384.34 | 5323.76 | 8265.221206 | 8299.736477 | 7068.766421 | 7022.225298 |
| 842.489 | 7695.05 | 7529.51 | 9567.471156 | 9434.385949 | 842.489 | 5525.38 | 5334.37 | 8458.592611 | 8316.151361 | 7197.883111 | 7036.588274 |
| 842.769 | 7578.53 | 7542.44 | 9505.304021 | 9450.516056 | 842.769 | 5390.41 | 5344.98 | 8417.367528 | 8332.57698 | 6999.895719 | 7050.962796 |
| 843.048 | 7443.88 | 7555.39 | 9440.180489 | 9466.657988 | 843.048 | 5398.43 | 5355.61 | 8378.107078 | 8349.013335 | 7145.31508 | 7065.348864 |
| 843.327 | 7582.94 | 7568.34 | 9394.628311 | 9482.799921 | 843.327 | 5332.23 | 5366.24 | 8291.759853 | 8365.481896 | 7070.579112 | 7079.746477 |
| 843.607 | 7576.72 | 7581.3 | 9482.8827 | 9498.965504 | 843.607 | 5390.17 | 5376.89 | 8414.887582 | 8381.961194 | 7003.913658 | 7094.167183 |
| 843.886 | 7634.52 | 7594.27 | 9539.586411 | 9515.131088 | 843.886 | 5494.01 | 5387.55 | 8496.554042 | 8398.451227 | 7262.851332 | 7108.599434 |
| 844.165 | 7553.54 | 7607.25 | 9623.394378 | 9531.308497 | 844.165 | 5532.24 | 5398.22 | 8479.280303 | 8414.951996 | 7205.133874 | 7123.054776 |
| 844.444 | 7639.2 | 7620.23 | 9592.813399 | 9547.497732 | 844.444 | 5518.56 | 5408.9 | 8573.529002 | 8431.484972 | 7277.572228 | 7137.510119 |
| 844.724 | 7717.62 | 7633.23 | 9522.285571 | 9563.698792 | 844.724 | 5467.13 | 5419.58 | 8542.255912 | 8448.017948 | 7348.278712 | 7151.988553 |
| 845.003 | 7606.62 | 7646.23 | 9572.106788 | 9579.911678 | 845.003 | 5362.65 | 5430.28 | 8510.177644 | 8464.572396 | 7092.931782 | 7166.490079 |
| 845.282 | 7722.96 | 7659.24 | 9603.72842 | 9596.13639 | 845.282 | 5415.61 | 5440.99 | 8457.733756 | 8481.148314 | 7239.10162 | 7180.991604 |
| 845.562 | 7674.32 | 7672.26 | 9538.285596 | 9612.372927 | 845.562 | 5492.21 | 5451.71 | 8362.797972 | 8497.734969 | 7117.201055 | 7195.516222 |
| 845.841 | 7671.52 | 7685.29 | 9552.559085 | 9628.609464 | 845.841 | 5468.4 | 5462.44 | 8457.87332 | 8514.332359 | 7190.909447 | 7210.052385 |
| 846.12 | 7644.02 | 7698.33 | 9774.489961 | 9644.869652 | 846.12 | 5456.57 | 5473.19 | 8634.0354 | 8530.961957 | 7286.25467 | 7224.61164 |
| 846.399 | 7787.39 | 7711.38 | 9688.789899 | 9661.129841 | 846.399 | 5499.15 | 5483.94 | 8488.99611 | 8547.591554 | 7171.131491 | 7239.18244 |
| 846.678 | 7734.56 | 7724.43 | 9649.588063 | 9677.401855 | 846.678 | 5596.45 | 5494.7 | 8499.839166 | 8564.242623 | 7262.782057 | 7253.764787 |
| 846.957 | 7798.19 | 7737.49 | 9626.409904 | 9693.69752 | 846.957 | 5557.67 | 5505.47 | 8600.078385 | 8580.904427 | 7313.087111 | 7268.358679 |
| 847.236 | 7900.44 | 7750.56 | 9808.169245 | 9709.993185 | 847.236 | 5635.45 | 5516.25 | 8514.182059 | 8597.587703 | 7326.064591 | 7282.975662 |
| 847.516 | 7863.13 | 7763.65 | 9826.534389 | 9726.28885 | 847.516 | 5605.54 | 5527.05 | 8591.178491 | 8614.29245 | 7424.273172 | 7297.604192 |
| 847.795 | 7890.56 | 7776.73 | 9872.263951 | 9742.608166 | 847.795 | 5672.17 | 5537.85 | 8736.539851 | 8631.007933 | 7304.416214 | 7312.244267 |
| 848.074 | 7842.9 | 7789.83 | 9819.226174 | 9758.939308 | 848.074 | 5620.71 | 5548.66 | 8729.593854 | 8647.734152 | 7336.628999 | 7326.907434 |
| 848.353 | 7743.56 | 7802.94 | 9836.160421 | 9775.27045 | 848.353 | 5746.69 | 5559.49 | 8647.068538 | 8664.481842 | 7433.913916 | 7341.582147 |
| 848.632 | 7794.32 | 7816.05 | 9713.446258 | 9791.625243 | 848.632 | 5668.48 | 5570.32 | 8811.077808 | 8681.240267 | 7459.891968 | 7356.268406 |
| 848.911 | 7755.53 | 7829.17 | 9848.872931 | 9807.980036 | 848.911 | 5549.13 | 5581.17 | 8767.662642 | 8698.020164 | 7461.289009 | 7370.96621 |
| 849.19 | 7800.75 | 7842.3 | 9867.616494 | 9824.346655 | 849.19 | 5718.57 | 5592.02 | 8684.675691 | 8714.810797 | 7435.969068 | 7385.687106 |
| 849.469 | 7798.2 | 7855.44 | 9803.710997 | 9840.725099 | 849.469 | 5633.31 | 5602.89 | 8720.039083 | 8731.612166 | 7428.799126 | 7400.419548 |
| 849.748 | 7792.41 | 7868.59 | 9805.342929 | 9857.115369 | 849.748 | 5520.15 | 5613.76 | 8797.829955 | 8748.445741 | 7435.472599 | 7415.175081 |
| 850.027 | 7837.66 | 7881.74 | 9741.733072 | 9873.517464 | 850.027 | 5549.8 | 5624.65 | 8819.687838 | 8765.279317 | 7446.30256 | 7429.930615 |
| 850.306 | 7910.57 | 7894.91 | 9917.863432 | 9889.91956 | 850.306 | 5653.33 | 5635.55 | 8646.16674 | 8782.134364 | 7481.41334 | 7444.70924 |
| 850.585 | 7810.69 | 7908.08 | 9848.423559 | 9906.345306 | 850.585 | 5673.99 | 5646.45 | 8739.384812 | 8799.000147 | 7564.219823 | 7459.49941 |
| 850.864 | 7884.91 | 7921.26 | 9736.742673 | 9922.771053 | 850.864 | 5678 | 5657.37 | 8706.436952 | 8815.887401 | 7329.389782 | 7474.312673 |
| 851.143 | 7989.72 | 7934.45 | 9769.58234 | 9939.208625 | 851.143 | 5627.59 | 5668.3 | 8803.262219 | 8832.796126 | 7631.012281 | 7489.137481 |
| 851.421 | 7930.25 | 7947.64 | 9982.868709 | 9955.658023 | 851.421 | 5758.59 | 5679.23 | 8811.968871 | 8849.704852 | 7435.41487 | 7503.973835 |
| 851.7 | 7868.73 | 7960.85 | 9903.294304 | 9972.119246 | 851.7 | 5645.32 | 5690.18 | 8737.280615 | 8866.645784 | 7549.418107 | 7518.821735 |
| 851.979 | 7825.99 | 7974.06 | 9845.218824 | 9988.592295 | 851.979 | 5636.86 | 5701.14 | 8839.183867 | 8883.586717 | 7630.307987 | 7533.692726 |
| 852.258 | 7980.17 | 7987.28 | 9854.561041 | 10005.07717 | 852.258 | 5686.75 | 5712.11 | 9052.362637 | 8900.549121 | 7820.998432 | 7548.575263 |
| 852.537 | 8005.05 | 8000.51 | 10036.6042 | 10021.56204 | 852.537 | 5705.35 | 5723.09 | 8981.81836 | 8917.532996 | 7520.772975 | 7563.469346 |
| 852.816 | 8070.28 | 8013.75 | 9977.984739 | 10038.05874 | 852.816 | 5647.45 | 5734.07 | 8987.690787 | 8934.527607 | 7485.800744 | 7578.374975 |
| 853.094 | 7873.81 | 8026.99 | 9980.98844 | 10054.5791 | 853.094 | 5689.92 | 5745.07 | 8900.602799 | 8951.54369 | 7565.362858 | 7593.303695 |
| 853.373 | 7966.27 | 8040.25 | 9893.159772 | 10071.09945 | 853.373 | 5751.98 | 5756.08 | 8864.208781 | 8968.559772 | 7509.365724 | 7608.243962 |
| 853.652 | 7945.15 | 8053.51 | 9990.413436 | 10087.6198 | 853.652 | 5853.27 | 5767.1 | 8987.841087 | 8985.608062 | 7663.37516 | 7623.195774 |
| 853.93 | 7941.55 | 8066.78 | 10041.52364 | 10104.1638 | 853.93 | 5723.96 | 5778.13 | 9048.433371 | 9002.667087 | 7655.95121 | 7638.170677 |
| 854.209 | 8058.51 | 8080.06 | 9934.833156 | 10120.71963 | 854.209 | 5845.8 | 5789.17 | 8868.65336 | 9019.736848 | 7637.731937 | 7653.157127 |
| 854.488 | 8096.48 | 8093.34 | 10169.27551 | 10137.27546 | 854.488 | 5831.31 | 5800.22 | 8994.02485 | 9036.817345 | 7636.819818 | 7668.155122 |
| 854.766 | 7977.27 | 8106.64 | 10053.12455 | 10153.84311 | 854.766 | 5782.9 | 5811.28 | 9122.928384 | 9053.919313 | 7436.384717 | 7683.164663 |
| 855.045 | 8083.89 | 8119.94 | 10062.38399 | 10170.42259 | 855.045 | 5787.1 | 5822.35 | 9203.585689 | 9071.042752 | 7774.688225 | 7698.197295 |
| 855.324 | 8067.98 | 8133.25 | 10153.52382 | 10187.01389 | 855.324 | 5798.31 | 5833.43 | 9055.196861 | 9088.176928 | 7686.212754 | 7713.241474 |
| 855.602 | 7936.82 | 8146.56 | 10019.50439 | 10203.61703 | 855.602 | 5851.25 | 5844.52 | 8902.44934 | 9105.321838 | 7585.325547 | 7728.297198 |
| 855.881 | 8093.49 | 8159.89 | 10121.06257 | 10220.22016 | 855.881 | 5866.02 | 5855.62 | 9057.386944 | 9122.488221 | 7881.60234 | 7743.364468 |
| 856.16 | 8151.39 | 8173.23 | 10233.74863 | 10236.84694 | 856.16 | 5834.7 | 5866.73 | 9114.146583 | 9139.665339 | 7793.969713 | 7758.454829 |
| 856.438 | 8095.61 | 8186.57 | 10150.37821 | 10253.47372 | 856.438 | 5832.55 | 5877.85 | 9187.70759 | 9156.863928 | 7833.975912 | 7773.556737 |
| 856.717 | 8029.1 | 8199.92 | 10142.15943 | 10270.11233 | 856.717 | 5804.94 | 5888.98 | 9064.816047 | 9174.073253 | 7787.596431 | 7788.67019 |
| 856.995 | 8129.88 | 8213.27 | 10269.0362 | 10286.76276 | 856.995 | 5929.25 | 5900.12 | 9139.804903 | 9191.293314 | 7811.634788 | 7803.795189 |
| 857.274 | 8214.77 | 8226.64 | 10344.5426 | 10303.41319 | 857.274 | 5971.18 | 5911.27 | 9094.543197 | 9208.534846 | 7926.376956 | 7818.94328 |
| 857.552 | 8120.23 | 8240.01 | 10274.84256 | 10320.08728 | 857.552 | 5872.9 | 5922.43 | 9327.464916 | 9225.797849 | 7927.612356 | 7834.102916 |
| 857.831 | 8296.4 | 8253.39 | 10193.90821 | 10336.76136 | 857.831 | 6004.28 | 5933.6 | 9114.17879 | 9243.060853 | 7858.660834 | 7849.274098 |
| 858.109 | 8217.69 | 8266.78 | 10368.18196 | 10353.44727 | 858.109 | 6044.24 | 5944.78 | 9243.994859 | 9260.345328 | 7843.651293 | 7864.468372 |
| 858.388 | 8202.46 | 8280.18 | 10346.76581 | 10370.145 | 858.388 | 5940.04 | 5955.97 | 9306.369268 | 9277.651274 | 8019.309106 | 7879.674192 |
| 858.666 | 8287.89 | 8293.58 | 10296.03402 | 10386.85457 | 858.666 | 5882.96 | 5967.17 | 9419.190726 | 9294.967956 | 7899.925526 | 7894.891557 |
| 858.945 | 8270.81 | 8307 | 10488.50735 | 10403.56413 | 858.945 | 6114.74 | 5978.38 | 9227.891311 | 9312.295374 | 7904.347568 | 7910.120468 |
| 859.223 | 8221.57 | 8320.42 | 10211.38644 | 10420.29734 | 859.223 | 5982.7 | 5989.6 | 9385.394747 | 9329.644263 | 7747.116853 | 7925.360925 |
| 859.501 | 8352.66 | 8333.84 | 10373.14871 | 10437.03055 | 859.501 | 5910.72 | 6000.83 | 9186.16165 | 9347.003888 | 7996.217505 | 7940.624474 |
| 859.78 | 8359.86 | 8347.28 | 10417.83762 | 10453.77559 | 859.78 | 6066.01 | 6012.07 | 9289.814821 | 9364.384984 | 7969.200331 | 7955.899568 |
| 860.058 | 8309.91 | 8360.72 | 10323.75321 | 10470.52063 | 860.058 | 6061.39 | 6023.32 | 9363.5476 | 9381.776816 | 7890.89671 | 7971.197754 |
| 860.337 | 8197.8 | 8374.17 | 10205.47364 | 10487.28931 | 860.337 | 6096.78 | 6034.58 | 9361.958716 | 9399.179384 | 8013.016645 | 7986.49594 |
| 860.615 | 8184.29 | 8387.63 | 10560.12313 | 10504.058 | 860.615 | 6103.85 | 6045.85 | 9237.47829 | 9416.603423 | 8147.386674 | 8001.817218 |
| 860.893 | 8362.07 | 8401.1 | 10453.89384 | 10520.83852 | 860.893 | 6095.27 | 6057.13 | 9394.938783 | 9434.038198 | 7888.922378 | 8017.150041 |
| 861.171 | 8269.09 | 8414.57 | 10410.41114 | 10537.63086 | 861.171 | 6045.22 | 6068.42 | 9253.85023 | 9451.494444 | 8066.161966 | 8032.494411 |
| 861.45 | 8411.08 | 8428.06 | 10550.93465 | 10554.43502 | 861.45 | 6069.7 | 6079.72 | 9549.99448 | 9468.961426 | 7915.385353 | 8047.861871 |
| 861.728 | 8385.14 | 8441.55 | 10478.36099 | 10571.25101 | 861.728 | 6114.96 | 6091.03 | 9480.545245 | 9486.439143 | 8037.586109 | 8063.229332 |
| 862.006 | 8339.53 | 8455.04 | 10414.53828 | 10588.06701 | 862.006 | 6102.32 | 6102.35 | 9469.337175 | 9503.938332 | 8030.312254 | 8078.619885 |
| 862.284 | 8371.27 | 8468.55 | 10545.28202 | 10604.89482 | 862.284 | 6102.7 | 6113.68 | 9439.363104 | 9521.448257 | 7907.222472 | 8094.033529 |
| 862.563 | 8505.24 | 8482.06 | 10601.98573 | 10621.73446 | 862.563 | 6244.12 | 6125.02 | 9533.321939 | 9538.979653 | 8240.087908 | 8109.447173 |
| 862.841 | 8421.3 | 8495.58 | 10650.60073 | 10638.58593 | 862.841 | 6236.26 | 6136.37 | 9527.213327 | 9556.511049 | 7965.51722 | 8124.883908 |
| 863.119 | 8483.98 | 8509.11 | 10571.4757 | 10655.44923 | 863.119 | 6069.55 | 6147.73 | 9466.352651 | 9574.074652 | 8184.287053 | 8140.33219 |
| 863.397 | 8285.12 | 8522.64 | 10645.4093 | 10672.31252 | 863.397 | 6210.56 | 6159.1 | 9495.478601 | 9591.648991 | 7923.825334 | 8155.792017 |
| 863.675 | 8424.77 | 8536.18 | 10597.0308 | 10689.18764 | 863.675 | 6148.72 | 6170.48 | 9490.76563 | 9609.234065 | 8147.363582 | 8171.274936 |
| 863.953 | 8518.46 | 8549.73 | 10581.01895 | 10706.07458 | 863.953 | 6208.72 | 6181.87 | 9616.523605 | 9626.829875 | 8126.788965 | 8186.757855 |
| 864.231 | 8404.42 | 8563.29 | 10633.74927 | 10722.96153 | 864.231 | 6214.49 | 6193.26 | 9661.076754 | 9644.447157 | 8076.079808 | 8202.263865 |
| 864.51 | 8422 | 8576.86 | 10615.70341 | 10739.87212 | 864.51 | 6167.87 | 6204.67 | 9617.865567 | 9662.075174 | 8081.921984 | 8217.781421 |
| 864.788 | 8411.59 | 8590.43 | 10485.07793 | 10756.78272 | 864.788 | 6094.16 | 6216.09 | 9559.098352 | 9679.724663 | 8354.783893 | 8233.322069 |
| 865.066 | 8490.98 | 8604.01 | 10695.67989 | 10773.70514 | 865.066 | 6305.45 | 6227.52 | 9517.282806 | 9697.384887 | 8262.60222 | 8248.862717 |
| 865.344 | 8489.08 | 8617.59 | 10602.45875 | 10790.63939 | 865.344 | 6220.61 | 6238.95 | 9582.770568 | 9715.055847 | 8217.862242 | 8264.426456 |
| 865.622 | 8591.23 | 8631.19 | 10781.64011 | 10807.57363 | 865.622 | 6325.42 | 6250.4 | 9528.308368 | 9732.748279 | 8308.496778 | 8280.001741 |
| 865.9 | 8569.62 | 8644.79 | 10544.40692 | 10824.51971 | 865.9 | 6180.97 | 6261.86 | 9697.631808 | 9750.451446 | 8083.295934 | 8295.588572 |
| 866.178 | 8417.5 | 8658.4 | 10570.15123 | 10841.47761 | 866.178 | 6303.98 | 6273.32 | 9690.073877 | 9768.176085 | 8340.224638 | 8311.198495 |
| 866.456 | 8527.86 | 8672.02 | 10605.95913 | 10858.44733 | 866.456 | 6263.14 | 6284.8 | 9791.118273 | 9785.911459 | 8113.061008 | 8326.819964 |
| 866.734 | 8523.37 | 8685.64 | 10729.78489 | 10875.42888 | 866.734 | 6342.74 | 6296.29 | 9770.505731 | 9803.657569 | 8243.424645 | 8342.452978 |
| 867.011 | 8501.58 | 8699.27 | 10680.16471 | 10892.41043 | 867.011 | 6251.51 | 6307.78 | 9661.903403 | 9821.414415 | 8041.777234 | 8358.097538 |
| 867.289 | 8499.73 | 8712.91 | 10640.005 | 10909.4038 | 867.289 | 6328.4 | 6319.29 | 10023.81454 | 9839.192732 | 8381.281506 | 8373.753644 |
| 867.567 | 8591.71 | 8726.56 | 10744.4841 | 10926.409 | 867.567 | 6401.6 | 6330.8 | 9830.947715 | 9856.981785 | 8351.008416 | 8389.432841 |
| 867.845 | 8589.84 | 8740.21 | 10713.19359 | 10943.42603 | 867.845 | 6377.11 | 6342.33 | 9817.936048 | 9874.792309 | 8184.841252 | 8405.123584 |
| 868.123 | 8689.72 | 8753.87 | 10789.66969 | 10960.44306 | 868.123 | 6220.55 | 6353.86 | 9690.556983 | 9892.613569 | 8191.872644 | 8420.825873 |
| 868.401 | 8551.88 | 8767.54 | 10696.79149 | 10977.47191 | 868.401 | 6244.98 | 6365.4 | 9815.96068 | 9910.4563 | 8415.029881 | 8436.539708 |
| 868.679 | 8621.2 | 8781.21 | 10775.10056 | 10994.51259 | 868.679 | 6268.12 | 6376.96 | 9878.141847 | 9928.299031 | 8355.996202 | 8452.265089 |
| 868.957 | 8619.93 | 8794.9 | 10795.61796 | 11011.56509 | 868.957 | 6376.73 | 6388.52 | 9857.829905 | 9946.17397 | 8430.836083 | 8468.013561 |
| 869.234 | 8681.66 | 8808.58 | 10830.82275 | 11028.61759 | 869.234 | 6388.91 | 6400.09 | 9864.389417 | 9964.048908 | 8394.039616 | 8483.773579 |
| 869.512 | 8737.19 | 8822.28 | 10904.96921 | 11045.68192 | 869.512 | 6394.03 | 6411.68 | 9875.586751 | 9981.945318 | 8335.721776 | 8499.545143 |
| 869.79 | 8682.6 | 8835.99 | 10720.80927 | 11062.75807 | 869.79 | 6253 | 6423.27 | 9726.575252 | 9999.852463 | 8518.60726 | 8515.328253 |
| 870.068 | 8835.89 | 8849.7 | 10819.52931 | 11079.83423 | 870.068 | 6279.06 | 6434.87 | 9964.252886 | 10017.77034 | 8414.614233 | 8531.134454 |
| 870.345 | 8640.44 | 8863.41 | 10899.02093 | 11096.93403 | 870.345 | 6328.38 | 6446.48 | 9890.509372 | 10035.7097 | 8485.00898 | 8546.952201 |
| 870.623 | 8635.14 | 8877.14 | 10898.66617 | 11114.03384 | 870.623 | 6323.19 | 6458.1 | 9826.052237 | 10053.65979 | 8476.153351 | 8562.781494 |
| 870.901 | 8827.81 | 8890.87 | 10814.69264 | 11131.14547 | 870.901 | 6404.39 | 6469.73 | 9937.424375 | 10071.63134 | 8383.752307 | 8578.622332 |
| 871.178 | 8745.93 | 8904.61 | 11073.59032 | 11148.2571 | 871.178 | 6401.25 | 6481.37 | 9888.254875 | 10089.61364 | 8673.009254 | 8594.474717 |
| 871.456 | 8728.4 | 8918.36 | 10900.66469 | 11165.38056 | 871.456 | 6514.68 | 6493.02 | 10048.02354 | 10107.60667 | 8558.855921 | 8610.350193 |
| 871.734 | 8838.66 | 8932.11 | 11114.58964 | 11182.51584 | 871.734 | 6447.67 | 6504.68 | 10035.03335 | 10125.62117 | 8536.814988 | 8626.225669 |
| 872.011 | 8783.91 | 8945.87 | 10965.8237 | 11199.66295 | 872.011 | 6374.06 | 6516.35 | 9964.338772 | 10143.63568 | 8643.463549 | 8642.135782 |
| 872.289 | 8808.68 | 8959.64 | 11079.17199 | 11216.82188 | 872.289 | 6512.43 | 6528.02 | 9984.382321 | 10161.68239 | 8595.306014 | 8658.045896 |
| 872.567 | 8799.41 | 8973.41 | 11031.94058 | 11233.98081 | 872.567 | 6392.8 | 6539.71 | 10102.22809 | 10179.7291 | 8520.327584 | 8673.967555 |
| 872.844 | 8893.1 | 8987.19 | 11013.49266 | 11251.15157 | 872.844 | 6421.04 | 6551.41 | 9824.624389 | 10197.79728 | 8675.757154 | 8689.912306 |
| 873.122 | 8872.78 | 9000.98 | 11130.66062 | 11268.32233 | 873.122 | 6623.04 | 6563.11 | 10150.98963 | 10215.87619 | 8555.750101 | 8705.857057 |
| 873.399 | 8827.51 | 9014.78 | 11038.21997 | 11285.51674 | 873.399 | 6565.45 | 6574.83 | 10174.79067 | 10233.97658 | 8641.985687 | 8721.824899 |
| 873.677 | 9017.18 | 9028.58 | 11139.69537 | 11302.71115 | 873.677 | 6469 | 6586.55 | 10288.51393 | 10252.0877 | 8469.399057 | 8737.804287 |
| 873.954 | 8847.66 | 9042.39 | 11190.93566 | 11319.91739 | 873.954 | 6703.36 | 6598.29 | 10207.70633 | 10270.20956 | 8626.699047 | 8753.806767 |
| 874.232 | 9063.63 | 9056.2 | 11157.69392 | 11337.12362 | 874.232 | 6555.05 | 6610.03 | 10320.25939 | 10288.3529 | 8701.781389 | 8769.809247 |
| 874.509 | 8921.38 | 9070.02 | 11070.58662 | 11354.34168 | 874.509 | 6675.26 | 6621.78 | 10411.25517 | 10306.49623 | 8779.080525 | 8785.834819 |
| 874.787 | 8986.53 | 9083.85 | 11257.73843 | 11371.57157 | 874.787 | 6794.55 | 6633.54 | 10346.0358 | 10324.67176 | 8847.616399 | 8801.871936 |
| 875.064 | 8961.67 | 9097.69 | 11307.25218 | 11388.81328 | 875.064 | 6622.82 | 6645.32 | 10270.86444 | 10342.8473 | 8752.479 | 8817.920599 |
| 875.342 | 9033.23 | 9111.53 | 11289.56109 | 11406.055 | 875.342 | 6547.78 | 6657.1 | 10167.29716 | 10361.04431 | 8711.987877 | 8833.980808 |
| 875.619 | 8878.23 | 9125.39 | 11186.82035 | 11423.32036 | 875.619 | 6563.91 | 6668.89 | 10366.47657 | 10379.25206 | 8833.588251 | 8850.064108 |
| 875.896 | 8979.83 | 9139.24 | 11099.3228 | 11440.5739 | 875.896 | 6721.46 | 6680.69 | 10184.5709 | 10397.47054 | 8948.873071 | 8866.147409 |
| 876.174 | 9008.15 | 9153.1 | 11316.34606 | 11457.85109 | 876.174 | 6655.9 | 6692.5 | 10299.07786 | 10415.71049 | 8971.837669 | 8882.253801 |
| 876.451 | 9032.04 | 9166.98 | 11350.81766 | 11475.12828 | 876.451 | 6697.15 | 6704.31 | 10436.38744 | 10433.96118 | 8819.005904 | 8898.371739 |
| 876.728 | 8967.48 | 9180.85 | 11292.25733 | 11492.41729 | 876.728 | 6765.1 | 6716.14 | 10399.57473 | 10452.23334 | 8923.276031 | 8914.501222 |
| 877.006 | 9011.18 | 9194.74 | 11218.83223 | 11509.71813 | 877.006 | 6599.95 | 6727.98 | 10387.31457 | 10470.5055 | 8986.731752 | 8930.653798 |
| 877.283 | 8991.27 | 9208.63 | 11358.06674 | 11527.01897 | 877.283 | 6773.43 | 6739.82 | 10298.49813 | 10488.79913 | 8926.127844 | 8946.806373 |
| 877.56 | 9165.25 | 9222.52 | 11302.66385 | 11544.33164 | 877.56 | 6804.14 | 6751.68 | 10610.63857 | 10507.11423 | 8929.868683 | 8962.98204 |
| 877.838 | 9148.61 | 9236.43 | 11324.91961 | 11561.6443 | 877.838 | 6711.64 | 6763.55 | 10663.80175 | 10525.42933 | 8860.963344 | 8979.169252 |
| 878.115 | 9224.85 | 9250.34 | 11545.45507 | 11578.98062 | 878.115 | 6706.09 | 6775.42 | 10517.00181 | 10543.7659 | 8977.795302 | 8995.368011 |
| 878.392 | 9001.99 | 9264.26 | 11356.00909 | 11596.31694 | 878.392 | 6663.18 | 6787.3 | 10395.00133 | 10562.11321 | 8923.795592 | 9011.578315 |
| 878.669 | 9082.06 | 9278.18 | 11554.44252 | 11613.66508 | 878.669 | 6793.31 | 6799.19 | 10267.5149 | 10580.48199 | 9024.001597 | 9027.811711 |
| 878.947 | 9189.91 | 9292.11 | 11494.80606 | 11631.01322 | 878.947 | 6655.93 | 6811.1 | 10529.21903 | 10598.85077 | 8849.844738 | 9044.045107 |
| 879.224 | 9217.99 | 9306.05 | 11523.04557 | 11648.37319 | 879.224 | 6746.19 | 6823.01 | 10524.3665 | 10617.25176 | 8951.251506 | 9060.301594 |
| 879.501 | 9219.08 | 9319.99 | 11411.15183 | 11665.74498 | 879.501 | 6785.71 | 6834.93 | 10531.02263 | 10635.65275 | 8970.683089 | 9076.569628 |
| 879.778 | 9204.43 | 9333.94 | 11526.40404 | 11683.11678 | 879.778 | 6887.36 | 6846.86 | 10787.12272 | 10654.07521 | 8914.558951 | 9092.849207 |
| 880.055 | 9211.51 | 9347.9 | 11425.95747 | 11700.51222 | 880.055 | 6687.71 | 6858.8 | 10617.05852 | 10672.5084 | 9135.534032 | 9109.140332 |
| 880.332 | 9236.02 | 9361.86 | 11588.86681 | 11717.89584 | 880.332 | 6744.4 | 6870.74 | 10669.54535 | 10690.95233 | 8937.46582 | 9125.454548 |
| 880.609 | 9355.22 | 9375.83 | 11603.66063 | 11735.30311 | 880.609 | 6832.23 | 6882.7 | 10803.76306 | 10709.407 | 9074.80312 | 9141.78031 |
| 880.886 | 9297.35 | 9389.81 | 11709.21586 | 11752.71038 | 880.886 | 6846.96 | 6894.66 | 10705.62803 | 10727.88314 | 9216.955019 | 9158.106073 |
| 881.163 | 9239.15 | 9403.79 | 11601.92227 | 11770.12948 | 881.163 | 6834 | 6906.64 | 10552.35446 | 10746.32707 | 9006.382705 | 9174.454926 |
| 881.441 | 9400.12 | 9417.78 | 11712.07765 | 11787.5604 | 881.441 | 6951.54 | 6918.62 | 10585.93573 | 10764.89983 | 9274.072095 | 9190.815326 |
| 881.717 | 9337.47 | 9431.77 | 11707.00447 | 11804.99132 | 881.717 | 6928.75 | 6930.62 | 10889.756 | 10783.36523 | 9382.371707 | 9207.187272 |
| 881.995 | 9370.25 | 9445.78 | 11530.76768 | 11822.43407 | 881.995 | 6858.86 | 6942.62 | 10801.72327 | 10801.93799 | 9198.320097 | 9223.582309 |
| 882.271 | 9410.79 | 9459.79 | 11729.92247 | 11839.90047 | 882.271 | 6843.85 | 6954.63 | 10717.287 | 10820.51075 | 9229.493759 | 9239.977346 |
| 882.548 | 9298.06 | 9473.8 | 11593.50245 | 11857.40235 | 882.548 | 6793.52 | 6966.65 | 10679.36851 | 10838.97615 | 8990.287859 | 9256.395474 |
| 882.825 | 9406.8 | 9487.82 | 11676.83739 | 11874.78597 | 882.825 | 6815.33 | 6978.68 | 10689.64258 | 10857.54891 | 9192.501013 | 9272.825149 |
| 883.102 | 9361.5 | 9501.85 | 11787.84422 | 11892.28784 | 883.102 | 7051.25 | 6990.72 | 10695.4828 | 10876.12166 | 9346.037072 | 9289.266369 |
| 883.379 | 9360.4 | 9515.88 | 11802.4843 | 11909.78972 | 883.379 | 6971.75 | 7002.77 | 10838.43936 | 10894.80178 | 9283.100912 | 9305.719135 |
| 883.656 | 9457.99 | 9529.92 | 11752.15458 | 11927.29159 | 883.656 | 7032.16 | 7014.82 | 10890.29279 | 10913.37454 | 9357.340411 | 9322.183447 |
| 883.933 | 9352.97 | 9543.97 | 11593.38419 | 11944.79347 | 883.933 | 6824.56 | 7026.89 | 10836.82901 | 10932.05465 | 9341.441843 | 9338.670851 |
| 884.21 | 9372.83 | 9558.02 | 11804.53013 | 11962.29534 | 884.21 | 6942.49 | 7038.96 | 10770.48239 | 10950.62741 | 9398.385732 | 9355.158254 |
| 884.487 | 9315.74 | 9572.08 | 11785.46727 | 11979.79722 | 884.487 | 6915.94 | 7051.05 | 10692.26209 | 10969.30753 | 9392.959206 | 9371.668749 |
| 884.764 | 9496.43 | 9586.15 | 11795.37712 | 11997.29909 | 884.764 | 6972.82 | 7063.14 | 10899.84756 | 10987.98765 | 9470.81254 | 9388.19079 |
| 885.04 | 9348.9 | 9600.22 | 11748.94985 | 12014.80097 | 885.04 | 6965.23 | 7075.24 | 10748.58156 | 11006.66776 | 9336.465603 | 9404.724377 |
| 885.317 | 9311.08 | 9614.3 | 11846.75931 | 12032.30284 | 885.317 | 7015.74 | 7087.35 | 10977.46666 | 11025.34788 | 9377.949665 | 9421.269509 |
| 885.594 | 9437.57 | 9628.38 | 11765.80131 | 12049.92298 | 885.594 | 6943.07 | 7099.47 | 10896.84156 | 11044.02799 | 9299.565224 | 9437.826188 |
| 885.871 | 9449.23 | 9642.47 | 11770.76806 | 12067.42485 | 885.871 | 7049.62 | 7111.6 | 11012.03561 | 11062.70811 | 9365.076097 | 9454.405958 |
| 886.148 | 9612.13 | 9656.57 | 11928.82892 | 12085.04498 | 886.148 | 6930.07 | 7123.73 | 11038.44543 | 11081.49558 | 9297.12906 | 9470.997273 |
| 886.424 | 9505.95 | 9670.67 | 11862.72386 | 12102.54686 | 886.424 | 7039.58 | 7135.88 | 11190.78499 | 11100.1757 | 9496.236394 | 9487.588589 |
| 886.701 | 9433.23 | 9684.78 | 11852.55385 | 12120.16699 | 886.701 | 7057.01 | 7148.03 | 11036.29829 | 11118.96317 | 9476.250613 | 9504.202996 |
| 886.978 | 9533.34 | 9698.9 | 11836.35279 | 12137.66886 | 886.978 | 7010.8 | 7160.2 | 11091.05035 | 11137.75064 | 9292.418373 | 9520.828949 |
| 887.254 | 9486.52 | 9713.02 | 11879.98923 | 12155.289 | 887.254 | 7036.06 | 7172.37 | 11084.17951 | 11156.53812 | 9471.008819 | 9537.466448 |
| 887.531 | 9395.43 | 9727.14 | 11849.4792 | 12172.90913 | 887.531 | 7155.73 | 7184.55 | 11145.15827 | 11175.32559 | 9495.901565 | 9554.115493 |
| 887.808 | 9547.61 | 9741.28 | 11977.66861 | 12190.52926 | 887.808 | 7125.72 | 7196.74 | 10899.95492 | 11194.11306 | 9475.938876 | 9570.787629 |
| 888.084 | 9591.21 | 9755.42 | 11962.29534 | 12208.14939 | 888.084 | 7136.79 | 7208.94 | 10967.16039 | 11213.00789 | 9434.54718 | 9587.459766 |
| 888.361 | 9625.86 | 9769.56 | 12061.86682 | 12225.76952 | 888.361 | 7061.03 | 7221.15 | 11108.22747 | 11231.79536 | 9314.136024 | 9604.154993 |
| 888.638 | 9677.31 | 9783.71 | 11965.01523 | 12243.38965 | 888.638 | 7079.42 | 7233.37 | 11079.4558 | 11250.69019 | 9622.351175 | 9620.861767 |
| 888.914 | 9646.27 | 9797.87 | 12087.4101 | 12261.00978 | 888.914 | 7209.02 | 7245.59 | 11003.44705 | 11269.58502 | 9531.381811 | 9637.568541 |
| 889.191 | 9593.68 | 9812.03 | 12086.81882 | 12278.62992 | 889.191 | 7290.33 | 7257.83 | 11047.35606 | 11288.47985 | 9573.801083 | 9654.298406 |
| 889.467 | 9612.17 | 9826.2 | 11986.30129 | 12296.25005 | 889.467 | 7188.45 | 7270.07 | 11258.74197 | 11307.26733 | 9497.633436 | 9671.051363 |
| 889.744 | 9663.19 | 9840.38 | 12134.3577 | 12313.87018 | 889.744 | 7118.67 | 7282.32 | 11249.93869 | 11326.16216 | 9583.199365 | 9687.80432 |
| 890.02 | 9776.16 | 9854.56 | 12035.14099 | 12331.60857 | 890.02 | 7285.04 | 7294.58 | 10997.43506 | 11345.16434 | 9563.964061 | 9704.568823 |
| 890.297 | 9584.1 | 9868.75 | 11960.28499 | 12349.2287 | 890.297 | 7142.57 | 7306.85 | 11108.44219 | 11364.05917 | 9668.176459 | 9721.356417 |
| 890.573 | 9643.25 | 9882.94 | 12095.33325 | 12366.96708 | 890.573 | 7121.41 | 7319.13 | 10999.47484 | 11383.06136 | 9588.6028 | 9738.144011 |
| 890.85 | 9627.47 | 9897.14 | 12028.75517 | 12384.58722 | 890.85 | 7281.7 | 7331.41 | 11228.35994 | 11401.95619 | 9693.323213 | 9754.954697 |
| 891.126 | 9601.2 | 9911.34 | 12082.20684 | 12402.3256 | 891.126 | 7270.26 | 7343.71 | 11016.75932 | 11420.95837 | 9633.943159 | 9771.776929 |
| 891.402 | 9808.04 | 9925.55 | 12182.13309 | 12419.94573 | 891.402 | 7248.7 | 7356.02 | 11134.20786 | 11439.96056 | 9641.909762 | 9788.610706 |
| 891.679 | 9605.5 | 9939.77 | 11991.74107 | 12437.68412 | 891.679 | 7332.27 | 7368.33 | 11031.57458 | 11458.96275 | 9635.605755 | 9805.45603 |
| 891.955 | 9709.27 | 9953.99 | 12044.24669 | 12455.42251 | 891.955 | 7215.64 | 7380.65 | 11341.94363 | 11477.96494 | 9576.352705 | 9822.312899 |
| 892.232 | 9646.59 | 9968.22 | 12085.39975 | 12473.1609 | 892.232 | 7216.3 | 7392.98 | 11254.6624 | 11496.96712 | 9701.786285 | 9839.181314 |
| 892.508 | 9763.77 | 9982.45 | 12270.11549 | 12490.89928 | 892.508 | 7358.51 | 7405.32 | 11352.14255 | 11515.96931 | 9815.142957 | 9856.07282 |
| 892.784 | 9764.93 | 9996.69 | 12158.36365 | 12508.63767 | 892.784 | 7112.75 | 7417.66 | 11188.85257 | 11535.07885 | 9631.299171 | 9872.964327 |
| 893.061 | 9886.49 | 10010.9 | 12379.2657 | 12526.37606 | 893.061 | 7283.59 | 7430.02 | 11342.69513 | 11554.08104 | 9876.658983 | 9889.867379 |
| 893.337 | 9757.49 | 10025.2 | 12123.24164 | 12544.11444 | 893.337 | 7306.77 | 7442.39 | 11236.30436 | 11573.19058 | 9916.884553 | 9906.793523 |
| 893.613 | 9764.48 | 10039.5 | 12214.41695 | 12561.85283 | 893.613 | 7272.11 | 7454.76 | 11187.99371 | 11592.30013 | 9812.083319 | 9923.731213 |
| 893.889 | 9849.36 | 10053.7 | 12190.52926 | 12579.59122 | 893.889 | 7366.3 | 7467.14 | 11300.82591 | 11611.40967 | 9759.596109 | 9940.680448 |
| 894.166 | 9871.62 | 10068 | 12196.91508 | 12597.32961 | 894.166 | 7421.65 | 7479.53 | 11409.68589 | 11630.51921 | 9747.346014 | 9957.64123 |
| 894.442 | 9906.04 | 10082.2 | 12216.19079 | 12615.18625 | 894.442 | 7348.71 | 7491.93 | 11295.56541 | 11649.62876 | 9793.27521 | 9974.613557 |
| 894.718 | 9894.09 | 10096.5 | 12381.39431 | 12632.92464 | 894.718 | 7413.26 | 7504.34 | 11459.39218 | 11668.7383 | 9936.858788 | 9991.59743 |
| 894.994 | 9827.69 | 10110.8 | 12164.3947 | 12650.66302 | 894.994 | 7506.19 | 7516.75 | 11344.41284 | 11687.9552 | 9952.757356 | 10008.59285 |
| 895.27 | 9825.18 | 10125.1 | 12240.315 | 12668.51967 | 895.27 | 7446.48 | 7529.18 | 11275.27494 | 11707.06475 | 9670.670352 | 10025.61136 |
| 895.547 | 9788.98 | 10139.4 | 12418.76317 | 12686.25805 | 895.547 | 7453.17 | 7541.61 | 11321.22373 | 11726.28165 | 9852.874633 | 10042.62987 |
| 895.823 | 9870.19 | 10153.7 | 12489.71672 | 12704.1147 | 895.823 | 7282.86 | 7554.05 | 11437.81342 | 11745.39119 | 9807.234083 | 10059.67147 |
| 896.099 | 9925.67 | 10168 | 12478.36416 | 12721.97134 | 896.099 | 7351.73 | 7566.5 | 11503.08647 | 11764.60809 | 9928.846002 | 10076.71307 |
| 896.375 | 9964.85 | 10182.3 | 12362.47336 | 12739.70973 | 896.375 | 7491.68 | 7578.96 | 11519.94152 | 11783.82499 | 9830.002402 | 10093.77777 |
| 896.651 | 9987.48 | 10196.6 | 12414.15119 | 12757.56637 | 896.651 | 7514.57 | 7591.43 | 11695.14812 | 11803.04189 | 10015.26632 | 10110.85401 |
| 896.927 | 9952.05 | 10211 | 12382.34035 | 12775.42301 | 896.927 | 7553.77 | 7603.9 | 11534.32735 | 11822.36615 | 9903.445241 | 10127.94179 |
| 897.203 | 9987.34 | 10225.3 | 12524.60222 | 12793.27966 | 897.203 | 7468.18 | 7616.38 | 11612.37588 | 11841.58305 | 10058.75935 | 10145.04112 |
| 897.479 | 9933.42 | 10239.6 | 12497.63987 | 12811.1363 | 897.479 | 7305.6 | 7628.88 | 11554.9399 | 11860.79995 | 9988.826437 | 10162.152 |
| 897.755 | 10011.6 | 10254 | 12603.2424 | 12828.99294 | 897.755 | 7441.8 | 7641.38 | 11686.66692 | 11880.12421 | 10005.99504 | 10179.27442 |
| 898.031 | 9942.07 | 10268.3 | 12491.84533 | 12846.84959 | 898.031 | 7536.52 | 7653.89 | 11611.62438 | 11899.44847 | 10071.02099 | 10196.40839 |
| 898.307 | 10061.1 | 10282.7 | 12648.53442 | 12864.70623 | 898.307 | 7532.44 | 7666.4 | 11643.2947 | 11918.77272 | 10023.87949 | 10213.5539 |
| 898.583 | 10047.7 | 10297 | 12554.52097 | 12882.56287 | 898.583 | 7447.31 | 7678.93 | 11861.87352 | 11938.09698 | 10044.85821 | 10230.72251 |
| 898.859 | 10108.4 | 10311.4 | 12594.25495 | 12900.53777 | 898.859 | 7527.57 | 7691.46 | 11737.23206 | 11957.42124 | 10065.76765 | 10247.89111 |
| 899.135 | 10024.3 | 10325.8 | 12568.47516 | 12918.39442 | 899.135 | 7625.6 | 7704 | 11701.58954 | 11976.7455 | 10154.02375 | 10265.08281 |
| 899.411 | 10082.4 | 10340.1 | 12728.83018 | 12936.25106 | 899.411 | 7738.33 | 7716.55 | 11760.20645 | 11996.06976 | 10279.30724 | 10282.27451 |
| 899.687 | 10147.8 | 10354.5 | 12711.44656 | 12954.22596 | 899.687 | 7590.76 | 7729.11 | 11927.68335 | 12015.39401 | 10240.59417 | 10299.4893 |
| 899.963 | 10210.5 | 10368.9 | 12652.20035 | 12972.0826 | 899.963 | 7465.73 | 7741.68 | 11807.7656 | 12034.82563 | 10221.69369 | 10316.70409 |
| 900.239 | 10140.5 | 10383.3 | 12511.47581 | 12989.93924 | 900.239 | 7590.95 | 7754.26 | 11821.07787 | 12054.14989 | 10252.4748 | 10333.94197 |
| 900.514 | 10120.9 | 10397.7 | 12636.2358 | 13007.91414 | 900.514 | 7731.57 | 7766.84 | 12012.38802 | 12073.5815 | 10244.47356 | 10351.19139 |
| 900.79 | 10185.2 | 10412.1 | 12668.87444 | 13025.88904 | 900.79 | 7710.51 | 7779.43 | 11885.27734 | 12093.01312 | 10207.3769 | 10368.45237 |
| 901.066 | 10233.4 | 10426.5 | 12752.95439 | 13043.74569 | 901.066 | 7747.91 | 7792.03 | 11968.80108 | 12112.44473 | 10271.23672 | 10385.72488 |
| 901.342 | 10223.2 | 10440.9 | 12684.36596 | 13061.72059 | 901.342 | 7629 | 7804.64 | 11842.11984 | 12131.87634 | 10190.64703 | 10403.00895 |
| 901.617 | 10281.2 | 10455.3 | 12893.32416 | 13079.69548 | 901.617 | 7702 | 7817.26 | 11967.40544 | 12151.30796 | 10332.4872 | 10420.30456 |
| 901.893 | 10314.4 | 10469.7 | 12937.31536 | 13097.67038 | 901.893 | 7673.72 | 7829.88 | 12081.95535 | 12170.84693 | 10340.95027 | 10437.61171 |
| 902.169 | 10392 | 10484.2 | 12820.24201 | 13115.64528 | 902.169 | 7674.49 | 7842.51 | 11893.8659 | 12190.27855 | 10309.33787 | 10454.93041 |
| 902.445 | 10393.3 | 10498.6 | 12792.57012 | 13133.50193 | 902.445 | 7788.38 | 7855.15 | 11964.5068 | 12209.81752 | 10157.90314 | 10472.27221 |
| 902.72 | 10209.3 | 10513 | 12727.05635 | 13151.47682 | 902.72 | 7737.98 | 7867.81 | 12237.73033 | 12229.24913 | 10391.8788 | 10489.614 |
| 902.996 | 10338.3 | 10527.5 | 12908.69743 | 13169.56998 | 902.996 | 7590.61 | 7880.46 | 12094.62347 | 12248.7881 | 10229.83348 | 10506.96734 |
| 903.272 | 10194.5 | 10541.9 | 12718.66018 | 13187.54488 | 903.272 | 7757.48 | 7893.13 | 11979.96621 | 12268.32708 | 10396.43939 | 10524.34377 |
| 903.547 | 10366.9 | 10556.4 | 12986.0368 | 13205.51978 | 903.547 | 7783.02 | 7905.8 | 12024.94879 | 12287.86605 | 10613.00397 | 10541.7202 |
| 903.823 | 10368 | 10570.8 | 12940.62653 | 13223.49468 | 903.823 | 7821.12 | 7918.48 | 12324.26007 | 12307.40502 | 10537.5868 | 10559.11972 |
| 904.098 | 10449.2 | 10585.3 | 12859.97599 | 13241.46958 | 904.098 | 7892.79 | 7931.17 | 11989.62834 | 12326.94399 | 10368.7641 | 10576.53079 |
| 904.374 | 10366.4 | 10599.8 | 12830.53027 | 13259.56273 | 904.374 | 7820.57 | 7943.87 | 12250.39846 | 12346.59032 | 10403.11286 | 10593.94186 |
| 904.65 | 10470.6 | 10614.2 | 13089.0377 | 13277.53763 | 904.65 | 7773.23 | 7956.58 | 12264.14015 | 12366.12929 | 10403.57469 | 10611.37601 |
| 904.925 | 10398.9 | 10628.7 | 13122.62238 | 13295.51253 | 904.925 | 7785.13 | 7969.29 | 12152.7036 | 12385.77562 | 10681.30893 | 10628.82172 |
| 905.201 | 10406 | 10643.2 | 12848.85994 | 13313.60568 | 905.201 | 7936.2 | 7982.02 | 12196.18318 | 12405.42195 | 10848.23812 | 10646.26742 |
| 905.476 | 10311.5 | 10657.7 | 12946.0663 | 13331.58058 | 905.476 | 7814.51 | 7994.75 | 12271.33307 | 12424.96092 | 10412.33796 | 10663.73622 |
| 905.752 | 10430.7 | 10672.2 | 12944.52897 | 13349.67374 | 905.752 | 7886.17 | 8007.48 | 12427.0007 | 12444.60725 | 10563.69186 | 10681.21656 |
| 906.027 | 10516.9 | 10686.7 | 13031.80184 | 13367.76689 | 906.027 | 7897.8 | 8020.23 | 12531.88848 | 12464.25358 | 10686.56227 | 10698.70845 |
| 906.303 | 10411.5 | 10701.2 | 13178.32092 | 13385.74179 | 906.303 | 7863.83 | 8032.99 | 12441.70861 | 12484.00726 | 10719.54862 | 10716.21189 |
| 906.578 | 10458.2 | 10715.7 | 13142.60763 | 13403.83495 | 906.578 | 7757.9 | 8045.75 | 12182.87091 | 12503.65359 | 10757.01475 | 10733.72687 |
| 906.854 | 10433.8 | 10730.2 | 13102.16411 | 13421.9281 | 906.854 | 7795.85 | 8058.52 | 12336.71348 | 12523.29992 | 10552.50398 | 10751.25339 |
| 907.129 | 10527.8 | 10744.7 | 13057.81814 | 13440.02126 | 907.129 | 7963.73 | 8071.3 | 12377.40177 | 12543.05361 | 10396.2662 | 10768.79146 |
| 907.404 | 10568.7 | 10759.2 | 12978.35016 | 13458.11441 | 907.404 | 8023.62 | 8084.08 | 12323.83064 | 12562.69993 | 10536.13203 | 10786.34108 |
| 907.68 | 10541.8 | 10773.8 | 13237.68539 | 13476.08931 | 907.68 | 7826.55 | 8096.88 | 12246.53361 | 12582.45362 | 10756.89929 | 10803.90224 |
| 907.955 | 10332.1 | 10788.3 | 13259.68099 | 13494.18247 | 907.955 | 8037.27 | 8109.68 | 12272.72871 | 12602.20731 | 10630.04557 | 10821.47495 |
| 908.231 | 10449.2 | 10802.8 | 13228.57968 | 13512.39388 | 908.231 | 7923.81 | 8122.49 | 12206.70416 | 12621.96099 | 10590.38575 | 10839.05921 |
| 908.506 | 10562 | 10817.4 | 13180.44952 | 13530.36878 | 908.506 | 7949.34 | 8135.31 | 12429.46991 | 12641.71468 | 10647.73374 | 10856.65501 |
| 908.781 | 10577.5 | 10831.9 | 13147.10136 | 13548.58019 | 908.781 | 8183.48 | 8148.14 | 12461.89172 | 12661.46836 | 10909.82342 | 10874.26235 |
| 909.056 | 10667.9 | 10846.5 | 13208.71269 | 13566.67334 | 909.056 | 8145.7 | 8160.97 | 12510.95387 | 12681.22205 | 10806.80024 | 10891.89279 |
| 909.332 | 10716.2 | 10861 | 13292.31962 | 13584.7665 | 909.332 | 8061.39 | 8173.81 | 12405.95873 | 12700.97573 | 10709.44605 | 10909.52323 |
| 909.607 | 10591.3 | 10875.6 | 13246.90935 | 13602.85965 | 909.607 | 8061.9 | 8186.66 | 12441.92332 | 12720.83678 | 10737.67553 | 10927.16521 |
| 909.882 | 10591.6 | 10890.1 | 13173.59068 | 13621.07106 | 909.882 | 8027.4 | 8199.52 | 12573.54299 | 12740.59046 | 10846.18296 | 10944.81874 |
| 910.157 | 10735.8 | 10904.7 | 13321.41057 | 13639.16422 | 910.157 | 7997.19 | 8212.38 | 12471.33914 | 12760.4515 | 10647.86074 | 10962.49536 |
| 910.433 | 10707.3 | 10919.3 | 13440.37603 | 13657.37563 | 910.433 | 8146.22 | 8225.26 | 12546.91846 | 12780.31255 | 10726.93793 | 10980.17198 |
| 910.708 | 10657.7 | 10933.9 | 13241.23306 | 13675.46878 | 910.708 | 7978.07 | 8238.14 | 12596.83946 | 12800.17359 | 10627.08985 | 10997.86015 |
| 910.983 | 10829 | 10948.4 | 13348.25467 | 13693.56194 | 910.983 | 8142.27 | 8251.03 | 12725.0237 | 12820.03463 | 10909.08449 | 11015.57141 |
| 911.258 | 10674.2 | 10963 | 13242.88865 | 13711.77335 | 911.258 | 8114.41 | 8263.92 | 12516.96586 | 12839.89568 | 10953.98611 | 11033.28267 |
| 911.533 | 10676.8 | 10977.6 | 13350.97455 | 13729.98476 | 911.533 | 8210.62 | 8276.83 | 12669.84221 | 12859.75672 | 11057.69049 | 11051.00547 |
| 911.808 | 10697.5 | 10992.2 | 13443.33242 | 13748.07792 | 911.808 | 8101.99 | 8289.74 | 12614.66072 | 12879.61776 | 10950.40691 | 11068.75137 |
| 912.083 | 10938.3 | 11006.8 | 13596.71034 | 13766.28933 | 912.083 | 8148.04 | 8302.66 | 12819.71256 | 12899.58616 | 10814.76684 | 11086.49726 |
| 912.358 | 10875.7 | 11021.4 | 13440.13951 | 13784.50074 | 912.358 | 8263.81 | 8315.59 | 12849.77252 | 12919.4472 | 10916.27752 | 11104.2547 |
| 912.633 | 10813.9 | 11036 | 13440.25777 | 13802.71215 | 912.633 | 8131.78 | 8328.52 | 12925.02977 | 12939.4156 | 11045.03629 | 11122.03524 |
| 912.909 | 10815.7 | 11050.7 | 13460.95255 | 13820.8053 | 912.909 | 8218.36 | 8341.47 | 12606.50159 | 12959.384 | 10877.11416 | 11139.81577 |
| 913.183 | 10816.4 | 11065.3 | 13510.50178 | 13839.01671 | 913.183 | 8181.54 | 8354.42 | 12604.67652 | 12979.3524 | 11148.87922 | 11157.60785 |
| 913.458 | 10888.9 | 11079.9 | 13595.7643 | 13857.22813 | 913.458 | 8228.35 | 8367.37 | 12941.13331 | 12999.3208 | 10986.89164 | 11175.42302 |
| 913.734 | 10913.1 | 11094.5 | 13411.28507 | 13875.43954 | 913.734 | 8177.61 | 8380.34 | 12908.60415 | 13019.2892 | 10829.83411 | 11193.23819 |
| 914.008 | 10883.2 | 11109.2 | 13449.48173 | 13893.65095 | 914.008 | 8276.3 | 8393.31 | 12782.99647 | 13039.2576 | 11113.02951 | 11211.07645 |
| 914.283 | 10900.1 | 11123.8 | 13593.75395 | 13911.86236 | 914.283 | 8346.11 | 8406.29 | 12814.13 | 13059.226 | 11162.35317 | 11228.91471 |
| 914.558 | 10837.8 | 11138.5 | 13578.14417 | 13930.19203 | 914.558 | 8338.92 | 8419.28 | 12860.83029 | 13079.30176 | 11160.04401 | 11246.76452 |
| 914.833 | 10960.7 | 11153.1 | 13753.28118 | 13948.40344 | 914.833 | 8319.67 | 8432.28 | 13036.35896 | 13099.27016 | 11304.78217 | 11264.63742 |
| 915.108 | 10981.2 | 11167.8 | 13875.7943 | 13966.61485 | 915.108 | 8374.03 | 8445.28 | 12947.03795 | 13119.34591 | 11045.35957 | 11282.51032 |
| 915.383 | 11098 | 11182.4 | 13755.64629 | 13984.82626 | 915.383 | 8432.36 | 8458.3 | 12873.06898 | 13139.42167 | 11260.03065 | 11300.39477 |
| 915.658 | 10946.4 | 11197.1 | 13701.24857 | 14003.15592 | 915.658 | 8360.53 | 8471.32 | 13108.28814 | 13159.39007 | 11301.14524 | 11318.29076 |
| 915.933 | 10888.7 | 11211.7 | 13807.08762 | 14021.36734 | 915.933 | 8276.81 | 8484.34 | 12894.11095 | 13179.57318 | 11302.48456 | 11336.20984 |
| 916.207 | 10964 | 11226.4 | 13765.22502 | 14039.57875 | 916.207 | 8416.81 | 8497.38 | 12963.46357 | 13199.54158 | 11311.88284 | 11354.12892 |
| 916.482 | 11127.6 | 11241.1 | 13783.90946 | 14057.90841 | 916.482 | 8437.41 | 8510.42 | 12986.00854 | 13219.7247 | 11274.32435 | 11372.05955 |
| 916.757 | 11017.5 | 11255.8 | 13661.51459 | 14076.11982 | 916.757 | 8328.63 | 8523.47 | 12974.73605 | 13239.80045 | 11286.60908 | 11390.01327 |
| 917.032 | 11030 | 11270.4 | 13969.92601 | 14094.44949 | 917.032 | 8448.77 | 8536.52 | 13239.04895 | 13259.87621 | 11147.20508 | 11407.96699 |
| 917.307 | 10981.7 | 11285.1 | 13668.7282 | 14112.6609 | 917.307 | 8464.5 | 8549.59 | 12917.08535 | 13280.05932 | 11471.71136 | 11425.93226 |
| 917.581 | 10980.7 | 11299.8 | 13799.40098 | 14130.99057 | 917.581 | 8340.92 | 8562.66 | 12990.62489 | 13300.13508 | 11368.42262 | 11443.90907 |
| 917.856 | 11088.9 | 11314.5 | 13863.61395 | 14149.32024 | 917.856 | 8523.31 | 8575.74 | 13186.44403 | 13320.31819 | 11456.73645 | 11461.89743 |
| 918.131 | 11144.7 | 11329.2 | 13767.94491 | 14167.6499 | 918.131 | 8341.63 | 8588.83 | 13102.92029 | 13340.50131 | 11109.47341 | 11479.89733 |
| 918.405 | 10937.7 | 11343.9 | 13785.32853 | 14185.86131 | 918.405 | 8563.98 | 8601.92 | 13075.00748 | 13360.68442 | 11491.35077 | 11497.90878 |
| 918.68 | 11114.8 | 11358.6 | 13827.19112 | 14204.19098 | 918.68 | 8404.68 | 8615.02 | 13054.28758 | 13380.86753 | 11255.50469 | 11515.94332 |
| 918.955 | 11023 | 11373.3 | 13919.19423 | 14222.52065 | 918.955 | 8420.47 | 8628.13 | 13299.59829 | 13401.05065 | 11208.20155 | 11533.96632 |
| 919.229 | 11251.4 | 11388 | 13885.01827 | 14240.85031 | 919.229 | 8569.62 | 8641.25 | 13220.15412 | 13421.23376 | 11545.13111 | 11552.03549 |
| 919.504 | 11253 | 11402.8 | 14021.01257 | 14259.17998 | 919.504 | 8445.97 | 8654.37 | 13168.83748 | 13441.41688 | 11493.8562 | 11570.04694 |
| 919.779 | 11274.6 | 11417.5 | 14001.85511 | 14277.50965 | 919.779 | 8468.75 | 8667.5 | 13356.39014 | 13461.70735 | 11580.55362 | 11588.17385 |
| 920.053 | 11163.8 | 11432.2 | 13900.50979 | 14295.83931 | 920.053 | 8585.43 | 8680.64 | 13273.08112 | 13481.89046 | 11548.68721 | 11606.1853 |
| 920.328 | 11249.7 | 11446.9 | 14037.45014 | 14314.16898 | 920.328 | 8630.23 | 8693.79 | 13456.98364 | 13502.18093 | 11568.54599 | 11624.31221 |
| 920.602 | 11139.8 | 11461.7 | 14121.29358 | 14332.49865 | 920.602 | 8693.08 | 8706.94 | 13216.82606 | 13522.36404 | 11753.74063 | 11642.43911 |
| 920.877 | 11135.6 | 11476.4 | 14107.5759 | 14350.82832 | 920.877 | 8592.18 | 8720.1 | 13155.73993 | 13542.65451 | 11424.54676 | 11660.56602 |
| 921.151 | 11144.6 | 11491.2 | 13978.44044 | 14369.15798 | 921.151 | 8654.26 | 8733.27 | 13555.85942 | 13562.94498 | 11525.39933 | 11678.69293 |
| 921.426 | 11138.1 | 11505.9 | 13970.28078 | 14387.6059 | 921.426 | 8715.59 | 8746.44 | 13406.74057 | 13583.23546 | 11369.20774 | 11696.81984 |
| 921.7 | 11298.7 | 11520.7 | 14032.60165 | 14405.93557 | 921.7 | 8741.19 | 8759.63 | 13482.64196 | 13603.52593 | 11593.13854 | 11714.94674 |
| 921.975 | 11257.1 | 11535.4 | 14019.12047 | 14424.26524 | 921.975 | 8616.95 | 8772.82 | 13371.42012 | 13623.8164 | 11700.39903 | 11733.07365 |
| 922.249 | 11376.5 | 11550.2 | 14154.40524 | 14442.59491 | 922.249 | 8694.75 | 8786.01 | 13339.74981 | 13644.21422 | 11743.23395 | 11751.31602 |
| 922.523 | 11405.6 | 11565 | 13936.57784 | 14461.04283 | 922.523 | 8600.79 | 8799.22 | 13470.08119 | 13664.50469 | 11818.05074 | 11769.44292 |
| 922.798 | 11277.1 | 11579.7 | 14018.41094 | 14479.37249 | 922.798 | 8690.02 | 8812.43 | 13558.43599 | 13684.90252 | 11805.2349 | 11787.68529 |
| 923.072 | 11395.4 | 11594.5 | 14224.17623 | 14497.82042 | 923.072 | 8694.4 | 8825.65 | 13512.27249 | 13705.19299 | 11716.56315 | 11805.81219 |
| 923.346 | 11366.3 | 11609.3 | 14105.09252 | 14516.15008 | 923.346 | 8708.63 | 8838.87 | 13446.57001 | 13725.59082 | 11678.92384 | 11824.05456 |
| 923.621 | 11482.3 | 11624 | 14195.67655 | 14534.59801 | 923.621 | 8689.68 | 8852.11 | 13448.82451 | 13745.98865 | 11686.77499 | 11842.29693 |
| 923.895 | 11221.5 | 11638.8 | 14245.81706 | 14553.04593 | 923.895 | 8705.44 | 8865.35 | 13407.70678 | 13766.38648 | 11800.27021 | 11860.53929 |
| 924.169 | 11433.2 | 11653.6 | 14185.15178 | 14571.3756 | 924.169 | 8789.91 | 8878.59 | 13546.62672 | 13786.7843 | 11771.75208 | 11878.78166 |
| 924.444 | 11437.1 | 11668.4 | 14261.78161 | 14589.82352 | 924.444 | 8859.06 | 8891.85 | 13606.42456 | 13807.18213 | 11960.8723 | 11897.02402 |
| 924.718 | 11560.2 | 11683.2 | 14280.46605 | 14608.15319 | 924.718 | 8740.65 | 8905.11 | 13440.02123 | 13827.57996 | 11841.14235 | 11915.26639 |
| 924.992 | 11365.9 | 11698 | 14250.31079 | 14626.60111 | 924.992 | 8739.23 | 8918.38 | 13500.89265 | 13847.97779 | 11675.69102 | 11933.50875 |
| 925.266 | 11510 | 11712.8 | 14341.36784 | 14645.04903 | 925.266 | 8834.76 | 8931.66 | 13722.1554 | 13868.48297 | 11987.65856 | 11951.86657 |
| 925.54 | 11472.3 | 11727.6 | 14468.13818 | 14663.49695 | 925.54 | 8777.14 | 8944.94 | 13747.06222 | 13888.8808 | 12065.93909 | 11970.10894 |
| 925.815 | 11647.2 | 11742.4 | 14320.90957 | 14681.94488 | 925.815 | 8908.06 | 8958.23 | 13706.80335 | 13909.38598 | 11879.4744 | 11988.46676 |
| 926.089 | 11407.2 | 11757.2 | 14538.97348 | 14700.3928 | 926.089 | 8752.51 | 8971.53 | 13768.74833 | 13929.89117 | 11908.80074 | 12006.70913 |
| 926.363 | 11559.6 | 11772.1 | 14518.63346 | 14718.84072 | 926.363 | 8821.09 | 8984.83 | 13709.59463 | 13950.39635 | 11913.3036 | 12025.06695 |
| 926.637 | 11506.5 | 11786.9 | 14406.52685 | 14737.28865 | 926.637 | 8879.52 | 8998.14 | 13694.88672 | 13970.79418 | 11936.27974 | 12043.42477 |
| 926.911 | 11583.8 | 11801.7 | 14373.29694 | 14755.73657 | 926.911 | 8910.5 | 9011.46 | 13639.16845 | 13991.29936 | 11927.62039 | 12061.7826 |
| 927.185 | 11632.7 | 11816.5 | 14461.98888 | 14774.18449 | 927.185 | 8916.78 | 9024.78 | 13825.9696 | 14011.80455 | 12141.56408 | 12080.14042 |
| 927.459 | 11648.1 | 11831.4 | 14560.85082 | 14792.63241 | 927.459 | 8953.61 | 9038.12 | 13800.95543 | 14032.41709 | 11979.34558 | 12098.49824 |
| 927.734 | 11559.9 | 11846.2 | 14527.62091 | 14811.08034 | 927.734 | 8778.62 | 9051.46 | 13623.60168 | 14052.92227 | 12089.83889 | 12116.85607 |
| 928.008 | 11591.1 | 11861 | 14352.4839 | 14829.52826 | 928.008 | 8831.73 | 9064.8 | 13671.37554 | 14073.42746 | 11816.31887 | 12135.32935 |
| 928.282 | 11724.1 | 11875.9 | 14518.16043 | 14847.97618 | 928.282 | 8951.73 | 9078.16 | 13584.09431 | 14094.04 | 12235.89327 | 12153.68717 |
| 928.556 | 11755.2 | 11890.7 | 14541.22034 | 14866.54236 | 928.556 | 8912.8 | 9091.52 | 13917.00833 | 14114.54519 | 12092.26351 | 12172.16045 |
| 928.83 | 11667.3 | 11905.6 | 14668.34545 | 14884.99028 | 928.83 | 9051.59 | 9104.88 | 14091.57079 | 14135.15773 | 12137.17668 | 12190.51828 |
| 929.104 | 11688 | 11920.5 | 14503.61496 | 14903.43821 | 929.104 | 8920.76 | 9118.26 | 13856.45899 | 14155.77027 | 12159.34461 | 12208.99156 |
| 929.377 | 11760.2 | 11935.3 | 14627.66541 | 14922.00438 | 929.377 | 9092.32 | 9131.64 | 13760.15977 | 14176.38281 | 11877.04979 | 12227.46484 |
| 929.651 | 11767.4 | 11950.2 | 14546.54185 | 14940.45231 | 929.651 | 8927.87 | 9145.02 | 13851.41321 | 14196.99535 | 12141.21771 | 12245.82266 |
| 929.925 | 11525.5 | 11965 | 14679.34325 | 14958.90023 | 929.925 | 9043.23 | 9158.42 | 14037.03344 | 14217.60789 | 12018.48584 | 12264.29594 |
| 930.199 | 11644.7 | 11979.9 | 14660.65881 | 14977.46641 | 930.199 | 9043.42 | 9171.82 | 13835.41702 | 14238.22043 | 12254.13564 | 12282.76922 |
| 930.473 | 11766.4 | 11994.8 | 14698.97373 | 14996.03259 | 930.473 | 8978.51 | 9185.23 | 14024.36532 | 14258.83298 | 12170.77496 | 12301.35796 |
| 930.747 | 11676.3 | 12009.6 | 14828.10919 | 15014.48051 | 930.747 | 9197.86 | 9198.65 | 14244.76921 | 14279.55287 | 12503.52493 | 12319.83124 |
| 931.021 | 11813.6 | 12024.5 | 14840.64432 | 15033.04669 | 931.021 | 9127.56 | 9212.07 | 14344.61121 | 14300.16542 | 12289.46579 | 12338.30453 |
| 931.294 | 11850.3 | 12039.4 | 14818.53046 | 15051.49461 | 931.294 | 9101.76 | 9225.5 | 14029.08902 | 14320.88531 | 12060.05073 | 12356.89327 |
| 931.568 | 11765.1 | 12054.3 | 14896.10634 | 15070.06079 | 931.568 | 9079.42 | 9238.93 | 14164.57354 | 14341.49786 | 12181.05072 | 12375.36655 |
| 931.842 | 11772 | 12069.2 | 14946.3651 | 15088.62697 | 931.842 | 9121.28 | 9252.37 | 14169.61932 | 14362.21775 | 12216.14995 | 12393.95529 |
| 932.116 | 11768.9 | 12084.1 | 14757.62866 | 15107.07489 | 932.116 | 9010.68 | 9265.82 | 14310.04226 | 14382.93765 | 12522.34459 | 12412.42857 |
| 932.39 | 11838.5 | 12099 | 14767.20739 | 15125.64107 | 932.39 | 9105.09 | 9279.28 | 14388.09079 | 14403.65755 | 12321.79403 | 12431.01731 |
| 932.663 | 11981.8 | 12113.9 | 14878.01318 | 15144.20725 | 932.663 | 9140.22 | 9292.74 | 14170.58553 | 14424.37745 | 12386.56597 | 12449.60605 |
| 932.937 | 11803.7 | 12128.8 | 14666.3351 | 15162.77343 | 932.937 | 9131.3 | 9306.21 | 14109.49941 | 14445.09735 | 12282.19193 | 12468.19478 |
| 933.211 | 11841.3 | 12143.7 | 14902.49216 | 15181.22135 | 933.211 | 9098.1 | 9319.69 | 13995.48629 | 14465.81725 | 12319.83124 | 12486.78352 |
| 933.484 | 11851.1 | 12158.6 | 14842.65467 | 15199.78753 | 933.484 | 9233.62 | 9333.17 | 14389.057 | 14486.53715 | 12479.50967 | 12505.37226 |
| 933.758 | 11955.7 | 12173.5 | 15034.46576 | 15218.35371 | 933.758 | 9224.94 | 9346.66 | 14216.64168 | 14507.3644 | 12426.6299 | 12523.961 |
| 934.032 | 11964.7 | 12188.4 | 14857.55491 | 15236.91989 | 934.032 | 9403.02 | 9360.16 | 14336.23736 | 14528.0843 | 12467.84841 | 12542.6652 |
| 934.305 | 11945.8 | 12203.3 | 14757.51041 | 15255.48607 | 934.305 | 9167.07 | 9373.66 | 14351.48206 | 14548.91156 | 12617.7129 | 12561.25394 |
| 934.579 | 11834.9 | 12218.3 | 15026.54261 | 15274.05224 | 934.579 | 9162.78 | 9387.17 | 14370.91367 | 14569.63145 | 12461.96005 | 12579.84268 |
| 934.852 | 11872.8 | 12233.2 | 14936.66812 | 15292.61842 | 934.852 | 9222.48 | 9400.68 | 14237.14686 | 14590.45871 | 12507.10413 | 12598.54688 |
| 935.126 | 12039.7 | 12248.1 | 14910.17879 | 15311.1846 | 935.126 | 9320.98 | 9414.21 | 14342.67878 | 14611.28597 | 12633.06882 | 12617.25107 |
| 935.4 | 12035.8 | 12263.1 | 14972.3814 | 15329.86904 | 935.4 | 9200.42 | 9427.74 | 14312.08204 | 14632.11322 | 12719.20049 | 12635.83981 |
| 935.673 | 11848.6 | 12278 | 14998.51596 | 15348.43521 | 935.673 | 9406.69 | 9441.27 | 14360.07061 | 14652.94048 | 12610.66997 | 12654.54401 |
| 935.947 | 12017.1 | 12292.9 | 14966.35035 | 15367.00139 | 935.947 | 9382.49 | 9454.82 | 14372.20195 | 14673.76773 | 12622.56214 | 12673.24821 |
| 936.22 | 12084.2 | 12307.9 | 14982.55141 | 15385.56757 | 936.22 | 9335.74 | 9468.36 | 14469.25267 | 14694.59499 | 12504.67951 | 12691.9524 |
| 936.494 | 12093.9 | 12322.8 | 15096.19535 | 15404.13375 | 936.494 | 9296.22 | 9481.92 | 14578.97151 | 14715.42224 | 12686.1795 | 12710.6566 |
| 936.767 | 12042 | 12337.8 | 14872.57341 | 15422.81818 | 936.767 | 9201.52 | 9495.48 | 14358.13819 | 14736.35686 | 12572.68428 | 12729.3608 |
| 937.04 | 12026.8 | 12352.7 | 15121.2656 | 15441.38436 | 937.04 | 9302.66 | 9509.05 | 14364.15018 | 14757.18411 | 12779.46957 | 12748.18045 |
| 937.314 | 11975.2 | 12367.7 | 15084.84278 | 15460.0688 | 937.314 | 9251.95 | 9522.63 | 14555.88976 | 14778.11872 | 12626.2568 | 12766.88465 |
| 937.587 | 11937.4 | 12382.6 | 15119.37351 | 15478.63498 | 937.587 | 9245.09 | 9536.21 | 14470.11153 | 14799.05334 | 12781.08599 | 12785.58885 |
| 937.861 | 11912 | 12397.6 | 15059.18125 | 15497.20116 | 937.861 | 9504.06 | 9549.8 | 14532.37858 | 14819.88059 | 12604.55069 | 12804.4085 |
| 938.134 | 12169.5 | 12412.5 | 15031.39111 | 15515.88559 | 938.134 | 9322.17 | 9563.39 | 14618.47888 | 14840.8152 | 12848.62892 | 12823.1127 |
| 938.407 | 12066.1 | 12427.5 | 15170.93309 | 15534.45177 | 938.407 | 9264.13 | 9576.99 | 14403.12077 | 14861.74982 | 12546.12894 | 12841.93236 |
| 938.681 | 12161.5 | 12442.5 | 15067.69567 | 15553.1362 | 938.681 | 9447.19 | 9590.6 | 14703.2909 | 14882.68443 | 12487.82265 | 12860.75201 |
| 938.954 | 12253.5 | 12457.5 | 15182.40391 | 15571.70238 | 938.954 | 9544.01 | 9604.21 | 14722.93723 | 14903.61904 | 12950.80926 | 12879.57167 |
| 939.227 | 12013.1 | 12472.4 | 15063.08369 | 15590.38682 | 939.227 | 9333.95 | 9617.83 | 14559.86197 | 14924.55365 | 12516.34077 | 12898.39132 |
| 939.5 | 11977.1 | 12487.4 | 15210.66707 | 15608.953 | 939.5 | 9417.94 | 9631.46 | 14864.64846 | 14945.59562 | 12653.50489 | 12917.21098 |
| 939.774 | 12270.9 | 12502.4 | 15321.94589 | 15627.63743 | 939.774 | 9363.07 | 9645.09 | 14483.10172 | 14966.53024 | 12886.49915 | 12936.03063 |
| 940.047 | 12142.8 | 12517.4 | 15031.50936 | 15646.32186 | 940.047 | 9404.34 | 9658.73 | 14608.92411 | 14987.46485 | 12753.72244 | 12954.85029 |
| 940.32 | 12332 | 12532.4 | 15498.73848 | 15664.88804 | 940.32 | 9570.05 | 9672.38 | 14591.85435 | 15008.50682 | 12820.34171 | 12973.66994 |
| 940.593 | 12160.9 | 12547.4 | 15088.2722 | 15683.57248 | 940.593 | 9463.38 | 9686.03 | 14687.50943 | 15029.54879 | 12743.56213 | 12992.60506 |
| 940.867 | 12094.8 | 12562.4 | 15393.84549 | 15702.25691 | 940.867 | 9294.89 | 9699.69 | 14597.54427 | 15050.4834 | 12757.30164 | 13011.42471 |
| 941.14 | 12240.9 | 12577.3 | 15275.58957 | 15720.94135 | 941.14 | 9498.79 | 9713.36 | 14767.16831 | 15071.52537 | 12847.47434 | 13030.35982 |
| 941.413 | 12207.2 | 12592.3 | 15483.83824 | 15739.62578 | 941.413 | 9506.89 | 9727.03 | 14707.90725 | 15092.56734 | 13144.54779 | 13049.17948 |
| 941.686 | 12375.3 | 12607.3 | 15300.77808 | 15758.19196 | 941.686 | 9619.14 | 9740.7 | 14818.27024 | 15113.60931 | 13097.90276 | 13068.11459 |
| 941.959 | 12362.2 | 12622.4 | 15188.67147 | 15776.87639 | 941.959 | 9576.17 | 9754.39 | 14733.56557 | 15134.65128 | 12961.31594 | 13087.04971 |
| 942.232 | 12439.3 | 12637.4 | 15513.87524 | 15795.56083 | 942.232 | 9784.73 | 9768.08 | 14910.38253 | 15155.69325 | 13106.79302 | 13105.98482 |
| 942.505 | 12516.8 | 12652.4 | 15459.59577 | 15814.24526 | 942.505 | 9719.97 | 9781.77 | 14936.57764 | 15176.84257 | 13067.19093 | 13124.91993 |
| 942.778 | 12462.5 | 12667.4 | 15468.22846 | 15832.9297 | 942.778 | 9690.3 | 9795.48 | 14745.58956 | 15197.88454 | 12868.25678 | 13143.85505 |
| 943.051 | 12416.4 | 12682.4 | 15494.1265 | 15851.61413 | 943.051 | 9572.58 | 9809.19 | 15081.50957 | 15218.92651 | 12827.73102 | 13162.79016 |
| 943.324 | 12347.7 | 12697.4 | 15434.64378 | 15870.29857 | 943.324 | 9811.3 | 9822.9 | 14742.90563 | 15240.07584 | 12908.898 | 13181.72527 |
| 943.597 | 12266.9 | 12712.4 | 15488.56847 | 15888.983 | 943.597 | 9669.11 | 9836.62 | 15107.91939 | 15261.11781 | 12817.45526 | 13200.66038 |
| 943.87 | 12359.4 | 12727.5 | 15425.1833 | 15907.66744 | 943.87 | 9724.06 | 9850.35 | 15254.03225 | 15282.26714 | 13043.29112 | 13219.71096 |
| 944.143 | 12468.7 | 12742.5 | 15323.95624 | 15926.35187 | 944.143 | 9673.81 | 9864.08 | 14826.32201 | 15303.41646 | 12969.74437 | 13238.64607 |
| 944.416 | 12408.6 | 12757.5 | 15557.9847 | 15945.03631 | 944.416 | 9618.62 | 9877.82 | 15001.42126 | 15324.56579 | 13239.22336 | 13257.69664 |
| 944.689 | 12488.3 | 12772.5 | 15647.74094 | 15963.72074 | 944.689 | 9555.65 | 9891.57 | 14933.89371 | 15345.60776 | 13143.27776 | 13276.63175 |
| 944.962 | 12456.6 | 12787.6 | 15538.94549 | 15982.52343 | 944.962 | 9717.61 | 9905.32 | 15235.45949 | 15366.86444 | 13138.54398 | 13295.68233 |
| 945.235 | 12444.2 | 12802.6 | 15382.01989 | 16001.20787 | 945.235 | 9820.1 | 9919.08 | 15087.52156 | 15388.01377 | 12971.82261 | 13314.7329 |
| 945.508 | 12354.4 | 12817.6 | 15537.99945 | 16019.8923 | 945.508 | 9703.05 | 9932.84 | 15077.21529 | 15409.16309 | 13151.24436 | 13333.78347 |
| 945.78 | 12660.1 | 12832.7 | 15715.73809 | 16038.57673 | 945.78 | 9887.58 | 9946.61 | 15084.19349 | 15430.31242 | 13008.76918 | 13352.83404 |
| 946.053 | 12466.9 | 12847.7 | 15672.57468 | 16057.37943 | 946.053 | 9817.94 | 9960.39 | 15198.42133 | 15451.46175 | 12991.79685 | 13371.88461 |
| 946.326 | 12605.5 | 12862.8 | 15738.32497 | 16076.06386 | 946.326 | 9958.32 | 9974.17 | 15257.03824 | 15472.71843 | 13220.51916 | 13390.93518 |
| 946.599 | 12665.5 | 12877.8 | 15756.77289 | 16094.74829 | 946.599 | 9758.06 | 9987.96 | 15462.84159 | 15493.86776 | 13441.50579 | 13409.98575 |
| 946.872 | 12693.4 | 12892.9 | 15699.65528 | 16113.55098 | 946.872 | 9922.13 | 10001.8 | 15373.95 | 15515.12444 | 13324.31591 | 13429.03632 |
| 947.144 | 12655 | 12907.9 | 15573.35796 | 16132.23542 | 947.144 | 9914.32 | 10015.6 | 15269.06223 | 15536.38112 | 13430.76819 | 13448.20235 |
| 947.417 | 12629.6 | 12923 | 15777.58593 | 16150.91985 | 947.417 | 9695.02 | 10029.4 | 15223.97229 | 15557.53045 | 13242.80256 | 13467.25292 |
| 947.69 | 12620.5 | 12938 | 15954.85155 | 16169.72254 | 947.69 | 9687.4 | 10043.2 | 15546.36532 | 15578.78713 | 13291.52584 | 13486.41895 |
| 947.963 | 12459.5 | 12953.1 | 15706.75064 | 16188.40698 | 947.963 | 9943.83 | 10057 | 15382.00178 | 15600.04382 | 13532.25578 | 13505.46953 |
| 948.235 | 12579.1 | 12968.2 | 15781.84314 | 16207.20967 | 948.235 | 9899.48 | 10070.8 | 15282.58921 | 15621.3005 | 13178.72336 | 13524.63555 |
| 948.508 | 12598.3 | 12983.2 | 15684.16376 | 16225.8941 | 948.508 | 9859 | 10084.6 | 15272.60501 | 15642.55718 | 13352.14129 | 13543.80158 |
| 948.78 | 12642.8 | 12998.3 | 15714.0825 | 16244.69679 | 948.78 | 9942.92 | 10098.5 | 15306.85189 | 15663.92123 | 13217.74817 | 13562.85215 |
| 949.053 | 12651 | 13013.4 | 15900.69034 | 16263.38123 | 949.053 | 9915.33 | 10112.3 | 15532.40892 | 15685.17791 | 13593.44853 | 13582.01818 |
| 949.326 | 12542.3 | 13028.5 | 15789.64803 | 16282.18392 | 949.326 | 9831.18 | 10126.2 | 15609.8133 | 15706.43459 | 13336.20809 | 13601.18421 |
| 949.598 | 12567.5 | 13043.5 | 15912.51593 | 16300.98661 | 949.598 | 9854.88 | 10140 | 15375.77507 | 15727.79863 | 13565.27677 | 13620.35024 |
| 949.871 | 12725 | 13058.6 | 15683.80899 | 16319.67104 | 949.871 | 9921.46 | 10153.9 | 15136.5837 | 15749.05532 | 13574.2825 | 13639.63173 |
| 950.143 | 12652 | 13073.7 | 15880.23206 | 16338.47374 | 950.143 | 9877.33 | 10167.8 | 15311.36088 | 15770.41936 | 13481.68518 | 13658.79776 |
| 950.416 | 12766.4 | 13088.8 | 15760.43882 | 16357.15817 | 950.416 | 9894 | 10181.6 | 15400.03775 | 15791.67604 | 13441.50579 | 13677.96379 |
| 950.688 | 12771.4 | 13103.9 | 15857.40867 | 16375.96086 | 950.688 | 9955.05 | 10195.5 | 15337.66334 | 15813.04008 | 13525.44376 | 13697.24528 |
| 950.961 | 12659.4 | 13119 | 15911.56988 | 16394.76355 | 950.961 | 10037.5 | 10209.4 | 15641.2689 | 15834.40412 | 13484.918 | 13716.4113 |
| 951.233 | 12648.1 | 13134 | 15786.69164 | 16413.56624 | 951.233 | 9995.59 | 10223.3 | 15359.77888 | 15855.76816 | 13192.00103 | 13735.57733 |
| 951.506 | 12773.7 | 13149.1 | 15809.86979 | 16432.25068 | 951.506 | 10042.8 | 10237.2 | 15553.23617 | 15877.1322 | 13342.2119 | 13754.85882 |
| 951.778 | 12759.5 | 13164.2 | 16030.06231 | 16451.05337 | 951.778 | 9849.34 | 10251.1 | 15448.45575 | 15898.49624 | 13432.03823 | 13774.14031 |
| 952.051 | 12654.4 | 13179.3 | 15808.21421 | 16469.85606 | 952.051 | 9908.8 | 10265 | 15594.46125 | 15919.86028 | 13190.03825 | 13793.4218 |
| 952.323 | 12655.5 | 13194.4 | 16034.55603 | 16488.65875 | 952.323 | 10167.6 | 10278.9 | 15395.74347 | 15941.22432 | 13466.90655 | 13812.70328 |
| 952.595 | 12778 | 13209.5 | 15888.27347 | 16507.46144 | 952.595 | 9822.79 | 10292.8 | 15710.29944 | 15962.58836 | 13681.19661 | 13831.86931 |
| 952.868 | 12945.5 | 13224.6 | 16013.74299 | 16526.26413 | 952.868 | 9955.12 | 10306.7 | 15925.33549 | 15984.05976 | 13524.05826 | 13851.26626 |
| 953.14 | 12749.2 | 13239.7 | 15866.27787 | 16545.06682 | 953.14 | 10057.3 | 10320.7 | 15487.31898 | 16005.4238 | 13528.33021 | 13870.54775 |
| 953.412 | 12941.5 | 13254.8 | 15893.83149 | 16563.75125 | 953.412 | 10033.5 | 10334.6 | 15520.81436 | 16026.8952 | 13616.8865 | 13889.82923 |
| 953.685 | 12715.2 | 13269.9 | 15795.56083 | 16582.55394 | 953.685 | 10188.8 | 10348.5 | 15723.82642 | 16048.25924 | 13579.47811 | 13909.11072 |
| 953.957 | 12646.2 | 13285.1 | 16078.54723 | 16601.35663 | 953.957 | 10136.1 | 10362.5 | 15632.14356 | 16069.73064 | 13567.81685 | 13928.39221 |
| 954.229 | 12917.4 | 13300.2 | 16074.76304 | 16620.15932 | 954.229 | 10130.6 | 10376.4 | 15849.97088 | 16091.20203 | 13513.20521 | 13947.78915 |
| 954.501 | 12799.7 | 13315.3 | 15865.80484 | 16638.96202 | 954.501 | 10078.4 | 10390.4 | 15437.6127 | 16112.67343 | 13368.18995 | 13967.07064 |
| 954.774 | 12922.1 | 13330.4 | 16037.15766 | 16657.76471 | 954.774 | 10311.9 | 10404.4 | 15603.15717 | 16134.14483 | 13666.99528 | 13986.46758 |
| 955.046 | 12907.9 | 13345.5 | 16243.98726 | 16676.5674 | 955.046 | 10183 | 10418.3 | 15585.01384 | 16155.61623 | 13534.10311 | 14005.74907 |
| 955.318 | 12919.7 | 13360.6 | 16146.18962 | 16695.37009 | 955.318 | 10094.5 | 10432.3 | 15311.68295 | 16177.08762 | 13617.23288 | 14025.14602 |
| 955.59 | 13062.4 | 13375.8 | 15886.49963 | 16714.29103 | 955.59 | 10141.9 | 10446.3 | 15636.00841 | 16198.55902 | 13596.56589 | 14044.54296 |
| 955.862 | 12945.7 | 13390.9 | 16236.65539 | 16733.09372 | 955.862 | 10312.2 | 10460.3 | 15843.42211 | 16220.03042 | 13850.91988 | 14063.93991 |
| 956.135 | 13003.5 | 13406 | 16360.35108 | 16751.89641 | 956.135 | 10165.7 | 10474.3 | 15704.39481 | 16241.50182 | 13700.36264 | 14083.33685 |
| 956.407 | 12993.5 | 13421.1 | 16371.46714 | 16770.6991 | 956.407 | 10247 | 10488.3 | 15815.29458 | 16263.08057 | 13683.1594 | 14102.7338 |
| 956.679 | 12863.4 | 13436.3 | 16398.666 | 16789.5018 | 956.679 | 10400.1 | 10502.3 | 15633.96863 | 16284.55197 | 13693.78154 | 14122.13074 |
| 956.951 | 13109 | 13451.4 | 16357.86771 | 16808.30449 | 956.951 | 10262.7 | 10516.3 | 15580.93427 | 16306.13072 | 13796.30825 | 14141.52769 |
| 957.223 | 12893.7 | 13466.5 | 16206.97316 | 16827.22543 | 957.223 | 10418.3 | 10530.3 | 15875.73656 | 16327.70948 | 14073.75384 | 14160.92463 |
| 957.495 | 13199.1 | 13481.7 | 16446.20487 | 16846.02812 | 957.495 | 10414.6 | 10544.3 | 15779.00792 | 16349.18087 | 13794.92275 | 14180.43704 |
| 957.767 | 13014.4 | 13496.8 | 16467.01791 | 16864.83081 | 957.767 | 10444 | 10558.3 | 15908.69516 | 16370.75963 | 13819.28439 | 14199.83398 |
| 958.039 | 12925.7 | 13512 | 16186.98791 | 16883.6335 | 958.039 | 10045.2 | 10572.4 | 16028.29084 | 16392.33838 | 13852.42084 | 14219.23093 |
| 958.311 | 13126.5 | 13527.1 | 16394.17227 | 16902.55445 | 958.311 | 10368.1 | 10586.4 | 16268.77049 | 16413.91714 | 13769.52199 | 14238.74333 |
| 958.583 | 12954 | 13542.3 | 16387.54994 | 16921.35714 | 958.583 | 10313.1 | 10600.4 | 15888.18997 | 16435.49589 | 13901.25958 | 14258.14028 |
| 958.855 | 13205.1 | 13557.4 | 16565.87986 | 16940.15983 | 958.855 | 10210 | 10614.5 | 16120.61785 | 16457.07465 | 14227.65936 | 14277.65268 |
| 959.127 | 13158.4 | 13572.5 | 16275.08856 | 16959.08078 | 959.127 | 10308.6 | 10628.5 | 16036.44997 | 16478.6534 | 14117.97426 | 14297.16508 |
| 959.399 | 13285.2 | 13587.7 | 16351.48189 | 16977.88347 | 959.399 | 10409 | 10642.6 | 16231.3029 | 16500.23216 | 14421.3979 | 14316.67749 |
| 959.67 | 13268.1 | 13602.9 | 16346.39688 | 16996.68616 | 959.67 | 10375.4 | 10656.6 | 16233.98683 | 16521.81091 | 14159.07731 | 14336.18989 |
| 959.942 | 13408.7 | 13618 | 16657.29168 | 17015.6071 | 959.942 | 10375.7 | 10670.7 | 16227.86748 | 16543.49702 | 14166.58208 | 14355.70229 |
| 960.214 | 13089 | 13633.2 | 16448.68825 | 17034.4098 | 960.214 | 10370.5 | 10684.8 | 16378.0599 | 16565.07578 | 14313.21375 | 14375.2147 |
| 960.486 | 13381.1 | 13648.3 | 16336.6999 | 17053.33074 | 960.486 | 10621 | 10698.9 | 15975.14913 | 16586.76189 | 14050.66224 | 14394.7271 |
| 960.758 | 13260.7 | 13663.5 | 16635.29608 | 17072.13343 | 960.758 | 10637.3 | 10713 | 16340.69967 | 16608.34064 | 14152.26528 | 14414.2395 |
| 961.03 | 13223.6 | 13678.6 | 16431.18637 | 17090.93612 | 961.03 | 10517.9 | 10727 | 16333.29204 | 16630.02675 | 14396.68989 | 14433.75191 |
| 961.301 | 13324.5 | 13693.8 | 16342.25792 | 17109.85707 | 961.301 | 10639.7 | 10741.1 | 16253.09637 | 16651.71286 | 14264.25955 | 14453.26431 |
| 961.573 | 13307.8 | 13709 | 16790.68435 | 17128.77802 | 961.573 | 10405 | 10755.2 | 16270.05878 | 16673.29162 | 14257.67844 | 14472.89217 |
| 961.845 | 13498.9 | 13724.1 | 16942.52495 | 17147.58071 | 961.845 | 10690.9 | 10769.3 | 16593.95481 | 16694.97773 | 14434.90649 | 14492.40457 |
| 962.117 | 13250.2 | 13739.3 | 16542.22868 | 17166.50165 | 962.117 | 10441.6 | 10783.5 | 16031.40419 | 16716.66384 | 14009.7901 | 14512.03244 |
| 962.388 | 13333.3 | 13754.5 | 16781.34214 | 17185.30434 | 962.388 | 10546.6 | 10797.6 | 16812.74835 | 16738.34995 | 14818.22707 | 14531.6603 |
| 962.66 | 13262.9 | 13769.6 | 16879.6128 | 17204.22529 | 962.66 | 10395.1 | 10811.7 | 16495.50845 | 16760.03606 | 14335.84352 | 14551.1727 |
| 962.932 | 13525.4 | 13784.8 | 16613.18223 | 17223.02798 | 962.932 | 10481.8 | 10825.8 | 16314.934 | 16781.82953 | 14250.52005 | 14570.80056 |
| 963.203 | 13440.1 | 13800 | 16637.54294 | 17241.94893 | 963.203 | 10700.2 | 10839.9 | 16465.23378 | 16803.51564 | 14405.11832 | 14590.42842 |
| 963.475 | 13533.6 | 13815.1 | 16635.88736 | 17260.75162 | 963.475 | 10591.2 | 10854.1 | 16538.88067 | 16825.20176 | 14404.65649 | 14610.05628 |
| 963.746 | 13546.3 | 13830.3 | 16785.24458 | 17279.67256 | 963.746 | 10654.2 | 10868.2 | 16335.86861 | 16846.88787 | 14640.19082 | 14629.68415 |
| 964.018 | 13279 | 13845.5 | 16588.585 | 17298.59351 | 964.018 | 10278.5 | 10882.4 | 16384.60868 | 16868.68134 | 14363.78435 | 14649.31201 |
| 964.29 | 13620.2 | 13860.7 | 16739.24303 | 17317.3962 | 964.29 | 10615.7 | 10896.5 | 16386.86318 | 16890.36745 | 14457.76717 | 14668.93987 |
| 964.561 | 13385.5 | 13875.9 | 16847.80196 | 17336.31715 | 964.561 | 10657.8 | 10910.7 | 16488.31553 | 16912.16092 | 14358.58874 | 14688.56773 |
| 964.833 | 13460.2 | 13891 | 17090.58135 | 17355.23809 | 964.833 | 10808.3 | 10924.8 | 16533.83489 | 16933.95438 | 14293.12405 | 14708.19559 |
| 965.104 | 13444.4 | 13906.2 | 16711.6894 | 17374.15904 | 965.104 | 10543.1 | 10939 | 16723.21262 | 16955.6405 | 14468.27385 | 14727.93891 |
| 965.376 | 13743.4 | 13921.4 | 17010.5221 | 17392.96173 | 965.376 | 10699.5 | 10953.2 | 16374.62448 | 16977.43396 | 14602.20514 | 14747.56677 |
| 965.647 | 13523.8 | 13936.6 | 16668.05297 | 17411.88268 | 965.647 | 10502.3 | 10967.4 | 16687.03331 | 16999.22743 | 14712.92937 | 14767.31009 |
| 965.919 | 13491.4 | 13951.8 | 16702.22893 | 17430.80362 | 965.919 | 10748.4 | 10981.5 | 16385.78961 | 17021.0209 | 14567.45228 | 14786.93795 |
| 966.19 | 13730.3 | 13967 | 17309.00003 | 17449.60631 | 966.19 | 10618.7 | 10995.7 | 16886.82467 | 17042.81437 | 14431.09637 | 14806.68127 |
| 966.461 | 13593.3 | 13982.2 | 16939.21378 | 17468.52726 | 966.461 | 10820.9 | 11009.9 | 16669.21205 | 17064.60784 | 14802.52478 | 14826.42459 |
| 966.733 | 13611 | 13997.4 | 16987.93522 | 17487.44821 | 966.733 | 10684.5 | 11024.1 | 16944.47537 | 17086.40131 | 14644.80914 | 14846.05245 |
| 967.004 | 13724.1 | 14012.6 | 16881.74141 | 17506.36915 | 967.004 | 11116.5 | 11038.3 | 16638.93738 | 17108.19477 | 14602.66697 | 14865.79577 |
| 967.275 | 13733.8 | 14027.8 | 16905.51085 | 17525.2901 | 967.275 | 10865.9 | 11052.5 | 16837.11838 | 17130.0956 | 14731.97994 | 14885.53909 |
| 967.547 | 13628.1 | 14043 | 16809.36879 | 17544.21105 | 967.547 | 10853.2 | 11066.7 | 16522.77712 | 17151.88907 | 14614.44369 | 14905.28241 |
| 967.818 | 13744.9 | 14058.1 | 16990.06383 | 17563.01374 | 967.818 | 10692.1 | 11081 | 16884.57017 | 17173.68254 | 14966.59061 | 14925.02573 |
| 968.09 | 13685.9 | 14073.4 | 16846.6194 | 17581.93468 | 968.09 | 10734.8 | 11095.2 | 16443.44031 | 17195.58336 | 14747.2204 | 14944.76905 |
| 968.361 | 13722.7 | 14088.5 | 17199.84982 | 17600.85563 | 968.361 | 10967.3 | 11109.4 | 17075.77296 | 17217.37683 | 15113.10682 | 14964.51237 |
| 968.632 | 13585.8 | 14103.8 | 17069.41355 | 17619.77657 | 968.632 | 11070.6 | 11123.6 | 16815.53963 | 17239.27766 | 14849.97802 | 14984.25569 |
| 968.903 | 13706.1 | 14119 | 17355.4746 | 17638.69752 | 968.903 | 10751 | 11137.9 | 16918.92441 | 17261.17848 | 14958.85492 | 15004.11446 |
| 969.174 | 13672.4 | 14134.2 | 17196.89342 | 17657.61847 | 969.174 | 10964.8 | 11152.1 | 16844.95544 | 17282.97195 | 14731.97994 | 15023.85778 |
| 969.446 | 13755 | 14149.4 | 17352.39995 | 17676.53941 | 969.446 | 10962.5 | 11166.4 | 16813.6072 | 17304.87278 | 14921.44653 | 15043.6011 |
| 969.717 | 13861.9 | 14164.6 | 17166.14688 | 17695.46036 | 969.717 | 10951.6 | 11180.6 | 16997.18765 | 17326.7736 | 14931.72229 | 15063.45988 |
| 969.988 | 13901.6 | 14179.8 | 17192.04493 | 17714.26305 | 969.988 | 11062.3 | 11194.9 | 16955.96257 | 17348.67443 | 14848.9389 | 15083.31866 |
| 970.259 | 13649.8 | 14195 | 17069.65006 | 17733.184 | 970.259 | 10989.4 | 11209.1 | 17255.27385 | 17370.57525 | 14795.5973 | 15103.06198 |
| 970.53 | 13741.9 | 14210.2 | 17213.56751 | 17752.10494 | 970.53 | 11227.2 | 11223.4 | 16835.40067 | 17392.47608 | 14780.58776 | 15122.92075 |
| 970.801 | 13911 | 14225.4 | 17294.92758 | 17771.02589 | 970.801 | 11244.8 | 11237.7 | 16989.99473 | 17414.3769 | 15395.28619 | 15142.77953 |
| 971.072 | 13929.6 | 14240.6 | 17079.11053 | 17789.94684 | 971.072 | 10962.9 | 11252 | 17100.67979 | 17436.27773 | 15008.27095 | 15162.63831 |
| 971.343 | 13833.3 | 14255.8 | 17245.02358 | 17808.86778 | 971.343 | 11031.4 | 11266.2 | 16904.64593 | 17458.28591 | 14978.36733 | 15182.49709 |
| 971.615 | 13996.4 | 14271 | 17411.40965 | 17827.78873 | 971.615 | 11225.3 | 11280.5 | 17147.8095 | 17480.18674 | 15406.48562 | 15202.2404 |
| 971.886 | 13918.9 | 14286.3 | 17137.29244 | 17846.70968 | 971.886 | 10965.4 | 11294.8 | 17078.56425 | 17502.08756 | 15249.57819 | 15222.09918 |
| 972.157 | 13927.9 | 14301.5 | 17216.16914 | 17865.63062 | 972.157 | 10874.8 | 11309.1 | 17311.95834 | 17524.09574 | 15028.12973 | 15242.07342 |
| 972.428 | 13853.9 | 14316.7 | 17348.14274 | 17884.55157 | 972.428 | 11167.7 | 11323.4 | 16886.71731 | 17545.99657 | 15228.44937 | 15261.93219 |
| 972.698 | 14015 | 14331.9 | 17382.3187 | 17903.59077 | 972.698 | 11182.5 | 11337.7 | 17045.28358 | 17568.00475 | 15274.2862 | 15281.79097 |
| 972.97 | 13887.5 | 14347.1 | 17325.67411 | 17922.51172 | 972.97 | 11185.1 | 11352 | 17484.69573 | 17590.01293 | 15309.26998 | 15301.64975 |
| 973.24 | 14244.4 | 14362.4 | 17384.4473 | 17941.43266 | 973.24 | 11278 | 11366.3 | 17269.76704 | 17611.91376 | 15320.23849 | 15321.62398 |
| 973.511 | 14089.8 | 14377.6 | 17464.03353 | 17960.35361 | 973.511 | 11436.3 | 11380.6 | 17584.96716 | 17633.92194 | 15157.09632 | 15341.48276 |
| 973.782 | 14063.4 | 14392.8 | 17475.14959 | 17979.27456 | 973.782 | 11154.3 | 11395 | 17471.49082 | 17655.93012 | 15573.20698 | 15361.457 |
| 974.053 | 14080.3 | 14408 | 17290.78862 | 17998.1955 | 974.053 | 11006.4 | 11409.3 | 17412.8739 | 17677.93831 | 15500.00661 | 15381.31577 |
| 974.324 | 14052.8 | 14423.2 | 17500.21984 | 18017.11645 | 974.324 | 11186.7 | 11423.6 | 17329.35017 | 17699.94649 | 15149.59155 | 15401.29001 |
| 974.595 | 14104.9 | 14438.5 | 17681.74267 | 18036.0374 | 974.595 | 11234.7 | 11438 | 17677.50888 | 17721.95467 | 15355.45318 | 15421.14879 |
| 974.866 | 14061.2 | 14453.7 | 17433.287 | 18054.95834 | 974.866 | 11462.9 | 11452.3 | 16993.64487 | 17743.96285 | 15182.26617 | 15441.12302 |
| 975.136 | 13855.5 | 14468.9 | 17629.94658 | 18073.99754 | 975.136 | 11312.7 | 11466.6 | 17349.42592 | 17765.97104 | 15445.51043 | 15461.09726 |
| 975.407 | 14005.4 | 14484.1 | 17648.63102 | 18092.91849 | 975.407 | 11276.6 | 11481 | 17230.15231 | 17787.97922 | 14931.83775 | 15481.07149 |
| 975.678 | 14008.3 | 14499.4 | 17436.12514 | 18111.83944 | 975.678 | 11278.3 | 11495.4 | 17119.89669 | 17810.09476 | 15319.08391 | 15501.04573 |
| 975.949 | 14119.9 | 14514.6 | 17694.51431 | 18130.76038 | 975.949 | 11064.7 | 11509.7 | 17225.64332 | 17832.10294 | 15470.3339 | 15521.01996 |
| 976.219 | 14319.9 | 14529.8 | 17742.64447 | 18149.68133 | 976.219 | 11152.9 | 11524.1 | 17460.11098 | 17854.21848 | 15419.30146 | 15540.9942 |
| 976.49 | 14104 | 14545.1 | 17580.63387 | 18168.60228 | 976.49 | 11235.3 | 11538.5 | 17529.35624 | 17876.22666 | 15404.061 | 15560.96843 |
| 976.761 | 14252.7 | 14560.3 | 17577.79572 | 18187.64148 | 976.761 | 11467.2 | 11552.8 | 17458.0712 | 17898.3422 | 15365.8444 | 15580.94267 |
| 977.031 | 14214.5 | 14575.5 | 17434.3513 | 18206.56243 | 977.031 | 11130.5 | 11567.2 | 17990.02507 | 17920.35038 | 15206.85873 | 15600.91691 |
| 977.302 | 14191.9 | 14590.8 | 17508.85253 | 18225.48337 | 977.302 | 11553.2 | 11581.6 | 17536.76387 | 17942.46592 | 15200.16216 | 15621.0066 |
| 977.573 | 14202.6 | 14606 | 17638.57927 | 18244.40432 | 977.573 | 11360.8 | 11596 | 17565.21347 | 17964.58146 | 15450.24421 | 15640.98083 |
| 977.843 | 14531.1 | 14621.2 | 17914.82508 | 18263.32526 | 977.843 | 11423.3 | 11610.4 | 17436.27773 | 17986.58964 | 15204.78048 | 15661.07053 |
| 978.114 | 14382.6 | 14636.5 | 18028.35076 | 18282.36447 | 978.114 | 11443.1 | 11624.8 | 17627.26581 | 18008.70518 | 15556.81195 | 15681.04476 |
| 978.385 | 14217.6 | 14651.7 | 17616.93843 | 18301.28541 | 978.385 | 11447.1 | 11639.2 | 17547.92899 | 18030.82072 | 15325.66502 | 15701.13446 |
| 978.655 | 14170.3 | 14667 | 17631.48391 | 18320.20636 | 978.655 | 11613.6 | 11653.6 | 17630.59387 | 18052.93626 | 15919.11917 | 15721.10869 |
| 978.926 | 14276.6 | 14682.2 | 17764.87658 | 18339.12731 | 978.926 | 11160.3 | 11668 | 17741.601 | 18075.0518 | 15630.24324 | 15741.19838 |
| 979.196 | 14053.4 | 14697.4 | 17909.85833 | 18358.16651 | 979.196 | 11478.7 | 11682.4 | 17690.177 | 18097.16734 | 15792.34628 | 15761.28808 |
| 979.466 | 14308.2 | 14712.7 | 17807.68522 | 18377.08745 | 979.466 | 11546.2 | 11696.8 | 17796.35306 | 18119.28288 | 15734.38636 | 15781.37777 |
| 979.737 | 14393.5 | 14727.9 | 17902.28996 | 18396.0084 | 979.737 | 11399.9 | 11711.2 | 18119.81967 | 18141.39842 | 15801.81384 | 15801.46746 |
| 1101.15 | 20883 | 21546.5 | 26557.9134 | 26871.40983 | 1101.15 | 18305.2 | 18532.7 | 28397.31877 | 28576.17551 | 25568.75308 | 25402.26263 |
| 1102.87 | 21323.5 | 21640.8 | 27028.92671 | 26988.60144 | 1102.87 | 18158.4 | 18632.4 | 28196.13178 | 28728.193 | 25312.08993 | 25544.27598 |
| 1104.6 | 21206.8 | 21734.9 | 26796.67209 | 27105.55654 | 1104.6 | 18787.4 | 18732.2 | 28731.62843 | 28880.31785 | 25890.65 | 25686.17387 |
| 1106.32 | 20963.2 | 21828.9 | 26684.92025 | 27222.39339 | 1106.32 | 18960.3 | 18831.9 | 28668.50252 | 29032.22799 | 25530.65193 | 25828.07176 |
| 1108.05 | 21374 | 21922.7 | 27115.25353 | 27338.87547 | 1108.05 | 18734.8 | 18931.6 | 28546.11555 | 29184.13813 | 25353.30843 | 25970.08511 |
| 1109.77 | 21553.3 | 22016.3 | 27330.95232 | 27455.23929 | 1109.77 | 18846.7 | 19031.3 | 29350.64882 | 29336.15562 | 26107.82651 | 26112.09846 |
| 1111.5 | 21539.8 | 22109.8 | 27065.7043 | 27571.48485 | 1111.5 | 19152 | 19130.9 | 29296.11147 | 29487.9584 | 26080.23205 | 26254.11181 |
| 1113.22 | 21981.1 | 22203.2 | 27050.56754 | 27687.37565 | 1113.22 | 19181.5 | 19230.6 | 29274.74743 | 29639.76118 | 26428.4534 | 26396.12516 |
| 1114.95 | 22063.5 | 22296.3 | 27595.84557 | 27803.14819 | 1114.95 | 19152.4 | 19330.2 | 29717.16557 | 29791.56396 | 26665.14232 | 26538.13851 |
| 1116.67 | 21811.8 | 22389.3 | 27624.34524 | 27918.68422 | 1116.67 | 19385.2 | 19429.8 | 29677.76556 | 29943.15203 | 26394.50875 | 26680.15186 |
| 1118.39 | 22086.2 | 22482.1 | 27325.15778 | 28033.98373 | 1118.39 | 19473.7 | 19529.3 | 30221.09927 | 30094.84745 | 26746.77113 | 26822.16521 |
| 1120.12 | 22350.5 | 22574.7 | 27792.85992 | 28149.04674 | 1120.12 | 19542.7 | 19628.8 | 30121.25727 | 30246.43552 | 26516.54786 | 26964.17856 |
| 1121.84 | 22335.3 | 22667.2 | 28089.20925 | 28263.99149 | 1121.84 | 19599.2 | 19728.3 | 29714.80372 | 30397.91623 | 26275.47154 | 27106.07645 |
| 1123.57 | 22477.8 | 22759.4 | 28163.23745 | 28378.58147 | 1123.57 | 19436.2 | 19827.7 | 30034.19075 | 30549.28958 | 26626.92572 | 27248.0898 |
| 1125.29 | 22522.4 | 22851.5 | 28138.16719 | 28493.0532 | 1125.29 | 19622 | 19927.1 | 30628.73375 | 30700.55557 | 27084.25488 | 27390.10315 |
| 1127.01 | 22792.4 | 22943.4 | 28251.57462 | 28607.17015 | 1127.01 | 19456.3 | 20026.4 | 30336.0786 | 30851.82157 | 26983.46004 | 27532.00104 |
| 1128.74 | 22802 | 23035.1 | 28415.12255 | 28721.16886 | 1128.74 | 19765.3 | 20125.7 | 30395.01759 | 31002.98021 | 27350.96288 | 27673.89893 |
| 1130.46 | 22531.2 | 23126.6 | 28584.34676 | 28834.81279 | 1130.46 | 20280.8 | 20225 | 30874.79596 | 31154.03149 | 27565.83023 | 27815.79682 |
| 1132.18 | 22769.2 | 23217.9 | 28845.10106 | 28948.33847 | 1132.18 | 19839.5 | 20324.2 | 31094.87779 | 31304.97541 | 27772.61553 | 27957.57925 |
| 1133.91 | 23104.4 | 23309.1 | 28693.02395 | 29061.62764 | 1133.91 | 20090.7 | 20423.3 | 31498.00328 | 31455.81198 | 28127.41798 | 28099.36169 |
| 1135.63 | 23219.2 | 23400 | 28849.94955 | 29174.56203 | 1135.63 | 20501.6 | 20522.3 | 31465.47411 | 31606.64855 | 27794.78346 | 28241.14412 |
| 1137.35 | 22979.2 | 23490.7 | 28949.40277 | 29287.25992 | 1137.35 | 19986.3 | 20621.3 | 30886.81995 | 31757.2704 | 27828.38174 | 28382.92655 |
| 1139.07 | 23202.6 | 23581.3 | 28946.80114 | 29399.83955 | 1139.07 | 20486.2 | 20720.3 | 31222.95467 | 31907.7849 | 28470.21281 | 28524.47807 |
| 1140.8 | 23312.1 | 23671.6 | 29341.7759 | 29512.06442 | 1140.8 | 20422.7 | 20819.2 | 31989.91299 | 32058.19203 | 28472.75288 | 28666.14505 |
| 1142.52 | 23351.3 | 23761.7 | 29341.06636 | 29624.05277 | 1142.52 | 20131.8 | 20918 | 31822.32873 | 32208.49182 | 27779.54301 | 28807.69656 |
| 1144.24 | 23651.8 | 23851.6 | 29311.85715 | 29735.68635 | 1144.24 | 20903.1 | 21016.7 | 32222.98501 | 32358.57688 | 28326.23667 | 28949.13262 |
| 1145.96 | 23858.6 | 23941.3 | 29819.05677 | 29847.20168 | 1145.96 | 20505 | 21115.3 | 32294.37741 | 32508.66195 | 28752.85401 | 29090.56868 |
| 1147.69 | 23717.7 | 24030.8 | 29753.54299 | 29958.36224 | 1147.69 | 20996.8 | 21213.9 | 32646.8304 | 32658.53231 | 29167.34826 | 29231.88928 |
| 1149.41 | 23505.2 | 24120.1 | 29948.547 | 30069.40454 | 1149.41 | 21194.9 | 21312.4 | 32643.8244 | 32808.2953 | 29024.41124 | 29373.20989 |
| 1151.13 | 23852 | 24209.2 | 30385.5026 | 30180.09208 | 1151.13 | 20566.3 | 21410.8 | 32449.61561 | 32957.84359 | 28953.0582 | 29514.41503 |
| 1152.85 | 23864.7 | 24298.1 | 30130.66111 | 30290.42485 | 1152.85 | 20982 | 21509.1 | 33057.9003 | 33107.39187 | 29113.19845 | 29655.50471 |
| 1154.57 | 24352.5 | 24386.7 | 30074.48955 | 30400.63936 | 1154.57 | 21409.3 | 21607.3 | 33101.91666 | 33256.61808 | 29633.56769 | 29796.47894 |
| 1156.3 | 24494.8 | 24475.1 | 30251.99168 | 30510.38085 | 1156.3 | 21262.9 | 21705.5 | 33355.1718 | 33405.8443 | 30096.78522 | 29937.45317 |
| 1158.02 | 24185.6 | 24563.3 | 30564.18729 | 30620.00408 | 1158.02 | 21797.9 | 21803.5 | 33422.91406 | 33554.85579 | 30277.82338 | 30078.31194 |
| 1159.74 | 24652.8 | 24651.3 | 30614.80082 | 30729.39081 | 1159.74 | 21914.2 | 21901.5 | 33206.69708 | 33703.65258 | 29882.3797 | 30219.05525 |
| 1161.46 | 24415.5 | 24739 | 30782.2512 | 30838.42276 | 1161.46 | 21415 | 21999.3 | 33508.47758 | 33852.34201 | 29655.73563 | 30359.6831 |
| 1163.18 | 24526.5 | 24826.5 | 30805.07459 | 30947.09995 | 1163.18 | 21587.4 | 22097.1 | 33784.92182 | 34000.92408 | 30507.81573 | 30500.31096 |
| 1164.9 | 24628.1 | 24913.8 | 30628.16374 | 31055.54062 | 1164.9 | 21839.6 | 22194.7 | 34032.91646 | 34149.18408 | 30341.20982 | 30640.70789 |
| 1166.62 | 24694.2 | 25000.9 | 30852.96824 | 31163.74478 | 1166.62 | 22141.4 | 22292.3 | 34390.84465 | 34297.33672 | 30823.36246 | 30781.10483 |
| 1168.34 | 24846 | 25087.7 | 30873.19 | 31271.59418 | 1168.34 | 21889.4 | 22389.7 | 34247.9525 | 34445.382 | 30912.03421 | 30921.27085 |
| 1170.06 | 24630.4 | 25174.3 | 30563.83252 | 31379.20706 | 1170.06 | 21827.1 | 22487 | 33983.6396 | 34593.21257 | 30317.31001 | 31061.43687 |
| 1171.78 | 24836.6 | 25260.7 | 30921.08364 | 31486.58343 | 1171.78 | 21900.4 | 22584.3 | 34036.45924 | 34740.82843 | 30691.2785 | 31201.37198 |
| 1173.5 | 24821.3 | 25346.8 | 31263.43452 | 31593.60503 | 1173.5 | 21967.2 | 22681.4 | 34469.42997 | 34888.22958 | 30842.87486 | 31341.19163 |
| 1175.22 | 24921.3 | 25432.7 | 31290.51512 | 31700.27187 | 1175.22 | 22252.5 | 22778.3 | 34731.48837 | 35035.52336 | 31120.66683 | 31481.01127 |
| 1176.94 | 25020.5 | 25518.3 | 31547.36697 | 31806.70219 | 1176.94 | 22539 | 22875.2 | 35282.22972 | 35182.49508 | 31852.09331 | 31620.6 |
| 1178.66 | 25076.8 | 25603.7 | 31459.14806 | 31912.77775 | 1178.66 | 22750.5 | 22971.9 | 35224.68637 | 35329.35944 | 31556.17444 | 31760.07328 |
| 1180.38 | 25254.3 | 25688.8 | 31590.88515 | 32018.61679 | 1180.38 | 22736.9 | 23068.6 | 35143.63185 | 35476.00908 | 31826.23072 | 31899.43109 |
| 1182.11 | 25619.6 | 25773.7 | 31912.65949 | 32124.10107 | 1182.11 | 22831 | 23165 | 35528.50665 | 35622.44401 | 32098.94253 | 32038.55799 |
| 1183.82 | 25475.7 | 25858.4 | 31827.51523 | 32229.34883 | 1183.82 | 22997.2 | 23261.4 | 35561.5726 | 35768.66423 | 32241.1868 | 32177.68489 |
| 1185.54 | 25560.8 | 25942.8 | 31909.11181 | 32334.24183 | 1185.54 | 22959.7 | 23357.6 | 35283.518 | 35914.66973 | 32064.6515 | 32316.58087 |
| 1187.26 | 25992.9 | 26027 | 32238.57279 | 32438.78006 | 1187.26 | 23295.5 | 23453.7 | 35349.22048 | 36060.46052 | 31957.85284 | 32455.3614 |
| 1188.98 | 25754.1 | 26110.9 | 32451.43344 | 32543.08177 | 1188.98 | 23227.2 | 23549.7 | 35736.24242 | 36206.03659 | 32178.95493 | 32593.91101 |
| 1190.7 | 25950.1 | 26194.5 | 32657.43524 | 32647.02872 | 1190.7 | 23149.6 | 23645.5 | 36270.12872 | 36351.39795 | 32682.35184 | 32732.34516 |
| 1192.42 | 26316.5 | 26278 | 32640.05162 | 32750.73916 | 1192.42 | 23601.9 | 23741.2 | 36711.47329 | 36496.5446 | 32875.62855 | 32870.66385 |
| 1194.14 | 26177.7 | 26361.1 | 32871.47845 | 32854.09483 | 1194.14 | 23603.7 | 23836.8 | 36695.4771 | 36641.47653 | 32719.52932 | 33008.75163 |
| 1195.86 | 26470.1 | 26444 | 32395.85316 | 32957.09573 | 1195.86 | 23668.2 | 23932.2 | 36491.49882 | 36786.0864 | 32957.48827 | 33146.72395 |
| 1197.58 | 26588.2 | 26526.6 | 32468.81706 | 33059.74187 | 1197.58 | 23908.2 | 24027.4 | 36583.07433 | 36930.5889 | 33147.07032 | 33284.58081 |
| 1199.3 | 26434.4 | 26609 | 32942.31374 | 33162.15149 | 1199.3 | 23735.8 | 24122.5 | 36827.20412 | 37074.76933 | 33511.80217 | 33422.20675 |
| 1201.02 | 26696.2 | 26691.1 | 32790.82792 | 33264.20634 | 1201.02 | 24010.3 | 24217.5 | 36883.02976 | 37218.6277 | 33465.38805 | 33559.60178 |
| 1202.73 | 26486.4 | 26773 | 32962.77202 | 33365.90643 | 1202.73 | 24097.8 | 24312.3 | 37311.70621 | 37362.3787 | 33448.87755 | 33696.88136 |
| 1204.45 | 26567.7 | 26854.5 | 33172.43975 | 33467.37001 | 1204.45 | 24451.8 | 24406.9 | 37580.5281 | 37505.80764 | 33745.83555 | 33833.93001 |
| 1206.17 | 26831.1 | 26935.9 | 33429.52811 | 33568.36056 | 1206.17 | 24751.4 | 24501.4 | 37454.27628 | 37649.02186 | 33881.61417 | 33970.86321 |
| 1207.89 | 26794.5 | 27016.9 | 33639.0776 | 33669.1146 | 1207.89 | 24422 | 24595.8 | 37324.26697 | 37792.02137 | 33779.31837 | 34107.56549 |
| 1209.61 | 26636.7 | 27097.7 | 33362.83178 | 33769.51387 | 1209.61 | 24275.2 | 24689.9 | 37345.84573 | 37934.6988 | 34080.66377 | 34244.15231 |
| 1211.32 | 26712 | 27178.2 | 33340.12664 | 33869.55837 | 1211.32 | 24186.9 | 24784 | 37616.17062 | 38077.16152 | 34025.35939 | 34380.50822 |
| 1213.04 | 27146.7 | 27258.5 | 33734.27361 | 33969.36637 | 1213.04 | 24357.9 | 24877.8 | 38085.75008 | 38219.30218 | 34005.26969 | 34516.63321 |
| 1214.76 | 27529 | 27338.5 | 34129.72139 | 34068.70133 | 1214.76 | 25092.2 | 24971.5 | 38469.87338 | 38361.22811 | 34801.69903 | 34652.52728 |
| 1216.48 | 27328.4 | 27418.2 | 34158.69409 | 34167.79979 | 1216.48 | 25228 | 25065 | 38681.58136 | 38502.83198 | 34823.52059 | 34788.3059 |
| 1218.19 | 27273.7 | 27497.6 | 34074.37762 | 34266.54348 | 1218.19 | 24886.3 | 25158.3 | 38775.51873 | 38644.22113 | 35101.54348 | 34923.73814 |
| 1219.91 | 27332.8 | 27576.8 | 34472.7818 | 34364.9324 | 1219.91 | 25183.4 | 25251.5 | 38200.72942 | 38785.28821 | 34881.48051 | 35059.17039 |
| 1221.63 | 27498.3 | 27655.7 | 34469.23412 | 34462.96656 | 1221.63 | 25721.2 | 25344.5 | 38237.76758 | 38926.14058 | 34564.54828 | 35194.25626 |
| 1223.35 | 27709.9 | 27734.3 | 34374.51113 | 34560.7642 | 1223.35 | 25474.7 | 25437.4 | 39155.56246 | 39066.67088 | 35427.94326 | 35329.11121 |
| 1225.06 | 27806.9 | 27812.6 | 34879.46389 | 34658.08882 | 1225.06 | 25296.5 | 25530 | 39127.11286 | 39206.98646 | 35864.60545 | 35463.8507 |
| 1226.78 | 27722.7 | 27890.7 | 34883.60285 | 34755.17692 | 1226.78 | 25470 | 25622.5 | 39602.59696 | 39346.87261 | 35987.68368 | 35598.35928 |
| 1228.5 | 27881.4 | 27968.5 | 34873.90586 | 34851.792 | 1228.5 | 25718.6 | 25714.8 | 39870.02321 | 39486.65141 | 35886.88884 | 35732.52149 |
| 1230.21 | 27989 | 28046 | 35045.96822 | 34948.17057 | 1230.21 | 25832.9 | 25806.9 | 39244.99083 | 39626.00078 | 35616.71711 | 35866.56823 |
| 1231.93 | 27770.2 | 28123.2 | 34949.94441 | 35044.07612 | 1231.93 | 25684.6 | 25898.8 | 39480.63942 | 39765.13544 | 35667.0568 | 36000.38407 |
| 1233.64 | 27741.2 | 28200.1 | 34779.77415 | 35139.74516 | 1233.64 | 25831.5 | 25990.6 | 39663.79044 | 39903.94802 | 35959.74285 | 36133.96898 |
| 1235.36 | 27890.8 | 28276.8 | 35184.32764 | 35235.05942 | 1235.36 | 26052.3 | 26082.1 | 39758.80137 | 40042.43853 | 36153.59684 | 36267.32298 |
| 1237.08 | 28206.5 | 28353.2 | 35291.34924 | 35329.90067 | 1237.08 | 26377.5 | 26173.5 | 39642.10433 | 40180.60698 | 36227.48997 | 36400.44606 |
| 1238.79 | 28156.4 | 28429.2 | 34914.11287 | 35424.5054 | 1238.79 | 25996.6 | 26264.7 | 39825.68478 | 40318.5607 | 36327.13023 | 36533.22277 |
| 1240.51 | 28208.9 | 28505 | 35082.03627 | 35518.75536 | 1240.51 | 25846.9 | 26355.7 | 39952.68809 | 40456.19236 | 36504.12735 | 36665.88402 |
| 1242.22 | 28398.1 | 28580.5 | 35400.73596 | 35612.65056 | 1242.22 | 26317.5 | 26446.5 | 40393.81795 | 40593.39459 | 36853.15691 | 36798.31436 |
| 1243.94 | 28406 | 28655.8 | 35308.02332 | 35706.19099 | 1243.94 | 26206 | 26537 | 41065.33591 | 40730.38211 | 37155.0796 | 36930.39832 |
| 1245.66 | 28554.5 | 28730.7 | 35628.26034 | 35799.2584 | 1245.66 | 26521.3 | 26627.4 | 41029.47867 | 40867.04755 | 37338.54237 | 37062.36682 |
| 1247.37 | 28528.2 | 28805.3 | 36147.52207 | 35892.08929 | 1247.37 | 26606.6 | 26717.6 | 40825.60776 | 41003.49828 | 37484.25038 | 37193.98895 |
| 1249.09 | 28853 | 28879.7 | 35979.12564 | 35984.56541 | 1249.09 | 26599.6 | 26807.6 | 41383.21995 | 41139.51959 | 37400.19695 | 37325.38016 |
| 1250.8 | 29090.3 | 28953.8 | 35993.31635 | 36076.56852 | 1250.8 | 27170.5 | 26897.4 | 41331.15181 | 41275.21882 | 37337.61871 | 37456.54046 |
| 1252.52 | 29030 | 29027.5 | 36286.70928 | 36168.33511 | 1252.52 | 26952.7 | 26987 | 41139.84166 | 41410.59598 | 37324.22558 | 37587.46984 |
| 1254.23 | 28992 | 29101 | 36229.11865 | 36259.62867 | 1254.23 | 27052.4 | 27076.4 | 41428.73931 | 41545.65107 | 37202.6483 | 37718.05284 |
| 1255.95 | 29269.3 | 29174.2 | 36549.11915 | 36350.68573 | 1255.95 | 27317.8 | 27165.6 | 41734.81408 | 41680.38409 | 37985.10722 | 37848.40493 |
| 1257.66 | 29338 | 29247.1 | 36719.0529 | 36441.26976 | 1257.66 | 27385.1 | 27254.6 | 41985.70736 | 41814.79504 | 38133.81713 | 37978.52611 |
| 1259.38 | 29212.3 | 29319.7 | 36698.35812 | 36531.61728 | 1259.38 | 27734.6 | 27343.3 | 41595.89414 | 41948.88391 | 37963.86294 | 38108.41637 |
| 1261.09 | 29251.3 | 29392 | 36785.51273 | 36621.49177 | 1261.09 | 27370.8 | 27431.9 | 41871.37217 | 42082.65072 | 38153.79137 | 38237.96025 |
| 1262.81 | 29295.7 | 29464 | 36356.59852 | 36711.0115 | 1262.81 | 27181.4 | 27520.2 | 42074.49159 | 42216.09545 | 38009.46886 | 38367.27322 |
| 1264.52 | 29468.8 | 29535.7 | 36748.14386 | 36800.17646 | 1264.52 | 27417.1 | 27608.3 | 42157.5859 | 42349.21812 | 38249.85243 | 38496.23982 |
| 1266.23 | 29468.5 | 29607.2 | 36996.48128 | 36888.98665 | 1266.23 | 27591.6 | 27696.2 | 42405.79525 | 42481.91135 | 38492.77608 | 38625.09095 |
| 1267.95 | 29519.9 | 29678.3 | 36667.37507 | 36977.44208 | 1267.95 | 27732.6 | 27783.9 | 42717.98937 | 42614.28252 | 38734.31423 | 38753.48026 |
| 1269.66 | 29694.8 | 29749.1 | 37049.22342 | 37065.42448 | 1269.66 | 27840.7 | 27871.4 | 43103.40095 | 42746.33161 | 38817.09762 | 38881.7541 |
| 1271.37 | 30028.5 | 29819.6 | 37162.63084 | 37153.17037 | 1271.37 | 27979.1 | 27958.7 | 42914.56001 | 42878.05864 | 38918.81612 | 39009.68158 |
| 1273.09 | 30287 | 29889.8 | 37301.22677 | 37240.44323 | 1273.09 | 28587.6 | 28045.7 | 43429.87355 | 43009.46359 | 39586.50978 | 39137.26267 |
| 1274.8 | 30134 | 29959.8 | 37511.13102 | 37327.47959 | 1274.8 | 28636.1 | 28132.5 | 43608.19351 | 43140.43911 | 39303.5222 | 39264.61286 |
| 1276.51 | 30284.2 | 30029.4 | 37715.35899 | 37414.04292 | 1276.51 | 28343.6 | 28219.1 | 43394.66046 | 43271.09257 | 39504.18822 | 39391.73212 |
| 1278.23 | 30413.5 | 30098.7 | 37806.88907 | 37500.25148 | 1278.23 | 28422.6 | 28305.5 | 43834.28732 | 43401.42395 | 39914.17961 | 39518.50502 |
| 1279.94 | 30312.9 | 30167.7 | 37797.07382 | 37586.10527 | 1279.94 | 28453.7 | 28391.6 | 43885.81868 | 43531.32591 | 39553.83517 | 39644.93153 |
| 1281.65 | 30525.5 | 30236.4 | 37900.90252 | 37671.48604 | 1281.65 | 28894.9 | 28477.5 | 44161.51142 | 43660.90579 | 40456.60133 | 39771.12714 |
| 1283.37 | 30787.2 | 30304.8 | 38005.44075 | 37756.6303 | 1283.37 | 28965.5 | 28563.2 | 44598.56172 | 43790.1636 | 40333.86946 | 39896.97636 |
| 1285.08 | 30637.7 | 30372.9 | 38274.11819 | 37841.30154 | 1285.08 | 28425.6 | 28648.7 | 44440.1028 | 43918.99199 | 39648.39527 | 40022.59468 |
| 1286.79 | 30564.5 | 30440.8 | 38198.67091 | 37925.618 | 1286.79 | 28949.6 | 28733.9 | 44121.78934 | 44047.60566 | 40404.87614 | 40147.86661 |
| 1288.5 | 30849.3 | 30508.2 | 38047.77636 | 38009.5797 | 1288.5 | 28865.6 | 28818.8 | 44405.96328 | 44175.68254 | 40571.02021 | 40272.90764 |
| 1290.22 | 30693.2 | 30575.5 | 38218.30139 | 38093.18664 | 1290.22 | 28807.7 | 28903.6 | 44291.8428 | 44303.43736 | 40465.95343 | 40397.60229 |
| 1291.93 | 30784.9 | 30642.3 | 38120.26724 | 38176.32054 | 1291.93 | 29063.7 | 28988.1 | 44188.35067 | 44430.8701 | 40638.44769 | 40521.95056 |
| 1293.64 | 30910.8 | 30708.9 | 38164.96798 | 38259.21794 | 1293.64 | 29058.8 | 29072.4 | 44529.42382 | 44557.87342 | 40570.44292 | 40646.06792 |
| 1295.35 | 30805.5 | 30775.2 | 38300.84402 | 38341.64231 | 1295.35 | 29374.1 | 29156.5 | 44508.81127 | 44684.55466 | 40791.08318 | 40769.8389 |
| 1297.07 | 31023.8 | 30841.2 | 38425.3675 | 38423.71192 | 1297.07 | 29514.5 | 29240.3 | 44842.36943 | 44810.91384 | 40977.66332 | 40893.26351 |
| 1298.78 | 30880.2 | 30906.8 | 38575.07949 | 38505.42676 | 1298.78 | 29459 | 29323.8 | 44977.53188 | 44936.84358 | 41096.58506 | 41016.45721 |
| 1300.49 | 30985.5 | 30972.2 | 38602.04184 | 38586.66857 | 1300.49 | 29545.8 | 29407.2 | 45517.32281 | 45062.3439 | 41507.38465 | 41139.30453 |
| 1302.2 | 31314.7 | 31037.3 | 38946.28481 | 38667.55562 | 1302.2 | 30132.7 | 29490.2 | 45877.39815 | 45187.52215 | 41873.502 | 41261.80547 |
| 1303.91 | 31395.3 | 31102 | 38992.28636 | 38748.20615 | 1303.91 | 30409.9 | 29573.1 | 45896.18562 | 45312.27097 | 41854.22051 | 41383.96004 |
| 1305.62 | 31432.6 | 31166.5 | 38989.68473 | 38828.38366 | 1305.62 | 30224.4 | 29655.7 | 46036.17913 | 45436.69772 | 41742.68807 | 41505.8837 |
| 1307.33 | 31608.7 | 31230.6 | 39230.21726 | 38908.08815 | 1307.33 | 30184.2 | 29738 | 46057.00639 | 45560.69504 | 41899.94188 | 41627.46098 |
| 1309.04 | 31888.5 | 31294.4 | 39322.81164 | 38987.55612 | 1309.04 | 30307.7 | 29820.1 | 46141.92576 | 45684.37028 | 42009.51153 | 41748.69189 |
| 1310.76 | 31915.6 | 31357.9 | 39334.51898 | 39066.55107 | 1310.76 | 30298.7 | 29902 | 46342.14655 | 45807.61611 | 42095.98958 | 41869.57642 |
| 1312.47 | 31872.1 | 31421.1 | 39386.43332 | 39145.19126 | 1312.47 | 30444.3 | 29983.6 | 46397.97218 | 45930.4325 | 42241.00483 | 41990.23004 |
| 1314.18 | 32041.2 | 31484 | 39650.9718 | 39223.47667 | 1314.18 | 30327.9 | 30065 | 46755.04152 | 46052.92682 | 42552.51054 | 42110.42183 |
| 1315.89 | 32104.1 | 31546.6 | 39584.39372 | 39301.28906 | 1315.89 | 30290.6 | 30146.1 | 46958.69772 | 46174.99172 | 42821.75861 | 42230.3827 |
| 1317.6 | 32043.1 | 31608.9 | 39700.99406 | 39378.74669 | 1317.6 | 30508.6 | 30227 | 46927.13477 | 46296.73454 | 43096.08684 | 42349.99719 |
| 1319.31 | 31746.9 | 31670.9 | 39780.58029 | 39455.84954 | 1319.31 | 30813 | 30307.6 | 46885.05083 | 46418.04794 | 42784.35022 | 42469.26532 |
| 1321.02 | 31942.9 | 31732.5 | 39689.16847 | 39532.59763 | 1321.02 | 30716.7 | 30387.9 | 46949.89445 | 46538.9319 | 42958.22998 | 42588.18706 |
| 1322.73 | 32083.4 | 31793.9 | 39767.69039 | 39608.99095 | 1322.73 | 30545.3 | 30468 | 47160.31415 | 46659.38644 | 43007.29963 | 42706.76244 |
| 1324.44 | 31885.1 | 31854.9 | 39821.37858 | 39684.91125 | 1324.44 | 30605.3 | 30547.9 | 46803.24481 | 46779.51891 | 42596.6155 | 42824.99144 |
| 1326.15 | 32057.6 | 31915.6 | 39954.65299 | 39760.47678 | 1326.15 | 30672.5 | 30627.5 | 46657.77609 | 46899.22195 | 42831.45709 | 42942.98952 |
| 1327.86 | 32016.6 | 31976.1 | 40097.74265 | 39835.68754 | 1327.86 | 30763.8 | 30706.8 | 46942.70153 | 47018.60292 | 43107.17081 | 43060.52577 |
| 1329.57 | 32248.5 | 32036.2 | 40207.72065 | 39910.54354 | 1329.57 | 31052.9 | 30785.9 | 47478.84233 | 47137.44711 | 43224.70706 | 43177.83111 |

| | | | | | | | | | | | | |
|---|---|---|---|---|---|---|---|---|---|---|---|---|
| 1331.28 | 32430.8 | 32096 | 40525.94732 | 39984.92651 | 1331.28 | 31348.6 | 30864.7 | 47737.46531 | 47255.96922 | 43703.97325 | 43294.67461 |
| 1332.98 | 32386.3 | 32155.5 | 40376.82661 | 40058.95471 | 1332.98 | 30992 | 30943.3 | 47396.82159 | 47374.06191 | 43589.78528 | 43411.2872 |
| 1334.69 | 32602.7 | 32214.6 | 40310.72155 | 40132.62815 | 1334.69 | 31313.1 | 31021.6 | 47629.67889 | 47491.83252 | 43326.19465 | 43527.43796 |
| 1336.4 | 32692.1 | 32273.5 | 40667.61791 | 40205.82856 | 1336.4 | 31364 | 31099.7 | 47552.38186 | 47609.06635 | 43511.27383 | 43643.3578 |
| 1338.11 | 32597.9 | 32332 | 40631.4316 | 40278.79246 | 1338.11 | 31376.3 | 31177.4 | 47807.78414 | 47725.97811 | 43955.44079 | 43758.8158 |
| 1339.82 | 32403.2 | 32390.3 | 40745.4303 | 40351.28333 | 1339.82 | 31576.5 | 31255 | 48460.94405 | 47842.46044 | 44298.23561 | 43874.04289 |
| 1341.53 | 32694.9 | 32448.2 | 40787.41115 | 40423.30119 | 1341.53 | 31330 | 31332.2 | 48289.49494 | 47958.51335 | 43902.67648 | 43988.80815 |
| 1343.24 | 32856.5 | 32505.8 | 40883.3167 | 40495.08253 | 1343.24 | 31433.5 | 31409.2 | 48504.7457 | 48074.24418 | 44359.42836 | 44103.3425 |
| 1344.95 | 32739.6 | 32563.1 | 41060.93708 | 40566.39084 | 1344.95 | 31671.9 | 31486 | 48535.87922 | 48189.43823 | 44668.16307 | 44217.41501 |
| 1346.65 | 32726.4 | 32620.1 | 40826.4356 | 40637.34439 | 1346.65 | 31262 | 31562.4 | 48464.59418 | 48304.3102 | 44539.65831 | 44331.14115 |
| 1348.36 | 32858.2 | 32676.8 | 41043.19869 | 40707.94317 | 1348.36 | 31308.5 | 31638.6 | 48391.26936 | 48418.75275 | 44472.46175 | 44444.52091 |
| 1350.07 | 33007.6 | 32733.2 | 41124.67702 | 40778.18719 | 1350.07 | 31617 | 31714.5 | 48232.27366 | 48532.76587 | 44303.7776 | 44557.66976 |
| 1351.78 | 33153.9 | 32789.3 | 41233.70897 | 40847.95818 | 1351.78 | 31996.2 | 31790.2 | 48664.27818 | 48646.34956 | 44641.9541 | 44670.35677 |
| 1353.49 | 33298.9 | 32845 | 41143.95273 | 40917.3744 | 1353.49 | 31876.5 | 31865.6 | 48583.54573 | 48759.50383 | 44486.54762 | 44782.58196 |
| 1355.19 | 33271.1 | 32900.5 | 41068.50546 | 40986.43585 | 1355.19 | 31665.5 | 31940.7 | 48823.38123 | 48872.33602 | 44693.33292 | 44894.57622 |
| 1356.9 | 33325.3 | 32955.6 | 41428.83123 | 41055.02428 | 1356.9 | 32177.8 | 32015.6 | 49079.85708 | 48984.63143 | 45292.90635 | 45006.22412 |
| 1358.61 | 33326.8 | 33010.4 | 41578.89799 | 41123.25795 | 1358.61 | 31863 | 32090.2 | 48604.80241 | 49096.60477 | 44994.4474 | 45117.41018 |
| 1360.32 | 33207.2 | 33064.9 | 41392.1719 | 41191.13684 | 1360.32 | 32494.5 | 32164.5 | 49020.48866 | 49208.04132 | 45007.60961 | 45228.36532 |
| 1362.02 | 33436.9 | 33119.1 | 41717.84869 | 41258.66097 | 1362.02 | 32595.2 | 32238.6 | 49538.16406 | 49319.1558 | 45571.16015 | 45338.85864 |
| 1363.73 | 33597.2 | 33173 | 41887.66418 | 41325.83033 | 1363.73 | 32453.7 | 32312.3 | 49325.81194 | 49429.84086 | 45460.89775 | 45449.00558 |
| 1365.44 | 33413 | 33226.6 | 41554.89204 | 41392.52667 | 1365.44 | 32613 | 32385.9 | 49663.02023 | 49540.09648 | 45189.68689 | 45558.80614 |
| 1367.14 | 33683.7 | 33279.8 | 41733.34021 | 41458.86823 | 1367.14 | 32365.8 | 32459.1 | 50024.59857 | 49649.92268 | 45523.82237 | 45668.14487 |
| 1368.85 | 33877.1 | 33332.8 | 41857.39067 | 41524.85503 | 1368.85 | 32554.8 | 32532 | 49821.69386 | 49759.31945 | 45848.02845 | 45777.25269 |
| 1370.56 | 33737.1 | 33385.4 | 41897.12466 | 41590.36881 | 1370.56 | 32478.8 | 32604.7 | 49761.14452 | 49868.28679 | 45573.23839 | 45885.89868 |
| 1372.26 | 33841.7 | 33437.8 | 42073.79899 | 41655.52782 | 1372.26 | 32552.4 | 32677.2 | 49746.32925 | 49976.82471 | 45763.28227 | 45994.19829 |
| 1373.97 | 33996.3 | 33489.8 | 42343.8955 | 41720.33206 | 1373.97 | 32608.8 | 32749.3 | 50185.84876 | 50084.93319 | 46356.04368 | 46102.15153 |
| 1375.67 | 33922.5 | 33541.5 | 42561.7229 | 41784.78154 | 1375.67 | 32772.5 | 32821.2 | 50485.2674 | 50192.61225 | 46452.68204 | 46209.75839 |
| 1377.38 | 33833.4 | 33592.9 | 42470.19282 | 41848.87624 | 1377.38 | 33037.5 | 32892.7 | 50359.55237 | 50299.86188 | 46587.30607 | 46316.90342 |
| 1379.09 | 34031.4 | 33644 | 42269.15776 | 41912.49792 | 1379.09 | 33128.2 | 32964.1 | 50639.86146 | 50406.68208 | 46669.51217 | 46423.70208 |
| 1380.79 | 34206.8 | 33694.8 | 42701.50139 | 41975.76484 | 1380.79 | 33470 | 33035.1 | 50687.63532 | 50513.18021 | 46558.0952 | 46530.15436 |
| 1382.5 | 34053.4 | 33745.3 | 42665.19682 | 42038.67699 | 1382.5 | 33437.6 | 33105.8 | 50738.20046 | 50619.14156 | 46679.32611 | 46636.26027 |
| 1384.2 | 34352.6 | 33795.5 | 42801.90066 | 42101.11611 | 1384.2 | 33255.4 | 33176.3 | 50921.35148 | 50724.67348 | 46805.17533 | 46741.90435 |
| 1385.91 | 34618.3 | 33845.3 | 43185.04983 | 42163.31872 | 1385.91 | 33802.1 | 33246.5 | 51541.98222 | 50829.77597 | 47399.89953 | 46847.31751 |
| 1387.61 | 34482.5 | 33894.9 | 42926.06937 | 42225.04831 | 1387.61 | 33853.5 | 33316.4 | 51603.06835 | 50934.44903 | 47519.39857 | 46952.15338 |
| 1389.32 | 34707.3 | 33944.1 | 42818.92951 | 42286.42313 | 1389.32 | 33548.6 | 33386.1 | 51096.12866 | 51038.69267 | 47395.97396 | 47056.75833 |
| 1391.02 | 34642.4 | 33993.1 | 42936.94892 | 42347.44318 | 1391.02 | 33772.2 | 33455.4 | 51321.68569 | 51142.50687 | 47604.83749 | 47160.90146 |
| 1392.73 | 34564.4 | 34041.7 | 43045.98087 | 42407.99021 | 1392.73 | 33523.4 | 33524.5 | 51301.71729 | 51245.89165 | 47243.91576 | 47264.6982 |
| 1394.43 | 34694.2 | 34090 | 43033.32749 | 42468.18247 | 1394.43 | 33511.8 | 33593.3 | 51047.71065 | 51348.847 | 47320.00259 | 47368.14858 |
| 1396.14 | 34811.4 | 34138 | 43155.24934 | 42528.01996 | 1396.14 | 34101.3 | 33661.9 | 51820.89568 | 51451.37293 | 48020.02449 | 47471.13712 |
| 1397.84 | 34690.4 | 34185.7 | 43289.35154 | 42587.50269 | 1397.84 | 34115.7 | 33730.1 | 51942.6385 | 51553.46942 | 47516.97395 | 47573.77929 |
| 1399.55 | 34665.2 | 34233.2 | 43099.78731 | 42646.63064 | 1399.55 | 33961.5 | 33798.1 | 51707.84877 | 51655.13649 | 47662.68196 | 47676.07509 |
| 1401.25 | 34713.6 | 34280.2 | 42915.78111 | 42705.28558 | 1401.25 | 33980.7 | 33865.7 | 51864.16054 | 51756.37413 | 48081.44815 | 47777.90905 |
| 1402.95 | 34743.5 | 34327 | 43186.23238 | 42763.704 | 1402.95 | 34091.8 | 33933.1 | 51898.40742 | 51857.18234 | 48052.46819 | 47879.39664 |
| 1404.66 | 34617.3 | 34373.5 | 43178.90052 | 42821.6494 | 1404.66 | 34147.2 | 34000.2 | 52028.20202 | 51957.45376 | 48637.03208 | 47980.53785 |
| 1406.36 | 34717.9 | 34419.7 | 43323.5275 | 42879.24003 | 1406.36 | 33953.3 | 34067.1 | 52257.08712 | 52057.40312 | 48794.17043 | 48081.21723 |
| 1408.07 | 34775.6 | 34465.6 | 43376.15138 | 42936.35764 | 1408.07 | 34142.6 | 34133.6 | 52396.11441 | 52156.92305 | 48468.69431 | 48181.55024 |
| 1409.77 | 34865.5 | 34511.1 | 43383.48325 | 42993.23873 | 1409.77 | 34393.8 | 34199.9 | 52342.22121 | 52255.90619 | 48507.25728 | 48281.53688 |
| 1411.47 | 35063.1 | 34556.4 | 43969.08654 | 43049.6468 | 1411.47 | 34483.1 | 34265.9 | 52522.90302 | 52354.4599 | 48735.17139 | 48381.06168 |
| 1413.17 | 35004.3 | 34601.4 | 43810.26885 | 43105.70011 | 1413.17 | 34835.3 | 34331.6 | 52520.21909 | 52452.69155 | 48316.28974 | 48480.24011 |
| 1414.88 | 35100.5 | 34646 | 43481.39915 | 43161.39864 | 1414.88 | 34550.1 | 34397 | 52505.94061 | 52550.3864 | 48535.65995 | 48579.07216 |
| 1416.58 | 34994.7 | 34690.4 | 43534.49605 | 43216.74241 | 1416.58 | 34689 | 34462.1 | 52814.48459 | 52647.65183 | 48773.50345 | 48677.44239 |
| 1418.28 | 35302.2 | 34734.4 | 43500.43835 | 43271.61316 | 1418.28 | 34836.7 | 34527 | 53106.06617 | 52744.48784 | 48788.16661 | 48775.46623 |
| 1419.99 | 35234.4 | 34778.2 | 43826.11514 | 43326.24739 | 1419.99 | 34717.1 | 34591.5 | 53399.04339 | 52840.89441 | 49345.59788 | 48873.02825 |
| 1421.69 | 35219.3 | 34821.6 | 44135.94564 | 43380.4086 | 1421.69 | 35190.5 | 34655.8 | 53688.15576 | 52936.87156 | 49564.3908 | 48970.24389 |
| 1423.39 | 35297.4 | 34864.8 | 44177.33521 | 43434.21504 | 1423.39 | 35259.7 | 34719.8 | 53655.8413 | 53032.41928 | 49463.36504 | 49067.11316 |
| 1425.09 | 35092 | 34907.6 | 43733.87553 | 43487.66671 | 1425.09 | 35202.3 | 34783.5 | 53396.14475 | 53127.53757 | 49373.76963 | 49163.5206 |
| 1426.8 | 35246.2 | 34950.2 | 43642.70022 | 43540.76362 | 1426.8 | 35251.8 | 34846.9 | 53282.88313 | 53222.11907 | 49235.56639 | 49259.58166 |
| 1428.5 | 35367.2 | 34992.4 | 43899.55206 | 43593.50576 | 1428.5 | 35382 | 34910.1 | 53921.87192 | 53316.37851 | 49534.37172 | 49355.29635 |
| 1430.2 | 35349.9 | 35034.4 | 43931.36291 | 43645.77487 | 1430.2 | 35507.9 | 34972.9 | 53586.38133 | 53410.10116 | 49511.51103 | 49450.5492 |
| 1431.9 | 35220.8 | 35076 | 43974.99934 | 43697.80747 | 1431.9 | 35350 | 35035.5 | 53535.92355 | 53503.39438 | 49572.3574 | 49545.34023 |
| 1433.6 | 35373.1 | 35117.4 | 44236.93619 | 43749.36705 | 1433.6 | 35395.4 | 35097.8 | 54274.21755 | 53596.25817 | 50233.12358 | 49639.90034 |
| 1435.3 | 35648.6 | 35158.4 | 44266.38191 | 43800.57186 | 1435.3 | 35787 | 35159.8 | 53895.67681 | 53688.7999 | 50088.80107 | 49733.88316 |
| 1437.01 | 35638.8 | 35199.2 | 43900.02509 | 43851.42191 | 1437.01 | 35710.6 | 35221.4 | 54140.02131 | 53780.69748 | 50218.57587 | 49827.63506 |
| 1438.71 | 35616.5 | 35239.6 | 44000.06959 | 43901.91718 | 1438.71 | 35390.1 | 35282.9 | 54169.43713 | 53872.27299 | 50195.25335 | 49920.92513 |
| 1440.41 | 35587.8 | 35279.8 | 44271.11215 | 43952.05769 | 1440.41 | 35616 | 35344 | 53968.89428 | 53963.41907 | 50130.2505 | 50013.75337 |
| 1442.11 | 35632.3 | 35319.6 | 44306.47067 | 44001.84343 | 1442.11 | 35649.9 | 35404.8 | 54564.83291 | 54054.13572 | 50256.67701 | 50106.35069 |
| 1443.81 | 35795.2 | 35359.2 | 44304.34206 | 44051.15615 | 1443.81 | 35811.8 | 35465.4 | 54490.00509 | 54144.31559 | 50391.30105 | 50198.37072 |
| 1445.51 | 35668.7 | 35398.5 | 44325.1551 | 44100.23235 | 1445.51 | 36004.4 | 35525.7 | 54448.88737 | 54234.17339 | 50770.00332 | 50290.15984 |
| 1447.21 | 35707.5 | 35437.4 | 44548.18576 | 44148.83553 | 1447.21 | 35960.8 | 35585.7 | 54888.62159 | 54323.49441 | 50909.13021 | 50381.37166 |
| 1448.91 | 35674.2 | 35476.1 | 44539.43482 | 44197.2022 | 1448.91 | 36034.3 | 35645.4 | 54809.49949 | 54412.38599 | 51158.2886 | 50472.35257 |
| 1450.61 | 35654.1 | 35514.5 | 44406.04215 | 44245.09585 | 1450.61 | 35959.4 | 35704.8 | 54598.65036 | 54500.84815 | 50939.03384 | 50562.87165 |
| 1452.31 | 35793.7 | 35552.6 | 44393.034 | 44292.63473 | 1452.31 | 35753.7 | 35763.9 | 54798.11965 | 54588.88088 | 50493.82776 | 50652.9289 |
| 1454.01 | 35565.5 | 35590.4 | 44202.76023 | 44339.81884 | 1454.01 | 35992.9 | 35822.8 | 55012.18948 | 54676.48418 | 50940.53479 | 50742.63977 |
| 1455.71 | 35623.3 | 35627.9 | 44551.14216 | 44386.64818 | 1455.71 | 35960.7 | 35881.3 | 54908.58999 | 54763.65805 | 51235.41454 | 50832.00427 |
| 1457.41 | 35919.5 | 35665.1 | 44611.09791 | 44433.12275 | 1457.41 | 36111.2 | 35939.6 | 54803.16542 | 54850.29514 | 50588.50333 | 50920.90693 |
| 1459.11 | 35796.7 | 35702 | 44604.59383 | 44479.24256 | 1459.11 | 36527.1 | 35997.5 | 54808.53327 | 54936.61016 | 50819.65026 | 51009.46322 |
| 1460.81 | 35874.3 | 35738.6 | 44576.09415 | 44525.0076 | 1460.81 | 36017.6 | 36055.2 | 55389.11986 | 55022.38839 | 51445.54812 | 51097.55768 |
| 1462.51 | 35981.5 | 35775 | 44351.1714 | 44570.41787 | 1462.51 | 36113.3 | 36112.6 | 55325.8866 | 55107.7372 | 51172.8363 | 51185.30577 |
| 1464.21 | 35861 | 35811 | 44839.21357 | 44615.47337 | 1464.21 | 36493.7 | 36169.8 | 55130.38952 | 55192.76393 | 50886.6159 | 51272.59202 |
| 1465.91 | 35878.2 | 35846.8 | 44648.82154 | 44660.17411 | 1465.91 | 36159.6 | 36226.6 | 55622.62131 | 55277.25388 | 51053.68364 | 51359.5319 |
| 1467.61 | 36014.3 | 35882.2 | 44374.82259 | 44704.52008 | 1467.61 | 36320.3 | 36283.1 | 55729.87094 | 55361.3144 | 51722.30096 | 51446.12541 |
| 1469.31 | 35874.8 | 35917.4 | 44529.50132 | 44748.39302 | 1469.31 | 36579.7 | 36339.4 | 55385.0403 | 55444.83814 | 51408.25518 | 51532.25708 |
| 1471.01 | 35979.7 | 35952.3 | 44519.56783 | 44792.02946 | 1471.01 | 36008.3 | 36395.3 | 55351.65227 | 55528.0398 | 51123.99757 | 51617.92692 |
| 1472.7 | 36168.2 | 35986.9 | 44632.14746 | 44835.31112 | 1472.7 | 36388.6 | 36451 | 55473.18038 | 55610.81204 | 51395.20843 | 51703.36585 |
| 1474.4 | 36259.8 | 36021.2 | 44976.86345 | 44878.23802 | 1474.4 | 36722.7 | 36506.4 | 55766.05024 | 55693.15485 | 51862.35153 | 51788.22748 |
| 1476.1 | 36335.4 | 36055.3 | 45179.67234 | 44920.69189 | 1476.1 | 36758.6 | 36561.5 | 55804.80612 | 55774.96087 | 51894.21794 | 51872.74275 |
| 1477.8 | 36364 | 36089 | 45058.22352 | 44962.90925 | 1477.8 | 37176.3 | 36616.4 | 55935.03014 | 55856.33747 | 51828.40687 | 51956.91163 |
| 1479.5 | 36295 | 36122.5 | 45150.8179 | 45004.77185 | 1479.5 | 37013 | 36670.9 | 56327.41993 | 55937.392 | 52371.6368 | 52040.61869 |
| 1481.19 | 36045.1 | 36155.7 | 45202.14097 | 45046.27967 | 1481.19 | 36908.6 | 36725.2 | 56046.89612 | 56017.90974 | 52320.4889 | 52123.97937 |
| 1482.89 | 36184.2 | 36188.5 | 45031.14292 | 45087.31448 | 1482.89 | 36677.6 | 36779.1 | 56117.96645 | 56097.99805 | 52450.72553 | 52206.99368 |
| 1484.59 | 36387.9 | 36221.2 | 45126.57544 | 45128.11277 | 1484.59 | 36830.3 | 36832.8 | 56552.1181 | 56177.65693 | 52713.73887 | 52289.54615 |
| 1486.29 | 36429.5 | 36253.5 | 45073.36028 | 45168.55629 | 1486.29 | 36698.6 | 36886.2 | 56195.58555 | 56256.88639 | 52220.27135 | 52371.6368 |
| 1487.98 | 36588.6 | 36285.5 | 45503.33878 | 45208.64504 | 1487.98 | 37142.1 | 36939.3 | 56553.6211 | 56335.68642 | 52354.54901 | 52453.38107 |
| 1489.68 | 36661.5 | 36317.3 | 45792.59275 | 45248.37903 | 1489.68 | 37314.8 | 36992.1 | 56877.8392 | 56413.94966 | 52930.91539 | 52534.77896 |
| 1491.38 | 36444.6 | 36348.8 | 45427.77325 | 45287.75825 | 1491.38 | 37125.2 | 37044.7 | 56728.50563 | 56491.89083 | 52806.10528 | 52615.71503 |
| 1493.08 | 36321.9 | 36380 | 45410.27138 | 45326.7827 | 1493.08 | 37281.1 | 37096.9 | 56893.19125 | 56569.29522 | 53126.15488 | 52696.18926 |
| 1494.77 | 36441.2 | 36410.9 | 45507.95077 | 45365.57064 | 1494.77 | 37326.2 | 37148.9 | 56739.45605 | 56646.37754 | 53021.66538 | 52776.31711 |
| 1496.47 | 36478.4 | 36441.5 | 45161.10617 | 45403.88556 | 1496.47 | 37339.4 | 37200.6 | 56284.79921 | 56722.92307 | 52346.12058 | 52856.0986 |
| 1498.17 | 36246.7 | 36471.9 | 45348.7783 | 45441.96396 | 1498.17 | 37308.2 | 37251.9 | 56732.69256 | 56799.14653 | 52851.01844 | 52935.41825 |
| 1499.86 | 36418.9 | 36502 | 45444.92036 | 45479.56935 | 1499.86 | 37563.1 | 37303.1 | 57255.19901 | 56874.83321 | 53385.3581 | 53014.39153 |
| 1501.56 | 36500.4 | 36531.8 | 45141.23917 | 45516.93821 | 1501.56 | 37427 | 37353.9 | 56926.04249 | 56950.09046 | 53400.71402 | 53093.01843 |
| 1503.25 | 36599.6 | 36561.3 | 45389.22183 | 45553.83406 | 1503.25 | 37213.2 | 37404.4 | 57037.26433 | 57024.91828 | 53540.9955 | 53171.06804 |
| 1504.95 | 36858.9 | 36590.6 | 45571.92722 | 45590.49339 | 1504.95 | 37492.8 | 37454.7 | 57372.75491 | 57099.31667 | 53264.01174 | 53248.88674 |
| 1506.65 | 36361.6 | 36619.6 | 45303.24978 | 45626.79796 | 1506.65 | 37625.7 | 37504.7 | 57224.70963 | 57173.39299 | 53073.15965 | 53326.2436 |
| 1508.34 | 36333.4 | 36648.3 | 45395.25288 | 45662.74776 | 1508.34 | 37474.8 | 37554.3 | 57169.63549 | 57246.93253 | 53424.72928 | 53403.13864 |
| 1510.04 | 36536.3 | 36676.7 | 45591.91246 | 45698.34279 | 1510.04 | 37622.9 | 37603.8 | 57224.70963 | 57320.04263 | 53508.5518 | 53479.80275 |
| 1511.73 | 36393.6 | 36704.9 | 45355.40063 | 45733.70131 | 1511.73 | 37662.4 | 37652.9 | 57075.37606 | 57392.72331 | 53155.59667 | 53555.88958 |
| 1513.43 | 36447.4 | 36732.8 | 45275.22312 | 45768.5868 | 1513.43 | 37118.9 | 37701.7 | 57188.74504 | 57464.86721 | 53123.96117 | 53631.63003 |
| 1515.12 | 36580.7 | 36760.4 | 45134.49859 | 45803.23579 | 1515.12 | 37229.6 | 37750.3 | 57187.3494 | 57536.68903 | 53528.0642 | 53707.02411 |
| 1516.82 | 36418.2 | 36787.8 | 45360.6039 | 45837.53 | 1516.82 | 37643.6 | 37798.6 | 57203.23823 | 57608.08143 | 53500.23882 | 53782.07182 |
| 1518.51 | 36402.3 | 36814.8 | 45472.35573 | 45871.46945 | 1518.51 | 37543 | 37846.6 | 57520.47813 | 57679.0444 | 53574.94015 | 53856.65769 |
| 1520.21 | 36356.6 | 36841.7 | 45321.10642 | 45905.05413 | 1520.21 | 37566.8 | 37894.3 | 57475.49555 | 57749.57794 | 53658.87812 | 53930.78173 |
| 1521.9 | 36395.8 | 36868.2 | 45623.36854 | 45938.28404 | 1521.9 | 37636.8 | 37941.7 | 57613.66399 | 57819.5747 | 53683.12431 | 54004.5594 |
| 1523.6 | 36686.2 | 36894.5 | 45608.23178 | 45971.27744 | 1523.6 | 37725.6 | 37988.8 | 57745.9278 | 57889.24938 | 53600.80275 | 54077.99069 |
| 1525.29 | 36466.2 | 36920.5 | 45669.25183 | 46003.79782 | 1525.29 | 37723.8 | 38035.7 | 57514.78821 | 57958.49464 | 53473.10619 | 54150.96015 |
| 1526.99 | 36511.2 | 36946.2 | 45787.74426 | 46036.08168 | 1526.99 | 37722.2 | 38082.3 | 57700.73051 | 58027.31047 | 54002.59661 | 54223.46778 |
| 1528.68 | 36426.4 | 36971.7 | 45739.37759 | 46068.12903 | 1528.68 | 37939 | 38128.6 | 57753.33543 | 58095.69687 | 53871.89815 | 54295.74449 |
| 1530.37 | 36492.9 | 36996.9 | 45781.94972 | 46099.70336 | 1530.37 | 38183.8 | 38174.6 | 57488.16368 | 58163.54648 | 54151.42198 | 54367.55937 |
| 1532.07 | 36696.5 | 37021.8 | 45658.25403 | 46131.04118 | 1532.07 | 38054.6 | 38220.4 | 57783.6101 | 58231.07403 | 54663.82462 | 54438.91242 |
| 1533.76 | 36672 | 37046.5 | 45656.95322 | 46162.02423 | 1533.76 | 38011.4 | 38265.8 | 58358.07734 | 58298.17214 | 54453.69105 | 54509.9191 |
| 1535.45 | 36654.5 | 37070.9 | 45802.88102 | 46192.65251 | 1535.45 | 38055.1 | 38311 | 58041.91102 | 58364.84083 | 54228.66339 | 54580.5794 |
| 1537.15 | 36524.6 | 37095.1 | 45549.81336 | 46222.92603 | 1537.15 | 37815.7 | 38355.9 | 58187.80916 | 58430.97274 | 54323.45442 | 54650.77787 |
| 1538.84 | 36645.5 | 37119 | 45872.77026 | 46252.96303 | 1538.84 | 38079 | 38400.6 | 58637.84965 | 58496.78257 | 54828.23683 | 54720.62996 |
| 1540.53 | 36870.5 | 37142.6 | 45866.73921 | 46282.64526 | 1540.53 | 37842.9 | 38444.9 | 58420.23704 | 58562.16298 | 54839.89808 | 54790.02022 |
| 1542.23 | 36804.8 | 37166 | 45951.64696 | 46311.97273 | 1542.23 | 38053 | 38489 | 58925.02959 | 58627.11395 | 55414.7635 | 54859.06411 |
| 1543.92 | 36854 | 37189.1 | 46138.37305 | 46340.94543 | 1543.92 | 38529.3 | 38532.8 | 58697.32542 | 58691.6355 | 55058.80647 | 54927.76163 |
| 1545.61 | 36804.5 | 37211.9 | 46011.01143 | 46369.68162 | 1545.61 | 38560.5 | 38576.3 | 58587.0698 | 58755.72762 | 54825.81221 | 54995.99731 |
| 1547.31 | 36901.7 | 37234.5 | 46386.47396 | 46398.06304 | 1547.31 | 38385.5 | 38619.6 | 59031.95715 | 58819.39032 | 54966.67098 | 55063.88662 |
| 1549 | 37106.7 | 37256.8 | 46270.93793 | 46426.08969 | 1549 | 38325.5 | 38662.5 | 58592.43765 | 58882.73094 | 55057.53643 | 55131.3141 |
| 1550.69 | 37161.8 | 37278.9 | 46437.324 | 46453.87983 | 1550.69 | 38740.4 | 38705.2 | 58652.4502 | 58945.53478 | 55588.75872 | 55198.3952 |
| 1552.38 | 37066.5 | 37300.7 | 46194.42635 | 46481.3152 | 1552.38 | 38641.9 | 38747.6 | 58831.73637 | 59007.90919 | 55272.05741 | 55265.12993 |
| 1554.07 | 36876 | 37322.3 | 46034.18959 | 46508.39581 | 1554.07 | 38862.6 | 38789.8 | 58530.06324 | 59069.96153 | 55220.1013 | 55331.40282 |
| 1555.77 | 36915.5 | 37343.6 | 46182.00948 | 46535.2399 | 1555.77 | 38829.9 | 38831.6 | 58453.51771 | 59131.47708 | 54886.19675 | 55397.32934 |
| 1557.46 | 36727.2 | 37364.7 | 46003.32479 | 46561.72922 | 1557.46 | 38381.7 | 38873.2 | 58658.46219 | 59192.67056 | 54977.75495 | 55462.90949 |
| 1559.15 | 36994.7 | 37385.5 | 46367.90778 | 46587.86378 | 1559.15 | 38768.2 | 38914.5 | 59172.16538 | 59253.32726 | 55562.20338 | 55528.02781 |
| 1560.84 | 37160.5 | 37406 | 46395.2249 | 46613.64357 | 1560.84 | 38590.3 | 38955.6 | 59248.38884 | 59313.66189 | 55670.50299 | 55592.79975 |
| 1562.53 | 37022.8 | 37426.3 | 46390.13989 | 46639.18685 | 1562.53 | 38671 | 38996.3 | 59362.93874 | 59373.56709 | 55992.28446 | 55657.10986 |
| 1564.22 | 36847 | 37446.4 | 46278.62456 | 46664.49361 | 1564.22 | 38977.3 | 39036.8 | 59401.90933 | 59433.04286 | 55874.97912 | 55721.0736 |
| 1565.91 | 36621.7 | 37466.2 | 45644.77286 | 46689.44561 | 1565.91 | 38747.8 | 39077.1 | 59319.02974 | 59492.0892 | 55431.274 | 55784.69096 |
| 1567.61 | 36379.7 | 37485.7 | 46045.66041 | 46714.04284 | 1567.61 | 38707.6 | 39117 | 58979.6743 | 59550.70612 | 55143.55264 | 55847.84649 |
| 1569.3 | 36556.3 | 37505 | 46022.83702 | 46738.28531 | 1569.3 | 38361.7 | 39156.7 | 58397.79943 | 59609.00096 | 55238.11275 | 55910.65565 |
| 1570.99 | 36677.8 | 37524.1 | 45885.5419 | 46762.29126 | 1570.99 | 38589.8 | 39196.1 | 58891.74893 | 59666.75902 | 55378.27877 | 55973.11843 |
| 1572.68 | 36717.4 | 37542.9 | 46295.88993 | 46786.0607 | 1572.68 | 38377 | 39235.2 | 59008.76804 | 59724.19501 | 55425.03927 | 56035.11938 |
| 1574.37 | 36788 | 37561.4 | 45744.58085 | 46809.47537 | 1574.37 | 38177.4 | 39274.1 | 58706.98755 | 59781.20157 | 55265.93813 | 56096.77395 |
| 1576.06 | 37046.9 | 37579.7 | 45912.74076 | 46832.53527 | 1576.06 | 38849.5 | 39312.7 | 59004.90319 | 59837.7787 | 55113.64902 | 56158.08216 |
| 1577.75 | 37109.6 | 37597.8 | 46217.84102 | 46855.24041 | 1577.75 | 38995.5 | 39351 | 59245.81227 | 59893.9264 | 55790.23294 | 56218.92853 |
| 1579.44 | 36824.4 | 37615.6 | 46087.16824 | 46877.70903 | 1579.44 | 38911.8 | 39389.1 | 59510.33989 | 59949.64468 | 55710.91329 | 56279.42852 |

Figure 5 Data

| Figure 5 A and B | T = 1645 K, n = 1.03, 1080 sec | T = 1645 K, n = 1.03, 1080 sec | T = 1640 K, n = 0.90, 9240 sec | T = 1640 K, n = 0.90, 9240 sec | T = 1288 K, n = 1.3, M = 71.68 MDa | T = 1288 K, n = 1.3, M = 71.68 MDa | T = 1289 K, n = 1.02, M = 57.26 MDa | T = 1289 K, n = 1.02, M = 57.26 MDa | Figure 5 C and D | T = 1649 K, n = 0.91, 480 sec | T = 1649 K, n = 0.91, 480 sec | T = 1641 K, n = 1.32, 4020 sec | T = 1641 K, n = 1.32, 4020 sec | T = 1728 K, n = 1.16, M = 14.22 Mda, 2640 sec |
|---|---|---|---|---|---|---|---|---|---|---|---|---|---|---|
| 600.015 | 149.404 | 136.555 | 347.2334056 | 205.1775191 | 177.26776 | 109.22189 | 0 | 83.4915 | 600.015 | 20.2581 | 47.8367 | 169.4643845 | 160.960087 | 387.666 |
| 600.306 | 157.375 | 137.187 | 117.5649733 | 206.1439088 | 84.0259 | 109.92661 | 53.47172 | 84.04116 | 600.306 | 30.1726 | 48.0596 | 98.15635224 | 161.6843267 | 430.754 |
| 600.597 | 134.669 | 137.82 | 268.2742856 | 207.1136892 | 239.67509 | 110.63448 | 152.52312 | 84.59383 | 600.597 | 87.739 | 48.2833 | 149.5102532 | 162.4107746 | 417.53 |
| 600.888 | 178.619 | 138.456 | 155.6746336 | 208.0868606 | 238.4206 | 111.34655 | 50.57477 | 85.14941 | 600.888 | 40.4685 | 48.5077 | 211.5817991 | 163.1394305 | 350.333 |
| 601.179 | 103.85 | 139.094 | 199.491079 | 209.0622924 | 356.44592 | 112.06177 | 302.44316 | 85.70812 | 601.179 | 110.51 | 48.7329 | 238.1148099 | 163.8702944 | 366.192 |
| 601.47 | 205.864 | 139.734 | 175.7585839 | 210.0422455 | 274.78499 | 112.78119 | 149.88404 | 86.26962 | 601.47 | -39.6328 | 48.9588 | 124.3274639 | 164.6033664 | 505.617 |
| 601.76 | 176.514 | 140.376 | 263.7260377 | 211.0255893 | 181.23608 | 113.50376 | 0 | 86.83426 | 601.76 | 39.0095 | 49.1855 | 217.5501533 | 165.3397505 | 364.362 |
| 602.052 | 156.612 | 141.021 | 272.9886847 | 212.012324 | 0 | 114.23054 | 177.37021 | 87.40192 | 602.052 | 9.44518 | 49.413 | 144.8545836 | 166.0772386 | 490.597 |
| 602.342 | 216.344 | 141.667 | 209.1515846 | 213.0013192 | 67.94567 | 114.96151 | 216.19328 | 87.9726 | 602.342 | 46.5257 | 49.6412 | 142.7072568 | 166.8169347 | 398.53 |
| 602.633 | 161.189 | 142.316 | 166.7276445 | 213.9948356 | 198.59946 | 115.69564 | 252.7664 | 88.54641 | 602.633 | 103.148 | 49.8702 | 235.3271491 | 167.5599429 | 427.248 |
| 602.924 | 191.703 | 142.966 | 250.617047 | 214.9906125 | 95.95367 | 116.43396 | 203.54033 | 89.12313 | 602.924 | 91.7197 | 50.1 | 268.5406096 | 168.3051592 | 434.616 |
| 603.215 | 81.2661 | 143.619 | 156.1086612 | 215.9909106 | 61.9828 | 117.17649 | 78.88833 | 89.70299 | 603.215 | 118.026 | 50.3306 | 202.5265936 | 169.0525835 | 355.748 |
| 603.506 | 147.639 | 144.274 | 254.5210351 | 216.9945995 | 121.26769 | 117.92217 | 308.68659 | 90.28598 | 603.506 | 62.8529 | 50.5619 | 187.7790399 | 169.8022158 | 343.65 |
| 603.797 | 214.341 | 144.932 | 272.2054004 | 218.0016792 | 257.47998 | 118.67205 | 182.05946 | 90.87199 | 603.797 | 52.2348 | 50.794 | 182.0657153 | 170.5540562 | 388.727 |
| 604.088 | 232.302 | 145.591 | 252.4198441 | 219.0110195 | 112.74945 | 119.42508 | 107.62621 | 91.46113 | 604.088 | 60.4832 | 51.0268 | 216.8391618 | 171.3081047 | 440.925 |
| 604.379 | 132.727 | 146.253 | 183.0511523 | 220.0248809 | 218.03668 | 120.18231 | 104.06438 | 92.05342 | 604.379 | -16.7463 | 51.2605 | 210.1311115 | 172.0654652 | 473.065 |
| 604.669 | 160.692 | 146.917 | 195.0705528 | 221.0421332 | 134.0493 | 120.94374 | 170.61105 | 92.64884 | 604.669 | 50.165 | 51.4949 | 116.5672786 | 172.8250338 | 441.893 |
| 604.96 | 202.832 | 147.582 | 212.1321337 | 222.061646 | 216.57841 | 121.70937 | 68.91209 | 93.24739 | 604.96 | 65.3743 | 51.7301 | 170.8974077 | 173.5857063 | 484.027 |
| 605.251 | 203.32 | 148.25 | 289.3099319 | 223.0856799 | 310.14005 | 122.47816 | 32.89408 | 93.84909 | 605.251 | 89.5198 | 51.9661 | 260.9647963 | 174.349691 | 430.124 |
| 605.542 | 221.58 | 148.921 | 234.4992502 | 224.1131047 | 320.38932 | 123.25219 | 94.10161 | 94.45403 | 605.542 | 56.2237 | 52.2028 | 156.7758357 | 175.1169877 | 468.426 |
| 605.832 | 166.188 | 149.593 | 151.7378674 | 225.1439203 | 147.8742 | 124.02833 | 125.47021 | 95.06211 | 605.832 | 39.0899 | 52.4404 | 130.7992531 | 175.8853885 | 406.077 |
| 606.123 | 124.519 | 150.268 | 186.0407437 | 226.1781268 | 237.41976 | 124.80971 | 60.43497 | 95.67333 | 606.123 | 104.461 | 52.6787 | 291.2812976 | 176.6559973 | 483.274 |
| 606.414 | 186.35 | 150.945 | 278.4208117 | 227.2157241 | 142.12497 | 125.59425 | 90.44453 | 96.2878 | 606.414 | 39.6827 | 52.9178 | 320.8912235 | 177.4299182 | 426.7 |
| 606.705 | 213.493 | 151.624 | 255.1799885 | 228.2567122 | 205.16061 | 126.38404 | 232.10327 | 96.90552 | 606.705 | 44.9343 | 53.1577 | 187.9435395 | 178.2060471 | 423.115 |
| 606.995 | 176.372 | 152.305 | 313.7059016 | 229.3010912 | 134.51416 | 127.17698 | 28.53368 | 97.52638 | 606.995 | 44.5935 | 53.3984 | 186.5182444 | 178.9832801 | 474.087 |
| 607.286 | 225.697 | 152.989 | 192.6913128 | 230.3499913 | 215.16577 | 127.97412 | 164.3107 | 98.1506 | 607.286 | 14.9616 | 53.6399 | 302.4650615 | 179.7649291 | 470.367 |
| 607.577 | 153.82 | 153.674 | 169.3397222 | 231.401152 | 216.10293 | 128.77546 | 247.53935 | 98.77797 | 607.577 | 51.7648 | 53.8821 | 164.9621742 | 180.5476822 | 478.447 |
| 607.868 | 200.623 | 154.362 | 219.0449279 | 232.4557035 | 185.37421 | 129.57995 | 104.85958 | 99.4087 | 607.868 | 100.605 | 54.1252 | 210.3971813 | 181.3337474 | 419.095 |
| 608.158 | 195.03 | 155.052 | 338.0114488 | 233.5147761 | 229.06178 | 130.3897 | 79.51 | 100.04268 | 608.158 | 57.8184 | 54.369 | 201.5274518 | 182.1209165 | 493.943 |
| 608.449 | 170.469 | 155.745 | 243.6206121 | 234.5772395 | 138.48885 | 131.20259 | 75.54011 | 100.67992 | 608.449 | 118.089 | 54.6137 | 213.0645035 | 182.9113978 | 443.095 |
| 608.74 | 177.714 | 156.439 | 298.7082125 | 235.6419636 | 276.71449 | 132.02074 | 100.62497 | 101.32052 | 608.74 | 7.03115 | 54.8591 | 235.2697398 | 183.7040871 | 429.28 |
| 609.03 | 179.57 | 157.136 | 236.0963362 | 236.7112087 | 153.38251 | 132.84203 | 170.81548 | 101.96438 | 609.03 | 41.1196 | 55.1053 | 238.8721704 | 184.4989844 | 424.665 |
| 609.321 | 245.287 | 157.835 | 282.7113557 | 237.7838447 | 261.89406 | 133.66753 | 47.61777 | 102.6116 | 609.321 | 69.3212 | 55.3524 | 173.9676987 | 185.2971938 | 461.136 |
| 609.612 | 170.943 | 158.536 | 297.9690092 | 238.8598715 | 190.12048 | 134.49723 | 169.38211 | 103.2622 | 609.612 | 33.0072 | 55.6002 | 184.0761432 | 186.0965072 | 499.294 |
| 609.902 | 195.06 | 159.24 | 218.5758616 | 239.9404195 | 204.5461 | 135.33113 | 165.66742 | 103.91616 | 609.902 | 45.6145 | 55.8488 | 199.8736238 | 186.8991327 | 498.797 |
| 610.193 | 201.447 | 159.945 | 259.4637508 | 241.023228 | 129.79124 | 136.16923 | 259.58712 | 104.57349 | 610.193 | 90.7446 | 56.0983 | 207.8490937 | 187.7039663 | 413.457 |
| 610.484 | 206.529 | 160.653 | 258.8014066 | 242.1094273 | 160.04661 | 137.01153 | 181.06516 | 105.23419 | 610.484 | 46.0879 | 56.3485 | 192.7681248 | 188.5121119 | 461.975 |
| 610.774 | 183.424 | 161.363 | 303.412439 | 243.2001478 | 209.82728 | 137.85803 | 44.50941 | 105.89826 | 610.774 | 57.7543 | 56.5996 | 214.7249557 | 189.3213616 | 473.66 |
| 611.065 | 169.918 | 162.076 | 273.4001067 | 244.2942591 | 156.99634 | 138.70873 | 155.4131 | 106.56581 | 611.065 | 83.228 | 56.8514 | 285.6309023 | 190.1339233 | 460.885 |
| 611.355 | 191.519 | 162.791 | 276.6055816 | 245.3917613 | 272.64004 | 139.56469 | 238.56275 | 107.23673 | 611.355 | 55.9792 | 57.1041 | 234.1403232 | 190.9486931 | 464.77 |
| 611.646 | 175.546 | 163.508 | 243.5460136 | 246.4926543 | 267.75048 | 140.42379 | 255.5832 | 107.91102 | 611.646 | 72.6201 | 57.3576 | 194.0587289 | 191.7656709 | 463.228 |
| 611.937 | 204.128 | 164.227 | 304.9835286 | 247.5969381 | 247.65949 | 141.2871 | 147.09627 | 108.58891 | 611.937 | 74.3799 | 57.6118 | 207.403068 | 192.5848568 | 513.991 |
| 612.227 | 201.101 | 164.948 | 309.5645546 | 248.7046128 | 309.04794 | 142.15461 | 82.80827 | 109.27017 | 612.227 | 35.0793 | 57.8669 | 195.6319632 | 193.4073547 | 447.526 |
| 612.518 | 204.444 | 165.672 | 278.5033222 | 249.8168086 | 208.64602 | 143.02737 | 163.41746 | 109.95492 | 612.518 | 51.7854 | 58.1228 | 280.7765085 | 194.2320607 | 477.14 |
| 612.808 | 164.762 | 166.398 | 272.542224 | 250.9323952 | 282.70995 | 143.90327 | 79.95929 | 110.64315 | 612.808 | 63.3223 | 58.3796 | 224.7241018 | 195.0589747 | 472.113 |
| 613.099 | 141.499 | 167.127 | 290.6900042 | 252.0513727 | 283.25865 | 144.78444 | 80.11471 | 111.33486 | 613.099 | 46.0864 | 58.6371 | 281.1121318 | 195.8880968 | 509.207 |
| 613.389 | 256.424 | 167.857 | 313.4595005 | 253.173741 | 202.21541 | 145.6698 | 79.19045 | 112.03017 | 613.389 | 61.6813 | 58.8954 | 195.4475912 | 196.720531 | 510.976 |
| 613.68 | 207.159 | 168.59 | 318.1083433 | 254.2995002 | 199.66822 | 146.55936 | 195.48137 | 112.72886 | 613.68 | 43.5927 | 59.1546 | 258.3040983 | 197.5540692 | 486.62 |
| 613.97 | 205.119 | 169.325 | 272.4122417 | 255.4297805 | 273.73958 | 147.45312 | 212.91185 | 113.43126 | 613.97 | 60.4439 | 59.4146 | 260.4757136 | 198.3909194 | 460.667 |
| 614.26 | 168.731 | 170.063 | 294.1136231 | 256.5634516 | 195.15438 | 148.35109 | 191.06277 | 114.13714 | 614.26 | 85.2276 | 59.6754 | 207.9440398 | 199.2310817 | 470.839 |
| 614.551 | 177.197 | 170.802 | 314.3852 | 257.7005136 | 206.91499 | 149.2543 | 131.6753 | 114.84651 | 614.551 | 64.725 | 59.937 | 210.5605769 | 200.0723481 | 486.592 |
| 614.841 | 187.882 | 171.544 | 232.5879462 | 258.8409664 | 229.24221 | 150.16067 | 218.82423 | 115.5596 | 614.841 | 62.4731 | 60.1995 | 210.4932314 | 200.9169265 | 455.431 |
| 615.132 | 187.268 | 172.289 | 319.7630736 | 259.98481 | 228.49185 | 151.07229 | 236.28492 | 116.2764 | 615.132 | 72.7478 | 60.4628 | 239.0753108 | 201.763713 | 511.394 |
| 615.422 | 177.395 | 173.036 | 249.4505978 | 261.1331748 | 278.51238 | 151.98811 | 70.89512 | 116.99657 | 615.422 | 56.6446 | 60.7269 | 272.8209109 | 202.6138115 | 544.332 |
| 615.713 | 215.307 | 173.784 | 321.6856804 | 262.2849304 | 232.60026 | 152.90918 | 69.65654 | 117.72023 | 615.713 | 53.7986 | 60.9918 | 272.7513574 | 203.4650141 | 490.242 |
| 616.003 | 204.483 | 174.536 | 381.8911899 | 263.4400768 | 204.66178 | 153.8334 | 191.02095 | 118.44737 | 616.003 | 58.1029 | 61.2576 | 243.023301 | 204.3195287 | 507.404 |
| 616.293 | 212.875 | 175.289 | 236.6897334 | 264.5986141 | 184.98364 | 154.76287 | 168.16943 | 119.17916 | 616.293 | 68.2829 | 61.5242 | 278.8025134 | 205.1762514 | 546.674 |
| 616.584 | 207.022 | 176.045 | 324.0366633 | 265.7616725 | 236.80207 | 155.69759 | 150.69481 | 119.91327 | 616.584 | 76.5072 | 61.7916 | 273.3342821 | 206.0351822 | 501.095 |
| 616.874 | 193.799 | 176.803 | 246.1174012 | 266.9281218 | 310.34701 | 156.63547 | 246.87028 | 120.65203 | 616.874 | 18.6171 | 62.0598 | 266.0510353 | 206.897425 | 483.5 |
| 617.165 | 197.262 | 177.564 | 323.4771746 | 268.0979619 | 256.82196 | 157.57859 | 179.77698 | 121.39427 | 617.165 | 41.9079 | 62.3289 | 285.6507748 | 207.7618758 | 509.184 |
| 617.455 | 174.588 | 178.327 | 289.0477069 | 269.2711928 | 289.6553 | 158.52592 | 192.34282 | 122.13999 | 617.455 | 63.3543 | 62.5988 | 189.2772006 | 208.6285347 | 511.339 |
| 617.745 | 205.498 | 179.092 | 341.1445857 | 270.4489449 | 243.76122 | 159.4785 | 170.63544 | 122.8892 | 617.745 | 66.6312 | 62.8696 | 216.7618801 | 209.4985057 | 558.605 |
| 618.036 | 217.886 | 179.859 | 345.8849812 | 271.6300878 | 193.84151 | 160.43527 | 154.19345 | 123.64306 | 618.036 | 53.731 | 63.1412 | 255.9503189 | 210.3695807 | 522.268 |
| 618.326 | 216.37 | 180.629 | 302.6393273 | 272.8146215 | 147.44011 | 161.39625 | 125.10199 | 124.4004 | 618.326 | 69.833 | 63.4136 | 255.5749507 | 211.2439678 | 461.358 |
| 618.616 | 199.928 | 181.402 | 325.2607569 | 274.0025461 | 189.72142 | 162.36248 | 166.00775 | 125.16007 | 618.616 | 51.8852 | 63.6869 | 253.1869467 | 212.1216669 | 500.387 |
| 618.907 | 232.076 | 182.176 | 281.3165898 | 275.1949918 | 268.79059 | 163.33397 | 193.36151 | 125.92554 | 618.907 | 63.0864 | 63.961 | 246.2757559 | 213.0015741 | 475.232 |
| 619.197 | 241.462 | 182.953 | 328.5521331 | 276.3908283 | 184.12502 | 164.3086 | 73.23231 | 126.69334 | 619.197 | 70.7796 | 64.236 | 232.1928247 | 213.8836893 | 542.544 |
| 619.487 | 166.484 | 183.732 | 376.2996933 | 277.591186 | 262.0384 | 165.28848 | 188.505 | 127.46578 | 619.487 | 48.906 | 64.5118 | 244.3304655 | 214.7680126 | 493.938 |
| 619.777 | 230.79 | 184.514 | 313.4708033 | 278.7938042 | 291.17512 | 166.27362 | 156.78839 | 128.24171 | 619.777 | 40.9798 | 64.7884 | 245.6784789 | 215.654544 | 519.068 |
| 620.067 | 214.486 | 185.298 | 338.9461905 | 280.0009435 | 231.17489 | 167.26295 | 84.06462 | 129.02112 | 620.067 | 82.8847 | 65.0659 | 236.6199613 | 216.5443874 | 499.93 |
| 620.358 | 184.324 | 186.084 | 324.3531418 | 281.2114737 | 273.29594 | 168.25754 | 194.78792 | 129.80518 | 620.358 | 35.0909 | 65.3442 | 206.5684258 | 217.4364388 | 493.863 |
| 620.648 | 188.278 | 186.873 | 287.2867302 | 282.426525 | 277.96898 | 169.25633 | 190.49825 | 130.59272 | 620.648 | 38.5213 | 65.6234 | 236.3097305 | 218.3318024 | 554.622 |
| 620.938 | 229.889 | 187.664 | 296.899764 | 283.6449672 | 239.17415 | 170.25932 | 185.2921 | 131.38375 | 620.938 | 49.2037 | 65.9035 | 242.7395668 | 219.229374 | 520.801 |
| 621.228 | 218.701 | 188.457 | 309.2503367 | 284.8668002 | 198.03165 | 171.26861 | 159.18471 | 132.17826 | 621.228 | 67.0176 | 66.1843 | 254.3550041 | 220.1291536 | 529.274 |
| 621.519 | 216.283 | 189.253 | 357.7664875 | 286.0931543 | 298.91647 | 172.28105 | 157.42493 | 132.97742 | 621.519 | 74.126 | 66.4661 | 237.6985928 | 221.0311413 | 537.004 |
| 621.809 | 212.151 | 190.051 | 311.3108376 | 287.3228992 | 283.09415 | 173.2998 | 193.02117 | 133.78122 | 621.809 | 32.5556 | 66.7487 | 252.1635163 | 221.936441 | 540.589 |
| 622.099 | 197.325 | 190.852 | 311.5063761 | 288.556035 | 248.40348 | 174.32274 | 164.40015 | 134.58735 | 622.099 | 58.1573 | 67.0321 | 253.385671 | 222.8439488 | 510.087 |
| 622.389 | 214.923 | 191.655 | 315.6691984 | 289.7936919 | 252.76449 | 175.34989 | 160.85156 | 135.39813 | 622.389 | 73.923 | 67.3164 | 211.0364112 | 223.7536646 | 574.186 |
| 622.679 | 230.471 | 192.46 | 289.6840547 | 291.0336094 | 283.12811 | 176.38228 | 161.53572 | 136.21355 | 622.679 | 60.674 | 67.6016 | 273.6809457 | 224.6666926 | 549.642 |
| 622.969 | 213.762 | 193.268 | 341.0304274 | 292.2791782 | 232.05898 | 177.41993 | 172.28835 | 137.03245 | 622.969 | 30.9903 | 67.8876 | 273.6445129 | 225.5819285 | 514.597 |
| 623.259 | 206.75 | 194.078 | 324.6074549 | 293.528138 | 233.79744 | 178.46178 | 130.92841 | 137.85484 | 623.259 | 78.7095 | 68.1745 | 238.5608356 | 226.5004766 | 548.848 |
| 623.549 | 207.72 | 194.89 | 336.6505913 | 294.7804885 | 291.26746 | 179.50888 | 107.62528 | 138.68072 | 623.549 | 95.5456 | 68.4622 | 266.4219873 | 227.4201286 | 541.055 |
| 623.84 | 238.493 | 195.705 | 330.5063877 | 296.0362299 | 295.17422 | 180.56123 | 187.84058 | 139.5124 | 623.84 | 70.2779 | 68.7508 | 279.9495944 | 228.3430927 | 542.824 |
| 624.13 | 204.716 | 196.523 | 325.9174497 | 297.2964924 | 304.50969 | 181.61883 | 205.18046 | 140.3464 | 624.13 | 69.1859 | 69.0403 | 266.4120511 | 229.269369 | 522.381 |
| 624.42 | 211.095 | 197.342 | 345.4306085 | 298.5601458 | 286.14548 | 182.68064 | 216.23858 | 141.18506 | 624.42 | 78.7572 | 69.3306 | 283.6602194 | 230.1978532 | 586.104 |
| 624.71 | 198.24 | 198.165 | 320.8684877 | 299.8283203 | 264.51979 | 183.74769 | 224.4439 | 142.02835 | 624.71 | 74.3301 | 69.6218 | 270.3434809 | 231.1285455 | 559.538 |
| 625 | 235.488 | 198.989 | 346.9497052 | 301.0998856 | 296.76303 | 184.81895 | 166.63384 | 142.87514 | 625 | 50.6022 | 69.9138 | 260.1533386 | 232.0614459 | 576.393 |
| 625.29 | 238.388 | 199.816 | 346.4116918 | 302.3748417 | 287.58676 | 185.89651 | 188.55843 | 143.7254 | 625.29 | 70.3433 | 70.2068 | 258.5569198 | 232.9976583 | 556.51 |
| 625.58 | 235.213 | 200.645 | 356.1761832 | 303.654319 | 271.43861 | 186.97826 | 172.73672 | 144.58148 | 625.58 | 80.2192 | 70.5006 | 279.6062429 | 233.9360787 | 545.439 |
| 625.87 | 225.861 | 201.477 | 304.6467051 | 304.9371871 | 284.62883 | 188.06527 | 202.43916 | 145.43988 | 625.87 | 71.4763 | 70.7952 | 251.2990655 | 234.8778113 | 566.375 |
| 626.16 | 248.379 | 202.311 | 319.4251198 | 306.2245764 | 302.04952 | 189.15753 | 224.25224 | 146.30292 | 626.16 | 85.7089 | 71.0908 | 264.5992436 | 235.8217519 | 571.823 |
| 626.45 | 222.583 | 203.148 | 327.9383909 | 307.5153565 | 280.49495 | 190.254 | 188.99982 | 147.17061 | 626.45 | 72.5084 | 71.3872 | 261.1546885 | 236.7679005 | 580.897 |
| 626.74 | 233.761 | 203.987 | 408.2221993 | 308.8095274 | 243.04482 | 191.35676 | 175.28984 | 148.04179 | 626.74 | 66.1301 | 71.6845 | 272.4069019 | 237.7173612 | 569.744 |
| 627.03 | 240.81 | 204.829 | 359.7896892 | 310.1082194 | 309.11162 | 192.46372 | 176.00421 | 148.91761 | 627.03 | 64.3961 | 71.9826 | 277.5074931 | 238.66903 | 562.806 |
| 627.32 | 226.288 | 205.673 | 400.3724028 | 311.4103023 | 222.47835 | 193.57594 | 171.91781 | 149.79808 | 627.32 | 71.9157 | 72.2816 | 248.6582399 | 239.6229068 | 592.423 |
| 627.61 | 234.218 | 206.519 | 332.2210229 | 312.7169063 | 348.11766 | 194.69341 | 201.39259 | 150.68087 | 627.61 | 89.3845 | 72.5816 | 253.2620203 | 240.5800957 | 557.027 |
| 627.9 | 208.583 | 207.368 | 328.9522523 | 314.0269012 | 286.03934 | 195.81613 | 186.94733 | 151.56947 | 627.9 | 68.9381 | 72.8824 | 288.3434895 | 241.5394926 | 620.349 |
| 628.19 | 240.839 | 208.219 | 346.6671352 | 315.3414171 | 274.32012 | 196.94409 | 249.38508 | 152.46155 | 628.19 | 72.2677 | 73.184 | 275.1426736 | 242.5010976 | 582.557 |
| 628.48 | 259.408 | 209.073 | 383.3176036 | 316.6593239 | 296.676 | 198.07732 | 193.6368 | 153.35828 | 628.48 | 94.4092 | 73.4866 | 286.4810009 | 243.4660146 | 609.807 |
| 628.77 | 240.175 | 209.929 | 346.025136 | 317.9817519 | 335.71813 | 199.21579 | 203.93061 | 154.25966 | 628.77 | 73.9434 | 73.79 | 274.9141406 | 244.4331397 | 613.918 |
| 629.059 | 246.463 | 210.787 | 368.2340132 | 319.3075706 | 258.6432 | 200.35951 | 225.72743 | 155.16452 | 629.059 | 78.0671 | 74.0943 | 270.2286624 | 245.4035769 | 640.251 |
| 629.349 | 254.601 | 211.649 | 355.5861769 | 320.6367803 | 236.31598 | 201.50743 | 187.98113 | 156.07403 | 629.349 | 64.4501 | 74.3995 | 262.0831727 | 246.3762221 | 609.328 |
| 629.639 | 237.932 | 212.512 | 343.2299528 | 321.970511 | 299.21577 | 202.66166 | 176.48045 | 156.98818 | 629.639 | 82.4732 | 74.7056 | 274.4802591 | 247.3521794 | 591.535 |
| 629.929 | 254.582 | 213.378 | 368.0949888 | 323.3076326 | 328.71759 | 203.82114 | 204.53811 | 157.90582 | 629.929 | 84.3971 | 75.0126 | 290.4908163 | 248.3292407 | 579.207 |
| 630.219 | 229.8 | 214.247 | 359.3081898 | 324.6492753 | 299.47792 | 204.98587 | 199.65488 | 158.82811 | 630.219 | 72.6881 | 75.3204 | 267.8340341 | 249.3107181 | 605.598 |
| 630.509 | 246.213 | 215.118 | 348.2947388 | 325.9954391 | 265.60129 | 206.15479 | 213.50077 | 159.75504 | 630.509 | 83.8479 | 75.6292 | 294.0259015 | 250.2932995 | 615.885 |
| 630.799 | 258.906 | 215.991 | 398.6668098 | 327.3449937 | 338.39056 | 207.33003 | 215.34185 | 160.68545 | 630.799 | 80.8193 | 75.9388 | 242.9647877 | 251.279193 | 612.19 |
| 631.088 | 270.165 | 216.867 | 380.3596601 | 328.6979392 | 256.48233 | 208.51051 | 211.74216 | 161.62051 | 631.088 | 79.9016 | 76.2494 | 275.0179188 | 252.2683986 | 596.138 |
| 631.378 | 253.681 | 217.745 | 375.1897581 | 330.0554058 | 280.3453 | 209.69624 | 208.86379 | 162.56022 | 631.378 | 101.787 | 76.5608 | 273.4413725 | 253.2598122 | 604.032 |
| 631.668 | 250.81 | 218.626 | 386.08566 | 331.4162633 | 317.73387 | 210.88828 | 258.12355 | 163.50458 | 631.668 | 99.832 | 76.8731 | 288.5576702 | 254.2534339 | 592.429 |
| 631.958 | 262.701 | 219.51 | 364.5922502 | 332.7816419 | 312.29243 | 212.08451 | 253.70262 | 164.45358 | 631.958 | 78.5473 | 77.1863 | 294.3118437 | 255.2492636 | 633.005 |
| 632.247 | 255.707 | 220.396 | 408.0097066 | 334.1515416 | 304.73044 | 213.286 | 244.50998 | 165.40606 | 632.247 | 72.3029 | 77.5004 | 302.4164845 | 256.2495095 | 625.92 |
| 632.537 | 265.489 | 221.284 | 394.8170752 | 335.5248321 | 303.14269 | 214.49379 | 226.46154 | 166.36319 | 632.537 | 73.5465 | 77.8154 | 277.8486366 | 257.2508593 | 642.091 |
| 632.827 | 259.028 | 222.175 | 383.3571634 | 336.9015135 | 319.5116 | 215.70683 | 207.47804 | 167.32497 | 632.827 | 83.5395 | 78.1313 | 292.8434917 | 258.2555212 | 648.156 |
| 633.117 | 265.123 | 223.068 | 408.6607481 | 338.282716 | 358.52613 | 216.92512 | 252.60146 | 168.2914 | 633.117 | 62.8426 | 78.4481 | 301.7871907 | 259.2623912 | 625.679 |
| 633.407 | 264.603 | 223.964 | 436.481467 | 339.6673094 | 332.26243 | 218.14866 | 203.31034 | 169.26247 | 633.407 | 51.0619 | 78.7659 | 297.7078218 | 260.2725732 | 626.045 |
| 633.696 | 287.013 | 224.862 | 393.3793587 | 341.0564238 | 310.13899 | 219.37745 | 217.50353 | 170.23818 | 633.696 | 96.784 | 79.0845 | 289.1516351 | 261.2849633 | 646.762 |
| 633.986 | 262.186 | 225.763 | 384.8536545 | 342.4500594 | 312.98017 | 220.61255 | 203.07802 | 171.21739 | 633.986 | 93.4791 | 79.404 | 308.3406777 | 262.3006654 | 648.224 |
| 634.276 | 254.981 | 226.667 | 408.172467 | 343.8470859 | 279.43999 | 221.85289 | 216.486 | 172.20123 | 634.276 | 64.4107 | 79.7244 | 331.3386033 | 263.3185757 | 637.544 |
| 634.565 | 287.891 | 227.572 | 382.3387809 | 345.2486334 | 323.1265 | 223.09849 | 225.02817 | 173.19089 | 634.565 | 94.5503 | 80.0457 | 285.1053869 | 264.3386939 | 639.182 |
| 634.855 | 257.798 | 228.481 | 398.6848943 | 346.6535718 | 325.84138 | 224.34934 | 207.35608 | 174.18403 | 634.855 | 83.8124 | 80.3679 | 316.7312607 | 265.3621242 | 691.389 |
| 635.144 | 263.187 | 229.392 | 396.582573 | 348.0630313 | 348.13039 | 225.60649 | 259.73697 | 175.18182 | 635.144 | 68.7679 | 80.6911 | 284.6350726 | 266.3877626 | 679.524 |
| 635.434 | 270.962 | 230.305 | 418.1121518 | 349.4758817 | 355.57776 | 226.86889 | 211.19506 | 176.18425 | 635.434 | 97.0039 | 81.0151 | 345.6235709 | 267.416713 | 660.018 |
| 635.724 | 281.214 | 231.221 | 394.0993472 | 350.8932531 | 314.10412 | 228.1376 | 236.90287 | 177.19133 | 635.724 | 85.96 | 81.34 | 295.2314958 | 268.4478715 | 653.399 |
| 636.013 | 291.493 | 232.14 | 435.513947 | 352.3151457 | 351.44387 | 229.41155 | 219.98232 | 178.20305 | 636.013 | 79.1933 | 81.6659 | 310.8975975 | 269.4823421 | 668.478 |
| 636.303 | 270.247 | 233.061 | 464.6559636 | 353.7404292 | 354.82209 | 230.69076 | 196.66385 | 179.21942 | 636.303 | 75.4717 | 81.9927 | 307.1383955 | 270.5190207 | 682.163 |
| 636.593 | 264.673 | 233.984 | 407.995013 | 355.1691034 | 330.68742 | 231.97521 | 256.80865 | 180.24044 | 636.593 | 71.0977 | 82.3203 | 311.5357234 | 271.5590114 | 657.136 |
| 636.882 | 293.527 | 234.91 | 423.000614 | 356.6034291 | 333.27813 | 233.26597 | 226.22806 | 181.26611 | 636.882 | 87.7075 | 82.6489 | 304.2855972 | 272.6012101 | 682.141 |
| 637.172 | 262.199 | 235.839 | 416.2302351 | 358.0411456 | 322.4759 | 234.56303 | 243.47502 | 182.29642 | 637.172 | 104.973 | 82.9784 | 303.7556657 | 273.6456169 | 667.796 |
| 637.461 | 272.488 | 236.77 | 398.0711521 | 359.482253 | 306.98047 | 235.86534 | 205.63579 | 183.33137 | 637.461 | 100.556 | 83.3089 | 327.1256474 | 274.6933358 | 700.718 |
| 637.751 | 268.443 | 237.704 | 440.2769482 | 360.9278815 | 285.42802 | 237.1729 | 233.04531 | 184.37098 | 637.751 | 81.2727 | 83.6402 | 318.350864 | 275.7432627 | 677.121 |
| 638.04 | 263.139 | 238.64 | 435.1658207 | 362.3780311 | 369.81233 | 238.48677 | 204.64265 | 185.41523 | 638.04 | 81.0131 | 83.9725 | 330.7799672 | 276.7965016 | 706.597 |
| 638.33 | 276.268 | 239.579 | 436.352615 | 363.8315715 | 363.30318 | 239.80588 | 247.94938 | 186.46412 | 638.33 | 96.5787 | 84.3057 | 309.6975233 | 277.8530527 | 675.9 |
| 638.619 | 268.385 | 240.52 | 447.8181782 | 365.2896331 | 381.57188 | 241.1313 | 232.84203 | 187.51766 | 638.619 | 86.6208 | 84.6398 | 325.547997 | 278.9107078 | 684.663 |
| 638.909 | 282.397 | 241.464 | 430.5339321 | 366.7510855 | 338.29928 | 242.46197 | 267.47417 | 188.57585 | 638.909 | 80.9089 | 84.9748 | 327.9183367 | 279.9727789 | 636.123 |
| 639.198 | 297.353 | 242.411 | 432.8770031 | 368.217059 | 381.43603 | 243.79894 | 258.91806 | 189.63985 | 639.198 | 93.0726 | 85.3107 | 334.7886786 | 281.0359541 | 702.25 |
| 639.488 | 267.996 | 243.36 | 422.8073361 | 369.6875537 | 340.02501 | 245.14116 | 235.4753 | 190.70849 | 639.488 | 89.5237 | 85.6476 | 308.215923 | 282.1024414 | 690.923 |
| 639.777 | 289.4 | 244.311 | 429.613884 | 371.1625694 | 321.73297 | 246.48968 | 188.99285 | 191.78062 | 639.777 | 87.1914 | 85.9854 | 346.9616481 | 283.1722407 | 670.389 |
| 640.067 | 290.889 | 245.266 | 423.5374972 | 372.640976 | 331.21384 | 247.8445 | 220.21347 | 192.85855 | 640.067 | 98.3804 | 86.3241 | 296.7705053 | 284.2442481 | 699.267 |
| 640.356 | 292.438 | 246.222 | 428.6282796 | 374.1227735 | 390.12939 | 249.20458 | 238.84037 | 193.94113 | 640.356 | 95.8096 | 86.6637 | 309.2106487 | 285.3195675 | 662.886 |
| 640.646 | 271.636 | 247.182 | 430.0354785 | 375.6090921 | 350.37829 | 250.57095 | 250.46069 | 195.02836 | 640.646 | 102.155 | 87.0043 | 324.1403663 | 286.3970951 | 686.933 |
| 640.935 | 276.959 | 248.144 | 446.0425079 | 377.101062 | 374.27522 | 251.94258 | 232.07191 | 196.12139 | 640.935 | 86.3682 | 87.3458 | 320.3049867 | 287.4768306 | 675.275 |
| 641.225 | 295.473 | 249.108 | 444.1187708 | 378.5952926 | 413.17089 | 253.32051 | 211.55282 | 197.21791 | 641.225 | 93.1499 | 87.6883 | 319.8666891 | 288.5598782 | 686.925 |
| 641.514 | 286.857 | 250.075 | 467.7868399 | 380.0940442 | 340.67667 | 254.70474 | 249.46522 | 198.32024 | 641.514 | 89.4451 | 88.0316 | 316.560137 | 289.6462379 | 725.728 |
| 641.803 | 307.041 | 251.045 | 440.4973528 | 381.597317 | 363.05696 | 256.09422 | 266.58441 | 199.42722 | 641.803 | 109.175 | 88.3759 | 288.0354668 | 290.7348057 | 673.448 |
| 642.093 | 272.12 | 252.017 | 475.6083795 | 383.1051109 | 374.4323 | 257.49001 | 249.90081 | 200.54 | 642.093 | 102.853 | 88.7211 | 336.1631886 | 291.8255815 | 712.924 |
| 642.382 | 313.233 | 252.992 | 429.8422006 | 384.6162956 | 399.69304 | 258.89209 | 208.10761 | 201.65626 | 642.382 | 94.2931 | 89.0673 | 299.8308601 | 292.9196693 | 719.057 |
| 642.671 | 303.141 | 253.969 | 442.9071104 | 386.1320015 | 370.88215 | 260.30048 | 255.9247 | 202.77834 | 642.671 | 92.7609 | 89.4144 | 344.8750426 | 294.0170693 | 688.094 |
| 642.961 | 294.284 | 254.95 | 424.6090029 | 387.6522285 | 367.90299 | 261.71412 | 296.93733 | 203.90506 | 642.961 | 96.8183 | 89.7624 | 344.2115241 | 295.1166772 | 694.045 |
| 643.25 | 295.11 | 255.932 | 458.2563166 | 389.1758463 | 375.98609 | 263.13406 | 270.23057 | 205.03759 | 643.25 | 97.9561 | 90.1114 | 331.3540597 | 296.2184933 | 706.787 |
| 643.539 | 299.314 | 256.917 | 432.4554086 | 390.7039852 | 381.73532 | 264.5603 | 228.96472 | 206.1736 | 643.539 | 109.228 | 90.4613 | 340.4081612 | 297.3236214 | 747.976 |
| 643.829 | 307.327 | 257.905 | 483.59494 | 392.2366453 | 370.72189 | 265.99284 | 260.89621 | 207.31542 | 643.829 | 92.319 | 90.8122 | 346.1910393 | 298.4309575 | 722.324 |
| 644.118 | 293.959 | 258.896 | 467.9970721 | 393.7726962 | 415.71065 | 267.43168 | 218.18305 | 208.46305 | 644.118 | 100.078 | 91.1639 | 318.303391 | 299.5416058 | 735.274 |
| 644.407 | 323.16 | 259.889 | 456.5337695 | 395.3132682 | 371.3364 | 268.87578 | 260.75334 | 209.61416 | 644.407 | 100.96 | 91.5167 | 331.3253551 | 300.6555661 | 711.939 |
| 644.697 | 307.396 | 260.884 | 423.5781873 | 396.8583614 | 412.18491 | 270.32617 | 248.78222 | 210.77225 | 644.697 | 93.8965 | 91.8703 | 375.5658081 | 301.7717344 | 716.771 |
| 644.986 | 295.923 | 261.882 | 472.8697104 | 398.4068454 | 388.42277 | 271.78287 | 236.43476 | 211.93381 | 644.986 | 107.547 | 92.225 | 319.1645298 | 302.8901108 | 733.619 |
| 645.275 | 307.096 | 262.883 | 489.2576442 | 399.9609808 | 401.29778 | 273.24587 | 324.5385 | 213.10119 | 645.275 | 102.891 | 92.5805 | 343.8637565 | 304.0117993 | 715.61 |
| 645.564 | 304.387 | 263.887 | 454.7773139 | 401.5173767 | 403.98188 | 274.71517 | 278.06883 | 214.27321 | 645.564 | 98.612 | 92.937 | 345.7980067 | 305.1367998 | 709.355 |
| 645.853 | 303.776 | 264.893 | 447.0224609 | 403.0794241 | 370.60833 | 276.19077 | 300.63808 | 215.45104 | 645.853 | 105.736 | 93.2945 | 355.0508326 | 306.2640084 | 735.538 |
| 646.143 | 312.343 | 265.902 | 480.3227786 | 404.6459925 | 390.23022 | 277.67268 | 274.19964 | 216.63352 | 646.143 | 92.1548 | 93.6529 | 354.3652336 | 307.393425 | 723.281 |
| 646.432 | 306.902 | 266.913 | 460.7836233 | 406.2159518 | 421.43441 | 279.16088 | 260.52683 | 217.8218 | 646.432 | 100.81 | 94.0122 | 321.8395801 | 308.5261537 | 756.209 |
| 646.721 | 316.539 | 267.927 | 495.241348 | 407.789302 | 406.31257 | 280.65434 | 253.49586 | 219.01357 | 646.721 | 100.234 | 94.3725 | 356.4551512 | 309.6621945 | 732.029 |
| 647.01 | 307.212 | 268.944 | 467.9111708 | 409.3683035 | 393.88862 | 282.15515 | 288.63909 | 220.21231 | 647.01 | 92.1744 | 94.7337 | 343.3636336 | 310.8004434 | 746.762 |
| 647.299 | 334.949 | 269.963 | 478.296186 | 410.9506959 | 442.84783 | 283.66226 | 281.81604 | 221.41569 | 647.299 | 92.1706 | 95.0959 | 334.1141199 | 311.9420043 | 732.659 |
| 647.588 | 306.252 | 270.985 | 461.9026008 | 412.5376095 | 356.24533 | 285.17567 | 315.41555 | 222.62372 | 647.588 | 102.114 | 95.4591 | 363.8311361 | 313.0857732 | 772.849 |
| 647.878 | 313.685 | 272.01 | 503.2256479 | 414.1290441 | 443.50055 | 286.69538 | 279.77749 | 223.83756 | 647.878 | 100.932 | 95.8232 | 359.6181802 | 314.2328542 | 737.072 |
| 648.167 | 315.253 | 273.037 | 468.8606062 | 415.7249999 | 426.28363 | 288.22035 | 268.83088 | 225.0572 | 648.167 | 106.785 | 96.1882 | 368.2847692 | 315.3821433 | 757.981 |
| 648.456 | 299.704 | 274.067 | 471.4613811 | 417.3243465 | 411.3125 | 289.75266 | 293.54904 | 226.2815 | 648.456 | 102.712 | 96.5542 | 350.2648882 | 316.5336404 | 760.157 |
| 648.745 | 316.427 | 275.099 | 479.0975547 | 418.9282142 | 399.47441 | 291.29128 | 278.1966 | 227.51044 | 648.745 | 116.571 | 96.9212 | 359.8765218 | 317.6895536 | 773.932 |
| 649.034 | 314.671 | 276.134 | 470.1434743 | 420.536603 | 427.23883 | 292.83725 | 281.42227 | 228.74634 | 649.034 | 101.232 | 97.2891 | 373.8291782 | 318.8476749 | 787.852 |
| 649.323 | 312.69 | 277.172 | 465.1069454 | 422.149513 | 392.99923 | 294.38847 | 254.81308 | 229.98574 | 649.323 | 102.355 | 97.658 | 351.128235 | 320.0080042 | 763.354 |
| 649.612 | 324.339 | 278.212 | 476.1961252 | 423.7658138 | 393.30489 | 295.94599 | 228.83462 | 231.2321 | 649.612 | 103.379 | 98.0279 | 346.2904015 | 321.1716456 | 764.644 |
| 649.901 | 299.131 | 279.255 | 471.755254 | 425.3866357 | 421.17651 | 297.51087 | 258.6207 | 232.48311 | 649.901 | 110.799 | 98.3987 | 349.8994562 | 322.337495 | 756.637 |
| 650.19 | 302.781 | 280.301 | 493.1333753 | 427.0119788 | 385.68453 | 299.08204 | 245.43807 | 233.73876 | 650.19 | 118.283 | 98.7704 | 365.0290022 | 323.5066565 | 771.456 |
| 650.479 | 328.807 | 281.35 | 481.0292037 | 428.6418429 | 398.38654 | 300.65952 | 316.32157 | 235.00022 | 650.479 | 104.385 | 99.1431 | 362.9136921 | 324.6791301 | 798.293 |
| 650.768 | 319.691 | 282.401 | 459.7313324 | 430.2750979 | 387.34021 | 302.2433 | 307.55057 | 236.26749 | 650.768 | 95.9702 | 99.5168 | 368.4150441 | 325.8538117 | 769.109 |
| 651.057 | 319.586 | 283.454 | 519.0925225 | 431.9140043 | 406.83049 | 303.83443 | 292.66508 | 237.54057 | 651.057 | 108.302 | 99.8915 | 348.7280867 | 327.0307014 | 773.075 |
| 651.346 | 319.786 | 284.511 | 497.6929259 | 433.5563016 | 437.21747 | 305.43081 | 285.01849 | 238.8183 | 651.346 | 98.8615 | 100.267 | 370.5402904 | 328.2120071 | 785.052 |
| 651.635 | 315.511 | 285.57 | 499.1272516 | 435.20312 | 438.53139 | 307.03454 | 310.57646 | 240.10183 | 651.635 | 94.7972 | 100.644 | 353.6365777 | 329.3944169 | 777.459 |
| 651.924 | 347.336 | 286.632 | 493.4894136 | 436.8544595 | 433.5771 | 308.64563 | 262.56538 | 241.39001 | 651.924 | 111.387 | 101.021 | 374.1239526 | 330.5812428 | 763.569 |
| 652.213 | 339.666 | 287.696 | 530.9107332 | 438.5091898 | 439.12574 | 310.26196 | 288.3185 | 242.68515 | 652.213 | 106.47 | 101.4 | 376.4059704 | 331.7702768 | 779.636 |
| 652.502 | 334.628 | 288.763 | 496.0540195 | 440.1695715 | 414.95923 | 311.88565 | 275.16375 | 243.98495 | 652.502 | 102.243 | 101.779 | 388.9289155 | 332.9615187 | 795.967 |
| 652.791 | 343.77 | 289.833 | 496.3072023 | 441.8333441 | 439.64579 | 313.51669 | 297.26373 | 245.28939 | 652.791 | 103.882 | 102.16 | 392.619668 | 334.1560728 | 801.781 |
| 653.08 | 318.167 | 290.906 | 521.368907 | 443.5016378 | 438.0729 | 315.15298 | 296.2009 | 246.6008 | 653.08 | 113.908 | 102.541 | 371.7227002 | 335.3528349 | 797.312 |
| 653.369 | 360.363 | 291.981 | 538.4101428 | 445.1744526 | 438.20344 | 316.79662 | 306.96282 | 247.91685 | 653.369 | 118.421 | 102.924 | 394.5130693 | 336.5529091 | 803.093 |
| 653.658 | 350.923 | 293.059 | 497.5177325 | 446.8506583 | 416.82293 | 318.44762 | 304.39111 | 249.23872 | 653.658 | 100.288 | 103.307 | 367.6819721 | 337.7562953 | 828.133 |
| 653.947 | 353.68 | 294.139 | 524.651241 | 448.5325153 | 411.09386 | 320.10386 | 308.78416 | 250.56639 | 653.947 | 112.058 | 103.691 | 382.853471 | 338.9618896 | 827.002 |
| 654.236 | 334.924 | 295.222 | 509.7032842 | 450.2177632 | 435.10329 | 321.76851 | 317.04639 | 251.89987 | 654.236 | 104.588 | 104.076 | 367.0570945 | 340.170796 | 796.407 |
| 654.524 | 358.724 | 296.308 | 513.472769 | 451.9075323 | 451.70357 | 323.43946 | 332.16651 | 253.238 | 654.524 | 108.022 | 104.463 | 397.0490127 | 341.3819104 | 813.209 |
| 654.813 | 344.444 | 297.397 | 503.479961 | 453.6018224 | 437.86169 | 325.11671 | 308.1964 | 254.58309 | 654.813 | 120.213 | 104.85 | 393.8650072 | 342.5963369 | 818.298 |
| 655.102 | 348.466 | 298.488 | 530.7344094 | 455.3006337 | 459.04692 | 326.80026 | 338.03011 | 255.93283 | 655.102 | 119.41 | 105.238 | 381.247116 | 343.8140755 | 821.541 |
| 655.391 | 348.047 | 299.582 | 535.4917591 | 457.0039661 | 455.28238 | 328.49117 | 318.70162 | 257.28838 | 655.391 | 105.928 | 105.627 | 401.5744074 | 345.0340221 | 812.35 |
| 655.68 | 353.493 | 300.679 | 541.027872 | 458.7118196 | 463.27314 | 330.18942 | 317.81418 | 258.64974 | 655.68 | 111.94 | 106.018 | 387.7056569 | 346.2561767 | 796.718 |
| 655.968 | 355.753 | 301.779 | 562.2567962 | 460.4230639 | 445.44915 | 331.89398 | 312.22937 | 260.0169 | 655.968 | 113.294 | 106.409 | 399.8245292 | 347.4827475 | 838.819 |
| 656.257 | 360.332 | 302.881 | 509.6512913 | 462.1388294 | 442.15796 | 333.60589 | 322.15612 | 261.38988 | 656.257 | 103.943 | 106.801 | 398.9335818 | 348.7104223 | 834.376 |
| 656.546 | 347.267 | 303.986 | 537.4053237 | 463.8602462 | 463.96725 | 335.3241 | 337.43538 | 262.7675 | 656.546 | 112.241 | 107.194 | 397.7401317 | 349.9425131 | 826.766 |
| 656.835 | 339.216 | 305.093 | 542.1344164 | 465.585054 | 463.47373 | 337.04966 | 365.16665 | 264.15209 | 656.835 | 125.187 | 107.588 | 387.1558529 | 351.1768121 | 801.654 |
| 657.124 | 349.441 | 306.204 | 527.3300052 | 467.3143828 | 452.04532 | 338.78258 | 333.29904 | 265.54248 | 657.124 | 97.372 | 107.983 | 412.0626361 | 352.413319 | 859.121 |
| 657.412 | 359.533 | 307.317 | 542.4034231 | 469.0471025 | 451.50935 | 340.52179 | 364.6091 | 266.93752 | 657.412 | 110.037 | 108.379 | 393.8175342 | 353.6531381 | 849.725 |
| 657.701 | 350.197 | 308.432 | 537.1272747 | 470.7854735 | 499.66619 | 342.26836 | 402.89785 | 268.33954 | 657.701 | 108.698 | 108.776 | 416.7580506 | 354.8962692 | 854.322 |
| 657.99 | 356.458 | 309.551 | 529.152017 | 472.5283657 | 428.03907 | 344.02123 | 326.86977 | 269.74619 | 657.99 | 119.717 | 109.175 | 382.1800163 | 356.1416084 | 864.078 |
| 658.278 | 367.455 | 310.672 | 542.1219833 | 474.2746488 | 488.55723 | 345.78145 | 301.1271 | 271.15982 | 658.278 | 117.274 | 109.574 | 396.423031 | 357.3902596 | 837.794 |
| 658.567 | 362.579 | 311.796 | 515.9243469 | 476.0254529 | 491.26469 | 347.54902 | 336.07403 | 272.57926 | 658.567 | 108.107 | 109.974 | 398.3407208 | 358.6422229 | 846.405 |
| 658.856 | 370.323 | 312.923 | 548.1995004 | 477.7819085 | 445.43217 | 349.32394 | 316.24258 | 274.00334 | 658.856 | 118.36 | 110.375 | 408.6909466 | 359.8963943 | 849.24 |
| 659.144 | 363.319 | 314.052 | 564.8801768 | 479.5417549 | 514.85808 | 351.10517 | 343.05969 | 275.43439 | 659.144 | 118.543 | 110.777 | 404.0043644 | 361.1538777 | 863.684 |
| 659.433 | 339.046 | 315.184 | 541.5534523 | 481.3061224 | 452.19603 | 352.89375 | 310.48353 | 276.87125 | 659.433 | 117.575 | 111.18 | 391.5962377 | 362.4135692 | 861.091 |
| 659.722 | 353.903 | 316.319 | 548.7126476 | 483.075011 | 495.75943 | 354.68968 | 324.93343 | 278.31276 | 659.722 | 123.708 | 111.584 | 422.1114633 | 363.6765727 | 874.15 |
| 660.01 | 368.836 | 317.456 | 564.348945 | 484.8484208 | 494.88807 | 356.49191 | 358.30759 | 279.76123 | 660.01 | 126.146 | 111.989 | 411.874952 | 364.9417843 | 859.676 |
| 660.299 | 364.239 | 318.597 | 580.3729286 | 486.6263517 | 484.32465 | 358.30254 | 351.17324 | 281.21551 | 660.299 | 110.466 | 112.395 | 395.3984967 | 366.210308 | 851.76 |
| 660.588 | 360.322 | 319.74 | 569.7245581 | 488.4076734 | 489.41054 | 360.11947 | 367.73371 | 282.6756 | 660.588 | 113.52 | 112.802 | 406.1285067 | 367.4821437 | 882.214 |
| 660.876 | 343.225 | 320.886 | 563.5939178 | 490.1946465 | 468.00031 | 361.94376 | 335.05069 | 284.14266 | 660.876 | 116.184 | 113.21 | 447.3174371 | 368.7561875 | 908.18 |
| 661.165 | 370.408 | 322.034 | 571.2707815 | 491.9850105 | 472.28384 | 363.7754 | 346.64429 | 285.61437 | 661.165 | 108.588 | 113.618 | 406.6871429 | 370.0335434 | 883.771 |
| 661.453 | 361.892 | 323.185 | 573.2657262 | 493.7810259 | 505.9153 | 365.61334 | 384.85987 | 287.09305 | 661.453 | 131.396 | 114.028 | 446.164836 | 371.3131073 | 901.875 |
| 661.742 | 377.355 | 324.34 | 557.3909396 | 495.5804321 | 491.83781 | 367.45968 | 363.94468 | 288.57753 | 661.742 | 111.992 | 114.439 | 420.2666391 | 372.5959833 | 861.253 |
| 662.03 | 362.449 | 325.496 | 593.0998846 | 497.3854897 | 517.92214 | 369.31232 | 369.87448 | 290.06782 | 662.03 | 128.366 | 114.851 | 404.124703 | 373.8821713 | 882.967 |
| 662.319 | 364.729 | 326.656 | 552.9240719 | 499.1939382 | 527.80207 | 371.17336 | 352.21865 | 291.56392 | 662.319 | 108.907 | 115.264 | 417.9205879 | 375.1705674 | 869.551 |
| 662.607 | 363.65 | 327.818 | 575.0945197 | 501.0069077 | 533.20849 | 373.04071 | 353.75773 | 293.06583 | 662.607 | 128.018 | 115.678 | 410.224436 | 376.4622756 | 894.682 |
| 662.896 | 368.915 | 328.983 | 590.7839403 | 502.8255287 | 472.2191 | 374.91646 | 352.69141 | 294.5747 | 662.896 | 116.085 | 116.093 | 438.277688 | 377.7572959 | 901.5 |
| 663.184 | 372.778 | 330.151 | 558.5460861 | 504.6475405 | 572.23364 | 376.79851 | 335.30972 | 296.08939 | 663.184 | 121.37 | 116.509 | 425.9070981 | 379.0545241 | 897.914 |
| 663.473 | 369.529 | 331.322 | 578.7690609 | 506.4740735 | 518.02934 | 378.68791 | 342.13276 | 297.60988 | 663.473 | 112.472 | 116.926 | 432.2861493 | 380.3550645 | 885.481 |
| 663.761 | 385.657 | 332.495 | 601.1180929 | 508.3051275 | 514.3359 | 380.58571 | 375.07831 | 299.13734 | 663.761 | 125.069 | 117.344 | 419.0577327 | 381.6578129 | 907.164 |
| 664.05 | 382.855 | 333.671 | 573.5607293 | 510.1407027 | 527.15678 | 382.48982 | 382.88985 | 300.66945 | 664.05 | 112.368 | 117.763 | 432.7619836 | 382.9638734 | 878.863 |
| 664.338 | 388.547 | 334.85 | 594.4155308 | 511.980799 | 523.69897 | 384.40127 | 371.85612 | 302.20852 | 664.338 | 135.768 | 118.183 | 438.6497441 | 384.273246 | 898.451 |
| 664.627 | 380.595 | 336.032 | 596.1539019 | 513.8254164 | 534.86842 | 386.32113 | 364.68808 | 303.75457 | 664.627 | 124.975 | 118.604 | 430.7261632 | 385.5848266 | 920.186 |
| 664.915 | 361.116 | 337.216 | 578.9397332 | 515.674555 | 522.18339 | 388.24729 | 399.45264 | 305.30642 | 664.915 | 131.148 | 119.026 | 444.651219 | 386.8997193 | 914.423 |
| 665.203 | 371.419 | 338.404 | 618.7843737 | 517.5282146 | 553.69325 | 390.18185 | 345.44672 | 306.86409 | 665.203 | 114.313 | 119.449 | 438.2865201 | 388.21682 | 916.862 |
| 665.492 | 383.879 | 339.594 | 588.5991085 | 519.3852651 | 534.79625 | 392.12377 | 386.50581 | 308.42756 | 665.492 | 124.746 | 119.873 | 448.3949646 | 389.5372328 | 889.868 |
| 665.78 | 388.257 | 340.787 | 592.0475936 | 521.247967 | 528.46647 | 394.07303 | 373.66816 | 309.998 | 665.78 | 121.045 | 120.298 | 423.930895 | 390.8609577 | 916.836 |
| 666.069 | 378.725 | 341.982 | 586.5815582 | 523.1151901 | 497.1434 | 396.02965 | 413.97107 | 311.57424 | 666.069 | 123.615 | 120.724 | 435.6125738 | 392.1868906 | 915.349 |
| 666.357 | 394.131 | 343.18 | 602.2461126 | 524.9869342 | 525.62104 | 397.99362 | 385.17117 | 313.15746 | 666.357 | 117.725 | 121.151 | 435.4270978 | 393.5161356 | 904.32 |
| 666.645 | 384.677 | 344.382 | 602.8813302 | 526.8631995 | 531.52098 | 399.96599 | 366.85325 | 314.74648 | 666.645 | 122.733 | 121.579 | 434.473221 | 394.8486927 | 944.533 |
| 666.934 | 389.846 | 345.586 | 604.7485532 | 528.7439859 | 551.73297 | 401.94466 | 388.24351 | 316.34248 | 666.934 | 122.316 | 122.008 | 436.994812 | 396.1834578 | 918.683 |
| 667.222 | 384.886 | 346.792 | 576.5017186 | 530.6281631 | 523.65439 | 403.93174 | 392.62843 | 317.94312 | 667.222 | 129.202 | 122.438 | 458.2837083 | 397.521535 | 941.885 |
| 667.51 | 383.762 | 348.002 | 579.1906554 | 532.5179917 | 516.97968 | 405.92721 | 383.54846 | 319.55189 | 667.51 | 127.846 | 122.869 | 449.6005589 | 398.8618202 | 939.825 |
| 667.799 | 397.101 | 349.214 | 593.2999442 | 534.4123415 | 524.40581 | 407.92899 | 387.96706 | 321.16647 | 667.799 | 124.824 | 123.301 | 444.8930003 | 400.2065215 | 949.316 |
| 668.087 | 404.349 | 350.429 | 601.2379026 | 536.3112124 | 536.6557 | 409.93917 | 384.20242 | 322.78686 | 668.087 | 126.219 | 123.734 | 441.2530329 | 401.5523269 | 938.166 |
| 668.375 | 393.146 | 351.647 | 617.7739032 | 538.2146043 | 565.61836 | 411.95671 | 374.73564 | 324.41421 | 668.375 | 129.211 | 124.168 | 462.8598882 | 402.9025483 | 941.951 |
| 668.663 | 392.875 | 352.868 | 630.2521975 | 540.1213872 | 574.13448 | 413.98159 | 399.76975 | 326.04854 | 668.663 | 121.391 | 124.603 | 462.1687691 | 404.2549778 | 926.641 |
| 668.952 | 403.186 | 354.091 | 633.9448232 | 542.0338214 | 583.76606 | 416.01487 | 387.78585 | 327.68867 | 668.952 | 139.449 | 125.039 | 446.255366 | 405.6096154 | 951.772 |
| 669.24 | 407.35 | 355.318 | 607.1763952 | 543.9507768 | 594.47808 | 418.05551 | 440.27361 | 329.33461 | 669.24 | 130.026 | 125.476 | 477.8127903 | 406.968669 | 944.511 |
| 669.528 | 399.411 | 356.547 | 615.2658112 | 545.8722533 | 577.1338 | 420.1035 | 427.13628 | 330.98868 | 669.528 | 120.061 | 125.915 | 452.8607421 | 408.3299307 | 943.942 |
| 669.816 | 383.852 | 357.779 | 614.5695586 | 547.7982509 | 586.46184 | 422.15989 | 376.42572 | 332.6474 | 669.816 | 138.449 | 126.354 | 467.1258372 | 409.6934005 | 936.966 |
| 670.105 | 403.107 | 359.013 | 591.0676406 | 549.7287696 | 555.06873 | 424.22364 | 407.32807 | 334.31425 | 670.105 | 129.62 | 126.794 | 470.3782922 | 411.0612863 | 964.202 |
| 670.393 | 388.392 | 360.251 | 629.2281635 | 551.6638094 | 591.5764 | 426.29578 | 403.46702 | 335.98691 | 670.393 | 128.973 | 127.235 | 474.4090841 | 412.4302762 | 936.887 |
| 670.681 | 433.561 | 361.491 | 650.5791581 | 553.6033704 | 593.1928 | 428.37528 | 417.31058 | 337.66537 | 670.681 | 142.863 | 127.677 | 469.408959 | 413.8036821 | 966.85 |
| 670.969 | 416.345 | 362.734 | 644.5185952 | 555.5474525 | 604.09267 | 430.46318 | 427.49637 | 339.35081 | 670.969 | 123.743 | 128.121 | 457.949189 | 415.1792961 | 949.35 |
| 671.257 | 401.886 | 363.98 | 633.8849183 | 557.4960557 | 599.34746 | 432.55843 | 411.47719 | 341.04321 | 671.257 | 126.056 | 128.565 | 486.7874021 | 416.5582222 | 938.79 |
| 671.545 | 417.166 | 365.229 | 630.2024651 | 559.44918 | 561.24992 | 434.66103 | 431.11349 | 342.74259 | 671.545 | 123.512 | 129.01 | 461.2999021 | 417.9393563 | 958.736 |
| 671.833 | 397.64 | 366.481 | 630.3640952 | 561.4068255 | 607.43905 | 436.77203 | 434.08478 | 344.44777 | 671.833 | 131.137 | 129.457 | 474.1021654 | 419.3249065 | 976.593 |
| 672.122 | 411.996 | 367.735 | 639.1768906 | 563.3701223 | 632.50026 | 438.89144 | 441.96717 | 346.15992 | 672.122 | 134.758 | 129.904 | 480.4227032 | 420.7115608 | 972.504 |
| 672.41 | 400.141 | 368.992 | 632.3590399 | 565.33681 | 566.86861 | 441.0182 | 410.92893 | 347.87904 | 672.41 | 130.388 | 130.352 | 491.0820558 | 422.1026311 | 987.665 |
| 672.698 | 407.043 | 370.252 | 661.6932041 | 567.3080188 | 580.7816 | 443.15336 | 396.10384 | 349.60397 | 672.698 | 128.075 | 130.802 | 478.2499839 | 423.4959095 | 985.903 |
| 672.986 | 412.805 | 371.515 | 629.1264383 | 569.284879 | 578.6791 | 445.29587 | 402.58423 | 351.33586 | 672.986 | 130.712 | 131.252 | 475.5716214 | 424.8925 | 982.266 |
| 673.274 | 434.031 | 372.781 | 649.8467364 | 571.2651301 | 565.32331 | 447.44678 | 457.87136 | 353.07473 | 673.274 | 142.746 | 131.703 | 487.5955477 | 426.2912985 | 969.083 |
| 673.562 | 429.057 | 374.05 | 650.6334115 | 573.2510325 | 635.72034 | 449.6061 | 448.12232 | 354.82057 | 673.562 | 130.415 | 132.156 | 477.4495664 | 427.6934091 | 987.112 |
| 673.85 | 408.697 | 375.321 | 672.234198 | 575.2403258 | 620.7577 | 451.77277 | 451.02159 | 356.57337 | 673.85 | 137.162 | 132.609 | 492.4554617 | 429.0988317 | 986.846 |
| 674.138 | 409.58 | 376.595 | 645.8658893 | 577.2352705 | 569.18125 | 453.94784 | 443.90118 | 358.33199 | 674.138 | 126.198 | 133.064 | 502.003062 | 430.5064624 | 995.585 |
| 674.426 | 420.602 | 377.872 | 643.3091953 | 579.2347364 | 606.80013 | 456.13026 | 433.84317 | 360.09757 | 674.426 | 134.916 | 133.519 | 487.5789873 | 431.9174052 | 1022.37 |
| 674.714 | 425.493 | 379.152 | 658.6165812 | 581.2387233 | 634.70571 | 458.32213 | 425.32888 | 361.87012 | 674.714 | 128.912 | 133.976 | 493.9536224 | 433.3316601 | 991.717 |
| 675.002 | 413.591 | 380.435 | 640.9740362 | 583.2472314 | 587.79806 | 460.52136 | 446.40785 | 363.64964 | 675.002 | 133.856 | 134.434 | 479.3661522 | 434.748123 | 988.569 |
| 675.29 | 416.724 | 381.721 | 639.9873015 | 585.2602605 | 627.23818 | 462.72898 | 431.02869 | 365.43497 | 675.29 | 133.027 | 134.892 | 488.0746942 | 436.1678979 | 1000.08 |
| 675.578 | 432.856 | 383.009 | 653.4014679 | 587.2778109 | 641.59905 | 464.94396 | 458.38361 | 367.22843 | 675.578 | 138.039 | 135.352 | 492.099966 | 437.590985 | 1009.49 |
| 675.866 | 428.861 | 384.3 | 647.4256761 | 589.2998823 | 600.46398 | 467.16734 | 464.00211 | 369.0277 | 675.866 | 124.926 | 135.813 | 498.1588501 | 439.0162801 | 1021.37 |
| 676.154 | 426.578 | 385.594 | 653.5111051 | 591.3264748 | 596.48505 | 469.39913 | 449.99361 | 370.8351 | 676.154 | 154.773 | 136.274 | 495.8216311 | 440.4448872 | 1008.65 |
| 676.442 | 418.86 | 386.891 | 657.8332969 | 593.3575885 | 654.99411 | 471.63931 | 475.93257 | 372.64831 | 676.442 | 134.078 | 136.737 | 480.685461 | 441.8768065 | 996.476 |
| 676.73 | 423.405 | 388.191 | 680.8491943 | 595.3943536 | 593.29788 | 473.8879 | 465.8583 | 374.46848 | 676.73 | 143.565 | 137.201 | 507.0705325 | 443.3109337 | 1028.36 |
| 677.018 | 439.648 | 389.493 | 675.4848841 | 597.4345095 | 645.58435 | 476.14489 | 453.12403 | 376.29563 | 677.018 | 141.834 | 137.666 | 476.9880843 | 444.7483731 | 994.807 |
| 677.305 | 422.668 | 390.799 | 671.0496643 | 599.4803168 | 618.90461 | 478.40923 | 437.28257 | 378.1309 | 677.305 | 141.166 | 138.132 | 498.6777414 | 446.1891245 | 1007 |
| 677.593 | 437.896 | 392.107 | 664.8455558 | 601.5306452 | 626.45917 | 480.68197 | 430.25277 | 379.97199 | 677.593 | 139.833 | 138.599 | 511.4601323 | 447.632084 | 1001.29 |
| 677.881 | 432.194 | 393.418 | 672.0872616 | 603.5854948 | 646.86962 | 482.96417 | 452.06584 | 381.82004 | 677.881 | 141.016 | 139.067 | 516.2405565 | 449.0783555 | 1028.01 |
| 678.169 | 441.005 | 394.733 | 688.1202874 | 605.6448654 | 625.98687 | 485.25371 | 434.57496 | 383.67507 | 678.169 | 134.674 | 139.536 | 510.3318196 | 450.5279392 | 1022.68 |
| 678.457 | 418.279 | 396.049 | 712.1421342 | 607.7087572 | 629.67182 | 487.55166 | 414.21268 | 385.53706 | 678.457 | 140.793 | 140.006 | 484.1156638 | 451.9797308 | 1031.9 |
| 678.745 | 426.17 | 397.369 | 694.6160082 | 609.7771702 | 638.70905 | 489.85801 | 493.12726 | 387.40718 | 678.745 | 137.733 | 140.477 | 525.3608995 | 453.4348346 | 1049.07 |
| 679.032 | 444.05 | 398.692 | 675.38768 | 611.8501042 | 615.78748 | 492.17277 | 479.28718 | 389.28312 | 679.032 | 148.55 | 140.949 | 511.1907504 | 454.8932504 | 1028.94 |
| 679.32 | 428.72 | 400.017 | 671.9787547 | 613.9286896 | 640.63218 | 494.49592 | 479.09901 | 391.16602 | 679.32 | 141.783 | 141.422 | 505.5712678 | 456.3538742 | 1050.83 |
| 679.608 | 432.537 | 401.346 | 697.5739517 | 616.0106659 | 629.60601 | 496.82748 | 433.55394 | 393.05705 | 679.608 | 148 | 141.896 | 516.4536332 | 457.8178102 | 1036.83 |
| 679.896 | 446.795 | 402.677 | 684.6288516 | 618.0982936 | 633.28353 | 499.16744 | 444.79791 | 394.95389 | 679.896 | 151.737 | 142.372 | 503.9262719 | 459.2850582 | 1044.8 |
| 680.183 | 437.099 | 404.011 | 688.6809064 | 620.1904424 | 659.35831 | 501.5158 | 464.55386 | 396.85886 | 680.183 | 155.806 | 142.848 | 501.2909664 | 460.7545142 | 1046.12 |
| 680.471 | 442.017 | 405.348 | 721.1651617 | 622.2871123 | 662.46588 | 503.87256 | 480.08518 | 398.7708 | 680.471 | 153.136 | 143.325 | 521.7661971 | 462.2272824 | 1050.52 |
| 680.759 | 432.605 | 406.687 | 709.0225606 | 624.3883034 | 664.22876 | 506.23878 | 489.43928 | 400.68971 | 680.759 | 144.341 | 143.804 | 525.1334705 | 463.7033626 | 1053.03 |
| 681.047 | 453.794 | 408.03 | 710.6004319 | 626.4940155 | 686.53475 | 508.61234 | 503.53606 | 402.61559 | 681.047 | 136.615 | 144.283 | 523.7942893 | 465.1816508 | 1059.73 |
| 681.334 | 450.73 | 409.376 | 708.6631315 | 628.6042488 | 652.24102 | 510.99536 | 489.18722 | 404.54844 | 681.334 | 156.074 | 144.763 | 516.0561845 | 466.6632511 | 1052.3 |
| 681.622 | 452.509 | 410.724 | 714.0805649 | 630.7201335 | 680.53823 | 513.38573 | 488.14065 | 406.48942 | 681.622 | 139.086 | 145.245 | 507.2526965 | 468.1481635 | 1005.82 |
| 681.91 | 457.352 | 412.075 | 750.3376958 | 632.839409 | 705.98258 | 515.78556 | 489.56821 | 408.4362 | 681.91 | 149.82 | 145.727 | 514.5315272 | 469.6352839 | 989.994 |
| 682.197 | 452.951 | 413.429 | 624.4855075 | 634.964336 | 666.90862 | 518.19378 | 462.98458 | 410.39112 | 682.197 | 131.043 | 146.211 | 467.4316519 | 471.1268205 | 994.079 |
| 682.485 | 424.978 | 414.786 | 630.3867008 | 637.093784 | 695.80229 | 520.61041 | 488.69704 | 412.35417 | 682.485 | 149.143 | 146.696 | 467.5619267 | 472.619461 | 996.784 |
| 682.773 | 456.824 | 416.146 | 668.0657243 | 639.2277532 | 650.18628 | 523.03649 | 501.89012 | 414.32303 | 682.773 | 182.167 | 147.181 | 492.421237 | 474.1165177 | 997.956 |
| 683.06 | 450.665 | 417.509 | 666.7399056 | 641.3673738 | 686.7364 | 525.46992 | 460.96694 | 416.30002 | 683.06 | 145.127 | 147.668 | 497.4633149 | 475.6157824 | 1045.46 |
| 683.348 | 464.256 | 418.875 | 694.2950086 | 643.5103852 | 652.02769 | 527.9128 | 480.21063 | 418.28397 | 683.348 | 127.923 | 148.156 | 481.788381 | 477.1183592 | 1089.42 |
| 683.636 | 445.002 | 420.243 | 676.6931537 | 645.659048 | 695.35654 | 530.36409 | 537.27496 | 420.2749 | 683.636 | 141.704 | 148.645 | 504.0907715 | 478.623144 | 1064.33 |
| 683.923 | 476.594 | 421.615 | 736.4182941 | 647.8111017 | 649.55585 | 532.82483 | 529.21949 | 422.27396 | 683.923 | 135.178 | 149.135 | 524.971179 | 480.1312409 | 1102.42 |
| 684.211 | 467.574 | 422.989 | 744.4127666 | 649.9688067 | 709.95408 | 535.29397 | 534.00863 | 424.27999 | 684.211 | 149.351 | 149.626 | 540.3149056 | 481.6426499 | 1098.16 |
| 684.498 | 467.274 | 424.366 | 751.391117 | 652.1310329 | 718.06796 | 537.77151 | 486.82459 | 426.29298 | 684.498 | 135.386 | 150.118 | 575.2285632 | 483.1573709 | 1076.18 |
| 684.786 | 447.515 | 425.746 | 748.7519126 | 654.2977802 | 711.56518 | 540.25745 | 501.30934 | 428.31411 | 684.786 | 140.35 | 150.611 | 581.6296948 | 484.6743 | 1114.76 |
| 685.073 | 474.625 | 427.129 | 744.2511365 | 656.4701789 | 700.24927 | 542.75285 | 532.9365 | 430.34221 | 685.073 | 153.025 | 151.105 | 532.7722131 | 486.1945411 | 1090.49 |
| 685.361 | 455.707 | 428.515 | 748.8830251 | 658.6470987 | 728.41806 | 545.25665 | 516.5433 | 432.37727 | 685.361 | 145.797 | 151.6 | 555.7083134 | 487.7180944 | 1081.03 |
| 685.649 | 448.815 | 429.904 | 751.2430503 | 660.8274094 | 708.17635 | 547.7699 | 489.72154 | 434.42047 | 685.649 | 138.053 | 152.096 | 567.2619254 | 489.2438557 | 1109.55 |
| 685.936 | 467.579 | 431.295 | 743.5006304 | 663.0133715 | 685.88734 | 550.29155 | 551.97343 | 436.47064 | 685.936 | 134.402 | 152.593 | 568.4597916 | 490.772929 | 1131.71 |
| 686.223 | 455.411 | 432.69 | 771.0772087 | 665.2038546 | 708.40136 | 552.8216 | 527.18907 | 438.52893 | 686.223 | 137.739 | 153.092 | 578.2127403 | 492.3053144 | 1150.04 |
| 686.511 | 446.894 | 434.087 | 738.950122 | 667.3999892 | 751.27913 | 555.36111 | 548.53403 | 440.5942 | 686.511 | 142.544 | 153.591 | 578.7426719 | 493.8399079 | 1117.08 |
| 686.798 | 466.608 | 435.488 | 784.1262946 | 669.5995147 | 724.82864 | 557.90902 | 549.907 | 442.6676 | 686.798 | 129.394 | 154.091 | 564.0602556 | 495.3778135 | 1137.64 |
| 687.086 | 470.219 | 436.891 | 760.9781544 | 671.8046915 | 741.8959 | 560.46638 | 539.99883 | 444.74797 | 687.086 | 134.635 | 154.593 | 545.1935881 | 496.9190311 | 1129.58 |
| 687.373 | 473.103 | 438.297 | 748.1969449 | 674.0143894 | 745.03532 | 563.03319 | 539.25427 | 446.8353 | 687.373 | 130.776 | 155.095 | 569.7802044 | 498.4624567 | 1142.8 |
| 687.661 | 472.972 | 439.706 | 746.4303169 | 676.2286085 | 773.76131 | 565.60736 | 563.17327 | 448.93193 | 687.661 | 126.012 | 155.599 | 566.7717387 | 500.0091945 | 1122.24 |
| 687.948 | 465.583 | 441.118 | 800.3582497 | 678.4473487 | 799.973 | 568.19203 | 504.54779 | 451.03437 | 687.948 | 119.146 | 156.104 | 565.7549326 | 501.5592443 | 1108.54 |
| 688.236 | 464.875 | 442.533 | 730.2797419 | 680.67061 | 729.59826 | 570.78509 | 549.94301 | 453.14494 | 688.236 | 138.64 | 156.609 | 555.8672929 | 503.1126062 | 1121.33 |
| 688.523 | 452.496 | 443.95 | 770.183157 | 682.8995227 | 742.88506 | 573.38762 | 581.15317 | 455.26364 | 688.523 | 139.45 | 157.116 | 543.4194214 | 504.6681761 | 1127.89 |
| 688.81 | 453.246 | 445.371 | 802.4910886 | 685.1329566 | 770.83416 | 575.99854 | 584.40556 | 457.38931 | 688.81 | 140.595 | 157.624 | 562.6482088 | 506.2270581 | 1133.29 |
| 689.098 | 490.916 | 446.795 | 772.2820875 | 687.3709115 | 733.34688 | 578.61892 | 589.81266 | 459.52311 | 689.098 | 151.85 | 158.133 | 574.6434304 | 507.7881482 | 1182.5 |
| 689.385 | 493.04 | 448.221 | 798.1372489 | 689.6133876 | 734.89218 | 581.24769 | 501.07006 | 461.66504 | 689.385 | 128.687 | 158.643 | 606.1401335 | 509.3536543 | 1170.29 |
| 689.672 | 492.154 | 449.65 | 783.357704 | 691.8615151 | 735.95033 | 583.88592 | 554.73099 | 463.81394 | 689.672 | 119.482 | 159.154 | 612.2244102 | 510.9213685 | 1177.9 |
| 689.96 | 467.349 | 451.082 | 755.774344 | 694.1130334 | 732.90643 | 586.53361 | 595.53454 | 465.96981 | 689.96 | 125.65 | 159.666 | 585.3535681 | 512.4912907 | 1198.89 |
| 690.247 | 463.513 | 452.518 | 774.2340815 | 696.3702032 | 774.55943 | 589.19074 | 568.97414 | 468.13497 | 690.247 | 131.982 | 160.179 | 552.0319132 | 514.0656291 | 1201.35 |
| 690.534 | 484.58 | 453.956 | 804.2961462 | 698.631894 | 795.33392 | 591.85628 | 583.64357 | 470.3071 | 690.534 | 124.779 | 160.693 | 582.3097737 | 515.6421755 | 1168.59 |
| 690.822 | 492.563 | 455.397 | 757.258402 | 700.8992362 | 756.63249 | 594.53127 | 581.43427 | 472.48736 | 690.822 | 145.119 | 161.208 | 579.115832 | 517.2220339 | 1172.35 |
| 691.109 | 483.001 | 456.841 | 771.4569829 | 703.1699693 | 790.52079 | 597.21466 | 581.04863 | 474.67459 | 691.109 | 127.936 | 161.725 | 596.9910859 | 518.8041004 | 1147.18 |
| 691.396 | 491.273 | 458.287 | 794.7170208 | 705.4463538 | 818.71929 | 599.90855 | 533.26987 | 476.86996 | 691.396 | 142.881 | 162.242 | 568.2875638 | 520.389479 | 1160.63 |
| 691.683 | 484.261 | 459.737 | 830.3163286 | 707.7272594 | 810.01001 | 602.61085 | 567.01457 | 479.07345 | 691.683 | 138.679 | 162.76 | 613.085549 | 521.9781697 | 1211.73 |
| 691.971 | 513.937 | 461.19 | 840.6064003 | 710.0126862 | 747.69608 | 605.32259 | 578.31895 | 481.28508 | 691.971 | 135.824 | 163.28 | 604.7556873 | 523.5701724 | 1220.74 |
| 692.258 | 495.865 | 462.645 | 824.3767057 | 712.3037643 | 786.09928 | 608.0438 | 556.85434 | 483.50367 | 692.258 | 141.912 | 163.8 | 620.6028489 | 525.1643832 | 1198.8 |
| 692.545 | 473.559 | 464.104 | 802.2333847 | 714.5982333 | 765.34814 | 610.77445 | 565.40348 | 485.7304 | 692.545 | 134.943 | 164.322 | 585.3193433 | 526.761906 | 1215.25 |
| 692.832 | 500.043 | 465.565 | 795.7387942 | 716.8983536 | 805.29346 | 613.5135 | 543.15249 | 487.96525 | 692.832 | 164.244 | 164.845 | 593.9053386 | 528.3616369 | 1221.45 |
| 693.119 | 489.211 | 467.029 | 839.0036628 | 719.2029951 | 754.68813 | 616.26306 | 515.55364 | 490.20824 | 693.119 | 156.852 | 165.368 | 581.1108035 | 529.9657839 | 1214.64 |
| 693.407 | 470.765 | 468.497 | 799.5806169 | 721.513288 | 851.3393 | 619.02102 | 575.62876 | 492.45936 | 693.407 | 147.597 | 165.893 | 586.7236619 | 531.5721389 | 1202.06 |
| 693.694 | 457.798 | 469.967 | 791.9116652 | 723.8269718 | 787.2901 | 621.78948 | 600.59781 | 494.71744 | 693.694 | 154.926 | 166.419 | 596.4523221 | 533.180702 | 1178.34 |
| 693.981 | 489.432 | 471.44 | 810.557899 | 726.1463069 | 791.63943 | 624.56634 | 589.26672 | 496.98366 | 693.981 | 123.933 | 166.946 | 623.2723792 | 534.7936812 | 1210.01 |
| 694.268 | 508.664 | 472.916 | 793.7314164 | 728.4701632 | 847.97701 | 627.35266 | 582.55402 | 499.25801 | 694.268 | 159.485 | 167.474 | 601.5959703 | 536.4088684 | 1225.63 |
| 694.555 | 493.255 | 474.395 | 839.3540497 | 730.7996708 | 810.27216 | 630.14843 | 553.88653 | 501.54049 | 694.555 | 159.291 | 168.003 | 603.6781598 | 538.0273678 | 1197.32 |
| 694.842 | 475.823 | 475.876 | 817.2841969 | 733.1325693 | 831.38628 | 632.95365 | 639.16767 | 503.8311 | 694.842 | 145.683 | 168.533 | 617.9576072 | 539.6480751 | 1205.95 |
| 695.129 | 490.275 | 477.361 | 828.4637992 | 735.4711192 | 847.77323 | 635.76937 | 564.82386 | 506.12985 | 695.129 | 149.067 | 169.065 | 624.7230668 | 541.2720945 | 1225.15 |
| 695.416 | 510.005 | 478.849 | 803.795432 | 737.8141903 | 815.90252 | 638.59349 | 566.69746 | 508.43556 | 695.416 | 164.913 | 169.597 | 591.4069321 | 542.899426 | 1210.68 |
| 695.703 | 476.426 | 480.339 | 811.5808026 | 740.1617824 | 830.02671 | 641.42707 | 630.78579 | 510.75056 | 695.703 | 129.526 | 170.13 | 635.1229737 | 544.5300696 | 1239.78 |
| 695.99 | 517.999 | 481.833 | 861.1334205 | 742.515026 | 785.40305 | 644.2701 | 612.76872 | 513.0737 | 695.99 | 149.593 | 170.665 | 602.1060295 | 546.1629212 | 1247 |
| 696.277 | 520.447 | 483.329 | 813.2457055 | 744.8727906 | 856.5844 | 647.12363 | 641.66852 | 515.4038 | 696.277 | 141.191 | 171.2 | 615.3123655 | 547.7990849 | 1270.21 |
| 696.564 | 508.499 | 484.829 | 840.0672566 | 747.2350764 | 850.41488 | 649.98556 | 596.84479 | 517.7432 | 696.564 | 160.584 | 171.737 | 624.3995878 | 549.4385607 | 1257.24 |
| 696.851 | 494.125 | 486.331 | 870.6300354 | 749.6018833 | 847.47075 | 652.858 | 613.69449 | 520.08956 | 696.851 | 165.472 | 172.274 | 633.8147051 | 551.0802445 | 1238.73 |
| 697.138 | 508.961 | 487.836 | 824.5925892 | 751.9743417 | 789.36606 | 655.73884 | 589.15289 | 522.44522 | 697.138 | 141.212 | 172.813 | 632.7360736 | 552.7252404 | 1245.52 |
| 697.425 | 508.257 | 489.345 | 862.3642957 | 754.3513211 | 867.48958 | 658.63018 | 642.4944 | 524.80785 | 697.425 | 145.767 | 173.353 | 649.5890012 | 554.3735483 | 1273.71 |
| 697.712 | 511.829 | 490.856 | 902.0043555 | 756.7328216 | 852.61078 | 661.53097 | 592.46916 | 527.17977 | 697.712 | 162.815 | 173.893 | 616.1403836 | 556.0251683 | 1239.1 |
| 697.999 | 510.479 | 492.37 | 860.7559069 | 759.1188433 | 816.63696 | 664.44227 | 621.76622 | 529.55982 | 697.999 | 163.064 | 174.435 | 637.3122534 | 557.6789964 | 1232.59 |
| 698.286 | 505.384 | 493.887 | 883.8983957 | 761.5105164 | 907.281 | 667.36196 | 554.02476 | 531.94685 | 698.286 | 145.204 | 174.978 | 614.831011 | 559.3361366 | 1213.73 |
| 698.573 | 537.939 | 495.407 | 842.2577398 | 763.9067106 | 837.47514 | 670.29216 | 637.65995 | 534.34316 | 698.573 | 159.522 | 175.523 | 640.8329862 | 560.9954847 | 1235.19 |
| 698.86 | 500.696 | 496.93 | 892.4681808 | 766.3074259 | 885.88669 | 673.23182 | 646.52039 | 536.74761 | 698.86 | 153.912 | 176.068 | 611.3202145 | 562.659249 | 1294.42 |
| 699.147 | 540.283 | 498.455 | 848.5500101 | 768.7126624 | 880.0377 | 676.18092 | 578.80332 | 539.16018 | 699.147 | 135.093 | 176.614 | 641.6378198 | 564.3252214 | 1277.71 |
| 699.434 | 520.65 | 499.984 | 848.3578625 | 771.1235502 | 877.76963 | 679.13948 | 633.90344 | 541.58089 | 699.434 | 139.761 | 177.161 | 666.8117769 | 565.9934018 | 1245.25 |
| 699.721 | 539.249 | 501.516 | 878.0378924 | 773.5389592 | 856.68629 | 682.10854 | 625.76201 | 544.01089 | 699.721 | 142.645 | 177.709 | 672.6366079 | 567.6659982 | 1307.31 |
| 700.008 | 515.351 | 503.05 | 927.7216228 | 775.9588892 | 907.74693 | 685.08705 | 577.66498 | 546.44786 | 700.008 | 148.4 | 178.259 | 634.6648037 | 569.3408028 | 1317.04 |
| 700.294 | 523.172 | 504.588 | 861.2215824 | 778.3844707 | 886.73788 | 688.07606 | 684.76268 | 548.89412 | 700.294 | 143.344 | 178.81 | 671.6551305 | 571.0189194 | 1281.02 |
| 700.581 | 504.716 | 506.128 | 879.5332532 | 780.8145733 | 890.47164 | 691.07453 | 610.99268 | 551.34851 | 700.581 | 151.466 | 179.361 | 641.1134083 | 572.699244 | 1321.56 |
| 700.868 | 516.645 | 507.672 | 838.6329309 | 783.2491971 | 945.04316 | 694.08245 | 685.94632 | 553.81103 | 700.868 | 156.466 | 179.914 | 646.7869881 | 574.3828808 | 1331.59 |
| 701.155 | 523.089 | 509.218 | 884.6251659 | 785.6883419 | 902.4095 | 697.09982 | 591.67117 | 556.28169 | 701.155 | 162.72 | 180.468 | 643.8447639 | 576.0698296 | 1306.35 |
| 701.442 | 522.165 | 510.768 | 897.5973927 | 788.1331382 | 888.19084 | 700.12874 | 609.67546 | 558.76163 | 701.442 | 158.154 | 181.022 | 666.5975962 | 577.7600904 | 1333.3 |
| 701.729 | 496.612 | 512.32 | 929.4091312 | 790.5824555 | 959.39024 | 703.16607 | 697.74784 | 561.24971 | 701.729 | 155.209 | 181.578 | 633.9063391 | 579.4525593 | 1340.89 |
| 702.015 | 539.667 | 513.875 | 887.0044059 | 793.036294 | 912.87953 | 706.2139 | 698.07308 | 563.74592 | 702.015 | 148.066 | 182.135 | 684.9332281 | 581.1483403 | 1287.63 |
| 702.302 | 538.062 | 515.433 | 949.1743425 | 795.4957839 | 928.67426 | 709.27118 | 714.30367 | 566.25142 | 702.302 | 176.659 | 182.693 | 656.1579445 | 582.8474334 | 1315.83 |
| 702.589 | 535.836 | 516.995 | 921.302761 | 797.9597949 | 933.50119 | 712.33896 | 670.21521 | 568.76389 | 702.589 | 154.522 | 183.252 | 697.6030085 | 584.5487345 | 1347.32 |
| 702.876 | 545.318 | 518.558 | 939.4177631 | 800.4283271 | 872.32288 | 715.41724 | 647.64246 | 571.28682 | 702.876 | 169.953 | 183.813 | 706.9872132 | 586.2533477 | 1367.53 |
| 703.162 | 554.084 | 520.125 | 934.1608295 | 802.9013803 | 922.0091 | 718.50393 | 679.29983 | 573.81671 | 703.162 | 165.807 | 184.374 | 698.6750159 | 587.9612729 | 1328.81 |
| 703.449 | 533.503 | 521.696 | 968.9192088 | 805.380085 | 931.27876 | 721.60217 | 673.60118 | 576.3559 | 703.449 | 157.067 | 184.936 | 689.3482204 | 589.6714062 | 1350.83 |
| 703.736 | 534.682 | 523.268 | 920.173611 | 807.8633108 | 944.42547 | 724.70986 | 715.40715 | 578.90322 | 703.736 | 164.452 | 185.5 | 672.2987766 | 591.3859556 | 1346.6 |
| 704.022 | 579.219 | 524.844 | 921.9605842 | 810.3510577 | 905.85139 | 727.82701 | 719.36926 | 581.45866 | 704.022 | 181.119 | 186.064 | 705.8887092 | 593.1027131 | 1341.47 |
| 704.309 | 551.028 | 526.423 | 900.4615229 | 812.844456 | 953.87662 | 730.95571 | 681.68104 | 584.02341 | 704.309 | 175.975 | 186.63 | 662.0468089 | 594.8216786 | 1325.08 |
| 704.596 | 543.689 | 528.005 | 917.0709917 | 815.3423754 | 908.49092 | 734.0928 | 669.08152 | 586.59744 | 704.596 | 177.991 | 187.196 | 658.52166 | 596.5439561 | 1338.41 |
| 704.882 | 588.51 | 529.59 | 874.2604957 | 817.844816 | 960.46431 | 737.24146 | 674.32948 | 589.17844 | 704.882 | 171.697 | 187.764 | 654.073547 | 598.2695458 | 1289.94 |
| 705.169 | 557.426 | 531.177 | 896.2376655 | 820.3529079 | 893.4699 | 740.39956 | 716.65468 | 591.76874 | 705.169 | 171.603 | 188.333 | 681.670837 | 599.9984475 | 1287.27 |
| 705.456 | 565.97 | 532.768 | 914.5063857 | 822.8643907 | 946.91748 | 743.56817 | 707.8709 | 594.36833 | 705.456 | 182.751 | 188.903 | 646.1311978 | 601.7306613 | 1306 |
| 705.742 | 574.343 | 534.361 | 897.5940018 | 825.3815249 | 947.03316 | 746.74623 | 730.40415 | 596.97605 | 705.742 | 181.091 | 189.474 | 672.8419564 | 603.4650831 | 1306.44 |
| 706.029 | 588.962 | 535.958 | 937.75173 | 827.9043105 | 974.59168 | 749.93479 | 718.64793 | 599.59306 | 706.029 | 176.013 | 190.046 | 657.879118 | 605.201713 | 1320.79 |
| 706.315 | 562.616 | 537.557 | 920.3035933 | 830.4316172 | 982.71405 | 753.13385 | 704.36645 | 602.2182 | 706.315 | 191.3 | 190.619 | 690.3153455 | 606.942759 | 1295.03 |
| 706.602 | 580.542 | 539.16 | 919.4174535 | 832.963445 | 1011.17683 | 756.34342 | 692.73103 | 604.85264 | 706.602 | 176.962 | 191.194 | 674.199906 | 608.686013 | 1356.46 |
| 706.888 | 585.111 | 540.765 | 925.2813476 | 835.499794 | 945.31168 | 759.56244 | 775.57752 | 607.4952 | 706.888 | 196.344 | 191.769 | 688.3093338 | 610.4325791 | 1341.89 |
| 707.175 | 611.894 | 542.373 | 956.5019496 | 838.0417943 | 983.45911 | 762.79196 | 708.62941 | 610.14706 | 707.175 | 193.309 | 192.345 | 689.051238 | 612.1813532 | 1370.18 |
| 707.462 | 577.411 | 543.985 | 953.6615553 | 840.5883158 | 968.64292 | 766.03198 | 749.69779 | 612.80705 | 707.462 | 207.016 | 192.923 | 683.1656856 | 613.9345435 | 1384.55 |
| 707.748 | 584.101 | 545.599 | 950.0367464 | 843.1393584 | 971.28776 | 769.28251 | 715.87062 | 615.47633 | 707.748 | 209.83 | 193.501 | 698.2731511 | 615.6899418 | 1402.48 |
| 708.034 | 616.774 | 547.216 | 928.8338186 | 845.6960524 | 995.98281 | 772.54354 | 731.96646 | 618.15375 | 708.034 | 198.834 | 194.081 | 744.6002096 | 617.4475481 | 1386.48 |
| 708.321 | 597.967 | 548.836 | 962.133006 | 848.2572675 | 981.01061 | 775.81402 | 723.90867 | 620.84161 | 708.321 | 192.525 | 194.662 | 702.2785505 | 619.2095705 | 1381.8 |
| 708.607 | 598.973 | 550.459 | 951.3071814 | 850.8230038 | 1043.36169 | 779.09605 | 751.59579 | 623.53645 | 708.607 | 179.021 | 195.244 | 703.1529376 | 620.973801 | 1390.14 |
| 708.894 | 608.899 | 552.085 | 928.7874771 | 853.3943914 | 1014.22179 | 782.38754 | 745.2304 | 626.24174 | 708.894 | 204.496 | 195.826 | 672.8088357 | 622.7402396 | 1343.62 |
| 709.18 | 614.206 | 553.714 | 948.4554843 | 855.9703002 | 1001.8074 | 785.68953 | 719.08584 | 628.95516 | 709.18 | 183.437 | 196.41 | 709.6710957 | 624.5110942 | 1374.02 |
| 709.467 | 590.924 | 555.346 | 950.9149742 | 858.55073 | 1050.23275 | 789.00202 | 762.3763 | 631.67672 | 709.467 | 182.314 | 196.995 | 702.349208 | 626.2841569 | 1389.4 |
| 709.753 | 598.443 | 556.981 | 929.7086555 | 861.1356811 | 1010.678 | 792.32501 | 771.12755 | 634.40872 | 709.753 | 194.728 | 197.582 | 699.6675335 | 628.0605316 | 1390.44 |
| 710.04 | 585.655 | 558.619 | 979.8953606 | 863.7262835 | 1013.23475 | 795.65851 | 717.94983 | 637.14886 | 710.04 | 195.192 | 198.169 | 733.8260386 | 629.8391144 | 1376.94 |
| 710.326 | 600.673 | 560.26 | 955.5932043 | 866.321407 | 1010.80218 | 799.00251 | 761.7386 | 639.89829 | 710.326 | 183.311 | 198.757 | 747.690373 | 631.6210093 | 1403.87 |
| 710.612 | 617.781 | 561.904 | 978.6339678 | 868.9221819 | 1099.74066 | 802.35701 | 751.05102 | 642.65585 | 710.612 | 212.501 | 199.347 | 686.7482437 | 633.4062162 | 1429.56 |
| 710.899 | 623.444 | 563.551 | 938.888792 | 871.527478 | 1061.25467 | 805.72201 | 734.29889 | 645.42387 | 710.899 | 204.42 | 199.937 | 713.1863084 | 635.1947352 | 1396.65 |
| 711.185 | 622.086 | 565.2 | 991.0071461 | 874.1372951 | 1055.58292 | 809.09752 | 767.22004 | 648.20002 | 711.185 | 196.774 | 200.529 | 722.0317494 | 636.9854623 | 1435.38 |
| 711.471 | 616.918 | 566.853 | 961.0106377 | 876.7516334 | 1029.90401 | 812.48353 | 737.75571 | 650.98545 | 711.471 | 204.49 | 201.121 | 727.2162464 | 638.7795014 | 1405.22 |
| 711.758 | 625.687 | 568.509 | 955.8667321 | 879.3716231 | 1024.38722 | 815.88004 | 781.58747 | 653.77902 | 711.758 | 189.665 | 201.715 | 705.1545332 | 640.5768526 | 1406.97 |
| 712.044 | 614.484 | 570.167 | 992.6901334 | 881.9972642 | 1013.291 | 819.28706 | 767.57315 | 656.58305 | 712.044 | 201.952 | 202.31 | 721.2478923 | 642.3764119 | 1440.51 |
| 712.33 | 610.882 | 571.829 | 1030.031203 | 884.6262962 | 1082.03766 | 822.70562 | 799.3095 | 659.3952 | 712.33 | 198.083 | 202.906 | 737.3710599 | 644.1792832 | 1409.73 |
| 712.616 | 621.727 | 573.494 | 997.8871621 | 887.2609795 | 1099.46472 | 826.13364 | 762.32868 | 662.21665 | 712.616 | 208.585 | 203.503 | 754.7241105 | 645.9854666 | 1406.55 |
| 712.903 | 648.108 | 575.161 | 1005.09496 | 889.900184 | 1085.29595 | 829.57321 | 810.69171 | 665.04739 | 712.903 | 206.536 | 204.101 | 713.626814 | 647.793858 | 1438.37 |
| 713.189 | 623.612 | 576.831 | 997.5119091 | 892.5450398 | 1067.33822 | 833.02329 | 767.81708 | 667.88742 | 713.189 | 205.228 | 204.7 | 750.9185395 | 649.6066656 | 1476.05 |
| 713.475 | 621.166 | 578.505 | 992.8189854 | 895.1932865 | 1067.25332 | 836.48386 | 847.21487 | 670.73675 | 713.475 | 208.448 | 205.301 | 752.2775931 | 651.4205771 | 1462.34 |
| 713.761 | 629.046 | 580.181 | 1001.573006 | 897.8483149 | 1049.09394 | 839.95494 | 813.05201 | 673.59537 | 713.761 | 200.192 | 205.902 | 707.5436413 | 653.2389048 | 1422.92 |
| 714.048 | 598.97 | 581.861 | 1014.228754 | 900.5067341 | 1053.56108 | 843.43652 | 771.19957 | 676.46212 | 714.048 | 205.861 | 206.504 | 745.672217 | 655.0594405 | 1422.57 |
| 714.334 | 617.966 | 583.543 | 1015.464151 | 903.1708048 | 1107.37163 | 846.92966 | 767.49533 | 679.33932 | 714.334 | 219.181 | 207.108 | 722.4192618 | 656.8832883 | 1476.34 |
| 714.62 | 636.051 | 585.228 | 1008.933391 | 905.8393965 | 1078.99164 | 850.4333 | 830.27341 | 682.22581 | 714.62 | 209.97 | 207.713 | 725.5524822 | 658.7093441 | 1421.74 |
| 714.906 | 622.494 | 586.916 | 995.1586655 | 908.5136397 | 1111.808 | 853.94743 | 784.96298 | 685.12044 | 714.906 | 203.977 | 208.318 | 730.0701487 | 660.5398161 | 1466.66 |
| 715.192 | 646.796 | 588.608 | 1071.842531 | 911.1924039 | 1165.13988 | 857.47208 | 827.76326 | 688.02436 | 715.192 | 212.454 | 208.925 | 744.0559258 | 662.372496 | 1464.5 |
| 715.479 | 623.505 | 590.302 | 1044.789273 | 913.8756893 | 1078.34423 | 861.00827 | 801.48512 | 690.93873 | 715.479 | 210.014 | 209.533 | 761.0082155 | 664.2073841 | 1488.48 |
| 715.765 | 646.963 | 591.999 | 1054.914324 | 916.5646261 | 1081.89969 | 864.55497 | 844.64316 | 693.86124 | 715.765 | 214.146 | 210.142 | 740.7250851 | 666.0455842 | 1471 |
| 716.051 | 664.417 | 593.699 | 1031.834 | 919.258084 | 1111.85045 | 868.11322 | 785.77608 | 696.7942 | 716.051 | 201.763 | 210.752 | 772.8665382 | 667.8870963 | 1470.86 |
| 716.337 | 636.135 | 595.402 | 1022.461716 | 921.9560631 | 1106.28907 | 871.68092 | 792.90114 | 699.73645 | 716.337 | 213.355 | 211.363 | 740.7924306 | 669.7319206 | 1460.02 |
| 716.623 | 649.218 | 597.108 | 1021.694256 | 924.6585632 | 1104.35745 | 875.26123 | 807.75178 | 702.68683 | 716.623 | 197.776 | 211.975 | 750.4714097 | 671.5789529 | 1446.44 |
| 716.909 | 639.837 | 598.817 | 1013.950706 | 927.3667148 | 1098.85976 | 878.85098 | 825.61669 | 705.64766 | 716.909 | 214.829 | 212.589 | 732.3124216 | 673.4292972 | 1496.8 |
| 717.195 | 657.832 | 600.529 | 1011.112572 | 930.0805177 | 1117.68777 | 882.45229 | 852.09462 | 708.61779 | 717.195 | 213.582 | 213.203 | 718.6766202 | 675.2829536 | 1495.26 |
| 717.481 | 650.512 | 602.244 | 1032.066838 | 932.7977115 | 1118.37764 | 886.06411 | 790.78476 | 711.59605 | 717.481 | 211.46 | 213.819 | 749.2724395 | 677.1388181 | 1483.27 |
| 717.767 | 623.486 | 603.962 | 1034.511634 | 935.5205567 | 1107.99782 | 889.68747 | 838.20808 | 714.58593 | 717.767 | 211.559 | 214.435 | 747.5170413 | 678.9990987 | 1470.45 |
| 718.053 | 650.787 | 605.683 | 1037.571303 | 938.2490533 | 1053.35731 | 893.32239 | 856.42843 | 717.58393 | 718.053 | 222.929 | 215.053 | 747.1129684 | 680.8604833 | 1521.21 |
| 718.339 | 672.932 | 607.407 | 1069.450858 | 940.9809408 | 1136.83418 | 896.96781 | 868.86766 | 720.59123 | 718.339 | 220.263 | 215.672 | 793.2026618 | 682.726284 | 1497.28 |
| 718.625 | 658.495 | 609.133 | 1078.536051 | 943.7184796 | 1133.93675 | 900.62373 | 860.09666 | 723.60782 | 718.625 | 222.628 | 216.292 | 771.5825582 | 684.5942928 | 1491.86 |
| 718.911 | 646.008 | 610.863 | 1066.531344 | 946.4616699 | 1193.9338 | 904.29121 | 865.71865 | 726.63487 | 718.911 | 212.537 | 216.912 | 764.2761268 | 686.4656136 | 1491.88 |
| 719.197 | 659.697 | 612.596 | 1060.718313 | 949.2093812 | 1137.18442 | 907.96919 | 830.68228 | 729.6712 | 719.197 | 219.94 | 217.535 | 754.1356656 | 688.3391424 | 1505.56 |
| 719.483 | 643.093 | 614.331 | 1072.068587 | 951.9616137 | 1159.12214 | 911.65872 | 871.25119 | 732.71683 | 719.483 | 208.471 | 218.158 | 754.5231781 | 690.2159834 | 1494.82 |
| 719.769 | 679.45 | 616.07 | 1063.49089 | 954.7183673 | 1121.19016 | 915.3598 | 936.68694 | 735.77176 | 719.769 | 217.256 | 218.782 | 749.8984212 | 692.0961364 | 1513.64 |
| 720.055 | 675.602 | 617.812 | 1051.663637 | 957.4807723 | 1165.7873 | 919.07139 | 919.04157 | 738.83713 | 720.055 | 200.17 | 219.407 | 735.8894596 | 693.9796015 | 1514.93 |
| 720.341 | 688.735 | 619.556 | 1081.208034 | 960.2476985 | 1134.70091 | 922.79453 | 829.48006 | 741.9118 | 720.341 | 231.141 | 220.034 | 774.8869023 | 695.8652746 | 1516.52 |
| 720.627 | 685.076 | 621.304 | 1091.902746 | 963.020276 | 1155.48177 | 926.52817 | 820.30368 | 744.9946 | 720.627 | 223.956 | 220.661 | 766.5471044 | 697.7542598 | 1518.52 |
| 720.913 | 696.672 | 623.054 | 1094.469612 | 965.7973746 | 1164.63044 | 930.27337 | 877.56547 | 748.08902 | 720.913 | 219.22 | 221.29 | 786.3532964 | 699.6454531 | 1500.18 |
| 721.199 | 652.607 | 624.808 | 1058.340203 | 968.5789944 | 1185.93136 | 934.03011 | 797.34529 | 751.19273 | 721.199 | 221.421 | 221.919 | 778.4805006 | 701.5399584 | 1502.07 |
| 721.485 | 651.336 | 626.564 | 1083.283228 | 971.3662656 | 1187.75685 | 937.79736 | 903.29303 | 754.30457 | 721.485 | 233.087 | 222.55 | 765.6892776 | 703.4377758 | 1527.55 |
| 721.77 | 666.225 | 628.323 | 1078.749674 | 974.1580579 | 1239.3057 | 941.57617 | 890.72487 | 757.42802 | 721.77 | 228.785 | 223.182 | 791.5300653 | 705.3389052 | 1499.91 |
| 722.056 | 680.885 | 630.086 | 1073.536821 | 976.9543713 | 1189.29578 | 945.36652 | 893.92848 | 760.55961 | 722.056 | 231.075 | 223.815 | 759.1821597 | 707.2422428 | 1529.55 |
| 722.342 | 671.894 | 631.851 | 1071.224268 | 979.7563361 | 1231.10161 | 949.16843 | 872.9738 | 763.70281 | 722.342 | 236.841 | 224.449 | 799.3675322 | 709.1488923 | 1527.36 |
| 722.628 | 705.856 | 633.619 | 1059.540561 | 982.5628221 | 1226.91997 | 952.98084 | 885.38282 | 766.85414 | 722.628 | 243.274 | 225.084 | 802.9136576 | 711.05775 | 1547.87 |
| 722.913 | 688.497 | 635.391 | 1099.78193 | 985.3738291 | 1180.18956 | 956.8048 | 841.16892 | 770.01593 | 722.913 | 224.349 | 225.72 | 769.2541714 | 712.9699197 | 1574.87 |
| 723.199 | 675.064 | 637.165 | 1116.425307 | 988.1904876 | 1145.33545 | 960.64032 | 896.47928 | 773.18701 | 723.199 | 235.946 | 226.358 | 791.6029309 | 714.8854015 | 1563.95 |
| 723.485 | 687.272 | 638.942 | 1094.851647 | 991.0116672 | 1198.59304 | 964.48739 | 836.20205 | 776.36854 | 723.485 | 221.456 | 226.996 | 747.0401029 | 716.8041953 | 1556.34 |
| 723.771 | 686.11 | 640.722 | 1088.236117 | 993.8373679 | 1212.29482 | 968.34496 | 912.94218 | 779.55937 | 723.771 | 235.063 | 227.636 | 788.3571001 | 718.7251972 | 1565.55 |
| 724.057 | 703.954 | 642.505 | 1106.507097 | 996.66872 | 1216.39156 | 972.21409 | 851.29778 | 782.75949 | 724.057 | 223.465 | 228.276 | 789.4821006 | 720.6495112 | 1549 |
| 724.342 | 714.981 | 644.291 | 1121.649462 | 999.5057235 | 1208.97286 | 976.09581 | 913.37312 | 785.97006 | 724.342 | 225.109 | 228.918 | 780.83428 | 722.5760332 | 1564.75 |
| 724.628 | 692.055 | 646.08 | 1110.699307 | 1002.346118 | 1190.10239 | 979.98804 | 939.10881 | 789.19109 | 724.628 | 218.535 | 229.561 | 797.9113245 | 724.5058673 | 1573.41 |
| 724.914 | 726.173 | 647.872 | 1084.959434 | 1005.192164 | 1227.6629 | 983.89182 | 939.02285 | 792.42141 | 724.914 | 224.878 | 230.204 | 786.0905386 | 726.4390135 | 1561.16 |
| 725.199 | 697.142 | 649.667 | 1091.421246 | 1008.04273 | 1259.54528 | 987.80716 | 890.18126 | 795.66102 | 725.199 | 226.557 | 230.849 | 762.3352525 | 728.3754717 | 1543.42 |
| 725.485 | 707.001 | 651.465 | 1101.657065 | 1010.898949 | 1258.31414 | 991.73405 | 968.53364 | 798.91109 | 725.485 | 228.301 | 231.495 | 799.5949612 | 730.314138 | 1572.87 |
| 725.771 | 691.932 | 653.266 | 1128.996284 | 1013.759688 | 1182.8535 | 995.67144 | 902.5264 | 802.17161 | 725.771 | 228.222 | 232.143 | 792.4430932 | 732.2561164 | 1580.56 |
| 726.056 | 707.953 | 655.07 | 1130.046314 | 1016.626079 | 1215.00122 | 999.62143 | 954.55882 | 805.44142 | 726.056 | 235.01 | 232.791 | 807.7029141 | 734.2003028 | 1606.65 |
| 726.342 | 734.402 | 656.877 | 1114.135359 | 1019.495861 | 1234.72075 | 1003.58298 | 908.39812 | 808.72169 | 726.342 | 216.904 | 233.44 | 789.0294507 | 736.1478013 | 1562.17 |
| 726.628 | 712.421 | 658.687 | 1113.254871 | 1022.371294 | 1204.83367 | 1007.55608 | 917.77662 | 812.01125 | 726.628 | 237.519 | 234.09 | 815.6143506 | 738.0986118 | 1588.23 |
| 726.913 | 708.987 | 660.499 | 1105.043384 | 1025.252378 | 1209.42923 | 1011.53968 | 896.6419 | 815.31126 | 726.913 | 244.459 | 234.742 | 811.2004624 | 740.0527344 | 1555.32 |
| 727.199 | 702.427 | 662.315 | 1148.556915 | 1028.137984 | 1249.16547 | 1015.53588 | 915.03184 | 818.62173 | 727.199 | 241.881 | 235.394 | 810.8946477 | 742.0090651 | 1582.98 |
| 727.484 | 717.394 | 664.134 | 1139.356433 | 1031.02811 | 1246.65011 | 1019.54364 | 919.80936 | 821.94149 | 727.484 | 232.174 | 236.048 | 813.8291437 | 743.9676038 | 1582.78 |
| 727.77 | 685.029 | 665.955 | 1140.599741 | 1033.923888 | 1280.83559 | 1023.56295 | 944.18834 | 825.2717 | 727.77 | 221.214 | 236.703 | 782.5532455 | 745.9305587 | 1578.56 |
| 728.055 | 708.788 | 667.78 | 1159.430211 | 1036.824188 | 1302.98557 | 1027.59381 | 958.68703 | 828.61121 | 728.055 | 244.143 | 237.359 | 817.8544155 | 747.8957215 | 1614.72 |
| 728.341 | 714.054 | 669.607 | 1145.245193 | 1039.729008 | 1286.03611 | 1031.63623 | 951.89767 | 831.96117 | 728.341 | 239.753 | 238.016 | 822.5719104 | 749.8641965 | 1597.79 |
| 728.626 | 710.973 | 671.438 | 1138.610448 | 1042.63948 | 1286.54555 | 1035.69125 | 1017.13827 | 835.32158 | 728.626 | 243.559 | 238.674 | 809.3920709 | 751.8348794 | 1616.44 |
| 728.912 | 732.302 | 673.271 | 1135.016157 | 1045.554473 | 1301.22376 | 1039.75677 | 932.54015 | 838.69245 | 728.912 | 247.344 | 239.333 | 805.6792379 | 753.8088745 | 1593.92 |
| 729.197 | 743.077 | 675.108 | 1137.627104 | 1048.473986 | 1292.86047 | 1043.83489 | 968.61843 | 842.07262 | 729.197 | 236.224 | 239.993 | 814.0455324 | 755.7861816 | 1553.53 |
| 729.483 | 739.944 | 676.947 | 1149.822829 | 1051.399152 | 1294.53737 | 1047.92352 | 993.98707 | 845.46439 | 729.483 | 257.716 | 240.654 | 844.0705713 | 757.7656968 | 1641.89 |
| 729.768 | 734.37 | 678.789 | 1155.315991 | 1054.328838 | 1365.86943 | 1052.0258 | 950.68848 | 848.8643 | 729.768 | 228.804 | 241.317 | 809.8910898 | 759.7485241 | 1609.29 |
| 730.054 | 723.003 | 680.635 | 1170.529563 | 1057.263046 | 1337.37269 | 1056.14278 | 932.30551 | 852.27582 | 730.054 | 233.117 | 241.98 | 809.4119433 | 761.7335594 | 1646.33 |
| 730.339 | 721.536 | 682.483 | 1167.014392 | 1060.202905 | 1330.9198 | 1060.25977 | 971.28539 | 855.69664 | 730.339 | 246.871 | 242.645 | 796.9099746 | 763.7230108 | 1630.47 |
| 730.625 | 717.327 | 684.334 | 1177.718146 | 1063.148415 | 1308.14363 | 1064.39776 | 981.88005 | 859.12907 | 730.625 | 245.484 | 243.31 | 845.5764602 | 765.7146702 | 1627.64 |
| 730.91 | 718.809 | 686.188 | 1147.73181 | 1066.097317 | 1276.20818 | 1068.54625 | 997.1628 | 862.5708 | 730.91 | 246.962 | 243.977 | 842.8760173 | 767.7085377 | 1622.16 |
| 731.195 | 733.219 | 688.045 | 1119.449937 | 1069.051869 | 1309.9479 | 1072.70524 | 967.69499 | 866.02298 | 731.195 | 267.536 | 244.645 | 827.1569224 | 769.7057172 | 1628.01 |
| 731.481 | 736.198 | 689.906 | 1135.694325 | 1072.012073 | 1269.63854 | 1076.88524 | 978.06314 | 869.48445 | 731.481 | 257.531 | 245.314 | 840.938455 | 771.7062089 | 1599.5 |
| 731.766 | 719.723 | 691.769 | 1195.373124 | 1074.975668 | 1281.89692 | 1081.06524 | 989.04112 | 872.95754 | 731.766 | 240.877 | 245.984 | 860.6717815 | 773.7089086 | 1634.51 |
| 732.052 | 763.372 | 693.635 | 1173.21963 | 1077.944915 | 1285.81323 | 1085.26624 | 980.34214 | 876.44108 | 732.052 | 252.07 | 246.655 | 827.8336892 | 775.7149203 | 1603.82 |
| 732.337 | 727.076 | 695.504 | 1197.678896 | 1080.919812 | 1332.64977 | 1089.46724 | 953.09757 | 879.93391 | 732.337 | 248.25 | 247.327 | 807.5958237 | 777.7242442 | 1611.04 |
| 732.622 | 746.889 | 697.375 | 1162.481968 | 1083.899231 | 1295.30153 | 1093.68925 | 971.14136 | 883.4372 | 732.622 | 247.471 | 248.001 | 850.0223652 | 779.7368801 | 1642.12 |
| 732.908 | 718.681 | 699.25 | 1173.0953 | 1086.883171 | 1344.18644 | 1097.92176 | 964.26836 | 886.9521 | 732.908 | 248.979 | 248.675 | 824.6287073 | 781.75062 | 1652.86 |
| 733.193 | 739.031 | 701.128 | 1202.05308 | 1089.871632 | 1357.03916 | 1102.16478 | 970.79521 | 890.4763 | 733.193 | 265.989 | 249.351 | 873.6849133 | 783.768776 | 1627.87 |
| 733.478 | 765.216 | 703.009 | 1176.407021 | 1092.865745 | 1367.93903 | 1106.41829 | 989.2258 | 894.01095 | 733.478 | 255.328 | 250.027 | 840.1170611 | 785.7891401 | 1650.4 |
| 733.763 | 756.93 | 704.893 | 1189.789539 | 1095.865508 | 1368.60767 | 1110.69281 | 998.7344 | 897.55605 | 733.763 | 285.687 | 250.705 | 820.5140095 | 787.8128162 | 1692.17 |
| 734.048 | 763.53 | 706.779 | 1193.101261 | 1098.868663 | 1341.08735 | 1114.96734 | 1025.70135 | 901.11161 | 734.048 | 261.387 | 251.383 | 854.5764645 | 789.8398044 | 1672.53 |
| 734.334 | 728.337 | 708.669 | 1170.292204 | 1101.877469 | 1333.58374 | 1119.26286 | 1011.24448 | 904.67762 | 734.334 | 259.82 | 252.063 | 855.7831628 | 791.8690007 | 1655.27 |
| 734.619 | 746.893 | 710.562 | 1216.882358 | 1104.891927 | 1322.93859 | 1123.55839 | 992.07397 | 908.25409 | 734.619 | 275.029 | 252.744 | 848.9525658 | 793.901509 | 1607.9 |
| 734.904 | 767.705 | 712.457 | 1154.050077 | 1107.909775 | 1343.13572 | 1127.87492 | 1058.42852 | 911.84101 | 734.904 | 264.845 | 253.426 | 838.3406864 | 795.9362254 | 1688.35 |
| 735.189 | 756.656 | 714.356 | 1228.806815 | 1110.934405 | 1343.36921 | 1132.21245 | 1052.16534 | 915.43955 | 735.189 | 253.301 | 254.109 | 848.5297246 | 797.9742539 | 1705.57 |
| 735.475 | 733.925 | 716.257 | 1166.381435 | 1113.962426 | 1361.12528 | 1136.54999 | 1030.69841 | 919.04738 | 735.475 | 248.925 | 254.793 | 853.3079408 | 800.0155944 | 1656.52 |
| 735.76 | 746.216 | 718.161 | 1205.636069 | 1116.996098 | 1304.06812 | 1140.89803 | 1029.91668 | 922.66566 | 735.76 | 256.867 | 255.479 | 888.9226537 | 802.059143 | 1675.07 |
| 736.045 | 791.609 | 720.069 | 1239.047154 | 1120.034292 | 1406.5078 | 1145.26707 | 1102.84454 | 926.29556 | 736.045 | 277.269 | 256.165 | 853.0142704 | 804.1060036 | 1685.12 |
| 736.33 | 775.391 | 721.979 | 1195.305307 | 1123.078137 | 1397.48649 | 1149.63612 | 1006.70507 | 929.93475 | 736.33 | 265.187 | 256.853 | 856.294326 | 806.1550723 | 1726.09 |
| 736.615 | 760.874 | 723.892 | 1226.828824 | 1126.126502 | 1376.02637 | 1154.02617 | 1037.9768 | 933.58556 | 736.615 | 263.669 | 257.541 | 858.8269572 | 808.2074531 | 1698.67 |
| 736.9 | 770.049 | 725.808 | 1275.803869 | 1129.179389 | 1372.38601 | 1158.42672 | 1049.62035 | 937.24682 | 736.9 | 261.54 | 258.231 | 849.2307799 | 810.2631459 | 1691.3 |
| 737.185 | 760.428 | 727.727 | 1213.050708 | 1132.235667 | 1365.63594 | 1162.84827 | 1043.65919 | 940.91853 | 737.185 | 273.732 | 258.921 | 899.5809022 | 812.3210468 | 1706.25 |
| 737.47 | 778.701 | 729.649 | 1246.540912 | 1135.298727 | 1359.37409 | 1167.26983 | 1047.71306 | 944.6007 | 737.47 | 274.623 | 259.613 | 832.7686768 | 814.3822598 | 1716.31 |
| 737.755 | 757.074 | 731.574 | 1226.715796 | 1138.373089 | 1427.21437 | 1171.71239 | 1024.7872 | 948.29332 | 737.755 | 264.429 | 260.306 | 862.3675624 | 816.4467848 | 1705.59 |
| 738.041 | 780.25 | 733.503 | 1178.13635 | 1141.436149 | 1438.88901 | 1176.15495 | 1080.32755 | 951.9964 | 738.041 | 248.756 | 261 | 838.1441701 | 818.5135179 | 1701.15 |
| 738.326 | 783.597 | 735.433 | 1202.686037 | 1144.521814 | 1395.40628 | 1180.61852 | 1131.64096 | 955.7111 | 738.326 | 266.123 | 261.695 | 842.5161054 | 820.583563 | 1748.37 |
| 738.611 | 783.223 | 737.367 | 1248.925804 | 1147.607479 | 1388.38027 | 1185.09259 | 1100.85245 | 959.43624 | 738.611 | 253.975 | 262.391 | 898.9803131 | 822.6558162 | 1707.76 |
| 738.896 | 793.876 | 739.304 | 1226.77231 | 1150.693144 | 1399.98061 | 1189.57716 | 1107.61858 | 963.17184 | 738.896 | 260.589 | 263.089 | 878.4189686 | 824.7313815 | 1736.12 |
| 739.181 | 761.336 | 741.244 | 1296.420181 | 1153.77881 | 1411.07152 | 1194.08273 | 1098.64199 | 966.91906 | 739.181 | 267.572 | 263.787 | 875.8620488 | 826.8102588 | 1700.42 |
| 739.466 | 752.592 | 743.186 | 1269.225638 | 1156.875778 | 1422.17305 | 1198.58831 | 1083.59969 | 970.67557 | 739.466 | 269.667 | 264.486 | 891.918975 | 828.8913442 | 1714.83 |
| 739.751 | 779.262 | 745.132 | 1232.344592 | 1159.984048 | 1428.18018 | 1203.11489 | 1093.88072 | 974.4437 | 739.751 | 265.673 | 265.187 | 926.0123427 | 830.9757417 | 1728.65 |
| 740.035 | 776.222 | 747.081 | 1244.370774 | 1163.092319 | 1432.64838 | 1207.65197 | 1067.05548 | 978.22344 | 740.035 | 279.764 | 265.889 | 900.8681942 | 833.0623472 | 1745.09 |
| 740.321 | 794.754 | 749.032 | 1196.379073 | 1166.20059 | 1459.95643 | 1212.19956 | 1110.00909 | 982.01247 | 740.321 | 271.795 | 266.591 | 883.6487306 | 835.1522648 | 1743.33 |
| 740.605 | 804.473 | 750.986 | 1257.640265 | 1169.320163 | 1416.4737 | 1216.76815 | 1105.34539 | 985.81312 | 740.605 | 275.801 | 267.295 | 906.0272988 | 837.2454945 | 1793.79 |
| 740.89 | 785.731 | 752.944 | 1313.442202 | 1172.439737 | 1437.99749 | 1221.33674 | 1126.9157 | 989.62422 | 740.89 | 270.938 | 268 | 902.6633374 | 839.3409322 | 1718.64 |
| 741.175 | 792.852 | 754.904 | 1283.896676 | 1175.570613 | 1488.09232 | 1225.92634 | 1135.55777 | 993.44694 | 741.175 | 286.193 | 268.706 | 881.7696817 | 841.4396819 | 1754.73 |
| 741.46 | 787.288 | 756.867 | 1228.671181 | 1178.70149 | 1463.21472 | 1230.52644 | 1091.12432 | 997.28011 | 741.46 | 283.188 | 269.413 | 874.8386185 | 843.5406398 | 1758.21 |
| 741.745 | 811.962 | 758.833 | 1254.780656 | 1181.843669 | 1443.665 | 1235.13704 | 1056.40971 | 1001.12374 | 741.745 | 283.172 | 270.121 | 898.6281294 | 845.6449097 | 1779.3 |
| 742.03 | 832.81 | 760.803 | 1277.024572 | 1184.985848 | 1486.10763 | 1239.75814 | 1106.78109 | 1004.97898 | 742.03 | 271.198 | 270.83 | 852.398225 | 847.7524916 | 1775.43 |
| 742.315 | 794.434 | 762.775 | 1287.072763 | 1188.128027 | 1470.97305 | 1244.40025 | 1078.66651 | 1008.84468 | 742.315 | 264.003 | 271.54 | 938.2636979 | 849.8622817 | 1807.52 |
| 742.6 | 810.971 | 764.75 | 1314.27861 | 1191.281509 | 1518.95583 | 1249.04236 | 1058.58301 | 1012.72199 | 742.6 | 292.2 | 272.252 | 921.1965897 | 851.9753838 | 1782.37 |
| 742.884 | 809.331 | 766.727 | 1279.03647 | 1194.434991 | 1471.76905 | 1253.70548 | 1134.75745 | 1016.60976 | 742.884 | 280.445 | 272.964 | 871.4647209 | 854.0906939 | 1784.56 |
| 743.169 | 815.688 | 768.708 | 1310.062664 | 1197.599776 | 1521.97001 | 1258.37909 | 1140.67796 | 1020.50914 | 743.169 | 288.339 | 273.678 | 919.5891306 | 856.2093161 | 1793.27 |
| 743.454 | 817.196 | 770.692 | 1252.305342 | 1200.764561 | 1546.16836 | 1263.07371 | 1134.66684 | 1024.41898 | 743.454 | 296.119 | 274.392 | 925.5883974 | 858.3312504 | 1793.13 |
| 743.739 | 843.876 | 772.679 | 1249.841331 | 1203.940648 | 1514.48763 | 1267.76834 | 1163.18217 | 1028.33927 | 743.739 | 287.106 | 275.108 | 935.6692413 | 860.4553927 | 1767.07 |
| 744.023 | 819.01 | 774.668 | 1282.902029 | 1207.116736 | 1575.04718 | 1272.48396 | 1109.48174 | 1032.27118 | 744.023 | 289.782 | 275.825 | 942.7051868 | 862.5817431 | 1823.46 |
| 744.308 | 840.013 | 776.661 | 1321.489798 | 1210.292824 | 1540.67066 | 1277.21009 | 1114.51597 | 1036.2147 | 744.308 | 291.528 | 276.543 | 946.7138982 | 864.7125096 | 1799.46 |
| 744.593 | 820.969 | 778.656 | 1305.665874 | 1213.480214 | 1535.07745 | 1281.94672 | 1072.90281 | 1040.16867 | 744.593 | 288.523 | 277.262 | 900.0600486 | 866.8443801 | 1780.17 |
| 744.878 | 827.487 | 780.655 | 1286.530229 | 1216.667604 | 1525.3875 | 1286.69386 | 1143.75727 | 1044.1331 | 744.878 | 269.547 | 277.982 | 937.8364406 | 868.9806667 | 1838.9 |
| 745.162 | 819.45 | 782.656 | 1327.728945 | 1219.866298 | 1574.41038 | 1291.46199 | 1161.21447 | 1048.11031 | 745.162 | 289.839 | 278.703 | 932.2721592 | 871.1191613 | 1808.27 |
| 745.447 | 819.535 | 784.66 | 1329.559999 | 1223.064991 | 1526.15165 | 1296.23013 | 1138.65102 | 1052.09681 | 745.447 | 278.411 | 279.425 | 956.9084566 | 873.2609681 | 1831.08 |
| 745.732 | 822.037 | 786.668 | 1285.354737 | 1226.274987 | 1529.50546 | 1301.01928 | 1168.00267 | 1056.09609 | 745.732 | 279.043 | 280.148 | 919.560426 | 875.4049828 | 1836.09 |
| 746.016 | 849.139 | 788.678 | 1307.519534 | 1229.484983 | 1570.14383 | 1305.81892 | 1179.72289 | 1060.10582 | 746.016 | 291.452 | 280.873 | 979.6127119 | 877.5512056 | 1831.56 |
| 746.301 | 845.964 | 790.691 | 1293.334516 | 1232.706282 | 1513.85083 | 1310.63958 | 1134.36135 | 1064.12601 | 746.301 | 287.766 | 281.598 | 920.4149406 | 879.7018445 | 1817.97 |
| 746.586 | 844.505 | 792.707 | 1313.329174 | 1235.916278 | 1508.00289 | 1315.46023 | 1216.81174 | 1068.15897 | 746.586 | 301.974 | 282.325 | 900.6197888 | 881.8546915 | 1831.58 |
| 746.87 | 817.83 | 794.726 | 1331.45887 | 1239.148879 | 1629.58898 | 1320.30188 | 1214.07044 | 1072.20239 | 746.87 | 297.189 | 283.052 | 974.1323363 | 884.0097465 | 1811.7 |
| 747.155 | 826.817 | 796.747 | 1378.150748 | 1242.381481 | 1544.99028 | 1325.15404 | 1233.72417 | 1076.25626 | 747.155 | 288.159 | 283.781 | 962.1867958 | 886.1681136 | 1821.56 |
| 747.439 | 833.709 | 798.772 | 1351.442225 | 1245.614083 | 1516.4617 | 1330.0167 | 1230.51824 | 1080.32174 | 747.439 | 278.353 | 284.511 | 944.6747658 | 888.3297927 | 1842.3 |
| 747.724 | 855.411 | 800.8 | 1391.533267 | 1248.857987 | 1609.2751 | 1334.90037 | 1212.02608 | 1084.39885 | 747.724 | 284.052 | 285.242 | 974.4502953 | 890.4936799 | 1821.29 |
| 748.009 | 848.565 | 802.83 | 1327.220319 | 1252.101891 | 1536.21307 | 1339.79454 | 1196.72824 | 1088.48756 | 748.009 | 286.957 | 285.974 | 927.033565 | 892.6597752 | 1803.35 |
| 748.293 | 840.48 | 804.864 | 1379.077578 | 1255.345796 | 1599.80803 | 1344.69921 | 1213.19927 | 1092.58674 | 748.293 | 284.042 | 286.707 | 977.6861898 | 894.8302865 | 1841.34 |
| 748.578 | 840.547 | 806.9 | 1326.643876 | 1258.601003 | 1588.5367 | 1349.61439 | 1205.46323 | 1096.69868 | 748.578 | 276.411 | 287.441 | 944.2982936 | 897.0019019 | 1854.91 |
| 748.862 | 839.967 | 808.939 | 1390.244747 | 1261.867513 | 1631.84962 | 1354.54006 | 1232.51614 | 1100.81993 | 748.862 | 287.482 | 288.176 | 948.6128197 | 899.1779334 | 1896.9 |
| 749.147 | 867.499 | 810.981 | 1383.022256 | 1265.134023 | 1634.37559 | 1359.48674 | 1222.35242 | 1104.95395 | 749.147 | 312.166 | 288.912 | 957.6669212 | 901.3561729 | 1875.55 |
| 749.431 | 859.895 | 813.027 | 1333.301227 | 1268.400533 | 1651.97245 | 1364.44393 | 1229.31021 | 1109.09958 | 749.431 | 305.708 | 289.649 | 926.345758 | 903.5366204 | 1854.57 |
| 749.716 | 859.171 | 815.075 | 1351.724795 | 1271.678346 | 1615.13365 | 1369.41161 | 1244.54998 | 1113.25567 | 749.716 | 287.795 | 290.388 | 949.3326434 | 905.7203801 | 1901.62 |
| 750 | 863.44 | 817.125 | 1399.433926 | 1274.956159 | 1655.01847 | 1374.3898 | 1224.45486 | 1117.42337 | 750 | 292.83 | 291.127 | 999.3151256 | 907.9074518 | 1867.9 |
| 750.284 | 833.384 | 819.179 | 1395.613579 | 1278.233971 | 1613.58411 | 1379.389 | 1178.41032 | 1121.60269 | 750.284 | 294.323 | 291.868 | 959.1882663 | 910.0967315 | 1902.01 |
| 750.569 | 856.066 | 821.236 | 1369.051992 | 1281.523087 | 1642.33557 | 1384.39869 | 1200.52656 | 1125.79247 | 750.569 | 306.703 | 292.61 | 975.4273566 | 912.2893234 | 1872.21 |
| 750.853 | 852.259 | 823.296 | 1357.975245 | 1284.823506 | 1653.81917 | 1389.41889 | 1280.86058 | 1129.99502 | 750.853 | 302.088 | 293.352 | 983.8919092 | 914.4841232 | 1910.03 |
| 751.138 | 891.788 | 825.358 | 1345.643888 | 1288.123924 | 1643.89572 | 1394.4496 | 1275.982 | 1134.20802 | 751.138 | 299.877 | 294.096 | 1003.766551 | 916.6811312 | 1875.59 |
| 751.422 | 860.796 | 827.424 | 1342.919912 | 1291.424342 | 1645.19055 | 1399.5013 | 1264.07593 | 1138.43381 | 751.422 | 307.786 | 294.841 | 988.5376425 | 918.8825552 | 1916.75 |
| 751.706 | 864.036 | 829.492 | 1372.454136 | 1294.736064 | 1676.95619 | 1404.56351 | 1265.10972 | 1142.67004 | 751.706 | 293.849 | 295.587 | 980.1393313 | 921.0850832 | 1902.65 |
| 751.991 | 867.616 | 831.563 | 1378.015115 | 1298.047785 | 1657.69303 | 1409.63622 | 1277.10872 | 1146.9179 | 751.991 | 304.678 | 296.334 | 956.6313466 | 923.2920274 | 1874.63 |
| 752.275 | 882.72 | 833.638 | 1399.241778 | 1301.370809 | 1618.2009 | 1414.71944 | 1360.00979 | 1151.17737 | 752.275 | 302.514 | 297.082 | 985.2774594 | 925.5011795 | 1925.15 |
| 752.559 | 869.461 | 835.715 | 1378.455924 | 1304.693833 | 1696.33609 | 1419.82366 | 1266.21321 | 1155.44845 | 752.559 | 308.143 | 297.831 | 965.775978 | 927.7125398 | 1876.84 |
| 752.844 | 906.952 | 837.794 | 1452.941395 | 1308.016857 | 1738.96976 | 1424.93838 | 1328.83332 | 1159.73115 | 752.844 | 309.018 | 298.582 | 950.2732719 | 929.9272121 | 1912.12 |
| 753.128 | 870.392 | 839.877 | 1383.847361 | 1311.351184 | 1697.23823 | 1430.0636 | 1316.7414 | 1164.03011 | 753.128 | 308.373 | 299.333 | 1004.347267 | 932.1451965 | 1946.57 |
| 753.412 | 878.121 | 841.963 | 1404.746243 | 1314.685511 | 1715.27024 | 1435.19933 | 1259.15088 | 1168.32791 | 753.412 | 306.329 | 300.085 | 1012.160446 | 934.3653889 | 1881.38 |
| 753.697 | 892.743 | 844.052 | 1425.814668 | 1318.03114 | 1698.71348 | 1440.35606 | 1318.10044 | 1172.64894 | 753.697 | 303.782 | 300.839 | 975.5211986 | 936.5877894 | 1894.65 |
| 753.981 | 882.645 | 846.143 | 1362.891965 | 1321.388073 | 1677.52931 | 1445.5233 | 1311.53758 | 1176.98159 | 753.981 | 310.081 | 301.593 | 990.6551607 | 938.8135019 | 1959.63 |
| 754.265 | 896.692 | 848.238 | 1406.712931 | 1324.733702 | 1703.28781 | 1450.70103 | 1320.70235 | 1181.31424 | 754.265 | 285.633 | 302.349 | 999.928963 | 941.0425265 | 1936.79 |
| 754.549 | 932.172 | 850.335 | 1446.668339 | 1328.090635 | 1748.70217 | 1455.88927 | 1359.88202 | 1185.67012 | 754.549 | 327.61 | 303.106 | 998.9397574 | 943.2737592 | 1931.37 |
| 754.834 | 874.995 | 852.435 | 1420.072844 | 1331.45887 | 1765.71531 | 1461.09852 | 1271.16149 | 1190.03761 | 754.834 | 303.922 | 303.863 | 978.243722 | 945.508304 | 1929.53 |
| 755.118 | 889.071 | 854.538 | 1410.521976 | 1334.827105 | 1704.35976 | 1466.31826 | 1262.48458 | 1194.41672 | 755.118 | 301.798 | 304.622 | 999.7644634 | 947.7450567 | 1914.88 |
| 755.402 | 885.556 | 856.644 | 1424.57136 | 1338.19534 | 1704.03074 | 1471.54851 | 1286.49418 | 1198.79583 | 755.402 | 300.337 | 305.382 | 1004.957793 | 949.9851216 | 1937.45 |
| 755.686 | 915.493 | 858.753 | 1408.216204 | 1341.574878 | 1753.24467 | 1476.78926 | 1340.61162 | 1203.19817 | 755.686 | 315.587 | 306.143 | 1009.710616 | 952.2273945 | 1913.38 |
| 755.97 | 917.324 | 860.865 | 1462.820044 | 1344.954417 | 1786.10348 | 1482.05102 | 1330.78475 | 1207.61213 | 755.97 | 317.193 | 306.905 | 968.0403314 | 954.4718755 | 1967.55 |
| 756.254 | 906.67 | 862.98 | 1488.545224 | 1348.345257 | 1813.67686 | 1487.32328 | 1392.05744 | 1212.02608 | 756.254 | 315.227 | 307.668 | 1029.569802 | 956.7207725 | 1953.54 |
| 756.538 | 924.691 | 865.097 | 1425.690337 | 1351.736098 | 1735.02161 | 1492.61655 | 1306.64738 | 1216.46327 | 756.538 | 308.738 | 308.433 | 988.4305522 | 958.9707736 | 1928.85 |
| 756.823 | 913.293 | 867.217 | 1483.097273 | 1355.126939 | 1816.5 | 1497.90981 | 1319.21554 | 1220.91208 | 756.823 | 324.938 | 309.198 | 1015.356596 | 961.2240868 | 1992.06 |
| 757.107 | 861.946 | 869.341 | 1432.618955 | 1358.529083 | 1721.69129 | 1503.22408 | 1345.44374 | 1225.3725 | 757.107 | 325.962 | 309.964 | 999.8119365 | 963.480712 | 1913.59 |
| 757.391 | 918.829 | 871.467 | 1392.222738 | 1361.942529 | 1731.46615 | 1508.54885 | 1313.41932 | 1229.84453 | 757.391 | 305.817 | 310.732 | 1001.358674 | 965.7395452 | 1985.43 |
| 757.675 | 900.31 | 873.596 | 1482.193048 | 1365.355976 | 1805.32418 | 1513.89463 | 1297.73816 | 1234.31657 | 757.675 | 311.937 | 311.5 | 1032.460137 | 968.0016906 | 1920.45 |
| 757.959 | 908.439 | 875.728 | 1462.435749 | 1368.769422 | 1785.06337 | 1519.2404 | 1348.12696 | 1238.81183 | 757.959 | 317.466 | 312.27 | 1012.193567 | 970.266044 | 1995.08 |
| 758.243 | 911.069 | 877.863 | 1470.732006 | 1372.194171 | 1813.71931 | 1524.60719 | 1438.31107 | 1243.31871 | 758.243 | 321.816 | 313.04 | 1038.01669 | 972.5337094 | 1999.26 |
| 758.527 | 931.908 | 880 | 1443.040139 | 1375.618921 | 1759.75063 | 1529.99497 | 1456.25728 | 1247.83721 | 758.527 | 317.719 | 313.812 | 1045.895006 | 974.803583 | 1950.83 |
| 758.811 | 923.011 | 882.141 | 1527.505985 | 1379.04367 | 1748.75524 | 1535.38276 | 1334.54823 | 1252.36733 | 758.811 | 323.608 | 314.585 | 1011.626098 | 977.0767685 | 1987.05 |
| 759.095 | 923.076 | 884.284 | 1480.87062 | 1382.479722 | 1838.57569 | 1540.79155 | 1403.12718 | 1256.90906 | 759.095 | 304.998 | 315.359 | 1004.607817 | 979.3521622 | 1996.47 |
| 759.379 | 915.842 | 886.431 | 1456.298327 | 1385.927077 | 1823.25007 | 1546.21085 | 1335.52395 | 1261.4624 | 759.379 | 321.863 | 316.134 | 1007.460615 | 981.6308679 | 1981.22 |
| 759.663 | 911.861 | 888.58 | 1485.70822 | 1389.374432 | 1846.69487 | 1551.65115 | 1408.44716 | 1266.02736 | 759.663 | 327.483 | 316.909 | 1063.563807 | 983.9117816 | 1988.3 |
| 759.947 | 948.431 | 890.732 | 1487.686211 | 1392.821786 | 1781.58221 | 1557.10195 | 1421.08502 | 1270.60394 | 759.947 | 311.382 | 317.687 | 1023.058268 | 986.1960075 | 1983.91 |
| 760.231 | 923.441 | 892.887 | 1456.23051 | 1396.280444 | 1872.51705 | 1562.56325 | 1380.94124 | 1275.19213 | 760.231 | 333.671 | 318.465 | 987.4534909 | 988.4824413 | 2007.26 |
| 760.515 | 943.051 | 895.044 | 1465.74747 | 1399.739102 | 1842.69365 | 1568.03506 | 1465.79375 | 1279.79194 | 760.515 | 327.18 | 319.244 | 1070.353555 | 990.7721873 | 2016.58 |
| 760.799 | 930.25 | 897.205 | 1489.845046 | 1403.197759 | 1824.89513 | 1573.52787 | 1361.54306 | 1284.40336 | 760.799 | 322.872 | 320.024 | 1035.527115 | 993.0641413 | 2019.99 |
| 761.082 | 954.25 | 899.369 | 1506.460166 | 1406.66772 | 1935.04007 | 1579.03119 | 1399.01523 | 1289.03802 | 761.082 | 328.641 | 320.805 | 1068.524187 | 995.3583033 | 2005.08 |
| 761.366 | 935.832 | 901.535 | 1500.469681 | 1410.148983 | 1862.24337 | 1584.545 | 1410.49152 | 1293.67267 | 761.366 | 336.895 | 321.588 | 1056.381026 | 997.6557774 | 2000.77 |
| 761.65 | 958.536 | 903.704 | 1469.714754 | 1413.630246 | 1940.18753 | 1590.06932 | 1387.32986 | 1298.31894 | 761.65 | 327.346 | 322.371 | 1048.800797 | 999.9565636 | 2050.64 |
| 761.934 | 938.541 | 905.876 | 1469.115706 | 1417.11151 | 1922.00692 | 1595.61465 | 1370.41744 | 1302.97683 | 761.934 | 317.255 | 323.156 | 1047.599619 | 1002.259558 | 2007.31 |
| 762.218 | 955.735 | 908.051 | 1511.704667 | 1420.604076 | 1877.87678 | 1601.17048 | 1489.25742 | 1307.65794 | 762.218 | 333.589 | 323.941 | 1042.883228 | 1004.565864 | 2010.97 |
| 762.502 | 935.482 | 910.229 | 1512.348926 | 1424.096642 | 1907.30749 | 1606.74731 | 1395.48407 | 1312.33906 | 762.502 | 325.264 | 324.728 | 1064.087114 | 1006.873274 | 2044.45 |
| 762.786 | 949.639 | 912.409 | 1560.71362 | 1427.589208 | 1848.49913 | 1612.33464 | 1444.69969 | 1317.04341 | 762.786 | 331.052 | 325.515 | 1055.495599 | 1009.185101 | 2052.07 |
| 763.069 | 951.668 | 914.593 | 1540.357272 | 1431.093077 | 1834.82919 | 1617.93248 | 1371.57901 | 1321.74776 | 763.069 | 336.985 | 326.304 | 1063.862998 | 1011.499135 | 2043.66 |
| 763.353 | 934.145 | 916.779 | 1476.168655 | 1434.608249 | 1869.35429 | 1623.54082 | 1448.02177 | 1326.47534 | 763.353 | 342.776 | 327.094 | 1033.058518 | 1013.815378 | 2037.36 |
| 763.637 | 959.95 | 918.968 | 1485.674312 | 1438.12342 | 1919.17317 | 1629.17016 | 1427.84534 | 1331.20292 | 763.637 | 319.351 | 327.885 | 1075.405569 | 1016.134932 | 2028.86 |
| 763.921 | 987.62 | 921.16 | 1515.140719 | 1441.638592 | 1887.15281 | 1634.81001 | 1396.22747 | 1335.95373 | 763.921 | 341.437 | 328.677 | 1095.237154 | 1018.456695 | 2034.08 |
| 764.204 | 954.322 | 923.355 | 1502.108587 | 1445.165066 | 1876.50766 | 1640.46036 | 1424.9995 | 1340.70454 | 764.204 | 326.734 | 329.47 | 1049.167333 | 1020.78177 | 2041.96 |
| 764.488 | 954.66 | 925.553 | 1523.267434 | 1448.691541 | 1940.3255 | 1646.13172 | 1427.84534 | 1345.47858 | 764.488 | 334.511 | 330.264 | 1078.612759 | 1023.109053 | 2100.82 |
| 764.772 | 975.86 | 927.754 | 1575.203813 | 1452.218015 | 1955.37517 | 1651.81357 | 1514.85837 | 1350.26424 | 764.772 | 346.472 | 331.059 | 1050.51093 | 1025.439647 | 2029.32 |
| 765.056 | 980.518 | 929.957 | 1543.94026 | 1455.755793 | 1918.86538 | 1657.50593 | 1478.78008 | 1355.06152 | 765.056 | 341.698 | 331.855 | 1073.246098 | 1027.77245 | 2076.22 |
| 765.339 | 947.766 | 932.163 | 1518.588074 | 1459.304873 | 1818.29365 | 1663.2193 | 1530.87639 | 1359.87041 | 765.339 | 324.358 | 332.652 | 1050.383968 | 1030.107461 | 2077.25 |
| 765.623 | 958.244 | 934.372 | 1630.813603 | 1462.853953 | 1960.11932 | 1668.94316 | 1477.85083 | 1364.69091 | 765.623 | 334.512 | 333.45 | 1069.612755 | 1032.445784 | 2076.11 |
| 765.907 | 970.38 | 936.584 | 1596.385265 | 1466.403033 | 1910.83111 | 1674.67753 | 1478.97755 | 1369.52303 | 765.907 | 331.506 | 334.25 | 1059.258113 | 1034.787419 | 2076.99 |
| 766.19 | 961.911 | 938.799 | 1510.314422 | 1469.963416 | 1804.9421 | 1680.43291 | 1534.47725 | 1374.36677 | 766.19 | 331.228 | 335.05 | 1042.663527 | 1037.130158 | 2065.67 |
| 766.474 | 927.445 | 941.016 | 1582.494121 | 1473.523799 | 1960.52263 | 1686.19879 | 1520.60813 | 1379.22212 | 766.474 | 341.59 | 335.851 | 1075.632998 | 1039.477314 | 2059.39 |
| 766.757 | 1002.77 | 943.237 | 1547.760608 | 1477.084182 | 1996.72463 | 1691.97517 | 1475.31861 | 1384.08909 | 766.757 | 349.345 | 336.654 | 1105.812601 | 1041.826677 | 2069.23 |
| 767.041 | 931.566 | 945.46 | 1602.89568 | 1480.655867 | 1988.44625 | 1697.77255 | 1522.98934 | 1388.97929 | 767.041 | 351.911 | 337.458 | 1131.547402 | 1044.178248 | 2139.45 |
| 767.324 | 983.387 | 947.686 | 1577.204409 | 1484.238856 | 1907.12706 | 1703.58044 | 1506.27439 | 1393.86949 | 767.324 | 363.261 | 338.262 | 1107.325114 | 1046.532028 | 2100.04 |
| 767.608 | 1015.33 | 949.915 | 1564.952171 | 1487.810541 | 1978.09827 | 1709.39883 | 1533.24598 | 1398.77131 | 767.608 | 355.603 | 339.068 | 1101.948516 | 1048.889119 | 2111.03 |
| 767.891 | 979.847 | 952.147 | 1569.676742 | 1491.404833 | 1934.84903 | 1715.22772 | 1498.36411 | 1403.69635 | 767.891 | 355.783 | 339.875 | 1125.199263 | 1051.249522 | 2158.61 |
| 768.175 | 997.867 | 954.381 | 1574.83082 | 1494.987821 | 1988.62668 | 1721.07762 | 1510.44441 | 1408.6214 | 768.175 | 339.634 | 340.682 | 1101.254085 | 1053.612134 | 2065.66 |
| 768.459 | 982.275 | 956.619 | 1590.892103 | 1498.593415 | 1989.40145 | 1726.94852 | 1507.4708 | 1413.56968 | 768.459 | 342.871 | 341.491 | 1044.459774 | 1055.976954 | 2118.71 |
| 768.742 | 958.179 | 958.859 | 1602.115786 | 1502.187707 | 2036.64129 | 1732.81942 | 1435.56977 | 1418.52957 | 768.742 | 345.727 | 342.301 | 1069.134713 | 1058.345085 | 2114.12 |
| 769.026 | 991.869 | 961.102 | 1579.657117 | 1505.793301 | 1988.19153 | 1738.71133 | 1561.73924 | 1423.48946 | 769.026 | 332.003 | 343.112 | 1105.867802 | 1060.715425 | 2081.28 |
| 769.309 | 1011.86 | 963.348 | 1612.310915 | 1509.398895 | 1960.95777 | 1744.62424 | 1619.75954 | 1428.47259 | 769.309 | 355.478 | 343.924 | 1128.68798 | 1063.087973 | 2072.45 |
| 769.592 | 994.698 | 965.596 | 1618.889146 | 1513.015792 | 2026.16596 | 1750.53716 | 1547.27772 | 1433.46733 | 769.592 | 342.644 | 344.737 | 1137.078563 | 1065.463832 | 2172.59 |
| 769.876 | 1015.19 | 967.848 | 1610.694614 | 1516.632689 | 2016.96422 | 1756.47107 | 1542.06228 | 1438.47369 | 769.876 | 337.134 | 345.551 | 1142.554522 | 1067.843004 | 2123.33 |
| 770.159 | 984.928 | 970.102 | 1627.298431 | 1520.260888 | 1944.04016 | 1762.426 | 1522.62926 | 1443.49166 | 770.159 | 353.975 | 346.366 | 1132.187736 | 1070.224384 | 2148.82 |
| 770.443 | 1006.22 | 972.359 | 1616.074748 | 1523.889088 | 2020.48784 | 1768.39142 | 1616.93693 | 1448.52124 | 770.443 | 342.036 | 347.182 | 1091.541985 | 1072.607972 | 2075.11 |
| 770.726 | 999.556 | 974.619 | 1551.309688 | 1527.528591 | 1983.05469 | 1774.36735 | 1527.04321 | 1453.57406 | 770.726 | 341.796 | 348 | 1085.340682 | 1074.993768 | 2121.03 |
| 771.01 | 1016.47 | 976.882 | 1683.190791 | 1531.15679 | 2099.96031 | 1780.35378 | 1525.31248 | 1458.62688 | 771.01 | 350.097 | 348.818 | 1130.454418 | 1077.382876 | 2167.96 |
| 771.293 | 1013.97 | 979.147 | 1629.321633 | 1534.807596 | 2025.18953 | 1786.36121 | 1612.10481 | 1463.69132 | 771.293 | 347.188 | 349.637 | 1108.230413 | 1079.775297 | 2188.47 |
| 771.576 | 1024.84 | 981.415 | 1637.550073 | 1538.458401 | 2050.48105 | 1792.37915 | 1534.87218 | 1468.77898 | 771.576 | 367.275 | 350.457 | 1047.807175 | 1082.169925 | 2148.63 |
| 771.86 | 985.493 | 983.687 | 1611.576232 | 1542.109206 | 2086.06748 | 1798.40759 | 1565.54918 | 1473.87827 | 771.86 | 347.606 | 351.279 | 1142.355798 | 1084.566761 | 2144.66 |
| 772.143 | 1043.9 | 985.96 | 1650.231818 | 1545.760012 | 2077.81033 | 1804.45704 | 1621.4322 | 1478.97755 | 772.143 | 356.168 | 352.101 | 1138.657317 | 1086.965806 | 2162.54 |
| 772.426 | 1019.98 | 988.237 | 1636.510215 | 1549.433422 | 2111.57127 | 1810.51698 | 1569.60305 | 1484.10006 | 772.426 | 356.851 | 352.925 | 1123.973797 | 1089.369266 | 2187.26 |
| 772.709 | 997.68 | 990.517 | 1629.604203 | 1553.095531 | 2058.16509 | 1816.59794 | 1618.34243 | 1489.23419 | 772.709 | 378.889 | 353.749 | 1135.102359 | 1091.77383 | 2165.96 |
| 772.993 | 1044.77 | 992.799 | 1683.190791 | 1556.768941 | 2193.89872 | 1822.68939 | 1609.8746 | 1494.37993 | 772.993 | 345.579 | 354.575 | 1131.525321 | 1094.181707 | 2185.28 |
| 773.276 | 1051.74 | 995.084 | 1625.908186 | 1560.442352 | 2106.91203 | 1828.79135 | 1691.45149 | 1499.53729 | 773.276 | 356.471 | 355.401 | 1172.396293 | 1096.591791 | 2182.23 |
| 773.559 | 1038.06 | 997.372 | 1651.520338 | 1564.127066 | 2069.69114 | 1834.91431 | 1570.47423 | 1504.70627 | 773.559 | 352.927 | 356.229 | 1093.456363 | 1099.004084 | 2176.2 |
| 773.842 | 1048.85 | 999.663 | 1641.585174 | 1567.81178 | 2162.18615 | 1841.04778 | 1663.94557 | 1509.88686 | 773.842 | 370.786 | 357.058 | 1141.306975 | 1101.419689 | 2199.79 |
| 774.126 | 1042.92 | 1001.96 | 1667.061691 | 1571.507796 | 2052.4339 | 1847.19174 | 1587.46796 | 1515.09068 | 774.126 | 365.374 | 357.888 | 1175.134273 | 1103.838606 | 2224.86 |
| 774.409 | 1032.08 | 1004.25 | 1618.595273 | 1575.203813 | 2117.02651 | 1853.35672 | 1628.55261 | 1520.29451 | 774.409 | 337.527 | 358.718 | 1126.104563 | 1106.25421 | 2223.62 |
| 774.692 | 1031.45 | 1006.55 | 1653.147941 | 1578.899829 | 2073.44826 | 1859.53219 | 1662.4123 | 1525.52156 | 774.692 | 366.158 | 359.55 | 1149.995644 | 1108.683063 | 2177.72 |
| 774.975 | 1037.84 | 1008.85 | 1655.182446 | 1582.607149 | 2176.81129 | 1865.72867 | 1615.76375 | 1530.74861 | 774.975 | 379.798 | 360.383 | 1133.898973 | 1111.111916 | 2193.51 |
| 775.258 | 991.195 | 1011.16 | 1689.339516 | 1586.314468 | 2151.13769 | 1871.93565 | 1564.48054 | 1535.9989 | 775.258 | 371.926 | 361.217 | 1118.961527 | 1113.540769 | 2220.22 |
| 775.542 | 1058.51 | 1013.47 | 1654.300827 | 1590.03309 | 2152.86766 | 1878.15314 | 1633.93066 | 1541.2608 | 775.542 | 384.692 | 362.052 | 1160.252028 | 1115.969622 | 2214.2 |
| 775.825 | 1060.27 | 1015.77 | 1684.886211 | 1593.751712 | 2077.18414 | 1884.38113 | 1651.17994 | 1546.53432 | 775.825 | 374.859 | 362.888 | 1129.030227 | 1118.398475 | 2247.13 |
| 776.108 | 1052.77 | 1018.09 | 1706.180692 | 1597.470334 | 2131.41816 | 1890.63012 | 1635.90533 | 1551.81945 | 776.108 | 369.07 | 363.725 | 1133.501525 | 1120.838368 | 2224.96 |
| 776.391 | 1059.06 | 1020.4 | 1676.454321 | 1601.200259 | 2139.29324 | 1896.90012 | 1686.39867 | 1557.1162 | 776.391 | 347.091 | 364.563 | 1176.006452 | 1123.278262 | 2208.05 |
| 776.674 | 1071.44 | 1022.72 | 1643.043235 | 1604.941487 | 2208.95901 | 1903.18062 | 1665.86216 | 1562.42456 | 776.674 | 373.084 | 365.402 | 1145.800353 | 1125.718155 | 2202.7 |
| 776.957 | 1042.38 | 1025.04 | 1654.23301 | 1608.671412 | 2158.29106 | 1909.47162 | 1634.44175 | 1567.75616 | 776.957 | 382.806 | 366.242 | 1199.478004 | 1128.169088 | 2223.66 |
| 777.24 | 1057.85 | 1027.37 | 1727.113483 | 1612.423943 | 2198.69594 | 1915.77313 | 1713.67228 | 1573.08775 | 777.24 | 373.215 | 367.084 | 1109.820208 | 1130.608981 | 2220.12 |
| 777.523 | 1049.7 | 1029.69 | 1636.442399 | 1616.16517 | 2210.76327 | 1922.09564 | 1686.4219 | 1578.44258 | 777.523 | 383.822 | 367.926 | 1134.186019 | 1133.059915 | 2257.32 |
| 777.806 | 1090.54 | 1032.02 | 1740.405579 | 1619.917701 | 2181.66157 | 1928.43915 | 1622.03621 | 1583.80902 | 777.806 | 370.462 | 368.769 | 1181.360969 | 1135.510848 | 2285.28 |
| 778.089 | 1084.66 | 1034.36 | 1715.901102 | 1623.681534 | 2173.54239 | 1934.78267 | 1735.30067 | 1589.18708 | 778.089 | 368.229 | 369.613 | 1144.122236 | 1137.961782 | 2267.36 |
| 778.372 | 1074.81 | 1036.69 | 1658.50547 | 1627.445368 | 2213.42721 | 1941.14719 | 1649.83252 | 1594.57675 | 778.372 | 357.086 | 370.459 | 1201.465247 | 1140.423756 | 2252.56 |
| 778.655 | 1046.61 | 1039.03 | 1705.683369 | 1631.209201 | 2271.07872 | 1947.53271 | 1656.97616 | 1599.97804 | 778.655 | 391.18 | 371.305 | 1162.150949 | 1142.885729 | 2253.62 |
| 778.938 | 1065.07 | 1041.37 | 1683.722023 | 1634.984337 | 2163.87367 | 1953.92874 | 1659.72908 | 1605.39095 | 778.938 | 373.204 | 372.152 | 1160.163706 | 1145.347703 | 2259.22 |
| 779.221 | 1058.2 | 1043.71 | 1688.050996 | 1638.759473 | 2232.22339 | 1960.33527 | 1622.15237 | 1610.81547 | 779.221 | 374.627 | 373.001 | 1184.882805 | 1147.809677 | 2272.65 |
| 779.504 | 1068.21 | 1046.06 | 1749.526941 | 1642.534609 | 2222.55466 | 1966.76281 | 1682.57711 | 1616.26322 | 779.504 | 382.616 | 373.85 | 1169.547911 | 1150.282691 | 2247.72 |
| 779.787 | 1108.48 | 1048.41 | 1731.159886 | 1646.321048 | 2273.49855 | 1973.20084 | 1775.7116 | 1621.71097 | 779.787 | 379.287 | 374.701 | 1149.101385 | 1152.744664 | 2284.24 |
| 780.07 | 1111.94 | 1050.76 | 1746.41867 | 1650.11879 | 2232.52056 | 1979.64938 | 1673.58658 | 1627.18196 | 780.07 | 376.057 | 375.552 | 1179.219162 | 1155.217678 | 2265.32 |
| 780.353 | 1082.86 | 1053.11 | 1763.847592 | 1653.905229 | 2324.42122 | 1986.11893 | 1812.76561 | 1632.66455 | 780.353 | 391.495 | 376.405 | 1218.886747 | 1157.701732 | 2283.93 |
| 780.635 | 1088.36 | 1055.47 | 1778.529933 | 1657.714273 | 2305.00947 | 1992.59898 | 1803.57761 | 1638.15877 | 780.635 | 389.579 | 377.258 | 1184.231431 | 1160.174746 | 2265.28 |
| 780.918 | 1095.09 | 1057.83 | 1739.546566 | 1661.512015 | 2255.76371 | 1999.10003 | 1670.15996 | 1643.6646 | 780.918 | 392.614 | 378.113 | 1194.300131 | 1162.658801 | 2296 |
| 781.201 | 1100.57 | 1060.19 | 1758.162282 | 1665.32106 | 2337.54989 | 2005.61159 | 1821.96522 | 1649.18205 | 781.201 | 389.642 | 378.968 | 1232.963054 | 1165.142855 | 2331.54 |
| 781.484 | 1109.35 | 1062.56 | 1748.634019 | 1669.141407 | 2211.04983 | 2012.14415 | 1701.33643 | 1654.72272 | 781.484 | 404.459 | 379.825 | 1241.684844 | 1167.626909 | 2327.84 |
| 781.767 | 1078.07 | 1064.93 | 1744.610222 | 1672.961755 | 2314.93292 | 2018.67671 | 1790.81198 | 1660.2634 | 781.767 | 385.932 | 380.683 | 1194.476775 | 1170.110963 | 2302.25 |
| 782.05 | 1060.05 | 1067.3 | 1760.89756 | 1676.782102 | 2332.37059 | 2025.24077 | 1758.26486 | 1665.82731 | 782.05 | 391.198 | 381.541 | 1217.539838 | 1172.606058 | 2315.32 |
| 782.332 | 1111.84 | 1069.67 | 1762.592981 | 1680.602449 | 2300.09551 | 2031.80484 | 1836.55451 | 1671.39122 | 782.332 | 397.268 | 382.401 | 1247.602413 | 1175.090112 | 2289.93 |
| 782.615 | 1129.54 | 1072.05 | 1746.520395 | 1684.445402 | 2277.61652 | 2038.38992 | 1806.58607 | 1676.97836 | 782.615 | 374.755 | 383.262 | 1216.678699 | 1177.596246 | 2295.15 |
| 782.898 | 1113.64 | 1074.43 | 1746.644726 | 1688.277052 | 2354.61609 | 2044.99599 | 1837.507 | 1682.58873 | 782.898 | 382.583 | 384.123 | 1164.292756 | 1180.091341 | 2316.59 |
| 783.181 | 1080.93 | 1076.81 | 1742.813076 | 1692.120005 | 2329.67481 | 2051.61257 | 1864.68768 | 1688.1991 | 783.181 | 383.192 | 384.986 | 1230.622523 | 1182.586435 | 2364.83 |
| 783.463 | 1113.08 | 1079.19 | 1752.703028 | 1695.962958 | 2376.51135 | 2058.23966 | 1826.83219 | 1693.82109 | 783.463 | 394.185 | 385.85 | 1193.207147 | 1185.09257 | 2344.46 |
| 783.746 | 1093.24 | 1081.58 | 1819.366959 | 1699.817214 | 2288.7605 | 2064.88775 | 1885.43328 | 1699.46631 | 783.746 | 379.333 | 386.715 | 1230.357557 | 1187.598705 | 2388.48 |
| 784.029 | 1110.07 | 1083.97 | 1757.879712 | 1703.67147 | 2335.35293 | 2071.54634 | 1730.01553 | 1705.11153 | 784.029 | 411.57 | 387.581 | 1206.113188 | 1190.104839 | 2315.96 |
| 784.311 | 1138.91 | 1086.36 | 1768.15396 | 1707.537028 | 2271.87472 | 2078.22593 | 1770.60071 | 1710.77997 | 784.311 | 393.64 | 388.448 | 1195.238551 | 1192.622014 | 2318.35 |
| 784.594 | 1131.45 | 1088.76 | 1781.570387 | 1711.402587 | 2404.04228 | 2084.90553 | 1819.29362 | 1716.46004 | 784.594 | 398.238 | 389.315 | 1253.3875 | 1195.128149 | 2359.42 |
| 784.877 | 1116.94 | 1091.15 | 1739.343115 | 1715.279448 | 2357.34371 | 2091.61663 | 1769.99669 | 1722.15172 | 784.877 | 398.707 | 390.184 | 1265.045994 | 1197.645324 | 2348.93 |
| 785.159 | 1122.87 | 1093.56 | 1823.899383 | 1719.145007 | 2371.97947 | 2098.33824 | 1875.88519 | 1727.85502 | 785.159 | 380.864 | 391.054 | 1231.053092 | 1200.162499 | 2342.98 |
| 785.442 | 1106.9 | 1095.96 | 1773.522791 | 1723.033171 | 2420.43984 | 2105.07034 | 1851.12057 | 1733.56993 | 785.442 | 380.742 | 391.925 | 1275.479021 | 1202.690714 | 2327.47 |
| 785.725 | 1128.93 | 1098.37 | 1747.447225 | 1726.910032 | 2382.76259 | 2111.82346 | 1920.67524 | 1739.30807 | 785.725 | 390.325 | 392.797 | 1241.5082 | 1205.207889 | 2366.82 |
| 786.007 | 1099.08 | 1100.77 | 1794.342554 | 1730.798197 | 2439.24663 | 2118.58707 | 1860.87774 | 1745.05783 | 786.007 | 397.491 | 393.67 | 1253.818069 | 1207.736104 | 2328.53 |
| 786.29 | 1139.33 | 1103.19 | 1813.704255 | 1734.697664 | 2370.21766 | 2125.36119 | 1844.04662 | 1750.80759 | 786.29 | 390.257 | 394.544 | 1261.579358 | 1210.264319 | 2401.36 |
| 786.572 | 1123.04 | 1105.6 | 1764.830936 | 1738.597131 | 2389.69308 | 2132.15631 | 1871.6687 | 1756.58058 | 786.572 | 392.296 | 395.419 | 1247.249126 | 1212.803574 | 2386.43 |
| 786.855 | 1126.13 | 1108.02 | 1809.12662 | 1742.496597 | 2450.12527 | 2138.97244 | 1793.18158 | 1762.36519 | 786.855 | 404.672 | 396.295 | 1216.899504 | 1215.331789 | 2357.16 |
| 787.137 | 1169.93 | 1110.44 | 1843.611472 | 1746.407367 | 2380.03497 | 2145.78857 | 1953.25721 | 1768.17303 | 787.137 | 393.153 | 397.172 | 1243.926013 | 1217.871045 | 2383.59 |
| 787.42 | 1155.65 | 1112.86 | 1787.707809 | 1750.318137 | 2406.86542 | 2152.63621 | 1903.61181 | 1773.98087 | 787.42 | 408.326 | 398.05 | 1238.394853 | 1220.4103 | 2363.01 |
| 787.702 | 1160.49 | 1115.28 | 1807.295566 | 1754.228907 | 2469.46273 | 2159.48384 | 1907.0849 | 1779.81194 | 787.702 | 407.979 | 398.929 | 1260.552616 | 1222.949556 | 2397.34 |
| 787.985 | 1099.9 | 1117.71 | 1830.206347 | 1758.150979 | 2455.50622 | 2166.35248 | 1866.83658 | 1785.65462 | 787.985 | 386.598 | 399.809 | 1250.925526 | 1225.488811 | 2377.05 |
| 788.267 | 1111.73 | 1120.14 | 1827.31283 | 1762.084355 | 2408.21331 | 2173.24213 | 1955.9172 | 1791.49731 | 788.267 | 409.702 | 400.69 | 1268.53471 | 1228.039107 | 2377.03 |
| 788.55 | 1119.38 | 1122.58 | 1821.1415 | 1766.006427 | 2411.60957 | 2180.14228 | 1899.85995 | 1797.37484 | 788.55 | 405.498 | 401.572 | 1273.425537 | 1230.589402 | 2444.83 |
| 788.832 | 1140.11 | 1125.01 | 1851.455617 | 1769.951106 | 2411.56712 | 2187.05293 | 1872.87673 | 1803.25237 | 788.832 | 403.006 | 402.455 | 1295.638501 | 1233.139698 | 2387.07 |
| 789.115 | 1184.55 | 1127.45 | 1833.020745 | 1773.884481 | 2471.98869 | 2193.98458 | 1818.32952 | 1809.14152 | 789.115 | 435.423 | 403.339 | 1249.313651 | 1235.689994 | 2441.18 |
| 789.397 | 1159.52 | 1129.89 | 1851.184349 | 1777.829159 | 2530.12841 | 2200.93724 | 1891.2295 | 1815.0539 | 789.397 | 411.422 | 404.225 | 1282.29085 | 1238.251329 | 2409.1 |
| 789.679 | 1150.84 | 1132.34 | 1872.241471 | 1781.78514 | 2411.80061 | 2207.8899 | 1955.48742 | 1820.96628 | 789.679 | 404.444 | 405.111 | 1268.269744 | 1240.801625 | 2434.85 |
| 789.962 | 1138.67 | 1134.78 | 1883.024345 | 1785.729818 | 2501.55738 | 2214.86357 | 1969.6237 | 1826.90189 | 789.962 | 408.384 | 405.998 | 1249.666938 | 1243.362961 | 2392.34 |
| 790.244 | 1158.94 | 1137.23 | 1853.343185 | 1789.697102 | 2410.37842 | 2221.85824 | 1934.6489 | 1832.84911 | 790.244 | 413.941 | 406.886 | 1269.671855 | 1245.924297 | 2416.51 |
| 790.526 | 1186.34 | 1139.68 | 1909.766776 | 1793.653083 | 2429.47177 | 2228.86341 | 1990.10213 | 1838.81957 | 790.526 | 407.657 | 407.775 | 1346.600253 | 1248.496673 | 2455.89 |
| 790.809 | 1170.34 | 1142.14 | 1844.142703 | 1797.620367 | 2475.89439 | 2235.88959 | 2002.35667 | 1844.79003 | 790.809 | 414.626 | 408.665 | 1242.954472 | 1251.058009 | 2399.76 |
| 791.091 | 1164.18 | 1144.59 | 1837.134966 | 1801.598953 | 2443.83158 | 2242.92627 | 1964.19918 | 1850.78372 | 791.091 | 412.312 | 409.556 | 1332.104417 | 1253.630385 | 2444.59 |
| 791.373 | 1149.49 | 1147.05 | 1861.899407 | 1805.57754 | 2499.62576 | 2249.97346 | 1930.07232 | 1856.78902 | 791.373 | 399.033 | 410.448 | 1295.649541 | 1256.202761 | 2460.78 |
| 791.655 | 1205.38 | 1149.52 | 1837.519261 | 1809.556127 | 2478.79183 | 2257.04165 | 1960.26146 | 1862.80594 | 791.655 | 405.778 | 411.341 | 1269.241285 | 1258.786177 | 2425.09 |
| 791.938 | 1174.57 | 1151.98 | 1854.778641 | 1813.546016 | 2580.04281 | 2264.12034 | 1993.36614 | 1868.83448 | 791.938 | 423.56 | 412.236 | 1301.390467 | 1261.358553 | 2436.82 |
| 792.22 | 1211.71 | 1154.45 | 1885.578778 | 1817.535905 | 2466.95799 | 2271.22003 | 1982.55194 | 1874.87463 | 792.22 | 381.881 | 413.131 | 1289.61053 | 1263.94197 | 2460.02 |
| 792.502 | 1165.06 | 1156.92 | 1901.527033 | 1821.525795 | 2500.87813 | 2278.33023 | 2014.19304 | 1880.93801 | 792.502 | 398.657 | 414.027 | 1275.964792 | 1266.525386 | 2493.47 |
| 792.784 | 1167.86 | 1159.39 | 1875.779248 | 1825.526987 | 2494.51014 | 2285.46143 | 2074.80364 | 1887.00139 | 792.784 | 412.098 | 414.924 | 1258.101682 | 1269.108803 | 2449.98 |
| 793.067 | 1187.64 | 1161.87 | 1924.17785 | 1829.528179 | 2549.11562 | 2292.60314 | 1922.41759 | 1893.08801 | 793.067 | 423.381 | 415.822 | 1258.035441 | 1271.692219 | 2424.56 |
| 793.349 | 1184.67 | 1164.35 | 1909.891107 | 1833.540674 | 2555.58974 | 2299.75535 | 1996.65337 | 1899.18624 | 793.349 | 415.701 | 416.721 | 1327.114228 | 1274.286675 | 2446.58 |
| 793.631 | 1163.64 | 1166.83 | 1872.151049 | 1837.553169 | 2558.33859 | 2306.93906 | 1956.41667 | 1905.29608 | 793.631 | 426.059 | 417.621 | 1293.187568 | 1276.881132 | 2462.12 |
| 793.913 | 1151.63 | 1169.31 | 1882.56093 | 1841.576967 | 2511.80984 | 2314.12278 | 2016.43487 | 1911.42916 | 793.913 | 448.287 | 418.522 | 1266.657869 | 1279.475589 | 2477.61 |
| 794.195 | 1211.23 | 1171.8 | 1866.737006 | 1845.600765 | 2575.98853 | 2321.3275 | 1974.53713 | 1917.56224 | 794.195 | 428.833 | 419.424 | 1286.464061 | 1282.070045 | 2468.06 |
| 794.477 | 1184.73 | 1174.29 | 1906.240302 | 1849.624563 | 2568.80332 | 2328.55323 | 1987.47699 | 1923.71855 | 794.477 | 432.68 | 420.327 | 1321.572027 | 1284.675542 | 2491.86 |
| 794.76 | 1240.63 | 1176.78 | 1852.585897 | 1853.659663 | 2557.60627 | 2335.77895 | 2007.44434 | 1929.88647 | 794.76 | 426.887 | 421.232 | 1313.413289 | 1287.281039 | 2467.29 |
| 795.042 | 1164 | 1179.27 | 1830.850607 | 1857.694764 | 2588.0028 | 2343.03619 | 2072.5502 | 1936.06601 | 795.042 | 422.803 | 422.137 | 1313.843858 | 1289.886536 | 2460.24 |
| 795.324 | 1180.83 | 1181.77 | 1910.603184 | 1861.729864 | 2590.92146 | 2350.29342 | 2029.28181 | 1942.26878 | 795.324 | 440.545 | 423.043 | 1289.665731 | 1292.492033 | 2496.02 |
| 795.606 | 1219.13 | 1184.27 | 1907.427096 | 1865.776268 | 2645.31467 | 2357.58216 | 1973.64272 | 1948.47155 | 795.606 | 433.951 | 423.95 | 1332.899314 | 1295.097529 | 2525.95 |
| 795.888 | 1215.37 | 1186.77 | 1918.57166 | 1869.833974 | 2669.70406 | 2364.8709 | 2032.18573 | 1954.69755 | 795.888 | 444.814 | 424.858 | 1310.885074 | 1297.714066 | 2548.54 |
| 796.17 | 1185.95 | 1189.28 | 1916.345008 | 1873.880378 | 2585.92259 | 2372.19115 | 2114.13432 | 1960.93517 | 796.17 | 422.81 | 425.767 | 1326.805101 | 1300.330604 | 2518.26 |
| 796.452 | 1218.42 | 1191.79 | 1972.960747 | 1877.938084 | 2614.47239 | 2379.5114 | 2067.60192 | 1967.1844 | 796.452 | 432.31 | 426.677 | 1308.997193 | 1302.9361 | 2530.43 |
| 796.734 | 1212.65 | 1194.3 | 1893.49074 | 1882.007093 | 2614.62098 | 2386.86316 | 2016.12125 | 1973.44525 | 796.734 | 427.967 | 427.588 | 1295.274173 | 1305.563678 | 2534.91 |
| 797.016 | 1214.52 | 1196.81 | 1951.496725 | 1886.076102 | 2626.63525 | 2394.21491 | 2105.05086 | 1979.72933 | 797.016 | 427.235 | 428.5 | 1281.672597 | 1308.180215 | 2520.34 |
| 797.298 | 1201.24 | 1199.32 | 1967.309346 | 1890.145111 | 2591.55825 | 2401.58767 | 2036.15829 | 1986.02503 | 797.298 | 422.083 | 429.413 | 1307.926289 | 1310.807792 | 2507.9 |
| 797.58 | 1205.4 | 1201.84 | 1977.16539 | 1894.225423 | 2695.68545 | 2408.98144 | 2129.22308 | 1992.32073 | 797.58 | 436.147 | 430.327 | 1345.54039 | 1313.435369 | 2544.1 |
| 797.862 | 1206.22 | 1204.36 | 1928.868513 | 1898.305734 | 2637.34408 | 2416.38571 | 2081.69174 | 1998.65127 | 797.862 | 436.259 | 431.242 | 1355.123319 | 1316.062947 | 2506.09 |
| 798.144 | 1228.72 | 1206.88 | 1934.587731 | 1902.397349 | 2682.10041 | 2423.80048 | 2078.06765 | 2004.98181 | 798.144 | 448.468 | 432.158 | 1345.705993 | 1318.690524 | 2588.64 |
| 798.426 | 1219.63 | 1209.41 | 2016.08094 | 1906.488964 | 2702.93434 | 2431.23626 | 2111.31171 | 2011.33559 | 798.426 | 436.525 | 433.075 | 1319.264617 | 1321.329142 | 2522.54 |
| 798.707 | 1191.04 | 1211.94 | 1997.069626 | 1910.580578 | 2641.86535 | 2438.69304 | 2146.92537 | 2017.68936 | 798.707 | 430.474 | 433.993 | 1319.264617 | 1323.967759 | 2575.87 |
| 798.989 | 1215.72 | 1214.47 | 1939.244486 | 1914.683496 | 2672.31494 | 2446.16032 | 2155.93914 | 2024.06637 | 798.989 | 434.502 | 434.911 | 1363.944471 | 1326.606377 | 2531.11 |
| 799.271 | 1275.32 | 1217 | 1986.851892 | 1918.786413 | 2678.33268 | 2453.63811 | 2106.27051 | 2030.45499 | 799.271 | 461.168 | 435.831 | 1347.251627 | 1329.244994 | 2576.1 |
| 799.553 | 1254.13 | 1219.54 | 2011.119009 | 1922.889331 | 2770.31825 | 2461.1369 | 2158.33197 | 2036.86684 | 799.553 | 444.16 | 436.752 | 1353.113995 | 1331.883612 | 2544.44 |
| 799.835 | 1219.94 | 1222.08 | 1957.396788 | 1927.003551 | 2678.09919 | 2468.65669 | 2134.97284 | 2043.2787 | 799.835 | 436.244 | 437.674 | 1355.509727 | 1334.53327 | 2520.92 |
| 800.117 | 1227.97 | 1224.62 | 2038.765665 | 1931.129074 | 2804.04735 | 2476.18699 | 2146.91376 | 2049.71378 | 800.117 | 441.921 | 438.597 | 1364.419201 | 1337.182927 | 2658.32 |
| 800.399 | 1261.57 | 1227.16 | 1979.324225 | 1935.243294 | 2774.5848 | 2483.72779 | 2175.63932 | 2056.16048 | 800.399 | 443.157 | 439.521 | 1362.255314 | 1339.832585 | 2555.55 |
| 800.68 | 1234.08 | 1229.71 | 1975.063068 | 1939.368817 | 2727.59967 | 2491.2896 | 2122.00975 | 2062.6188 | 800.68 | 453.802 | 440.445 | 1343.553146 | 1342.482243 | 2572.39 |
| 800.962 | 1232.67 | 1232.26 | 2047.028014 | 1943.505643 | 2756.18132 | 2498.86191 | 2119.66338 | 2069.10034 | 800.962 | 459.091 | 441.371 | 1335.957461 | 1345.131901 | 2615.89 |
| 801.244 | 1244.43 | 1234.81 | 2027.37244 | 1947.642469 | 2729.87092 | 2506.45522 | 2176.75443 | 2075.58189 | 801.244 | 438.081 | 442.298 | 1336.652996 | 1347.792599 | 2594.21 |
| 801.525 | 1279.61 | 1237.37 | 2018.49974 | 1951.779295 | 2746.36401 | 2514.05904 | 2210.81159 | 2082.08667 | 801.525 | 453.385 | 443.225 | 1312.419667 | 1350.453297 | 2592.78 |
| 801.807 | 1273.69 | 1239.92 | 2054.736526 | 1955.927423 | 2822.20673 | 2521.68386 | 2140.28121 | 2088.60306 | 801.807 | 441.652 | 444.154 | 1322.665011 | 1353.113995 | 2621.44 |
| 802.089 | 1232.19 | 1242.48 | 2024.478922 | 1960.075552 | 2704.08058 | 2529.31919 | 2146.37944 | 2095.14269 | 802.089 | 467.24 | 445.084 | 1395.928049 | 1355.785733 | 2610.61 |
| 802.371 | 1267.44 | 1245.04 | 1987.936961 | 1964.234983 | 2813.20664 | 2536.97551 | 2148.29602 | 2101.68232 | 802.371 | 424.313 | 446.014 | 1392.174367 | 1358.446431 | 2600.48 |
| 802.652 | 1258.48 | 1247.61 | 2032.164828 | 1968.383112 | 2788.42456 | 2544.65285 | 2188.98574 | 2108.24517 | 802.652 | 460.387 | 446.946 | 1374.653505 | 1361.118169 | 2627.52 |
| 802.934 | 1244.33 | 1250.17 | 2008.756724 | 1972.553846 | 2820.95436 | 2552.33018 | 2232.34705 | 2114.81964 | 802.934 | 467.225 | 447.878 | 1367.775435 | 1363.789908 | 2615.21 |
| 803.216 | 1233.95 | 1252.74 | 1983.528868 | 1976.713278 | 2824.42491 | 2560.03902 | 2273.57109 | 2121.40573 | 803.216 | 461.568 | 448.812 | 1381.630937 | 1366.461646 | 2632.72 |
| 803.497 | 1279.73 | 1255.32 | 2061.947714 | 1980.895315 | 2837.40499 | 2567.74786 | 2250.65336 | 2128.01505 | 803.497 | 449.435 | 449.746 | 1396.391739 | 1369.133384 | 2632.3 |
| 803.779 | 1273.37 | 1257.89 | 2010.248694 | 1985.066049 | 2811.83752 | 2575.48821 | 2213.66905 | 2134.62437 | 803.779 | 453.708 | 450.682 | 1399.085558 | 1371.816163 | 2615.64 |
| 804.06 | 1268.1 | 1260.47 | 1997.386104 | 1989.248086 | 2839.05005 | 2583.23907 | 2182.01633 | 2141.25692 | 804.06 | 472.867 | 451.618 | 1387.460184 | 1374.498941 | 2621.93 |
| 804.342 | 1266.25 | 1263.05 | 1955.215347 | 1993.430123 | 2866.3581 | 2591.00042 | 2178.68263 | 2147.90109 | 804.342 | 443.7 | 452.555 | 1449.053688 | 1377.18172 | 2648.93 |
| 804.624 | 1281.71 | 1265.63 | 2042.088689 | 1997.623463 | 2887.45736 | 2598.78278 | 2181.73755 | 2154.55687 | 804.624 | 454.973 | 453.494 | 1408.171676 | 1379.864498 | 2651.71 |
| 804.905 | 1288.53 | 1268.22 | 1990.841781 | 2001.816803 | 2937.99795 | 2606.57564 | 2197.95304 | 2161.23589 | 804.905 | 459.073 | 454.433 | 1399.869415 | 1382.558317 | 2693.47 |
| 805.187 | 1279.87 | 1270.8 | 2056.002439 | 2006.010143 | 2793.73121 | 2614.38951 | 2306.7803 | 2167.92651 | 805.187 | 454.997 | 455.373 | 1399.328443 | 1385.241096 | 2709.65 |
| 805.468 | 1298.92 | 1273.39 | 2096.771649 | 2010.214785 | 2943.41074 | 2622.21388 | 2320.41711 | 2174.62876 | 805.468 | 468.057 | 456.314 | 1439.007068 | 1387.934914 | 2639.32 |
| 805.75 | 1271.34 | 1275.99 | 2079.806142 | 2014.419428 | 2852.82613 | 2630.05925 | 2305.93236 | 2181.34262 | 805.75 | 475.79 | 457.257 | 1462.721506 | 1390.628733 | 2700.04 |
| 806.031 | 1277.57 | 1278.58 | 2055.143426 | 2018.635373 | 2785.36793 | 2637.91513 | 2240.22248 | 2188.0681 | 806.031 | 458.11 | 458.2 | 1424.080663 | 1393.322552 | 2667.34 |
| 806.313 | 1287.21 | 1281.18 | 2070.187457 | 2022.851319 | 2871.5374 | 2645.79201 | 2204.8876 | 2194.8168 | 806.313 | 467.854 | 459.144 | 1430.947693 | 1396.027411 | 2681.9 |
| 806.594 | 1280.37 | 1283.78 | 2058.703809 | 2027.067264 | 2897.90086 | 2653.67939 | 2295.86157 | 2201.57713 | 806.594 | 451.78 | 460.089 | 1408.392481 | 1398.72123 | 2688.72 |
| 806.875 | 1255.81 | 1286.38 | 2044.06668 | 2031.294512 | 2852.3273 | 2661.58778 | 2427.9318 | 2208.34907 | 806.875 | 458.417 | 461.035 | 1424.610594 | 1401.426089 | 2639.6 |
| 807.157 | 1299.07 | 1288.99 | 2098.060169 | 2035.521761 | 2881.46084 | 2669.51717 | 2268.94805 | 2215.14424 | 807.157 | 463.739 | 461.982 | 1427.260252 | 1404.130948 | 2686.51 |
| 807.438 | 1305.75 | 1291.6 | 2067.949502 | 2039.760312 | 2917.63101 | 2677.44657 | 2301.65779 | 2221.93941 | 807.438 | 461.001 | 462.93 | 1378.075979 | 1406.846847 | 2713.02 |
| 807.72 | 1333.38 | 1294.2 | 2084.214235 | 2043.998863 | 2833.0323 | 2685.40747 | 2218.18754 | 2228.75781 | 807.72 | 478.309 | 463.879 | 1443.014676 | 1409.551706 | 2662.38 |
| 808.001 | 1318.25 | 1296.82 | 2077.963785 | 2048.237414 | 3007.86537 | 2693.37887 | 2283.02625 | 2235.58783 | 808.001 | 465.879 | 464.828 | 1402.695716 | 1412.267605 | 2677.83 |
| 808.283 | 1299.06 | 1299.43 | 2036.516407 | 2052.487268 | 2890.49277 | 2701.36078 | 2363.33703 | 2242.44108 | 808.283 | 463.433 | 465.779 | 1425.537975 | 1414.983504 | 2704.45 |
| 808.564 | 1314.48 | 1302.05 | 2050.03456 | 2056.737122 | 2971.0478 | 2709.36369 | 2218.59409 | 2249.29432 | 808.564 | 487.931 | 466.731 | 1477.338785 | 1417.699404 | 2650.82 |
| 808.845 | 1311.86 | 1304.67 | 2104.367133 | 2060.998278 | 2903.16506 | 2717.38761 | 2324.93561 | 2256.1708 | 808.845 | 476.332 | 467.684 | 1413.548273 | 1420.415303 | 2691.66 |
| 809.126 | 1308.29 | 1307.3 | 2105.836497 | 2065.259435 | 2938.29512 | 2725.41153 | 2354.11418 | 2263.0589 | 809.126 | 440.123 | 468.637 | 1450.212913 | 1423.142242 | 2725.55 |
| 809.408 | 1336.97 | 1309.92 | 2091.018523 | 2069.520592 | 2984.86633 | 2733.46695 | 2305.20057 | 2269.97023 | 809.408 | 471.534 | 469.592 | 1425.946464 | 1425.869182 | 2745.16 |
| 809.689 | 1284.6 | 1312.55 | 2123.468869 | 2073.793051 | 2999.13062 | 2741.53288 | 2334.75085 | 2276.88155 | 809.689 | 489.928 | 470.547 | 1419.521043 | 1428.585081 | 2678.7 |
| 809.97 | 1323.01 | 1315.18 | 2087.17557 | 2078.065511 | 2991.0645 | 2749.61981 | 2404.82822 | 2283.81611 | 809.97 | 457.785 | 471.504 | 1424.069623 | 1431.323061 | 2689.83 |
| 810.252 | 1327.58 | 1317.81 | 2054.6348 | 2082.33797 | 2977.88277 | 2757.71724 | 2346.90085 | 2290.76229 | 810.252 | 476.323 | 472.461 | 1448.424394 | 1434.05 | 2717.96 |
| 810.533 | 1352.48 | 1320.45 | 2143.056627 | 2086.621732 | 2958.73636 | 2765.82518 | 2266.83399 | 2297.73169 | 810.533 | 467.111 | 473.42 | 1403.965344 | 1436.78798 | 2778.22 |
| 810.814 | 1338.08 | 1323.09 | 2096.715135 | 2090.905495 | 2984.27198 | 2773.95412 | 2406.15241 | 2304.7011 | 810.814 | 478.905 | 474.379 | 1403.258769 | 1439.514919 | 2692.76 |
| 811.095 | 1329.83 | 1325.72 | 2117.014969 | 2095.20056 | 2932.72313 | 2782.10407 | 2343.83431 | 2311.69374 | 811.095 | 486.267 | 475.339 | 1431.455544 | 1442.252899 | 2698.41 |
| 811.376 | 1337.89 | 1328.37 | 2075.703225 | 2099.495625 | 2983.14697 | 2790.26452 | 2324.84268 | 2318.69799 | 811.376 | 493.944 | 476.3 | 1428.883168 | 1444.990879 | 2723.52 |
| 811.658 | 1335.62 | 1331.01 | 2143.565253 | 2103.79069 | 3101.67643 | 2798.44597 | 2374.10476 | 2325.72547 | 811.658 | 490.067 | 477.262 | 1464.388582 | 1447.739899 | 2748.25 |
| 811.939 | 1372.82 | 1333.66 | 2146.549193 | 2108.097058 | 3066.16429 | 2806.63793 | 2446.51689 | 2332.75296 | 811.939 | 489.527 | 478.226 | 1473.839028 | 1450.477879 | 2704.23 |
| 812.22 | 1329.38 | 1336.31 | 2154.246402 | 2112.403426 | 3035.24772 | 2814.85089 | 2374.81332 | 2339.80367 | 812.22 | 479.522 | 479.19 | 1467.877298 | 1453.226899 | 2766.25 |
| 812.501 | 1343.18 | 1338.96 | 2098.139288 | 2116.721096 | 3139.61902 | 2823.07436 | 2435.30776 | 2346.866 | 812.501 | 490.652 | 480.154 | 1451.339017 | 1455.975919 | 2756.37 |
| 812.782 | 1342.12 | 1341.62 | 2108.639592 | 2121.038767 | 2938.34819 | 2831.31882 | 2442.57917 | 2353.95156 | 812.782 | 469.762 | 481.12 | 1436.832141 | 1458.724939 | 2798.1 |
| 813.063 | 1299.48 | 1344.28 | 2153.398691 | 2125.356438 | 3101.34742 | 2839.5738 | 2472.79154 | 2361.03713 | 813.063 | 482.63 | 482.087 | 1450.279154 | 1461.484999 | 2780.2 |
| 813.344 | 1342.03 | 1346.94 | 2134.862095 | 2129.674108 | 3050.17003 | 2847.84977 | 2511.40205 | 2368.14592 | 813.344 | 491.582 | 483.055 | 1497.520345 | 1464.234019 | 2760.99 |
| 813.625 | 1366.91 | 1349.6 | 2095.155348 | 2134.014385 | 3103.79909 | 2856.13625 | 2381.28325 | 2375.26633 | 813.625 | 486.357 | 484.024 | 1457.510512 | 1466.994079 | 2727.82 |
| 813.907 | 1363.31 | 1352.27 | 2174.987045 | 2138.343358 | 3185.59588 | 2864.44374 | 2497.17285 | 2382.40997 | 813.907 | 481.698 | 484.993 | 1503.25023 | 1469.754139 | 2728.7 |
| 814.187 | 1353.94 | 1354.93 | 2130.747875 | 2142.683634 | 3076.23632 | 2872.77223 | 2442.42817 | 2389.55361 | 814.187 | 494.35 | 485.964 | 1448.678319 | 1472.514199 | 2794.66 |
| 814.468 | 1375.39 | 1357.6 | 2119.987606 | 2147.023911 | 3101.4111 | 2881.11122 | 2462.45359 | 2396.7321 | 814.468 | 484.45 | 486.935 | 1476.345163 | 1475.27426 | 2787.45 |
| 814.75 | 1355.98 | 1360.28 | 2173.981095 | 2151.37549 | 3202.82128 | 2889.46071 | 2581.08449 | 2403.91059 | 814.75 | 464.331 | 487.908 | 1509.929575 | 1478.04536 | 2764.73 |
| 815.03 | 1371.69 | 1362.95 | 2194.04357 | 2155.727069 | 3199.20214 | 2897.83121 | 2486.50966 | 2411.10069 | 815.03 | 505.686 | 488.881 | 1468.815719 | 1480.81646 | 2726.45 |
| 815.312 | 1366.5 | 1365.63 | 2112.900749 | 2160.078648 | 3196.62311 | 2906.22272 | 2369.05194 | 2418.31402 | 815.312 | 507.624 | 489.856 | 1500.468089 | 1483.587561 | 2835.84 |
| 815.592 | 1336.71 | 1368.31 | 2118.925143 | 2164.44153 | 3129.41962 | 2914.62472 | 2408.18515 | 2425.53897 | 815.592 | 503.38 | 490.831 | 1471.796584 | 1486.358661 | 2814.85 |
| 815.873 | 1383.28 | 1370.99 | 2197.163144 | 2168.804412 | 3178.96256 | 2923.04774 | 2492.34073 | 2432.77554 | 815.873 | 491.286 | 491.807 | 1478.255125 | 1489.129762 | 2783.34 |
| 816.154 | 1382.72 | 1373.68 | 2180.095912 | 2173.167294 | 3100.82737 | 2931.48125 | 2484.53499 | 2440.03534 | 816.154 | 490.176 | 492.784 | 1479.491631 | 1491.911902 | 2840.61 |
| 816.435 | 1378.35 | 1376.36 | 2228.426697 | 2177.541478 | 3162.50131 | 2939.93577 | 2481.57299 | 2447.30675 | 816.435 | 480.663 | 493.762 | 1459.310071 | 1494.683003 | 2850.63 |
| 816.716 | 1374.25 | 1379.05 | 2185.374321 | 2181.915663 | 3123.5823 | 2948.40079 | 2587.32211 | 2454.58978 | 816.716 | 501.711 | 494.741 | 1503.901604 | 1497.465144 | 2817.71 |
| 816.997 | 1352.45 | 1381.74 | 2122.112533 | 2186.30115 | 3204.04181 | 2956.88682 | 2517.63967 | 2461.88442 | 816.997 | 503.965 | 495.721 | 1518.055193 | 1500.247284 | 2843.66 |
| 817.278 | 1374.64 | 1384.44 | 2254.739621 | 2190.686638 | 3132.77343 | 2965.39385 | 2509.27638 | 2469.2023 | 817.278 | 487.626 | 496.702 | 1473.629264 | 1503.040465 | 2836.64 |
| 817.559 | 1399.78 | 1387.14 | 2198.994198 | 2195.072125 | 3228.39936 | 2973.90088 | 2598.03176 | 2476.53179 | 817.559 | 502.462 | 497.683 | 1500.887618 | 1505.822606 | 2842.66 |
| 817.84 | 1406.35 | 1389.84 | 2233.659894 | 2199.468916 | 3211.33316 | 2982.43942 | 2598.42669 | 2483.8729 | 817.84 | 496.643 | 498.666 | 1523.332428 | 1508.615787 | 2877.13 |
| 818.12 | 1399.4 | 1392.54 | 2249.291671 | 2203.865706 | 3270.48114 | 2990.98847 | 2433.87903 | 2491.23724 | 818.12 | 523.818 | 499.649 | 1530.707309 | 1511.408968 | 2813.36 |
| 818.401 | 1368.92 | 1395.24 | 2195.716385 | 2208.273799 | 3145.54124 | 2999.55851 | 2513.70195 | 2498.60158 | 818.401 | 495.594 | 500.634 | 1531.568447 | 1514.202149 | 2844.92 |
| 818.682 | 1386.13 | 1397.95 | 2238.452283 | 2212.681892 | 3225.4807 | 3008.13906 | 2488.14747 | 2506.00076 | 818.682 | 513.133 | 501.619 | 1531.976936 | 1516.99533 | 2822.4 |
| 818.963 | 1366.22 | 1400.66 | 2203.56053 | 2217.089985 | 3291.24078 | 3016.74062 | 2504.13064 | 2513.39995 | 818.963 | 491.573 | 502.605 | 1479.41435 | 1519.799551 | 2852.66 |
| 819.243 | 1407.93 | 1403.37 | 2199.08462 | 2221.509381 | 3188.0051 | 3025.35267 | 2651.20833 | 2520.81075 | 819.243 | 513.543 | 503.593 | 1527.925168 | 1522.592732 | 2881.38 |
| 819.524 | 1390.59 | 1406.08 | 2242.317841 | 2225.928777 | 3377.20859 | 3033.98574 | 2550.39587 | 2528.24478 | 819.524 | 525.217 | 504.581 | 1518.154555 | 1525.396953 | 2871.64 |
| 819.805 | 1400.61 | 1408.8 | 2253.586736 | 2230.348173 | 3286.89993 | 3042.6293 | 2602.48057 | 2535.69043 | 819.805 | 512.511 | 505.57 | 1543.282143 | 1528.201174 | 2936.98 |
| 820.086 | 1365.79 | 1411.52 | 2266.652776 | 2234.778872 | 3309.87775 | 3051.29387 | 2655.7849 | 2543.15931 | 820.086 | 535.875 | 506.56 | 1573.764248 | 1531.016435 | 2888.82 |
| 820.366 | 1452.46 | 1414.24 | 2316.520742 | 2239.20957 | 3367.79459 | 3059.97945 | 2611.48271 | 2550.6398 | 820.366 | 518.038 | 507.551 | 1520.053476 | 1533.820657 | 2905.85 |
| 820.647 | 1423.66 | 1416.96 | 2308.394027 | 2243.640269 | 3373.73804 | 3068.66502 | 2666.93595 | 2558.13191 | 820.647 | 504.349 | 508.542 | 1521.069178 | 1536.635918 | 2942.59 |
| 820.928 | 1424.84 | 1419.69 | 2269.128089 | 2248.082271 | 3406.72421 | 3077.3821 | 2638.43108 | 2565.63564 | 820.928 | 497.702 | 509.535 | 1565.914637 | 1539.451179 | 2880.06 |
| 821.208 | 1440.56 | 1422.42 | 2297.430308 | 2252.524272 | 3410.08863 | 3086.10969 | 2675.50832 | 2573.1626 | 821.208 | 527.576 | 510.529 | 1543.359425 | 1542.266441 | 2934.18 |
| 821.489 | 1447.11 | 1425.15 | 2304.065053 | 2256.977576 | 3344.042 | 3094.85828 | 2719.31103 | 2580.70117 | 821.489 | 518.951 | 511.523 | 1624.703919 | 1545.081702 | 2915.99 |
| 821.769 | 1412 | 1427.88 | 2368.750994 | 2261.430881 | 3360.23791 | 3103.61737 | 2633.61058 | 2588.25136 | 821.769 | 536.113 | 512.519 | 1589.264746 | 1547.896964 | 2876.8 |
| 822.05 | 1431.91 | 1430.61 | 2349.841404 | 2265.884185 | 3489.13657 | 3112.39747 | 2609.77521 | 2595.81317 | 822.05 | 499.435 | 513.515 | 1573.168075 | 1550.723265 | 2919.7 |
| 822.331 | 1446.4 | 1433.35 | 2322.23996 | 2270.348792 | 3415.39529 | 3121.19857 | 2675.74063 | 2603.3982 | 822.331 | 517.062 | 514.512 | 1552.975474 | 1553.549567 | 2930 |
| 822.611 | 1434.33 | 1436.09 | 2334.842585 | 2274.813399 | 3386.81364 | 3129.99967 | 2699.29722 | 2610.99486 | 822.611 | 539.943 | 515.51 | 1576.391825 | 1556.375869 | 2959.46 |
| 822.892 | 1419.15 | 1438.84 | 2306.020438 | 2279.289309 | 3335.5089 | 3138.83228 | 2768.44534 | 2618.60312 | 822.892 | 528.594 | 516.509 | 1576.038537 | 1559.20217 | 2968.97 |
| 823.172 | 1428.11 | 1441.58 | 2321.369644 | 2283.765219 | 3421.35997 | 3147.67539 | 2568.28402 | 2626.23462 | 823.172 | 510.155 | 517.509 | 1583.678384 | 1562.028472 | 2945.46 |
| 823.453 | 1449.49 | 1444.33 | 2222.809204 | 2288.241129 | 3427.13361 | 3156.53951 | 2694.9065 | 2633.87774 | 823.453 | 543.756 | 518.51 | 1544.099121 | 1564.865814 | 2939 |
| 823.733 | 1433.32 | 1447.08 | 2367.338143 | 2292.717039 | 3494.94205 | 3165.41413 | 2673.55689 | 2641.53247 | 823.733 | 517.058 | 519.512 | 1586.69237 | 1567.692115 | 2994.61 |
| 824.014 | 1478.31 | 1449.83 | 2331.180477 | 2297.204252 | 3440.13491 | 3174.30976 | 2718.27724 | 2649.21043 | 824.014 | 509.551 | 520.514 | 1615.098909 | 1570.529457 | 2978.24 |
| 824.294 | 1417.12 | 1452.58 | 2309.953813 | 2301.702767 | 3452.12796 | 3183.21588 | 2778.81814 | 2656.90001 | 824.294 | 532.414 | 521.518 | 1575.298841 | 1573.366799 | 2936.4 |
| 824.574 | 1494.64 | 1455.34 | 2351.604642 | 2306.18998 | 3481.36763 | 3192.14302 | 2665.01936 | 2664.6012 | 824.574 | 540.638 | 522.522 | 1568.796139 | 1576.215181 | 2962.02 |
| 824.855 | 1512.99 | 1458.1 | 2340.087086 | 2310.699798 | 3496.38546 | 3201.09115 | 2702.86324 | 2672.31401 | 824.855 | 532.02 | 523.528 | 1613.630557 | 1579.052523 | 2973.23 |
| 825.135 | 1451.38 | 1460.86 | 2387.57016 | 2315.198314 | 3448.9758 | 3210.04979 | 2814.95451 | 2680.05005 | 825.135 | 528.47 | 524.534 | 1575.420284 | 1581.900905 | 2981.38 |
| 825.416 | 1487.05 | 1463.62 | 2332.853292 | 2319.708132 | 3494.34771 | 3219.01894 | 2818.71799 | 2687.7977 | 825.416 | 529.271 | 525.541 | 1637.13523 | 1584.749287 | 2982.27 |
| 825.696 | 1452.82 | 1466.39 | 2396.849761 | 2324.21795 | 3451.08785 | 3228.00908 | 2714.01428 | 2695.55698 | 825.696 | 502.216 | 526.549 | 1569.403353 | 1587.59767 | 3002.01 |
| 825.976 | 1451.6 | 1469.16 | 2407.519607 | 2328.727769 | 3498.3277 | 3237.02024 | 2749.24463 | 2703.33948 | 825.976 | 535.681 | 527.558 | 1601.188206 | 1590.446052 | 2949.44 |
| 826.257 | 1492.2 | 1471.93 | 2366.637369 | 2333.24889 | 3524.90343 | 3246.05239 | 2884.04455 | 2711.1336 | 826.257 | 520.185 | 528.568 | 1664.349424 | 1593.294434 | 3010.64 |
| 826.537 | 1434.05 | 1474.7 | 2313.661133 | 2337.781314 | 3470.37224 | 3255.08455 | 2735.1432 | 2718.93933 | 826.537 | 547.84 | 529.578 | 1570.253451 | 1596.153856 | 3019.56 |
| 826.817 | 1500.37 | 1477.48 | 2412.684988 | 2342.302435 | 3447.47933 | 3264.14822 | 2798.92488 | 2726.7683 | 826.817 | 552.28 | 530.59 | 1604.621721 | 1599.013279 | 2999.09 |
| 827.098 | 1498.49 | 1480.26 | 2400.839651 | 2346.834859 | 3606.75328 | 3273.22238 | 2795.20786 | 2734.60888 | 827.098 | 542.788 | 531.602 | 1651.112175 | 1601.872701 | 3001.74 |
| 827.378 | 1491.01 | 1483.04 | 2358.725407 | 2351.367283 | 3616.51753 | 3282.31756 | 2758.95533 | 2742.46108 | 827.378 | 561.449 | 532.616 | 1598.869755 | 1604.732123 | 3004.42 |
| 827.658 | 1455.49 | 1485.82 | 2452.77603 | 2355.91101 | 3532.09925 | 3291.42323 | 2735.96791 | 2750.32489 | 827.658 | 540.086 | 533.63 | 1607.492184 | 1607.591546 | 3010.84 |
| 827.939 | 1481.46 | 1488.6 | 2392.441668 | 2360.454736 | 3462.79433 | 3300.54991 | 2676.65827 | 2758.21193 | 827.939 | 549.44 | 534.645 | 1596.672748 | 1610.462008 | 3051.13 |
| 828.219 | 1491.03 | 1491.39 | 2374.458909 | 2365.009766 | 3527.34449 | 3309.68709 | 2864.25144 | 2766.11059 | 828.219 | 538.611 | 535.661 | 1583.656304 | 1613.332471 | 3065.14 |
| 828.499 | 1492.16 | 1494.18 | 2375.329225 | 2369.564795 | 3489.48681 | 3318.84528 | 2822.88801 | 2774.02087 | 828.499 | 543.854 | 536.678 | 1598.063818 | 1616.191893 | 3034.12 |
| 828.779 | 1537.04 | 1496.97 | 2380.076402 | 2374.119825 | 3635.02714 | 3328.02447 | 2951.68263 | 2781.95437 | 828.779 | 538.534 | 537.696 | 1621.568491 | 1619.073396 | 2972.18 |
| 829.059 | 1482.61 | 1499.76 | 2420.95864 | 2378.674854 | 3559.46037 | 3337.21417 | 2898.4596 | 2789.8995 | 829.059 | 562.173 | 538.714 | 1574.989715 | 1621.943859 | 3016.77 |
| 829.34 | 1525.42 | 1502.56 | 2430.351269 | 2383.241187 | 3679.5606 | 3346.41437 | 2870.1522 | 2797.86785 | 829.34 | 560.445 | 539.734 | 1605.449739 | 1624.814321 | 3055.76 |
| 829.62 | 1524.92 | 1505.36 | 2409.350661 | 2387.807519 | 3633.85968 | 3355.64607 | 2934.72374 | 2805.8362 | 829.62 | 560.902 | 540.754 | 1654.622972 | 1627.695824 | 3014.79 |
| 829.9 | 1508.74 | 1508.16 | 2429.774826 | 2392.385154 | 3679.21036 | 3364.88828 | 2857.78151 | 2813.82779 | 829.9 | 552.915 | 541.776 | 1637.654121 | 1630.577327 | 3012.23 |
| 830.18 | 1510.72 | 1510.96 | 2424.473812 | 2396.962789 | 3622.80061 | 3374.14099 | 2985.9721 | 2821.83099 | 830.18 | 564.153 | 542.798 | 1656.37837 | 1633.45883 | 3072.43 |
| 830.46 | 1503.69 | 1513.77 | 2463.321545 | 2401.540425 | 3614.33119 | 3383.41471 | 2837.97678 | 2829.85742 | 830.46 | 567.325 | 543.821 | 1644.9186 | 1636.340333 | 3066.75 |
| 830.74 | 1501.27 | 1516.58 | 2513.969404 | 2406.129363 | 3685.45098 | 3392.70943 | 2879.08465 | 2837.89547 | 830.74 | 527.424 | 544.845 | 1641.087636 | 1639.221836 | 3108.88 |
| 831.02 | 1555.55 | 1519.38 | 2475.901564 | 2410.7183 | 3630.41035 | 3402.01465 | 2978.02698 | 2845.94513 | 831.02 | 535.995 | 545.87 | 1668.456393 | 1642.103338 | 3068.39 |
| 831.3 | 1527.73 | 1522.2 | 2435.934854 | 2415.307238 | 3597.92301 | 3411.34088 | 2941.47245 | 2854.01803 | 831.3 | 538.449 | 546.895 | 1696.531726 | 1644.995882 | 3106.73 |
| 831.58 | 1534.18 | 1525.01 | 2463.332848 | 2419.907479 | 3689.50527 | 3420.67761 | 2994.96264 | 2862.10254 | 831.58 | 559.318 | 547.922 | 1649.732145 | 1647.888425 | 3056.88 |
| 831.86 | 1532.52 | 1527.83 | 2472.725477 | 2424.50772 | 3620.63549 | 3430.03534 | 3007.51918 | 2870.19866 | 831.86 | 577.961 | 548.949 | 1633.878359 | 1650.780968 | 3117.55 |
| 832.14 | 1476.09 | 1530.65 | 2408.469043 | 2429.119263 | 3637.45759 | 3439.40358 | 2975.82 | 2878.3064 | 832.14 | 547.062 | 549.977 | 1689.808219 | 1653.673511 | 3092.47 |
| 832.42 | 1550.11 | 1533.47 | 2470.103227 | 2433.730807 | 3686.75642 | 3448.80332 | 2990.18859 | 2886.43738 | 832.42 | 570.913 | 551.006 | 1667.83814 | 1656.577094 | 3102.13 |
| 832.7 | 1547.3 | 1536.29 | 2499.004493 | 2438.342351 | 3786.55338 | 3458.20307 | 2982.06924 | 2894.57997 | 832.7 | 546.56 | 552.037 | 1684.564105 | 1659.469637 | 3024.51 |
| 832.98 | 1546.46 | 1539.12 | 2489.668378 | 2442.953894 | 3816.85438 | 3467.63432 | 2918.42695 | 2902.74579 | 832.98 | 551.556 | 553.067 | 1678.127644 | 1662.373221 | 3120.51 |
| 833.26 | 1548.81 | 1541.95 | 2481.225185 | 2447.57674 | 3822.63864 | 3477.07608 | 2987.5983 | 2910.92322 | 833.26 | 574.845 | 554.099 | 1662.141376 | 1665.276804 | 3138.14 |
| 833.54 | 1536.66 | 1544.78 | 2502.949172 | 2452.21089 | 3708.38635 | 3486.52834 | 2993.06928 | 2919.11227 | 833.54 | 564.969 | 555.132 | 1662.273858 | 1668.180387 | 3144.58 |
| 833.82 | 1552.94 | 1547.61 | 2507.119906 | 2456.833736 | 3776.72546 | 3496.0016 | 3009.43577 | 2927.31294 | 833.82 | 562.143 | 556.165 | 1661.545203 | 1671.083971 | 3097.64 |
| 834.1 | 1533.77 | 1550.44 | 2477.483957 | 2461.467885 | 3731.21558 | 3505.49587 | 2956.59606 | 2935.53684 | 834.1 | 551.717 | 557.199 | 1688.95812 | 1673.998594 | 3125.72 |
| 834.38 | 1543.31 | 1553.28 | 2493.059219 | 2466.113337 | 3725.08109 | 3515.00064 | 2996.06613 | 2943.77235 | 834.38 | 559.569 | 558.235 | 1661.754967 | 1676.902178 | 3093.21 |
| 834.66 | 1534.63 | 1556.12 | 2468.453017 | 2470.747486 | 3766.28196 | 3524.52641 | 3072.06749 | 2952.01948 | 834.66 | 541.565 | 559.271 | 1686.904636 | 1679.816801 | 3152.41 |
| 834.939 | 1531.55 | 1558.96 | 2457.907502 | 2475.392938 | 3815.47465 | 3534.06269 | 3014.46536 | 2960.28984 | 834.939 | 561.361 | 560.308 | 1703.100669 | 1682.731425 | 3143.7 |
| 835.219 | 1549.86 | 1561.81 | 2513.980707 | 2480.049693 | 3787.88005 | 3543.61997 | 3030.34399 | 2968.57182 | 835.219 | 581.681 | 561.345 | 1690.172547 | 1685.657088 | 3196.02 |
| 835.499 | 1553.92 | 1564.65 | 2554.862945 | 2484.695145 | 3841.45604 | 3553.19826 | 3109.21443 | 2976.87703 | 835.499 | 552.508 | 562.384 | 1644.9186 | 1688.571712 | 3112.93 |
| 835.779 | 1548.66 | 1567.5 | 2495.218054 | 2489.3519 | 3782.50971 | 3562.78705 | 3049.44016 | 2985.19385 | 835.779 | 588.429 | 563.423 | 1681.351395 | 1691.486336 | 3075.07 |
| 836.059 | 1566.61 | 1570.35 | 2567.612506 | 2494.019957 | 3785.08875 | 3572.39685 | 3070.67361 | 2993.52229 | 836.059 | 542.666 | 564.464 | 1673.33618 | 1694.411999 | 3184.81 |
| 836.338 | 1573.06 | 1573.2 | 2557.112203 | 2498.688015 | 3877.91276 | 3582.02765 | 2948.4767 | 3001.86235 | 836.338 | 576.501 | 565.505 | 1680.346733 | 1697.337663 | 3176.13 |
| 836.618 | 1591.35 | 1576.06 | 2521.395346 | 2503.356072 | 3815.12442 | 3591.66895 | 3050.29972 | 3010.22564 | 836.618 | 559.349 | 566.547 | 1687.533929 | 1700.263327 | 3174.86 |
| 836.898 | 1593.4 | 1578.92 | 2524.514919 | 2508.02413 | 3946.76131 | 3601.32076 | 3047.0241 | 3018.60054 | 836.898 | 576.891 | 567.59 | 1708.698071 | 1703.188991 | 3143.37 |
| 837.178 | 1592.76 | 1581.78 | 2537.953952 | 2512.70349 | 3734.96208 | 3611.00407 | 3198.24858 | 3026.98706 | 837.178 | 568.503 | 568.633 | 1716.415199 | 1706.114655 | 3161.07 |
| 837.457 | 1582.14 | 1584.64 | 2521.16929 | 2517.382851 | 3795.8931 | 3620.69789 | 3010.35341 | 3035.39681 | 837.457 | 582.677 | 569.678 | 1708.19022 | 1709.051359 | 3139.06 |
| 837.737 | 1560.38 | 1587.5 | 2535.015223 | 2522.073514 | 3773.91293 | 3630.4022 | 3124.3845 | 3043.81817 | 837.737 | 563.827 | 570.723 | 1708.101898 | 1711.988063 | 3212.39 |
| 838.017 | 1592.28 | 1590.37 | 2514.184158 | 2526.764177 | 3761.54842 | 3640.12753 | 3034.52563 | 3052.26277 | 838.017 | 546.185 | 571.77 | 1699.711315 | 1714.924767 | 3143.45 |
| 838.296 | 1582.13 | 1593.24 | 2511.57321 | 2531.45484 | 3844.91598 | 3649.87385 | 3135.39616 | 3060.71898 | 838.296 | 571.321 | 572.817 | 1672.000311 | 1717.861471 | 3126.11 |
| 838.576 | 1622.81 | 1596.11 | 2556.049739 | 2536.145503 | 3942.69641 | 3659.63069 | 3100.73498 | 3069.18681 | 838.576 | 590.482 | 573.865 | 1714.505238 | 1720.798175 | 3160.02 |
| 838.856 | 1595.48 | 1598.98 | 2568.550639 | 2540.847469 | 3922.66909 | 3669.40852 | 3191.73219 | 3077.66625 | 838.856 | 591.068 | 574.913 | 1686.860475 | 1723.734879 | 3232.89 |
| 839.135 | 1622.38 | 1601.85 | 2576.225242 | 2545.560738 | 3950.92173 | 3679.19686 | 3204.98567 | 3086.16892 | 839.135 | 604.191 | 575.963 | 1714.295473 | 1726.682623 | 3246.68 |
| 839.415 | 1584.45 | 1604.73 | 2565.589305 | 2550.262704 | 3962.936 | 3689.0062 | 3194.25279 | 3094.68321 | 839.415 | 586.58 | 577.014 | 1707.627168 | 1729.630368 | 3274.71 |
| 839.694 | 1610.05 | 1607.61 | 2605.827283 | 2554.975973 | 3922.93443 | 3698.83655 | 3166.44486 | 3103.22074 | 839.694 | 594.639 | 578.065 | 1720.466968 | 1732.578112 | 3212.84 |
| 839.974 | 1575.47 | 1610.49 | 2507.967616 | 2559.700545 | 3945.69998 | 3708.6774 | 3145.57149 | 3111.75826 | 839.974 | 605.665 | 579.117 | 1802.915486 | 1735.514816 | 3229.72 |
| 840.254 | 1594.57 | 1613.37 | 2615.389454 | 2564.413813 | 3953.42647 | 3718.53926 | 3225.81258 | 3120.33063 | 840.254 | 597.926 | 580.17 | 1746.400493 | 1738.473601 | 3213.85 |
| 840.533 | 1603.28 | 1616.26 | 2567.533387 | 2569.138385 | 3965.91834 | 3728.42212 | 3274.73781 | 3128.903 | 840.533 | 545.508 | 581.224 | 1712.164707 | 1741.421345 | 3279.16 |
| 840.812 | 1637.84 | 1619.15 | 2639.984353 | 2573.874259 | 3991.91033 | 3738.31548 | 3182.91589 | 3137.4986 | 840.812 | 589.192 | 582.279 | 1764.473368 | 1744.380129 | 3233.46 |
| 841.092 | 1618.72 | 1622.04 | 2619.187196 | 2578.598831 | 3934.3968 | 3748.22985 | 3149.7996 | 3146.10581 | 841.092 | 593.085 | 583.334 | 1694.378879 | 1747.327874 | 3234.26 |
| 841.371 | 1604.43 | 1624.93 | 2671.858257 | 2583.334705 | 4032.72913 | 3758.15472 | 3221.74709 | 3154.73626 | 841.371 | 595.857 | 584.39 | 1757.286171 | 1750.286658 | 3226.56 |
| 841.651 | 1600.09 | 1627.83 | 2629.517958 | 2588.081882 | 4029.61943 | 3768.1006 | 3236.98686 | 3163.37832 | 841.651 | 590.821 | 585.448 | 1715.487819 | 1753.256483 | 3205.09 |
| 841.93 | 1624.27 | 1630.72 | 2583.425128 | 2592.817757 | 4089.25562 | 3778.06748 | 3183.64768 | 3172.032 | 841.93 | 609.345 | 586.506 | 1734.201027 | 1756.215268 | 3271.25 |
| 842.21 | 1602.52 | 1633.62 | 2647.704167 | 2597.564934 | 3995.99646 | 3788.04486 | 3310.88579 | 3180.70891 | 842.21 | 595.769 | 587.564 | 1757.032245 | 1759.174052 | 3280.26 |
| 842.489 | 1684.41 | 1636.52 | 2653.389477 | 2602.323414 | 4135.8056 | 3798.03275 | 3226.81153 | 3189.38582 | 842.489 | 579.07 | 588.624 | 1771.00919 | 1762.143877 | 3279.07 |
| 842.769 | 1632.17 | 1639.42 | 2603.19373 | 2607.081894 | 4132.2183 | 3808.05214 | 3277.24679 | 3198.09758 | 842.769 | 594.584 | 589.684 | 1700.130844 | 1765.113702 | 3266.49 |
| 843.048 | 1651.25 | 1642.33 | 2629.40493 | 2611.840374 | 4002.57671 | 3818.08204 | 3225.90551 | 3206.80933 | 843.048 | 598.745 | 590.746 | 1761.282738 | 1768.072486 | 3271.25 |
| 843.327 | 1632.37 | 1645.24 | 2609.658933 | 2616.598854 | 3977.65666 | 3828.12243 | 3271.74096 | 3215.54432 | 843.327 | 610.718 | 591.808 | 1792.857827 | 1771.042311 | 3287.55 |
| 843.607 | 1643.95 | 1648.15 | 2599.48641 | 2621.368637 | 4157.83884 | 3838.19434 | 3337.17206 | 3224.30254 | 843.607 | 586.769 | 592.871 | 1777.40149 | 1774.023176 | 3322.07 |
| 843.886 | 1611.14 | 1651.06 | 2720.980238 | 2626.13842 | 4073.40995 | 3848.26625 | 3301.67456 | 3233.06076 | 843.886 | 611.455 | 593.935 | 1780.481717 | 1776.993001 | 3343.15 |
| 844.165 | 1674.22 | 1653.97 | 2623.43705 | 2630.908202 | 4178.75767 | 3858.36966 | 3284.27428 | 3241.85383 | 844.165 | 614.937 | 594.999 | 1782.347518 | 1779.973866 | 3248.68 |
| 844.444 | 1614.41 | 1656.89 | 2665.641715 | 2635.689288 | 4150.33523 | 3868.48357 | 3253.13265 | 3250.64689 | 844.444 | 611.96 | 596.065 | 1776.131862 | 1782.943691 | 3279.95 |
| 844.724 | 1649.4 | 1659.81 | 2674.751774 | 2640.470374 | 4128.49303 | 3878.6185 | 3317.92489 | 3259.46319 | 844.724 | 596.125 | 597.131 | 1774.409584 | 1785.924556 | 3362.68 |
| 845.003 | 1631.87 | 1662.73 | 2682.166413 | 2645.262762 | 4143.78681 | 3888.76392 | 3284.22781 | 3268.29111 | 845.003 | 615.634 | 598.198 | 1799.349488 | 1788.905421 | 3338.36 |
| 845.282 | 1675.27 | 1665.65 | 2662.194361 | 2650.05515 | 4137.35515 | 3898.93035 | 3327.69367 | 3277.13064 | 845.282 | 613.956 | 599.266 | 1827.192976 | 1791.897326 | 3258.79 |
| 845.562 | 1651.69 | 1668.57 | 2654.937961 | 2654.847539 | 4028.30338 | 3909.10728 | 3255.16539 | 3285.9934 | 845.562 | 594.477 | 600.335 | 1765.599472 | 1794.878191 | 3295.76 |
| 845.841 | 1709.88 | 1671.5 | 2720.030803 | 2659.639927 | 4322.73785 | 3919.30521 | 3315.68306 | 3294.86777 | 845.841 | 611.778 | 601.404 | 1762.640688 | 1797.859056 | 3350.89 |
| 846.12 | 1649.58 | 1674.43 | 2695.09682 | 2664.443618 | 4121.92339 | 3929.52416 | 3310.95548 | 3303.76538 | 846.12 | 593.849 | 602.474 | 1799.967742 | 1800.850961 | 3301.71 |
| 846.399 | 1681.47 | 1677.36 | 2665.969497 | 2669.247309 | 4271.57108 | 3939.7641 | 3399.60632 | 3312.6746 | 846.399 | 627.266 | 603.545 | 1785.825193 | 1803.842866 | 3340.73 |
| 846.678 | 1671.81 | 1680.29 | 2697.71907 | 2674.062303 | 4235.40091 | 3950.01455 | 3370.83429 | 3321.59544 | 846.678 | 588.65 | 604.617 | 1802.6174 | 1806.834772 | 3297.48 |
| 846.957 | 1676.27 | 1683.23 | 2689.852319 | 2678.865995 | 4122.08259 | 3960.2755 | 3290.39574 | 3330.53951 | 846.957 | 625.041 | 605.69 | 1812.741301 | 1809.826677 | 3362.15 |
| 847.236 | 1674.11 | 1686.16 | 2723.85115 | 2683.692292 | 4201.37464 | 3970.56796 | 3410.32759 | 3339.4952 | 847.236 | 611.331 | 606.764 | 1779.477055 | 1812.829622 | 3402.11 |
| 847.516 | 1668.31 | 1689.1 | 2654.960567 | 2688.507286 | 4271.95316 | 3980.86042 | 3498.89712 | 3348.4625 | 847.516 | 632.051 | 607.838 | 1832.326688 | 1815.821528 | 3364.37 |
| 847.795 | 1671.88 | 1692.04 | 2725.365726 | 2693.333582 | 4295.84372 | 3991.18438 | 3440.79551 | 3357.45303 | 847.795 | 599.618 | 608.913 | 1811.019023 | 1818.824473 | 3360.93 |
| 848.074 | 1726.23 | 1694.99 | 2688.699433 | 2698.159879 | 4296.03476 | 4001.51885 | 3447.26544 | 3366.45518 | 848.074 | 612.35 | 609.989 | 1850.697648 | 1821.816378 | 3421.24 |
| 848.353 | 1736.03 | 1697.93 | 2734.950502 | 2702.997479 | 4255.6405 | 4011.87433 | 3453.83991 | 3375.48056 | 848.353 | 630.925 | 611.066 | 1823.295771 | 1824.819324 | 3394.62 |
| 848.632 | 1659.61 | 1700.88 | 2725.727415 | 2707.823776 | 4177.88738 | 4022.2403 | 3358.82368 | 3384.50594 | 848.632 | 586.964 | 612.144 | 1804.759206 | 1827.83331 | 3406.24 |
| 848.911 | 1696.94 | 1703.83 | 2720.776788 | 2712.661375 | 4270.33993 | 4032.62728 | 3410.04881 | 3393.55456 | 848.911 | 642.248 | 613.222 | 1789.026863 | 1830.836255 | 3387.29 |
| 849.19 | 1701.14 | 1706.78 | 2717.261616 | 2717.510278 | 4331.45138 | 4043.03527 | 3434.07003 | 3402.6264 | 849.19 | 597.745 | 614.301 | 1829.776392 | 1833.839201 | 3341.21 |
| 849.469 | 1723.54 | 1709.73 | 2748.660802 | 2722.35918 | 4326.43128 | 4053.45376 | 3396.31909 | 3411.70986 | 849.469 | 609.827 | 615.381 | 1800.8068 | 1836.853186 | 3327.09 |
| 849.748 | 1702.9 | 1712.69 | 2761.319941 | 2727.208082 | 4272.56873 | 4063.89325 | 3395.83123 | 3420.80493 | 849.748 | 623.74 | 616.462 | 1773.813411 | 1839.867172 | 3403.18 |
| 850.027 | 1675.63 | 1715.65 | 2733.661983 | 2732.056985 | 4279.50983 | 4074.34325 | 3426.90316 | 3429.92324 | 850.027 | 593.686 | 617.544 | 1807.773192 | 1842.881158 | 3367.15 |
| 850.306 | 1686.43 | 1718.61 | 2759.534098 | 2736.91719 | 4278.19378 | 4084.81425 | 3447.1609 | 3439.05316 | 850.306 | 637.132 | 618.626 | 1849.549463 | 1845.895144 | 3391.74 |
| 850.585 | 1717.02 | 1721.57 | 2798.381832 | 2741.777395 | 4346.01284 | 4095.30626 | 3399.01392 | 3448.19469 | 850.585 | 639.783 | 619.709 | 1831.410348 | 1848.909129 | 3371.65 |
| 850.864 | 1716.6 | 1724.53 | 2678.583425 | 2746.648903 | 4328.38413 | 4105.80877 | 3424.95173 | 3457.35946 | 850.864 | 648.9 | 620.793 | 1823.914024 | 1851.923115 | 3404.49 |
| 851.143 | 1761.29 | 1727.5 | 2683.613172 | 2751.509108 | 4250.11096 | 4116.34278 | 3486.50319 | 3466.53585 | 851.143 | 648.417 | 621.878 | 1878.993786 | 1854.948141 | 3406.14 |
| 851.421 | 1715.51 | 1730.47 | 2795.940426 | 2756.391919 | 4302.8485 | 4126.8768 | 3487.06075 | 3475.73546 | 851.421 | 624.513 | 622.963 | 1834.501615 | 1857.962127 | 3365.63 |
| 851.7 | 1697.64 | 1733.44 | 2770.260458 | 2761.263427 | 4327.58813 | 4137.43182 | 3626.61148 | 3484.94669 | 851.7 | 610.313 | 624.05 | 1826.994251 | 1860.987153 | 3408.05 |
| 851.979 | 1750.66 | 1736.41 | 2759.827971 | 2766.146238 | 4393.41189 | 4148.00785 | 3533.98808 | 3494.16954 | 851.979 | 630.992 | 625.137 | 1870.349277 | 1864.012179 | 3346.61 |
| 852.258 | 1736.74 | 1739.38 | 2829.057639 | 2771.029049 | 4361.31723 | 4158.60488 | 3608.96726 | 3503.404 | 852.258 | 646.362 | 626.225 | 1811.427512 | 1867.037205 | 3492.35 |
| 852.537 | 1719.18 | 1742.36 | 2807.017173 | 2775.91186 | 4366.60266 | 4169.21241 | 3580.8341 | 3512.6617 | 852.537 | 652.796 | 627.313 | 1814.827906 | 1870.073271 | 3396.17 |
| 852.816 | 1703.95 | 1745.34 | 2794.199795 | 2780.805973 | 4315.37221 | 4179.83045 | 3584.35365 | 3521.94262 | 852.816 | 624.205 | 628.403 | 1860.711147 | 1873.098297 | 3481.87 |
| 853.094 | 1727.24 | 1748.32 | 2792.673916 | 2785.700087 | 4391.91541 | 4190.47999 | 3548.06628 | 3531.22355 | 853.094 | 615.24 | 629.493 | 1888.565674 | 1876.134363 | 3437.64 |
| 853.373 | 1766.99 | 1751.3 | 2749.033795 | 2790.605503 | 4231.22988 | 4201.14004 | 3506.27307 | 3540.53932 | 853.373 | 623.187 | 630.584 | 1833.441752 | 1879.17043 | 3425 |
| 853.652 | 1702.6 | 1754.29 | 2788.763147 | 2795.499617 | 4486.56492 | 4211.82109 | 3570.04313 | 3549.85509 | 853.652 | 638.03 | 631.676 | 1891.656942 | 1882.206496 | 3469.46 |
| 853.93 | 1739.92 | 1757.27 | 2831.532952 | 2800.405033 | 4315.27669 | 4222.51264 | 3460.426 | 3559.19409 | 853.93 | 663.101 | 632.769 | 1847.197892 | 1885.242562 | 3482.72 |
| 854.209 | 1740.85 | 1760.26 | 2822.535921 | 2805.321753 | 4407.23042 | 4233.2252 | 3603.70536 | 3568.54471 | 854.209 | 658.146 | 633.862 | 1887.671415 | 1888.278628 | 3432.82 |
| 854.488 | 1739.26 | 1763.25 | 2799.410387 | 2810.227169 | 4437.41467 | 4243.94826 | 3630.00325 | 3577.91856 | 854.488 | 654.828 | 634.956 | 1871.309778 | 1891.314694 | 3500.8 |
| 854.766 | 1708.76 | 1766.24 | 2784.004667 | 2815.143888 | 4459.93612 | 4254.69233 | 3680.97284 | 3587.30403 | 854.766 | 649.784 | 636.051 | 1895.388543 | 1894.361801 | 3472.02 |
| 855.045 | 1772.44 | 1769.24 | 2791.735784 | 2820.07191 | 4502.19832 | 4265.4574 | 3636.18279 | 3596.70111 | 855.045 | 626.333 | 637.147 | 1851.337982 | 1897.397867 | 3456.4 |
| 855.324 | 1724.35 | 1772.23 | 2804.338409 | 2824.98863 | 4405.56413 | 4276.24348 | 3649.64536 | 3606.12142 | 855.324 | 626.568 | 638.244 | 1924.744544 | 1900.444974 | 3474.35 |
| 855.602 | 1763.65 | 1775.23 | 2878.722154 | 2829.916652 | 4483.70994 | 4287.04006 | 3637.41405 | 3615.55335 | 855.602 | 620.325 | 639.341 | 1867.235929 | 1903.49208 | 3473.24 |
| 855.881 | 1761.36 | 1778.23 | 2789.237864 | 2834.844674 | 4625.4507 | 4297.85764 | 3692.39105 | 3625.00851 | 855.881 | 638.902 | 640.439 | 1880.705023 | 1906.539186 | 3454.32 |
| 856.16 | 1764.52 | 1781.23 | 2812.340793 | 2839.783999 | 4416.25173 | 4308.68573 | 3717.67837 | 3634.46367 | 856.16 | 653.076 | 641.538 | 1931.90966 | 1909.586293 | 3489.94 |
| 856.438 | 1770.23 | 1784.24 | 2828.492498 | 2844.723323 | 4513.83051 | 4319.53482 | 3583.23854 | 3643.95368 | 856.438 | 639.617 | 642.637 | 1911.694979 | 1912.64444 | 3452.29 |
| 856.717 | 1759.19 | 1787.25 | 2821.066557 | 2849.662648 | 4510.15831 | 4330.39441 | 3620.33901 | 3653.44369 | 856.717 | 628.18 | 643.738 | 1905.446203 | 1915.691546 | 3451.15 |
| 856.995 | 1782.22 | 1790.25 | 2800.608484 | 2854.613276 | 4545.58554 | 4341.28551 | 3705.90008 | 3662.95693 | 856.995 | 652.135 | 644.839 | 1903.381678 | 1918.749693 | 3508.86 |
| 857.274 | 1770.15 | 1793.27 | 2886.814961 | 2859.563903 | 4680.71421 | 4352.17662 | 3638.84278 | 3672.4934 | 857.274 | 663.706 | 645.941 | 1953.91286 | 1921.80784 | 3505.74 |
| 857.552 | 1790.52 | 1796.28 | 2853.245637 | 2864.514531 | 4578.03043 | 4363.09922 | 3716.54004 | 3682.04148 | 857.552 | 671.578 | 647.043 | 1899.80464 | 1924.865986 | 3522.72 |
| 857.831 | 1712.68 | 1799.29 | 2831.905945 | 2869.465159 | 4630.7892 | 4374.03234 | 3692.66982 | 3691.60118 | 857.831 | 641.041 | 648.147 | 1908.956999 | 1927.924133 | 3553.49 |
| 858.109 | 1798.09 | 1802.31 | 2854.274192 | 2874.427089 | 4505.97666 | 4384.98645 | 3759.88974 | 3701.1725 | 858.109 | 651.987 | 649.251 | 1886.821316 | 1930.99332 | 3522.15 |
| 858.388 | 1744.81 | 1805.33 | 2920.022596 | 2879.38902 | 4527.98867 | 4395.95107 | 3763.57191 | 3710.76705 | 858.388 | 666.861 | 650.356 | 1897.331626 | 1934.051467 | 3496.52 |
| 858.666 | 1806.95 | 1808.35 | 2883.70669 | 2884.362253 | 4710.94092 | 4406.9367 | 3768.93835 | 3720.38483 | 858.666 | 660.069 | 651.462 | 1912.865244 | 1937.120654 | 3522.68 |
| 858.945 | 1837.17 | 1811.37 | 2873.014239 | 2889.335486 | 4705.65549 | 4417.94333 | 3691.5431 | 3730.00261 | 858.945 | 646.453 | 652.568 | 1869.145891 | 1940.1788 | 3495.7 |
| 859.223 | 1799.9 | 1814.39 | 2893.607946 | 2894.308719 | 4544.37562 | 4428.96046 | 3834.88054 | 3739.65523 | 859.223 | 644.87 | 653.675 | 1933.576736 | 1943.247987 | 3532.93 |
| 859.501 | 1769.93 | 1817.42 | 2957.740049 | 2899.281953 | 4608.98946 | 4439.99859 | 3815.6566 | 3749.30786 | 859.501 | 656.806 | 654.783 | 1905.655967 | 1946.317174 | 3602.93 |
| 859.78 | 1798.77 | 1820.45 | 2932.726946 | 2904.266489 | 4604.17101 | 4451.05774 | 3865.92925 | 3758.98372 | 859.78 | 644.098 | 655.892 | 1953.526451 | 1949.386361 | 3538.19 |
| 860.058 | 1808.63 | 1823.47 | 2872.121317 | 2909.251025 | 4797.33327 | 4462.12738 | 3834.81085 | 3768.67119 | 860.058 | 662.091 | 657.002 | 1967.702121 | 1952.466588 | 3614.08 |
| 860.337 | 1808.8 | 1826.51 | 2893.811396 | 2914.235561 | 4538.58075 | 4473.21803 | 3810.29016 | 3778.3819 | 860.337 | 649.752 | 658.112 | 1926.157694 | 1955.535775 | 3572.71 |
| 860.615 | 1821.51 | 1829.54 | 2933.439023 | 2919.2314 | 4624.94126 | 4484.32968 | 3804.81918 | 3788.10422 | 860.615 | 665.321 | 659.223 | 1944.142247 | 1958.616002 | 3611.56 |
| 860.893 | 1824.99 | 1832.58 | 2894.2296 | 2924.227238 | 4634.75858 | 4495.45184 | 3822.49823 | 3797.83815 | 860.893 | 681 | 660.335 | 1904.364259 | 1961.685189 | 3563.2 |
| 861.171 | 1836.16 | 1835.61 | 2950.562769 | 2929.223077 | 4728.64392 | 4506.595 | 3872.16686 | 3807.59532 | 861.171 | 657.403 | 661.447 | 1941.801716 | 1964.765417 | 3609.76 |
| 861.45 | 1818.8 | 1838.65 | 2969.743625 | 2934.230219 | 4767.22331 | 4517.74867 | 3828.49192 | 3817.36411 | 861.45 | 690.944 | 662.56 | 1977.439613 | 1967.845644 | 3621.24 |
| 861.728 | 1832.28 | 1841.69 | 2971.190384 | 2939.237361 | 4682.35928 | 4528.92334 | 3930.69825 | 3827.15612 | 861.728 | 676.383 | 663.674 | 1927.592926 | 1970.925871 | 3659.18 |
| 862.006 | 1815.16 | 1844.74 | 2915.24151 | 2944.244502 | 4651.36841 | 4540.11901 | 3854.7782 | 3836.95975 | 862.006 | 667.485 | 664.789 | 1951.583369 | 1974.017138 | 3609.33 |
| 862.284 | 1851.96 | 1847.78 | 2909.985707 | 2949.251644 | 4813.60348 | 4551.32519 | 3775.00174 | 3846.775 | 862.284 | 642.524 | 665.905 | 1987.695997 | 1977.097366 | 3611.77 |
| 862.563 | 1798.38 | 1850.83 | 3006.455129 | 2954.270088 | 4737.6334 | 4562.55237 | 3846.68207 | 3856.61347 | 862.563 | 655.159 | 667.021 | 1942.166044 | 1980.188633 | 3611.53 |
| 862.841 | 1856.73 | 1853.88 | 2886.023765 | 2959.288533 | 4804.20008 | 4573.80056 | 3934.32235 | 3866.46357 | 862.841 | 689.838 | 668.138 | 1900.014404 | 1983.26886 | 3601.24 |
| 863.119 | 1840.21 | 1856.93 | 2946.798936 | 2964.31828 | 4813.8582 | 4585.05925 | 3974.51258 | 3876.33689 | 863.119 | 689.328 | 669.255 | 1968.927587 | 1986.360128 | 3638.35 |
| 863.397 | 1830.79 | 1859.98 | 3015.836455 | 2969.348027 | 4771.22453 | 4596.33894 | 3943.09218 | 3886.22183 | 863.397 | 694.86 | 670.374 | 1910.447432 | 1989.451395 | 3634.96 |
| 863.675 | 1859.26 | 1863.04 | 2977.158264 | 2974.366472 | 4821.97738 | 4607.63964 | 3967.90326 | 3896.11839 | 863.675 | 678.833 | 671.493 | 1949.132436 | 1992.542662 | 3613.08 |
| 863.953 | 1870.08 | 1866.1 | 3000.193376 | 2979.407522 | 4768.18912 | 4618.95084 | 3939.87464 | 3906.03817 | 863.953 | 713.408 | 672.613 | 1917.071576 | 1995.64497 | 3611.52 |
| 864.231 | 1848.83 | 1869.15 | 3003.79897 | 2984.448572 | 4804.66707 | 4630.28305 | 3806.66607 | 3915.96958 | 864.231 | 663.118 | 673.734 | 1974.834116 | 1998.736238 | 3644.41 |
| 864.51 | 1802.47 | 1872.21 | 2904.017827 | 2989.489622 | 4681.55266 | 4641.62576 | 3877.48684 | 3925.92421 | 864.51 | 676.621 | 674.855 | 1956.982047 | 2001.838545 | 3635.84 |
| 864.788 | 1840.02 | 1875.28 | 2934.411064 | 2994.530672 | 4845.31605 | 4652.98947 | 3889.85754 | 3935.89046 | 864.788 | 674.121 | 675.977 | 1955.646178 | 2004.929813 | 3632.62 |
| 865.066 | 1864.56 | 1878.34 | 2961.017862 | 2999.571722 | 4811.28978 | 4664.37419 | 3969.45977 | 3945.86833 | 865.066 | 653.437 | 677.1 | 1970.572583 | 2008.03212 | 3662.65 |
| 865.344 | 1837.86 | 1881.41 | 2988.483672 | 3004.624075 | 4850.26185 | 4675.77992 | 3814.82027 | 3955.86942 | 865.344 | 686.367 | 678.224 | 1985.697713 | 2011.134428 | 3641.25 |
| 865.622 | 1826.15 | 1884.47 | 2866.029107 | 3009.687731 | 4971.77365 | 4687.19614 | 3894.37604 | 3965.88214 | 865.622 | 677.152 | 679.348 | 2009.125104 | 2014.236736 | 3622.11 |
| 865.9 | 1829.2 | 1887.54 | 2947.477104 | 3014.740083 | 4809.54919 | 4698.62287 | 3936.75002 | 3975.91808 | 865.9 | 685.688 | 680.473 | 2014.998512 | 2017.339043 | 3674.61 |
| 866.178 | 1888.45 | 1890.62 | 2962.758493 | 3019.803739 | 4942.33233 | 4710.08111 | 3867.16051 | 3985.95402 | 866.178 | 678.203 | 681.599 | 2002.357437 | 2020.452391 | 3601.37 |
| 866.456 | 1835.55 | 1893.69 | 3053.406972 | 3024.867395 | 5036.7165 | 4721.54985 | 4103.78344 | 3996.02482 | 866.456 | 664.269 | 682.725 | 1994.883194 | 2023.554699 | 3667.31 |
| 866.734 | 1864.38 | 1896.77 | 3001.17672 | 3029.93105 | 4995.34581 | 4733.03959 | 4012.11253 | 4006.10722 | 866.734 | 704.939 | 683.853 | 1991.008069 | 2026.668047 | 3668.54 |
| 867.011 | 1865.5 | 1899.84 | 2946.471154 | 3035.006009 | 5001.78809 | 4744.53984 | 3905.37608 | 4016.20124 | 867.011 | 692.095 | 684.981 | 2020.032862 | 2029.781395 | 3654.38 |
| 867.289 | 1853.58 | 1902.92 | 2983.3522 | 3040.080967 | 4891.81297 | 4756.06109 | 4131.03382 | 4026.30688 | 867.289 | 674.109 | 686.109 | 1987.729117 | 2032.894743 | 3637.46 |
| 867.567 | 1876.99 | 1906 | 3087.586648 | 3045.155926 | 4858.25368 | 4767.59285 | 4061.90893 | 4036.43575 | 867.567 | 686.439 | 687.239 | 1987.972003 | 2036.008091 | 3671.84 |
| 867.845 | 1890.54 | 1909.09 | 2958.3504 | 3050.242187 | 4958.23107 | 4779.15611 | 4019.92988 | 4046.57624 | 867.845 | 673.751 | 688.369 | 2021.412892 | 2039.121438 | 3616.45 |
| 868.123 | 1875.05 | 1912.17 | 2965.199899 | 3055.328448 | 5066.39132 | 4790.72987 | 4111.64725 | 4056.73995 | 868.123 | 685.085 | 689.5 | 2004.013473 | 2042.234786 | 3725.69 |
| 868.401 | 1901.83 | 1915.26 | 2972.806685 | 3060.41471 | 5045.23899 | 4802.31414 | 4043.20769 | 4066.91529 | 868.401 | 648.386 | 690.631 | 2002.456799 | 2045.348134 | 3693.37 |
| 868.679 | 1854.38 | 1918.34 | 2998.825737 | 3065.500971 | 4862.38225 | 4813.91941 | 4059.70195 | 4077.11385 | 868.679 | 692.318 | 691.763 | 2048.185476 | 2048.472522 | 3653.89 |
| 868.957 | 1902.31 | 1921.43 | 2945.408691 | 3070.598535 | 5039.47596 | 4825.54569 | 4059.92265 | 4087.32403 | 868.957 | 697.796 | 692.896 | 1990.544379 | 2051.596911 | 3694.6 |
| 869.234 | 1902.85 | 1924.53 | 3050.355215 | 3075.696099 | 4991.06865 | 4837.19297 | 4096.25648 | 4097.54582 | 869.234 | 707.336 | 694.03 | 2039.430565 | 2054.721299 | 3703.87 |
| 869.512 | 1867.36 | 1927.62 | 3051.225531 | 3080.804966 | 4974.71354 | 4848.85076 | 4101.92493 | 4107.79085 | 869.512 | 688.168 | 695.164 | 2009.41215 | 2057.834647 | 3693.95 |
| 869.79 | 1896.6 | 1930.71 | 3069.140473 | 3085.913833 | 5073.88431 | 4860.52955 | 4090.95973 | 4118.04749 | 869.79 | 673.551 | 696.3 | 2020.39719 | 2060.970075 | 3747.39 |
| 870.068 | 1957.01 | 1933.81 | 3061.82756 | 3091.0227 | 5117.22908 | 4872.21884 | 3961.23587 | 4128.32736 | 870.068 | 714.292 | 697.435 | 2049.587587 | 2064.094463 | 3736.48 |
| 870.345 | 1854.68 | 1936.91 | 3048.546767 | 3096.131567 | 5065.45735 | 4883.92914 | 4115.41073 | 4138.61885 | 870.345 | 698.246 | 698.572 | 2046.341756 | 2067.218851 | 3756.3 |
| 870.623 | 1871.8 | 1940.01 | 3064.381993 | 3101.240433 | 4975.881 | 4895.66044 | 4061.59531 | 4148.92196 | 870.623 | 684.15 | 699.709 | 2010.184967 | 2070.35428 | 3731.2 |
| 870.901 | 1906.91 | 1943.12 | 3054.288591 | 3106.360603 | 5072.86544 | 4907.40224 | 4093.45711 | 4159.24829 | 870.901 | 707.418 | 700.847 | 2046.915848 | 2073.478668 | 3762.36 |
| 871.178 | 1931.95 | 1946.22 | 3085.834713 | 3111.480773 | 5111.1052 | 4919.16505 | 4086.61547 | 4169.58624 | 871.178 | 713.228 | 701.986 | 2050.139599 | 2076.614096 | 3769.84 |
| 871.456 | 1918.43 | 1949.33 | 3049.530111 | 3116.612245 | 5135.4309 | 4930.94887 | 4096.91858 | 4179.94743 | 871.456 | 711.482 | 703.125 | 2003.748507 | 2079.749525 | 3717.94 |
| 871.734 | 1929.32 | 1952.43 | 3092.299916 | 3121.743718 | 5064.84177 | 4942.74319 | 4172.16492 | 4190.32023 | 871.734 | 691.633 | 704.265 | 2096.188443 | 2082.884953 | 3716.73 |
| 872.011 | 1911 | 1955.55 | 3076.656837 | 3126.87519 | 5012.59244 | 4954.55851 | 4076.7073 | 4200.70464 | 872.011 | 698.698 | 705.406 | 2060.528465 | 2086.020381 | 3742.04 |
| 872.289 | 1897.45 | 1958.66 | 3062.844812 | 3132.006663 | 5111.72077 | 4966.39484 | 4175.57993 | 4211.11229 | 872.289 | 679.96 | 706.548 | 2021.32457 | 2089.15581 | 3762.8 |
| 872.567 | 1900.99 | 1961.77 | 3067.761531 | 3137.149438 | 5220.61334 | 4978.24167 | 4164.12688 | 4221.53155 | 872.567 | 720.902 | 707.69 | 2083.006396 | 2092.302278 | 3774.94 |
| 872.844 | 1914.12 | 1964.88 | 3134.210709 | 3142.280911 | 5027.72703 | 4990.1095 | 4155.49643 | 4231.97404 | 872.844 | 698.793 | 708.832 | 2064.646475 | 2095.437707 | 3747.15 |
| 873.122 | 1926.25 | 1968 | 3175.409426 | 3147.434989 | 5220.91051 | 5001.98784 | 4284.08196 | 4242.42815 | 873.122 | 718.374 | 709.976 | 2108.752237 | 2098.584175 | 3861.07 |
| 873.399 | 1952 | 1971.12 | 3086.46767 | 3152.577764 | 5210.60499 | 5013.88718 | 4241.24335 | 4252.89387 | 873.399 | 724.098 | 711.12 | 2017.272802 | 2101.730644 | 3795.55 |
| 873.677 | 1958.15 | 1974.24 | 3166.751479 | 3157.731842 | 5137.53234 | 5025.80753 | 4260.44406 | 4263.38283 | 873.677 | 735.992 | 712.265 | 2024.96785 | 2104.877113 | 3830.22 |
| 873.954 | 1950.26 | 1977.36 | 3126.999521 | 3162.88592 | 5313.51165 | 5037.74888 | 4295.68602 | 4273.8834 | 873.954 | 706.315 | 713.411 | 2058.861389 | 2108.012541 | 3837.96 |
| 874.232 | 1937.04 | 1980.49 | 3093.057204 | 3168.039998 | 5195.88432 | 5049.70074 | 4183.85029 | 4284.4072 | 874.232 | 708.917 | 714.557 | 2091.220335 | 2111.17005 | 3841.58 |
| 874.509 | 1986.42 | 1983.62 | 3139.0144 | 3173.205379 | 5298.70608 | 5061.6736 | 4255.46094 | 4294.94262 | 874.509 | 721.268 | 715.704 | 2120.697778 | 2114.316519 | 3869.6 |
| 874.787 | 1995.57 | 1986.74 | 3116.702668 | 3178.37076 | 5261.97341 | 5073.65696 | 4336.67774 | 4305.50127 | 874.787 | 719.883 | 716.852 | 2047.412659 | 2117.462987 | 3795.49 |
| 875.064 | 1961.78 | 1989.87 | 3174.45999 | 3183.536141 | 5215.14748 | 5085.66133 | 4288.1823 | 4316.07153 | 875.064 | 736.578 | 718 | 2123.579281 | 2120.620496 | 3851.07 |
| 875.342 | 1959.51 | 1993 | 3126.863887 | 3188.712825 | 5295.15062 | 5097.6867 | 4243.48517 | 4326.65341 | 875.342 | 734.502 | 719.149 | 2117.59547 | 2123.766965 | 3797.19 |
| 875.619 | 1976.53 | 1996.14 | 3176.528403 | 3193.878205 | 5358.39534 | 5109.72257 | 4423.33068 | 4337.25853 | 875.619 | 723.703 | 720.299 | 2087.201687 | 2126.924474 | 3858.06 |
| 875.896 | 1970.88 | 1999.27 | 3133.713386 | 3199.066192 | 5307.44083 | 5121.78996 | 4273.76724 | 4347.87525 | 875.896 | 711.478 | 721.449 | 2062.305944 | 2130.081983 | 3848.31 |
| 876.174 | 1934.45 | 2002.41 | 3237.60875 | 3204.242876 | 5264.69042 | 5133.85734 | 4373.94083 | 4358.5036 | 876.174 | 694.835 | 722.601 | 2101.686483 | 2133.239491 | 3833.61 |
| 876.451 | 1997.42 | 2005.55 | 3222.078698 | 3209.430862 | 5295.61761 | 5145.94573 | 4313.57416 | 4369.15517 | 876.451 | 732.983 | 723.753 | 2113.587863 | 2136.397 | 3892.48 |
| 876.728 | 1974.73 | 2008.68 | 3154.363607 | 3214.618849 | 5314.29703 | 5158.06562 | 4405.37285 | 4379.82998 | 876.728 | 718.768 | 724.905 | 2121.6914 | 2139.554509 | 3868.7 |
| 877.006 | 1984.01 | 2011.83 | 3190.736026 | 3219.806835 | 5328.26415 | 5170.18552 | 4397.21865 | 4390.5164 | 877.006 | 737.878 | 726.058 | 2145.53832 | 2142.723058 | 3885.75 |
| 877.283 | 1969.22 | 2014.97 | 3150.012028 | 3224.994822 | 5274.4016 | 5182.32642 | 4360.90804 | 4401.21444 | 877.283 | 735.564 | 727.212 | 2096.762536 | 2145.880567 | 3868.78 |
| 877.56 | 2001.57 | 2018.11 | 3173.838336 | 3230.194111 | 5302.73914 | 5194.48832 | 4457.78278 | 4411.93571 | 877.56 | 727.74 | 728.366 | 2156.070709 | 2149.049116 | 3874.42 |
| 877.838 | 2020.17 | 2021.26 | 3258.609357 | 3235.3934 | 5428.46442 | 5206.67123 | 4544.44734 | 4422.66859 | 877.838 | 752.861 | 729.522 | 2082.719349 | 2152.217665 | 3868.35 |
| 878.115 | 2052.88 | 2024.41 | 3244.119164 | 3240.592689 | 5602.43781 | 5218.86464 | 4418.49856 | 4433.41309 | 878.115 | 763.178 | 730.678 | 2124.374178 | 2155.375174 | 3876.71 |
| 878.392 | 1983.85 | 2027.56 | 3245.645042 | 3245.803282 | 5389.25885 | 5231.07906 | 4456.37728 | 4444.18082 | 878.392 | 763.359 | 731.834 | 2156.324635 | 2158.543723 | 3844.63 |
| 878.669 | 2020.08 | 2030.71 | 3223.977569 | 3251.013874 | 5497.19622 | 5243.31448 | 4445.60955 | 4454.96017 | 878.669 | 737.909 | 732.991 | 2118.555971 | 2161.723313 | 3907.78 |
| 878.947 | 2016.55 | 2033.87 | 3244.650396 | 3256.224466 | 5425.93845 | 5255.56041 | 4421.03078 | 4465.76275 | 878.947 | 752.957 | 734.149 | 2092.158755 | 2164.891862 | 3940.13 |
| 879.224 | 2030.96 | 2037.02 | 3210.029911 | 3261.446361 | 5526.39344 | 5267.82734 | 4454.50716 | 4476.57694 | 879.224 | 747.023 | 735.308 | 2138.417364 | 2168.060411 | 3880.39 |
| 879.501 | 2001.09 | 2040.18 | 3219.931166 | 3266.668256 | 5373.98629 | 5280.10477 | 4572.12749 | 4487.41437 | 879.501 | 761.543 | 736.467 | 2162.485089 | 2171.22896 | 3934.46 |
| 879.778 | 1954.54 | 2043.34 | 3294.993079 | 3271.890151 | 5399.904 | 5292.40321 | 4485.83464 | 4498.26341 | 879.778 | 738.248 | 737.627 | 2134.177912 | 2174.408549 | 3903.78 |
| 880.055 | 2048.42 | 2046.5 | 3220.078102 | 3277.112045 | 5543.9054 | 5304.72265 | 4476.66987 | 4509.12407 | 880.055 | 745.447 | 738.787 | 2123.159752 | 2177.588139 | 3897.44 |
| 880.332 | 2025.57 | 2049.66 | 3310.975243 | 3282.33394 | 5402.88634 | 5317.0631 | 4485.27708 | 4520.00795 | 880.332 | 762.934 | 739.948 | 2119.936001 | 2180.767728 | 3918.11 |
| 880.609 | 2027.81 | 2052.82 | 3230.860976 | 3287.567138 | 5620.96865 | 5329.41405 | 4568.44532 | 4530.90346 | 880.609 | 768.131 | 741.11 | 2192.161257 | 2183.936277 | 4014.74 |
| 880.886 | 2038.21 | 2055.99 | 3266.419594 | 3292.800336 | 5527.74133 | 5341.7755 | 4567.20245 | 4541.82219 | 880.886 | 752.925 | 742.273 | 2080.89771 | 2187.115867 | 3930.84 |
| 881.163 | 2042.14 | 2059.15 | 3229.391612 | 3298.044836 | 5578.39866 | 5354.16846 | 4579.14336 | 4552.75255 | 881.163 | 740.928 | 743.436 | 2123.513039 | 2190.295456 | 3882.55 |
| 881.441 | 2074.47 | 2062.32 | 3307.98 | 3303.278034 | 5422.74384 | 5366.57193 | 4513.77034 | 4563.69451 | 881.441 | 755.916 | 744.6 | 2143.25299 | 2193.486086 | 3963.52 |
| 881.717 | 2057.57 | 2065.49 | 3286.674217 | 3308.533837 | 5539.56456 | 5378.99639 | 4575.5425 | 4574.65971 | 881.717 | 768.99 | 745.764 | 2175.910022 | 2196.665675 | 3933.18 |
| 881.995 | 2019.66 | 2068.66 | 3260.090024 | 3313.778338 | 5606.87417 | 5391.43137 | 4675.06561 | 4585.63652 | 881.995 | 771.582 | 746.929 | 2162.705894 | 2199.856304 | 4006.25 |
| 882.271 | 2065.58 | 2071.84 | 3261.954987 | 3319.022838 | 5561.02467 | 5403.88734 | 4587.18141 | 4596.63657 | 882.271 | 751.788 | 748.095 | 2163.688476 | 2203.035894 | 3923.25 |
| 882.548 | 2001.54 | 2075.01 | 3245.317261 | 3324.278642 | 5562.25582 | 5416.36432 | 4632.55224 | 4607.64823 | 882.548 | 738.206 | 749.262 | 2172.995399 | 2206.226523 | 3993.2 |
| 882.825 | 2054.18 | 2078.19 | 3347.74326 | 3329.534445 | 5604.07226 | 5428.8518 | 4792.47686 | 4618.68312 | 882.825 | 734.619 | 750.429 | 2169.915171 | 2209.417153 | 4016.55 |
| 883.102 | 2081.11 | 2081.37 | 3277.021623 | 3334.790248 | 5665.07757 | 5441.36029 | 4581.47811 | 4629.72963 | 883.102 | 710.573 | 751.596 | 2163.037101 | 2212.607783 | 4001.36 |
| 883.379 | 2051.9 | 2084.55 | 3323.928255 | 3340.057354 | 5605.4626 | 5453.88978 | 4604.75593 | 4640.78775 | 883.379 | 754.515 | 752.765 | 2188.440695 | 2215.798412 | 4005.01 |
| 883.656 | 2083.93 | 2087.73 | 3287.567138 | 3345.32446 | 5536.10462 | 5466.42978 | 4720.85461 | 4651.86911 | 883.656 | 727.634 | 753.934 | 2158.102114 | 2218.989042 | 4015.06 |
| 883.933 | 2071.49 | 2090.91 | 3266.00139 | 3350.591567 | 5715.95778 | 5478.99078 | 4663.07823 | 4662.96208 | 883.933 | 756.792 | 755.104 | 2170.124936 | 2222.179671 | 4005.72 |
| 884.21 | 2062.76 | 2094.1 | 3335.626656 | 3355.858673 | 5653.81685 | 5491.56228 | 4602.32825 | 4674.07828 | 884.21 | 771.536 | 756.274 | 2226.463285 | 2225.381341 | 3941.09 |
| 884.487 | 2017.38 | 2097.29 | 3276.309546 | 3361.137082 | 5741.47218 | 5504.16529 | 4578.26057 | 4685.2061 | 884.487 | 751.38 | 757.445 | 2238.099699 | 2228.571971 | 4065.16 |
| 884.764 | 2051.83 | 2100.47 | 3265.108469 | 3366.415491 | 5765.30968 | 5516.7683 | 4712.75848 | 4696.35715 | 884.764 | 744.044 | 758.617 | 2161.314824 | 2231.773641 | 4022.55 |
| 885.04 | 2073.75 | 2103.66 | 3294.699207 | 3371.693899 | 5663.63416 | 5529.40282 | 4710.27273 | 4707.51981 | 885.04 | 777.529 | 759.789 | 2173.713014 | 2234.96427 | 3976.1 |
| 885.317 | 2084.51 | 2106.86 | 3354.626667 | 3376.983611 | 5644.57265 | 5542.04784 | 4714.26852 | 4718.69409 | 885.317 | 759.176 | 760.962 | 2211.238793 | 2238.16594 | 4012.7 |
| 885.594 | 2079.09 | 2110.05 | 3375.356008 | 3382.26202 | 5659.56926 | 5554.71387 | 4650.39392 | 4729.8916 | 885.594 | 761.645 | 762.135 | 2227.710832 | 2241.36761 | 4038.35 |
| 885.871 | 2081.03 | 2113.24 | 3304.71349 | 3387.563035 | 5743.14909 | 5567.3904 | 4616.813 | 4741.10073 | 885.871 | 782.312 | 763.309 | 2247.583266 | 2244.56928 | 4063.57 |
| 886.148 | 2096.9 | 2116.44 | 3285.001402 | 3392.852746 | 5618.71863 | 5580.09843 | 4690.31699 | 4752.33309 | 886.148 | 792.332 | 764.484 | 2207.46303 | 2247.77095 | 4039.09 |
| 886.424 | 2109.26 | 2119.64 | 3347.17812 | 3398.142458 | 5710.95891 | 5592.80647 | 4774.89073 | 4763.57706 | 886.424 | 791.477 | 765.66 | 2205.078338 | 2250.98366 | 4076.67 |
| 886.701 | 2112.41 | 2122.84 | 3412.689165 | 3403.443473 | 5864.438 | 5605.54601 | 4786.75034 | 4774.83265 | 886.701 | 777.822 | 766.836 | 2263.35977 | 2254.18533 | 4047.78 |
| 886.978 | 2121.75 | 2126.04 | 3387.38219 | 3408.744487 | 5826.41051 | 5618.29606 | 4852.08851 | 4786.11147 | 886.978 | 782.491 | 768.013 | 2233.054309 | 2257.39804 | 4103.74 |
| 887.254 | 2103.95 | 2129.24 | 3393.779576 | 3414.056804 | 5652.36282 | 5631.06711 | 4786.55287 | 4797.41353 | 887.254 | 785.606 | 769.19 | 2215.191199 | 2260.599709 | 4090.07 |
| 887.531 | 2097.01 | 2132.45 | 3353.157303 | 3419.357819 | 5962.68542 | 5643.84866 | 4799.53919 | 4808.71558 | 887.531 | 792.03 | 770.368 | 2174.41959 | 2263.81242 | 4107.61 |
| 887.808 | 2123.51 | 2135.65 | 3408.3941 | 3424.670136 | 5934.80425 | 5656.65122 | 4772.37013 | 4820.04086 | 887.808 | 762.116 | 771.546 | 2226.209359 | 2267.02513 | 4089.59 |
| 888.084 | 2146.66 | 2138.86 | 3427.382809 | 3429.982454 | 5844.69724 | 5669.47478 | 4834.03775 | 4831.38938 | 888.084 | 811.13 | 772.726 | 2191.443641 | 2270.23784 | 4110.64 |
| 888.361 | 2061.76 | 2142.07 | 3373.219778 | 3435.306074 | 5827.0367 | 5682.30885 | 4945.80379 | 4842.74951 | 888.361 | 792.269 | 773.905 | 2231.873003 | 2273.45055 | 4056.07 |
| 888.638 | 2141.49 | 2145.28 | 3415.944373 | 3440.618391 | 5793.77458 | 5695.16392 | 4868.36207 | 4854.13287 | 888.638 | 766.395 | 775.086 | 2275.824202 | 2276.66326 | 4038.68 |
| 888.914 | 2081.17 | 2148.49 | 3445.128209 | 3445.942011 | 5861.0842 | 5708.03999 | 4709.85456 | 4865.52785 | 888.914 | 796.271 | 776.267 | 2283.287404 | 2279.87597 | 4134.2 |
| 889.191 | 2145.97 | 2151.71 | 3401.126398 | 3451.276934 | 5861.42382 | 5720.92657 | 4689.88721 | 4876.93444 | 889.191 | 788.203 | 777.449 | 2235.86957 | 2283.08868 | 4072.18 |
| 889.467 | 2077.43 | 2154.92 | 3349.201322 | 3456.600554 | 5813.58963 | 5733.83416 | 4836.11696 | 4888.36427 | 889.467 | 794.45 | 778.631 | 2243.906865 | 2286.31243 | 4142.72 |
| 889.744 | 2117.45 | 2158.14 | 3407.953291 | 3461.935477 | 5947.60391 | 5746.75224 | 4943.03925 | 4899.80571 | 889.744 | 793.726 | 779.814 | 2213.237076 | 2289.52514 | 4163.88 |
| 890.02 | 2130.08 | 2161.36 | 3441.0592 | 3467.2704 | 5842.0333 | 5759.70183 | 4837.78962 | 4911.25877 | 890.02 | 776.249 | 780.998 | 2305.511409 | 2292.748891 | 4133.32 |
| 890.297 | 2083.81 | 2164.58 | 3432.26562 | 3472.605323 | 5845.60999 | 5772.65143 | 4866.42226 | 4922.73505 | 890.297 | 788.564 | 782.182 | 2217.940219 | 2295.972641 | 4124.36 |
| 890.573 | 2120.87 | 2167.8 | 3458.533334 | 3477.940246 | 5963.23731 | 5785.63253 | 4818.54244 | 4934.23457 | 890.573 | 775.631 | 783.366 | 2210.63158 | 2299.196391 | 4116.94 |
| 890.85 | 2155.55 | 2171.02 | 3384.217405 | 3483.286472 | 5961.81513 | 5798.62413 | 4942.81856 | 4945.74571 | 890.85 | 784.288 | 784.552 | 2261.350446 | 2302.420142 | 4113.49 |
| 891.126 | 2111.51 | 2174.24 | 3399.849181 | 3488.632697 | 5941.23592 | 5811.63674 | 4920.79524 | 4957.26846 | 891.126 | 810.159 | 785.738 | 2259.065116 | 2305.643892 | 4145.52 |
| 891.402 | 2143.34 | 2177.47 | 3441.172228 | 3493.990226 | 6040.16259 | 5824.67035 | 4869.44233 | 4968.81444 | 891.402 | 819.074 | 786.925 | 2304.959397 | 2308.867642 | 4124.49 |
| 891.679 | 2140.53 | 2180.7 | 3416.803386 | 3499.336452 | 6049.65089 | 5837.71447 | 4939.2177 | 4980.37204 | 891.679 | 795.896 | 788.112 | 2242.29499 | 2312.091393 | 4181.23 |
| 891.955 | 2141.16 | 2183.93 | 3475.498841 | 3504.69398 | 6056.34789 | 5850.76909 | 4894.76451 | 4991.94125 | 891.955 | 801.247 | 789.3 | 2274.455212 | 2315.326183 | 4080.27 |
| 892.232 | 2124.54 | 2187.16 | 3485.965236 | 3510.051509 | 5988.92153 | 5863.85521 | 4986.0521 | 5003.53369 | 892.232 | 777.916 | 790.488 | 2263.050643 | 2318.549933 | 4070.42 |
| 892.508 | 2216.24 | 2190.39 | 3423.664187 | 3515.409037 | 6016.1659 | 5876.95184 | 4997.28446 | 5015.14937 | 892.508 | 799.76 | 791.677 | 2359.321543 | 2321.784724 | 4169.71 |
| 892.784 | 2144.45 | 2193.62 | 3513.351927 | 3520.777869 | 6036.35241 | 5890.06948 | 4929.63476 | 5026.77666 | 892.784 | 815.465 | 792.867 | 2310.468477 | 2325.008474 | 4204.16 |
| 893.061 | 2132.57 | 2196.86 | 3432.966394 | 3526.135397 | 6009.75546 | 5903.19761 | 5054.23612 | 5038.41557 | 893.061 | 805.067 | 794.057 | 2248.72041 | 2328.243265 | 4098.46 |
| 893.337 | 2160.58 | 2200.1 | 3395.192427 | 3531.504229 | 5969.40365 | 5916.34675 | 4858.77914 | 5050.07771 | 893.337 | 791.421 | 795.248 | 2327.713333 | 2331.478055 | 4141.49 |
| 893.613 | 2192.84 | 2203.34 | 3431.078825 | 3536.884363 | 6101.13606 | 5929.5169 | 5072.62373 | 5061.75146 | 893.613 | 810.398 | 796.439 | 2273.207664 | 2334.712846 | 4196.87 |
| 893.889 | 2175.83 | 2206.57 | 3497.652334 | 3542.253194 | 6041.56355 | 5942.69755 | 4939.71717 | 5073.43683 | 893.889 | 794.958 | 797.631 | 2233.970649 | 2337.947637 | 4153.11 |
| 894.166 | 2165.83 | 2209.82 | 3500.071134 | 3547.633328 | 5973.511 | 5955.8992 | 5023.41973 | 5085.14543 | 894.166 | 786.417 | 798.824 | 2299.59384 | 2341.182427 | 4205.9 |
| 894.442 | 2154.32 | 2213.06 | 3497.968813 | 3553.013462 | 6026.5351 | 5969.12186 | 5021.66576 | 5096.87727 | 894.442 | 803.147 | 800.017 | 2328.47511 | 2344.428258 | 4185.19 |
| 894.718 | 2196.48 | 2216.3 | 3521.636882 | 3558.393597 | 6117.28952 | 5982.35502 | 5172.69278 | 5108.62071 | 894.718 | 802.327 | 801.211 | 2322.380897 | 2347.663049 | 4180.56 |
| 894.994 | 2209.26 | 2219.55 | 3512.990238 | 3563.785033 | 6155.25333 | 5995.60919 | 5141.71377 | 5120.37578 | 894.994 | 810.212 | 802.406 | 2292.152718 | 2350.897839 | 4166.45 |
| 895.27 | 2186.74 | 2222.8 | 3502.049124 | 3569.17647 | 6124.70823 | 6008.87386 | 5054.77044 | 5132.15407 | 895.27 | 787.193 | 803.601 | 2325.637768 | 2354.14367 | 4194.43 |
| 895.547 | 2110.74 | 2226.05 | 3543.519108 | 3574.567907 | 6144.76738 | 6022.15953 | 5066.46743 | 5143.94398 | 895.547 | 804.164 | 804.796 | 2331.577418 | 2357.389501 | 4149.56 |
| 895.823 | 2127.84 | 2229.3 | 3395.520208 | 3579.959344 | 6029.9632 | 6035.46621 | 5182.7868 | 5155.74551 | 895.823 | 820.956 | 805.993 | 2328.165983 | 2360.624291 | 4201.85 |
| 896.099 | 2212.88 | 2232.55 | 3542.366222 | 3585.362084 | 6096.44498 | 6048.79389 | 5043.01538 | 5167.57027 | 896.099 | 789.601 | 807.189 | 2318.84802 | 2363.870122 | 4210 |
| 896.375 | 2185.63 | 2235.8 | 3510.413199 | 3590.764824 | 6069.99661 | 6062.13208 | 5075.67866 | 5179.41826 | 896.375 | 822.305 | 808.387 | 2275.360511 | 2367.115953 | 4227.9 |
| 896.651 | 2196.48 | 2239.05 | 3545.1015 | 3596.167564 | 6288.56713 | 6075.49127 | 5149.79829 | 5191.27786 | 896.651 | 806.157 | 809.585 | 2322.06073 | 2370.361784 | 4261.02 |
| 896.927 | 2142.41 | 2242.31 | 3492.893854 | 3601.570303 | 6283.65317 | 6088.86096 | 5067.1876 | 5203.14908 | 896.927 | 798.761 | 810.783 | 2297.73908 | 2373.607614 | 4258.51 |
| 897.203 | 2222.31 | 2245.57 | 3482.845662 | 3606.984346 | 6138.91945 | 6102.25166 | 5144.96616 | 5215.04354 | 897.203 | 809.693 | 811.982 | 2308.547475 | 2376.864485 | 4204.9 |
| 897.479 | 2157.5 | 2248.83 | 3539.370979 | 3612.387086 | 6270.76861 | 6115.66336 | 5169.68432 | 5226.9496 | 897.479 | 787.403 | 813.182 | 2403.405224 | 2380.110316 | 4232.11 |
| 897.755 | 2233.15 | 2252.09 | 3592.313308 | 3617.801128 | 6152.36651 | 6129.08557 | 5200.59364 | 5238.86729 | 897.755 | 833.083 | 814.383 | 2268.239556 | 2383.356147 | 4176.6 |
| 898.031 | 2208.84 | 2255.35 | 3624.356754 | 3623.226474 | 6248.71415 | 6142.52878 | 5150.10029 | 5250.8082 | 898.031 | 817.219 | 815.583 | 2294.415967 | 2386.613018 | 4200.74 |
| 898.307 | 2229.22 | 2258.61 | 3549.204418 | 3628.640516 | 6249.47831 | 6155.993 | 5235.59167 | 5262.77235 | 898.307 | 811.15 | 816.785 | 2369.633128 | 2389.858849 | 4272.29 |
| 898.583 | 2178.33 | 2261.88 | 3564.237146 | 3634.065861 | 6299.35026 | 6169.46772 | 5233.6867 | 5274.74811 | 898.583 | 848.109 | 817.987 | 2319.709159 | 2393.11572 | 4307.12 |
| 898.859 | 2252.26 | 2265.14 | 3521.354312 | 3639.491207 | 6304.03073 | 6182.96344 | 5178.30315 | 5286.73549 | 898.859 | 833.409 | 819.189 | 2392.939076 | 2396.372591 | 4310.41 |
| 899.135 | 2264.61 | 2268.41 | 3539.766577 | 3644.916552 | 6452.78689 | 6196.48017 | 5275.06173 | 5298.7461 | 899.135 | 830.479 | 820.393 | 2298.004045 | 2399.629462 | 4285.83 |
| 899.411 | 2229.52 | 2271.68 | 3597.388266 | 3650.3532 | 6466.83892 | 6210.0074 | 5231.37518 | 5310.76832 | 899.411 | 814.487 | 821.596 | 2343.821045 | 2402.886333 | 4238.78 |
| 899.687 | 2302.32 | 2274.95 | 3642.192577 | 3655.789848 | 6410.79002 | 6223.55564 | 5090.93004 | 5322.81378 | 899.687 | 817.132 | 822.801 | 2333.697144 | 2406.143204 | 4243.53 |
| 899.963 | 2224.7 | 2278.22 | 3599.954002 | 3661.226497 | 6451.78924 | 6237.11438 | 5278.8136 | 5334.87085 | 899.963 | 841.329 | 824.006 | 2338.058039 | 2409.400075 | 4308.6 |
| 900.239 | 2247.93 | 2281.5 | 3616.772573 | 3666.663145 | 6369.56792 | 6250.69412 | 5224.18507 | 5346.93954 | 900.239 | 845.626 | 825.211 | 2354.000147 | 2412.656946 | 4341.63 |
| 900.514 | 2227.15 | 2284.77 | 3637.377583 | 3672.099793 | 6499.14583 | 6264.29487 | 5397.40965 | 5359.03145 | 900.514 | 830 | 826.417 | 2420.661121 | 2415.924857 | 4243.48 |
| 900.79 | 2219.44 | 2288.05 | 3626.752948 | 3677.547744 | 6367.85918 | 6277.91662 | 5217.52929 | 5371.13499 | 900.79 | 829.479 | 827.624 | 2356.506281 | 2419.181729 | 4338.47 |
| 901.066 | 2215.76 | 2291.32 | 3660.367484 | 3682.995695 | 6372.62456 | 6291.54888 | 5374.09699 | 5383.26175 | 901.066 | 846.802 | 828.831 | 2404.752134 | 2422.44964 | 4251.2 |
| 901.342 | 2255.1 | 2294.6 | 3598.563758 | 3688.443646 | 6379.41707 | 6305.19164 | 5198.37504 | 5395.40014 | 901.342 | 860.279 | 830.039 | 2319.78644 | 2425.706511 | 4334.23 |
| 901.617 | 2300.12 | 2297.88 | 3653.676224 | 3693.891597 | 6411.0129 | 6318.8659 | 5370.11281 | 5407.56175 | 901.617 | 852.355 | 831.247 | 2420.296793 | 2428.974422 | 4309.46 |
| 901.893 | 2287.26 | 2301.16 | 3721.606069 | 3699.350851 | 6425.447 | 6332.55067 | 5358.65975 | 5419.73498 | 901.893 | 877.941 | 832.456 | 2398.392955 | 2432.242333 | 4304.05 |
| 902.169 | 2288.13 | 2304.45 | 3734.208694 | 3704.810104 | 6210.99444 | 6346.24594 | 5346.18452 | 5431.91982 | 902.169 | 842.205 | 833.665 | 2398.006547 | 2435.510245 | 4322.17 |
| 902.445 | 2258.21 | 2307.73 | 3654.376998 | 3710.269358 | 6463.06058 | 6359.97272 | 5247.93913 | 5444.1279 | 902.445 | 844.332 | 834.875 | 2366.365216 | 2438.778156 | 4427.99 |
| 902.72 | 2273.74 | 2311.02 | 3665.566773 | 3715.728612 | 6463.19855 | 6373.71 | 5419.97891 | 5456.3592 | 902.72 | 851.445 | 836.086 | 2416.156703 | 2442.046067 | 4339.2 |
| 902.996 | 2272.07 | 2314.3 | 3649.697638 | 3721.199168 | 6427.12391 | 6387.45779 | 5341.23624 | 5468.59051 | 902.996 | 855.167 | 837.297 | 2438.49111 | 2445.313979 | 4354.54 |
| 903.272 | 2244.45 | 2317.59 | 3605.277623 | 3726.669725 | 6570.66894 | 6401.22658 | 5497.38577 | 5480.84505 | 903.272 | 850.005 | 838.508 | 2442.112309 | 2448.58189 | 4352.22 |
| 903.547 | 2285.49 | 2320.88 | 3682.475766 | 3732.140282 | 6532.31243 | 6415.01637 | 5333.58151 | 5493.12282 | 903.547 | 866.306 | 839.721 | 2411.630204 | 2451.849801 | 4426 |
| 903.823 | 2350.02 | 2324.18 | 3683.685166 | 3737.610838 | 6664.52245 | 6428.82717 | 5412.38225 | 5505.4122 | 903.823 | 866.742 | 840.933 | 2438.888558 | 2455.128753 | 4339.44 |
| 904.098 | 2306.9 | 2327.47 | 3694.773215 | 3743.092697 | 6593.63614 | 6442.64847 | 5423.92824 | 5517.72482 | 904.098 | 829.156 | 842.147 | 2391.901293 | 2458.396664 | 4318.11 |
| 904.374 | 2278.55 | 2330.76 | 3603.152696 | 3748.563254 | 6501.22604 | 6456.49078 | 5269.17259 | 5530.04905 | 904.374 | 832.141 | 843.36 | 2366.53082 | 2461.675615 | 4415.94 |
| 904.65 | 2232.26 | 2334.06 | 3714.372275 | 3754.045113 | 6541.64153 | 6470.34359 | 5334.37138 | 5542.3849 | 904.65 | 847.949 | 844.575 | 2376.500158 | 2464.954567 | 4393.07 |
| 904.925 | 2333.34 | 2337.35 | 3757.786341 | 3759.538276 | 6618.67294 | 6484.2174 | 5440.9452 | 5554.74398 | 904.925 | 851.244 | 845.79 | 2397.763661 | 2468.222478 | 4397.82 |
| 905.201 | 2350.89 | 2340.65 | 3757.413349 | 3765.020135 | 6729.04076 | 6498.11222 | 5478.96331 | 5567.11467 | 905.201 | 859.92 | 847.005 | 2381.733232 | 2471.50143 | 4416.33 |
| 905.476 | 2318.24 | 2343.95 | 3665.544167 | 3770.513297 | 6656.15916 | 6512.01754 | 5538.72596 | 5579.5086 | 905.476 | 844.676 | 848.221 | 2446.528405 | 2474.780381 | 4423.54 |
| 905.752 | 2277.22 | 2347.25 | 3720.747056 | 3776.006459 | 6814.42485 | 6525.94387 | 5467.33602 | 5591.91414 | 905.752 | 854.586 | 849.438 | 2437.287723 | 2478.059333 | 4387.02 |
| 906.027 | 2289.76 | 2350.55 | 3777.306282 | 3781.499621 | 6644.93027 | 6539.8912 | 5592.48331 | 5604.3313 | 906.027 | 875.187 | 850.655 | 2440.390031 | 2481.338284 | 4358.62 |
| 906.303 | 2313.99 | 2353.86 | 3637.716667 | 3786.992784 | 6730.05964 | 6553.84903 | 5439.3887 | 5616.77169 | 906.303 | 874.377 | 851.872 | 2432.021529 | 2484.617236 | 4470.75 |
| 906.578 | 2307.48 | 2357.16 | 3744.143858 | 3792.497249 | 6689.50618 | 6567.82787 | 5490.64868 | 5629.23531 | 906.578 | 871.589 | 853.091 | 2468.586806 | 2487.907228 | 4342.02 |
| 906.854 | 2296.27 | 2360.47 | 3704.199753 | 3798.001714 | 6782.0967 | 6581.81722 | 5558.13576 | 5641.69893 | 906.854 | 848.803 | 854.309 | 2494.045602 | 2491.186179 | 4462.66 |
| 907.129 | 2260.17 | 2363.78 | 3796.181962 | 3803.506179 | 6710.39317 | 6595.82756 | 5571.73771 | 5654.19739 | 907.129 | 879.557 | 855.529 | 2495.215867 | 2494.465131 | 4359.15 |
| 907.404 | 2321.88 | 2367.08 | 3678.293729 | 3809.010644 | 6604.40865 | 6609.85891 | 5452.5028 | 5666.69586 | 907.404 | 889.787 | 856.748 | 2411.950371 | 2497.755122 | 4395.72 |
| 907.68 | 2367.41 | 2370.39 | 3751.12899 | 3814.526411 | 6810.45547 | 6623.91127 | 5547.10086 | 5679.21756 | 907.68 | 868.865 | 857.969 | 2447.23498 | 2501.034074 | 4402.65 |
| 907.955 | 2337.43 | 2373.7 | 3797.9452 | 3820.030876 | 6817.46026 | 6637.97413 | 5436.4267 | 5691.76249 | 907.955 | 881.299 | 859.189 | 2475.807124 | 2504.324066 | 4440.99 |
| 908.231 | 2290.12 | 2377.02 | 3795.322949 | 3825.546644 | 6708.91792 | 6652.04749 | 5626.54047 | 5704.31904 | 908.231 | 880.186 | 860.411 | 2525.444046 | 2507.614057 | 4465.85 |
| 908.506 | 2287.07 | 2380.33 | 3709.127775 | 3831.073715 | 6723.23528 | 6666.15236 | 5648.61025 | 5716.8872 | 908.506 | 885.709 | 861.633 | 2449.321586 | 2510.904049 | 4437.08 |
| 908.781 | 2344.02 | 2383.65 | 3684.465059 | 3836.589482 | 6751.66834 | 6680.26773 | 5659.02952 | 5729.47859 | 908.781 | 858.489 | 862.855 | 2493.736475 | 2514.183001 | 4461.92 |
| 909.056 | 2388.94 | 2386.96 | 3782.968986 | 3842.116553 | 6957.37553 | 6694.39361 | 5612.95013 | 5742.0816 | 909.056 | 872.791 | 864.078 | 2419.325252 | 2517.472993 | 4456.29 |
| 909.332 | 2325.43 | 2390.28 | 3769.586467 | 3847.632321 | 6904.967 | 6708.54049 | 5596.27002 | 5754.70784 | 909.332 | 857.092 | 865.302 | 2398.105909 | 2520.762984 | 4461.88 |
| 909.607 | 2369.39 | 2393.6 | 3726.014162 | 3853.159391 | 6806.8894 | 6722.70837 | 5614.51825 | 5767.34569 | 909.607 | 858.144 | 866.525 | 2431.855925 | 2524.064016 | 4463.9 |
| 909.882 | 2379.57 | 2396.92 | 3810.186135 | 3858.697765 | 6896.61433 | 6736.89726 | 5635.01991 | 5780.00678 | 909.882 | 882.042 | 867.75 | 2488.293636 | 2527.354008 | 4520.17 |
| 910.157 | 2414.93 | 2400.24 | 3763.087356 | 3864.224835 | 6837.22224 | 6751.09665 | 5637.72637 | 5792.67948 | 910.157 | 868.444 | 868.975 | 2427.759996 | 2530.644 | 4485.79 |
| 910.433 | 2351.67 | 2403.57 | 3902.959541 | 3869.763209 | 6821.81171 | 6765.30655 | 5732.8936 | 5805.3638 | 910.433 | 903.484 | 870.201 | 2511.003411 | 2533.933992 | 4517.34 |
| 910.708 | 2318.51 | 2406.89 | 3734.87556 | 3875.301582 | 6879.98327 | 6779.54795 | 5624.36834 | 5818.07135 | 910.708 | 870.864 | 871.427 | 2482.111101 | 2537.235024 | 4531.65 |
| 910.983 | 2348.55 | 2410.22 | 3788.360423 | 3880.839955 | 6886.92437 | 6793.79985 | 5680.83214 | 5830.79052 | 910.983 | 895.859 | 872.653 | 2431.535758 | 2540.525015 | 4497.43 |
| 911.258 | 2313.25 | 2413.54 | 3839.68645 | 3886.389632 | 6884.62129 | 6808.06226 | 5763.48929 | 5843.53291 | 911.258 | 889.517 | 873.88 | 2532.200674 | 2543.826047 | 4535.43 |
| 911.533 | 2383.91 | 2416.87 | 3782.336029 | 3891.928005 | 6847.98414 | 6822.35617 | 5699.34753 | 5856.28692 | 911.533 | 910.793 | 875.108 | 2529.26397 | 2547.127079 | 4503.57 |
| 911.808 | 2362.99 | 2420.2 | 3849.36165 | 3897.477681 | 6967.18223 | 6836.66059 | 5809.34798 | 5869.06417 | 911.808 | 859.483 | 876.336 | 2552.856964 | 2550.417071 | 4467.94 |
| 912.083 | 2384.83 | 2423.53 | 3844.851831 | 3903.027357 | 6906.37857 | 6850.97551 | 5758.8314 | 5881.85303 | 912.083 | 898.223 | 877.565 | 2549.500731 | 2553.718103 | 4475.88 |
| 912.358 | 2383.19 | 2426.86 | 3852.75249 | 3908.588336 | 6990.87114 | 6865.31143 | 5917.72223 | 5894.6535 | 912.358 | 881.576 | 878.794 | 2522.286538 | 2557.019135 | 4509.69 |
| 912.633 | 2424.64 | 2430.19 | 3867.796521 | 3914.138013 | 7079.20573 | 6879.66836 | 5915.78242 | 5907.47721 | 912.633 | 880.087 | 880.023 | 2504.390307 | 2560.320167 | 4570.62 |
| 912.909 | 2398.54 | 2433.53 | 3825.885728 | 3919.698992 | 6995.84878 | 6894.0463 | 5853.98702 | 5920.31253 | 912.909 | 900.769 | 881.254 | 2478.180775 | 2563.621199 | 4556.36 |
| 913.183 | 2393.86 | 2436.86 | 3798.272981 | 3925.259971 | 7066.024 | 6908.43474 | 5726.78375 | 5933.17108 | 913.183 | 891.464 | 882.484 | 2511.930792 | 2566.922231 | 4492.55 |
| 913.458 | 2398.97 | 2440.2 | 3836.103462 | 3930.82095 | 6970.23887 | 6922.83368 | 5889.55422 | 5946.04125 | 913.458 | 880.334 | 883.715 | 2507.194528 | 2570.223263 | 4484.14 |
| 913.734 | 2340.67 | 2443.54 | 3829.649561 | 3936.381929 | 6962.46992 | 6937.26412 | 5962.10573 | 5958.93465 | 913.734 | 885.158 | 884.947 | 2537.455828 | 2573.535335 | 4554.3 |
| 914.008 | 2406.75 | 2446.88 | 3925.090429 | 3941.95421 | 6925.56744 | 6951.70507 | 5934.13518 | 5971.83967 | 914.008 | 884.693 | 886.179 | 2544.510542 | 2576.836367 | 4523.11 |
| 914.283 | 2440.26 | 2450.22 | 3819.115349 | 3947.515189 | 7248.89134 | 6966.15652 | 5841.17493 | 5984.7563 | 914.283 | 897.972 | 887.412 | 2527.111123 | 2580.137399 | 4612.13 |
| 914.558 | 2467.3 | 2453.56 | 3847.214117 | 3953.098774 | 6940.1289 | 6980.62898 | 5841.09362 | 5997.69616 | 914.558 | 908.678 | 888.645 | 2548.562311 | 2583.449472 | 4641.15 |
| 914.833 | 2466.97 | 2456.9 | 3967.204672 | 3958.671056 | 7233.68247 | 6995.12244 | 5900.32195 | 6010.65926 | 914.833 | 918.08 | 889.879 | 2595.958064 | 2586.750504 | 4577.12 |
| 915.108 | 2419.69 | 2460.25 | 3913.538964 | 3964.243338 | 7071.43678 | 7009.62641 | 5948.60832 | 6023.62235 | 915.108 | 886.488 | 891.113 | 2565.02331 | 2590.062576 | 4550.88 |
| 915.383 | 2378.84 | 2463.59 | 3838.409234 | 3969.826922 | 7038.0367 | 7024.15138 | 6076.48529 | 6036.60868 | 915.383 | 906.558 | 892.347 | 2580.711492 | 2593.374648 | 4557.39 |
| 915.658 | 2453.3 | 2466.94 | 3900.393804 | 3975.410507 | 7190.31648 | 7038.69735 | 5950.03704 | 6049.61823 | 915.658 | 926.638 | 893.583 | 2585.977687 | 2596.67568 | 4586.68 |
| 915.933 | 2398.09 | 2470.29 | 3882.105869 | 3980.994091 | 7200.32483 | 7053.25383 | 5885.15188 | 6062.63941 | 915.933 | 907.576 | 894.818 | 2558.211481 | 2599.987752 | 4576.34 |
| 916.207 | 2430.47 | 2473.64 | 3983.107715 | 3986.577676 | 7157.97772 | 7067.83131 | 6048.53798 | 6075.68381 | 916.207 | 933.589 | 896.054 | 2479.273759 | 2603.299825 | 4683.94 |
| 916.482 | 2449.3 | 2476.99 | 3942.937554 | 3992.161261 | 7090.6681 | 7082.4298 | 5987.84607 | 6088.72821 | 916.482 | 912.761 | 897.291 | 2547.292683 | 2606.611897 | 4601.84 |
| 916.757 | 2469.29 | 2480.34 | 3894.663283 | 3997.756148 | 7172.77268 | 7097.03879 | 5955.86811 | 6101.80746 | 916.757 | 938.555 | 898.528 | 2521.02795 | 2609.923969 | 4644.65 |
| 917.032 | 2491.05 | 2483.69 | 3881.054709 | 4003.351035 | 7234.80748 | 7111.66878 | 5932.16052 | 6114.89833 | 917.032 | 945.345 | 899.766 | 2571.625374 | 2613.236041 | 4600.48 |
| 917.307 | 2433.83 | 2487.04 | 3925.33909 | 4008.945923 | 7224.03497 | 7126.30928 | 5869.99342 | 6128.00081 | 917.307 | 911.5 | 901.004 | 2612.363862 | 2616.559154 | 4599.86 |
| 917.581 | 2432.3 | 2490.4 | 3931.544329 | 4014.54081 | 7283.78791 | 7140.97078 | 5965.30004 | 6141.11491 | 917.581 | 909.885 | 902.242 | 2588.837109 | 2619.871226 | 4655.27 |
| 917.856 | 2457.59 | 2493.75 | 3946.023219 | 4020.147 | 7449.21759 | 7155.65329 | 6064.77669 | 6154.25224 | 917.856 | 930.674 | 903.481 | 2552.150389 | 2623.183298 | 4709.37 |
| 918.131 | 2438.59 | 2497.11 | 3890.842936 | 4025.753191 | 7282.57799 | 7170.3463 | 5957.40138 | 6167.4128 | 918.131 | 885.801 | 904.721 | 2573.447013 | 2626.506411 | 4650.15 |
| 918.405 | 2458.62 | 2500.47 | 3950.239165 | 4031.359381 | 7342.36277 | 7185.06031 | 6114.53825 | 6180.58498 | 918.405 | 903.906 | 905.961 | 2548.473989 | 2629.818483 | 4661.94 |
| 918.68 | 2379.76 | 2503.83 | 3939.218932 | 4036.965571 | 7334.96529 | 7199.79533 | 6018.3953 | 6193.76877 | 918.68 | 924.793 | 907.201 | 2554.95461 | 2633.130555 | 4663.91 |
| 918.955 | 2417.52 | 2507.19 | 3949.244518 | 4042.571761 | 7302.88126 | 7214.54086 | 6089.82009 | 6206.97579 | 918.955 | 882.713 | 908.442 | 2597.393296 | 2636.453668 | 4620.22 |
| 919.229 | 2475.74 | 2510.55 | 3985.153523 | 4048.189254 | 7304.03811 | 7229.29688 | 6144.739 | 6220.19443 | 919.229 | 929.537 | 909.684 | 2561.39107 | 2639.77678 | 4685.84 |
| 919.504 | 2494.93 | 2513.91 | 4047.477178 | 4053.795444 | 7376.47395 | 7244.08441 | 6122.61114 | 6233.4363 | 919.504 | 911.418 | 910.926 | 2656.73459 | 2643.088853 | 4697.35 |
| 919.779 | 2477.27 | 2517.28 | 3971.827518 | 4059.412937 | 7289.80566 | 7258.88245 | 6013.76064 | 6246.68979 | 919.779 | 939.982 | 912.168 | 2594.809879 | 2646.411965 | 4753.28 |
| 920.053 | 2487 | 2520.64 | 4024.611607 | 4065.03043 | 7241.71674 | 7273.69099 | 6235.73621 | 6259.95489 | 920.053 | 937.863 | 913.411 | 2671.914921 | 2649.735077 | 4698.34 |
| 920.328 | 2476.3 | 2524.01 | 3957.122572 | 4070.659226 | 7342.47952 | 7288.52053 | 6202.55022 | 6273.24323 | 920.328 | 932.206 | 914.654 | 2536.042678 | 2653.05819 | 4692.82 |
| 920.602 | 2500.97 | 2527.38 | 3996.614565 | 4076.276719 | 7451.96643 | 7303.37108 | 6207.30103 | 6286.54317 | 920.602 | 908.08 | 915.898 | 2582.488971 | 2656.381302 | 4751.64 |
| 920.877 | 2489.66 | 2530.74 | 4001.033961 | 4081.905515 | 7441.92624 | 7318.24263 | 6074.40609 | 6299.86635 | 920.877 | 932.688 | 917.142 | 2611.690408 | 2659.704415 | 4679.56 |
| 921.151 | 2502.29 | 2534.11 | 3997.541395 | 4087.534311 | 7366.07291 | 7333.12468 | 6204.36227 | 6313.20115 | 921.151 | 918.519 | 918.387 | 2620.621962 | 2663.027527 | 4682.62 |
| 921.426 | 2463.28 | 2537.49 | 4016.010174 | 4093.163107 | 7408.89762 | 7348.01724 | 6115.47912 | 6326.55918 | 921.426 | 960.342 | 919.632 | 2605.287068 | 2666.35064 | 4792.3 |
| 921.7 | 2506.57 | 2540.86 | 3931.273062 | 4098.803205 | 7600.41481 | 7362.9413 | 6211.85438 | 6339.92882 | 921.7 | 950.038 | 920.878 | 2582.764977 | 2669.673752 | 4747.77 |
| 921.975 | 2501.29 | 2544.23 | 3993.427174 | 4104.432001 | 7421.7185 | 7377.86537 | 6208.76461 | 6353.31008 | 921.975 | 935.974 | 922.124 | 2645.341062 | 2673.007905 | 4704.76 |
| 922.249 | 2521.51 | 2547.61 | 4062.622933 | 4110.0721 | 7521.15461 | 7392.82094 | 6231.05509 | 6366.71457 | 922.249 | 903.623 | 923.37 | 2630.778984 | 2676.331018 | 4716.93 |
| 922.523 | 2516.17 | 2550.98 | 4037.010782 | 4115.712198 | 7444.47344 | 7407.78702 | 6177.35582 | 6380.14229 | 922.523 | 920.175 | 924.617 | 2667.940435 | 2679.66517 | 4717.72 |
| 922.798 | 2533.57 | 2554.36 | 4031.641951 | 4121.352297 | 7606.55992 | 7422.7741 | 6275.47343 | 6393.58163 | 922.798 | 923.911 | 925.865 | 2580.689411 | 2682.988283 | 4754.44 |
| 923.072 | 2503 | 2557.73 | 4026.578295 | 4126.992395 | 7550.8931 | 7437.77168 | 6313.45669 | 6407.03258 | 923.072 | 942.938 | 927.113 | 2633.770889 | 2686.311395 | 4725.18 |
| 923.346 | 2461.01 | 2561.11 | 4087.421283 | 4132.643797 | 7615.11425 | 7452.79027 | 6320.32156 | 6420.49515 | 923.346 | 946.808 | 928.361 | 2559.823356 | 2689.645548 | 4759.88 |
| 923.621 | 2484.61 | 2564.49 | 4117.791914 | 4138.295198 | 7779.38708 | 7467.82986 | 6276.54208 | 6433.98095 | 923.621 | 913.163 | 929.61 | 2633.240958 | 2692.979701 | 4736.26 |
| 923.895 | 2503.08 | 2567.87 | 4151.462964 | 4143.9466 | 7712.48076 | 7482.87996 | 6366.91204 | 6447.48998 | 923.895 | 935.889 | 930.86 | 2656.811872 | 2696.313853 | 4673.7 |
| 924.169 | 2497.67 | 2571.26 | 4040.616376 | 4149.598001 | 7529.762 | 7497.95106 | 6336.26988 | 6461.01063 | 924.169 | 944.261 | 932.109 | 2615.02456 | 2699.636966 | 4783.93 |
| 924.444 | 2528.51 | 2574.64 | 4025.301078 | 4155.249402 | 7566.02769 | 7513.03266 | 6291.28237 | 6474.54289 | 924.444 | 939.417 | 933.36 | 2617.243649 | 2702.971119 | 4796.52 |
| 924.718 | 2529.63 | 2578.02 | 4111.258894 | 4160.900804 | 7460.53138 | 7528.13527 | 6153.0326 | 6488.09838 | 924.718 | 994.942 | 934.61 | 2634.024815 | 2706.305271 | 4775.79 |
| 924.992 | 2494.16 | 2581.41 | 4053.219001 | 4166.563508 | 7432.65021 | 7543.25888 | 6248.15337 | 6501.66549 | 924.992 | 932.197 | 935.861 | 2669.795195 | 2709.639424 | 4801.28 |
| 925.266 | 2496.97 | 2584.79 | 4102.985242 | 4172.226212 | 7774.01674 | 7558.39299 | 6278.77229 | 6515.24422 | 925.266 | 940.531 | 937.113 | 2647.527029 | 2712.973577 | 4803.11 |
| 925.54 | 2533.92 | 2588.18 | 4023.594355 | 4177.888917 | 7671.09946 | 7573.54811 | 6300.59814 | 6528.84617 | 925.54 | 962.933 | 938.365 | 2664.584201 | 2716.30773 | 4787.45 |
| 925.815 | 2541.96 | 2591.57 | 4089.444484 | 4183.551621 | 7622.56479 | 7588.71374 | 6471.54604 | 6542.47136 | 925.815 | 904.856 | 939.617 | 2650.640377 | 2719.641882 | 4819.33 |
| 926.089 | 2571.12 | 2594.96 | 4153.124476 | 4189.225628 | 7604.90424 | 7603.90036 | 6479.54924 | 6556.10816 | 926.089 | 958.833 | 940.87 | 2685.119049 | 2722.976035 | 4714 |
| 926.363 | 2550.25 | 2598.35 | 4155.283311 | 4194.888332 | 7690.29894 | 7619.108 | 6538.60334 | 6569.75658 | 926.363 | 992.246 | 942.123 | 2661.448773 | 2726.310188 | 4826.33 |
| 926.637 | 2541.3 | 2601.74 | 4159.001933 | 4200.562339 | 7898.24554 | 7634.32613 | 6560.97513 | 6583.42823 | 926.637 | 985.713 | 943.377 | 2626.782417 | 2729.655381 | 4799.1 |
| 926.911 | 2577.82 | 2605.13 | 4182.387432 | 4206.236346 | 7635.26893 | 7649.56527 | 6526.58112 | 6597.1115 | 926.911 | 968.709 | 944.631 | 2724.256703 | 2732.989533 | 4790.94 |
| 927.185 | 2588.22 | 2608.53 | 4133.683655 | 4211.921656 | 7741.70982 | 7664.81491 | 6435.10767 | 6610.80638 | 927.185 | 967.113 | 945.886 | 2687.503741 | 2736.334726 | 4862.89 |
| 927.459 | 2533.06 | 2611.92 | 4135.446892 | 4217.595663 | 7967.68844 | 7680.08556 | 6485.68232 | 6624.52449 | 927.459 | 973.711 | 947.141 | 2652.572419 | 2739.668879 | 4774.15 |
| 927.734 | 2566.4 | 2615.32 | 4178.081064 | 4223.280973 | 7724.90895 | 7695.37721 | 6570.05859 | 6638.26584 | 927.734 | 973.73 | 948.397 | 2671.075863 | 2743.014072 | 4826.39 |
| 928.008 | 2643.87 | 2618.72 | 4093.886486 | 4228.95498 | 7675.70564 | 7710.67937 | 6400.31872 | 6652.00718 | 928.008 | 967.124 | 949.653 | 2717.776082 | 2746.348225 | 4869.27 |
| 928.282 | 2546.53 | 2622.11 | 4210.237538 | 4234.64029 | 7811.74706 | 7726.00253 | 6557.97829 | 6665.78338 | 928.282 | 962.751 | 950.909 | 2696.225532 | 2749.693418 | 4851.91 |
| 928.556 | 2604.85 | 2625.51 | 4204.032299 | 4240.336902 | 7812.20343 | 7741.33619 | 6569.65204 | 6679.55957 | 928.556 | 993.56 | 952.166 | 2709.529022 | 2753.027571 | 4831.79 |
| 928.83 | 2602.73 | 2628.91 | 4125.217855 | 4246.022212 | 7866.01292 | 7756.69086 | 6623.21192 | 6693.35899 | 928.83 | 941.328 | 953.423 | 2704.362189 | 2756.372764 | 4877.63 |
| 929.104 | 2564.9 | 2632.31 | 4169.683082 | 4251.707522 | 7875.20405 | 7772.06653 | 6686.00627 | 6707.18164 | 929.104 | 970.161 | 954.681 | 2723.936536 | 2759.717957 | 4931.18 |
| 929.377 | 2632.42 | 2635.71 | 4187.044187 | 4257.404134 | 7942.59857 | 7787.45271 | 6577.01638 | 6721.01591 | 929.377 | 999.587 | 955.939 | 2735.296944 | 2763.063149 | 4878.6 |
| 929.651 | 2619.36 | 2639.12 | 4143.743149 | 4263.100747 | 7967.50801 | 7802.85989 | 6701.09503 | 6734.8618 | 929.651 | 993.477 | 957.197 | 2736.97506 | 2766.408342 | 4886.81 |
| 929.925 | 2514.37 | 2642.52 | 4129.987638 | 4268.79736 | 7955.72724 | 7818.27757 | 6580.75663 | 6748.73092 | 929.925 | 991.664 | 958.456 | 2710.721368 | 2769.753535 | 4847.88 |
| 930.199 | 2647.14 | 2645.93 | 4172.859169 | 4274.493972 | 8067.97361 | 7833.71626 | 6597.76198 | 6762.61165 | 930.199 | 971.399 | 959.715 | 2753.89975 | 2773.098728 | 4944.01 |
| 930.473 | 2594.47 | 2649.33 | 4225.12333 | 4280.201888 | 8000.67461 | 7849.17595 | 6762.58842 | 6776.51561 | 930.473 | 945.176 | 960.975 | 2716.572695 | 2776.443921 | 4932.59 |
| 930.747 | 2619.04 | 2652.74 | 4213.232781 | 4285.8985 | 8142.90359 | 7864.64615 | 6720.52806 | 6790.43119 | 930.747 | 963.784 | 962.235 | 2802.410568 | 2779.789114 | 4920.4 |
| 931.021 | 2643.42 | 2656.15 | 4205.490361 | 4291.606416 | 8084.05278 | 7880.13735 | 6546.95501 | 6804.35839 | 931.021 | 977.763 | 963.496 | 2712.068277 | 2783.134307 | 4858.25 |
| 931.294 | 2619.65 | 2659.55 | 4257.664099 | 4297.314331 | 8128.98954 | 7895.63905 | 6698.1098 | 6818.30882 | 931.294 | 1000.47 | 964.757 | 2794.560957 | 2786.4795 | 5000.06 |
| 931.568 | 2591.11 | 2662.97 | 4267.554051 | 4303.033549 | 8125.39162 | 7911.16176 | 6851.87812 | 6832.27086 | 931.568 | 1006.1 | 966.018 | 2776.466002 | 2789.824693 | 4961.21 |
| 931.842 | 2675.62 | 2666.38 | 4289.04068 | 4308.741465 | 7843.91601 | 7926.70547 | 6764.64439 | 6846.25613 | 931.842 | 993.761 | 967.28 | 2773.440976 | 2793.180926 | 4905.56 |
| 932.116 | 2641.39 | 2669.79 | 4289.379764 | 4314.44938 | 8193.24252 | 7942.25969 | 6826.41656 | 6860.25302 | 932.116 | 999.842 | 968.543 | 2720.105572 | 2796.526119 | 4945.35 |
| 932.39 | 2633.35 | 2673.2 | 4273.047214 | 4320.168598 | 7893.01318 | 7957.82441 | 6798.72479 | 6874.27314 | 932.39 | 974.974 | 969.805 | 2757.476788 | 2799.871312 | 4922.96 |
| 932.663 | 2640.68 | 2676.61 | 4265.632575 | 4325.887817 | 8182.18345 | 7973.42063 | 6893.74101 | 6888.30488 | 932.663 | 969.024 | 971.068 | 2770.206185 | 2803.227546 | 4933.81 |
| 932.937 | 2616.73 | 2680.03 | 4275.782492 | 4331.607035 | 8056.64921 | 7989.02736 | 6775.00558 | 6902.34823 | 932.937 | 1003.87 | 972.332 | 2762.765063 | 2806.572739 | 4975.71 |
| 933.211 | 2606.08 | 2683.45 | 4257.449346 | 4337.326253 | 7952.33098 | 8004.64459 | 6859.50962 | 6916.41481 | 933.211 | 1010.65 | 973.596 | 2776.532243 | 2809.928972 | 4927.77 |
| 933.484 | 2681.16 | 2686.86 | 4252.656957 | 4343.056774 | 8149.08053 | 8020.28283 | 6861.18228 | 6930.49301 | 933.484 | 1005.35 | 974.86 | 2764.123013 | 2813.274165 | 4897.74 |
| 933.758 | 2595.11 | 2690.28 | 4223.100128 | 4348.775992 | 8253.86575 | 8035.94207 | 6869.63849 | 6944.59444 | 933.758 | 994.939 | 976.125 | 2762.080568 | 2816.630398 | 4904.19 |
| 934.032 | 2655.43 | 2693.7 | 4212.656338 | 4354.506513 | 7993.13916 | 8051.61181 | 6911.87308 | 6958.70749 | 934.032 | 988.676 | 977.39 | 2813.660573 | 2819.986631 | 4992.28 |
| 934.305 | 2624.89 | 2697.12 | 4238.596271 | 4360.237035 | 8020.98848 | 8067.30256 | 6710.8522 | 6972.83215 | 934.305 | 977.644 | 978.655 | 2777.846032 | 2823.342864 | 4960.26 |
| 934.579 | 2658.16 | 2700.54 | 4287.130506 | 4365.967556 | 8134.20067 | 8083.01431 | 6943.31672 | 6986.98005 | 934.579 | 978.777 | 979.921 | 2741.600921 | 2826.688057 | 4997.63 |
| 934.852 | 2697.26 | 2703.96 | 4255.324419 | 4371.698077 | 8068.5255 | 8098.73657 | 6798.34147 | 7001.13955 | 934.852 | 1015.26 | 981.187 | 2786.44638 | 2830.04429 | 4970.34 |
| 935.126 | 2673.15 | 2707.38 | 4343.463675 | 4377.4399 | 8304.57615 | 8114.46932 | 6988.16484 | 7015.32229 | 935.126 | 1034.22 | 982.454 | 2789.372043 | 2833.400524 | 4970.55 |
| 935.4 | 2700.76 | 2710.8 | 4404.984831 | 4383.170421 | 8155.15135 | 8130.22309 | 6806.99515 | 7029.51665 | 935.4 | 1018.71 | 983.721 | 2800.66621 | 2836.756757 | 4987.4 |
| 935.673 | 2627.15 | 2714.23 | 4286.768816 | 4388.912245 | 8273.9992 | 8145.99786 | 6901.56998 | 7043.73424 | 935.673 | 1004.64 | 984.989 | 2826.34581 | 2840.11299 | 5009.88 |
| 935.947 | 2657.96 | 2717.66 | 4235.578422 | 4394.654069 | 8257.43182 | 8161.79363 | 6894.96066 | 7057.96344 | 935.947 | 984.228 | 986.257 | 2799.429703 | 2843.469223 | 5051.29 |
| 936.22 | 2654.58 | 2721.08 | 4238.709299 | 4400.395893 | 8193.25313 | 8177.5894 | 7004.48487 | 7072.20426 | 936.22 | 1068.1 | 987.525 | 2819.291096 | 2846.825457 | 5024.9 |
| 936.494 | 2671.74 | 2724.51 | 4333.539814 | 4406.14902 | 8306.87924 | 8193.41668 | 7188.05901 | 7086.46831 | 936.494 | 1004.99 | 988.794 | 2825.087222 | 2850.18169 | 5058.38 |
| 936.767 | 2733.11 | 2727.94 | 4362.972313 | 4411.890843 | 8248.46358 | 8209.25447 | 6962.88913 | 7100.74397 | 936.767 | 1033.58 | 990.063 | 2838.997926 | 2853.537923 | 5038.67 |
| 937.04 | 2669.53 | 2731.36 | 4336.772416 | 4417.64397 | 8349.21574 | 8225.11325 | 6927.22901 | 7115.04287 | 937.04 | 989.695 | 991.332 | 2820.671126 | 2856.894156 | 5035.82 |
| 937.314 | 2698.21 | 2734.8 | 4364.45298 | 4423.397097 | 8462.01401 | 8240.98254 | 7077.60555 | 7129.35338 | 937.314 | 1005.84 | 992.602 | 2814.88604 | 2860.26143 | 5056.96 |
| 937.587 | 2687.35 | 2738.22 | 4329.176932 | 4429.150223 | 8395.71265 | 8256.87284 | 6927.04316 | 7143.67551 | 937.587 | 1038.51 | 993.872 | 2801.306544 | 2863.617663 | 4978.73 |
| 937.861 | 2672.7 | 2741.66 | 4406.397681 | 4434.90335 | 8321.66358 | 8272.77364 | 6915.49718 | 7158.02087 | 937.861 | 995.289 | 995.143 | 2798.281518 | 2866.973896 | 4995.8 |
| 938.134 | 2709.17 | 2745.09 | 4391.884882 | 4440.656477 | 8227.29002 | 8288.69544 | 7086.14307 | 7172.37785 | 938.134 | 1021.17 | 996.414 | 2883.313452 | 2870.330129 | 5022.61 |
| 938.407 | 2743.84 | 2748.52 | 4344.333991 | 4446.420906 | 8376.16293 | 8304.63825 | 6840.41345 | 7186.75805 | 938.407 | 971.233 | 997.685 | 2805.844083 | 2873.697403 | 4987.8 |
| 938.681 | 2722.97 | 2751.96 | 4385.9057 | 4452.174033 | 8458.44793 | 8320.59156 | 6984.5756 | 7201.14988 | 938.681 | 975.208 | 998.957 | 2827.438794 | 2877.053636 | 5061 |
| 938.954 | 2676.77 | 2755.39 | 4372.794448 | 4457.938462 | 8458.28873 | 8336.56588 | 7066.11764 | 7215.56493 | 938.954 | 1018.38 | 1000.23 | 2779.976798 | 2880.420909 | 5048.9 |
| 939.227 | 2727.09 | 2758.83 | 4302.909219 | 4463.702892 | 8365.26306 | 8352.5507 | 6912.36094 | 7229.9916 | 939.227 | 993.091 | 1001.5 | 2828.907146 | 2883.777143 | 5076.6 |
| 939.5 | 2695.61 | 2762.26 | 4478.475655 | 4469.467321 | 8588.92798 | 8368.55652 | 6966.53646 | 7244.42989 | 939.5 | 998.223 | 1002.77 | 2791.270965 | 2887.144416 | 5088.86 |
| 939.774 | 2683.01 | 2765.7 | 4388.053232 | 4475.243053 | 8501.17712 | 8384.57284 | 7018.44691 | 7258.8914 | 939.774 | 992.125 | 1004.05 | 2837.065884 | 2890.500649 | 5010.89 |
| 940.047 | 2761.23 | 2769.14 | 4404.204937 | 4481.007483 | 8567.75442 | 8400.61018 | 6942.85209 | 7273.36453 | 940.047 | 1015.29 | 1005.32 | 2809.995213 | 2893.867923 | 5047.24 |
| 940.32 | 2752.71 | 2772.58 | 4501.533373 | 4486.783215 | 8626.28683 | 8416.66851 | 7083.04168 | 7287.84928 | 940.32 | 1052.78 | 1006.6 | 2890.798736 | 2897.235196 | 5060.28 |
| 940.593 | 2701.96 | 2776.02 | 4461.408423 | 4492.558947 | 8446.23201 | 8432.73735 | 6974.16795 | 7302.35726 | 940.593 | 1025.19 | 1007.87 | 2845.500628 | 2900.591429 | 5159.66 |
| 940.867 | 2724.97 | 2779.46 | 4414.750453 | 4498.334679 | 8500.40235 | 8448.82719 | 7139.3661 | 7316.88847 | 940.867 | 1016.9 | 1009.14 | 2859.34509 | 2903.958703 | 5101.77 |
| 941.14 | 2759.09 | 2782.91 | 4484.104451 | 4504.110412 | 8604.08379 | 8464.92754 | 7347.5887 | 7331.4313 | 941.14 | 1016.13 | 1010.42 | 2844.032276 | 2907.325976 | 5133.57 |
| 941.413 | 2660.45 | 2786.35 | 4427.420894 | 4509.886144 | 8543.54546 | 8481.04889 | 7106.15688 | 7345.98574 | 941.413 | 1025.06 | 1011.7 | 2839.605139 | 2910.69325 | 5003.94 |
| 941.686 | 2754.88 | 2789.79 | 4426.426248 | 4515.661876 | 8569.0917 | 8497.18074 | 7333.98675 | 7360.5518 | 941.686 | 1044.92 | 1012.97 | 2814.223625 | 2914.060523 | 5172.02 |
| 941.959 | 2813.43 | 2793.24 | 4400.10202 | 4521.448911 | 8313.96893 | 8513.3336 | 7178.17407 | 7375.14109 | 941.959 | 1021.72 | 1014.25 | 2867.790874 | 2917.416756 | 5050.48 |
| 942.232 | 2742.6 | 2796.68 | 4492.762398 | 4527.235946 | 8609.17817 | 8529.50747 | 7147.67131 | 7389.75361 | 942.232 | 1038.25 | 1015.52 | 2900.481027 | 2920.78403 | 5082.94 |
| 942.505 | 2751.87 | 2800.13 | 4448.61365 | 4533.022981 | 8571.45847 | 8545.69183 | 7443.41803 | 7404.37774 | 942.505 | 1024.24 | 1016.8 | 2834.416226 | 2924.151303 | 5222.48 |
| 942.778 | 2729.32 | 2803.57 | 4491.021766 | 4538.810016 | 8535.5218 | 8561.8867 | 7082.8326 | 7419.01349 | 942.778 | 1069.65 | 1018.08 | 2928.203072 | 2927.518577 | 5120.84 |
| 943.051 | 2734.73 | 2807.02 | 4387.894993 | 4544.597051 | 8551.49483 | 8578.11308 | 7166.9301 | 7433.67248 | 943.051 | 1038.31 | 1019.36 | 2827.229029 | 2930.88585 | 5063.41 |
| 943.324 | 2786.05 | 2810.47 | 4392.64217 | 4550.384086 | 8669.59975 | 8594.33945 | 7328.67838 | 7448.34308 | 943.324 | 1031.23 | 1020.63 | 2916.500416 | 2934.253124 | 5089.8 |
| 943.597 | 2752.27 | 2813.92 | 4467.692781 | 4556.182424 | 8719.95991 | 8610.59734 | 7385.0841 | 7463.02529 | 943.597 | 1002.34 | 1021.91 | 2923.720734 | 2937.631437 | 5145.39 |
| 943.87 | 2702.84 | 2817.37 | 4467.670175 | 4561.980762 | 8672.64577 | 8626.86572 | 7344.90548 | 7477.73074 | 943.87 | 1042.23 | 1023.19 | 2919.87873 | 2940.998711 | 5105.27 |
| 944.143 | 2780.85 | 2820.82 | 4445.030661 | 4567.7791 | 8661.77774 | 8643.14461 | 7250.23772 | 7492.4478 | 944.143 | 1044.89 | 1024.47 | 2877.660849 | 2944.365984 | 5165.51 |
| 944.416 | 2809.78 | 2824.28 | 4555.052144 | 4573.577438 | 8720.73469 | 8659.4445 | 7513.44894 | 7507.18809 | 944.416 | 1034.25 | 1025.75 | 2900.492067 | 2947.733258 | 5082.44 |
| 944.689 | 2777.42 | 2827.73 | 4447.845059 | 4579.375776 | 8643.98983 | 8675.7654 | 7308.83881 | 7521.94 | 944.689 | 1020.34 | 1027.03 | 3009.393003 | 2951.100531 | 5125.67 |
| 944.962 | 2776.15 | 2831.18 | 4531.327561 | 4585.174114 | 8620.87404 | 8692.0968 | 7270.37931 | 7536.70352 | 944.962 | 1024.84 | 1028.31 | 2924.670195 | 2954.478845 | 5078.03 |
| 945.235 | 2765.24 | 2834.64 | 4567.846917 | 4590.983754 | 8774.38497 | 8708.44921 | 7225.24079 | 7551.49028 | 945.235 | 1069.29 | 1029.59 | 2967.671933 | 2957.846118 | 5118.3 |
| 945.508 | 2754.24 | 2838.09 | 4547.061062 | 4596.782092 | 8756.71381 | 8724.81212 | 7312.62552 | 7566.28865 | 945.508 | 1033.36 | 1030.87 | 2875.839209 | 2961.213392 | 5230.39 |
| 945.78 | 2809.74 | 2841.55 | 4439.921794 | 4602.591733 | 8813.76036 | 8741.19603 | 7571.03946 | 7581.11025 | 945.78 | 1054.01 | 1032.15 | 2958.773498 | 2964.591705 | 5183.96 |
| 946.053 | 2826.56 | 2845.01 | 4430.325715 | 4608.401373 | 8661.78836 | 8757.59045 | 7373.2245 | 7595.94347 | 946.053 | 1024.54 | 1033.43 | 2964.779389 | 2967.958979 | 5250.92 |
| 946.326 | 2799.96 | 2848.46 | 4503.466152 | 4614.211014 | 8802.82864 | 8774.00587 | 7497.61677 | 7610.79992 | 946.326 | 1092.07 | 1034.71 | 2858.362508 | 2971.326252 | 5175.53 |
| 946.599 | 2808.56 | 2851.92 | 4486.692793 | 4620.031957 | 8856.42586 | 8790.43179 | 7385.81589 | 7625.66798 | 946.599 | 1079.33 | 1036 | 2904.775681 | 2974.704566 | 5241.87 |
| 946.872 | 2769.99 | 2855.38 | 4442.56665 | 4625.841598 | 8701.36539 | 8806.87872 | 7414.18137 | 7640.54766 | 946.872 | 1043.81 | 1037.28 | 2921.976376 | 2978.071839 | 5126.31 |
| 947.144 | 2764.52 | 2858.84 | 4641.156896 | 4631.651239 | 8853.09329 | 8823.34666 | 7348.83158 | 7655.43896 | 947.144 | 1068.42 | 1038.56 | 2950.559559 | 2981.450153 | 5279.91 |
| 947.417 | 2822.84 | 2862.3 | 4597.765436 | 4637.472182 | 8927.97019 | 8839.82509 | 7662.25736 | 7670.3651 | 947.417 | 1046.03 | 1039.85 | 2928.247233 | 2984.828467 | 5214.43 |
| 947.69 | 2810.26 | 2865.76 | 4569.689274 | 4643.293126 | 8992.48851 | 8856.32453 | 7444.05689 | 7685.29125 | 947.69 | 1067.94 | 1041.13 | 2903.550214 | 2988.19574 | 5099.29 |
| 947.963 | 2794.43 | 2869.23 | 4570.435259 | 4649.114069 | 8895.11138 | 8872.83448 | 7430.90795 | 7700.24062 | 947.963 | 1047.57 | 1042.41 | 2889.319343 | 2991.574054 | 5154.11 |
| 948.235 | 2787.09 | 2872.69 | 4494.050917 | 4654.935013 | 8904.82256 | 8889.36543 | 7449.22587 | 7715.20161 | 948.235 | 1032.44 | 1043.7 | 2907.734465 | 2994.941327 | 5298 |
| 948.508 | 2861.78 | 2876.15 | 4596.962937 | 4660.755956 | 8896.1621 | 8905.90688 | 7550.78172 | 7730.18583 | 948.508 | 1023.23 | 1044.98 | 2983.680282 | 2998.319641 | 5143.5 |
| 948.78 | 2817.84 | 2879.62 | 4578.403735 | 4666.588202 | 8815.43726 | 8922.46933 | 7675.16238 | 7745.18167 | 948.78 | 1097.15 | 1046.27 | 2849.143907 | 3001.697955 | 5288.94 |
| 949.053 | 2831.05 | 2883.08 | 4583.105701 | 4672.409146 | 9045.22605 | 8939.04229 | 7706.67571 | 7760.20074 | 949.053 | 1024.89 | 1047.55 | 2921.976376 | 3005.076268 | 5256.72 |
| 949.326 | 2839.86 | 2886.55 | 4645.824953 | 4678.241392 | 8852.33974 | 8955.63626 | 7606.90867 | 7775.21981 | 949.326 | 1055.67 | 1048.84 | 2949.521777 | 3008.443542 | 5267.93 |
| 949.598 | 2893.24 | 2890.02 | 4597.188993 | 4684.073638 | 8774.82012 | 8972.25123 | 7561.4333 | 7790.27373 | 949.598 | 1117.83 | 1050.12 | 2902.733236 | 3011.821856 | 5227.44 |
| 949.871 | 2810.2 | 2893.48 | 4617.873122 | 4689.905884 | 9023.00178 | 8988.8767 | 7475.34952 | 7805.32764 | 949.871 | 1074.45 | 1051.41 | 3010.728872 | 3015.200169 | 5182.08 |
| 950.143 | 2841.92 | 2896.95 | 4589.536995 | 4695.738131 | 8959.92687 | 9005.51267 | 7614.81894 | 7820.40479 | 950.143 | 1061.55 | 1052.7 | 2948.230068 | 3018.578483 | 5273.06 |
| 950.416 | 2819.35 | 2900.42 | 4485.042583 | 4701.570377 | 9034.78255 | 9022.16965 | 7521.99808 | 7835.50517 | 950.416 | 1075.02 | 1053.98 | 2961.986208 | 3021.956797 | 5202.62 |
| 950.688 | 2779.06 | 2903.89 | 4521.799298 | 4707.413926 | 8975.64519 | 9038.84763 | 7540.98971 | 7850.61716 | 950.688 | 1063.27 | 1055.27 | 2928.986929 | 3025.33511 | 5293.86 |
| 950.961 | 2883.01 | 2907.36 | 4502.923617 | 4713.246172 | 9010.52053 | 9055.53612 | 7600.32258 | 7865.74077 | 950.961 | 1043.6 | 1056.56 | 2871.323751 | 3028.702384 | 5280.03 |
| 951.233 | 2809.66 | 2910.83 | 4612.334749 | 4719.089721 | 8952.14732 | 9072.24561 | 7635.94786 | 7880.876 | 951.233 | 1070.21 | 1057.85 | 2930.488401 | 3032.080697 | 5291.64 |
| 951.506 | 2828.35 | 2914.3 | 4576.244899 | 4724.93327 | 8952.22161 | 9088.9656 | 7785.0699 | 7896.03446 | 951.506 | 1079.41 | 1059.13 | 2919.746247 | 3035.459011 | 5251.86 |
| 951.778 | 2866.05 | 2917.78 | 4562.240727 | 4730.77682 | 9039.42057 | 9105.7066 | 7596.83788 | 7911.21614 | 951.778 | 1064.17 | 1060.42 | 2897.842409 | 3038.837325 | 5314.74 |
| 952.051 | 2852.87 | 2921.25 | 4510.95991 | 4736.620369 | 9104.16177 | 9122.45811 | 7788.29906 | 7926.40945 | 952.051 | 1028.02 | 1061.71 | 2980.964382 | 3042.215638 | 5333.37 |
| 952.323 | 2872.25 | 2924.72 | 4584.654185 | 4742.47522 | 9377.34839 | 9139.23061 | 7740.1637 | 7941.61437 | 952.323 | 1067.29 | 1063 | 2930.598804 | 3045.593952 | 5278.83 |
| 952.595 | 2820.23 | 2928.2 | 4553.571477 | 4748.318769 | 8906.43579 | 9156.02412 | 7628.28151 | 7956.84252 | 952.595 | 1095.27 | 1064.29 | 2941.296797 | 3048.983306 | 5281.2 |
| 952.868 | 2896.33 | 2931.68 | 4602.682155 | 4754.173621 | 9165.35812 | 9172.82814 | 7635.51808 | 7972.07067 | 952.868 | 1074.19 | 1065.58 | 3044.202882 | 3052.36162 | 5405.71 |
| 953.14 | 2830.94 | 2935.15 | 4588.610166 | 4760.01717 | 9052.16716 | 9189.64265 | 7829.12816 | 7987.33367 | 953.14 | 1074.97 | 1066.87 | 3015.774262 | 3055.739933 | 5225.03 |
| 953.412 | 2896.21 | 2938.63 | 4569.463218 | 4765.872022 | 9140.33193 | 9206.47818 | 7734.9018 | 8002.60828 | 953.412 | 1108.13 | 1068.16 | 3071.549558 | 3059.118247 | 5217.26 |
| 953.685 | 2890.45 | 2942.11 | 4633.131906 | 4771.726874 | 9334.82085 | 9223.3347 | 7982.55962 | 8017.89451 | 953.685 | 1109.86 | 1069.45 | 2959.413832 | 3062.496561 | 5261.93 |
| 953.957 | 2906.24 | 2945.58 | 4688.018316 | 4777.581726 | 9263.43572 | 9240.20173 | 7746.14577 | 8033.19236 | 953.957 | 1066.31 | 1070.74 | 2914.215086 | 3065.874875 | 5291.36 |
| 954.229 | 2936.11 | 2949.06 | 4714.670325 | 4783.447881 | 9029.21057 | 9257.07926 | 7830.74274 | 8048.51343 | 954.229 | 1096.08 | 1072.03 | 3019.196736 | 3069.253188 | 5238.67 |
| 954.501 | 2870.94 | 2952.54 | 4578.810636 | 4789.302733 | 9276.9677 | 9273.9778 | 7824.62128 | 8063.85774 | 954.501 | 1078.28 | 1073.32 | 3028.481579 | 3072.642542 | 5362.51 |
| 954.774 | 2973.99 | 2956.02 | 4662.236623 | 4795.157584 | 9243.11123 | 9290.89734 | 7764.01068 | 8079.20205 | 954.774 | 1090.47 | 1074.62 | 3025.092225 | 3076.020856 | 5364.34 |
| 955.046 | 2979.85 | 2959.5 | 4672.804744 | 4801.023739 | 9282.18944 | 9307.82738 | 7836.20211 | 8094.56959 | 955.046 | 1081.9 | 1075.91 | 2979.253145 | 3079.39917 | 5385.58 |
| 955.318 | 2856.76 | 2962.99 | 4689.815462 | 4806.889894 | 9369.42025 | 9324.77843 | 7778.80905 | 8109.96036 | 955.318 | 1088.26 | 1077.2 | 3056.523791 | 3082.777483 | 5350.58 |
| 955.59 | 2969.37 | 2966.47 | 4766.663218 | 4812.756048 | 9295.60467 | 9341.73999 | 7818.29074 | 8125.36274 | 955.59 | 1110.23 | 1078.49 | 2957.073301 | 3086.166837 | 5394.37 |
| 955.862 | 2872.38 | 2969.95 | 4704.475197 | 4818.622203 | 9169.8051 | 9358.72254 | 7999.19327 | 8140.77675 | 955.862 | 1119.68 | 1079.79 | 2963.498721 | 3089.545151 | 5381.54 |
| 956.135 | 2892.03 | 2973.43 | 4773.139724 | 4824.488358 | 9179.66486 | 9375.7156 | 7711.49621 | 8156.21398 | 956.135 | 1101.03 | 1081.08 | 2977.575029 | 3092.923465 | 5329.45 |
| 956.407 | 2870.62 | 2976.92 | 4758.807771 | 4830.354512 | 9424.63069 | 9392.72967 | 7951.38315 | 8171.65121 | 956.407 | 1089.31 | 1082.37 | 2995.890788 | 3096.312818 | 5366.51 |
| 956.679 | 2861.11 | 2980.4 | 4669.798198 | 4836.23197 | 9260.52768 | 9409.75423 | 8083.40692 | 8187.12329 | 956.679 | 1125.6 | 1083.67 | 2984.905748 | 3099.691132 | 5394.37 |
| 956.951 | 2974.87 | 2983.89 | 4737.660226 | 4842.098125 | 9308.61659 | 9426.7998 | 7847.03953 | 8202.60699 | 956.951 | 1131.89 | 1084.96 | 3005.440596 | 3103.080486 | 5381.54 |
| 957.223 | 2936.7 | 2987.38 | 4835.372957 | 4847.975582 | 9401.85453 | 9443.85588 | 7996.38228 | 8218.1023 | 957.223 | 1112.55 | 1086.25 | 2983.183471 | 3106.4588 | 5386.67 |
| 957.495 | 2933.14 | 2990.86 | 4803.645989 | 4853.853039 | 9503.48759 | 9460.93296 | 7875.95095 | 8233.60923 | 957.495 | 1083.16 | 1087.55 | 3000.903057 | 3109.837113 | 5364.22 |
| 957.767 | 2937.13 | 2994.35 | 4773.80659 | 4859.730497 | 9495.73987 | 9478.03104 | 7896.5107 | 8249.13939 | 957.767 | 1096.99 | 1088.84 | 3033.008078 | 3113.226467 | 5406.07 |
| 958.039 | 2993.11 | 2997.84 | 4731.805375 | 4865.607954 | 9442.01529 | 9495.13963 | 7910.62374 | 8264.69278 | 958.039 | 1088.29 | 1090.14 | 3086.718849 | 3116.604781 | 5365.04 |
| 958.311 | 2929.25 | 3001.33 | 4846.223648 | 4871.485412 | 9758.4087 | 9512.25872 | 8145.17909 | 8280.24617 | 958.311 | 1174.21 | 1091.43 | 3061.381497 | 3119.994135 | 5487.01 |
| 958.583 | 2952.21 | 3004.82 | 4833.835776 | 4877.374172 | 9541.69551 | 9529.39881 | 8063.3118 | 8295.82279 | 958.583 | 1147.64 | 1092.73 | 3094.314535 | 3123.372449 | 5453.68 |
| 958.855 | 2970.2 | 3008.31 | 4801.747118 | 4883.25163 | 9478.03687 | 9546.54941 | 8044.83126 | 8311.42264 | 958.855 | 1139.84 | 1094.03 | 2987.908694 | 3126.761803 | 5410.16 |
| 959.127 | 2968.36 | 3011.8 | 4837.147497 | 4889.14039 | 9486.46384 | 9563.72101 | 8013.32955 | 8327.03411 | 959.127 | 1085.45 | 1095.32 | 2957.603233 | 3130.151157 | 5481.02 |
| 959.399 | 3045.45 | 3015.29 | 4810.879783 | 4895.02915 | 9528.57746 | 9580.91362 | 8265.30841 | 8342.65719 | 959.399 | 1141.9 | 1096.62 | 3062.397199 | 3133.52947 | 5426.4 |
| 959.67 | 3032.22 | 3018.78 | 4904.851286 | 4900.906608 | 9649.83453 | 9598.11673 | 8154.56455 | 8358.29189 | 959.67 | 1127.47 | 1097.92 | 3113.944083 | 3136.918824 | 5436.8 |
| 959.942 | 3004.24 | 3022.27 | 4727.905908 | 4906.795368 | 9639.06202 | 9615.33034 | 8086.91486 | 8373.94982 | 959.942 | 1172.25 | 1099.21 | 3121.970338 | 3140.297138 | 5450.1 |
| 960.214 | 2988.62 | 3025.77 | 4753.314608 | 4912.695431 | 9674.83949 | 9632.56496 | 8153.25198 | 8389.63099 | 960.214 | 1145.8 | 1100.51 | 3074.894751 | 3143.686492 | 5431.99 |
| 960.486 | 2974.22 | 3029.26 | 4736.507341 | 4918.584191 | 9556.71334 | 9649.81008 | 8265.63365 | 8405.31215 | 960.486 | 1119.17 | 1101.81 | 3010.463906 | 3147.064806 | 5560.05 |
| 960.758 | 3001.28 | 3032.75 | 4873.101713 | 4924.472952 | 9630.32727 | 9667.0867 | 8205.19728 | 8421.01654 | 960.758 | 1118.45 | 1103.11 | 3068.292687 | 3150.454159 | 5515.48 |
| 961.03 | 2953.52 | 3036.25 | 4725.588833 | 4930.361712 | 9807.85612 | 9684.36333 | 8195.32396 | 8436.74417 | 961.03 | 1122.68 | 1104.41 | 3103.444814 | 3153.843513 | 5552.72 |
| 961.301 | 2995.54 | 3039.74 | 4765.996353 | 4936.261775 | 9780.16599 | 9701.66096 | 8548.80059 | 8452.48341 | 961.301 | 1138.37 | 1105.7 | 3116.494379 | 3157.221827 | 5588.8 |
| 961.573 | 3001.61 | 3043.24 | 4800.933317 | 4942.161838 | 9514.97119 | 9718.9691 | 8244.09818 | 8468.23426 | 961.573 | 1147.37 | 1107 | 3175.570707 | 3160.611181 | 5394.21 |
| 961.845 | 2982.36 | 3046.74 | 4946.750776 | 4948.061901 | 9852.58061 | 9736.29823 | 8357.00255 | 8484.00835 | 961.845 | 1177.97 | 1108.3 | 3124.189427 | 3164.000535 | 5650.56 |
| 962.117 | 2945.97 | 3050.23 | 4717.789899 | 4953.961964 | 9547.85123 | 9753.64838 | 8166.7494 | 8499.79406 | 962.117 | 1160.6 | 1109.6 | 2992.843682 | 3167.378849 | 5547.93 |
| 962.388 | 3004.16 | 3053.73 | 4862.477078 | 4959.862027 | 9936.72294 | 9771.00903 | 8390.90871 | 8515.59138 | 962.388 | 1153.24 | 1110.9 | 3051.158234 | 3170.768203 | 5550.24 |
| 962.66 | 2977.51 | 3057.23 | 4902.274247 | 4965.76209 | 9809.15094 | 9788.39068 | 8320.14601 | 8531.41193 | 962.66 | 1154.22 | 1112.2 | 3108.192118 | 3174.157556 | 5567.5 |
| 962.932 | 2996.59 | 3060.73 | 4821.832199 | 4971.662153 | 9916.47274 | 9805.78283 | 8524.03597 | 8547.24409 | 962.932 | 1165.38 | 1113.5 | 3135.251748 | 3177.53587 | 5572.45 |
| 963.203 | 3011.53 | 3064.23 | 4876.017836 | 4977.573519 | 10065.18646 | 9823.18549 | 8290.63058 | 8563.08787 | 963.203 | 1158.08 | 1114.8 | 3070.169528 | 3180.925224 | 5561.24 |
| 963.475 | 2954.48 | 3067.73 | 4755.54126 | 4983.473582 | 9690.93989 | 9840.60915 | 8357.45556 | 8578.95489 | 963.475 | 1127.9 | 1116.1 | 3169.476494 | 3184.314578 | 5533.29 |
| 963.746 | 3007.58 | 3071.23 | 4925.411084 | 4989.384948 | 9865.59253 | 9858.05382 | 8244.53958 | 8594.83352 | 963.746 | 1121.05 | 1117.4 | 3139.92177 | 3187.703932 | 5538.8 |
| 964.018 | 3055.19 | 3074.73 | 5014.183298 | 4995.296314 | 9901.97496 | 9875.50899 | 8070.98977 | 8610.73538 | 964.018 | 1137.75 | 1118.71 | 3179.512073 | 3191.093286 | 5510.45 |
| 964.29 | 3074.57 | 3078.23 | 4955.691293 | 5001.20768 | 9961.75974 | 9892.97466 | 8549.56723 | 8626.64885 | 964.29 | 1211.14 | 1120.01 | 3155.54371 | 3194.4716 | 5603.36 |
| 964.561 | 2967.42 | 3081.73 | 5004.869788 | 5007.119046 | 9915.83595 | 9910.46134 | 8345.56111 | 8642.57394 | 964.561 | 1133.8 | 1121.31 | 3239.659305 | 3197.860954 | 5607.03 |
| 964.833 | 3036.42 | 3085.24 | 4890.157642 | 5013.030412 | 9897.17774 | 9927.95852 | 8376.94667 | 8658.51065 | 964.833 | 1155.42 | 1122.61 | 3129.742668 | 3201.250307 | 5667.04 |
| 965.104 | 3008.06 | 3088.74 | 4931.107697 | 5018.941778 | 9881.30023 | 9945.4767 | 8518.03067 | 8674.48221 | 965.104 | 1192.57 | 1123.91 | 3193.080529 | 3204.639661 | 5662.69 |
| 965.376 | 3079.16 | 3092.24 | 4989.497976 | 5024.864446 | 9803.88674 | 9963.01589 | 8627.43872 | 8690.45376 | 965.376 | 1151.09 | 1125.22 | 3126.949487 | 3208.029015 | 5590.39 |
| 965.647 | 3098.83 | 3095.75 | 4912.062474 | 5030.775812 | 10077.55096 | 9980.56558 | 8358.61713 | 8706.44855 | 965.647 | 1150.27 | 1126.52 | 3225.417395 | 3211.407329 | 5628.72 |
| 965.919 | 3143.49 | 3099.25 | 4887.671026 | 5036.698481 | 9858.93799 | 9998.12578 | 8524.63999 | 8722.45495 | 965.919 | 1164.17 | 1127.82 | 3210.137701 | 3214.796683 | 5679.13 |
| 966.19 | 3095.64 | 3102.76 | 4960.110689 | 5042.609847 | 10183.75837 | 10015.70698 | 8689.66389 | 8738.47296 | 966.19 | 1136.15 | 1129.13 | 3186.279741 | 3218.186037 | 5616.93 |
| 966.461 | 3093.53 | 3106.26 | 4965.083922 | 5048.532515 | 10120.98063 | 10033.30918 | 8563.21565 | 8754.51421 | 966.461 | 1196.52 | 1130.43 | 3121.959298 | 3221.575391 | 5620.31 |
| 966.733 | 3157.06 | 3109.77 | 4972.408138 | 5054.455184 | 10125.61865 | 10050.91139 | 8640.57605 | 8770.56708 | 966.733 | 1205.51 | 1131.73 | 3152.76157 | 3224.964745 | 5731.98 |
| 967.004 | 3089.02 | 3113.28 | 4996.132721 | 5060.377853 | 9914.1272 | 10068.5451 | 8384.54332 | 8786.64317 | 967.004 | 1157.02 | 1133.04 | 3233.587173 | 3228.354099 | 5585.6 |
| 967.275 | 3061.92 | 3116.78 | 5047.718714 | 5066.300521 | 10259.17655 | 10086.18931 | 8603.22003 | 8802.73088 | 967.275 | 1198.66 | 1134.34 | 3199.428668 | 3231.743453 | 5597.94 |
| 967.547 | 3059.3 | 3120.29 | 5051.539061 | 5072.234493 | 10047.24996 | 10103.84403 | 8614.30139 | 8818.83021 | 967.547 | 1134.72 | 1135.65 | 3197.110217 | 3235.121766 | 5690.69 |
| 967.818 | 3084.69 | 3123.8 | 4990.775193 | 5078.157162 | 10372.00666 | 10121.51975 | 8574.0763 | 8834.95277 | 967.818 | 1198.57 | 1136.95 | 3098.65335 | 3238.51112 | 5583.97 |
| 968.09 | 3063.56 | 3127.31 | 4894.147532 | 5084.07983 | 10246.99247 | 10139.20598 | 8424.00177 | 8851.08694 | 968.09 | 1157.09 | 1138.26 | 3200.488531 | 3241.900474 | 5720.43 |
| 968.361 | 3048.74 | 3130.82 | 4956.369461 | 5090.013802 | 9908.83116 | 10156.91321 | 8560.76474 | 8867.23273 | 968.361 | 1176.1 | 1139.56 | 3335.632119 | 3245.289828 | 5708.55 |
| 968.632 | 3097.29 | 3134.33 | 5026.017332 | 5095.947773 | 10206.25858 | 10174.63094 | 8747.75389 | 8883.40175 | 968.632 | 1143.04 | 1140.87 | 3220.946097 | 3248.679182 | 5665.23 |
| 968.903 | 3120.76 | 3137.84 | 4888.258771 | 5101.881745 | 10311.56386 | 10192.36968 | 8701.97651 | 8899.58239 | 968.903 | 1170.56 | 1142.17 | 3233.995662 | 3252.068536 | 5693.85 |
| 969.174 | 3143.41 | 3141.35 | 4880.674591 | 5107.804413 | 10450.12063 | 10210.11892 | 8556.98964 | 8915.77464 | 969.174 | 1207.53 | 1143.48 | 3200.632054 | 3255.45789 | 5739.21 |
| 969.446 | 3126.89 | 3144.86 | 5016.941181 | 5113.749688 | 10215.25867 | 10227.88916 | 8718.48239 | 8931.99012 | 969.446 | 1144.09 | 1144.78 | 3186.390143 | 3258.847244 | 5864.68 |
| 969.717 | 3017.31 | 3148.37 | 4944.501518 | 5119.683659 | 10207.97794 | 10245.66991 | 8872.68048 | 8948.21722 | 969.717 | 1204.36 | 1146.09 | 3198.722092 | 3262.236598 | 5566.87 |
| 969.988 | 3093.7 | 3151.89 | 5174.016232 | 5125.617631 | 10263.9207 | 10263.47166 | 8571.02138 | 8964.46755 | 969.988 | 1161.49 | 1147.4 | 3143.233842 | 3265.625952 | 5701.44 |
| 970.259 | 3064.53 | 3155.4 | 5113.308878 | 5131.551602 | 10344.46512 | 10281.27341 | 8684.27422 | 8980.7295 | 970.259 | 1147.51 | 1148.7 | 3230.517986 | 3269.004265 | 5673.29 |
| 970.53 | 3073.04 | 3158.91 | 5038.574746 | 5137.496876 | 10276.62484 | 10299.10667 | 8735.76651 | 8997.00306 | 970.53 | 1218.1 | 1150.01 | 3304.366157 | 3272.393619 | 5640.41 |
| 970.801 | 3076.51 | 3162.43 | 5167.867508 | 5143.430848 | 10512.57996 | 10316.95044 | 8769.48682 | 9013.29986 | 970.801 | 1163.45 | 1151.32 | 3194.085191 | 3275.782973 | 5744.19 |
| 971.072 | 3144.5 | 3165.94 | 5051.561666 | 5149.376122 | 10177.22057 | 10334.8152 | 8801.39508 | 9029.60826 | 971.072 | 1200.45 | 1152.62 | 3238.875448 | 3279.172327 | 5816.49 |
| 971.343 | 3122.9 | 3169.46 | 5048.50991 | 5155.321396 | 10592.63617 | 10352.69047 | 8869.55587 | 9045.92829 | 971.343 | 1178.88 | 1153.93 | 3318.608067 | 3282.561681 | 5730.26 |
| 971.615 | 3118.86 | 3172.97 | 5015.935232 | 5161.266671 | 10609.87218 | 10370.57625 | 9232.09273 | 9062.27154 | 971.615 | 1156.59 | 1155.24 | 3158.690179 | 3285.951035 | 5803.11 |
| 971.886 | 3031.8 | 3176.49 | 5081.197615 | 5167.211945 | 10310.4813 | 10388.48302 | 8807.92309 | 9078.6148 | 971.886 | 1225.75 | 1156.55 | 3193.455897 | 3289.340389 | 5631.16 |
| 972.157 | 3159.21 | 3180 | 5141.226801 | 5173.157219 | 10467.16561 | 10406.4003 | 8758.7307 | 9094.9929 | 972.157 | 1219.87 | 1157.86 | 3290.76458 | 3292.729743 | 5702.83 |
| 972.428 | 3145.1 | 3183.52 | 5123.809182 | 5179.102494 | 10418.33376 | 10424.33859 | 8704.27641 | 9111.37101 | 972.428 | 1212.68 | 1159.17 | 3302.080827 | 3296.119097 | 5738.01 |
| 972.698 | 3175.14 | 3187.04 | 5005.005422 | 5185.047768 | 10472.5041 | 10442.28738 | 9096.49133 | 9127.77234 | 972.698 | 1206.78 | 1160.47 | 3269.324432 | 3299.508451 | 5813.74 |
| 972.97 | 3206.36 | 3190.55 | 5064.808551 | 5190.993042 | 10302.62745 | 10460.25717 | 8781.83428 | 9144.19691 | 972.97 | 1172.38 | 1161.78 | 3281.082289 | 3302.897805 | 5804.03 |
| 973.24 | 3182.52 | 3194.07 | 5197.367823 | 5196.949619 | 10480.24121 | 10478.23746 | 8963.34083 | 9160.63309 | 973.24 | 1202.82 | 1163.09 | 3322.041582 | 3306.287159 | 5790.95 |
| 973.511 | 3111.61 | 3197.59 | 5136.242265 | 5202.894893 | 10554.79971 | 10496.23876 | 8915.19386 | 9177.08089 | 973.511 | 1202.33 | 1164.4 | 3264.378404 | 3309.665472 | 5721.91 |
| 973.782 | 3163.61 | 3201.11 | 5049.56107 | 5208.851471 | 10700.44618 | 10514.27157 | 8882.77451 | 9193.5403 | 973.782 | 1245.67 | 1165.71 | 3226.510378 | 3313.054826 | 5860.84 |
| 974.053 | 3088.9 | 3204.63 | 4983.541399 | 5214.808048 | 10370.31914 | 10532.33589 | 8788.98954 | 9210.02294 | 974.053 | 1196.57 | 1167.02 | 3283.643625 | 3316.44418 | 5779 |
| 974.324 | 3131.55 | 3208.15 | 5082.282685 | 5220.764625 | 10486.78962 | 10550.29518 | 8826.07839 | 9226.5172 | 974.324 | 1248.3 | 1168.33 | 3358.794544 | 3319.833534 | 5716.84 |
| 974.595 | 3238.64 | 3211.67 | 5070.064355 | 5226.721202 | 10602.67636 | 10568.35949 | 8941.51498 | 9243.02308 | 974.595 | 1201.84 | 1169.64 | 3204.474058 | 3323.222888 | 5870.3 |
| 974.866 | 3122.37 | 3215.19 | 5108.267828 | 5232.677779 | 10723.2648 | 10586.42381 | 8832.83871 | 9259.55218 | 974.866 | 1174.16 | 1170.95 | 3261.750827 | 3326.612242 | 5811.03 |
| 975.136 | 3232.34 | 3218.71 | 5267.558228 | 5238.634356 | 10595.24704 | 10604.48812 | 9115.47134 | 9276.09291 | 975.136 | 1233.38 | 1172.26 | 3311.354629 | 3330.001596 | 5665.45 |
| 975.407 | 3149.33 | 3222.23 | 5115.106024 | 5244.590933 | 10592.05244 | 10622.65746 | 9244.69573 | 9292.65686 | 975.407 | 1257.61 | 1173.58 | 3164.320702 | 3333.39095 | 5829.77 |
| 975.678 | 3109 | 3225.75 | 5214.152485 | 5250.54751 | 10454.38718 | 10640.72178 | 9090.6951 | 9309.23243 | 975.678 | 1175.34 | 1174.89 | 3256.440471 | 3336.780304 | 5797.78 |
| 975.949 | 3222.97 | 3229.27 | 5141.034654 | 5256.51539 | 10637.29698 | 10658.89112 | 9025.67055 | 9325.81961 | 975.949 | 1223.39 | 1176.2 | 3344.265587 | 3340.169658 | 5894.56 |
| 976.219 | 3224.9 | 3232.8 | 5145.386233 | 5262.471967 | 10734.83331 | 10677.06046 | 8886.05013 | 9342.41842 | 976.219 | 1151.43 | 1177.51 | 3261.110493 | 3343.559012 | 5790.1 |
| 976.49 | 3159.71 | 3236.32 | 5170.003737 | 5268.439847 | 10563.8741 | 10695.2298 | 9002.28819 | 9359.04045 | 976.49 | 1189.41 | 1178.82 | 3295.412521 | 3346.948365 | 5842.29 |
| 976.761 | 3254.6 | 3239.84 | 5079.163111 | 5274.407727 | 10754.0434 | 10713.39914 | 9014.53112 | 9375.6741 | 976.761 | 1222.79 | 1180.13 | 3362.989835 | 3350.326679 | 5815.26 |
| 977.031 | 3184.11 | 3243.36 | 5224.539761 | 5280.375607 | 10630.71673 | 10731.56848 | 8813.38246 | 9392.33097 | 977.031 | 1249.74 | 1181.44 | 3254.762355 | 3353.716033 | 5872.38 |
| 977.302 | 3316.56 | 3246.89 | 5226.257787 | 5286.343487 | 10762.42792 | 10749.84284 | 9559.42247 | 9408.99947 | 977.302 | 1205.65 | 1182.76 | 3306.397561 | 3357.105387 | 5797.21 |
| 977.573 | 3198.97 | 3250.41 | 5369.814686 | 5292.311367 | 10765.29351 | 10768.01218 | 9051.60835 | 9425.67958 | 977.573 | 1258.68 | 1184.07 | 3333.39095 | 3360.494741 | 5740.53 |
| 977.843 | 3166.84 | 3253.94 | 5232.090033 | 5298.279247 | 10987.64237 | 10786.28655 | 9233.90477 | 9442.38292 | 977.843 | 1237.57 | 1185.38 | 3290.863942 | 3363.884095 | 5864.82 |
| 978.114 | 3161.79 | 3257.46 | 5255.825919 | 5304.247126 | 10830.88377 | 10804.56091 | 9281.26188 | 9459.09788 | 978.114 | 1264.89 | 1186.69 | 3345.060484 | 3367.273449 | 5780.17 |
| 978.385 | 3164.96 | 3260.99 | 5095.778231 | 5310.215006 | 10707.76936 | 10822.83528 | 9052.64215 | 9475.82445 | 978.385 | 1226.74 | 1188.01 | 3337.343356 | 3370.662803 | 5893.2 |
| 978.655 | 3194.22 | 3264.51 | 5142.515321 | 5316.182886 | 11142.49058 | 10841.10964 | 9314.61049 | 9492.56264 | 978.655 | 1209.19 | 1189.32 | 3262.534684 | 3374.041116 | 5837.66 |
| 978.926 | 3254.47 | 3268.04 | 5169.382083 | 5322.162069 | 10705.54057 | 10859.38401 | 9153.51268 | 9509.32406 | 978.926 | 1227.89 | 1190.63 | 3288.159083 | 3377.43047 | 5932.31 |
| 979.196 | 3172.55 | 3271.57 | 5250.875291 | 5328.129949 | 10810.40008 | 10877.65837 | 9061.26098 | 9526.0971 | 979.196 | 1204.46 | 1191.95 | 3389.640977 | 3380.819824 | 5821.77 |
| 979.466 | 3216.62 | 3275.09 | 5275.594521 | 5334.109131 | 10837.57016 | 10896.03776 | 9333.85766 | 9542.89337 | 979.466 | 1208.1 | 1193.26 | 3346.462595 | 3384.209178 | 5777.46 |
| 979.737 | 3214.7 | 3278.62 | 5167.098917 | 5340.088314 | 11154.37749 | 10914.41716 | 9381.54001 | 9559.70125 | 979.737 | 1229.14 | 1194.58 | 3316.676025 | 3387.598532 | 5811.6 |
| 1101.15 | 4866.28 | 4876.14 | 7758.526429 | 8086.499859 | 20408.33553 | 20602.87646 | 18888.59902 | 18640.37202 | 1101.15 | 1669.81 | 1795.08 | 4430.702565 | 4879.190268 | 8146.69 |
| 1102.87 | 4880.78 | 4898.51 | 7550.679188 | 8125.551042 | 20258.26331 | 20759.57389 | 18739.68605 | 18790.3304 | 1102.87 | 1694.8 | 1803.57 | 4536.876561 | 4899.44911 | 8257.26 |
| 1104.6 | 4873.99 | 4920.85 | 8002.000105 | 8164.568318 | 20478.91404 | 20916.69142 | 18844.45945 | 18940.7534 | 1104.6 | 1686 | 1812.05 | 4541.623864 | 4919.66379 | 8189.23 |
| 1106.32 | 4969.28 | 4943.16 | 8329.408394 | 8203.551685 | 20613.27856 | 21074.22906 | 18808.56701 | 19091.75719 | 1106.32 | 1828.56 | 1820.52 | 4691.373691 | 4939.845351 | 8290.31 |
| 1108.05 | 5021.36 | 4965.44 | 8463.685692 | 8242.501143 | 21169.20378 | 21232.08176 | 19235.55926 | 19243.2256 | 1108.05 | 1948.65 | 1828.98 | 4884.191497 | 4959.960669 | 8363.82 |
| 1109.77 | 5217.64 | 4987.7 | 8298.144841 | 8281.416693 | 21613.47698 | 21390.45959 | 19533.6175 | 19395.15864 | 1109.77 | 1951.98 | 1837.44 | 4985.905235 | 4980.042867 | 8558.76 |
| 1111.5 | 5096.63 | 5009.92 | 7883.998843 | 8320.298335 | 21360.77405 | 21549.25753 | 19218.71653 | 19547.55631 | 1111.5 | 1915.01 | 1845.88 | 5144.575576 | 5000.080905 | 8567.96 |
| 1113.22 | 5031.43 | 5032.11 | 7853.074375 | 8359.134765 | 21498.21642 | 21708.37053 | 20167.60109 | 19700.4186 | 1113.22 | 1813.62 | 1854.32 | 4762.318278 | 5020.06374 | 8688.58 |
| 1114.95 | 5180.74 | 5054.26 | 8043.187518 | 8397.925985 | 21708.35998 | 21867.90364 | 19907.75842 | 19853.74553 | 1114.95 | 1729.93 | 1862.74 | 4982.990612 | 5040.013456 | 8884.42 |
| 1116.67 | 5152.4 | 5076.39 | 8294.821817 | 8436.683296 | 21836.88717 | 22027.75183 | 19822.61551 | 20007.65323 | 1116.67 | 1640.76 | 1871.16 | 5308.335468 | 5059.90797 | 8471.5 |
| 1118.39 | 4982.83 | 5098.47 | 8568.440068 | 8475.395395 | 22351.10208 | 22188.12514 | 20057.60064 | 20161.90941 | 1118.39 | 1644.71 | 1879.56 | 5018.606429 | 5079.747282 | 8980.22 |
| 1120.12 | 5225.36 | 5120.53 | 8480.244298 | 8514.062284 | 22291.98594 | 22348.81352 | 19997.54759 | 20316.63021 | 1120.12 | 1780.7 | 1887.96 | 5109.412409 | 5099.542434 | 8670.97 |
| 1121.84 | 5232.61 | 5142.55 | 8217.634977 | 8552.695264 | 22451.82241 | 22509.81698 | 20707.38155 | 20471.81564 | 1121.84 | 1937.29 | 1896.34 | 5113.000488 | 5119.282385 | 8480.67 |
| 1123.57 | 5134.93 | 5164.53 | 8264.598123 | 8591.27173 | 22832.20347 | 22671.24054 | 20792.52445 | 20627.4657 | 1123.57 | 1950.39 | 1904.71 | 5255.408554 | 5138.978174 | 8947.21 |
| 1125.29 | 5458.55 | 5186.48 | 8477.475111 | 8629.814288 | 22692.63843 | 22833.0842 | 20942.13436 | 20783.58038 | 1125.29 | 1820.62 | 1913.07 | 5168.058169 | 5158.629803 | 8757.36 |
| 1127.01 | 5517.6 | 5208.39 | 8628.966577 | 8668.300331 | 22952.55841 | 22995.24293 | 20997.77344 | 20940.15969 | 1127.01 | 2017.99 | 1921.42 | 4848.542559 | 5178.22623 | 8900.19 |
| 1128.74 | 5216.71 | 5230.27 | 8745.306327 | 8706.741164 | 23256.84204 | 23157.71674 | 20955.84085 | 21097.20363 | 1128.74 | 1985.66 | 1929.76 | 5077.495073 | 5197.767457 | 9055.9 |
| 1130.46 | 4891 | 5252.11 | 8847.189791 | 8745.136785 | 23420.92382 | 23320.61065 | 20872.67261 | 21254.59604 | 1130.46 | 2014.4 | 1938.08 | 5235.999811 | 5217.264522 | 8927.06 |
| 1132.18 | 5047.61 | 5273.9 | 8751.20639 | 8783.475892 | 23032.68891 | 23483.81964 | 21347.75376 | 21412.45307 | 1132.18 | 1780.47 | 1946.4 | 5221.117566 | 5236.695346 | 8796.38 |
| 1133.91 | 5119.42 | 5295.66 | 8720.417555 | 8821.769788 | 23260.13216 | 23647.44873 | 21682.05291 | 21570.77473 | 1133.91 | 1801.11 | 1954.7 | 5215.133755 | 5256.093049 | 8971.53 |
| 1135.63 | 5391.06 | 5317.39 | 8462.544109 | 8860.00717 | 24148.67855 | 23811.28786 | 21369.9397 | 21729.44487 | 1135.63 | 1916.22 | 1962.99 | 5209.635715 | 5275.42451 | 8869.8 |
| 1137.35 | 5303.1 | 5339.07 | 8845.867364 | 8898.199341 | 23693.68591 | 23975.5471 | 21073.97228 | 21888.69578 | 1137.35 | 1988.76 | 1971.26 | 5287.944144 | 5294.700771 | 8849.89 |
| 1139.07 | 5167.03 | 5360.71 | 8891.711532 | 8936.334997 | 23866.57675 | 24140.12141 | 21292.11467 | 22048.17901 | 1139.07 | 2209.89 | 1979.52 | 5224.849167 | 5313.93287 | 8920 |
| 1140.8 | 5287.95 | 5382.31 | 8683.163517 | 8974.41414 | 24435.45021 | 24305.0108 | 22142.84677 | 22208.24303 | 1140.8 | 1899 | 1987.77 | 5407.587233 | 5333.109769 | 9137.27 |
| 1142.52 | 5416.05 | 5403.87 | 9127.906202 | 9012.436769 | 24528.31668 | 24470.21526 | 22457.51543 | 22368.53935 | 1142.52 | 1810.32 | 1996.01 | 5479.967052 | 5352.220426 | 9085.53 |
| 1144.24 | 5627.22 | 5425.39 | 8903.11606 | 9050.402883 | 24938.73341 | 24635.7348 | 22430.10244 | 22529.41647 | 1144.24 | 1873.17 | 2004.23 | 5455.755804 | 5371.286921 | 9072.69 |
| 1145.96 | 5420.24 | 5446.86 | 8901.62409 | 9088.312484 | 24731.03092 | 24801.56942 | 22430.33475 | 22690.64205 | 1145.96 | 1895.77 | 2012.44 | 5173.666611 | 5390.298216 | 9067.06 |
| 1147.69 | 5492.14 | 5468.3 | 9069.809796 | 9126.165571 | 24656.73774 | 24967.7191 | 22520.24008 | 22852.2161 | 1147.69 | 2197.96 | 2020.64 | 5141.208303 | 5409.25431 | 9135.9 |
| 1149.41 | 5531.63 | 5489.68 | 8962.500986 | 9163.962143 | 24943.5094 | 25134.18387 | 22833.28255 | 23014.13862 | 1149.41 | 2067.39 | 2028.82 | 5469.445703 | 5428.144162 | 9293.2 |
| 1151.13 | 5311.74 | 5511.03 | 8963.40521 | 9201.690899 | 24809.56942 | 25300.85868 | 23090.45361 | 23176.52577 | 1151.13 | 1851.7 | 2036.98 | 5363.967242 | 5446.989853 | 9050.59 |
| 1152.85 | 5635.86 | 5532.33 | 8996.646753 | 9239.363141 | 25895.62953 | 25467.9536 | 23172.22797 | 23339.37754 | 1152.85 | 2070.59 | 2045.14 | 5520.915305 | 5465.769302 | 9306.35 |
| 1154.57 | 5620.98 | 5553.59 | 9015.285075 | 9276.978869 | 25762.53861 | 25635.25856 | 23651.25844 | 23502.46163 | 1154.57 | 1808.28 | 2053.28 | 5386.35685 | 5484.493551 | 9540.75 |
| 1156.3 | 5566.63 | 5574.8 | 9162.673624 | 9314.52678 | 25784.50816 | 25802.8786 | 23884.73353 | 23666.01035 | 1156.3 | 2015.76 | 2061.4 | 5374.543793 | 5503.162598 | 9493.03 |
| 1158.02 | 5542.79 | 5595.97 | 9352.131205 | 9352.018177 | 26289.27723 | 25970.81372 | 24112.0523 | 23829.90754 | 1158.02 | 2103.43 | 2069.51 | 5519.391752 | 5521.776444 | 9594.82 |
| 1159.74 | 5432.2 | 5617.09 | 9273.576725 | 9389.430454 | 26310.82225 | 26138.95888 | 24605.2539 | 23994.15319 | 1159.74 | 1926.29 | 2077.6 | 5486.469754 | 5540.324049 | 9536.83 |
| 1161.46 | 5684.28 | 5638.16 | 9081.982915 | 9426.79752 | 26403.05192 | 26307.41912 | 24176.05468 | 24158.86348 | 1161.46 | 2060.63 | 2085.68 | 5331.729739 | 5558.816452 | 9481.14 |
| 1163.18 | 5557.4 | 5659.19 | 9383.812961 | 9464.085467 | 26171.36334 | 26476.08941 | 23946.76124 | 24323.80607 | 1163.18 | 2143.15 | 2093.74 | 5300.331294 | 5577.242614 | 9467.49 |
| 1164.9 | 5451.56 | 5680.17 | 9590.823795 | 9501.305597 | 26744.48213 | 26645.07478 | 24556.23575 | 24489.2133 | 1164.9 | 2042.14 | 2101.78 | 5137.719587 | 5595.624615 | 9443.76 |
| 1166.62 | 5637.79 | 5701.1 | 9432.188957 | 9538.457909 | 26729.6235 | 26814.27019 | 25055.24518 | 24654.96899 | 1166.62 | 2022.43 | 2109.81 | 5433.907167 | 5613.940374 | 9558.3 |
| 1168.34 | 5865.48 | 5721.98 | 9531.291932 | 9575.553708 | 26734.82402 | 26983.78069 | 24620.35428 | 24820.957 | 1168.34 | 2131.55 | 2117.83 | 5500.137572 | 5632.189892 | 9408.66 |
| 1170.06 | 5815.3 | 5742.82 | 9263.302477 | 9612.570388 | 27041.01804 | 27153.50123 | 24699.10856 | 24987.40964 | 1170.06 | 2101.49 | 2125.82 | 5596.121426 | 5650.384209 | 9381.49 |
| 1171.78 | 5400.08 | 5763.6 | 9090.267869 | 9649.51925 | 27064.89799 | 27323.43182 | 24971.38001 | 25154.09459 | 1171.78 | 2175.74 | 2133.81 | 5949.983224 | 5668.512285 | 9562.81 |
| 1173.5 | 5452.09 | 5784.34 | 9414.488768 | 9686.400296 | 27179.84014 | 27493.67749 | 25028.18066 | 25321.24416 | 1173.5 | 2031.89 | 2141.77 | 5687.633982 | 5686.585159 | 9685.66 |
| 1175.22 | 5877.57 | 5805.02 | 9639.753628 | 9723.213524 | 27811.01374 | 27664.02818 | 25370.95926 | 25488.62605 | 1175.22 | 2139.49 | 2149.72 | 5752.19731 | 5704.602832 | 9713.17 |
| 1176.94 | 6104.79 | 5825.66 | 9714.815542 | 9759.947634 | 27721.4374 | 27834.69395 | 25409.75561 | 25656.35641 | 1176.94 | 2196.4 | 2157.65 | 5584.84934 | 5722.543223 | 9623.3 |
| 1178.66 | 5802.4 | 5846.25 | 9604.71494 | 9796.602623 | 27706.79103 | 28005.67479 | 25452.38514 | 25824.43524 | 1178.66 | 1968.36 | 2165.56 | 5531.337293 | 5740.428413 | 9719.61 |
| 1180.38 | 5816.4 | 5866.78 | 9879.463471 | 9833.189796 | 28175.47484 | 28176.76066 | 25529.16476 | 25992.86254 | 1180.38 | 1949.11 | 2173.46 | 5622.90505 | 5758.258402 | 9826.35 |
| 1182.11 | 5991.5 | 5887.26 | 9790.815588 | 9869.709152 | 28016.06291 | 28348.05657 | 26309.73818 | 26161.63831 | 1182.11 | 2126.76 | 2181.34 | 5369.012632 | 5776.01111 | 9764.66 |
| 1183.82 | 5687.86 | 5907.69 | 9732.481823 | 9906.149388 | 28435.07642 | 28519.56254 | 26580.49958 | 26330.6464 | 1183.82 | 2173.01 | 2189.2 | 5661.170525 | 5793.708616 | 9671.47 |
| 1185.54 | 5750.51 | 5928.07 | 9856.055366 | 9942.510505 | 28410.55968 | 28691.38359 | 25938.26887 | 26499.88679 | 1185.54 | 2148.19 | 2197.05 | 5820.271435 | 5811.35092 | 9874.12 |
| 1187.26 | 5857.88 | 5948.4 | 9660.211701 | 9978.792502 | 28437.94202 | 28863.30965 | 26419.73863 | 26669.59182 | 1187.26 | 2286.08 | 2204.87 | 5634.243377 | 5828.915944 | 9943.06 |
| 1188.98 | 6111.68 | 5968.67 | 10007.65986 | 10015.00668 | 29168.45621 | 29035.44577 | 26546.93028 | 26839.52916 | 1188.98 | 2332.65 | 2212.68 | 5881.02588 | 5846.425765 | 9992.69 |
| 1190.7 | 5894.06 | 5988.89 | 10240.26024 | 10051.14174 | 29111.0382 | 29207.79194 | 26779.59227 | 27009.69881 | 1190.7 | 2117.42 | 2220.47 | 5961.078666 | 5863.869346 | 10018.4 |
| 1192.42 | 6031.42 | 6009.05 | 9920.933454 | 10087.18638 | 29111.56886 | 29380.34816 | 27202.63518 | 27180.21693 | 1192.42 | 2073.05 | 2228.24 | 6009.909651 | 5881.246685 | 9898.69 |
| 1194.14 | 6151.33 | 6029.16 | 10067.54211 | 10123.1632 | 29414.26049 | 29553.0094 | 27311.24175 | 27351.08352 | 1194.14 | 2245.96 | 2236 | 5970.131664 | 5898.557783 | 10045.7 |
| 1195.86 | 5806.68 | 6049.22 | 10192.59632 | 10159.0496 | 29269.17653 | 29725.8807 | 27302.06537 | 27522.18243 | 1195.86 | 2164.59 | 2243.73 | 5974.614002 | 5915.813679 | 9735.51 |
| 1197.58 | 5966.14 | 6069.22 | 9912.28681 | 10194.86818 | 30093.61852 | 29898.96204 | 27339.81632 | 27693.51365 | 1197.58 | 2363.01 | 2251.45 | 6091.485991 | 5933.003334 | 9834.95 |
| 1199.3 | 5997.6 | 6089.17 | 10153.53383 | 10230.59634 | 29904.27705 | 30072.14841 | 27610.34541 | 27865.19334 | 1199.3 | 2314.71 | 2259.15 | 5821.254016 | 5950.115707 | 9971.61 |
| 1201.02 | 6067.53 | 6109.06 | 10380.81059 | 10266.24538 | 29778.40318 | 30245.64986 | 28298.5742 | 28037.10534 | 1201.02 | 2353.93 | 2266.83 | 5934.427525 | 5967.172879 | 9852.97 |
| 1202.73 | 6162.9 | 6128.89 | 9992.480196 | 10301.8153 | 30231.27316 | 30419.1513 | 28354.79408 | 28209.24966 | 1202.73 | 2416.91 | 2274.49 | 5905.170887 | 5984.15277 | 10062.4 |
| 1204.45 | 5957.91 | 6148.67 | 10175.14479 | 10337.2948 | 30339.4228 | 30592.8628 | 28041.40314 | 28381.74244 | 1204.45 | 2185.95 | 2282.13 | 5801.072456 | 6001.077459 | 10111.5 |
| 1206.17 | 5991.89 | 6168.39 | 10442.33175 | 10372.69518 | 31012.51898 | 30766.78434 | 28002.026 | 28554.46754 | 1206.17 | 2288.77 | 2289.75 | 6076.062774 | 6017.935907 | 10202.8 |
| 1207.89 | 6007.44 | 6188.06 | 10012.14707 | 10408.00514 | 30882.61205 | 30940.81091 | 28181.95282 | 28727.42496 | 1207.89 | 2272.99 | 2297.35 | 5938.987144 | 6034.717073 | 10270.6 |
| 1209.61 | 6113.93 | 6207.66 | 10317.23233 | 10443.23597 | 30916.89304 | 31114.94251 | 28712.90536 | 28900.61468 | 1209.61 | 2054.99 | 2304.93 | 5922.2943 | 6051.443037 | 10193.5 |
| 1211.32 | 6198.92 | 6227.21 | 10450.63931 | 10478.37639 | 31215.97615 | 31289.17913 | 29354.20683 | 29074.03673 | 1211.32 | 2091.37 | 2312.49 | 5982.585055 | 6068.102761 | 10184.5 |
| 1213.04 | 6164.77 | 6246.71 | 10508.87135 | 10513.42638 | 31302.89916 | 31463.6258 | 29803.26886 | 29247.69108 | 1213.04 | 2321.97 | 2320.03 | 5974.205513 | 6084.685203 | 10237.1 |
| 1214.76 | 6389.45 | 6266.14 | 10638.1415 | 10548.39725 | 31526.73389 | 31638.17749 | 29575.48545 | 29421.6939 | 1214.76 | 2269.91 | 2327.56 | 5840.607559 | 6101.201403 | 10330.4 |
| 1216.48 | 6623.35 | 6285.52 | 10534.83389 | 10583.2777 | 32212.14151 | 31812.93924 | 29594.30285 | 29595.81289 | 1216.48 | 2458.77 | 2335.06 | 5787.526081 | 6117.651362 | 10312.5 |
| 1218.19 | 6463.2 | 6304.83 | 10648.30272 | 10618.06773 | 32261.06887 | 31987.70099 | 30117.47289 | 29770.28034 | 1218.19 | 2389.64 | 2342.54 | 6032.332381 | 6134.046119 | 10284.2 |
| 1219.91 | 6351.81 | 6324.09 | 10656.87025 | 10652.76733 | 32295.88053 | 32162.67278 | 30110.03886 | 29944.86395 | 1219.91 | 2341.7 | 2350 | 6034.672912 | 6150.352555 | 10381.5 |
| 1221.63 | 6344.24 | 6343.29 | 10675.79114 | 10687.37651 | 32646.33206 | 32337.64458 | 30303.3237 | 30119.67987 | 1221.63 | 2517.63 | 2357.44 | 6239.281695 | 6166.603789 | 10469.2 |
| 1223.35 | 6260.95 | 6362.42 | 10788.84177 | 10721.90658 | 32430.88184 | 32512.82642 | 30398.22378 | 30294.7281 | 1223.35 | 2250.72 | 2364.86 | 6339.880369 | 6182.788782 | 10433.2 |
| 1225.06 | 6583.49 | 6381.5 | 11094.71822 | 10756.33491 | 32907.95017 | 32688.11329 | 30726.36662 | 30470.00865 | 1225.06 | 2240.82 | 2372.26 | 5920.947391 | 6198.896494 | 10243.2 |
| 1226.78 | 6408.96 | 6400.52 | 10941.29398 | 10790.67283 | 33211.59699 | 32863.40017 | 30951.01379 | 30645.52152 | 1226.78 | 2281.86 | 2379.64 | 6218.90141 | 6214.937964 | 10304.5 |
| 1228.5 | 6372.8 | 6419.48 | 10787.65498 | 10824.92032 | 33366.97587 | 33038.89709 | 31015.94542 | 30821.15053 | 1228.5 | 2210.61 | 2386.99 | 6171.649179 | 6230.913192 | 10458.9 |
| 1230.21 | 6400.55 | 6438.37 | 10897.26956 | 10859.07739 | 33634.1129 | 33214.39401 | 31402.97974 | 30997.01187 | 1230.21 | 2271.04 | 2394.33 | 6029.605441 | 6246.822179 | 10402.4 |
| 1231.93 | 6502.48 | 6457.21 | 10774.14813 | 10893.14404 | 33437.12985 | 33389.99596 | 31473.25458 | 31173.10551 | 1231.93 | 2417.07 | 2401.65 | 6285.705907 | 6262.653884 | 10652 |
| 1233.64 | 6537.51 | 6475.98 | 10784.47889 | 10927.10896 | 33367.93106 | 33565.70293 | 31404.37362 | 31349.43147 | 1233.64 | 2403.49 | 2408.94 | 6278.794717 | 6278.419348 | 10781.6 |
| 1235.36 | 6365.75 | 6494.69 | 11007.53971 | 10960.98346 | 33738.65401 | 33741.51492 | 31489.51652 | 31525.87359 | 1235.36 | 2352.93 | 2416.21 | 6347.034445 | 6294.107531 | 10674.2 |
| 1237.08 | 6593.58 | 6513.34 | 11105.63673 | 10994.76754 | 33802.65228 | 33917.32692 | 31626.11687 | 31702.54802 | 1237.08 | 2296.04 | 2423.46 | 6414.832564 | 6309.740512 | 10560.8 |
| 1238.79 | 6792.11 | 6531.93 | 11006.97457 | 11028.44989 | 34171.14643 | 34093.24394 | 31884.68182 | 31879.33861 | 1238.79 | 2257.45 | 2430.69 | 6300.135502 | 6325.296211 | 10511.3 |
| 1240.51 | 6513.86 | 6550.46 | 10936.27553 | 11062.04182 | 34530.61927 | 34269.26599 | 32076.45663 | 32056.36151 | 1240.51 | 2407.26 | 2437.9 | 6103.000962 | 6340.785669 | 10581.2 |
| 1242.22 | 6709.44 | 6568.92 | 11069.84076 | 11095.54333 | 34594.29914 | 34445.28804 | 32257.08039 | 32233.50056 | 1242.22 | 2339.61 | 2445.08 | 6374.624007 | 6356.197845 | 10725.6 |
| 1243.94 | 6804.4 | 6587.32 | 10992.84606 | 11128.93181 | 34527.96594 | 34621.31008 | 32108.98052 | 32410.87194 | 1243.94 | 2443.68 | 2452.25 | 6423.962843 | 6371.54378 | 10912.5 |
| 1245.66 | 6519.45 | 6605.66 | 11220.11152 | 11162.24117 | 35172.83071 | 34797.54218 | 32574.65297 | 32588.35946 | 1245.66 | 2640.43 | 2459.39 | 6387.717732 | 6386.812433 | 10598.6 |
| 1247.37 | 6639.89 | 6623.93 | 11184.63202 | 11195.4375 | 35427.44404 | 34973.66925 | 32800.92633 | 32765.96315 | 1247.37 | 2534.51 | 2466.51 | 6433.203525 | 6402.025885 | 10584.4 |
| 1249.09 | 6641.56 | 6642.14 | 10864.7514 | 11228.54341 | 35256.25134 | 35149.90135 | 32932.18347 | 32943.79914 | 1249.09 | 2629.62 | 2473.6 | 6491.407674 | 6417.151014 | 10868.2 |
| 1250.8 | 6614.16 | 6660.29 | 11380.90519 | 11261.54759 | 35672.71766 | 35326.23847 | 33421.66805 | 33121.7513 | 1250.8 | 2491.22 | 2480.68 | 6559.97861 | 6432.220943 | 10788.2 |
| 1252.52 | 6653.1 | 6678.37 | 11297.27575 | 11294.46136 | 35724.29835 | 35502.47057 | 33129.41765 | 33299.81961 | 1252.52 | 2397.3 | 2487.73 | 6622.256608 | 6447.20255 | 10563.7 |
| 1254.23 | 6695.88 | 6696.39 | 11345.41439 | 11327.21688 | 35743.93297 | 35678.80769 | 33293.89562 | 33478.12024 | 1254.23 | 2535.34 | 2494.76 | 6580.458257 | 6462.128956 | 10839.7 |
| 1255.95 | 6822.01 | 6714.34 | 11640.19149 | 11359.99501 | 35993.87645 | 35855.24984 | 33615.30137 | 33656.42086 | 1255.95 | 2477.16 | 2501.77 | 6549.368938 | 6476.978079 | 11096.4 |
| 1257.66 | 6768.29 | 6732.23 | 11507.60961 | 11392.54708 | 35985.0674 | 36031.58696 | 33839.48391 | 33834.9538 | 1257.66 | 2474.3 | 2508.75 | 6619.452387 | 6491.760962 | 10906 |
| 1259.38 | 6749.86 | 6750.05 | 11531.11944 | 11425.09915 | 36120.06871 | 36208.02911 | 34268.33467 | 34013.60289 | 1259.38 | 2505.13 | 2515.71 | 6626.992872 | 6506.466563 | 11066.4 |
| 1261.09 | 6601.86 | 6767.81 | 11281.93785 | 11457.5382 | 36528.68118 | 36384.47125 | 34875.48604 | 34192.36815 | 1261.09 | 2490.91 | 2522.65 | 6572.542404 | 6521.094882 | 11067.5 |
| 1262.81 | 6581.69 | 6785.5 | 11361.91648 | 11489.86421 | 36785.62944 | 36560.80838 | 34626.67827 | 34371.24955 | 1262.81 | 2547.32 | 2529.57 | 6597.360865 | 6535.656959 | 10675.4 |
| 1264.52 | 6791.89 | 6803.13 | 11163.22451 | 11521.96417 | 37325.74083 | 36737.25052 | 34266.01153 | 34550.24712 | 1264.52 | 2538.72 | 2536.46 | 6619.849836 | 6550.152796 | 10883.3 |
| 1266.23 | 6709.32 | 6820.69 | 11373.55837 | 11554.06413 | 37405.1284 | 36913.69267 | 34983.97646 | 34729.477 | 1266.23 | 2503.63 | 2543.33 | 6622.079964 | 6564.57135 | 11105.5 |
| 1267.95 | 6876.31 | 6838.19 | 11392.88616 | 11586.05106 | 37332.32108 | 37090.13482 | 34953.65954 | 34908.70688 | 1267.95 | 2498.47 | 2550.17 | 6526.460439 | 6578.923663 | 11076.5 |
| 1269.66 | 6827.93 | 6855.62 | 11456.40792 | 11618.038 | 37857.04317 | 37266.57696 | 34818.45307 | 35088.05291 | 1269.66 | 2400.01 | 2556.99 | 6574.67317 | 6593.198695 | 11258.3 |
| 1271.37 | 6710.35 | 6872.98 | 11616.11652 | 11649.79887 | 38096.5856 | 37442.91409 | 35290.04952 | 35267.39895 | 1271.37 | 2531.69 | 2563.79 | 6667.068946 | 6607.396444 | 11082.6 |
| 1273.09 | 6879.31 | 6890.28 | 11683.59425 | 11681.44672 | 38042.98838 | 37619.35623 | 35906.60959 | 35446.9773 | 1273.09 | 2692.41 | 2570.57 | 6540.371142 | 6621.527952 | 11231.3 |
| 1274.8 | 6805.26 | 6907.51 | 11987.63965 | 11713.09457 | 38062.62301 | 37795.69336 | 36228.36381 | 35626.6718 | 1274.8 | 2517.64 | 2577.32 | 6543.694255 | 6635.593219 | 11270.4 |
| 1276.51 | 6914 | 6924.68 | 11826.34865 | 11744.51636 | 38268.20284 | 37972.1355 | 36180.0426 | 35806.36631 | 1276.51 | 2639.29 | 2584.05 | 6693.565524 | 6649.581204 | 11025.6 |
| 1278.23 | 7084.37 | 6941.78 | 12006.4023 | 11775.93815 | 38780.29509 | 38148.47262 | 36611.56496 | 35986.17697 | 1278.23 | 2659.02 | 2590.76 | 6876.877682 | 6663.491908 | 11234.1 |
| 1279.94 | 7054.1 | 6958.81 | 12007.87166 | 11807.13389 | 39123.52956 | 38324.70472 | 36867.57446 | 36166.10379 | 1279.94 | 2610.94 | 2597.44 | 6807.07024 | 6677.33637 | 11134.5 |
| 1281.65 | 7139.19 | 6975.77 | 12012.50581 | 11838.32962 | 38901.28683 | 38500.93682 | 36604.36324 | 36346.03061 | 1281.65 | 2657.37 | 2604.09 | 6705.654587 | 6691.10355 | 11227.2 |
| 1283.37 | 7240.42 | 6992.66 | 11952.26188 | 11869.41233 | 39158.87189 | 38677.16892 | 37004.9879 | 36526.07359 | 1283.37 | 2507.66 | 2610.73 | 6896.606593 | 6704.804489 | 11430 |
| 1285.08 | 7018.43 | 7009.49 | 11819.34091 | 11900.38201 | 39536.81189 | 38853.40101 | 37199.08585 | 36706.23272 | 1285.08 | 2550.87 | 2617.34 | 6836.923051 | 6718.428146 | 11239.8 |
| 1286.79 | 7131.81 | 7026.25 | 11939.37668 | 11931.23866 | 39138.17593 | 39029.52808 | 37050.52135 | 36886.39185 | 1286.79 | 2802.26 | 2623.92 | 6743.688217 | 6731.974522 | 11456.4 |
| 1288.5 | 7144.87 | 7042.95 | 11835.39089 | 11961.86926 | 39354.58134 | 39205.55013 | 37354.03896 | 37066.66714 | 1288.5 | 2797.53 | 2630.48 | 6832.495914 | 6745.454656 | 11493.5 |
| 1290.22 | 6995.14 | 7059.57 | 12044.83183 | 11992.49985 | 39751.83757 | 39381.57218 | 37249.14941 | 37246.94243 | 1290.22 | 2743.53 | 2637.02 | 6550.285278 | 6758.857508 | 11449.7 |
| 1291.93 | 7255.6 | 7076.13 | 12075.57545 | 12023.01742 | 39642.20207 | 39557.59423 | 37599.94282 | 37427.33388 | 1291.93 | 2736.49 | 2643.53 | 6687.305707 | 6772.194119 | 11439.3 |
| 1293.64 | 7134 | 7092.62 | 11806.7948 | 12053.53499 | 40201.31128 | 39733.40622 | 38209.30118 | 37607.72532 | 1293.64 | 2717.27 | 2650.02 | 6956.411577 | 6785.453448 | 11347.1 |
| 1295.35 | 7158.47 | 7109.04 | 12250.54284 | 12083.8265 | 40451.89155 | 39909.32324 | 37781.37968 | 37788.11677 | 1295.35 | 2682.67 | 2656.49 | 6899.775142 | 6798.646536 | 11452.3 |
| 1297.07 | 7146.54 | 7125.39 | 12377.69937 | 12114.00498 | 40759.04077 | 40085.03022 | 38120.90588 | 37968.62437 | 1297.07 | 2653.3 | 2662.93 | 6947.513143 | 6811.762342 | 11478.3 |
| 1298.78 | 7096.39 | 7141.67 | 12130.84616 | 12144.07044 | 40893.40529 | 40260.73719 | 38576.58885 | 38149.24813 | 1298.78 | 2752.68 | 2669.34 | 6890.501339 | 6824.800866 | 11463.3 |
| 1300.49 | 7241.34 | 7157.89 | 12482.58938 | 12174.02287 | 41247.35921 | 40436.33913 | 38600.16867 | 38329.75574 | 1300.49 | 2622.82 | 2675.73 | 6946.243515 | 6837.773149 | 11595.5 |
| 1302.2 | 7161.83 | 7174.03 | 12621.3878 | 12203.86227 | 41445.61585 | 40611.94108 | 38494.81449 | 38510.3795 | 1302.2 | 2696.57 | 2682.1 | 6982.65423 | 6850.668151 | 11459.4 |
| 1303.91 | 7157.63 | 7190.11 | 12415.45073 | 12233.70167 | 41379.17653 | 40787.33298 | 39081.87075 | 38691.00326 | 1303.91 | 2821.55 | 2688.44 | 6991.287698 | 6863.48587 | 11492.4 |
| 1305.62 | 7264.63 | 7206.11 | 12352.60715 | 12263.31501 | 41391.38184 | 40962.72487 | 39445.20909 | 38871.62702 | 1305.62 | 2793.65 | 2694.76 | 6954.126247 | 6876.237348 | 11751 |
| 1307.33 | 7289.1 | 7222.05 | 12296.20617 | 12292.81532 | 41695.66546 | 41137.90672 | 39394.21627 | 39052.36693 | 1307.33 | 2876.19 | 2701.05 | 7054.19499 | 6888.911545 | 11635.1 |
| 1309.04 | 7328.06 | 7237.92 | 12428.67501 | 12322.31564 | 42070.846 | 41313.08857 | 39978.48477 | 39233.10685 | 1309.04 | 2685.89 | 2707.32 | 7049.778894 | 6901.5195 | 11621.8 |
| 1310.76 | 7251.58 | 7253.72 | 12647.83636 | 12351.5899 | 42179.10177 | 41488.16539 | 40116.479 | 39413.73061 | 1310.76 | 2649.11 | 2713.56 | 7078.04191 | 6914.050173 | 11805.1 |
| 1312.47 | 7316.48 | 7269.45 | 12619.01421 | 12380.86416 | 42744.47284 | 41663.13718 | 40275.72992 | 39594.47053 | 1312.47 | 2620.11 | 2719.78 | 7124.388841 | 6926.503565 | 11846.3 |
| 1314.18 | 7297.61 | 7285.11 | 12714.86198 | 12409.91236 | 42952.28147 | 41838.00395 | 40199.18261 | 39775.21044 | 1314.18 | 2596.47 | 2725.97 | 6987.997706 | 6938.890715 | 11873.6 |
| 1315.89 | 7183.29 | 7300.7 | 12530.06116 | 12438.84754 | 42679.62551 | 42012.7657 | 40217.65154 | 39955.95036 | 1315.89 | 2616.8 | 2732.14 | 7073.979102 | 6951.200583 | 11851 |
| 1317.6 | 7450.07 | 7316.22 | 12788.55626 | 12467.78271 | 43101.82302 | 42187.42242 | 40839.43867 | 40136.69028 | 1317.6 | 2646.51 | 2738.28 | 6936.241057 | 6963.444211 | 11840.4 |
| 1319.31 | 7469.86 | 7331.67 | 12828.79423 | 12496.49183 | 43012.03441 | 42361.86909 | 40451.47509 | 40317.43019 | 1319.31 | 2749.63 | 2744.4 | 6903.020973 | 6975.610556 | 11816 |
| 1321.02 | 7291.37 | 7347.06 | 12663.66028 | 12525.20095 | 42931.16098 | 42536.31576 | 40547.65289 | 40498.05395 | 1321.02 | 2683.07 | 2750.49 | 7115.236481 | 6987.69962 | 11566.4 |
| 1322.73 | 7291.69 | 7362.37 | 12790.25168 | 12553.68401 | 43264.73734 | 42710.55238 | 40537.54725 | 40678.79387 | 1322.73 | 2776.64 | 2756.56 | 7142.925405 | 6999.722442 | 11645.5 |
| 1324.44 | 7397.17 | 7377.61 | 12555.15338 | 12582.16708 | 43534.63384 | 42884.68398 | 41144.69863 | 40859.41763 | 1324.44 | 2671.49 | 2762.61 | 7108.137607 | 7011.667982 | 11803.5 |
| 1326.15 | 7439.29 | 7392.78 | 12704.80249 | 12610.42408 | 43466.28412 | 43058.71055 | 41447.86777 | 41040.15755 | 1326.15 | 2837.72 | 2768.62 | 7107.331669 | 7023.547282 | 11947.9 |
| 1327.86 | 7430.49 | 7407.88 | 12690.3349 | 12638.56806 | 43762.07709 | 43232.52707 | 40904.02182 | 41220.78131 | 1327.86 | 2816.79 | 2774.61 | 7061.028899 | 7035.338259 | 11803.1 |
| 1329.57 | 7453.01 | 7422.91 | 12840.43612 | 12666.71204 | 44138.21283 | 43406.23856 | 41728.5025 | 41401.40507 | 1329.57 | 2702.89 | 2780.58 | 7181.058397 | 7047.074035 | 11752.4 |
| 1331.28 | 7353.66 | 7437.87 | 13117.01571 | 12694.62996 | 44300.59649 | 43579.84503 | 41656.36915 | 41581.91267 | 1331.28 | 2866.46 | 2786.52 | 7175.317472 | 7058.732529 | 11978.9 |
| 1332.98 | 7610.5 | 7452.76 | 12929.95432 | 12722.54789 | 44379.45339 | 43753.24145 | 41395.82954 | 41762.53643 | 1332.98 | 2765.95 | 2792.44 | 7060.465847 | 7070.313742 | 11737.4 |
| 1334.69 | 7661.93 | 7467.58 | 12752.0482 | 12750.23975 | 44317.15325 | 43926.53285 | 42136.56119 | 41943.04403 | 1334.69 | 2893.6 | 2798.33 | 7235.497825 | 7081.817673 | 11906 |
| 1336.4 | 7506.32 | 7482.33 | 12618.5621 | 12777.93162 | 44618.46515 | 44099.71921 | 42105.31502 | 42123.43548 | 1336.4 | 2896.79 | 2804.2 | 7151.647196 | 7093.255362 | 11940.1 |
| 1338.11 | 7611.48 | 7497.01 | 12874.90967 | 12805.39743 | 45003.19767 | 44272.69553 | 42301.96842 | 42303.94308 | 1338.11 | 2770.6 | 2810.04 | 7147.385663 | 7104.61577 | 11790.9 |
| 1339.82 | 7544.6 | 7511.61 | 12915.3737 | 12832.75022 | 44953.10284 | 44445.56683 | 42780.30195 | 42484.21837 | 1339.82 | 2798.33 | 2815.85 | 7165.12733 | 7115.909936 | 11963.8 |
| 1341.53 | 7444.94 | 7526.15 | 12839.30584 | 12860.103 | 45267.78751 | 44618.12305 | 42510.81827 | 42664.60982 | 1341.53 | 2813.63 | 2821.64 | 7300.469642 | 7127.126821 | 12237.7 |
| 1343.24 | 7404.44 | 7540.62 | 12968.27082 | 12887.22972 | 45786.46001 | 44790.67927 | 42942.80526 | 42844.76895 | 1343.24 | 2822.25 | 2827.4 | 7327.131823 | 7138.277464 | 12136.6 |
| 1344.95 | 7763.21 | 7555.01 | 12866.43257 | 12914.35645 | 45769.79711 | 44963.02543 | 43502.21621 | 43025.04424 | 1344.95 | 2806.32 | 2833.14 | 7349.952001 | 7149.350826 | 11920.6 |
| 1346.65 | 7690.33 | 7569.34 | 13131.37027 | 12941.25712 | 45602.95587 | 45135.16155 | 43023.99883 | 43205.20337 | 1346.65 | 2906.2 | 2838.85 | 7362.007944 | 7160.346905 | 11996.3 |
| 1348.36 | 7413.74 | 7583.59 | 13160.75755 | 12968.04476 | 45536.51654 | 45307.08762 | 43180.11351 | 43385.24635 | 1348.36 | 2809.1 | 2844.54 | 7251.208087 | 7171.276744 | 11887.1 |
| 1350.07 | 7510.35 | 7597.77 | 12786.63478 | 12994.83241 | 46075.67273 | 45478.90867 | 43495.82758 | 43565.28933 | 1350.07 | 2646.8 | 2850.2 | 7045.197194 | 7182.129301 | 11933.5 |
| 1351.78 | 7544.59 | 7611.89 | 13163.01811 | 13021.39399 | 45851.41347 | 45650.51966 | 43441.46622 | 43745.21614 | 1351.78 | 2857.38 | 2855.84 | 7154.042928 | 7192.915616 | 12106 |
| 1353.49 | 7409.99 | 7625.93 | 13204.83848 | 13047.84255 | 46119.61184 | 45821.9206 | 43829.4298 | 43925.02681 | 1353.49 | 2944.86 | 2861.45 | 7239.41711 | 7203.624649 | 12137.7 |
| 1355.19 | 7596.69 | 7639.9 | 13049.76403 | 13074.17808 | 46508.37742 | 45993.21652 | 43693.64254 | 44104.83747 | 1355.19 | 2867.44 | 2867.03 | 6944.753083 | 7214.267442 | 12039.1 |
| 1356.9 | 7703.49 | 7653.8 | 13201.10856 | 13100.51361 | 46565.15863 | 46164.19736 | 44196.60131 | 44284.53198 | 1356.9 | 2914.56 | 2872.59 | 7120.524757 | 7224.832952 | 12216.4 |
| 1358.61 | 7518.84 | 7667.62 | 13265.42151 | 13126.62309 | 47022.27393 | 46335.07318 | 44748.34591 | 44464.11033 | 1358.61 | 2920.49 | 2878.12 | 7304.433088 | 7235.321181 | 12260.1 |
| 1360.32 | 7527.9 | 7681.38 | 13265.42151 | 13152.61954 | 46676.5984 | 46505.73895 | 44529.62274 | 44643.68867 | 1360.32 | 2916.38 | 2883.63 | 7328.887222 | 7245.743168 | 12021 |
| 1362.02 | 7580 | 7695.07 | 13066.15309 | 13178.50295 | 46937.26131 | 46676.19466 | 45030.02605 | 44823.03471 | 1362.02 | 2889.97 | 2889.11 | 7286.857025 | 7256.098914 | 12072.7 |
| 1363.73 | 7599.54 | 7708.68 | 13464.35084 | 13204.27334 | 47445.6389 | 46846.33531 | 44947.20628 | 45002.38075 | 1363.73 | 2926.18 | 2894.57 | 7331.249833 | 7266.377378 | 12158.8 |
| 1365.44 | 7671.75 | 7722.23 | 13184.26738 | 13229.93071 | 47439.90772 | 47016.37092 | 44825.474 | 45181.61062 | 1365.44 | 2903.2 | 2899.99 | 7328.544974 | 7276.578561 | 12205.8 |
| 1367.14 | 7765.56 | 7735.7 | 13356.63513 | 13255.47504 | 47730.18177 | 47186.30151 | 45151.40987 | 45360.72435 | 1367.14 | 2890.82 | 2905.4 | 7276.700003 | 7286.713502 | 12034.8 |
| 1368.85 | 7807.81 | 7749.1 | 13466.04626 | 13280.90635 | 47748.64893 | 47355.81201 | 45417.17653 | 45539.72191 | 1368.85 | 2847.52 | 2910.78 | 7312.558706 | 7296.782201 | 12194 |
| 1370.56 | 7771.25 | 7762.44 | 13110.00797 | 13306.22463 | 47881.20918 | 47525.21747 | 45201.12496 | 45718.71948 | 1370.56 | 2806.97 | 2916.13 | 7166.275515 | 7306.773619 | 12416.5 |
| 1372.26 | 7744.72 | 7775.69 | 13465.59415 | 13331.42988 | 48077.7677 | 47694.41289 | 45595.12869 | 45897.48473 | 1372.26 | 3001.32 | 2921.45 | 7255.93331 | 7316.687756 | 12355.1 |
| 1373.97 | 7835.72 | 7788.88 | 13464.91598 | 13356.5221 | 48300.11656 | 47863.39825 | 45885.7529 | 46076.13382 | 1373.97 | 3143.33 | 2926.76 | 7411.125976 | 7326.53565 | 12408.5 |
| 1375.67 | 7966.26 | 7802 | 13468.53287 | 13381.50129 | 48557.70162 | 48032.06855 | 45918.16063 | 46254.78292 | 1375.67 | 2956.26 | 2932.03 | 7542.714606 | 7336.306264 | 12102.6 |
| 1377.38 | 7956.29 | 7815.05 | 13513.51803 | 13406.36746 | 48708.6229 | 48200.63381 | 45832.66926 | 46433.1997 | 1377.38 | 2754.54 | 2937.28 | 7589.602509 | 7346.021676 | 12259 |
| 1379.09 | 7980.25 | 7828.02 | 13325.66545 | 13431.1206 | 49046.76299 | 48368.884 | 46420.19014 | 46611.50032 | 1379.09 | 2851.52 | 2942.5 | 7391.915957 | 7355.648765 | 12416.7 |
| 1380.79 | 8168.37 | 7840.93 | 13521.20394 | 13455.76071 | 49178.47417 | 48536.92414 | 46609.52566 | 46789.68479 | 1380.79 | 2872.97 | 2947.7 | 7600.764192 | 7365.209614 | 12205.5 |
| 1382.5 | 8001.69 | 7853.76 | 13719.68115 | 13480.28779 | 49356.35326 | 48704.64921 | 46543.20015 | 46967.7531 | 1382.5 | 2955.24 | 2952.87 | 7271.080521 | 7374.704221 | 12339.1 |
| 1384.2 | 7894.8 | 7866.53 | 13660.68052 | 13504.58882 | 49821.00402 | 48872.26925 | 47057.77459 | 47145.70526 | 1384.2 | 2986.14 | 2958.02 | 7165.502698 | 7384.121547 | 12407.8 |
| 1385.91 | 7930.39 | 7879.22 | 13646.77807 | 13528.88984 | 50302.63606 | 49039.57421 | 47501.02878 | 47323.4251 | 1385.91 | 3076.08 | 2963.14 | 7461.76756 | 7393.472631 | 12308 |
| 1387.61 | 8064.46 | 7891.84 | 13746.12971 | 13553.07784 | 49856.45247 | 49206.5641 | 47885.50766 | 47501.14494 | 1387.61 | 3040.19 | 2968.23 | 7404.954481 | 7402.757473 | 12390.5 |
| 1389.32 | 7887.91 | 7904.39 | 13667.4622 | 13577.03978 | 50227.91836 | 49373.44896 | 47946.25764 | 47678.63247 | 1389.32 | 3041.65 | 2973.3 | 7457.881395 | 7411.965034 | 12489.5 |
| 1391.02 | 7871.43 | 7916.87 | 13527.75956 | 13601.00172 | 50921.07369 | 49540.01875 | 47734.27156 | 47856.00384 | 1391.02 | 3149.86 | 2978.34 | 7398.860268 | 7421.106353 | 12399.6 |
| 1392.73 | 8093.98 | 7929.28 | 13703.40512 | 13624.73761 | 50408.13237 | 49706.27347 | 47873.77582 | 48033.1429 | 1392.73 | 2893.28 | 2983.35 | 7499.547264 | 7430.170391 | 12447.4 |
| 1394.43 | 8185.43 | 7941.61 | 13798.00958 | 13648.47349 | 50608.40555 | 49872.31813 | 47994.57885 | 48210.28195 | 1394.43 | 3011.54 | 2988.34 | 7412.097517 | 7439.168187 | 12367.8 |
| 1396.14 | 8142.56 | 7953.88 | 13957.3791 | 13671.98332 | 51054.58914 | 50038.15274 | 48108.18016 | 48387.1887 | 1396.14 | 3235.74 | 2993.31 | 7252.853083 | 7448.088702 | 12373 |
| 1397.84 | 7941.37 | 7966.08 | 13802.5307 | 13695.49315 | 50856.86316 | 50203.67228 | 48587.79142 | 48563.86313 | 1397.84 | 3058.21 | 2998.25 | 7538.33163 | 7456.954015 | 12339.9 |
| 1399.55 | 7877.24 | 7978.2 | 14029.15189 | 13718.77693 | 50978.70397 | 50368.98177 | 48371.856 | 48740.4214 | 1399.55 | 3148.9 | 3003.16 | 7637.848361 | 7465.742047 | 12303.5 |
| 1401.25 | 7980.55 | 7990.26 | 14009.93713 | 13742.0607 | 51181.20594 | 50533.97618 | 48588.48836 | 48916.86352 | 1401.25 | 3169.01 | 3008.05 | 7599.527685 | 7474.452797 | 12366.1 |
| 1402.95 | 8173.27 | 8002.24 | 13846.83768 | 13765.11842 | 51463.30774 | 50698.65552 | 49045.79752 | 49093.07332 | 1402.95 | 2945.23 | 3012.91 | 7568.990379 | 7483.097305 | 12390.4 |
| 1404.66 | 7964.38 | 8014.16 | 13900.86508 | 13788.06311 | 51307.18594 | 50863.12481 | 49113.63307 | 49269.16697 | 1404.66 | 3071.85 | 3017.74 | 7569.696954 | 7491.675573 | 12460 |
| 1406.36 | 8081.81 | 8026 | 13873.28624 | 13811.0078 | 51491.64528 | 51027.38404 | 49341.53263 | 49445.14446 | 1406.36 | 2918.33 | 3022.55 | 7655.545867 | 7500.187598 | 12589.3 |
| 1408.07 | 8200.65 | 8037.77 | 14041.69801 | 13833.72643 | 51913.52439 | 51191.22318 | 49476.62294 | 49620.77347 | 1408.07 | 3147.69 | 3027.33 | 7518.448157 | 7508.622342 | 12597 |
| 1409.77 | 7980.6 | 8049.47 | 13850.79367 | 13856.33204 | 52223.64534 | 51354.85227 | 49921.73564 | 49796.40249 | 1409.77 | 3269.25 | 3032.09 | 7550.961666 | 7516.990845 | 12558.5 |
| 1411.47 | 8108.94 | 8061.1 | 13882.21546 | 13878.82462 | 52526.97376 | 51518.16628 | 49936.13907 | 49971.7992 | 1411.47 | 3080.29 | 3036.82 | 7474.739843 | 7525.293106 | 12597.5 |
| 1413.17 | 7965.45 | 8072.66 | 14109.51482 | 13901.20417 | 52296.24038 | 51681.27024 | 49577.09853 | 50146.96359 | 1413.17 | 3244.01 | 3041.53 | 7438.759698 | 7533.529126 | 12547.9 |
| 1414.88 | 7907.15 | 8084.16 | 13750.08569 | 13923.47069 | 52545.22866 | 51844.05913 | 50226.29866 | 50321.89567 | 1414.88 | 3086.36 | 3046.2 | 7697.344219 | 7541.687863 | 12506.8 |
| 1416.58 | 8054.99 | 8095.58 | 13782.52474 | 13945.62418 | 52582.69364 | 52006.53294 | 50532.37172 | 50496.71159 | 1416.58 | 3024.62 | 3050.86 | 7675.274777 | 7549.78036 | 12554.1 |
| 1418.28 | 8212.41 | 8106.93 | 14192.47739 | 13967.66465 | 52820.85634 | 52168.69167 | 50513.55432 | 50671.2952 | 1418.28 | 3161.97 | 3055.49 | 7782.574878 | 7557.806615 | 12681.3 |
| 1419.99 | 8049.3 | 8118.21 | 14090.8652 | 13989.59208 | 53747.07998 | 52330.64036 | 50766.19527 | 50845.76265 | 1419.99 | 2855.03 | 3060.09 | 7867.286646 | 7565.766629 | 12436.4 |
| 1421.69 | 8083.33 | 8129.42 | 14183.2091 | 14011.40649 | 54187.63852 | 52492.27397 | 51324.90928 | 51019.99779 | 1421.69 | 3056.66 | 3064.67 | 7729.912929 | 7573.649361 | 12618.3 |
| 1423.39 | 8160.16 | 8140.56 | 14273.17941 | 14033.10788 | 53936.73985 | 52653.48748 | 51441.9953 | 51194.00062 | 1423.39 | 3210.59 | 3069.22 | 7693.800301 | 7581.465851 | 12726.4 |
| 1425.09 | 7991.27 | 8151.63 | 14053.00081 | 14054.69623 | 53690.0865 | 52814.49093 | 51086.67177 | 51367.77113 | 1425.09 | 2969.78 | 3073.74 | 7595.16679 | 7589.2161 | 12541.4 |
| 1426.8 | 8187.5 | 8162.63 | 14014.68431 | 14076.17155 | 53753.2357 | 52975.17932 | 51354.52926 | 51541.42548 | 1426.8 | 3183.2 | 3078.24 | 7604.429552 | 7596.900108 | 12634.9 |
| 1428.5 | 8262.47 | 8173.56 | 14055.14834 | 14097.53385 | 53757.0565 | 53135.55263 | 51487.29643 | 51714.84752 | 1428.5 | 3152.34 | 3082.72 | 7648.182026 | 7604.517874 | 12755 |
| 1430.2 | 8272.09 | 8184.42 | 14067.80748 | 14118.78312 | 54022.8138 | 53295.71589 | 51770.02198 | 51888.03725 | 1430.2 | 3266.53 | 3087.16 | 7825.698058 | 7612.069399 | 12734.6 |
| 1431.9 | 8080.57 | 8195.21 | 14164.10736 | 14139.80633 | 54366.04827 | 53455.45904 | 51922.18734 | 52060.99466 | 1431.9 | 3125.27 | 3091.59 | 7702.963701 | 7619.543642 | 12485.3 |
| 1433.6 | 8126.76 | 8205.93 | 14334.55363 | 14160.82955 | 54960.49982 | 53614.88713 | 52381.81963 | 52233.71976 | 1433.6 | 3043.57 | 3095.98 | 7570.86722 | 7626.951643 | 12575.2 |
| 1435.3 | 8315.25 | 8216.58 | 14418.19437 | 14181.73973 | 55038.82605 | 53774.00014 | 52580.21538 | 52406.32871 | 1435.3 | 3166.54 | 3100.35 | 7801.608253 | 7634.304444 | 12695.4 |
| 1437.01 | 8406.8 | 8227.17 | 14364.39303 | 14202.42386 | 55084.46329 | 53932.90309 | 52438.27182 | 52578.58918 | 1437.01 | 3052.74 | 3104.7 | 7700.446526 | 7641.579962 | 12696.5 |
| 1438.71 | 8440.73 | 8237.68 | 14467.47459 | 14223.10799 | 55408.80607 | 54091.38595 | 52522.48547 | 52750.7335 | 1438.71 | 3050.58 | 3109.02 | 7797.34672 | 7648.78924 | 12677.6 |
| 1440.41 | 8250.79 | 8248.12 | 14346.87368 | 14243.67909 | 55140.71384 | 54249.55373 | 52461.50317 | 52922.52934 | 1440.41 | 3231.76 | 3113.31 | 7734.847917 | 7655.932275 | 12651.5 |
| 1442.11 | 8337.75 | 8258.5 | 14089.50886 | 14264.02414 | 55384.18319 | 54407.40644 | 52762.69764 | 53094.20903 | 1442.11 | 3224.6 | 3117.58 | 7654.3204 | 7663.00907 | 12688.5 |
| 1443.81 | 8500.49 | 8268.81 | 14370.27048 | 14284.36918 | 55785.04794 | 54564.94407 | 53237.89494 | 53265.65641 | 1443.81 | 3020.31 | 3121.82 | 7787.576107 | 7670.019623 | 12697.9 |
| 1445.51 | 8209.1 | 8279.04 | 14384.39899 | 14304.48817 | 55483.41765 | 54722.16663 | 53340.22905 | 53436.75531 | 1445.51 | 3370.38 | 3126.04 | 7693.126847 | 7676.963934 | 12802.4 |
| 1447.21 | 8199.66 | 8289.21 | 14341.22228 | 14324.60716 | 55952.10146 | 54878.96908 | 53800.32597 | 53607.73806 | 1447.21 | 3147.54 | 3130.23 | 7740.467399 | 7683.842004 | 12665.5 |
| 1448.91 | 8252.2 | 8299.31 | 14600.50858 | 14344.50009 | 56325.0532 | 55035.56149 | 53915.90194 | 53778.37234 | 1448.91 | 3242.85 | 3134.4 | 7823.313366 | 7690.653833 | 12772.6 |
| 1450.61 | 8246.31 | 8309.34 | 14338.62264 | 14364.39303 | 56203.42466 | 55191.73379 | 53594.61235 | 53948.89046 | 1450.61 | 3654.42 | 3138.54 | 7767.626392 | 7697.39942 | 12967.4 |
| 1452.31 | 8497.44 | 8319.3 | 14226.04672 | 14384.0599 | 56651.20024 | 55347.59102 | 54009.05967 | 54119.06012 | 1452.31 | 3264.99 | 3142.66 | 7790.821938 | 7704.078766 | 12715.3 |
| 1454.01 | 8371.65 | 8329.19 | 14484.99393 | 14403.61375 | 56333.54385 | 55503.13318 | 53607.15728 | 54288.99745 | 1454.01 | 3001.93 | 3146.75 | 7886.860993 | 7710.69187 | 12919.6 |
| 1455.71 | 8201.81 | 8339.02 | 14441.93025 | 14423.1676 | 56291.19674 | 55658.25523 | 54246.94871 | 54458.70248 | 1455.71 | 2989.27 | 3150.81 | 7806.189953 | 7717.238733 | 12833.7 |
| 1457.41 | 8435.86 | 8348.77 | 14389.82433 | 14442.49539 | 56864.20939 | 55813.16724 | 54960.26736 | 54628.17519 | 1457.41 | 3163.55 | 3154.85 | 7837.091587 | 7723.719354 | 12617 |
| 1459.11 | 8487.89 | 8358.46 | 14590.90119 | 14461.71016 | 56934.78791 | 55967.65914 | 54656.40128 | 54797.41559 | 1459.11 | 3105.72 | 3158.87 | 7849.710582 | 7730.133734 | 12697.2 |
| 1460.81 | 8298.08 | 8368.08 | 14610.22899 | 14480.92492 | 57072.97322 | 56121.83597 | 54745.72583 | 54966.30751 | 1460.81 | 3246.13 | 3162.86 | 7716.852324 | 7736.492913 | 12749 |
| 1462.51 | 8281.45 | 8377.63 | 14446.79046 | 14499.91363 | 56796.17807 | 56275.5927 | 54575.44002 | 55135.08329 | 1462.51 | 3359.37 | 3166.82 | 7760.163189 | 7742.77481 | 12653.4 |
| 1464.21 | 8344.48 | 8387.12 | 14576.32058 | 14518.78931 | 57089.95452 | 56429.03435 | 55085.25204 | 55303.51059 | 1464.21 | 3122.3 | 3170.76 | 7691.658495 | 7749.001506 | 12713.4 |
| 1465.91 | 8324.77 | 8396.53 | 14509.97313 | 14537.66499 | 57563.30818 | 56582.16093 | 55285.85476 | 55471.58942 | 1465.91 | 3262.28 | 3174.67 | 7690.190143 | 7755.16196 | 12807.9 |
| 1467.61 | 8484.35 | 8405.88 | 14556.99279 | 14556.31462 | 57551.31514 | 56734.86741 | 55507.94648 | 55639.55209 | 1467.61 | 3248.7 | 3178.56 | 7643.545125 | 7761.256173 | 12762 |
| 1469.31 | 8333.62 | 8415.16 | 14500.81786 | 14574.85121 | 57587.93106 | 56887.25881 | 55249.38154 | 55807.16629 | 1469.31 | 3229.62 | 3182.43 | 7600.333623 | 7767.284144 | 12798.2 |
| 1471.01 | 8268.8 | 8424.38 | 14513.36397 | 14593.27478 | 57451.86842 | 57039.33513 | 55590.65009 | 55974.54818 | 1471.01 | 3187.57 | 3186.26 | 7573.384395 | 7773.245874 | 12661.1 |
| 1472.7 | 8454 | 8433.52 | 14790.28264 | 14611.58532 | 57898.68881 | 57190.99136 | 56265.52086 | 56141.5816 | 1472.7 | 3160.57 | 3190.08 | 7820.928674 | 7779.141363 | 12743.5 |
| 1474.4 | 8396.18 | 8442.6 | 14718.28378 | 14629.89586 | 58601.18386 | 57342.33251 | 56377.72829 | 56308.38271 | 1474.4 | 3336.37 | 3193.86 | 7855.672312 | 7784.98165 | 13040.4 |
| 1476.1 | 8383.14 | 8451.61 | 14822.94774 | 14647.98035 | 58275.03681 | 57493.25356 | 56596.45147 | 56474.9515 | 1476.1 | 3217.13 | 3197.63 | 7785.246616 | 7790.755696 | 12982.9 |
| 1477.8 | 8438.72 | 8460.55 | 14752.53127 | 14665.95181 | 58646.82109 | 57643.85954 | 56685.65986 | 56641.28798 | 1477.8 | 3206.09 | 3201.37 | 7939.224855 | 7796.463501 | 12811.3 |
| 1479.5 | 8511.22 | 8469.43 | 14665.49969 | 14683.81023 | 58856.32785 | 57794.04541 | 56926.10435 | 56807.27598 | 1479.5 | 3204.84 | 3205.08 | 7944.016319 | 7802.116104 | 12871.4 |
| 1481.19 | 8434.49 | 8478.24 | 14685.73171 | 14701.55563 | 58517.6571 | 57943.91621 | 56707.96196 | 56972.91552 | 1481.19 | 3139.91 | 3208.77 | 7981.939547 | 7807.702466 | 12781.5 |
| 1482.89 | 8327.53 | 8486.98 | 14647.64126 | 14719.18801 | 58742.97769 | 58093.47194 | 56797.40266 | 57138.32275 | 1482.89 | 3259.28 | 3212.43 | 7864.868833 | 7813.222586 | 12845 |
| 1484.59 | 8610.94 | 8495.66 | 14685.84474 | 14736.70735 | 59124.10168 | 58242.50254 | 57343.57174 | 57303.49766 | 1484.59 | 3387.23 | 3216.07 | 7932.799435 | 7818.676465 | 12913.2 |
| 1486.29 | 8545.1 | 8504.27 | 14706.18978 | 14754.2267 | 59205.82418 | 58391.32309 | 57530.11949 | 57468.3241 | 1486.29 | 3231.95 | 3219.68 | 7935.658857 | 7824.075143 | 12796.3 |
| 1487.98 | 8386 | 8512.82 | 14857.53431 | 14771.51998 | 59554.57757 | 58539.61851 | 57753.6051 | 57632.91823 | 1487.98 | 3194.33 | 3223.27 | 7865.277322 | 7829.407579 | 12951.7 |
| 1489.68 | 8706.56 | 8521.29 | 14926.59444 | 14788.70024 | 60014.983 | 58687.59886 | 58008.8015 | 57797.16388 | 1489.68 | 3145.26 | 3226.84 | 7947.935605 | 7834.673774 | 13193.1 |
| 1491.38 | 8625.19 | 8529.7 | 15141.46072 | 14805.76748 | 60005.00649 | 58835.26413 | 58382.59395 | 57961.06107 | 1491.38 | 3141.52 | 3230.38 | 7839.012589 | 7839.884768 | 13018.7 |
| 1493.08 | 8610.82 | 8538.05 | 15011.25243 | 14822.72168 | 60076.8586 | 58982.50931 | 58402.3406 | 58124.72594 | 1493.08 | 3258.46 | 3233.89 | 7744.905576 | 7845.02952 | 13033.6 |
| 1494.77 | 8761.67 | 8546.33 | 15135.24418 | 14839.56286 | 60337.52152 | 59129.33438 | 58467.03991 | 58288.1585 | 1494.77 | 3146.22 | 3237.38 | 7752.346698 | 7850.119071 | 13060.7 |
| 1496.47 | 8582.92 | 8554.54 | 15260.81832 | 14856.40403 | 60125.67983 | 59275.84438 | 58352.04472 | 58451.24259 | 1496.47 | 3109.11 | 3240.85 | 7807.603104 | 7855.142381 | 12764.3 |
| 1498.17 | 8589.84 | 8562.69 | 14916.19586 | 14873.01915 | 60272.24966 | 59421.93427 | 58194.30384 | 58613.97821 | 1498.17 | 3251.6 | 3244.29 | 8005.212374 | 7860.099449 | 12885.8 |
| 1499.86 | 8514.64 | 8570.77 | 15026.62424 | 14889.52125 | 60410.11656 | 59567.60407 | 58780.77932 | 58776.48152 | 1499.86 | 3241.07 | 3247.71 | 7739.727703 | 7865.001316 | 12776.5 |
| 1501.56 | 8542.83 | 8578.79 | 14921.84726 | 14905.91031 | 60399.92779 | 59712.85376 | 58641.15889 | 58938.63635 | 1501.56 | 3224.53 | 3251.1 | 7716.333433 | 7869.836941 | 12864.5 |
| 1503.25 | 8834.55 | 8586.74 | 14926.70747 | 14922.18635 | 60394.30273 | 59857.78839 | 59139.23907 | 59100.44272 | 1503.25 | 3329.05 | 3254.47 | 7727.241191 | 7874.617366 | 12825 |
| 1504.95 | 8610.27 | 8594.63 | 15085.39882 | 14938.34935 | 60582.47673 | 60002.30291 | 59752.08213 | 59262.01677 | 1504.95 | 3187.69 | 3257.81 | 7739.816025 | 7879.342589 | 12816.5 |
| 1506.65 | 8525.27 | 8602.45 | 14726.19574 | 14954.51236 | 61053.70774 | 60146.50235 | 59460.1802 | 59423.24235 | 1506.65 | 3182.95 | 3261.13 | 7859.746161 | 7884.00157 | 12931.3 |
| 1508.34 | 8574.37 | 8610.21 | 14828.26005 | 14970.44931 | 61245.5964 | 60290.2817 | 59181.40398 | 59584.11946 | 1508.34 | 3193.23 | 3264.43 | 7777.010596 | 7888.59431 | 12783.2 |
| 1510.04 | 8571.56 | 8617.9 | 15016.7908 | 14986.27324 | 61109.63989 | 60433.64094 | 59097.88727 | 59744.76426 | 1510.04 | 3206.96 | 3267.7 | 7753.384481 | 7893.131849 | 12692.3 |
| 1511.73 | 8532.21 | 8625.52 | 14874.60155 | 15002.09716 | 61240.71428 | 60576.58009 | 59515.35466 | 59904.94443 | 1511.73 | 3348.42 | 3270.95 | 7741.792228 | 7897.603147 | 12735 |
| 1513.43 | 8602 | 8633.09 | 15038.83127 | 15017.69503 | 61234.77082 | 60719.09913 | 59894.02569 | 60064.89228 | 1513.43 | 3265.78 | 3274.17 | 7834.143843 | 7902.030283 | 12775.4 |
| 1515.12 | 8799.69 | 8640.59 | 15018.03411 | 15033.17987 | 61104.12096 | 60861.3031 | 60058.96829 | 60224.60783 | 1515.12 | 3367.16 | 3277.37 | 7848.175989 | 7906.380138 | 12807.7 |
| 1516.82 | 8771.9 | 8648.02 | 14805.54142 | 15048.55168 | 61197.30583 | 61003.08697 | 60228.55716 | 60383.85874 | 1516.82 | 3155.21 | 3280.54 | 7999.857857 | 7910.685832 | 12814.6 |
| 1518.51 | 8633.75 | 8655.39 | 14853.4653 | 15063.92349 | 61641.26063 | 61144.45073 | 60626.85868 | 60542.87735 | 1518.51 | 3101.97 | 3283.69 | 8056.030603 | 7914.925285 | 12756.5 |
| 1520.21 | 8543.7 | 8662.7 | 14868.61106 | 15079.06925 | 61367.11881 | 61285.3944 | 60475.97105 | 60701.54748 | 1520.21 | 3308.67 | 3286.82 | 7929.597765 | 7919.098496 | 12939.3 |
| 1521.9 | 8607.94 | 8669.94 | 15018.03411 | 15094.215 | 61450.85783 | 61425.91797 | 60576.56281 | 60859.86914 | 1521.9 | 3118.46 | 3289.92 | 7883.681403 | 7923.227546 | 12747.6 |
| 1523.6 | 8410.39 | 8677.12 | 15187.01102 | 15109.1347 | 62260.97185 | 61566.12646 | 60640.10055 | 61017.84233 | 1523.6 | 3238.25 | 3293 | 7918.182156 | 7927.279314 | 12864.8 |
| 1525.29 | 8637.68 | 8684.24 | 15093.87592 | 15124.0544 | 61962.84395 | 61705.80983 | 61005.52972 | 61175.58321 | 1525.29 | 3222.45 | 3296.06 | 7912.783478 | 7931.286922 | 12935.9 |
| 1526.99 | 8788.58 | 8691.29 | 15226.00569 | 15138.74805 | 62695.26853 | 61845.17812 | 61399.18497 | 61332.85946 | 1526.99 | 3245.15 | 3299.09 | 7869.704458 | 7935.239328 | 12831.8 |
| 1528.68 | 8645.54 | 8698.28 | 15152.42444 | 15153.44169 | 62594.44208 | 61984.12631 | 61188.01199 | 61489.9034 | 1528.68 | 3362.18 | 3302.1 | 7886.805791 | 7939.125493 | 12878.7 |
| 1530.37 | 8615.81 | 8705.2 | 14985.59507 | 15167.90928 | 62469.84181 | 62122.6544 | 61477.70694 | 61646.59887 | 1530.37 | 3336.01 | 3305.08 | 7964.639489 | 7942.956456 | 12786.6 |
| 1532.07 | 8551.96 | 8712.07 | 15092.74564 | 15182.37687 | 62728.8066 | 62260.76239 | 61851.15092 | 61802.94587 | 1532.07 | 3263.33 | 3308.04 | 7972.831348 | 7946.721178 | 12569.5 |
| 1533.76 | 8801.17 | 8718.87 | 15032.16262 | 15196.73143 | 62658.33421 | 62398.45028 | 61810.03143 | 61958.94439 | 1533.76 | 3202.62 | 3310.98 | 7889.808737 | 7950.441739 | 12722.2 |
| 1535.45 | 8599.49 | 8725.6 | 14867.48078 | 15210.85993 | 62624.79615 | 62535.82309 | 61879.72548 | 62114.59445 | 1535.45 | 3199.74 | 3313.89 | 7703.857961 | 7954.096059 | 12863.6 |
| 1537.15 | 8493.12 | 8732.28 | 14932.01978 | 15224.98843 | 62499.13455 | 62672.67078 | 62220.76173 | 62269.89604 | 1537.15 | 3222.61 | 3316.78 | 7796.463501 | 7957.695178 | 12755.6 |
| 1538.84 | 8683.86 | 8738.89 | 15178.98603 | 15239.00391 | 63132.00628 | 62809.09837 | 62547.62685 | 62424.84915 | 1538.84 | 3240.91 | 3319.65 | 7879.265307 | 7961.239095 | 12961.1 |
| 1540.53 | 8741.36 | 8745.44 | 15344.45906 | 15252.90636 | 63261.06414 | 62945.21089 | 62729.8768 | 62579.4538 | 1540.53 | 3228.31 | 3322.49 | 7798.207859 | 7964.727811 | 12863.8 |
| 1542.23 | 8777.63 | 8751.93 | 15414.53644 | 15266.69578 | 63741.95325 | 63080.79827 | 63306.59511 | 62733.70997 | 1542.23 | 3337.49 | 3325.31 | 7713.043441 | 7968.150286 | 12833.9 |
| 1543.92 | 8870.07 | 8758.36 | 15409.45018 | 15280.37217 | 63450.93627 | 63215.96556 | 63243.87046 | 62887.61768 | 1543.92 | 3369.4 | 3328.11 | 7652.863089 | 7971.528599 | 12846.3 |
| 1545.61 | 8821.56 | 8764.72 | 15140.10438 | 15293.93553 | 63583.07199 | 63350.81777 | 63204.60948 | 63041.17692 | 1545.61 | 3401.3 | 3330.88 | 7915.278572 | 7974.840672 | 12742.1 |
| 1547.31 | 8829.64 | 8771.03 | 14938.80147 | 15307.38587 | 63940.5283 | 63485.14486 | 63720.69395 | 63194.38768 | 1547.31 | 3433.74 | 3333.63 | 7914.594078 | 7978.097543 | 12847.2 |
| 1549 | 8898.75 | 8777.27 | 15034.64923 | 15320.72317 | 63923.33474 | 63619.15687 | 64040.47351 | 63347.24997 | 1549 | 3424.34 | 3336.36 | 7871.978748 | 7981.310253 | 13061.4 |
| 1550.69 | 8728.26 | 8783.45 | 15064.60166 | 15333.94745 | 64603.86023 | 63752.64376 | 64058.59397 | 63499.7638 | 1550.69 | 3299.25 | 3339.06 | 7967.609314 | 7984.456721 | 13149.3 |
| 1552.38 | 8776.6 | 8789.57 | 15157.73676 | 15347.0587 | 64303.50353 | 63885.81557 | 63715.23459 | 63651.92915 | 1552.38 | 3285.49 | 3341.74 | 8119.280142 | 7987.547989 | 13089.8 |
| 1554.07 | 8655.42 | 8795.63 | 15193.7927 | 15360.16995 | 64437.54965 | 64018.46226 | 64107.26365 | 63803.74604 | 1554.07 | 3338.4 | 3344.4 | 8018.350261 | 7990.584055 | 12831.8 |
| 1555.77 | 8737.87 | 8801.63 | 15114.89913 | 15373.05515 | 64728.88504 | 64150.68884 | 64301.59391 | 63955.21445 | 1555.77 | 3293.7 | 3347.03 | 7897.702509 | 7993.57596 | 12846.6 |
| 1557.46 | 8750.4 | 8807.56 | 15163.38816 | 15385.94035 | 64264.87108 | 64282.49533 | 64346.77888 | 64106.21824 | 1557.46 | 3401.79 | 3349.64 | 7899.226062 | 7996.501624 | 12814.7 |
| 1559.15 | 8832.25 | 8813.44 | 15119.64631 | 15398.59948 | 64772.08121 | 64413.98674 | 64844.04597 | 64256.98971 | 1559.15 | 3385.59 | 3352.23 | 7987.382385 | 7999.372087 | 12704.7 |
| 1560.84 | 8866.38 | 8819.26 | 15022.89432 | 15411.25862 | 65029.87853 | 64544.95302 | 64549.0078 | 64407.29656 | 1560.84 | 3307.61 | 3354.8 | 7892.049906 | 8002.187348 | 12904.4 |
| 1562.53 | 8761.35 | 8825.01 | 15218.99795 | 15423.69171 | 65335.43575 | 64675.4992 | 64860.0756 | 64557.37109 | 1562.53 | 3389.1 | 3357.34 | 7867.562652 | 8004.958449 | 12844.4 |
| 1564.22 | 9068.12 | 8830.71 | 15274.0426 | 15436.12479 | 65354.22131 | 64805.62529 | 65364.5444 | 64706.98099 | 1564.22 | 3271.55 | 3359.86 | 7898.751332 | 8007.663308 | 12789.7 |
| 1565.91 | 8819.18 | 8836.35 | 15238.77785 | 15448.44484 | 64956.43442 | 64935.33127 | 65064.39534 | 64856.24243 | 1565.91 | 3296.47 | 3362.36 | 7928.802867 | 8010.324006 | 12873 |
| 1567.61 | 8783.54 | 8841.92 | 15306.14256 | 15460.65187 | 64900.29001 | 65064.61715 | 65063.93071 | 65005.15539 | 1567.61 | 3405.56 | 3364.84 | 7884.376938 | 8012.929502 | 12834.2 |
| 1569.3 | 8690.22 | 8847.44 | 15364.8041 | 15472.74587 | 64667.75236 | 65193.37791 | 64933.25436 | 65153.60373 | 1569.3 | 3449.62 | 3367.29 | 7866.259903 | 8015.468758 | 12843.9 |
| 1570.99 | 8370.63 | 8852.9 | 15406.39842 | 15484.72684 | 65085.28002 | 65321.82359 | 64827.20324 | 65301.81976 | 1570.99 | 3399.25 | 3369.72 | 8034.137805 | 8017.963852 | 12797.1 |
| 1572.68 | 8728.23 | 8858.3 | 15387.40971 | 15496.59478 | 65203.72456 | 65449.74415 | 65281.95695 | 65449.57115 | 1572.68 | 3384.41 | 3372.13 | 7933.616412 | 8020.414786 | 12783 |
| 1574.37 | 8765.52 | 8863.64 | 15168.02231 | 15508.3497 | 65290.11692 | 65577.34963 | 65436.5616 | 65596.97408 | 1574.37 | 3163.87 | 3374.51 | 7815.099427 | 8022.799478 | 12895.6 |
| 1576.06 | 8662.07 | 8868.92 | 15194.92298 | 15519.99159 | 65681.53582 | 65704.42999 | 65580.13135 | 65744.02854 | 1576.06 | 3136.43 | 3376.88 | 7814.867582 | 8025.140009 | 12922 |
| 1577.75 | 8971.86 | 8874.14 | 15345.47631 | 15531.63347 | 66214.85469 | 65831.09024 | 66346.18517 | 65890.61837 | 1577.75 | 3339.65 | 3379.22 | 7972.389738 | 8027.425339 | 12749.7 |
| 1579.44 | 9013.61 | 8879.3 | 15541.91902 | 15543.0493 | 66560.84863 | 65957.22537 | 66027.21871 | 66036.97588 | 1579.44 | 3230.3 | 3381.54 | 7762.205634 | 8029.655467 | 12784 |

| T = 1728 K, n = 1.16, M = 14.22 Mda, 2640 sec | T = 1726 K, n = 1.04, M = 14.17 MDa, 4380 sec | T = 1726 K, n = 1.04, M = 14.17 MDa, 4380 sec | Figure 5 E and F | T = 1643 K, n = 1.00, 2040 sec | T = 1643 K, n = 1.00, 2040 sec | T = 1646 K, n = 0.78, 14280 sec | T = 1646 K, n = 0.78, 14280 sec |
|---|---|---|---|---|---|---|---|
| 350.447 | 381.64899 | 298.91928 | 600.015 | 214.365 | 211.425 | 297.827269 | 263.6348297 |
| 351.927 | 272.0704 | 300.19842 | 600.306 | 304.831 | 212.406 | 302.511035 | 264.8839743 |
| 353.412 | 254.29889 | 301.48258 | 600.597 | 177.121 | 213.391 | 230.1027729 | 266.1362788 |
| 354.901 | 351.87665 | 302.77074 | 600.888 | 224.248 | 214.379 | 305.1989078 | 267.3938494 |
| 356.394 | 300.88564 | 304.06192 | 601.179 | 239.471 | 215.371 | 285.1778359 | 268.655633 |
| 357.891 | 306.32023 | 305.35711 | 601.47 | 221.531 | 216.365 | 226.1257321 | 269.9216295 |
| 359.393 | 329.7681 | 306.65632 | 601.76 | 321.449 | 217.364 | 226.2015655 | 271.191839 |
| 360.899 | 319.38049 | 307.95953 | 602.052 | 238.751 | 218.365 | 284.3204971 | 272.4673148 |
| 362.41 | 332.62535 | 309.26676 | 602.342 | 342.365 | 219.37 | 326.1689506 | 273.7459502 |
| 363.925 | 335.19868 | 310.577 | 602.633 | 320.226 | 220.378 | 317.7893594 | 275.0298519 |
| 365.444 | 301.28092 | 311.89226 | 602.924 | 206.291 | 221.39 | 261.0196305 | 276.3179665 |
| 366.968 | 394.70925 | 313.21053 | 603.215 | 287.337 | 222.405 | 208.2838173 | 277.6102941 |
| 368.496 | 381.69012 | 314.5328 | 603.506 | 195.538 | 223.423 | 203.7527714 | 278.9078879 |
| 370.028 | 361.66633 | 315.8591 | 603.797 | 299.848 | 224.445 | 288.4102348 | 280.2096947 |
| 371.565 | 400.6906 | 317.1894 | 604.088 | 261.344 | 225.47 | 297.6935073 | 281.5146612 |
| 373.106 | 339.75843 | 318.52372 | 604.379 | 241.709 | 226.498 | 327.0947501 | 282.8259471 |
| 374.652 | 387.72264 | 319.86205 | 604.669 | 248.574 | 227.53 | 244.5363975 | 284.1403928 |
| 376.202 | 307.90135 | 321.20339 | 604.96 | 371.054 | 228.565 | 337.9052231 | 285.4601047 |
| 377.756 | 306.63525 | 322.54975 | 605.251 | 322.979 | 229.604 | 334.9961695 | 286.7829763 |
| 379.315 | 395.75162 | 323.89911 | 605.542 | 228.507 | 230.646 | 295.7850333 | 288.1121673 |
| 380.879 | 264.35844 | 325.2535 | 605.832 | 337.8 | 231.691 | 283.958182 | 289.4445181 |
| 382.446 | 395.92318 | 326.61089 | 606.123 | 296.702 | 232.74 | 353.3320547 | 290.7821351 |
| 384.019 | 329.70891 | 327.97229 | 606.414 | 286.473 | 233.793 | 238.8057092 | 292.1239651 |
| 385.595 | 354.53023 | 329.33771 | 606.705 | 266.498 | 234.849 | 317.3638498 | 293.470008 |
| 387.176 | 344.6613 | 330.70714 | 606.995 | 253.059 | 235.908 | 269.0706102 | 294.8213171 |
| 388.762 | 361.36736 | 332.08059 | 607.286 | 329.613 | 236.97 | 320.2181348 | 296.1768392 |
| 390.352 | 350.07682 | 333.45805 | 607.577 | 357.186 | 238.037 | 238.617179 | 297.5365743 |
| 391.946 | 409.61448 | 334.83952 | 607.868 | 257.408 | 239.106 | 336.2031845 | 298.9015756 |
| 393.545 | 349.12173 | 336.225 | 608.158 | 293.82 | 240.179 | 239.9316246 | 300.2707898 |
| 395.148 | 389.24055 | 337.61349 | 608.449 | 358.908 | 241.256 | 294.4400436 | 301.644217 |
| 396.756 | 305.68216 | 339.007 | 608.74 | 247.016 | 242.336 | 237.2279531 | 303.0218572 |
| 398.368 | 370.78082 | 340.40452 | 609.03 | 324.635 | 243.419 | 358.9837498 | 304.4047636 |
| 399.985 | 401.83531 | 341.80506 | 609.321 | 286.578 | 244.506 | 386.1816108 | 305.7929362 |
| 401.606 | 325.26654 | 343.2106 | 609.612 | 367.904 | 245.597 | 310.3387275 | 307.1853217 |
| 403.232 | 352.55183 | 344.61916 | 609.902 | 292.367 | 246.691 | 348.170117 | 308.5819203 |
| 404.862 | 313.10719 | 346.03274 | 610.193 | 276.578 | 247.788 | 364.9745887 | 309.9827318 |
| 406.497 | 362.57226 | 347.44932 | 610.484 | 243.754 | 248.889 | 384.1920372 | 311.3888095 |
| 408.136 | 446.38045 | 348.87092 | 610.774 | 274.926 | 249.994 | 285.4264009 | 312.8001534 |
| 409.78 | 414.411 | 350.29553 | 611.065 | 365.699 | 251.102 | 343.3736542 | 314.214657 |
| 411.428 | 384.5875 | 351.72516 | 611.355 | 322.88 | 252.213 | 319.0595689 | 315.6354801 |
| 413.081 | 403.41843 | 353.15779 | 611.646 | 350.824 | 253.328 | 353.5111058 | 317.0594629 |
| 414.738 | 371.27642 | 354.59444 | 611.937 | 332.822 | 254.447 | 325.0040652 | 318.4887119 |
| 416.4 | 355.85352 | 356.03611 | 612.227 | 255.751 | 255.569 | 296.7561222 | 319.9232272 |
| 418.066 | 344.65729 | 357.48078 | 612.518 | 304.121 | 256.695 | 277.404912 | 321.3619554 |
| 419.737 | 417.11475 | 358.93047 | 612.808 | 291.279 | 257.824 | 256.3853669 | 322.8048965 |
| 421.413 | 411.85573 | 360.38317 | 613.099 | 222.057 | 258.957 | 362.6606166 | 324.2531039 |
| 423.093 | 362.96653 | 361.84089 | 613.389 | 294.752 | 260.094 | 276.3274457 | 325.7055242 |
| 424.777 | 405.36373 | 363.30262 | 613.68 | 328.193 | 261.234 | 339.2144026 | 327.1632108 |
| 426.466 | 419.38911 | 364.76736 | 613.97 | 312.699 | 262.377 | 350.4777696 | 328.6261635 |
| 428.16 | 432.28584 | 366.23712 | 614.26 | 344.987 | 263.524 | 345.9646288 | 330.092276 |
| 429.858 | 411.66913 | 367.71088 | 614.551 | 321.775 | 264.675 | 333.5184713 | 331.5647079 |
| 431.561 | 407.40634 | 369.18766 | 614.841 | 248.392 | 265.829 | 343.9560968 | 333.0413528 |
| 433.268 | 336.35442 | 370.66946 | 615.132 | 310.915 | 266.987 | 315.4058734 | 334.5222107 |
| 434.98 | 407.95311 | 372.15527 | 615.422 | 325.644 | 268.149 | 407.8573255 | 336.0083348 |
| 436.697 | 396.90636 | 373.64509 | 615.713 | 259.274 | 269.314 | 367.8857488 | 337.4986718 |
| 438.418 | 388.26239 | 375.13892 | 616.003 | 313.563 | 270.483 | 379.9622184 | 338.9942751 |
| 440.143 | 471.2098 | 376.63677 | 616.293 | 346.277 | 271.655 | 418.7141412 | 340.4951445 |
| 441.873 | 392.67366 | 378.13863 | 616.584 | 334.165 | 272.831 | 328.4639643 | 342.000227 |
| 443.608 | 377.71325 | 379.6445 | 616.874 | 276.325 | 274.011 | 409.7805449 | 343.5105756 |
| 445.348 | 467.86197 | 381.15439 | 617.165 | 336.531 | 275.194 | 369.9353572 | 345.0251372 |
| 447.092 | 432.82659 | 382.66929 | 617.455 | 294.505 | 276.381 | 399.0911946 | 346.544965 |
| 448.841 | 429.15772 | 384.1872 | 617.745 | 314.505 | 277.572 | 356.9773243 | 348.0690058 |
| 450.594 | 389.32984 | 385.71013 | 618.036 | 346.817 | 278.766 | 378.1095662 | 349.5983128 |
| 452.352 | 417.45184 | 387.23607 | 618.326 | 307.145 | 279.964 | 407.0652877 | 351.132886 |
| 454.114 | 417.72774 | 388.76702 | 618.616 | 325.105 | 281.165 | 404.2373337 | 352.6716721 |
| 455.881 | 369.75751 | 390.30199 | 618.907 | 325.129 | 282.371 | 370.3513877 | 354.2157245 |
| 457.653 | 422.35671 | 391.84097 | 619.197 | 306.237 | 283.579 | 320.4835518 | 355.7650431 |
| 459.429 | 370.11767 | 393.38396 | 619.487 | 344.527 | 284.792 | 361.0512633 | 357.3185746 |
| 461.211 | 391.95534 | 394.93097 | 619.777 | 316.086 | 286.008 | 435.5681153 | 358.8773724 |
| 462.996 | 385.12925 | 396.48299 | 620.067 | 359.529 | 287.228 | 385.8645851 | 360.4403831 |
| 464.786 | 425.6584 | 398.03802 | 620.358 | 334.993 | 288.452 | 425.1778857 | 362.00866 |
| 466.581 | 468.9234 | 399.59807 | 620.648 | 357.793 | 289.679 | 379.8821721 | 363.5822031 |
| 468.381 | 427.83043 | 401.16113 | 620.938 | 331.243 | 290.91 | 424.4258711 | 365.1599592 |
| 470.185 | 392.07874 | 402.7292 | 621.228 | 361.421 | 292.145 | 395.3711449 | 366.7429816 |
| 471.994 | 436.84861 | 404.30129 | 621.519 | 357.426 | 293.384 | 395.9504278 | 368.3312701 |
| 473.808 | 443.8312 | 405.87739 | 621.809 | 346.521 | 294.626 | 388.3860458 | 369.9248248 |
| 475.626 | 414.3779 | 407.4575 | 622.099 | 384.618 | 295.872 | 415.0888832 | 371.5225925 |
| 477.449 | 442.11365 | 409.04263 | 622.389 | 341.038 | 297.122 | 406.1310623 | 373.1256264 |
| 479.277 | 437.91305 | 410.63177 | 622.679 | 377.92 | 298.375 | 393.7965485 | 374.7328733 |
| 481.109 | 402.53156 | 412.22393 | 622.969 | 311.905 | 299.632 | 431.7954034 | 376.3464396 |
| 482.946 | 400.05154 | 413.82109 | 623.259 | 343.154 | 300.893 | 368.2343718 | 377.9642189 |
| 484.788 | 428.57484 | 415.42228 | 623.549 | 329.608 | 302.158 | 451.1718915 | 379.5862111 |
| 486.634 | 484.97735 | 417.02747 | 623.84 | 349.629 | 303.426 | 434.0788315 | 381.2145229 |
| 488.485 | 467.89307 | 418.63768 | 624.13 | 325.025 | 304.698 | 391.3625069 | 382.8470475 |
| 490.341 | 466.04408 | 420.2519 | 624.42 | 374.994 | 305.974 | 459.4503714 | 384.4848384 |
| 492.202 | 437.11447 | 421.86914 | 624.71 | 366.21 | 307.254 | 414.9372164 | 386.1278955 |
| 494.067 | 439.98977 | 423.49139 | 625 | 373.073 | 308.537 | 406.5586785 | 387.7762188 |
| 495.937 | 466.45241 | 425.11865 | 625.29 | 365.466 | 309.825 | 431.0370694 | 389.4287551 |
| 497.811 | 470.96902 | 426.74893 | 625.58 | 400.192 | 311.116 | 456.2095466 | 391.0865575 |
| 499.69 | 474.37504 | 428.38422 | 625.87 | 337.569 | 312.411 | 454.2599962 | 392.7496262 |
| 501.575 | 455.42773 | 430.02252 | 626.16 | 399.229 | 313.709 | 447.2253947 | 394.4179611 |
| 503.463 | 464.09277 | 431.66584 | 626.45 | 372.59 | 315.012 | 447.666703 | 396.090509 |
| 505.357 | 473.36377 | 433.31317 | 626.74 | 369.158 | 316.318 | 474.6317976 | 397.7693763 |
| 507.255 | 463.77875 | 434.96552 | 627.03 | 387.064 | 317.628 | 441.4104472 | 399.4524566 |
| 509.158 | 453.00389 | 436.62188 | 627.32 | 387.692 | 318.942 | 408.2701963 | 401.140803 |
| 511.066 | 476.35144 | 438.28125 | 627.61 | 386.039 | 320.26 | 493.9156004 | 402.8344157 |
| 512.978 | 480.09254 | 439.94563 | 627.9 | 427.09 | 321.581 | 471.4826049 | 404.5332946 |
| 514.895 | 496.79058 | 441.61503 | 628.19 | 366.626 | 322.907 | 489.2971354 | 406.2363865 |
| 516.817 | 517.75341 | 443.28744 | 628.48 | 383.447 | 324.236 | 467.5940364 | 407.9457978 |
| 518.744 | 502.71274 | 444.96487 | 628.77 | 412.294 | 325.569 | 512.9719029 | 409.6594221 |
| 520.676 | 470.73827 | 446.64631 | 629.059 | 375.478 | 326.906 | 475.5344258 | 411.3783126 |
| 522.612 | 469.59257 | 448.33177 | 629.349 | 413.943 | 328.247 | 450.3945991 | 413.1024693 |
| 524.553 | 514.05745 | 450.02224 | 629.639 | 391.415 | 329.591 | 508.1796531 | 414.8318922 |
| 526.499 | 522.35029 | 451.71572 | 629.929 | 415.352 | 330.94 | 487.6130019 | 416.5665813 |
| 528.449 | 514.14975 | 453.41421 | 630.219 | 393.935 | 332.292 | 465.7003078 | 418.3065367 |
| 530.404 | 520.409 | 455.11773 | 630.509 | 408.7 | 333.649 | 476.6371699 | 420.050705 |
| 532.365 | 509.96722 | 456.82425 | 630.799 | 432.127 | 335.009 | 490.907542 | 421.8011927 |
| 534.329 | 511.41289 | 458.53579 | 631.088 | 410.928 | 336.373 | 476.045248 | 423.5558934 |
| 536.299 | 513.27893 | 460.25134 | 631.378 | 405.343 | 337.741 | 489.2771238 | 425.3169136 |
| 538.274 | 510.47686 | 461.9709 | 631.668 | 392.577 | 339.113 | 461.0144354 | 427.0821467 |
| 540.253 | 521.71623 | 463.69548 | 631.958 | 383.175 | 340.488 | 494.602314 | 428.8526461 |
| 542.237 | 512.25662 | 465.42408 | 632.247 | 414.085 | 341.868 | 499.4830361 | 430.6284116 |
| 544.226 | 520.30366 | 467.15668 | 632.537 | 414.587 | 343.251 | 487.3886614 | 432.4094434 |
| 546.219 | 539.5981 | 468.8933 | 632.827 | 426 | 344.639 | 522.9671669 | 434.1967946 |
| 548.218 | 510.44175 | 470.63494 | 633.117 | 433.758 | 346.03 | 504.2500082 | 435.9883588 |
| 550.221 | 553.86226 | 472.38059 | 633.407 | 414.87 | 347.425 | 521.6727328 | 437.7851891 |
| 552.229 | 547.67122 | 474.13025 | 633.696 | 431.375 | 348.825 | 544.1078348 | 439.5872857 |
| 554.242 | 544.07458 | 475.88493 | 633.986 | 430.632 | 350.228 | 518.7204962 | 441.3946485 |
| 556.259 | 508.55565 | 477.64262 | 634.276 | 431.176 | 351.635 | 490.1344626 | 443.2072775 |
| 558.282 | 550.20542 | 479.40632 | 634.565 | 408.149 | 353.046 | 483.1230324 | 445.0251728 |
| 560.309 | 573.20184 | 481.17304 | 634.855 | 433.159 | 354.461 | 585.3854312 | 446.8483342 |
| 562.341 | 554.34181 | 482.94477 | 635.144 | 421.193 | 355.88 | 501.5842534 | 448.6767618 |
| 564.378 | 558.13809 | 484.72052 | 635.434 | 434.984 | 357.302 | 546.470256 | 450.5104557 |
| 566.42 | 548.92428 | 486.50028 | 635.724 | 438.674 | 358.729 | 553.1320099 | 452.3494157 |
| 568.467 | 537.58458 | 488.28505 | 636.013 | 432.176 | 360.16 | 536.8004437 | 454.193642 |
| 570.518 | 534.05015 | 490.07384 | 636.303 | 424.679 | 361.595 | 544.2163187 | 456.0431344 |
| 572.575 | 565.21098 | 491.86664 | 636.593 | 436.263 | 363.033 | 562.6017062 | 457.8989464 |
| 574.636 | 536.00347 | 493.66446 | 636.882 | 425.633 | 364.476 | 514.8761639 | 459.7589712 |
| 576.702 | 537.69895 | 495.46629 | 637.172 | 421.025 | 365.923 | 517.1353674 | 461.6242623 |
| 578.773 | 535.27411 | 497.27214 | 637.461 | 456.127 | 367.373 | 543.1862483 | 463.4958729 |
| 580.848 | 568.32907 | 499.083 | 637.751 | 460.457 | 368.828 | 519.8906478 | 465.3716964 |
| 582.929 | 544.15183 | 500.89787 | 638.04 | 466.896 | 370.287 | 556.0115727 | 467.2538394 |
| 585.014 | 593.42929 | 502.71776 | 638.33 | 439.568 | 371.749 | 520.9386233 | 469.1401953 |
| 587.105 | 591.25927 | 504.54066 | 638.619 | 455.349 | 373.216 | 514.6444508 | 471.0328707 |
| 589.2 | 559.89779 | 506.36857 | 638.909 | 429.991 | 374.687 | 512.0619021 | 472.9308123 |
| 591.3 | 548.50693 | 508.2015 | 639.198 | 438.95 | 376.161 | 554.4854255 | 474.83402 |
| 593.404 | 564.83577 | 510.03744 | 639.488 | 445.694 | 377.64 | 547.5740533 | 476.7424941 |
| 595.514 | 584.39606 | 511.8794 | 639.777 | 473.124 | 379.123 | 561.0502812 | 478.6572875 |
| 597.629 | 613.4862 | 513.72437 | 640.067 | 450.597 | 380.61 | 576.9647636 | 480.5762939 |
| 599.748 | 553.66662 | 515.57436 | 640.356 | 460.088 | 382.1 | 569.2508212 | 482.5005666 |
| 601.873 | 581.95316 | 517.42836 | 640.646 | 441.366 | 383.595 | 553.272091 | 484.4311586 |
| 604.002 | 599.11369 | 519.28637 | 640.935 | 448.989 | 385.094 | 543.9045591 | 486.3670169 |
| 606.136 | 599.95942 | 521.1494 | 641.225 | 475.022 | 386.597 | 579.3745807 | 488.3081414 |
| 608.275 | 604.90242 | 523.01745 | 641.514 | 447.972 | 388.104 | 609.6858243 | 490.2545321 |
| 610.419 | 603.65037 | 524.8885 | 641.803 | 467.45 | 389.615 | 547.7288798 | 492.2061891 |
| 612.568 | 605.98091 | 526.76557 | 642.093 | 462.264 | 391.13 | 539.5283398 | 494.1631122 |
| 614.721 | 557.77793 | 528.64566 | 642.382 | 479.959 | 392.649 | 551.9334208 | 496.1263548 |
| 616.88 | 581.73144 | 530.53076 | 642.671 | 466.787 | 394.172 | 570.9044107 | 498.0948635 |
| 619.044 | 575.25849 | 532.41987 | 642.961 | 468.685 | 395.7 | 566.762011 | 500.0686385 |
| 621.212 | 581.78662 | 534.314 | 643.25 | 485.272 | 397.231 | 575.7693343 | 502.0476797 |
| 623.385 | 592.81731 | 536.21114 | 643.539 | 468.98 | 398.766 | 592.2125441 | 504.0319871 |
| 625.563 | 602.80965 | 538.1143 | 643.829 | 487.277 | 400.306 | 550.953906 | 506.022614 |
| 627.747 | 593.72224 | 540.02147 | 644.118 | 456.343 | 401.849 | 613.4332583 | 508.0185071 |
| 629.935 | 613.74604 | 541.93265 | 644.407 | 435.551 | 403.397 | 576.0874133 | 510.0196664 |
| 632.128 | 593.36709 | 543.84785 | 644.697 | 479.9 | 404.949 | 600.0728871 | 512.0260918 |
| 634.326 | 615.99732 | 545.76806 | 644.986 | 466.596 | 406.505 | 600.26879 | 514.0377835 |
| 636.528 | 638.82419 | 547.69329 | 645.275 | 494.013 | 408.065 | 574.2495065 | 516.0557947 |
| 638.736 | 620.70455 | 549.62254 | 645.564 | 488.832 | 409.629 | 597.9748296 | 518.079072 |
| 640.949 | 612.92337 | 551.55579 | 645.853 | 501.017 | 411.197 | 592.7328455 | 520.1076156 |
| 643.166 | 614.61184 | 553.49306 | 646.143 | 486.953 | 412.769 | 622.8513458 | 522.1424786 |
| 645.389 | 615.41945 | 555.43535 | 646.432 | 487.292 | 414.346 | 586.0837305 | 524.1826078 |
| 647.616 | 579.62161 | 557.38265 | 646.721 | 535.042 | 415.926 | 589.3287682 | 526.2280033 |
| 649.848 | 598.3141 | 559.33396 | 647.01 | 476.629 | 417.511 | 623.0240775 | 528.2786649 |
| 652.086 | 598.67527 | 561.28929 | 647.299 | 510.356 | 419.1 | 605.8667698 | 530.335646 |
| 654.328 | 630.29258 | 563.24963 | 647.588 | 488.871 | 420.693 | 637.9811632 | 532.3978933 |
| 656.575 | 616.22105 | 565.21399 | 647.878 | 472.89 | 422.29 | 590.9360151 | 534.4654068 |
| 658.827 | 650.22508 | 567.18236 | 648.167 | 492.6 | 423.892 | 625.3612209 | 536.5381865 |
| 661.084 | 636.96216 | 569.15575 | 648.456 | 514.778 | 425.497 | 620.4183575 | 538.6172857 |
| 663.346 | 634.16712 | 571.13415 | 648.745 | 500.119 | 427.107 | 610.0576186 | 540.701651 |
| 665.613 | 619.85681 | 573.11656 | 649.034 | 486.763 | 428.72 | 624.6766138 | 542.7923359 |
| 667.885 | 637.5621 | 575.10299 | 649.323 | 526.413 | 430.338 | 635.7935802 | 544.8872337 |
| 670.162 | 656.32783 | 577.09443 | 649.612 | 510.689 | 431.96 | 640.5500198 | 546.9884509 |
| 672.443 | 661.61494 | 579.08989 | 649.901 | 524.57 | 433.587 | 640.6658764 | 549.0959876 |
| 674.73 | 657.40431 | 581.09036 | 650.19 | 525.427 | 435.217 | 590.756964 | 551.2087905 |
| 677.022 | 644.66007 | 583.09485 | 650.479 | 511.579 | 436.852 | 612.7339058 | 553.3268596 |
| 679.318 | 643.94977 | 585.10435 | 650.768 | 507.429 | 438.49 | 635.0068086 | 555.4501949 |
| 681.62 | 625.30344 | 587.11787 | 651.057 | 526.562 | 440.134 | 628.7652981 | 557.5798497 |
| 683.927 | 638.81315 | 589.1354 | 651.346 | 510.448 | 441.781 | 630.0734244 | 559.7147707 |
| 686.238 | 663.37563 | 591.15794 | 651.635 | 533.534 | 443.432 | 631.9703127 | 561.8560111 |
| 688.555 | 642.00849 | 593.1845 | 651.924 | 504.941 | 445.088 | 646.668301 | 564.0025177 |
| 690.876 | 649.01516 | 595.21607 | 652.213 | 542.65 | 446.747 | 681.3810417 | 566.1542906 |
| 693.203 | 662.18578 | 597.25266 | 652.502 | 526.073 | 448.411 | 666.6862132 | 568.3123829 |
| 695.534 | 674.72335 | 599.29226 | 652.791 | 532.036 | 450.08 | 673.1836614 | 570.4757414 |
| 697.87 | 679.63926 | 601.33788 | 653.08 | 520.472 | 451.752 | 662.5585589 | 572.6454193 |
| 700.212 | 672.00355 | 603.38651 | 653.369 | 516.592 | 453.429 | 652.7244409 | 574.8203635 |
| 702.558 | 693.78203 | 605.44116 | 653.658 | 533.366 | 455.11 | 687.6889064 | 577.0005738 |
| 704.909 | 681.68689 | 607.49882 | 653.947 | 539.63 | 456.795 | 673.3827241 | 579.1871037 |
| 707.266 | 667.82002 | 609.5615 | 654.236 | 527.889 | 458.484 | 695.6587866 | 581.3788997 |
| 709.627 | 702.06985 | 611.62919 | 654.524 | 570.936 | 460.177 | 647.7615659 | 583.5770152 |
| 711.993 | 703.80847 | 613.70089 | 654.813 | 559.33 | 461.875 | 662.9008624 | 585.7803969 |
| 714.365 | 693.3416 | 615.77761 | 655.102 | 544.95 | 463.577 | 645.8130687 | 587.9890448 |
| 716.741 | 691.07828 | 617.85834 | 655.391 | 549.666 | 465.284 | 693.6355092 | 590.2040121 |
| 719.122 | 670.51273 | 619.94309 | 655.68 | 556.674 | 466.994 | 685.0220984 | 592.4252989 |
| 721.508 | 689.79513 | 622.03285 | 655.968 | 526.06 | 468.709 | 680.7406708 | 594.6518519 |
| 723.899 | 710.84925 | 624.12763 | 656.257 | 557.073 | 470.428 | 692.645462 | 596.8836711 |
| 726.296 | 699.20257 | 626.22642 | 656.546 | 552.583 | 472.151 | 703.5412476 | 599.1218098 |
| 728.697 | 678.39624 | 628.33023 | 656.835 | 553.27 | 473.879 | 711.8839753 | 601.3652147 |
| 731.103 | 685.07184 | 630.43805 | 657.124 | 555.876 | 475.611 | 702.4111193 | 603.614939 |
| 733.514 | 688.26718 | 632.54988 | 657.412 | 567.78 | 477.347 | 704.5723713 | 605.8699295 |
| 735.93 | 726.35744 | 634.66673 | 657.701 | 584.676 | 479.087 | 703.7266182 | 608.1312395 |
| 738.351 | 717.74858 | 636.7886 | 657.99 | 558.727 | 480.832 | 709.4341351 | 610.3988689 |
| 740.778 | 711.87156 | 638.91448 | 658.278 | 578.873 | 482.58 | 705.5982287 | 612.6707113 |
| 743.209 | 714.05563 | 641.04437 | 658.567 | 566.058 | 484.334 | 709.5004893 | 614.9499264 |
| 745.645 | 708.54078 | 643.17928 | 658.856 | 575.753 | 486.091 | 711.8492184 | 617.2333545 |
| 748.086 | 733.82059 | 645.3192 | 659.144 | 565.682 | 487.853 | 710.0271102 | 619.5241552 |
| 750.533 | 715.20334 | 647.46314 | 659.433 | 580.875 | 489.619 | 686.0289975 | 621.8202222 |
| 752.984 | 725.48863 | 649.61109 | 659.722 | 565.265 | 491.389 | 714.5149732 | 624.1215554 |
| 755.44 | 769.77192 | 651.76406 | 660.01 | 584.597 | 493.164 | 734.6171447 | 626.429208 |
| 757.901 | 733.72227 | 653.92204 | 660.299 | 594.45 | 494.943 | 706.5008569 | 628.7421268 |
| 760.367 | 734.60713 | 656.08404 | 660.588 | 574.477 | 496.726 | 705.372835 | 631.0624183 |
| 762.839 | 747.63228 | 658.25105 | 660.876 | 562.867 | 498.514 | 756.0200698 | 633.3869228 |
| 765.315 | 731.27636 | 660.42208 | 661.165 | 583.284 | 500.306 | 708.5336134 | 635.7177468 |
| 767.796 | 742.21375 | 662.59812 | 661.453 | 598.223 | 502.102 | 741.7391653 | 638.0548901 |
| 770.283 | 743.22703 | 664.77817 | 661.742 | 590.65 | 503.903 | 725.3665227 | 640.3972997 |
| 772.774 | 729.10031 | 666.96224 | 662.03 | 602.459 | 505.708 | 731.272049 | 642.7460288 |
| 775.27 | 753.55645 | 669.15233 | 662.319 | 596.21 | 507.517 | 774.4286287 | 645.100024 |
| 777.772 | 766.02079 | 671.34542 | 662.607 | 591.563 | 509.331 | 767.1538881 | 647.4603387 |
| 780.278 | 755.29207 | 673.54454 | 662.896 | 598.906 | 511.149 | 709.1423871 | 649.8269729 |
| 782.789 | 730.37745 | 675.74767 | 663.184 | 612.577 | 512.971 | 756.7183691 | 652.1988733 |
| 785.306 | 768.45968 | 677.95481 | 663.473 | 619.068 | 514.797 | 768.1702663 | 654.5770931 |
| 787.827 | 757.14807 | 680.16697 | 663.761 | 593.346 | 516.628 | 741.3252413 | 656.9605791 |
| 790.354 | 757.4952 | 682.38314 | 664.05 | 612.418 | 518.464 | 782.3490064 | 659.3503846 |
| 792.885 | 739.57722 | 684.60433 | 664.338 | 590.478 | 520.303 | 750.8539192 | 661.7465095 |
| 795.422 | 774.67479 | 686.83053 | 664.627 | 615.439 | 522.147 | 755.1953815 | 664.1479006 |
| 797.963 | 749.66586 | 689.06075 | 664.915 | 602.544 | 523.996 | 773.4459541 | 666.5556112 |
| 800.51 | 771.17747 | 691.29498 | 665.203 | 619.784 | 525.848 | 760.2498886 | 668.9696413 |
| 803.062 | 752.30841 | 693.53422 | 665.492 | 619.097 | 527.705 | 739.8443834 | 671.3889375 |
| 805.618 | 764.34436 | 695.77849 | 665.78 | 619.168 | 529.567 | 760.4299929 | 673.8145532 |
| 808.18 | 755.1155 | 698.02676 | 666.069 | 597.449 | 531.433 | 776.2791744 | 676.2454351 |
| 810.747 | 750.09425 | 700.28005 | 666.357 | 606.199 | 533.303 | 776.7468137 | 678.6826364 |
| 813.318 | 782.77901 | 702.53736 | 666.645 | 605.99 | 535.178 | 800.4257942 | 681.1261572 |
| 815.895 | 775.50347 | 704.79968 | 666.934 | 606.797 | 537.056 | 782.5133122 | 683.5749443 |
| 818.477 | 781.06948 | 707.06601 | 667.222 | 617.08 | 538.94 | 737.3513602 | 686.0300507 |
| 821.064 | 780.72035 | 709.33736 | 667.51 | 648.797 | 540.827 | 788.8569871 | 688.4914766 |
| 823.656 | 793.39035 | 711.61373 | 667.799 | 629.517 | 542.72 | 773.2532109 | 690.9581687 |
| 826.253 | 788.9871 | 713.89411 | 668.087 | 617.288 | 544.616 | 802.0488397 | 693.4322336 |
| 828.855 | 805.49151 | 716.1785 | 668.375 | 629.387 | 546.517 | 800.3941969 | 695.9105113 |
| 831.462 | 783.96084 | 718.46891 | 668.663 | 630.524 | 548.423 | 785.5066251 | 698.3961618 |
| 834.074 | 790.26223 | 720.76233 | 668.952 | 645.638 | 550.332 | 768.8696189 | 700.8870785 |
| 836.691 | 808.46012 | 723.06077 | 669.24 | 633.745 | 552.246 | 812.5717777 | 703.3843146 |
| 839.313 | 798.46176 | 725.36422 | 669.528 | 627.784 | 554.165 | 791.5838299 | 705.886817 |
| 841.94 | 807.00942 | 727.67169 | 669.816 | 654.782 | 556.088 | 804.0489457 | 708.396692 |
| 844.572 | 796.66595 | 729.98417 | 670.105 | 605.28 | 558.015 | 806.1038203 | 710.9118332 |
| 847.21 | 807.7147 | 732.30167 | 670.393 | 651.598 | 559.947 | 821.5411843 | 713.4322407 |
| 849.852 | 802.64329 | 734.62319 | 670.681 | 652.37 | 561.883 | 796.1285679 | 715.9600208 |
| 852.499 | 806.96528 | 736.94871 | 670.969 | 641.198 | 563.824 | 796.1001304 | 718.4930672 |
| 855.152 | 795.54934 | 739.28026 | 671.257 | 654.224 | 565.769 | 814.9415716 | 721.032433 |
| 857.809 | 790.8702 | 741.61481 | 671.545 | 639.45 | 567.718 | 856.5393535 | 723.577065 |
| 860.472 | 814.33312 | 743.95538 | 671.833 | 654.908 | 569.672 | 822.9714865 | 726.1290697 |
| 863.139 | 803.00045 | 746.29997 | 672.122 | 648.511 | 571.631 | 818.7711585 | 728.6863406 |
| 865.812 | 847.67702 | 748.64857 | 672.41 | 652.351 | 573.594 | 817.1281014 | 731.2499309 |
| 868.489 | 820.28839 | 751.00219 | 672.698 | 664.444 | 575.561 | 838.3635611 | 733.8198407 |
| 871.172 | 826.03198 | 753.35982 | 672.986 | 666.825 | 577.533 | 799.9908053 | 736.3950168 |
| 873.86 | 836.94128 | 755.72346 | 673.274 | 667.512 | 579.509 | 845.4824219 | 738.9765122 |
| 876.552 | 830.5496 | 758.09012 | 673.562 | 650.141 | 581.489 | 874.0947866 | 741.5643271 |
| 879.25 | 828.86816 | 760.4618 | 673.85 | 667.587 | 583.474 | 846.676798 | 744.1584615 |
| 881.953 | 840.66934 | 762.83849 | 674.138 | 669.431 | 585.464 | 832.1167844 | 746.7578621 |
| 884.661 | 848.95716 | 765.2202 | 674.426 | 690.62 | 587.458 | 860.7702256 | 749.3635821 |
| 887.374 | 832.838 | 767.60592 | 674.714 | 662.559 | 589.456 | 856.0222119 | 751.9756216 |
| 890.092 | 846.92358 | 769.99565 | 675.002 | 681.262 | 591.459 | 842.6755327 | 754.5939805 |
| 892.815 | 821.85346 | 772.3904 | 675.29 | 672.655 | 593.466 | 840.0066182 | 757.2186589 |
| 895.543 | 826.33696 | 774.79016 | 675.578 | 684.259 | 595.478 | 841.2210059 | 759.8486035 |
| 898.276 | 844.85288 | 777.19394 | 675.866 | 681.67 | 597.495 | 858.7216704 | 762.4859208 |
| 901.014 | 845.92936 | 779.60274 | 676.154 | 693.9 | 599.515 | 880.6617488 | 765.1285042 |
| 903.757 | 878.99135 | 782.01554 | 676.442 | 677.545 | 601.54 | 841.2009943 | 767.7774072 |
| 906.506 | 867.50919 | 784.43337 | 676.73 | 684.405 | 603.57 | 866.7652775 | 770.4315763 |
| 909.259 | 847.91479 | 786.85621 | 677.018 | 701.997 | 605.605 | 846.2502351 | 773.0931182 |
| 912.017 | 858.45891 | 789.28306 | 677.305 | 681.775 | 607.643 | 874.0968931 | 775.7599262 |
| 914.781 | 877.28483 | 791.71493 | 677.593 | 699.199 | 609.686 | 886.0427607 | 778.4330537 |
| 917.549 | 862.56118 | 794.15081 | 677.881 | 691.931 | 611.734 | 877.1228566 | 781.1125007 |
| 920.323 | 889.09003 | 796.59171 | 678.169 | 695.582 | 613.786 | 859.1103166 | 783.7982671 |
| 923.101 | 881.2306 | 799.03662 | 678.457 | 698.48 | 615.843 | 868.0607648 | 786.4903529 |
| 925.885 | 865.34921 | 801.48655 | 678.745 | 711.591 | 617.904 | 898.1876911 | 789.187705 |
| 928.673 | 872.71604 | 803.94149 | 679.032 | 711.03 | 619.97 | 902.5091419 | 791.8924297 |
| 931.467 | 880.81927 | 806.40045 | 679.32 | 713.246 | 622.04 | 858.3698877 | 794.6024207 |
| 934.266 | 890.97312 | 808.86442 | 679.608 | 695.81 | 624.114 | 934.7341257 | 797.3187311 |
| 937.07 | 868.04894 | 811.33241 | 679.896 | 694.693 | 626.194 | 898.2940685 | 800.0413609 |
| 939.879 | 882.48365 | 813.80541 | 680.183 | 702.217 | 628.277 | 896.222342 | 802.7703103 |
| 942.692 | 891.01325 | 816.28343 | 680.471 | 716.336 | 630.365 | 887.0486066 | 805.5045258 |
| 945.511 | 880.43502 | 818.76546 | 680.759 | 710.987 | 632.458 | 847.2508148 | 808.246114 |
| 948.335 | 902.45829 | 821.25251 | 681.047 | 727.797 | 634.555 | 834.2780365 | 810.9929684 |
| 951.164 | 894.8286 | 823.74357 | 681.334 | 727.546 | 636.657 | 814.359129 | 813.7471955 |
| 953.999 | 871.78704 | 826.23965 | 681.622 | 730.418 | 638.763 | 829.9355208 | 816.5066888 |
| 956.838 | 868.46429 | 828.73974 | 681.91 | 732.556 | 640.874 | 830.6748965 | 819.2725016 |
| 959.682 | 817.98795 | 831.24485 | 682.197 | 733.954 | 642.989 | 835.9432116 | 822.0446338 |
| 962.531 | 817.60772 | 833.75497 | 682.485 | 736.218 | 645.109 | 848.7801218 | 824.8230855 |
| 965.385 | 842.70493 | 836.2691 | 682.773 | 758.695 | 647.234 | 860.0782457 | 827.6078566 |
| 968.245 | 844.16265 | 838.78826 | 683.06 | 776.317 | 649.362 | 861.1999482 | 830.3989472 |
| 971.109 | 865.12046 | 841.31142 | 683.348 | 758.419 | 651.496 | 853.8809714 | 833.195304 |
| 973.978 | 901.89948 | 843.8396 | 683.636 | 750.476 | 653.634 | 894.5634863 | 835.9990334 |
| 976.853 | 920.77757 | 846.3718 | 683.923 | 736.588 | 655.777 | 897.4999242 | 838.8080291 |
| 979.732 | 922.96765 | 848.91001 | 684.211 | 740.387 | 657.923 | 914.452903 | 841.6243975 |
| 982.617 | 962.77748 | 851.45123 | 684.498 | 737.678 | 660.075 | 887.216072 | 844.4460321 |
| 985.506 | 953.32188 | 853.99848 | 684.786 | 745.537 | 662.231 | 882.7650725 | 847.2739861 |
| 988.401 | 937.70534 | 856.54873 | 685.073 | 748.455 | 664.392 | 908.9318099 | 850.1093128 |
| 991.301 | 946.64126 | 859.105 | 685.361 | 757.355 | 666.557 | 966.7337158 | 852.9499057 |
| 994.205 | 946.26103 | 861.66528 | 685.649 | 744.935 | 668.727 | 945.258117 | 855.7968181 |
| 997.115 | 952.01164 | 864.22958 | 685.936 | 751.414 | 670.901 | 963.9299863 | 858.65005 |
| 1000.03 | 899.90302 | 866.7999 | 686.223 | 751.249 | 673.08 | 855.051123 | 861.5096012 |
| 1002.95 | 923.93579 | 869.37323 | 686.511 | 755.569 | 675.264 | 919.0229189 | 864.375472 |
| 1005.87 | 950.91108 | 871.95257 | 686.798 | 779.377 | 677.452 | 942.3174661 | 867.2476622 |
| 1008.8 | 944.48027 | 874.53593 | 687.086 | 754.7 | 679.644 | 941.3210994 | 870.1251186 |
| 1011.74 | 911.01698 | 877.12331 | 687.373 | 775.186 | 681.841 | 976.1928797 | 873.0099476 |
| 1014.68 | 883.05049 | 879.71569 | 687.661 | 755.331 | 684.043 | 934.4666023 | 875.9010962 |
| 1017.62 | 940.06097 | 882.3131 | 687.948 | 777.865 | 686.249 | 897.8137903 | 878.7985642 |
| 1020.57 | 963.38645 | 884.91452 | 688.236 | 764.195 | 688.46 | 900.9766752 | 881.7012984 |
| 1023.53 | 929.69041 | 887.52095 | 688.523 | 754.267 | 690.676 | 925.5193138 | 884.6114052 |
| 1026.49 | 944.9538 | 890.1324 | 688.81 | 764.252 | 692.896 | 931.8461369 | 887.5278316 |
| 1029.45 | 936.39611 | 892.74787 | 689.098 | 801.089 | 695.12 | 921.8466599 | 890.4505774 |
| 1032.42 | 945.05713 | 895.36834 | 689.385 | 790.545 | 697.349 | 989.0213639 | 893.3785894 |
| 1035.4 | 965.60061 | 897.99284 | 689.672 | 785.279 | 699.583 | 921.4274697 | 896.313974 |
| 1038.38 | 946.86799 | 900.62134 | 689.96 | 790.981 | 701.822 | 923.5971477 | 899.2556782 |
| 1041.37 | 987.99307 | 903.25587 | 690.247 | 806.548 | 704.064 | 958.1940319 | 902.2026485 |
| 1044.36 | 961.06192 | 905.89441 | 690.534 | 810.62 | 706.312 | 925.023237 | 905.1569916 |
| 1047.35 | 941.1204 | 908.53696 | 690.822 | 796.031 | 708.564 | 963.9763289 | 908.1176541 |
| 1050.35 | 969.46913 | 911.18452 | 691.109 | 798.061 | 710.821 | 965.397152 | 911.0835828 |
| 1053.36 | 965.43407 | 913.83711 | 691.396 | 795.904 | 713.082 | 940.8724185 | 914.0568841 |
| 1056.37 | 956.52123 | 916.49471 | 691.683 | 775.901 | 715.348 | 953.7651504 | 917.036505 |
| 1059.38 | 990.51824 | 919.15532 | 691.971 | 802.404 | 717.618 | 919.73807 | 920.021392 |
| 1062.4 | 984.42352 | 921.82195 | 692.258 | 783.909 | 719.893 | 978.5900579 | 923.0136518 |
| 1065.43 | 971.38834 | 924.49259 | 692.545 | 811.653 | 722.173 | 930.6001519 | 926.012231 |
| 1068.46 | 969.8935 | 927.16825 | 692.832 | 823.2 | 724.457 | 951.3532269 | 929.0171296 |
| 1071.49 | 1006.27623 | 929.84792 | 693.119 | 801.824 | 726.746 | 967.6131726 | 932.0272944 |
| 1074.54 | 993.08354 | 932.5316 | 693.407 | 809.179 | 729.039 | 970.5991129 | 935.044832 |
| 1077.58 | 992.45752 | 935.22131 | 693.694 | 848.267 | 731.337 | 953.4744557 | 938.068689 |
| 1080.63 | 1037.40697 | 937.91502 | 693.981 | 832.616 | 733.64 | 973.0879231 | 941.0988654 |
| 1083.69 | 986.97779 | 940.61275 | 694.268 | 840.02 | 735.947 | 965.6172795 | 944.1353613 |
| 1086.75 | 1001.58405 | 943.3165 | 694.555 | 805.406 | 738.259 | 981.087294 | 947.1781767 |
| 1089.81 | 981.8211 | 946.02326 | 694.842 | 816.538 | 740.576 | 1005.879551 | 950.2273115 |
| 1092.88 | 983.40322 | 948.73504 | 695.129 | 820.579 | 742.897 | 955.1764943 | 953.2827657 |
| 1095.96 | 982.36586 | 951.45183 | 695.416 | 838.767 | 745.222 | 1000.594384 | 956.3445394 |
| 1099.04 | 1004.50049 | 954.17363 | 695.703 | 847.189 | 747.553 | 1017.562108 | 959.4136858 |
| 1102.12 | 1026.30103 | 956.89945 | 695.99 | 829.259 | 749.888 | 1031.947284 | 962.4880984 |
| 1105.22 | 1014.90415 | 959.62929 | 696.277 | 831.685 | 752.227 | 988.0734464 | 965.5688304 |
| 1108.31 | 1034.28687 | 962.36414 | 696.564 | 878.651 | 754.571 | 977.4557165 | 968.6569352 |
| 1111.41 | 1022.7997 | 965.104 | 696.851 | 886.542 | 756.92 | 1021.810885 | 971.7503061 |
| 1114.52 | 1020.20129 | 967.84788 | 697.138 | 861.363 | 759.274 | 1019.913997 | 974.8510497 |
| 1117.63 | 1024.67577 | 970.59678 | 697.425 | 843.646 | 761.631 | 1057.675872 | 977.9581128 |
| 1120.74 | 1037.99889 | 973.35069 | 697.712 | 852.374 | 763.994 | 1026.840115 | 981.0704421 |
| 1123.87 | 993.44371 | 976.10761 | 697.999 | 845.571 | 766.361 | 1013.971607 | 984.1901441 |
| 1126.99 | 1046.06499 | 978.87055 | 698.286 | 841.921 | 768.733 | 1069.609101 | 987.3161656 |
| 1130.12 | 1030.05317 | 981.63751 | 698.573 | 844.525 | 771.11 | 990.3716198 | 990.4485064 |
| 1133.26 | 1056.44859 | 984.40948 | 698.86 | 841.873 | 773.491 | 1049.738643 | 993.5871668 |
| 1136.4 | 1051.08122 | 987.18546 | 699.147 | 867.138 | 775.877 | 1024.375529 | 996.7321466 |
| 1139.54 | 1069.68141 | 989.96646 | 699.434 | 867.774 | 778.267 | 1031.034123 | 999.8844991 |
| 1142.69 | 1099.86909 | 992.75147 | 699.721 | 896.988 | 780.662 | 1056.264529 | 1003.042118 |
| 1145.85 | 1073.09245 | 995.5415 | 700.008 | 871.895 | 783.062 | 1029.967189 | 1006.207109 |
| 1149.01 | 1090.36835 | 998.33654 | 700.294 | 873.633 | 785.466 | 1073.906327 | 1009.377367 |
| 1152.18 | 1083.04465 | 1001.1356 | 700.581 | 895.504 | 787.875 | 1041.320082 | 1012.554997 |
| 1155.35 | 1053.73982 | 1003.93867 | 700.868 | 900.466 | 790.289 | 1042.213231 | 1015.738947 |
| 1158.52 | 1045.99476 | 1006.74776 | 701.155 | 886.988 | 792.707 | 1040.751331 | 1018.929216 |
| 1161.7 | 1084.57962 | 1009.55685 | 701.442 | 879.104 | 795.13 | 1079.27786 | 1022.125805 |
| 1164.89 | 1063.76225 | 1012.37597 | 701.729 | 890.167 | 797.558 | 1049.19201 | 1025.328713 |
| 1168.08 | 1073.75459 | 1015.1951 | 702.015 | 901.681 | 799.99 | 1078.603786 | 1028.53794 |
| 1171.27 | 1092.4852 | 1018.02425 | 702.302 | 920.32 | 802.427 | 1102.628229 | 1031.75454 |
| 1174.47 | 1133.42769 | 1020.85341 | 702.589 | 891.87 | 804.868 | 1120.838779 | 1034.976407 |
| 1177.68 | 1102.95909 | 1023.69259 | 702.876 | 880.225 | 807.314 | 1108.473721 | 1038.205646 |
| 1180.89 | 1131.17038 | 1026.53178 | 703.162 | 903.564 | 809.765 | 1158.702819 | 1041.441205 |
| 1184.11 | 1173.10608 | 1029.381 | 703.449 | 901.721 | 812.221 | 1152.720406 | 1044.683083 |
| 1187.33 | 1143.58054 | 1032.23022 | 703.736 | 934.084 | 814.681 | 1120.891441 | 1047.93128 |
| 1190.55 | 1104.49406 | 1035.07944 | 704.022 | 899.301 | 817.146 | 1164.948542 | 1051.185797 |
| 1193.78 | 1127.1273 | 1037.93869 | 704.309 | 957.386 | 819.615 | 1125.409848 | 1054.44242 |
| 1197.02 | 1097.14026 | 1040.79794 | 704.596 | 914.678 | 822.089 | 1092.327525 | 1057.718002 |
| 1200.26 | 1089.93695 | 1043.66723 | 704.882 | 956.66 | 824.568 | 1117.721183 | 1060.993584 |
| 1203.5 | 1086.63627 | 1046.54655 | 705.169 | 946.471 | 827.051 | 1111.359603 | 1064.269166 |
| 1206.75 | 1096.70887 | 1049.41583 | 705.456 | 914.6 | 829.539 | 1146.411488 | 1067.55528 |
| 1210.01 | 1156.14119 | 1052.30518 | 705.742 | 983.688 | 832.032 | 1111.443863 | 1070.851926 |
| 1213.27 | 1084.06796 | 1055.1845 | 706.029 | 924.136 | 834.529 | 1115.393519 | 1074.148573 |
| 1216.53 | 1143.33976 | 1058.07385 | 706.315 | 934.243 | 837.031 | 1202.064781 | 1077.455752 |
| 1219.8 | 1151.96768 | 1060.97323 | 706.602 | 917.3 | 839.538 | 1194.060143 | 1080.773464 |
| 1223.08 | 1174.63102 | 1063.87261 | 706.888 | 954.443 | 842.049 | 1195.450423 | 1084.091175 |
| 1226.36 | 1167.47787 | 1066.78203 | 707.175 | 946.208 | 844.565 | 1219.348477 | 1087.419419 |
| 1229.64 | 1157.80658 | 1069.68141 | 707.462 | 970.071 | 847.086 | 1219.401139 | 1090.747663 |
| 1232.93 | 1132.8458 | 1072.60086 | 707.748 | 976.461 | 849.611 | 1202.886309 | 1094.086439 |
| 1236.22 | 1164.64872 | 1075.5203 | 708.034 | 954.942 | 852.141 | 1200.042556 | 1097.435748 |
| 1239.53 | 1174.01904 | 1078.43975 | 708.321 | 969.948 | 854.676 | 1182.6746 | 1100.785056 |
| 1242.83 | 1179.0754 | 1081.36923 | 708.607 | 997.297 | 857.215 | 1242.382874 | 1104.144897 |
| 1246.14 | 1180.79095 | 1084.29871 | 708.894 | 961.225 | 859.759 | 1218.821856 | 1107.504738 |
| 1249.45 | 1160.75612 | 1087.22819 | 709.18 | 992.476 | 862.308 | 1207.057146 | 1110.875112 |
| 1252.77 | 1179.18576 | 1090.17773 | 709.467 | 965.025 | 864.861 | 1217.715953 | 1114.256018 |
| 1256.1 | 1191.49559 | 1093.11725 | 709.753 | 986.803 | 867.419 | 1241.287502 | 1117.636924 |
| 1259.42 | 1205.5912 | 1096.06679 | 710.04 | 965.839 | 869.982 | 1263.27919 | 1121.038895 |
| 1262.76 | 1181.29257 | 1099.02637 | 710.326 | 988.904 | 872.549 | 1222.402878 | 1124.430333 |
| 1266.1 | 1205.32033 | 1101.97591 | 710.612 | 983.116 | 875.121 | 1244.257644 | 1127.832304 |
| 1269.44 | 1164.36781 | 1104.94552 | 710.899 | 973.215 | 877.698 | 1222.581929 | 1131.244807 |
| 1272.79 | 1211.58058 | 1107.91513 | 711.185 | 977.644 | 880.279 | 1213.187013 | 1134.667843 |
| 1276.14 | 1209.23299 | 1110.88474 | 711.471 | 979.475 | 882.865 | 1250.661354 | 1138.090878 |
| 1279.5 | 1226.36844 | 1113.86438 | 711.758 | 1016.45 | 885.456 | 1213.029027 | 1141.513914 |
| 1282.87 | 1234.09343 | 1116.84402 | 712.044 | 981.452 | 888.051 | 1275.865428 | 1144.958014 |
| 1286.23 | 1230.52188 | 1119.82367 | 712.33 | 996.481 | 890.651 | 1274.243436 | 1148.402115 |
| 1289.61 | 1195.6691 | 1122.81334 | 712.616 | 980.377 | 893.256 | 1270.051534 | 1151.846215 |
| 1292.99 | 1254.09817 | 1125.81305 | 712.903 | 1004.21 | 895.865 | 1291.811508 | 1155.300848 |
| 1296.37 | 1217.69035 | 1128.81275 | 713.189 | 1028.09 | 898.479 | 1289.936738 | 1158.766013 |
| 1299.76 | 1229.51863 | 1131.81246 | 713.475 | 1008.21 | 901.098 | 1242.951624 | 1162.231179 |
| 1303.15 | 1252.32242 | 1134.8222 | 713.761 | 1000.37 | 903.722 | 1273.379778 | 1165.706876 |
| 1306.55 | 1226.45873 | 1137.84197 | 714.048 | 1013.65 | 906.35 | 1302.912676 | 1169.193106 |
| 1309.95 | 1236.3708 | 1140.85171 | 714.334 | 1014.64 | 908.983 | 1300.542882 | 1172.679336 |
| 1313.36 | 1255.95417 | 1143.88151 | 714.62 | 1005.57 | 911.62 | 1307.325759 | 1176.176099 |
| 1316.77 | 1249.0719 | 1146.90129 | 714.906 | 1031.6 | 914.262 | 1298.963019 | 1179.683394 |
| 1320.19 | 1232.51834 | 1149.93109 | 715.192 | 1047.94 | 916.909 | 1274.559408 | 1183.190689 |
| 1323.61 | 1251.70041 | 1152.97093 | 715.479 | 1013.46 | 919.561 | 1277.887652 | 1186.708516 |
| 1327.04 | 1261.05067 | 1156.01076 | 715.765 | 1009.68 | 922.217 | 1294.886974 | 1190.226344 |
| 1330.47 | 1269.97956 | 1159.06063 | 716.051 | 1034.13 | 924.878 | 1320.14371 | 1193.754703 |
| 1333.91 | 1289.28202 | 1162.10047 | 716.337 | 1009.34 | 927.543 | 1337.627523 | 1197.293596 |
| 1337.35 | 1266.60865 | 1165.16037 | 716.623 | 1049.63 | 930.213 | 1301.048438 | 1200.832488 |
| 1340.8 | 1245.30973 | 1168.22027 | 716.909 | 1028.28 | 932.888 | 1315.941276 | 1204.381912 |
| 1344.25 | 1264.863 | 1171.28017 | 717.195 | 1038.09 | 935.568 | 1282.01636 | 1207.931337 |
| 1347.7 | 1274.60452 | 1174.35011 | 717.481 | 1052.31 | 938.252 | 1334.615252 | 1211.491294 |
| 1351.16 | 1252.40268 | 1177.42004 | 717.767 | 1041.77 | 940.941 | 1304.482006 | 1215.061783 |
| 1354.63 | 1320.57328 | 1180.48998 | 718.053 | 1041.87 | 943.635 | 1298.057231 | 1218.632273 |
| 1358.1 | 1290.91731 | 1183.56994 | 718.339 | 1056.23 | 946.333 | 1359.450691 | 1222.213295 |
| 1361.58 | 1297.60897 | 1186.65994 | 718.625 | 1047.31 | 949.036 | 1313.276574 | 1225.804849 |
| 1365.06 | 1285.89105 | 1189.74994 | 718.911 | 1003.36 | 951.744 | 1316.288846 | 1229.396403 |
| 1368.54 | 1325.76006 | 1192.83994 | 719.197 | 1053.92 | 954.456 | 1405.793327 | 1232.99849 |
| 1372.03 | 1298.98341 | 1195.93997 | 719.483 | 1056.63 | 957.173 | 1356.48055 | 1236.611109 |
| 1375.53 | 1309.29679 | 1199.04 | 719.769 | 1058.44 | 959.895 | 1366.876046 | 1240.223728 |
| 1379.03 | 1282.46998 | 1202.15007 | 720.055 | 1089.94 | 962.622 | 1357.923491 | 1243.836347 |
| 1382.53 | 1300.2776 | 1205.26013 | 720.341 | 1101.74 | 965.353 | 1384.76009 | 1247.470031 |
| 1386.04 | 1321.79724 | 1208.38023 | 720.627 | 1045.46 | 968.089 | 1362.241782 | 1251.103715 |
| 1389.56 | 1308.79516 | 1211.50032 | 720.913 | 1053.78 | 970.829 | 1377.471657 | 1254.737399 |
| 1393.08 | 1309.35698 | 1214.62042 | 721.199 | 1077.52 | 973.574 | 1416.26255 | 1258.381616 |
| 1396.6 | 1302.81581 | 1217.76058 | 721.485 | 1093.98 | 976.324 | 1438.138381 | 1262.036364 |
| 1400.13 | 1315.96837 | 1220.89071 | 721.77 | 1108.94 | 979.079 | 1393.501996 | 1265.701646 |
| 1403.66 | 1317.77422 | 1224.03087 | 722.056 | 1100.84 | 981.838 | 1363.45301 | 1269.366927 |
| 1407.2 | 1338.73204 | 1227.17103 | 722.342 | 1089.86 | 984.602 | 1385.255114 | 1273.04274 |
| 1410.74 | 1340.50778 | 1230.32123 | 722.628 | 1096.11 | 987.371 | 1382.92745 | 1276.718554 |
| 1414.29 | 1329.52223 | 1233.47142 | 722.913 | 1111.34 | 990.144 | 1388.709747 | 1280.4049 |
| 1417.85 | 1338.56148 | 1236.63165 | 723.199 | 1108.77 | 992.922 | 1407.194139 | 1284.091246 |
| 1421.4 | 1299.37468 | 1239.79187 | 723.485 | 1071.72 | 995.705 | 1381.568768 | 1287.798657 |
| 1424.97 | 1332.24103 | 1242.9521 | 723.771 | 1102.1 | 998.492 | 1383.706848 | 1291.495535 |
| 1428.53 | 1356.14839 | 1246.12236 | 724.057 | 1117.17 | 1001.28 | 1431.429231 | 1295.213479 |
| 1432.11 | 1355.32573 | 1249.30265 | 724.342 | 1109.61 | 1004.08 | 1425.899712 | 1298.931422 |
| 1435.68 | 1305.45435 | 1252.47291 | 724.628 | 1119.48 | 1006.88 | 1431.850528 | 1302.659898 |
| 1439.26 | 1339.23366 | 1255.66323 | 724.914 | 1115.47 | 1009.69 | 1401.62249 | 1306.388373 |
| 1442.85 | 1365.89995 | 1258.85356 | 725.199 | 1116.32 | 1012.5 | 1422.002718 | 1310.127382 |
| 1446.44 | 1373.47446 | 1262.04388 | 725.485 | 1112.78 | 1015.32 | 1424.562095 | 1313.876922 |
| 1450.04 | 1350.23927 | 1265.2342 | 725.771 | 1101.85 | 1018.13 | 1395.239845 | 1317.626463 |
| 1453.64 | 1335.88282 | 1268.43456 | 726.056 | 1096.11 | 1020.96 | 1456.770227 | 1321.386536 |
| 1457.24 | 1388.17202 | 1271.64495 | 726.342 | 1130.72 | 1023.79 | 1418.937784 | 1325.146609 |
| 1460.86 | 1373.64501 | 1274.85534 | 726.628 | 1134.19 | 1026.62 | 1434.452035 | 1328.917214 |
| 1464.47 | 1357.21184 | 1278.06572 | 726.913 | 1148.68 | 1029.46 | 1488.588659 | 1332.698352 |
| 1468.09 | 1366.00027 | 1281.28615 | 727.199 | 1121.12 | 1032.31 | 1424.404109 | 1336.479489 |
| 1471.71 | 1363.83326 | 1284.50657 | 727.484 | 1129.41 | 1035.15 | 1491.211231 | 1340.281692 |
| 1475.34 | 1380.13602 | 1287.73702 | 727.77 | 1114.91 | 1038.01 | 1484.933911 | 1344.073362 |
| 1478.98 | 1397.09088 | 1290.96747 | 728.055 | 1141.64 | 1040.87 | 1470.525564 | 1347.875565 |
| 1482.62 | 1378.03923 | 1294.20796 | 728.341 | 1131.72 | 1043.73 | 1431.050064 | 1351.6883 |
| 1486.26 | 1366.71258 | 1297.44845 | 728.626 | 1142.73 | 1046.59 | 1512.202339 | 1355.511567 |
| 1489.91 | 1393.05783 | 1300.68893 | 728.912 | 1174.68 | 1049.47 | 1482.206015 | 1359.334835 |
| 1493.56 | 1363.94362 | 1303.93945 | 729.197 | 1118 | 1052.34 | 1494.497345 | 1363.168635 |
| 1497.22 | 1409.11981 | 1307.18997 | 729.483 | 1161.87 | 1055.22 | 1443.530978 | 1367.002435 |
| 1500.88 | 1390.69017 | 1310.45052 | 729.768 | 1173.08 | 1058.11 | 1468.80878 | 1370.846767 |
| 1504.55 | 1396.78991 | 1313.71107 | 730.054 | 1168.26 | 1061 | 1467.23945 | 1374.701632 |
| 1508.22 | 1383.76777 | 1316.98165 | 730.339 | 1157.27 | 1063.89 | 1532.266594 | 1378.556496 |
| 1511.9 | 1377.92887 | 1320.25224 | 730.625 | 1189.27 | 1066.79 | 1491.03218 | 1382.421894 |
| 1515.58 | 1398.10416 | 1323.53285 | 730.91 | 1171.39 | 1069.7 | 1493.55996 | 1386.287291 |
| 1519.27 | 1425.3724 | 1326.81347 | 731.195 | 1155.17 | 1072.61 | 1515.130351 | 1390.16322 |
| 1522.96 | 1452.46005 | 1330.09408 | 731.481 | 1156.45 | 1075.52 | 1476.486912 | 1394.049682 |
| 1526.65 | 1409.28033 | 1333.38473 | 731.766 | 1153.71 | 1078.44 | 1512.191806 | 1397.946677 |
| 1530.35 | 1411.38714 | 1336.67538 | 732.052 | 1158.48 | 1081.36 | 1487.303704 | 1401.843671 |
| 1534.06 | 1392.57627 | 1339.97606 | 732.337 | 1164.97 | 1084.29 | 1494.613202 | 1405.740665 |
| 1537.77 | 1375.83209 | 1343.27674 | 732.622 | 1160.25 | 1087.22 | 1501.533 | 1409.658725 |
| 1541.48 | 1424.76042 | 1346.58746 | 732.908 | 1184.11 | 1090.16 | 1515.878152 | 1413.576784 |
| 1545.2 | 1416.64415 | 1349.89817 | 733.193 | 1163.84 | 1093.1 | 1545.052948 | 1417.494843 |
| 1548.93 | 1432.92684 | 1353.20888 | 733.478 | 1164.88 | 1096.05 | 1559.124258 | 1421.423435 |
| 1552.65 | 1462.84365 | 1356.52963 | 733.763 | 1175.02 | 1099 | 1527.042515 | 1425.362559 |
| 1556.39 | 1432.47538 | 1359.86041 | 734.048 | 1180.94 | 1101.96 | 1551.46719 | 1429.312215 |
| 1560.12 | 1433.75954 | 1363.18115 | 734.334 | 1180.05 | 1104.92 | 1528.127354 | 1433.261872 |
| 1563.87 | 1426.96756 | 1366.52196 | 734.619 | 1176.13 | 1107.88 | 1485.723842 | 1437.211528 |
| 1567.62 | 1415.53055 | 1369.85274 | 734.904 | 1211.32 | 1110.85 | 1558.702961 | 1441.182249 |
| 1571.37 | 1443.95252 | 1373.19355 | 735.189 | 1189.75 | 1113.83 | 1580.947426 | 1445.142438 |
| 1575.12 | 1455.08856 | 1376.54439 | 735.475 | 1164.54 | 1116.81 | 1518.606049 | 1449.123692 |
| 1578.88 | 1423.61672 | 1379.89524 | 735.76 | 1196.34 | 1119.79 | 1498.00464 | 1453.104946 |
| 1582.65 | 1458.92096 | 1383.24608 | 736.045 | 1175.1 | 1122.78 | 1560.914769 | 1457.096732 |
| 1586.42 | 1522.79766 | 1386.60696 | 736.33 | 1204.26 | 1125.77 | 1596.472209 | 1461.088518 |
| 1590.2 | 1446.09947 | 1389.96783 | 736.615 | 1252.32 | 1128.77 | 1588.246391 | 1465.101369 |
| 1593.98 | 1482.82832 | 1393.33874 | 736.9 | 1202.31 | 1131.77 | 1564.980282 | 1469.103687 |
| 1597.76 | 1474.52144 | 1396.70965 | 737.185 | 1203.53 | 1134.78 | 1539.681415 | 1473.127071 |
| 1601.55 | 1474.72209 | 1400.09059 | 737.47 | 1209.82 | 1137.79 | 1548.412789 | 1477.139922 |
| 1605.34 | 1478.39397 | 1403.4615 | 737.755 | 1230.32 | 1140.81 | 1574.459457 | 1481.173838 |
| 1609.14 | 1484.2429 | 1406.85247 | 738.041 | 1187.45 | 1143.83 | 1590.100097 | 1485.207753 |
| 1612.94 | 1474.15024 | 1410.24344 | 738.326 | 1225.24 | 1146.85 | 1605.161453 | 1489.252202 |
| 1616.75 | 1483.42024 | 1413.63441 | 738.611 | 1234.6 | 1149.88 | 1574.606911 | 1493.307182 |
| 1620.56 | 1441.19359 | 1417.03542 | 738.896 | 1240.06 | 1152.92 | 1569.224846 | 1497.362163 |
| 1624.38 | 1498.21812 | 1420.43643 | 739.181 | 1235.98 | 1155.96 | 1629.649324 | 1501.427676 |
| 1628.2 | 1519.196 | 1423.83743 | 739.466 | 1247.63 | 1159 | 1608.426503 | 1505.493189 |
| 1632.03 | 1485.99858 | 1427.24847 | 739.751 | 1220.43 | 1162.05 | 1633.125021 | 1509.569235 |
| 1635.86 | 1484.87494 | 1430.66954 | 740.035 | 1259.42 | 1165.11 | 1637.274794 | 1513.64528 |
| 1639.69 | 1523.4598 | 1434.08058 | 740.321 | 1252.54 | 1168.16 | 1617.568641 | 1517.74239 |
| 1643.53 | 1571.29459 | 1437.51168 | 740.605 | 1229.31 | 1171.23 | 1656.570182 | 1521.828968 |
| 1647.38 | 1520.69084 | 1440.93275 | 740.89 | 1255.24 | 1174.3 | 1636.221552 | 1525.936611 |
| 1651.23 | 1504.22757 | 1444.36385 | 741.175 | 1278.48 | 1177.37 | 1664.996116 | 1530.044254 |
| 1655.08 | 1521.22256 | 1447.80499 | 741.46 | 1283.16 | 1180.44 | 1611.765279 | 1534.162429 |
| 1658.94 | 1532.71977 | 1451.24612 | 741.745 | 1275.42 | 1183.53 | 1697.446493 | 1538.280604 |
| 1662.8 | 1546.77525 | 1454.68726 | 742.03 | 1324.63 | 1186.61 | 1675.749714 | 1542.409312 |
| 1666.67 | 1536.04051 | 1458.13843 | 742.315 | 1278.4 | 1189.7 | 1617.6529 | 1546.548551 |
| 1670.54 | 1569.35833 | 1461.59963 | 742.6 | 1277.26 | 1192.8 | 1650.76682 | 1550.687791 |
| 1674.42 | 1529.21844 | 1465.05079 | 742.884 | 1253.44 | 1195.9 | 1680.247056 | 1554.837564 |
| 1678.3 | 1544.89918 | 1468.51199 | 743.169 | 1314.67 | 1199.01 | 1670.599362 | 1558.997869 |
| 1682.18 | 1551.56073 | 1471.98323 | 743.454 | 1332.36 | 1202.11 | 1697.404364 | 1563.158173 |
| 1686.07 | 1591.45984 | 1475.45446 | 743.739 | 1260 | 1205.23 | 1633.90442 | 1567.329011 |
| 1689.97 | 1542.682 | 1478.92569 | 744.023 | 1312.01 | 1208.35 | 1737.090512 | 1571.499848 |
| 1693.87 | 1570.52209 | 1482.40696 | 744.308 | 1334.64 | 1211.47 | 1676.444853 | 1575.681218 |
| 1697.77 | 1583.05264 | 1485.88822 | 744.593 | 1294.04 | 1214.6 | 1700.342908 | 1579.87312 |
| 1701.68 | 1543.8859 | 1489.37952 | 744.878 | 1324.31 | 1217.73 | 1650.429783 | 1584.065022 |
| 1705.59 | 1588.31968 | 1492.87082 | 745.162 | 1326.04 | 1220.87 | 1697.930985 | 1588.267456 |
| 1709.51 | 1619.7213 | 1496.36212 | 745.447 | 1294.96 | 1224.01 | 1716.109937 | 1592.469891 |
| 1713.43 | 1567.35183 | 1499.86345 | 745.732 | 1285.72 | 1227.16 | 1682.96442 | 1596.69339 |
| 1717.36 | 1574.96647 | 1503.37481 | 746.016 | 1292.05 | 1230.31 | 1727.421753 | 1600.906357 |
| 1721.29 | 1586.32322 | 1506.87614 | 746.301 | 1319.2 | 1233.46 | 1654.505828 | 1605.140389 |
| 1725.22 | 1598.4926 | 1510.39753 | 746.586 | 1318.71 | 1236.63 | 1693.644291 | 1609.37442 |
| 1729.16 | 1592.63364 | 1513.9089 | 746.87 | 1342.84 | 1239.79 | 1712.234007 | 1613.618985 |
| 1733.11 | 1576.80241 | 1517.43029 | 747.155 | 1335.94 | 1242.96 | 1727.748258 | 1617.863549 |
| 1737.06 | 1605.27455 | 1520.96172 | 747.439 | 1333.03 | 1246.13 | 1738.723037 | 1622.118645 |
| 1741.01 | 1591.5702 | 1524.48311 | 747.724 | 1307.74 | 1249.31 | 1699.352861 | 1626.373742 |
| 1744.97 | 1637.61921 | 1528.02457 | 748.009 | 1324.92 | 1252.5 | 1738.617712 | 1630.639371 |
| 1748.93 | 1646.50798 | 1531.556 | 748.293 | 1367.43 | 1255.69 | 1737.417017 | 1634.915532 |
| 1752.9 | 1603.45867 | 1535.09746 | 748.578 | 1319.19 | 1258.88 | 1740.387159 | 1639.202226 |
| 1756.87 | 1623.81454 | 1538.64895 | 748.862 | 1386.95 | 1262.08 | 1740.302899 | 1643.48892 |
| 1760.84 | 1634.15801 | 1542.20045 | 749.147 | 1363.53 | 1265.28 | 1806.909906 | 1647.775614 |
| 1764.82 | 1633.95736 | 1545.75194 | 749.431 | 1350.34 | 1268.48 | 1795.471701 | 1652.083372 |
| 1768.81 | 1628.03821 | 1549.31346 | 749.716 | 1363 | 1271.7 | 1798.283856 | 1656.391131 |
| 1772.8 | 1628.80068 | 1552.87499 | 750 | 1381.5 | 1274.91 | 1771.362998 | 1660.69889 |
| 1776.79 | 1671.74966 | 1556.44654 | 750.284 | 1361.46 | 1278.13 | 1769.741005 | 1665.027713 |
| 1780.79 | 1641.02022 | 1560.0181 | 750.569 | 1371.01 | 1281.36 | 1746.906725 | 1669.346004 |
| 1784.79 | 1619.55074 | 1563.59969 | 750.853 | 1380.41 | 1284.59 | 1784.244144 | 1673.68536 |
| 1788.8 | 1613.79211 | 1567.17125 | 751.138 | 1393.13 | 1287.82 | 1839.223362 | 1678.024716 |
| 1792.81 | 1669.78329 | 1570.76287 | 751.422 | 1399.69 | 1291.06 | 1803.307819 | 1682.374604 |
| 1796.83 | 1671.01729 | 1574.34446 | 751.706 | 1347.58 | 1294.31 | 1775.51277 | 1686.724493 |
| 1800.85 | 1634.50915 | 1577.94612 | 751.991 | 1399.09 | 1297.55 | 1785.139399 | 1691.084913 |
| 1804.87 | 1685.31355 | 1581.53774 | 752.275 | 1391.07 | 1300.81 | 1801.053882 | 1695.445334 |
| 1808.9 | 1651.65463 | 1585.13939 | 752.559 | 1355.09 | 1304.06 | 1824.57277 | 1699.82682 |
| 1812.93 | 1648.03291 | 1588.74105 | 752.844 | 1414.19 | 1307.33 | 1828.985852 | 1704.208305 |
| 1816.97 | 1635.57259 | 1592.35273 | 753.128 | 1374.91 | 1310.59 | 1843.826028 | 1708.589791 |
| 1821.01 | 1662.61008 | 1595.96442 | 753.412 | 1344.47 | 1313.86 | 1826.79511 | 1712.981809 |
| 1825.06 | 1697.18195 | 1599.58614 | 753.697 | 1413.42 | 1317.14 | 1771.952813 | 1717.384359 |
| 1829.11 | 1689.50712 | 1603.20786 | 753.981 | 1404.68 | 1320.42 | 1825.899854 | 1721.78691 |
| 1833.17 | 1687.06923 | 1606.82958 | 754.265 | 1372.96 | 1323.7 | 1841.845934 | 1726.199993 |
| 1837.23 | 1661.07511 | 1610.46133 | 754.549 | 1412.4 | 1326.99 | 1855.464349 | 1730.623608 |
| 1841.29 | 1688.04238 | 1614.09308 | 754.834 | 1408.24 | 1330.29 | 1842.604268 | 1735.047223 |
| 1845.36 | 1691.6039 | 1617.73487 | 755.118 | 1415.13 | 1333.59 | 1809.827385 | 1739.481371 |
| 1849.43 | 1697.4428 | 1621.37665 | 755.402 | 1402.53 | 1336.89 | 1857.444444 | 1743.915518 |
| 1853.51 | 1696.96124 | 1625.02847 | 755.686 | 1453.49 | 1340.2 | 1887.672481 | 1748.360198 |
| 1857.59 | 1672.18105 | 1628.68029 | 755.97 | 1453.7 | 1343.51 | 1820.191284 | 1752.815411 |
| 1861.68 | 1718.07958 | 1632.3321 | 756.254 | 1422.97 | 1346.83 | 1856.0331 | 1757.270623 |
| 1865.77 | 1727.05863 | 1635.99395 | 756.538 | 1399.99 | 1350.15 | 1876.223744 | 1761.736368 |
| 1869.86 | 1694.24244 | 1639.6558 | 756.823 | 1458.79 | 1353.47 | 1850.598373 | 1766.212646 |
| 1873.96 | 1698.13504 | 1643.31765 | 757.107 | 1411.47 | 1356.81 | 1871.32617 | 1770.688923 |
| 1878.06 | 1691.82462 | 1646.98953 | 757.391 | 1397.12 | 1360.14 | 1925.262678 | 1775.175733 |
| 1882.17 | 1718.78185 | 1650.67145 | 757.675 | 1442.66 | 1363.48 | 1912.823894 | 1779.662542 |
| 1886.28 | 1743.72256 | 1654.34333 | 757.959 | 1471.27 | 1366.83 | 1917.437092 | 1784.159885 |
| 1890.4 | 1748.26726 | 1658.02525 | 758.243 | 1445.25 | 1370.17 | 1917.363366 | 1788.667759 |
| 1894.52 | 1701.6464 | 1661.71719 | 758.527 | 1500.03 | 1373.53 | 1930.181317 | 1793.175634 |
| 1898.64 | 1734.20175 | 1665.40914 | 758.811 | 1474.4 | 1376.89 | 1908.284422 | 1797.694041 |
| 1902.78 | 1725.01201 | 1669.10109 | 759.095 | 1444.19 | 1380.25 | 1946.35911 | 1802.22298 |
| 1906.91 | 1729.32597 | 1672.80307 | 759.379 | 1452.96 | 1383.62 | 1913.045074 | 1806.75192 |
| 1911.05 | 1755.64112 | 1676.50504 | 759.663 | 1446.5 | 1386.99 | 1910.633151 | 1811.291391 |
| 1915.19 | 1747.0433 | 1680.21706 | 759.947 | 1491.33 | 1390.36 | 1863.511116 | 1815.830863 |
| 1919.34 | 1718.04948 | 1683.92907 | 760.231 | 1484.21 | 1393.74 | 1873.54851 | 1820.380868 |
| 1923.48 | 1716.12325 | 1687.64108 | 760.515 | 1509.54 | 1397.13 | 1936.974726 | 1824.941404 |
| 1927.64 | 1793.92501 | 1691.36312 | 760.799 | 1485.33 | 1400.52 | 1987.40394 | 1829.501941 |
| 1931.8 | 1818.22364 | 1695.08517 | 761.082 | 1482.15 | 1403.91 | 1935.637109 | 1834.07301 |
| 1935.96 | 1811.96338 | 1698.81725 | 761.366 | 1494.29 | 1407.31 | 1932.993473 | 1838.654611 |
| 1940.13 | 1785.90907 | 1702.54932 | 761.65 | 1501.63 | 1410.72 | 1949.424044 | 1843.236213 |
| 1944.3 | 1765.82407 | 1706.2814 | 761.934 | 1474.58 | 1414.12 | 1931.45574 | 1847.828347 |
| 1948.48 | 1741.58564 | 1710.02351 | 762.218 | 1473.75 | 1417.54 | 2039.002252 | 1852.420481 |
| 1952.66 | 1774.11089 | 1713.76562 | 762.502 | 1480.82 | 1420.95 | 1931.329351 | 1857.023147 |
| 1956.85 | 1766.53638 | 1717.51776 | 762.786 | 1477.51 | 1424.38 | 1960.6516 | 1861.636346 |
| 1961.03 | 1803.71669 | 1721.2699 | 763.069 | 1531.87 | 1427.8 | 1995.324318 | 1866.249545 |
| 1965.23 | 1757.89843 | 1725.02204 | 763.353 | 1467.56 | 1431.23 | 1964.864567 | 1870.873276 |
| 1969.42 | 1765.34252 | 1728.78422 | 763.637 | 1541.67 | 1434.67 | 1997.114829 | 1875.497007 |
| 1973.63 | 1754.74823 | 1732.54639 | 763.921 | 1550 | 1438.11 | 1978.177543 | 1880.141803 |
| 1977.83 | 1818.17347 | 1736.3186 | 764.204 | 1474.91 | 1441.55 | 1987.930561 | 1884.776067 |
| 1982.04 | 1769.36553 | 1740.0908 | 764.488 | 1524.58 | 1445 | 2003.918771 | 1889.431395 |
| 1986.26 | 1785.65826 | 1743.86301 | 764.772 | 1507.94 | 1448.45 | 1995.103137 | 1894.086723 |
| 1990.48 | 1817.5314 | 1747.64525 | 765.056 | 1531.51 | 1451.91 | 1967.771515 | 1898.742052 |
| 1994.7 | 1794.06546 | 1751.42749 | 765.339 | 1518.47 | 1455.37 | 1996.061587 | 1903.407913 |
| 1998.93 | 1819.27704 | 1755.21976 | 765.623 | 1501.57 | 1458.84 | 2040.940217 | 1908.084306 |
| 2003.16 | 1802.01117 | 1759.01203 | 765.907 | 1574.9 | 1462.31 | 2060.667435 | 1912.771232 |
| 2007.39 | 1815.03331 | 1762.8043 | 766.19 | 1527.75 | 1465.79 | 1990.605795 | 1917.458157 |
| 2011.63 | 1815.71552 | 1766.60661 | 766.474 | 1542.14 | 1469.27 | 2039.402484 | 1922.145083 |
| 2015.87 | 1845.05044 | 1770.40891 | 766.757 | 1597.21 | 1472.75 | 2032.946112 | 1926.853074 |
| 2020.12 | 1854.25021 | 1774.22125 | 767.041 | 1574.68 | 1476.24 | 1999.90592 | 1931.561064 |
| 2024.37 | 1813.88961 | 1778.03358 | 767.324 | 1595.25 | 1479.74 | 2045.469157 | 1936.269055 |
| 2028.63 | 1825.60753 | 1781.84592 | 767.608 | 1554.69 | 1483.24 | 2022.856057 | 1940.987577 |
| 2032.89 | 1866.48982 | 1785.66829 | 767.891 | 1561.51 | 1486.74 | 2071.673811 | 1945.716633 |
| 2037.16 | 1834.24548 | 1789.49066 | 768.175 | 1589.12 | 1490.25 | 2050.124485 | 1950.456221 |
| 2041.42 | 1835.01798 | 1793.31303 | 768.459 | 1523.95 | 1493.76 | 2070.673231 | 1955.195808 |
| 2045.7 | 1888.56124 | 1797.14543 | 768.742 | 1588.06 | 1497.27 | 2032.3247 | 1959.935396 |
| 2049.97 | 1872.108 | 1800.98786 | 769.026 | 1548.59 | 1500.8 | 2034.062549 | 1964.685516 |
| 2054.26 | 1801.14838 | 1804.82026 | 769.309 | 1569.58 | 1504.32 | 2032.97771 | 1969.446169 |
| 2058.54 | 1860.33992 | 1808.6627 | 769.592 | 1567.21 | 1507.85 | 2097.478233 | 1974.217354 |
| 2062.83 | 1849.04337 | 1812.51516 | 769.876 | 1578.94 | 1511.38 | 2092.843969 | 1978.988539 |
| 2067.12 | 1908.55594 | 1816.36763 | 770.159 | 1573.28 | 1514.92 | 2077.371848 | 1983.770256 |
| 2071.42 | 1889.70494 | 1820.2201 | 770.443 | 1527.8 | 1518.46 | 2095.603463 | 1988.551974 |
| 2075.72 | 1908.46565 | 1824.08259 | 770.726 | 1606.24 | 1522.01 | 2072.084575 | 1993.344224 |
| 2080.03 | 1903.68017 | 1827.94509 | 771.01 | 1567.77 | 1525.57 | 2082.100904 | 1998.136473 |
| 2084.34 | 1872.57953 | 1831.80759 | 771.293 | 1599.84 | 1529.12 | 2115.172694 | 2002.949788 |
| 2088.65 | 1868.86752 | 1835.68012 | 771.576 | 1626.06 | 1532.68 | 2124.030457 | 2007.763103 |
| 2092.97 | 1865.99823 | 1839.55265 | 771.86 | 1598 | 1536.25 | 2124.441221 | 2012.576417 |
| 2097.29 | 1896.96845 | 1843.43522 | 772.143 | 1582.83 | 1539.82 | 2145.590315 | 2017.400265 |
| 2101.62 | 1899.95812 | 1847.31778 | 772.426 | 1618.89 | 1543.39 | 2107.220719 | 2022.234644 |
| 2105.95 | 1921.45769 | 1851.20035 | 772.709 | 1598.8 | 1546.97 | 2187.30922 | 2027.069024 |
| 2110.28 | 1934.93129 | 1855.09294 | 772.993 | 1570.23 | 1550.55 | 2107.115395 | 2031.913936 |
| 2114.62 | 1920.02305 | 1858.98554 | 773.276 | 1658.38 | 1554.14 | 2130.823866 | 2036.758847 |
| 2118.96 | 1898.06199 | 1862.87813 | 773.559 | 1603.86 | 1557.73 | 2213.650795 | 2041.614292 |
| 2123.31 | 1915.99 | 1866.78076 | 773.842 | 1627.72 | 1561.33 | 2124.978374 | 2046.480269 |
| 2127.66 | 1962.49047 | 1870.69342 | 774.126 | 1655.12 | 1564.93 | 2157.755257 | 2051.346245 |
| 2132.01 | 1913.41166 | 1874.59605 | 774.409 | 1627.46 | 1568.54 | 2212.97672 | 2056.222755 |
| 2136.37 | 1932.42318 | 1878.50871 | 774.692 | 1646.18 | 1572.15 | 2178.388262 | 2061.109796 |
| 2140.74 | 1910.56244 | 1882.43141 | 774.975 | 1673.05 | 1575.76 | 2180.410486 | 2065.996838 |
| 2145.1 | 1961.7882 | 1886.34407 | 775.258 | 1609.32 | 1579.38 | 2138.217623 | 2070.894412 |
| 2149.47 | 1985.96644 | 1890.27679 | 775.542 | 1656.15 | 1583 | 2190.827047 | 2075.791986 |
| 2153.85 | 1904.17176 | 1894.19949 | 775.825 | 1673.69 | 1586.63 | 2233.420142 | 2080.700092 |
| 2158.22 | 1968.08859 | 1898.13221 | 776.108 | 1637.52 | 1590.26 | 2194.260615 | 2085.608199 |
| 2162.61 | 1984.15056 | 1902.07497 | 776.391 | 1634.46 | 1593.9 | 2214.430194 | 2090.53737 |
| 2166.99 | 1966.052 | 1906.0077 | 776.674 | 1623.73 | 1597.54 | 2204.455995 | 2095.466541 |
| 2171.39 | 1952.12694 | 1909.95046 | 776.957 | 1643.33 | 1601.18 | 2195.766751 | 2100.395713 |
| 2175.78 | 1964.22609 | 1913.90325 | 777.24 | 1660.52 | 1604.83 | 2185.876811 | 2105.335416 |
| 2180.18 | 1965.46008 | 1917.85604 | 777.523 | 1682.11 | 1608.49 | 2121.713325 | 2110.285652 |
| 2184.58 | 1936.77727 | 1921.80883 | 777.806 | 1689.77 | 1612.14 | 2187.203895 | 2115.235888 |
| 2188.99 | 1971.89089 | 1925.77165 | 778.089 | 1663.89 | 1615.81 | 2202.402174 | 2120.196657 |
| 2193.4 | 1988.45449 | 1929.72444 | 778.372 | 1716.68 | 1619.47 | 2211.080885 | 2125.157425 |
| 2197.81 | 1987.28069 | 1933.6973 | 778.655 | 1666.74 | 1623.14 | 2233.630791 | 2130.128726 |
| 2202.23 | 1979.8968 | 1937.66012 | 778.938 | 1681.56 | 1626.82 | 2193.807721 | 2135.11056 |
| 2206.65 | 1983.59878 | 1941.64301 | 779.221 | 1662.28 | 1630.5 | 2265.322834 | 2140.092393 |
| 2211.08 | 2021.01987 | 1945.61587 | 779.504 | 1734.43 | 1634.19 | 2266.4814 | 2145.084759 |
| 2215.51 | 1993.20988 | 1949.59876 | 779.787 | 1681.94 | 1637.87 | 2249.861246 | 2150.087657 |
| 2219.94 | 2026.91896 | 1953.58164 | 780.07 | 1718.96 | 1641.57 | 2232.998846 | 2155.090555 |
| 2224.38 | 2012.54244 | 1957.57456 | 780.353 | 1723.07 | 1645.27 | 2265.838923 | 2160.103986 |
| 2228.82 | 1981.7528 | 1961.56749 | 780.635 | 1704.07 | 1648.97 | 2322.20842 | 2165.117416 |
| 2233.27 | 2008.95081 | 1965.56041 | 780.918 | 1709.96 | 1652.67 | 2253.705578 | 2170.14138 |
| 2237.72 | 2063.35687 | 1969.56336 | 781.201 | 1714.27 | 1656.38 | 2248.923861 | 2175.165343 |
| 2242.17 | 2012.32172 | 1973.56631 | 781.484 | 1720.6 | 1660.1 | 2292.31742 | 2180.199838 |
| 2246.63 | 2060.31703 | 1977.56927 | 781.767 | 1718.26 | 1663.82 | 2293.296934 | 2185.244866 |
| 2251.09 | 2032.2462 | 1981.58225 | 782.05 | 1687.17 | 1667.54 | 2247.39666 | 2190.289894 |
| 2255.55 | 1989.16679 | 1985.59524 | 782.332 | 1681.46 | 1671.27 | 2296.41453 | 2195.345454 |
| 2260.02 | 2029.05587 | 1989.61826 | 782.615 | 1724.52 | 1675 | 2327.148123 | 2200.411547 |
| 2264.49 | 2057.13674 | 1993.64127 | 782.898 | 1753.95 | 1678.74 | 2284.428639 | 2205.477639 |
| 2268.97 | 2059.1332 | 1997.66429 | 783.181 | 1756.47 | 1682.48 | 2298.815921 | 2210.554265 |
| 2273.45 | 2035.77762 | 2001.69734 | 783.463 | 1758.52 | 1686.23 | 2296.930618 | 2215.63089 |
| 2277.94 | 2010.55601 | 2005.73039 | 783.746 | 1732.62 | 1689.98 | 2366.750012 | 2220.718047 |
| 2282.42 | 2060.307 | 2009.76344 | 784.029 | 1768.46 | 1693.73 | 2349.466316 | 2225.815737 |
| 2286.92 | 2076.34891 | 2013.80653 | 784.311 | 1773.02 | 1697.49 | 2289.547394 | 2230.913427 |
| 2291.41 | 2040.54304 | 2017.84961 | 784.594 | 1790.6 | 1701.25 | 2347.686337 | 2236.021649 |
| 2295.91 | 2085.96001 | 2021.90273 | 784.877 | 1744.85 | 1705.02 | 2312.982022 | 2241.129872 |
| 2300.41 | 2068.97505 | 2025.94581 | 785.159 | 1713.02 | 1708.79 | 2386.645748 | 2246.248627 |
| 2304.92 | 2072.93787 | 2030.00896 | 785.442 | 1784.51 | 1712.57 | 2356.596762 | 2251.377914 |
| 2309.43 | 2040.18188 | 2034.06207 | 785.725 | 1773.4 | 1716.35 | 2313.350657 | 2256.507201 |
| 2313.95 | 2051.52859 | 2038.12522 | 786.007 | 1757.27 | 1720.13 | 2339.892348 | 2261.647021 |
| 2318.47 | 2065.38343 | 2042.1984 | 786.29 | 1757.32 | 1723.92 | 2268.282443 | 2266.78684 |
| 2322.99 | 2119.09724 | 2046.26155 | 786.572 | 1757.42 | 1727.71 | 2346.00115 | 2271.937192 |
| 2327.51 | 2114.5124 | 2050.33473 | 786.855 | 1791.27 | 1731.51 | 2434.399728 | 2277.098077 |
| 2332.05 | 2131.91873 | 2054.41794 | 787.137 | 1783.66 | 1735.31 | 2350.993516 | 2282.258961 |
| 2336.58 | 2117.19107 | 2058.50116 | 787.42 | 1824.33 | 1739.12 | 2352.415392 | 2287.430378 |
| 2341.12 | 2139.88451 | 2062.58437 | 787.702 | 1805.63 | 1742.93 | 2374.217496 | 2292.601795 |
| 2345.66 | 2030.6711 | 2066.66758 | 787.985 | 1820.02 | 1746.74 | 2339.892348 | 2297.783744 |
| 2350.2 | 2159.02645 | 2070.76083 | 788.267 | 1785 | 1750.56 | 2398.800158 | 2302.976226 |
| 2354.75 | 2111.64312 | 2074.85407 | 788.55 | 1787.04 | 1754.39 | 2398.947612 | 2308.168708 |
| 2359.3 | 2128.62808 | 2078.95735 | 788.832 | 1809.64 | 1758.21 | 2398.54738 | 2313.371722 |
| 2363.86 | 2155.93645 | 2083.06063 | 789.115 | 1857.71 | 1762.05 | 2390.669132 | 2318.574736 |
| 2368.42 | 2170.76443 | 2087.16391 | 789.397 | 1868.69 | 1765.88 | 2378.167153 | 2323.788282 |
| 2372.98 | 2135.62071 | 2091.27722 | 789.679 | 1837.09 | 1769.72 | 2393.913116 | 2329.012361 |
| 2377.55 | 2117.49205 | 2095.39053 | 789.962 | 1832.84 | 1773.57 | 2396.377702 | 2334.23644 |
| 2382.12 | 2107.49971 | 2099.50384 | 790.244 | 1815.88 | 1777.41 | 2327.87486 | 2339.471052 |
| 2386.7 | 2168.72784 | 2103.62718 | 790.526 | 1816.78 | 1781.27 | 2430.692317 | 2344.705663 |
| 2391.28 | 2129.5912 | 2107.75052 | 790.809 | 1830.53 | 1785.12 | 2453.716182 | 2349.950807 |
| 2395.86 | 2105.34273 | 2111.87387 | 791.091 | 1858.62 | 1788.98 | 2491.127328 | 2355.195951 |
| 2400.45 | 2234.95214 | 2116.00724 | 791.373 | 1831.99 | 1792.85 | 2467.892815 | 2360.462159 |
| 2405.03 | 2133.25305 | 2120.14062 | 791.655 | 1835.86 | 1796.72 | 2425.120669 | 2365.717835 |
| 2409.63 | 2161.49443 | 2124.28402 | 791.938 | 1842.36 | 1800.6 | 2449.239904 | 2370.994576 |
| 2414.22 | 2130.6747 | 2128.4174 | 792.22 | 1855.64 | 1804.47 | 2420.707586 | 2376.271317 |
| 2418.82 | 2134.32652 | 2132.57084 | 792.502 | 1849.06 | 1808.36 | 2449.945576 | 2381.548059 |
| 2423.43 | 2154.54194 | 2136.71425 | 792.784 | 1885.83 | 1812.24 | 2458.06607 | 2386.835332 |
| 2428.04 | 2133.37344 | 2140.86769 | 793.067 | 1843.89 | 1816.13 | 2498.45789 | 2392.133138 |
| 2432.65 | 2173.08193 | 2145.02113 | 793.349 | 1820.97 | 1820.03 | 2485.03959 | 2397.430944 |
| 2437.26 | 2152.71603 | 2149.1846 | 793.631 | 1867.36 | 1823.93 | 2459.867113 | 2402.739282 |
| 2441.88 | 2147.54931 | 2153.34807 | 793.913 | 1860.36 | 1827.83 | 2467.187143 | 2408.058153 |
| 2446.51 | 2175.40946 | 2157.51155 | 794.195 | 1843.55 | 1831.74 | 2469.851845 | 2413.377023 |
| 2451.13 | 2192.99637 | 2161.68505 | 794.477 | 1889.64 | 1835.65 | 2459.487946 | 2418.695894 |
| 2455.76 | 2157.59181 | 2165.85856 | 794.76 | 1867.74 | 1839.57 | 2488.768066 | 2424.025297 |
| 2460.39 | 2179.40238 | 2170.03206 | 795.042 | 1863.45 | 1843.49 | 2512.908366 | 2429.365233 |
| 2465.03 | 2220.67594 | 2174.2156 | 795.324 | 1932.39 | 1847.41 | 2522.629788 | 2434.715701 |
| 2469.67 | 2195.57472 | 2178.39914 | 795.606 | 1936.9 | 1851.34 | 2533.046348 | 2440.066169 |
| 2474.32 | 2220.76623 | 2182.58267 | 795.888 | 1896.7 | 1855.28 | 2563.516631 | 2445.416637 |
| 2478.96 | 2233.21652 | 2186.77624 | 796.17 | 1894.53 | 1859.21 | 2528.338358 | 2450.78817 |
| 2483.61 | 2174.1554 | 2190.96981 | 796.452 | 1883.94 | 1863.15 | 2548.823909 | 2456.14917 |
| 2488.27 | 2241.96483 | 2195.16338 | 796.734 | 1943.42 | 1867.1 | 2561.378551 | 2461.531235 |
| 2492.93 | 2227.36759 | 2199.36699 | 797.016 | 1920.68 | 1871.05 | 2576.471505 | 2466.9133 |
| 2497.59 | 2225.54169 | 2203.57059 | 797.298 | 1899.49 | 1875 | 2521.513351 | 2472.295366 |
| 2502.26 | 2238.76447 | 2207.78422 | 797.58 | 1881.5 | 1878.96 | 2585.065957 | 2477.687963 |
| 2506.93 | 2284.18144 | 2211.98783 | 797.862 | 1926.42 | 1882.92 | 2585.03436 | 2483.091093 |
| 2511.6 | 2233.62785 | 2216.20146 | 798.144 | 1928.49 | 1886.89 | 2547.244047 | 2488.494223 |
| 2516.27 | 2287.40186 | 2220.42513 | 798.426 | 1981.16 | 1890.86 | 2579.831346 | 2493.907886 |
| 2520.95 | 2277.73057 | 2224.6488 | 798.707 | 1896.65 | 1894.84 | 2543.852608 | 2499.332081 |
| 2525.64 | 2261.1469 | 2228.87246 | 798.989 | 1963.1 | 1898.82 | 2613.545613 | 2504.756276 |
| 2530.32 | 2283.0879 | 2233.09613 | 799.271 | 1931 | 1902.8 | 2564.959572 | 2510.191003 |
| 2535.01 | 2227.53815 | 2237.32983 | 799.553 | 1954.95 | 1906.79 | 2604.150697 | 2515.62573 |
| 2539.71 | 2226.81581 | 2241.56353 | 799.835 | 1969.9 | 1910.78 | 2561.220564 | 2521.060457 |
| 2544.41 | 2258.56856 | 2245.80726 | 800.117 | 1996.98 | 1914.77 | 2579.704957 | 2526.51625 |
| 2549.11 | 2245.16519 | 2250.051 | 800.399 | 1966.85 | 1918.77 | 2569.43585 | 2531.972042 |
| 2553.81 | 2286.26819 | 2254.29473 | 800.68 | 1977.95 | 1922.78 | 2608.795493 | 2537.427834 |
| 2558.52 | 2270.54732 | 2258.53846 | 800.962 | 2000.81 | 1926.78 | 2625.721088 | 2542.904691 |
| 2563.23 | 2310.44643 | 2262.79223 | 801.244 | 1981.39 | 1930.8 | 2632.0932 | 2548.371015 |
| 2567.94 | 2258.19736 | 2267.04599 | 801.525 | 1939.9 | 1934.81 | 2667.324136 | 2553.858405 |
| 2572.66 | 2297.15342 | 2271.30979 | 801.807 | 1976.17 | 1938.83 | 2648.839743 | 2559.345794 |
| 2577.38 | 2295.73884 | 2275.57359 | 802.089 | 1938.29 | 1942.86 | 2636.990774 | 2564.833183 |
| 2582.11 | 2312.20211 | 2279.83738 | 802.371 | 2023.7 | 1946.88 | 2640.234759 | 2570.331105 |
| 2586.84 | 2325.50516 | 2284.11121 | 802.652 | 1975.19 | 1950.92 | 2611.007301 | 2575.83956 |
| 2591.57 | 2330.28061 | 2288.37501 | 802.934 | 1984.17 | 1954.95 | 2656.486278 | 2581.348014 |
| 2596.3 | 2310.60695 | 2292.65887 | 803.216 | 1973.99 | 1958.99 | 2610.196305 | 2586.867 |
| 2601.04 | 2344.97817 | 2296.9327 | 803.497 | 1985.07 | 1963.04 | 2642.278048 | 2592.385987 |
| 2605.78 | 2328.33432 | 2301.21656 | 803.779 | 1993.93 | 1967.09 | 2641.245871 | 2597.915506 |
| 2610.53 | 2313.90763 | 2305.50043 | 804.06 | 2049.87 | 1971.14 | 2705.504148 | 2603.445025 |
| 2615.28 | 2335.58779 | 2309.79432 | 804.342 | 1977.7 | 1975.2 | 2646.069718 | 2608.995609 |
| 2620.03 | 2289.30803 | 2314.07818 | 804.624 | 2037.86 | 1979.26 | 2683.575656 | 2614.53566 |
| 2624.79 | 2323.9401 | 2318.38211 | 804.905 | 2044.4 | 1983.32 | 2658.171465 | 2620.086244 |
| 2629.55 | 2354.3485 | 2322.67601 | 805.187 | 2000.07 | 1987.39 | 2676.023912 | 2625.647361 |
| 2634.31 | 2329.27737 | 2326.97993 | 805.468 | 2030.84 | 1991.47 | 2673.864767 | 2631.219009 |
| 2639.07 | 2356.45531 | 2331.28386 | 805.75 | 2067.63 | 1995.54 | 2704.819541 | 2636.790658 |
| 2643.84 | 2357.0372 | 2335.59782 | 806.031 | 2031.01 | 1999.62 | 2712.150104 | 2642.362307 |
| 2648.62 | 2350.15493 | 2339.90175 | 806.313 | 2064.51 | 2003.71 | 2707.220932 | 2647.944488 |
| 2653.39 | 2353.60609 | 2344.22574 | 806.594 | 2029.75 | 2007.8 | 2734.057531 | 2653.537202 |
| 2658.17 | 2399.2839 | 2348.5397 | 806.875 | 2038.12 | 2011.89 | 2757.070863 | 2659.129915 |
| 2662.95 | 2382.15849 | 2352.86369 | 807.157 | 2057.94 | 2015.99 | 2659.045656 | 2664.733161 |
| 2667.74 | 2384.77696 | 2357.18768 | 807.438 | 2079.58 | 2020.09 | 2777.693336 | 2670.346939 |
| 2672.53 | 2350.30541 | 2361.51167 | 807.72 | 2071.26 | 2024.2 | 2742.020039 | 2675.950185 |
| 2677.32 | 2391.52881 | 2365.8457 | 808.001 | 2091.76 | 2028.31 | 2761.167973 | 2681.574496 |
| 2682.12 | 2374.64417 | 2370.17972 | 808.283 | 2093.95 | 2032.42 | 2744.895389 | 2687.198807 |
| 2686.92 | 2437.91892 | 2374.52378 | 808.564 | 2095.18 | 2036.54 | 2746.527913 | 2692.83365 |
| 2691.72 | 2423.08091 | 2378.85781 | 808.845 | 2054.56 | 2040.66 | 2689.115707 | 2698.468494 |
| 2696.53 | 2398.03988 | 2383.20186 | 809.126 | 2081.62 | 2044.79 | 2659.382693 | 2704.113869 |
| 2701.34 | 2424.08415 | 2387.55595 | 809.408 | 2081.24 | 2048.91 | 2746.886016 | 2709.759245 |
| 2706.15 | 2429.16058 | 2391.90001 | 809.689 | 2110.3 | 2053.05 | 2807.310493 | 2715.415153 |
| 2710.96 | 2358.37151 | 2396.2541 | 809.97 | 2141.43 | 2057.19 | 2729.181022 | 2721.081593 |
| 2715.78 | 2395.78257 | 2400.61822 | 810.252 | 2081.43 | 2061.33 | 2799.442778 | 2726.748034 |
| 2720.6 | 2407.27978 | 2404.97231 | 810.533 | 2120.69 | 2065.47 | 2763.316587 | 2732.425007 |
| 2725.43 | 2396.01332 | 2409.33643 | 810.814 | 2082.31 | 2069.62 | 2800.390695 | 2738.10198 |
| 2730.26 | 2443.42675 | 2413.71059 | 811.095 | 2093.08 | 2073.78 | 2804.803778 | 2743.789485 |
| 2735.09 | 2411.56364 | 2418.07471 | 811.376 | 2040.91 | 2077.93 | 2766.760687 | 2749.47699 |
| 2739.93 | 2427.45506 | 2422.44886 | 811.658 | 2094.55 | 2082.09 | 2816.126127 | 2755.175028 |
| 2744.76 | 2443.95847 | 2426.82302 | 811.939 | 2011.16 | 2086.26 | 2729.465397 | 2760.883598 |
| 2749.61 | 2429.96318 | 2431.2072 | 812.22 | 2105.72 | 2090.43 | 2802.844748 | 2766.592168 |
| 2754.45 | 2412.4164 | 2435.59139 | 812.501 | 2130.84 | 2094.6 | 2812.534572 | 2772.300738 |
| 2759.3 | 2455.35535 | 2439.97558 | 812.782 | 2170.77 | 2098.78 | 2792.912679 | 2778.019841 |
| 2764.15 | 2405.08267 | 2444.35976 | 813.063 | 2131.31 | 2102.96 | 2820.307496 | 2783.749476 |
| 2769 | 2419.19834 | 2448.75398 | 813.344 | 2089.58 | 2107.15 | 2858.814014 | 2789.479111 |
| 2773.86 | 2429.83276 | 2453.1482 | 813.625 | 2090.66 | 2111.34 | 2864.806959 | 2795.219278 |
| 2778.72 | 2433.9862 | 2457.55246 | 813.907 | 2145.01 | 2115.53 | 2837.475336 | 2800.969978 |
| 2783.59 | 2477.63745 | 2461.94667 | 814.187 | 2175.93 | 2119.73 | 2808.574383 | 2806.720678 |
| 2788.45 | 2446.32613 | 2466.35093 | 814.468 | 2141.77 | 2123.93 | 2906.209891 | 2812.471378 |
| 2793.32 | 2477.48696 | 2470.76521 | 814.75 | 2098.2 | 2128.13 | 2814.388278 | 2818.243142 |
| 2798.2 | 2497.47163 | 2475.16946 | 815.03 | 2156.15 | 2132.34 | 2856.876049 | 2824.004375 |
| 2803.07 | 2489.25504 | 2479.58375 | 815.312 | 2110.44 | 2136.55 | 2838.865615 | 2829.786672 |
| 2807.95 | 2481.40966 | 2484.00806 | 815.592 | 2175.19 | 2140.77 | 2874.117616 | 2835.558436 |
| 2812.84 | 2468.63833 | 2488.42235 | 815.873 | 2190.24 | 2144.99 | 2928.212111 | 2841.351266 |
| 2817.72 | 2552.14856 | 2492.84666 | 816.154 | 2200.33 | 2149.22 | 2933.415125 | 2847.144095 |
| 2822.61 | 2491.90361 | 2497.27098 | 816.435 | 2178.65 | 2153.44 | 2886.261493 | 2852.936925 |
| 2827.51 | 2478.40995 | 2501.70533 | 816.716 | 2161.14 | 2157.68 | 2883.997023 | 2858.740287 |
| 2832.4 | 2511.32646 | 2506.13968 | 816.997 | 2146.53 | 2161.91 | 2914.69902 | 2864.554181 |
| 2837.3 | 2535.05324 | 2510.57403 | 817.278 | 2167.72 | 2166.15 | 2905.240909 | 2870.368075 |
| 2842.2 | 2529.65578 | 2515.00838 | 817.559 | 2161.48 | 2170.4 | 2899.700857 | 2876.192502 |
| 2847.11 | 2516.79416 | 2519.45276 | 817.84 | 2182.72 | 2174.64 | 2922.714189 | 2882.016929 |
| 2852.01 | 2474.12609 | 2523.89714 | 818.12 | 2184.93 | 2178.9 | 2938.323232 | 2887.851888 |
| 2856.92 | 2580.9317 | 2528.34152 | 818.401 | 2167.17 | 2183.15 | 2936.553785 | 2893.686847 |
| 2861.84 | 2528.44185 | 2532.79594 | 818.682 | 2212.41 | 2187.41 | 2944.748006 | 2899.532339 |
| 2866.76 | 2567.36781 | 2537.25035 | 818.963 | 2177.52 | 2191.67 | 2932.509337 | 2905.37783 |
| 2871.68 | 2575.34362 | 2541.70477 | 819.243 | 2222.11 | 2195.94 | 2928.64394 | 2911.233854 |
| 2876.6 | 2555.04794 | 2546.16921 | 819.524 | 2243.16 | 2200.21 | 2913.677375 | 2917.100411 |
| 2881.53 | 2540.81188 | 2550.62363 | 819.805 | 2207.23 | 2204.49 | 3011.260221 | 2922.966967 |
| 2886.45 | 2565.53187 | 2555.09811 | 820.086 | 2241.66 | 2208.76 | 2970.752544 | 2928.844056 |
| 2891.39 | 2511.36659 | 2559.56255 | 820.366 | 2217.78 | 2213.05 | 2984.739594 | 2934.721145 |
| 2896.32 | 2633.77269 | 2564.03703 | 820.647 | 2231.98 | 2217.33 | 2988.267954 | 2940.608766 |
| 2901.26 | 2604.76884 | 2568.51151 | 820.928 | 2262.57 | 2221.62 | 2935.089779 | 2946.496387 |
| 2906.2 | 2621.68357 | 2572.98599 | 821.208 | 2258.3 | 2225.92 | 3022.424583 | 2952.394541 |
| 2911.15 | 2597.88657 | 2577.4705 | 821.489 | 2220.45 | 2230.22 | 2985.224086 | 2958.303227 |
| 2916.09 | 2591.02436 | 2581.95501 | 821.769 | 2217.46 | 2234.52 | 2985.445266 | 2964.211913 |
| 2921.04 | 2629.57912 | 2586.43952 | 822.05 | 2266.43 | 2238.82 | 2996.957198 | 2970.120599 |
| 2926 | 2591.45576 | 2590.92404 | 822.331 | 2291.35 | 2243.13 | 2989.479182 | 2976.039818 |
| 2930.95 | 2615.99516 | 2595.41858 | 822.611 | 2263.77 | 2247.44 | 3082.722672 | 2981.969569 |
| 2935.91 | 2688.48975 | 2599.91312 | 822.892 | 2303.21 | 2251.76 | 3057.8767 | 2987.89932 |
| 2940.87 | 2626.07779 | 2604.4177 | 823.172 | 2299.3 | 2256.08 | 3011.102235 | 2993.839603 |
| 2945.84 | 2589.32887 | 2608.91224 | 823.453 | 2293.39 | 2260.41 | 3028.744034 | 2999.779886 |
| 2950.81 | 2636.75233 | 2613.41682 | 823.733 | 2278.84 | 2264.73 | 3127.653964 | 3005.730702 |
| 2955.78 | 2638.34749 | 2617.9214 | 824.014 | 2286.23 | 2269.07 | 3021.992754 | 3011.681518 |
| 2960.75 | 2646.27314 | 2622.43601 | 824.294 | 2252.39 | 2273.4 | 3075.423707 | 3017.642866 |
| 2965.73 | 2624.46256 | 2626.95062 | 824.574 | 2313.25 | 2277.74 | 3119.27016 | 3023.614746 |
| 2970.71 | 2702.65559 | 2631.46522 | 824.855 | 2270.27 | 2282.08 | 3086.819782 | 3029.586627 |
| 2975.69 | 2632.35811 | 2635.97983 | 825.135 | 2328.73 | 2286.43 | 3089.347562 | 3035.558508 |
| 2980.68 | 2646.63431 | 2640.50447 | 825.416 | 2355.87 | 2290.78 | 3176.071486 | 3041.540921 |
| 2985.67 | 2625.48588 | 2645.02912 | 825.696 | 2335.52 | 2295.14 | 3069.736202 | 3047.533866 |
| 2990.66 | 2642.88217 | 2649.55376 | 825.976 | 2320.33 | 2299.49 | 3065.944531 | 3053.526811 |
| 2995.65 | 2722.84091 | 2654.08843 | 826.257 | 2315.01 | 2303.86 | 3136.153625 | 3059.530289 |
| 3000.65 | 2638.6886 | 2658.62311 | 826.537 | 2292.77 | 2308.22 | 3116.447472 | 3065.533767 |
| 3005.65 | 2667.84294 | 2663.15778 | 826.817 | 2365.36 | 2312.59 | 3080.521397 | 3071.547777 |
| 3010.65 | 2676.09966 | 2667.69245 | 827.098 | 2343.48 | 2316.96 | 3191.722658 | 3077.561788 |
| 3015.66 | 2676.23008 | 2672.23716 | 827.378 | 2374.27 | 2321.34 | 3099.648267 | 3083.58633 |
| 3020.67 | 2676.72167 | 2676.78187 | 827.658 | 2339.87 | 2325.72 | 3157.418575 | 3089.610873 |
| 3025.68 | 2677.14303 | 2681.32657 | 827.939 | 2364.92 | 2330.11 | 3118.174788 | 3095.645948 |
| 3030.7 | 2694.54936 | 2685.88131 | 828.219 | 2341.42 | 2334.49 | 3151.941718 | 3101.691555 |
| 3035.71 | 2717.06221 | 2690.43605 | 828.499 | 2328.28 | 2338.89 | 3104.703827 | 3107.737163 |
| 3040.73 | 2733.11415 | 2694.99079 | 828.779 | 2403.16 | 2343.28 | 3171.479352 | 3113.78277 |
| 3045.76 | 2746.4473 | 2699.54553 | 829.059 | 2417.38 | 2347.68 | 3147.949932 | 3119.83891 |
| 3050.78 | 2765.15784 | 2704.1103 | 829.34 | 2402.64 | 2352.08 | 3159.261748 | 3125.905583 |
| 3055.81 | 2695.39209 | 2708.67507 | 829.62 | 2373.72 | 2356.49 | 3249.924796 | 3131.972255 |
| 3060.84 | 2691.35904 | 2713.23984 | 829.9 | 2389.75 | 2360.9 | 3189.131683 | 3138.04946 |
| 3065.88 | 2708.52458 | 2717.81464 | 830.18 | 2413.54 | 2365.31 | 3163.032354 | 3144.126665 |
| 3070.91 | 2709.70841 | 2722.37942 | 830.46 | 2404.36 | 2369.73 | 3140.31393 | 3150.214402 |
| 3075.95 | 2717.19263 | 2726.96425 | 830.74 | 2378.52 | 2374.15 | 3299.04799 | 3156.302139 |
| 3080.99 | 2727.60633 | 2731.53906 | 831.02 | 2424.46 | 2378.58 | 3244.374212 | 3162.400408 |
| 3086.04 | 2767.13424 | 2736.11386 | 831.3 | 2495.45 | 2383 | 3262.953396 | 3168.498678 |
| 3091.09 | 2713.31007 | 2740.6987 | 831.58 | 2399.35 | 2387.44 | 3220.813195 | 3174.60748 |
| 3096.14 | 2745.58451 | 2745.29356 | 831.86 | 2412.21 | 2391.87 | 3252.852808 | 3180.716282 |
| 3101.19 | 2729.65295 | 2749.8784 | 832.14 | 2436.75 | 2396.31 | 3239.750481 | 3186.835617 |
| 3106.25 | 2804.03364 | 2754.47327 | 832.42 | 2442.22 | 2400.75 | 3231.893297 | 3192.965483 |
| 3111.31 | 2801.30481 | 2759.06814 | 832.7 | 2462.4 | 2405.2 | 3227.37489 | 3199.09535 |
| 3116.37 | 2772.9631 | 2763.66301 | 832.98 | 2422.18 | 2409.65 | 3197.030996 | 3205.225217 |
| 3121.43 | 2810.19358 | 2768.26791 | 833.26 | 2506.62 | 2414.11 | 3231.619455 | 3211.365616 |
| 3126.5 | 2813.70494 | 2772.86278 | 833.54 | 2405.11 | 2418.56 | 3255.928274 | 3217.516548 |
| 3131.57 | 2803.66244 | 2777.47771 | 833.82 | 2478.87 | 2423.02 | 3284.850291 | 3223.66748 |
| 3136.64 | 2791.06167 | 2782.08261 | 834.1 | 2411.7 | 2427.49 | 3236.00094 | 3229.828944 |
| 3141.72 | 2781.00914 | 2786.69754 | 834.38 | 2409.78 | 2431.96 | 3297.3944 | 3235.990408 |
| 3146.8 | 2784.53053 | 2791.30245 | 834.66 | 2464.04 | 2436.43 | 3254.611721 | 3242.151872 |
| 3151.88 | 2793.18855 | 2795.91738 | 834.939 | 2433.61 | 2440.9 | 3246.291112 | 3248.334401 |
| 3156.96 | 2770.2744 | 2800.54235 | 835.219 | 2440.48 | 2445.38 | 3274.349471 | 3254.506397 |
| 3162.05 | 2810.94601 | 2805.16731 | 835.499 | 2525.06 | 2449.86 | 3327.011558 | 3260.699459 |
| 3167.14 | 2764.53583 | 2809.79228 | 835.779 | 2379.28 | 2454.35 | 3334.752884 | 3266.881988 |
| 3172.23 | 2816.56419 | 2814.41724 | 836.059 | 2452.11 | 2458.84 | 3333.826032 | 3273.085581 |
| 3177.32 | 2820.60727 | 2819.04221 | 836.338 | 2478.55 | 2463.33 | 3330.792695 | 3279.278643 |
| 3182.42 | 2791.20212 | 2823.67721 | 836.618 | 2481.16 | 2467.83 | 3312.571614 | 3285.492769 |
| 3187.52 | 2790.53998 | 2828.31221 | 836.898 | 2491 | 2472.33 | 3282.733276 | 3291.706895 |
| 3192.62 | 2830.11805 | 2832.9472 | 837.178 | 2497.17 | 2476.84 | 3337.470248 | 3297.921021 |
| 3197.73 | 2829.92743 | 2837.59224 | 837.457 | 2444.86 | 2481.34 | 3250.577806 | 3304.14568 |
| 3202.83 | 2829.25526 | 2842.23727 | 837.737 | 2482.37 | 2485.85 | 3274.454796 | 3310.370338 |
| 3207.94 | 2735.6022 | 2846.8823 | 838.017 | 2476.44 | 2490.37 | 3383.517976 | 3316.605529 |
| 3213.06 | 2831.04104 | 2851.52733 | 838.296 | 2508.62 | 2494.89 | 3331.129733 | 3322.84072 |
| 3218.17 | 2830.73003 | 2856.18239 | 838.576 | 2465.8 | 2499.41 | 3312.824392 | 3329.086444 |
| 3223.29 | 2874.37125 | 2860.83746 | 838.856 | 2540.25 | 2503.93 | 3378.641467 | 3335.3427 |
| 3228.41 | 2839.23756 | 2865.49252 | 839.135 | 2506.18 | 2508.46 | 3334.521171 | 3341.598956 |
| 3233.53 | 2815.62114 | 2870.14758 | 839.415 | 2538.6 | 2512.99 | 3368.498749 | 3347.855211 |
| 3238.66 | 2829.24523 | 2874.81268 | 839.694 | 2513.62 | 2517.53 | 3328.296513 | 3354.122 |
| 3243.78 | 2884.9254 | 2879.47777 | 839.974 | 2527.71 | 2522.07 | 3432.746494 | 3360.39932 |
| 3248.92 | 2884.51407 | 2884.14287 | 840.254 | 2560.75 | 2526.61 | 3426.574498 | 3366.676641 |
| 3254.05 | 2879.548 | 2888.80797 | 840.533 | 2501.16 | 2531.16 | 3407.531888 | 3372.953962 |
| 3259.18 | 2886.99209 | 2893.48309 | 840.812 | 2508.43 | 2535.71 | 3412.208281 | 3379.252347 |
| 3264.32 | 2889.25943 | 2898.15822 | 841.092 | 2556.64 | 2540.26 | 3421.929702 | 3385.5402 |
| 3269.46 | 2901.31845 | 2902.83335 | 841.371 | 2556.08 | 2544.82 | 3415.726108 | 3391.838586 |
| 3274.61 | 2870.57898 | 2907.50848 | 841.651 | 2541.23 | 2549.38 | 3403.024013 | 3398.147504 |
| 3279.75 | 2949.89564 | 2912.19364 | 841.93 | 2552.23 | 2553.94 | 3426.700887 | 3404.456422 |
| 3284.9 | 2928.96792 | 2916.8788 | 842.21 | 2590.36 | 2558.51 | 3481.237743 | 3410.775872 |
| 3290.05 | 2895.37923 | 2921.56396 | 842.489 | 2585.85 | 2563.08 | 3527.569847 | 3417.095322 |
| 3295.2 | 2909.05348 | 2926.25916 | 842.769 | 2577.48 | 2567.65 | 3438.950088 | 3423.414773 |
| 3300.36 | 2905.0104 | 2930.94432 | 843.048 | 2614.22 | 2572.23 | 3403.792879 | 3429.744756 |
| 3305.52 | 2901.81004 | 2935.63951 | 843.327 | 2498.88 | 2576.81 | 3418.675185 | 3436.085271 |
| 3310.68 | 2980.61505 | 2940.3347 | 843.607 | 2610.7 | 2581.4 | 3408.469273 | 3442.425786 |
| 3315.84 | 2935.83013 | 2945.03993 | 843.886 | 2583.86 | 2585.98 | 3504.208945 | 3448.776833 |
| 3321.01 | 2938.8499 | 2949.74515 | 844.165 | 2616.03 | 2590.57 | 3560.146613 | 3455.127881 |
| 3326.17 | 2977.74576 | 2954.44035 | 844.444 | 2619.78 | 2595.17 | 3526.969499 | 3461.489461 |
| 3331.34 | 2994.30936 | 2959.15561 | 844.724 | 2633.23 | 2599.77 | 3472.495837 | 3467.851041 |
| 3336.52 | 2952.22317 | 2963.86083 | 845.003 | 2598.51 | 2604.37 | 3478.47825 | 3474.212621 |
| 3341.69 | 2951.83191 | 2968.57609 | 845.282 | 2640.28 | 2608.97 | 3441.53053 | 3480.595266 |
| 3346.87 | 2980.71537 | 2973.29135 | 845.562 | 2576.19 | 2613.58 | 3542.673333 | 3486.967378 |
| 3352.05 | 3007.25124 | 2978.00661 | 845.841 | 2571.94 | 2618.19 | 3507.305476 | 3493.350023 |
| 3357.23 | 2921.71445 | 2982.72186 | 846.12 | 2631.37 | 2622.81 | 3560.209808 | 3499.743201 |
| 3362.42 | 2981.68852 | 2987.44715 | 846.399 | 2585.93 | 2627.43 | 3598.252899 | 3506.136378 |
| 3367.61 | 3000.94082 | 2992.17245 | 846.678 | 2639.65 | 2632.05 | 3560.87335 | 3512.540087 |
| 3372.8 | 3043.52863 | 2996.89774 | 846.957 | 2623.76 | 2636.67 | 3522.040328 | 3518.943797 |
| 3377.99 | 2951.83191 | 3001.62303 | 847.236 | 2622.1 | 2641.3 | 3582.549065 | 3525.358039 |
| 3383.18 | 2957.57048 | 3006.35835 | 847.516 | 2692.9 | 2645.93 | 3575.976837 | 3531.772281 |
| 3388.38 | 2999.18514 | 3011.09367 | 847.795 | 2632.84 | 2650.57 | 3638.760576 | 3538.197056 |
| 3393.58 | 2977.10368 | 3015.829 | 848.074 | 2714.05 | 2655.21 | 3656.760477 | 3544.62183 |
| 3398.78 | 3057.88509 | 3020.56432 | 848.353 | 2648.24 | 2659.85 | 3527.369731 | 3551.046605 |
| 3403.98 | 3007.53215 | 3025.30968 | 848.632 | 2627.28 | 2664.5 | 3514.636039 | 3557.492444 |
| 3409.19 | 3028.99159 | 3030.045 | 848.911 | 2675.25 | 2669.14 | 3602.465866 | 3563.927751 |
| 3414.4 | 2955.99538 | 3034.80039 | 849.19 | 2655.01 | 2673.8 | 3568.677871 | 3570.373591 |
| 3419.61 | 3030.53659 | 3039.54574 | 849.469 | 2702.11 | 2678.45 | 3621.171439 | 3576.829962 |
| 3424.82 | 3011.77588 | 3044.2911 | 849.748 | 2696.63 | 2683.11 | 3537.122749 | 3583.286334 |
| 3430.04 | 3087.87212 | 3049.04649 | 850.027 | 2732.83 | 2687.77 | 3596.020027 | 3589.753238 |
| 3435.26 | 3075.20112 | 3053.80187 | 850.306 | 2719.25 | 2692.44 | 3635.916823 | 3596.220143 |
| 3440.47 | 3104.43572 | 3058.55726 | 850.585 | 2660.83 | 2697.11 | 3594.29271 | 3602.687047 |
| 3445.7 | 2995.12199 | 3063.32268 | 850.864 | 2734.88 | 2701.78 | 3575.545007 | 3609.164483 |
| 3450.92 | 3037.69977 | 3068.0881 | 851.143 | 2653.17 | 2706.45 | 3623.709752 | 3615.652452 |
| 3456.15 | 3111.63903 | 3072.85352 | 851.421 | 2739.46 | 2711.13 | 3611.80812 | 3622.140421 |
| 3461.38 | 3048.75554 | 3077.61894 | 851.7 | 2676.18 | 2715.81 | 3589.84803 | 3628.638923 |
| 3466.61 | 3026.65403 | 3082.38436 | 851.979 | 2675.82 | 2720.5 | 3653.790335 | 3635.137424 |
| 3471.84 | 3044.09045 | 3087.15982 | 852.258 | 2693.55 | 2725.19 | 3713.403817 | 3641.635926 |
| 3477.08 | 3049.11671 | 3091.93527 | 852.537 | 2701.42 | 2729.88 | 3662.13201 | 3648.14496 |
| 3482.32 | 3096.02852 | 3096.71072 | 852.816 | 2754.03 | 2734.57 | 3664.522868 | 3654.664526 |
| 3487.56 | 3095.65731 | 3101.48618 | 853.094 | 2737.59 | 2739.27 | 3618.717386 | 3661.184092 |
| 3492.8 | 3054.17308 | 3106.27166 | 853.373 | 2675.18 | 2743.97 | 3649.872276 | 3667.703658 |
| 3498.05 | 3134.90432 | 3111.04712 | 853.652 | 2759.1 | 2748.68 | 3647.365561 | 3674.233757 |
| 3503.29 | 3080.55846 | 3115.8326 | 853.93 | 2749.44 | 2753.39 | 3708.148141 | 3680.774388 |
| 3508.54 | 3102.7302 | 3120.62812 | 854.209 | 2697.13 | 2758.1 | 3662.216269 | 3687.315019 |
| 3513.79 | 3101.57647 | 3125.4136 | 854.488 | 2764.04 | 2762.81 | 3767.003288 | 3693.85565 |
| 3519.05 | 3136.32893 | 3130.20912 | 854.766 | 2758.91 | 2767.53 | 3797.189196 | 3700.406814 |
| 3524.3 | 3096.97157 | 3135.00464 | 855.045 | 2736.03 | 2772.25 | 3744.000489 | 3706.957977 |
| 3529.56 | 3102.04799 | 3139.80016 | 855.324 | 2783.36 | 2776.97 | 3700.627995 | 3713.519673 |
| 3534.82 | 3145.73938 | 3144.59568 | 855.602 | 2716.95 | 2781.7 | 3702.839802 | 3720.081369 |
| 3540.08 | 3066.28226 | 3149.40123 | 855.881 | 2769.1 | 2786.43 | 3761.38951 | 3726.653598 |
| 3545.35 | 3134.59331 | 3154.19674 | 856.16 | 2816.5 | 2791.16 | 3741.809746 | 3733.225826 |
| 3550.61 | 3119.75529 | 3159.00229 | 856.438 | 2813.56 | 2795.9 | 3805.994297 | 3739.808587 |
| 3555.88 | 3126.01555 | 3163.81788 | 856.717 | 2767.34 | 2800.64 | 3740.272013 | 3746.391348 |
| 3561.16 | 3132.66708 | 3168.62343 | 856.995 | 2758.07 | 2805.38 | 3808.785387 | 3752.984641 |
| 3566.43 | 3124.53075 | 3173.43901 | 857.274 | 2838.56 | 2810.13 | 3726.959038 | 3759.577934 |
| 3571.7 | 3138.70662 | 3178.25459 | 857.552 | 2815.94 | 2814.88 | 3753.279549 | 3766.18176 |
| 3576.98 | 3188.31716 | 3183.07018 | 857.831 | 2751.6 | 2819.63 | 3782.759784 | 3772.785585 |
| 3582.26 | 3167.05836 | 3187.88576 | 858.109 | 2796.66 | 2824.39 | 3811.850321 | 3779.399943 |
| 3587.54 | 3162.65411 | 3192.71137 | 858.388 | 2788.37 | 2829.15 | 3784.192193 | 3786.014301 |
| 3592.83 | 3130.89133 | 3197.52696 | 858.666 | 2861.96 | 2833.91 | 3783.855156 | 3792.628659 |
| 3598.11 | 3175.11443 | 3202.35257 | 858.945 | 2834.97 | 2838.67 | 3795.082713 | 3799.25355 |
| 3603.4 | 3143.88337 | 3207.18822 | 859.223 | 2781.66 | 2843.44 | 3813.061549 | 3805.888973 |
| 3608.69 | 3155.21002 | 3212.01384 | 859.501 | 2821.71 | 2848.21 | 3855.001634 | 3812.524396 |
| 3613.98 | 3229.3098 | 3216.84948 | 859.78 | 2807.28 | 2852.99 | 3796.251811 | 3819.159818 |
| 3619.28 | 3217.15046 | 3221.68513 | 860.058 | 2834.47 | 2857.77 | 3869.199333 | 3825.805774 |
| 3624.57 | 3224.23338 | 3226.52078 | 860.337 | 2859.39 | 2862.55 | 3805.151703 | 3832.462261 |
| 3629.87 | 3156.03269 | 3231.35643 | 860.615 | 2850.18 | 2867.33 | 3884.703051 | 3839.118749 |
| 3635.17 | 3236.80406 | 3236.19207 | 860.893 | 2827.29 | 2872.12 | 3839.118749 | 3845.775237 |
| 3640.47 | 3240.91737 | 3241.03776 | 861.171 | 2875.59 | 2876.91 | 3925.126468 | 3852.442257 |
| 3645.78 | 3244.63941 | 3245.88344 | 861.45 | 2875.96 | 2881.7 | 3833.673489 | 3859.109277 |
| 3651.08 | 3233.53347 | 3250.72912 | 861.728 | 2830.3 | 2886.5 | 3855.191218 | 3865.78683 |
| 3656.39 | 3230.61402 | 3255.58483 | 862.006 | 2871.81 | 2891.3 | 3854.064249 | 3872.464382 |
| 3661.7 | 3215.95659 | 3260.43051 | 862.284 | 2894.2 | 2896.1 | 3818.011785 | 3879.141935 |
| 3667.01 | 3250.94983 | 3265.28622 | 862.563 | 2922.4 | 2900.91 | 3969.731255 | 3885.840552 |
| 3672.33 | 3205.56296 | 3270.14193 | 862.841 | 2927.94 | 2905.72 | 3929.065592 | 3892.528637 |
| 3677.64 | 3206.39566 | 3274.99765 | 863.119 | 2838.99 | 2910.53 | 3910.981432 | 3899.227254 |
| 3682.96 | 3240.10474 | 3279.85336 | 863.397 | 2821.2 | 2915.34 | 3854.496078 | 3905.936404 |
| 3688.28 | 3254.64178 | 3284.7191 | 863.675 | 2890.61 | 2920.16 | 3871.347946 | 3912.645554 |
| 3693.61 | 3311.09445 | 3289.58485 | 863.953 | 2893.18 | 2924.98 | 3862.542845 | 3919.354704 |
| 3698.93 | 3238.74032 | 3294.4506 | 864.231 | 2921.04 | 2929.81 | 3966.466206 | 3926.074386 |
| 3704.26 | 3255.37415 | 3299.31634 | 864.51 | 2831.33 | 2934.63 | 3914.625648 | 3932.794068 |
| 3709.58 | 3291.48099 | 3304.18209 | 864.788 | 2902.64 | 2939.46 | 3918.006554 | 3939.524283 |
| 3714.91 | 3312.11776 | 3309.05786 | 865.066 | 2902.82 | 2944.3 | 3832.63078 | 3946.254497 |
| 3720.24 | 3232.20919 | 3313.93364 | 865.344 | 2901.43 | 2949.13 | 3862.816688 | 3952.995244 |
| 3725.58 | 3332.33318 | 3318.80942 | 865.622 | 2970.67 | 2953.97 | 3975.608344 | 3959.735991 |
| 3730.91 | 3279.72294 | 3323.6852 | 865.9 | 2924 | 2958.81 | 3931.835618 | 3966.487271 |
| 3736.25 | 3248.77279 | 3328.56097 | 866.178 | 2939.33 | 2963.66 | 3966.26609 | 3973.23855 |
| 3741.59 | 3312.64949 | 3333.44678 | 866.456 | 2996.81 | 2968.51 | 4021.350632 | 3980.000362 |
| 3746.93 | 3286.12365 | 3338.33259 | 866.734 | 2893.62 | 2973.36 | 3920.323686 | 3986.762174 |
| 3752.28 | 3243.2449 | 3343.2184 | 867.011 | 2966.75 | 2978.21 | 3949.656468 | 3993.523986 |
| 3757.62 | 3306.24877 | 3348.10421 | 867.289 | 2965.32 | 2983.07 | 4038.960834 | 4000.29633 |
| 3762.97 | 3287.18709 | 3352.99002 | 867.567 | 2914.81 | 2987.93 | 3977.546309 | 4007.068675 |
| 3768.32 | 3317.20422 | 3357.88587 | 867.845 | 2977.57 | 2992.79 | 4042.415467 | 4013.851551 |
| 3773.67 | 3291.88228 | 3362.77168 | 868.123 | 2942.47 | 2997.66 | 4029.88189 | 4020.634428 |
| 3779.02 | 3323.01302 | 3367.67755 | 868.401 | 2952.32 | 3002.53 | 3951.025682 | 4027.427837 |
| 3784.38 | 3354.35444 | 3372.57339 | 868.679 | 2926.07 | 3007.4 | 3952.679272 | 4034.221246 |
| 3789.73 | 3327.31695 | 3377.46924 | 868.957 | 2954.82 | 3012.28 | 4038.023449 | 4041.025188 |
| 3795.09 | 3316.3615 | 3382.37511 | 869.234 | 2946.57 | 3017.15 | 4006.584183 | 4047.829129 |
| 3800.45 | 3350.6023 | 3387.27095 | 869.512 | 2984.75 | 3022.03 | 4041.930976 | 4054.633071 |
| 3805.81 | 3291.71173 | 3392.17683 | 869.79 | 2972.28 | 3026.92 | 4030.829808 | 4061.447545 |
| 3811.18 | 3315.1877 | 3397.0827 | 870.068 | 2991.22 | 3031.8 | 4083.260181 | 4068.272551 |
| 3816.54 | 3324.13666 | 3401.99861 | 870.345 | 2933.35 | 3036.7 | 4032.620319 | 4075.087025 |
| 3821.91 | 3370.26593 | 3406.90449 | 870.623 | 2957.09 | 3041.59 | 4014.852131 | 4081.922564 |
| 3827.28 | 3397.15293 | 3411.82039 | 870.901 | 2964.68 | 3046.48 | 4012.45074 | 4088.758103 |
| 3832.65 | 3293.99913 | 3416.7363 | 871.178 | 3005.04 | 3051.38 | 4047.09186 | 4095.593642 |
| 3838.02 | 3369.3931 | 3421.65221 | 871.456 | 2995.49 | 3056.28 | 4107.095041 | 4102.42918 |
| 3843.4 | 3359.73184 | 3426.56812 | 871.734 | 2996.95 | 3061.19 | 4040.150997 | 4109.275252 |
| 3848.77 | 3441.62685 | 3431.49406 | 872.011 | 3006.32 | 3066.09 | 4069.568039 | 4116.131855 |
| 3854.15 | 3421.78263 | 3436.42 | 872.289 | 2960.53 | 3071 | 4117.195629 | 4122.988459 |
| 3859.53 | 3425.605 | 3441.3359 | 872.567 | 2984.38 | 3075.92 | 4129.22365 | 4129.845062 |
| 3864.91 | 3378.10128 | 3446.26184 | 872.844 | 3044.96 | 3080.83 | 4131.151082 | 4136.712198 |
| 3870.29 | 3403.2928 | 3451.19782 | 873.122 | 3012.27 | 3085.75 | 4147.444732 | 4143.579335 |
| 3875.68 | 3433.77143 | 3456.12376 | 873.399 | 3117.68 | 3090.67 | 4114.214955 | 4150.457003 |
| 3881.07 | 3353.56187 | 3461.05973 | 873.677 | 3043.53 | 3095.6 | 4138.028751 | 4157.334671 |
| 3886.45 | 3439.72068 | 3465.98567 | 873.954 | 3095.15 | 3100.52 | 4098.131954 | 4164.222872 |
| 3891.84 | 3443.3725 | 3470.92164 | 874.232 | 3047.06 | 3105.45 | 4254.390896 | 4171.111073 |
| 3897.24 | 3432.30669 | 3475.85761 | 874.509 | 3097.62 | 3110.39 | 4132.688815 | 4177.999274 |
| 3902.63 | 3443.38253 | 3480.80362 | 874.787 | 3116.55 | 3115.32 | 4237.391575 | 4184.898007 |
| 3908.02 | 3469.15593 | 3485.73959 | 875.064 | 3041.27 | 3120.26 | 4132.488699 | 4191.796741 |
| 3913.42 | 3470.78119 | 3490.6856 | 875.342 | 3150.63 | 3125.2 | 4207.605899 | 4198.706006 |
| 3918.82 | 3464.36041 | 3495.6316 | 875.619 | 3107.04 | 3130.15 | 4192.154843 | 4205.615272 |
| 3924.22 | 3463.39729 | 3500.57761 | 875.896 | 3032.09 | 3135.09 | 4190.469656 | 4212.53507 |
| 3929.62 | 3455.08038 | 3505.52361 | 876.174 | 3101.41 | 3140.04 | 4218.264705 | 4219.454868 |
| 3935.03 | 3467.00898 | 3510.46962 | 876.451 | 3059.41 | 3144.99 | 4205.067586 | 4226.374666 |
| 3940.43 | 3458.09012 | 3515.42565 | 876.728 | 3084.97 | 3149.95 | 4205.541545 | 4233.304997 |
| 3945.84 | 3524.99662 | 3520.38169 | 877.006 | 3134.05 | 3154.91 | 4207.43738 | 4240.235328 |
| 3951.24 | 3449.34181 | 3525.33773 | 877.283 | 3028.91 | 3159.87 | 4204.161799 | 4247.176191 |
| 3956.66 | 3519.23799 | 3530.29377 | 877.56 | 3133.99 | 3164.83 | 4226.648509 | 4254.117054 |
| 3962.07 | 3450.46545 | 3535.2498 | 877.838 | 3117.11 | 3169.8 | 4292.012691 | 4261.068449 |
| 3967.48 | 3509.28579 | 3540.20584 | 878.115 | 3136.41 | 3174.77 | 4268.87297 | 4268.019844 |
| 3972.89 | 3542.5434 | 3545.17191 | 878.392 | 3156.28 | 3179.74 | 4262.648312 | 4274.97124 |
| 3978.31 | 3469.8281 | 3550.13798 | 878.669 | 3192.45 | 3184.72 | 4350.646658 | 4281.933167 |
| 3983.73 | 3562.81902 | 3555.10405 | 878.947 | 3159.84 | 3189.69 | 4284.324026 | 4288.905628 |
| 3989.15 | 3530.50445 | 3560.07012 | 879.224 | 3109.14 | 3194.67 | 4331.351269 | 4295.867556 |
| 3994.57 | 3508.92462 | 3565.03619 | 879.501 | 3089.25 | 3199.66 | 4195.957045 | 4302.840016 |
| 3999.99 | 3562.34749 | 3570.01229 | 879.778 | 3168.2 | 3204.64 | 4280.521824 | 4309.823008 |
| 4005.42 | 3522.50857 | 3574.97836 | 880.055 | 3169.43 | 3209.63 | 4397.294734 | 4316.806001 |
| 4010.84 | 3601.52426 | 3579.95446 | 880.332 | 3142.74 | 3214.62 | 4306.347311 | 4323.788994 |
| 4016.27 | 3571.14596 | 3584.93057 | 880.609 | 3172.65 | 3219.61 | 4343.358225 | 4330.782519 |
| 4021.7 | 3573.99518 | 3589.90667 | 880.886 | 3138.8 | 3224.61 | 4315.784357 | 4337.786576 |
| 4027.13 | 3567.55434 | 3594.8928 | 881.163 | 3187.85 | 3229.61 | 4335.869676 | 4344.780101 |
| 4032.56 | 3581.72018 | 3599.86891 | 881.441 | 3209.45 | 3234.61 | 4431.293376 | 4351.784159 |
| 4037.99 | 3522.49854 | 3604.85504 | 881.717 | 3202.35 | 3239.62 | 4326.769668 | 4358.798749 |
| 4043.43 | 3561.54489 | 3609.83114 | 881.995 | 3242.27 | 3244.62 | 4294.835379 | 4365.813338 |
| 4048.87 | 3626.39474 | 3614.81728 | 882.271 | 3174.75 | 3249.63 | 4328.918281 | 4372.827928 |
| 4054.3 | 3631.0498 | 3619.81344 | 882.548 | 3188.6 | 3254.65 | 4389.52181 | 4379.853051 |
| 4059.74 | 3588.79307 | 3624.79958 | 882.825 | 3179.74 | 3259.66 | 4422.182836 | 4386.878173 |
| 4065.18 | 3605.08579 | 3629.78571 | 883.102 | 3195.22 | 3264.68 | 4373.259757 | 4393.913828 |
| 4070.63 | 3622.63257 | 3634.78188 | 883.379 | 3207.93 | 3269.7 | 4316.342575 | 4400.949482 |
| 4076.07 | 3584.9406 | 3639.77805 | 883.656 | 3242.18 | 3274.72 | 4383.391943 | 4407.985137 |
| 4081.52 | 3567.6045 | 3644.77422 | 883.933 | 3235.36 | 3279.75 | 4312.403451 | 4415.031324 |
| 4086.96 | 3634.62136 | 3649.77038 | 884.21 | 3281.68 | 3284.78 | 4400.338602 | 4422.077511 |
| 4092.41 | 3590.50862 | 3654.76655 | 884.487 | 3223.64 | 3289.81 | 4357.756039 | 4429.134231 |
| 4097.86 | 3638.53402 | 3659.77275 | 884.764 | 3249.84 | 3294.84 | 4451.273372 | 4436.190951 |
| 4103.31 | 3624.43841 | 3664.76892 | 885.04 | 3284.07 | 3299.88 | 4421.477164 | 4443.24767 |
| 4108.77 | 3655.50895 | 3669.77512 | 885.317 | 3261.8 | 3304.92 | 4399.453879 | 4450.314922 |
| 4114.22 | 3649.08818 | 3674.78132 | 885.594 | 3271.75 | 3309.96 | 4423.404596 | 4457.382174 |
| 4119.68 | 3632.32393 | 3679.78752 | 885.871 | 3251.99 | 3315.01 | 4459.572917 | 4464.459958 |
| 4125.13 | 3679.31599 | 3684.79372 | 886.148 | 3213.68 | 3320.05 | 4463.122341 | 4471.537743 |
| 4130.59 | 3658.78957 | 3689.79992 | 886.424 | 3199.88 | 3325.1 | 4439.487597 | 4478.62606 |
| 4136.05 | 3648.68688 | 3694.81615 | 886.701 | 3315.85 | 3330.15 | 4485.493196 | 4485.703844 |
| 4141.51 | 3699.83238 | 3699.83238 | 886.978 | 3291.62 | 3335.21 | 4448.629735 | 4492.802693 |
| 4146.97 | 3654.68629 | 3704.84861 | 887.254 | 3275.16 | 3340.27 | 4510.370765 | 4499.901542 |
| 4152.44 | 3678.97489 | 3709.86484 | 887.531 | 3292.07 | 3345.32 | 4463.269795 | 4507.000392 |
| 4157.9 | 3721.1313 | 3714.88108 | 887.808 | 3309.73 | 3350.39 | 4508.601319 | 4514.099241 |
| 4163.37 | 3699.61167 | 3719.89731 | 888.084 | 3309.53 | 3355.45 | 4498.58499 | 4521.208623 |
| 4168.84 | 3631.84237 | 3724.91354 | 888.361 | 3345.32 | 3360.52 | 4480.690413 | 4528.318004 |
| 4174.31 | 3713.99822 | 3729.93981 | 888.638 | 3321.27 | 3365.59 | 4502.977008 | 4535.437918 |
| 4179.78 | 3673.30654 | 3734.96607 | 888.914 | 3301.89 | 3370.66 | 4509.791482 | 4542.557832 |
| 4185.25 | 3694.1841 | 3739.99233 | 889.191 | 3341.66 | 3375.74 | 4525.885016 | 4549.688279 |
| 4190.72 | 3678.86453 | 3745.0186 | 889.467 | 3285.3 | 3380.82 | 4457.392706 | 4556.818725 |
| 4196.2 | 3671.55086 | 3750.04486 | 889.744 | 3343.54 | 3385.9 | 4446.007163 | 4563.949172 |
| 4201.67 | 3711.44997 | 3755.07113 | 890.02 | 3321.69 | 3390.98 | 4521.830035 | 4571.090151 |
| 4207.15 | 3785.42937 | 3760.10743 | 890.297 | 3268.39 | 3396.06 | 4484.239838 | 4578.23113 |
| 4212.63 | 3769.3072 | 3765.13369 | 890.573 | 3357.62 | 3401.15 | 4537.007248 | 4585.372108 |
| 4218.11 | 3685.43579 | 3770.16999 | 890.85 | 3356.23 | 3406.24 | 4500.702006 | 4592.52362 |
| 4223.59 | 3753.03454 | 3775.20628 | 891.126 | 3310.22 | 3411.33 | 4536.75447 | 4599.675131 |
| 4229.07 | 3720.67984 | 3780.24258 | 891.402 | 3299.06 | 3416.43 | 4561.474054 | 4606.837175 |
| 4234.56 | 3754.34879 | 3785.27888 | 891.679 | 3360.49 | 3421.53 | 4569.994779 | 4613.999218 |
| 4240.04 | 3720.45913 | 3790.32521 | 891.955 | 3389.29 | 3426.63 | 4468.767717 | 4621.161262 |
| 4245.53 | 3749.82415 | 3795.36151 | 892.232 | 3343.62 | 3431.73 | 4601.476174 | 4628.333838 |
| 4251.02 | 3751.02805 | 3800.40783 | 892.508 | 3399.36 | 3436.84 | 4604.930807 | 4635.516947 |
| 4256.5 | 3748.22899 | 3805.45416 | 892.784 | 3366.84 | 3441.94 | 4578.346986 | 4642.689523 |
| 4261.99 | 3708.84153 | 3810.49046 | 893.061 | 3374.89 | 3447.05 | 4633.852825 | 4649.872632 |
| 4267.48 | 3740.59428 | 3815.54682 | 893.337 | 3301.62 | 3452.16 | 4584.982409 | 4657.066273 |
| 4272.98 | 3786.65333 | 3820.59315 | 893.613 | 3375.39 | 3457.28 | 4570.753113 | 4664.249381 |
| 4278.47 | 3683.0581 | 3825.63948 | 893.889 | 3331.31 | 3462.4 | 4600.359738 | 4671.453555 |
| 4283.97 | 3798.86284 | 3830.69584 | 894.166 | 3418.49 | 3467.52 | 4617.306398 | 4678.647196 |
| 4289.46 | 3771.79525 | 3835.74217 | 894.442 | 3362.25 | 3472.64 | 4684.124053 | 4685.851369 |
| 4294.96 | 3836.82568 | 3840.79854 | 894.718 | 3364.94 | 3477.76 | 4675.982494 | 4693.055542 |
| 4300.46 | 3717.78046 | 3845.8549 | 894.994 | 3394.72 | 3482.89 | 4607.047823 | 4700.270248 |
| 4305.96 | 3808.01244 | 3850.91126 | 895.27 | 3355.93 | 3488.02 | 4604.541108 | 4707.484954 |
| 4311.46 | 3781.09534 | 3855.96762 | 895.547 | 3374.64 | 3493.15 | 4619.634062 | 4714.710192 |
| 4316.96 | 3757.62941 | 3861.02398 | 895.823 | 3400.7 | 3498.29 | 4698.416543 | 4721.924898 |
| 4322.46 | 3808.8351 | 3866.09038 | 896.099 | 3469.86 | 3503.42 | 4660.62623 | 4729.160669 |
| 4327.97 | 3786.80381 | 3871.14674 | 896.375 | 3391.77 | 3508.56 | 4619.591932 | 4736.385907 |
| 4333.47 | 3763.05697 | 3876.21314 | 896.651 | 3449.03 | 3513.7 | 4657.329583 | 4743.621678 |
| 4338.98 | 3793.39514 | 3881.27953 | 896.927 | 3401.73 | 3518.85 | 4573.954968 | 4750.867981 |
| 4344.49 | 3798.96316 | 3886.34592 | 897.203 | 3466.94 | 3523.99 | 4704.936109 | 4758.114284 |
| 4350 | 3824.00419 | 3891.41232 | 897.479 | 3429.9 | 3529.14 | 4644.890798 | 4765.360587 |
| 4355.51 | 3814.26267 | 3896.47871 | 897.755 | 3521.1 | 3534.29 | 4704.514812 | 4772.60689 |
| 4361.02 | 3863.54213 | 3901.55514 | 898.031 | 3427.64 | 3539.45 | 4691.728458 | 4779.863725 |
| 4366.53 | 3862.65928 | 3906.62153 | 898.307 | 3518.02 | 3544.6 | 4731.51993 | 4787.120561 |
| 4372.04 | 3873.65486 | 3911.69796 | 898.583 | 3467.02 | 3549.76 | 4773.238835 | 4794.387929 |
| 4377.56 | 3859.76993 | 3916.76436 | 898.859 | 3485.37 | 3554.92 | 4793.039779 | 4801.655297 |
| 4383.07 | 3838.41081 | 3921.84078 | 899.135 | 3471.75 | 3560.08 | 4793.366284 | 4808.922665 |
| 4388.59 | 3860.26152 | 3926.91721 | 899.411 | 3529.6 | 3565.25 | 4725.653374 | 4816.200565 |
| 4394.11 | 3877.57755 | 3932.00367 | 899.687 | 3533.73 | 3570.41 | 4816.569199 | 4823.478465 |
| 4399.63 | 3893.17803 | 3937.0801 | 899.963 | 3468.09 | 3575.58 | 4722.967607 | 4830.766898 |
| 4405.15 | 3882.00187 | 3942.15652 | 900.239 | 3498.8 | 3580.75 | 4724.884507 | 4838.044798 |
| 4410.67 | 3847.10896 | 3947.24298 | 900.514 | 3470.02 | 3585.93 | 4819.181239 | 4845.343763 |
| 4416.19 | 3886.12521 | 3952.31941 | 900.79 | 3408.78 | 3591.11 | 4732.246667 | 4852.632196 |
| 4421.71 | 3875.41054 | 3957.40587 | 901.066 | 3486.8 | 3596.28 | 4831.240857 | 4859.931161 |
| 4427.24 | 3903.72215 | 3962.49233 | 901.342 | 3505.6 | 3601.46 | 4783.392085 | 4867.230126 |
| 4432.76 | 3943.07951 | 3967.57879 | 901.617 | 3558.73 | 3606.65 | 4862.480006 | 4874.539624 |
| 4438.29 | 3920.15533 | 3972.66525 | 901.893 | 3564.95 | 3611.83 | 4814.589105 | 4881.849122 |
| 4443.81 | 3895.17449 | 3977.75171 | 902.169 | 3538.7 | 3617.02 | 4834.400582 | 4889.169152 |
| 4449.34 | 3962.10106 | 3982.8482 | 902.445 | 3522.01 | 3622.21 | 4791.818019 | 4896.489182 |
| 4454.87 | 3974.50119 | 3987.93466 | 902.72 | 3532.28 | 3627.4 | 4836.791441 | 4903.809211 |
| 4460.4 | 3949.67084 | 3993.03115 | 902.996 | 3564.54 | 3632.6 | 4887.947391 | 4911.129241 |
| 4465.93 | 3888.61326 | 3998.11761 | 903.272 | 3563.73 | 3637.79 | 4871.400964 | 4918.459804 |
| 4471.47 | 3912.24975 | 4003.2141 | 903.547 | 3594.28 | 3642.99 | 4923.789207 | 4925.790366 |
| 4477 | 4000.79628 | 4008.31059 | 903.823 | 3563.03 | 3648.19 | 4903.472174 | 4933.131461 |
| 4482.53 | 3947.21288 | 4013.40708 | 904.098 | 3552.97 | 3653.4 | 4842.88971 | 4940.472556 |
| 4488.07 | 3965.49203 | 4018.50358 | 904.374 | 3538.7 | 3658.6 | 4907.632479 | 4947.813651 |
| 4493.6 | 3937.27071 | 4023.6101 | 904.65 | 3585.64 | 3663.81 | 4939.535171 | 4955.165278 |
| 4499.14 | 4001.89985 | 4028.70659 | 904.925 | 3628.48 | 3669.02 | 4945.422792 | 4962.516905 |
| 4504.68 | 3965.74285 | 4033.81312 | 905.201 | 3633.47 | 3674.23 | 4965.044685 | 4969.868532 |
| 4510.22 | 4088.71076 | 4038.90961 | 905.476 | 3597.36 | 3679.45 | 4925.337472 | 4977.230692 |
| 4515.76 | 3967.7694 | 4044.01613 | 905.752 | 3589.17 | 3684.66 | 4996.146914 | 4984.592852 |
| 4521.3 | 3956.09162 | 4049.12266 | 906.027 | 3648.42 | 3689.88 | 5034.653431 | 4991.955011 |
| 4526.84 | 4039.01997 | 4054.22918 | 906.303 | 3619.61 | 3695.1 | 4891.317765 | 4999.327704 |
| 4532.38 | 3985.83787 | 4059.33571 | 906.578 | 3668.41 | 3700.33 | 4934.753453 | 5006.700396 |
| 4537.93 | 4014.86179 | 4064.44223 | 906.854 | 3639.55 | 3705.55 | 5006.942641 | 5014.08362 |
| 4543.47 | 4006.80572 | 4069.55879 | 907.129 | 3612.73 | 3710.78 | 4902.334673 | 5021.466845 |
| 4549.02 | 4016.3466 | 4074.66531 | 907.404 | 3673.33 | 3716.01 | 4988.068549 | 5028.850069 |
| 4554.56 | 4071.36463 | 4079.77184 | 907.68 | 3685.06 | 3721.24 | 4973.923513 | 5036.243826 |
| 4560.11 | 3992.72014 | 4084.88839 | 907.955 | 3617.58 | 3726.47 | 4973.070387 | 5043.637583 |
| 4565.66 | 4054.72077 | 4090.00495 | 908.231 | 3550.73 | 3731.71 | 4946.307515 | 5051.03134 |
| 4571.2 | 4083.3133 | 4095.12151 | 908.506 | 3653.66 | 3736.95 | 4859.731045 | 5058.425097 |
| 4576.76 | 3997.23475 | 4100.23806 | 908.781 | 3702.55 | 3742.19 | 4982.602225 | 5065.829386 |
| 4582.31 | 4154.8548 | 4105.35462 | 909.056 | 3702.2 | 3747.43 | 4993.808717 | 5073.244208 |
| 4587.86 | 4021.36283 | 4110.47118 | 909.332 | 3730.7 | 3752.68 | 5040.720103 | 5080.648497 |
| 4593.41 | 4028.73669 | 4115.58773 | 909.607 | 3663.67 | 3757.92 | 5109.875955 | 5088.063319 |
| 4598.96 | 4052.73434 | 4120.71432 | 909.882 | 3726.23 | 3763.17 | 5060.205075 | 5095.488673 |
| 4604.52 | 4049.55405 | 4125.83088 | 910.157 | 3646.07 | 3768.42 | 4981.264608 | 5102.903495 |
| 4610.07 | 4041.3475 | 4130.95747 | 910.433 | 3665.62 | 3773.68 | 5133.415908 | 5110.328849 |
| 4615.63 | 4107.89283 | 4136.07403 | 910.708 | 3676.85 | 3778.93 | 5026.754118 | 5117.764735 |
| 4621.18 | 4066.87009 | 4141.20061 | 910.983 | 3706.79 | 3784.19 | 5094.024667 | 5125.19009 |
| 4626.74 | 4082.77154 | 4146.3272 | 911.258 | 3734.76 | 3789.45 | 5111.392623 | 5132.625976 |
| 4632.3 | 4085.1994 | 4151.45379 | 911.533 | 3798.33 | 3794.71 | 5063.343736 | 5140.072395 |
| 4637.86 | 4084.40683 | 4156.58038 | 911.808 | 3745.66 | 3799.97 | 5060.710631 | 5147.508282 |
| 4643.41 | 4153.82145 | 4161.70697 | 912.083 | 3688.98 | 3805.24 | 5094.235315 | 5154.954701 |
| 4648.98 | 4160.98463 | 4166.84359 | 912.358 | 3743.31 | 3810.5 | 5187.489338 | 5162.411652 |
| 4654.54 | 4116.16962 | 4171.97018 | 912.633 | 3762.11 | 3815.77 | 5061.521627 | 5169.868604 |
| 4660.1 | 4110.07991 | 4177.1068 | 912.909 | 3741.13 | 3821.04 | 5100.417844 | 5177.325555 |
| 4665.66 | 4098.96394 | 4182.23339 | 913.183 | 3711.88 | 3826.32 | 5091.7286 | 5184.782506 |
| 4671.22 | 4134.76981 | 4187.37002 | 913.458 | 3807.64 | 3831.59 | 5146.444508 | 5192.24999 |
| 4676.79 | 4104.22095 | 4192.50664 | 913.734 | 3756.42 | 3836.87 | 5091.654873 | 5199.717474 |
| 4682.35 | 4154.40334 | 4197.64326 | 914.008 | 3750.44 | 3842.15 | 5052.895578 | 5207.184958 |
| 4687.92 | 4105.51514 | 4202.76985 | 914.283 | 3810.86 | 3847.43 | 5168.478325 | 5214.662974 |
| 4693.48 | 4202.1679 | 4207.9165 | 914.558 | 3763.43 | 3852.71 | 5157.714194 | 5222.14099 |
| 4699.05 | 4173.90645 | 4213.05312 | 914.833 | 3813.18 | 3858 | 5184.529728 | 5229.629539 |
| 4704.62 | 4145.87574 | 4218.18975 | 915.108 | 3767.85 | 3863.29 | 5246.607796 | 5237.107555 |
| 4710.19 | 4087.75768 | 4223.32637 | 915.383 | 3808.51 | 3868.58 | 5295.025318 | 5244.596104 |
| 4715.76 | 4238.07409 | 4228.47302 | 915.658 | 3785.69 | 3873.87 | 5147.10805 | 5252.095185 |
| 4721.32 | 4216.97582 | 4233.60964 | 915.933 | 3781.81 | 3879.16 | 5230.998753 | 5259.583734 |
| 4726.9 | 4228.39276 | 4238.7563 | 916.207 | 3834.72 | 3884.46 | 5182.275791 | 5267.082815 |
| 4732.47 | 4226.01507 | 4243.89292 | 916.482 | 3785.94 | 3889.75 | 5225.869466 | 5274.592428 |
| 4738.04 | 4220.18621 | 4249.03957 | 916.757 | 3823.45 | 3895.05 | 5229.418891 | 5282.102042 |
| 4743.61 | 4167.93713 | 4254.18623 | 917.032 | 3837.76 | 3900.35 | 5247.039625 | 5289.611655 |
| 4749.18 | 4226.3662 | 4259.33288 | 917.307 | 3784.39 | 3905.66 | 5215.842605 | 5297.121269 |
| 4754.76 | 4210.22397 | 4264.47954 | 917.581 | 3821.09 | 3910.96 | 5278.320904 | 5304.630882 |
| 4760.33 | 4192.02508 | 4269.62619 | 917.856 | 3831.25 | 3916.27 | 5304.862595 | 5312.151028 |
| 4765.91 | 4228.17205 | 4274.78288 | 918.131 | 3827.98 | 3921.58 | 5222.614949 | 5319.681707 |
| 4771.48 | 4229.06494 | 4279.92953 | 918.405 | 3945.37 | 3926.89 | 5265.502952 | 5327.201852 |
| 4777.06 | 4263.76723 | 4285.07619 | 918.68 | 3824.12 | 3932.2 | 5355.270744 | 5334.732531 |
| 4782.64 | 4214.65832 | 4290.23287 | 918.955 | 3905.69 | 3937.51 | 5262.564408 | 5342.273742 |
| 4788.21 | 4290.65424 | 4295.37953 | 919.229 | 3862.84 | 3942.83 | 5295.309693 | 5349.80442 |
| 4793.79 | 4208.18738 | 4300.53621 | 919.504 | 3914.61 | 3948.15 | 5247.597843 | 5357.345631 |
| 4799.37 | 4281.42437 | 4305.6929 | 919.779 | 3875.15 | 3953.47 | 5366.993325 | 5364.886841 |
| 4804.95 | 4255.12928 | 4310.84959 | 920.053 | 3885.79 | 3958.79 | 5229.692734 | 5372.438585 |
| 4810.53 | 4227.62026 | 4315.99624 | 920.328 | 3888.72 | 3964.12 | 5282.691857 | 5379.990328 |
| 4816.11 | 4290.94518 | 4321.16296 | 920.602 | 3919.7 | 3969.44 | 5359.69436 | 5387.542071 |
| 4821.69 | 4269.91713 | 4326.31965 | 920.877 | 3894.69 | 3974.77 | 5309.854961 | 5395.093814 |
| 4827.27 | 4255.55064 | 4331.47633 | 921.151 | 3858.69 | 3980.1 | 5336.154407 | 5402.65609 |
| 4832.85 | 4236.82003 | 4336.63302 | 921.426 | 3847.33 | 3985.43 | 5307.716881 | 5410.218365 |
| 4838.44 | 4307.30813 | 4341.78971 | 921.7 | 3871.22 | 3990.76 | 5417.348812 | 5417.791173 |
| 4844.02 | 4291.02544 | 4346.95643 | 921.975 | 3841.43 | 3996.1 | 5366.034875 | 5425.353449 |
| 4849.6 | 4240.32136 | 4352.11311 | 922.249 | 3893.76 | 4001.44 | 5329.224077 | 5432.926257 |
| 4855.19 | 4281.95609 | 4357.27983 | 922.523 | 3925.22 | 4006.78 | 5300.333656 | 5440.509597 |
| 4860.77 | 4278.67547 | 4362.43652 | 922.798 | 3925.49 | 4012.12 | 5378.16822 | 5448.092938 |
| 4866.36 | 4267.26856 | 4367.60324 | 923.072 | 3974.3 | 4017.46 | 5339.587975 | 5455.665746 |
| 4871.94 | 4320.23997 | 4372.76996 | 923.346 | 3948.41 | 4022.8 | 5469.789717 | 5463.259618 |
| 4877.53 | 4355.50408 | 4377.93667 | 923.621 | 3941.95 | 4028.15 | 5334.964244 | 5470.842959 |
| 4883.12 | 4331.0349 | 4383.09336 | 923.895 | 3977.83 | 4033.5 | 5404.983754 | 5478.436832 |
| 4888.7 | 4248.42759 | 4388.26008 | 924.169 | 4028.03 | 4038.85 | 5390.84925 | 5486.030705 |
| 4894.29 | 4274.40164 | 4393.43683 | 924.444 | 3974.25 | 4044.2 | 5431.89408 | 5493.63511 |
| 4899.88 | 4337.30519 | 4398.60355 | 924.718 | 3996.06 | 4049.55 | 5409.259915 | 5501.239515 |
| 4905.47 | 4408.59589 | 4403.77027 | 924.992 | 3979.24 | 4054.91 | 5369.879207 | 5508.84392 |
| 4911.06 | 4338.34857 | 4408.93699 | 925.266 | 3958.92 | 4060.26 | 5491.317978 | 5516.448326 |
| 4916.65 | 4403.98095 | 4414.10371 | 925.54 | 4092.85 | 4065.62 | 5394.64092 | 5524.063263 |
| 4922.24 | 4307.87998 | 4419.28046 | 925.815 | 3988.07 | 4070.98 | 5492.961035 | 5531.678201 |
| 4927.83 | 4348.42116 | 4424.44718 | 926.089 | 3985.92 | 4076.34 | 5500.87088 | 5539.293139 |
| 4933.42 | 4412.22764 | 4429.62393 | 926.363 | 3953.88 | 4081.71 | 5504.904796 | 5546.918609 |
| 4939.01 | 4342.75282 | 4434.80068 | 926.637 | 4058.88 | 4087.07 | 5498.480022 | 5554.544079 |
| 4944.6 | 4404.27189 | 4439.9674 | 926.911 | 3973.06 | 4092.44 | 5458.446304 | 5562.169549 |
| 4950.19 | 4384.23706 | 4445.14415 | 927.185 | 4042.94 | 4097.81 | 5573.41817 | 5569.795019 |
| 4955.79 | 4409.82988 | 4450.3209 | 927.459 | 4026.5 | 4103.18 | 5553.996393 | 5577.431021 |
| 4961.38 | 4332.22877 | 4455.49766 | 927.734 | 3966.14 | 4108.55 | 5466.703719 | 5585.067024 |
| 4966.97 | 4359.25623 | 4460.67441 | 928.008 | 4018.97 | 4113.93 | 5508.148781 | 5592.713559 |
| 4972.57 | 4506.69332 | 4465.85116 | 928.282 | 4070.48 | 4119.3 | 5525.327153 | 5600.349561 |
| 4978.16 | 4437.28873 | 4471.02791 | 928.556 | 4101.57 | 4124.68 | 5563.8758 | 5607.996096 |
| 4983.76 | 4375.33827 | 4476.20466 | 928.83 | 4057.04 | 4130.06 | 5553.76468 | 5615.653164 |
| 4989.35 | 4378.34801 | 4481.38141 | 929.104 | 4031.68 | 4135.44 | 5543.864208 | 5623.299699 |
| 4994.95 | 4465.22915 | 4486.5682 | 929.377 | 4088.82 | 4140.82 | 5581.201627 | 5630.956766 |
| 5000.54 | 4499.44988 | 4491.74495 | 929.651 | 4022.68 | 4146.2 | 5609.460102 | 5638.613833 |
| 5006.14 | 4414.64546 | 4496.9217 | 929.925 | 4040.72 | 4151.59 | 5588.942954 | 5646.281433 |
| 5011.74 | 4517.60864 | 4502.10849 | 930.199 | 4018.64 | 4156.98 | 5485.735797 | 5653.9385 |
| 5017.33 | 4460.08249 | 4507.28524 | 930.473 | 4124.1 | 4162.37 | 5609.312648 | 5661.6061 |
| 5022.93 | 4526.1864 | 4512.47202 | 930.747 | 4115.6 | 4167.76 | 5637.097165 | 5669.284232 |
| 5028.53 | 4472.93408 | 4517.64877 | 931.021 | 4104.58 | 4173.15 | 5750.331183 | 5676.951832 |
| 5034.13 | 4486.70865 | 4522.83556 | 931.294 | 4114.33 | 4178.55 | 5628.629102 | 5684.629964 |
| 5039.73 | 4502.04829 | 4528.02234 | 931.568 | 4124.6 | 4183.94 | 5632.894731 | 5692.308096 |
| 5045.32 | 4497.18255 | 4533.20912 | 931.842 | 4140.71 | 4189.34 | 5628.260467 | 5699.996761 |
| 5050.92 | 4604.72053 | 4538.39591 | 932.116 | 4137.21 | 4194.74 | 5666.819647 | 5707.674893 |
| 5056.52 | 4504.49621 | 4543.57266 | 932.39 | 4072.54 | 4200.14 | 5566.730086 | 5715.363558 |
| 5062.12 | 4557.81876 | 4548.76948 | 932.663 | 4088 | 4205.54 | 5648.482708 | 5723.062755 |
| 5067.72 | 4526.00582 | 4553.95626 | 932.937 | 4121.01 | 4210.94 | 5639.667075 | 5730.751419 |
| 5073.32 | 4512.49209 | 4559.14305 | 933.211 | 4073.66 | 4216.35 | 5609.259986 | 5738.450616 |
| 5078.92 | 4557.55792 | 4564.32983 | 933.484 | 4137.33 | 4221.76 | 5629.545422 | 5746.149813 |
| 5084.52 | 4546.77302 | 4569.51661 | 933.758 | 4114.39 | 4227.16 | 5663.365014 | 5753.859543 |
| 5090.12 | 4537.95448 | 4574.7034 | 934.032 | 4133.18 | 4232.58 | 5798.516992 | 5761.55874 |
| 5095.73 | 4514.2277 | 4579.90021 | 934.305 | 4162.23 | 4237.99 | 5689.854043 | 5769.268469 |
| 5101.33 | 4591.02622 | 4585.087 | 934.579 | 4175.03 | 4243.4 | 5652.242781 | 5776.988731 |
| 5106.93 | 4520.21708 | 4590.27378 | 934.852 | 4185.49 | 4248.82 | 5679.911441 | 5784.69846 |
| 5112.53 | 4563.49713 | 4595.4706 | 935.126 | 4238.54 | 4254.23 | 5696.784374 | 5792.418722 |
| 5118.13 | 4524.88218 | 4600.65738 | 935.4 | 4152.69 | 4259.65 | 5492.150039 | 5800.138984 |
| 5123.74 | 4557.17668 | 4605.8542 | 935.673 | 4149.16 | 4265.07 | 5708.190982 | 5807.859246 |
| 5129.34 | 4563.8884 | 4611.04098 | 935.947 | 4247.25 | 4270.49 | 5633.168574 | 5815.59004 |
| 5134.94 | 4580.62255 | 4616.2378 | 936.22 | 4172.71 | 4275.91 | 5656.181905 | 5823.320835 |
| 5140.54 | 4561.99226 | 4621.43462 | 936.494 | 4168.34 | 4281.34 | 5754.417761 | 5831.051629 |
| 5146.15 | 4487.38083 | 4626.63143 | 936.767 | 4211.35 | 4286.76 | 5839.835665 | 5838.792955 |
| 5151.75 | 4632.90172 | 4631.81822 | 937.04 | 4254.4 | 4292.19 | 5784.350891 | 5846.52375 |
| 5157.35 | 4606.5966 | 4637.01503 | 937.314 | 4176.39 | 4297.62 | 5733.837418 | 5854.265076 |
| 5162.96 | 4597.21625 | 4642.21185 | 937.587 | 4296 | 4303.05 | 5824.637387 | 5862.006403 |
| 5168.56 | 4489.08635 | 4647.40867 | 937.861 | 4215.27 | 4308.48 | 5779.600771 | 5869.758262 |
| 5174.17 | 4597.50719 | 4652.60548 | 938.134 | 4198.86 | 4313.91 | 5826.143522 | 5877.510121 |
| 5179.77 | 4634.6574 | 4657.8023 | 938.407 | 4199.31 | 4319.35 | 5914.057609 | 5885.26198 |
| 5185.38 | 4575.21505 | 4662.99912 | 938.681 | 4123.29 | 4324.78 | 5835.422582 | 5893.013839 |
| 5190.98 | 4588.43784 | 4668.19593 | 938.954 | 4283.46 | 4330.22 | 5892.065922 | 5900.776231 |
| 5196.59 | 4626.21007 | 4673.39275 | 939.227 | 4133.57 | 4335.66 | 5784.582604 | 5908.52809 |
| 5202.19 | 4589.47118 | 4678.5996 | 939.5 | 4260.54 | 4341.1 | 5879.679799 | 5916.301014 |
| 5207.8 | 4642.17172 | 4683.79641 | 939.774 | 4178.49 | 4346.55 | 5751.131647 | 5924.063406 |
| 5213.4 | 4718.46861 | 4688.99323 | 940.047 | 4188.35 | 4351.99 | 5732.32075 | 5931.836329 |
| 5219.01 | 4599.82469 | 4694.20008 | 940.32 | 4204.34 | 4357.43 | 5844.196086 | 5939.598721 |
| 5224.61 | 4582.00703 | 4699.3969 | 940.593 | 4252.78 | 4362.88 | 5802.698362 | 5947.382177 |
| 5230.22 | 4547.46526 | 4704.59371 | 940.867 | 4245.24 | 4368.33 | 5808.301608 | 5955.155101 |
| 5235.82 | 4560.40714 | 4709.80056 | 941.14 | 4293.39 | 4373.78 | 5952.279751 | 5962.938558 |
| 5241.43 | 4614.76303 | 4714.99738 | 941.413 | 4211.29 | 4379.23 | 5923.231345 | 5970.722014 |
| 5247.04 | 4675.05814 | 4720.20423 | 941.686 | 4284.48 | 4384.68 | 6016.200992 | 5978.50547 |
| 5252.64 | 4715.72975 | 4725.40104 | 941.959 | 4289.18 | 4390.13 | 5838.634969 | 5986.288927 |
| 5258.25 | 4598.84151 | 4730.60789 | 942.232 | 4339.66 | 4395.59 | 5943.285067 | 5994.082915 |
| 5263.86 | 4616.58894 | 4735.80471 | 942.505 | 4238.15 | 4401.04 | 5847.387408 | 6001.876904 |
| 5269.46 | 4699.35677 | 4741.01156 | 942.778 | 4201.33 | 4406.5 | 5920.313865 | 6009.670893 |
| 5275.07 | 4608.37235 | 4746.21841 | 943.051 | 4347.93 | 4411.96 | 5906.948227 | 6017.464882 |
| 5280.68 | 4684.06729 | 4751.42526 | 943.324 | 4315.03 | 4417.42 | 5943.569442 | 6025.269403 |
| 5286.28 | 4695.67485 | 4756.62207 | 943.597 | 4244.47 | 4422.88 | 5957.861932 | 6033.073924 |
| 5291.89 | 4662.20655 | 4761.82892 | 943.87 | 4275.06 | 4428.35 | 6005.268342 | 6040.878445 |
| 5297.49 | 4658.58483 | 4767.03577 | 944.143 | 4290.68 | 4433.81 | 5980.896329 | 6048.693499 |
| 5303.1 | 4662.95899 | 4772.24262 | 944.416 | 4333.69 | 4439.28 | 6002.951211 | 6056.508552 |
| 5308.71 | 4735.94516 | 4777.44947 | 944.689 | 4277.28 | 4444.74 | 5932.078575 | 6064.323606 |
| 5314.32 | 4708.86754 | 4782.65632 | 944.962 | 4319.61 | 4450.21 | 5969.152684 | 6072.13866 |
| 5319.92 | 4670.05194 | 4787.85313 | 945.235 | 4368.78 | 4455.68 | 6014.884439 | 6079.953713 |
| 5325.53 | 4787.80297 | 4793.07002 | 945.508 | 4313.68 | 4461.15 | 6064.723838 | 6087.779299 |
| 5331.14 | 4659.54795 | 4798.26683 | 945.78 | 4370.14 | 4466.63 | 6010.09219 | 6095.604885 |
| 5336.74 | 4734.66101 | 4803.48371 | 946.053 | 4364.62 | 4472.1 | 6056.771863 | 6103.430471 |
| 5342.35 | 4765.45064 | 4808.69056 | 946.326 | | 4477.58 | 6094.172476 | 6111.256057 |
| 5347.96 | 4711.08472 | 4813.89741 | 946.599 | | 4483.05 | 6106.906169 | 6119.092176 |
| 5353.56 | 4767.91863 | 4819.10426 | 946.872 | | 4488.53 | 6080.975358 | 6126.928294 |
| 5359.17 | 4704.53352 | 4824.31111 | 947.144 | 4385.81 | 4494.01 | 6017.433284 | 6134.764413 |
| 5364.78 | 4810.22553 | 4829.51796 | 947.417 | 4396.56 | 4499.49 | 6027.712924 | 6142.611063 |
| 5370.38 | 4732.99562 | 4834.72481 | 947.69 | 4403.75 | 4504.97 | 6039.698814 | 6150.447182 |
| 5375.99 | 4820.37838 | 4839.94169 | 947.963 | 4269.88 | 4510.45 | 6104.873412 | 6158.293833 |
| 5381.6 | 4725.46124 | 4845.14854 | 948.235 | 4298.07 | 4515.94 | 6084.145615 | 6166.140484 |
| 5387.2 | 4774.03843 | 4850.35539 | 948.508 | 4403.01 | 4521.43 | 6013.515225 | 6173.997667 |
| 5392.81 | 4775.98473 | 4855.56224 | 948.78 | 4397.15 | 4526.91 | 6065.03981 | 6181.844318 |
| 5398.42 | 4795.22699 | 4860.77912 | 949.053 | | 4532.4 | 5999.359656 | 6189.701501 |
| 5404.02 | 4815.09127 | 4865.98597 | 949.326 | | 4537.89 | 6091.44458 | 6197.558684 |
| 5409.63 | 4746.94074 | 4871.19281 | 949.598 | | 4543.38 | 6036.528557 | 6205.415867 |
| 5415.24 | 4766.02249 | 4876.4097 | 949.871 | 4388.07 | 4548.87 | 6050.778917 | 6213.283583 |
| 5420.84 | 4718.49871 | 4881.61655 | 950.143 | 4449.46 | 4554.36 | 6035.23307 | 6221.151299 |
| 5426.45 | 4734.33997 | 4886.83343 | 950.416 | 4503.52 | 4559.86 | 6045.112477 | 6229.019015 |
| 5432.06 | 4840.08214 | 4892.04028 | 950.688 | 4466.81 | 4565.35 | 6106.379548 | 6236.88673 |
| 5437.66 | 4746.37893 | 4897.24712 | 950.961 | 4414.96 | 4570.85 | 6141.484095 | 6244.764978 |
| 5443.27 | 4787.21106 | 4902.46401 | 951.233 | 4456.41 | 4576.35 | 6074.887621 | 6252.632694 |
| 5448.87 | 4835.36688 | 4907.67085 | 951.506 | 4404.3 | 4581.85 | 6092.381966 | 6260.510942 |
| 5454.48 | 4839.03877 | 4912.88774 | 951.778 | 4386.04 | 4587.35 | 6195.747108 | 6268.38919 |
| 5460.08 | 4832.39728 | 4918.10462 | 952.051 | 4402.95 | 4592.85 | 6211.335086 | 6276.277971 |
| 5465.69 | 4762.93249 | 4923.31147 | 952.323 | 4394.66 | 4598.35 | 6069.073726 | 6284.156219 |
| 5471.29 | 4851.48905 | 4928.52835 | 952.595 | 4414.11 | 4603.86 | 6148.604009 | 6292.044999 |
| 5476.9 | 4847.98772 | 4933.7352 | 952.868 | 4495.16 | 4609.36 | 6271.56998 | 6299.93378 |
| 5482.5 | 4866.98921 | 4938.95208 | 953.14 | 4390.3 | 4614.87 | 6125.780261 | 6307.833093 |
| 5488.11 | 4790.03018 | 4944.15893 | 953.412 | 4534.83 | 4620.38 | 6246.292179 | 6315.721873 |
| 5493.71 | 4853.94701 | 4949.37581 | 953.685 | 4496.92 | 4625.89 | 6220.403497 | 6323.621186 |
| 5499.32 | 4896.57495 | 4954.59269 | 953.957 | 4533.27 | 4631.4 | 6109.318092 | 6331.520499 |
| 5504.92 | 4885.26836 | 4959.79954 | 954.229 | 4484.53 | 4636.91 | 6278.911075 | 6339.419812 |
| 5510.52 | 4863.62834 | 4965.01642 | 954.501 | 4444.29 | 4642.42 | 6199.781024 | 6347.329657 |
| 5516.13 | 4932.32062 | 4970.2333 | 954.774 | 4541.49 | 4647.93 | 6227.744592 | 6355.22897 |
| 5521.73 | 4883.61301 | 4975.44015 | 955.046 | 4490.78 | 4653.45 | 6207.195846 | 6363.138816 |
| 5527.34 | 4828.13348 | 4980.65703 | 955.318 | 4487.02 | 4658.96 | 6302.756468 | 6371.048661 |
| 5532.94 | 4948.00136 | 4985.87391 | 955.59 | 4488.16 | 4664.48 | 6227.565541 | 6378.969039 |
| 5538.54 | 4936.0226 | 4991.08076 | 955.862 | 4527.15 | 4670 | 6266.061526 | 6386.878884 |
| 5544.15 | 4929.38111 | 4996.29764 | 956.135 | 4529.48 | 4675.52 | 6281.902282 | 6394.799262 |
| 5549.75 | 4927.45487 | 5001.51453 | 956.407 | 4590.36 | 4681.03 | 6190.733678 | 6402.71964 |
| 5555.35 | 4860.27749 | 5006.72137 | 956.679 | 4583.65 | 4686.56 | 6291.950208 | 6410.640018 |
| 5560.95 | 4892.06034 | 5011.93826 | 956.951 | 4562.56 | 4692.08 | 6254.697048 | 6418.560395 |
| 5566.56 | 4868.06269 | 5017.15514 | 957.223 | 4563.56 | 4697.6 | 6429.956471 | 6426.491306 |
| 5572.16 | 4867.04941 | 5022.37202 | 957.495 | 4541.31 | 4703.13 | 6306.874643 | 6434.422216 |
| 5577.76 | 4992.15424 | 5027.57887 | 957.767 | 4613.81 | 4708.65 | 6436.960528 | 6442.353126 |
| 5583.36 | 4979.58356 | 5032.79575 | 958.039 | 4564.87 | 4714.18 | 6272.012342 | 6450.284036 |
| 5588.96 | 5066.37441 | 5038.01263 | 958.311 | 4578.77 | 4719.71 | 6364.613354 | 6458.225479 |
| 5594.56 | 5152.06169 | 5043.22951 | 958.583 | 4533.56 | 4725.23 | 6254.739177 | 6466.156389 |
| 5600.16 | 5033.97958 | 5048.43636 | 958.855 | 4591.14 | 4730.76 | 6298.01688 | 6474.097831 |
| 5605.76 | 4892.29109 | 5053.65324 | 959.127 | 4695.27 | 4736.3 | 6419.339794 | 6482.039274 |
| 5611.36 | 4925.70923 | 5058.87012 | 959.399 | 4583.86 | 4741.83 | 6438.392937 | 6489.980717 |
| 5616.96 | 5022.60277 | 5064.08701 | 959.67 | 4590.38 | 4747.36 | 6307.29594 | 6497.932692 |
| 5622.56 | 4980.83762 | 5069.30389 | 959.942 | 4638.46 | 4752.89 | 6436.518167 | 6505.884667 |
| 5628.16 | 4954.8134 | 5074.51074 | 960.214 | 4564.73 | 4758.43 | 6462.364719 | 6513.836642 |
| 5633.76 | 4975.63077 | 5079.72762 | 960.486 | 4674.05 | 4763.96 | 6421.288291 | 6521.788617 |
| 5639.36 | 5055.77009 | 5084.9445 | 960.758 | 4724.66 | 4769.5 | 6377.052139 | 6529.740592 |
| 5644.95 | 5024.95036 | 5090.16138 | 961.03 | 4672.21 | 4775.04 | 6375.27216 | 6537.703099 |
| 5650.55 | 5136.35085 | 5095.36823 | 961.301 | 4636.8 | 4780.58 | 6444.775582 | 6545.655074 |
| 5656.15 | 5165.91652 | 5100.58511 | 961.573 | 4683.68 | 4786.12 | 6553.006701 | 6553.617582 |
| 5661.75 | 5138.99942 | 5105.80199 | 961.845 | 4727.65 | 4791.66 | 6481.849691 | 6561.590622 |
| 5667.34 | 5109.89524 | 5111.01887 | 962.117 | 4702.36 | 4797.2 | 6429.861679 | 6569.553129 |
| 5672.94 | 5117.07848 | 5116.22572 | 962.388 | 4785.93 | 4802.74 | 6409.797424 | 6577.515636 |
| 5678.54 | 5129.05724 | 5121.4426 | 962.66 | 4694.96 | 4808.29 | 6365.982568 | 6585.488676 |
| 5684.13 | 5168.55506 | 5126.65949 | 962.932 | 4716.96 | 4813.83 | 6458.225479 | 6593.461716 |
| 5689.73 | 5117.43965 | 5131.87637 | 963.203 | 4796.03 | 4819.38 | 6621.046117 | 6601.434756 |
| 5695.32 | 5044.94506 | 5137.08322 | 963.475 | 4753.78 | 4824.93 | 6493.340558 | 6609.418328 |
| 5700.92 | 4999.53813 | 5142.3001 | 963.746 | 4861.7 | 4830.47 | 6580.496311 | 6617.391368 |
| 5706.51 | 5089.24843 | 5147.51698 | 964.018 | 4738.97 | 4836.02 | 6467.736251 | 6625.37494 |
| 5712.1 | 5019.27199 | 5152.73386 | 964.29 | 4686.64 | 4841.57 | 6494.3938 | 6633.358513 |
| 5717.7 | 5154.45944 | 5157.94071 | 964.561 | 4689.81 | 4847.12 | 6407.238047 | 6641.342085 |
| 5723.29 | 5031.80254 | 5163.15759 | 964.833 | 4582.98 | 4852.67 | 6603.162072 | 6649.325657 |
| 5728.88 | 5069.50454 | 5168.37447 | 965.104 | 4671.22 | 4858.23 | 6514.689768 | 6657.319762 |
| 5734.48 | 5266.20103 | 5173.58132 | 965.376 | 4811.55 | 4863.78 | 6666.177525 | 6665.303334 |
| 5740.07 | 5213.88173 | 5178.7982 | 965.647 | 4785.87 | 4869.33 | 6476.520287 | 6673.297439 |
| 5745.66 | 5192.51258 | 5184.01508 | 965.919 | 4744.4 | 4874.89 | 6578.516216 | 6681.302076 |
| 5751.25 | 5152.69373 | 5189.22193 | 966.19 | 4877.65 | 4880.44 | 6712.130461 | 6689.296181 |
| 5756.84 | 5218.26592 | 5194.43881 | 966.461 | 4744.84 | 4886 | 6540.283541 | 6697.290285 |
| 5762.43 | 5021.52929 | 5199.6557 | 966.733 | 4790.7 | 4891.56 | 6606.079552 | 6705.294922 |
| 5768.02 | 5121.63322 | 5204.86254 | 967.004 | 4722.03 | 4897.12 | 6592.966693 | 6713.29956 |
| 5773.61 | 5142.63117 | 5210.07943 | 967.275 | 4819.06 | 4902.68 | 6808.997103 | 6721.304197 |
| 5779.2 | 5043.87159 | 5215.29631 | 967.547 | 4759.18 | 4908.24 | 6520.135027 | 6729.308834 |
| 5784.79 | 5224.87731 | 5220.50316 | 967.818 | 4884.48 | 4913.8 | 6489.696341 | 6737.313471 |
| 5790.37 | 5065.17051 | 5225.72004 | 968.09 | 4832.39 | 4919.36 | 6658.552055 | 6745.32864 |
| 5795.96 | 5255.82746 | 5230.92689 | 968.361 | 4845.15 | 4924.93 | 6781.265248 | 6753.34381 |
| 5801.55 | 5249.37659 | 5236.14377 | 968.632 | 4842.55 | 4930.49 | 6809.850229 | 6761.35898 |
| 5807.14 | 5181.33642 | 5241.35062 | 968.903 | 4935.06 | 4936.05 | 6751.38478 | 6769.374149 |
| 5812.72 | 5177.67457 | 5246.5675 | 969.174 | 4757.39 | 4941.62 | 6781.981452 | 6777.389319 |
| 5818.31 | 5130.92328 | 5251.77435 | 969.446 | 4895.7 | 4947.19 | 6566.561922 | 6785.41502 |
| 5823.89 | 5220.22225 | 5256.99123 | 969.717 | 4803.26 | 4952.75 | 6654.00205 | 6793.43019 |
| 5829.48 | 5176.64122 | 5262.19808 | 969.988 | 4762.89 | 4958.32 | 6807.828004 | 6801.455892 |
| 5835.06 | 5111.44024 | 5267.41496 | 970.259 | 4862.56 | 4963.89 | 6751.532234 | 6809.481594 |
| 5840.65 | 5151.97139 | 5272.62181 | 970.53 | 4837.65 | 4969.46 | 6835.012173 | 6817.517828 |
| 5846.23 | 5206.29719 | 5277.83869 | 970.801 | 4866.2 | 4975.03 | 6812.177893 | 6825.54353 |
| 5851.81 | 5280.7481 | 5283.04554 | 971.072 | 4799.4 | 4980.6 | 6794.957391 | 6833.579765 |
| 5857.39 | 5246.20633 | 5288.25239 | 971.343 | 4838.82 | 4986.17 | 6657.519878 | 6841.605467 |
| 5862.97 | 5284.00866 | 5293.46927 | 971.615 | 4882.61 | 4991.75 | 6742.042526 | 6849.641701 |
| 5868.56 | 5241.66162 | 5298.67612 | 971.886 | 4898.51 | 4997.32 | 6708.128143 | 6857.677935 |
| 5874.14 | 5319.87472 | 5303.88297 | 972.157 | 4853 | 5002.9 | 6941.905676 | 6865.724702 |
| 5879.72 | 5390.423 | 5309.08982 | 972.428 | 4838.52 | 5008.47 | 6825.037974 | 6873.760936 |
| 5885.29 | 5312.83193 | 5314.3067 | 972.698 | 4844.68 | 5014.05 | 6805.5846 | 6881.807703 |
| 5890.87 | 5303.29105 | 5319.51355 | 972.97 | 4798.02 | 5019.62 | 6886.157591 | 6889.843938 |
| 5896.45 | 5339.98981 | 5324.7204 | 973.24 | 4908.55 | 5025.2 | 6784.530297 | 6897.890704 |
| 5902.03 | 5380.00931 | 5329.92724 | 973.511 | 4977.53 | 5030.78 | 6797.927532 | 6905.948003 |
| 5907.61 | 5318.82131 | 5335.13409 | 973.782 | 4941.81 | 5036.36 | 6881.27055 | 6913.99477 |
| 5913.18 | 5320.54689 | 5340.35098 | 974.053 | 4874.29 | 5041.94 | 6783.635042 | 6922.041537 |
| 5918.76 | 5176.24996 | 5345.55782 | 974.324 | 4878.47 | 5047.52 | 6871.212091 | 6930.098836 |
| 5924.33 | 5217.99504 | 5350.76467 | 974.595 | 4856.93 | 5053.1 | 6841.953036 | 6938.156135 |
| 5929.91 | 5262.49905 | 5355.97152 | 974.866 | 4946.18 | 5058.68 | 6850.400035 | 6946.213435 |
| 5935.48 | 5392.17869 | 5361.17837 | 975.136 | 4883.6 | 5064.26 | 6931.394324 | 6954.270734 |
| 5941.06 | 5300.25122 | 5366.38522 | 975.407 | 4887.85 | 5069.85 | 6859.35259 | 6962.328033 |
| 5946.63 | 5440.05361 | 5371.59207 | 975.678 | 5018.4 | 5075.43 | 6865.145419 | 6970.395865 |
| 5952.2 | 5221.43618 | 5376.79892 | 975.949 | 4902.44 | 5081.02 | 6863.723543 | 6978.453164 |
| 5957.77 | 5340.7924 | 5382.00577 | 976.219 | 4924.41 | 5086.6 | 6834.359163 | 6986.520995 |
| 5963.34 | 5364.45899 | 5387.20258 | 976.49 | 5017.68 | 5092.19 | 6914.447664 | 6994.588827 |
| 5968.91 | 5274.95937 | 5392.40943 | 976.761 | 4973.71 | 5097.77 | 6960.674444 | 7002.656659 |
| 5974.48 | 5322.61358 | 5397.61628 | 977.031 | 4945.87 | 5103.36 | 6930.225225 | 7010.735023 |
| 5980.05 | 5342.11669 | 5402.82313 | 977.302 | 4928.72 | 5108.95 | 7048.230428 | 7018.802854 |
| 5985.62 | 5341.03318 | 5408.02998 | 977.573 | 4981.9 | 5114.54 | 6949.489016 | 7026.881218 |
| 5991.19 | 5462.78717 | 5413.2268 | 977.843 | 5019.88 | 5120.13 | 6948.204061 | 7034.94905 |
| 5996.76 | 5356.52331 | 5418.43364 | 978.114 | 4979.97 | 5125.72 | 7112.436044 | 7043.027414 |
| 6002.32 | 5372.806 | 5423.64049 | 978.385 | 5065.23 | 5131.31 | 7080.164717 | 7051.105778 |
| 6007.89 | 5417.39027 | 5428.83731 | 978.655 | 5082.29 | 5136.9 | 7004.66835 | 7059.194674 |
| 6013.45 | 5358.03821 | 5434.04416 | 978.926 | 4940.89 | 5142.49 | 6962.117385 | 7067.273038 |
| 6019.02 | 5414.98248 | 5439.25101 | 979.196 | 5062.6 | 5148.09 | 7002.098441 | 7075.351403 |
| 6024.58 | 5364.6496 | 5444.44782 | 979.466 | 4957.08 | 5153.68 | 7060.974653 | 7083.440299 |
| 6030.15 | 5339.14708 | 5449.64464 | 979.737 | 5061.48 | 5159.27 | 7032.863631 | 7091.529195 |
| 8431.5 | 7504.96551 | 7722.58972 | 1101.15 | 7602.33 | 7700.15 | 10591.29346 | 10838.17332 |
| 8463.47 | 7593.86317 | 7753.24893 | 1102.87 | 7563.48 | 7735.84 | 10677.44863 | 10891.88865 |
| 8495.34 | 7284.54223 | 7783.84795 | 1104.6 | 7738.38 | 7771.49 | 10666.81089 | 10945.60398 |
| 8527.13 | 7637.0128 | 7814.37674 | 1106.32 | 7697.3 | 7807.09 | 11014.69663 | 10999.3193 |
| 8558.83 | 7680.74431 | 7844.82527 | 1108.05 | 7510.64 | 7842.65 | 11013.74872 | 11052.92931 |
| 8590.44 | 7598.5082 | 7875.20357 | 1109.77 | 7793.42 | 7878.15 | 10907.58195 | 11106.53931 |
| 8621.95 | 7925.69696 | 7905.50161 | 1111.5 | 7958.79 | 7913.62 | 11002.16306 | 11160.04399 |
| 8653.38 | 7911.92239 | 7935.72943 | 1113.22 | 7971.19 | 7949.03 | 11200.48847 | 11213.54867 |
| 8684.72 | 7829.22478 | 7965.87698 | 1114.95 | 8093.13 | 7984.4 | 11251.67602 | 11267.05335 |
| 8715.96 | 7911.13986 | 7995.95431 | 1116.67 | 7854.21 | 8019.71 | 11009.43043 | 11320.45271 |
| 8747.11 | 8052.82835 | 8025.95138 | 1118.39 | 7693.66 | 8054.98 | 11247.35773 | 11373.85206 |
| 8778.17 | 7825.99433 | 8055.86819 | 1120.12 | 7931.03 | 8090.19 | 11111.2789 | 11427.14609 |
| 8809.13 | 7963.87049 | 8085.71477 | 1121.84 | 7995.3 | 8125.35 | 11099.48259 | 11480.44012 |
| 8840 | 8151.98923 | 8115.47106 | 1123.57 | 8253.6 | 8160.46 | 11261.1552 | 11533.62883 |
| 8870.77 | 8018.20631 | 8145.15712 | 1125.29 | 8410.7 | 8195.51 | 11378.27568 | 11586.81754 |
| 8901.45 | 7898.88018 | 8174.75289 | 1127.01 | 8421.99 | 8230.51 | 11386.28031 | 11640.00625 |
| 8932.03 | 7855.73056 | 8204.27843 | 1128.74 | 8205.78 | 8265.45 | 11489.07671 | 11693.08963 |
| 8962.51 | 8073.58552 | 8233.71368 | 1130.46 | 8166.55 | 8300.34 | 11479.38688 | 11746.17301 |
| 8992.9 | 8138.55576 | 8263.06867 | 1132.18 | 8283.37 | 8335.16 | 11493.395 | 11799.04575 |
| 9023.18 | 8203.63635 | 8292.3434 | 1133.91 | 8320.11 | 8369.93 | 11941.02273 | 11852.02381 |
| 9053.37 | 8173.03734 | 8321.52784 | 1135.63 | 8345.1 | 8404.64 | 11658.33265 | 11904.89654 |
| 9083.46 | 8015.07619 | 8350.63202 | 1137.35 | 8558.7 | 8439.29 | 11674.02595 | 11957.66395 |
| 9113.44 | 8131.82397 | 8379.65594 | 1139.07 | 8569.62 | 8473.87 | 11955.55747 | 12010.43136 |
| 9143.33 | 8383.43818 | 8408.58957 | 1140.8 | 8401.81 | 8508.4 | 11904.36992 | 12063.09345 |
| 9173.11 | 8482.49873 | 8437.44293 | 1142.52 | 8425.23 | 8542.86 | 11932.49147 | 12115.75553 |
| 9202.8 | 8453.50491 | 8466.20601 | 1144.24 | 8510.15 | 8577.26 | 11986.31213 | 12168.3123 |
| 9232.38 | 8499.91509 | 8494.87879 | 1145.96 | 8477.69 | 8611.59 | 12263.63067 | 12220.76373 |
| 9261.86 | 8532.95199 | 8523.47131 | 1147.69 | 8414.95 | 8645.86 | 12085.63282 | 12273.21517 |
| 9291.23 | 8306.76005 | 8551.97355 | 1149.41 | 8913.24 | 8680.06 | 11862.66155 | 12325.56128 |
| 9320.5 | 8369.81409 | 8580.38548 | 1151.13 | 8716.75 | 8714.2 | 12256.04733 | 12377.80207 |
| 9349.67 | 8407.99765 | 8608.70713 | 1152.85 | 8683.94 | 8748.26 | 12243.30311 | 12430.04286 |
| 9378.73 | 8437.11186 | 8636.93849 | 1154.57 | 8860.82 | 8782.26 | 12445.42019 | 12482.17833 |
| 9407.69 | 8553.50851 | 8665.08958 | 1156.3 | 8590.1 | 8816.19 | 12435.94102 | 12534.20847 |
| 9436.54 | 8597.62126 | 8693.14035 | 1158.02 | 8944.02 | 8850.05 | 12377.17013 | 12586.23861 |
| 9465.28 | 8700.83525 | 8721.10083 | 1159.74 | 9033.03 | 8883.84 | 12498.71422 | 12638.16343 |
| 9493.92 | 8570.51354 | 8748.97101 | 1161.46 | 8869.36 | 8917.56 | 12549.79645 | 12689.98292 |
| 9522.45 | 8746.72374 | 8776.75091 | 1163.18 | 8972.75 | 8951.21 | 12936.02019 | 12741.69709 |
| 9550.87 | 8754.58919 | 8804.44051 | 1164.9 | 9044.4 | 8984.78 | 12572.01985 | 12793.41126 |
| 9579.19 | 8771.81494 | 8832.02979 | 1166.62 | 8942.19 | 9018.28 | 12428.14703 | 12845.0201 |
| 9607.4 | 8728.59508 | 8859.5388 | 1168.34 | 8827.7 | 9051.71 | 12552.85085 | 12896.52362 |
| 9635.5 | 8604.71421 | 8886.93746 | 1170.06 | 8833.52 | 9085.06 | 12469.22346 | 12948.02714 |
| 9663.48 | 8817.262 | 8914.25586 | 1171.78 | 8976.96 | 9118.34 | 12809.31521 | 12999.32002 |
| 9691.36 | 8731.83557 | 8941.47394 | 1173.5 | 9088.2 | 9151.54 | 13015.85591 | 13050.61289 |
| 9719.13 | 8964.93987 | 8968.60172 | 1175.22 | 9005.47 | 9184.66 | 12943.60353 | 13101.80044 |
| 9746.79 | 8837.08615 | 8995.62918 | 1176.94 | 9016.27 | 9217.7 | 12869.13934 | 13152.88266 |
| 9774.34 | 8777.9548 | 9022.55632 | 1178.66 | 9137.04 | 9250.67 | 13075.68004 | 13203.96488 |
| 9801.78 | 8954.50611 | 9049.39316 | 1180.38 | 9057.73 | 9283.56 | 13144.56205 | 13254.83646 |
| 9829.1 | 8892.17441 | 9076.12968 | 1182.11 | 9193.42 | 9316.37 | 13195.22298 | 13305.70803 |
| 9856.32 | 8961.41848 | 9102.7759 | 1183.82 | 9347.84 | 9349.09 | 13333.30297 | 13356.47429 |
| 9883.42 | 8950.57338 | 9129.3218 | 1185.54 | 9107.46 | 9381.74 | 13388.91413 | 13407.13521 |
| 9910.41 | 9128.5794 | 9155.76738 | 1187.26 | 9250.14 | 9414.31 | 13429.35861 | 13457.69081 |
| 9937.28 | 9085.42977 | 9182.11263 | 1188.98 | 9516.37 | 9446.79 | 13311.39554 | 13508.14109 |
| 9964.05 | 8936.19686 | 9208.36759 | 1190.7 | 9553.21 | 9479.19 | 13400.28914 | 13558.48605 |
| 9990.69 | 9191.93441 | 9234.52222 | 1192.42 | 9439.65 | 9511.51 | 13044.18811 | 13608.831 |
| 10017.2 | 9264.12802 | 9260.5665 | 1194.14 | 9486.55 | 9543.75 | 13435.46741 | 13658.96531 |
| 10043.6 | 9093.20493 | 9286.52048 | 1195.86 | 9489.74 | 9575.9 | 13583.9745 | 13708.99429 |
| 10070 | 9401.17149 | 9312.37414 | 1197.58 | 9616.08 | 9607.96 | 13363.10971 | 13759.02327 |
| 10096.1 | 9347.60816 | 9338.12748 | 1199.3 | 9688.01 | 9639.94 | 13528.15269 | 13808.94693 |
| 10122.2 | 9211.66827 | 9363.78049 | 1201.02 | 9725.56 | 9671.84 | 13558.38072 | 13858.65994 |
| 10148.2 | 9352.05254 | 9389.33318 | 1202.73 | 9909.17 | 9703.64 | 13717.10425 | 13908.37295 |
| 10174 | 9216.12268 | 9414.78554 | 1204.45 | 9818.19 | 9735.36 | 13852.86711 | 13957.98063 |
| 10199.8 | 9224.0082 | 9440.12754 | 1206.17 | 9712.56 | 9766.99 | 13807.78836 | 14007.37767 |
| 10225.4 | 9369.02747 | 9465.37926 | 1207.89 | 9631.94 | 9798.54 | 13686.45492 | 14056.77471 |
| 10250.9 | 9403.35856 | 9490.52061 | 1209.61 | 9638.5 | 9829.99 | 13997.16123 | 14106.06642 |
| 10276.3 | 9316.22661 | 9515.56164 | 1211.32 | 9801.73 | 9861.36 | 13932.91348 | 14155.25281 |
| 10301.5 | 9447.83248 | 9540.50235 | 1213.04 | 9848.77 | 9892.63 | 13958.19128 | 14204.22855 |
| 10326.7 | 9488.37367 | 9565.34273 | 1214.76 | 9910.97 | 9923.82 | 14287.53997 | 14253.20429 |
| 10351.7 | 9428.43972 | 9590.07276 | 1216.48 | 9967.82 | 9954.91 | 14197.27715 | 14302.07471 |
| 10376.6 | 9469.21166 | 9614.70246 | 1218.19 | 9976.02 | 9985.91 | 14158.83383 | 14350.73447 |
| 10401.4 | 9619.40768 | 9639.23183 | 1219.91 | 10120.1 | 10016.8 | 14370.43009 | 14399.39424 |
| 10426.1 | 9651.94296 | 9663.65085 | 1221.63 | 10270.1 | 10047.6 | 14290.80502 | 14447.84336 |
| 10450.7 | 9600.22561 | 9687.96954 | 1223.35 | 10155.8 | 10078.4 | 14478.59802 | 14496.29248 |
| 10475.1 | 9751.60546 | 9712.18791 | 1225.06 | 10024 | 10109 | 14700.7267 | 14544.53095 |
| 10499.4 | 9842.46949 | 9736.29592 | 1226.78 | 10221.9 | 10139.5 | 14583.18492 | 14592.6641 |
| 10523.6 | 9650.73907 | 9760.30361 | 1228.5 | 10019.7 | 10170 | 14749.17582 | 14640.79724 |
| 10547.7 | 9610.73963 | 9784.20094 | 1230.21 | 10065.4 | 10200.3 | 14871.24653 | 14688.71974 |
| 10571.7 | 9694.34016 | 9807.99795 | 1231.93 | 10239.2 | 10230.6 | 14797.09832 | 14736.53692 |
| 10595.5 | 9751.09381 | 9831.68459 | 1233.64 | 10071.8 | 10260.8 | 14707.46745 | 14784.24877 |
| 10619.3 | 9787.94305 | 9855.27092 | 1235.36 | 10311.3 | 10290.8 | 14614.36088 | 14831.74997 |
| 10642.9 | 9697.91172 | 9878.74688 | 1237.08 | 10410 | 10320.8 | 14728.32163 | 14879.25117 |
| 10666.4 | 9720.42457 | 9902.12253 | 1238.79 | 10385.1 | 10350.6 | 14832.38191 | 14926.54172 |
| 10689.8 | 9886.95344 | 9925.38781 | 1240.51 | 10282.4 | 10380.4 | 14739.90729 | 14973.83228 |
| 10713 | 10047.21202 | 9948.55277 | 1242.22 | 10385.1 | 10410.1 | 15068.30806 | 15020.91218 |
| 10736.1 | 10142.72108 | 9971.59734 | 1243.94 | 10436.4 | 10439.7 | 15213.23412 | 15067.88676 |
| 10759.2 | 10017.73664 | 9994.55162 | 1245.66 | 10305.1 | 10469.1 | 15214.70866 | 15114.75602 |
| 10782.1 | 9987.31821 | 10017.38551 | 1247.37 | 10281.3 | 10498.5 | 15283.16937 | 15161.51995 |
| 10804.8 | 10132.187 | 10040.08898 | 1249.09 | 10348.5 | 10527.8 | 15267.1601 | 15208.07324 |
| 10827.5 | 9861.44088 | 10062.76234 | 1250.8 | 10570 | 10557 | 15276.84992 | 15254.62652 |
| 10850 | 10084.03117 | 10085.23506 | 1252.52 | 10614.1 | 10586.1 | 15298.02008 | 15300.96916 |
| 10872.4 | 10013.40262 | 10107.70778 | 1254.23 | 10694 | 10615 | 15205.33481 | 15347.20647 |
| 10894.7 | 10059.35131 | 10129.97986 | 1255.95 | 10828.8 | 10643.9 | 15288.22493 | 15393.23313 |
| 10916.9 | 10246.25612 | 10152.1516 | 1257.66 | 10588.6 | 10672.7 | 15531.10247 | 15439.2598 |
| 10939 | 10267.42462 | 10174.22302 | 1259.38 | 10678.2 | 10701.4 | 15571.96825 | 15485.07581 |
| 10960.9 | 10220.27203 | 10196.19412 | 1261.09 | 10587.4 | 10730 | 15480.12557 | 15530.89183 |
| 10982.7 | 10065.27046 | 10218.06489 | 1262.81 | 10610.9 | 10758.4 | 15471.69964 | 15576.39187 |
| 11004.4 | 10383.19925 | 10239.83534 | 1264.52 | 10862.1 | 10786.8 | 15574.70668 | 15621.89191 |
| 11026 | 10331.33141 | 10261.50546 | 1266.23 | 10904.2 | 10815.1 | 15618.41621 | 15667.28663 |
| 11047.4 | 10371.05997 | 10283.07526 | 1267.95 | 10887.8 | 10843.2 | 15927.22669 | 15712.4707 |
| 11068.7 | 10439.88268 | 10304.54473 | 1269.66 | 10887.3 | 10871.3 | 15918.06348 | 15757.54945 |
| 11089.9 | 10132.98959 | 10325.81356 | 1271.37 | 10896.1 | 10899.3 | 15676.34451 | 15802.41754 |
| 11111 | 10216.0584 | 10347.08238 | 1273.09 | 10861.8 | 10927.1 | 16088.16202 | 15847.28564 |
| 11132 | 10514.02259 | 10368.15056 | 1274.8 | 11135.1 | 10954.9 | 15832.85623 | 15891.94309 |
| 11152.8 | 10439.38105 | 10389.21873 | 1276.51 | 11288.6 | 10982.5 | 15826.53678 | 15936.49521 |
| 11173.5 | 10634.71313 | 10410.08626 | 1278.23 | 11134.4 | 11010 | 16156.41209 | 15980.94202 |
| 11194.1 | 10629.59658 | 10430.85346 | 1279.94 | 11014.8 | 11037.5 | 16119.54863 | 16025.17817 |
| 11214.6 | 10419.71742 | 10451.52034 | 1281.65 | 10968.7 | 11064.8 | 15953.34708 | 16069.309 |
| 11235 | 10573.31445 | 10472.08689 | 1283.37 | 11073.9 | 11092 | 16013.27654 | 16113.3345 |
| 11255.2 | 10541.31089 | 10492.55312 | 1285.08 | 11120.5 | 11119.2 | 16190.74777 | 16157.14936 |
| 11275.4 | 10667.82027 | 10512.91902 | 1286.79 | 11167.8 | 11146.2 | 16331.77683 | 16200.96421 |
| 11295.4 | 10726.00856 | 10533.1846 | 1288.5 | 11261.8 | 11173.1 | 16504.92977 | 16244.56842 |
| 11315.2 | 10588.66412 | 10553.24952 | 1290.22 | 11120.5 | 11199.9 | 16199.06838 | 16287.96198 |
| 11335 | 10650.96573 | 10573.31445 | 1291.93 | 11327.7 | 11226.6 | 16310.50135 | 16331.25021 |
| 11354.6 | 10677.15046 | 10593.27906 | 1293.64 | 11410.4 | 11253.2 | 16246.6749 | 16374.43312 |
| 11374.1 | 10635.31508 | 10613.04301 | 1295.35 | 11364.3 | 11279.6 | 16239.40753 | 16417.51071 |
| 11393.5 | 10680.66182 | 10632.70664 | 1297.07 | 11312.6 | 11306 | 16566.64974 | 16460.37765 |
| 11412.8 | 10752.29362 | 10652.26995 | 1298.78 | 11137.4 | 11332.3 | 16673.86975 | 16503.13926 |
| 11432 | 10898.06532 | 10671.73293 | 1300.49 | 11358.3 | 11358.5 | 16811.21247 | 16545.79555 |
| 11451 | 10903.08155 | 10691.19591 | 1302.2 | 11355.1 | 11384.5 | 16709.04802 | 16588.24119 |
| 11469.9 | 10813.89295 | 10710.45824 | 1303.91 | 11506.6 | 11410.4 | 16688.61513 | 16630.58151 |
| 11488.7 | 10851.41436 | 10729.51992 | 1305.62 | 11761.2 | 11436.3 | 16837.54351 | 16672.8165 |
| 11507.4 | 10773.36179 | 10748.5816 | 1307.33 | 11560.5 | 11462 | 16812.79233 | 16714.84085 |
| 11525.9 | 10926.0559 | 10767.54296 | 1309.04 | 11644.6 | 11487.6 | 16865.66506 | 16756.75987 |
| 11544.3 | 11097.00909 | 10786.40399 | 1310.76 | 11688.1 | 11513.1 | 16987.41981 | 16798.46824 |
| 11562.6 | 10876.89682 | 10805.06438 | 1312.47 | 11584.8 | 11538.5 | 17214.07743 | 16840.07129 |
| 11580.8 | 10864.55689 | 10823.62444 | 1314.18 | 11638.5 | 11563.8 | 17229.24411 | 16881.56901 |
| 11598.9 | 10875.09098 | 10842.1845 | 1315.89 | 11640.7 | 11589 | 17170.36789 | 16922.96141 |
| 11616.9 | 10790.41698 | 10860.54391 | 1317.6 | 11623.6 | 11614 | 17320.34952 | 16964.03784 |
| 11634.7 | 10965.28283 | 10878.80299 | 1319.31 | 11881.7 | 11639 | 17096.11435 | 17005.11427 |
| 11652.4 | 10981.63575 | 10896.96175 | 1321.02 | 11732.3 | 11663.9 | 17102.4338 | 17046.08537 |
| 11670 | 10975.11465 | 10915.02019 | 1322.73 | 11621.7 | 11688.6 | 17140.24518 | 17086.7405 |
| 11687.5 | 11017.35132 | 10932.9783 | 1324.44 | 11770.4 | 11713.2 | 17100.74862 | 17127.39563 |
| 11704.8 | 11029.18963 | 10950.83608 | 1326.15 | 11669.2 | 11737.8 | 17169.9466 | 17167.84011 |
| 11722 | 11146.56946 | 10968.59355 | 1327.86 | 11798.8 | 11762.1 | 17143.61556 | 17208.17927 |
| 11739.1 | 11036.91463 | 10986.15036 | 1329.57 | 11859.5 | 11786.5 | 17225.97906 | 17248.30778 |
| 11756.1 | 10982.2377 | 11003.70717 | 1331.28 | 11883.3 | 11810.6 | 17394.70838 | 17288.33097 |
| 11773 | 11205.761 | 11021.06333 | 1332.98 | 11914.7 | 11834.7 | 17511.19692 | 17328.14351 |
| 11789.8 | 11249.3019 | 11038.4195 | 1334.69 | 11956.3 | 11858.7 | 17572.91688 | 17367.95604 |
| 11806.4 | 11277.99475 | 11055.57501 | 1336.4 | 11986.4 | 11882.5 | 17551.11478 | 17407.45261 |
| 11822.9 | 11378.21906 | 11072.6302 | 1338.11 | 11877.6 | 11906.3 | 17572.70623 | 17446.94917 |
| 11839.3 | 11209.37269 | 11089.58507 | 1339.82 | 11874.9 | 11929.9 | 17603.14492 | 17486.12976 |
| 11855.6 | 11215.89379 | 11106.43961 | 1341.53 | 11990.5 | 11953.4 | 17720.2654 | 17525.31036 |
| 11871.8 | 11308.59376 | 11123.19382 | 1343.24 | 11841 | 11976.8 | 17816.53169 | 17564.2803 |
| 11887.8 | 11250.30514 | 11139.84771 | 1344.95 | 11969.1 | 12000.1 | 17667.60331 | 17603.0396 |
| 11903.8 | 11164.42725 | 11156.40128 | 1346.65 | 12040.7 | 12023.3 | 17848.33959 | 17641.79889 |
| 11919.6 | 11238.46684 | 11172.85452 | 1348.36 | 11997.2 | 12046.4 | 17825.6949 | 17680.24221 |
| 11935.2 | 11248.9006 | 11189.20744 | 1350.07 | 12189.1 | 12069.4 | 17851.28867 | 17718.68554 |
| 11950.8 | 11501.01643 | 11205.3597 | 1351.78 | 12189.9 | 12092.2 | 17928.80726 | 17756.91821 |
| 11966.3 | 11382.0314 | 11221.51197 | 1353.49 | 12108.2 | 12115 | 17697.41005 | 17794.94024 |
| 11981.6 | 11128.0094 | 11237.46359 | 1355.19 | 12279.2 | 12137.6 | 17895.94612 | 17832.85694 |
| 11996.9 | 11359.55868 | 11253.41521 | 1356.9 | 12241.9 | 12160.1 | 17901.42298 | 17870.66832 |
| 12012 | 11304.6811 | 11269.16618 | 1358.61 | 12235 | 12182.5 | 17817.47961 | 17908.26905 |
| 12027 | 11312.10512 | 11284.81682 | 1360.32 | 12237.4 | 12204.8 | 17913.32461 | 17945.65913 |
| 12041.8 | 11365.47783 | 11300.36714 | 1362.02 | 12055.7 | 12227 | 18071.31087 | 17983.04921 |
| 12056.6 | 11377.11549 | 11315.81714 | 1363.73 | 12318.7 | 12249.1 | 18178.2149 | 18020.22864 |
| 12071.2 | 11235.35677 | 11331.16681 | 1365.44 | 12326.4 | 12271 | 18061.19975 | 18057.19743 |
| 12085.8 | 11369.59114 | 11346.41615 | 1367.14 | 12227.6 | 12292.9 | 18080.36875 | 18094.06089 |
| 12100.2 | 11608.16315 | 11361.56517 | 1368.85 | 12363.2 | 12314.6 | 18202.43946 | 18130.7137 |
| 12114.5 | 11514.96155 | 11376.61387 | 1370.56 | 12354.9 | 12336.2 | 18348.52409 | 18167.26119 |
| 12128.7 | 11404.0025 | 11391.56224 | 1372.26 | 12468.9 | 12357.7 | 18476.59828 | 18203.70335 |
| 12142.7 | 11466.40442 | 11406.41029 | 1373.97 | 12469.8 | 12379.1 | 18382.33315 | 18239.93487 |
| 12156.7 | 11630.33489 | 11421.05769 | 1375.67 | 12531.8 | 12400.4 | 18328.82847 | 18276.06106 |
| 12170.5 | 11569.53816 | 11435.70508 | 1377.38 | 12560 | 12421.6 | 18307.55299 | 18311.9766 |
| 12184.2 | 11496.20084 | 11450.25216 | 1379.09 | 12457 | 12442.6 | 18399.81696 | 18347.68149 |
| 12197.9 | 11649.8982 | 11464.59858 | 1380.79 | 12582.1 | 12463.6 | 18542.21524 | 18383.38639 |
| 12211.3 | 11625.71996 | 11478.94501 | 1382.5 | 12492.4 | 12484.4 | 18650.17252 | 18418.88064 |
| 12224.7 | 11430.28755 | 11493.09078 | 1384.2 | 12602 | 12505.1 | 19014.38351 | 18454.16423 |
| 12238 | 11738.98648 | 11507.13623 | 1385.91 | 12708.9 | 12525.8 | 18872.09055 | 18489.34251 |
| 12251.1 | 11845.63158 | 11521.18168 | 1387.61 | 12587.9 | 12546.3 | 18622.05096 | 18524.31013 |
| 12264.2 | 11638.16021 | 11535.02648 | 1389.32 | 12694.6 | 12566.6 | 18962.14272 | 18559.17243 |
| 12277.1 | 11755.54005 | 11548.77096 | 1391.02 | 12628.2 | 12586.9 | 18850.81507 | 18593.92941 |
| 12289.9 | 11716.61408 | 11562.41511 | 1392.73 | 12777.1 | 12607.1 | 18786.56732 | 18628.47574 |
| 12302.6 | 11578.06575 | 11575.95894 | 1394.43 | 12891.5 | 12627.1 | 18927.06977 | 18662.81142 |
| 12315.2 | 11808.00983 | 11589.50276 | 1396.14 | 12756.4 | 12647.1 | 19006.48419 | 18697.04178 |
| 12327.7 | 11856.46664 | 11602.84594 | 1397.84 | 12684 | 12666.9 | 18947.81863 | 18731.16681 |
| 12340.1 | 11695.54591 | 11616.08879 | 1399.55 | 12757 | 12686.6 | 18951.50498 | 18765.08119 |
| 12352.3 | 11763.66634 | 11629.23132 | 1401.25 | 12868 | 12706.2 | 19030.70875 | 18798.78493 |
| 12364.5 | 11813.52769 | 11642.27352 | 1402.95 | 12850.4 | 12725.7 | 18889.15307 | 18832.48866 |
| 12376.5 | 11710.29363 | 11655.2154 | 1404.66 | 12952.4 | 12745.1 | 18896.10446 | 18865.87642 |
| 12388.5 | 11788.94815 | 11668.05696 | 1406.36 | 12989 | 12764.4 | 18932.23065 | 18899.15886 |
| 12400.3 | 11837.10398 | 11680.79819 | 1408.07 | 12902.6 | 12783.5 | 19003.21914 | 18932.33598 |
| 12412 | 11756.64362 | 11693.33877 | 1409.77 | 12945.1 | 12802.6 | 18943.39501 | 18965.40777 |
| 12423.6 | 11845.83222 | 11705.87935 | 1411.47 | 12974 | 12821.5 | 19071.46921 | 18998.16358 |
| 12435 | 11789.24913 | 11718.3196 | 1413.17 | 12927.1 | 12840.4 | 19270.74254 | 19030.9194 |
| 12446.4 | 11971.33835 | 11730.65953 | 1414.88 | 12966.6 | 12859.1 | 19272.11176 | 19063.46457 |
| 12457.7 | 12009.86302 | 11742.89914 | 1416.58 | 13102 | 12877.7 | 19240.83048 | 19095.79909 |
| 12468.8 | 11971.539 | 11755.03842 | 1418.28 | 13081.7 | 12896.2 | 19219.97629 | 19128.02829 |
| 12479.9 | 11921.27636 | 11766.97706 | 1419.99 | 12987.6 | 12914.5 | 19391.54937 | 19160.04684 |
| 12490.8 | 11896.1952 | 11778.91569 | 1421.69 | 13209.3 | 12932.8 | 19488.02631 | 19191.96006 |
| 12501.7 | 11999.52958 | 11790.754 | 1423.39 | 13205.8 | 12951 | 19629.47667 | 19223.76796 |
| 12512.4 | 11845.12995 | 11802.49198 | 1425.09 | 12904.7 | 12969 | 19637.16534 | 19255.36521 |
| 12523 | 11811.62152 | 11814.02931 | 1426.8 | 12936.6 | 12987 | 19420.51352 | 19286.75182 |
| 12533.5 | 11859.77735 | 11825.56665 | 1428.5 | 13047.4 | 13004.8 | 19348.36646 | 19318.0331 |
| 12543.9 | 11926.09194 | 11837.00366 | 1430.2 | 13056.8 | 13022.5 | 19458.11425 | 19349.20905 |
| 12554.2 | 11928.60006 | 11848.34034 | 1431.9 | 13079.6 | 13040.1 | 19541.10969 | 19380.17436 |
| 12564.4 | 12056.91527 | 11859.47638 | 1433.6 | 13050.9 | 13057.6 | 19479.7057 | 19410.92902 |
| 12574.5 | 11958.39647 | 11870.61241 | 1435.3 | 13158.5 | 13075 | 19673.6075 | 19441.57835 |
| 12584.5 | 11969.53251 | 11881.64812 | 1437.01 | 13267.2 | 13092.3 | 19693.51377 | 19472.12236 |
| 12594.3 | 11987.49062 | 11892.58351 | 1438.71 | 13316.4 | 13109.4 | 19546.27058 | 19502.45572 |
| 12604.1 | 11991.50361 | 11903.41857 | 1440.41 | 13287.7 | 13126.5 | 19730.06126 | 19532.68376 |
| 12613.7 | 12089.92208 | 11914.05298 | 1442.11 | 13243.7 | 13143.4 | 19671.60634 | 19562.70115 |
| 12623.3 | 12179.01036 | 11924.68739 | 1443.81 | 13346.1 | 13160.3 | 19638.7452 | 19592.61321 |
| 12632.7 | 12225.86197 | 11935.22148 | 1445.51 | 13303.1 | 13177 | 19839.38775 | 19622.31463 |
| 12642.1 | 12065.44287 | 11945.65524 | 1447.21 | 13351.1 | 13193.6 | 19820.95602 | 19651.91072 |
| 12651.3 | 12078.78605 | 11955.98868 | 1448.91 | 13200.6 | 13210.1 | 19795.36225 | 19681.40149 |
| 12660.5 | 12208.70646 | 11966.2218 | 1450.61 | 13131.9 | 13226.5 | 19815.58449 | 19710.57629 |
| 12669.5 | 11970.33511 | 11976.35459 | 1452.31 | 13276.5 | 13242.8 | 19854.8704 | 19739.75108 |
| 12678.4 | 12013.27405 | 11986.38705 | 1454.01 | 13267.7 | 13259 | 19872.24889 | 19768.71523 |
| 12687.3 | 12090.12273 | 11996.31919 | 1455.71 | 13482 | 13275.1 | 19990.63326 | 19797.46873 |
| 12696 | 11944.75232 | 12006.151 | 1457.41 | 13489.5 | 13291 | 20043.82197 | 19826.1169 |
| 12704.6 | 12190.74835 | 12015.88249 | 1459.11 | 13337.6 | 13306.9 | 20111.54541 | 19854.65975 |
| 12713.1 | 12076.87988 | 12025.51366 | 1460.81 | 13225.7 | 13322.6 | 20068.3625 | 19882.99196 |
| 12721.5 | 12135.06817 | 12035.0445 | 1462.51 | 13199 | 13338.3 | 19883.6239 | 19911.11351 |
| 12729.9 | 12202.58666 | 12044.47502 | 1464.21 | 13475.3 | 13353.8 | 19920.80334 | 19939.12974 |
| 12738.1 | 11992.80783 | 12053.90553 | 1465.91 | 13455 | 13369.2 | 19886.6783 | 19967.04065 |
| 12746.2 | 12104.97078 | 12063.1354 | 1467.61 | 13482 | 13384.5 | 19976.83579 | 19994.7409 |
| 12754.2 | 11992.50685 | 12072.26494 | 1469.31 | 13536.6 | 13399.8 | 20095.74679 | 20022.33584 |
| 12762.1 | 12081.89611 | 12081.39449 | 1471.01 | 13259.4 | 13414.9 | 19879.51626 | 20049.72012 |
| 12770 | 12307.52623 | 12090.32338 | 1472.7 | 13437.5 | 13429.9 | 19852.2373 | 20076.99908 |
| 12777.7 | 12287.8626 | 12099.25227 | 1474.4 | 13560 | 13444.8 | 20307.65902 | 20104.17272 |
| 12785.3 | 12150.21719 | 12107.98052 | 1476.1 | 13443.1 | 13459.5 | 20347.36623 | 20131.13571 |
| 12792.8 | 12136.07142 | 12116.70876 | 1477.8 | 13465 | 13474.2 | 20199.17512 | 20157.88805 |
| 12800.2 | 12111.39155 | 12125.33668 | 1479.5 | 13548 | 13488.8 | 20212.23532 | 20184.53506 |
| 12807.5 | 12142.79317 | 12133.86427 | 1481.19 | 13484.3 | 13503.2 | 20107.54309 | 20211.07675 |
| 12814.8 | 12153.9292 | 12142.29154 | 1482.89 | 13567.6 | 13517.6 | 20240.67285 | 20237.4078 |
| 12821.9 | 12114.20064 | 12150.61849 | 1484.59 | 13619.3 | 13531.9 | 20506.08976 | 20263.52819 |
| 12828.9 | 12295.1863 | 12158.84511 | 1486.29 | 13603 | 13546 | 20297.9692 | 20289.64859 |
| 12835.8 | 12252.8493 | 12166.97141 | 1487.98 | 13632.2 | 13560 | 20321.77246 | 20315.45301 |
| 12842.7 | 12451.59242 | 12174.99738 | 1489.68 | 13437.7 | 13574 | 20775.40367 | 20341.25743 |
| 12849.4 | 12345.54927 | 12182.92303 | 1491.38 | 13502.5 | 13587.8 | 20617.20676 | 20366.85121 |
| 12856 | 12132.66038 | 12190.84867 | 1493.08 | 13643.6 | 13601.5 | 20536.94974 | 20392.23433 |
| 12862.6 | 12224.3571 | 12198.57367 | 1494.77 | 13716.1 | 13615.1 | 20622.68362 | 20417.51213 |
| 12869 | 12119.7185 | 12206.29867 | 1496.47 | 13836.1 | 13628.7 | 20383.17645 | 20442.68461 |
| 12875.4 | 12089.82176 | 12213.82302 | 1498.17 | 13776.3 | 13642.1 | 20533.57937 | 20467.64644 |
| 12881.7 | 12092.12922 | 12221.34736 | 1499.86 | 13610.5 | 13655.4 | 20573.28658 | 20492.39762 |
| 12887.8 | 12328.29343 | 12228.77139 | 1501.56 | 13600.4 | 13668.6 | 20477.65224 | 20517.1488 |
| 12893.9 | 12252.74898 | 12236.09509 | 1503.25 | 13512 | 13681.7 | 20571.6014 | 20541.58401 |
| 12899.9 | 12090.02241 | 12243.31846 | 1504.95 | 13695.7 | 13694.7 | 20561.70092 | 20566.01922 |
| 12905.8 | 12168.37595 | 12250.44151 | 1506.65 | 13790.9 | 13707.6 | 20476.38835 | 20590.24378 |
| 12911.5 | 12161.6542 | 12257.56456 | 1508.34 | 13674.5 | 13720.4 | 20377.2783 | 20614.25769 |
| 12917.2 | 12122.42727 | 12264.48696 | 1510.04 | 13569.6 | 13733 | 20509.46014 | 20638.2716 |
| 12922.8 | 12046.68216 | 12271.40936 | 1511.73 | 13600.6 | 13745.6 | 20693.77744 | 20661.96954 |
| 12928.4 | 12221.04639 | 12278.13111 | 1513.43 | 13792.8 | 13758.1 | 20686.51007 | 20685.56215 |
| 12933.8 | 12280.43858 | 12284.85286 | 1515.12 | 13583.6 | 13770.5 | 20647.0135 | 20709.04944 |
| 12939.1 | 12120.1198 | 12291.47429 | 1516.82 | 13650.6 | 13782.8 | 20662.39083 | 20732.43141 |
| 12944.3 | 12121.92564 | 12297.99539 | 1518.51 | 13701.5 | 13794.9 | 20557.59328 | 20755.60273 |
| 12949.5 | 12229.17269 | 12304.41617 | 1520.21 | 13682.5 | 13807 | 20615.94287 | 20778.5634 |
| 12954.5 | 12190.24672 | 12310.83695 | 1521.9 | 13924.1 | 13819 | 20714.84227 | 20801.52407 |
| 12959.5 | 12127.94512 | 12317.05707 | 1523.6 | 13761 | 13830.9 | 20545.37568 | 20824.16876 |
| 12964.4 | 12192.65452 | 12323.2772 | 1525.29 | 13861.5 | 13842.6 | 20586.97873 | 20846.81346 |
| 12969.1 | 12272.51293 | 12329.29668 | 1526.99 | 13948.6 | 13854.3 | 20889.57507 | 20869.24751 |
| 12973.8 | 12283.64897 | 12335.31616 | 1528.68 | 13866.5 | 13865.9 | 20826.16992 | 20891.47091 |
| 12978.4 | 12185.63179 | 12341.23531 | 1530.37 | 13704.9 | 13877.3 | 20627.73918 | 20913.58899 |
| 12982.9 | 12084.7052 | 12347.05414 | 1532.07 | 13782.5 | 13888.7 | 20722.10964 | 20935.60174 |
| 12987.4 | 12321.47136 | 12352.87297 | 1533.76 | 13942.6 | 13900 | 20875.98826 | 20957.40384 |
| 12991.7 | 12195.8649 | 12358.49115 | 1535.45 | 13711.6 | 13911.1 | 20751.38976 | 20979.10062 |
| 12996 | 12072.96722 | 12364.10933 | 1537.15 | 13606 | 13922.2 | 20644.3804 | 21000.69208 |
| 13000.1 | 12356.08336 | 12369.62719 | 1538.84 | 13751.3 | 13933.2 | 20958.24643 | 21022.07288 |
| 13004.2 | 12320.56844 | 12375.04472 | 1540.53 | 13787.9 | 13944 | 20740.96267 | 21043.34837 |
| 13008.2 | 12305.21877 | 12380.36192 | 1542.23 | 13911 | 13954.8 | 20807.84352 | 21064.4132 |
| 13012.1 | 12243.11781 | 12385.5788 | 1543.92 | 13890.8 | 13965.5 | 20900.73944 | 21085.37271 |
| 13015.9 | 12172.18829 | 12390.69536 | 1545.61 | 14012.6 | 13976 | 20986.57864 | 21106.2269 |
| 13019.6 | 12239.8071 | 12395.81192 | 1547.31 | 14081.3 | 13986.5 | 21191.85545 | 21126.87043 |
| 13023.3 | 12379.96062 | 12400.82815 | 1549 | 13829.5 | 13996.9 | 21012.80436 | 21147.40865 |
| 13026.8 | 12336.41973 | 12405.74406 | 1550.69 | 14013.1 | 14007.2 | 20834.17456 | 21167.84154 |
| 13030.3 | 12354.77914 | 12410.55964 | 1552.38 | 13873.9 | 14017.4 | 20897.68503 | 21188.06378 |
| 13033.7 | 12213.42172 | 12415.37522 | 1554.07 | 13815.3 | 14027.4 | 20827.11784 | 21208.1807 |
| 13037 | 12167.27238 | 12419.99016 | 1555.77 | 13841.4 | 14037.4 | 20847.12943 | 21228.08696 |
| 13040.2 | 12189.04283 | 12424.60509 | 1557.46 | 13712.1 | 14047.3 | 21047.14004 | 21247.88791 |
| 13043.3 | 12230.97853 | 12429.1197 | 1559.15 | 13987.8 | 14057.1 | 21173.84502 | 21267.58353 |
| 13046.4 | 12432.33009 | 12433.53398 | 1560.84 | 13994.5 | 14066.8 | 21022.49418 | 21287.0685 |
| 13049.4 | 12071.663 | 12437.84794 | 1562.53 | 13839.7 | 14076.4 | 21010.59255 | 21306.44815 |
| 13052.2 | 12151.32076 | 12442.1619 | 1564.22 | 14009.3 | 14085.9 | 21159.09963 | 21325.72247 |
| 13055 | 12253.45125 | 12446.37554 | 1565.91 | 14036.1 | 14095.3 | 21057.56713 | 21344.78615 |
| 13057.8 | 12061.53021 | 12450.48885 | 1567.61 | 13995.5 | 14104.6 | 21039.03007 | 21363.7445 |
| 13060.4 | 12102.66331 | 12454.50183 | 1569.3 | 13845.4 | 14113.9 | 21116.23269 | 21382.59753 |
| 13063 | 12063.03508 | 12458.4145 | 1570.99 | 13931.6 | 14123 | 21224.82191 | 21401.2399 |
| 13065.4 | 12171.38569 | 12462.32716 | 1572.68 | 13889.3 | 14132 | 21235.56498 | 21419.77696 |
| 13067.8 | 12472.96157 | 12466.13949 | 1574.37 | 13816.3 | 14141 | 21339.62526 | 21438.10336 |
| 13070.2 | 12397.51744 | 12469.8515 | 1576.06 | 13987.5 | 14149.8 | 21173.95034 | 21456.32445 |
| 13072.4 | 12289.0665 | 12473.46319 | 1577.75 | 13972.8 | 14158.6 | 21195.01517 | 21474.4402 |
| 13074.5 | 12384.97686 | 12477.07488 | 1579.44 | 14113.8 | 14167.2 | 21322.7734 | 21492.45064 |

Figure 6 Data

| mass, Graphite | Temperature | n parameter | mass, Carbon Black | Temperature | n parameter |
|---|---|---|---|---|---|
| 43.83 | 1650.78504 | 0.76969 | 64.47 | 1900.53103 | 0.70479 |
| 33.56 | 1645.05179 | 1.22532 | 84.04 | 1887.70973 | 0.65437 |
| 6.305 | 1649.39388 | 0.91398 | 30.49 | 1915.55633 | 0.70108 |
| 14.17 | 1654.95504 | 1.06588 | 44.01 | 1913.28912 | 0.56109 |
| 4.77 | 1660.11889 | 1.44904 | 38.04 | 1893.33062 | 0.34217 |

Figure 7 Data

| Time (sec) | Fz | Q | Time (sec) | T |
|---|---|---|---|---|
| 64 | 2878.41107 | 48 | 64 | 1653.55029 |
| 96 | 2883.22152 | 48 | 96 | 1664.84927 |
| 129 | 2882.17569 | 48 | 129 | 1648.75295 |
| 161 | 2884.16222 | 48 | 161 | 1659.50226 |
| 193 | 2883.65114 | 48 | 193 | 1640.06897 |
| 226 | 2883.15681 | 48 | 225 | 1645.27027 |
| 258 | 2883.49124 | 48 | 258 | 1640.88382 |
| 290 | 2884.23229 | 48 | 290 | 1661.18262 |
| 323 | 2883.41213 | 48 | 322 | 1661.33464 |
| 355 | 2884.29104 | 48 | 355 | 1647.31352 |
| 387 | 2883.26564 | 48 | 387 | 1650.78504 |
| 419 | 2885.33663 | 48 | 419 | 1665.54281 |
| 452 | 2883.61134 | 48 | 451 | 1642.27059 |
| 484 | 2884.54437 | 48 | 484 | 1655.74934 |
| 516 | 2884.82375 | 48 | 516 | 1656.5718 |
| 549 | 2883.21173 | 48 | 548 | 1649.48712 |
| 581 | 2884.26546 | 48 | 581 | 1676.69152 |
| 613 | 2884.62592 | 48 | 613 | 1665.27849 |
| 646 | 2884.29419 | 48 | 645 | 1670.64972 |
| 678 | 2884.45752 | 48 | 677 | 1653.42288 |
| 710 | 2884.82261 | 48 | 710 | 1657.09384 |
| 743 | 2885.31018 | 48 | 742 | 1652.92506 |
| 775 | 2884.50918 | 48 | 774 | 1652.04804 |
| 807 | 2885.61606 | 48 | 807 | 1636.14503 |
| 839 | 2885.68476 | 48 | 839 | 1655.47979 |
| 872 | 2884.67587 | 48 | 871 | 1652.28959 |
| 904 | 2884.84005 | 48 | 904 | 1646.57859 |
| 936 | 2884.90088 | 48 | 936 | 1653.92713 |
| 968 | 2884.90346 | 48 | 968 | 1657.70949 |
| 1001 | 2885.43694 | 48 | 1000 | 1671.93587 |
| 1033 | 2883.42452 | 48 | 1033 | 1655.73503 |
| 1065 | 2883.74257 | 48 | 1065 | 1642.25601 |
| 1097 | 2885.34337 | 48 | 1097 | 1633.0965 |
| 1130 | 2884.9275 | 48 | 1129 | 1643.77902 |
| 1162 | 2885.37296 | 48 | 1162 | 1664.34155 |
| 1194 | 2884.80847 | 48 | 1194 | 1657.65527 |
| 1227 | 2884.56697 | 48 | 1226 | 1652.6298 |
| 1259 | 2885.32619 | 48 | 1258 | 1635.85011 |
| 1291 | 2885.31647 | 48 | 1291 | 1664.7349 |
| 1323 | 2885.82743 | 48 | 1323 | 1663.18166 |
| 1356 | 2884.71602 | 48 | 1355 | 1667.68381 |
| 1388 | 2885.77292 | 48 | 1388 | 1674.68672 |
| 1420 | 2886.14804 | 48 | 1420 | 1672.28856 |
| 1453 | 2884.40342 | 48 | 1452 | 1665.81456 |
| 1485 | 2884.64874 | 48 | 1484 | 1677.42301 |
| 1517 | 2886.3044 | 48 | 1517 | 1657.77623 |
| 1549 | 2885.83374 | 48 | 1549 | 1662.51502 |
| 1582 | 2885.10867 | 48 | 1581 | 1649.3068 |
| 1614 | 2885.4564 | 48 | 1614 | 1646.74986 |
| 1646 | 2886.25831 | 48 | 1646 | 1657.21189 |
| 1679 | 2885.22961 | 48 | 1678 | 1653.7265 |
| 1711 | 2886.3743 | 48 | 1711 | 1631.14618 |
| 1743 | 2886.05651 | 48 | 1743 | 1678.55941 |
| 1776 | 2886.22442 | 48 | 1775 | 1656.42473 |
| 1808 | 2884.87573 | 48 | 1807 | 1679.42804 |
| 1840 | 2884.88635 | 48 | 1840 | 1648.91827 |
| 1872 | 2885.60239 | 48 | 1872 | 1672.38712 |
| 1905 | 2888.68691 | 48 | 1904 | 1656.68733 |
| 1937 | 2886.0077 | 48 | 1937 | 1661.10625 |
| 1969 | 2886.46963 | 48 | 1969 | 1648.49679 |
| 2002 | 2887.15832 | 48 | 2001 | 1666.54736 |
| 2034 | 2886.11223 | 48 | 2034 | 1634.17643 |
| 2066 | 2886.8397 | 48 | 2066 | 1651.86261 |
| 2098 | 2886.6529 | 48 | 2098 | 1656.85022 |
| 2131 | 2887.78075 | 48 | 2130 | 1649.88402 |
| 2163 | 2885.14425 | 48 | 2163 | 1662.63238 |
| 2195 | 2886.57115 | 48 | 2195 | 1667.23674 |
| 2228 | 2885.89827 | 48 | 2227 | 1665.67165 |
| 2260 | 2886.31104 | 48 | 2260 | 1665.54426 |
| 2292 | 2887.85608 | 48 | 2292 | 1661.2625 |
| 2325 | 2887.65812 | 48 | 2324 | 1648.46958 |
| 2357 | 2887.41596 | 48 | 2356 | 1658.60042 |
| 2389 | 2887.92269 | 48 | 2389 | 1657.80707 |
| 2421 | 2886.99694 | 48 | 2421 | 1657.9008 |
| 2454 | 2886.10035 | 48 | 2453 | 1658.30694 |
| 2486 | 2887.0879 | 48 | 2486 | 1669.00597 |
| 2518 | 2887.75319 | 48 | 2518 | 1649.2693 |
| 2550 | 2886.79395 | 48 | 2550 | 1644.20179 |
| 2583 | 2886.40536 | 48 | 2582 | 1654.61123 |
| 2615 | 2887.40106 | 48 | 2615 | 1651.81666 |
| 2647 | 2887.03325 | 48 | 2647 | 1662.87985 |
| 2679 | 2886.46208 | 48 | 2679 | 1662.10978 |
| 2712 | 2887.43128 | 48 | 2711 | 1675.12116 |
| 2744 | 2886.86975 | 48 | 2744 | 1663.98389 |
| 2776 | 2887.36511 | 48 | 2776 | 1644.55401 |
| 2809 | 2887.05781 | 48 | 2808 | 1648.35273 |
| 2841 | 2887.39681 | 48 | 2841 | 1642.24355 |
| 2873 | 2887.58157 | 48 | 2873 | 1649.74008 |
| 2906 | 2887.43941 | 48 | 2905 | 1651.67028 |
| 2938 | 2886.59916 | 48 | 2938 | 1677.06456 |
| 2970 | 2887.42044 | 48 | 2970 | 1649.47722 |
| 2999 | 2949.23992 | 49 | 3002 | 1667.27672 |
| 3035 | 2948.30803 | 49 | 3035 | 1651.23157 |
| 3067 | 2949.46836 | 49 | 3067 | 1643.11548 |
| 3099 | 2950.74944 | 49 | 3099 | 1652.75818 |
| 3132 | 2949.00506 | 49 | 3131 | 1650.89009 |
| 3164 | 2949.83769 | 49 | 3164 | 1658.03155 |
| 3196 | 2949.35049 | 49 | 3196 | 1628.58362 |
| 3229 | 2948.71609 | 49 | 3228 | 1646.58733 |
| 3261 | 2949.69953 | 49 | 3261 | 1657.02262 |
| 3293 | 2949.94125 | 49 | 3293 | 1642.247 |
| 3325 | 2949.16862 | 49 | 3325 | 1644.30617 |
| 3354 | 3011.93544 | 50 | 3357 | 1658.67879 |
| 3390 | 3011.71817 | 50 | 3390 | 1653.93258 |
| 3422 | 3012.72823 | 50 | 3422 | 1637.25877 |
| 3455 | 3012.90866 | 50 | 3454 | 1692.30053 |
| 3487 | 3014.0095 | 50 | 3487 | 1645.6198 |
| 3519 | 3013.27701 | 50 | 3519 | 1665.61332 |
| 3552 | 3010.41055 | 50 | 3551 | 1654.15346 |
| 3584 | 3012.68865 | 50 | 3584 | 1666.53729 |
| 3616 | 3012.82275 | 50 | 3616 | 1659.92726 |
| 3649 | 3011.75805 | 50 | 3648 | 1671.06097 |
| 3681 | 3011.21228 | 50 | 3681 | 1662.34007 |
| 3713 | 3011.97654 | 50 | 3713 | 1645.19036 |
| 3746 | 3012.41402 | 50 | 3745 | 1651.00929 |
| 3778 | 3012.12887 | 50 | 3778 | 1650.83721 |
| 3810 | 3011.68474 | 50 | 3810 | 1647.57731 |
| 3842 | 3011.42265 | 50 | 3842 | 1668.80359 |
| 3875 | 3012.03375 | 50 | 3874 | 1659.38759 |
| 3907 | 3012.01381 | 50 | 3907 | 1642.41937 |
| 3939 | 3012.29284 | 50 | 3939 | 1678.22909 |
| 3971 | 3013.39668 | 50 | 3971 | 1644.84073 |
| 4004 | 3011.86791 | 50 | 4004 | 1662.09403 |
| 4036 | 3012.38999 | 50 | 4036 | 1662.34972 |
| 4069 | 3012.71319 | 50 | 4068 | 1660.67226 |
| 4101 | 3013.56371 | 50 | 4101 | 1661.70326 |
| 4133 | 3012.96053 | 50 | 4133 | 1641.88896 |
| 4166 | 3012.14143 | 50 | 4165 | 1663.32639 |
| 4198 | 3012.52624 | 50 | 4197 | 1655.15272 |
| 4230 | 3012.66563 | 50 | 4230 | 1669.65441 |
| 4262 | 3013.03546 | 50 | 4262 | 1656.58308 |
| 4295 | 3012.87698 | 50 | 4294 | 1653.2116 |
| 4327 | 3012.22872 | 50 | 4327 | 1658.42467 |
| 4359 | 3011.0706 | 50 | 4359 | 1643.20791 |
| 4391 | 3012.24867 | 50 | 4391 | 1666.44968 |
| 4424 | 3012.56655 | 50 | 4424 | 1649.44482 |
| 4456 | 3012.60189 | 50 | 4456 | 1643.32986 |
| 4488 | 3012.42587 | 50 | 4488 | 1656.27548 |
| 4521 | 3013.61499 | 50 | 4520 | 1637.88266 |
| 4553 | 3013.16818 | 50 | 4553 | 1642.5987 |
| 4585 | 3012.68179 | 50 | 4585 | 1647.45686 |
| 4618 | 3013.59199 | 50 | 4617 | 1652.22555 |
| 4650 | 3013.45335 | 50 | 4650 | 1641.73382 |
| 4682 | 3012.52515 | 50 | 4682 | 1659.10029 |
| 4715 | 3013.65456 | 50 | 4714 | 1650.14081 |
| 4747 | 3013.6254 | 50 | 4747 | 1667.24577 |
| 4779 | 3014.30241 | 50 | 4779 | 1649.46493 |
| 4812 | 3012.83145 | 50 | 4811 | 1651.38013 |
| 4844 | 3012.95507 | 50 | 4843 | 1637.7866 |
| 4876 | 3013.33215 | 50 | 4876 | 1670.00017 |
| 4908 | 3014.7535 | 50 | 4908 | 1669.11744 |
| 4941 | 3013.96869 | 50 | 4940 | 1651.79129 |
| 4973 | 3012.94762 | 50 | 4973 | 1657.93448 |
| 5005 | 3013.81272 | 50 | 5005 | 1643.70174 |
| 5037 | 3014.72539 | 50 | 5037 | 1647.45327 |
| 5070 | 3014.18657 | 50 | 5069 | 1653.69545 |
| 5102 | 3014.3037 | 50 | 5102 | 1663.17854 |
| 5134 | 3013.46199 | 50 | 5134 | 1654.80672 |
| 5167 | 3012.10704 | 50 | 5166 | 1655.03958 |
| 5199 | 3014.00166 | 50 | 5199 | 1662.66015 |
| 5231 | 3015.04813 | 50 | 5231 | 1660.88067 |
| 5263 | 3015.79684 | 50 | 5263 | 1642.72509 |
| 5296 | 3013.38513 | 50 | 5295 | 1661.79939 |
| 5328 | 3013.47165 | 50 | 5328 | 1658.228 |
| 5360 | 3014.26839 | 50 | 5360 | 1637.15785 |
| 5393 | 3013.21372 | 50 | 5392 | 1659.56233 |
| 5425 | 3014.94257 | 50 | 5425 | 1664.96419 |
| 5457 | 3012.97433 | 50 | 5457 | 1633.46203 |
| 5490 | 3013.39047 | 50 | 5489 | 1658.55941 |
| 5522 | 3014.58526 | 50 | 5521 | 1655.62288 |
| 5554 | 3014.76571 | 50 | 5554 | 1643.87348 |
| 5586 | 3014.30041 | 50 | 5586 | 1654.01645 |
| 5619 | 3012.49498 | 50 | 5618 | 1669.51724 |
| 5651 | 3013.32953 | 50 | 5651 | 1658.20514 |
| 5683 | 3013.58486 | 50 | 5683 | 1643.2817 |
| 5716 | 3013.41019 | 50 | 5715 | 1666.57557 |
| 5748 | 3013.39444 | 50 | 5748 | 1656.87648 |
| 5780 | 3012.89584 | 50 | 5780 | 1656.79784 |
| 5812 | 3014.84813 | 50 | 5812 | 1643.05395 |
| 5845 | 3014.37752 | 50 | 5844 | 1644.66324 |
| 5877 | 3012.74562 | 50 | 5877 | 1665.04638 |
| 5909 | 3014.63343 | 50 | 5909 | 1648.2721 |
| 5942 | 3014.89445 | 50 | 5941 | 1654.35414 |
| 5974 | 3014.34915 | 50 | 5974 | 1642.31384 |
| 6006 | 3014.36948 | 50 | 6006 | 1655.4225 |
| 6039 | 3014.02958 | 50 | 6038 | 1660.72249 |
| 6071 | 3014.44499 | 50 | 6071 | 1668.23832 |
| 6103 | 3014.50436 | 50 | 6103 | 1637.69106 |
| 6135 | 3016.32049 | 50 | 6135 | 1637.66367 |
| 6168 | 3015.44036 | 50 | 6167 | 1664.14039 |
| 6200 | 3013.23236 | 50 | 6200 | 1644.93439 |
| 6232 | 3014.18221 | 50 | 6232 | 1667.77129 |
| 6265 | 3014.76761 | 50 | 6265 | 1641.01065 |
| 6297 | 3015.45143 | 50 | 6297 | 1654.8692 |
| 6330 | 3013.30339 | 50 | 6329 | 1648.32867 |
| 6362 | 3014.75224 | 50 | 6361 | 1661.55891 |
| 6394 | 3014.72357 | 50 | 6393 | 1647.9946 |
| 6426 | 3014.5891 | 50 | 6426 | 1646.86034 |
| 6459 | 3015.01983 | 50 | 6458 | 1625.54001 |
| 6491 | 3014.5363 | 50 | 6491 | 1639.87749 |
| 6523 | 3015.47962 | 50 | 6523 | 1643.87079 |
| 6556 | 3014.62886 | 50 | 6555 | 1620.48327 |
| 6588 | 3016.40759 | 50 | 6587 | 1665.68022 |
| 6620 | 3015.96661 | 50 | 6620 | 1645.58622 |
| 6653 | 3014.04516 | 50 | 6652 | 1656.65924 |
| 6685 | 3013.7699 | 50 | 6684 | 1667.63532 |
| 6717 | 3015.42293 | 50 | 6717 | 1643.0482 |
| 6749 | 3015.65647 | 50 | 6749 | 1647.98939 |
| 6782 | 3015.29771 | 50 | 6781 | 1659.31647 |
| 6814 | 3014.03663 | 50 | 6814 | 1641.67943 |
| 6846 | 3015.13457 | 50 | 6846 | 1647.8956 |
| 6878 | 3014.49522 | 50 | 6878 | 1655.16271 |
| 6911 | 3015.44154 | 50 | 6910 | 1646.46644 |
| 6943 | 3013.82611 | 50 | 6943 | 1648.08414 |
| 6975 | 3013.79082 | 50 | 6975 | 1643.19084 |
| 7008 | 3014.62615 | 50 | 7008 | 1633.15708 |
| 7040 | 3015.1103 | 50 | 7040 | 1654.28071 |
| 7072 | 3014.68406 | 50 | 7232 | 1652.84448 |
| 7105 | | 50 | 7338 | 1659.20422 |
| 7136 | 3014.16378 | 50 | 7370 | 1631.67236 |
| 7161 | | -- | 7403 | 1639.54761 |
| 7189 | 3204.95765 | 53 | 7435 | 1638.29508 |
| 7233 | 3203.63256 | 53 | 7467 | 1645.17032 |
| 7265 | 3203.12287 | 53 | 7500 | 1659.89954 |
| 7306 | 3203.5269 | 53 | 7532 | 1633.20089 |
| 7338 | 3202.43963 | 53 | 7564 | 1668.0867 |
| 7371 | 3204.07755 | 53 | 7597 | 1657.70074 |
| 7403 | 3202.90438 | 53 | 7629 | 1670.32655 |
| 7435 | 3202.75777 | 53 | 7661 | 1646.48433 |
| 7467 | 3202.95045 | 53 | 7693 | 1641.01913 |
| 7500 | 3203.28644 | 53 | 7726 | 1644.88005 |
| 7532 | 3202.70897 | 53 | 7758 | 1652.42196 |
| 7565 | 3202.66082 | 53 | 7790 | 1660.23598 |
| 7597 | 3202.75905 | 53 | 7822 | 1654.25177 |
| 7629 | 3203.4285 | 53 | 7855 | 1663.81114 |
| 7661 | 3203.56479 | 53 | 7887 | 1646.58326 |
| 7694 | 3203.63951 | 53 | 7919 | 1637.59962 |
| 7726 | 3203.76663 | 53 | 7952 | 1652.85268 |
| 7758 | 3203.16178 | 53 | 7984 | 1667.59471 |
| 7790 | 3204.18301 | 53 | 8016 | 1653.21497 |
| 7823 | 3203.66875 | 53 | 8660 | 1656.80886 |
| 7855 | 3204.09746 | 53 | 8692 | 1653.08469 |
| 7887 | 3203.25101 | 53 | 8724 | 1656.19896 |
| 7920 | 3202.46536 | 53 | 8757 | 1651.55089 |
| 7952 | 3203.17398 | 53 | 8789 | 1629.87193 |
| 7984 | 3204.53348 | 53 | 8821 | 1664.54077 |
| 8017 | 3205.0968 | 53 | 8854 | 1642.87438 |
| 8049 | | 53 | 8886 | 1626.01551 |
| 8080 | 3203.04873 | 53 | 8918 | 1662.58351 |
| 8112 | 3202.84017 | 53 | 8950 | 1644.5073 |
| 8145 | 3201.95424 | 53 | 8983 | 1636.4733 |
| 8177 | 3203.77862 | 53 | 9015 | 1652.08694 |
| 8209 | 3205.43094 | 53 | 9048 | 1634.36724 |
| 8241 | 3204.33244 | 53 | 9080 | 1647.88788 |
| 8274 | 3203.90446 | 53 | 9112 | 1648.86236 |
| 8306 | 3203.68334 | 53 | 9144 | 1659.72628 |
| 8338 | | 53 | 9177 | 1652.87042 |
| 8370 | 3205.00204 | 53 | 9209 | 1653.01088 |
| 8402 | 3205.32552 | 53 | 9241 | 1636.04113 |
| 8434 | 3203.34668 | 53 | 9564 | 1648.28593 |
| 8467 | 3203.85979 | 53 | 9929 | 1639.25776 |
| 8499 | 3204.13644 | 53 | 9961 | 1658.12892 |
| 8531 | 3203.7276 | 53 | 9994 | 1643.13546 |
| 8564 | 3204.38873 | 53 | 10026 | 1636.52113 |
| 8600 | 3141.03861 | 52 | 10058 | 1650.32203 |
| 8632 | 3143.16208 | 52 | 10091 | 1643.2534 |
| 8661 | 3142.58048 | 52 | 10123 | 1638.27652 |
| 8693 | 3141.60777 | 52 | 10156 | 1647.29098 |
| 8725 | 3141.77866 | 52 | 10188 | 1629.69652 |
| 8757 | 3141.68864 | 52 | 10220 | 1623.26849 |
| 8790 | 3141.85765 | 52 | 10252 | 1634.69246 |
| 8822 | 3142.81446 | 52 | 10285 | 1630.02692 |
| 8854 | 3143.24214 | 52 | 10317 | 1645.89668 |
| 8886 | 3142.65983 | 52 | 10349 | 1655.18222 |
| 8919 | 3142.71147 | 52 | 10382 | 1640.64158 |
| 8951 | 3142.63043 | 52 | 10414 | 1637.88065 |
| 8983 | 3143.32463 | 52 | 10446 | 1636.77375 |
| 9016 | 3142.51749 | 52 | 10479 | 1630.11101 |
| 9048 | 3142.9693 | 52 | 10511 | 1659.50354 |
| 9080 | 3142.63483 | 52 | 10544 | 1657.26782 |
| 9113 | 3142.56058 | 52 | 10576 | 1646.25922 |
| 9145 | 3141.38378 | 52 | 10608 | 1628.11818 |
| 9177 | 3142.35551 | 52 | 10833 | 1641.82097 |
| 9209 | 3141.97335 | 52 | 10865 | 1629.40378 |
| 9242 | 3142.82123 | 52 | 10898 | 1628.2219 |
| 9274 | | 52 | 10930 | 1664.09672 |
| 9608 | 3143.01803 | 52 | 10962 | 1647.91439 |
| 9640 | | 52 | 10995 | 1624.47566 |
| 9671 | 3143.30611 | 52 | 11027 | 1645.43308 |
| 9704 | 3142.81442 | 52 | 11059 | 1653.79215 |
| 9736 | 3143.18679 | 52 | 11092 | 1659.23182 |
| 9768 | 3142.26732 | 52 | 11124 | 1634.06946 |
| 9800 | 3143.25083 | 52 | 11156 | 1639.21983 |
| 9833 | 3143.19866 | 52 | 11189 | 1629.14474 |
| 9869 | 3080.62392 | 51 | 11221 | 1635.54825 |
| 9901 | 3080.67693 | 51 | 11253 | 1646.79213 |
| 9933 | 3079.47246 | 51 | 11286 | 1643.21222 |
| 9962 | 3081.16776 | 51 | 11318 | 1644.34928 |
| 9994 | 3081.22779 | 51 | 11351 | 1660.84981 |
| 10027 | 3081.0821 | 51 | 11383 | 1629.34884 |
| 10059 | 3082.77058 | 51 | 11415 | 1647.86841 |
| 10091 | 3081.09283 | 51 | 11447 | 1631.93182 |
| 10124 | 3081.84003 | 51 | 11480 | 1629.49381 |
| 10156 | 3081.95027 | 51 | 11705 | 1658.28215 |
| 10188 | 3081.12707 | 51 | 11737 | 1650.42959 |
| 10221 | 3081.15843 | 51 | 11769 | 1651.23298 |
| 10253 | 3082.22329 | 51 | 11802 | 1646.50156 |
| 10285 | 3082.48715 | 51 | 11834 | 1640.34139 |
| 10318 | 3081.84882 | 51 | 11866 | 1631.24798 |
| 10350 | 3081.94193 | 51 | 11899 | 1629.22622 |
| 10382 | 3081.32832 | 51 | 11931 | 1636.88383 |
| 10414 | 3081.54039 | 51 | 11964 | 1650.0103 |
| 10447 | 3081.06429 | 51 | 11996 | 1645.82213 |
| 10479 | 3082.49013 | 51 | 12028 | 1633.63068 |
| 10511 | 3082.13305 | 51 | 12060 | 1637.72513 |
| 10544 | 3082.4176 | 51 | 12093 | 1647.19594 |
| 10576 | 3082.12061 | 51 | 12125 | 1647.54032 |
| 10609 | 3082.3047 | 51 | 12158 | 1655.77591 |
| 10641 | | 51 | 12190 | 1637.38369 |
| 10672 | 3081.70985 | 51 | 13380 | 1675.74579 |
| 10705 | 3080.93051 | 51 | 13413 | 1629.94506 |
| 10737 | 3080.78543 | 51 | 13445 | 1644.63919 |
| 10773 | 3018.57682 | 50 | 13477 | 1656.81789 |
| 10805 | 3019.74343 | 50 | 13510 | 1645.61622 |
| 10837 | 3020.58203 | 50 | 13542 | 1665.20436 |
| 10866 | 3021.21346 | 50 | 13574 | 1653.7958 |
| 10898 | 3020.36586 | 50 | 13606 | 1643.13656 |
| 10930 | 3020.56191 | 50 | 13639 | 1659.82598 |
| 10963 | 3018.44024 | 50 | 13671 | 1667.34367 |
| 10995 | 3020.0202 | 50 | 13703 | 1668.97594 |
| 11027 | 3019.67575 | 50 | 13736 | 1648.54644 |
| 11060 | 3020.27845 | 50 | 13768 | 1652.19594 |
| 11092 | 3019.85356 | 50 | 13800 | 1652.6665 |
| 11124 | 3020.57118 | 50 | 13833 | 1649.75035 |
| 11157 | 3020.70594 | 50 | 13865 | 1648.39139 |
| 11189 | 3020.18604 | 50 | 13897 | 1662.20162 |
| 11222 | 3019.18627 | 50 | 13930 | 1635.35591 |
| 11254 | 3020.81412 | 50 | 13962 | 1643.8972 |
| 11286 | 3020.17653 | 50 | 13994 | 1632.45006 |
| 11319 | 3020.71265 | 50 | 14027 | 1636.49061 |
| 11351 | 3019.7894 | 50 | 14059 | 1663.34177 |
| 11383 | 3021.02196 | 50 | 14091 | 1631.24348 |
| 11416 | 3020.00664 | 50 | 14124 | 1633.49676 |
| 11448 | 3020.84357 | 50 | 14156 | 1660.9302 |
| 11480 | 3020.92797 | 50 | 14188 | 1657.36862 |
| 11513 | | 50 | 14221 | 1651.17128 |
| 11544 | 3020.46384 | 50 | 14253 | 1635.30855 |
| 11576 | 3019.69378 | 50 | 14285 | 1641.22847 |
| 11608 | 3020.1632 | 50 | 14318 | 1662.93142 |
| 11637 | 3081.59707 | 51 | 14350 | 1654.74933 |
| 11669 | 3081.98125 | 51 | 14383 | 1655.14854 |
| 11701 | 3082.68167 | 51 | 14415 | 1638.29646 |
| 11737 | 3082.45962 | 51 | 14447 | 1653.29801 |
| 11770 | 3083.74404 | 51 | 14480 | 1650.68727 |
| 11802 | 3083.39362 | 51 | 14512 | 1660.80851 |
| 11834 | 3083.28322 | 51 | 14544 | 1654.06385 |
| 11867 | 3082.72233 | 51 | 14867 | 1629.10942 |
| 11899 | 3083.76294 | 51 | 14899 | 1637.82913 |
| 11931 | 3083.307 | 51 | 14931 | 1650.86576 |
| 11964 | 3082.35445 | 51 | 14964 | 1641.70845 |
| 11996 | 3082.02848 | 51 | 14996 | 1645.6309 |
| 12028 | 3083.91235 | 51 | 15028 | 1641.88093 |
| 12061 | 3083.334 | 51 | 15061 | 1637.96623 |
| 12093 | 3082.82384 | 51 | 15093 | 1653.81407 |
| 12126 | 3083.26215 | 51 | 15125 | 1642.84416 |
| 12158 | 3084.26528 | 51 | 15158 | 1654.15138 |
| 12190 | 3083.3558 | 51 | 15190 | 1654.75034 |
| 12223 | | 51 | 15222 | 1624.93581 |
| 12254 | 3084.16685 | 51 | 15255 | 1634.60818 |
| 12286 | 3083.50876 | 51 | 15287 | 1640.93983 |
| 12319 | 3083.20861 | 51 | 15319 | 1682.8266 |
| 12351 | 3082.99974 | 51 | 15352 | 1633.57753 |
| 12383 | 3082.17613 | 51 | 15384 | 1652.98285 |
| 12415 | 3083.18105 | 51 | 15417 | 1644.60267 |
| 12448 | 3082.96113 | 51 | 15642 | 1644.48849 |
| 12480 | 3084.52447 | 51 | 15674 | 1657.91834 |
| 12512 | 3082.63428 | 51 | 15706 | 1637.25536 |
| 12544 | 3083.55011 | 51 | 15739 | 1652.93243 |
| 12577 | 3081.80929 | 51 | 15771 | 1674.31529 |
| 12609 | 3083.36933 | 51 | 15804 | 1638.79583 |
| 12641 | 3083.39623 | 51 | 15836 | 1645.90625 |
| 12674 | | 51 | 15868 | 1651.49227 |
| 12705 | 3084.23051 | 51 | 15901 | 1644.57117 |
| 12737 | | 51 | 15933 | 1637.71179 |
| 12769 | 3084.55941 | 51 | 15965 | 1640.02813 |
| 12801 | 3086.01744 | 51 | 15998 | 1650.78691 |
| 12833 | | 51 | 16030 | 1619.10181 |
| 12865 | 3084.62092 | 51 | 16062 | 1642.49213 |
| 12885 | | 51 | 16514 | 1659.43575 |
| 12929 | 3085.74802 | 51 | 16547 | 1642.85125 |
| 12962 | | 51 | 16579 | 1633.13597 |
| 12993 | 3083.89337 | 51 | 16611 | 1658.24863 |
| 13025 | 3084.70781 | 51 | 16644 | 1646.57912 |
| 13058 | 3084.53236 | 51 | 16676 | 1660.20006 |
| 13090 | 3084.73203 | 51 | 16708 | 1657.02726 |
| 13122 | 3086.30322 | 51 | 16741 | 1637.45981 |
| 13155 | 3084.39005 | 51 | 16773 | 1644.38478 |
| 13187 | 3085.90236 | 51 | 16805 | 1648.60889 |
| 13219 | 3085.38251 | 51 | 16838 | 1627.60425 |
| 13251 | 3084.99009 | 51 | 16870 | 1665.45527 |
| 13287 | 3022.85432 | 50 | 16903 | 1668.93356 |
| 13320 | 3022.10583 | 50 | 16935 | 1658.92032 |
| 13352 | 3023.20336 | 50 | 17128 | 1650.12039 |
| 13384 | 3023.81159 | 50 | 17160 | 1644.86766 |
| 13413 | 3023.87917 | 50 | 17193 | 1656.36597 |
| 13445 | 3023.40699 | 50 | 17225 | 1634.55071 |
| 13478 | 3023.98436 | 50 | 17257 | 1644.78875 |
| 13510 | 3023.29124 | 50 | 17290 | 1666.13933 |
| 13542 | 3024.16053 | 50 | 17322 | 1644.46808 |
| 13575 | 3021.53089 | 50 | 17354 | 1674.52714 |
| 13607 | 3024.10028 | 50 | 17387 | 1642.63042 |
| 13639 | 3023.91889 | 50 | 17419 | 1668.23209 |
| 13672 | 3023.65996 | 50 | 17452 | 1649.83364 |
| 13704 | 3023.10142 | 50 | 17484 | 1648.9611 |
| 13736 | 3025.1577 | 50 | 17517 | 1644.65381 |
| 13768 | 3024.02857 | 50 | 17549 | 1643.45432 |
| 13801 | 3023.40025 | 50 | 17582 | 1637.052 |
| 13833 | 3024.10042 | 50 | 17614 | 1666.45917 |
| 13862 | 3086.17172 | 51 | 17646 | 1645.229 |
| 13894 | 3087.53215 | 51 | 17678 | 1652.978 |
| 13930 | 3087.48808 | 51 | 18162 | 1653.75969 |
| 13963 | 3086.67765 | 51 | 18195 | 1649.40257 |
| 13995 | 3086.69113 | 51 | 18227 | 1650.6465 |
| 14027 | 3085.6965 | 51 | 18260 | 1662.21619 |
| 14059 | 3086.19857 | 51 | 18292 | 1624.34517 |
| 14092 | 3087.81844 | 51 | 18325 | 1658.08764 |
| 14124 | 3086.19358 | 51 | 18357 | 1655.87287 |
| 14157 | 3086.75728 | 51 | 18390 | 1641.87524 |
| 14189 | 3087.60926 | 51 | 18422 | 1657.94848 |
| 14221 | 3086.96751 | 51 | 18455 | 1648.90036 |
| 14254 | 3086.50214 | 51 | 18709 | 1639.26055 |
| 14286 | 3086.76003 | 51 | 18741 | 1650.99918 |
| 14318 | 3087.70275 | 51 | 18774 | 1639.7116 |
| 14351 | 3087.23764 | 51 | 18807 | 1641.55327 |
| 14383 | 3087.57434 | 51 | 18839 | 1641.61641 |
| 14415 | 3087.70657 | 51 | 18871 | 1656.09092 |
| 14448 | 3087.2132 | 51 | 18904 | 1622.45573 |
| 14480 | 3087.28124 | 51 | 18936 | 1644.72771 |
| 14512 | 3086.57843 | 51 | 18969 | 1637.83916 |
| 14545 | 3086.30776 | 51 | 19001 | 1639.81321 |
| 14577 | 3086.05296 | 51 | 19034 | 1643.88216 |
| 14609 | 3087.33126 | 51 | 19066 | 1657.39789 |
| 14642 | 3086.80346 | 51 | 19098 | 1638.72409 |
| 14674 | 3085.65627 | 51 | 19131 | 1669.02906 |
| 14706 | 3086.95169 | 51 | 19164 | 1656.94142 |
| 14739 | 3086.28894 | 51 | 19196 | 1645.33672 |
| 14767 | 3150.68146 | 52 | 19229 | 1644.51161 |
| 14803 | | 52 | 19455 | 1627.56783 |
| 14831 | 3150.45449 | 52 | 19487 | 1669.51589 |
| 14863 | 3149.94405 | 52 | 19520 | 1652.65217 |
| 14899 | 3150.16588 | 52 | 19552 | 1638.21381 |
| 14932 | 3149.31725 | 52 | 19585 | 1629.91683 |
| 14964 | 3150.80853 | 52 | 19617 | 1652.03947 |
| 14996 | 3150.43231 | 52 | 19650 | 1638.07709 |
| 15029 | 3149.74384 | 52 | 19682 | 1657.545 |
| 15061 | 3150.0695 | 52 | 19715 | 1651.643 |
| 15093 | 3149.42938 | 52 | 19747 | 1653.13972 |
| 15126 | 3150.1392 | 52 | 19780 | 1636.74253 |
| 15158 | 3150.35309 | 52 | 19812 | 1650.79008 |
| 15191 | 3151.44996 | 52 | 19844 | 1658.83231 |
| 15223 | 3150.21852 | 52 | 19877 | 1630.83422 |
| 15255 | 3151.58428 | 52 | 19909 | 1654.22306 |
| 15287 | 3150.88901 | 52 | 19942 | 1623.44634 |
| 15320 | 3149.76067 | 52 | 19974 | 1635.64236 |
| 15352 | 3151.77722 | 52 | 20006 | 1673.75341 |
| 15384 | 3151.63836 | 52 | 20039 | 1648.62711 |
| 15417 | 3149.95152 | 52 | 20072 | 1642.93028 |
| 15449 | | 52 | 20200 | 1645.66989 |
| 15481 | 3151.3872 | 52 | 20233 | 1654.88332 |
| 15513 | 3150.9978 | 52 | 20265 | 1633.13211 |
| 15545 | 3150.51193 | 52 | 20298 | 1652.19119 |
| 15581 | 3088.48514 | 51 | 20330 | 1665.06014 |
| 15614 | 3089.3693 | 51 | 20363 | 1638.38415 |
| 15646 | 3089.21473 | 51 | 20395 | 1623.66991 |
| 15675 | 3088.30949 | 51 | 20428 | 1619.49515 |
| 15707 | 3088.91956 | 51 | 20461 | 1645.08482 |
| 15739 | 3088.88769 | 51 | 20493 | 1651.31609 |
| 15772 | 3089.06663 | 51 | 20526 | 1636.9203 |
| 15804 | 3088.17775 | 51 | 20559 | 1632.8109 |
| 15836 | 3088.11966 | 51 | 20591 | 1666.73529 |
| 15869 | 3090.11459 | 51 | 20624 | 1659.4792 |
| 15901 | 3089.08904 | 51 | 20657 | 1647.76647 |
| 15933 | 3088.82926 | 51 | 20690 | 1630.23489 |
| 15966 | 3090.15932 | 51 | 20722 | 1655.97844 |
| 15998 | 3089.68417 | 51 | 20755 | 1661.61561 |
| 16031 | 3089.78415 | 51 | 20788 | 1647.07615 |
| 16063 | 3090.04902 | 51 | 21209 | 1651.13116 |
| 16095 | | 51 | 21241 | 1638.51901 |
| 16127 | 3089.87889 | 51 | 21273 | 1630.43942 |
| 16159 | 3088.70396 | 51 | 21306 | 1630.00058 |
| 16191 | 3088.36642 | 51 | 21339 | 1650.95925 |
| 16224 | 3089.99259 | 51 | 21372 | 1641.53115 |
| 16256 | 3088.08987 | 51 | 21404 | 1637.88403 |
| 16288 | 3090.63159 | 51 | 21437 | 1669.88896 |
| 16321 | 3088.24439 | 51 | 21470 | 1646.73267 |
| 16353 | 3088.5139 | 51 | 21502 | 1648.86858 |
| 16385 | 3090.65714 | 51 | 21536 | 1663.23432 |
| 16418 | 3089.2081 | 51 | 21568 | 1652.39577 |
| 16446 | 3152.0735 | 52 | 21601 | 1652.80666 |
| 16478 | 3153.4919 | 52 | 21762 | 1658.80316 |
| 16511 | 3152.72178 | 52 | 21796 | 1628.41994 |
| 16547 | 3153.33857 | 52 | 21828 | 1652.12461 |
| 16579 | 3152.25657 | 52 | 21861 | 1638.03439 |
| 16612 | 3154.65507 | 52 | 21894 | 1652.60669 |
| 16644 | 3153.25787 | 52 | 21927 | 1652.5734 |
| 16677 | 3152.73026 | 52 | 21960 | 1636.53093 |
| 16709 | 3154.08292 | 52 | 22121 | 1653.25606 |
| 16741 | 3152.77048 | 52 | 22153 | 1660.87773 |
| 16774 | 3153.89205 | 52 | 22186 | 1652.42058 |
| 16806 | 3153.83276 | 52 | 22219 | 1641.45765 |
| 16838 | 3153.99251 | 52 | 22252 | 1643.21841 |
| 16871 | 3154.20515 | 52 | 22285 | 1645.18727 |
| 16903 | 3153.51046 | 52 | 22318 | 1639.48787 |
| 16935 | 3153.85184 | 52 | 22350 | 1625.87985 |
| 16968 | | 52 | 22383 | 1676.39809 |
| 16999 | 3154.55394 | 52 | 22416 | 1648.8877 |
| 17031 | 3154.23534 | 52 | 22771 | 1627.536 |
| 17067 | 3090.40212 | 51 | 22803 | 1621.12279 |
| 17100 | 3090.42417 | 51 | 22837 | 1663.42668 |
| 17132 | 3091.87465 | 51 | 22869 | 1631.53698 |
| 17161 | 3090.67245 | 51 | 22902 | 1634.87897 |
| 17193 | 3091.17148 | 51 | 22935 | 1643.24073 |
| 17225 | 3090.97236 | 51 | 22968 | 1649.72794 |
| 17258 | 3091.39412 | 51 | 23001 | 1651.5398 |
| 17290 | 3091.32105 | 51 | 23034 | 1670.74428 |
| 17322 | 3091.49874 | 51 | 23067 | 1781.07827 |
| 17355 | 3092.50758 | 51 | | |
| 17387 | 3092.36992 | 51 | | |
| 17420 | 3092.60456 | 51 | | |
| 17452 | 3091.55782 | 51 | | |
| 17484 | 3092.10162 | 51 | | |
| 17517 | 3091.17677 | 51 | | |
| 17549 | 3091.83473 | 51 | | |
| 17582 | 3093.22984 | 51 | | |
| 17614 | 3091.09324 | 51 | | |
| 17647 | 3091.12403 | 51 | | |
| 17679 | 3093.41807 | 51 | | |
| 17711 | | 51 | | |

| | | |
|---|---|---|
| 17742 | 3090.87998 | 51 |
| 17775 | 3091.62968 | 51 |
| 17807 | 3091.48958 | 51 |
| 17839 | 3091.55629 | 51 |
| 17872 | 3091.20923 | 51 |
| 17904 | 3091.89945 | 51 |
| 17936 | 3091.75175 | 51 |
| 17969 | 3092.38445 | 51 |
| 18001 | 3092.51584 | 51 |
| 18033 | 3094.28742 | 51 |
| 18066 | 3090.55657 | 51 |
| 18098 | 3091.10821 | 51 |
| 18130 | 3092.13273 | 51 |
| 18163 | 3091.72898 | 51 |
| 18195 | 3092.55034 | 51 |
| 18227 | 3092.2248 | 51 |
| 18260 | 3093.03441 | 51 |
| 18292 | 3093.28976 | 51 |
| 18325 | 3093.21782 | 51 |
| 18357 | 3092.65335 | 51 |
| 18390 | 3092.75597 | 51 |
| 18423 | 3093.79947 | 51 |
| 18455 | 3093.26876 | 51 |
| 18487 | | 51 |
| 18519 | | 51 |
| 18550 | | 51 |
| 18581 | 3092.63694 | 51 |
| 18610 | 3155.86728 | 52 |
| 18646 | | 52 |
| 18673 | 3156.2958 | 52 |
| 18706 | 3156.83943 | 52 |
| 18742 | 3156.60788 | 52 |
| 18775 | 3156.08992 | 52 |
| 18807 | 3156.42697 | 52 |
| 18839 | 3157.25708 | 52 |
| 18872 | 3156.91184 | 52 |
| 18904 | 3157.93776 | 52 |
| 18937 | 3157.09999 | 52 |
| 18969 | 3156.14718 | 52 |
| 19002 | 3157.33826 | 52 |
| 19034 | 3156.22292 | 52 |
| 19066 | 3158.52057 | 52 |
| 19099 | 3156.7606 | 52 |
| 19132 | 3161.16834 | 52 |
| 19164 | 3156.78963 | 52 |
| 19197 | 3157.71986 | 52 |
| 19229 | 3157.27873 | 52 |
| 19261 | 3157.71479 | 52 |
| 19294 | 3155.24833 | 52 |
| 19326 | 3156.11078 | 52 |
| 19358 | 3156.06827 | 52 |
| 19387 | 3218.80368 | 53 |
| 19419 | 3219.70426 | 53 |
| 19451 | 3219.13788 | 53 |
| 19488 | 3219.69745 | 53 |
| 19520 | 3221.77589 | 53 |
| 19553 | 3219.41663 | 53 |
| 19585 | 3220.21457 | 53 |
| 19617 | 3220.89453 | 53 |
| 19650 | 3220.28571 | 53 |
| 19683 | 3220.02986 | 53 |
| 19715 | 3221.0316 | 53 |
| 19747 | 3220.94079 | 53 |
| 19780 | 3221.06812 | 53 |
| 19812 | 3220.99139 | 53 |
| 19845 | 3221.06926 | 53 |
| 19877 | 3222.00561 | 53 |
| 19910 | 3220.62486 | 53 |
| 19942 | 3220.60544 | 53 |
| 19974 | 3221.08493 | 53 |
| 20007 | 3221.92559 | 53 |
| 20039 | 3221.91545 | 53 |
| 20072 | 3225.63183 | 53 |
| 20105 | | 53 |
| 20140 | 3157.56453 | 52 |
| 20172 | 3158.45823 | 52 |
| 20204 | 3160.1609 | 52 |
| 20233 | 3157.84819 | 52 |
| 20266 | 3158.59398 | 52 |
| 20298 | 3160.08382 | 52 |
| 20331 | 3163.54407 | 52 |
| 20364 | 3160.35251 | 52 |
| 20396 | 3164.28298 | 52 |
| 20429 | 3159.10244 | 52 |
| 20458 | 3223.55918 | 53 |
| 20490 | 3223.11422 | 53 |
| 20527 | 3221.21423 | 53 |
| 20559 | 3222.1724 | 53 |
| 20592 | 3226.07932 | 53 |
| 20624 | 3226.80674 | 53 |
| 20657 | 3222.00486 | 53 |
| 20690 | 3226.59527 | 53 |
| 20723 | 3227.67205 | 53 |
| 20756 | 3227.09182 | 53 |
| 20788 | 3228.30761 | 53 |
| 20821 | 3221.33451 | 53 |
| 20853 | 3222.12648 | 53 |
| 20886 | 3220.83361 | 53 |
| 20918 | 3221.08399 | 53 |
| 20950 | 3220.93892 | 53 |
| 20983 | 3221.59194 | 53 |
| 21015 | 3222.25149 | 53 |
| 21047 | 3221.15504 | 53 |
| 21079 | 3222.20765 | 53 |
| 21116 | 3159.11723 | 52 |
| 21148 | 3158.85154 | 52 |
| 21180 | 3160.99734 | 52 |
| 21213 | 3158.68941 | 52 |
| 21241 | 3161.16553 | 52 |
| 21273 | 3163.30469 | 52 |
| 21306 | 3164.5763 | 52 |
| 21339 | 3165.13768 | 52 |
| 21372 | 3164.0458 | 52 |
| 21405 | 3164.80846 | 52 |
| 21438 | 3164.51639 | 52 |
| 21470 | 3164.39563 | 52 |
| 21503 | 3164.80762 | 52 |
| 21536 | 3166.56048 | 52 |
| 21569 | 3165.67086 | 52 |
| 21602 | | 52 |
| 21634 | 3160.65535 | 52 |
| 21663 | 3221.79061 | 53 |
| 21695 | 3221.44713 | 53 |
| 21727 | 3222.56677 | 53 |
| 21759 | 3228.86278 | 53 |
| 21796 | 3228.66505 | 53 |
| 21829 | 3228.28698 | 53 |
| 21862 | 3229.33436 | 53 |
| 21894 | 3227.79142 | 53 |
| 21927 | 3224.7115 | 53 |
| 21960 | 3229.07356 | 53 |
| 21992 | | 53 |
| 22024 | 3225.47664 | 53 |
| 22052 | 3286.33242 | 54 |
| 22085 | 3286.52817 | 54 |
| 22117 | 3291.92851 | 54 |
| 22153 | 3291.80414 | 54 |
| 22187 | 3292.1331 | 54 |
| 22219 | 3291.99971 | 54 |
| 22252 | 3291.797 | 54 |
| 22285 | 3292.08194 | 54 |
| 22318 | 3292.19473 | 54 |
| 22350 | 3294.62131 | 54 |
| 22384 | 3292.43612 | 54 |
| 22416 | 3293.09335 | 54 |
| 22449 | | 54 |
| 22480 | 3286.99871 | 54 |
| 22513 | 3287.40284 | 54 |
| 22545 | 3287.89117 | 54 |
| 22577 | 3287.83869 | 54 |
| 22609 | 3288.0459 | 54 |
| 22642 | 3286.89489 | 54 |
| 22674 | 3285.54135 | 54 |
| 22719 | | 54 |
| 22742 | 3225.30021 | 53 |
| 22775 | 3229.91036 | 53 |
| 22804 | 3230.98478 | 53 |
| 22837 | 3230.42169 | 53 |
| 22870 | 3230.43929 | 53 |
| 22903 | 3226.51128 | 53 |

Figure 8 Data

| Carbon Black | 532 nm, T = 1526 K, n = 0.87, 820 sec, M = 37.62 Mda | 532 nm, T = 1526 K, n = 0.87, 820 sec, M = 37.62 MDa | 10.6 um, T = 1528 K, n= 0.77, 2764 sec, M = 37.62 MDa | 10.6 um, T = 1528 K, n= 0.77, 2764 sec, M = 37.62 Mda | Carbon Black | 532 nm, T = 1895 K, n = 0.34, 4024 Sec, M = 37.62 MDa | 532 nm, T = 1895 K, n = 0.34, 4024 Sec, M = 37.62 MDa | 10.6 um, T = 1894 K, n = 0.26, 4591 sec, M = 37.62 MDa | 10.6 um, T = 1894 K, n = 0.26, 4591 sec, M = 37.62 MDa | Graphite |
|---|---|---|---|---|---|---|---|---|---|---|
| 600.015 | 239.001 | 179.726 | 239.92422 | 141.82196 | 600.015 | 2032.48 | 1548.95 | 1452.29915 | 1260.19588 | 600.015 |
| 600.306 | 301.93 | 180.669 | 202.06456 | 142.57209 | 600.306 | 1935.21 | 1555.19 | 1687.10093 | 1265.3257 | 600.306 |
| 600.597 | 201.229 | 181.616 | 230.86463 | 143.32522 | 600.597 | 2077.97 | 1561.44 | 1672.20145 | 1270.46552 | 600.597 |
| 600.888 | 399.739 | 182.567 | 315.29348 | 144.08136 | 600.888 | 2120.86 | 1567.71 | 1689.54084 | 1275.61534 | 600.888 |
| 601.179 | 279.806 | 183.521 | 168.53166 | 144.84049 | 601.179 | 1980.72 | 1573.99 | 1641.65252 | 1280.78516 | 601.179 |
| 601.47 | 305.788 | 184.48 | 111.62564 | 145.60362 | 601.47 | 1870.85 | 1580.3 | 1561.29533 | 1285.96497 | 601.47 |
| 601.76 | 384.644 | 185.442 | 226.35183 | 146.36876 | 601.76 | 1983.63 | 1586.62 | 1609.59364 | 1291.16479 | 601.76 |
| 602.052 | 187.348 | 186.408 | 228.37319 | 147.13689 | 602.052 | 1988 | 1592.95 | 1785.0175 | 1296.37461 | 602.052 |
| 602.342 | 199.213 | 187.378 | 239.97923 | 147.90903 | 602.342 | 1986.21 | 1599.31 | 1851.44518 | 1301.59443 | 602.342 |
| 602.633 | 258.487 | 188.351 | 246.51138 | 148.68316 | 602.633 | 1984.44 | 1605.68 | 1659.15191 | 1306.83424 | 602.633 |
| 602.924 | 254.079 | 189.329 | 153.03593 | 149.4613 | 602.924 | 2026.59 | 1612.07 | 1630.8629 | 1312.09406 | 602.924 |
| 603.215 | 209.734 | 190.31 | 235.31341 | 150.24244 | 603.215 | 2091.11 | 1618.48 | 1620.91325 | 1317.36388 | 603.215 |
| 603.506 | 260.35 | 191.296 | 286.24637 | 151.02658 | 603.506 | 2064.25 | 1624.9 | 1802.90688 | 1322.64369 | 603.506 |
| 603.797 | 283.213 | 192.285 | 370.83526 | 151.81372 | 603.797 | 1970.46 | 1631.34 | 1864.79471 | 1327.9435 | 603.797 |
| 604.088 | 335.876 | 193.278 | 240.83838 | 152.60385 | 604.088 | 2206.61 | 1637.8 | 1636.17271 | 1333.25332 | 604.088 |
| 604.379 | 283.422 | 194.275 | 213.38755 | 153.39799 | 604.379 | 2084.03 | 1644.28 | 1681.82111 | 1338.58313 | 604.379 |
| 604.669 | 292.799 | 195.276 | 199.87217 | 154.19413 | 604.669 | 2037.66 | 1650.77 | 1773.6779 | 1343.92295 | 604.669 |
| 604.96 | 276.678 | 196.281 | 266.17384 | 154.99428 | 604.96 | 1976.97 | 1657.28 | 1873.04442 | 1349.28276 | 604.96 |
| 605.251 | 263.204 | 197.29 | 195.29337 | 155.79742 | 605.251 | 2105.08 | 1663.81 | 1813.38651 | 1354.65257 | 605.251 |
| 605.542 | 311.508 | 198.303 | 223.36531 | 156.60356 | 605.542 | 2118.61 | 1670.36 | 1662.52179 | 1360.04238 | 605.542 |
| 605.832 | 265.303 | 199.32 | 210.8431 | 157.4127 | 605.832 | 2033.46 | 1676.92 | 1851.46517 | 1365.44219 | 605.832 |
| 606.123 | 242.979 | 200.341 | 221.74302 | 158.22584 | 606.123 | 2059.18 | 1683.51 | 1772.28795 | 1370.852 | 606.123 |
| 606.414 | 324.75 | 201.365 | 228.20316 | 159.04199 | 606.414 | 2142.38 | 1690.1 | 1758.77842 | 1376.29181 | 606.414 |
| 606.705 | 277.616 | 202.394 | 278.68904 | 159.86113 | 606.705 | 2139.81 | 1696.72 | 1617.03338 | 1381.73162 | 606.705 |
| 606.995 | 261.797 | 203.427 | 231.2717 | 160.68328 | 606.995 | 2161.24 | 1703.36 | 1711.91006 | 1387.19143 | 606.995 |
| 607.286 | 313.787 | 204.464 | 251.99935 | 161.50842 | 607.286 | 2012.7 | 1710.01 | 1856.75499 | 1392.66124 | 607.286 |
| 607.577 | 297.999 | 205.505 | 299.14964 | 162.33757 | 607.577 | 2018.32 | 1716.68 | 1885.97397 | 1398.15105 | 607.577 |
| 607.868 | 284.605 | 206.549 | 244.88909 | 163.16871 | 607.868 | 2178.87 | 1723.37 | 1733.01932 | 1403.66085 | 607.868 |
| 608.158 | 264.86 | 207.598 | 285.57726 | 164.00486 | 608.158 | 2093.02 | 1730.07 | 1839.76558 | 1409.18066 | 608.158 |
| 608.449 | 276.38 | 208.651 | 198.17687 | 164.84301 | 608.449 | 2165.1 | 1736.79 | 1890.8238 | 1414.71047 | 608.449 |
| 608.74 | 297.375 | 209.708 | 228.86828 | 165.68516 | 608.74 | 2253.34 | 1743.53 | 1809.35665 | 1420.26027 | 608.74 |
| 609.03 | 348.066 | 210.769 | 213.52858 | 166.52931 | 609.03 | 2091.83 | 1750.29 | 1890.8738 | 1425.82008 | 609.03 |
| 609.321 | 272.146 | 211.834 | 273.19708 | 167.37845 | 609.321 | 2142.63 | 1757.07 | 1835.29574 | 1431.39988 | 609.321 |
| 609.612 | 332.82 | 212.904 | 278.421 | 168.2296 | 609.612 | 2046.29 | 1763.86 | 1780.23767 | 1436.99969 | 609.612 |
| 609.902 | 356.388 | 213.977 | 275.20343 | 169.08476 | 609.902 | 2168.42 | 1770.67 | 1823.81614 | 1442.59949 | 609.902 |
| 610.193 | 279.477 | 215.054 | 262.45619 | 169.94291 | 610.193 | 2139.29 | 1777.5 | 1889.72384 | 1448.22929 | 610.193 |
| 610.484 | 278.631 | 216.136 | 270.68363 | 170.80506 | 610.484 | 2240.34 | 1784.35 | 1838.93561 | 1453.8591 | 610.484 |
| 610.774 | 366.803 | 217.221 | 245.4792 | 171.66921 | 610.774 | 2246.52 | 1791.21 | 1891.93376 | 1459.5189 | 610.774 |
| 611.065 | 328.75 | 218.311 | 260.54185 | 172.53736 | 611.065 | 2285.79 | 1798.1 | 1943.56195 | 1465.1887 | 611.065 |
| 611.355 | 314.909 | 219.405 | 239.2291 | 173.40952 | 611.355 | 2189.32 | 1805 | 1886.44395 | 1470.8685 | 611.355 |
| 611.646 | 407.154 | 220.503 | 255.45495 | 174.28467 | 611.646 | 2376.47 | 1811.91 | 1973.6709 | 1476.5683 | 611.646 |
| 611.937 | 437.584 | 221.605 | 223.78138 | 175.16282 | 611.937 | 2304.89 | 1818.85 | 1788.57738 | 1482.2781 | 611.937 |
| 612.227 | 303.63 | 222.712 | 263.61339 | 176.04398 | 612.227 | 2250.01 | 1825.81 | 1876.4943 | 1488.0079 | 612.227 |
| 612.518 | 397.552 | 223.822 | 211.76027 | 176.92914 | 612.518 | 2248.8 | 1832.78 | 1733.18932 | 1493.7477 | 612.518 |
| 612.808 | 329.387 | 224.937 | 283.42188 | 177.81829 | 612.808 | 2142.47 | 1839.77 | 1777.59776 | 1499.5075 | 612.808 |
| 613.099 | 282.599 | 226.056 | 283.68992 | 178.70945 | 613.099 | 2473.22 | 1846.78 | 2039.08861 | 1505.2773 | 613.099 |
| 613.389 | 320.338 | 227.179 | 203.92589 | 179.60461 | 613.389 | 2274.65 | 1853.8 | 1943.32196 | 1511.06709 | 613.389 |
| 613.68 | 294.182 | 228.307 | 279.77523 | 180.50376 | 613.68 | 2369.62 | 1860.85 | 1995.42013 | 1516.86689 | 613.68 |
| 613.97 | 346.499 | 229.439 | 301.45905 | 181.40592 | 613.97 | 2433.64 | 1867.91 | 2015.56943 | 1522.68669 | 613.97 |
| 614.26 | 304.05 | 230.574 | 275.44547 | 182.31108 | 614.26 | 2262.51 | 1874.99 | 1904.01334 | 1528.52648 | 614.26 |
| 614.551 | 322.448 | 231.715 | 250.12702 | 183.22024 | 614.551 | 2345.41 | 1882.09 | 1984.44052 | 1534.36628 | 614.551 |
| 614.841 | 329.669 | 232.859 | 220.62883 | 184.1324 | 614.841 | 2322.74 | 1889.2 | 2042.57848 | 1540.23607 | 614.841 |
| 615.132 | 303.783 | 234.008 | 283.67992 | 185.04856 | 615.132 | 2289.21 | 1896.34 | 1898.88351 | 1546.11587 | 615.132 |
| 615.422 | 363.772 | 235.161 | 286.4134 | 185.96773 | 615.422 | 2396.61 | 1903.49 | 1981.56062 | 1552.00566 | 615.422 |
| 615.713 | 329.607 | 236.318 | 255.05288 | 186.88989 | 615.713 | 2211.39 | 1910.66 | 2038.51863 | 1557.91545 | 615.713 |
| 616.003 | 386.549 | 237.48 | 279.11812 | 187.81605 | 616.003 | 2373.26 | 1917.85 | 2060.82784 | 1563.83525 | 616.003 |
| 616.293 | 302.726 | 238.645 | 273.50613 | 188.74622 | 616.293 | 2347.54 | 1925.05 | 1990.35031 | 1569.78504 | 616.293 |
| 616.584 | 370.858 | 239.815 | 265.9218 | 189.67938 | 616.584 | 2522.27 | 1932.28 | 2019.1193 | 1575.73483 | 616.584 |
| 616.874 | 363.348 | 240.99 | 293.10358 | 190.61554 | 616.874 | 2419.43 | 1939.52 | 2056.318 | 1581.70462 | 616.874 |
| 617.165 | 349.728 | 242.169 | 282.81377 | 191.55571 | 617.165 | 2460.19 | 1946.78 | 2033.05882 | 1587.69441 | 617.165 |
| 617.455 | 350.639 | 243.352 | 260.8519 | 192.49988 | 617.455 | 2397.24 | 1954.06 | 1971.83096 | 1593.6942 | 617.455 |
| 617.745 | 378.354 | 244.54 | 291.92437 | 193.44704 | 617.745 | 2381.84 | 1961.36 | 2082.32709 | 1599.70399 | 617.745 |
| 618.036 | 369.047 | 245.732 | 249.05683 | 194.39721 | 618.036 | 2485.98 | 1968.67 | 2015.47943 | 1605.73378 | 618.036 |
| 618.326 | 376.337 | 246.928 | 284.09899 | 195.35138 | 618.326 | 2311.35 | 1976.01 | 2017.02938 | 1611.78357 | 618.326 |
| 618.616 | 372.671 | 248.129 | 305.27472 | 196.30955 | 618.616 | 2360.18 | 1983.36 | 1973.4609 | 1617.84335 | 618.616 |
| 618.907 | 309.139 | 249.334 | 298.62255 | 197.27072 | 618.907 | 2332.01 | 1990.73 | 2059.1579 | 1623.92314 | 618.907 |
| 619.197 | 352.142 | 250.543 | 318.73109 | 198.23589 | 619.197 | 2394.36 | 1998.12 | 2020.60925 | 1630.01293 | 619.197 |
| 619.487 | 343.727 | 251.757 | 287.5456 | 199.20406 | 619.487 | 2347.53 | 2005.53 | 2039.66859 | 1636.12271 | 619.487 |
| 619.777 | 358.242 | 252.975 | 268.29221 | 200.17623 | 619.777 | 2466.23 | 2012.95 | 2098.70652 | 1642.2525 | 619.777 |
| 620.067 | 330.511 | 254.198 | 293.07858 | 201.1524 | 620.067 | 2438.51 | 2020.4 | 2033.05882 | 1648.38229 | 620.067 |
| 620.358 | 327.686 | 255.425 | 296.0571 | 202.13057 | 620.358 | 2417.68 | 2027.86 | 1993.78019 | 1654.54207 | 620.358 |
| 620.648 | 382.596 | 256.657 | 272.05187 | 203.11374 | 620.648 | 2405.31 | 2035.34 | 2089.11685 | 1660.71185 | 620.648 |
| 620.938 | 362.792 | 257.893 | 290.29909 | 204.09992 | 620.938 | 2484.82 | 2042.84 | 2035.46873 | 1666.89164 | 620.938 |
| 621.228 | 379.056 | 259.134 | 265.84278 | 205.09009 | 621.228 | 2443.6 | 2050.35 | 2129.96542 | 1673.09142 | 621.228 |
| 621.519 | 398.067 | 260.379 | 306.19088 | 206.08427 | 621.519 | 2484.39 | 2057.89 | 2068.74757 | 1679.3112 | 621.519 |
| 621.809 | 342.177 | 261.628 | 271.72282 | 207.08144 | 621.809 | 2474.68 | 2065.44 | 2048.81826 | 1685.54098 | 621.809 |
| 622.099 | 314.964 | 262.882 | 326.21841 | 208.08262 | 622.099 | 2437.46 | 2073.01 | 2125.95556 | 1691.79077 | 622.099 |
| 622.389 | 305.127 | 264.141 | 296.42516 | 209.08679 | 622.389 | 2556.43 | 2080.61 | 2033.5488 | 1698.05055 | 622.389 |
| 622.679 | 375.637 | 265.404 | 288.93685 | 210.09497 | 622.679 | 2538.96 | 2088.22 | 2167.45411 | 1704.33033 | 622.679 |
| 622.969 | 340.918 | 266.671 | 313.31314 | 211.10715 | 622.969 | 2546.89 | 2095.84 | 2094.83665 | 1710.62011 | 622.969 |
| 623.259 | 383.205 | 267.944 | 281.47753 | 212.12233 | 623.259 | 2516.05 | 2103.49 | 2015.36944 | 1716.92989 | 623.259 |
| 623.549 | 361.86 | 269.22 | 330.23411 | 213.14151 | 623.549 | 2525.49 | 2111.15 | 2116.2359 | 1723.25966 | 623.549 |
| 623.84 | 367.8 | 270.502 | 295.37698 | 214.16469 | 623.84 | 2525.7 | 2118.84 | 2171.76396 | 1729.59944 | 623.84 |
| 624.13 | 372.06 | 271.787 | 308.94237 | 215.19087 | 624.13 | 2569.2 | 2126.54 | 2174.39387 | 1735.94922 | 624.13 |
| 624.42 | 350.149 | 273.078 | 296.72122 | 216.22105 | 624.42 | 2625.11 | 2134.26 | 2180.00367 | 1742.319 | 624.42 |
| 624.71 | 366.618 | 274.373 | 291.73434 | 217.25523 | 624.71 | 2527.49 | 2142 | 2103.00637 | 1748.70877 | 624.71 |
| 625 | 375.737 | 275.672 | 336.77526 | 218.29341 | 625 | 2640.18 | 2149.75 | 2198.21303 | 1755.10855 | 625 |
| 625.29 | 377.176 | 276.976 | 277.37081 | 219.3346 | 625.29 | 2614.89 | 2157.53 | 2062.93777 | 1761.52832 | 625.29 |
| 625.58 | 383.901 | 278.285 | 307.4411 | 220.37978 | 625.58 | 2584.71 | 2165.32 | 2184.68351 | 1767.9681 | 625.58 |
| 625.87 | 402.32 | 279.599 | 297.59137 | 221.42897 | 625.87 | 2592.53 | 2173.14 | 2115.17594 | 1774.41787 | 625.87 |
| 626.16 | 413.642 | 280.917 | 334.73391 | 222.48115 | 626.16 | 2647.02 | 2180.97 | 2226.66204 | 1780.87765 | 626.16 |
| 626.45 | 411.841 | 282.24 | 278.434 | 223.53834 | 626.45 | 2729.77 | 2188.82 | 2174.81385 | 1787.35742 | 626.45 |
| 626.74 | 408.957 | 283.567 | 316.11363 | 224.59852 | 626.74 | 2664.59 | 2196.68 | 2223.03217 | 1793.85719 | 626.74 |
| 627.03 | 412.374 | 284.899 | 328.73785 | 225.66171 | 627.03 | 2658.36 | 2204.57 | 2242.25149 | 1800.36696 | 627.03 |
| 627.32 | 401.18 | 286.236 | 333.5747 | 226.7299 | 627.32 | 2810.02 | 2212.48 | 2284.07003 | 1806.89674 | 627.32 |
| 627.61 | 433.467 | 287.577 | 327.35361 | 227.80109 | 627.61 | 2719.95 | 2220.4 | 2281.36012 | 1813.43651 | 627.61 |
| 627.9 | 449.241 | 288.923 | 356.8588 | 228.87728 | 627.9 | 2683.54 | 2228.35 | 2276.2803 | 1819.99628 | 627.9 |
| 628.19 | 467.031 | 290.274 | 312.556 | 229.95647 | 628.19 | 2797.74 | 2236.31 | 2318.56882 | 1826.57605 | 628.19 |
| 628.48 | 420.76 | 291.629 | 341.74714 | 231.03866 | 628.48 | 2677.23 | 2244.29 | 2305.59927 | 1833.16582 | 628.48 |
| 628.77 | 382.37 | 292.989 | 334.92494 | 232.12585 | 628.77 | 2698.54 | 2252.29 | 2229.50194 | 1839.77558 | 628.77 |
| 629.059 | 404.66 | 294.354 | 349.80256 | 233.21604 | 629.059 | 2689.72 | 2260.31 | 2333.59829 | 1846.39535 | 629.059 |
| 629.349 | 397.736 | 295.724 | 358.60511 | 234.31023 | 629.349 | 2709.79 | 2268.34 | 2304.7493 | 1853.03512 | 629.349 |
| 629.639 | 387.312 | 297.098 | 330.12009 | 235.40943 | 629.639 | 2826.99 | 2276.4 | 2337.99814 | 1859.68489 | 629.639 |
| 629.929 | 411.961 | 298.478 | 322.86082 | 236.51162 | 629.929 | 2756.26 | 2284.47 | 2370.26701 | 1866.35465 | 629.929 |
| 630.219 | 399.038 | 299.861 | 332.62353 | 237.61682 | 630.219 | 2834.7 | 2292.57 | 2358.34743 | 1873.04442 | 630.219 |
| 630.509 | 398.062 | 301.25 | 338.94665 | 238.72701 | 630.509 | 2827.77 | 2300.68 | 2341.898 | 1879.74418 | 630.509 |
| 630.799 | 465.689 | 302.644 | 337.43538 | 239.84121 | 630.799 | 2792.82 | 2308.81 | 2331.74836 | 1886.46395 | 630.799 |
| 631.088 | 464.278 | 304.042 | 333.90176 | 240.9584 | 631.088 | 2860.3 | 2316.96 | 2415.05544 | 1893.19371 | 631.088 |
| 631.378 | 407.404 | 305.445 | 354.44737 | 242.0796 | 631.378 | 2829.81 | 2325.13 | 2332.39834 | 1899.94348 | 631.378 |
| 631.668 | 427.282 | 306.853 | 374.4649 | 243.2058 | 631.668 | 2843.46 | 2333.31 | 2421.54521 | 1906.70324 | 631.668 |
| 631.958 | 424.616 | 308.265 | 375.51208 | 244.335 | 631.958 | 2889.57 | 2341.52 | 2389.53634 | 1913.483 | 631.958 |
| 632.247 | 410.847 | 309.683 | 345.37178 | 245.4682 | 632.247 | 2910.68 | 2349.74 | 2438.79461 | 1920.28277 | 632.247 |
| 632.537 | 454.602 | 311.105 | 337.30836 | 246.6054 | 632.537 | 2870.59 | 2357.99 | 2444.8544 | 1927.09253 | 632.537 |
| 632.827 | 426.942 | 312.533 | 369.70106 | 247.7466 | 632.827 | 2901.73 | 2366.25 | 2389.36634 | 1933.92229 | 632.827 |
| 633.117 | 419.348 | 313.965 | 346.48897 | 248.8908 | 633.117 | 2974.96 | 2374.53 | 2468.45357 | 1940.77205 | 633.117 |
| 633.407 | 493.828 | 315.402 | 375.85114 | 250.04 | 633.407 | 3002.43 | 2382.83 | 2508.73216 | 1947.63181 | 633.407 |
| 633.696 | 484.813 | 316.844 | 353.08513 | 251.1932 | 633.696 | 2979.22 | 2391.15 | 2497.89254 | 1954.50157 | 633.696 |
| 633.986 | 465.243 | 318.29 | 351.06378 | 252.35041 | 633.986 | 2966.84 | 2399.49 | 2517.78185 | 1961.40133 | 633.986 |
| 634.276 | 447.827 | 319.742 | 360.2614 | 253.51061 | 634.276 | 2920.58 | 2407.85 | 2464.10372 | 1968.30108 | 634.276 |
| 634.565 | 464.389 | 321.199 | 370.4762 | 254.67582 | 634.565 | 3063.22 | 2416.22 | 2467.5336 | 1975.23084 | 634.565 |
| 634.855 | 448.654 | 322.66 | 353.00612 | 255.84502 | 634.855 | 2965.01 | 2424.62 | 2584.26952 | 1982.1706 | 634.855 |
| 635.144 | 434.01 | 324.126 | 361.08254 | 257.01723 | 635.144 | 3037.94 | 2433.03 | 2456.65399 | 1989.12036 | 635.144 |
| 635.434 | 444.757 | 325.598 | 375.97916 | 258.19444 | 635.434 | 2952.35 | 2441.47 | 2450.38421 | 1996.10011 | 635.434 |
| 635.724 | 438.347 | 327.074 | 340.958 | 259.37564 | 635.724 | 3021.69 | 2449.92 | 2509.67213 | 2003.08987 | 635.724 |
| 636.013 | 489.309 | 328.555 | 342.96735 | 260.55985 | 636.013 | 3002.19 | 2458.39 | 2564.05023 | 2010.08962 | 636.013 |
| 636.303 | 453.106 | 330.041 | 346.82703 | 261.74906 | 636.303 | 3043.73 | 2466.88 | 2527.7715 | 2017.10938 | 636.303 |
| 636.593 | 477.008 | 331.532 | 411.64344 | 262.94227 | 636.593 | 3017.64 | 2475.39 | 2509.52213 | 2024.14913 | 636.593 |
| 636.882 | 468.516 | 333.029 | 372.92063 | 264.13848 | 636.882 | 3057.04 | 2483.91 | 2641.37752 | 2031.19888 | 636.882 |
| 637.172 | 478.071 | 334.53 | 397.17289 | 265.33969 | 637.172 | 3054.4 | 2492.46 | 2466.23365 | 2038.26863 | 637.172 |
| 637.461 | 478.925 | 336.036 | 353.11114 | 266.54491 | 637.461 | 2972.92 | 2501.03 | 2545.45088 | 2045.34839 | 637.461 |
| 637.751 | 437.716 | 337.547 | 330.22111 | 267.75412 | 637.751 | 3031.11 | 2509.61 | 2506.27225 | 2052.45814 | 637.751 |
| 638.04 | 476.199 | 339.063 | 384.04158 | 268.96733 | 638.04 | 3142.11 | 2518.22 | 2605.28878 | 2059.56789 | 638.04 |
| 638.33 | 452.502 | 340.584 | 385.98092 | 270.18455 | 638.33 | 3132.02 | 2526.84 | 2602.27889 | 2066.70764 | 638.33 |
| 638.619 | 467.104 | 342.111 | 393.36222 | 271.40576 | 638.619 | 3081.49 | 2535.48 | 2579.1897 | 2073.84739 | 638.619 |
| 638.909 | 483.2 | 343.642 | 384.44065 | 272.63098 | 638.909 | 3118.66 | 2544.14 | 2580.30966 | 2081.01714 | 638.909 |
| 639.198 | 468.121 | 345.178 | 365.21627 | 273.86119 | 639.198 | 3155.5 | 2552.82 | 2582.19959 | 2088.19689 | 639.198 |
| 639.488 | 470.354 | 346.719 | 386.32198 | 275.09441 | 639.488 | 3069.54 | 2561.53 | 2598.29903 | 2095.39663 | 639.488 |
| 639.777 | 461.577 | 348.266 | 398.58914 | 276.33263 | 639.777 | 3095.26 | 2570.24 | 2588.40937 | 2102.60638 | 639.777 |
| 640.067 | 467.212 | 349.817 | 394.9615 | 277.57385 | 640.067 | 3069.53 | 2578.98 | 2463.73374 | 2109.83613 | 640.067 |
| 640.356 | 490.384 | 351.374 | 403.66804 | 278.82007 | 640.356 | 3138.91 | 2587.74 | 2616.47839 | 2117.08587 | 640.356 |
| 640.646 | 473.416 | 352.936 | 390.3797 | 280.07029 | 640.646 | 3164.39 | 2596.51 | 2625.93806 | 2124.34562 | 640.646 |
| 640.935 | 485.263 | 354.503 | 362.96487 | 281.32451 | 640.935 | 3078.43 | 2605.31 | 2650.5172 | 2131.62537 | 640.935 |
| 641.225 | 484.991 | 356.075 | 410.8663 | 282.58373 | 641.225 | 3130.11 | 2614.12 | 2626.61803 | 2138.91511 | 641.225 |
| 641.514 | 473.846 | 357.652 | 409.41205 | 283.84595 | 641.514 | 3201.29 | 2622.96 | 2645.69737 | 2146.22485 | 641.514 |
| 641.803 | 481.363 | 359.234 | 425.42086 | 285.11317 | 641.803 | 3152.86 | 2631.81 | 2536.2812 | 2153.5546 | 641.803 |
| 642.093 | 472.752 | 360.821 | 354.19733 | 286.3834 | 642.093 | 3241.95 | 2640.68 | 2655.15704 | 2160.89434 | 642.093 |
| 642.382 | 487.916 | 362.414 | 398.67416 | 287.65862 | 642.382 | 3165.42 | 2649.57 | 2672.04644 | 2168.25408 | 642.382 |
| 642.671 | 491.234 | 364.012 | 411.69745 | 288.93785 | 642.671 | 3140.85 | 2658.49 | 2657.71695 | 2175.63382 | 642.671 |
| 642.961 | 467.559 | 365.615 | 375.49208 | 290.22207 | 642.961 | 3185.5 | 2667.41 | 2652.78712 | 2183.02357 | 642.961 |
| 643.25 | 524.694 | 367.223 | 387.29516 | 291.5103 | 643.25 | 3154.2 | 2676.36 | 2670.94648 | 2190.43331 | 643.25 |
| 643.539 | 487.801 | 368.836 | 392.63009 | 292.80153 | 643.539 | 3343.12 | 2685.33 | 2637.86764 | 2197.85305 | 643.539 |
| 643.829 | 508.66 | 370.454 | 393.29021 | 294.09775 | 643.829 | 3232.8 | 2694.32 | 2598.49902 | 2205.30279 | 643.829 |
| 644.118 | 535.891 | 372.078 | 392.62509 | 295.39798 | 644.118 | 3212.26 | 2703.33 | 2713.305 | 2212.75253 | 644.118 |
| 644.407 | 489.199 | 373.707 | 416.68033 | 296.70321 | 644.407 | 3205.18 | 2712.35 | 2710.14511 | 2220.23226 | 644.407 |
| 644.697 | 499.964 | 375.341 | 379.06771 | 298.01244 | 644.697 | 3228.54 | 2721.4 | 2657.34696 | 2227.722 | 644.697 |
| 644.986 | 522.464 | 376.98 | 433.54329 | 299.32567 | 644.986 | 3293.51 | 2730.46 | 2710.81509 | 2235.22174 | 644.986 |
| 645.275 | 498.592 | 378.625 | 410.8413 | 300.64291 | 645.275 | 3204.29 | 2739.55 | 2708.60516 | 2242.74148 | 645.275 |
| 645.564 | 505.868 | 380.275 | 392.6881 | 301.96514 | 645.564 | 3236.51 | 2748.65 | 2699.0055 | 2250.28121 | 645.564 |
| 645.853 | 510.978 | 381.93 | 392.37805 | 303.29037 | 645.853 | 3298.3 | 2757.77 | 2695.07564 | 2257.84095 | 645.853 |
| 646.143 | 500.586 | 383.591 | 413.36274 | 304.62061 | 646.143 | 3254.2 | 2766.91 | 2710.66509 | 2265.41068 | 646.143 |
| 646.432 | 520.723 | 385.256 | 414.16688 | 305.95584 | 646.432 | 3296.77 | 2776.07 | 2719.69478 | 2273.00042 | 646.432 |
| 646.721 | 519.694 | 386.927 | 431.40692 | 307.29508 | 646.721 | 3350.04 | 2785.26 | 2750.4137 | 2280.60015 | 646.721 |
| 647.01 | 529.453 | 388.603 | 391.21384 | 308.63831 | 647.01 | 3234 | 2794.45 | 2724.10462 | 2288.21988 | 647.01 |
| 647.299 | 515.286 | 390.285 | 410.32021 | 309.98555 | 647.299 | 3273.74 | 2803.67 | 2768.53307 | 2295.85962 | 647.299 |
| 647.588 | 515.022 | 391.972 | 419.46182 | 311.33779 | 647.588 | 3303.79 | 2812.91 | 2691.79575 | 2303.50935 | 647.588 |
| 647.878 | 518.182 | 393.664 | 425.42887 | 312.69403 | 647.878 | 3361.06 | 2822.17 | 2772.36293 | 2311.17908 | 647.878 |
| 648.167 | 498.498 | 395.362 | 431.39892 | 314.05427 | 648.167 | 3362.96 | 2831.45 | 2771.79295 | 2318.86881 | 648.167 |
| 648.456 | 531.353 | 397.065 | 428.62943 | 315.41951 | 648.456 | 3362.18 | 2840.74 | 2816.80138 | 2326.56854 | 648.456 |
| 648.745 | 536.938 | 398.773 | 438.77521 | 316.78875 | 648.745 | 3247.75 | 2850.06 | 2787.02242 | 2334.28827 | 648.745 |
| 649.034 | 512.832 | 400.487 | 425.27984 | 318.16199 | 649.034 | 3369.82 | 2859.39 | 2808.33167 | 2342.028 | 649.034 |
| 649.323 | 498.565 | 402.206 | 408.14882 | 319.54023 | 649.323 | 3389.71 | 2868.75 | 2739.98407 | 2349.77773 | 649.323 |
| 649.612 | 515.554 | 403.93 | 419.22977 | 320.92248 | 649.612 | 3388.12 | 2878.12 | 2829.94092 | 2357.54746 | 649.612 |
| 649.901 | 541.129 | 405.66 | 418.00256 | 322.30872 | 649.901 | 3379.78 | 2887.52 | 2831.30087 | 2365.32718 | 649.901 |
| 650.19 | 504.037 | 407.395 | 427.72027 | 323.69996 | 650.19 | 3335.2 | 2896.93 | 2858.50992 | 2373.12691 | 650.19 |
| 650.479 | 572.528 | 409.136 | 431.95801 | 325.09621 | 650.479 | 3419.58 | 2906.36 | 2755.45352 | 2380.94664 | 650.479 |
| 650.768 | 516.941 | 410.882 | 434.83352 | 326.49546 | 650.768 | 3394.56 | 2915.81 | 2824.6711 | 2388.77636 | 650.768 |
| 651.057 | 522.347 | 412.634 | 444.64725 | 327.8997 | 651.057 | 3436.19 | 2925.29 | 2835.30073 | 2396.62609 | 651.057 |
| 651.346 | 519.158 | 414.39 | 433.12922 | 329.30895 | 651.346 | 3415.21 | 2934.78 | 2842.61047 | 2404.49581 | 651.346 |
| 651.635 | 553.519 | 416.153 | 437.46998 | 330.7222 | 651.635 | 3467.18 | 2944.29 | 2925.48757 | 2412.37554 | 651.635 |
| 651.924 | 561.098 | 417.921 | 432.07003 | 332.14045 | 651.924 | 3478.2 | 2953.82 | 2902.82836 | 2420.27526 | 651.924 |
| 652.213 | 524.176 | 419.694 | 463.47456 | 333.5617 | 652.213 | 3512.96 | 2963.36 | 2941.887 | 2428.19498 | 652.213 |
| 652.502 | 537.504 | 421.473 | 418.90672 | 334.98895 | 652.502 | 3450.01 | 2972.93 | 2883.10905 | 2436.1247 | 652.502 |
| 652.791 | 551.32 | 423.257 | 452.3546 | 336.4202 | 652.791 | 3449.55 | 2982.52 | 2889.62883 | 2444.07443 | 652.791 |
| 653.08 | 554.872 | 425.047 | 413.1057 | 337.85545 | 653.08 | 3482.45 | 2992.13 | 2943.16695 | 2452.04415 | 653.08 |
| 653.369 | 552.495 | 426.842 | 444.3822 | 339.29571 | 653.369 | 3468.53 | 3001.76 | 2924.7576 | 2460.02387 | 653.369 |
| 653.658 | 563.578 | 428.643 | 466.11503 | 340.73996 | 653.658 | 3569.35 | 3011.4 | 2979.67567 | 2468.02359 | 653.658 |
| 653.947 | 544.387 | 430.449 | 467.7013 | 342.18822 | 653.947 | 3562.97 | 3021.07 | 3001.21492 | 2476.04331 | 653.947 |
| 654.236 | 576.667 | 432.261 | 458.08161 | 343.64247 | 654.236 | 3583.05 | 3030.75 | 2933.67728 | 2484.07303 | 654.236 |
| 654.524 | 587.629 | 434.078 | 454.40396 | 345.09973 | 654.524 | 3561.25 | 3040.46 | 2956.64648 | 2492.12274 | 654.524 |
| 654.813 | 586.306 | 435.901 | 466.57011 | 346.56299 | 654.813 | 3585.98 | 3050.18 | 2971.40596 | 2500.18246 | 654.813 |
| 655.102 | 570.43 | 437.729 | 463.32653 | 348.02925 | 655.102 | 3564.72 | 3059.93 | 2984.36551 | 2508.26218 | 655.102 |
| 655.391 | 601.87 | 439.563 | 466.87716 | 349.5015 | 655.391 | 3545.94 | 3069.69 | 2985.87546 | 2516.36189 | 655.391 |
| 655.68 | 580.905 | 441.403 | 479.76543 | 350.97676 | 655.68 | 3588.99 | 3079.47 | 2984.03552 | 2524.48161 | 655.68 |
| 655.968 | 584.551 | 443.248 | 476.48185 | 352.45802 | 655.968 | 3655.69 | 3089.28 | 3006.38474 | 2532.61133 | 655.968 |
| 656.257 | 592.626 | 445.099 | 479.45937 | 353.94329 | 656.257 | 3628.01 | 3099.1 | 3002.23488 | 2540.76104 | 656.257 |
| 656.546 | 593.943 | 446.955 | 446.38055 | 355.43255 | 656.546 | 3636.99 | 3108.94 | 3059.63287 | 2548.92075 | 656.546 |
| 656.835 | 589.249 | 448.817 | 478.25416 | 356.92681 | 656.835 | 3662.49 | 3118.8 | 3025.59407 | 2557.10047 | 656.835 |
| 657.124 | 576.137 | 450.685 | 474.4745 | 358.42607 | 657.124 | 3602.09 | 3128.68 | 3047.12331 | 2565.30018 | 657.124 |
| 657.412 | 586.555 | 452.558 | 464.8578 | 359.92934 | 657.412 | 3622.76 | 3138.58 | 3031.14387 | 2573.51989 | 657.412 |
| 657.701 | 576.666 | 454.437 | 464.96282 | 361.4376 | 657.701 | 3674.5 | 3148.5 | 2999.43498 | 2581.74961 | 657.701 |
| 657.99 | 599.655 | 456.321 | 457.9886 | 362.94987 | 657.99 | 3692.21 | 3158.44 | 3063.67273 | 2589.98932 | 657.99 |
| 658.278 | 590.833 | 458.211 | 481.52474 | 364.46714 | 658.278 | 3633.15 | 3168.4 | 3118.59081 | 2598.25903 | 658.278 |
| 658.567 | 580.404 | 460.107 | 473.11426 | 365.98841 | 658.567 | 3647.34 | 3178.38 | 3076.17229 | 2606.53874 | 658.567 |
| 658.856 | 593.471 | 462.008 | 512.59821 | 367.51567 | 658.856 | 3715.45 | 3188.38 | 3102.93136 | 2614.83845 | 658.856 |
| 659.144 | 599.726 | 463.915 | 497.66758 | 369.04594 | 659.144 | 3797.99 | 3198.4 | 3152.47962 | 2623.14816 | 659.144 |
| 659.433 | 604.719 | 465.828 | 470.16974 | 370.58221 | 659.433 | 3710.63 | 3208.43 | 3117.17086 | 2631.47786 | 659.433 |
| 659.722 | 602.217 | 467.746 | 490.45031 | 372.12249 | 659.722 | 3708.52 | 3218.49 | 3102.79136 | 2639.82757 | 659.722 |
| 660.01 | 613.121 | 469.671 | 480.52656 | 373.66776 | 660.01 | 3716.4 | 3228.57 | 3136.80017 | 2648.18728 | 660.01 |
| 660.299 | 615.689 | 471.6 | 479.73842 | 375.21703 | 660.299 | 3762.08 | 3238.66 | 3134.50025 | 2656.56699 | 660.299 |
| 660.588 | 617.971 | 473.536 | 494.23297 | 376.7713 | 660.588 | 3756.53 | 3248.78 | 3131.13037 | 2664.96669 | 660.588 |
| 660.876 | 612.859 | 475.477 | 501.51125 | 378.33058 | 660.876 | 3842.88 | 3258.92 | 3153.36959 | 2673.3864 | 660.876 |
| 661.165 | 634.465 | 477.424 | 505.743 | 379.89385 | 661.165 | 3824.43 | 3269.07 | 3258.7859 | 2681.8161 | 661.165 |
| 661.453 | 614.397 | 479.377 | 500.94415 | 381.46213 | 661.453 | 3853.74 | 3279.25 | 3177.87873 | 2690.26581 | 661.453 |
| 661.742 | 632.03 | 481.336 | 497.48255 | 383.03541 | 661.742 | 3772.32 | 3289.44 | 3192.44822 | 2698.72551 | 661.742 |
| 662.03 | 599.647 | 483.3 | 479.86645 | 384.61268 | 662.03 | 3811.39 | 3299.65 | 3143.16995 | 2707.20521 | 662.03 |
| 662.319 | 627.145 | 485.27 | 530.53436 | 386.19496 | 662.319 | 3861.29 | 3309.89 | 3143.48994 | 2715.70492 | 662.319 |
| 662.607 | 602.572 | 487.246 | 507.60833 | 387.78224 | 662.607 | 3863.77 | 3320.14 | 3232.9768 | 2724.21462 | 662.607 |
| 662.896 | 634.67 | 489.227 | 498.98481 | 389.37452 | 662.896 | 3886.3 | 3330.41 | 3244.5964 | 2732.74432 | 662.896 |
| 663.184 | 624.1 | 491.215 | 511.32198 | 390.9708 | 663.184 | 3916.24 | 3340.71 | 3191.37826 | 2741.29402 | 663.184 |
| 663.473 | 641.305 | 493.208 | 516.42488 | 392.57208 | 663.473 | 3976.79 | 3351.02 | 3257.85593 | 2749.85372 | 663.473 |
| 663.761 | 629.748 | 495.207 | 521.02769 | 394.17837 | 663.761 | 3881.8 | 3361.35 | 3273.1654 | 2758.44342 | 663.761 |
| 664.05 | 637.637 | 497.212 | 513.7284 | 395.78865 | 664.05 | 3834.92 | 3371.7 | 3182.41857 | 2767.03312 | 664.05 |
| 664.338 | 652.361 | 499.222 | 517.30103 | 397.40493 | 664.338 | 3912.1 | 3382.07 | 3265.85565 | 2775.65282 | 664.338 |
| 664.627 | 625.494 | 501.239 | 539.18288 | 399.02522 | 664.627 | 3900.43 | 3392.47 | 3270.2655 | 2784.28251 | 664.627 |
| 664.915 | 643.614 | 503.261 | 526.90372 | 400.64951 | 664.915 | 3921.51 | 3402.88 | 3199.06799 | 2792.93221 | 664.915 |
| 665.203 | 622.338 | 505.289 | 546.6492 | 402.27979 | 665.203 | 3926.2 | 3413.31 | 3288.93484 | 2801.59191 | 665.203 |
| 665.492 | 659.452 | 507.324 | 520.23755 | 403.91408 | 665.492 | 3919.72 | 3423.76 | 3230.74688 | 2810.2716 | 665.492 |
| 665.78 | 634.26 | 509.363 | 517.07799 | 405.55337 | 665.78 | 3972.48 | 3434.23 | 3270.88548 | 2818.9713 | 665.78 |
| 666.069 | 633.422 | 511.409 | 513.99345 | 407.19766 | 666.069 | 3949.76 | 3444.72 | 3315.34392 | 2827.69099 | 666.069 |
| 666.357 | 654.05 | 513.461 | 522.09688 | 408.84695 | 666.357 | 4010.85 | 3455.22 | 3256.10599 | 2836.42069 | 666.357 |
| 666.645 | 636.654 | 515.518 | 511.02693 | 410.50124 | 666.645 | 4035.46 | 3465.75 | 3230.83688 | 2845.17038 | 666.645 |
| 666.934 | 636.073 | 517.582 | 531.23649 | 412.15953 | 666.934 | 3954.55 | 3476.3 | 3341.31301 | 2853.93008 | 666.934 |
| 667.222 | 639.221 | 519.651 | 519.09135 | 413.82282 | 667.222 | 4036.37 | 3486.87 | 3355.30252 | 2862.71977 | 667.222 |
| 667.51 | 664.62 | 521.727 | 530.39434 | 415.49112 | 667.51 | 3934.86 | 3497.46 | 3298.01453 | 2871.50946 | 667.51 |
| 667.799 | 641.539 | 523.808 | 542.05839 | 417.16441 | 667.799 | 3983.93 | 3508.07 | 3315.07393 | 2880.32915 | 667.799 |
| 668.087 | 631.577 | 525.895 | 522.51195 | 418.84271 | 668.087 | 3986.81 | 3518.69 | 3348.47276 | 2889.15884 | 668.087 |
| 668.375 | 661.129 | 527.988 | 521.18972 | 420.525 | 668.375 | 4046.79 | 3529.34 | 3340.96302 | 2898.00853 | 668.375 |
| 668.663 | 683 | 530.087 | 534.69609 | 422.2133 | 668.663 | 4031.2 | 3540.01 | 3320.20375 | 2906.87822 | 668.663 |
| 668.952 | 669.323 | 532.192 | 551.83311 | 423.9056 | 668.952 | 4016.69 | 3550.69 | 3329.60342 | 2915.75791 | 668.952 |
| 669.24 | 683.094 | 534.303 | 523.85819 | 425.6029 | 669.24 | 4072.46 | 3561.4 | 3364.33221 | 2924.6576 | 669.24 |
| 669.528 | 658.542 | 536.42 | 545.83806 | 427.3052 | 669.528 | 4062.78 | 3572.12 | 3414.95043 | 2933.57729 | 669.528 |
| 669.816 | 698.673 | 538.543 | 535.82329 | 429.0125 | 669.816 | 4095.49 | 3582.87 | 3365.97215 | 2942.50697 | 669.816 |
| 670.105 | 681.192 | 540.672 | 551.34802 | 430.7248 | 670.105 | 4041.18 | 3593.63 | 3431.04987 | 2951.45666 | 670.105 |
| 670.393 | 694.763 | 542.807 | 566.51769 | 432.4411 | 670.393 | 4044.91 | 3604.42 | 3407.2507 | 2960.42635 | 670.393 |
| 670.681 | 683.633 | 544.948 | 557.28407 | 434.1634 | 670.681 | 4150.56 | 3615.22 | 3396.89107 | 2969.40603 | 670.681 |
| 670.969 | 688.505 | 547.095 | 541.28825 | 435.89071 | 670.969 | 4091.69 | 3626.05 | 3394.43115 | 2978.40572 | 670.969 |
| 671.257 | 682.596 | 549.248 | 556.907 | 437.62201 | 671.257 | 4153.94 | 3636.89 | 3426.70002 | 2987.4254 | 671.257 |
| 671.545 | 683.079 | 551.407 | 540.18706 | 439.35832 | 671.545 | 4178.16 | 3647.76 | 3443.86942 | 2996.45509 | 671.545 |
| 671.833 | 702.488 | 553.572 | 561.93789 | 441.10062 | 671.833 | 4203.63 | 3658.64 | 3491.86774 | 3005.50477 | 671.833 |
| 672.122 | 694.17 | 555.743 | 568.58106 | 442.84693 | 672.122 | 4171.02 | 3669.54 | 3461.28881 | 3014.57445 | 672.122 |
| 672.41 | 688.07 | 557.92 | 553.06933 | 444.59824 | 672.41 | 4135.51 | 3680.47 | 3413.90047 | 3023.65413 | 672.41 |
| 672.698 | 705.428 | 560.103 | 568.48204 | 446.35455 | 672.698 | 4155.5 | 3691.41 | 3488.49786 | 3032.75381 | 672.698 |
| 672.986 | 694.758 | 562.293 | 574.29206 | 448.11586 | 672.986 | 4210.66 | 3702.37 | 3470.54849 | 3041.8735 | 672.986 |
| 673.274 | 709.089 | 564.488 | 562.35596 | 449.88217 | 673.274 | 4243.11 | 3713.35 | 3497.06756 | 3051.01318 | 673.274 |
| 673.562 | 687.427 | 566.69 | 587.66742 | 451.65448 | 673.562 | 4232.19 | 3724.36 | 3512.967 | 3060.16285 | 673.562 |
| 673.85 | 714.315 | 568.897 | 573.65895 | 453.43079 | 673.85 | 4256.25 | 3735.38 | 3480.02815 | 3069.33253 | 673.85 |
| 674.138 | 705.228 | 571.111 | 566.36067 | 455.21211 | 674.138 | 4162.3 | 3746.42 | 3485.67796 | 3078.51221 | 674.138 |
| 674.426 | 695.262 | 573.331 | 571.15751 | 456.99842 | 674.426 | 4247.93 | 3757.48 | 3482.85806 | 3087.71189 | 674.426 |
| 674.714 | 718.952 | 575.557 | 595.42678 | 458.78974 | 674.714 | 4243.55 | 3768.56 | 3556.00549 | 3096.93157 | 674.714 |
| 675.002 | 721.206 | 577.789 | 571.57158 | 460.58605 | 675.002 | 4180.73 | 3779.66 | 3557.09546 | 3106.17124 | 675.002 |
| 675.29 | 714.258 | 580.027 | 591.32906 | 462.38737 | 675.29 | 4256.17 | 3790.78 | 3530.31639 | 3115.42092 | 675.29 |
| 675.578 | 720.796 | 582.271 | 597.01306 | 464.19369 | 675.578 | 4208.75 | 3801.92 | 3529.82641 | 3124.6906 | 675.578 |
| 675.866 | 715.306 | 584.521 | 572.08067 | 466.00501 | 675.866 | 4262.58 | 3813.08 | 3586.99441 | 3133.98027 | 675.866 |
| 676.154 | 712.385 | 586.778 | 584.02878 | 467.82133 | 676.154 | 4218.99 | 3824.26 | 3525.90655 | 3143.27994 | 676.154 |
| 676.442 | 743.693 | 589.041 | 583.29065 | 469.64265 | 676.442 | 4304.36 | 3835.46 | 3502.38737 | 3152.59962 | 676.442 |
| 676.73 | 740.09 | 591.31 | 597.66618 | 471.46997 | 676.73 | 4243.17 | 3846.68 | 3561.4053 | 3161.93929 | 676.73 |
| 677.018 | 709.686 | 593.585 | 588.37254 | 473.30129 | 677.018 | 4308.83 | 3857.92 | 3550.54569 | 3171.28896 | 677.018 |
| 677.305 | 723.676 | 595.866 | 601.46184 | 475.13861 | 677.305 | 4304.56 | 3869.18 | 3554.73554 | 3180.65864 | 677.305 |
| 677.593 | 741.827 | 598.154 | 575.01519 | 476.97994 | 677.593 | 4354.65 | 3880.45 | 3601.08392 | 3190.04831 | 677.593 |
| 677.881 | 757.986 | 600.448 | 619.12795 | 478.82726 | 677.881 | 4344.3 | 3891.75 | 3649.92221 | 3199.44798 | 677.881 |
| 678.169 | 725.651 | 602.748 | 586.28817 | 480.67859 | 678.169 | 4337.47 | 3903.07 | 3648.26226 | 3208.86765 | 678.169 |
| 678.457 | 730.109 | 605.054 | 612.93086 | 482.53592 | 678.457 | 4365.3 | 3914.41 | 3704.71029 | 3218.30732 | 678.457 |
| 678.745 | 723.307 | 607.366 | 591.42008 | 484.39824 | 678.745 | 4289.73 | 3925.76 | 3621.26321 | 3227.76699 | 678.745 |
| 679.032 | 744.897 | 609.685 | 587.45338 | 486.26557 | 679.032 | 4393.34 | 3937.14 | 3646.52232 | 3237.23666 | 679.032 |
| 679.32 | 751.079 | 612.01 | 599.43749 | 488.1369 | 679.32 | 4357.03 | 3948.54 | 3590.92427 | 3246.72632 | 679.32 |
| 679.608 | 745.835 | 614.341 | 593.12338 | 490.01523 | 679.608 | 4447.77 | 3959.95 | 3665.28167 | 3256.22599 | 679.608 |
| 679.896 | 760.358 | 616.678 | 584.7259 | 491.89756 | 679.896 | 4371.24 | 3971.39 | 3686.98091 | 3265.74566 | 679.896 |
| 680.183 | 755.795 | 619.022 | 595.96488 | 493.78489 | 680.183 | 4445.71 | 3982.85 | 3690.60078 | 3275.28532 | 680.183 |
| 680.471 | 716.082 | 621.372 | 608.16602 | 495.67823 | 680.471 | 4448.04 | 3994.32 | 3725.89955 | 3284.84499 | 680.471 |
| 680.759 | 753.112 | 623.728 | 598.02924 | 497.57556 | 680.759 | 4456.43 | 4005.82 | 3733.31929 | 3294.41465 | 680.759 |
| 681.047 | 755.988 | 626.091 | 590.64794 | 499.4789 | 681.047 | 4411.81 | 4017.33 | 3629.85291 | 3304.00432 | 681.047 |
| 681.334 | 789.339 | 628.46 | 645.99368 | 501.38723 | 681.334 | 4470.4 | 4028.87 | 3677.33125 | 3313.61398 | 681.334 |
| 681.622 | 770.126 | 630.835 | 593.98153 | 503.30057 | 681.622 | 4433.72 | 4040.42 | 3691.03077 | 3323.23364 | 681.622 |
| 681.91 | 779.508 | 633.216 | 622.29951 | 505.21991 | 681.91 | 4442.91 | 4052 | 3722.18968 | 3332.87331 | 681.91 |
| 682.197 | 798.928 | 635.604 | 638.59838 | 507.14325 | 682.197 | 4486.95 | 4063.59 | 3736.53917 | 3342.53297 | 682.197 |
| 682.485 | 730.353 | 637.998 | 623.28568 | 509.07259 | 682.485 | 4481.13 | 4075.2 | 3759.45837 | 3352.20263 | 682.485 |
| 682.773 | 755.497 | 640.398 | 659.188 | 511.00693 | 682.773 | 4516.06 | 4086.84 | 3764.76818 | 3361.89229 | 682.773 |
| 683.06 | 746.73 | 642.805 | 629.42076 | 512.94627 | 683.06 | 4467.07 | 4098.49 | 3756.75847 | 3371.60195 | 683.06 |
| 683.348 | 752.482 | 645.218 | 618.67687 | 514.89061 | 683.348 | 4512.98 | 4110.16 | 3760.68833 | 3381.33161 | 683.348 |
| 683.636 | 770.992 | 647.638 | 628.11553 | 516.84095 | 683.636 | 4540.95 | 4121.86 | 3854.48504 | 3391.07127 | 683.636 |
| 683.923 | 806.445 | 650.064 | 637.87025 | 518.7953 | 683.923 | 4564.18 | 4133.57 | 3820.16625 | 3400.83093 | 683.923 |
| 684.211 | 801.884 | 652.496 | 638.83642 | 520.75564 | 684.211 | 4509.98 | 4145.3 | 3788.52735 | 3410.60059 | 684.211 |
| 684.498 | 797.876 | 654.934 | 650.30544 | 522.72099 | 684.498 | 4622.83 | 4157.05 | 3793.58718 | 3420.39024 | 684.498 |
| 684.786 | 787.168 | 657.379 | 668.63767 | 524.69233 | 684.786 | 4603.33 | 4168.82 | 3787.35739 | 3430.1999 | 684.786 |
| 685.073 | 792.629 | 659.831 | 635.38181 | 526.66768 | 685.073 | 4603.34 | 4180.61 | 3822.76615 | 3440.02955 | 685.073 |
| 685.361 | 784.547 | 662.288 | 631.69116 | 528.64903 | 685.361 | 4654.01 | 4192.43 | 3829.60591 | 3449.86921 | 685.361 |
| 685.649 | 831.22 | 664.753 | 641.91396 | 530.63538 | 685.649 | 4627.11 | 4204.26 | 3826.93601 | 3459.72886 | 685.649 |
| 685.936 | 811.08 | 667.223 | 629.6028 | 532.62773 | 685.936 | 4652.69 | 4216.11 | 3843.07544 | 3469.59852 | 685.936 |
| 686.223 | 860.862 | 669.7 | 665.65714 | 534.62408 | 686.223 | 4714.87 | 4227.98 | 3902.71335 | 3479.49817 | 686.223 |
| 686.511 | 823.375 | 672.184 | 675.0988 | 536.62643 | 686.511 | 4761.15 | 4239.87 | 3929.51242 | 3489.40783 | 686.511 |
| 686.798 | 854.116 | 674.674 | 669.20176 | 538.63379 | 686.798 | 4768.37 | 4251.78 | 3911.41305 | 3499.32748 | 686.798 |
| 687.086 | 824.431 | 677.17 | 665.84317 | 540.64714 | 687.086 | 4722.32 | 4263.71 | 3891.78374 | 3509.27713 | 687.086 |
| 687.373 | 776.338 | 679.673 | 656.71757 | 542.6645 | 687.373 | 4713.27 | 4275.65 | 3996.70006 | 3519.23678 | 687.373 |
| 687.661 | 839.616 | 682.182 | 651.2466 | 544.68885 | 687.661 | 4789.22 | 4287.62 | 3930.84237 | 3529.21643 | 687.661 |
| 687.948 | 807.275 | 684.698 | 658.88495 | 546.71721 | 687.948 | 4712.61 | 4299.61 | 3936.71216 | 3539.20608 | 687.948 |
| 688.236 | 842.527 | 687.22 | 667.47446 | 548.75157 | 688.236 | 4795.48 | 4311.62 | 3948.37176 | 3549.21573 | 688.236 |
| 688.523 | 823.401 | 689.749 | 673.51052 | 550.79093 | 688.523 | 4732.77 | 4323.65 | 3966.71111 | 3559.24538 | 688.523 |
| 688.81 | 871.525 | 692.284 | 682.95618 | 552.83529 | 688.81 | 4791.42 | 4335.69 | 3955.10152 | 3569.28503 | 688.81 |
| 689.098 | 847.604 | 694.826 | 657.76275 | 554.88565 | 689.098 | 4714.81 | 4347.76 | 3979.14068 | 3579.35468 | 689.098 |
| 689.385 | 859.142 | 697.374 | 681.45092 | 556.94001 | 689.385 | 4790.56 | 4359.85 | 3987.97037 | 3589.43432 | 689.385 |
| 689.672 | 839.528 | 699.928 | 678.70144 | 559.00137 | 689.672 | 4822.91 | 4371.95 | 3950.48168 | 3599.52397 | 689.672 |
| 689.96 | 845.379 | 702.49 | 707.99159 | 561.06674 | 689.96 | 4852.94 | 4384.08 | 4030.24889 | 3609.63362 | 689.96 |
| 690.247 | 859.545 | 705.057 | 686.26177 | 563.1381 | 690.247 | 4881.79 | 4396.23 | 4082.94704 | 3619.76326 | 690.247 |
| 690.534 | 859.287 | 707.632 | 699.31406 | 565.21547 | 690.534 | 4865.4 | 4408.39 | 4054.83803 | 3629.91291 | 690.534 |
| 690.822 | 868.117 | 710.212 | 673.08645 | 567.29683 | 690.822 | 4922.31 | 4420.58 | 4060.70782 | 3640.07255 | 690.822 |
| 691.109 | 879.563 | 712.8 | 710.64606 | 569.3852 | 691.109 | 4893.45 | 4432.78 | 4012.9495 | 3650.25219 | 691.109 |
| 691.396 | 853.168 | 715.394 | 703.22675 | 571.47757 | 691.396 | 4912.68 | 4445.01 | 4076.56727 | 3660.45184 | 691.396 |
| 691.683 | 882.066 | 717.994 | 683.79433 | 573.57594 | 691.683 | 4983.92 | 4457.25 | 4095.36661 | 3670.66148 | 691.683 |
| 691.971 | 885.217 | 720.601 | 704.94505 | 575.67931 | 691.971 | 4987.51 | 4469.51 | 4113.70597 | 3680.89112 | 691.971 |
| 692.258 | 889.792 | 723.215 | 697.2717 | 577.78868 | 692.258 | 4983.79 | 4481.8 | 4144.89488 | 3691.14076 | 692.258 |
| 692.545 | 873.141 | 725.835 | 705.73319 | 579.90305 | 692.545 | 4937.23 | 4494.1 | 4102.03638 | 3701.4004 | 692.545 |
| 692.832 | 847.695 | 728.462 | 723.44331 | 582.02242 | 692.832 | 4942.03 | 4506.42 | 4132.99529 | 3711.68004 | 692.832 |
| 693.119 | 894.657 | 731.095 | 708.69071 | 584.1478 | 693.119 | 5000.65 | 4518.77 | 4146.68481 | 3721.97968 | 693.119 |
| 693.407 | 866.48 | 733.735 | 710.42802 | 586.27917 | 693.407 | 5026.9 | 4531.13 | 4218.03231 | 3732.29932 | 693.407 |
| 693.694 | 903.711 | 736.381 | 702.61864 | 588.41455 | 693.694 | 4954.75 | 4543.51 | 4145.03487 | 3742.62896 | 693.694 |
| 693.981 | 915.819 | 739.035 | 713.45855 | 590.55592 | 693.981 | 4964.97 | 4555.91 | 4236.09168 | 3752.9786 | 693.981 |
| 694.268 | 911.654 | 741.694 | 695.74644 | 592.7033 | 694.268 | 4976.46 | 4568.33 | 4151.57464 | 3763.33823 | 694.268 |
| 694.555 | 900.213 | 744.361 | 699.01301 | 594.85568 | 694.555 | 5030.97 | 4580.77 | 4197.34304 | 3773.71787 | 694.555 |
| 694.842 | 891.423 | 747.034 | 723.05124 | 597.01306 | 694.842 | 5058.9 | 4593.23 | 4172.14392 | 3784.11751 | 694.842 |
| 695.129 | 883.406 | 749.714 | 731.02564 | 599.17644 | 695.129 | 5049.02 | 4605.71 | 4164.1542 | 3794.52714 | 695.129 |
| 695.416 | 900.23 | 752.4 | 727.52003 | 601.34582 | 695.416 | 5074.64 | 4618.21 | 4261.66079 | 3804.96678 | 695.416 |
| 695.703 | 890.422 | 755.093 | 704.28994 | 603.52021 | 695.703 | 5109.78 | 4630.73 | 4179.84365 | 3815.41641 | 695.703 |
| 695.99 | 916.055 | 757.793 | 726.05777 | 605.69959 | 695.99 | 5139.74 | 4643.27 | 4229.7819 | 3825.87605 | 695.99 |
| 696.277 | 916.867 | 760.499 | 718.95652 | 607.88497 | 696.277 | 5072.72 | 4655.82 | 4224.1421 | 3836.35568 | 696.277 |
| 696.564 | 926.717 | 763.212 | 711.82727 | 610.07536 | 696.564 | 5087.17 | 4668.4 | 4326.9385 | 3846.85531 | 696.564 |
| 696.851 | 891.777 | 765.932 | 705.36213 | 612.27175 | 696.851 | 5097.26 | 4681 | 4285.84994 | 3857.37494 | 696.851 |
| 697.138 | 893.98 | 768.658 | 719.15055 | 614.47313 | 697.138 | 5075.16 | 4693.62 | 4241.08151 | 3867.90457 | 697.138 |
| 697.425 | 909.277 | 771.391 | 735.3024 | 616.68052 | 697.425 | 5151.65 | 4706.25 | 4254.50104 | 3878.4542 | 697.425 |
| 697.712 | 915.819 | 774.131 | 741.0274 | 618.89291 | 697.712 | 5167.44 | 4718.91 | 4291.53974 | 3889.02383 | 697.712 |
| 697.999 | 924.974 | 776.877 | 746.98045 | 621.1113 | 697.999 | 5146.07 | 4731.58 | 4357.17744 | 3899.60346 | 697.999 |
| 698.286 | 930.845 | 779.631 | 747.30951 | 623.33469 | 698.286 | 5165.48 | 4744.28 | 4240.08154 | 3910.20309 | 698.286 |
| 698.573 | 908.347 | 782.39 | 752.22037 | 625.56409 | 698.573 | 5230.9 | 4756.99 | 4249.34122 | 3920.82272 | 698.573 |
| 698.86 | 959.13 | 785.157 | 754.15871 | 627.79848 | 698.86 | 5192.27 | 4769.73 | 4359.79735 | 3931.46235 | 698.86 |
| 699.147 | 950.614 | 787.931 | 796.05509 | 630.03887 | 699.147 | 5256.7 | 4782.48 | 4341.74798 | 3942.11198 | 699.147 |
| 699.434 | 920.21 | 790.711 | 747.28751 | 632.28527 | 699.434 | 5220.79 | 4795.25 | 4297.14954 | 3952.7716 | 699.434 |
| 699.721 | 948.451 | 793.497 | 770.5256 | 634.53666 | 699.721 | 5282.63 | 4808.05 | 4378.57669 | 3963.46123 | 699.721 |
| 700.008 | 922.051 | 796.291 | 778.495 | 636.79306 | 700.008 | 5256.51 | 4820.86 | 4397.14604 | 3974.16085 | 700.008 |
| 700.294 | 959.624 | 799.091 | 778.02792 | 639.05546 | 700.294 | 5317.71 | 4833.69 | 4390.40628 | 3984.88048 | 700.294 |
| 700.581 | 937.786 | 801.899 | 756.07205 | 641.32386 | 700.581 | 5317.43 | 4846.54 | 4393.63617 | 3995.6101 | 700.581 |
| 700.868 | 973.918 | 804.713 | 775.74051 | 643.59726 | 700.868 | 5383.44 | 4859.41 | 4390.52628 | 4006.35973 | 700.868 |
| 701.155 | 978.334 | 807.533 | 783.96396 | 645.87666 | 701.155 | 5270.43 | 4872.3 | 4454.32404 | 4017.12935 | 701.155 |
| 701.442 | 965.554 | 810.361 | 776.90672 | 648.16206 | 701.442 | 5371.93 | 4885.21 | 4455.03402 | 4027.90897 | 701.442 |
| 701.729 | 964.242 | 813.195 | 773.14006 | 650.45146 | 701.729 | 5269.82 | 4898.14 | 4411.53554 | 4038.71859 | 701.729 |
| 702.015 | 968.837 | 816.036 | 797.88641 | 652.74787 | 702.015 | 5297.39 | 4911.09 | 4444.93437 | 4049.53821 | 702.015 |
| 702.302 | 974.887 | 818.884 | 766.27985 | 655.04927 | 702.302 | 5358.28 | 4924.06 | 4443.25443 | 4060.36783 | 702.302 |
| 702.589 | 984.967 | 821.739 | 785.03615 | 657.35668 | 702.589 | 5342.74 | 4937.05 | 4484.22299 | 4071.21745 | 702.589 |
| 702.876 | 999.174 | 824.6 | 810.43062 | 659.66909 | 702.876 | 5384.73 | 4950.05 | 4499.44246 | 4082.08707 | 702.876 |
| 703.162 | 998.238 | 827.468 | 792.17841 | 661.98749 | 703.162 | 5405.74 | 4963.08 | 4513.95195 | 4092.97669 | 703.162 |
| 703.449 | 981.51 | 830.344 | 782.68674 | 664.3119 | 703.449 | 5412.67 | 4976.13 | 4565.61014 | 4103.87631 | 703.449 |
| 703.736 | 987.236 | 833.226 | 780.52235 | 666.64131 | 703.736 | 5466.41 | 4989.19 | 4559.47036 | 4114.79593 | 703.736 |
| 704.022 | 968.981 | 836.115 | 790.58312 | 668.97672 | 704.022 | 5516.32 | 5002.28 | 4518.76179 | 4125.72555 | 704.022 |
| 704.309 | 983.838 | 839.01 | 757.88137 | 671.31814 | 704.309 | 5447.86 | 5015.38 | 4593.35917 | 4136.67516 | 704.309 |
| 704.596 | 1003.12 | 841.913 | 779.48317 | 673.66455 | 704.596 | 5545.12 | 5028.51 | 4576.29977 | 4147.64478 | 704.596 |
| 704.882 | 1044.75 | 844.822 | 816.14462 | 676.01696 | 704.882 | 5527.66 | 5041.65 | 4619.22827 | 4158.63439 | 704.882 |
| 705.169 | 978.536 | 847.738 | 791.15222 | 678.37438 | 705.169 | 5517.91 | 5054.81 | 4519.35176 | 4169.63401 | 705.169 |
| 705.456 | 977.022 | 850.662 | 815.05843 | 680.73779 | 705.456 | 5509.2 | 5067.99 | 4593.24918 | 4180.65362 | 705.456 |
| 705.742 | 1042.02 | 853.591 | 780.07628 | 683.10721 | 705.742 | 5568.71 | 5081.19 | 4609.97859 | 4191.69324 | 705.742 |
| 706.029 | 1032.19 | 856.528 | 828.68783 | 685.48163 | 706.029 | 5704.08 | 5094.42 | 4659.09687 | 4202.74285 | 706.029 |
| 706.315 | 1049.95 | 859.472 | 844.9967 | 687.86205 | 706.315 | 5662.03 | 5107.66 | 4700.14543 | 4213.81246 | 706.315 |
| 706.602 | 1023.49 | 862.423 | 834.7939 | 690.24847 | 706.602 | 5592.04 | 5120.92 | 4726.9445 | 4224.89207 | 706.602 |
| 706.888 | 1002.38 | 865.38 | 814.46433 | 692.64089 | 706.888 | 5614.19 | 5134.19 | 4633.13778 | 4236.00169 | 706.888 |
| 707.175 | 1023.56 | 868.345 | 827.15056 | 695.03831 | 707.175 | 5630.73 | 5147.49 | 4610.52857 | 4247.1213 | 707.175 |
| 707.462 | 1034.11 | 871.316 | 832.03042 | 697.44173 | 707.462 | 5618.27 | 5160.81 | 4688.99582 | 4258.25091 | 707.462 |
| 707.748 | 1034.69 | 874.295 | 822.39872 | 699.85016 | 707.748 | 5639.14 | 5174.15 | 4718.88478 | 4269.41052 | 707.748 |
| 708.034 | 1050.5 | 877.28 | 850.61469 | 702.26558 | 708.034 | 5663.79 | 5187.51 | 4652.73709 | 4280.57012 | 708.034 |
| 708.321 | 1026.32 | 880.272 | 851.2178 | 704.68601 | 708.321 | 5709.19 | 5200.88 | 4709.7351 | 4291.75973 | 708.321 |
| 708.607 | 1045.33 | 883.271 | 848.55833 | 707.11144 | 708.607 | 5643.42 | 5214.28 | 4691.83573 | 4302.95934 | 708.607 |
| 708.894 | 1050.32 | 886.277 | 806.05785 | 709.54386 | 708.894 | 5654.36 | 5227.69 | 4726.62451 | 4314.17895 | 708.894 |
| 709.18 | 1056.67 | 889.29 | 837.71142 | 711.98129 | 709.18 | 5696.55 | 5241.12 | 4688.61584 | 4325.41855 | 709.18 |
| 709.467 | 1048.2 | 892.31 | 825.92834 | 714.42472 | 709.467 | 5685.3 | 5254.58 | 4757.32343 | 4336.66816 | 709.467 |
| 709.753 | 1059.89 | 895.337 | 824.19804 | 716.87315 | 709.753 | 5789.98 | 5268.05 | 4721.95467 | 4347.93777 | 709.753 |
| 710.04 | 1073.38 | 898.371 | 849.63952 | 719.32859 | 710.04 | 5787.93 | 5281.54 | 4771.96292 | 4359.21737 | 710.04 |
| 710.326 | 1078.14 | 901.412 | 881.19007 | 721.78902 | 710.326 | 5784.67 | 5295.05 | 4789.14232 | 4370.52698 | 710.326 |
| 710.612 | 1038.82 | 904.46 | 864.40312 | 724.25545 | 710.612 | 5744.47 | 5308.58 | 4836.93065 | 4381.84658 | 710.612 |
| 710.899 | 1071.78 | 907.515 | 851.98993 | 726.72689 | 710.899 | 5836.48 | 5322.13 | 4850.35018 | 4393.17618 | 710.899 |
| 711.185 | 1066.73 | 910.577 | 896.37274 | 729.20532 | 711.185 | 5805.94 | 5335.7 | 4927.44748 | 4404.52578 | 711.185 |
| 711.471 | 1055.33 | 913.645 | 886.38598 | 731.68876 | 711.471 | 5789.7 | 5349.29 | 4810.74156 | 4415.89539 | 711.471 |
| 711.758 | 1077.25 | 916.721 | 849.39347 | 734.1782 | 711.758 | 5882.77 | 5362.9 | 4849.15022 | 4427.28499 | 711.758 |
| 712.044 | 1095.61 | 919.804 | 886.31797 | 736.67264 | 712.044 | 5885.9 | 5376.52 | 4890.31878 | 4438.68459 | 712.044 |
| 712.33 | 1077.83 | 922.894 | 885.02174 | 739.17408 | 712.33 | 5921.17 | 5390.17 | 4934.50723 | 4450.10419 | 712.33 |
| 712.616 | 1068.93 | 925.991 | 847.36912 | 741.68052 | 712.616 | 5962.34 | 5403.84 | 5021.41419 | 4461.53379 | 712.616 |
| 712.903 | 1137.46 | 929.095 | 905.60937 | 744.19296 | 712.903 | 5961.69 | 5417.52 | 4959.24636 | 4472.99339 | 712.903 |
| 713.189 | 1122.58 | 932.206 | 904.25013 | 746.7114 | 713.189 | 6008.87 | 5431.22 | 4987.72537 | 4484.45299 | 713.189 |
| 713.475 | 1119.01 | 935.324 | 885.9119 | 749.23585 | 713.475 | 5930.11 | 5444.95 | 4895.75859 | 4495.94258 | 713.475 |
| 713.761 | 1116.41 | 938.449 | 908.28684 | 751.76529 | 713.761 | 5946.85 | 5458.69 | 5007.92466 | 4507.44218 | 713.761 |
| 714.048 | 1110.77 | 941.581 | 883.88754 | 754.30074 | 714.048 | 5972.23 | 5472.45 | 4994.06514 | 4518.96178 | 714.048 |
| 714.334 | 1115.84 | 944.72 | 912.20853 | 756.84219 | 714.334 | 5920.82 | 5486.23 | 4912.338 | 4530.49137 | 714.334 |
| 714.62 | 1117.5 | 947.866 | 926.98413 | 759.38964 | 714.62 | 5995.24 | 5500.03 | 5002.13486 | 4542.04097 | 714.62 |
| 714.906 | 1085.97 | 951.019 | 904.14211 | 761.94308 | 714.906 | 5987.03 | 5513.85 | 4929.18742 | 4553.61057 | 714.906 |
| 715.192 | 1138.14 | 954.179 | 917.88653 | 764.50153 | 715.192 | 6068 | 5527.69 | 5051.92312 | 4565.20016 | 715.192 |
| 715.479 | 1143.1 | 957.347 | 904.69121 | 767.06699 | 715.479 | 6063.72 | 5541.54 | 5015.51439 | 4576.79975 | 715.479 |
| 715.765 | 1129.12 | 960.521 | 891.87195 | 769.63744 | 715.765 | 6084.05 | 5555.42 | 5060.72281 | 4588.40935 | 715.765 |
| 716.051 | 1147.75 | 963.703 | 928.91647 | 772.21389 | 716.051 | 6039.28 | 5569.31 | 5016.14437 | 4600.04894 | 716.051 |
| 716.337 | 1116.28 | 966.891 | 944.02113 | 774.79635 | 716.337 | 6046.72 | 5583.23 | 4975.4458 | 4611.69853 | 716.337 |
| 716.623 | 1137.1 | 970.087 | 908.47587 | 777.3848 | 716.623 | 6012.98 | 5597.16 | 4999.57495 | 4623.36812 | 716.623 |
| 716.909 | 1146.44 | 973.29 | 938.46215 | 779.97826 | 716.909 | 6046.07 | 5611.11 | 5011.09455 | 4635.04771 | 716.909 |
| 717.195 | 1177.27 | 976.5 | 904.15611 | 782.57872 | 717.195 | 6085.07 | 5625.09 | 5043.49341 | 4646.7473 | 717.195 |
| 717.481 | 1148.09 | 979.717 | 922.02025 | 785.18417 | 717.481 | 6111.48 | 5639.08 | 5149.22971 | 4658.46689 | 717.481 |
| 717.767 | 1138.83 | 982.941 | 931.15386 | 787.79663 | 717.767 | 6177.99 | 5653.09 | 5055.52299 | 4670.19648 | 717.767 |
| 718.053 | 1136.26 | 986.172 | 946.76661 | 790.41409 | 718.053 | 6171.82 | 5667.12 | 5044.70337 | 4681.94607 | 718.053 |
| 718.339 | 1164.42 | 989.41 | 915.29907 | 793.03756 | 718.339 | 6111.59 | 5681.16 | 5134.20024 | 4693.70566 | 718.339 |
| 718.625 | 1204.12 | 992.656 | 948.53092 | 795.66702 | 718.625 | 6160.91 | 5695.23 | 5133.20027 | 4705.49525 | 718.625 |
| 718.911 | 1168.09 | 995.909 | 917.25142 | 798.30248 | 718.911 | 6228.29 | 5709.32 | 5194.80811 | 4717.28483 | 718.911 |
| 719.197 | 1179.77 | 999.168 | 931.55393 | 800.94295 | 719.197 | 6169.02 | 5723.42 | 5106.00122 | 4729.10442 | 719.197 |
| 719.483 | 1190.37 | 1002.43 | 910.11416 | 803.59041 | 719.483 | 6190.22 | 5737.55 | 5134.00024 | 4740.93401 | 719.483 |
| 719.769 | 1147.68 | 1005.71 | 940.55552 | 806.24288 | 719.769 | 6231 | 5751.69 | 5209.6376 | 4752.78359 | 719.769 |
| 720.055 | 1174.29 | 1008.99 | 931.32689 | 808.90235 | 720.055 | 6236.2 | 5765.85 | 5169.68899 | 4764.64318 | 720.055 |
| 720.341 | 1140.39 | 1012.28 | 957.59752 | 811.56682 | 720.341 | 6381.53 | 5780.03 | 5255.39599 | 4776.52276 | 720.341 |
| 720.627 | 1141.65 | 1015.58 | 966.53609 | 814.23729 | 720.627 | 6270.39 | 5794.23 | 5209.83759 | 4788.42234 | 720.627 |
| 720.913 | 1179.44 | 1018.88 | 950.1012 | 816.91376 | 720.913 | 6334.18 | 5808.45 | 5206.16772 | 4800.34193 | 720.913 |
| 721.199 | 1193.24 | 1022.19 | 975.05959 | 819.59723 | 721.199 | 6247.33 | 5822.69 | 5258.0959 | 4812.26151 | 721.199 |
| 721.485 | 1213.82 | 1025.5 | 960.30299 | 822.2857 | 721.485 | 6323.38 | 5836.95 | 5294.12464 | 4824.21109 | 721.485 |
| 721.77 | 1191.63 | 1028.83 | 950.84933 | 824.98018 | 721.77 | 6400.29 | 5851.22 | 5330.58336 | 4836.17067 | 721.77 |
| 722.056 | 1222.41 | 1032.16 | 948.10685 | 827.67965 | 722.056 | 6366.67 | 5865.52 | 5349.10271 | 4848.15025 | 722.056 |
| 722.342 | 1188.33 | 1035.5 | 963.11649 | 830.38613 | 722.342 | 6322.31 | 5879.83 | 5305.36424 | 4860.14983 | 722.342 |
| 722.628 | 1210.65 | 1038.85 | 975.48066 | 833.09861 | 722.628 | 6367.91 | 5894.16 | 5341.87297 | 4872.15941 | 722.628 |
| 722.913 | 1215.59 | 1042.2 | 971.24592 | 835.81708 | 722.913 | 6334.87 | 5908.51 | 5255.63599 | 4884.18899 | 722.913 |
| 723.199 | 1258.36 | 1045.56 | 997.35251 | 838.54156 | 723.199 | 6354.78 | 5922.88 | 5262.39575 | 4896.22857 | 723.199 |
| 723.485 | 1224.44 | 1048.92 | 971.87103 | 841.27204 | 723.485 | 6391.71 | 5937.27 | 5301.37438 | 4908.28815 | 723.485 |
| 723.771 | 1184.85 | 1052.3 | 1013.80841 | 844.00753 | 723.771 | 6523.67 | 5951.68 | 5370.99195 | 4920.36772 | 723.771 |
| 724.057 | 1201.61 | 1055.68 | 975.01458 | 846.75001 | 724.057 | 6395.39 | 5966.11 | 5245.74633 | 4932.4573 | 724.057 |
| 724.342 | 1215.89 | 1059.07 | 981.45871 | 849.49749 | 724.342 | 6414.01 | 5980.55 | 5377.59172 | 4944.56688 | 724.342 |
| 724.628 | 1238.47 | 1062.47 | 968.2894 | 852.25198 | 724.628 | 6557.17 | 5995.02 | 5396.68105 | 4956.69645 | 724.628 |
| 724.914 | 1194.12 | 1065.87 | 970.19373 | 855.01246 | 724.914 | 6411.23 | 6009.5 | 5335.50319 | 4968.83603 | 724.914 |
| 725.199 | 1209.09 | 1069.28 | 986.08553 | 857.77795 | 725.199 | 6454.75 | 6024.01 | 5336.43316 | 4980.9956 | 725.199 |
| 725.485 | 1249.85 | 1072.7 | 979.6594 | 860.55044 | 725.485 | 6487.59 | 6038.53 | 5415.0304 | 4993.17517 | 725.485 |
| 725.771 | 1198.66 | 1076.13 | 1019.97949 | 863.32793 | 725.771 | 6339.58 | 6053.06 | 5300.44442 | 5005.36475 | 725.771 |
| 726.056 | 1209.71 | 1079.56 | 994.80706 | 866.11242 | 726.056 | 6486.02 | 6067.62 | 5478.27819 | 5017.56432 | 726.056 |
| 726.342 | 1223.89 | 1083 | 1004.6868 | 868.90291 | 726.342 | 6565.33 | 6082.2 | 5391.61122 | 5029.79389 | 726.342 |
| 726.628 | 1216.68 | 1086.44 | 1032.78175 | 871.6984 | 726.628 | 6469.69 | 6096.8 | 5386.41141 | 5042.03346 | 726.628 |
| 726.913 | 1239.13 | 1089.9 | 995.13112 | 874.50089 | 726.913 | 6622.29 | 6111.41 | 5474.09834 | 5054.29303 | 726.913 |
| 727.199 | 1258.83 | 1093.36 | 1041.83334 | 877.30839 | 727.199 | 6503.22 | 6126.05 | 5416.95034 | 5066.56261 | 727.199 |
| 727.484 | 1217.63 | 1096.83 | 1003.63662 | 880.12288 | 727.484 | 6550.61 | 6140.7 | 5402.58084 | 5078.85217 | 727.484 |
| 727.77 | 1284.16 | 1100.31 | 991.28044 | 882.94338 | 727.77 | 6634.76 | 6155.37 | 5401.12089 | 5091.15174 | 727.77 |
| 728.055 | 1277.87 | 1103.79 | 1022.94001 | 885.76988 | 728.055 | 6607.59 | 6170.06 | 5404.08079 | 5103.47131 | 728.055 |
| 728.341 | 1245.21 | 1107.28 | 1008.12741 | 888.60137 | 728.341 | 6694.01 | 6184.77 | 5420.20022 | 5115.81088 | 728.341 |
| 728.626 | 1296.72 | 1110.78 | 1007.59731 | 891.43987 | 728.626 | 6671.09 | 6199.49 | 5492.90768 | 5128.16045 | 728.626 |
| 728.912 | 1261.87 | 1114.29 | 1013.98844 | 894.28437 | 728.912 | 6562.52 | 6214.24 | 5519.14676 | 5140.53002 | 728.912 |
| 729.197 | 1281.11 | 1117.8 | 1025.23042 | 897.13488 | 729.197 | 6534.75 | 6229.01 | 5529.54639 | 5152.91958 | 729.197 |
| 729.483 | 1244.83 | 1121.32 | 1001.51624 | 899.99138 | 729.483 | 6588.02 | 6243.79 | 5507.71716 | 5165.31915 | 729.483 |
| 729.768 | 1264.2 | 1124.85 | 1025.84052 | 902.85388 | 729.768 | 6642.84 | 6258.59 | 5468.56853 | 5177.73871 | 729.768 |
| 730.054 | 1268.81 | 1128.38 | 1044.74385 | 905.72239 | 730.054 | 6677.59 | 6273.41 | 5576.35476 | 5190.16828 | 730.054 |
| 730.339 | 1268.71 | 1131.92 | 1022.21989 | 908.59689 | 730.339 | 6660.11 | 6288.25 | 5623.7731 | 5202.61784 | 730.339 |
| 730.625 | 1319.67 | 1135.47 | 1009.84771 | 911.4774 | 730.625 | 6653.93 | 6303.11 | 5551.26563 | 5215.0874 | 730.625 |
| 730.91 | 1248.29 | 1139.03 | 1027.49082 | 914.36391 | 730.91 | 6713.45 | 6317.98 | 5662.66173 | 5227.56697 | 730.91 |
| 731.195 | 1296.56 | 1142.59 | 1039.73297 | 917.25742 | 731.195 | 6622.3 | 6332.88 | 5637.9026 | 5240.06653 | 731.195 |
| 731.481 | 1321.58 | 1146.16 | 1025.44045 | 920.15593 | 731.481 | 6790.77 | 6347.79 | 5633.16277 | 5252.58609 | 731.481 |
| 731.766 | 1297.96 | 1149.74 | 1048.77456 | 923.06144 | 731.766 | 6765.32 | 6362.72 | 5632.84278 | 5265.11565 | 731.766 |
| 732.052 | 1288.85 | 1153.33 | 1054.84563 | 925.97195 | 732.052 | 6668.48 | 6377.67 | 5603.87379 | 5277.66521 | 732.052 |
| 732.337 | 1300.52 | 1156.92 | 1053.00531 | 928.88946 | 732.337 | 6758.44 | 6392.64 | 5558.30539 | 5290.22477 | 732.337 |
| 732.622 | 1365.7 | 1160.52 | 1040.84316 | 931.81298 | 732.622 | 6821.89 | 6407.63 | 5652.87208 | 5302.80433 | 732.622 |
| 732.908 | 1341.83 | 1164.13 | 1043.59365 | 934.74249 | 732.908 | 6771.42 | 6422.63 | 5606.5637 | 5315.39389 | 732.908 |
| 733.193 | 1333.54 | 1167.75 | 1068.80809 | 937.67801 | 733.193 | 6748.4 | 6437.66 | 5622.45314 | 5328.01345 | 733.193 |
| 733.478 | 1317.88 | 1171.37 | 1083.41066 | 940.61953 | 733.478 | 6939.77 | 6452.7 | 5736.27916 | 5340.63301 | 733.478 |
| 733.763 | 1333.85 | 1175 | 1074.5791 | 943.56705 | 733.763 | 6770.04 | 6467.76 | 5632.06281 | 5353.28257 | 733.763 |
| 734.048 | 1372.59 | 1178.64 | 1058.84633 | 946.52157 | 734.048 | 6874.69 | 6482.84 | 5681.95106 | 5365.94212 | 734.048 |
| 734.334 | 1321.41 | 1182.28 | 1074.79914 | 949.48109 | 734.334 | 6820.52 | 6497.94 | 5692.96067 | 5378.61168 | 734.334 |
| 734.619 | 1317.93 | 1185.93 | 1041.39326 | 952.44661 | 734.619 | 6860.53 | 6513.06 | 5681.35108 | 5391.30124 | 734.619 |
| 734.904 | 1377.06 | 1189.6 | 1077.4196 | 955.41913 | 734.904 | 6831.02 | 6528.19 | 5713.03997 | 5404.01079 | 734.904 |
| 735.189 | 1354.96 | 1193.26 | 1082.60051 | 958.39766 | 735.189 | 6910.9 | 6543.35 | 5802.38684 | 5416.74034 | 735.189 |
| 735.475 | 1324.16 | 1196.94 | 1047.63436 | 961.38218 | 735.475 | 6945.43 | 6558.52 | 5700.21042 | 5429.4699 | 735.475 |
| 735.76 | 1318.53 | 1200.62 | 1075.63929 | 964.37271 | 735.76 | 6884 | 6573.71 | 5777.30772 | 5442.22945 | 735.76 |
| 736.045 | 1382.87 | 1204.31 | 1062.11691 | 967.36924 | 736.045 | 6952.39 | 6588.92 | 5736.90913 | 5454.999 | 736.045 |
| 736.33 | 1343.26 | 1208.01 | 1088.35153 | 970.37276 | 736.33 | 6930.05 | 6604.14 | 5741.39898 | 5467.78856 | 736.33 |
| 736.615 | 1396.66 | 1211.71 | 1104.83443 | 973.38129 | 736.615 | 6944.01 | 6619.39 | 5772.44789 | 5480.58811 | 736.615 |
| 736.9 | 1366.11 | 1215.42 | 1087.76142 | 976.39682 | 736.9 | 7087.33 | 6634.65 | 5794.9471 | 5493.40766 | 736.9 |
| 737.185 | 1362.81 | 1219.14 | 1113.68598 | 979.41836 | 737.185 | 6937.06 | 6649.93 | 5732.1693 | 5506.24721 | 737.185 |
| 737.47 | 1347.55 | 1222.87 | 1106.23467 | 982.44589 | 737.47 | 7004.41 | 6665.23 | 5748.89871 | 5519.09676 | 737.47 |
| 737.755 | 1341.64 | 1226.6 | 1124.43788 | 985.47942 | 737.755 | 6943.34 | 6680.55 | 5789.03731 | 5531.95631 | 737.755 |
| 738.041 | 1379.54 | 1230.34 | 1105.71458 | 988.51896 | 738.041 | 7004.23 | 6695.89 | 5786.72739 | 5544.84586 | 738.041 |
| 738.326 | 1349.09 | 1234.09 | 1172.05626 | 991.56549 | 738.326 | 7112.41 | 6711.24 | 5945.84182 | 5557.74541 | 738.326 |
| 738.611 | 1384.75 | 1237.85 | 1114.59614 | 994.61803 | 738.611 | 7065.79 | 6726.62 | 5863.4747 | 5570.65496 | 738.611 |
| 738.896 | 1370.86 | 1241.61 | 1121.07728 | 997.67557 | 738.896 | 6991.87 | 6742.01 | 5998.13999 | 5583.5845 | 738.896 |
| 739.181 | 1406.75 | 1245.38 | 1102.77406 | 1000.73611 | 739.181 | 7138.93 | 6757.42 | 5924.05258 | 5596.53405 | 739.181 |
| 739.466 | 1386.44 | 1249.16 | 1098.61333 | 1003.80665 | 739.466 | 7101.01 | 6772.85 | 5925.74252 | 5609.4936 | 739.466 |
| 739.751 | 1409.91 | 1252.95 | 1105.94462 | 1006.88719 | 739.751 | 7096.38 | 6788.29 | 5965.27114 | 5622.47314 | 739.751 |
| 740.035 | 1421.34 | 1256.74 | 1147.84199 | 1009.96773 | 740.035 | 7278.75 | 6803.76 | 5999.82993 | 5635.46269 | 740.035 |
| 740.321 | 1443.51 | 1260.54 | 1149.05221 | 1013.05828 | 740.321 | 7240.49 | 6819.24 | 6015.74937 | 5648.47223 | 740.321 |
| 740.605 | 1390.78 | 1264.35 | 1171.42614 | 1016.15882 | 740.605 | 7226.08 | 6834.74 | 6087.33686 | 5661.50177 | 740.605 |
| 740.89 | 1415.51 | 1268.16 | 1148.02203 | 1019.25937 | 740.89 | 7166.55 | 6850.26 | 6010.64955 | 5674.54132 | 740.89 |
| 741.175 | 1431 | 1271.99 | 1128.92867 | 1022.36991 | 741.175 | 7177.85 | 6865.79 | 5921.34268 | 5687.60086 | 741.175 |
| 741.46 | 1445.67 | 1275.82 | 1130.34892 | 1025.48046 | 741.46 | 7248.09 | 6881.35 | 5992.59018 | 5700.6704 | 741.46 |
| 741.745 | 1464 | 1279.66 | 1161.34437 | 1028.60101 | 741.745 | 7199.88 | 6896.92 | 6051.87811 | 5713.75994 | 741.745 |
| 742.03 | 1426.43 | 1283.5 | 1140.75075 | 1031.73156 | 742.03 | 7268.28 | 6912.51 | 6005.33974 | 5726.85949 | 742.03 |
| 742.315 | 1413.52 | 1287.36 | 1199.07101 | 1034.86211 | 742.315 | 7285.75 | 6928.12 | 5996.18006 | 5739.97903 | 742.315 |
| 742.6 | 1416.76 | 1291.22 | 1158.20382 | 1038.00266 | 742.6 | 7362.02 | 6943.75 | 6104.40627 | 5753.11857 | 742.6 |
| 742.884 | 1454.92 | 1295.09 | 1150.92254 | 1041.14322 | 742.884 | 7320.6 | 6959.39 | 6133.07526 | 5766.26811 | 742.884 |
| 743.169 | 1429.23 | 1298.96 | 1194.81026 | 1044.29377 | 743.169 | 7266.61 | 6975.06 | 6117.50581 | 5779.43765 | 743.169 |
| 743.454 | 1455.34 | 1302.85 | 1179.44756 | 1047.45433 | 743.454 | 7394.89 | 6990.74 | 6195.62307 | 5792.61718 | 743.454 |
| 743.739 | 1487.15 | 1306.74 | 1197.56074 | 1050.61488 | 743.739 | 7392.72 | 7006.44 | 6160.5543 | 5805.81672 | 743.739 |
| 744.023 | 1512.31 | 1310.63 | 1215.4939 | 1053.78544 | 744.023 | 7363.85 | 7022.15 | 6138.06509 | 5819.02626 | 744.023 |
| 744.308 | 1489.81 | 1314.54 | 1135.51983 | 1056.966 | 744.308 | 7420.44 | 7037.89 | 6168.73401 | 5832.2558 | 744.308 |
| 744.593 | 1463.13 | 1318.45 | 1211.21315 | 1060.14656 | 744.593 | 7457.45 | 7053.64 | 6135.12519 | 5845.50533 | 744.593 |
| 744.878 | 1490.81 | 1322.37 | 1182.87816 | 1063.33712 | 744.878 | 7418.93 | 7069.41 | 6202.68283 | 5858.76487 | 744.878 |
| 745.162 | 1473.66 | 1326.3 | 1180.86781 | 1066.52768 | 745.162 | 7471.15 | 7085.2 | 6255.72097 | 5872.0344 | 745.162 |
| 745.447 | 1432.75 | 1330.24 | 1214.26368 | 1069.72825 | 745.447 | 7410.09 | 7101.01 | 6201.86285 | 5885.33394 | 745.447 |
| 745.732 | 1498.82 | 1334.18 | 1207.92257 | 1072.93881 | 745.732 | 7488.03 | 7116.83 | 6146.3648 | 5898.64347 | 745.732 |
| 746.016 | 1488.55 | 1338.13 | 1199.67111 | 1076.14938 | 746.016 | 7399.59 | 7132.68 | 6274.08033 | 5911.96301 | 746.016 |
| 746.301 | 1460.8 | 1342.09 | 1208.1326 | 1079.36994 | 746.301 | 7400.8 | 7148.54 | 6275.13029 | 5925.30254 | 746.301 |
| 746.586 | 1472.89 | 1346.06 | 1182.86816 | 1082.60051 | 746.586 | 7465.9 | 7164.41 | 6174.8638 | 5938.65207 | 746.586 |
| 746.87 | 1501.84 | 1350.03 | 1214.67376 | 1085.83108 | 746.87 | 7488.53 | 7180.31 | 6238.70157 | 5952.0216 | 746.87 |
| 747.155 | 1472.15 | 1354.01 | 1202.89168 | 1089.07165 | 747.155 | 7517.7 | 7196.23 | 6133.12526 | 5965.41113 | 747.155 |
| 747.439 | 1519.35 | 1358 | 1249.09981 | 1092.31222 | 747.439 | 7476.21 | 7212.16 | 6267.57055 | 5978.81066 | 747.439 |
| 747.724 | 1519.09 | 1362 | 1201.48143 | 1095.56279 | 747.724 | 7566.32 | 7228.11 | 6369.49699 | 5992.23019 | 747.724 |
| 748.009 | 1511.31 | 1366 | 1193.08996 | 1098.82337 | 748.009 | 7470.69 | 7244.07 | 6226.422 | 6005.65972 | 748.009 |
| 748.293 | 1485.57 | 1370.01 | 1227.64604 | 1102.09394 | 748.293 | 7562.9 | 7260.06 | 6308.67912 | 6019.10925 | 748.293 |
| 748.578 | 1501.07 | 1374.03 | 1229.59638 | 1105.36452 | 748.578 | 7604.37 | 7276.06 | 6397.81599 | 6032.56878 | 748.578 |
| 748.862 | 1568.06 | 1378.06 | 1198.31088 | 1108.63509 | 748.862 | 7561.64 | 7292.08 | 6395.99606 | 6046.04831 | 748.862 |
| 749.147 | 1491.43 | 1382.09 | 1252.92049 | 1111.91567 | 749.147 | 7659.47 | 7308.12 | 6419.80522 | 6059.54784 | 749.147 |
| 749.431 | 1519.72 | 1386.13 | 1212.51337 | 1115.20625 | 749.431 | 7796.45 | 7324.18 | 6331.63831 | 6073.05736 | 749.431 |
| 749.716 | 1547.31 | 1390.18 | 1220.0247 | 1118.50683 | 749.716 | 7603.76 | 7340.25 | 6326.89848 | 6086.57689 | 749.716 |
| 750 | 1548.32 | 1394.24 | 1238.768 | 1121.80741 | 750 | 7728.54 | 7356.34 | 6344.82785 | 6100.11642 | 750 |
| 750.284 | 1532.15 | 1398.3 | 1246.71939 | 1125.118 | 750.284 | 7676.73 | 7372.45 | 6426.82498 | 6113.67594 | 750.284 |
| 750.569 | 1541.29 | 1402.37 | 1272.23388 | 1128.42858 | 750.569 | 7720.91 | 7388.58 | 6330.17836 | 6127.24547 | 750.569 |
| 750.853 | 1564.85 | 1406.45 | 1239.3881 | 1131.74916 | 750.853 | 7675.26 | 7404.72 | 6380.89659 | 6140.83499 | 750.853 |
| 751.138 | 1529.55 | 1410.54 | 1223.68534 | 1135.07975 | 751.138 | 7700.55 | 7420.88 | 6558.62036 | 6154.43452 | 751.138 |
| 751.422 | 1606.62 | 1414.63 | 1244.37898 | 1138.41033 | 751.422 | 7806.32 | 7437.06 | 6441.95445 | 6168.05404 | 751.422 |
| 751.706 | 1528.98 | 1418.74 | 1256.12105 | 1141.75092 | 751.706 | 7834 | 7453.26 | 6496.20255 | 6181.68356 | 751.706 |
| 751.991 | 1557.18 | 1422.85 | 1277.30478 | 1145.09151 | 751.991 | 7754.47 | 7469.47 | 6378.89666 | 6195.33308 | 751.991 |
| 752.275 | 1545.73 | 1426.96 | 1284.45604 | 1148.4521 | 752.275 | 7710.68 | 7485.71 | 6538.10108 | 6208.99261 | 752.275 |
| 752.559 | 1606.7 | 1431.09 | 1278.21494 | 1151.80269 | 752.559 | 7914.4 | 7501.96 | 6494.31262 | 6222.67213 | 752.559 |
| 752.844 | 1587.49 | 1435.22 | 1259.72168 | 1155.17328 | 752.844 | 7766.49 | 7518.22 | 6562.30024 | 6236.37165 | 752.844 |
| 753.128 | 1568.18 | 1439.36 | 1321.68259 | 1158.54388 | 753.128 | 7805.88 | 7534.51 | 6501.32237 | 6250.08117 | 753.128 |
| 753.412 | 1562.35 | 1443.51 | 1300.52886 | 1161.92447 | 753.412 | 7838.75 | 7550.81 | 6493.22265 | 6263.80069 | 753.412 |
| 753.697 | 1618.96 | 1447.66 | 1289.66695 | 1165.30507 | 753.697 | 7983.25 | 7567.13 | 6630.24786 | 6277.54021 | 753.697 |
| 753.981 | 1570.65 | 1451.83 | 1288.8068 | 1168.69566 | 753.981 | 7897.19 | 7583.47 | 6536.16115 | 6291.29972 | 753.981 |
| 754.265 | 1591.5 | 1456 | 1296.70819 | 1172.08626 | 754.265 | 7891.87 | 7599.82 | 6602.49883 | 6305.06924 | 754.265 |
| 754.549 | 1655.13 | 1460.18 | 1325.84332 | 1175.49686 | 754.549 | 7827.2 | 7616.19 | 6437.6546 | 6318.84876 | 754.549 |
| 754.834 | 1637.01 | 1464.36 | 1281.33549 | 1178.90746 | 754.834 | 7966.76 | 7632.58 | 6565.53012 | 6332.64828 | 754.834 |
| 755.118 | 1627.86 | 1468.56 | 1270.6236 | 1182.31806 | 755.118 | 7886.12 | 7648.99 | 6501.90235 | 6346.46779 | 755.118 |
| 755.402 | 1576.88 | 1472.76 | 1298.84857 | 1185.73866 | 755.402 | 7897.93 | 7665.41 | 6662.20674 | 6360.29731 | 755.402 |
| 755.686 | 1623.68 | 1476.96 | 1293.72767 | 1189.16927 | 755.686 | 7990.9 | 7681.85 | 6581.94955 | 6374.14682 | 755.686 |
| 755.97 | 1610.67 | 1481.18 | 1262.29213 | 1192.59987 | 755.97 | 7923.27 | 7698.31 | 6541.20097 | 6388.00634 | 755.97 |
| 756.254 | 1578.93 | 1485.4 | 1319.43219 | 1196.05048 | 756.254 | 7960.79 | 7714.79 | 6735.59417 | 6401.88585 | 756.254 |
| 756.538 | 1639.72 | 1489.64 | 1315.48149 | 1199.49108 | 756.538 | 7988.1 | 7731.28 | 6617.23831 | 6415.77537 | 756.538 |
| 756.823 | 1639.44 | 1493.87 | 1305.38972 | 1202.94169 | 756.823 | 7997.07 | 7747.79 | 6610.96853 | 6429.68488 | 756.823 |
| 757.107 | 1655.3 | 1498.12 | 1289.957 | 1206.4023 | 757.107 | 8066.48 | 7764.32 | 6647.43725 | 6443.60439 | 757.107 |
| 757.391 | 1626.02 | 1502.38 | 1318.52203 | 1209.87291 | 757.391 | 8162.12 | 7780.86 | 6720.08471 | 6457.5439 | 757.391 |
| 757.675 | 1649.76 | 1506.64 | 1311.24075 | 1213.34352 | 757.675 | 7991.97 | 7797.43 | 6718.97475 | 6471.49341 | 757.675 |
| 757.959 | 1642.64 | 1510.91 | 1334.65487 | 1216.82413 | 757.959 | 8050.63 | 7814.01 | 6783.72248 | 6485.46293 | 757.959 |
| 758.243 | 1614.31 | 1515.18 | 1322.47273 | 1220.31475 | 758.243 | 8039.07 | 7830.6 | 6743.1239 | 6499.44244 | 758.243 |
| 758.527 | 1636.26 | 1519.47 | 1325.43325 | 1223.80536 | 758.527 | 8078.23 | 7847.22 | 6806.98167 | 6513.44195 | 758.527 |
| 758.811 | 1690.68 | 1523.76 | 1354.67839 | 1227.30598 | 758.811 | 8118.69 | 7863.85 | 6767.09306 | 6527.45146 | 758.811 |
| 759.095 | 1677.99 | 1528.06 | 1352.07794 | 1230.80659 | 759.095 | 8225.76 | 7880.5 | 6755.82346 | 6541.48096 | 759.095 |
| 759.379 | 1686.44 | 1532.37 | 1385.93389 | 1234.31721 | 759.379 | 8165.97 | 7897.16 | 6848.68021 | 6555.53047 | 759.379 |
| 759.663 | 1693.36 | 1536.69 | 1301.64906 | 1237.83783 | 759.663 | 8258.43 | 7913.85 | 6786.36239 | 6569.57998 | 759.663 |
| 759.947 | 1644.98 | 1541.01 | 1360.53942 | 1241.35845 | 759.947 | 8281.29 | 7930.55 | 6732.35428 | 6583.65949 | 759.947 |
| 760.231 | 1671.29 | 1545.34 | 1405.06726 | 1244.88907 | 760.231 | 8260.92 | 7947.26 | 6820.99118 | 6597.74899 | 760.231 |
| 760.515 | 1657.19 | 1549.68 | 1359.22919 | 1248.4297 | 760.515 | 8173.72 | 7964 | 6712.06499 | 6611.8485 | 760.515 |
| 760.799 | 1677.76 | 1554.03 | 1369.85106 | 1251.97032 | 760.799 | 8130.56 | 7980.75 | 6860.12981 | 6625.96801 | 760.799 |
| 761.082 | 1696.14 | 1558.38 | 1341.15601 | 1255.52094 | 761.082 | 8214.42 | 7997.52 | 6804.42176 | 6640.09751 | 761.082 |
| 761.366 | 1684.96 | 1562.74 | 1380.93301 | 1259.07157 | 761.366 | 8272.29 | 8014.3 | 6818.51126 | 6654.24702 | 761.366 |
| 761.65 | 1685.44 | 1567.11 | 1361.46959 | 1262.63219 | 761.65 | 8267.09 | 8031.1 | 6854.32001 | 6668.40652 | 761.65 |
| 761.934 | 1696.49 | 1571.49 | 1349.90755 | 1266.20282 | 761.934 | 8344.94 | 8047.92 | 7016.10435 | 6682.58602 | 761.934 |
| 762.218 | 1694.14 | 1575.87 | 1382.0232 | 1269.78345 | 762.218 | 8283.84 | 8064.76 | 6817.15131 | 6696.77553 | 762.218 |
| 762.502 | 1712.97 | 1580.26 | 1379.1427 | 1273.36408 | 762.502 | 8308.24 | 8081.61 | 6840.69049 | 6710.98503 | 762.502 |
| 762.786 | 1757.94 | 1584.66 | 1392.38503 | 1276.94471 | 762.786 | 8350.78 | 8098.48 | 6924.30756 | 6725.20453 | 762.786 |
| 763.069 | 1697.83 | 1589.07 | 1373.61172 | 1280.54535 | 763.069 | 8434.54 | 8115.36 | 6997.295 | 6739.44403 | 763.069 |
| 763.353 | 1727.44 | 1593.48 | 1380.67297 | 1284.14598 | 763.353 | 8433.45 | 8132.27 | 6936.94712 | 6753.69353 | 763.353 |
| 763.637 | 1726.58 | 1597.91 | 1375.12199 | 1287.74661 | 763.637 | 8432.87 | 8149.19 | 6893.52864 | 6767.96303 | 763.637 |
| 763.921 | 1747.1 | 1602.34 | 1407.23764 | 1291.36725 | 763.921 | 8379.08 | 8166.13 | 7005.39472 | 6782.24253 | 763.921 |
| 764.204 | 1719.99 | 1606.78 | 1424.30064 | 1294.98789 | 764.204 | 8501.52 | 8183.08 | 6969.94596 | 6796.54203 | 764.204 |
| 764.488 | 1729.6 | 1611.22 | 1406.89758 | 1298.60853 | 764.488 | 8517.89 | 8200.05 | 7036.65363 | 6810.85153 | 764.488 |
| 764.772 | 1760.49 | 1615.68 | 1450.89532 | 1302.23916 | 764.772 | 8458.6 | 8217.04 | 7059.08284 | 6825.17103 | 764.772 |
| 765.056 | 1756.62 | 1620.14 | 1422.95041 | 1305.87981 | 765.056 | 8510.78 | 8234.04 | 7048.7932 | 6839.51053 | 765.056 |
| 765.339 | 1746.04 | 1624.61 | 1396.63578 | 1309.53045 | 765.339 | 8397.36 | 8251.07 | 7093.54163 | 6853.87003 | 765.339 |
| 765.623 | 1759.08 | 1629.08 | 1424.94076 | 1313.18109 | 765.623 | 8524.42 | 8268.1 | 7077.86218 | 6868.23952 | 765.623 |
| 765.907 | 1762.99 | 1633.57 | 1398.40609 | 1316.84173 | 765.907 | 8386.99 | 8285.16 | 7016.85432 | 6882.61902 | 765.907 |
| 766.19 | 1736.7 | 1638.06 | 1408.00778 | 1320.50238 | 766.19 | 8564.02 | 8302.23 | 7120.72068 | 6897.02852 | 766.19 |
| 766.474 | 1790.93 | 1642.56 | 1424.25064 | 1324.17302 | 766.474 | 8621.46 | 8319.32 | 7091.8117 | 6911.43801 | 766.474 |
| 766.757 | 1804.94 | 1647.07 | 1439.69335 | 1327.85367 | 766.757 | 8598.57 | 8336.42 | 7098.81145 | 6925.86751 | 766.757 |
| 767.041 | 1754.53 | 1651.58 | 1416.39925 | 1331.53432 | 767.041 | 8393.59 | 8353.54 | 7049.14319 | 6940.307 | 767.041 |
| 767.324 | 1807.07 | 1656.1 | 1469.10853 | 1335.22497 | 767.324 | 8509.4 | 8370.68 | 7112.99095 | 6954.76649 | 767.324 |
| 767.608 | 1840.14 | 1660.63 | 1415.16904 | 1338.92562 | 767.608 | 8582.64 | 8387.84 | 7142.64992 | 6969.23599 | 767.608 |
| 767.891 | 1820.61 | 1665.17 | 1421.50015 | 1342.62627 | 767.891 | 8623.95 | 8405.01 | 7123.60058 | 6983.72548 | 767.891 |
| 768.175 | 1795.48 | 1669.72 | 1455.59615 | 1346.33692 | 768.175 | 8649.74 | 8422.2 | 7195.91805 | 6998.22497 | 768.175 |
| 768.459 | 1767.92 | 1674.27 | 1470.20872 | 1350.05758 | 768.459 | 8636.03 | 8439.4 | 7169.30898 | 7012.74446 | 768.459 |
| 768.742 | 1793.08 | 1678.83 | 1478.0101 | 1353.77823 | 768.742 | 8587.5 | 8456.62 | 7302.28433 | 7027.27395 | 768.742 |
| 769.026 | 1804.81 | 1683.4 | 1465.17784 | 1357.50889 | 769.026 | 8630.39 | 8473.86 | 7149.40968 | 7041.82345 | 769.026 |
| 769.309 | 1741.79 | 1687.98 | 1456.21626 | 1361.23955 | 769.309 | 8560.63 | 8491.11 | 7189.82826 | 7056.38294 | 769.309 |
| 769.592 | 1823.94 | 1692.56 | 1462.32734 | 1364.98021 | 769.592 | 8696.81 | 8508.39 | 7224.14706 | 7070.95243 | 769.592 |
| 769.876 | 1821.06 | 1697.15 | 1500.33402 | 1368.73087 | 769.876 | 8655.22 | 8525.67 | 7202.22783 | 7085.54191 | 769.876 |
| 770.159 | 1859.35 | 1701.76 | 1514.02643 | 1372.48153 | 770.159 | 8796.48 | 8542.98 | 7346.46278 | 7100.1514 | 770.159 |
| 770.443 | 1786 | 1706.36 | 1492.32261 | 1376.25219 | 770.443 | 8774.29 | 8560.3 | 7368.622 | 7114.77089 | 770.443 |
| 770.726 | 1810.56 | 1710.98 | 1490.29226 | 1380.01285 | 770.726 | 8760.29 | 8577.63 | 7300.34439 | 7129.40038 | 770.726 |
| 771.01 | 1845.8 | 1715.6 | 1500.7541 | 1383.79352 | 771.01 | 8698.88 | 8594.99 | 7351.76259 | 7144.04987 | 771.01 |
| 771.293 | 1822.49 | 1720.23 | 1460.79707 | 1387.57418 | 771.293 | 8794.49 | 8612.36 | 7314.79389 | 7158.70935 | 771.293 |
| 771.576 | 1862.63 | 1724.87 | 1502.54441 | 1391.35485 | 771.576 | 8957.36 | 8629.74 | 7354.69249 | 7173.37884 | 771.576 |
| 771.86 | 1882.99 | 1729.52 | 1527.69884 | 1395.15552 | 771.86 | 8726.2 | 8647.14 | 7304.29426 | 7188.06833 | 771.86 |
| 772.143 | 1924.91 | 1734.17 | 1497.21348 | 1398.94618 | 772.143 | 8939.75 | 8664.56 | 7509.22708 | 7202.77781 | 772.143 |
| 772.426 | 1842.94 | 1738.83 | 1487.1017 | 1402.75685 | 772.426 | 8945.84 | 8682 | 7384.53145 | 7217.49729 | 772.426 |
| 772.709 | 1858.15 | 1743.5 | 1505.99502 | 1406.56752 | 772.709 | 8894.02 | 8699.45 | 7392.75116 | 7232.22678 | 772.709 |
| 772.993 | 1881.2 | 1748.18 | 1454.29592 | 1410.3882 | 772.993 | 8981.92 | 8716.92 | 7402.24083 | 7246.97626 | 772.993 |
| 773.276 | 1895.11 | 1752.86 | 1530.43932 | 1414.21887 | 773.276 | 8989.96 | 8734.4 | 7443.32939 | 7261.73575 | 773.276 |
| 773.559 | 1899.72 | 1757.56 | 1542.96153 | 1418.04954 | 773.559 | 8822.62 | 8751.9 | 7501.02737 | 7276.51523 | 773.559 |
| 773.842 | 1909.8 | 1762.26 | 1569.88626 | 1421.89022 | 773.842 | 8969.8 | 8769.42 | 7565.95509 | 7291.30471 | 773.842 |
| 774.126 | 1888.65 | 1766.96 | 1498.38368 | 1425.7309 | 774.126 | 9004.54 | 8786.95 | 7439.00954 | 7306.11419 | 774.126 |
| 774.409 | 1929.39 | 1771.68 | 1525.72849 | 1429.58157 | 774.409 | 9005.8 | 8804.5 | 7506.42718 | 7320.93367 | 774.409 |
| 774.692 | 1923.41 | 1776.4 | 1548.91257 | 1433.44225 | 774.692 | 8910.61 | 8822.06 | 7525.36652 | 7335.76315 | 774.692 |
| 774.975 | 1893.35 | 1781.13 | 1503.5846 | 1437.31293 | 774.975 | 9086.3 | 8839.64 | 7465.06863 | 7350.61263 | 774.975 |
| 775.258 | 1905.79 | 1785.87 | 1534.40002 | 1441.18362 | 775.258 | 9005.49 | 8857.24 | 7524.02656 | 7365.47211 | 775.258 |
| 775.542 | 1933.14 | 1790.62 | 1513.88641 | 1445.0543 | 775.542 | 9044.36 | 8874.85 | 7569.45497 | 7380.35159 | 775.542 |
| 775.825 | 1920.62 | 1795.37 | 1551.66306 | 1448.94498 | 775.825 | 9209.57 | 8892.48 | 7648.19221 | 7395.24107 | 775.825 |
| 776.108 | 1893.58 | 1800.13 | 1521.06767 | 1452.83567 | 776.108 | 9057.27 | 8910.12 | 7543.52588 | 7410.14055 | 776.108 |
| 776.391 | 1905.75 | 1804.9 | 1547.86239 | 1456.72635 | 776.391 | 9121.84 | 8927.79 | 7521.30666 | 7425.06003 | 776.391 |
| 776.674 | 1926.72 | 1809.68 | 1589.4097 | 1460.63704 | 776.674 | 9158.99 | 8945.46 | 7593.43413 | 7439.9995 | 776.674 |
| 776.957 | 1928.97 | 1814.47 | 1610.73345 | 1464.54773 | 776.957 | 9166.14 | 8963.16 | 7565.14512 | 7454.94898 | 776.957 |
| 777.24 | 1878.41 | 1819.26 | 1567.56586 | 1468.45842 | 777.24 | 9090.35 | 8980.87 | 7638.04257 | 7469.90846 | 777.24 |
| 777.523 | 1902.75 | 1824.06 | 1629.09668 | 1472.37911 | 777.523 | 9202.87 | 8998.59 | 7715.00988 | 7484.87793 | 777.523 |
| 777.806 | 2019.65 | 1828.87 | 1599.50148 | 1476.3098 | 777.806 | 9241.21 | 9016.33 | 7687.17085 | 7499.86741 | 777.806 |
| 778.089 | 1944.65 | 1833.68 | 1582.78853 | 1480.25049 | 778.089 | 9214.07 | 9034.09 | 7740.73897 | 7514.87688 | 778.089 |
| 778.372 | 1968.96 | 1838.51 | 1639.74856 | 1484.19118 | 778.372 | 9208.14 | 9051.86 | 7684.59094 | 7529.88636 | 778.372 |
| 778.655 | 1922.08 | 1843.34 | 1576.08736 | 1488.14188 | 778.655 | 9260.97 | 9069.65 | 7780.35759 | 7544.92583 | 778.655 |
| 778.938 | 1952.53 | 1848.18 | 1598.67133 | 1492.09257 | 778.938 | 9116.55 | 9087.46 | 7726.24948 | 7559.9653 | 778.938 |
| 779.221 | 1966.82 | 1853.03 | 1580.94821 | 1496.05327 | 779.221 | 9197.57 | 9105.28 | 7686.65087 | 7575.02478 | 779.221 |
| 779.504 | 1963.41 | 1857.88 | 1623.69573 | 1500.02397 | 779.504 | 9249.14 | 9123.11 | 7740.67898 | 7590.10425 | 779.504 |
| 779.787 | 1981.81 | 1862.74 | 1594.77064 | 1504.00467 | 779.787 | 9381.64 | 9140.97 | 7849.16518 | 7605.19372 | 779.787 |
| 780.07 | 1993.13 | 1867.61 | 1603.86224 | 1507.98537 | 780.07 | 9246.51 | 9158.83 | 7855.33496 | 7620.29319 | 780.07 |
| 780.353 | 1951.69 | 1872.49 | 1557.05401 | 1511.96607 | 780.353 | 9262.85 | 9176.72 | 7767.35804 | 7635.41266 | 780.353 |
| 780.635 | 1945.74 | 1877.38 | 1582.55849 | 1515.96678 | 780.635 | 9418.34 | 9194.62 | 7791.4772 | 7650.54213 | 780.635 |
| 780.918 | 2014.14 | 1882.27 | 1599.35145 | 1519.96748 | 780.918 | 9475.86 | 9212.53 | 7793.32713 | 7665.6816 | 780.918 |
| 781.201 | 2035.37 | 1887.17 | 1609.97332 | 1523.96818 | 781.201 | 9551.61 | 9230.47 | 7915.90284 | 7680.84107 | 781.201 |
| 781.484 | 1975.1 | 1892.08 | 1603.6222 | 1527.98889 | 781.484 | 9439.58 | 9248.41 | 7850.60513 | 7696.01054 | 781.484 |
| 781.767 | 2011.34 | 1897 | 1601.48182 | 1532.0096 | 781.767 | 9415.17 | 9266.37 | 7854.95498 | 7711.19001 | 781.767 |
| 782.05 | 2013.12 | 1901.92 | 1649.42026 | 1536.03031 | 782.05 | 9389.6 | 9284.35 | 7909.12308 | 7726.38948 | 782.05 |
| 782.332 | 2031.54 | 1906.86 | 1647.95 | 1540.07102 | 782.332 | 9525.73 | 9302.35 | 7892.79365 | 7741.60894 | 782.332 |
| 782.615 | 2062.04 | 1911.8 | 1640.68872 | 1544.11173 | 782.615 | 9354.24 | 9320.36 | 7946.60177 | 7756.82841 | 782.615 |
| 782.898 | 2049.12 | 1916.74 | 1647.39991 | 1548.15244 | 782.898 | 9438.45 | 9338.38 | 7805.88669 | 7772.07788 | 782.898 |
| 783.181 | 1971.59 | 1921.7 | 1693.66805 | 1552.20315 | 783.181 | 9508.07 | 9356.42 | 7799.56691 | 7787.32734 | 783.181 |
| 783.463 | 2033.27 | 1926.66 | 1626.84629 | 1556.26387 | 783.463 | 9442.13 | 9374.48 | 7824.68604 | 7802.59681 | 783.463 |
| 783.746 | 2020.21 | 1931.63 | 1682.026 | 1560.33458 | 783.746 | 9515.92 | 9392.55 | 7994.56009 | 7817.87627 | 783.746 |
| 784.029 | 2004.45 | 1936.61 | 1633.1374 | 1564.4053 | 784.029 | 9457.67 | 9410.64 | 7899.69341 | 7833.17574 | 784.029 |
| 784.311 | 2083.75 | 1941.6 | 1657.0516 | 1568.48602 | 784.311 | 9552.53 | 9428.74 | 7974.89078 | 7848.4852 | 784.311 |
| 784.594 | 2056.32 | 1946.59 | 1645.06949 | 1572.56674 | 784.594 | 9546.05 | 9446.86 | 7997.68998 | 7863.80467 | 784.594 |
| 784.877 | 2046.54 | 1951.59 | 1683.49626 | 1576.65746 | 784.877 | 9614.24 | 9464.99 | 7996.30003 | 7879.14413 | 784.877 |
| 785.159 | 2043.27 | 1956.6 | 1647.30989 | 1580.75818 | 785.159 | 9464.72 | 9483.14 | 7935.09217 | 7894.49359 | 785.159 |
| 785.442 | 2055.12 | 1961.62 | 1673.89457 | 1584.8689 | 785.442 | 9687.35 | 9501.31 | 8045.6783 | 7909.85305 | 785.442 |
| 785.725 | 2050.77 | 1966.64 | 1666.22322 | 1588.96962 | 785.725 | 9642.23 | 9519.49 | 8011.4195 | 7925.23251 | 785.725 |
| 786.007 | 2112.14 | 1971.68 | 1753.55859 | 1593.09035 | 786.007 | 9424.21 | 9537.68 | 7998.36995 | 7940.62198 | 786.007 |
| 786.29 | 2106.06 | 1976.72 | 1626.57624 | 1597.22107 | 786.29 | 9594.05 | 9555.89 | 7984.84043 | 7956.03144 | 786.29 |
| 786.572 | 2067.65 | 1981.77 | 1675.85491 | 1601.3518 | 786.572 | 9710.18 | 9574.12 | 8073.69732 | 7971.4509 | 786.572 |
| 786.855 | 2110.2 | 1986.82 | 1666.57328 | 1605.48253 | 786.855 | 9646.97 | 9592.36 | 8056.77791 | 7986.88036 | 786.855 |
| 787.137 | 2105.6 | 1991.89 | 1685.27657 | 1609.63326 | 787.137 | 9748.47 | 9610.61 | 8175.93374 | 8002.32982 | 787.137 |
| 787.42 | 2100.77 | 1996.96 | 1699.96916 | 1613.78399 | 787.42 | 9722.45 | 9628.89 | 8131.52529 | 8017.78927 | 787.42 |
| 787.702 | 2137.3 | 2002.04 | 1703.97986 | 1617.93472 | 787.702 | 9699.8 | 9647.17 | 8150.83462 | 8033.25873 | 787.702 |
| 787.985 | 2066.95 | 2007.12 | 1710.34098 | 1622.10545 | 787.985 | 9743.13 | 9665.48 | 7995.58005 | 8048.74819 | 787.985 |
| 788.267 | 2094.54 | 2012.22 | 1721.95302 | 1626.26619 | 788.267 | 9657.14 | 9683.8 | 8205.39271 | 8064.24765 | 788.267 |
| 788.55 | 2106.88 | 2017.32 | 1684.51644 | 1630.44692 | 788.55 | 9844.9 | 9702.13 | 8177.35369 | 8079.7571 | 788.55 |
| 788.832 | 2164.18 | 2022.43 | 1681.38589 | 1634.62766 | 788.832 | 9814.9 | 9720.48 | 8279.09013 | 8095.28656 | 788.832 |
| 789.115 | 2085.94 | 2027.55 | 1792.54545 | 1638.81839 | 789.115 | 9892.59 | 9738.84 | 8154.84448 | 8110.82602 | 789.115 |
| 789.397 | 2125.99 | 2032.68 | 1713.45153 | 1643.00913 | 789.397 | 9876.41 | 9757.22 | 8156.78441 | 8126.37547 | 789.397 |
| 789.679 | 2141.73 | 2037.81 | 1681.82596 | 1647.21987 | 789.679 | 9875.64 | 9775.61 | 8199.8129 | 8141.94493 | 789.679 |
| 789.962 | 2139.1 | 2042.95 | 1781.39348 | 1651.42061 | 789.962 | 9945.67 | 9794.02 | 8235.51165 | 8157.52438 | 789.962 |
| 790.244 | 2124.22 | 2048.1 | 1701.99951 | 1655.64136 | 790.244 | 9868.14 | 9812.44 | 8226.10198 | 8173.12384 | 790.244 |
| 790.526 | 2124.83 | 2053.26 | 1742.93672 | 1659.8621 | 790.526 | 9983 | 9830.88 | 8282.84999 | 8188.73329 | 790.526 |
| 790.809 | 2107.87 | 2058.42 | 1722.3731 | 1664.09284 | 790.809 | 9904.8 | 9849.33 | 8246.16128 | 8204.35274 | 790.809 |
| 791.091 | 2147.03 | 2063.59 | 1761.82004 | 1668.32359 | 791.091 | 9992.67 | 9867.8 | 8275.51025 | 8219.98219 | 791.091 |
| 791.373 | 2146.59 | 2068.77 | 1705.65016 | 1672.56433 | 791.373 | 9874.51 | 9886.29 | 8216.33232 | 8235.63165 | 791.373 |
| 791.655 | 2180.89 | 2073.96 | 1799.18662 | 1676.81508 | 791.655 | 9985.64 | 9904.79 | 8298.79944 | 8251.2911 | 791.655 |
| 791.938 | 2185.63 | 2079.16 | 1770.27153 | 1681.07583 | 791.938 | 10039.6 | 9923.3 | 8348.2677 | 8266.97055 | 791.938 |
| 792.22 | 2145.19 | 2084.36 | 1732.0648 | 1685.33658 | 792.22 | 10016.6 | 9941.83 | 8356.09743 | 8282.65 | 792.22 |
| 792.502 | 2140.43 | 2089.57 | 1744.92707 | 1689.59733 | 792.502 | 10036.1 | 9960.37 | 8245.5713 | 8298.34945 | 792.502 |
| 792.784 | 2145.52 | 2094.79 | 1797.51632 | 1693.87808 | 792.784 | 9977.32 | 9978.93 | 8285.20991 | 8314.0689 | 792.784 |
| 793.067 | 2253.49 | 2100.01 | 1752.12833 | 1698.15884 | 793.067 | 10060.8 | 9997.5 | 8435.95463 | 8329.79835 | 793.067 |
| 793.349 | 2217.78 | 2105.25 | 1732.39486 | 1702.43959 | 793.349 | 10012.6 | 10016.1 | 8408.94558 | 8345.5378 | 793.349 |
| 793.631 | 2194.97 | 2110.49 | 1798.59651 | 1706.73035 | 793.631 | 10139.8 | 10034.7 | 8492.38266 | 8361.28725 | 793.631 |
| 793.913 | 2254.23 | 2115.74 | 1809.62845 | 1711.0311 | 793.913 | 10124.6 | 10053.3 | 8464.62363 | 8377.0567 | 793.913 |
| 794.195 | 2220.29 | 2121 | 1810.72865 | 1715.34186 | 794.195 | 10116.7 | 10071.9 | 8381.81653 | 8392.83614 | 794.195 |
| 794.477 | 2221.37 | 2126.26 | 1857.39686 | 1719.65262 | 794.477 | 10228.9 | 10090.6 | 8437.28459 | 8408.62559 | 794.477 |
| 794.76 | 2280.33 | 2131.53 | 1799.24663 | 1723.97338 | 794.76 | 10120.4 | 10109.2 | 8453.904 | 8424.43504 | 794.76 |
| 795.042 | 2227.75 | 2136.82 | 1847.81517 | 1728.30414 | 795.042 | 10143.6 | 10127.9 | 8484.50293 | 8440.25448 | 795.042 |
| 795.324 | 2220.26 | 2142.1 | 1765.5407 | 1732.6349 | 795.324 | 10151.6 | 10146.6 | 8515.66184 | 8456.08393 | 795.324 |
| 795.606 | 2282.91 | 2147.4 | 1835.89308 | 1736.97567 | 795.606 | 10097.3 | 10165.3 | 8516.48181 | 8471.93337 | 795.606 |
| 795.888 | 2216.63 | 2152.7 | 1862.11769 | 1741.31643 | 795.888 | 10350.4 | 10184 | 8619.51821 | 8487.78282 | 795.888 |
| 796.17 | 2190.99 | 2158.01 | 1822.10065 | 1745.6672 | 796.17 | 10244.2 | 10202.8 | 8493.38262 | 8503.66226 | 796.17 |
| 796.452 | 2230.61 | 2163.33 | 1794.23574 | 1750.02797 | 796.452 | 10236 | 10221.5 | 8578.95963 | 8519.54171 | 796.452 |
| 796.734 | 2256.68 | 2168.66 | 1804.42754 | 1754.38873 | 796.734 | 10197 | 10240.3 | 8558.53034 | 8535.44115 | 796.734 |
| 797.016 | 2226.72 | 2173.99 | 1860.37738 | 1758.7595 | 797.016 | 10343.3 | 10259.1 | 8630.04784 | 8551.35059 | 797.016 |
| 797.298 | 2308.65 | 2179.34 | 1861.97767 | 1763.14027 | 797.298 | 10289.2 | 10277.9 | 8642.73739 | 8567.27004 | 797.298 |
| 797.58 | 2283.2 | 2184.69 | 1854.07628 | 1767.53105 | 797.58 | 10351.7 | 10296.7 | 8608.99857 | 8583.20948 | 797.58 |
| 797.862 | 2290.66 | 2190.05 | 1847.17506 | 1771.92182 | 797.862 | 10519 | 10315.5 | 8687.12584 | 8599.15892 | 797.862 |
| 798.144 | 2245.68 | 2195.41 | 1881.02102 | 1776.31259 | 798.144 | 10241.3 | 10334.3 | 8612.69844 | 8615.11836 | 798.144 |
| 798.426 | 2295.61 | 2200.78 | 1852.54601 | 1780.72337 | 798.426 | 10197.1 | 10353.2 | 8689.70575 | 8631.0978 | 798.426 |
| 798.707 | 2257.44 | 2206.16 | 1860.97749 | 1785.13414 | 798.707 | 10333.1 | 10372 | 8697.93546 | 8647.08724 | 798.707 |
| 798.989 | 2291.28 | 2211.55 | 1923.38847 | 1789.54492 | 798.989 | 10387.6 | 10390.9 | 8639.06752 | 8663.08668 | 798.989 |
| 799.271 | 2306.62 | 2216.95 | 1858.53706 | 1793.9757 | 799.271 | 10430.3 | 10409.8 | 8746.58376 | 8679.09612 | 799.271 |
| 799.553 | 2321.76 | 2222.35 | 1886.86204 | 1798.40648 | 799.553 | 10441 | 10428.7 | 8737.91406 | 8695.12556 | 799.553 |
| 799.835 | 2327.32 | 2227.76 | 1905.45532 | 1802.83726 | 799.835 | 10413.2 | 10447.7 | 8828.61088 | 8711.165 | 799.835 |
| 800.117 | 2325.01 | 2233.18 | 1909.96611 | 1807.27804 | 800.117 | 10507.1 | 10466.6 | 8704.48523 | 8727.21444 | 800.117 |
| 800.399 | 2354.12 | 2238.61 | 1872.21947 | 1811.72882 | 800.399 | 10362 | 10485.5 | 8732.48425 | 8743.28387 | 800.399 |
| 800.68 | 2324.31 | 2244.05 | 1852.21595 | 1816.18961 | 800.68 | 10556.3 | 10504.5 | 8701.61533 | 8759.36331 | 800.68 |
| 800.962 | 2299.53 | 2249.49 | 1895.23352 | 1820.65039 | 800.962 | 10591 | 10523.5 | 8848.1102 | 8775.45275 | 800.962 |
| 801.244 | 2352.18 | 2254.94 | 1899.0842 | 1825.12118 | 801.244 | 10641.8 | 10542.5 | 8794.84207 | 8791.55218 | 801.244 |
| 801.525 | 2346 | 2260.4 | 1919.5378 | 1829.59197 | 801.525 | 10516.7 | 10561.5 | 8820.54117 | 8807.67162 | 801.525 |
| 801.807 | 2350.08 | 2265.86 | 1918.05753 | 1834.07276 | 801.807 | 10728 | 10580.5 | 8900.43837 | 8823.79105 | 801.807 |
| 802.089 | 2328.22 | 2271.34 | 1922.10825 | 1838.56355 | 802.089 | 10678 | 10599.6 | 8968.68598 | 8839.94049 | 802.089 |
| 802.371 | 2331.79 | 2276.82 | 1884.8917 | 1843.06434 | 802.371 | 10699 | 10618.6 | 8882.77899 | 8856.08992 | 802.371 |
| 802.652 | 2329.87 | 2282.31 | 1962.21531 | 1847.56513 | 802.652 | 10521 | 10637.7 | 8947.15673 | 8872.25936 | 802.652 |
| 802.934 | 2434.83 | 2287.8 | 1982.43886 | 1852.06592 | 802.934 | 10851.6 | 10656.8 | 8899.14842 | 8888.43879 | 802.934 |
| 803.216 | 2402.37 | 2293.31 | 1921.72818 | 1856.58672 | 803.216 | 10665.4 | 10675.9 | 9043.37337 | 8904.62822 | 803.216 |
| 803.497 | 2384.15 | 2298.82 | 1956.40428 | 1861.10751 | 803.497 | 10740.6 | 10695 | 8960.88625 | 8920.82766 | 803.497 |
| 803.779 | 2392.19 | 2304.34 | 1918.61763 | 1865.63831 | 803.779 | 10768.8 | 10714.1 | 9031.29379 | 8937.04709 | 803.779 |
| 804.06 | 2425.46 | 2309.87 | 1921.75819 | 1870.16911 | 804.06 | 10802 | 10733.2 | 9047.36323 | 8953.27652 | 804.06 |
| 804.342 | 2357.61 | 2315.4 | 1912.90663 | 1874.70991 | 804.342 | 10867.8 | 10752.4 | 9057.03289 | 8969.51595 | 804.342 |
| 804.624 | 2378.42 | 2320.94 | 1973.07722 | 1879.26071 | 804.624 | 10776.3 | 10771.5 | 9060.47277 | 8985.77538 | 804.624 |
| 804.905 | 2415.98 | 2326.49 | 1965.5859 | 1883.81151 | 804.905 | 10951.1 | 10790.7 | 9191.31819 | 9002.04481 | 804.905 |
| 805.187 | 2443.05 | 2332.05 | 2038.3187 | 1888.37231 | 805.187 | 10964.7 | 10809.9 | 9141.26994 | 9018.32424 | 805.187 |
| 805.468 | 2374.54 | 2337.62 | 1973.78734 | 1892.93311 | 805.468 | 10922.7 | 10829.1 | 9132.66024 | 9034.61367 | 805.468 |
| 805.75 | 2452.58 | 2343.19 | 1997.82157 | 1897.51392 | 805.75 | 10872 | 10848.3 | 9066.63255 | 9050.9131 | 805.75 |
| 806.031 | 2386.73 | 2348.77 | 1993.92088 | 1902.09473 | 806.031 | 10866.8 | 10867.6 | 9165.97907 | 9067.23253 | 806.031 |
| 806.313 | 2414.9 | 2354.36 | 1961.01509 | 1906.67553 | 806.313 | 10927 | 10886.8 | 9207.07763 | 9083.56196 | 806.313 |
| 806.594 | 2452.07 | 2359.96 | 2027.14673 | 1911.26634 | 806.594 | 10926 | 10906.1 | 9178.74863 | 9099.90139 | 806.594 |
| 806.875 | 2478.94 | 2365.56 | 1981.38868 | 1915.86715 | 806.875 | 10931.4 | 10925.4 | 9199.25791 | 9116.26081 | 806.875 |
| 807.157 | 2458.46 | 2371.17 | 1984.81928 | 1920.46796 | 807.157 | 10927.3 | 10944.6 | 9216.03732 | 9132.62024 | 807.157 |
| 807.438 | 2469.13 | 2376.79 | 2043.58963 | 1925.08877 | 807.438 | 11084.3 | 10963.9 | 9203.79775 | 9148.99967 | 807.438 |
| 807.72 | 2487.47 | 2382.42 | 1967.77628 | 1929.69958 | 807.72 | 11064.7 | 10983.2 | 9113.8509 | 9165.38909 | 807.72 |
| 808.001 | 2438.11 | 2388.05 | 2013.17427 | 1934.3304 | 808.001 | 11184.1 | 11002.6 | 9141.73992 | 9181.78852 | 808.001 |
| 808.283 | 2481.66 | 2393.7 | 1953.03369 | 1938.96121 | 808.283 | 11055.6 | 11021.9 | 9238.12655 | 9198.20794 | 808.283 |
| 808.564 | 2481.21 | 2399.35 | 2072.95479 | 1943.60203 | 808.564 | 11215.8 | 11041.3 | 9258.12585 | 9214.63737 | 808.564 |
| 808.845 | 2500.78 | 2405.01 | 2019.18533 | 1948.24285 | 808.845 | 11182.4 | 11060.6 | 9278.15515 | 9231.07679 | 808.845 |
| 809.126 | 2500.46 | 2410.67 | 2074.125 | 1952.89366 | 809.126 | 11009.2 | 11080 | 9240.58646 | 9247.52622 | 809.126 |
| 809.408 | 2446.59 | 2416.34 | 2023.35606 | 1957.55449 | 809.408 | 11169.5 | 11099.4 | 9235.62663 | 9263.99564 | 809.408 |
| 809.689 | 2490.91 | 2422.03 | 2050.79089 | 1962.21531 | 809.689 | 10997.1 | 11118.9 | 9257.68586 | 9280.46506 | 809.689 |
| 809.97 | 2517.27 | 2427.71 | 2031.64752 | 1966.88613 | 809.97 | 11089.1 | 11138.3 | 9250.41612 | 9296.95449 | 809.97 |
| 810.252 | 2542.92 | 2433.41 | 2035.70824 | 1971.55695 | 810.252 | 11184.6 | 11157.7 | 9306.92414 | 9313.45391 | 810.252 |
| 810.533 | 2504.74 | 2439.11 | 1991.1704 | 1976.24777 | 810.533 | 11140.3 | 11177.2 | 9288.21479 | 9329.97333 | 810.533 |
| 810.814 | 2508.18 | 2444.82 | 2055.04164 | 1980.9386 | 810.814 | 11177.9 | 11196.6 | 9397.95095 | 9346.49275 | 810.814 |
| 811.095 | 2579.14 | 2450.55 | 2056.62192 | 1985.62943 | 811.095 | 11196.3 | 11216.1 | 9279.21511 | 9363.03217 | 811.095 |
| 811.376 | 2480.13 | 2456.27 | 2058.72229 | 1990.33025 | 811.376 | 11194.6 | 11235.6 | 9260.95575 | 9379.58159 | 811.376 |
| 811.658 | 2537.54 | 2462.01 | 2046.43013 | 1995.04108 | 811.658 | 11295.6 | 11255.1 | 9362.4122 | 9396.14101 | 811.658 |
| 811.939 | 2523.16 | 2467.75 | 2060.12253 | 1999.76191 | 811.939 | 11291 | 11274.6 | 9288.84477 | 9412.72043 | 811.939 |
| 812.22 | 2502.59 | 2473.5 | 2088.76758 | 2004.48274 | 812.22 | 11351.2 | 11294.2 | 9401.43083 | 9429.29985 | 812.22 |
| 812.501 | 2532.73 | 2479.26 | 2050.8209 | 2009.21358 | 812.501 | 11315.5 | 11313.7 | 9333.50321 | 9445.89927 | 812.501 |
| 812.782 | 2563.4 | 2485.02 | 2091.91813 | 2013.94441 | 812.782 | 11309.1 | 11333.3 | 9427.21993 | 9462.50869 | 812.782 |
| 813.063 | 2545.41 | 2490.79 | 2116.23241 | 2018.68524 | 813.063 | 11368.7 | 11352.8 | 9540.10597 | 9479.12811 | 813.063 |
| 813.344 | 2557.91 | 2496.57 | 2081.16624 | 2023.43608 | 813.344 | 11351.6 | 11372.4 | 9554.75546 | 9495.75753 | 813.344 |
| 813.625 | 2555.14 | 2502.36 | 2094.43857 | 2028.18691 | 813.625 | 11248.9 | 11392 | 9520.33667 | 9512.40694 | 813.625 |
| 813.907 | 2618.74 | 2508.16 | 2080.28608 | 2032.94775 | 813.907 | 11363.7 | 11411.6 | 9425.49999 | 9529.06636 | 813.907 |
| 814.187 | 2632.9 | 2513.96 | 2096.53894 | 2037.71859 | 814.187 | 11352.8 | 11431.3 | 9477.52816 | 9545.73578 | 814.187 |
| 814.468 | 2591.96 | 2519.77 | 2156.93957 | 2042.48943 | 814.468 | 11404.6 | 11450.9 | 9589.62424 | 9562.41519 | 814.468 |
| 814.75 | 2615.88 | 2525.59 | 2131.59511 | 2047.27027 | 814.75 | 11368.2 | 11470.5 | 9624.38302 | 9579.10461 | 814.75 |
| 815.03 | 2603.71 | 2531.41 | 2083.51665 | 2052.06112 | 815.03 | 11458.5 | 11490.2 | 9603.39376 | 9595.80402 | 815.03 |
| 815.312 | 2610.8 | 2537.25 | 2098.13922 | 2056.85196 | 815.312 | 11448.8 | 11509.9 | 9629.55284 | 9612.52344 | 815.312 |
| 815.592 | 2613.34 | 2543.09 | 2081.20625 | 2061.6528 | 815.592 | 11438.9 | 11529.6 | 9700.85035 | 9629.25285 | 815.592 |
| 815.873 | 2632.74 | 2548.94 | 2124.34384 | 2066.45365 | 815.873 | 11464 | 11549.3 | 9563.13517 | 9645.99227 | 815.873 |
| 816.154 | 2570.77 | 2554.8 | 2179.8136 | 2071.2645 | 816.154 | 11564.3 | 11569 | 9610.49351 | 9662.74168 | 816.154 |
| 816.435 | 2542.91 | 2560.66 | 2155.87939 | 2076.08534 | 816.435 | 11464.1 | 11588.7 | 9779.90758 | 9679.50109 | 816.435 |
| 816.716 | 2646.44 | 2566.53 | 2173.30245 | 2080.91619 | 816.716 | 11516.2 | 11608.5 | 9627.66291 | 9696.28051 | 816.716 |
| 816.997 | 2708.48 | 2572.41 | 2137.58617 | 2085.74704 | 816.997 | 11666.9 | 11628.2 | 9754.02848 | 9713.06992 | 816.997 |
| 817.278 | 2604.79 | 2578.3 | 2182.91414 | 2090.5879 | 817.278 | 11630.9 | 11648 | 9732.79923 | 9729.86933 | 817.278 |
| 817.559 | 2676.04 | 2584.19 | 2163.12066 | 2095.42875 | 817.559 | 11709.9 | 11667.8 | 9781.55752 | 9746.67874 | 817.559 |
| 817.84 | 2616.32 | 2590.09 | 2163.04065 | 2100.2796 | 817.84 | 11706.5 | 11687.6 | 9750.52861 | 9763.49815 | 817.84 |
| 818.12 | 2676.14 | 2596.01 | 2118.35278 | 2105.14046 | 818.12 | 11705.1 | 11707.4 | 9756.4584 | 9780.32756 | 818.12 |
| 818.401 | 2680.98 | 2601.92 | 2132.28523 | 2110.00131 | 818.401 | 11667.3 | 11727.2 | 9840.51546 | 9797.17697 | 818.401 |
| 818.682 | 2686.89 | 2607.85 | 2203.1277 | 2114.87217 | 818.682 | 11746.8 | 11747 | 9725.71947 | 9814.03638 | 818.682 |
| 818.963 | 2654.26 | 2613.78 | 2193.56602 | 2119.75303 | 818.963 | 11711.3 | 11766.9 | 9851.99505 | 9830.89579 | 818.963 |
| 819.243 | 2684.86 | 2619.72 | 2180.99381 | 2124.63389 | 819.243 | 11825.4 | 11786.8 | 9917.59276 | 9847.7752 | 819.243 |
| 819.524 | 2752.99 | 2625.67 | 2190.84554 | 2129.52475 | 819.524 | 11669 | 11806.6 | 9806.69664 | 9864.67461 | 819.524 |
| 819.805 | 2706.43 | 2631.62 | 2193.51601 | 2134.41561 | 819.805 | 11822.5 | 11826.5 | 9875.87422 | 9881.57402 | 819.805 |
| 820.086 | 2692.08 | 2637.59 | 2207.05839 | 2139.32647 | 820.086 | 11567.7 | 11846.4 | 9890.31371 | 9898.48343 | 820.086 |
| 820.366 | 2696.73 | 2643.56 | 2234.78327 | 2144.22734 | 820.366 | 11771.4 | 11866.3 | 9862.1347 | 9915.41283 | 820.366 |
| 820.647 | 2729.84 | 2649.53 | 2154.69918 | 2149.1482 | 820.647 | 11918.7 | 11886.2 | 9784.59741 | 9932.35224 | 820.647 |
| 820.928 | 2719.71 | 2655.52 | 2209.83888 | 2154.06907 | 820.928 | 11804.1 | 11906.2 | 9902.92327 | 9949.30165 | 820.928 |
| 821.208 | 2672.01 | 2661.51 | 2240.0342 | 2158.99994 | 821.208 | 11892.7 | 11926.1 | 9972.74083 | 9966.26105 | 821.208 |
| 821.489 | 2742.2 | 2667.51 | 2208.31861 | 2163.9308 | 821.489 | 11900.3 | 11946.1 | 9950.5216 | 9983.23046 | 821.489 |
| 821.769 | 2739.45 | 2673.52 | 2268.95929 | 2168.87167 | 821.769 | 12041.3 | 11966.1 | 9964.04113 | 10000.24986 | 821.769 |
| 822.05 | 2760.35 | 2679.54 | 2207.50847 | 2173.82254 | 822.05 | 11962.7 | 11986 | 10004.9497 | 10017.24927 | 822.05 |
| 822.331 | 2777.52 | 2685.56 | 2204.95802 | 2178.77341 | 822.331 | 12074.3 | 12006 | 10126.84543 | 10034.24867 | 822.331 |
| 822.611 | 2730.98 | 2691.59 | 2245.03508 | 2183.73429 | 822.611 | 12027.8 | 12026.1 | 10017.74925 | 10051.24808 | 822.611 |
| 822.892 | 2742.54 | 2697.63 | 2266.7589 | 2188.70516 | 822.892 | 12096.2 | 12046.1 | 10026.84893 | 10068.24748 | 822.892 |
| 823.172 | 2848.81 | 2703.68 | 2238.50393 | 2193.67604 | 823.172 | 12211 | 12066.1 | 10124.14552 | 10085.24689 | 823.172 |
| 823.453 | 2805.81 | 2709.73 | 2277.83085 | 2198.65691 | 823.453 | 12111.7 | 12086.2 | 10022.84907 | 10102.34629 | 823.453 |
| 823.733 | 2786.16 | 2715.79 | 2271.03965 | 2203.63779 | 823.733 | 12093.5 | 12106.2 | 10184.04343 | 10119.44569 | 823.733 |
| 824.014 | 2823.18 | 2721.86 | 2234.50322 | 2208.62867 | 824.014 | 12126.5 | 12126.3 | 10091.74666 | 10136.44509 | 824.014 |
| 824.294 | 2857.26 | 2727.94 | 2309.02634 | 2213.62955 | 824.294 | 12243.5 | 12146.4 | 10091.64666 | 10153.5445 | 824.294 |
| 824.574 | 2824.39 | 2734.02 | 2298.49448 | 2218.64043 | 824.574 | 12133.5 | 12166.5 | 10254.44096 | 10170.6439 | 824.574 |
| 824.855 | 2857.31 | 2740.11 | 2298.97457 | 2223.65131 | 824.855 | 12254.2 | 12186.6 | 10128.24538 | 10187.7433 | 824.855 |
| 825.135 | 2802.44 | 2746.21 | 2292.42341 | 2228.6722 | 825.135 | 12214.6 | 12206.7 | 10164.04413 | 10204.8427 | 825.135 |
| 825.416 | 2794.74 | 2752.32 | 2325.61926 | 2233.69308 | 825.416 | 12244.2 | 12226.9 | 10263.54064 | 10222.0421 | 825.416 |
| 825.696 | 2826.2 | 2758.43 | 2299.29462 | 2238.72396 | 825.696 | 12348.5 | 12247 | 10349.73763 | 10239.1415 | 825.696 |
| 825.976 | 2846.65 | 2764.55 | 2313.84718 | 2243.75485 | 825.976 | 12230.9 | 12267.2 | 10234.44166 | 10256.3409 | 825.976 |
| 826.257 | 2857.56 | 2770.68 | 2346.70297 | 2248.80574 | 826.257 | 12243.5 | 12287.3 | 10316.53879 | 10273.4403 | 826.257 |
| 826.537 | 2856.24 | 2776.82 | 2304.99563 | 2253.85663 | 826.537 | 12235.9 | 12307.5 | 10271.84035 | 10290.6397 | 826.537 |
| 826.817 | 2811.51 | 2782.97 | 2343.27236 | 2258.90752 | 826.817 | 12266.3 | 12327.7 | 10336.03811 | 10307.83909 | 826.817 |
| 827.098 | 2887.02 | 2789.12 | 2316.01757 | 2263.96841 | 827.098 | 12468.5 | 12347.9 | 10301.9393 | 10325.03849 | 827.098 |
| 827.378 | 2848.66 | 2795.28 | 2334.35079 | 2269.0393 | 827.378 | 12220.9 | 12368.1 | 10272.64033 | 10342.23789 | 827.378 |
| 827.658 | 2867.58 | 2801.44 | 2345.10268 | 2274.12019 | 827.658 | 12404.8 | 12388.4 | 10381.9365 | 10359.43729 | 827.658 |
| 827.939 | 2901.28 | 2807.62 | 2341.13199 | 2279.20109 | 827.939 | 12420.1 | 12408.6 | 10240.74144 | 10376.63668 | 827.939 |
| 828.219 | 2825.82 | 2813.8 | 2366.91652 | 2284.29198 | 828.219 | 12447.8 | 12428.9 | 10272.64033 | 10393.93608 | 828.219 |
| 828.499 | 2898.46 | 2819.99 | 2364.31607 | 2289.38288 | 828.499 | 12436 | 12449.1 | 10337.33806 | 10411.13548 | 828.499 |
| 828.779 | 2850.41 | 2826.18 | 2401.80266 | 2294.48378 | 828.779 | 12440.6 | 12469.4 | 10371.83685 | 10428.43487 | 828.779 |
| 829.059 | 2943.9 | 2832.39 | 2351.64384 | 2299.59468 | 829.059 | 12418.6 | 12489.7 | 10435.23463 | 10445.63427 | 829.059 |
| 829.34 | 2927.95 | 2838.6 | 2371.8774 | 2304.70558 | 829.34 | 12454.5 | 12510 | 10477.83314 | 10462.93366 | 829.34 |
| 829.62 | 2914.23 | 2844.82 | 2321.3285 | 2309.82648 | 829.62 | 12524.1 | 12530.3 | 10365.23708 | 10480.23306 | 829.62 |
| 829.9 | 2910.44 | 2851.05 | 2368.4968 | 2314.94738 | 829.9 | 12490.9 | 12550.7 | 10470.3334 | 10497.53245 | 829.9 |
| 830.18 | 2983.85 | 2857.28 | 2328.1397 | 2320.08828 | 830.18 | 12653.5 | 12571 | 10395.13604 | 10514.83185 | 830.18 |
| 830.46 | 2878.93 | 2863.52 | 2374.22781 | 2325.22919 | 830.46 | 12557.9 | 12591.4 | 10529.53133 | 10532.23124 | 830.46 |
| 830.74 | 3012.08 | 2869.77 | 2426.83707 | 2330.37009 | 830.74 | 12633.7 | 12611.7 | 10501.7323 | 10549.53063 | 830.74 |
| 831.02 | 2980.35 | 2876.03 | 2407.47366 | 2335.521 | 831.02 | 12512.7 | 12632.1 | 10505.63217 | 10566.93002 | 831.02 |
| 831.3 | 2960.72 | 2882.29 | 2373.76773 | 2340.68191 | 831.3 | 12644.7 | 12652.5 | 10556.2304 | 10584.22942 | 831.3 |
| 831.58 | 2971.78 | 2888.56 | 2387.68018 | 2345.84282 | 831.58 | 12715.9 | 12672.9 | 10610.02851 | 10601.62881 | 831.58 |
| 831.86 | 2974.13 | 2894.84 | 2381.33906 | 2351.01373 | 831.86 | 12626.5 | 12693.3 | 10420.23516 | 10619.0282 | 831.86 |
| 832.14 | 2988.33 | 2901.13 | 2421.04605 | 2356.19464 | 832.14 | 12666.2 | 12713.7 | 10579.22959 | 10636.42759 | 832.14 |
| 832.42 | 3027.57 | 2907.42 | 2391.32082 | 2361.37555 | 832.42 | 12743.6 | 12734.2 | 10620.32815 | 10653.82698 | 832.42 |
| 832.7 | 2944.69 | 2913.72 | 2443.36998 | 2366.56646 | 832.7 | 12741.7 | 12754.6 | 10613.2284 | 10671.22637 | 832.7 |
| 832.98 | 3005.61 | 2920.03 | 2371.3173 | 2371.75738 | 832.98 | 12908.8 | 12775.1 | 10682.52597 | 10688.62576 | 832.98 |
| 833.26 | 3039.56 | 2926.35 | 2397.19185 | 2376.96829 | 833.26 | 12695.9 | 12795.5 | 10626.52793 | 10706.02515 | 833.26 |
| 833.54 | 3027.09 | 2932.67 | 2461.84323 | 2382.17921 | 833.54 | 12803.8 | 12816 | 10713.52489 | 10723.52454 | 833.54 |
| 833.82 | 3018.23 | 2939 | 2404.19308 | 2387.39013 | 833.82 | 12884.6 | 12836.5 | 10671.72635 | 10740.92393 | 833.82 |
| 834.1 | 2993.29 | 2945.34 | 2473.6353 | 2392.61105 | 834.1 | 12811.6 | 12857 | 10662.62667 | 10758.42332 | 834.1 |
| 834.38 | 3001.95 | 2951.68 | 2438.00903 | 2397.84197 | 834.38 | 12927.2 | 12877.5 | 10785.82236 | 10775.9227 | 834.38 |
| 834.66 | 3029.61 | 2958.04 | 2431.97797 | 2403.07289 | 834.66 | 12850.4 | 12898 | 10710.325 | 10793.42209 | 834.66 |
| 834.939 | 3059.96 | 2964.4 | 2444.59019 | 2408.31381 | 834.939 | 12785.7 | 12918.6 | 10648.62716 | 10810.82148 | 834.939 |
| 835.219 | 3006.5 | 2970.77 | 2435.16853 | 2413.56473 | 835.219 | 12886.2 | 12939.1 | 10689.72572 | 10828.42087 | 835.219 |
| 835.499 | 2932.55 | 2977.14 | 2474.06538 | 2418.81566 | 835.499 | 12878.7 | 12959.7 | 10731.22427 | 10845.92025 | 835.499 |
| 835.779 | 2977.97 | 2983.52 | 2462.97343 | 2424.07658 | 835.779 | 12966.7 | 12980.3 | 10868.01948 | 10863.41964 | 835.779 |
| 836.059 | 3086.47 | 2989.91 | 2537.31651 | 2429.33751 | 836.059 | 12836.3 | 13000.9 | 10789.62222 | 10880.91903 | 836.059 |
| 836.338 | 3095.87 | 2996.31 | 2498.40966 | 2434.60844 | 836.338 | 12928.9 | 13021.5 | 10861.01972 | 10898.51841 | 836.338 |
| 836.618 | 3074.61 | 3002.72 | 2458.48264 | 2439.88937 | 836.618 | 12954.7 | 13042.1 | 10819.32118 | 10916.0178 | 836.618 |
| 836.898 | 3122.18 | 3009.13 | 2505.96099 | 2445.1803 | 836.898 | 12974.6 | 13062.7 | 10916.81777 | 10933.61718 | 836.898 |
| 837.178 | 3079.85 | 3015.55 | 2452.32155 | 2450.47123 | 837.178 | 13105.9 | 13083.3 | 10897.51845 | 10951.21657 | 837.178 |
| 837.457 | 3049.45 | 3021.98 | 2475.3456 | 2455.76216 | 837.457 | 13074.2 | 13104 | 10868.91945 | 10968.81595 | 837.457 |
| 837.737 | 3091.93 | 3028.41 | 2477.40597 | 2461.07309 | 837.737 | 13195.4 | 13124.6 | 10873.0193 | 10986.41533 | 837.737 |
| 838.017 | 3086.2 | 3034.85 | 2511.65199 | 2466.37403 | 838.017 | 12926 | 13145.3 | 10863.61963 | 11004.01472 | 838.017 |
| 838.296 | 3105.99 | 3041.3 | 2565.54148 | 2471.69496 | 838.296 | 13036.9 | 13165.9 | 11059.91276 | 11021.6141 | 838.296 |
| 838.576 | 3156.81 | 3047.76 | 2518.40318 | 2477.0159 | 838.576 | 13166.7 | 13186.6 | 10976.71567 | 11039.21349 | 838.576 |
| 838.856 | 3142.25 | 3054.22 | 2499.89993 | 2482.34684 | 838.856 | 13149.7 | 13207.3 | 10862.11969 | 11056.91287 | 838.856 |
| 839.135 | 3102.51 | 3060.7 | 2514.5225 | 2487.67777 | 839.135 | 13081.1 | 13228 | 11086.51183 | 11074.51225 | 839.135 |
| 839.415 | 3108.99 | 3067.17 | 2483.54705 | 2493.01871 | 839.415 | 13067.7 | 13248.7 | 11019.71417 | 11092.21163 | 839.415 |
| 839.694 | 3137.9 | 3073.66 | 2533.2858 | 2498.36966 | 839.694 | 13125.1 | 13269.5 | 11193.0081 | 11109.91101 | 839.694 |
| 839.974 | 3111.41 | 3080.15 | 2607.61888 | 2503.7206 | 839.974 | 13181.5 | 13290.2 | 11002.01479 | 11127.61039 | 839.974 |
| 840.254 | 3111.98 | 3086.66 | 2558.77029 | 2509.08154 | 840.254 | 13232.5 | 13311 | 11195.40802 | 11145.20977 | 840.254 |
| 840.533 | 3152.49 | 3093.16 | 2616.86051 | 2514.45249 | 840.533 | 13259.3 | 13331.7 | 11125.21047 | 11162.90915 | 840.533 |
| 840.812 | 3131.91 | 3099.68 | 2543.70763 | 2519.82343 | 840.812 | 13302.5 | 13352.5 | 11046.21324 | 11180.60853 | 840.812 |
| 841.092 | 3124.82 | 3106.2 | 2622.44149 | 2525.20438 | 841.092 | 13382.6 | 13373.3 | 11054.81294 | 11198.40791 | 841.092 |
| 841.371 | 3213.76 | 3112.73 | 2542.09735 | 2530.58533 | 841.371 | 13225.5 | 13394.1 | 11048.81315 | 11216.10729 | 841.371 |
| 841.651 | 3146.6 | 3119.27 | 2596.47692 | 2535.97627 | 841.651 | 13438.2 | 13414.9 | 11147.3097 | 11233.80667 | 841.651 |
| 841.93 | 3197.44 | 3125.81 | 2570.69238 | 2541.37722 | 841.93 | 13406.1 | 13435.7 | 11265.80555 | 11251.60605 | 841.93 |
| 842.21 | 3180.74 | 3132.37 | 2571.96261 | 2546.77818 | 842.21 | 13328.7 | 13456.5 | 11075.71221 | 11269.40543 | 842.21 |
| 842.489 | 3205.31 | 3138.93 | 2613.47991 | 2552.18913 | 842.489 | 13445.4 | 13477.4 | 11271.00537 | 11287.10481 | 842.489 |
| 842.769 | 3204.69 | 3145.49 | 2585.07491 | 2557.61008 | 842.769 | 13406.6 | 13498.2 | 11108.31107 | 11304.90418 | 842.769 |
| 843.048 | 3242.36 | 3152.07 | 2621.39131 | 2563.03104 | 843.048 | 13363.4 | 13519.1 | 11304.80419 | 11322.70356 | 843.048 |
| 843.327 | 3192.71 | 3158.65 | 2650.42641 | 2568.46199 | 843.327 | 13299.6 | 13540 | 11295.10453 | 11340.50294 | 843.327 |
| 843.607 | 3236.88 | 3165.24 | 2655.33728 | 2573.89295 | 843.607 | 13516.8 | 13560.8 | 11346.10274 | 11358.30231 | 843.607 |
| 843.886 | 3291.54 | 3171.83 | 2635.17373 | 2579.3339 | 843.886 | 13672.5 | 13581.7 | 11376.30168 | 11376.10169 | 843.886 |
| 844.165 | 3232.18 | 3178.44 | 2614.65012 | 2584.78486 | 844.165 | 13563.9 | 13602.6 | 11374.20176 | 11394.00106 | 844.165 |
| 844.444 | 3239.57 | 3185.05 | 2670.38993 | 2590.23582 | 844.444 | 13543.7 | 13623.5 | 11479.79806 | 11411.80044 | 844.444 |
| 844.724 | 3224.19 | 3191.66 | 2624.54186 | 2595.69678 | 844.724 | 13581.2 | 13644.5 | 11348.80265 | 11429.59982 | 844.724 |
| 845.003 | 3222.64 | 3198.29 | 2666.17919 | 2601.16775 | 845.003 | 13697.1 | 13665.4 | 11369.20193 | 11447.49919 | 845.003 |
| 845.282 | 3233.22 | 3204.92 | 2609.36919 | 2606.63871 | 845.282 | 13581.2 | 13686.4 | 11367.10201 | 11465.39856 | 845.282 |
| 845.562 | 3293.41 | 3211.56 | 2667.44941 | 2612.11967 | 845.562 | 13571.6 | 13707.3 | 11423.00005 | 11483.29794 | 845.562 |
| 845.841 | 3246.54 | 3218.21 | 2723.89934 | 2617.60064 | 845.841 | 13692.8 | 13728.3 | 11482.29797 | 11501.09731 | 845.841 |
| 846.12 | 3251.77 | 3224.86 | 2759.79566 | 2623.0916 | 846.12 | 13710.3 | 13749.2 | 11570.29489 | 11519.09668 | 846.12 |
| 846.399 | 3250.24 | 3231.53 | 2695.8144 | 2628.59257 | 846.399 | 13665.4 | 13770.2 | 11599.69386 | 11536.99606 | 846.399 |
| 846.678 | 3241.8 | 3238.19 | 2684.77246 | 2634.09354 | 846.678 | 13734.5 | 13791.2 | 11512.7969 | 11554.89543 | 846.678 |
| 846.957 | 3263.99 | 3244.87 | 2692.43381 | 2639.60451 | 846.957 | 13762.6 | 13812.2 | 11467.0985 | 11572.7948 | 846.957 |
| 847.236 | 3348.23 | 3251.55 | 2732.95094 | 2645.11548 | 847.236 | 13779.7 | 13833.3 | 11558.19531 | 11590.69418 | 847.236 |
| 847.516 | 3291.42 | 3258.24 | 2697.21465 | 2650.64645 | 847.516 | 13844.6 | 13854.3 | 11604.49369 | 11608.69355 | 847.516 |
| 847.795 | 3331.9 | 3264.94 | 2707.49646 | 2656.16743 | 847.795 | 13860.8 | 13875.3 | 11609.49352 | 11626.59292 | 847.795 |
| 848.074 | 3407.36 | 3271.65 | 2712.52734 | 2661.7084 | 848.074 | 13794.4 | 13896.4 | 11702.29027 | 11644.59229 | 848.074 |
| 848.353 | 3350.18 | 3278.36 | 2727.08991 | 2667.24938 | 848.353 | 13855 | 13917.5 | 11656.09189 | 11662.59166 | 848.353 |
| 848.632 | 3334 | 3285.08 | 2711.32713 | 2672.79035 | 848.632 | 13957.7 | 13938.5 | 11875.6842 | 11680.59103 | 848.632 |
| 848.911 | 3358.72 | 3291.8 | 2719.16851 | 2678.34133 | 848.911 | 13925.9 | 13959.6 | 11673.79127 | 11698.5904 | 848.911 |
| 849.19 | 3449.56 | 3298.54 | 2732.64088 | 2683.90231 | 849.19 | 13916.1 | 13980.7 | 11635.6926 | 11716.58977 | 849.19 |
| 849.469 | 3314.32 | 3305.28 | 2721.21887 | 2689.47329 | 849.469 | 13881.2 | 14001.8 | 11644.69229 | 11734.58914 | 849.469 |
| 849.748 | 3382.76 | 3312.03 | 2727.29994 | 2695.04427 | 849.748 | 13994.6 | 14022.9 | 11790.18719 | 11752.58851 | 849.748 |
| 850.027 | 3360.7 | 3318.78 | 2702.13551 | 2700.61525 | 850.027 | 13950.1 | 14044 | 11635.7926 | 11770.58788 | 850.027 |
| 850.306 | 3357.84 | 3325.55 | 2713.30748 | 2706.20623 | 850.306 | 13877.2 | 14065.2 | 11730.18929 | 11788.68724 | 850.306 |
| 850.585 | 3378.82 | 3332.32 | 2748.73371 | 2711.79721 | 850.585 | 14003.6 | 14086.3 | 11581.3945 | 11806.68661 | 850.585 |
| 850.864 | 3445.79 | 3339.09 | 2761.23591 | 2717.3882 | 850.864 | 13976.8 | 14107.5 | 11809.8865 | 11824.78598 | 850.864 |
| 851.143 | 3409.86 | 3345.88 | 2755.37488 | 2722.98918 | 851.143 | 13979.6 | 14128.6 | 11667.69148 | 11842.78535 | 851.143 |
| 851.421 | 3384.59 | 3352.67 | 2794.6518 | 2728.60017 | 851.421 | 14095.3 | 14149.8 | 11821.68609 | 11860.88472 | 851.421 |
| 851.7 | 3365.12 | 3359.47 | 2794.77182 | 2734.21116 | 851.7 | 14017.4 | 14171 | 11818.5862 | 11878.98408 | 851.7 |
| 851.979 | 3393.02 | 3366.27 | 2795.18189 | 2739.83215 | 851.979 | 14117.5 | 14192.2 | 11778.88759 | 11897.08345 | 851.979 |
| 852.258 | 3468.45 | 3373.09 | 2759.69564 | 2745.46314 | 852.258 | 14079.8 | 14213.4 | 11670.59138 | 11915.18282 | 852.258 |
| 852.537 | 3405.64 | 3379.91 | 2804.43352 | 2751.09413 | 852.537 | 14116.6 | 14234.6 | 11886.48382 | 11933.28218 | 852.537 |
| 852.816 | 3443.45 | 3386.73 | 2804.30349 | 2756.73512 | 852.816 | 14075.3 | 14255.8 | 11879.78405 | 11951.48154 | 852.816 |
| 853.094 | 3459.49 | 3393.57 | 2791.00115 | 2762.37612 | 853.094 | 14126.7 | 14277 | 11813.58637 | 11969.58091 | 853.094 |
| 853.373 | 3418.28 | 3400.41 | 2786.65039 | 2768.03711 | 853.373 | 14096.7 | 14298.3 | 11859.38477 | 11987.68028 | 853.373 |
| 853.652 | 3372.68 | 3407.26 | 2818.48599 | 2773.68811 | 853.652 | 14001.7 | 14319.5 | 11865.28456 | 12005.87964 | 853.652 |
| 853.93 | 3565.02 | 3414.11 | 2837.3193 | 2779.3491 | 853.93 | 14189.7 | 14340.8 | 11929.8823 | 12023.97901 | 853.93 |
| 854.209 | 3393.39 | 3420.98 | 2824.68708 | 2785.0201 | 854.209 | 14158.2 | 14362.1 | 11950.28159 | 12042.17837 | 854.209 |
| 854.488 | 3511.45 | 3427.85 | 2882.00717 | 2790.7011 | 854.488 | 14389.4 | 14383.3 | 12149.07463 | 12060.37773 | 854.488 |
| 854.766 | 3490.48 | 3434.72 | 2831.14822 | 2796.3821 | 854.766 | 14351.3 | 14404.6 | 11950.78157 | 12078.57709 | 854.766 |
| 855.045 | 3520.75 | 3441.61 | 2830.29807 | 2802.0731 | 855.045 | 14317.4 | 14425.9 | 12035.07862 | 12096.77646 | 855.045 |
| 855.324 | 3482.68 | 3448.5 | 2849.87151 | 2807.7641 | 855.324 | 14434.7 | 14447.2 | 11942.78185 | 12114.97582 | 855.324 |
| 855.602 | 3467.09 | 3455.4 | 2829.75797 | 2813.46511 | 855.602 | 14309.2 | 14468.6 | 11932.18222 | 12133.17518 | 855.602 |
| 855.881 | 3515.87 | 3462.3 | 2788.27067 | 2819.17611 | 855.881 | 14272.3 | 14489.9 | 12015.37931 | 12151.37455 | 855.881 |
| 856.16 | 3425.34 | 3469.22 | 2907.79171 | 2824.88712 | 856.16 | 14461 | 14511.2 | 12143.57482 | 12169.6739 | 856.16 |
| 856.438 | 3459.3 | 3476.14 | 2860.15332 | 2830.59812 | 856.438 | 14289.1 | 14532.6 | 12031.27875 | 12187.87327 | 856.438 |
| 856.717 | 3502.4 | 3483.06 | 2859.55322 | 2836.32913 | 856.717 | 14254.6 | 14553.9 | 12090.37668 | 12206.07263 | 856.717 |
| 856.995 | 3487.7 | 3490 | 2897.99998 | 2842.06014 | 856.995 | 14380.9 | 14575.3 | 12109.97599 | 12224.37199 | 856.995 |
| 857.274 | 3551.49 | 3496.94 | 2883.44742 | 2847.79115 | 857.274 | 14376.7 | 14596.7 | 12188.37325 | 12242.67135 | 857.274 |
| 857.552 | 3529.06 | 3503.89 | 2874.69588 | 2853.53216 | 857.552 | 14547 | 14618.1 | 12021.57909 | 12260.97071 | 857.552 |
| 857.831 | 3535.17 | 3510.84 | 2910.00209 | 2859.28317 | 857.831 | 14503.5 | 14639.5 | 12087.57678 | 12279.17007 | 857.831 |
| 858.109 | 3544.39 | 3517.8 | 2885.21773 | 2865.04418 | 858.109 | 14606 | 14660.9 | 12307.86907 | 12297.46943 | 858.109 |
| 858.388 | 3533.65 | 3524.77 | 2905.69134 | 2870.8052 | 858.388 | 14471.5 | 14682.3 | 12230.07179 | 12315.76879 | 858.388 |
| 858.666 | 3561.55 | 3531.75 | 2938.05703 | 2876.56621 | 858.666 | 14614.8 | 14703.7 | 12217.87222 | 12334.06815 | 858.666 |
| 858.945 | 3592 | 3538.73 | 2864.98417 | 2882.34723 | 858.945 | 14664.4 | 14725.1 | 12233.37167 | 12352.4675 | 858.945 |
| 859.223 | 3609.54 | 3545.72 | 2924.76469 | 2888.11824 | 859.223 | 14428.4 | 14746.6 | 12254.67093 | 12370.76686 | 859.223 |
| 859.501 | 3590.43 | 3552.72 | 2944.50817 | 2893.90926 | 859.501 | 14523 | 14768 | 12065.07757 | 12389.06622 | 859.501 |
| 859.78 | 3523.61 | 3559.72 | 2888.02823 | 2899.70028 | 859.78 | 14535.3 | 14789.5 | 12203.9727 | 12407.46558 | 859.78 |
| 860.058 | 3622.93 | 3566.74 | 2950.70926 | 2905.5013 | 860.058 | 14672.3 | 14811 | 12192.5731 | 12425.76494 | 860.058 |
| 860.337 | 3579.7 | 3573.75 | 2980.5145 | 2911.30232 | 860.337 | 14860.6 | 14832.5 | 12248.77114 | 12444.16429 | 860.337 |
| 860.615 | 3637.03 | 3580.78 | 2921.56413 | 2917.11335 | 860.615 | 14727.8 | 14853.9 | 12200.97281 | 12462.56365 | 860.615 |
| 860.893 | 3607.55 | 3587.81 | 2884.22756 | 2922.92437 | 860.893 | 14601.3 | 14875.4 | 12224.27199 | 12480.96301 | 860.893 |
| 861.171 | 3564.27 | 3594.85 | 2976.71383 | 2928.74539 | 861.171 | 14829.4 | 14897 | 12389.36621 | 12499.36236 | 861.171 |
| 861.45 | 3607.24 | 3601.9 | 2952.26953 | 2934.57642 | 861.45 | 14613.9 | 14918.5 | 12359.66725 | 12517.66172 | 861.45 |
| 861.728 | 3657.5 | 3608.95 | 2950.60924 | 2940.40744 | 861.728 | 14659.5 | 14940 | 12370.16688 | 12536.16107 | 861.728 |
| 862.006 | 3632.83 | 3616.01 | 2924.15458 | 2946.24847 | 862.006 | 14901 | 14961.5 | 12446.26422 | 12554.56043 | 862.006 |
| 862.284 | 3596.68 | 3623.08 | 2992.72665 | 2952.0895 | 862.284 | 14764.1 | 14983.1 | 12438.46449 | 12572.95978 | 862.284 |
| 862.563 | 3657.49 | 3630.16 | 2991.62646 | 2957.94053 | 862.563 | 14755.3 | 15004.6 | 12463.06363 | 12591.35914 | 862.563 |
| 862.841 | 3610.35 | 3637.24 | 2986.05548 | 2963.80156 | 862.841 | 14699.8 | 15026.2 | 12547.96066 | 12609.85849 | 862.841 |
| 863.119 | 3686.83 | 3644.32 | 3020.8216 | 2969.66259 | 863.119 | 14693.1 | 15047.8 | 12389.06622 | 12628.25785 | 863.119 |
| 863.397 | 3662.4 | 3651.42 | 2971.81297 | 2975.53363 | 863.397 | 14901.1 | 15069.3 | 12610.05849 | 12646.7572 | 863.397 |
| 863.675 | 3644.12 | 3658.52 | 2956.99036 | 2981.41466 | 863.675 | 14790.2 | 15090.9 | 12435.3646 | 12665.15656 | 863.675 |
| 863.953 | 3612.41 | 3665.63 | 3013.80036 | 2987.28569 | 863.953 | 14909.1 | 15112.5 | 12544.0608 | 12683.65591 | 863.953 |
| 864.231 | 3735.18 | 3672.75 | 2976.43378 | 2993.17673 | 864.231 | 14983.8 | 15134.1 | 12603.75871 | 12702.15526 | 864.231 |
| 864.51 | 3665.21 | 3679.87 | 3029.76317 | 2999.06777 | 864.51 | 14808.1 | 15155.8 | 12524.66148 | 12720.65461 | 864.51 |
| 864.788 | 3699.48 | 3687 | 2994.18691 | 3004.96881 | 864.788 | 14944 | 15177.4 | 12540.66092 | 12739.15397 | 864.788 |
| 865.066 | 3713.39 | 3694.14 | 3095.32471 | 3010.87985 | 865.066 | 14990.9 | 15199 | 12582.05947 | 12757.65332 | 865.066 |
| 865.344 | 3687.56 | 3701.28 | 3018.23114 | 3016.79089 | 865.344 | 14856.1 | 15220.6 | 12664.0566 | 12776.15267 | 865.344 |
| 865.622 | 3730.14 | 3708.43 | 3092.93429 | 3022.70193 | 865.622 | 14998.3 | 15242.3 | 12591.85912 | 12794.65202 | 865.622 |
| 865.9 | 3734.39 | 3715.59 | 3077.1215 | 3028.63297 | 865.9 | 15198.8 | 15264 | 12658.85678 | 12813.25137 | 865.9 |
| 866.178 | 3744.84 | 3722.76 | 3017.9811 | 3034.55401 | 866.178 | 15058.7 | 15285.6 | 12624.058 | 12831.75072 | 866.178 |
| 866.456 | 3697.04 | 3729.93 | 3047.83635 | 3040.49506 | 866.456 | 14952.9 | 15307.3 | 12645.15726 | 12850.35007 | 866.456 |
| 866.734 | 3796.27 | 3737.11 | 2945.36832 | 3046.4361 | 866.734 | 14988.5 | 15329 | 12573.95975 | 12868.84942 | 866.734 |
| 867.011 | 3779.74 | 3744.29 | 3031.96356 | 3052.37715 | 867.011 | 15156 | 15350.7 | 12685.15586 | 12887.44877 | 867.011 |
| 867.289 | 3754.05 | 3751.49 | 2987.8858 | 3058.3382 | 867.289 | 14995.9 | 15372.4 | 12697.55542 | 12906.04812 | 867.289 |
| 867.567 | 3815.13 | 3758.68 | 3095.36471 | 3064.28925 | 867.567 | 15086.6 | 15394.1 | 12707.15509 | 12924.54747 | 867.567 |
| 867.845 | 3756.24 | 3765.89 | 3033.75387 | 3070.2603 | 867.845 | 15022 | 15415.8 | 12609.85849 | 12943.14682 | 867.845 |
| 868.123 | 3776.46 | 3773.11 | 3056.67791 | 3076.23135 | 868.123 | 15172.4 | 15437.6 | 12593.55906 | 12961.74617 | 868.123 |
| 868.401 | 3742.37 | 3780.32 | 3027.50277 | 3082.2024 | 868.401 | 15024 | 15459.3 | 12698.0554 | 12980.34552 | 868.401 |
| 868.679 | 3803.11 | 3787.55 | 3029.05304 | 3088.18345 | 868.679 | 15259.3 | 15481 | 12536.46106 | 12999.04487 | 868.679 |
| 868.957 | 3772.3 | 3794.78 | 3063.54912 | 3094.1745 | 868.957 | 15396.6 | 15502.8 | 12818.0512 | 13017.64421 | 868.957 |
| 869.234 | 3780 | 3802.03 | 3132.91132 | 3100.16556 | 869.234 | 15162.2 | 15524.6 | 12809.35151 | 13036.24356 | 869.234 |
| 869.512 | 3796.46 | 3809.27 | 3110.54739 | 3106.16661 | 869.512 | 15239.9 | 15546.3 | 12789.25221 | 13054.84291 | 869.512 |
| 869.79 | 3814.62 | 3816.53 | 3109.04712 | 3112.17767 | 869.79 | 15206.2 | 15568.1 | 12723.9545 | 13073.54226 | 869.79 |
| 870.068 | 3811.79 | 3823.79 | 3133.96151 | 3118.18873 | 870.068 | 15250.5 | 15589.9 | 12772.25281 | 13092.14161 | 870.068 |
| 870.345 | 3832.78 | 3831.06 | 3104.64635 | 3124.20979 | 870.345 | 15217.9 | 15611.7 | 12824.35098 | 13110.84095 | 870.345 |
| 870.623 | 3843.98 | 3838.33 | 3074.13098 | 3130.23085 | 870.623 | 15224.4 | 15633.5 | 12840.25043 | 13129.5403 | 870.623 |
| 870.901 | 3754.49 | 3845.61 | 3113.36788 | 3136.26191 | 870.901 | 15323.3 | 15655.3 | 12821.85107 | 13148.13965 | 870.901 |
| 871.178 | 3821.46 | 3852.9 | 3126.72023 | 3142.29297 | 871.178 | 15478.2 | 15677.1 | 13012.3444 | 13166.83899 | 871.178 |
| 871.456 | 3795.54 | 3860.2 | 3135.75182 | 3148.33404 | 871.456 | 15305.8 | 15699 | 12814.85132 | 13185.53834 | 871.456 |
| 871.734 | 3851.57 | 3867.5 | 3116.2884 | 3154.3851 | 871.734 | 15522.4 | 15720.8 | 12893.04858 | 13204.23768 | 871.734 |
| 872.011 | 3828.61 | 3874.81 | 3148.60408 | 3160.43617 | 872.011 | 15412.2 | 15742.6 | 12965.74603 | 13222.93703 | 872.011 |
| 872.289 | 3844.31 | 3882.12 | 3191.22158 | 3166.49723 | 872.289 | 15626.4 | 15764.5 | 13010.74446 | 13241.63637 | 872.289 |
| 872.567 | 3859.98 | 3889.44 | 3175.64884 | 3172.5583 | 872.567 | 15492 | 15786.4 | 12958.1463 | 13260.43571 | 872.567 |
| 872.844 | 3879.92 | 3896.77 | 3180.62972 | 3178.62937 | 872.844 | 15434.9 | 15808.2 | 13044.54327 | 13279.13506 | 872.844 |
| 873.122 | 3904.43 | 3904.11 | 3147.27385 | 3184.71044 | 873.122 | 15569.5 | 15830.1 | 13024.04399 | 13297.8344 | 873.122 |
| 873.399 | 3874.32 | 3911.45 | 3163.02662 | 3190.79151 | 873.399 | 15568.8 | 15852 | 13037.54352 | 13316.63375 | 873.399 |
| 873.677 | 3908.39 | 3918.8 | 3227.27793 | 3196.88258 | 873.677 | 15525.9 | 15873.9 | 12995.64499 | 13335.43309 | 873.677 |
| 873.954 | 4026.76 | 3926.16 | 3183.41021 | 3202.97365 | 873.954 | 15547.7 | 15895.8 | 12966.646 | 13354.13243 | 873.954 |
| 874.232 | 3924.23 | 3933.52 | 3139.48248 | 3209.07472 | 874.232 | 15703 | 15917.7 | 13140.23992 | 13372.93178 | 874.232 |
| 874.509 | 3926.66 | 3940.89 | 3210.47497 | 3215.1758 | 874.509 | 15584.3 | 15939.6 | 13157.23933 | 13391.63112 | 874.509 |
| 874.787 | 3985.91 | 3948.27 | 3296.55012 | 3221.28687 | 874.787 | 15621.2 | 15961.6 | 13289.13471 | 13410.43046 | 874.787 |
| 875.064 | 3920.82 | 3955.65 | 3281.73751 | 3227.40795 | 875.064 | 15715.7 | 15983.5 | 13284.33488 | 13429.2298 | 875.064 |
| 875.342 | 3947.44 | 3963.04 | 3276.66662 | 3233.52903 | 875.342 | 15714.8 | 16005.4 | 13039.84344 | 13448.02915 | 875.342 |
| 875.619 | 3942.88 | 3970.44 | 3260.44376 | 3239.66011 | 875.619 | 15791.3 | 16027.4 | 13140.7399 | 13466.82849 | 875.619 |
| 875.896 | 3938.4 | 3977.84 | 3256.33304 | 3245.79119 | 875.896 | 15747.2 | 16049.4 | 13142.23985 | 13485.62783 | 875.896 |
| 876.174 | 4042.78 | 3985.25 | 3255.22285 | 3251.93227 | 876.174 | 15795.5 | 16071.3 | 13322.33355 | 13504.52717 | 876.174 |
| 876.451 | 3936.03 | 3992.67 | 3200.48321 | 3258.08335 | 876.451 | 15851.1 | 16093.3 | 13326.73339 | 13523.32651 | 876.451 |
| 876.728 | 3963.73 | 4000.09 | 3318.23393 | 3264.23443 | 876.728 | 15816.6 | 16115.3 | 13368.83192 | 13542.12585 | 876.728 |
| 877.006 | 4013.33 | 4007.52 | 3245.15107 | 3270.39552 | 877.006 | 15883.5 | 16137.3 | 13226.5369 | 13561.02519 | 877.006 |
| 877.283 | 3978.73 | 4014.96 | 3264.81453 | 3276.5566 | 877.283 | 16037.6 | 16159.3 | 13284.03489 | 13579.82453 | 877.283 |
| 877.56 | 4008.14 | 4022.4 | 3245.3211 | 3282.72769 | 877.56 | 15954.8 | 16181.3 | 13350.83255 | 13598.72387 | 877.56 |
| 877.838 | 4051 | 4029.85 | 3299.45063 | 3288.89877 | 877.838 | 15981.8 | 16203.3 | 13369.53189 | 13617.62321 | 877.838 |
| 878.115 | 4031.27 | 4037.31 | 3308.44221 | 3295.07986 | 878.115 | 15886.7 | 16225.3 | 13386.13131 | 13636.42255 | 878.115 |
| 878.392 | 4065.47 | 4044.77 | 3221.60693 | 3301.27095 | 878.392 | 15987.7 | 16247.3 | 13355.63238 | 13655.32189 | 878.392 |
| 878.669 | 4023.63 | 4052.24 | 3337.17727 | 3307.46204 | 878.669 | 16179 | 16269.4 | 13426.12991 | 13674.22123 | 878.669 |
| 878.947 | 4039.04 | 4059.72 | 3304.3815 | 3313.66313 | 878.947 | 16167 | 16291.4 | 13497.42742 | 13693.12056 | 878.947 |
| 879.224 | 4056.96 | 4067.2 | 3300.05073 | 3319.86422 | 879.224 | 15948.2 | 16313.5 | 13494.12753 | 13712.0199 | 879.224 |
| 879.501 | 4060.51 | 4074.69 | 3281.49747 | 3326.07531 | 879.501 | 16170.5 | 16335.5 | 13464.82856 | 13730.91924 | 879.501 |
| 879.778 | 4094.39 | 4082.19 | 3276.63661 | 3332.28641 | 879.778 | 16067.8 | 16357.6 | 13603.32371 | 13749.81858 | 879.778 |
| 880.055 | 4131.22 | 4089.69 | 3293.18953 | 3338.5075 | 880.055 | 16106.4 | 16379.7 | 13553.72545 | 13768.81791 | 880.055 |
| 880.332 | 4112.8 | 4097.2 | 3388.84636 | 3344.7386 | 880.332 | 16104.8 | 16401.8 | 13622.02305 | 13787.71725 | 880.332 |
| 880.609 | 4125.85 | 4104.72 | 3344.1785 | 3350.96969 | 880.609 | 16208.3 | 16423.8 | 13615.42328 | 13806.61659 | 880.609 |
| 880.886 | 4126.13 | 4112.24 | 3353.36012 | 3357.21079 | 880.886 | 16238.6 | 16445.9 | 13507.32707 | 13825.61593 | 880.886 |
| 881.163 | 4149.42 | 4119.77 | 3391.24678 | 3363.45189 | 881.163 | 16380.5 | 16468 | 13628.52283 | 13844.51526 | 881.163 |
| 881.441 | 4099.81 | 4127.31 | 3410.05009 | 3369.70299 | 881.441 | 16231.6 | 16490.2 | 13797.8169 | 13863.5146 | 881.441 |
| 881.717 | 4210.96 | 4134.85 | 3386.87601 | 3375.96409 | 881.717 | 16266.3 | 16512.3 | 13605.72362 | 13882.51393 | 881.717 |
| 881.995 | 4120.22 | 4142.4 | 3359.58121 | 3382.22519 | 881.995 | 16352.2 | 16534.4 | 13771.81781 | 13901.41327 | 881.995 |
| 882.271 | 4129.19 | 4149.95 | 3416.23118 | 3388.4863 | 882.271 | 16531.2 | 16556.5 | 13659.42174 | 13920.41261 | 882.271 |
| 882.548 | 4210.34 | 4157.52 | 3381.64509 | 3394.7574 | 882.548 | 16498.9 | 16578.7 | 13795.31699 | 13939.41194 | 882.548 |
| 882.825 | 4227.87 | 4165.09 | 3379.26467 | 3401.03851 | 882.825 | 16407.2 | 16600.8 | 13787.61726 | 13958.41128 | 882.825 |
| 883.102 | 4150.54 | 4172.66 | 3418.30154 | 3407.31961 | 883.102 | 16281 | 16623 | 13698.72037 | 13977.41061 | 883.102 |
| 883.379 | 4169.55 | 4180.24 | 3340.49785 | 3413.61072 | 883.379 | 16400.2 | 16645.1 | 13712.41989 | 13996.40995 | 883.379 |
| 883.656 | 4107.45 | 4187.83 | 3441.29559 | 3419.91183 | 883.656 | 16442.9 | 16667.3 | 13728.51933 | 14015.40928 | 883.656 |
| 883.933 | 4178.35 | 4195.43 | 3430.52369 | 3426.21294 | 883.933 | 16456.8 | 16689.5 | 13822.51603 | 14034.40862 | 883.933 |
| 884.21 | 4161.44 | 4203.03 | 3459.01871 | 3432.51404 | 884.21 | 16489.2 | 16711.7 | 13998.50987 | 14053.50795 | 884.21 |
| 884.487 | 4169.92 | 4210.64 | 3444.51616 | 3438.82516 | 884.487 | 16365.3 | 16733.8 | 13810.81644 | 14072.50728 | 884.487 |
| 884.764 | 4233.27 | 4218.25 | 3393.82724 | 3445.14627 | 884.764 | 16498.7 | 16756 | 13683.5209 | 14091.50662 | 884.764 |
| 885.04 | 4242.1 | 4225.87 | 3429.57353 | 3451.46738 | 885.04 | 16494.5 | 16778.2 | 13900.11332 | 14110.60595 | 885.04 |
| 885.317 | 4232.37 | 4233.5 | 3423.55247 | 3457.79849 | 885.317 | 16547.7 | 16800.5 | 13871.9143 | 14129.70528 | 885.317 |
| 885.594 | 4269.86 | 4241.14 | 3411.44034 | 3464.12961 | 885.594 | 16742.1 | 16822.7 | 13980.11052 | 14148.70461 | 885.594 |
| 885.871 | 4346.98 | 4248.78 | 3479.92239 | 3470.47072 | 885.871 | 16669.6 | 16844.9 | 14026.80888 | 14167.80394 | 885.871 |
| 886.148 | 4256.17 | 4256.43 | 3432.8241 | 3476.82184 | 886.148 | 16734.5 | 16867.1 | 14040.5084 | 14186.90328 | 886.148 |
| 886.424 | 4279.69 | 4264.08 | 3540.88312 | 3483.17296 | 886.424 | 16537.3 | 16889.4 | 13912.91287 | 14206.00261 | 886.424 |
| 886.701 | 4201.43 | 4271.74 | 3476.84185 | 3489.53408 | 886.701 | 16733.8 | 16911.6 | 14016.20925 | 14225.00194 | 886.701 |
| 886.978 | 4250.61 | 4279.41 | 3477.38194 | 3495.8952 | 886.978 | 16795.4 | 16933.9 | 14046.50819 | 14244.10127 | 886.978 |
| 887.254 | 4286.66 | 4287.08 | 3521.31967 | 3502.25632 | 887.254 | 16693.6 | 16956.1 | 14090.60665 | 14263.2006 | 887.254 |
| 887.531 | 4239.81 | 4294.76 | 3448.30682 | 3508.63744 | 887.531 | 16862.8 | 16978.4 | 14180.20351 | 14282.39993 | 887.531 |
| 887.808 | 4357.33 | 4302.45 | 3563.5171 | 3515.01856 | 887.808 | 16810.1 | 17000.7 | 14198.10288 | 14301.49926 | 887.808 |
| 888.084 | 4323.73 | 4310.14 | 3330.38607 | 3521.39969 | 888.084 | 16779.7 | 17023 | 14220.10211 | 14320.59859 | 888.084 |
| 888.361 | 4313.34 | 4317.84 | 3520.55954 | 3527.79081 | 888.361 | 16909.3 | 17045.2 | 14317.39871 | 14339.69793 | 888.361 |
| 888.638 | 4267.36 | 4325.54 | 3523.84012 | 3534.19194 | 888.638 | 16838.3 | 17067.5 | 14208.10253 | 14358.89725 | 888.638 |
| 888.914 | 4272.16 | 4333.26 | 3530.91136 | 3540.59306 | 888.914 | 16807.6 | 17089.8 | 14227.20186 | 14377.99659 | 888.914 |
| 889.191 | 4313.95 | 4340.98 | 3506.66709 | 3546.99419 | 889.191 | 16977.6 | 17112.2 | 14358.79726 | 14397.09592 | 889.191 |
| 889.467 | 4377.19 | 4348.7 | 3536.96243 | 3553.41532 | 889.467 | 16983.7 | 17134.5 | 14175.40368 | 14416.29524 | 889.467 |
| 889.744 | 4306.54 | 4356.43 | 3576.38936 | 3559.82645 | 889.744 | 16814.2 | 17156.8 | 14070.00737 | 14435.49457 | 889.744 |
| 890.02 | 4323.79 | 4364.17 | 3488.92397 | 3566.25758 | 890.02 | 17050.1 | 17179.1 | 14302.69922 | 14454.5939 | 890.02 |
| 890.297 | 4356.66 | 4371.91 | 3139.11241 | 3572.68871 | 890.297 | 16890 | 17201.5 | 14202.00275 | 14473.79323 | 890.297 |
| 890.573 | 4305.55 | 4379.66 | 3195.10227 | 3579.11984 | 890.573 | 16903 | 17223.8 | 14137.20502 | 14492.99256 | 890.573 |
| 890.85 | 4298.65 | 4387.42 | 3216.02595 | 3585.56098 | 890.85 | 17007.2 | 17246.1 | 14182.90342 | 14512.19189 | 890.85 |
| 891.126 | 4295.11 | 4395.18 | 3516.14876 | 3592.01211 | 891.126 | 16913.5 | 17268.5 | 14259.30074 | 14531.39121 | 891.126 |
| 891.402 | 4347.01 | 4402.95 | 3615.32622 | 3598.46325 | 891.402 | 17006 | 17290.8 | 14289.0997 | 14550.59054 | 891.402 |
| 891.679 | 4394.54 | 4410.73 | 3626.20813 | 3604.91438 | 891.679 | 16820.1 | 17313.2 | 14323.59849 | 14569.78987 | 891.679 |
| 891.955 | 4367.33 | 4418.51 | 3621.72734 | 3611.38552 | 891.955 | 17061.1 | 17335.6 | 14225.10194 | 14588.9892 | 891.955 |
| 892.232 | 4447.68 | 4426.3 | 3551.29495 | 3617.84666 | 892.232 | 17079.7 | 17358 | 14333.99813 | 14608.18853 | 892.232 |
| 892.508 | 4402.43 | 4434.09 | 3609.86525 | 3624.3278 | 892.508 | 17094.5 | 17380.4 | 14425.9949 | 14627.38785 | 892.508 |
| 892.784 | 4441.44 | 4441.9 | 3608.06494 | 3630.80894 | 892.784 | 17146.4 | 17402.8 | 14437.4945 | 14646.68718 | 892.784 |
| 893.061 | 4419.59 | 4449.7 | 3472.18103 | 3637.29008 | 893.061 | 17095.8 | 17425.2 | 14545.9907 | 14665.88651 | 893.061 |
| 893.337 | 4527.43 | 4457.52 | 3476.93186 | 3643.78122 | 893.337 | 17165.8 | 17447.6 | 14590.68914 | 14685.18583 | 893.337 |
| 893.613 | 4419.89 | 4465.34 | 3486.52355 | 3650.28237 | 893.613 | 17100.7 | 17470 | 14603.38869 | 14704.38516 | 893.613 |
| 893.889 | 4433.2 | 4473.16 | 3681.61788 | 3656.78351 | 893.889 | 17258.2 | 17492.4 | 14585.38932 | 14723.68448 | 893.889 |
| 894.166 | 4497.66 | 4480.99 | 3695.19027 | 3663.28465 | 894.166 | 17312.3 | 17514.8 | 14467.69344 | 14742.88381 | 894.166 |
| 894.442 | 4470.47 | 4488.83 | 3684.12832 | 3669.7958 | 894.442 | 17425.9 | 17537.2 | 14553.19045 | 14762.18313 | 894.442 |
| 894.718 | 4459.21 | 4496.68 | 3615.07617 | 3676.31695 | 894.718 | 17126.8 | 17559.7 | 14565.49002 | 14781.48246 | 894.718 |
| 894.994 | 4454.63 | 4504.53 | 3755.51088 | 3682.8381 | 894.994 | 17311.3 | 17582.1 | 14556.69033 | 14800.68179 | 894.994 |
| 895.27 | 4465.4 | 4512.39 | 3590.67188 | 3689.36925 | 895.27 | 17529.6 | 17604.6 | 14590.08916 | 14819.98111 | 895.27 |
| 895.547 | 4520.83 | 4520.25 | 3692.32977 | 3695.90039 | 895.547 | 17315.3 | 17627 | 14435.99455 | 14839.28043 | 895.547 |
| 895.823 | 4418.24 | 4528.12 | 3661.96442 | 3702.44155 | 895.823 | 17223.2 | 17649.5 | 14634.18762 | 14858.57976 | 895.823 |
| 896.099 | 4522.33 | 4536 | 3769.50335 | 3708.9927 | 896.099 | 17507.2 | 17672 | 14694.38551 | 14877.87908 | 896.099 |
| 896.375 | 4563.81 | 4543.88 | 3690.28941 | 3715.54385 | 896.375 | 17556.1 | 17694.4 | 14604.68865 | 14897.2784 | 896.375 |
| 896.651 | 4543.8 | 4551.77 | 3687.89899 | 3722.095 | 896.651 | 17592 | 17716.9 | 14597.08891 | 14916.57773 | 896.651 |
| 896.927 | 4545.12 | 4559.67 | 3666.53523 | 3728.65616 | 896.927 | 17391.7 | 17739.4 | 14835.38057 | 14935.87705 | 896.927 |
| 897.203 | 4542.4 | 4567.57 | 3704.00182 | 3735.22732 | 897.203 | 17456.7 | 17761.9 | 14636.48753 | 14955.17638 | 897.203 |
| 897.479 | 4504.54 | 4575.48 | 3750.17995 | 3741.79847 | 897.479 | 17404.4 | 17784.4 | 14777.2826 | 14974.4757 | 897.479 |
| 897.755 | 4537.18 | 4583.39 | 3724.49543 | 3748.37963 | 897.755 | 17530.5 | 17806.9 | 14614.5883 | 14993.87502 | 897.755 |
| 898.031 | 4582.74 | 4591.31 | 3763.99238 | 3754.96079 | 898.031 | 17656.3 | 17829.4 | 14769.08289 | 15013.17435 | 898.031 |
| 898.307 | 4570.63 | 4599.24 | 3739.73811 | 3761.55195 | 898.307 | 17600.9 | 17851.9 | 14808.8815 | 15032.57367 | 898.307 |
| 898.583 | 4577.74 | 4607.17 | 3685.90864 | 3768.14311 | 898.583 | 17556.7 | 17874.4 | 14890.57864 | 15051.97299 | 898.583 |
| 898.859 | 4594.4 | 4615.11 | 3737.19766 | 3774.74427 | 898.859 | 17725.9 | 17897 | 14779.88251 | 15071.27231 | 898.859 |
| 899.135 | 4620.99 | 4623.05 | 3686.09867 | 3781.34543 | 899.135 | 17610.6 | 17919.5 | 15008.8745 | 15090.67163 | 899.135 |
| 899.411 | 4591.91 | 4631.01 | 3781.28542 | 3787.95659 | 899.411 | 17551.3 | 17942 | 14913.67783 | 15110.07095 | 899.411 |
| 899.687 | 4639.23 | 4638.96 | 3686.98883 | 3794.57776 | 899.687 | 17700.1 | 17964.6 | 14831.88069 | 15129.37028 | 899.687 |
| 899.963 | 4618.57 | 4646.93 | 3684.21834 | 3801.19892 | 899.963 | 17698.2 | 17987.1 | 14744.18376 | 15148.7696 | 899.963 |
| 900.239 | 4672.9 | 4654.9 | 3800.18875 | 3807.82009 | 900.239 | 17716.9 | 18009.7 | 14941.97684 | 15168.16892 | 900.239 |
| 900.514 | 4670.26 | 4662.87 | 3796.47809 | 3814.45126 | 900.514 | 17753.2 | 18032.2 | 14965.876 | 15187.56824 | 900.514 |
| 900.79 | 4605.9 | 4670.85 | 3702.41154 | 3821.09243 | 900.79 | 17656.3 | 18054.8 | 14861.17967 | 15206.96756 | 900.79 |
| 901.066 | 4637.13 | 4678.84 | 3847.6571 | 3827.73359 | 901.066 | 17718.4 | 18077.4 | 15046.27319 | 15226.36688 | 901.066 |
| 901.342 | 4750.52 | 4686.84 | 3742.40858 | 3834.37476 | 901.342 | 17928 | 18100 | 15000.0748 | 15245.7662 | 901.342 |
| 901.617 | 4643.78 | 4694.84 | 3825.54321 | 3841.03594 | 901.617 | 17729.7 | 18122.5 | 14983.0754 | 15265.26552 | 901.617 |
| 901.893 | 4705.9 | 4702.84 | 3952.87562 | 3847.68711 | 901.893 | 17940.2 | 18145.1 | 14987.57524 | 15284.66484 | 901.893 |
| 902.169 | 4737.61 | 4710.85 | 3834.97487 | 3854.35828 | 902.169 | 17835 | 18167.7 | 14918.77765 | 15304.06416 | 902.169 |
| 902.445 | 4679.03 | 4718.87 | 3795.22787 | 3861.01945 | 902.445 | 17764.8 | 18190.3 | 15141.56985 | 15323.46348 | 902.445 |
| 902.72 | 4749.1 | 4726.9 | 3861.65956 | 3867.70063 | 902.72 | 17969.1 | 18212.9 | 15129.97026 | 15342.9628 | 902.72 |
| 902.996 | 4757.71 | 4734.93 | 3788.19664 | 3874.3818 | 902.996 | 17906.2 | 18235.5 | 15237.76648 | 15362.36212 | 902.996 |
| 903.272 | 4676.92 | 4742.97 | 3835.88503 | 3881.06298 | 903.272 | 17860.5 | 18258.2 | 15090.97162 | 15381.86144 | 903.272 |
| 903.547 | 4722.72 | 4751.01 | 3795.49792 | 3887.75416 | 903.547 | 18058 | 18280.8 | 15132.87015 | 15401.36075 | 903.547 |
| 903.823 | 4724.15 | 4759.06 | 3811.09067 | 3894.44533 | 903.823 | 17931 | 18303.4 | 15310.16395 | 15420.76007 | 903.823 |
| 904.098 | 4747.65 | 4767.11 | 3828.42372 | 3901.14651 | 904.098 | 18053.4 | 18326 | 15229.16678 | 15440.25939 | 904.098 |
| 904.374 | 4750.15 | 4775.17 | 3924.24058 | 3907.85769 | 904.374 | 18191.1 | 18348.7 | 15137.56999 | 15459.75871 | 904.374 |
| 904.65 | 4729.72 | 4783.24 | 3883.29337 | 3914.56888 | 904.65 | 18124.3 | 18371.3 | 15176.66862 | 15479.15803 | 904.65 |
| 904.925 | 4726.27 | 4791.31 | 3902.7568 | 3921.28006 | 904.925 | 18162.9 | 18394 | 15353.36243 | 15498.65735 | 904.925 |
| 905.201 | 4872.26 | 4799.39 | 3900.70644 | 3928.01124 | 905.201 | 18093 | 18416.6 | 15378.56155 | 15518.15666 | 905.201 |
| 905.476 | 4761.4 | 4807.48 | 3895.9556 | 3934.73242 | 905.476 | 18179.3 | 18439.3 | 15201.16776 | 15537.65598 | 905.476 |
| 905.752 | 4852.21 | 4815.57 | 3880.65291 | 3941.46361 | 905.752 | 18104.2 | 18461.9 | 15250.96602 | 15557.1553 | 905.752 |
| 906.027 | 4786.93 | 4823.66 | 3926.52098 | 3948.20479 | 906.027 | 18257 | 18484.6 | 15261.26566 | 15576.65462 | 906.027 |
| 906.303 | 4765.26 | 4831.77 | 3946.57451 | 3954.94598 | 906.303 | 18294.6 | 18507.3 | 15430.35974 | 15596.15393 | 906.303 |
| 906.578 | 4732.67 | 4839.88 | 3840.02576 | 3961.69717 | 906.578 | 18102.3 | 18530 | 15120.1706 | 15615.65325 | 906.578 |
| 906.854 | 4798 | 4847.99 | 3890.34461 | 3968.44836 | 906.854 | 18109.2 | 18552.7 | 15410.76043 | 15635.25256 | 906.854 |
| 907.129 | 4740.01 | 4856.11 | 3947.42466 | 3975.20955 | 907.129 | 18413.4 | 18575.3 | 15497.65738 | 15654.75188 | 907.129 |
| 907.404 | 4772.86 | 4864.24 | 3967.38817 | 3981.97074 | 907.404 | 18210.5 | 18598 | 15309.96395 | 15674.2512 | 907.404 |
| 907.68 | 4888.2 | 4872.37 | 3897.90594 | 3988.74193 | 907.68 | 18143.9 | 18620.7 | 15323.56348 | 15693.85051 | 907.68 |
| 907.955 | 4781.75 | 4880.51 | 4029.24906 | 3995.51312 | 907.955 | 18395 | 18643.4 | 15401.06076 | 15713.34983 | 907.955 |
| 908.231 | 4836.24 | 4888.66 | 4001.54418 | 4002.29431 | 908.231 | 18333.7 | 18666.2 | 15269.26538 | 15732.94914 | 908.231 |
| 908.506 | 4823.34 | 4896.81 | 3995.92319 | 4009.07551 | 908.506 | 18234 | 18688.9 | 15480.65798 | 15752.44846 | 908.506 |
| 908.781 | 4878.12 | 4904.96 | 3996.46329 | 4015.8667 | 908.781 | 18384 | 18711.6 | 15574.85468 | 15772.04778 | 908.781 |
| 909.056 | 4856.88 | 4913.13 | 3955.5961 | 4022.6679 | 909.056 | 18428.3 | 18734.3 | 15518.45665 | 15791.54709 | 909.056 |
| 909.332 | 4850.75 | 4921.29 | 4002.2043 | 4029.4691 | 909.332 | 18325.9 | 18757 | 15421.56005 | 15811.14641 | 909.332 |
| 909.607 | 4848.53 | 4929.47 | 4069.31611 | 4036.27029 | 909.607 | 18312.8 | 18779.8 | 15413.16034 | 15830.74572 | 909.607 |
| 909.882 | 4829.12 | 4937.65 | 3960.54697 | 4043.08149 | 909.882 | 18479.1 | 18802.5 | 15555.85534 | 15850.34503 | 909.882 |
| 910.157 | 4864.53 | 4945.84 | 3967.69822 | 4049.90269 | 910.157 | 18374 | 18825.3 | 15491.25761 | 15869.84435 | 910.157 |
| 910.433 | 4886.6 | 4954.03 | 3940.8235 | 4056.72389 | 910.433 | 18321.1 | 18848 | 15512.15687 | 15889.44366 | 910.433 |
| 910.708 | 4799.24 | 4962.23 | 3971.00881 | 4063.54509 | 910.708 | 18418.5 | 18870.8 | 15598.55385 | 15909.04298 | 910.708 |
| 910.983 | 4935.9 | 4970.43 | 4046.02201 | 4070.37629 | 910.983 | 18360.8 | 18893.5 | 15660.65168 | 15928.64229 | 910.983 |
| 911.258 | 4943.46 | 4978.64 | 3978.10006 | 4077.2175 | 911.258 | 18351.4 | 18916.3 | 15440.95937 | 15948.24161 | 911.258 |
| 911.533 | 4930.63 | 4986.85 | 3992.52259 | 4084.0587 | 911.533 | 18345.5 | 18939.1 | 15710.54993 | 15967.84092 | 911.533 |
| 911.808 | 4933.75 | 4995.07 | 3948.88491 | 4090.89991 | 911.808 | 18606.9 | 18961.8 | 15627.15285 | 15987.44023 | 911.808 |
| 912.083 | 4997.87 | 5003.3 | 4040.99112 | 4097.75111 | 912.083 | 18677.8 | 18984.6 | 15831.94568 | 16007.03955 | 912.083 |
| 912.358 | 4962.37 | 5011.54 | 4106.04257 | 4104.61232 | 912.358 | 18823.7 | 19007.4 | 15897.0434 | 16026.73886 | 912.358 |
| 912.633 | 4903.13 | 5019.77 | 4051.43296 | 4111.47353 | 912.633 | 18724 | 19030.2 | 15772.94774 | 16046.33817 | 912.633 |
| 912.909 | 5043.12 | 5028.02 | 4001.61419 | 4118.33473 | 912.909 | 18686.6 | 19053 | 15763.64807 | 16065.93749 | 912.909 |
| 913.183 | 4963.12 | 5036.27 | 4074.56703 | 4125.21594 | 913.183 | 18629.2 | 19075.8 | 15825.5459 | 16085.6368 | 913.183 |
| 913.458 | 4960.85 | 5044.53 | 4091.60003 | 4132.08715 | 913.458 | 18659.2 | 19098.6 | 15693.55052 | 16105.23611 | 913.458 |
| 913.734 | 4948.19 | 5052.79 | 4066.76566 | 4138.96836 | 913.734 | 18632.2 | 19121.4 | 15920.64257 | 16124.93542 | 913.734 |
| 914.008 | 5052.28 | 5061.06 | 4082.3284 | 4145.85958 | 914.008 | 18657.8 | 19144.2 | 15708.65 | 16144.53473 | 914.008 |
| 914.283 | 5017.26 | 5069.33 | 4038.72072 | 4152.75079 | 914.283 | 18827.4 | 19167 | 15694.45049 | 16164.23404 | 914.283 |
| 914.558 | 4925 | 5077.61 | 4123.08557 | 4159.652 | 914.558 | 18724.5 | 19189.8 | 15683.35088 | 16183.83336 | 914.558 |
| 914.833 | 5053.82 | 5085.89 | 4101.02169 | 4166.55322 | 914.833 | 18830.2 | 19212.7 | 15801.44675 | 16203.53267 | 914.833 |
| 915.108 | 5082.41 | 5094.19 | 4188.6071 | 4173.46444 | 915.108 | 18932.9 | 19235.5 | 15828.24581 | 16223.13198 | 915.108 |
| 915.383 | 5108.54 | 5102.48 | 4184.45637 | 4180.37565 | 915.383 | 18854 | 19258.3 | 15953.84141 | 16242.83129 | 915.383 |
| 915.658 | 5068.99 | 5110.78 | 4145.84958 | 4187.28687 | 915.658 | 19083.9 | 19281.2 | 15994.53999 | 16262.5306 | 915.658 |
| 915.933 | 5089.56 | 5119.09 | 4085.30892 | 4194.21809 | 915.933 | 18869.9 | 19304 | 15987.04025 | 16282.22991 | 915.933 |
| 916.207 | 5043.56 | 5127.41 | 4225.07352 | 4201.13931 | 916.207 | 19022.5 | 19326.8 | 15906.44307 | 16301.92922 | 916.207 |
| 916.482 | 5071.28 | 5135.73 | 4130.13681 | 4208.07052 | 916.482 | 18983.4 | 19349.7 | 15978.74054 | 16321.52854 | 916.482 |
| 916.757 | 5098.69 | 5144.05 | 4167.62341 | 4215.01175 | 916.757 | 18953.7 | 19372.5 | 15868.5444 | 16341.32784 | 916.757 |
| 917.032 | 5087.95 | 5152.38 | 4197.82872 | 4221.95297 | 917.032 | 18900.9 | 19395.4 | 16003.63967 | 16360.92716 | 917.032 |
| 917.307 | 5082.79 | 5160.72 | 4216.76205 | 4228.90419 | 917.307 | 19040.9 | 19418.3 | 15902.24322 | 16380.62647 | 917.307 |
| 917.581 | 5097.43 | 5169.06 | 4171.42408 | 4235.85541 | 917.581 | 18871.7 | 19441.1 | 15949.54156 | 16400.42577 | 917.581 |
| 917.856 | 5043.51 | 5177.41 | 4207.33039 | 4242.81664 | 917.856 | 19001 | 19464 | 16088.03671 | 16420.12508 | 917.856 |
| 918.131 | 4979.28 | 5185.76 | 4136.84799 | 4249.77786 | 918.131 | 19070.7 | 19486.9 | 15982.44041 | 16439.82439 | 918.131 |
| 918.405 | 5073.09 | 5194.12 | 4216.01192 | 4256.74909 | 918.405 | 19035.1 | 19509.8 | 16102.03622 | 16459.5237 | 918.405 |
| 918.68 | 5172.46 | 5202.49 | 4228.45411 | 4263.72032 | 918.68 | 19246.6 | 19532.6 | 16292.42956 | 16479.22302 | 918.68 |
| 918.955 | 5158.16 | 5210.86 | 4171.52409 | 4270.69154 | 918.955 | 19205.6 | 19555.5 | 16102.5362 | 16499.02232 | 918.955 |
| 919.229 | 5110.61 | 5219.23 | 4229.84436 | 4277.68278 | 919.229 | 19287.9 | 19578.4 | 16094.73648 | 16518.72163 | 919.229 |
| 919.504 | 5209.45 | 5227.62 | 4200.17914 | 4284.664 | 919.504 | 19269.7 | 19601.3 | 16158.63424 | 16538.42094 | 919.504 |
| 919.779 | 5175.59 | 5236 | 4237.04562 | 4291.66524 | 919.779 | 19280.2 | 19624.2 | 16190.83311 | 16558.22025 | 919.779 |
| 920.053 | 5111.53 | 5244.4 | 4330.77212 | 4298.65647 | 920.053 | 19024.5 | 19647.1 | 16255.43085 | 16577.91956 | 920.053 |
| 920.328 | 5098.6 | 5252.8 | 4225.20354 | 4305.6577 | 920.328 | 19356 | 19670 | 16207.33253 | 16597.71887 | 920.328 |
| 920.602 | 5078.33 | 5261.2 | 4291.77526 | 4312.66893 | 920.602 | 19254.2 | 19692.9 | 16252.43096 | 16617.41818 | 920.602 |
| 920.877 | 5211.69 | 5269.61 | 4288.94476 | 4319.68017 | 920.877 | 19315.4 | 19715.9 | 16345.12771 | 16637.21748 | 920.877 |
| 921.151 | 5245.94 | 5278.03 | 4225.81365 | 4326.7014 | 921.151 | 19414 | 19738.8 | 16241.53134 | 16657.01679 | 921.151 |
| 921.426 | 5239.1 | 5286.45 | 4291.40519 | 4333.72264 | 921.426 | 19326.6 | 19761.7 | 16049.93805 | 16676.7161 | 921.426 |
| 921.7 | 5312.33 | 5294.87 | 4192.54779 | 4340.74387 | 921.7 | 19336.2 | 19784.6 | 16330.52822 | 16696.51541 | 921.7 |
| 921.975 | 5270.79 | 5303.31 | 4243.64679 | 4347.78511 | 921.975 | 19306.5 | 19807.6 | 16278.93003 | 16716.31471 | 921.975 |
| 922.249 | 5141.25 | 5311.74 | 4307.74807 | 4354.81635 | 922.249 | 19564.3 | 19830.5 | 16332.22816 | 16736.11402 | 922.249 |
| 922.523 | 5198.51 | 5320.19 | 4260.72979 | 4361.85759 | 922.523 | 19517.4 | 19853.4 | 16462.32361 | 16755.81333 | 922.523 |
| 922.798 | 5142.94 | 5328.64 | 4267.08091 | 4368.90883 | 922.798 | 19509.5 | 19876.4 | 16392.52605 | 16775.61264 | 922.798 |
| 923.072 | 5194.52 | 5337.09 | 4272.00178 | 4375.96007 | 923.072 | 19627 | 19899.3 | 16497.12239 | 16795.41194 | 923.072 |
| 923.346 | 5228.8 | 5345.55 | 4261.81998 | 4383.01131 | 923.346 | 19619.9 | 19922.3 | 16284.02985 | 16815.21125 | 923.346 |
| 923.621 | 5230.64 | 5354.02 | 4409.05589 | 4390.08255 | 923.621 | 19534 | 19945.2 | 16483.22288 | 16835.01056 | 923.621 |
| 923.895 | 5264.36 | 5362.49 | 4367.6086 | 4397.1438 | 923.895 | 19543.2 | 19968.2 | 16376.8266 | 16854.80986 | 923.895 |
| 924.169 | 5257.32 | 5370.97 | 4274.9623 | 4404.21504 | 924.169 | 19593.5 | 19991.2 | 16418.22515 | 16874.60917 | 924.169 |
| 924.444 | 5268.78 | 5379.45 | 4308.11813 | 4411.29629 | 924.444 | 19576.1 | 20014.1 | 16635.31755 | 16894.40848 | 924.444 |
| 924.718 | 5303.24 | 5387.94 | 4426.19891 | 4418.37753 | 924.718 | 19615.6 | 20037.1 | 16537.22098 | 16914.20779 | 924.718 |
| 924.992 | 5349.13 | 5396.43 | 4361.99761 | 4425.45878 | 924.992 | 19577 | 20060.1 | 16569.31986 | 16934.10709 | 924.992 |
| 925.266 | 5311.7 | 5404.93 | 4332.70246 | 4432.55003 | 925.266 | 19574.3 | 20083 | 16650.41702 | 16953.9064 | 925.266 |
| 925.54 | 5284.91 | 5413.43 | 4393.05308 | 4439.64128 | 925.54 | 19741.9 | 20106 | 16673.11623 | 16973.7057 | 925.54 |
| 925.815 | 5271.48 | 5421.94 | 4369.20888 | 4446.74253 | 925.815 | 19576 | 20129 | 16704.91511 | 16993.60501 | 925.815 |
| 926.089 | 5342.59 | 5430.45 | 4384.86164 | 4453.85378 | 926.089 | 19705.8 | 20152 | 16596.7189 | 17013.40431 | 926.089 |
| 926.363 | 5320.68 | 5438.98 | 4307.15796 | 4460.96503 | 926.363 | 19903.7 | 20175 | 16536.42101 | 17033.20362 | 926.363 |
| 926.637 | 5339.05 | 5447.5 | 4376.25012 | 4468.07628 | 926.637 | 19798.8 | 20198 | 16606.61855 | 17053.10292 | 926.637 |
| 926.911 | 5370.36 | 5456.03 | 4379.57071 | 4475.19753 | 926.911 | 19841.7 | 20221 | 16618.81813 | 17072.90223 | 926.911 |
| 927.185 | 5243.78 | 5464.57 | 4416.14714 | 4482.31879 | 927.185 | 19677.4 | 20244 | 16740.71386 | 17092.80153 | 927.185 |
| 927.459 | 5387.66 | 5473.11 | 4423.67847 | 4489.45004 | 927.459 | 19937.8 | 20267 | 16752.91343 | 17112.60084 | 927.459 |
| 927.734 | 5436.78 | 5481.66 | 4382.04114 | 4496.5813 | 927.734 | 20214.6 | 20290 | 16764.51303 | 17132.50014 | 927.734 |
| 928.008 | 5446.74 | 5490.21 | 4424.9987 | 4503.72255 | 928.008 | 19903 | 20313 | 16990.40512 | 17152.29945 | 928.008 |
| 928.282 | 5384.06 | 5498.77 | 4412.67653 | 4510.86381 | 928.282 | 19769.1 | 20336 | 16860.20968 | 17172.19875 | 928.282 |
| 928.556 | 5470.66 | 5507.33 | 4450.40317 | 4518.01507 | 928.556 | 19998 | 20359 | 16990.40512 | 17192.09806 | 928.556 |
| 928.83 | 5416.67 | 5515.9 | 4469.99662 | 4525.16633 | 928.83 | 19937.5 | 20382 | 16973.8057 | 17211.89736 | 928.83 |
| 929.104 | 5451.08 | 5524.48 | 4446.68252 | 4532.32759 | 929.104 | 20074.6 | 20405.1 | 16814.11129 | 17231.79667 | 929.104 |
| 929.377 | 5517.03 | 5533.06 | 4460.54495 | 4539.48885 | 929.377 | 20055.7 | 20428.1 | 16941.60683 | 17251.69597 | 929.377 |
| 929.651 | 5480.16 | 5541.64 | 4460.2249 | 4546.65011 | 929.651 | 19915 | 20451.1 | 16900.90825 | 17271.59527 | 929.651 |
| 929.925 | 5401.46 | 5550.23 | 4484.50917 | 4553.83137 | 929.925 | 20082.8 | 20474.1 | 17053.20292 | 17291.49458 | 929.925 |
| 930.199 | 5449.25 | 5558.83 | 4488.0998 | 4561.00263 | 930.199 | 19974 | 20497.2 | 16903.50816 | 17311.39388 | 930.199 |
| 930.473 | 5425.91 | 5567.43 | 4404.19504 | 4568.1839 | 930.473 | 20298.2 | 20520.2 | 17020.60406 | 17331.19319 | 930.473 |
| 930.747 | 5530.14 | 5576.04 | 4458.87466 | 4575.37516 | 930.747 | 20353.2 | 20543.3 | 16848.0101 | 17351.09249 | 930.747 |
| 931.021 | 5615.16 | 5584.65 | 4579.06581 | 4582.56643 | 931.021 | 20197.7 | 20566.3 | 17183.99834 | 17370.99179 | 931.021 |
| 931.294 | 5553.5 | 5593.27 | 4488.97996 | 4589.75769 | 931.294 | 20190.8 | 20589.4 | 16996.90489 | 17390.89109 | 931.294 |
| 931.568 | 5379 | 5601.89 | 4475.29755 | 4596.95896 | 931.568 | 20053.9 | 20612.4 | 16764.91301 | 17410.89039 | 931.568 |
| 931.842 | 5541.84 | 5610.52 | 4583.57661 | 4604.16023 | 931.842 | 20419.2 | 20635.5 | 17129.50025 | 17430.7897 | 931.842 |
| 932.116 | 5555.18 | 5619.15 | 4621.99337 | 4611.3715 | 932.116 | 20472 | 20658.5 | 17247.79611 | 17450.689 | 932.116 |
| 932.39 | 5563.15 | 5627.79 | 4490.71026 | 4618.58277 | 932.39 | 20190.7 | 20681.6 | 16925.90738 | 17470.5883 | 932.39 |
| 932.663 | 5535.47 | 5636.43 | 4545.86997 | 4625.80404 | 932.663 | 20208.3 | 20704.6 | 17219.99708 | 17490.48761 | 932.663 |
| 932.937 | 5491.04 | 5645.08 | 4533.12773 | 4633.03531 | 932.937 | 20297.6 | 20727.7 | 17163.19907 | 17510.38691 | 932.937 |
| 933.211 | 5535.9 | 5653.73 | 4584.51677 | 4640.25658 | 933.211 | 20270.1 | 20750.8 | 17140.99984 | 17530.38621 | 933.211 |
| 933.484 | 5607.71 | 5662.39 | 4587.86736 | 4647.48785 | 933.484 | 20596 | 20773.8 | 17062.80258 | 17550.28551 | 933.484 |
| 933.758 | 5575.77 | 5671.06 | 4612.93177 | 4654.72913 | 933.758 | 20478.9 | 20796.9 | 17178.19854 | 17570.18482 | 933.758 |
| 934.032 | 5569.8 | 5679.73 | 4575.79524 | 4661.9704 | 934.032 | 20421.2 | 20820 | 17209.49745 | 17590.18412 | 934.032 |
| 934.305 | 5584.15 | 5688.4 | 4729.30225 | 4669.22168 | 934.305 | 20455.6 | 20843.1 | 17284.69481 | 17610.08342 | 934.305 |
| 934.579 | 5614.07 | 5697.08 | 4520.14544 | 4676.47295 | 934.579 | 20479.2 | 20866.2 | 17314.29378 | 17629.98272 | 934.579 |
| 934.852 | 5554.72 | 5705.77 | 4512.79415 | 4683.72423 | 934.852 | 20305.2 | 20889.2 | 17202.1977 | 17649.98202 | 934.852 |
| 935.126 | 5497.18 | 5714.46 | 4617.54258 | 4690.98551 | 935.126 | 20545.4 | 20912.3 | 17264.09553 | 17669.88133 | 935.126 |
| 935.4 | 5502.45 | 5723.15 | 4603.09004 | 4698.24678 | 935.4 | 20378.3 | 20935.4 | 17117.60066 | 17689.88063 | 935.4 |
| 935.673 | 5634.63 | 5731.85 | 4643.37713 | 4705.51806 | 935.673 | 20293.2 | 20958.5 | 17299.59429 | 17709.77993 | 935.673 |
| 935.947 | 5694.48 | 5740.56 | 4625.08391 | 4712.79935 | 935.947 | 20556.8 | 20981.6 | 17310.29392 | 17729.77923 | 935.947 |
| 936.22 | 5565.12 | 5749.27 | 4716.28996 | 4720.07063 | 936.22 | 20600.7 | 21004.7 | 17324.69341 | 17749.77853 | 936.22 |
| 936.494 | 5667.42 | 5757.99 | 4704.75793 | 4727.36191 | 936.494 | 20680.7 | 21027.8 | 17553.5854 | 17769.67783 | 936.494 |
| 936.767 | 5706.95 | 5766.71 | 4648.51803 | 4734.64319 | 936.767 | 20677.2 | 21050.9 | 17394.19098 | 17789.67713 | 936.767 |
| 937.04 | 5612.83 | 5775.43 | 4671.98216 | 4741.93447 | 937.04 | 20742.3 | 21074 | 17400.59076 | 17809.67643 | 937.04 |
| 937.314 | 5724.58 | 5784.17 | 4720.05062 | 4749.23576 | 937.314 | 20687.8 | 21097.2 | 17473.5882 | 17829.57574 | 937.314 |
| 937.587 | 5742.09 | 5792.9 | 4716.43999 | 4756.53704 | 937.587 | 20762.1 | 21120.3 | 17427.58981 | 17849.57504 | 937.587 |
| 937.861 | 5628.88 | 5801.65 | 4696.0664 | 4763.84833 | 937.861 | 20601.6 | 21143.4 | 17608.08349 | 17869.57433 | 937.861 |
| 938.134 | 5694.86 | 5810.39 | 4726.7318 | 4771.15962 | 938.134 | 20848.8 | 21166.5 | 17365.69198 | 17889.57363 | 938.134 |
| 938.407 | 5717.98 | 5819.14 | 4664.12078 | 4778.4709 | 938.407 | 20763 | 21189.6 | 17380.39146 | 17909.57293 | 938.407 |
| 938.681 | 5645.52 | 5827.9 | 4694.3561 | 4785.79219 | 938.681 | 20769.1 | 21212.7 | 17584.8843 | 17929.47224 | 938.681 |
| 938.954 | 5721.8 | 5836.66 | 4719.19047 | 4793.11348 | 938.954 | 20723.9 | 21235.9 | 17469.08836 | 17949.47154 | 938.954 |
| 939.227 | 5739.1 | 5845.43 | 4744.12486 | 4800.44477 | 939.227 | 20897.9 | 21259 | 17553.28541 | 17969.47084 | 939.227 |
| 939.5 | 5695.68 | 5854.2 | 4640.54663 | 4807.77606 | 939.5 | 21000.8 | 21282.1 | 17605.48358 | 17989.47014 | 939.5 |
| 939.774 | 5602.79 | 5862.98 | 4691.24555 | 4815.11735 | 939.774 | 20870.2 | 21305.2 | 17438.68942 | 18009.46944 | 939.774 |
| 940.047 | 5566.41 | 5871.76 | 4762.04801 | 4822.45864 | 940.047 | 20829.1 | 21328.4 | 17349.79253 | 18029.46874 | 940.047 |
| 940.32 | 5860.74 | 5880.55 | 4656.92951 | 4829.80994 | 940.32 | 20990.6 | 21351.5 | 17748.97856 | 18049.46804 | 940.32 |
| 940.593 | 5711.6 | 5889.34 | 4684.41435 | 4837.16123 | 940.593 | 20972 | 21374.7 | 17633.4826 | 18069.46734 | 940.593 |
| 940.867 | 5802.55 | 5898.14 | 4666.16114 | 4844.51252 | 940.867 | 20947.1 | 21397.8 | 17688.08069 | 18089.46664 | 940.867 |
| 941.14 | 5881.01 | 5906.94 | 4730.76251 | 4851.87382 | 941.14 | 21101 | 21421 | 17780.17746 | 18109.56593 | 941.14 |
| 941.413 | 5734.36 | 5915.75 | 4725.14152 | 4859.23512 | 941.413 | 20777 | 21444.1 | 17599.48379 | 18129.56523 | 941.413 |
| 941.686 | 5761.92 | 5924.56 | 4711.49912 | 4866.60641 | 941.686 | 20915.5 | 21467.2 | 17502.38719 | 18149.56453 | 941.686 |
| 941.959 | 5832.27 | 5933.38 | 4793.37352 | 4873.97771 | 941.959 | 21185 | 21490.4 | 17745.97866 | 18169.56383 | 941.959 |
| 942.232 | 5734.15 | 5942.2 | 4791.24315 | 4881.35901 | 942.232 | 21013.8 | 21513.5 | 17545.38569 | 18189.56313 | 942.232 |
| 942.505 | 5797.51 | 5951.03 | 4741.95448 | 4888.74031 | 942.505 | 20896.6 | 21536.7 | 17765.67797 | 18209.66243 | 942.505 |
| 942.778 | 5783.09 | 5959.86 | 4720.87077 | 4896.13161 | 942.778 | 21139.9 | 21559.8 | 17824.47591 | 18229.66173 | 942.778 |
| 943.051 | 5943.22 | 5968.7 | 4759.17751 | 4903.52291 | 943.051 | 21118 | 21583 | 17781.37742 | 18249.66103 | 943.051 |
| 943.324 | 5798.99 | 5977.54 | 4761.25787 | 4910.91421 | 943.324 | 21163.7 | 21606.2 | 17723.87944 | 18269.66033 | 943.324 |
| 943.597 | 5745.18 | 5986.39 | 4842.79222 | 4918.31551 | 943.597 | 21028.5 | 21629.3 | 17775.57763 | 18289.75962 | 943.597 |
| 943.87 | 5791.71 | 5995.24 | 4787.59251 | 4925.72682 | 943.87 | 21207.7 | 21652.5 | 17815.87622 | 18309.75892 | 943.87 |
| 944.143 | 5855.18 | 6004.1 | 4794.47372 | 4933.13812 | 944.143 | 21058.1 | 21675.7 | 17967.87089 | 18329.85822 | 944.143 |
| 944.416 | 5854.35 | 6012.96 | 4794.50372 | 4940.54942 | 944.416 | 21189.7 | 21698.8 | 17869.37434 | 18349.85752 | 944.416 |
| 944.689 | 5932.3 | 6021.83 | 4937.87895 | 4947.96073 | 944.689 | 21285.8 | 21722 | 18076.06711 | 18369.95682 | 944.689 |
| 944.962 | 5901.74 | 6030.7 | 4882.52921 | 4955.39204 | 944.962 | 21254.6 | 21745.2 | 17989.57013 | 18389.95611 | 944.962 |
| 945.235 | 5879.78 | 6039.58 | 4934.95844 | 4962.81334 | 945.235 | 20979.8 | 21768.3 | 17696.3804 | 18410.05541 | 945.235 |
| 945.508 | 5982.92 | 6048.46 | 4892.49097 | 4970.24465 | 945.508 | 21286.3 | 21791.5 | 17960.47115 | 18430.05471 | 945.508 |
| 945.78 | 5960.99 | 6057.35 | 4849.97349 | 4977.67596 | 945.78 | 21408.4 | 21814.7 | 17949.07155 | 18450.15401 | 945.78 |
| 946.053 | 5931.51 | 6066.24 | 4844.32249 | 4985.11727 | 946.053 | 21215.8 | 21837.9 | 18060.66764 | 18470.15331 | 946.053 |
| 946.326 | 5952.72 | 6075.13 | 4799.30457 | 4992.56858 | 946.326 | 21370.5 | 21861.1 | 18213.1623 | 18490.2526 | 946.326 |
| 946.599 | 5965.9 | 6084.03 | 4866.87646 | 5000.00989 | 946.599 | 21489.7 | 21884.3 | 18209.56243 | 18510.3519 | 946.599 |
| 946.872 | 5975.22 | 6092.94 | 4886.88998 | 5007.4612 | 946.872 | 21263.3 | 21907.4 | 17997.36986 | 18530.3512 | 946.872 |
| 947.144 | 6017.18 | 6101.85 | 4886.68995 | 5014.92251 | 947.144 | 21372.2 | 21930.6 | 18148.66456 | 18550.4505 | 947.144 |
| 947.417 | 5970.48 | 6110.77 | 4945.64032 | 5022.38383 | 947.417 | 21291.7 | 21953.8 | 18135.56502 | 18570.54979 | 947.417 |
| 947.69 | 5875.16 | 6119.69 | 4878.19845 | 5029.84514 | 947.69 | 21535.4 | 21977 | 18102.86617 | 18590.54909 | 947.69 |
| 947.963 | 6140.58 | 6128.61 | 4924.72664 | 5037.31645 | 947.963 | 21462.9 | 22000.2 | 18070.26731 | 18610.64839 | 947.963 |
| 948.235 | 5976.67 | 6137.54 | 4868.67678 | 5044.79777 | 948.235 | 21338.6 | 22023.4 | 18206.26255 | 18630.74768 | 948.235 |
| 948.508 | 6099.85 | 6146.48 | 4946.48047 | 5052.26908 | 948.508 | 21333.5 | 22046.6 | 17979.87047 | 18650.84698 | 948.508 |
| 948.78 | 5987.92 | 6155.42 | 4969.09445 | 5059.7504 | 948.78 | 21602.7 | 22069.8 | 18194.56296 | 18670.84628 | 948.78 |
| 949.053 | 6008.1 | 6164.36 | 4996.79932 | 5067.24172 | 949.053 | 21542.9 | 22093 | 18235.56152 | 18690.94558 | 949.053 |
| 949.326 | 6026.5 | 6173.31 | 4939.22919 | 5074.73304 | 949.326 | 21487.1 | 22116.2 | 18240.26136 | 18711.04487 | 949.326 |
| 949.598 | 6103.23 | 6182.27 | 4947.2506 | 5082.23436 | 949.598 | 21806.8 | 22139.4 | 18346.05765 | 18731.14417 | 949.598 |
| 949.871 | 5974.64 | 6191.22 | 5019.54333 | 5089.72568 | 949.871 | 21756.7 | 22162.6 | 18139.1649 | 18751.24347 | 949.871 |
| 950.143 | 6115.72 | 6200.19 | 4953.30167 | 5097.237 | 950.143 | 21581.9 | 22185.8 | 18180.46345 | 18771.34276 | 950.143 |
| 950.416 | 6150.49 | 6209.16 | 5027.02464 | 5104.73832 | 950.416 | 21835 | 22209 | 18263.56054 | 18791.44206 | 950.416 |
| 950.688 | 6168.48 | 6218.13 | 4990.80827 | 5112.25964 | 950.688 | 21608.7 | 22232.2 | 18271.06028 | 18811.54135 | 950.688 |
| 950.961 | 6138 | 6227.11 | 5066.49159 | 5119.77096 | 950.961 | 21697.6 | 22255.4 | 18505.65206 | 18831.64065 | 950.961 |
| 951.233 | 6099.33 | 6236.09 | 4937.33886 | 5127.29229 | 951.233 | 21718.2 | 22278.6 | 18331.15817 | 18851.73995 | 951.233 |
| 951.506 | 6154.97 | 6245.07 | 5049.75864 | 5134.81361 | 951.506 | 21872.4 | 22301.8 | 18495.55242 | 18871.83924 | 951.506 |
| 951.778 | 6149.04 | 6254.07 | 5051.52895 | 5142.34494 | 951.778 | 21890.6 | 22325.1 | 18609.54843 | 18891.93854 | 951.778 |
| 952.051 | 6218.39 | 6263.06 | 5086.51511 | 5149.88626 | 952.051 | 21669 | 22348.3 | 18365.05699 | 18912.03784 | 952.051 |
| 952.323 | 6091.06 | 6272.06 | 5006.321 | 5157.41759 | 952.323 | 21853.9 | 22371.5 | 18364.85699 | 18932.13713 | 952.323 |
| 952.595 | 6121.62 | 6281.07 | 5037.5865 | 5164.95892 | 952.595 | 21675.1 | 22394.7 | 18406.85552 | 18952.23643 | 952.595 |
| 952.868 | 6135.55 | 6290.08 | 5029.65511 | 5172.51024 | 952.868 | 21930.2 | 22417.9 | 18650.54699 | 18972.33572 | 952.868 |
| 953.14 | 6141.66 | 6299.1 | 5115.01013 | 5180.06157 | 953.14 | 21786.6 | 22441.2 | 18440.25435 | 18992.43502 | 953.14 |
| 953.412 | 6251.09 | 6308.11 | 5051.89902 | 5187.6129 | 953.412 | 21908.9 | 22464.4 | 18673.64618 | 19012.53432 | 953.412 |
| 953.685 | 6187.29 | 6317.14 | 4950.99126 | 5195.17423 | 953.685 | 21875.1 | 22487.6 | 18520.45155 | 19032.63361 | 953.685 |
| 953.957 | 6232.16 | 6326.17 | 5114.37001 | 5202.73556 | 953.957 | 21986.5 | 22510.8 | 18571.04977 | 19052.83291 | 953.957 |
| 954.229 | 6207.34 | 6335.2 | 5109.55917 | 5210.3069 | 954.229 | 22195.8 | 22534 | 18661.44661 | 19072.9322 | 954.229 |
| 954.501 | 6244.4 | 6344.24 | 5111.15945 | 5217.87823 | 954.501 | 22050.8 | 22557.3 | 18609.44843 | 19093.0315 | 954.501 |
| 954.774 | 6272.46 | 6353.28 | 5176.94102 | 5225.44956 | 954.774 | 22167.2 | 22580.5 | 18669.34633 | 19113.13079 | 954.774 |
| 955.046 | 6175.56 | 6362.33 | 5188.01297 | 5233.03089 | 955.046 | 21872.5 | 22603.7 | 18590.04911 | 19133.33009 | 955.046 |
| 955.318 | 6266.3 | 6371.38 | 5166.97927 | 5240.61223 | 955.318 | 21942.5 | 22627 | 18607.5485 | 19153.42938 | 955.318 |
| 955.59 | 6266.85 | 6380.44 | 5109.7492 | 5248.20356 | 955.59 | 22301.1 | 22650.2 | 18689.54563 | 19173.52868 | 955.59 |
| 955.862 | 6215.93 | 6389.5 | 5128.29246 | 5255.7949 | 955.862 | 22108.9 | 22673.4 | 18564.25001 | 19193.72797 | 955.862 |
| 956.135 | 6285.84 | 6398.56 | 5146.26563 | 5263.38624 | 956.135 | 22168.1 | 22696.7 | 18795.54191 | 19213.82727 | 956.135 |
| 956.407 | 6256.12 | 6407.64 | 5147.97593 | 5270.98757 | 956.407 | 22123.3 | 22719.9 | 18733.64408 | 19233.92657 | 956.407 |
| 956.679 | 6275.27 | 6416.71 | 5218.42832 | 5278.58891 | 956.679 | 22250.9 | 22743.1 | 18961.83609 | 19254.12586 | 956.679 |
| 956.951 | 6317.39 | 6425.79 | 5116.35036 | 5286.20025 | 956.951 | 22062.2 | 22766.4 | 18802.54167 | 19274.22515 | 956.951 |
| 957.223 | 6399.73 | 6434.87 | 5178.28126 | 5293.81159 | 957.223 | 22265.8 | 22789.6 | 18674.94614 | 19294.32445 | 957.223 |
| 957.495 | 6218.15 | 6443.96 | 5149.23615 | 5301.43293 | 957.495 | 22280.4 | 22812.8 | 18852.53992 | 19314.52374 | 957.495 |
| 957.767 | 6298.49 | 6453.05 | 5217.02808 | 5309.05427 | 957.767 | 22456.4 | 22836.1 | 18768.14287 | 19334.62304 | 957.767 |
| 958.039 | 6360.85 | 6462.15 | 5140.40459 | 5316.67561 | 958.039 | 22317.1 | 22859.3 | 18755.14333 | 19354.82233 | 958.039 |
| 958.311 | 6318.83 | 6471.25 | 5191.07351 | 5324.30696 | 958.311 | 22251.4 | 22882.6 | 18847.54009 | 19374.92163 | 958.311 |
| 958.583 | 6312.38 | 6480.36 | 5175.39075 | 5331.9383 | 958.583 | 22457.9 | 22905.8 | 19059.13269 | 19395.12092 | 958.583 |
| 958.855 | 6448.97 | 6489.47 | 5338.71949 | 5339.56964 | 958.855 | 22599.1 | 22929 | 18841.24031 | 19415.22022 | 958.855 |
| 959.127 | 6480.1 | 6498.59 | 5286.02022 | 5347.21099 | 959.127 | 22357.1 | 22952.3 | 18781.14242 | 19435.41951 | 959.127 |
| 959.399 | 6502.35 | 6507.7 | 5297.14218 | 5354.86233 | 959.399 | 22422.6 | 22975.5 | 19102.03118 | 19455.51881 | 959.399 |
| 959.67 | 6366.97 | 6516.83 | 5224.48939 | 5362.50368 | 959.67 | 22604.6 | 22998.8 | 18816.34119 | 19475.7181 | 959.67 |
| 959.942 | 6350.03 | 6525.96 | 5236.71154 | 5370.16503 | 959.942 | 22862.2 | 23022 | 18999.23478 | 19495.8174 | 959.942 |
| 960.214 | 6251.56 | 6535.09 | 5338.87952 | 5377.81637 | 960.214 | 22510.8 | 23045.3 | 19194.92793 | 19516.01669 | 960.214 |
| 960.486 | 6409.46 | 6544.23 | 5388.44824 | 5385.47772 | 960.486 | 22549.3 | 23068.5 | 19193.52798 | 19536.11598 | 960.486 |
| 960.758 | 6408.39 | 6553.37 | 5332.73844 | 5393.13907 | 960.758 | 22731.3 | 23091.8 | 19255.42581 | 19556.31528 | 960.758 |
| 961.03 | 6544.89 | 6562.52 | 5336.41909 | 5400.81042 | 961.03 | 22600.8 | 23115 | 19092.13153 | 19576.51457 | 961.03 |
| 961.301 | 6496.27 | 6571.67 | 5297.78229 | 5408.48177 | 961.301 | 22659.6 | 23138.3 | 18978.8355 | 19596.61387 | 961.301 |
| 961.573 | 6384.11 | 6580.82 | 5304.94355 | 5416.16312 | 961.573 | 22836 | 23161.5 | 19386.12124 | 19616.81316 | 961.573 |
| 961.845 | 6521.31 | 6589.98 | 5308.0541 | 5423.84447 | 961.845 | 22801.2 | 23184.8 | 19194.42795 | 19636.91246 | 961.845 |
| 962.117 | 6491.9 | 6599.15 | 5360.84339 | 5431.52583 | 962.117 | 22680.6 | 23208 | 19476.11809 | 19657.11175 | 962.117 |
| 962.388 | 6541.26 | 6608.31 | 5319.08604 | 5439.21718 | 962.388 | 22826.4 | 23231.3 | 19265.72545 | 19677.31104 | 962.388 |
| 962.66 | 6407.8 | 6617.48 | 5287.21043 | 5446.90853 | 962.66 | 22734.3 | 23254.5 | 19279.62497 | 19697.51033 | 962.66 |
| 962.932 | 6539.66 | 6626.66 | 5389.85849 | 5454.59989 | 962.932 | 22736 | 23277.8 | 19338.5229 | 19717.60963 | 962.932 |
| 963.203 | 6493.89 | 6635.84 | 5372.70547 | 5462.30124 | 963.203 | 22582.9 | 23301 | 19143.52973 | 19737.80892 | 963.203 |
| 963.475 | 6603.62 | 6645.03 | 5412.15242 | 5470.0126 | 963.475 | 22934.9 | 23324.3 | 19575.01462 | 19758.00822 | 963.475 |
| 963.746 | 6531.01 | 6654.22 | 5351.34171 | 5477.71395 | 963.746 | 22834.1 | 23347.5 | 19196.42788 | 19778.10751 | 963.746 |
| 964.018 | 6428.54 | 6663.41 | 5331.33819 | 5485.42531 | 964.018 | 22821.4 | 23370.8 | 19511.91683 | 19798.3068 | 964.018 |
| 964.29 | 6527.65 | 6672.61 | 5579.02178 | 5493.14667 | 964.29 | 23237 | 23394 | 19482.61786 | 19818.5061 | 964.29 |
| 964.561 | 6475.86 | 6681.81 | 5455.66007 | 5500.86803 | 964.561 | 22916.4 | 23417.3 | 19566.61492 | 19838.70539 | 964.561 |
| 964.833 | 6603.92 | 6691.02 | 5337.21923 | 5508.58939 | 964.833 | 23125.2 | 23440.6 | 19605.31356 | 19858.80469 | 964.833 |
| 965.104 | 6533.9 | 6700.23 | 5473.4332 | 5516.32075 | 965.104 | 22926 | 23463.8 | 19513.91676 | 19879.00398 | 965.104 |
| 965.376 | 6608.26 | 6709.45 | 5490.5062 | 5524.04211 | 965.376 | 23030.9 | 23487.1 | 19282.12488 | 19899.20327 | 965.376 |
| 965.647 | 6636.99 | 6718.67 | 5509.06947 | 5531.78347 | 965.647 | 23247.2 | 23510.3 | 19592.01403 | 19919.40256 | 965.647 |
| 965.919 | 6649.58 | 6727.89 | 5496.03718 | 5539.52483 | 965.919 | 23190.5 | 23533.6 | 19488.21766 | 19939.60186 | 965.919 |
| 966.19 | 6682.13 | 6737.12 | 5504.57868 | 5547.26619 | 966.19 | 23150.7 | 23556.8 | 19579.31447 | 19959.70115 | 966.19 |
| 966.461 | 6718.05 | 6746.35 | 5499.27775 | 5555.00756 | 966.461 | 23101.1 | 23580.1 | 19630.0127 | 19979.90045 | 966.461 |
| 966.733 | 6642.73 | 6755.59 | 5362.31365 | 5562.75892 | 966.733 | 23373.8 | 23603.4 | 19629.31272 | 20000.09974 | 966.733 |
| 967.004 | 6556.81 | 6764.83 | 5427.25507 | 5570.52029 | 967.004 | 23257.1 | 23626.6 | 19589.9141 | 20020.29903 | 967.004 |
| 967.275 | 6637.73 | 6774.07 | 5568.20988 | 5578.27165 | 967.275 | 23141.8 | 23649.9 | 19616.21318 | 20040.49832 | 967.275 |
| 967.547 | 6665.47 | 6783.32 | 5494.55692 | 5586.03302 | 967.547 | 23341.9 | 23673.1 | 19843.10524 | 20060.69762 | 967.547 |
| 967.818 | 6765.41 | 6792.58 | 5483.53498 | 5593.80438 | 967.818 | 23026.8 | 23696.4 | 19418.6201 | 20080.89691 | 967.818 |
| 968.09 | 6706.51 | 6801.84 | 5551.34691 | 5601.57575 | 968.09 | 23173.5 | 23719.7 | 19781.50739 | 20101.0962 | 968.09 |
| 968.361 | 6690.38 | 6811.1 | 5581.79227 | 5609.34712 | 968.361 | 23104.5 | 23742.9 | 19729.20922 | 20121.1955 | 968.361 |
| 968.632 | 6697.7 | 6820.36 | 5532.14353 | 5617.12849 | 968.632 | 23193.7 | 23766.2 | 19716.40967 | 20141.39479 | 968.632 |
| 968.903 | 6729.72 | 6829.63 | 5559.87841 | 5624.89985 | 968.903 | 23297.5 | 23789.4 | 19581.61439 | 20161.59408 | 968.903 |
| 969.174 | 6630.18 | 6838.91 | 5627.60033 | 5632.69123 | 969.174 | 23621.8 | 23812.7 | 19678.21101 | 20181.79338 | 969.174 |
| 969.446 | 6655.9 | 6848.19 | 5433.39615 | 5640.4826 | 969.446 | 23558.2 | 23836 | 19600.61373 | 20201.99267 | 969.446 |
| 969.717 | 6815.99 | 6857.47 | 5481.22457 | 5648.27397 | 969.717 | 23248.4 | 23859.2 | 20050.79796 | 20222.19196 | 969.717 |
| 969.988 | 6696.21 | 6866.76 | 5436.75675 | 5656.06534 | 969.988 | 23359.8 | 23882.5 | 19830.10569 | 20242.39126 | 969.988 |
| 970.259 | 6735.65 | 6876.05 | 5599.96547 | 5663.86671 | 970.259 | 23501 | 23905.7 | 20035.3985 | 20262.59055 | 970.259 |
| 970.53 | 6795.28 | 6885.34 | 5554.1374 | 5671.66808 | 970.53 | 23416.6 | 23929 | 19685.71075 | 20282.78984 | 970.53 |
| 970.801 | 6769.23 | 6894.64 | 5591.32395 | 5679.47946 | 970.801 | 23588.3 | 23952.2 | 20024.99887 | 20302.98913 | 970.801 |
| 971.072 | 6682.49 | 6903.94 | 5575.78121 | 5687.29083 | 971.072 | 23714.8 | 23975.5 | 19834.50554 | 20323.18843 | 971.072 |
| 971.343 | 6799.46 | 6913.25 | 5618.06865 | 5695.10221 | 971.343 | 23503.7 | 23998.8 | 20019.29907 | 20343.38772 | 971.343 |
| 971.615 | 6806.7 | 6922.56 | 5518.67116 | 5702.92359 | 971.615 | 23474.3 | 24022 | 20014.39924 | 20363.58701 | 971.615 |
| 971.886 | 6760.65 | 6931.88 | 5711.5551 | 5710.74496 | 971.886 | 23561.7 | 24045.3 | 19987.0002 | 20383.78631 | 971.886 |
| 972.157 | 6939.09 | 6941.2 | 5718.95641 | 5718.56634 | 972.157 | 23626 | 24068.5 | 20224.49188 | 20403.9856 | 972.157 |
| 972.428 | 6810.34 | 6950.52 | 5636.96198 | 5726.39772 | 972.428 | 23490.4 | 24091.8 | 20187.29318 | 20424.18489 | 972.428 |
| 972.698 | 6726.11 | 6959.85 | 5548.97649 | 5734.2391 | 972.698 | 23752.9 | 24115.1 | 20175.69359 | 20444.38418 | 972.698 |
| 972.97 | 6813.77 | 6969.18 | 5621.82931 | 5742.07047 | 972.97 | 23559.6 | 24138.3 | 19795.6069 | 20464.58348 | 972.97 |
| 973.24 | 6818.65 | 6978.52 | 5677.47911 | 5749.91185 | 973.24 | 23742 | 24161.6 | 20090.59657 | 20484.78277 | 973.24 |
| 973.511 | 6827.5 | 6987.86 | 5560.46852 | 5757.75323 | 973.511 | 23755.4 | 24184.8 | 20132.4951 | 20504.98206 | 973.511 |
| 973.782 | 6854.29 | 6997.2 | 5647.16377 | 5765.60462 | 973.782 | 23609.5 | 24208.1 | 20110.49587 | 20525.18135 | 973.782 |
| 974.053 | 6804.6 | 7006.55 | 5636.56191 | 5773.456 | 974.053 | 23813.9 | 24231.3 | 20115.19571 | 20545.38065 | 974.053 |
| 974.324 | 6842.11 | 7015.9 | 5690.17134 | 5781.30738 | 974.324 | 24019.6 | 24254.6 | 20163.69401 | 20565.57994 | 974.324 |
| 974.595 | 6710 | 7025.25 | 5617.2185 | 5789.16876 | 974.595 | 23729.3 | 24277.9 | 19967.70087 | 20585.77923 | 974.595 |
| 974.866 | 7039.93 | 7034.61 | 5658.48577 | 5797.03015 | 974.866 | 23902.7 | 24301.1 | 20176.89355 | 20605.97853 | 974.866 |
| 975.136 | 6868.49 | 7043.98 | 5664.9069 | 5804.90153 | 975.136 | 23833.9 | 24324.4 | 20017.49913 | 20626.17782 | 975.136 |
| 975.407 | 6949.34 | 7053.35 | 5595.35466 | 5812.77292 | 975.407 | 23958.2 | 24347.6 | 19937.30194 | 20646.37711 | 975.407 |
| 975.678 | 7182.84 | 7062.72 | 5704.49386 | 5820.6443 | 975.678 | 23981.6 | 24370.9 | 20195.89288 | 20666.5764 | 975.678 |
| 975.949 | 6923.35 | 7072.09 | 5782.98767 | 5828.51569 | 975.949 | 23720.5 | 24394.1 | 19914.90272 | 20686.7757 | 975.949 |
| 976.219 | 6986.02 | 7081.47 | 5765.33457 | 5836.39707 | 976.219 | 23852 | 24417.4 | 20351.28744 | 20707.07499 | 976.219 |
| 976.49 | 6984.15 | 7090.85 | 5685.80057 | 5844.28846 | 976.49 | 23620.3 | 24440.7 | 20212.79229 | 20727.27428 | 976.49 |
| 976.761 | 6989.97 | 7100.24 | 5817.35372 | 5852.16985 | 976.761 | 24171.9 | 24463.9 | 20124.19539 | 20747.47357 | 976.761 |
| 977.031 | 7021.01 | 7109.63 | 5779.79711 | 5860.06124 | 977.031 | 23917.9 | 24487.2 | 20323.8884 | 20767.67286 | 977.031 |
| 977.302 | 6889.39 | 7119.03 | 5668.30749 | 5867.96263 | 977.302 | 23930.4 | 24510.4 | 20245.69114 | 20787.87216 | 977.302 |
| 977.573 | 6949.97 | 7128.43 | 5725.31753 | 5875.85402 | 977.573 | 24062.7 | 24533.7 | 20092.5965 | 20808.07145 | 977.573 |
| 977.843 | 6976.63 | 7137.83 | 5679.6995 | 5883.75541 | 977.843 | 23932.9 | 24556.9 | 20260.9906 | 20828.27074 | 977.843 |
| 978.114 | 7058.71 | 7147.23 | 5891.6568 | 5891.6668 | 978.114 | 23999.3 | 24580.2 | 20280.18993 | 20848.47004 | 978.114 |
| 978.385 | 7132.81 | 7156.65 | 5757.02311 | 5899.56819 | 978.385 | 24042.6 | 24603.4 | 20493.98245 | 20868.66933 | 978.385 |
| 978.655 | 7016.65 | 7166.06 | 5767.68498 | 5907.48958 | 978.655 | 24084.5 | 24626.7 | 20106.796 | 20888.96862 | 978.655 |
| 978.926 | 6997.27 | 7175.48 | 5786.72833 | 5915.40098 | 978.926 | 24349.9 | 24649.9 | 20503.28212 | 20909.16791 | 978.926 |
| 979.196 | 7041.89 | 7184.9 | 5754.30263 | 5923.32237 | 979.196 | 24302 | 24673.2 | 20389.98609 | 20929.3672 | 979.196 |
| 979.466 | 7080.23 | 7194.33 | 5764.41441 | 5931.24376 | 979.466 | 24230.4 | 24696.4 | 20489.98259 | 20949.5665 | 979.466 |
| 979.737 | 7000.18 | 7203.76 | 5774.67621 | 5939.17516 | 979.737 | 24211.2 | 24719.6 | 20440.58432 | 20969.76579 | 979.737 |
| 1101.15 | 11649.6 | 11765.2 | 9601.67968 | 9805.91562 | 1101.15 | 34297.6 | 34994.3 | 29888.9535 | 29979.85032 | 1101.15 |
| 1102.87 | 11479.9 | 11833.4 | 9648.978 | 9864.19587 | 1102.87 | 34480.9 | 35135.1 | 29807.95634 | 30104.44596 | 1102.87 |
| 1104.6 | 11645.9 | 11901.6 | 9835.49082 | 9922.52614 | 1104.6 | 34820 | 35275.7 | 29742.55863 | 30228.9416 | 1104.6 |
| 1106.32 | 11694 | 11969.8 | 9758.56729 | 9980.89641 | 1106.32 | 34885.4 | 35415.9 | 30134.04492 | 30353.13725 | 1106.32 |
| 1108.05 | 11664.6 | 12038.1 | 9848.26307 | 10039.26668 | 1108.05 | 35013.1 | 35556 | 30067.34726 | 30477.13291 | 1108.05 |
| 1109.77 | 11939.3 | 12106.4 | 10166.58909 | 10097.77698 | 1109.77 | 35420.1 | 35695.7 | 29926.65218 | 30600.92857 | 1109.77 |
| 1111.5 | 11787.6 | 12174.8 | 9925.24662 | 10156.28728 | 1111.5 | 35250.4 | 35835.2 | 30275.73996 | 30724.62424 | 1111.5 |
| 1113.22 | 11842 | 12243.1 | 10130.88281 | 10214.79757 | 1113.22 | 35240.8 | 35974.5 | 30419.43493 | 30848.01992 | 1113.22 |
| 1114.95 | 12433.6 | 12311.5 | 10113.57976 | 10273.30787 | 1114.95 | 35552.9 | 36113.4 | 30520.93137 | 30971.21561 | 1114.95 |
| 1116.67 | 12296.4 | 12379.9 | 10130.08266 | 10331.91818 | 1116.67 | 35728.1 | 36252.1 | 30656.32663 | 31094.2113 | 1116.67 |
| 1118.39 | 12217.9 | 12448.4 | 10341.61989 | 10390.5285 | 1118.39 | 35999.3 | 36390.5 | 30797.62169 | 31217.007 | 1118.39 |
| 1120.12 | 12237 | 12516.8 | 10263.40613 | 10449.23883 | 1120.12 | 35921.1 | 36528.6 | 30950.81632 | 31339.60271 | 1120.12 |
| 1121.84 | 12337.6 | 12585.3 | 10153.28675 | 10507.84914 | 1121.84 | 36198.7 | 36666.5 | 31209.30727 | 31461.99843 | 1121.84 |
| 1123.57 | 12452.2 | 12653.8 | 10460.24077 | 10566.55947 | 1123.57 | 36502.9 | 36804 | 31226.30668 | 31584.09415 | 1123.57 |
| 1125.29 | 12504.3 | 12722.3 | 10387.32793 | 10625.26981 | 1125.29 | 36297.6 | 36941.3 | 31149.50937 | 31705.98988 | 1125.29 |
| 1127.01 | 12695.1 | 12790.8 | 10321.81641 | 10684.08016 | 1127.01 | 36465.1 | 37078.3 | 31532.99594 | 31827.68562 | 1127.01 |
| 1128.74 | 12778.5 | 12859.3 | 10731.78855 | 10742.8905 | 1128.74 | 36605.9 | 37215 | 31447.19894 | 31949.18137 | 1128.74 |
| 1130.46 | 12654.4 | 12927.8 | 10645.07329 | 10801.60084 | 1130.46 | 36612.3 | 37351.3 | 31559.99499 | 32070.37712 | 1130.46 |
| 1132.18 | 12758.4 | 12996.3 | 10756.39288 | 10860.41119 | 1132.18 | 36892.9 | 37487.4 | 31782.98719 | 32191.37289 | 1132.18 |
| 1133.91 | 12983.4 | 13064.8 | 10920.42175 | 10919.32155 | 1133.91 | 36960.7 | 37623.2 | 31753.48822 | 32312.16866 | 1133.91 |
| 1135.63 | 12716.1 | 13133.4 | 10849.90934 | 10978.1319 | 1135.63 | 37032.2 | 37758.7 | 31969.08067 | 32432.66444 | 1135.63 |
| 1137.35 | 12783.4 | 13201.9 | 10908.8197 | 11037.04227 | 1137.35 | 37334.8 | 37893.9 | 32083.87665 | 32552.96023 | 1137.35 |
| 1139.07 | 12968.2 | 13270.4 | 10879.21449 | 11095.85262 | 1139.07 | 37369.9 | 38028.8 | 32127.57512 | 32673.05602 | 1139.07 |
| 1140.8 | 13030 | 13338.9 | 10941.62548 | 11154.76299 | 1140.8 | 37561.1 | 38163.3 | 32385.56609 | 32792.85183 | 1140.8 |
| 1142.52 | 13181 | 13407.4 | 11002.43618 | 11213.67335 | 1142.52 | 37948.6 | 38297.6 | 32361.66693 | 32912.34764 | 1142.52 |
| 1144.24 | 13129.2 | 13476 | 10983.63287 | 11272.58372 | 1144.24 | 37836.2 | 38431.5 | 32347.86741 | 33031.64347 | 1144.24 |
| 1145.96 | 13195.6 | 13544.4 | 11203.97164 | 11331.49409 | 1145.96 | 37935.5 | 38565.1 | 32611.35818 | 33150.7393 | 1145.96 |
| 1147.69 | 13421.8 | 13612.9 | 11144.76123 | 11390.40445 | 1147.69 | 38114.2 | 38698.4 | 32804.85141 | 33269.53514 | 1147.69 |
| 1149.41 | 13341.1 | 13681.4 | 11232.37664 | 11449.31482 | 1149.41 | 38184.8 | 38831.4 | 33015.94402 | 33388.13099 | 1149.41 |
| 1151.13 | 13177.6 | 13749.9 | 11313.89099 | 11508.22519 | 1151.13 | 38344 | 38964 | 33093.6413 | 33506.42684 | 1151.13 |
| 1152.85 | 13288 | 13818.3 | 11415.90894 | 11567.13555 | 1152.85 | 38567.8 | 39096.3 | 33201.13753 | 33624.42271 | 1152.85 |
| 1154.57 | 13641.5 | 13886.7 | 11655.15104 | 11626.04592 | 1154.57 | 38623 | 39228.3 | 33239.43619 | 33742.21859 | 1154.57 |
| 1156.3 | 13794.8 | 13955.1 | 11609.24296 | 11684.95629 | 1156.3 | 38696.5 | 39359.9 | 33479.62778 | 33859.71447 | 1156.3 |
| 1158.02 | 13456 | 14023.5 | 11603.24191 | 11743.86665 | 1158.02 | 38996.7 | 39491.3 | 33514.52656 | 33977.01037 | 1158.02 |
| 1159.74 | 13857.3 | 14091.8 | 11663.75256 | 11802.77702 | 1159.74 | 39255.4 | 39622.2 | 33754.71815 | 34093.90627 | 1159.74 |
| 1161.46 | 14111.6 | 14160.1 | 11720.36252 | 11861.68739 | 1161.46 | 39310.2 | 39752.9 | 33862.61437 | 34210.70219 | 1161.46 |
| 1163.18 | 14218.8 | 14228.4 | 11776.97248 | 11920.49774 | 1163.18 | 39312.8 | 39883.2 | 33793.2168 | 34327.09811 | 1163.18 |
| 1164.9 | 14344.8 | 14296.7 | 11787.8744 | 11979.4081 | 1164.9 | 39652.9 | 40013.2 | 33725.11919 | 34443.29404 | 1164.9 |
| 1166.62 | 14082.7 | 14364.9 | 11819.78001 | 12038.31847 | 1166.62 | 39787.2 | 40142.8 | 33914.51256 | 34559.18998 | 1166.62 |
| 1168.34 | 14368 | 14433.1 | 11887.892 | 12097.12882 | 1168.34 | 39638.6 | 40272 | 34094.80624 | 34674.78594 | 1168.34 |
| 1170.06 | 14370.1 | 14501.3 | 12016.71467 | 12155.93917 | 1170.06 | 39916.4 | 40401 | 34315.9985 | 34790.0819 | 1170.06 |
| 1171.78 | 14320.7 | 14569.5 | 11968.20613 | 12214.74952 | 1171.78 | 40019.8 | 40529.5 | 34672.28602 | 34905.17787 | 1171.78 |
| 1173.5 | 14356.4 | 14637.5 | 11981.70851 | 12273.55987 | 1173.5 | 40050.4 | 40657.7 | 34695.38521 | 35019.97385 | 1173.5 |
| 1175.22 | 14537.8 | 14705.6 | 12095.42852 | 12332.37022 | 1175.22 | 40368.4 | 40785.6 | 34696.78517 | 35134.46984 | 1175.22 |
| 1176.94 | 14830 | 14773.6 | 12049.62046 | 12391.08055 | 1176.94 | 40416.1 | 40913.1 | 34681.28571 | 35248.66584 | 1176.94 |
| 1178.66 | 14710.8 | 14841.6 | 12145.43732 | 12449.8909 | 1178.66 | 40454.1 | 41040.3 | 34598.48861 | 35362.56186 | 1178.66 |
| 1180.38 | 14708.2 | 14909.5 | 12264.15821 | 12508.60123 | 1180.38 | 40896.1 | 41167.1 | 34921.07731 | 35476.15788 | 1180.38 |
| 1182.11 | 14857.4 | 14977.4 | 12189.84514 | 12567.21154 | 1182.11 | 40873.2 | 41293.5 | 35135.7698 | 35589.55391 | 1182.11 |
| 1183.82 | 14944.9 | 15045.2 | 12449.49083 | 12625.92188 | 1183.82 | 40920.8 | 41419.6 | 35238.06621 | 35702.54995 | 1183.82 |
| 1185.54 | 14835.8 | 15113 | 12573.31262 | 12684.53219 | 1185.54 | 40956.4 | 41545.2 | 35609.55321 | 35815.346 | 1185.54 |
| 1187.26 | 15025.6 | 15180.7 | 12668.82943 | 12743.1425 | 1187.26 | 41127.8 | 41670.6 | 35450.05879 | 35927.74207 | 1187.26 |
| 1188.98 | 15146.6 | 15248.4 | 12743.94264 | 12801.75282 | 1188.98 | 41345.2 | 41795.6 | 35629.95249 | 36039.93814 | 1188.98 |
| 1190.7 | 15100.2 | 15316 | 12580.81394 | 12860.26311 | 1190.7 | 41198.2 | 41920.1 | 35919.74235 | 36151.83422 | 1190.7 |
| 1192.42 | 15030.8 | 15383.6 | 12681.13159 | 12918.77341 | 1192.42 | 41493.7 | 42044.4 | 35855.54459 | 36263.33032 | 1192.42 |
| 1194.14 | 15042.8 | 15451.1 | 12725.13934 | 12977.28371 | 1194.14 | 41564.8 | 42168.2 | 36056.93754 | 36374.62642 | 1194.14 |
| 1195.86 | 15209.5 | 15518.6 | 12895.16926 | 13035.69399 | 1195.86 | 41657.9 | 42291.7 | 36064.93726 | 36485.52254 | 1195.86 |
| 1197.58 | 15439 | 15586 | 12994.58675 | 13094.10427 | 1197.58 | 42025.8 | 42414.8 | 36056.93754 | 36596.21866 | 1197.58 |
| 1199.3 | 15435.1 | 15653.3 | 12928.87519 | 13152.41453 | 1199.3 | 42263.5 | 42537.5 | 36261.43038 | 36706.5148 | 1199.3 |
| 1201.02 | 15403.5 | 15720.6 | 12979.28406 | 13210.82481 | 1201.02 | 42370.4 | 42659.8 | 36393.92574 | 36816.61094 | 1201.02 |
| 1202.73 | 15685.5 | 15787.8 | 13128.11025 | 13269.03505 | 1202.73 | 42474.7 | 42781.8 | 36647.41687 | 36926.3071 | 1202.73 |
| 1204.45 | 15750.5 | 15854.9 | 13309.24212 | 13327.34531 | 1204.45 | 42742.2 | 42903.4 | 36908.60772 | 37035.70327 | 1204.45 |
| 1206.17 | 15682.9 | 15922 | 13259.53338 | 13385.45554 | 1206.17 | 42692.6 | 43024.6 | 36935.0068 | 37144.79945 | 1206.17 |
| 1207.89 | 15821.5 | 15989 | 13399.15795 | 13443.66578 | 1207.89 | 42778.8 | 43145.4 | 37160.5989 | 37253.59564 | 1207.89 |
| 1209.61 | 15828.7 | 16055.9 | 13505.37664 | 13501.77601 | 1209.61 | 42943.5 | 43265.8 | 37186.89798 | 37362.09185 | 1209.61 |
| 1211.32 | 15929.8 | 16122.8 | 13497.7753 | 13559.78621 | 1211.32 | 43243.6 | 43385.8 | 37081.70166 | 37470.28806 | 1211.32 |
| 1213.04 | 16084.9 | 16189.6 | 13652.20248 | 13617.79642 | 1213.04 | 43268.6 | 43505.5 | 37470.68804 | 37578.08428 | 1213.04 |
| 1214.76 | 16061.4 | 16256.3 | 13548.38421 | 13675.80663 | 1214.76 | 43272.3 | 43624.8 | 37585.28403 | 37685.58052 | 1214.76 |
| 1216.48 | 16312.2 | 16322.9 | 13652.3025 | 13733.6168 | 1216.48 | 43481 | 43743.6 | 37745.07844 | 37792.77677 | 1216.48 |
| 1218.19 | 16280.2 | 16389.5 | 13707.11214 | 13791.527 | 1218.19 | 43515.3 | 43862.1 | 38017.5689 | 37899.67302 | 1218.19 |
| 1219.91 | 16350.8 | 16455.9 | 13527.38051 | 13849.33717 | 1219.91 | 43894.4 | 43980.2 | 37689.48038 | 38006.26929 | 1219.91 |
| 1221.63 | 16515.5 | 16522.3 | 13659.20371 | 13907.04732 | 1221.63 | 43992.6 | 44097.8 | 37848.57481 | 38112.46557 | 1221.63 |
| 1223.35 | 16391.7 | 16588.6 | 13821.93235 | 13964.75748 | 1223.35 | 44174.2 | 44215.1 | 38085.06653 | 38218.36187 | 1223.35 |
| 1225.06 | 16565.7 | 16654.8 | 13868.14048 | 14022.36762 | 1225.06 | 44365.8 | 44332 | 37999.96951 | 38323.95817 | 1225.06 |
| 1226.78 | 16622.6 | 16721 | 13972.85891 | 14079.97776 | 1226.78 | 44253.2 | 44448.5 | 38230.76143 | 38429.25448 | 1226.78 |
| 1228.5 | 16819 | 16787 | 14230.50425 | 14137.48788 | 1228.5 | 44386.9 | 44564.6 | 38370.45654 | 38534.15081 | 1228.5 |
| 1230.21 | 16838.9 | 16853 | 14227.90379 | 14194.89798 | 1230.21 | 44511.8 | 44680.2 | 38542.45052 | 38638.74715 | 1230.21 |
| 1231.93 | 16759.6 | 16918.9 | 14280.41303 | 14252.30808 | 1231.93 | 44769.7 | 44795.5 | 38929.63696 | 38742.9435 | 1231.93 |
| 1233.64 | 16892.6 | 16984.7 | 14186.39648 | 14309.61817 | 1233.64 | 44771.1 | 44910.4 | 38800.54148 | 38846.93986 | 1233.64 |
| 1235.36 | 17036 | 17050.4 | 14141.28855 | 14366.82824 | 1235.36 | 45048.3 | 45024.9 | 38925.43711 | 38950.53623 | 1235.36 |
| 1237.08 | 16922.1 | 17116 | 14332.72223 | 14424.0383 | 1237.08 | 45204.3 | 45138.9 | 38959.13593 | 39053.83261 | 1237.08 |
| 1238.79 | 16995.4 | 17181.5 | 14255.00856 | 14481.14835 | 1238.79 | 45111 | 45252.6 | 38874.1389 | 39156.72901 | 1238.79 |
| 1240.51 | 17160.9 | 17246.9 | 14470.14642 | 14538.2584 | 1240.51 | 45370 | 45365.8 | 39251.12571 | 39259.32542 | 1240.51 |
| 1242.22 | 17164.4 | 17312.2 | 14608.37074 | 14595.16842 | 1242.22 | 45633.6 | 45478.7 | 39466.71816 | 39361.52184 | 1242.22 |
| 1243.94 | 17341.1 | 17377.4 | 14550.66059 | 14652.07843 | 1243.94 | 45620 | 45591.1 | 39523.01619 | 39463.51827 | 1243.94 |
| 1245.66 | 17259.8 | 17442.5 | 14822.00834 | 14708.98845 | 1245.66 | 45772.4 | 45703.1 | 39778.30725 | 39565.01472 | 1245.66 |
| 1247.37 | 17391.7 | 17507.5 | 14742.49434 | 14765.69843 | 1247.37 | 46042.8 | 45814.7 | 39922.4022 | 39666.31117 | 1247.37 |
| 1249.09 | 17546.6 | 17572.5 | 14801.00464 | 14822.40841 | 1249.09 | 45938.3 | 45925.9 | 39842.20501 | 39767.20764 | 1249.09 |
| 1250.8 | 17664.5 | 17637.3 | 14843.61214 | 14879.01837 | 1250.8 | 45902.4 | 46036.7 | 39794.80667 | 39867.80411 | 1250.8 |
| 1252.52 | 17730.5 | 17702 | 14826.10906 | 14935.62833 | 1252.52 | 46290.3 | 46147.1 | 40071.49698 | 39968.00061 | 1252.52 |
| 1254.23 | 17627.2 | 17766.6 | 14940.22914 | 14992.03826 | 1254.23 | 46413.7 | 46257.1 | 40242.09101 | 40067.89711 | 1254.23 |
| 1255.95 | 17912.6 | 17831.1 | 15036.046 | 15048.44819 | 1255.95 | 46581.6 | 46366.6 | 40294.68917 | 40167.39362 | 1255.95 |
| 1257.66 | 17978.5 | 17895.5 | 15181.27156 | 15104.75809 | 1257.66 | 46775.4 | 46475.7 | 40483.38256 | 40266.59015 | 1257.66 |
| 1259.38 | 17983.8 | 17959.7 | 15170.66969 | 15160.96799 | 1259.38 | 46877.6 | 46584.4 | 40772.77243 | 40365.38669 | 1259.38 |
| 1261.09 | 17851.4 | 18023.9 | 15084.15447 | 15217.17788 | 1261.09 | 46916.2 | 46692.7 | 40734.77376 | 40463.88324 | 1261.09 |
| 1262.81 | 18081.1 | 18087.9 | 15264.18615 | 15273.18773 | 1262.81 | 47163 | 46800.6 | 40769.37255 | 40562.07981 | 1262.81 |
| 1264.52 | 18381.2 | 18151.9 | 15347.40079 | 15329.19759 | 1264.52 | 47226.4 | 46908 | 40872.56893 | 40659.87638 | 1264.52 |
| 1266.23 | 18337 | 18215.7 | 15304.69328 | 15385.10743 | 1266.23 | 47361.8 | 47015.1 | 40903.06787 | 40757.27297 | 1266.23 |
| 1267.95 | 18360 | 18279.4 | 15412.81231 | 15440.91725 | 1267.95 | 47358.1 | 47121.7 | 41202.75737 | 40854.36957 | 1267.95 |
| 1269.66 | 18473.9 | 18343 | 15574.24071 | 15496.62705 | 1269.66 | 47514.9 | 47227.9 | 41235.95621 | 40951.16618 | 1269.66 |
| 1271.37 | 18516.3 | 18406.5 | 15629.75048 | 15552.23684 | 1271.37 | 47866.7 | 47333.6 | 41285.45448 | 41047.56281 | 1271.37 |
| 1273.09 | 18456.5 | 18469.8 | 15713.66525 | 15607.74661 | 1273.09 | 47868.3 | 47439 | 41493.94718 | 41143.55945 | 1273.09 |
| 1274.8 | 18635.5 | 18533.1 | 15920.30161 | 15663.25638 | 1274.8 | 47941.3 | 47543.9 | 41544.8454 | 41239.2561 | 1274.8 |
| 1276.51 | 18649.8 | 18596.2 | 15863.89168 | 15718.66613 | 1276.51 | 48055.8 | 47648.4 | 41763.63774 | 41334.65276 | 1276.51 |
| 1278.23 | 18715.3 | 18659.2 | 15794.97956 | 15773.87584 | 1278.23 | 48090.1 | 47752.5 | 41733.63879 | 41429.54943 | 1278.23 |
| 1279.94 | 18781.2 | 18722.1 | 15995.01476 | 15829.08556 | 1279.94 | 48018.3 | 47856.1 | 41734.83874 | 41524.24612 | 1279.94 |
| 1281.65 | 18860.1 | 18784.8 | 16021.71946 | 15884.19526 | 1281.65 | 48129.2 | 47959.3 | 41923.23215 | 41618.54282 | 1281.65 |
| 1283.37 | 18999.7 | 18847.4 | 15891.19649 | 15939.20494 | 1283.37 | 48396.2 | 48062.1 | 41918.13233 | 41712.43953 | 1283.37 |
| 1285.08 | 19041.1 | 18909.9 | 16041.32291 | 15994.01458 | 1285.08 | 48434.3 | 48164.5 | 42062.52727 | 41805.93625 | 1285.08 |
| 1286.79 | 19071.7 | 18972.3 | 16031.82124 | 16048.82423 | 1286.79 | 48583.2 | 48266.5 | 42377.01626 | 41899.23299 | 1286.79 |
| 1288.5 | 19106.4 | 19034.5 | 16062.72668 | 16103.63387 | 1288.5 | 48806.6 | 48368 | 42229.12144 | 41992.02974 | 1288.5 |
| 1290.22 | 19295.6 | 19096.7 | 16387.98391 | 16158.24348 | 1290.22 | 48895.9 | 48469.1 | 42270.71998 | 42084.5265 | 1290.22 |
| 1291.93 | 19287.7 | 19158.7 | 16425.69055 | 16212.75308 | 1291.93 | 48968.2 | 48569.7 | 42430.71438 | 42176.72327 | 1291.93 |
| 1293.64 | 19442 | 19220.5 | 16396.38539 | 16267.06263 | 1293.64 | 49107.4 | 48670 | 42549.71021 | 42268.42006 | 1293.64 |
| 1295.35 | 19401.2 | 19282.2 | 16509.70533 | 16321.37219 | 1295.35 | 49273.4 | 48769.8 | 42774.70233 | 42359.91686 | 1295.35 |
| 1297.07 | 19385 | 19343.8 | 16613.02352 | 16375.68175 | 1297.07 | 49352.6 | 48869.2 | 42916.99735 | 42450.91367 | 1297.07 |
| 1298.78 | 19531.2 | 19405.3 | 16540.81081 | 16429.69125 | 1298.78 | 49394.3 | 48968.1 | 42958.69589 | 42541.6105 | 1298.78 |
| 1300.49 | 19422 | 19466.6 | 16600.92139 | 16483.70076 | 1300.49 | 49453.8 | 49066.6 | 43115.39041 | 42632.00733 | 1300.49 |
| 1302.2 | 19521.7 | 19527.8 | 16659.63172 | 16537.61024 | 1302.2 | 49509.9 | 49164.7 | 43104.59078 | 42722.00418 | 1302.2 |
| 1303.91 | 19612 | 19588.8 | 16723.04288 | 16591.41971 | 1303.91 | 49605.6 | 49262.4 | 43281.88458 | 42811.60104 | 1303.91 |
| 1305.62 | 19624.1 | 19649.7 | 16704.93969 | 16645.12917 | 1305.62 | 49754.8 | 49359.6 | 43295.78409 | 42900.89792 | 1305.62 |
| 1307.33 | 19838.1 | 19710.5 | 16835.66269 | 16698.7386 | 1307.33 | 49885.5 | 49456.4 | 43148.18926 | 42989.7948 | 1307.33 |
| 1309.04 | 19868.3 | 19771.1 | 16940.68118 | 16752.148 | 1309.04 | 49697.8 | 49552.8 | 43294.28414 | 43078.29171 | 1309.04 |
| 1310.76 | 19799 | 19831.6 | 16913.07632 | 16805.5574 | 1310.76 | 49940.5 | 49648.7 | 43245.28586 | 43166.48862 | 1310.76 |
| 1312.47 | 19938.4 | 19892 | 16932.17968 | 16858.76676 | 1312.47 | 50280 | 49744.2 | 43461.57829 | 43254.28554 | 1312.47 |
| 1314.18 | 20128.1 | 19952.2 | 16841.9638 | 16911.97612 | 1314.18 | 50222.1 | 49839.3 | 43673.77086 | 43341.68248 | 1314.18 |
| 1315.89 | 20277.3 | 20012.2 | 16859.16683 | 16964.98545 | 1315.89 | 50465.6 | 49933.9 | 43716.76935 | 43428.77943 | 1315.89 |
| 1317.6 | 20174.3 | 20072.2 | 17058.50191 | 17017.89476 | 1317.6 | 50675.4 | 50028.1 | 43960.76081 | 43515.57639 | 1317.6 |
| 1319.31 | 20250.4 | 20132 | 17228.53183 | 17070.70406 | 1319.31 | 50675.2 | 50121.9 | 44126.05502 | 43601.87337 | 1319.31 |
| 1321.02 | 20328.5 | 20191.6 | 17160.01977 | 17123.41333 | 1321.02 | 50916.8 | 50215.3 | 44184.05299 | 43687.87036 | 1321.02 |
| 1322.73 | 20333.5 | 20251.1 | 17304.14514 | 17176.02259 | 1322.73 | 50812.6 | 50308.2 | 44208.95212 | 43773.46737 | 1322.73 |
| 1324.44 | 20562 | 20310.4 | 17366.65614 | 17228.53183 | 1324.44 | 50897.5 | 50400.7 | 44297.04903 | 43858.76438 | 1324.44 |
| 1326.15 | 20450.2 | 20369.6 | 17246.73503 | 17280.84104 | 1326.15 | 50835.3 | 50492.7 | 44099.65594 | 43943.66141 | 1326.15 |
| 1327.86 | 20614.3 | 20428.6 | 17353.75387 | 17333.05022 | 1327.86 | 50833.6 | 50584.3 | 44347.94725 | 44028.25844 | 1327.86 |
| 1329.57 | 20560 | 20487.5 | 17417.76513 | 17385.15939 | 1329.57 | 50926.7 | 50675.5 | 44478.74267 | 44112.3555 | 1329.57 |

| | | | | | | | | | | | |
|---|---|---|---|---|---|---|---|---|---|---|---|
| 1331.28 | 20491.2 | 20546.3 | 17393.26082 | 17437.16855 | 1331.28 | 50959.6 | 50766.3 | 44226.25151 | 44196.15257 | 1331.28 |
| 1332.98 | 20687 | 20604.9 | 17464.57337 | 17489.07768 | 1332.98 | 51106.2 | 50856.6 | 44586.63889 | 44279.64964 | 1332.98 |
| 1334.69 | 20756.5 | 20663.3 | 17441.56932 | 17540.8868 | 1334.69 | 51195.6 | 50946.5 | 44672.43589 | 44362.64674 | 1334.69 |
| 1336.4 | 20805.3 | 20721.6 | 17492.97837 | 17592.49588 | 1336.4 | 51330 | 51035.9 | 44527.54096 | 44445.44384 | 1336.4 |
| 1338.11 | 20928.7 | 20779.7 | 17718.41804 | 17644.00494 | 1338.11 | 51439.7 | 51125 | 44858.32938 | 44527.74096 | 1338.11 |
| 1339.82 | 21091.9 | 20837.7 | 17769.72707 | 17695.41399 | 1339.82 | 51504.7 | 51213.6 | 44840.03002 | 44609.73809 | 1339.82 |
| 1341.53 | 20919.1 | 20895.5 | 17792.03099 | 17746.72302 | 1341.53 | 51613.4 | 51301.7 | 44818.93076 | 44691.33523 | 1341.53 |
| 1343.24 | 20804 | 20953.2 | 17775.92816 | 17797.93203 | 1343.24 | 51657.2 | 51389.4 | 44859.02936 | 44772.63238 | 1343.24 |
| 1344.95 | 20899.4 | 21010.7 | 17818.43564 | 17848.94101 | 1344.95 | 51638.8 | 51476.7 | 44858.42938 | 44853.42955 | 1344.95 |
| 1346.65 | 21016 | 21068.1 | 17914.45254 | 17899.84997 | 1346.65 | 51612 | 51563.6 | 45065.02214 | 44934.02673 | 1346.65 |
| 1348.36 | 20969.6 | 21125.3 | 17876.7459 | 17950.65891 | 1348.36 | 51673.9 | 51650 | 45089.62128 | 45014.12393 | 1348.36 |
| 1350.07 | 21035.3 | 21182.3 | 17968.06197 | 18001.36783 | 1350.07 | 51861 | 51736 | 45047.42276 | 45093.92113 | 1350.07 |
| 1351.78 | 21376.8 | 21239.2 | 18006.56875 | 18051.87672 | 1351.78 | 51864.1 | 51821.6 | 45340.01252 | 45173.31835 | 1351.78 |
| 1353.49 | 21380.5 | 21295.9 | 18005.36854 | 18102.28559 | 1353.49 | 51954.6 | 51906.7 | 45392.51068 | 45252.31559 | 1353.49 |
| 1355.19 | 21313.8 | 21352.5 | 18228.00772 | 18152.59444 | 1355.19 | 52317.5 | 51991.4 | 45418.40977 | 45331.01283 | 1355.19 |
| 1356.9 | 21472.9 | 21408.9 | 18321.72421 | 18202.70326 | 1356.9 | 52493.4 | 52075.7 | 45633.00226 | 45409.31009 | 1356.9 |
| 1358.61 | 21685.1 | 21465.1 | 18312.82264 | 18252.71206 | 1358.61 | 52600.4 | 52159.6 | 45748.69821 | 45487.30736 | 1358.61 |
| 1360.32 | 21624.6 | 21521.2 | 18418.44123 | 18302.72086 | 1360.32 | 52531.7 | 52243 | 45905.9927 | 45564.90464 | 1360.32 |
| 1362.02 | 21618.6 | 21577.1 | 18403.43859 | 18352.42961 | 1362.02 | 52496.4 | 52325.9 | 45960.39079 | 45642.10194 | 1362.02 |
| 1363.73 | 21757.3 | 21632.9 | 18435.04415 | 18402.13836 | 1363.73 | 52669.7 | 52408.5 | 46071.6869 | 45718.99925 | 1363.73 |
| 1365.44 | 21731.9 | 21688.5 | 18564.867 | 18451.64707 | 1365.44 | 52851.7 | 52490.6 | 46127.48494 | 45795.49657 | 1365.44 |
| 1367.14 | 21638 | 21743.9 | 18574.76874 | 18501.05577 | 1367.14 | 52880.2 | 52572.3 | 46050.98762 | 45871.5939 | 1367.14 |
| 1368.85 | 21835.3 | 21799.2 | 18616.27604 | 18550.26443 | 1368.85 | 53012.1 | 52653.6 | 46216.08184 | 45947.29125 | 1368.85 |
| 1370.56 | 22024.7 | 21854.3 | 18663.48435 | 18599.37307 | 1370.56 | 53271.5 | 52734.4 | 46127.28495 | 46022.68861 | 1370.56 |
| 1372.26 | 22078.9 | 21909.2 | 18668.78528 | 18648.38169 | 1372.26 | 53211.8 | 52814.8 | 46381.67604 | 46097.68599 | 1372.26 |
| 1373.97 | 22095.7 | 21964 | 18735.49702 | 18697.19028 | 1373.97 | 53252.1 | 52894.8 | 46491.5722 | 46172.38337 | 1373.97 |
| 1375.67 | 22022.8 | 22018.5 | 18783.20542 | 18745.99887 | 1375.67 | 53423.5 | 52974.3 | 46565.56961 | 46246.68077 | 1375.67 |
| 1377.38 | 22061 | 22073 | 18932.33166 | 18794.50741 | 1377.38 | 53478.4 | 53053.4 | 46675.06577 | 46320.57818 | 1377.38 |
| 1379.09 | 22342.5 | 22127.2 | 18973.73895 | 18843.01594 | 1379.09 | 53424.4 | 53132.1 | 46534.07071 | 46394.17561 | 1379.09 |
| 1380.79 | 22393.9 | 22181.3 | 18803.20894 | 18891.32444 | 1380.79 | 53491.1 | 53210.4 | 46682.56551 | 46467.37304 | 1380.79 |
| 1382.5 | 22429.8 | 22235.3 | 19015.5463 | 18939.53293 | 1382.5 | 53491.3 | 53288.2 | 46772.06238 | 46540.1705 | 1382.5 |
| 1384.2 | 22489.1 | 22289 | 19044.35137 | 18987.54138 | 1384.2 | 53720.8 | 53365.6 | 46964.55564 | 46612.66796 | 1384.2 |
| 1385.91 | 22521.8 | 22342.6 | 19183.07579 | 19035.44981 | 1385.91 | 54039.7 | 53442.6 | 47052.35256 | 46684.76543 | 1385.91 |
| 1387.61 | 22561.9 | 22396 | 19292.19499 | 19083.25822 | 1387.61 | 53964.8 | 53519.2 | 47112.75045 | 46756.56292 | 1387.61 |
| 1389.32 | 22589.1 | 22449.2 | 19339.4033 | 19130.8666 | 1389.32 | 54093.3 | 53595.3 | 47328.74288 | 46827.86042 | 1389.32 |
| 1391.02 | 22733.6 | 22502.3 | 19338.90321 | 19178.37496 | 1391.02 | 54112 | 53671 | 47428.3394 | 46898.85794 | 1391.02 |
| 1392.73 | 22698 | 22555.2 | 19342.50384 | 19225.7833 | 1392.73 | 54116.2 | 53746.3 | 47425.5395 | 46969.55546 | 1392.73 |
| 1394.43 | 22689.8 | 22607.9 | 19318.3996 | 19272.99161 | 1394.43 | 54263.2 | 53821.2 | 47412.43995 | 47039.853 | 1394.43 |
| 1396.14 | 22793 | 22660.5 | 19408.51546 | 19320.0999 | 1396.14 | 54440.8 | 53895.6 | 47407.34013 | 47109.75055 | 1396.14 |
| 1397.84 | 22912.2 | 22712.9 | 19542.23899 | 19367.00815 | 1397.84 | 54301.5 | 53969.6 | 47584.03395 | 47179.34812 | 1397.84 |
| 1399.55 | 22908.6 | 22765.1 | 19560.54221 | 19413.81639 | 1399.55 | 54487.4 | 54043.2 | 47775.72723 | 47248.54569 | 1399.55 |
| 1401.25 | 22945.3 | 22817.1 | 19584.04635 | 19460.42459 | 1401.25 | 54849.9 | 54116.3 | 47960.62076 | 47317.34328 | 1401.25 |
| 1402.95 | 23048.8 | 22869 | 19568.74365 | 19507.0328 | 1402.95 | 55012.1 | 54189.1 | 47921.62213 | 47385.84089 | 1402.95 |
| 1404.66 | 23002.3 | 22920.6 | 19707.56808 | 19553.34094 | 1404.66 | 54971.4 | 54261.4 | 48064.11714 | 47453.9385 | 1404.66 |
| 1406.36 | 23126.5 | 22972.2 | 19886.79963 | 19599.64909 | 1406.36 | 55134.6 | 54333.3 | 48290.30922 | 47521.63613 | 1406.36 |
| 1408.07 | 23305.6 | 23023.5 | 19845.79241 | 19645.75721 | 1408.07 | 55273.9 | 54404.8 | 48235.91112 | 47589.03377 | 1408.07 |
| 1409.77 | 23366.6 | 23074.6 | 19899.40184 | 19691.66529 | 1409.77 | 55279 | 54475.8 | 48298.40893 | 47656.13142 | 1409.77 |
| 1411.47 | 23243.5 | 23125.6 | 19997.91918 | 19737.47335 | 1411.47 | 55238.4 | 54546.4 | 48226.71144 | 47722.72909 | 1411.47 |
| 1413.17 | 23645.8 | 23176.4 | 20021.9234 | 19783.18139 | 1413.17 | 55330.6 | 54616.7 | 48663.99613 | 47789.02677 | 1413.17 |
| 1414.88 | 23692 | 23227 | 20080.4337 | 19828.6894 | 1414.88 | 55565.4 | 54686.5 | 48701.89481 | 47855.02446 | 1414.88 |
| 1416.58 | 23505.8 | 23277.5 | 20084.33439 | 19874.09739 | 1416.58 | 55319.1 | 54755.8 | 48374.00629 | 47920.62216 | 1416.58 |
| 1418.28 | 23707.8 | 23327.7 | 20060.83025 | 19919.30535 | 1418.28 | 55377 | 54824.8 | 48575.19924 | 47985.81988 | 1418.28 |
| 1419.99 | 23667.9 | 23377.8 | 20172.14984 | 19964.41328 | 1419.99 | 55716.6 | 54893.3 | 48837.29007 | 48050.71761 | 1419.99 |
| 1421.69 | 23801.5 | 23427.7 | 20341.37962 | 20009.32119 | 1421.69 | 55777.8 | 54961.5 | 48770.09242 | 48115.21535 | 1421.69 |
| 1423.39 | 23612.9 | 23477.5 | 20208.95632 | 20054.12907 | 1423.39 | 55746.7 | 55029.2 | 48740.69345 | 48179.4131 | 1423.39 |
| 1425.09 | 23588.8 | 23527 | 20203.7554 | 20098.83694 | 1425.09 | 55697 | 55096.5 | 48726.09396 | 48243.21087 | 1425.09 |
| 1426.8 | 23778.7 | 23576.4 | 20282.66929 | 20143.34477 | 1426.8 | 55662.3 | 55163.4 | 48841.88991 | 48306.60865 | 1426.8 |
| 1428.5 | 23918.2 | 23625.6 | 20330.97779 | 20187.65257 | 1428.5 | 55952.4 | 55229.8 | 48963.28565 | 48369.70644 | 1428.5 |
| 1430.2 | 24001.5 | 23674.6 | 20460.30055 | 20231.86035 | 1430.2 | 56007 | 55295.9 | 48971.38537 | 48432.40424 | 1430.2 |
| 1431.9 | 23963.4 | 23723.4 | 20407.09119 | 20275.96811 | 1431.9 | 55717 | 55361.5 | 48915.68732 | 48494.80206 | 1431.9 |
| 1433.6 | 23950.6 | 23772.1 | 20551.11653 | 20319.87584 | 1433.6 | 56192.9 | 55426.8 | 49199.0774 | 48556.79989 | 1433.6 |
| 1435.3 | 23958.3 | 23820.5 | 20658.13536 | 20363.68355 | 1435.3 | 56121.6 | 55491.6 | 49215.27683 | 48618.49773 | 1435.3 |
| 1437.01 | 24104.5 | 23868.8 | 20596.02443 | 20407.29122 | 1437.01 | 56120 | 55556 | 48985.38488 | 48679.79558 | 1437.01 |
| 1438.71 | 24182.2 | 23916.9 | 20638.83197 | 20450.79888 | 1438.71 | 56201.3 | 55620 | 49267.77499 | 48740.69345 | 1438.71 |
| 1440.41 | 24249.6 | 23964.9 | 20732.94853 | 20494.1065 | 1440.41 | 56092.4 | 55683.6 | 49425.86946 | 48801.29133 | 1440.41 |
| 1442.11 | 24188.2 | 24012.6 | 20687.44052 | 20537.3141 | 1442.11 | 56422.6 | 55746.7 | 49295.67402 | 48861.58922 | 1442.11 |
| 1443.81 | 24357.2 | 24060.2 | 20792.75905 | 20580.32167 | 1443.81 | 56340.2 | 55809.5 | 49423.46954 | 48921.48712 | 1443.81 |
| 1445.51 | 24507.1 | 24107.5 | 20851.76944 | 20623.22922 | 1445.51 | 56437.5 | 55871.9 | 49470.86788 | 48980.98504 | 1445.51 |
| 1447.21 | 24334.9 | 24154.7 | 20824.76469 | 20665.93674 | 1447.21 | 56539.7 | 55933.8 | 49374.67125 | 49040.18296 | 1447.21 |
| 1448.91 | 24300.4 | 24201.8 | 20852.36954 | 20708.54423 | 1448.91 | 56424.3 | 55995.4 | 49631.26227 | 49099.0809 | 1448.91 |
| 1450.61 | 24525.2 | 24248.6 | 20967.58982 | 20750.9517 | 1450.61 | 56574 | 56056.5 | 49714.05937 | 49157.47886 | 1450.61 |
| 1452.31 | 24492.6 | 24295.2 | 21057.90571 | 20793.25914 | 1452.31 | 56712.2 | 56117.2 | 49646.66173 | 49215.67682 | 1452.31 |
| 1454.01 | 24517.8 | 24341.7 | 21074.80869 | 20835.36655 | 1454.01 | 56798.8 | 56177.6 | 49882.95345 | 49273.47479 | 1454.01 |
| 1455.71 | 24610.2 | 24388 | 21143.6208 | 20877.37394 | 1455.71 | 56767.3 | 56237.5 | 49835.65511 | 49330.87278 | 1455.71 |
| 1457.41 | 24638.1 | 24434.1 | 21066.9073 | 20919.1813 | 1457.41 | 56846.3 | 56297 | 49814.25586 | 49387.97079 | 1457.41 |
| 1459.11 | 24764.1 | 24480 | 21072.80834 | 20960.88864 | 1459.11 | 56971.3 | 56356.2 | 49771.85734 | 49444.7688 | 1459.11 |
| 1460.81 | 24927.9 | 24525.7 | 21363.35947 | 21002.39595 | 1460.81 | 56935.5 | 56414.9 | 49877.05366 | 49501.16682 | 1460.81 |
| 1462.51 | 24977.3 | 24571.2 | 21361.75918 | 21043.80323 | 1462.51 | 56902.9 | 56473.2 | 50035.24812 | 49557.16486 | 1462.51 |
| 1464.21 | 24918.7 | 24616.6 | 21391.9645 | 21085.01048 | 1464.21 | 57047.7 | 56531.1 | 50001.14932 | 49612.86291 | 1464.21 |
| 1465.91 | 24936.9 | 24661.8 | 21445.47392 | 21126.0177 | 1465.91 | 56920.8 | 56588.7 | 50045.34777 | 49668.26097 | 1465.91 |
| 1467.61 | 24820.7 | 24706.7 | 21287.54613 | 21167.02492 | 1467.61 | 57160.6 | 56645.8 | 49949.65112 | 49723.25905 | 1467.61 |
| 1469.31 | 24912.3 | 24751.5 | 21391.26438 | 21207.73208 | 1469.31 | 57261.5 | 56702.5 | 50102.44577 | 49777.85713 | 1469.31 |
| 1471.01 | 24984.3 | 24796.2 | 21389.06399 | 21248.33923 | 1471.01 | 57105.6 | 56758.8 | 50194.44255 | 49832.25523 | 1471.01 |
| 1472.7 | 24831.6 | 24840.6 | 21330.35366 | 21288.84635 | 1472.7 | 57201.1 | 56814.8 | 50055.9474 | 49886.15334 | 1472.7 |
| 1474.4 | 25021.6 | 24884.8 | 21492.78224 | 21329.15345 | 1474.4 | 57180.8 | 56870.3 | 50285.13937 | 49939.85146 | 1474.4 |
| 1476.1 | 25139.5 | 24928.9 | 21533.68944 | 21369.2605 | 1476.1 | 57201.4 | 56925.4 | 50319.03819 | 49993.1496 | 1476.1 |
| 1477.8 | 25135.6 | 24972.8 | 21641.50841 | 21409.26755 | 1477.8 | 57267.8 | 56980.2 | 50201.6423 | 50046.04774 | 1477.8 |
| 1479.5 | 25204.1 | 25016.5 | 21655.81093 | 21449.17457 | 1479.5 | 57290.5 | 57034.6 | 50292.8391 | 50098.6459 | 1479.5 |
| 1481.19 | 25114.9 | 25060 | 21450.97488 | 21488.88156 | 1481.19 | 57268.7 | 57088.5 | 50458.13332 | 50150.94407 | 1481.19 |
| 1482.89 | 25062.2 | 25103.3 | 21616.60403 | 21528.38851 | 1482.89 | 57426.2 | 57142.1 | 50413.03489 | 50202.84225 | 1482.89 |
| 1484.59 | 25184.8 | 25146.4 | 21562.59453 | 21567.79544 | 1484.59 | 57290.6 | 57195.3 | 50289.03924 | 50254.44045 | 1484.59 |
| 1486.29 | 25174.1 | 25189.3 | 21452.07508 | 21607.00234 | 1486.29 | 57410.7 | 57248.1 | 50351.53705 | 50305.73865 | 1486.29 |
| 1487.98 | 25343.8 | 25232.1 | 21728.52373 | 21646.10922 | 1487.98 | 57606.5 | 57300.5 | 50432.03423 | 50356.63687 | 1487.98 |
| 1489.68 | 25381.5 | 25274.7 | 21575.99689 | 21685.01607 | 1489.68 | 57316.4 | 57352.5 | 50395.7355 | 50407.1351 | 1489.68 |
| 1491.38 | 25345.7 | 25317.1 | 21552.49275 | 21723.8229 | 1491.38 | 57450.3 | 57404.1 | 50547.1302 | 50457.43334 | 1491.38 |
| 1493.08 | 25487.6 | 25359.2 | 21961.76477 | 21762.42969 | 1493.08 | 57641.8 | 57455.4 | 50709.42452 | 50507.33159 | 1493.08 |
| 1494.77 | 25378.8 | 25401.3 | 21909.85564 | 21800.93647 | 1494.77 | 57555.7 | 57506.2 | 50654.72643 | 50556.92986 | 1494.77 |
| 1496.47 | 25404.7 | 25443.1 | 21880.05039 | 21839.24321 | 1496.47 | 57521.6 | 57556.7 | 50691.52514 | 50606.12813 | 1496.47 |
| 1498.17 | 25581 | 25484.7 | 21953.56333 | 21877.34992 | 1498.17 | 57520.4 | 57606.8 | 50693.62507 | 50655.02642 | 1498.17 |
| 1499.86 | 25473.9 | 25526.2 | 21772.03138 | 21915.35661 | 1499.86 | 57368.4 | 57656.5 | 50577.92912 | 50703.52472 | 1499.86 |
| 1501.56 | 25516.2 | 25567.4 | 21823.94052 | 21953.26328 | 1501.56 | 57601.9 | 57705.8 | 50623.62752 | 50751.82303 | 1501.56 |
| 1503.25 | 25685.4 | 25608.5 | 21997.07099 | 21990.86989 | 1503.25 | 57772.5 | 57754.7 | 50710.92447 | 50799.72136 | 1503.25 |
| 1504.95 | 25641.7 | 25649.4 | 22051.48056 | 22028.47651 | 1504.95 | 57707.5 | 57803.3 | 50646.02674 | 50847.21969 | 1504.95 |
| 1506.65 | 25645.2 | 25690.1 | 22127.59395 | 22065.88309 | 1506.65 | 57892.8 | 57851.5 | 50721.62409 | 50894.41804 | 1506.65 |
| 1508.34 | 25668.3 | 25730.6 | 22070.18385 | 22103.08964 | 1508.34 | 57877.3 | 57899.3 | 50898.01791 | 50941.41639 | 1508.34 |
| 1510.04 | 25751.9 | 25771 | 22029.17664 | 22140.09615 | 1510.04 | 57711.1 | 57946.7 | 50741.52339 | 50987.91477 | 1510.04 |
| 1511.73 | 25883.5 | 25811.1 | 22183.50379 | 22177.00265 | 1511.73 | 57842 | 57993.8 | 50809.921 | 51034.21315 | 1511.73 |
| 1513.43 | 25797.7 | 25851.1 | 22170.70154 | 22213.80913 | 1513.43 | 57686.9 | 58040.4 | 50901.3178 | 51080.11154 | 1513.43 |
| 1515.12 | 25770.4 | 25890.8 | 22270.41909 | 22250.41557 | 1515.12 | 57656.7 | 58086.7 | 50932.8167 | 51125.70994 | 1515.12 |
| 1516.82 | 25907.7 | 25930.4 | 22365.33579 | 22286.82197 | 1516.82 | 57877 | 58132.7 | 51077.41163 | 51170.90836 | 1516.82 |
| 1518.51 | 25886 | 25969.8 | 22156.69908 | 22323.12836 | 1518.51 | 57863.2 | 58178.2 | 51162.00867 | 51215.90678 | 1518.51 |
| 1520.21 | 25841.2 | 26009 | 22131.29461 | 22359.23472 | 1520.21 | 57893.7 | 58223.4 | 50976.01518 | 51260.50522 | 1520.21 |
| 1521.9 | 25865.5 | 26048.1 | 22407.54322 | 22395.24105 | 1521.9 | 57886.7 | 58268.2 | 51059.41226 | 51304.70368 | 1521.9 |
| 1523.6 | 26058.3 | 26086.9 | 22330.7297 | 22431.14737 | 1523.6 | 57691.8 | 58312.6 | 51038.513 | 51348.70213 | 1523.6 |
| 1525.29 | 26128.6 | 26125.5 | 22334.43035 | 22466.75364 | 1525.29 | 57726.3 | 58356.7 | 50795.92149 | 51392.30061 | 1525.29 |
| 1526.99 | 26035.2 | 26164 | 22404.54269 | 22502.25989 | 1526.99 | 57836.1 | 58400.4 | 51026.31342 | 51435.59909 | 1526.99 |
| 1528.68 | 26162.7 | 26202.3 | 22523.46362 | 22537.66612 | 1528.68 | 57888.6 | 58443.7 | 51072.31181 | 51478.59759 | 1528.68 |
| 1530.37 | 26136.9 | 26240.4 | 22709.29632 | 22572.87231 | 1530.37 | 57949.8 | 58486.7 | 51092.81109 | 51521.29609 | 1530.37 |
| 1532.07 | 26102 | 26278.3 | 22560.97022 | 22607.87847 | 1532.07 | 58051.8 | 58529.2 | 51096.91095 | 51563.59461 | 1532.07 |
| 1533.76 | 26215.9 | 26316 | 22473.85489 | 22642.78462 | 1533.76 | 58171.1 | 58571.5 | 51216.60676 | 51605.69314 | 1533.76 |
| 1535.45 | 26348.9 | 26353.5 | 22632.38279 | 22677.49072 | 1535.45 | 58139.6 | 58613.3 | 51110.01049 | 51647.39168 | 1535.45 |
| 1537.15 | 26449.8 | 26390.9 | 22720.39827 | 22712.09681 | 1537.15 | 58289.7 | 58654.8 | 51247.50568 | 51688.79023 | 1537.15 |
| 1538.84 | 26405 | 26428 | 22680.7913 | 22746.50287 | 1538.84 | 58218.3 | 58696 | 51413.49987 | 51729.78879 | 1538.84 |
| 1540.53 | 26310 | 26465 | 22719.29808 | 22780.80891 | 1540.53 | 57993 | 58736.8 | 51302.40376 | 51770.58736 | 1540.53 |
| 1542.23 | 26338.5 | 26501.8 | 22722.69868 | 22814.91491 | 1542.23 | 58203.2 | 58777.2 | 51543.09533 | 51810.98595 | 1542.23 |
| 1543.92 | 26424.5 | 26538.4 | 22720.69833 | 22848.92089 | 1543.92 | 58102.4 | 58817.2 | 51360.30173 | 51851.08454 | 1543.92 |
| 1545.61 | 26483.8 | 26574.8 | 22727.89959 | 22882.72684 | 1545.61 | 58047 | 58856.9 | 51567.69447 | 51890.98315 | 1545.61 |
| 1547.31 | 26469.5 | 26611 | 22711.29667 | 22916.33275 | 1547.31 | 58258.5 | 58896.3 | 51586.7938 | 51930.38177 | 1547.31 |
| 1549 | 26592.7 | 26647.1 | 22733.50058 | 22949.83865 | 1549 | 58215.1 | 58935.3 | 51411.29994 | 51969.5804 | 1549 |
| 1550.69 | 26598.8 | 26683 | 23016.75043 | 22983.14451 | 1550.69 | 58296.9 | 58973.9 | 51528.99582 | 52008.47903 | 1550.69 |
| 1552.38 | 26528.1 | 26718.6 | 22838.91913 | 23016.35036 | 1552.38 | 58356 | 59012.2 | 51365.10156 | 52046.97769 | 1552.38 |
| 1554.07 | 26295.1 | 26754.1 | 22813.41464 | 23049.35616 | 1554.07 | 58226.1 | 59050.1 | 51421.1996 | 52085.27634 | 1554.07 |
| 1555.77 | 26516.5 | 26789.4 | 22908.53138 | 23082.26195 | 1555.77 | 58409.4 | 59087.7 | 51426.19942 | 52123.17502 | 1555.77 |
| 1557.46 | 26713.7 | 26824.5 | 22814.81489 | 23114.96771 | 1557.46 | 58536.6 | 59124.9 | 51529.89579 | 52160.7737 | 1557.46 |
| 1559.15 | 26718.8 | 26859.5 | 23022.35141 | 23147.57345 | 1559.15 | 58465.6 | 59161.7 | 51577.59412 | 52198.0724 | 1559.15 |
| 1560.84 | 26837.7 | 26894.2 | 22996.24682 | 23179.97915 | 1560.84 | 58330.5 | 59198.2 | 51473.39777 | 52235.0711 | 1560.84 |
| 1562.53 | 26728.2 | 26928.8 | 23070.95997 | 23212.18482 | 1562.53 | 58411.3 | 59234.4 | 51783.18692 | 52271.86981 | 1562.53 |
| 1564.22 | 26630.5 | 26963.2 | 22949.33856 | 23244.29047 | 1564.22 | 58333.6 | 59270.2 | 51590.49367 | 52308.16854 | 1564.22 |
| 1565.91 | 26645.3 | 26997.4 | 23069.85977 | 23276.19608 | 1565.91 | 58357.4 | 59305.7 | 51493.39707 | 52344.26728 | 1565.91 |
| | | | | | | | | | | 1567.61 |
| | | | | | | | | | | 1569.3 |
| | | | | | | | | | | 1570.99 |
| | | | | | | | | | | 1572.68 |
| | | | | | | | | | | 1574.37 |
| | | | | | | | | | | 1576.06 |
| | | | | | | | | | | 1577.75 |
| | | | | | | | | | | 1579.44 |

| 532 nm, T = 1930 K, n = 0.87, 3000 sec | 532 nm, T = 1930 K, n = 0.87, 3000 sec | CO2, T = 1942 K, n = 0.78, 9780 sec | CO2, T = 1942 K, n = 0.78, 9780 sec |
|---|---|---|---|
| 977.72 | 921.819 | 1058.66267 | 976.25422 |
| 1056.48 | 925.191 | 1095.45131 | 979.82879 |
| 904.111 | 928.57 | 1018.91719 | 983.41186 |
| 966.996 | 931.957 | 1057.41763 | 987.00343 |
| 1017.27 | 935.352 | 1038.96749 | 990.60245 |
| 1050.17 | 938.755 | 1188.18579 | 994.21103 |
| 1107.48 | 942.166 | 1008.56244 | 997.82706 |
| 947.461 | 945.584 | 1049.53382 | 1001.45266 |
| 986.25 | 949.011 | 1064.68158 | 1005.0857 |
| 1073.74 | 952.445 | 1041.97322 | 1008.7283 |
| 937.415 | 955.888 | 988.90341 | 1012.37836 |
| 988.048 | 959.338 | 1178.63803 | 1016.03691 |
| 1002.66 | 962.796 | 1066.88246 | 1019.70398 |
| 1054.54 | 966.262 | 1120.62848 | 1023.37954 |
| 1169.28 | 969.736 | 1150.23929 | 1027.06362 |
| 1045.95 | 973.218 | 1151.50453 | 1030.75514 |
| 1091.88 | 976.708 | 1122.21268 | 1034.45622 |
| 1028.52 | 980.206 | 1150.78154 | 1038.16581 |
| 1024.29 | 983.711 | 1181.8596 | 1041.88285 |
| 972.92 | 987.225 | 1192.94904 | 1045.60945 |
| 1117.16 | 990.746 | 1092.35733 | 1049.3435 |
| 1186.26 | 994.276 | 1149.26113 | 1053.08711 |
| 1089.04 | 997.813 | 1213.70319 | 1056.83817 |
| 1020.55 | 1001.36 | 1046.39518 | 1060.5988 |
| 1076.11 | 1004.91 | 1046.63546 | 1064.36261 |
| 1111.46 | 1008.47 | 1135.77943 | 1068.14769 |
| 1196.68 | 1012.04 | 1090.19898 | 1071.93278 |
| 1094.87 | 1015.62 | 1138.46939 | 1075.71786 |
| 1087.47 | 1019.2 | 1129.09175 | 1079.5242 |
| 1064.93 | 1022.8 | 1176.91561 | 1083.33055 |
| 1097.88 | 1026.4 | 1229.94927 | 1087.15816 |
| 1037.68 | 1030.01 | 1198.83931 | 1090.98577 |
| 1088.98 | 1033.62 | 1273.64782 | 1094.81338 |
| 1143.23 | 1037.25 | 1096.68465 | 1098.66225 |
| 1054.5 | 1040.88 | 1125.56184 | 1102.51113 |
| 1048.9 | 1044.52 | 1212.2891 | 1106.38126 |
| 1109.06 | 1048.17 | 1216.92476 | 1110.2514 |
| 1141.68 | 1051.83 | 1270.80901 | 1114.12154 |
| 1181.43 | 1055.49 | 1274.88116 | 1118.01295 |
| 1212.54 | 1059.16 | 1218.06241 | 1121.91498 |
| 1133.93 | 1062.85 | 1258.35864 | 1125.81702 |
| 1137.95 | 1066.53 | 1227.47196 | 1129.72968 |
| 1150.82 | 1070.23 | 1254.29712 | 1133.65298 |
| 1159.82 | 1073.93 | 1190.11023 | 1137.57629 |
| 1084.08 | 1077.65 | 1222.32595 | 1141.52085 |
| 1139.4 | 1081.37 | 1261.25061 | 1145.46541 |
| 1130.28 | 1085.1 | 1258.50749 | 1149.42061 |
| 1134.47 | 1088.83 | 1312.90208 | 1153.38644 |
| 1206.4 | 1092.58 | 1263.48339 | 1157.35227 |
| 1144.28 | 1096.33 | 1218.95552 | 1161.33936 |
| 1209.72 | 1100.09 | 1317.37826 | 1165.32646 |
| 1231.68 | 1103.86 | 1277.27341 | 1169.32418 |
| 1209.69 | 1107.64 | 1320.5892 | 1173.33254 |
| 1242.66 | 1111.42 | 1303.04599 | 1177.3409 |
| 1180.22 | 1115.21 | 1194.52261 | 1181.37052 |
| 1122.61 | 1119.01 | 1219.88053 | 1185.40014 |
| 1144.67 | 1122.82 | 1314.66703 | 1189.4404 |
| 1203.68 | 1126.64 | 1220.4972 | 1193.49129 |
| 1180.41 | 1130.46 | 1326.22429 | 1197.54217 |
| 1273.5 | 1134.3 | 1285.85364 | 1201.61432 |
| 1203.37 | 1138.14 | 1269.56503 | 1205.68648 |
| 1160.22 | 1141.98 | 1236.43494 | 1209.76926 |
| 1228.82 | 1145.84 | 1206.90918 | 1213.86268 |
| 1177.15 | 1149.71 | 1229.26881 | 1217.96672 |
| 1219.34 | 1153.58 | 1312.91271 | 1222.07077 |
| 1269.64 | 1157.46 | 1290.96775 | 1226.18545 |
| 1260.83 | 1161.35 | 1332.96514 | 1230.31077 |
| 1238.03 | 1165.24 | 1215.10665 | 1234.44671 |
| 1283.64 | 1169.15 | 1302.96093 | 1238.59329 |
| 1314.75 | 1173.06 | 1259.26238 | 1242.7505 |
| 1250.83 | 1176.98 | 1299.74999 | 1246.90771 |
| 1213.92 | 1180.91 | 1296.83675 | 1251.07555 |
| 1221.39 | 1184.85 | 1253.39338 | 1255.25402 |
| 1201.48 | 1188.79 | 1241.14503 | 1259.4325 |
| 1238.12 | 1192.75 | 1345.60688 | 1263.63224 |
| 1180.99 | 1196.71 | 1337.17551 | 1267.83198 |
| 1300.23 | 1200.67 | 1327.65964 | 1272.04235 |
| 1232.56 | 1204.65 | 1378.71571 | 1276.26335 |
| 1323.99 | 1208.64 | 1397.87502 | 1280.49498 |
| 1195.9 | 1212.63 | 1371.37945 | 1284.73725 |
| 1311.6 | 1216.63 | 1381.75653 | 1288.97952 |
| 1246.48 | 1220.64 | 1396.56725 | 1293.23242 |
| 1339.63 | 1224.66 | 1402.77649 | 1297.49595 |
| 1265.81 | 1228.68 | 1362.86302 | 1301.77012 |
| 1350.87 | 1232.72 | 1386.36029 | 1306.04428 |
| 1336.48 | 1236.76 | 1365.31907 | 1310.33971 |
| 1316.88 | 1240.81 | 1402.23424 | 1314.63514 |
| 1358.6 | 1244.86 | 1417.651 | 1318.9412 |
| 1322.66 | 1248.93 | 1378.1522 | 1323.25789 |
| 1317.52 | 1253 | 1432.55742 | 1327.57459 |
| 1355.69 | 1257.08 | 1407.69922 | 1331.91254 |
| 1360.91 | 1261.17 | 1465.8151 | 1336.2505 |
| 1339.81 | 1265.27 | 1422.03149 | 1340.59909 |
| 1376.7 | 1269.37 | 1420.17085 | 1344.95831 |
| 1385.63 | 1273.49 | 1402.01097 | 1349.32817 |
| 1397.06 | 1277.61 | 1414.81219 | 1353.69802 |
| 1416.22 | 1281.74 | 1447.36814 | 1358.07851 |
| 1378.31 | 1285.88 | 1434.19479 | 1362.46963 |
| 1426.86 | 1290.02 | 1457.20297 | 1366.87138 |
| 1355.45 | 1294.18 | 1506.50472 | 1371.28376 |
| 1357.82 | 1298.34 | 1451.21702 | 1375.69615 |
| 1453.63 | 1302.51 | 1489.04656 | 1380.1298 |
| 1379.68 | 1306.69 | 1574.69996 | 1384.56344 |
| 1422.65 | 1310.87 | 1497.24403 | 1389.00772 |
| 1498.58 | 1315.07 | 1479.44564 | 1393.452 |
| 1447.51 | 1319.27 | 1450.81299 | 1397.91755 |
| 1405.36 | 1323.48 | 1536.58335 | 1402.38309 |
| 1408.31 | 1327.69 | 1519.13583 | 1406.85927 |
| 1419.03 | 1331.92 | 1545.9291 | 1411.34608 |
| 1456.29 | 1336.15 | 1532.30918 | 1415.84352 |
| 1447.15 | 1340.4 | 1612.21054 | 1420.34096 |
| 1503.02 | 1344.65 | 1580.32442 | 1424.85967 |
| 1477.31 | 1348.91 | 1571.30828 | 1429.37838 |
| 1529.18 | 1353.17 | 1591.33943 | 1433.90771 |
| 1482.35 | 1357.45 | 1559.45332 | 1438.44769 |
| 1517.74 | 1361.73 | 1585.73624 | 1442.98766 |
| 1493.79 | 1366.02 | 1574.381 | 1447.54889 |
| 1542.7 | 1370.32 | 1607.88322 | 1452.11013 |
| 1553.28 | 1374.62 | 1541.88884 | 1456.68199 |
| 1533.58 | 1378.94 | 1599.04782 | 1461.26449 |
| 1515.33 | 1383.26 | 1618.51546 | 1465.84699 |
| 1555.76 | 1387.59 | 1598.26103 | 1470.45076 |
| 1542.21 | 1391.93 | 1660.11946 | 1475.05452 |
| 1519.8 | 1396.27 | 1610.03093 | 1479.66892 |
| 1571.84 | 1400.63 | 1588.92591 | 1484.29395 |
| 1551.97 | 1404.99 | 1620.25915 | 1488.92961 |
| 1520.58 | 1409.36 | 1588.98971 | 1493.56527 |
| 1563.5 | 1413.74 | 1671.82556 | 1498.21156 |
| 1512.59 | 1418.13 | 1584.1733 | 1502.87912 |
| 1531.89 | 1422.52 | 1609.56311 | 1507.53604 |
| 1585.84 | 1426.92 | 1649.92313 | 1512.21423 |
| 1527.84 | 1431.33 | 1663.74505 | 1516.89242 |
| 1555.56 | 1435.75 | 1674.56869 | 1521.59188 |
| 1564.6 | 1440.18 | 1659.43899 | 1526.29133 |
| 1604.35 | 1444.61 | 1631.24227 | 1531.00142 |
| 1638.25 | 1449.06 | 1724.37014 | 1535.72214 |
| 1601.03 | 1453.51 | 1738.69178 | 1540.44286 |
| 1569.45 | 1457.96 | 1716.68303 | 1545.17421 |
| 1579.33 | 1462.43 | 1643.22481 | 1549.92682 |
| 1637.06 | 1466.91 | 1663.10712 | 1554.66881 |
| 1631.77 | 1471.39 | 1678.08796 | 1559.43205 |
| 1610.15 | 1475.88 | 1719.88333 | 1564.20593 |
| 1598.15 | 1480.38 | 1715.6198 | 1568.97981 |
| 1653.84 | 1484.88 | 1759.95628 | 1573.76433 |
| 1643.53 | 1489.4 | 1679.73596 | 1578.55947 |
| 1574.88 | 1493.92 | 1691.7823 | 1583.36525 |
| 1617.15 | 1498.45 | 1647.09495 | 1588.17102 |
| 1612.44 | 1502.99 | 1754.61889 | 1592.99807 |
| 1600.34 | 1507.54 | 1691.85672 | 1597.82511 |
| 1657.95 | 1512.09 | 1715.48158 | 1602.66278 |
| 1694.91 | 1516.65 | 1691.24005 | 1607.51109 |
| 1664.45 | 1521.22 | 1749.41972 | 1612.35939 |
| 1630.75 | 1525.8 | 1751.34416 | 1617.22896 |
| 1687.42 | 1530.39 | 1754.96976 | 1622.09853 |
| 1643.14 | 1534.98 | 1798.49819 | 1626.97874 |
| 1685.11 | 1539.59 | 1776.51069 | 1631.86957 |
| 1678.2 | 1544.2 | 1742.93405 | 1636.76041 |
| 1651.48 | 1548.82 | 1783.52798 | 1641.6725 |
| 1709.51 | 1553.44 | 1772.80004 | 1646.5846 |
| 1710.43 | 1558.08 | 1790.63032 | 1651.50734 |
| 1739.32 | 1562.72 | 1766.51638 | 1656.4407 |
| 1662.02 | 1567.37 | 1754.81027 | 1661.37406 |
| 1675.05 | 1572.03 | 1782.57108 | 1666.32869 |
| 1716.31 | 1576.69 | 1767.55834 | 1671.28332 |
| 1732.71 | 1581.37 | 1828.71504 | 1676.24858 |
| 1703.36 | 1586.05 | 1807.5356 | 1681.22447 |
| 1743.19 | 1590.74 | 1836.44469 | 1686.20037 |
| 1702.31 | 1595.43 | 1817.18968 | 1691.19752 |
| 1751.92 | 1600.14 | 1835.19008 | 1696.19468 |
| 1741.23 | 1604.86 | 1815.09513 | 1701.20247 |
| 1734.6 | 1609.58 | 1834.35013 | 1706.22089 |
| 1694.89 | 1614.31 | 1874.41245 | 1711.23931 |
| 1769.87 | 1619.04 | 1832.89351 | 1716.279 |
| 1739.4 | 1623.79 | 1883.26911 | 1721.31869 |
| 1800.6 | 1628.54 | 1907.25547 | 1726.36901 |
| 1774.1 | 1633.3 | 1841.2292 | 1731.42996 |
| 1747.21 | 1638.07 | 1879.4734 | 1736.49091 |
| 1776.35 | 1642.85 | 1878.67598 | 1741.57312 |
| 1794.2 | 1647.63 | 1884.34297 | 1746.65534 |
| 1806.33 | 1652.43 | 1923.14005 | 1751.74819 |
| 1805.22 | 1657.23 | 1890.92433 | 1756.85166 |
| 1840.76 | 1662.04 | 1894.2097 | 1761.95514 |
| 1768.34 | 1666.85 | 1904.35286 | 1767.07989 |
| 1808.82 | 1671.68 | 1958.29027 | 1772.20463 |
| 1793.07 | 1676.51 | 1972.11219 | 1777.34001 |
| 1834.86 | 1681.35 | 1913.83683 | 1782.48602 |
| 1840.04 | 1686.2 | 1956.28077 | 1787.63203 |
| 1815.68 | 1691.06 | 1945.53157 | 1792.78867 |
| 1906.26 | 1695.92 | 2002.78623 | 1797.96657 |
| 1869.11 | 1700.79 | 1965.32882 | 1803.14448 |
| 1849.85 | 1705.67 | 1927.2441 | 1808.32239 |
| 1902.38 | 1710.56 | 1960.23597 | 1813.52156 |
| 1883.37 | 1715.45 | 1975.65273 | 1818.72073 |
| 1911.98 | 1720.36 | 2020.26565 | 1823.93053 |
| 1873.36 | 1725.27 | 2009.59087 | 1829.15096 |
| 1863.24 | 1730.19 | 1978.35332 | 1834.38203 |
| 1865.36 | 1735.12 | 1975.79095 | 1839.6131 |
| 1923.15 | 1740.05 | 2010.20754 | 1844.86543 |
| 1900.14 | 1744.99 | 2041.80659 | 1850.11776 |
| 1930.87 | 1749.94 | 1991.88818 | 1855.38072 |
| 1906.92 | 1754.9 | 2006.44373 | 1860.64369 |
| 1936.93 | 1759.87 | 1997.42758 | 1865.92791 |
| 1934.41 | 1764.84 | 2007.36873 | 1871.21214 |
| 1934.5 | 1769.83 | 2020.20186 | 1876.507 |
| 1922.67 | 1774.81 | 2056.52162 | 1881.81249 |
| 1936.84 | 1779.81 | 2055.28828 | 1887.11799 |
| 1955.27 | 1784.82 | 2023.62544 | 1892.44474 |
| 1920.22 | 1789.83 | 2077.00997 | 1897.7715 |
| 1949.03 | 1794.85 | 2039.18042 | 1903.10889 |
| 1930.66 | 1799.88 | 2089.00314 | 1908.45691 |
| 1961.94 | 1804.92 | 2092.0865 | 1913.80493 |
| 1989.73 | 1809.96 | 2084.81404 | 1919.17422 |
| 2021.61 | 1815.01 | 2125.25911 | 1924.54351 |
| 2005.74 | 1820.07 | 2094.60634 | 1929.92342 |
| 1959.32 | 1825.14 | 2078.53038 | 1935.30334 |
| 2006.29 | 1830.22 | 2091.68247 | 1940.70453 |
| 1996.01 | 1835.3 | 2113.78692 | 1946.10571 |
| 1969.63 | 1840.39 | 2124.82319 | 1951.51752 |
| 2032.38 | 1845.49 | 2143.25951 | 1956.93997 |
| 2025.17 | 1850.6 | 2119.17747 | 1962.37305 |
| 2033.08 | 1855.71 | 2107.26935 | 1967.80613 |
| 2063.82 | 1860.83 | 2146.54488 | 1973.24984 |
| 2012.11 | 1865.96 | 2138.58132 | 1978.70419 |
| 2035.17 | 1871.1 | 2159.23978 | 1984.16916 |
| 2059.09 | 1876.25 | 2187.24513 | 1989.63414 |
| 2055.4 | 1881.4 | 2188.96755 | 1995.12038 |
| 2037.32 | 1886.56 | 2153.01992 | 2000.60662 |
| 2055.47 | 1891.73 | 2178.68617 | 2006.10349 |
| 2038.42 | 1896.91 | 2173.8485 | 2011.60037 |
| 2024.66 | 1902.09 | 2151.51014 | 2017.1185 |
| 2023.49 | 1907.28 | 2150.96789 | 2022.63664 |
| 2049.18 | 1912.48 | 2194.1242 | 2028.16541 |
| 2068.67 | 1917.69 | 2155.17827 | 2033.70481 |
| 2069.39 | 1922.9 | 2203.2254 | 2039.24422 |
| 2066.76 | 1928.13 | 2169.39358 | 2044.80488 |
| 2087.62 | 1933.36 | 2242.55409 | 2050.36555 |
| 2129.58 | 1938.59 | 2220.01372 | 2055.93685 |
| 2117.27 | 1943.84 | 2226.73331 | 2061.50815 |
| 2140.68 | 1949.09 | 2240.02362 | 2067.10071 |
| 2097.98 | 1954.36 | 2191.77447 | 2072.69327 |
| 2113.82 | 1959.62 | 2255.74871 | 2078.29647 |
| 2126.88 | 1964.9 | 2227.92412 | 2083.9103 |
| 2114.87 | 1970.18 | 2276.64108 | 2089.52412 |
| 2145 | 1975.48 | 2246.15843 | 2095.14858 |
| 2101.15 | 1980.78 | 2217.48325 | 2100.79431 |
| 2123.34 | 1986.08 | 2269.33673 | 2106.4294 |
| 2136.1 | 1991.4 | 2270.84651 | 2112.08576 |
| 2161.93 | 1996.72 | 2273.35572 | 2117.74211 |
| 2171.31 | 2002.05 | 2296.44897 | 2123.4091 |
| 2168.81 | 2007.39 | 2295.52396 | 2129.08672 |
| 2215.9 | 2012.73 | 2320.12699 | 2134.77498 |
| 2179.94 | 2018.09 | 2351.0562 | 2140.47386 |
| 2172.33 | 2023.45 | 2322.58304 | 2146.17275 |
| 2182.49 | 2028.81 | 2294.85413 | 2151.88227 |
| 2205.55 | 2034.19 | 2329.85549 | 2157.60242 |
| 2199.84 | 2039.57 | 2359.22177 | 2163.32257 |
| 2199.78 | 2044.96 | 2363.21949 | 2169.06398 |
| 2210.81 | 2050.36 | 2351.48149 | 2174.8054 |
| 2185.94 | 2055.77 | 2353.17202 | 2180.55745 |
| 2203.11 | 2061.18 | 2344.70875 | 2186.30949 |
| 2250.26 | 2066.6 | 2391.74582 | 2192.07217 |
| 2244.71 | 2072.03 | 2339.53084 | 2197.85612 |
| 2227.2 | 2077.47 | 2331.69487 | 2203.62943 |
| 2215.93 | 2082.91 | 2379.11471 | 2209.424 |
| 2232.81 | 2088.36 | 2358.25423 | 2215.22921 |
| 2259.4 | 2093.82 | 2376.8713 | 2221.03442 |
| 2191.49 | 2099.29 | 2367.13216 | 2226.85026 |
| 2251.36 | 2104.76 | 2405.84418 | 2232.67673 |
| 2232.44 | 2110.24 | 2342.76305 | 2238.50321 |
| 2282.54 | 2115.73 | 2451.82866 | 2244.34031 |
| 2299.76 | 2121.23 | 2434.28545 | 2250.18805 |
| 2254.42 | 2126.73 | 2405.20625 | 2256.04642 |
| 2285.49 | 2132.24 | 2455.95398 | 2261.91542 |
| 2268.76 | 2137.76 | 2423.14285 | 2267.78442 |
| 2275.28 | 2143.29 | 2448.53266 | 2273.66406 |
| 2295.16 | 2148.82 | 2455.58185 | 2279.55432 |
| 2308.81 | 2154.37 | 2413.11664 | 2285.44459 |
| 2289.44 | 2159.91 | 2424.29114 | 2291.34549 |
| 2356.96 | 2165.47 | 2462.81178 | 2297.26765 |
| 2288.17 | 2171.03 | 2468.46813 | 2303.17918 |
| 2349.74 | 2176.61 | 2446.16167 | 2309.11198 |
| 2316.36 | 2182.19 | 2469.61642 | 2315.04477 |
| 2324.25 | 2187.77 | 2447.38438 | 2320.9882 |
| 2355.8 | 2193.37 | 2462.20574 | 2326.94226 |
| 2340.91 | 2198.97 | 2479.94033 | 2332.90695 |
| 2300.58 | 2204.57 | 2482.84293 | 2338.87164 |
| 2305.99 | 2210.19 | 2523.84089 | 2344.84697 |
| 2267.72 | 2215.82 | 2547.10425 | 2350.83292 |
| 2354.13 | 2221.45 | 2511.57127 | 2356.82951 |
| 2377.63 | 2227.08 | 2543.54245 | 2362.8261 |
| 2362.76 | 2232.73 | 2489.30734 | 2368.83332 |
| 2426.69 | 2238.38 | 2553.4836 | 2374.85117 |
| 2434.29 | 2244.04 | 2554.75947 | 2380.86903 |
| 2459.99 | 2249.71 | 2567.1992 | 2386.90815 |
| 2471.18 | 2255.39 | 2526.27567 | 2392.94726 |
| 2400.52 | 2261.07 | 2607.02761 | 2398.99701 |
| 2461.35 | 2266.76 | 2586.73065 | 2405.04676 |
| 2421.49 | 2272.45 | 2579.13922 | 2411.10715 |
| 2437.5 | 2278.16 | 2590.35624 | 2417.17816 |
| 2468.97 | 2283.87 | 2633.69329 | 2423.25981 |
| 2450.64 | 2289.59 | 2575.71563 | 2429.35209 |
| 2436.2 | 2295.32 | 2600.16981 | 2435.44437 |
| 2480.05 | 2301.05 | 2581.2763 | 2441.54728 |
| 2504.24 | 2306.79 | 2594.20512 | 2447.66082 |
| 2493.63 | 2312.54 | 2671.05502 | 2453.77436 |
| 2488.72 | 2318.3 | 2611.93971 | 2459.89854 |
| 2503.28 | 2324.06 | 2677.72144 | 2466.03335 |
| 2490.18 | 2329.83 | 2672.5329 | 2472.17879 |
| 2520.93 | 2335.61 | 2695.91322 | 2478.33486 |
| 2517.1 | 2341.39 | 2702.45205 | 2484.49093 |
| 2515.61 | 2347.18 | 2684.69619 | 2490.65764 |
| 2576.61 | 2352.98 | 2664.66504 | 2496.82434 |
| 2579.87 | 2358.79 | 2725.22633 | 2503.01231 |
| 2538.78 | 2364.6 | 2678.85909 | 2509.20028 |
| 2555.85 | 2370.42 | 2668.29063 | 2515.39888 |
| 2535.06 | 2376.25 | 2711.63831 | 2521.59748 |
| 2540.85 | 2382.09 | 2713.78603 | 2527.81735 |
| 2573.25 | 2387.93 | 2717.3372 | 2534.03722 |
| 2540.86 | 2393.78 | 2737.9744 | 2540.25708 |
| 2578.33 | 2399.64 | 2729.84073 | 2546.49821 |
| 2544.84 | 2405.5 | 2770.30707 | 2552.73934 |
| 2558.36 | 2411.37 | 2790.67846 | 2558.99111 |
| 2574.26 | 2417.25 | 2750.38223 | 2565.2535 |
| 2638.89 | 2423.14 | 2734.43386 | 2571.5159 |
| 2608.19 | 2429.03 | 2777.5051 | 2577.78892 |
| 2596.12 | 2434.93 | 2766.90475 | 2584.07258 |
| 2668.88 | 2440.84 | 2794.72935 | 2590.36687 |
| 2586.78 | 2446.75 | 2732.3393 | 2596.66117 |
| 2608.58 | 2452.67 | 2783.02324 | 2602.96609 |
| 2639.39 | 2458.6 | 2783.17209 | 2609.28165 |
| 2632.21 | 2464.54 | 2751.59431 | 2615.5972 |
| 2649.96 | 2470.48 | 2768.61654 | 2621.93402 |
| 2681.32 | 2476.43 | 2836.06753 | 2628.27084 |
| 2652.12 | 2482.39 | 2842.9466 | 2634.60767 |
| 2650.97 | 2488.35 | 2844.42448 | 2640.96575 |
| 2663.76 | 2494.32 | 2843.30809 | 2647.32384 |
| 2685.28 | 2500.3 | 2873.65253 | 2653.69255 |
| 2690.35 | 2506.29 | 2880.76551 | 2660.06127 |
| 2683.67 | 2512.28 | 2884.73134 | 2666.44062 |
| 2694.16 | 2518.28 | 2869.4209 | 2672.8306 |
| 2765.19 | 2524.29 | 2859.73492 | 2679.23122 |
| 2705.36 | 2530.3 | 2954.79786 | 2685.63183 |
| 2685.26 | 2536.32 | 2842.21297 | 2692.04308 |
| 2729.87 | 2542.35 | 2904.16709 | 2698.46496 |
| 2724.22 | 2548.38 | 2917.4893 | 2704.89747 |
| 2735.92 | 2554.43 | 2904.81566 | 2711.32998 |
| 2727.74 | 2560.47 | 2946.82368 | 2717.77312 |
| 2765.41 | 2566.53 | 2900.57339 | 2724.21627 |
| 2737.15 | 2572.59 | 2932.67215 | 2730.68067 |
| 2786.49 | 2578.66 | 2926.40976 | 2737.14508 |
| 2788.13 | 2584.74 | 2964.36689 | 2743.62012 |
| 2779.65 | 2590.82 | 3001.21827 | 2750.09516 |
| 2796.76 | 2596.91 | 2994.29667 | 2756.58083 |
| 2818.85 | 2603.01 | 2969.40657 | 2763.07714 |
| 2857.71 | 2609.11 | 2951.79957 | 2769.58407 |
| 2801.87 | 2615.22 | 2951.76767 | 2776.09101 |
| 2861.77 | 2621.34 | 2957.11569 | 2782.60858 |
| 2852.85 | 2627.47 | 3010.40453 | 2789.13678 |
| 2853.15 | 2633.6 | 3022.51466 | 2795.66498 |
| 2843.49 | 2639.74 | 2997.06106 | 2802.20382 |
| 2878.29 | 2645.89 | 3001.70735 | 2808.75328 |
| 2853.51 | 2652.04 | 3000.36769 | 2815.31338 |
| 2812.15 | 2658.2 | 3028.11786 | 2821.87348 |
| 2833.31 | 2664.36 | 3091.1671 | 2828.44421 |
| 2831.76 | 2670.54 | 3019.69712 | 2835.01494 |
| 2847.37 | 2676.72 | 3124.15897 | 2841.60693 |
| 2866.52 | 2682.91 | 3059.35541 | 2848.19893 |
| 2868.86 | 2689.1 | 3072.60319 | 2854.79092 |
| 2903.65 | 2695.3 | 3086.82914 | 2861.40418 |
| 2915.88 | 2701.51 | 3071.32732 | 2868.01744 |
| 2918.42 | 2707.72 | 3061.58818 | 2874.64133 |
| 2922.46 | 2713.94 | 3123.07448 | 2881.26522 |
| 2916.14 | 2720.17 | 3122.76614 | 2887.89975 |
| 2903.97 | 2726.41 | 3149.3893 | 2894.5449 |
| 2967.9 | 2732.65 | 3161.89282 | 2901.20069 |
| 2925.92 | 2738.89 | 3095.40936 | 2907.85648 |
| 2941.97 | 2745.15 | 3155.5241 | 2914.5229 |
| 2939.91 | 2751.41 | 3090.89066 | 2921.19995 |
| 2898.36 | 2757.68 | 3134.9826 | 2927.87701 |
| 2956.62 | 2763.95 | 3164.91238 | 2934.56469 |
| 2994.3 | 2770.24 | 3157.06578 | 2941.25238 |
| 2929.74 | 2776.53 | 3207.31379 | 2947.96133 |
| 3033.03 | 2782.82 | 3213.93768 | 2954.67028 |
| 2996.07 | 2789.12 | 3203.029 | 2961.37923 |
| 2980.59 | 2795.43 | 3141.12804 | 2968.10944 |
| 3049.84 | 2801.75 | 3213.92705 | 2974.83965 |
| 3031.75 | 2808.07 | 3249.65141 | 2981.5805 |
| 3015.75 | 2814.4 | 3256.06266 | 2988.32135 |
| 3048.42 | 2820.73 | 3238.6364 | 2995.07282 |
| 2991.75 | 2827.07 | 3226.40931 | 3001.83494 |
| 3034.19 | 2833.42 | 3267.386 | 3008.59705 |
| 3080.97 | 2839.78 | 3251.43763 | 3015.36979 |
| 3073.32 | 2846.14 | 3240.33756 | 3022.15316 |
| 2999.52 | 2852.51 | 3244.66489 | 3028.94717 |
| 3045.39 | 2858.89 | 3268.48112 | 3035.74118 |
| 3103.65 | 2865.27 | 3276.2533 | 3042.53519 |
| 3087.39 | 2871.66 | 3301.50489 | 3049.35046 |
| 3075.61 | 2878.05 | 3272.75529 | 3056.16573 |
| 3124.04 | 2884.45 | 3339.39823 | 3062.99164 |
| 3093.71 | 2890.86 | 3313.71071 | 3069.81754 |
| 3131.9 | 2897.27 | 3281.64385 | 3076.66471 |
| 3115.09 | 2903.7 | 3329.81857 | 3083.50125 |
| 3103.2 | 2910.12 | 3281.44184 | 3090.35905 |
| 3149.91 | 2916.56 | 3311.03139 | 3097.21685 |
| 3085.29 | 2923 | 3230.19439 | 3104.08528 |
| 3151.65 | 2929.44 | 3364.34149 | 3110.95371 |
| 3133.84 | 2935.9 | 3347.32989 | 3117.83278 |
| 3100.27 | 2942.36 | 3344.40602 | 3124.72248 |
| 3169.93 | 2948.82 | 3360.09922 | 3131.62281 |
| 3203.12 | 2955.3 | 3411.81448 | 3138.52314 |
| 3183.94 | 2961.78 | 3374.24011 | 3145.4341 |
| 3160.21 | 2968.26 | 3403.61702 | 3152.34506 |
| 3192.22 | 2974.76 | 3385.90369 | 3159.26666 |
| 3215.67 | 2981.26 | 3466.36855 | 3166.19888 |
| 3201.38 | 2987.76 | 3428.76229 | 3173.13111 |
| 3193.74 | 2994.27 | 3364.86247 | 3180.07397 |
| 3196.36 | 3000.79 | 3404.20179 | 3187.02746 |
| 3219.41 | 3007.31 | 3408.74176 | 3193.98095 |
| 3202.17 | 3013.84 | 3421.4792 | 3200.94507 |
| 3214.56 | 3020.38 | 3452.65295 | 3207.91983 |
| 3226.09 | 3026.93 | 3468.12288 | 3214.89459 |
| 3200.91 | 3033.48 | 3471.87606 | 3221.87997 |
| 3211.53 | 3040.03 | 3446.97533 | 3228.87599 |
| 3236.59 | 3046.59 | 3394.37759 | 3235.87201 |
| 3256.49 | 3053.16 | 3455.77883 | 3242.87867 |
| 3218.99 | 3059.74 | 3478.20225 | 3249.89595 |
| 3277.09 | 3066.32 | 3450.82421 | 3256.91324 |
| 3262.6 | 3072.91 | 3509.71624 | 3263.94115 |
| 3258.08 | 3079.5 | 3510.82199 | 3270.96907 |
| 3325.45 | 3086.1 | 3456.10843 | 3278.00762 |
| 3260.39 | 3092.71 | 3500.42365 | 3285.0568 |
| 3289.09 | 3099.32 | 3485.2089 | 3292.10598 |
| 3254.67 | 3105.94 | 3501.69952 | 3299.1658 |
| 3340.2 | 3112.57 | 3472.06744 | 3306.23624 |
| 3288.14 | 3119.2 | 3503.94292 | 3313.30669 |
| 3295.02 | 3125.84 | 3550.40586 | 3320.38777 |
| 3281.79 | 3132.48 | 3541.50666 | 3327.47948 |
| 3293.56 | 3139.13 | 3465.82631 | 3334.57119 |
| 3331.12 | 3145.79 | 3589.20293 | 3341.67353 |
| 3277.94 | 3152.45 | 3522.91086 | 3348.77587 |
| 3360.2 | 3159.12 | 3523.1235 | 3355.89948 |
| 3324.27 | 3165.8 | 3556.25359 | 3363.01246 |
| 3281.21 | 3172.48 | 3521.85827 | 3370.1467 |
| 3302.61 | 3179.17 | 3536.81784 | 3377.28093 |
| 3394.54 | 3185.86 | 3528.30141 | 3384.41517 |
| 3300.84 | 3192.56 | 3578.54942 | 3391.56005 |
| 3323.67 | 3199.27 | 3563.00507 | 3398.71555 |
| 3356.76 | 3205.98 | 3547.36503 | 3405.88169 |
| 3352.16 | 3212.7 | 3644.58632 | 3413.04782 |
| 3398.64 | 3219.42 | 3604.07745 | 3420.22459 |
| 3317.26 | 3226.15 | 3595.04004 | 3427.40136 |
| 3375.93 | 3232.89 | 3649.79613 | 3434.58876 |
| 3392.8 | 3239.63 | 3572.44651 | 3441.78679 |
| 3376.39 | 3246.38 | 3619.78128 | 3448.98483 |
| 3443.24 | 3253.13 | 3599.76076 | 3456.19349 |
| 3396.2 | 3259.89 | 3610.88209 | 3463.40216 |
| 3475.02 | 3266.66 | 3647.26565 | 3470.62145 |
| 3435.17 | 3273.43 | 3657.13238 | 3477.85138 |
| 3398.97 | 3280.21 | 3636.01673 | 3485.08131 |
| 3385.29 | 3286.99 | 3616.99564 | 3492.32188 |
| 3391.52 | 3293.79 | 3695.36595 | 3499.57307 |
| 3435.12 | 3300.58 | 3631.64688 | 3506.82426 |
| 3439.75 | 3307.39 | 3625.71408 | 3514.07546 |
| 3435.39 | 3314.19 | 3706.74245 | 3521.34792 |
| 3454.07 | 3321.01 | 3683.69174 | 3528.62038 |
| 3499.77 | 3327.83 | 3709.57063 | 3535.89283 |
| 3464.83 | 3334.65 | 3682.35207 | 3543.18656 |
| 3448.85 | 3341.49 | 3668.3494 | 3550.46965 |
| 3464.76 | 3348.32 | 3712.88789 | 3557.774 |
| 3522.95 | 3355.17 | 3718.63994 | 3565.07836 |
| 3461.11 | 3362.02 | 3729.82507 | 3572.38272 |
| 3544.24 | 3368.87 | 3709.13471 | 3579.70834 |
| 3503.24 | 3375.74 | 3763.93332 | 3587.02332 |
| 3489.49 | 3382.6 | 3733.45067 | 3594.35958 |
| 3518.34 | 3389.48 | 3729.29346 | 3601.69583 |
| 3497.63 | 3396.36 | 3760.83934 | 3609.03208 |
| 3542.79 | 3403.24 | 3785.95271 | 3616.3896 |
| 3511.7 | 3410.13 | 3721.298 | 3623.73648 |
| 3532.72 | 3417.03 | 3766.7934 | 3631.10463 |
| 3510.39 | 3423.93 | 3789.674 | 3638.47278 |
| 3599.47 | 3430.84 | 3816.41411 | 3645.84093 |
| 3601.89 | 3437.75 | 3777.9041 | 3653.21971 |
| 3585.09 | 3444.67 | 3822.61271 | 3660.60912 |
| 3595.63 | 3451.6 | 3816.67992 | 3668.00917 |
| 3615.83 | 3458.53 | 3818.35981 | 3675.39858 |
| 3557.32 | 3465.47 | 3774.12965 | 3682.80926 |
| 3653.34 | 3472.41 | 3835.15877 | 3690.21994 |
| 3596.5 | 3479.36 | 3860.25087 | 3697.64125 |
| 3637.21 | 3486.31 | 3876.90098 | 3705.06256 |
| 3599.57 | 3493.27 | 3842.80335 | 3712.4945 |
| 3613.36 | 3500.24 | 3853.53129 | 3719.92644 |
| 3629.56 | 3507.21 | 3860.22961 | 3727.36902 |
| 3681.03 | 3514.18 | 3934.02805 | 3734.82223 |
| 3660.56 | 3521.17 | 3987.39132 | 3742.27543 |
| 3643.8 | 3528.15 | 3924.98001 | 3749.73927 |
| 3703.1 | 3535.15 | 3860.30404 | 3757.20311 |
| 3651.43 | 3542.15 | 3957.42964 | 3764.67758 |
| 3717.8 | 3549.15 | 3966.6478 | 3772.16269 |
| 3702.85 | 3556.16 | 3959.34344 | 3779.64779 |
| 3673.59 | 3563.18 | 3928.62687 | 3787.13289 |
| 3690.27 | 3570.2 | 3947.77555 | 3794.63926 |
| 3688.66 | 3577.23 | 3975.51509 | 3802.135 |
| 3721.3 | 3584.26 | 3978.42833 | 3809.652 |
| 3738.55 | 3591.3 | 3995.50372 | 3817.169 |
| 3787.64 | 3598.34 | 3969.41218 | 3824.686 |
| 3757.32 | 3605.39 | 3989.93242 | 3832.21363 |
| 3736.97 | 3612.45 | 3996.30114 | 3839.7519 |
| 3760.65 | 3619.51 | 4006.13597 | 3847.29016 |
| 3758.75 | 3626.57 | 4038.5856 | 3854.83906 |
| 3783.74 | 3633.64 | 4026.34788 | 3862.38796 |
| 3782.41 | 3640.72 | 4031.65337 | 3869.94749 |
| 3791.39 | 3647.8 | 3978.84299 | 3877.50702 |
| 3818.37 | 3654.89 | 4042.89166 | 3885.07718 |
| 3786.26 | 3661.98 | 3982.00076 | 3892.65797 |
| 3794.93 | 3669.08 | 4016.73632 | 3900.23877 |
| 3768.97 | 3676.19 | 4033.57781 | 3907.83019 |
| 3827.24 | 3683.3 | 4019.80904 | 3915.42162 |
| 3826.17 | 3690.41 | 4088.06809 | 3923.02368 |
| 3823.55 | 3697.53 | 4082.14592 | 3930.62573 |
| 3833.27 | 3704.66 | 4130.95858 | 3938.23843 |
| 3848.84 | 3711.79 | 4116.14786 | 3945.85112 |
| 3886.58 | 3718.92 | 4105.9409 | 3953.47444 |
| 3812.36 | 3726.07 | 4158.24093 | 3961.10839 |
| 3857.1 | 3733.21 | 4103.35726 | 3968.74235 |
| 3893.32 | 3740.36 | 4162.57889 | 3976.3763 |
| 3901.45 | 3747.52 | 4151.26618 | 3984.02089 |
| 3884.38 | 3754.69 | 4170.3936 | 3991.67611 |
| 3909.55 | 3761.85 | 4146.8219 | 3999.33133 |
| 3917.04 | 3769.03 | 4091.91696 | 4006.99718 |
| 3844.83 | 3776.21 | 4135.70057 | 4014.67367 |
| 3883.03 | 3783.39 | 4070.87574 | 4022.35015 |
| 3940.8 | 3790.58 | 4225.03273 | 4030.02664 |
| 3942.59 | 3797.77 | 4204.43806 | 4037.71375 |
| 4003.07 | 3804.97 | 4197.65468 | 4045.40087 |
| 3913.97 | 3812.18 | 4173.5939 | 4053.10925 |
| 3960.86 | 3819.39 | 4237.3874 | 4060.807 |
| 3914.11 | 3826.6 | 4157.9645 | 4068.51538 |
| 3990.15 | 3833.82 | 4150.46876 | 4076.23439 |
| 3950.82 | 3841.05 | 4155.54034 | 4083.95341 |
| 3917.01 | 3848.28 | 4202.55615 | 4091.68305 |
| 3990.92 | 3855.52 | 4192.27477 | 4099.4127 |
| 3958.62 | 3862.76 | 4267.42351 | 4107.15298 |
| 3904.59 | 3870 | 4182.15286 | 4114.89325 |
| 4000.46 | 3877.26 | 4172.74332 | 4122.64416 |
| 3983.37 | 3884.51 | 4292.68573 | 4130.39507 |
| 4063.32 | 3891.77 | 4333.02449 | 4138.15662 |
| 3976.89 | 3899.04 | 4252.37887 | 4145.92879 |
| 3979.63 | 3906.31 | 4290.14462 | 4153.69033 |
| 4081.16 | 3913.59 | 4267.52983 | 4161.47314 |
| 3994.16 | 3920.87 | 4291.44176 | 4169.25595 |
| 4005.03 | 3928.16 | 4349.84471 | 4177.03875 |
| 4079.32 | 3935.45 | 4227.88217 | 4184.83219 |
| 3963.21 | 3942.75 | 4283.46757 | 4192.63626 |
| 4055.26 | 3950.05 | 4328.63337 | 4200.44033 |
| 4080.08 | 3957.36 | 4268.90139 | 4208.2444 |
| 4066.49 | 3964.67 | 4361.6465 | 4216.06974 |
| 4043.29 | 3971.98 | 4365.05945 | 4223.88444 |
| 4068.65 | 3979.3 | 4319.40457 | 4231.70978 |
| 4098.23 | 3986.63 | 4394.60648 | 4239.54575 |
| 4097.69 | 3993.96 | 4334.22593 | 4247.38172 |
| 4098.95 | 4001.3 | 4334.5449 | 4255.21768 |
| 4103.07 | 4008.64 | 4442.06884 | 4263.07492 |
| 4115.82 | 4015.99 | 4389.48173 | 4270.92152 |
| 4125.42 | 4023.34 | 4339.65901 | 4278.77875 |
| 4084.51 | 4030.7 | 4381.52881 | 4286.64661 |
| 4113.01 | 4038.06 | 4351.96052 | 4294.51448 |
| 4160.15 | 4045.42 | 4404.49447 | 4302.39298 |
| 4160.37 | 4052.79 | 4436.15731 | 4310.27147 |
| 4158.42 | 4060.17 | 4430.40526 | 4318.1606 |
| 4199.31 | 4067.55 | 4432.89321 | 4326.04973 |
| 4126.38 | 4074.93 | 4356.60682 | 4333.93886 |
| 4126.73 | 4082.32 | 4390.02398 | 4341.84925 |
| 4191.05 | 4089.72 | 4477.37854 | 4349.74902 |
| 4190.52 | 4097.12 | 4398.48725 | 4357.65941 |
| 4197.6 | 4104.52 | 4492.72088 | 4365.58043 |
| 4200.72 | 4111.93 | 4411.61808 | 4373.50146 |
| 4159.29 | 4119.35 | 4450.39389 | 4381.43312 |
| 4211.36 | 4126.77 | 4457.49623 | 4389.36478 |
| 4220.6 | 4134.19 | 4444.12086 | 4397.29644 |
| 4231.86 | 4141.62 | 4494.01801 | 4405.23873 |
| 4252.48 | 4149.05 | 4481.16362 | 4413.19165 |
| 4217.51 | 4156.49 | 4511.69944 | 4421.14457 |
| 4332.21 | 4163.93 | 4509.36035 | 4429.10813 |
| 4240.28 | 4171.38 | 4524.8409 | 4437.07168 |
| 4284.75 | 4178.83 | 4600.69137 | 4445.03524 |
| 4271.89 | 4186.28 | 4528.29638 | 4453.00942 |
| 4254.86 | 4193.74 | 4533.77199 | 4460.99424 |
| 4277.25 | 4201.21 | 4505.73475 | 4468.96843 |
| 4299.58 | 4208.68 | 4454.23213 | 4476.96388 |
| 4270.09 | 4216.15 | 4558.91726 | 4484.95933 |
| 4269.58 | 4223.64 | 4619.6912 | 4492.95479 |
| 4304.62 | 4231.12 | 4570.75096 | 4500.96087 |
| 4262.36 | 4238.61 | 4555.84454 | 4508.96695 |
| 4346.92 | 4246.1 | 4602.90288 | 4516.98367 |
| 4304.52 | 4253.6 | 4578.22543 | 4525.00039 |
| 4288.55 | 4261.1 | 4569.07106 | 4533.02773 |
| 4344.12 | 4268.61 | 4586.51858 | 4541.05508 |
| 4386.01 | 4276.12 | 4618.7662 | 4549.09306 |
| 4355.53 | 4283.63 | 4653.86325 | 4557.13104 |
| 4336.03 | 4291.15 | 4707.93887 | 4565.17966 |
| 4327.25 | 4298.68 | 4662.49664 | 4573.22827 |
| 4398.21 | 4306.21 | 4664.65498 | 4581.27688 |
| 4419.91 | 4313.74 | 4624.32686 | 4589.33613 |
| 4342.19 | 4321.28 | 4719.124 | 4597.40601 |
| 4406.17 | 4328.82 | 4653.67187 | 4605.47588 |
| 4404.63 | 4336.37 | 4699.8371 | 4613.54576 |
| 4427.56 | 4343.92 | 4720.69757 | 4621.62627 |
| 4443.22 | 4351.47 | 4658.39259 | 4629.70678 |
| 4397.33 | 4359.03 | 4705.06817 | 4637.79792 |
| 4429.5 | 4366.6 | 4717.35905 | 4645.88906 |
| 4460.73 | 4374.17 | 4686.27035 | 4653.99084 |
| 4452.78 | 4381.74 | 4685.78126 | 4662.09261 |
| 4452.21 | 4389.32 | 4700.36871 | 4670.19439 |
| 4467.11 | 4396.9 | 4730.9258 | 4678.30679 |
| 4486.95 | 4404.48 | 4789.87099 | 4686.42983 |
| 4414.52 | 4412.07 | 4784.88446 | 4694.55287 |
| 4526.42 | 4419.67 | 4764.41738 | 4702.67591 |
| 4455.59 | 4427.27 | 4721.96281 | 4710.80958 |
| 4486.38 | 4434.87 | 4763.19467 | 4718.94325 |
| 4527.72 | 4442.48 | 4869.2301 | 4727.08755 |
| 4551.26 | 4450.09 | 4786.20286 | 4735.23186 |
| 4566.33 | 4457.71 | 4799.85467 | 4743.37616 |
| 4578.9 | 4465.33 | 4778.83471 | 4751.5311 |
| 4531.4 | 4472.95 | 4885.05089 | 4759.69666 |
| 4590.01 | 4480.58 | 4852.68632 | 4767.86223 |
| 4534.13 | 4488.21 | 4857.53463 | 4776.0278 |
| 4575.46 | 4495.85 | 4913.56658 | 4784.204 |
| 4602.68 | 4503.49 | 4911.07863 | 4792.3802 |
| 4565.99 | 4511.14 | 4903.76365 | 4800.56703 |
| 4615.94 | 4518.78 | 4856.1099 | 4808.75386 |
| 4624.93 | 4526.44 | 5001.58034 | 4816.9407 |
| 4632.44 | 4534.1 | 4923.21003 | 4825.13816 |
| 4580.58 | 4541.76 | 4949.84382 | 4833.33563 |
| 4629.91 | 4549.42 | 4928.81323 | 4841.54372 |
| 4649.37 | 4557.09 | 4954.5539 | 4849.75182 |
| 4641.28 | 4564.77 | 4851.30413 | 4857.97055 |
| 4681.81 | 4572.45 | 4896.60814 | 4866.18928 |
| 4611.8 | 4580.13 | 4923.22066 | 4874.40801 |
| 4704.17 | 4587.81 | 5027.00205 | 4882.63737 |
| 4670.41 | 4595.5 | 4972.72442 | 4890.86673 |
| 4589.42 | 4603.2 | 4967.46145 | 4899.10672 |
| 4520 | 4610.9 | 4957.76484 | 4907.34672 |
| 4725.2 | 4618.6 | 4891.72794 | 4915.59734 |
| 4580.86 | 4626.31 | 4946.72857 | 4923.84797 |
| 4710.74 | 4634.02 | 4956.31886 | 4932.09859 |
| 4685.53 | 4641.73 | 5020.99483 | 4940.35985 |
| 4647.75 | 4649.45 | 4987.35439 | 4948.62111 |
| 4758.1 | 4657.17 | 5061.43991 | 4956.893 |
| 4735.12 | 4664.9 | 4906.33665 | 4965.16489 |
| 4643.34 | 4672.63 | 4979.05061 | 4973.43678 |
| 4725.84 | 4680.36 | 5002.73926 | 4981.7193 |
| 4711.86 | 4688.1 | 5083.57625 | 4990.00182 |
| 4777.84 | 4695.84 | 5059.40915 | 4998.29498 |
| 4687.15 | 4703.58 | 5017.41176 | 5006.58813 |
| 4707.76 | 4711.33 | 5064.78907 | 5014.89192 |
| 4736.1 | 4719.09 | 5069.10576 | 5023.18507 |
| 4832.6 | 4726.84 | 5103.1077 | 5031.49949 |
| 4796.09 | 4734.6 | 5132.59092 | 5039.80328 |
| 4699.5 | 4742.37 | 5093.04959 | 5048.12833 |
| 4792.68 | 4750.14 | 5034.35957 | 5056.44275 |
| 4834.14 | 4757.91 | 5044.74728 | 5064.7678 |
| 4786.13 | 4765.69 | 5107.58387 | 5073.09285 |
| 4756.83 | 4773.47 | 5061.8971 | 5081.42854 |
| 4766.64 | 4781.25 | 5149.76201 | 5089.76422 |
| 4829.83 | 4789.04 | 5110.82671 | 5098.09991 |
| 4826.35 | 4796.83 | 5119.83222 | 5106.44622 |
| 4807.29 | 4804.62 | 5104.66 | 5114.79254 |
| 4841.3 | 4812.42 | 5122.25638 | 5123.14949 |
| 4853.64 | 4820.22 | 5136.84382 | 5131.50643 |
| 4891.64 | 4828.03 | 5174.14176 | 5139.86338 |
| 4786.83 | 4835.84 | 5127.44491 | 5148.23096 |
| 4827.98 | 4843.65 | 5166.98625 | 5156.59854 |
| 4899.25 | 4851.47 | 5160.70259 | 5164.97676 |
| 4873.03 | 4859.29 | 5179.07512 | 5173.35497 |
| 4934.96 | 4867.11 | 5193.96027 | 5181.73318 |
| 4823.31 | 4874.94 | 5116.86583 | 5190.11139 |
| 4903.75 | 4882.77 | 5200.46721 | 5198.50024 |
| 4883.66 | 4890.6 | 5273.00041 | 5206.89972 |
| 4912.27 | 4898.44 | 5226.0484 | 5215.29919 |
| 4814.08 | 4906.28 | 5129.95413 | 5223.69867 |
| 4990.76 | 4914.13 | 5189.67547 | 5232.09815 |
| 4957.51 | 4921.98 | 5275.84986 | 5240.50826 |
| 4958.29 | 4929.83 | 5207.61208 | 5248.929 |
| 4934.11 | 4937.69 | 5340.65342 | 5257.33911 |
| 4893.01 | 4945.55 | 5275.79669 | 5265.75985 |
| 4972.08 | 4953.41 | 5259.53998 | 5274.19122 |
| 5019.75 | 4961.28 | 5268.46044 | 5282.61197 |
| 4914.29 | 4969.15 | 5321.74928 | 5291.05397 |
| 4974.75 | 4977.02 | 5311.85065 | 5299.48535 |
| 4966.7 | 4984.9 | 5254.91496 | 5307.92735 |
| 4965.46 | 4992.78 | 5237.29732 | 5316.36936 |
| 5099.16 | 5000.66 | 5371.40188 | 5324.822 |
| 4940.82 | 5008.55 | 5397.04687 | 5333.27464 |
| 5053.7 | 5016.44 | 5301.97329 | 5341.72727 |
| 5009.9 | 5024.33 | 5394.71841 | 5350.19054 |
| 4959.55 | 5032.23 | 5355.24086 | 5358.65382 |
| 4982.42 | 5040.13 | 5335.84764 | 5367.11709 |
| 5029.52 | 5048.04 | 5361.37567 | 5375.59099 |
| 5090.48 | 5055.94 | 5314.34923 | 5384.06489 |
| 5109.56 | 5063.85 | 5407.14751 | 5392.53879 |
| 5085.19 | 5071.77 | 5428.78413 | 5401.02333 |
| 5072 | 5079.69 | 5391.47557 | 5409.50787 |
| 5095.84 | 5087.61 | 5357.33542 | 5417.9924 |
| 5049.26 | 5095.53 | 5461.99928 | 5426.48757 |
| 5093.7 | 5103.46 | 5447.63511 | 5434.99337 |
| 5092.49 | 5111.39 | 5438.64023 | 5443.48854 |
| 5144.05 | 5119.32 | 5451.6541 | 5451.99433 |
| 5095.96 | 5127.26 | 5463.28578 | 5460.50013 |
| 5138.28 | 5135.2 | 5373.1881 | 5469.01657 |
| 5098.08 | 5143.14 | 5477.60742 | 5477.52237 |
| 5111.38 | 5151.09 | 5442.01065 | 5486.04943 |
| 5089.19 | 5159.04 | 5429.12437 | 5494.56586 |
| 5136.03 | 5166.99 | 5516.89359 | 5503.09293 |
| 5207.9 | 5174.95 | 5538.29631 | 5511.61999 |
| 5179.31 | 5182.91 | 5495.80984 | 5520.15769 |
| 5184.04 | 5190.87 | 5585.24832 | 5528.69538 |
| 5189.76 | 5198.84 | 5510.80131 | 5537.23308 |
| 5167.63 | 5206.8 | 5507.46278 | 5545.77078 |
| 5180.49 | 5214.78 | 5512.45994 | 5554.31911 |
| 5223.74 | 5222.75 | 5564.56859 | 5562.86743 |
| 5178 | 5230.73 | 5541.73052 | 5571.42639 |
| 5232.59 | 5238.71 | 5561.29386 | 5579.98536 |
| 5159.27 | 5246.69 | 5519.9982 | 5588.54432 |
| 5206.8 | 5254.68 | 5612.17981 | 5597.11391 |
| 5235.14 | 5262.67 | 5608.27777 | 5605.67287 |
| 5151.31 | 5270.66 | 5563.38841 | 5614.2531 |
| 5311.89 | 5278.66 | 5598.64495 | 5622.82269 |
| 5226.32 | 5286.66 | 5579.25173 | 5631.40292 |
| 5305.69 | 5294.66 | 5638.73917 | 5639.98314 |
| 5182.6 | 5302.67 | 5670.89109 | 5648.56337 |
| 5236.95 | 5310.68 | 5569.20425 | 5657.15422 |
| 5206.43 | 5318.69 | 5648.05302 | 5665.74508 |
| 5218.73 | 5326.7 | 5624.02413 | 5674.33594 |
| 5249.95 | 5334.72 | 5652.1996 | 5682.93743 |
| 5247.16 | 5342.74 | 5707.54045 | 5691.53892 |
| 5285.74 | 5350.76 | 5587.47046 | 5700.14041 |
| 5267.89 | 5358.79 | 5632.14717 | 5708.75253 |
| 5291.98 | 5366.82 | 5642.02453 | 5717.36465 |
| 5286.79 | 5374.85 | 5644.9803 | 5725.97678 |
| 5325.82 | 5382.88 | 5645.19294 | 5734.59953 |
| 5341.23 | 5390.92 | 5697.19528 | 5743.21165 |
| 5316.71 | 5398.96 | 5695.63234 | 5751.84504 |
| 5329.66 | 5407 | 5724.88165 | 5760.46779 |
| 5342.12 | 5415.05 | 5631.34975 | 5769.10118 |
| 5361.29 | 5423.1 | 5633.90149 | 5777.73457 |
| 5384.12 | 5431.15 | 5624.3431 | 5786.36795 |
| 5384.26 | 5439.2 | 5621.27038 | 5795.01197 |
| 5410.23 | 5447.26 | 5704.64848 | 5803.65599 |
| 5415.69 | 5455.32 | 5768.29313 | 5812.30001 |
| 5432.54 | 5463.38 | 5772.5354 | 5820.94403 |
| 5374.08 | 5471.45 | 5744.14729 | 5829.59868 |
| 5411.1 | 5479.52 | 5777.90468 | 5838.25333 |
| 5404.62 | 5487.59 | 5756.85283 | 5846.91862 |
| 5406.33 | 5495.66 | 5900.04797 | 5855.57327 |
| 5459.4 | 5503.74 | 5822.37938 | 5864.23855 |
| 5486.22 | 5511.81 | 5905.88507 | 5872.91447 |
| 5464.64 | 5519.9 | 5782.31707 | 5881.57975 |
| 5371.35 | 5527.98 | 5816.07446 | 5890.25566 |
| 5498.39 | 5536.07 | 5874.28603 | 5898.93158 |
| 5518.79 | 5544.16 | 5859.2095 | 5907.6075 |
| 5426.2 | 5552.25 | 5897.01778 | 5916.29404 |
| 5552.17 | 5560.34 | 5839.91196 | 5924.98059 |
| 5457.8 | 5568.44 | 5880.83549 | 5933.66714 |
| 5616.7 | 5576.54 | 5962.70381 | 5942.36432 |
| 5507.59 | 5584.64 | 5863.09027 | 5951.0615 |
| 5446.02 | 5592.75 | 5877.79467 | 5959.75868 |
| 5646.44 | 5600.85 | 5919.30297 | 5968.45586 |
| 5587.84 | 5608.96 | 5968.0412 | 5977.16367 |
| 5613.84 | 5617.08 | 5933.57145 | 5985.86085 |
| 5651.11 | 5625.19 | 6007.97593 | 5994.56867 |
| 5606.48 | 5633.31 | 6025.05133 | 6003.28711 |
| 5608.09 | 5641.43 | 5964.92595 | 6012.00556 |
| 5678.37 | 5649.55 | 5993.9201 | 6020.724 |
| 5705.7 | 5657.68 | 6013.37712 | 6029.44244 |
| 5591.79 | 5665.81 | 6032.12177 | 6038.16089 |
| 5646.08 | 5673.93 | 5975.61136 | 6046.88997 |
| 5592.84 | 5682.07 | 5967.13746 | 6055.61904 |
| 5670.51 | 5690.2 | 6098.09488 | 6064.34812 |
| 5673.77 | 5698.34 | 6152.10671 | 6073.08783 |
| 5636.06 | 5706.48 | 6112.86308 | 6081.82754 |
| 5722.94 | 5714.62 | 6033.59965 | 6090.56725 |
| 5661.6 | 5722.77 | 6131.87354 | 6099.30696 |
| 5779.68 | 5730.91 | 6072.7157 | 6108.04667 |
| 5623.23 | 5739.06 | 6127.17408 | 6116.79701 |
| 5719.44 | 5747.21 | 6124.84562 | 6125.54735 |
| 5738.77 | 5755.36 | 6172.15913 | 6134.30832 |
| 5794.32 | 5763.52 | 6112.85244 | 6143.05866 |
| 5717.35 | 5771.68 | 6119.45507 | 6151.81964 |
| 5742.37 | 5779.84 | 6111.18318 | 6160.58061 |
| 5792.69 | 5788 | 6184.81151 | 6169.34159 |
| 5807.74 | 5796.17 | 6150.86274 | 6178.11319 |
| 5818.88 | 5804.34 | 6113.32026 | 6186.8848 |
| 5834.98 | 5812.51 | 6087.22872 | 6195.6564 |
| 5722.76 | 5820.68 | 6128.55628 | 6204.42801 |
| 5901.41 | 5828.85 | 6197.62337 | 6213.21025 |
| 5731.47 | 5837.03 | 6118.56196 | 6221.98185 |
| 5762.44 | 5845.21 | 6158.93261 | 6230.76409 |
| 5808.46 | 5853.39 | 6152.63832 | 6239.55696 |
| 5827.57 | 5861.57 | 6258.66311 | 6248.3392 |
| 5821.17 | 5869.76 | 6256.96195 | 6257.13207 |
| 5806.33 | 5877.94 | 6226.20286 | 6265.92494 |
| 5859.6 | 5886.13 | 6164.46138 | 6274.71781 |
| 5798.79 | 5894.32 | 6322.63736 | 6283.51068 |
| 5877.72 | 5902.52 | 6236.59056 | 6292.31418 |
| 5835.79 | 5910.71 | 6250.43375 | 6301.11769 |
| 5890.26 | 5918.91 | 6243.23572 | 6309.92119 |
| 5859.16 | 5927.11 | 6273.77154 | 6318.72469 |
| 5892.64 | 5935.31 | 6292.27166 | 6327.53883 |
| 5942.4 | 5943.52 | 6193.23225 | 6336.35296 |
| 6002.39 | 5951.73 | 6343.29582 | 6345.1671 |
| 5867.27 | 5959.93 | 6236.2397 | 6353.98123 |
| 5968.52 | 5968.14 | 6393.79901 | 6362.806 |
| 5882.78 | 5976.36 | 6329.09114 | 6371.62013 |
| 5939.27 | 5984.57 | 6307.39071 | 6380.4449 |
| 5914.96 | 5992.79 | 6356.77751 | 6389.26967 |
| 5846.68 | 6001.01 | 6330.0693 | 6398.10507 |
| 6002.81 | 6009.23 | 6442.65419 | 6406.92984 |
| 5913.8 | 6017.45 | 6376.20263 | 6415.76524 |
| 5966.9 | 6025.67 | 6346.13463 | 6424.60063 |
| 5993.67 | 6033.9 | 6368.32414 | 6433.43603 |
| 5946.38 | 6042.13 | 6342.57283 | 6442.28207 |
| 5957.53 | 6050.36 | 6368.02643 | 6451.11747 |
| 5877.27 | 6058.59 | 6431.16073 | 6459.9635 |
| 5920.02 | 6066.82 | 6425.74892 | 6468.80953 |
| 5973.75 | 6075.06 | 6358.70195 | 6477.65556 |
| 6015.25 | 6083.3 | 6510.16898 | 6486.51222 |
| 5981.35 | 6091.53 | 6409.01376 | 6495.35826 |
| 6072.52 | 6099.78 | 6434.21219 | 6504.21492 |
| 6002.43 | 6108.02 | 6409.31146 | 6513.07158 |
| 6042.66 | 6116.26 | 6414.96782 | 6521.93888 |
| 6013.84 | 6124.51 | 6461.6115 | 6530.79554 |
| 6132.71 | 6132.76 | 6426.44001 | 6539.66284 |
| 5977.73 | 6141.01 | 6442.27143 | 6548.53014 |
| 5967.56 | 6149.26 | 6394.30936 | 6557.39743 |
| 6045.69 | 6157.52 | 6452.12753 | 6566.26473 |
| 6025.76 | 6165.77 | 6484.7898 | 6575.14266 |
| 6104.04 | 6174.03 | 6530.47658 | 6584.00995 |
| 6093.85 | 6182.29 | 6628.26138 | 6592.88788 |
| 6118.65 | 6190.55 | 6504.4382 | 6601.76581 |
| 6091.67 | 6198.81 | 6558.046 | 6610.64374 |
| 6080.02 | 6207.08 | 6616.96993 | 6619.5323 |
| 6106.81 | 6215.34 | 6518.73857 | 6628.41023 |
| 6123.74 | 6223.61 | 6501.72697 | 6637.29879 |
| 6149.57 | 6231.88 | 6562.12878 | 6646.18735 |
| 6172.59 | 6240.15 | 6589.1028 | 6655.07591 |
| 6215.2 | 6248.42 | 6545.35109 | 6663.9751 |
| 6220.16 | 6256.7 | 6555.24972 | 6672.86366 |
| 6249.08 | 6264.98 | 6677.68007 | 6681.76286 |
| 6139.28 | 6273.25 | 6672.27889 | 6690.66205 |
| 6191.47 | 6281.53 | 6632.04646 | 6699.56124 |
| 6230.53 | 6289.81 | 6544.41546 | 6708.46044 |
| 6207.85 | 6298.1 | 6650.99313 | 6717.37026 |
| 6234.09 | 6306.38 | 6646.62327 | 6726.26945 |
| 6190.26 | 6314.66 | 6684.22954 | 6735.17928 |
| 6276.33 | 6322.95 | 6677.73323 | 6744.0891 |
| 6258.62 | 6331.24 | 6658.69087 | 6752.99893 |
| 6237.06 | 6339.53 | 6651.59916 | 6761.91939 |
| 6290.35 | 6347.82 | 6732.58501 | 6770.82921 |
| 6270.09 | 6356.11 | 6755.69952 | 6779.74967 |
| 6281.85 | 6364.41 | 6618.96879 | 6788.67013 |
| 6296.37 | 6372.71 | 6674.60735 | 6797.57995 |
| 6287.67 | 6381 | 6795.67678 | 6806.51104 |
| 6270.91 | 6389.3 | 6746.40693 | 6815.4315 |
| 6352.57 | 6397.6 | 6814.09184 | 6824.36259 |
| 6309.21 | 6405.9 | 6778.35684 | 6833.28305 |
| 6427.87 | 6414.21 | 6854.83462 | 6842.21414 |
| 6353.86 | 6422.51 | 6730.08643 | 6851.14523 |
| 6393.4 | 6430.82 | 6827.41404 | 6860.07631 |
| 6420.64 | 6439.13 | 6788.88277 | 6869.0074 |
| 6454.14 | 6447.44 | 6878.95919 | 6877.94913 |
| 6395.2 | 6455.75 | 6879.27816 | 6886.88022 |
| 6282.46 | 6464.06 | 6773.35969 | 6895.82194 |
| 6394.12 | 6472.37 | 6813.54959 | 6904.76366 |
| 6407.59 | 6480.69 | 6845.11674 | 6913.70538 |
| 6390.22 | 6489 | 6880.5434 | 6922.6471 |
| 6396.62 | 6497.32 | 6826.39335 | 6931.59946 |
| 6444.19 | 6505.64 | 6909.6226 | 6940.54118 |
| 6426.84 | 6513.96 | 6929.01582 | 6949.49353 |
| 6443.28 | 6522.28 | 6814.24069 | 6958.44589 |
| 6396.74 | 6530.6 | 6838.65233 | 6967.39824 |
| 6388.07 | 6538.92 | 6849.57165 | 6976.3506 |
| 6468.78 | 6547.25 | 6827.58416 | 6985.30295 |
| 6406.75 | 6555.57 | 6863.468 | 6994.26594 |
| 6462.34 | 6563.9 | 6871.79305 | 7003.21829 |
| 6508.48 | 6572.23 | 6980.43338 | 7012.18128 |
| 6436.07 | 6580.56 | 6900.11737 | 7021.13363 |
| 6555.87 | 6588.89 | 6917.68184 | 7030.09662 |
| 6439.08 | 6597.22 | 7037.51793 | 7039.05961 |
| 6440.6 | 6605.55 | 6931.74831 | 7048.03322 |
| 6556.68 | 6613.89 | 7025.23768 | 7056.99621 |
| 6402.64 | 6622.22 | 6937.69174 | 7065.96983 |
| 6510.82 | 6630.56 | 6933.7578 | 7074.93282 |
| 6464.22 | 6638.9 | 6989.47079 | 7083.90643 |
| 6522.53 | 6647.24 | 6962.58183 | 7092.88005 |
| 6605.85 | 6655.58 | 6955.01167 | 7101.85367 |
| 6521.16 | 6663.92 | 6987.7909 | 7110.82729 |
| 6538.46 | 6672.26 | 7024.61038 | 7119.80091 |
| 6592.03 | 6680.6 | 6960.79562 | 7128.78516 |
| 6635.5 | 6688.95 | 7024.419 | 7137.75878 |
| 6562.65 | 6697.29 | 7004.19646 | 7146.74303 |
| 6598.12 | 6705.64 | 7039.64438 | 7155.72728 |
| 6635.17 | 6713.99 | 7083.39609 | 7164.71153 |
| 6636.12 | 6722.34 | 7003.33525 | 7173.69578 |
| 6650.41 | 6730.68 | 7085.61823 | 7182.68003 |
| 6588.14 | 6739.03 | 7139.88523 | 7191.66428 |
| 6637.05 | 6747.39 | 7098.75969 | 7200.65917 |
| 6626.29 | 6755.74 | 7093.56052 | 7209.64342 |
| 6678.45 | 6764.09 | 7110.74223 | 7218.6383 |
| 6657.34 | 6772.45 | 7137.72688 | 7227.62255 |
| 6656.54 | 6780.8 | 7128.95528 | 7236.61744 |
| 6752.4 | 6789.16 | 7216.07593 | 7245.61232 |
| 6699.96 | 6797.52 | 7233.41713 | 7254.6072 |
| 6787.54 | 6805.87 | 7119.64143 | 7263.61272 |
| 6777.28 | 6814.23 | 7165.6578 | 7272.6076 |
| 6764.38 | 6822.59 | 7186.52891 | 7281.60248 |
| 6677.44 | 6830.95 | 7195.71517 | 7290.608 |
| 6754.84 | 6839.31 | 7221.20067 | 7299.60288 |
| 6755.38 | 6847.68 | 7213.69431 | 7308.6084 |
| 6733.17 | 6856.04 | 7166.32763 | 7317.61391 |
| 6723.86 | 6864.4 | 7228.05847 | 7326.61943 |
| 6692.14 | 6872.77 | 7155.52527 | 7335.62494 |
| 6714.84 | 6881.13 | 7227.85646 | 7344.63046 |
| 6737.19 | 6889.5 | 7237.88267 | 7353.63598 |
| 6797 | 6897.87 | 7240.17924 | 7362.64149 |
| 6698.4 | 6906.24 | 7318.12426 | 7371.65764 |
| 6794.83 | 6914.6 | 7320.89928 | 7380.66315 |
| 6862.74 | 6922.98 | 7331.72291 | 7389.6793 |
| 6719.29 | 6931.35 | 7221.77482 | 7398.69545 |
| 6825.1 | 6939.72 | 7236.46858 | 7407.70096 |
| 6811.3 | 6948.09 | 7291.06519 | 7416.71711 |
| 6803.31 | 6956.46 | 7323.35533 | 7425.73326 |
| 6808.53 | 6964.83 | 7316.11477 | 7434.76004 |
| 6911.72 | 6973.21 | 7365.86306 | 7443.77619 |
| 6876.54 | 6981.58 | 7321.3671 | 7452.79234 |
| 6782.22 | 6989.96 | 7227.85646 | 7461.80848 |
| 6849.9 | 6998.33 | 7279.68868 | 7470.83526 |
| 6836.95 | 7006.71 | 7222.86994 | 7479.85141 |
| 6799.29 | 7015.09 | 7406.53142 | 7488.87819 |
| 6833.95 | 7023.47 | 7342.55717 | 7497.89434 |
| 6856.91 | 7031.85 | 7311.72365 | 7506.92112 |
| 6892.16 | 7040.22 | 7299.88995 | 7515.9479 |
| 6823.27 | 7048.6 | 7202.98763 | 7524.97468 |
| 6883.96 | 7056.98 | 7407.69033 | 7534.00146 |
| 6911.82 | 7065.36 | 7419.56656 | 7543.02824 |
| 6878.34 | 7073.75 | 7347.61812 | 7552.05502 |
| 6878.35 | 7082.13 | 7490.16469 | 7561.09243 |
| 6835.55 | 7090.51 | 7361.12108 | 7570.11921 |
| 6875.19 | 7098.9 | 7353.46586 | 7579.14599 |
| 6889.65 | 7107.28 | 7396.42015 | 7588.1834 |
| 6976.77 | 7115.66 | 7445.40292 | 7597.21018 |
| 6968.63 | 7124.05 | 7382.18357 | 7606.24759 |
| 7021.9 | 7132.43 | 7423.79819 | 7615.28501 |
| 7087.63 | 7140.82 | 7429.56087 | 7624.31179 |
| 7002.34 | 7149.2 | 7467.53927 | 7633.3492 |
| 7030.86 | 7157.59 | 7441.39456 | 7642.38661 |
| 6969.68 | 7165.98 | 7512.6519 | 7651.42402 |
| 7068.32 | 7174.37 | 7503.73144 | 7660.46144 |
| 7086.64 | 7182.76 | 7455.66304 | 7669.49885 |
| 7041.48 | 7191.14 | 7533.38479 | 7678.53626 |
| 7076.34 | 7199.53 | 7593.04234 | 7687.57367 |
| 7028.33 | 7207.92 | 7584.53654 | 7696.61108 |
| 7069.93 | 7216.31 | 7514.37433 | 7705.65913 |
| 7014.61 | 7224.7 | 7602.38809 | 7714.69654 |
| 7024.45 | 7233.09 | 7559.95478 | 7723.73395 |
| 7189.82 | 7241.48 | 7650.12689 | 7732.782 |
| 7138.92 | 7249.87 | 7543.4429 | 7741.81941 |
| 7037.73 | 7258.27 | 7585.65293 | 7750.86745 |
| 7138.29 | 7266.66 | 7614.78529 | 7759.9155 |
| 7109.51 | 7275.05 | 7622.88706 | 7768.95291 |
| 7067.15 | 7283.44 | 7567.13155 | 7778.00095 |
| 7197.12 | 7291.84 | 7591.19233 | 7787.049 |
| 7201.74 | 7300.23 | 7611.00021 | 7796.09704 |
| 7299.03 | 7308.62 | 7636.9429 | 7805.14509 |
| 7160.81 | 7317.02 | 7623.12097 | 7814.19313 |
| 7163.16 | 7325.41 | 7729.03945 | 7823.23054 |
| 7148.87 | 7333.8 | 7616.3057 | 7832.27859 |
| 7260.84 | 7342.2 | 7697.23839 | 7841.33727 |
| 7280.96 | 7350.59 | 7748.4752 | 7850.38531 |
| 7245.3 | 7358.99 | 7838.94501 | 7859.43335 |
| 7238.86 | 7367.38 | 7771.46212 | 7868.4814 |
| 7278.82 | 7375.78 | 7734.56822 | 7877.52944 |
| 7203.47 | 7384.17 | 7758.13991 | 7886.57749 |
| 7277.32 | 7392.57 | 7729.486 | 7895.63616 |
| 7202.92 | 7400.96 | 7669.40316 | 7904.68421 |
| 7276.15 | 7409.36 | 7763.40288 | 7913.73225 |
| 7347.64 | 7417.76 | 7736.58834 | 7922.79093 |
| 7375.3 | 7426.15 | 7810.05719 | 7931.83897 |
| 7285.37 | 7434.55 | 7883.31339 | 7940.89765 |
| 7281.06 | 7442.94 | 7819.36041 | 7949.9457 |
| 7325.7 | 7451.34 | 7863.10148 | 7959.00437 |
| 7347.98 | 7459.74 | 7831.65129 | 7968.05242 |
| 7358.81 | 7468.14 | 7880.15561 | 7977.11109 |
| 7278.34 | 7476.53 | 7842.73009 | 7986.16977 |
| 7536.63 | 7484.93 | 7881.52717 | 7995.21781 |
| 7323.67 | 7493.33 | 7804.87928 | 8004.27649 |
| 7367.62 | 7501.72 | 7915.14634 | 8013.32454 |
| 7432.24 | 7510.12 | 7948.25517 | 8022.38321 |
| 7418.19 | 7518.52 | 7811.83277 | 8031.44189 |
| 7401.8 | 7526.91 | 7796.11831 | 8040.50057 |
| 7435.12 | 7535.31 | 7938.72867 | 8049.55924 |
| 7356.96 | 7543.71 | 7895.52984 | 8058.60729 |
| 7459.03 | 7552.11 | 7948.14885 | 8067.66596 |
| 7423.08 | 7560.5 | 7761.08505 | 8076.72464 |
| 7446.89 | 7568.9 | 7942.89651 | 8085.78332 |
| 7479.62 | 7577.3 | 7988.8491 | 8094.84199 |
| 7454.63 | 7585.69 | 7995.29224 | 8103.89004 |
| 7499.41 | 7594.09 | 8020.03349 | 8112.94872 |
| 7502.93 | 7602.49 | 7971.81623 | 8122.00739 |
| 7535.22 | 7610.89 | 7982.01256 | 8131.06607 |
| 7470.57 | 7619.28 | 8030.43183 | 8140.12475 |
| 7547.47 | 7627.68 | 7999.72589 | 8149.18342 |
| 7478.55 | 7636.08 | 7929.09585 | 8158.2421 |
| 7557.91 | 7644.47 | 8127.27036 | 8167.30078 |
| 7535.4 | 7652.87 | 7980.53468 | 8176.35945 |
| 7493.68 | 7661.26 | 7971.63549 | 8185.41813 |
| 7514.65 | 7669.66 | 8037.17267 | 8194.47681 |
| 7566.48 | 7678.06 | 8054.48197 | 8203.53548 |
| 7508.09 | 7686.45 | 8097.07477 | 8212.59416 |
| 7558.36 | 7694.85 | 8066.79412 | 8221.65284 |
| 7532.04 | 7703.24 | 8039.87326 | 8230.71151 |
| 7529.57 | 7711.64 | 8096.61758 | 8239.77019 |
| 7571.3 | 7720.03 | 8077.69218 | 8248.82887 |
| 7614.19 | 7728.43 | 8044.01984 | 8257.88754 |
| 7546.35 | 7736.82 | 8122.63469 | 8266.94622 |
| 7536.89 | 7745.22 | 8092.91756 | 8276.0049 |
| 7525.28 | 7753.61 | 8135.85058 | 8285.06357 |
| 7552.67 | 7762 | 8199.7823 | 8294.11162 |
| 7543.48 | 7770.4 | 8086.38936 | 8303.17029 |
| 7610.12 | 7778.79 | 8085.20918 | 8312.22897 |
| 7597.23 | 7787.18 | 8246.2346 | 8321.28765 |
| 7609.78 | 7795.58 | 8186.97044 | 8330.34632 |
| 7600.35 | 7803.97 | 8149.24722 | 8339.405 |
| 7598.92 | 7812.36 | 8171.03269 | 8348.46368 |
| 7740.94 | 7820.75 | 8159.45417 | 8357.52235 |
| 7698.99 | 7829.14 | 8144.23943 | 8366.58103 |
| 7705.59 | 7837.53 | 8172.72322 | 8375.62907 |
| 7687.11 | 7845.93 | 8257.69616 | 8384.68775 |
| 7737.21 | 7854.31 | 8176.48704 | 8393.74643 |
| 7731.78 | 7862.7 | 8280.83194 | 8402.8051 |
| 7716.78 | 7871.1 | 8357.53299 | 8411.86378 |
| 7687.78 | 7879.48 | 8179.91062 | 8420.91183 |
| 7729.3 | 7887.87 | 8313.46231 | 8429.9705 |
| 7766.4 | 7896.26 | 8363.604 | 8439.02918 |
| 7800.33 | 7904.65 | 8205.36423 | 8448.08786 |
| 7758.76 | 7913.04 | 8213.26399 | 8457.1359 |
| 7848.57 | 7921.42 | 8290.52855 | 8466.19458 |
| 7649.16 | 7929.81 | 8265.42581 | 8475.25325 |
| 7857.32 | 7938.2 | 8390.60991 | 8484.3013 |
| 7768.3 | 7946.58 | 8244.26763 | 8493.35997 |
| 7725.67 | 7954.97 | 8371.42934 | 8502.40802 |
| 7725.44 | 7963.35 | 8351.88726 | 8511.4667 |
| 7724.65 | 7971.74 | 8334.1314 | 8520.51474 |
| 7740.9 | 7980.12 | 8432.70299 | 8529.57342 |
| 7888.53 | 7988.5 | 8366.07068 | 8538.62146 |
| 7839.81 | 7996.89 | 8363.97613 | 8547.66951 |
| 7862.59 | 8005.27 | 8412.91637 | 8556.72818 |
| 7890.52 | 8013.65 | 8524.42741 | 8565.77623 |
| 7815.83 | 8022.03 | 8438.65705 | 8574.82427 |
| 7865.28 | 8030.41 | 8420.25263 | 8583.88295 |
| 7930.31 | 8038.79 | 8333.536 | 8592.93099 |
| 7902.19 | 8047.17 | 8392.54498 | 8601.97904 |
| 7859.49 | 8055.55 | 8391.88578 | 8611.02708 |
| 7848.37 | 8063.93 | 8497.55971 | 8620.07513 |
| 7884.77 | 8072.31 | 8451.5965 | 8629.12317 |
| 7978.18 | 8080.68 | 8509.82933 | 8638.17121 |
| 8081.27 | 8089.06 | 8519.84491 | 8647.21926 |
| 7995.14 | 8097.43 | 8547.41433 | 8656.2673 |
| 8049.38 | 8105.81 | 8486.48091 | 8665.31535 |
| 8003.21 | 8114.18 | 8522.10958 | 8674.35276 |
| 7971.88 | 8122.56 | 8513.27418 | 8683.4008 |
| 7935.69 | 8130.93 | 8580.59758 | 8692.44885 |
| 7922.5 | 8139.3 | 8496.76229 | 8701.49689 |
| 7985.21 | 8147.67 | 8564.27708 | 8710.53431 |
| 7980.78 | 8156.05 | 8548.41376 | 8719.58235 |
| 8001.02 | 8164.42 | 8609.67679 | 8728.61976 |
| 8023.05 | 8172.78 | 8511.45606 | 8737.66781 |
| 7983.85 | 8181.15 | 8709.17338 | 8746.70522 |
| 8095.46 | 8189.52 | 8498.3465 | 8755.74263 |
| 7975.06 | 8197.89 | 8600.43736 | 8764.79068 |
| 8053.53 | 8206.26 | 8571.06045 | 8773.82809 |
| 7962.76 | 8214.62 | 8646.04971 | 8782.8655 |
| 8007.86 | 8222.99 | 8562.32075 | 8791.90291 |
| 8101.7 | 8231.35 | 8680.23239 | 8800.94032 |
| 8054.57 | 8239.72 | 8598.52356 | 8809.97774 |
| 8050.74 | 8248.08 | 8719.06137 | 8819.01515 |
| 8132.25 | 8256.44 | 8615.18429 | 8828.05256 |
| 8019.23 | 8264.8 | 8490.84013 | 8837.08997 |
| 8120.63 | 8273.16 | 8625.01912 | 8846.12739 |
| 8064.68 | 8281.52 | 8656.53311 | 8855.15417 |
| 8106.07 | 8289.88 | 8680.99792 | 8864.19158 |
| 8081.86 | 8298.24 | 8617.18315 | 8873.22899 |
| 8097.26 | 8306.59 | 8563.44776 | 8882.25577 |
| 8113.98 | 8314.95 | 8688.23848 | 8891.28255 |
| 8123.36 | 8323.31 | 8680.27492 | 8900.31996 |
| 8119.68 | 8331.66 | 8772.69044 | 8909.34674 |
| 8225.86 | 8340.01 | 8769.89416 | 8918.37352 |
| 8209.83 | 8348.36 | 8824.57582 | 8927.4003 |
| 8165.85 | 8356.72 | 8717.67918 | 8936.42708 |
| 8251.79 | 8365.07 | 8706.23888 | 8945.45386 |
| 8135.01 | 8373.42 | 8735.67958 | 8954.48064 |
| 8195.62 | 8381.77 | 8796.71932 | 8963.50742 |
| 8264.24 | 8390.11 | 8840.14143 | 8972.5342 |
| 8149.2 | 8398.46 | 8952.65189 | 8981.56098 |
| 8296.62 | 8406.81 | 8743.73882 | 8990.57713 |
| 8220.38 | 8415.15 | 8749.36328 | 8999.60391 |
| 8242.63 | 8423.5 | 8796.23024 | 9008.62006 |
| 8228.42 | 8431.84 | 8804.27885 | 9017.6362 |
| 8184.23 | 8440.18 | 8759.93174 | 9026.66298 |
| 8258.93 | 8448.52 | 8901.38319 | 9035.67913 |
| 8334.01 | 8456.86 | 8932.42936 | 9044.69528 |
| 8373.63 | 8465.2 | 8888.4969 | 9053.71143 |
| 8274.25 | 8473.54 | 8971.42845 | 9062.72757 |
| 8254.87 | 8481.87 | 8949.73866 | 9071.74372 |
| 8350.1 | 8490.21 | 8935.86357 | 9080.75987 |
| 8185.14 | 8498.54 | 8797.82508 | 9089.76539 |
| 8311.19 | 8506.88 | 9050.11773 | 9098.78153 |
| 8336.85 | 8515.21 | 8859.54528 | 9107.78705 |
| 8206.69 | 8523.54 | 8984.98457 | 9116.8032 |
| 8357.51 | 8531.87 | 8926.83679 | 9125.80871 |
| 8414.94 | 8540.2 | 8962.78443 | 9134.81423 |
| 8340.12 | 8548.53 | 9005.96199 | 9143.81974 |
| 8444.23 | 8556.86 | 9010.08731 | 9152.83589 |
| 8461.66 | 8565.18 | 8954.57633 | 9161.83077 |
| 8336.12 | 8573.51 | 9023.26066 | 9170.83629 |
| 8427.68 | 8581.83 | 9008.71575 | 9179.8418 |
| 8422.31 | 8590.15 | 9029.95898 | 9188.84732 |
| 8512.22 | 8598.47 | 9193.6212 | 9197.8422 |
| 8409.45 | 8606.79 | 8943.84839 | 9206.84772 |
| 8387.51 | 8615.11 | 9024.2282 | 9215.8426 |
| 8420.47 | 8623.43 | 9002.17691 | 9224.83748 |
| 8442.25 | 8631.75 | 9056.28443 | 9233.83237 |
| 8422.52 | 8640.06 | 9081.37654 | 9242.82725 |
| 8421.52 | 8648.38 | 9067.31007 | 9251.82213 |
| 8523.51 | 8656.69 | 9014.97814 | 9260.81702 |
| 8497.12 | 8665 | 9042.28176 | 9269.8119 |
| 8508.76 | 8673.31 | 9141.71456 | 9278.79615 |
| 8520.61 | 8681.62 | 9080.62165 | 9287.79103 |
| 8557.35 | 8689.93 | 9212.11068 | 9296.77529 |
| 8510.59 | 8698.23 | 9169.05007 | 9305.77017 |
| 8430.29 | 8706.54 | 9045.02488 | 9314.75442 |
| 8508.26 | 8714.84 | 9093.58236 | 9323.73867 |
| 8469.61 | 8723.15 | 9153.25055 | 9332.72292 |
| 8528.09 | 8731.45 | 9092.84874 | 9341.70717 |
| 8544.16 | 8739.75 | 9289.88559 | 9350.69142 |
| 8606.51 | 8748.05 | 9126.37222 | 9359.66504 |
| 8442.49 | 8756.34 | 9111.77414 | 9368.64929 |
| 8556.9 | 8764.64 | 9158.73679 | 9377.62291 |
| 8619.68 | 8772.93 | 9366.55474 | 9386.59653 |
| 8522.11 | 8781.23 | 9226.06019 | 9395.57015 |
| 8531.85 | 8789.52 | 9243.74162 | 9404.5544 |
| 8539.8 | 8797.81 | 9191.3778 | 9413.52802 |
| 8741.71 | 8806.1 | 9278.33897 | 9422.49101 |
| 8670.19 | 8814.39 | 9264.59147 | 9431.46462 |
| 8657.41 | 8822.67 | 9190.76112 | 9440.43824 |
| 8681.47 | 8830.96 | 9371.38178 | 9449.40123 |
| 8651.52 | 8839.24 | 9257.72303 | 9458.36422 |
| 8742.85 | 8847.52 | 9279.39156 | 9467.3272 |
| 8739.07 | 8855.81 | 9297.26437 | 9476.30082 |
| 8655.77 | 8864.08 | 9445.20149 | 9485.26381 |
| 8684.64 | 8872.36 | 9449.94347 | 9494.21616 |
| 8772.42 | 8880.64 | 9349.5963 | 9503.17915 |
| 8762.05 | 8888.91 | 9278.25391 | 9512.14213 |
| 8725.12 | 8897.19 | 9314.43545 | 9521.09449 |
| 8675.44 | 8905.46 | 9312.27711 | 9530.04684 |
| 8779.16 | 8913.73 | 9415.23981 | 9538.9992 |
| 8797.95 | 8922 | 9337.12467 | 9547.95155 |
| 8759.73 | 8930.27 | 9338.60256 | 9556.90391 |
| 8755.31 | 8938.53 | 9349.40492 | 9565.85626 |
| 8836.13 | 8946.8 | 9391.84886 | 9574.80861 |
| 8806.52 | 8955.06 | 9368.19211 | 9583.75034 |
| 8884.67 | 8963.32 | 9538.77592 | 9592.70269 |
| 8847.58 | 8971.58 | 9514.35364 | 9601.64441 |
| 8915.24 | 8979.84 | 9551.7579 | 9610.58613 |
| 8943.53 | 8988.1 | 9530.87616 | 9619.52786 |
| 8775.1 | 8996.35 | 9402.92767 | 9628.46958 |
| 8840.59 | 9004.6 | 9520.02063 | 9637.40067 |
| 8945.82 | 9012.86 | 9472.11171 | 9646.34239 |
| 8768.79 | 9021.11 | 9526.43188 | 9655.27348 |
| 8963.4 | 9029.36 | 9513.69444 | 9664.2152 |
| 8802.72 | 9037.6 | 9465.43466 | 9673.14629 |
| 8938.69 | 9045.85 | 9565.35654 | 9682.07738 |
| 8883.74 | 9054.09 | 9598.59296 | 9691.00847 |
| 8953.77 | 9062.34 | 9615.42381 | 9699.92893 |
| 8756.23 | 9070.58 | 9548.99351 | 9708.86002 |
| 8995.79 | 9078.82 | 9667.64942 | 9717.78047 |
| 8862.77 | 9087.05 | 9551.17312 | 9726.70093 |
| 8983.66 | 9095.29 | 9577.2434 | 9735.62139 |
| 8923.91 | 9103.52 | 9528.52643 | 9744.54185 |
| 9018.19 | 9111.76 | 9625.87531 | 9753.4623 |
| 8885.67 | 9119.99 | 9629.69229 | 9762.38276 |
| 8851.16 | 9128.22 | 9640.60097 | 9771.29259 |
| 8913.94 | 9136.44 | 9577.03075 | 9780.20241 |
| 8829.33 | 9144.67 | 9577.8707 | 9789.12287 |
| 8909.11 | 9152.89 | 9744.89271 | 9798.03269 |
| 9038.91 | 9161.12 | 9667.45804 | 9806.94252 |
| 8876.37 | 9169.34 | 9678.87707 | 9815.84171 |
| 8999.29 | 9177.56 | 9756.71577 | 9824.75154 |
| 8926.37 | 9185.77 | 9643.98203 | 9833.65073 |
| 8944.92 | 9193.99 | 9651.4884 | 9842.54992 |
| 9066.5 | 9202.2 | 9673.35894 | 9851.45975 |
| 8946.86 | 9210.41 | 9633.62622 | 9860.34831 |
| 9065.83 | 9218.62 | 9668.93592 | 9869.2475 |
| 9095.48 | 9226.83 | 9760.26694 | 9878.1467 |
| 8993.4 | 9235.04 | 9623.68507 | 9887.03526 |
| 8962.07 | 9243.24 | 9650.48897 | 9895.93445 |
| 9059.51 | 9251.44 | 9693.24124 | 9904.82301 |
| 9047.73 | 9259.65 | 9671.26438 | 9913.71157 |
| 8968.01 | 9267.85 | 9717.67415 | 9922.5895 |
| 9040.08 | 9276.04 | 9807.67615 | 9931.47806 |
| 9023.07 | 9284.24 | 9711.59251 | 9940.35599 |
| 9139.25 | 9292.43 | 9733.76075 | 9949.24455 |
| 9016.64 | 9300.62 | 9698.58926 | 9958.12248 |
| 9115.75 | 9308.81 | 9723.69201 | 9967.00041 |
| 9125.92 | 9317 | 9709.83818 | 9975.87833 |
| 9144.78 | 9325.19 | 9753.58989 | 9984.74563 |
| 9094.89 | 9333.37 | 9792.88669 | 9993.62356 |
| 9090.68 | 9341.56 | 9774.57795 | 10002.49086 |
| 9293.38 | 9349.74 | 9632.8288 | 10011.35815 |
| 9086.22 | 9357.91 | 9772.10064 | 10020.22545 |
| 9230.7 | 9366.09 | 9852.6293 | 10029.09274 |
| 9183.22 | 9374.27 | 9834.91597 | 10037.94941 |
| 9150.74 | 9382.44 | 9765.99773 | 10046.8167 |
| 9166.44 | 9390.61 | 9782.9668 | 10055.67337 |
| 9231.66 | 9398.78 | 9848.84422 | 10064.53003 |
| 9171.34 | 9406.95 | 9963.1728 | 10073.3867 |
| 9160.94 | 9415.11 | 9906.58796 | 10082.23273 |
| 9188.85 | 9423.27 | 9835.13925 | 10091.08939 |
| 9270.99 | 9431.43 | 9859.01928 | 10099.93542 |
| 9298.08 | 9439.59 | 9869.3113 | 10108.78145 |
| 9343.57 | 9447.75 | 9895.23272 | 10117.62749 |
| 9186.81 | 9455.91 | 9965.4481 | 10126.47352 |
| 9255.26 | 9464.06 | 9908.03395 | 10135.31955 |
| 9275.02 | 9472.21 | 9890.55453 | 10144.15495 |
| 9262.04 | 9480.36 | 9973.81568 | 10152.99035 |
| 9172.11 | 9488.51 | 9909.95839 | 10161.82575 |
| 9297.09 | 9496.65 | 9915.67854 | 10170.66115 |
| 9198.16 | 9504.8 | 9997.12157 | 10179.49655 |
| 9235.74 | 9512.94 | 9875.62685 | 10188.32131 |
| 9274.94 | 9521.08 | 10079.09621 | 10197.15671 |
| 9273.83 | 9529.21 | 10044.84974 | 10205.98148 |
| 9373.51 | 9537.35 | 10040.20344 | 10214.80625 |
| 9289.23 | 9545.48 | 10006.9564 | 10223.62038 |
| 9286.27 | 9553.61 | 10014.08001 | 10232.44515 |
| 9321.66 | 9561.74 | 9968.14869 | 10241.25929 |
| 9348.35 | 9569.87 | 10026.64733 | 10250.07342 |
| 9276.85 | 9577.99 | 10005.2765 | 10258.88756 |
| 9393.31 | 9586.11 | 9953.33797 | 10267.70169 |
| 9396.65 | 9594.23 | 10008.33859 | 10276.51583 |
| 9309.83 | 9602.35 | 10104.18832 | 10285.31933 |
| 9350.15 | 9610.47 | 10110.48261 | 10294.12283 |
| 9418.65 | 9618.58 | 10020.09786 | 10302.92633 |
| 9361.02 | 9626.69 | 10019.28981 | 10311.72984 |
| 9410.98 | 9634.8 | 10130.13101 | 10320.52271 |
| 9444.85 | 9642.91 | 10280.85378 | 10329.32621 |
| 9381.5 | 9651.01 | 10042.53191 | 10338.11908 |
| 9342.16 | 9659.12 | 10104.4435 | 10346.91195 |
| 9387.19 | 9667.22 | 10095.39545 | 10355.70482 |
| 9501.37 | 9675.32 | 10077.97983 | 10364.48706 |
| 9460.68 | 9683.41 | 10096.40552 | 10373.27993 |
| 9353.15 | 9691.51 | 10129.87584 | 10382.06217 |
| 9430.84 | 9699.6 | 10145.2926 | 10390.84441 |
| 9498.77 | 9707.69 | 10204.37601 | 10399.61601 |
| 9533.76 | 9715.78 | 10013.90989 | 10408.39825 |
| 9410.66 | 9723.86 | 10061.06392 | 10417.16986 |
| 9455.32 | 9731.95 | 10192.65927 | 10425.95209 |
| 9519.28 | 9740.03 | 10277.07933 | 10434.71307 |
| 9451.38 | 9748.1 | 10291.26276 | 10443.48467 |
| 9490.15 | 9756.18 | 10206.26855 | 10452.25628 |
| 9613.8 | 9764.26 | 10179.81551 | 10461.01725 |
| 9651.91 | 9772.33 | 10285.72335 | 10469.77823 |
| 9637.62 | 9780.4 | 10374.84287 | 10478.5392 |
| 9618.89 | 9788.47 | 10182.52674 | 10487.30017 |
| 9494.91 | 9796.53 | 10283.04403 | 10496.05052 |
| 9567.08 | 9804.59 | 10335.92884 | 10504.80086 |
| 9661.32 | 9812.65 | 10345.58292 | 10513.5512 |
| 9590.17 | 9820.71 | 10357.01259 | 10522.30154 |
| 9702.18 | 9828.77 | 10371.53624 | 10531.05188 |
| 9670.38 | 9836.82 | 10340.78777 | 10539.79159 |
| 9653.58 | 9844.87 | 10443.30393 | 10548.5313 |
| 9766.09 | 9852.92 | 10492.10595 | 10557.27101 |
| 9764.78 | 9860.97 | 10515.51816 | 10566.01072 |
| 9811.37 | 9869.01 | 10366.53908 | 10574.7398 |
| 9733.82 | 9877.05 | 10572.61335 | 10583.4795 |
| 9801.1 | 9885.09 | 10577.49355 | 10592.20858 |
| 9840.77 | 9893.13 | 10447.76947 | 10600.93766 |
| 9746.34 | 9901.16 | 10550.41321 | 10609.6561 |
| 9911.47 | 9909.19 | 10505.83219 | 10618.37455 |
| 9792.98 | 9917.23 | 10471.06473 | 10627.10362 |
| 9869.08 | 9925.25 | 10496.75224 | 10635.8646 |
| 9780.12 | 9933.28 | 10497.36891 | 10644.47672 |
| 9735.3 | 9941.3 | 10391.57803 | 10653.30149 |
| 9739.86 | 9949.32 | 10629.28324 | 10661.91361 |
| 9933.88 | 9957.34 | 10691.04597 | 10670.63205 |
| 9850.34 | 9965.35 | 10560.50321 | 10679.3505 |
| 9856.17 | 9973.37 | 10470.94777 | 10688.06894 |
| 9982.13 | 9981.38 | 10565.69175 | 10696.78739 |
| 9943.32 | 9989.39 | 10410.64165 | 10705.50583 |
| 9790.45 | 9997.39 | 10533.92259 | 10714.11796 |
| 10015.3 | 10005.4 | 10686.58043 | 10722.8364 |
| 9852.79 | 10013.4 | 10680.20108 | 10731.55484 |
| 9847.35 | 10021.4 | 10558.27044 | 10740.27329 |
| 9849.39 | 10029.4 | 10556.66497 | 10748.88541 |
| 9773.67 | 10037.4 | 10558.61067 | 10757.60386 |
| 9813.91 | 10045.4 | 10592.55945 | 10766.3223 |
| 9925.31 | 10053.4 | 10490.94704 | 10774.93442 |
| 9993.39 | 10061.4 | 10520.61101 | 10783.65287 |
| 10067.7 | 10069.3 | 10694.12933 | 10792.37131 |
| 9981.61 | 10077.3 | 10630.16571 | 10800.98344 |
| 9753.23 | 10085.3 | 10592.77209 | 10809.70188 |
| 9963.56 | 10093.3 | 10774.29649 | 10818.314 |
| 10044 | 10101.2 | 10730.49162 | 10827.03245 |
| 10015.2 | 10109.2 | 10597.17384 | 10835.64457 |
| 10060.9 | 10117.2 | 10912.94102 | 10844.36301 |
| 9856.92 | 10125.1 | 10755.26476 | 10852.97514 |
| 10043.4 | 10133.1 | 10757.49753 | 10861.58726 |
| 9964.02 | 10141.1 | 10784.2908 | 10870.3057 |
| 10121.7 | 10149 | 10810.33981 | 10878.91782 |
| 10155.1 | 10157 | 10908.36916 | 10887.52995 |
| 9946.14 | 10164.9 | 10555.60175 | 10896.24839 |
| 10017.4 | 10172.9 | 10825.65025 | 10904.86051 |
| 9984.28 | 10180.8 | 10830.32844 | 10913.47264 |
| 10086 | 10188.8 | 10682.75282 | 10922.08476 |
| 10105 | 10196.7 | 10722.09214 | 10930.8032 |
| 9936.39 | 10204.6 | 10854.78262 | 10939.41533 |
| 10117.1 | 10212.6 | 10860.73668 | 10948.02745 |
| 9984.41 | 10220.5 | 10784.82242 | 10956.63957 |
| 10051.6 | 10228.5 | 10935.0561 | 10965.25169 |
| 10033.8 | 10236.4 | 10778.76203 | 10973.86381 |
| 10107.4 | 10244.3 | 10688.49423 | 10982.47594 |
| 10103.3 | 10252.2 | 10909.32606 | 10991.08806 |
| 10139.6 | 10260.1 | 10826.39451 | 10999.70018 |
| 10146.5 | 10268.1 | 10676.58611 | 11008.3123 |
| 10016.6 | 10276 | 10730.17265 | 11016.92443 |
| 10251.6 | 10283.9 | 10839.47218 | 11025.53655 |
| 10197.8 | 10291.8 | 10853.61307 | 11034.14867 |
| 10177.6 | 10299.7 | 10890.40065 | 11042.76079 |
| 9992.38 | 10307.6 | 11005.97321 | 11051.37291 |
| 10290.7 | 10315.5 | 11060.19768 | 11059.87871 |
| 10071.6 | 10323.4 | 10945.79468 | 11068.49084 |
| 10120.9 | 10331.3 | 10847.1274 | 11077.10296 |
| 10247.3 | 10339.2 | 10874.66493 | 11085.71508 |
| 10211.7 | 10347.1 | 10858.29126 | 11094.22088 |
| 10229 | 10355 | 10838.83424 | 11102.833 |
| 10260.7 | 10362.9 | 11112.08306 | 11111.44512 |
| 10331.3 | 10370.8 | 10925.27443 | 11120.05725 |
| 10253 | 10378.6 | 10966.20859 | 11128.56305 |
| 10339.9 | 10386.5 | 11007.67437 | 11137.17517 |
| 10201.9 | 10394.4 | 11123.99118 | 11145.68097 |
| 10090 | 10402.3 | 11160.67244 | 11154.29309 |
| 13717.8 | 13642.9 | 14445.29332 | 14700.99892 |
| 13730.9 | 13683.9 | 14343.33005 | 14746.29231 |
| 13631.5 | 13724.7 | 14439.33926 | 14791.26672 |
| 13820.5 | 13765.3 | 14578.19644 | 14836.13482 |
| 13910.9 | 13805.8 | 14775.53099 | 14880.79026 |
| 13658 | 13846.1 | 14833.05146 | 14925.23307 |
| 13531.2 | 13886.3 | 14749.80095 | 14969.56955 |
| 13658.7 | 13926.2 | 14706.63402 | 15013.69338 |
| 14004.2 | 13966 | 14912.26172 | 15057.7109 |
| 13743.6 | 14005.6 | 14759.05101 | 15101.40944 |
| 13748.8 | 14045.1 | 14841.87623 | 15145.00167 |
| 14226.2 | 14084.4 | 14933.31358 | 15188.48757 |
| 14032.7 | 14123.5 | 15031.13027 | 15231.6545 |
| 13935.2 | 14162.4 | 15236.65166 | 15274.71511 |
| 14182.9 | 14201.1 | 15164.77765 | 15317.56308 |
| 14101.4 | 14239.7 | 15113.53021 | 15360.30472 |
| 14133.4 | 14278.1 | 15185.40422 | 15402.7274 |
| 14171.6 | 14316.3 | 15042.71943 | 15445.04375 |
| 13981 | 14354.3 | 15222.08548 | 15487.14746 |
| 14251.3 | 14392.2 | 15511.38899 | 15529.14485 |
| 14108.3 | 14429.9 | 15448.1271 | 15570.92959 |
| 14052 | 14467.4 | 15416.76197 | 15612.50168 |
| 14262.2 | 14504.7 | 15447.1702 | 15653.86114 |
| 14227.8 | 14541.9 | 15740.30132 | 15695.00794 |
| 14454 | 14578.8 | 15671.51067 | 15736.04842 |
| 14521.5 | 14615.6 | 15528.82588 | 15776.87626 |
| 14578.1 | 14652.2 | 15545.83748 | 15817.49146 |
| 14258 | 14688.7 | 15569.44107 | 15857.89401 |
| 14470.2 | 14724.9 | 15692.13723 | 15898.19023 |
| 14844.2 | 14761 | 15964.32282 | 15938.27381 |
| 14573.7 | 14796.9 | 16008.23402 | 15978.03843 |
| 14836.4 | 14832.6 | 15947.2049 | 16017.80304 |
| 15107.4 | 14868.1 | 15989.09597 | 16057.24869 |
| 15030.3 | 14903.5 | 15784.21252 | 16096.58801 |
| 14907.5 | 14938.6 | 15820.78745 | 16135.60837 |
| 14979.3 | 14973.6 | 16005.36331 | 16174.5224 |
| 14873.7 | 15008.4 | 15796.86489 | 16213.33011 |
| 14643.3 | 15043 | 15971.97804 | 16251.81885 |
| 15078.5 | 15077.5 | 16156.44758 | 16290.20128 |
| 15037.4 | 15111.7 | 15877.13838 | 16328.26473 |
| 14847.4 | 15145.8 | 16228.00262 | 16366.22186 |
| 14959.6 | 15179.7 | 16245.75847 | 16403.96635 |
| 14896.9 | 15213.4 | 16118.70309 | 16441.60451 |
| 15135.1 | 15246.9 | 16557.60235 | 16478.92371 |
| 15162.4 | 15280.2 | 16481.47545 | 16516.13658 |
| 14896.4 | 15313.4 | 16125.61405 | 16553.13681 |
| 15162.8 | 15346.3 | 16107.00761 | 16589.92439 |
| 15247.9 | 15379.1 | 16389.61281 | 16626.49933 |
| 15190.8 | 15411.7 | 16483.17661 | 16662.86163 |
| 15520 | 15444.1 | 16668.81569 | 16699.1176 |
| 15471.1 | 15476.4 | 16691.35605 | 16735.16092 |
| 15666 | 15508.4 | 16653.50525 | 16770.9916 |
| 15579.9 | 15540.3 | 16690.08018 | 16806.60964 |
| 15415.5 | 15571.9 | 16678.27839 | 16842.01503 |
| 15643.4 | 15603.4 | 16641.06551 | 16877.20778 |
| 15563 | 15634.8 | 16819.79363 | 16912.2942 |
| 15476.9 | 15665.9 | 16900.4924 | 16947.16798 |
| 15456 | 15696.8 | 16948.23121 | 16981.82911 |
| 15695.4 | 15727.6 | 17045.94158 | 17016.2776 |
| 15862.7 | 15758.1 | 16983.21131 | 17050.51345 |
| 15666.7 | 15788.5 | 17169.16935 | 17084.53665 |
| 15704.7 | 15818.7 | 16969.60203 | 17118.45352 |
| 15756.1 | 15848.8 | 16964.2859 | 17152.15775 |
| 15485.2 | 15878.6 | 17002.77465 | 17185.54302 |
| 15708.5 | 15908.2 | 17013.83219 | 17218.92828 |
| 15951.5 | 15937.7 | 17140.78125 | 17251.99458 |
| 15754.5 | 15967 | 17071.35266 | 17284.84823 |
| 15930.3 | 15996.1 | 17236.57782 | 17317.59556 |
| 16152.2 | 16025 | 17379.15628 | 17350.13024 |
| 16099.9 | 16053.7 | 17368.63036 | 17382.34596 |
| 16089.3 | 16082.3 | 17567.55975 | 17414.45535 |
| 16177.6 | 16110.6 | 17491.43284 | 17446.45842 |
| 16188.2 | 16138.8 | 17373.20222 | 17478.14253 |
| 16133.4 | 16166.8 | 17429.44682 | 17509.72031 |
| 16262.9 | 16194.6 | 17564.0511 | 17540.97912 |
| 16245.9 | 16222.3 | 17488.77478 | 17572.13161 |
| 16245.6 | 16249.7 | 17580.1058 | 17603.07146 |
| 16586.4 | 16277 | 17563.94478 | 17633.79866 |
| 16327.2 | 16304 | 17350.23656 | 17664.41954 |
| 16099.8 | 16331 | 17587.44205 | 17694.72145 |
| 16519 | 16357.7 | 17789.98641 | 17724.91704 |
| 16444.7 | 16384.2 | 17607.11172 | 17754.89999 |
| 16352.2 | 16410.5 | 17743.20451 | 17784.67028 |
| 16688 | 16436.7 | 17969.56511 | 17814.33426 |
| 16551.2 | 16462.7 | 17865.47538 | 17843.67927 |
| 16492.8 | 16488.5 | 17981.26058 | 17872.91796 |
| 16659.9 | 16514.1 | 18114.27002 | 17901.83768 |
| 16566 | 16539.6 | 18044.09718 | 17930.75739 |
| 16569.3 | 16564.8 | 17829.00677 | 17959.35815 |
| 16830.6 | 16589.9 | 17814.9722 | 17987.74625 |
| 16687.7 | 16614.8 | 17924.59069 | 18016.02804 |
| 16783.6 | 16639.5 | 18192.20441 | 18043.99085 |
| 16804.8 | 16664.1 | 18308.94651 | 18071.84735 |
| 16514.9 | 16688.4 | 18262.48358 | 18099.4912 |
| 16632.5 | 16712.6 | 18105.33893 | 18127.02872 |
| 16685.2 | 16736.6 | 18267.16177 | 18154.24728 |
| 16858.7 | 16760.4 | 18438.55364 | 18181.35952 |
| 16880.2 | 16784.1 | 18289.59582 | 18208.25911 |
| 16946.7 | 16807.5 | 18450.6744 | 18234.94606 |
| 17092.3 | 16830.8 | 18332.76275 | 18261.52668 |
| 16909.6 | 16853.9 | 18380.82052 | 18287.78834 |
| 16983.6 | 16876.8 | 18641.73593 | 18313.94367 |
| 17313.7 | 16899.6 | 18620.57775 | 18339.88636 |
| 17223.8 | 16922.2 | 18705.31678 | 18365.6164 |
| 16942.7 | 16944.6 | 18609.9455 | 18391.24013 |
| 16989.7 | 16966.8 | 18485.97347 | 18416.54488 |
| 17132.1 | 16988.8 | 18668.63552 | 18441.74331 |
| 17123.9 | 17010.7 | 18673.42003 | 18466.7291 |
| 17117.2 | 17032.4 | 18660.9803 | 18491.60856 |
| 17244.7 | 17053.9 | 18663.85101 | 18516.16906 |
| 17169.8 | 17075.2 | 18673.63268 | 18540.62323 |
| 17207.6 | 17096.4 | 18788.56729 | 18564.86476 |
| 17355.2 | 17117.4 | 18869.26607 | 18588.89365 |
| 17344.8 | 17138.2 | 18670.76197 | 18612.81621 |
| 17373 | 17158.9 | 18674.27061 | 18636.52612 |
| 17290.8 | 17179.3 | 18745.50668 | 18660.0234 |
| 17348.4 | 17199.6 | 18713.50361 | 18683.30802 |
| 17392.3 | 17219.8 | 18938.37569 | 18706.48633 |
| 17570.7 | 17239.7 | 19040.12632 | 18729.45199 |
| 17613.3 | 17259.5 | 19031.08891 | 18752.205 |
| 17587 | 17279.1 | 19112.85091 | 18774.74537 |
| 17602.7 | 17298.5 | 19189.19046 | 18797.17942 |
| 17524.3 | 17317.8 | 19175.36854 | 18819.2945 |
| 17636.7 | 17336.9 | 19238.31145 | 18841.40958 |
| 17640.9 | 17355.8 | 19375.99909 | 18863.20569 |
| 17612.3 | 17374.5 | 19323.47577 | 18884.89548 |
| 17673.5 | 17393.1 | 19220.98089 | 18906.37262 |
| 17456 | 17411.5 | 19132.09528 | 18927.63712 |
| 17574.3 | 17429.8 | 19014.71524 | 18948.7953 |
| 17519.8 | 17447.9 | 19088.18409 | 18969.74083 |
| 17477.5 | 17465.8 | 19180.36569 | 18990.47372 |
| 17599.4 | 17483.5 | 19203.96929 | 19010.99396 |
| 17575.6 | 17501.1 | 19221.29985 | 19031.40788 |
| 17689 | 17518.5 | 19139.85682 | 19051.60915 |
| 17582.1 | 17535.7 | 19300.72276 | 19071.7041 |
| 17679.5 | 17552.8 | 19407.04526 | 19091.58641 |
| 17783 | 17569.7 | 19264.89208 | 19111.25607 |
| 17793.5 | 17586.5 | 19458.08005 | 19130.71309 |
| 17815.2 | 17603 | 19467.96805 | 19150.06378 |
| 17753 | 17619.4 | 19377.91289 | 19169.20183 |
| 17695.1 | 17635.7 | 19443.93916 | 19188.23356 |
| 17894.3 | 17651.8 | 19367.17432 | 19206.94632 |
| 17919 | 17667.7 | 19374.29793 | 19225.55275 |
| 17747.9 | 17683.5 | 19411.82977 | 19244.05287 |
| 17903.3 | 17699 | 19382.91005 | 19262.34034 |
| 17702.3 | 17714.5 | 19406.19468 | 19280.41516 |
| 17876.5 | 17729.8 | 19404.17455 | 19298.38366 |
| 17870.2 | 17744.9 | 19451.7007 | 19316.0332 |
| 17712.4 | 17759.8 | 19510.39072 | 19333.68273 |
| 18093.2 | 17774.6 | 19428.20343 | 19351.0133 |
| 17955.9 | 17789.3 | 19623.62418 | 19368.23754 |
| 17789.3 | 17803.7 | 19671.36298 | 19385.35547 |
| 17937.1 | 17818 | 19655.94622 | 19402.26074 |
| 17959.3 | 17832.2 | 19413.74357 | 19418.95338 |
| 17961.7 | 17846.2 | 19348.24892 | 19435.53968 |
| 17998.1 | 17860 | 19524.85058 | 19451.91335 |
| 17982.4 | 17873.7 | 19634.46908 | 19468.07437 |
| 18030 | 17887.3 | 19688.4809 | 19484.12907 |
| 18135.2 | 17900.6 | 19680.08143 | 19499.97112 |
| 18138.8 | 17913.8 | 19754.50717 | 19515.70685 |
| 18044.6 | 17926.9 | 19699.00683 | 19531.22993 |
| 18098.1 | 17939.8 | 19780.55619 | 19546.54037 |
| 18071 | 17952.6 | 19741.8548 | 19561.74449 |
| 18110 | 17965.2 | 19895.91609 | 19576.84228 |
| 18191.2 | 17977.6 | 19907.2926 | 19591.72743 |
| 18246.8 | 17989.9 | 19836.3755 | 19606.39994 |
| 18306.8 | 18002 | 20187.23973 | 19620.96612 |
| 18231.7 | 18014 | 20066.98899 | 19635.31966 |
| 18379.9 | 18025.9 | 20086.97762 | 19649.56687 |
| 18409.3 | 18037.6 | 20065.28783 | 19663.60144 |
| 18237.5 | 18049.1 | 19990.86208 | 19677.42336 |
| 18340.2 | 18060.5 | 19935.68071 | 19691.13897 |
| 18263.8 | 18071.7 | 19811.60235 | 19704.74825 |
| 18190.4 | 18082.8 | 19963.11191 | 19718.14488 |
| 18225.2 | 18093.8 | 19738.34615 | 19731.43519 |
| 18222.1 | 18104.6 | 19899.7437 | 19744.51286 |
| 18359.4 | 18115.2 | 20085.59543 | 19757.37788 |
| 18347.4 | 18125.7 | 19771.09348 | 19770.13658 |
| 18394.7 | 18136.1 | 20006.0662 | 19782.78896 |
| 18361.6 | 18146.3 | 20044.12965 | 19795.22869 |
| 18476.2 | 18156.3 | 19967.68378 | 19807.45578 |
| 18441.2 | 18166.2 | 20310.57383 | 19819.57654 |
| 18437.3 | 18176 | 20285.05643 | 19831.59098 |
| 18435.3 | 18185.6 | 20226.04745 | 19843.39278 |
| 18148.7 | 18195.1 | 20168.20801 | 19855.08826 |
| 18339.3 | 18204.5 | 20021.80193 | 19866.57108 |
| 18269.8 | 18213.7 | 20036.36811 | 19877.94759 |
| 18519.2 | 18222.7 | 20079.21608 | 19889.11145 |
| 18730 | 18231.7 | 20225.40951 | 19900.16899 |
| 18486.4 | 18240.4 | 20234.3406 | 19911.01389 |
| 18519.2 | 18249.1 | 20224.02732 | 19921.85878 |
| 18440 | 18257.5 | 20059.97171 | 19932.38471 |
| 18521.2 | 18265.9 | 20092.93168 | 19942.80431 |
| 18537 | 18274.1 | 20167.88904 | 19953.1176 |
| 18354.3 | 18282.2 | 20234.87221 | 19963.21823 |
| 18530.8 | 18290.2 | 20208.61056 | 19973.21255 |
| 18461.8 | 18298 | 20042.21585 | 19983.10054 |
| 18252.3 | 18305.6 | 20045.51185 | 19992.77589 |
| 18464.1 | 18313.2 | 20094.42019 | 20002.23859 |
| 18443.9 | 18320.5 | 20083.78794 | 20011.70129 |
| 18425.9 | 18327.8 | 20312.91292 | 20020.95135 |
| 18532.7 | 18334.9 | 20441.3505 | 20029.98876 |
| 18400.8 | 18341.9 | 20219.24281 | 20038.91985 |
| 18407.5 | 18348.8 | 20249.86368 | 20047.74462 |
| 18582.3 | 18355.5 | 20257.5189 | 20056.46306 |
| 18471.6 | 18362.1 | 20076.87698 | 20064.96886 |
| 18338.1 | 18368.5 | 20097.07826 | 20073.26202 |
| 18234.9 | 18374.9 | 20165.97523 | 20081.55517 |
| 18406.5 | 18381.1 | 20264.32354 | 20089.63568 |
| 18540.8 | 18387.1 | 20403.60601 | 20097.60987 |
| 18423.2 | 18393.1 | 20215.73416 | 20105.37141 |
| 18415.6 | 18398.9 | 20192.13057 | 20113.02663 |
| 18446.2 | 18404.6 | 20203.93237 | 20120.46921 |
| 18322.5 | 18410.1 | 20172.03562 | 20127.91178 |
| 18437.8 | 18415.5 | 20300.47319 | 20135.14171 |
| 18423.5 | 18420.8 | 20249.11943 | 20142.26532 |
| 18356 | 18426 | 20029.03186 | 20149.17628 |
| 18391.3 | 18431 | 20207.8663 | 20155.98092 |
| 18408.8 | 18435.9 | 20179.37187 | 20162.67924 |
| 18531.8 | 18440.7 | 20093.25065 | 20169.27123 |
| 18468.3 | 18445.4 | 20303.3439 | 20175.65058 |
| 18361.3 | 18449.9 | 20152.15331 | 20181.92361 |
| 18599.3 | 18454.3 | 20198.50992 | 20188.09031 |
| 18515.5 | 18458.6 | 20151.83434 | 20194.04437 |
| 18247.5 | 18462.8 | 20188.09031 | 20199.89211 |
| 18460 | 18466.8 | 20311.63705 | 20205.63353 |
| 18494.6 | 18470.8 | 20151.30273 | 20211.26862 |
| 18527.7 | 18474.6 | 20302.17435 | 20216.69107 |
| 18774.4 | 18478.2 | 20309.5106 | 20222.00719 |
| 18580.6 | 18481.8 | 20063.48035 | 20227.21699 |
| 18320.2 | 18485.2 | 20200.3174 | 20232.32047 |
| 18550.9 | 18488.6 | 20191.38631 | 20237.21131 |
| 18563.1 | 18491.8 | 20298.02778 | 20242.10214 |
| 18307.7 | 18494.8 | 20288.24611 | 20246.78033 |
| 18401.2 | 18497.8 | 20111.53812 | 20251.24588 |
| 18509 | 18500.7 | 20291.22314 | 20255.71142 |
| 18478.9 | 18503.4 | 20077.19595 | 20259.96432 |
| 18417.6 | 18506 | 19922.49672 | 20264.21722 |
| 18209.8 | 18508.5 | 20055.71881 | 20268.25748 |
| 18303.6 | 18510.9 | 19994.68969 | 20272.19141 |
| 18353.4 | 18513.2 | 20125.1474 | 20275.9127 |
| 18234.8 | 18515.3 | 20054.23029 | 20279.63398 |
| 18344 | 18517.4 | 20066.77635 | 20283.14263 |
| 18310.8 | 18519.3 | 19999.04892 | 20286.54495 |
| 18356.8 | 18521.1 | 19882.62578 | 20289.84094 |
| 18370.6 | 18522.8 | 20085.17014 | 20293.03062 |
| 18232.8 | 18524.4 | 20073.26202 | 20296.00765 |
| 18224.3 | 18525.9 | 20090.06097 | 20298.98468 |
| 18320.7 | 18527.2 | 19954.92508 | 20301.74906 |
| 18386.9 | 18528.5 | 20094.42019 | 20304.40712 |
| 18365.5 | 18529.7 | 20061.67287 | 20306.95886 |
| 18486.6 | 18530.7 | 19928.5571 | 20309.40428 |
| 18400.7 | 18531.6 | 20064.11828 | 20311.74338 |
| 18298.6 | 18532.5 | 20049.44578 | 20313.97615 |
| 18347.1 | 18533.2 | 20046.36243 | 20315.99628 |
| 18329.9 | 18533.8 | 19978.31603 | 20318.0164 |
| 18524.4 | 18534.3 | 19975.87061 | 20319.82389 |
| 18237.7 | 18534.7 | 19918.88175 | 20321.52505 |
| 18143 | 18535 | 19872.3125 | 20323.11988 |
| 18243.4 | 18535.2 | 19970.44816 | 20324.6084 |
| 18238.1 | 18535.2 | 20031.68992 | 20325.99059 |
| 18426.8 | 18535.2 | 20105.90302 | 20327.26646 |
| 18431.4 | 18535.1 | 20183.09316 | 20328.43601 |
| 18471 | 18534.8 | 20085.38278 | 20329.49923 |
| 18197.3 | 18534.5 | 20123.12727 | 20330.34981 |
| 18328.2 | 18534.1 | 20085.91439 | 20331.20039 |
| 18260.2 | 18533.5 | 20055.39984 | 20331.83833 |
| 18113.2 | 18532.9 | 19930.47091 | 20332.47626 |
| 18462.4 | 18532.2 | 20061.3539 | 20332.90155 |
| 18244.9 | 18531.3 | 20126.42327 | 20333.32684 |
| 18233.4 | 18530.4 | 19974.80739 | 20333.53949 |
| 18495.3 | 18529.3 | 19986.60918 | 20333.64581 |
| 18194.5 | 18528.2 | 20090.91155 | 20333.75213 |
| 17896.6 | 18527 | 20047.6383 | 20333.64581 |
| 18111 | 18525.6 | 20010.85071 | 20333.43317 |
| 17987.6 | 18524.2 | 19934.61748 | 20333.22052 |
| 18121 | 18522.7 | 19915.47943 | 20332.79523 |
| 18294.1 | 18521 | 19923.98523 | 20332.26362 |
| 17918.5 | 18519.3 | 19726.54436 | 20331.62568 |
| 18144.8 | 18517.5 | 19751.3175 | 20330.98775 |
| 18219.8 | 18515.6 | 19932.91632 | 20330.13717 |